%% file: ms.tex
\PassOptionsToPackage{table}{xcolor}
\documentclass[12pt,a4paper,oneside,table]{book}
\pdfoutput=1 
\usepackage{dcolumn}
\usepackage{amsmath,amsfonts}
\usepackage{graphicx}
\usepackage{appendix}
\usepackage{pdfpages}
\usepackage{float}

\newcommand\Tstrut{\rule{0pt}{2.6ex}}       
\newcommand\Bstrut{\rule[-1.1ex]{0pt}{0pt}} 
\newcommand{\term}[1]{\textbf{#1}}
\DeclareMathOperator{\sinc}{sinc}
\DeclareMathOperator{\rect}{rect}

\DeclareMathOperator{\Hz}{Hz}

\DeclareMathOperator{\m}{m}
\DeclareMathOperator{\ms}{ms}
\DeclareMathOperator{\rad}{rad}
\DeclareMathOperator{\s}{s}

\DeclareMathOperator{\const}{const}

\DeclareMathOperator{\CF}{CF}
\DeclareMathOperator{\FWHM}{FWHM}
\DeclareMathOperator{\ERB}{ERB}
\DeclareMathOperator{\FFR}{FFR}
\DeclareMathOperator{\Depth}{Depth}
\DeclareMathOperator{\of}{of}
\DeclareMathOperator{\Field}{Field}
\DeclareMathOperator{\Focus}{Focus}
\renewcommand{\Re}{\operatorname{Re}}
\renewcommand{\Im}{\operatorname{Im}}
\usepackage{bm}
\usepackage{setspace}
\usepackage{subcaption}
\usepackage{multirow}
\usepackage{mathrsfs}
\usepackage{caption}
\usepackage{tocloft}
\usepackage[stable]{footmisc}
\usepackage{dblfloatfix}
\usepackage{longtable}
\usepackage[T1]{fontenc}
\usepackage{etoolbox}
\makeatletter
\patchcmd{\@maketitle}{\HUGE}{\Huge}{}{}
\makeatother

\newcolumntype{P}[1]{>{\raggedright\arraybackslash}p{#1}}
\usepackage{hyperref}
\hypersetup{
    colorlinks=true, 
    linktoc=all,     
    linkcolor=blue,  
		citecolor=blue
}
\usepackage{cleveref}
\crefname{chapter}{\S}{\S\S}
\Crefname{chapter}{\S}{\S\S}
\crefformat{chapter}{\S#2#1#3}

\crefname{section}{\S}{\S\S}
\Crefname{section}{\S}{\S\S}
\crefformat{section}{\S#2#1#3}

\crefname{subsection}{\S}{\S\S}
\Crefname{subsection}{\S}{\S\S}
\crefname{subsection}{\S\hspace*{-0.1cm}}{\S\hspace*{-0.1cm}}

\crefname{paragraph}{\S}{\S\S}
\Crefname{paragraph}{\S}{\S\S}
\crefformat{paragraph}{\S#2#1#3}

\Crefname{appsec}{Appendix}{appendices}

\usepackage{natbib} 
\usepackage[hmargin=2cm, top=2.5cm, bottom=2.5cm]{geometry}

\defcitealias{Akhmanov1968}{Serge\v{i} Aleksandrovich Akhmanov et al. (1968)}
\defcitealias{Bell2005}{Andrew Bell (2005)}
\defcitealias{Bekesy1960}{Georg von  B{\'e}k{\'e}sy (1960)}
\defcitealias{Bregman}{Albert S. Bregman (1990)}
\defcitealias{Couch}{Leon W. Couch II (2013)}
\defcitealias{Dirac1963}{Paul Dirac (1963)}
\defcitealias{Duffieux1983}{Pierre-Michel Duffieux (1946 / 1983)}
\defcitealias{Goodman}{Joseph W. Goodman (2017)}
\defcitealias{Hamilton}{William R. Hamilton (1839)}
\defcitealias{Helmholtz}{Helmholtz (1945)}
\defcitealias{Hopkins1951}{Harold Hopkins (1951)}
\defcitealias{Kolner}{Brian Kolner (1994a)}
\defcitealias{KolnerNazarathy}{Brian Kolner and Moshe Nazarathy (1989)}
\defcitealias{Landauer1996}{Rolf Landauer (1996)}
\defcitealias{Lyon2018}{Richard F. Lyon (2018)}
\defcitealias{Marr2010}{David Marr (2010)}
\defcitealias{Mayer1874}{Alfred M. Mayer (1874)}
\defcitealias{Michelson1890}{Albert Michelson (1890)}
\defcitealias{Moore2013}{Brian C. J. Moore (2013)}
\defcitealias{Ohm}{Georg Simon Ohm (1843)}
\defcitealias{Pickles}{James O. Pickles (2012)}
\defcitealias{Rayleigh1907Duplex}{Lord Rayleigh (1907b)}
\defcitealias{Rayleigh1945}{Lord Rayleigh (1945)}
\defcitealias{Resnikoff}{Howard L. Resnikoff (1989)}
\defcitealias{diFrancia1955}{Giuliano Toraldo di Francia, 1955}
\defcitealias{Ville1948}{Jean Ville (1948)}
\defcitealias{Zernike1938}{Frits Zernike (1938)}
\defcitealias{Tournois}{Pierre Tournois (1964)}

\usepackage{fancyheadings} 
\renewcommand{\chaptermark}[1]{\markboth{\MakeUppercase{\thechapter.\ #1}}{}}
\graphicspath{ {images/} } 

\DeclareGraphicsExtensions{.jpg,.png,.pdf}
\makeatletter
\renewcommand\subsubsection{\@startsection{paragraph}{4}{\z@}%
            {-2.5ex\@plus -1ex \@minus -.25ex}%
            {1.25ex \@plus .25ex}%
            {\normalfont\normalsize\bfseries}}
\makeatother
\usepackage[utf8]{inputenc}
\usepackage{chngcntr}
\usepackage[table]{xcolor}
\counterwithout{footnote}{chapter} 
\cftsetindents{section}{1em}{3em}
\cftsetindents{subsection}{2em}{4em}

\begin{document}
\pagenumbering{roman} %
\includepdf{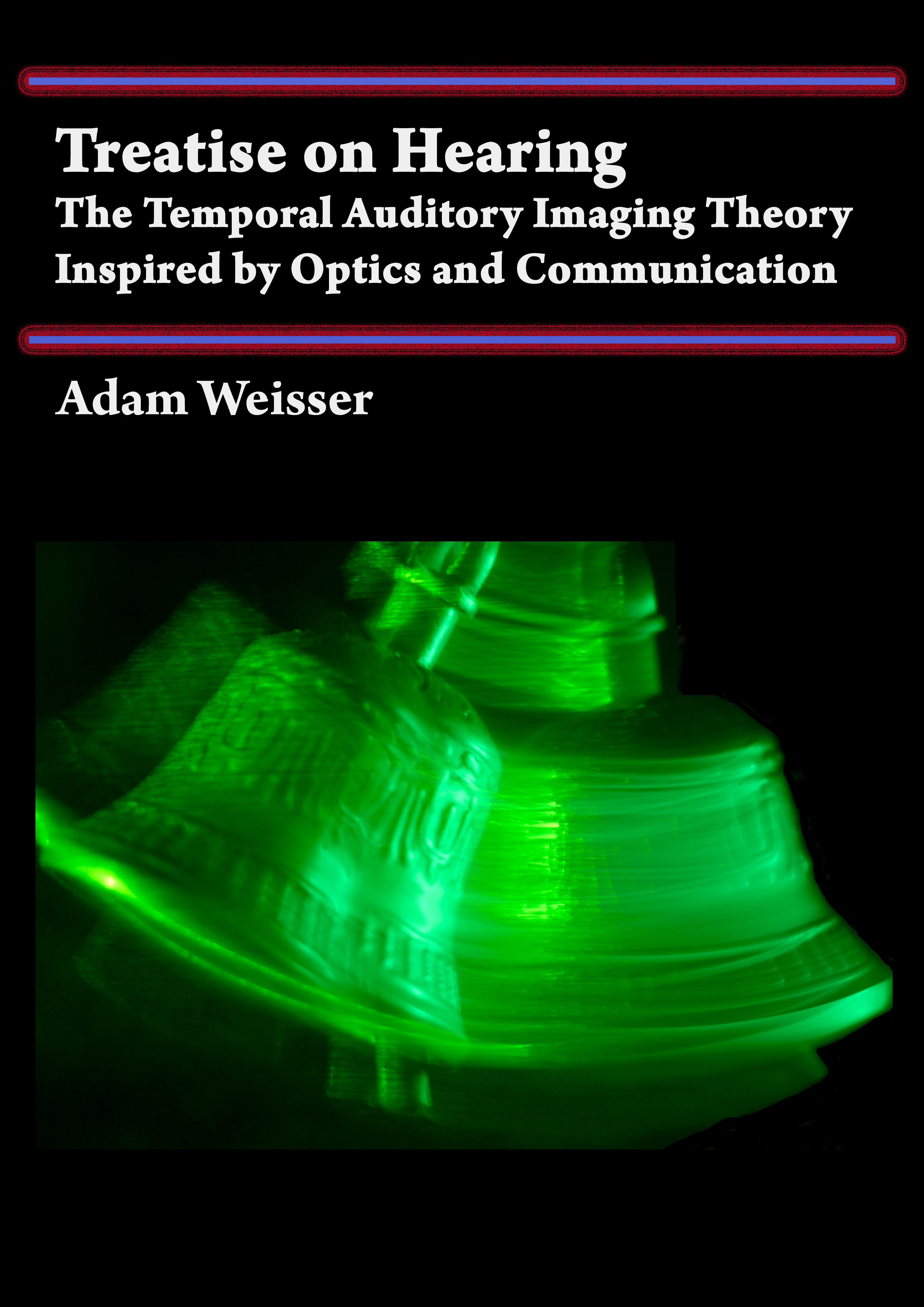}
\newpage
\thispagestyle{empty}
~\\

\newpage
\title{Treatise on Hearing:\\
The Temporal Auditory Imaging Theory\\Inspired by Optics and Communication}

\author{Adam Weisser, PhD}

\date{\today} 
\maketitle 

\pagebreak
\thispagestyle{empty}
\hspace{0pt}
\vfill
\begin{changemargin}{1cm}{1cm} 
\centering

\Large
\vspace{2cm}
\large
All photographs and illustrations are by Adam Weisser,\\
unless credited otherwise.\\
\vspace{10cm}
v0 published September 2021\\
v1, v2, v3 published November 2021\\
v4 April 2022\\
v5 May 2022\\
v6 May 2023\\
v7 this version May 2024\\
\vspace{2cm}
Contact the author at \url{weisser@f-m.fm}\\
\vspace{2cm}
\textbf{\textcopyright 2024 Adam Weisser} \\
Haifa, Israel\\
All rights reserved.\\

\end{changemargin} 
\vfill
\hspace{0pt}

\chapter*{Abstract}
Contrary to traditional thinking about hearing in which the broadband audio spectrum is taken as a whole, modern hearing science has gradually uncovered how the channel-based temporal envelope and its own spectrum are often prioritized by the auditory system. This is achieved through various processing mechanisms at different stages between the auditory brainstem and cortex, which operate on the temporal envelopes both within single auditory channels and between channels of different frequencies. Without loss of generality, it is possible to formulate the temporal envelope as a complex function that varies slowly around a fast center carrier frequency. The complex envelope includes all frequency and amplitude modulations, and hence includes the signal onset and offset cues, by definition. Tracking the transformations that the complex envelope undergoes between the acoustic source and the listener's brain should therefore be one of the key points of hearing theory. However, no systematic treatment of the complex envelope transformations relevant to hearing exists. Rather, only fragmentary treatments are available that primarily rely on empirical findings that pertain to particular stages of hearing. 

The new theory of mammalian hearing that is presented here attempts to bridge this gap in the science by consulting the two disciplines that offer the most extensive analytical tools that deal with complex envelope transformations. The first one is imaging optics, which deals with the spatial envelope that propagates between an object and an image and undergoes diffraction and refraction---as is the basis for vision. The second is communication theory, which devises various types of temporal modulations to transfer information between a receiver and a transmitter, over a noisy channel. 

Drawing from optical physics, it is argued that an auditory image is formed in the midbrain (inferior colliculus) of an object that is located in the acoustical environment of the listener. Using the space-time duality, it is shown that the ear is a temporal imaging system that comprises three transformations of the envelope functions: cochlear group-delay dispersion, cochlear time lensing, and neural group-delay dispersion. These elements are analogous to the familiar transformations from the visual system of diffraction between the object and the eye, spatial lensing by the crystalline lens, and second diffraction between the lens and the retina. However, unlike the eye, it is established that the human auditory system is naturally defocused, so that coherent stimuli do not react to the defocus, whereas completely incoherent stimuli are impacted by the defocus and may be blurred by design. It is argued that the auditory system can use this differential focusing to enhance or degrade the images of real-world acoustical objects that are partially coherent, predominantly. In addition to the imaging transformations, the corresponding inverse-domain modulation transfer functions are derived and interpreted with consideration to the nonuniform neural sampling operation of the auditory nerve. These ideas are used to rigorously initiate the concepts of sharpness and blur in auditory imaging, auditory aberrations, and auditory depth of field.  

In parallel, ideas from communication theory are invoked to show that the organ of Corti functions as a multichannel phase-locked loop (PLL) that constitutes the point of entry for auditory phase locking. It provides an anchor for a dual coherent and noncoherent auditory detection further downstream in the auditory brain. Phase locking enables conservation of coherence between the mechanical and neural domains. 

Combining the logic of both imaging and phase locking, it is speculated that the auditory system should be able to dynamically adjust the proportion of coherent and noncoherent processing that comprises the final image or detected product. This can be the basis for auditory accommodation, in analogy to the accommodation of the eye. Such a function may be achieved primarily through the olivocochlear efferent bundle, although additional accommodative brainstem circuits are considered as well.

The hypothetical effect of dispersion and synchronization anomalies in hearing impairments is considered. While much evidence is still lacking to make it less speculative, it is concluded that impairments as a result of accommodation dysfunction and excessive higher-order aberrations may have a role in known hearing-impairment effects. 

\thispagestyle{empty}

\setcounter{secnumdepth}{-2}

\chapter*{Summary}
\label{InformalSummary}
Below is an informal summary that provides a concise and broad overview of the main ideas found in this work. 

\section*{Vision and hearing}
Out of the five traditional senses---vision, hearing, touch, taste, and smell---hearing and vision superficially share the most in common---something that has led to recurrent juxtapositions and comparisons over millennia. For start, the peripheral organs themselves, the eyes and the ears, are placed in proximity and at similar height on the human face, they both come in pairs, and both provide near-nonstop information from the distance about the immediate and remote environments. On top of that, both are central for communication and both are used expressively in many art forms. 

As the understanding of the senses has matured over the last two centuries, additional characteristics have stood out, the main one being that both hearing and vision are based on physical wave stimuli that radiate toward the body---sound or light waves--- albeit at very different characteristic speeds, wavelengths, and frequencies. With the advent of psychophysics, analogies between visual and auditory perception were made clear too, which occasionally turned out to have parallels in the respective structure of the relevant brain area or its presumed method of processing the stimuli.

The human eye works like a camera, complete with an object, a lens, a pupil to limit the amount of light, and a screen that is the retina, on which an optical image appears upside down. Seeing the image simplifies the understanding of the eye, as it is intuitively clear that the ideal image is a demagnified replica of the object, which has to be as sharp and free of aberrations (various distortions in the two-dimensional image) as possible. The eye achieves focus using a variable focal length of its lens, which is controlled by accommodation. Once the image hits the retina, it is transduced by photoreceptors that also filter the light into three broad frequency ranges, which form the basis for color perception. After initial signal processing in the retina, a neural image is sent to the visual cortex through the optic nerve and through the thalamus. 

A quick inspection of the ear does not reveal an equally obvious mechanism of operation and certainly nothing that looks (or sounds) like a lens or an image, let alone a two-dimensional one. Optics does not apply here, so an intricate combination of principles from physics, engineering, and biology must be invoked to explain its operation. The ear also does not work as a tape recorder, which might be thought of as the associative analog to the camera for sound. It does not ``record''  the incoming sound as a broadband signal, let alone ``play'' it back as is. Rather, after a series of complex mechanical transformations, the organ of Corti in the cochlea filters the broadband sound into numerous narrowband auditory channels that are sometimes independent of each other, and yet interact in other cases, according to complex rules that must be uncovered through experiment. The sound is finally perceived as broadband, as though the input has been perceptually resynthesized, following signal processing in different brain areas related to the auditory system.

\section*{Hearing points}
The present theory attempts to show how the principles of optical imaging apply to hearing notwithstanding. The recipe for doing so is straightforward as long as several tropes of hearing science are shed. For the sake of this summary, let us treat the following statements as correct. They will all be properly demonstrated and motivated in the main text:
\begin{enumerate}
	\item Time is for hearing what space is for vision. 
	\item Wave physics does not stop at the auditory periphery. 
	\item Constant frequency is the exception, not the rule.
	\item The ear is not a lowpass receiver, but rather a (multichannel) bandpass system. 
  \item Acoustic source coherence propagates in space according to the wave equation.
	\item Information arriving to the auditory brain is discretized and the rules of sampling theory must apply. 
\end{enumerate}
Each one of these statements on its own may not be particularly novel or controversial, at least in some contexts of hearing theory, but once their totality is internalized, new ways to understand hearing inevitably arise.

\section*{Inferring the auditory image}
The visual image is spatial, since it is distributed over the area of the retina, where a two-dimensional projection of three dimensional objects appears as a pattern of light. In imaging analysis, it is customary to ``freeze'' the progress of time at an arbitrary moment and look at a single still image, which contains much of the information from the object and its environment, all simultaneously available within same image. The passage of time entails movement of the object(s) and observer, which can be understood as incremental changes to the reference still image. In contradistinction, it is exceedingly difficult to make sense of a ``frozen'' image of sound---for example, a particular geometrical configuration of the traveling wave in the cochlea---which carries limited information about the auditory scene on the whole. Here, it is necessary to let time pass and hear how the different sounds develop and interact in order to be able to say something meaningful about the acoustic situation they represent and how they are distributed in space. Hence, we intuitively arrive at the space-time analogy between vision and hearing, which was epitomized in Point 1.

Further pushing the logic of Point 1 entails that if the visual image occurs in space, then a hypothetical auditory image must occur in time. Then, we should expect that just as the optical object-image pair in vision can be represented as a spatial envelope that propagates between points in space (e.g., between the object origin and the image origin), so should the acoustical object-image pair of hearing be representable by a temporal envelope, between reference points in time. Both object types should have a center frequency that carries the envelope, as there is no physical difference in the way that the information about the objects is borne by waves.

Another important substitution is to find the temporal equivalent of diffraction, which in optics determines how different parts of the light waves interfere and change their shape, as a result of scattering by various boundaries and objects on their path to the screen. Diffraction is really a general term for wave propagation from the object, which may or may not encounter scattering obstacles on its way to the screen. Switching between the spatial and temporal dimensions (Point 1), diffraction is replaced by group-velocity dispersion, in which the shape of the temporal envelope is impacted by differential changes to its constituent frequencies. We recall that the cochlea itself is inherently a dispersive path, so that the acoustical signal that arrives into the cochlea is automatically dispersed. Note that while we often talk about dispersion, we are actually concerned with its derivative---the group-velocity dispersion---that goes by different names, such as group-delay dispersion and phase curvature. 

Next, if there is any chance for us to construct an imaging system within the auditory system, we should be also looking for a temporal aperture---something that limits the duration of signal that can be processed at one point in time (really, sampled) for a given chunk of acoustical input. Here, the neurons that transduce the inner hair cell motion produce spikes that are limited in time by definition, so there is a time window in effect that continuously truncates the signal into manageable chunks. 

While the above steps are relatively simple endeavors, completing the identification of the temporal imaging system in the ear requires bolder steps---borderline speculative. First, we require an additional dispersive section in the auditory brain, regardless of the acoustical signal representation that is now fully neural (Point 2). Current science has it that there is no neural dispersion in the auditory brain, whereas the present work claims otherwise, as can be demonstrated by several measurements. While the precise magnitude of dispersion is difficult to ascertain, it is readily evident how no cochlear measurement of the group delay based on otoacoustic emissions has ever matched the auditory brainstem response measurements that include the brainstem as well. This discrepancy translates to a non-zero group-velocity dispersion of the path difference, which is mostly neural.

The second speculative step is to identify a lens. A temporal imaging system requires a ``time lens'', rather than a spatial lens, although it is not strictly necessary (this is because a pinhole camera---the simplest imaging device possible---does not have a lens, but still produces a sharp image, as long as the aperture is very small). A time lens performs the same mathematical operation as the spatial lens in the eye, only over one dimension of time instead of over two dimensions of space, and where frequency is a variable rather than a constant (Point 3). Indeed, an inspection of the dynamic properties of the organ of Corti that were recorded in several state-of-the-art physiological measurements reveals a phase dependence in time, frequency, and space that is symmetrical in shape. Such phase response can be readily attributed to a time lens, which is modeled using a quadratic phase function---a form of phase modulation. 

\section*{Putting the system together}
We have now identified the four necessary elements of a basic temporal imaging system in the ear: cochlear dispersion, cochlear time lens, temporal aperture, and neural dispersion. The values of the different elements can be roughly estimated for humans following an analysis of available physiological data from literature. These estimates can also be cross-validated using human psychoacoustic data from other sources.

Although we are dealing with sound, now we are in the conceptual realm of optics, which has devised a number of powerful analytical tools to characterize the image and objectively assess its quality. For example, it is possible to compute whether the above combination of elements produces a focused image (putatively, inside the brain)---a temporal envelope carried by a center frequency that is propagated to the midbrain or thereabout, where the ``auditory retina resides''. Surprisingly, the answer is a definite ``no''. The auditory image is defocused, unlike the optical image that appears on the retina in normal conditions. However, in vision, we know how information about the object is superior when the image is focused (for instance, try to read a blurred text from the distance). 

Why should hearing be any different than vision and be defocused and not sharply focused? Why would we want to hear sounds that are blurry rather than sharp? To be able to answer these questions requires us to revisit the idea of group-velocity dispersion and establish its relevance to realistic acoustic signals. This will help us establish the meaning of sharpness and blur in hearing. 

\section*{Spatial blur}
In spatial imaging, there are two general domains of blur. In the geometrical one, light ``rays'' are traced following refraction. Every object can be thought of as a collection of point sources in a continuum, from which light rays diverge in all directions, each carries the information about the point it emanated from. The goal of imaging is to collect the rays so they form the same pattern of light in another region in space as they do at the object position. This generally includes a linear scaling factor---magnification---which does not have direct bearing on the fidelity of the image that is otherwise a one-to-one mapping of the object in space. Deviations from the one-to-one mapping in two dimensions---when the rays do not converge exactly where they should in order to reconstruct the object---are called aberrations. An out-of-focus imaging system has a ``defocus'' aberration, which entails a lack of convergence of the rays coming from different directions on the screen, so that information from different points of the object is ``mixed'' at the position of the image, in a way that is visible. The geometrical form of blur is the most dominant one when the wavelength of the light is much smaller than the object and also smaller than the different obstacles in the optical path to the image. 

On the other extreme, when the wavelength of the light is comparable to the details of the object or the aperture, or other things that scatter the light on the way to the screen, then the effect of diffraction may be visible in the image as different interference patterns that can distort the details of the object. These may be thought of as diffraction blur, although the underlying mechanism is very different from geometrical blur. Effects here, if they are visible at all under normal conditions, tend to appear along edges and around very fine details of the image. 

With these rough definitions in mind, we note that sharpness is simply the absence of visible blur, or blur that is quantified to be below a certain threshold. 

\section*{Auditory blur and coherence}
Back to hearing, how does the auditory image---really, the temporal envelope carried by high frequency---ever becomes blurred? First, we have to transform the two types of blur to temporally-relevant phenomena (invoking Point 1 again). An example of geometrical blur is relatively easy to see, since it manifests in reverberation. Here, the information from the source contained within a point in time (or rather, an infinitesimally short interval) arrives to the receiver mixed with information originating in other points in time. The mixing is asynchronous, so it adds up randomly and does not interfere. A corollary is that direct or free-field (anechoic) sound suffers from no geometrical blur at the input to the ear. 

The spatial diffraction blur can be analogized to temporal dispersion blur. Group velocity dispersion entails that every component of the temporal envelope propagates at somewhat different velocity. Temporal obstacles in time (i.e., those that can be expressed as filters or time windows) that are approximately proportional to the period of the carrier wave may impose a differential amount of delay to the different components of the envelope spectrum. The result is similar to interference in time and is, therefore, an analogous form of blur of the temporal envelope to that of diffraction blur of the spatial envelope. 

It appears that in order to know whether the different types of blur ever apply in reality, it is necessary to know how much bandwidth the acoustic source occupies. If the bandwidth is very narrow (with the extreme being that of a pure tone), then group-velocity dispersion is unlikely to have any effect, because only a single velocity is relevant per given frequency. However, as the bandwidth becomes wider, more and more frequencies are subjected to differential group-velocity dispersion, so the envelope may become blurry if the dispersion is high or if it is accumulated over large distances. But, there is an effective limit imposed on the bandwidth here, because the auditory system analyzes the acoustic signal in parallel bandpass filters, each with a finite bandwidth, that together cover the entire audio spectrum. Therefore, the maximum relevant bandwidth of a signal has to be related to the auditory filter in which it is being analyzed. Either way, it is realistic signals that are naturally modulated in frequency and do not have constant frequency (Point 3) that may experience the effects of group-velocity dispersion most strongly. 

Another effect of the bandwidth that is related to geometrical blur has to do with the degree of randomness of the signal. Unlike deterministic signals, a truly random signal does not interfere with a delayed copy of itself. Therefore, the notion of geometric blur as in reverberation applies best to random signals that do not interfere, and only mix in energy without consideration of the signal phase. In general, the more random a signal is, the broader its bandwidth is going to be, since its amplitude and phase cannot point to one frequency at all times. This reasoning is best captured by the concept of (degree of) coherence (also called correlation in the context of hearing and acoustics). It describes the ability of a signal to interfere with itself. A completely random signal, such as white noise (broadband spectrum), does not interfere with itself and is considered incoherent. A deterministic signal, such as a pure tone, can interfere with itself and is considered coherent. Roughly translating these terms into more intuitive understanding, a coherent source sounds more tonal---like a melodic musical instrument---whereas an incoherent sound source is more like noise. However, the vast majority of real-world sounds are neither completely tonal nor are they completely noise-like, so they can be classified as partially coherent. 

All in all, it appears that the auditory defocus is applied differentially to different types of signals. Coherent signals that are largely unaffected by dispersion, are also unaffected by defocus. In contrast, partially coherent signals are made more incoherent by defocus, whereas incoherent signals remain incoherent also after defocus. 

\section*{The modulation transfer function}
Returning to standard imaging theory, it should not come as a great surprise that the imaging process and quality differ depending on the kind of light that illuminates the object: coherence matters. This becomes immediately apparent when analyzing the spatial frequencies of the object, which make the spectrum of the spatial envelope that is being imaged. A major result in imaging optics is that whatever diffractive and geometrical blurring (or other aberration) effect beyond magnification (i.e., linear scaling) exist in the system, they can all be expressed through its ``pupil function'', which then leads to the derivation of the so-called modulation transfer functions (it is different for coherent and incoherent illumination).

In exact analogy to spatial optics, we obtain similar, degree-of-coherence-dependent modulation transfer functions in the temporal domain that incorporate the effects of dispersion in the aperture and blur due to defocus. Such functions are nothing new in hearing science, but so far they have been obtained only empirically without explanation for why the coherent and incoherent functions are different, whereas the present theory contains the first derivation of these functions from the basic principles of auditory imaging. Once these functions become available, all sorts of predictions may be offered to explain different auditory effects that have been also measured only empirically until now. 

The theoretical modulation transfer functions that have been obtained here are only good as first approximation and there are clear discrepancies from experiment in several cases. It is argued that a major reason for the discrepancy is the discrete nature of the transduction. Irregular spiking in the auditory nerve further downstream in the auditory system is tantamount to repeatedly sampling the original signal at nonuniform intervals, which tends to degrade the possible image at the output of the system. In very subtle contexts it may also create perceivable artifacts, which are not captured by the analytically derived (continuous) modulation transfer function (Point 6). 

\section*{Coherence conservation and the phase locked loop}
It is necessary to take a small detour in the auditory imaging account and introduce another element to the discussion that is borrowed from communication and control theories. In the brief mention of coherent signals above we sidestepped an important question: while we know that coherence propagates in space according to the wave equation (Point 5), do we also know for certain that it is conserved in the ear? Notably, does the transduction between mechanical to neural information conserve the degree of coherence of the original signal? A big clue seems to suggest that the answer is yes: signals are known to phase lock in the auditory nerve, following transduction by the inner hair cells. This applies to coherent (tonal) signals, and to a lesser degree to other signals, where incoherent signals only lock to the slow envelope phase and not to the random carrier. Phase locking to coherent signals is special in hearing compared to vision, where the phase of the light wave changes too rapidly to be tracked by a biological system. 

Phase locking is a hallmark of coherent reception---a form of information transfer in communication theory, which tracks the minute variations in the phase of the (complex) temporal envelope of a signal. Realizing this form of reception requires an oscillator, which is an active and nonlinear component within the receiver. (In the complementary noncoherent reception, which only tracks the slowly-varying envelope magnitude, an oscillator is not strictly necessary.) A closer inspection of the cochlear mechanics and transduction suggests that phase locking first emerges at the cochlea, before it becomes manifest in the auditory nerve. Therefore, it seems reasonable to look for the components of a classical coherent detector (Point 4) that provides phase locking, i.e, a phase-locked loop (PLL). For a PLL to be constructed, we require a phase detector, a filter, an oscillator, and a feedback loop that returns the output to the phase detector. All these components can be identified within the organ of Corti and the outer hair cells. 

Conventional theory holds that the outer hair cells perform cycle-by-cycle amplification for the incoming signal, but theory and experiment are still in disagreement as for how this process exactly works. The PLL model does not necessarily clash with this standard amplification model, and might even interact with it, by incorporating amplification into its own feedback loop. A similar argument may be made about the time lens and the PLL---their function may not necessarily be in conflict. 

As it currently appears in the main text, the auditory PLL model is strictly qualitative and, accordingly, speculative. However, it does provide the missing link for coherence conservation between the outside world and the brain---a link that has been glossed over until now and is critical for the understanding of how the auditory system handles different kinds of stimuli according to their degree of coherence. 

\section*{Auditory imaging concepts}
With the imaging system specified, we can now turn to explore some of the hallmark concepts of imaging theory and apply them to hearing, beyond those of auditory sharpness and blur. 

First, we calculate the temporal resolving power between two pulses---analogous to the resolving power between two object points, as is mandatory information in telescopy, for example---using the estimated temporal modulation transfer function. The predictions are comparable to empirical findings from relevant studies, especially at the 1000--8000 Hz range. 

We also elaborate on the auditory analogs to monochromatic and polychromatic images. It is argued that the various pitch types can be thought of as the quintessential monochromatic image (pure tone, unresolved complex tone) or polychromatic image (resolved complex tone, interrupted pitch)---all of which highlight different periodicities in the acoustic object. 

This understanding can then be used to hypothesize deviations from perfect imaging---various monochromatic and polychromatic aberrations. Monochromatic aberrations relate to the variation of the group delay within the auditory channel, whereas polychromatic aberrations to variations between channels. Several examples are given to the different types, based on known phenomena from the psychoacoustic literature, as well as a new effect. 

Finally, we can hypothesize about the auditory depth of field that is temporal rather than spatial. It should be most clearly observable between objects of different degree of coherence. For example, it is argued that forward masking can be readily recast as small depth of field, in the case of   incoherent (broadband noise) masker and coherent (pure tone) probe, since the boundary between them is effectively blurred as a result of forward masking. However, when the masker and probe are of the same type (e.g., incoherent and incoherent), then the depth of field is large, as the forward masking becomes longer.  

\section*{Auditory accommodation and impairments}
An even bigger leap in applying the analogy between optical / visual imaging and acoustical / auditory imaging is the search for auditory accommodation. In vision, accommodation is an unconscious mechanical process that varies the focal length of the lens to match the distance of the object, so to bring its image to sharp focus on the retina for arbitrary distance of the object. Accommodation involves several ocular muscles that are fed by an efferent nerve from the midbrain, which together with the output from the retina form a feedback loop. 

We have already stated that the ear is defocused, so what could possibly be the use of auditory accommodation in this context? One attractive answer is to control the degree of coherence that enters the neural system, which as was argued above, is captured by the degree of phase locking. This is interesting because of much converging empirical evidence that shows how the hearing system can process sound either according to its slowly-varying temporal envelope (its magnitude), or using phase-locking to track the fast variation of the carrier phase (the so-called temporal fine structure of the stimulus). The difference between the two processing schemes runs throughout the very physiology of the auditory brainstem, which appears to have dedicated parts for each type of processing. Coming in full circle, this differentiation is akin to the types of imaging that exist in optics: coherent, incoherent, or a mixture of the two---partially coherent. It is also akin to the detection schemes that are used in standard communication engineering: either coherent or noncoherent. Applying a variable stage in coherence processing may be achieved, for example, by the medial olivocochlear reflex---an efferent nerve that innervates the outer hair cells and whose function is not well understood. But additional mechanisms to achieve the same function may exist. Once again, this involves considerable speculation at present, but evidence to support this and related possibilities does exist and is discussed in depth in the main text. 

Although there are many uncertainties about the specifics of this system, the penultimate chapter is dedicated for hypothesizing what happens when things go wrong: what is the effect of faulty imaging that may translate into hearing impairments? There are several possible answers here, with dysfunction in hypothetical auditory accommodation being the most attractive candidate. Nevertheless, evidence here is difficult to gather and much work has to be done to uncover the basics before turning to these more challenging, yet important, questions.

\pagebreak
\thispagestyle{empty}
\hspace{0pt}
\vfill
\begin{changemargin}{2cm}{2cm} 
\centering 
\Large
\textit{``With the four-dimensional space curved, any section that we make in it also has to be curved, because in general we cannot give a meaning to a flat section in a curved space.''}\\
\vspace{0.35cm}
\textsc{\citetalias{Dirac1963}}\\

\vspace{1.5cm}

\textit{``Information is physical.''}\\
\vspace{0.35cm}
\textsc{\citetalias{Landauer1996}}\\
\end{changemargin} 
\vfill
\hspace{0pt}
\pagebreak

\addtocontents{toc}{\protect\thispagestyle{empty}}
\tableofcontents
\thispagestyle{empty}

\include{lists}

\chapter{Preface}
This was never intended to be a book-long piece. My original wish upon starting to write it was to introduce into hearing the ``paraxial'' dispersion equation---perhaps in an article format. Naively, I was convinced that the beauty of the equation---or rather, the beauty in the analogy to vision---would be self-evident and immediately applicable. But as I was trying to come up with a way to introduce these ideas, it had quickly become clear that there is no straightforward way to motivate them at the present state of hearing science. At least, not without taking many steps back and venture into several scientific corners that have somehow escaped rigorous treatment. Wherever I looked, it had seemed as if some aspects of the science froze in time and have left me with an intuition about acoustic waves that would have been excellent at the turn of the previous century. It has taken considerably more research to be able to assort all the pieces that would be required to motivate the need for a dispersion equation in hearing. Some of these pieces were in plain sight, while some are in vogue in contemporary research, and yet others have been long forgotten. The price for picking up these pieces may be considered steep, though: let go of static frequencies and stationary signals, embrace coherence as a fundamental descriptor of sound, and learn to accept that the auditory system must know what it is doing much better than what we would like to think it should want to do. 

Therefore, this is an invitation for the daring reader. I expect that a fair share of the topics that are touched upon will elicit resistance in some of the readers. Nevertheless, some of the topics---mainly in the introductory chapters---may appear timely and not nearly as controversial as the more advanced chapters. In more than one occasion I have resorted to indirect and unconfirmed methods that can associate the new ideas with an unwelcome, yet unavoidable, air of speculation. Also, the very broad scope of the present theory may simply be overwhelming, as there should be numerous points of contention that may be deemed worthy of rebuttal. This is all well, as far as I am concerned, as long as the ideas will be seriously considered and stir a long-needed discussion in the community.

~\\

My quest for a hearing theory has been initially triggered by a simmering dissatisfaction with the traditional presentation of auditory science. It had struck me as a bewildering collection of facts, which had to be learned by rote instead of through internalization of deeper concepts. Many results had to be empirically measured rather than derived based on higher-level principles. There did not seem to be a theory that is general enough to offer the necessary predictive power that may spare a direct experiment. Quite the contrary---predictions often seem to be proven wrong by experiment, which had made it difficult to develop an intuition for the inner workings of the ear and its inherent logic. 

This is why stumbling across the temporal imaging theory in optics seemed like a possible key to my unrest. This theory is based on the space-time duality principle as formulated by \citetalias{Akhmanov1968} and \citet{Akhmanov1969} and was later developed into a temporal imaging theory by \citetalias{KolnerNazarathy}, and greatly elaborated by \citet{Kolner} and his subsequent work. Here was an elegant theory that is completely analogous to the classical single-lens imaging theory from optics, only with the dimensions reversed, as the temporal envelope is used instead of spatial envelope (expressing the distribution of light of the optical object and image). Knowing that vision rests on spatial imaging that is neatly formulated using the paraxial equation and a double Fourier transform, there was an immediate allure of having a paraxial equation and a double Fourier transform expressed in time and frequency coordinates that can, rather organically, represent hearing.

Alas, the basic elements in the temporal imaging theory are group-velocity dispersion, time-lens curvature, and aperture time. What do these concepts have to do with hearing? Everything, it seems. But the reasoning behind it, which is the main subject of this treatise, has taken me the better part of the last four years to arrive at. Apart from my own time-consuming ignorance of many of the associated disciplines---some of them are routinely alluded to in auditory research---there appeared to be several gaps in the fundamentals of acoustics and hearing science, which had to be retraced and patched up in order to be able to tackle the idea of imaging with a degree of rigor that I thought the topic deserves. 

There were two principal ``culprits'' that underpin the gaps in auditory science. The first one is the over-reliance on pure tones---a mathematically degenerate signal with no curvature, which carries little-to-no information and is not encountered in nature. This deficiency is addressed in several chapters that adopt the complex envelope and constant carrier formulation as the most general representation of waves, signals, objects, images, and communication functions. In turn, it opens the door for unifying the auditory concepts of temporal envelope and temporal fine structure with mathematically related concepts in acoustics, optics, and communication. 

The second ``culprit'' is acoustic coherence theory, or the lack thereof. As a scalar wave theory, linear acoustics of plane waves is completely compatible with scalar wave theory in optics, which is also where classical coherence theory was developed. The main developments in coherence theory gathered momentum in the 1950s, at a point in which acoustics and optics may have been practiced by different scholars. Acoustics, and hearing too, imported a m\'elange of coherence-related concepts from several disciplines---each with its own jargon---that hardly coalesce into a consistent understanding. The two chapters about hearing-relevant coherence theory, while not adding anything new to the science, are a first attempt to unify and revive these ideas in a manner that is consistent with wave physics, room acoustics, hearing (phase locking), communication engineering, and neuroscience. Optical coherence theory provides a bridge that can be applied in sound, using Fourier optics, alongside some of the most insightful tools from imaging. The proposed amalgamation of the different coherence theories attempt to connect concepts of coherence with synchronization that manifest both in the mechanical and in the neural parts of the auditory system, and is thought to generally characterize perception throughout the brain. 

With the availability of these introductory chapters, the motivation for the temporal imaging theory should be in place. I have put substantial effort in exploring some of the potential implications of temporal imaging---temporal modulation transfer functions, aberrations, accommodation, and dispersive hearing impairments. Thus, it is my hope that interested readers will be able to follow the wildly different approach to hearing that is presented in the advanced chapters of this work, despite the effort that it may require. While I cannot foresee the correctness of some of the hypotheses put forth, I will feel greatly rewarded to know that these ideas will have influenced future researchers in solving some of the more persistent challenges in our understanding of hearing and hearing impairments. 

\chapter{Preface to v6}
While not exactly a second edition, v6 contains more significant updates to the theory compared to the one that first appeared in September 2021 and its subsequent public versions on arXiv (v1--v5; \url{https://arxiv.org/abs/2111.04338}). 

The most substantial update is the addition of two more supporting studies to the section about the time lens (\cref{PhaseModEvidence}). Although human-relevant curvature values are difficult to extract from these animal data (\cref{TimeLensExtrapolation}), the current total of five different studies helps to push the idea that the cochlea contains a time lens away from the realm of speculation. Still, the data appear to be clustered in two curvature ranges of the putative time lens, which may be difficult to accept without the interaction with auditory accommodation---itself a speculative idea, albeit completely in line with the system physiology and the precedent of accommodation in vision. The addition of more data points to the time lens curvature had a slight cascade effect on many of the quantitative predictions in this work, which were corrected accordingly. It also led to a correction of an error in the related formula of the octave stretch effect in \cref{TransChromAb}. Unfortunately, this update produces an octave stretch effect prediction that is more limited and somewhat messier than in the previous editions, although still relevant. Throughout, the large-curvature time-lens estimates have been used, whereas the small-curvature estimates were no longer viable in most contexts. This is unlike the previous versions of the text, where the difference between the two curvatures informed us of the extreme values of the system curvature range. 

Another major addition to this work is the summary, which aims to provide a briefer and lighter exposition to the ideas of the temporal auditory imaging theory and is more accessible than the technical abstract and long introductory chapters. The summary was constructed in a rather deductive manner, which rests on six points that were themselves gathered from the text (although without explicit reference to them later). Other changes are the additions of a couple dozen recent references (more than 1600 in total), a figure of the peripheral ear, and minor corrections to some figures and text. 

\section*{Preface to v7}
With the exception of a clearer presentation of the results of the experiments in \cref{Aliasing}, the changes in this version (v7) are all relatively minor: the usual correction of typos and references, as well as the inclusion of several new references from the past couple of years and a few ones that were overlooked before (notably, a mention of an imaging system in lizards with a parietal eye that is part of their pineal complex, the hypothesis that color vision in cephalopods is based on defocus, and the term ``contrast sensitivity function'' that is the correct visual equivalent to temporal modulation transfer function). However, to ease the future search and increase the visibility of this manuscript, this may be the final version that will appear on arXiv, for which I am grateful. Future updates may appear exclusively on the new web version to be found at \url{http://www.hearingtheory.org}.

\chapter{About the author}
Adam Weisser (born 1978) is an independent researcher with a broad academic and industrial background in hearing science and acoustical engineering. He holds a PhD in Hearing Science from Macquarie University in Sydney (topic: Complex Acoustic Environments, supervised by J\"org Buchholz and co-supervised by Gitte Keidser), MSc in engineering acoustics from the Technical University of Denmark (topic: Small Room Acoustics, supervised by Jens Holdger Rindel and co-supervised by Jan Voetmann), and BA Cum Laude in Physics from the Technion in Haifa. In between he spent eight years in the hearing aid industry in Denmark, primarily in the research and development of remote-microphone wireless technologies for the hearing impaired population. Before that, he also worked in high-frequency microwave chip design (MMIC) in Israel. This industrial experience involved basic acquaintance with communication engineering. Along with a previous training in the fundamentals of optical physics and a long standing interest in photography and cinematography, as well as extensive experience in music performance and production, these influences coalesced into a way of thinking about hearing that informed the ideas presented in this manuscript.

~\\

\begin{figure}[H]
		\centering
		\includegraphics[width=0.35\linewidth]{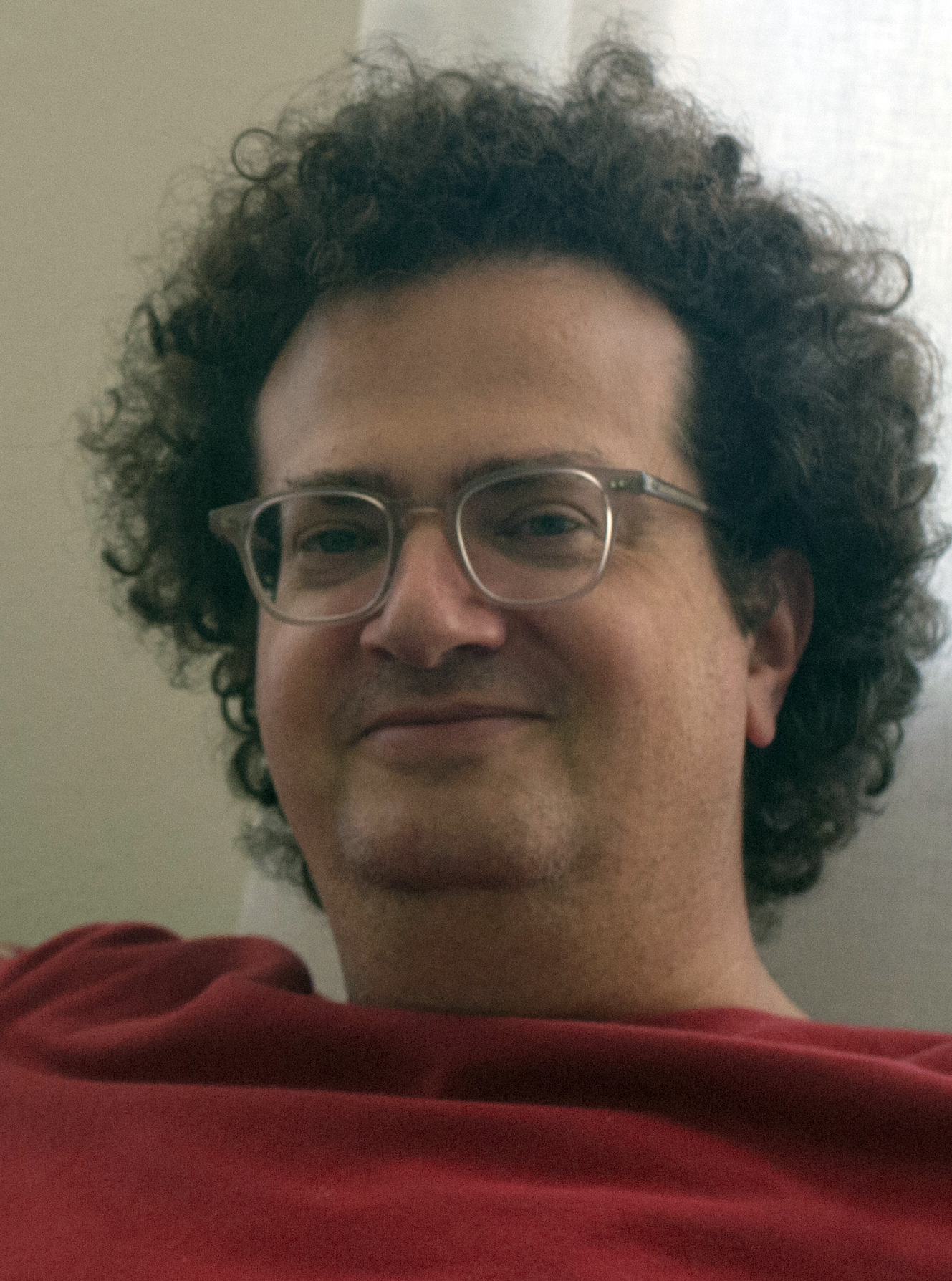}	
\end{figure}

\chapter{Acknowledgments}
The kernel of this work emerged from a long-standing interest I have had in exploring analogies that exist between psychoacoustics and imaging and Fourier optics to improve my intuitive understanding of hearing. Early during my PhD studies I had realized that such analogies tend to be fragmentary, where they appear in print, or altogether nonexistent. Fortunately, I had the chance to delve into this question in the context of a journal club presentation  I titled ``Optical Hearing'', on 5 October, 2017 at Macquarie University. Thus, I am indebted to the large group of participants in my talk and for this group for sharing their own presentations that had undoubtedly provided me with background knowledge and motivation that had helped me to crystallize my own ideas. 

I would like to thank J\"org Buchholz, for giving me the necessary freedom to explore these uncharted territories while working on my PhD during that period in 2017--2018, and for providing the opportunity to be immersed in this research in the first place.

A big thanks goes to Nicholas Haywood for his insightful suggestion to explore aliasing as a proxy for discrete processing. While he was unavailable for direct cooperation, the results garnered from this original suggestion have gone a long way. 

I am also thankful to David McAlpine for hearing out the idea in its initial and very raw form (along with J\"org Buchholz), and for getting me to pay attention early on to the inferior colliculus rather than to the auditory cortex as the main auditory hub. 

Many thanks for Gojko Obradovic for reviewing the text and providing invaluable comments throughout. 

Special thanks to Fabrice Bardy and Macarena Paz Bowen for audiological help in two measurements and for their general encouragement in these early stages. 

Thanks also to Jody Ghani for her remarkable patience with transforming some abstract ideas into original technical drawings.

~\\

An early source of inspiration was a presentation given by James B. Lee, who provocatively tied together several ideas in optics, nuclear physics, and concert hall acoustics. He held quite unlikely talks in the 2016 meeting of the Acoustical Society of America in Honolulu (\citealp{Lee2016a}; \citealp{Lee2016b}), which attempted to conceptually link optics and acoustics in a way that was both fresh and insightful. 

Sometimes people say things that resonate and can be later recognized in a completely different context in life. Such was something that my friend Ofer Meir said, who has also taken it upon himself to ensure that this work would see the light of day much sooner than it would have been without him. Thus, I am deeply grateful for his friendship and support. 

The presence of close and loving friends throughout the process has been indispensable. Kelly Miles, who provided early inspiration and ongoing enthusiasm, Ophir Ilzetzki for invigorating discussions and sincere curiosity along the discovery process, Dotan Perlstein for his unyielding encouragement and for making sure that my feet remain on firm scientific ground, Jan Tom\'a\v{s} Matys for his unflinching confidence and moral support of this work, Augusto Bravo for his endless openness for radical thinking and many fascinating conversations about science, Michael Yang for his firm friendship and taking this project seriously all along, and for Diogo Flores, who drove me with his excitement during the early parts of the work. 

For their valuable advice at key moments along the way I am indebted to Timothy Beechey, Andrew Bell, Yaniv Ganor, Ami Goren, Barak Mann, Yehuda Spira, and Eduardo Vistisen. 

For keeping things close at different points throughout the writing process and for their sincere support and friendship, I am grateful to Fadwa AlNafjan, Camilla Althoehn, Emily Arday, Javier Badajoz-Davila, Jerome Barkhan, John Beerends, Isabelle Boisvert, Ammalia Duvall, Jack Garzonio,  Gady Goldsobel, Julia Gutz and Paul Springthorpe, Sharon Israel, Brent Kirkwood, Emilija Klovait\.e, R\`emi Marchand, Jason Mikiel-Hunter, Juan Carlos Negrete, Allie O'Connor, Anders Pedersen, Heivet Hernandez Perez, Claudiu Pop, Matthieu Recugnat, Mariana Reis, Mariana Roslyng-Jensen, Ana Ruediger, Jeremy Rutman, Greg Stewart, Kramer Thompson, Lindsey Van Yper, Sarah Verhulst, Jay and Jane Woo, Jaime Undurraga, Gil Zilberstein and St\`ephanie \`Ethier, and the late Liviu Sigler.

I am grateful to my sister, Oriyan Miller, and my brother-in-law, Roee Miller, for their continuous engagement, constructive questions, and much encouragement at numerous points throughout the discovery process. 

And, finally, a huge thanks to my mother, Vivian Savitri, whose involvement and trust in this work have been essential from the early development of the ideas, which could have not been nearly as peaceful and resolute without her ongoing dedication to this project. 

~\\ 

This work is standing on the shoulders of numerous scholars in hearing, acoustics, optics, communication engineering, information theory, signal processing, physics, neuroscience, and biology, without whom this treatise could have not been written. The comprehensive bibliography of this treatise is a testimony of my deep appreciation for their work that is frequently studded with unmistakable ingenuity. Of special help were the textbooks by \citetalias{Moore2013} about psychological acoustics and by \citetalias{Pickles} about the physiology of hearing, as well as the entire Springer series of handbooks on auditory sciences, edited by Richard R. Fay and Arthur N. Popper, which has been an indispensable source of knowledge of all things auditory. Two additional texts that provided an exceptionally useful overview of the historical and current state of hearing science from original perspectives were by \citetalias{Bell2005} and \citetalias{Lyon2018}. A book by \citetalias{Resnikoff} provided a unique point of view that related perception (mainly vision) to information theory and served as early inspiration. Finally, non-hearing texts that were used extensively were books about communication theory by \citetalias{Couch} and Fourier optics by \citetalias{Goodman}. I was fortunate enough to attend a semester of Stephen G. Lipson's optical physics class in the Technion back in 1998, which evidently stuck for longer than I had realized at the time. The textbook of that course has thus remained an important reference throughout my writing \citep{Lipson}.

~\\ 

I could not have carried out this research without access to the Macquarie University Library, and to a lesser extent the Technion and Haifa University libraries. My gratitude goes to all the librarians that have worked behind the scenes.

~\\ 

\chapter{About the text}
This treatise is not intended to be an introduction to hearing science. It assumes that the reader is well-versed in basic hearing phenomena and has at least some acquaintance with wave physics, Fourier analysis, and linear signal processing. Readers that are already familiar with Fourier and geometrical imaging optics, as well as photography, astronomy, or microscopy enthusiasts, are likely to find certain optics-inspired passages relatively easy to follow---bordering on trivial. The same goes for readers with background in communication and radar engineering, who are going to find some sections relatively straightforward. While several chapters contain mathematical derivations that can be outside the comfort zone of the less mathematically-inclined readers, they are encouraged to gloss over them and focus on the qualitative descriptions that may be sufficient to develop the necessary insight. Nevertheless, a few topics should undoubtedly benefit from mathematical understanding---mostly those that introduce the basic space-time duality equations, the analytic signal, modulation and demodulation, coherence, and the various modulation transfer functions.

Some specific conclusions and derivations may raise interest among non-specialists as well. The derivation of the modulation transfer functions from the temporal imaging equations has not appeared in the optics literature previously, which has focused primarily on time-domain solutions. 

There are several implications of the ideas expressed in this work that may also be of interest to perception, vision, and neuroscience specialists. If proven correct, then auditory imaging as presented here suggests that certain imaging principles are biologically common to both hearing and vision. This begs the question of whether additional sensory inputs are processed in a similar fashion, only with less obvious dimensional substitutions. 

For the neuroscientist, the idea that the brainstem performs neural processing in hearing in part to achieve a function that is performed analogically in the eye may be curious as well. This suggests that biological computation is both analog and digital and that the segregation between the mechanical and neural domains may be at least somewhat contrived. It also underlines the significance of sampling considerations, which are usually taken for granted in the discussion of neural coding.

The scope of this work has been limited on purpose and largely excludes an in-depth treatment of some topics that have already received much attention in hearing science. Major topics that are only mentioned in passing are binaural processing, intensity and dynamic range compression, and lateral inhibition, as well as across-channel frequency weighting that is achieved by different segments of the auditory processing chain, which possibly contains the spectrotemporal modulated class of signals. The theory also formally deals with the auditory system up to the inferior colliculus, so higher-level effects (like attention or speech perception) are mostly avoided. Finally, in order to limit the scope of the literature reviewed, I tended to ignore most of the mathematical models of the associated ear parts, such as the cochlea, the auditory nerve, or of the complete system. These omissions notwithstanding, there was still much left to be explored in this work. 

In the text, I have striven to remain agnostic about the particular cochlear mechanics that transduces the signal, as numerous publications and models have been exclusively dedicated to this problem and, for all I can tell, the jury is still out as for which one is (the most) correct. Experimental data are still being reported regarding the cochlear mechanics and it is not unusual that they contradict predictions made based on different cochlear models or on classical observations done with outdated methods. However, I have posited two new functions of the organ of Corti (of a phase-locked loop and a time lens), which has made it almost impossible to retain this agnostic approach all throughout. 

\section{Chapter overview}
The heart of this work---the temporal imaging theory---is contained in chapters \crefrange{temporaltheory}{TemporalSampling}. Confident readers are encouraged to skim them before committing to the various introductory chapters. Results and implications are presented in the final chapters \crefrange{AudImageFun}{GeneralDisc}, which are more qualitative in nature and likely have relevance to a wider audience within the auditory research community. 

The introductory chapters survey a range of topics that are not necessarily new, but they attempt to tackle several acoustical issues in a fresh manner that is especially pertinent to hearing as a communication system that is embedded in a realistic world of arbitrary stimuli. Notable among them are chapters \cref{PhysicalSignals} about physical signals, and \cref{IntroCoh} and \cref{CoherenceTheory} about synchronization and coherence. Chapter \cref{PLLChapter} may be considered a standalone text that is introductory in spirit, but presents a novel hypothesis regarding the auditory phase-locked loop (PLL). This chapter was required for the assumption of coherence conservation between the external world and the auditory brain, but I believe that it may have far-reaching consequences beyond it, which are only superficially explored in the present work. 

Appendices \cref{ExCohere}, \cref{Aliasing}, and \cref{PsychoEstimation} feature results of small-scale measurements, which were necessary to corroborate some of the claims in the text and may be interesting in their own right, although only the first two may be understood without reference to the main text. 

Some of the sections refer to audio demos, which can be found in the supplementary directories and are printed in \textsc{small caps}. They are found in \url{https://zenodo.org/record/5656125}.

Below is an overview of the individual chapters. 

~\\

Chapter \cref{HearingTheory} motivates the treatise and provides a brief review of current and historical hearing theories with emphasis on visual analogies. It dwells on existing attempts to define the acoustic object, the auditory image and object, and the inconsistencies and shortcomings they bring about. Using various physiological, functional, and physical considerations, it makes the case that a correct analogy between the ear and the eye has it that the cochlea of the inner ear is at an analogous level to the lens, whereas the inferior colliculus of the auditory midbrain is at an analogous level to the retina. A temporal imaging theory is then motivated using four additional perspectives: the prominence of direct versus reflected sound in hearing (unlike light in vision), imaging mathematics analogies between spatial and temporal equations, insights from communication about the physical transfer of information, and signal coherence propagation from the acoustic environment into the listener's brain. 

Chapter \cref{EarAnatomy} reviews the anatomical structure and physiology of the mammalian ear, with emphasis on humans, from the external ear to the auditory cortex. The review is deliberately high level in that it tends to neglect low-level details (e.g., cellular, biochemical) in order to crystallize a systemic perspective, where attainable. It is intended mainly for reference and for highlighting possible roles that have been attributed to the different components of the auditory system. The chapter concludes with a comparative section about some of the major differences between the auditory systems of humans and other mammals.

Chapter \cref{InfoSourceChannel} presents a novel point of view on known aspects of real acoustic sources and environments. The idea behind this chapter is to highlight how the acoustics of realistic sounds and environments diverge from classical linear descriptions. For this purpose, a general formalism is adopted for the representation of waves, which allows for straightforward incorporation of the concepts of dispersion, instantaneous envelope and phase, and group delay. The overarching difference can be boiled down to that between constant Fourier frequency representation and time-dependent complex envelope representation, which facilitates amplitude and frequency modulation. The effects of dispersion and other realistic acoustic signal degradations in realistic environments---both outdoors and indoors---are emphasized. 

A very short introduction to physical optics is provided in Chapter \cref{ChapterImaging}, which revolves around spatial imaging. Several basic concepts in geometrical, wave, and Fourier optics are presented, as they provide the basis for the analogy with hearing in later chapters. The optics of the eye and the main elements in its peripheral physiology are presented. Finally, notable links and differences between imaging and Fourier analysis in acoustics and optics are mentioned.

Chapter \cref{InfoTransfer} introduces a few basic information- and communication-theoretic concepts in a qualitative manner. Information theory is not applied directly in the work, but the physical propagation of information is taken to be the unifying element across the different stages of auditory processing. Several historical connections between information and hearing are briefly mentioned and it is argued that conservation of information---over the various signal transformations---has been taken as an implicit assumption of hearing theory. Actual communication systems can be described using generalized receivers and transmitters that deal with modulated signals. It is argued that hearing can be viewed as a communication system by assigning the appropriate roles of transmitter, channel, and receiver to the acoustic source, environment, and ear, respectively, and by recognizing that the intentional transfer of information is optional. The communication approach to information transfer is very similar to that used in simple spatial imaging, but there are some important differences between them that are highlighted as well. 

Chapter \cref{PhysicalSignals} deals with the mathematical basis of physical and communication signals, which are used in all the theories that are relevant to this work. It begins from the analytic signal and the narrowband approximation, which gives rise to the important concept of instantaneous frequency. It then explores the roles of the temporal envelope and amplitude modulation in hearing and briefly reviews the role of phase in hearing, with emphasis on linear frequency modulation. Auditory phase perception has been a contentious topic, which gave rise to the concept of temporal fine structure as a proxy of auditory phase locking. However, several authors have indicated that the common way of applying these concepts in hearing has been inconsistent with the mathematics of broadband signals and with certain psychoacoustic observations. It is shown that the emphasis that has been put on the (mathematically) real envelope has led to auditory theory that treats hearing as a baseband (i.e., with low-pass characteristics, as though the system is capable of detecting sound down to 0 Hz) rather than a bandpass system. It is argued that a correct treatment of the system as bandpass is critical for embracing modulation and demodulation phenomena in hearing as a reality, rather than a metaphor. The existence of auditory demodulation along with a two-dimensional (carrier and modulation) spectrum are considered. 

Chapter \cref{IntroCoh} is an exposition of the concepts of coherence and synchronization that are found in six different scientific fields that have some bearing on hearing: acoustics, optics, communication, neuroscience, auditory neuroscience and physiology, and psychoacoustics. While the essence of coherence as a concept may be shared between all six fields, it is obfuscated by the use of different jargons, sometimes for narrowly defined purposes. A standardized jargon is then proposed, which is used throughout the work. It largely adheres to the jargon used in optics of coherent and incoherent illumination and imaging, which overlaps with coherent and noncoherent detection in communication. 

Chapter \cref{CoherenceTheory} draws heavily on optical coherence theory and summarizes its most important concepts that include interference, the mutual coherence function, partial coherence, temporal and spatial coherence, coherence time,  coherence propagation according to the wave function, spectral coherence, the effect of narrowband filtering, and nonstationary coherence. These ideas are then linked to data gathered about known sound sources and to the theory of room acoustics, which has used coherence rather sporadically. Other topics in hearing that relate to coherence are briefly mentioned as well, such as binaural hearing and coincidence detection. It is argued that partial coherence and coherence time have central roles in auditory perception. Appendix \cref{ExCohere} provides several quantitative figures from typical acoustical sources to substantiate some of the main claims of the chapter. 

Chapter \cref{PLLChapter} introduces the concept of synchronization from nonlinear dynamical system point of view. It focuses on the phase-locked loop (PLL)---one of the most important circuits in communication engineering, control theory, and general electronics. It is shown how a PLL can conserve the degree of coherence of an input signal at the output. It is then hypothesized and demonstrated how the phase locking that characterizes the mammalian low-frequency hearing can be the result of an auditory PLL, which may be assembled from known functions of the organ of Corti and the outer hair cells: a phase detector from the distorting mechanoelectrical transduction channels, a loop filter from the outer-hair cell membrane, and the self-oscillating hair bundle as the voltage controlled oscillator. This hypothetical feedback process may be additionally amplified by the somatic motility of the outer hair cells and feed into the inner hair cell transduction path. Available evidence that supports this idea, as well as known gaps in the model, are discussed at length. The usefulness and likelihood of having dual coherent and noncoherent detection within hearing are discussed as well.

Chapter \cref{temporaltheory} derives the paratonal (originally, the ``paraxial'') dispersion equation that was first introduced in nonlinear optics and has been applied to scalar electromagnetic plane waves, but can just as well apply to pressure waves. This equation employs the space-time duality principle, which analogizes the spatial envelope to the temporal envelope of the wave field. The general solution of the wave equation requires the narrowband approximation---the decomposition of the wave into a fast moving carrier and a slow-moving complex envelope. Only the complex envelope is considered in the solution that has a fixed carrier. The solution requires the group-delay dispersion (or group-velocity dispersion), which is a fundamental property of the medium. It is expressed using the derivative of the standard (phase-velocity) dispersion and has not appeared in this name in acoustics before. The reciprocal operation to the group-velocity dispersion of the medium---that of a time lens, or a phase modulator---is also presented. Both group-delay dispersion and time lensing rely on quadratic phase transformations that can produce linear frequency modulation. 

Chapter \cref{paramestimate} goes through the signal transmission chain of the human ear and attempts to estimate its frequency-dependent dispersion parameters. The passive cochlea is known to be group-delay dispersive, and the magnitude of this dispersion is estimated to be much larger than that of the outer and middle ears. The group-delay dispersion associated with the neural pathways between the inner hair cells and the inferior colliculus---a quantity that has been considered to be negligible before---is estimated as well. It is speculated that the organ of Corti functions also as a time lens, and a physical principle of phase modulation is hypothesized, which has to do with the active change of stiffness that is caused by the outer hair cell electromotility. The time-lens curvature values for humans are roughly estimated based on animal data. The uncertainty in these estimates is very large and we derive approximate lower and upper bounds that may pertain to humans. Appendix \cref{PsychoEstimation} offers an alternative derivation of the dispersion parameters using strictly psychoacoustic data instead of physiological data. The results are only partially consistent with the physiological estimates, because they introduce group-delay absorption to the parameters, which is largely ignored in this work and in optics, although it may be physically justifiable. Alternative explanations for the discrepancy are discussed.

Chapter \cref{ImagingEqs} introduces the imaging equations, which require the three dispersive components to be in cascade---cochlear group-delay dispersion, time-lens curvature, and neural group-delay dispersion. The image of a pulse is computed and it is shown that it is inherently defocused in humans, based on the estimated parameters from Chapter \cref{paramestimate}. The same parameters are used to model available psychoacoustic data of the cochlear curvature in humans and high correspondence is found above 1 kHz. At lower frequencies, additional constraints of modulation bandwidth had to be introduced in order to obtain better estimates, although some uncertainty remains. The results also reveal the existence and the durations of the temporal aperture---a short sampling window that is associated with the different auditory channels. The estimates show close correspondence to additional human and animal data. It suggests that at high frequencies the aperture stop is determined by the auditory nerve, but at low frequencies it is determined by the cochlear filters. 

Chapter \cref{ImpFun} begins from the derivation of the impulse response of the defocused temporal imaging system for a single channel. The equations are then used to further derive the modulation domain transfer functions, which have not been previously introduced in optical temporal imaging, but are completely analogous to the spatial modulation and optical transfer functions from Fourier optics. Predictions are compared to available observations of the auditory temporal modulation transfer function. Interesting predictions and discrepancies are highlighted, where the dependence of the results on the degree of coherence of the stimulus is argued to be key. 

The effect of sampling of continuous signals by the neural system is explored in Chapter \cref{TemporalSampling}. While sampling has been invoked several times in hearing models, the consequences of discretization have not been fully considered before. The significance of nonuniform sampling is discussed with respect to the modulation transfer functions, which are thought to degrade (lose modulation bandwidth) upon repeated resampling that occurs downstream, within the auditory pathways. The tradeoff between nonuniform sampling noise and aliasing from undersampling, as is known to take place on the retina, is hypothesized in the context of hearing. A psychoacoustic experiment that is interpreted as demonstrating the existence of auditory sampling is presented in Appendix \cref{Aliasing}, with emphasis on the effects of aliasing.

Chapter \cref{AudImageFun} explores in greater depth the idea of an auditory image based on all the previous findings and the principle of space-time duality. The concepts of sharpness, blur, focus, defocus, and depth of field are discussed, and a simple computation of the system temporal acuity is presented, based on its impulse response or the modulation frequency discrimination. A formal presentation of polychromatic images is made and pitch is discussed as a special case in auditory imaging that manifests in different ways. A subset of image aberrations from optics that can be relevant to hearing are discussed with speculations about the most significant auditory aberrations in humans. Ideas from masking theory are extrapolated to examine how supra-threshold stimuli sound in the presence of other sounds. Furthermore, nonsimultaneous masking is analogized to the auditory depth of field that applies temporally and is exaggerated by the signal processing of the auditory system. Most of the imaging effects considered are well-known auditory phenomena that are reinterpreted in light of the concepts of temporal imaging. Seven rules of thumb for auditory imaging are proposed that epitomize some of the analyses in the chapter. 

Chapter \cref{accommodation} takes these ideas a step further and hypothesizes what an auditory accommodation function that is roughly analogous to accommodation in vision could be like. Different mechanisms are proposed for parameters that can be shifted within the system. The operation of the olivocochlear efferent bundle appears to be key in accommodating the PLL gain and/or the time-lens curvature. The plausibility of other mechanisms of accommodation is discussed. The coherence of the stimulus is a recurrent key parameter in the analysis that the system is hypothesized to react to. The idea that the system may be combining coherent and incoherent imaging products in different amounts is considered and complements the earlier discussion made in the context of coherent and noncoherent communication detection in Chapter \cref{PLLChapter}.

Chapter \cref{impairments} brings together the ideas of auditory imaging and coherence and sets to find out if they can be used to shed light on known hearing impairments. Evidence for dispersive shifts in hearing-impaired individuals is examined and their effects are considered. Additionally, the possibility of aberration and accommodation impairments is explored, also with analogy to eye disorders. Out of the different impairments considered, accommodation disorders and excessive higher-order aberrations appear to have the highest likelihood to be detrimental, but more conclusive relations to known hearing disorders cannot be made before the entire temporal imaging theory is elucidated and more relevant data become available. 

The treatise closes in Chapter \cref{GeneralDisc}, where a functional model of the hearing system is presented that encompasses all of its standard parts, as well as the imaging components and the PLL, which are roughly mapped to the different auditory organs. The chapter concludes with a general discussion that highlights open questions, limitations, weaknesses, and merits of the ideas presented in this treatise. It also highlights several topics that can benefit from future experimentation. Finally, certain novel ideas and issues that appear in this work are mentioned, which may find interest outside of hearing research alone.


\newpage
\pagenumbering{arabic} 
\setcounter{page}{1}
\setcounter{secnumdepth}{3}

\chapter{Background}
\label{HearingTheory}
Several fundamental aspects of hearing come in pairs: place coding and temporal coding of pitch, localization using interaural time difference cues or interaural level difference cues, signal identification using temporal-envelope cues or temporal-fine-structure cues, and discrete signal sampling in time or continuous temporal integration. When one mechanism is unavailable, the other seems to cover for it, within certain ranges of signal parameters that are particular for these cues. Present auditory models largely consider a single signal path that functionally combines a predetermined weighting of one or two types of signal processing. This is the case even though the auditory signal splits into three parallel pathways (two in non-mammalian vertebrates) in the auditory brainstem that converge only at the inferior colliculus. Can it be that our sense of hearing is set to do everything twice, so that two signal processing outputs are combined to obtain a superior output to either of the two on their own? 

The theory put forth in this treatise portrays hearing as both a communication system and an imaging system. Both imaging and communication theories provide a built-in duality that distinguishes between coherent and incoherent (or noncoherent) imaging or detection. While most engineered systems tend to stick to either coherence or incoherence, we shall argue that hearing makes the most of both, which also matches the acoustic environment best, as it tends to be partially coherent. Therefore, this work will employ the notion of degree of coherence in a manner that is germane to hearing with emphasis on analogies to vision, since they are numerous and they can provide novel intuition for the development of alternative concepts in hearing. Namely, the theory highlights some parallels with vision that are simpler to grasp if the spatial and temporal envelope dimensions are swapped. However, unlike vision, whose image can be visually seen on the retina before being demodulated by the photoreceptors\footnote{Christoph Scheiner was the first to directly observe the retinal image in 1619 \citep[p. 57]{LeGrand1980}.}, the auditory image resides deep inside the brain and cannot be listened to---at least not without sophisticated instrumentation and signal processing. Moreover, auditory processing includes various phase and across-channel interactions that give rise to auditory phenomena, such as harmony, that do not have analogs in vision.

This introductory chapter surveys some of the overarching historical themes in hearing theory. Then, it examines more closely the various comparisons that have been made between hearing and vision, in attempt to garner insight about hearing that is more readily found in vision. This background is used to motivate the main themes that have driven this work---temporal imaging being the foremost one.

A historical review of hearing theory is found in \citet[pp. 399--436]{Boring1942} including some interesting notes about discoveries and ideas about the physiology of hearing, which anticipated Helmholtz's work and provided the context for its development. An even more comprehensive review of historical theories was given by \citet[pp. 3--94]{Wever1949}, where he delineated the so-called ``place theories'' and ``frequency theories'', which are now referred to as temporal theories. A vivid account of several hearing theories has been presented in \citet[Chapter 2 and throughout]{Lyon2018}. A recent report on how the understanding of tonotopy has developed is found in \citet{Ruben2020}, where the origin of the frequency analysis and tonotopy of the cochlea is traced back to Guichard Joseph Du Verney in 1683, who mistakenly switched the frequency mapping between the cochlear base and apex. A more detailed (but still brief) overview of the anatomy and physiology of the ear is provided in \cref{EarAnatomy}. 

\section{The scope and state of hearing theory}
\label{HearingIntro}
Although most scholars did not attempt to formally define the scope of hearing theory, it seems to have meant different things for those who did. For example, \citet{Fletcher1930} delineated several aspects that such a theory should be able to account for: the auditory bandwidth and dynamic range, the just-noticeable differences in pitch and intensity, distortion products, masking phenomena, loudness, binaural effects, frequency selectivity of complex tones, effects of pitch and loudness and their relations to the physical signal. Two decades later, \citet[p. 1034]{Licklider1951Stevens} suggested a more general scope: ``\textit{The principal tasks of auditory theory are (1) to explain the psychophysics of hearing in terms of aural mechanics and neurophysiology, (2) to give audition its proper setting in a general theory of communication, and (3) to provide a calculus of response to auditory stimulation.}'' Although the role of the brain in hearing was recognized from anatomical data and was being crudely incorporated in early theories \citep[e.g,][]{Fletcher1930,Wever1930Present}, hearing theory was regularly considered about equivalent to a mechanical explanation of how the cochlea transduces the acoustical waves to neural spiking patterns as late as 1975 \citep{Bekesy1956,Schroeder1975}. More recently, \citet{Cariani2012Poepel} proposed a more cautious scope: ``\textit{A full theory of audition thus should explain the relations between sounds, neuronal responses, auditory functions, and auditory experience.}'' This definition seems to be general enough to be inclusive of just about anything auditory.

Ideally, a complete theory of hearing should be able to explain what hearing does and how it is facilitated by the ears, for a given listener, acoustic environment, stimulus, and situation. Furthermore, it should enable derivation of particular auditory effects, motivate observed behaviors and responses, and help resolve ostensible contradictions in the empirical science. In other words, the ideal hearing theory should be able to reduce the complexity of observations that had seemed disparate before the theory was introduced. At present, no such theory exists\footnote{Acknowledgment that hearing theory does not exist has been seldom made. See for example, \citet[p. 57]{NordmarkTobias} and \citet{Bialek1985}. A passage that may be interpreted as saying essentially that was provided by Reinier Plomp---one of the most prominent psychoacousticians of the second half of the 20th century. In his book, ``The Intelligent Ear'', he candidly admitted \citep[p. 9]{Plomp2002}: ``\textit{Many investigators have assumed that full knowledge of how sinusoidal tones are heard will be sufficient to explain the perception of everyday sounds. A long time ago, a well-known Dutch composer visited me, presuming that current laboratory knowledge of tone perception could help him to a better understanding of how musical sounds are perceived. I had to disappoint him.}'' Plomp then went on to suggest that the reason for this lack of knowledge is the discrepancy between the perfect laboratory-based stimuli and sounds encountered in the real world. In the context of comparing visual and auditory processing models, \citet{Schonwiesner2009} stated: ``\textit{In auditory neuroscience there is no consensus yet about the suitable set of low-level features.}'' In the narrow context of masking phenomena, \citet{Durlach2006} pointed at a ``\textit{conceptual chaos}'' and that: ``\textit{Not only is there no overarching conceptual structure available to organize the area and provide it with scientific elegance, but there are few definitions that evidence even a modest degree of scientific stability (varying across both individuals and time).}'' As a final example, in the context of a theory of the auditory thalamus and cortex, \citet[p. 679]{Winer2010winer32} stated: ``\textit{There is no global theory of auditory forebrain function since the facts available cannot support such an edifice.''} }. Instead, various ``part-theories'' (an expression coined by \citealp{Licklider1959}) are available, which attempt to account for local phenomena within hearing, such as the cochlear function, pitch perception, sound localization, auditory scene analysis, auditory attention, speech perception, music perception, etc. 

All the part-theories notwithstanding, it is not a given that once they reach maturity, they will coalesce into a grand unified theory of hearing. Hearing is a complex biological system, and biology is presently best understood through evolution theory \citep{Dobzhansky1973}, as \citet{Lettvin1959} noted with regards to vision: ``\textit{...since the purpose of a frog's vision is to get him food and allow him to evade predators no matter how bright or dim it is about him, it is not enough to know the reaction of his visual system to points of light.}'' Therefore, any account of hearing phenomena beyond evolution that results in a more compact and less complex theory than the present state of knowledge may be seen as a boon. 

\section{Elements of hearing theory}
\label{ElementsHearing}
Hearing theory has made a slow progress over about two and a half millennia from the external to the internal hearing organs---from the outer ear to the brain \citep[pp. 399--400]{Boring1942}. With the monumental work of Hermann von Helmholtz, it became abundantly clear that the most critical sensory transformation of sound occurs in the cochlea, where it is transduced to neural information \citep[first published in 1863]{Helmholtz}. Helmholtz hypothesized that the capability of the ear to analyze complex tones is due to the radial fibers that make up the basilar membrane of the cochlea, which locally resonate in sympathetic vibration with incoming tones, just like tuning forks. Almost a century later, the collected works of \citetalias{Bekesy1960} were published, where what may have been the most influential model of cochlear mechanics since Helmholtz's was laid out. B\'ek\'esy showed that the unique spectral analytical property of the cochlea corresponds to a \term{traveling wave} that propagates along the basilar membrane. The traveling wave peaks at a place along the cochlea that is mapped to a specific frequency, as a result of the elastic properties and geometry of the basilar membrane. 

By the mid-20th century, the maturation of electronic engineering and signal processing precipitated the transformation of empirical science as a whole, including acoustics and neurophysiology. Using a mixture of synthesized pure tones and white noise maskers, it was found that the frequency range of the ear is internally covered by a bank of bandpass filters (also referred to as channels) with overlapping flanks. The response of these filters provides a robust explanation for simultaneous masking phenomena \citep{Fletcher1940}, loudness summation \citep{Zwicker1957} and other important effects. 

The significance of the sound information transfer to neural signals was hypothesized by \citet{Rutherford1886} in his \term{telephone theory} where he noted: ``\textit{that simple and complex vibrations of nerve energy arrive in the sensory cells of the brain, and there produce, not sound again of course, but the sensations of sound...}'' Direct physiological recordings of auditory nerve fibers of cats showed that the spiking patterns are spectrally tuned \citep{Galambos1943,Kiang1965}, in a way that corresponds to the incoming sound, if amplified and played back \citep{Wever1930Action}. The spiking pattern was hypothesized to follow the \term{volley principle}. This principle predicts that the auditory nerve fiber bundle, which innervates a given hair cell in the cochlea, tracks the incoming stimulus together, so that all fibers spike in tandem at a certain phase of the stimulus \citep{Wever1930Present}. This principle can account for how low-intensity stimuli can be heard even though each individual fiber experiences refractory periods and fires stochastically (see also \citealp[pp. 166--441]{Wever1949}). Indeed, once the neural and mechanical recording techniques became sufficiently precise, it became possible to establish a close correspondence between the cochlear mechanical response of the traveling wave and the auditory nerve fibers \citep{Sellick1982,Khanna1982}. 

Things became more complicated as it had gradually become evident that the cochlea is unmistakably nonlinear, since the dynamic range of the mechanical traveling wave response is compressed relatively to the acoustic input \citep{Rhode1971}. Dispelling any doubt that the cochlea must contain an active mechanism needed the discovery of \term{otoacoustic emissions} from the ear by \citet{Kemp1978}, which had soon after received the name \term{the cochlear amplifier} by \citet{Davis1983}. This idea resurrected a much earlier model of active hearing that was proposed by \citet{Gold1948} but was prematurely rejected at the time. The discovery of electromotility of the outer hair cells by \citet{Brownell1985}, along with converging physiological evidence using various methods and models, has led to the conclusion that the outer hair cells are the main cause of nonlinearity in the cochlea. However, as the complexity of the organ of Corti is very high, and since handling it requires delicate methods and tools if it is to remain intact during experimentation, much work has been carried out through modeling and indirect measurements of cochlear mechanics. Thus, the exact mechanism of amplification (as well as other cochlear effects and the very function of various cochlear features) are still being debated \citep[e.g.,][]{Ashmore2010}. Importantly, the same organ appears to be the main cause for a host of other nonlinear phenomena, including the generation of audible distortion products in the cochlea \citep{Avan2013}.

The above observations, which follow the auditory signal in the auditory nerve channels, can partially account for most phenomena on Fletcher's list (\cref{HearingIntro}) that underscore the significance of the cochlea in hearing sensation. Binaural effects such as localization, however, are strictly dependent on neural structures in the auditory brain to work, as was hypothesized by \citetalias{Rayleigh1907Duplex} and has largely been confirmed since \citep{Grothe2010}. But binaural effects are not the only phenomena that extend beyond the auditory nerve. Another important case is the detection of periodicity across the cochlear bandpass channels, which gives rise to the effects of the missing fundamental and periodicity pitch \citep{Schouten1940}. The highly influential \term{duplex pitch model} by \citet{Licklider1951} maintained that the auditory system carries out an autocorrelation operation, centrally, using delay lines and coincidence detectors on top of the cochlear bandpass filtering. The latter can give rise to the standard tonal pitch, whereas the former can give rise to periodicity pitch. Other notable phenomena that require central processing are the processing of amplitude and frequency modulation, broadband sound processing, various adaptations that improve detection of specific sounds in noise and reverberation, forward masking, and also disorders such as tinnitus. Therefore, it is clear that peripheral theories of sound only cover a restricted range of auditory phenomena. 

Probably the most influential higher-level theory of hearing in the last years is that of \term{auditory scene analysis}, which was synthesized by \citetalias{Bregman}. The main idea behind it is that the auditory system utilizes different acoustic cues in a given stimulus, which enables the mental organization of different sound elements into auditory streams---the auditory counterpart to visual objects---that are mentally localized in space. Different cues may be available to the listener that enable auditory streaming, such as grouping based on common onset time of sounds, common fundamental frequency (harmonic content), cues based on spectral range, location in space, and others. Importantly, this theory provides a framework for an intermediate stepping stone in the auditory information processing of meaningful sounds such as speech, vocalizations, or music. Scene analysis is also related to other cognitive factors such as memory, attention, and decision-making, and it generally opens a much broader perspective for hearing theory. However, while auditory scene analysis provides a powerful framework for understanding sound processing and demystify several auditory illusions, it was originally framed without a clear physiological substrate that can realize the necessary signal processing. Thus, scene analysis does not readily connect with the peripheral auditory output mentioned above, which was traditionally considered the essence of hearing theory. Auditory scene analysis is thought to take place at the level of the cortex \citep{Christison2015}, although there are indications that basic processing of relevant cues begins in the brainstem \citep{Masterton1992,Pressnitzer2008,Felix2018}. 

Harnessing auditory scene analysis to its core, \citet{Cariani2012Poepel} outlined the scope for an auditory theory that should take basic auditory (and cognitive) attributes as multidimensional variables of auditory information. The basic auditory attributes are loudness, duration, location, pitch, and timbre, which are processed in the auditory cortex, go through scene analysis, and culminate in conscious perception. Such a theory should draw on the three levels for understanding information processing in the visual cortex that were laid out by David Marr and could be readily generalized to other sensory modalities: 1. ``\textit{Computational theory---What is the goal of the computation, why is it appropriate, and what is the logic of the strategy by which it can be carried out}.'' 2. ``\textit{Representation and algorithm---How can this computational theory be implemented? In particular, what is the representation for the input and output, and what is the algorithm for the transformation?}'' 3. ``\textit{Hardware implementation---How can the representation and algorithm be realized physically?}'' \citep[p. 10]{Marr2010}. In the theoretical framework outlined by \citet{Cariani2012Poepel}, there is emphasis on elucidating the different neural codes that represent the auditory stimulus, which can then be processed in a broad range of perceptual operations using different neurocomputational architectures that lead toward decision making, which sometimes leads also to action. 

The above high-level theoretical frameworks leave a substantial gap in the understanding of the subcortical processing between the auditory nerve and the cortex. It is known that important processing takes place, probably gradually, in the brainstem and midbrain, which reflects the animal's specific needs and ecology (\citealp[e.g.,][]{Casseday1996,Felix2018}; see \cref{CentralNeuroanatomy}). Such processing is likely directed by the auditory cortex via the descending efferent pathways \citep{Cariani2012Poepel}. But relevant theories tend to be too general, probably due to the complexity of the involved circuitry (a complexity that relates both to their apparent algorithms and hardware, re Marr), as well as the large number and variety of processes (apparent computational goals) implicated by them. This was implied by \citet[p. 67]{Winer2010Winer2}, who stressed that the auditory system is an artificial construct made up of streams, although in reality it is integrated with many other sensory modalities and processes. Moreover, he underscored that even with the auditory scene analysis theoretical framework, when it is applied to the physiological circuitry, a significant number of auditory nuclei in the midbrain and forebrain are excluded from the processing chain, and their functions within the complete auditory process are poorly understood.  

At present, it remains unknown whether the different auditory brain functions can be explained using a compact theory from which lower-level effects may be unambiguously derived, or whether each auditory phenomenon has to be accounted for using its own part-theory. In this sense, the combined part-theories of hearing---covering the entire cascade of periphery, brainstem, and perception---are incomplete.

\section{Hearing theoretical development and vision}
\label{HearingTheoryVision}
One approach that has been repeatedly invoked to produce insights about hearing is comparison to vision, which is by far the most studied sense in humans. Vision is also touted as the most dominant of all our senses and humans are often said to be ``visual creatures''. The analogy between the two senses is natural, given that hearing and vision are intuitively juxtaposed---both organs are placed on the same level of the face, both come in pairs, both are based on wave physics, both sensory experiences are ubiquitous and elemental in the human experience, and both can give rise to rich aesthetics and culture. 

The cross-inspiration between hearing and vision can be traced back at least to ancient Greece, when various analogies between hearing and vision were conceived \citep{Darrigol2010A,Darrigol2010B}. 
At that time, light was thought to emanate from the eyes (originally due to Empedocles) and there was generally no distinction between sensation and perception \citep{Hamlyn1961}. A more fertile correspondence between the nature of light and sound began only around 1000 A.D., when it was proposed by Ibn al-Haitham---widely considered as the father of modern optics---that light is an entity that is independent of the beholder. Many of the most prominent scholars that studied acoustics and hearing from the Renaissance until the end of the 19th century contributed significantly to both sound and light theories. Notable figures such as Isaac Newton, Thomas Young, Hermann von Helmholtz, and Lord Rayleigh critically advanced both fields. Interestingly, Helmholtz, who is best known in hearing for his influential place-theory of pitch perception \citep{Helmholtz}, is also credited with laying the foundations to the optics and the physiology of the eye \citep{Helmholtz1867}---still relevant today as well (e.g., \citealp{LeGrand1980,Charman2008}; see also \citealp{Wade2021} for a historical review). 

Perhaps the most epigrammatic of all the hearing-vision analogies is best captured by the aphorism ``\textit{architecture is frozen music}\footnote{Attributed to both Johann Wolfgang von Goethe and Friedrich Wilhelm Joseph von Schelling.}.'' It relates to the fact that visual perception unfolds in space, whereas hearing unfolds in time \citep[e.g.,][]{Hirsh1952, Massaro1972b,Jones1976, Welch1980, Kubovy1988, Naatanen1999}. \citet{Hirsh1952} maintained that the perceived visual dimensions are determined by objects, which are not meaningful auditory entities, since hearing is concerned with events. Using these dimensions for comparison, visual acuity---``\textit{a measure of the \textsl{interval of space} between two visual stimuli that are perceived as two}''---becomes the fundamental descriptor of visual capacity, whereas temporal acuity is the analogous one for hearing. This point was elaborated in \citet{Julesz1972}, where it was stressed that visual events do exist---making the comparison asymmetrical between the two senses. They further emphasized the transient nature of the auditory environment compared to the stable one of vision, but they implied that the two senses can be ultimately seen as complementary, as they both detect both space and time of the animal's environment. \citet{Handel1988} pushed back against the adage that visual space and auditory time are analogous, as neither sensory object can exist in space or in time only. Nevertheless, the observation that vision is largely spatial and hearing is largely temporal has been robust. For reasons that will become clearer later, we shall borrow from optics and refer to it as the \term{space-time duality}.

\begin{figure} 
		\centering
		\includegraphics[width=0.9\linewidth]{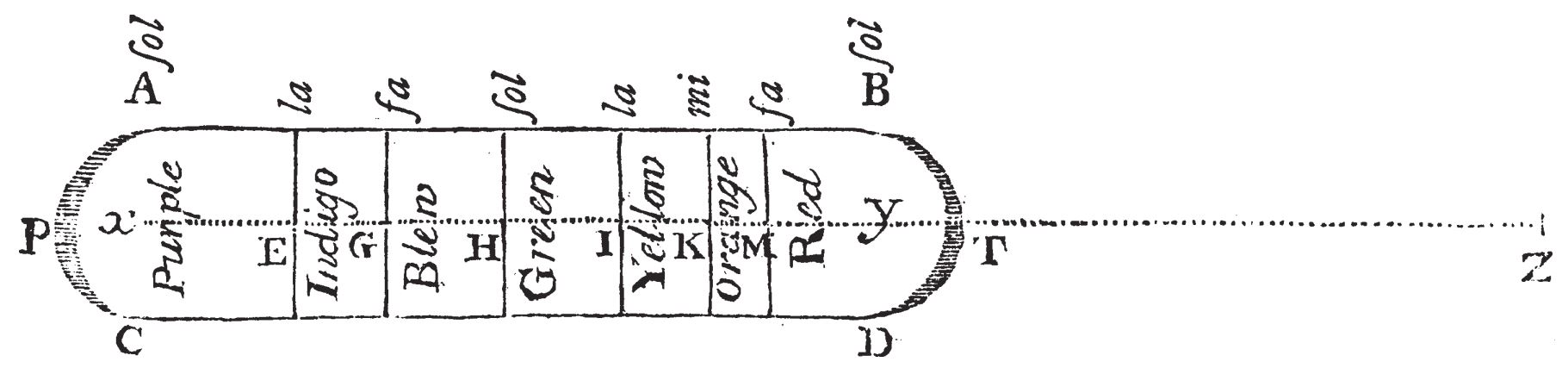}	
		\caption{Isaac Newton's color-pitch map from his second paper on color and light, read at
the Royal Society in 1675.}
		\label{NewtonColors}
\end{figure}

Other comparisons between hearing and vision were usually made by emphasizing specific dimensions. For example, a popular analogy that appeared early is between pitch and color. Marin Mersenne wrote in 1632 that the lowest notes are akin to black, the highest notes to white, and the colors are anything in between \citep{Darrigol2010B}. An influential analogy between color and pitch was drawn circa 1665 by Isaac Newton, who tried to impose the natural musical intervals of the seven diatonic notes on the seven primary colors that constitute white light \citep{Pesic2006} (Figure \ref{NewtonColors}). A similar analogy was independently proposed by Robert Hooke in 1672. Leonhard Euler was the first to relate colors to the frequency of light waves---drawing from sound wave theory instead of the other way round \citep{Moller2008}. Helmholtz too continued this line of thinking, having recognized that both pitch and color relate to the wave frequency, but he eventually abandoned the analogy with musical intervals, seeing that the spectra of light and sound cannot be made to match \citep{Pesic2013Helm}. 

Helmholtz made additional comparisons between the ear and the eye throughout his book from which  two points stand out. First, the eye is much slower than the ear in its ability to resolve changes \citep[p. 173]{Helmholtz}. Second, the ear has an analytic ability to decompose complex tones to their harmonics, which the eye does not possess. He additionally compared vibrations in the spatial-frequency domain (he used water waves as an object) to the audio frequency domain \citep[pp. 29 and 128]{Helmholtz}. As will be seen later in this work, this is confusing the carrier and modulation domains of the two systems---not an uncommon mistake in hearing and vision that is still occasionally encountered in literature.

Several commonalities between the two senses were contrasted by \citet{Harris1948}, who compared certain aspects in the psychophysical and neural coding in both hearing and vision. He found some that were directly comparable and differ in value (e.g., sensitivity, internal noise, neural adaptation, integration time, dynamic range, lateral inhibition), but others that were not directly comparable (e.g., frequency required for tonal modulation to sound continuous, binaural/binocular summation, quantum-effect threshold). While most comparisons are dated, they are valuable in showing the general properties that both the ear and the eye have as physical detectors, whose outputs manifest neurally. This is also the underlying thinking behind the work of \citet{Jacobson1950,Jacobson1951}, who was the first to attempt to quantify the informational channel capacity (in bits per second) of the eye and the ear. More narrowly, \citet{Stevens1957} focused his comparison on psychophysical characteristics of the two senses and on the similarity of the power laws that describe them. He contrasted loudness in hearing and brightness in vision as the psychophysical counterparts of intensity, and are comparable with respect to their growth-rate dependence on frequency (wavelength), the effect of masking (glare), the degree of adaptation to baseline signal level, and their pathologies---loudness recruitment in hearing and a rare genetic disorder of congenital stationary night blindness in vision. Stevens also mentioned loudness dependence on bandwidth as an auditory phenomenon with no analog in vision.

The scope of comparison was further broadened by \citet{Julesz1972}, who noted the omnidirectionality of hearing, its high alertness, and the fact that it does not cease (as vision does when the eyes are closed). Their discussion is driven by perception-general logic, where hearing is dominated by events and less so by objects, which are not as spatially well-defined as in vision. They began developing the figure-ground concept from Gestalt psychology relevant to hearing, which Bregman greatly elaborated and brought to maturity in his theory of auditory scene analysis that in itself was largely inspired by the analogy to vision. \citet[p. 6]{Bregman} wrote: \textit{``In vision, you can describe the problem of scene analysis in terms of the correct grouping of regions. Most people know that the retina of the eye acts something like a sensitive photographic film and that it records, in the form of neural impulses, the `image' that has been written onto it by the light. This image has regions. Therefore, it is possible to imagine some process that groups them. But what about the sense of hearing? What are the basic parts that must be grouped to make a sound?}'' Bregman preferred the term \term{auditory stream}---``\textit{the perceptual unit that represents a single happening}''---to auditory event (Ibid., pp. 10--11), which is the physical occurrence that can be composed of smaller subunits (e.g., footsteps, notes in a melody).

Auditory objects have entered the jargon of auditory scene analysis notwithstanding, as subunits of the auditory stream. For example, \citet{Shinn2008} defined the auditory object to be \textit{``a perceptual entity that, correctly or not, is perceived as coming from one physical source.''} As such, it is taken to be the basic unit of auditory attention. Just as in vision, the listener can attend to only one auditory object at a time, and the different objects effectively compete for attention in a process that combines bottom-up feature extraction with top-down enhancement and control. These high-level similarities probably have physiological correlates, and significant similarities indeed exist between the visual and auditory cortices, which suggest that on an advanced processing level the sensory information may become modality-invariant \citep{Rauschecker2015}. Projections from both the visual and auditory cortices split into the ``what'' and the ``where'' processing paths, which relate to the perceived object attributes that are processed in each path (see \cref{OrgPrinciples}). In a similar vein, it was shown that the human auditory cortex strongly responds to modulated stimuli, which are mathematically analogous to patterns that evoke similar response in vision \citep{Schonwiesner2009}. Such high-level processing similarities have been revisited several times \citep[e.g.,][]{Massaro1972b,Shamma2001,Kubovy2001}, as, for example, in the case of scene analysis processing similarities \citep{Handel2006}, or in the context of units that are tuned to spatiotemporal modulation in vision and spectrotemporal modulation in hearing \citep{Shamma2001,Schonwiesner2009}.

It can be seen that several authors have put vision and hearing on equal footing once perceptual analysis begins. However, vision (or visual perception and the analysis thereof) enjoys the existence of a well-defined optical image on the retina that is amenable for processing in the central nervous system, whereas hearing does not. Hearing has its cochlear frequency map (tonotopy) represented throughout the auditory brain (\cref{OrgPrinciples}), but it constitutes a spectral representation and not an image in any intuitive way, as is the retinotopic map between the eye and the visual cortex. Therefore, several attempts have been made to model the \term{auditory image}, which can feed into higher-order processing such as required by scene analysis.


\section{Objects and images in hearing research}
\label{ObjImg}
The roles of the object and the image are well-defined in optical imaging---the underlying physical basis of visual sensation and perception. They can be expressed mathematically in unambiguous terms using the laws of optics and projective geometry, which enable complete prediction of the real optical image on the retina, prior to transduction \citep[e.g.,][]{LeGrand1980}. However, a precise definition of a visual object that applies to the perceptual experience that corresponds to the optical object is not as straightforward \citep{Feldman2003}. Similarly, the concept of an object that applies equally intuitively to hearing has been elusive and controversial. Most hearing theories overlook this discussion, or rather dodge it by referring to the acoustical source as the de-facto object of hearing. Some hearing models relate to the auditory image without accounting for the object that produced it in the first place. Yet other higher-level hearing models posit the existence of a perceptual auditory object that may or may not follow an auditory image and may not have a definitive physical object it corresponds to. Different models of these three imaging stages (object, image, mental object) are reviewed below. It will be seen that they are generally inconsistent among one another, and, when contrasted, they portray rather confused conceptualizations of both the object and the image of hearing. The main features of all auditory image and object models, including the one presented in this work, are summarized in Table \ref{tab:AudImagingModels}.

\subsection{The acoustic object}
\label{AcousticSource}
A few accounts of the acoustical object---distinct from the auditory object---are described below. Unfortunately, these two concepts are often conflated in a way that is counterproductive for the understanding of hearing as a physical process. The main complication in arriving at a definition of an acoustic object is that, unlike optical objects, acoustic objects continuously change in time. Also, one-dimensional time signals arriving at the ears do not map to two-dimensional spatial images as in vision. Auditory objects are coupled to auditory events, which modulate the active duration of the acoustic sources. Does a spoken word constitute an object? Or should the mouth or the person that utters the word should be considered the object? What happens to the object when the source ceases to emit sound? Possible answers are inconsistent between the visual and auditory domains: in optics and vision we will never refer to the reflected light pattern from the object as the object itself, whereas in hearing the object is sometimes taken to be the actual transient sound---e.g., the word---and not the vocal cords, mouth, or person that produces it. 

Acoustic sources are routinely presented in a naive manner within physical acoustics, which sidesteps any philosophical challenges regarding objecthood. An acoustic source is anything that creates pressure vibrations, which can acoustically radiate through the medium that surrounds it. For example, it can be a rigid body like a string, a plate, a larynx, a loudspeaker, or a locus in a fluid that undergoes disturbance. Vibration and radiation require an investment of energy, so the source has to be forced in order to vibrate. The acoustic source vibrations have the potential to become the object of sensation if they vibrate at frequencies that are within the animal's hearing range, and arrive to the hearing organ at a level that is above their hearing threshold. A passive acoustic object may be subjected to echolocation---targeted radiation by echolocating animals---whose reflection from the object contains information about its geometry and position. 

Another practical way to sidestep the problem of objecthood is to consider auditory stimuli, acoustic signals, or simply ``sounds'', as the entities that are to be sensed by the ear. This terminology dissociates the radiated vibrations from their particular source, assuming that an arbitrary method can be devised to produce the same vibrational pattern, such as a mechanical instrument, a person, a loudspeaker, an array of loudspeakers, or even a direct electrical stimulation of the cochlea. This useful approach runs the risk of losing touch with the geometrical and mechanical acoustics of real sources, which the animal may have adapted to associate with the sounds, possibly using various imperfections in the signals and multimodal cues.

\subsection{The auditory image}
\label{TheAuditoryImage}
Despite the substantial commonalities between hearing and vision, there is an ostensible asymmetry between them at the level of the cortex, due to the different peripheral extent associated with each sense. While the visual periphery produces an optical image, the cochlea does not produce an obvious image, but rather a multiband filtered version of the sound stimulus---often compared to a spectrogram or to the output of a Fourier analysis. Several models have attempted to make up for the missing link by hypothesizing an auditory image, which may have some analogous properties to the visual image and may therefore provide a gateway for further processing. All the models skip the acoustical transformations that take place in the outer and middle ear and rather consider the image to emerge either somewhere at the cochlea or on the various auditory nuclei in the brain, where it necessarily manifests in the neural domain. 

The simplest conceptualization of an auditory image is due to \citet{Kemp2002}, who suggested that the traveling wave in the cochlea is itself an image of the acoustic object. This image represents a size mapping of larger objects---dominated by low frequencies that are mapped to the apical cochlear region, and smaller objects to the basal region of the cochlea. This is the only auditory image model that relates to the concept of image sharpness. According to Kemp, cochlear filters that are actively sharpened by the outer hair cells produce sharper images of the intensity envelopes. Therefore, it is also one of few auditory-image models that relates the image to the physical properties of the object\footnote{This model may be a distant relative of a historical theory by J. R. Ewald, who posited that the basilar membrane vibrates with standing waves (rather than traveling waves), which give rise to an ``acoustic image'' that faithfully represents the sound and some of its characteristic effects \citep[pp. 45--52]{Wever1949}.}.

There have been two prominent attempts to define auditory images that are not strictly spatial as in vision, but are instead composed of some combination of the temporal and spectral dimensions of the acoustic stimulus. Early incarnations of the model can be found in \citet{Lyon1984}, who highlighted the parallel two-dimensional laminae that are found along the auditory neural pathways as a likely target area for an image. The laminae are particularly attractive for imaging because they are tonotopically organized along one dimension found throughout the auditory brain, including the primary auditory cortex (\cref{CentralNeuroanatomy}). Related to Lyon's is the \term{auditory image model} by Patterson and colleagues \citep{Patterson1992, Patterson1995, Patterson1996, Patterson2001}, who proposed that sustained auditory images exist following cochlear and auditory brainstem processing, which includes multichannel compression, half-wave rectification, suppression, phase-alignment, and adaptation. \citet{Lyon2018} further elaborated these ideas and called his model the \term{stabilized auditory image}. These two models consider the second laminar dimension to correspond to periodicity. They consider sound information to be coded in the temporal patterns of the neural spiking, through synchronization. Such processing readily produces correlational responses that are in line with Licklider's autocorrelation pitch and Jeffress's binaural localization models \citep{Licklider1951,Jeffress1948}. The stabilization of the image is related to its ``movie-like'' property of being anchored with zero lag-time, which can be readily obtained using autocorrelation of the neural activity pattern of each auditory channel. 

The resultant image from Lyon's and Patterson's models can be visualized by plotting the time series of the neural activity on the x-axis with all the parallel auditory channels on the y-axis. Pure tones and other periodic stimuli tend to appear as stable patterns on these plots (using a long integration time constant of 200 ms), whereas transient sounds and random noise decay much more rapidly and do not form stable images. Fine details are tracked through faster temporal integration (time constant in the order of 10 ms). This may be achieved by generating short pulses (pulselets) from the stimulus through ``strobing''\footnote{\citet{Patterson1996} also used the term ``quantization'', which in information theory is reserved for amplitude steps in the dynamic range, whereas ``discretization'' is used for generating samples, or symbols, from a continuous sequence. (The symbols are anyway quantized, assuming a finite dynamic range). The concept of strobing was used with no mechanism to explain it \citep{Patterson1995, Patterson2001}, in what can be thought of as a rough ``sample-and-hold'' processing (\cref{FlatTop}) through the various temporal integration stages.}.  
These models were criticized by \citet{Carlyon2003} as they fail to account for across-channel information (produced by the relative delay between channels) that is sometimes used by listeners and has to be extracted using a summary measure of the spectrogram (or image). 

A similar idea to the image was presented by \citet{Carney2018}, who referred to auditory ``\term{fluctuation profiles}'' that appear in the inferior colliculus, which correspond to the dynamic changes in the low-frequency temporal envelopes of the coded signals across all channels. Here, the system optimizes the cochlear gain through its efferent system, in order to maintain an adequate level for coding in the auditory nerve that would otherwise be limited in dynamic range. While ocular focus, blur, and accommodation were invoked as motivation, none of these terms was employed in a more rigorous analogy, where acoustical and optical factors were compared. 

A more ecologically motivated auditory imaging model was proposed by Simmons and colleagues, specifically for echolocating bats \citep{Simmons1980,Simmons1989,Saillant1993,Simmons1996,Simmons2014}. These bat species produce periodic frequency-modulated vocalizations during flight, which are reflected from objects in their environments and are neurally processed to obtain information about the distance, shape, and movement of a remote target. This intricate biosonar process is usually compared to man-made sonar and radar systems\footnote{In radar (RAdio Detection And Ranging), electromagnetic radiation is mainly used to detect and evaluate the distance and sometimes shape of targets of interest \citep{Levanon}. Sonar (SOund NAvigation Ranging) systems achieve the same using sound waves in underwater environments.}, but here the returning echoes are used to construct an image of the remote target. In contrast to the above-mentioned general-purpose auditory imaging models, bats (and probably other echolocating animals) appear to be able to use the stored knowledge of the probing signal in order to obtain an image of the reflecting object and perform exceptionally fast and precise information processing on it. The bat echolocation system extracts information in at least two time scales: individual echoes merge if they are received with less than 0.3--0.5 ms separation. Thus, target (object) features are detected through direct comparison between transmitted and received reflected frequency chirps, which are on the order of 10 ns (!) or longer. The target and image in this case are much more similar to those that are familiar from visual imaging, as reflected waves are used to produce a spatial image of the target, which endows the animal with a superior three-dimensional model of the remote object compared to that achievable with passive hearing.

It is common in the audio jargon to talk about a \textbf{sound image}, especially in the spatial sense, but without defining it precisely. Typically, it relates to where vibrating objects are localized within the listener's mental geometrical space (including inside the listener's head) and how large they are (e.g., \citealp{Sayers1964, Heyser1974,Altman1977,Toole}; \citealp[pp. 245--282]{Moore2013}). This is the context in which stereo or phantom images are discussed in audio engineering. However, this terminology is sometimes used loosely with respect to the object-image pair. For example, \citet[p. 2]{Moore2013} implies that the image is formed in space before it arrives to the ear: ``\textit{The sound wave generally weakens as it moves away from the source, and also may be subject to reflections and refractions caused by walls or objects in its path. Thus, the sound ``image'' reaching the ear differs somewhat from that initially generated.}'' This careful wording is not unique, as another classic example can illustrate: ``\textit{The sonic image, if one could speak of such, is smeared in space behind the physical loudspeaker}'' \citep{Heyser1971}. \citet{Heyser1974} explained later: ``\textit{The subjective sound image, or illusion of sonic presence, is the final form of this figure when we want to study subjective properties.}'' These examples show a conflation between objects and images that hint that all sound images are purely mental or subjective \citep[e.g.,][]{Whitworth1961}---something that is not the case in vision, where a real optical image appears on the retina.

An early theory considered the \term{preperceptual auditory image}---a sustained version of the sensory information about the stimulus that can serve as the input for perception, in tasks such as pattern recognition, detection, and short-term memorization \citep{Massaro1970, Massaro1972b}. Accessing the preperceptual image is likened to a sequential readout process, which is dependent on the complexity of the stimulus. Conceptually proximate, for \citet{McAdams1984}, the auditory image is a metaphor of the internal form of sound objects that are automatically perceived as though they belong together and can be taken as fundamental units in music, for example. He defined (Ibid., p. 11): ``\textit{the auditory image is a psychological representation of a sound entity exhibiting an internal coherence in its acoustic behavior.}'' He added: ``\textit{...there is an identity relation between a percept and its image, where the image notion serves as a kind of bridge between the percept and its interpretation (the concept or schema).}'' 

\citet{McAdams1984} retained a relatively loose usage of the concept of coherence (a coherent stimulus can be a complex tone whose harmonics are modulated in phase, for example), which nevertheless affords the auditory system with the ability to stabilize and organize its image using a variety of cues and principles, in the spirit of auditory scene analysis\footnote{Note that McAdams sometimes used the term ``behavioral coherence'' instead. See \cref{IntroCoh} for a breakdown of the different definitions of coherence found in literature relating to hearing.}. \citet{McAdams1982} noted different mechanisms for separation and grouping of sounds, which result in contiguous images that are also invariant to transformations (such as retaining a sung voice identity, despite vibrato). Later, \citet{Yost1991} drew a stronger parallel between the auditory image and the acoustic source. Following various segregation and grouping processes, listeners can easily identify different simultaneous sources in a mix of sounds, which may seem inseparable just by looking at their neural patterns. It should be noted that \citet{Bregman} himself generally reserved the word ``image'' for vision and only rarely did he use it to designate sound percepts that are distinctly separate, when they do not fuse with other sounds in the same stream.

\subsection{The perceived auditory object}
In the hearing research literature, the basic perceptual unit that is extracted from the auditory scene is often taken to be the ``auditory object'' (e.g., \citealp{Kubovy2001,Shinn2008,Bizley,Nelken2014}). As such, it is dependent on attention, context, familiarity, and other higher-level cognitive factors, which may help to segregate it from other objects and from its background. \citet{Griffiths} suggested that the time scale over which the auditory object is formed need not be fixed and it can be drawn from patterns in the frequency-time plane at different time scales, which can also facilitate the separation of the object from its background. 

Discussions about the auditory object do not usually state whether it arises after an auditory imaging step, or what the relation is between the putative image and object. For example, in \citet{Yost1989Mod} the authors suggested that the auditory system uses binding of comodulated bandpass-filtered stimuli to form auditory objects, with no mention of intermediate images. In contrast, \citet{Griffiths} drew on the auditory image model of \citet{Patterson1995} and employed a building block of a ``\textit{two-dimensional frequency-time object image in the auditory nerve}'' that ''\textit{...might correspond to a sound source or an event.}''. Other auditory object references are almost indistinguishable from those of auditory images. So, according to \citet{Bizley}, auditory objects are fundamental, stable, perceptual units of hearing, which have neural correlates in the auditory pathways, starting from the cochlear nucleus but more prominent in the cortex---very similar properties to the auditory image models by Patterson and Lyon. In another parallel definition, \citet{Griffiths} emphasized the role of memory and familiarity with the objects that is likely required in forming them---elements that overlap with the image models of Massaro and McAdams. 

Treating the acoustic source as an object creates several difficulties, especially if compared to the more familiar visual object. \citet{Hirsh1952} and \citet{Julesz1972} downplayed the importance of auditory objects and preferred to relate to events instead. Bregman, who preferred to use the term ``auditory stream'' instead of ``auditory object'', observed that auditory objects are transparent and add up in loudness, whereas visual objects can block each other and generally have little effect on each other's brightness \citep[p. 121]{Bregman}. 

Another difficulty in the comparison between auditory and visual objects is that vision is mostly concerned with reflected light from surfaces, whereas hearing is concerned with the source itself---not with its reflections \citep{Kubovy2001}. This neat distinction has been disrupted due to technological shifts over the last century that allowed for light sources to convey information directly (e.g., using traffic lights, or electronic displays) and for sounds to be reproduced using loudspeakers without a clear relationship to natural acoustic sources that are being reproduced. For some authors, these advances weakened the obvious differences between optical and acoustical objects, as there was little left in the acoustic source dimensions, shape, or other visible properties that could be described by listening, which undermines the very notion of acoustic object. Without a doubt, these difficulties have created some confusion in the field, which tends to focus on mental objects that are elusive and do not exactly correspond to the acoustic sources. 


\begin{table}
\footnotesize\sf\centering
\begin{tabular}{P{4cm}P{2.5cm}P{2.5cm}P{2.5cm}P{4cm}}
\textbf{Model}   & \textbf{Object}  & \textbf{Image}  & \textbf{Physiology}  & \textbf{Main features} \\
\hline
Standard vision and optics & External spatial objects, reflected light, spatial modulation envelopes & Scaled spatial modulation envelope, overlapping color channels & Image on retina & Accommodated focus according to distance; pupil control; binocular integration in cortex; reconstructed 3D shapes \\
\hline\hline
``Preperceptual image'' \citep{Massaro1970,Massaro1972b}; Auditory image \citep{McAdams1984}; \citet{Yost1991}; Auditory object in \citet{Griffiths} & Any sound stimulus, acoustic source & Undefined & Psychological; unspecified & Coherent input to perception prior to scene analysis \\
\hline
\citet{Simmons1980} & Spatial objects under echolocation & Reconstructed objects & Neural; unspecified & Specific to bats; requires comparison between emitted and received sounds \\
\hline
``Auditory Image Model'' \citep{Patterson1992}; ``Stabilized Auditory Image'' \citep{Lyon2018}; Auditory object in \citet{Bizley} & Temporal, periodic, broadband & Filtered, processed, autocorrelated & Neural, somewhere past the brainstem & Discrete; stabilized; dual time constant \\
\hline
\citet{Kemp2002} & Spatial acoustic & Traveling wave  & In cochlea & Image sharpness stems from filter sharpness; size is mapped to frequency\\
\hline
\citet{Carney2018} & Low-frequency envelopes & ``Fluctuation profile'' & Inferior colliculus & Dynamic range optimization using the efferents \\
\hline
``Sound image'' (audio) & Spatial distribution of sounds & The perceived spatial distribution of sounds & Unspecified & Binaural \\
\hline
``Temporal auditory image'' (This work) & Narrowband temporal envelopes and the entirety thereof that relate to the acoustic source & Scaled temporal envelopes in parallel frequency channels and the entirety thereof & Inferior colliculus & Cochlear dispersion; cochlear time lens; neural dispersion; neural aperture; blur and focus; aberrations; depth of field; accommodation; coherent and incoherent dual processing in the brainstem \\
\hline
\end{tabular}
\caption{Comparison of auditory imaging models and some of their key features. Conceptually similar models are grouped together.}
\label{tab:AudImagingModels}
\end{table}

\subsection{Discussion}
\label{AuditoryImageDiscusion}
From the above overviews about the acoustic object, its auditory image, and the resultant auditory object, it can be seen that definitions are both inconsistent among one another and are in many cases rather vague and even confusing. 

Perhaps the most problematic aspect of these concepts is that, much like the word ``sound'' itself, the term ``auditory object'' is used to refer both to the acoustical entity that evokes the auditory sensation, as well as its mental representation at the level of perception. The following quotation from \citet[p. 891]{Griffiths} illustrates the confusion between image and object aptly: ``\textit{Operationally, an auditory object might be defined as an acoustic experience that produces a two-dimensional image with frequency and time dimensions}.'' According to this definition, the ``acoustic experience'' assumes the role of the reflecting surface (the object) in vision. Is the auditory object equivalent to the auditory image? Moreover, is the resultant auditory object external or internal to the perceiver? According to many perception models it is both, giving a hint of the dreaded \term{homunculus fallacy}\footnote{The homunculus (``a little man'', in Latin) fallacy suggests that a recursive chain of images is formed in the brain---each one is the object of a successive observer, ad absurdum \citep{Atteneave,Dennet,Nizami2017}.}$^,$\footnote{A remnant of the homunculus fallacy may be inferred from the very choice of terminology that associates a mentally perceived entity---the visual or auditory object---with the word ``object''. We generally assign objecthood and objectivity to elements in the reality that is external to us and does not depend on our knowledge, whereas ``subjective'' are exactly these mental entities that are perceived or thought within the mind. Evidently, this oxymoronic reversal of the meanings of objective/subjective has already happened around the mid-18th century, whereupon these words had received the meanings that are in modern use \citep{Daston1994}.}. But this is clearly an unhelpful conclusion if we aim at constructing a physical understanding of hearing with tractable cause and effect.

The problem of associating the acoustic source with objecthood is that an object is defined from the answer to the question---What do we perceive as an auditory whole (i.e., a single and coherent percept) \citep[e.g.,][]{Nudds2010}? This is akin to reverse-engineering auditory perception in an all-inclusive manner, where phantom perception from within the system (e.g., tinnitus) produces messy answers. These approaches prescribe that perception becomes intermingled with physics and that objects cannot exist independently of the perceiver's brain. It is not the same as asking---What kind of acoustic radiation can be sensed by the auditory system?---which was answered in \cref{AcousticSource} based on the physical system only. The difference between these two questions may have been a critical motivation for studying auditory scene analysis, wherein a scene is a collection of sound sources that are nevertheless perceived as separate entities by the listener. The inability to disentangle the two---the external sounds from the perceptual experience that they evoke---is a common thread in the early development of acoustics as a science, whose only method of observing (and thus measuring) sound was through listening \citep{Hunt1992}. It is also a source for a centuries-long debate in the philosophy of perception between the so-called \term{naive realistic view}, which states that we directly perceive sensory inputs, and \term{indirect realism}, which states that perception is forever indirect and therefore we can never directly access the world through our senses \citep[e.g.][]{Searle2015}. 

The auditory image is also riddled with problems. First, all auditory image models appear to have been developed without a direct anatomical analogy to the eye's image or its mathematical and physical imaging principles as are known in optics (see \cref{ChapterImaging}). This also means that the acoustical object that is imaged in each model is not always well-defined (with the exception of the echolocation imaging model; \cref{TheAuditoryImage}). Second, and arguably a consequence of the first, some of the most critical concepts in optical imaging---e.g., focus, sharpness, blur, depth of field, aberrations---do not have meaningful correlates with these putative auditory images. These fundamental and intuitive concepts in optics, which have major implications in vision (e.g., focus accommodation, refractive errors of vision), are completely foreign to hearing science and remain inaccessible even with the available auditory image models. 

~\\

While it is arguable whether vision theory is doing significantly better than auditory theory, it is undeniably better with respect to the visual periphery---the eye---its function, its optics, and its mechanics. Roughly, the eye produces an optical image of an object at a distance. The image that appears on the retina can automatically be made sharp by focusing the lens using accommodation, as well as by appropriately controlling the level of light by closing and opening the pupil. Hearing models that drew parallels to vision sometimes hypothesized the existence of an auditory image and/or an auditory object, but ignored all the other elements of visual imaging: the lens, its focus, the degree of image sharpness, accommodation, and the pupil---none of which have obvious analogs in hearing. 

The common invocation in hearing of Marr's highly influential three levels of analysis (\cref{ElementsHearing}) is telling, because his theory takes for granted the sharp optical image that is formed on the retina, which serves as an input to post-retinal central information processing in vision. While nowhere stated explicitly, it is implied that the optics of the eye does not process information and does not execute any algorithm. This mistake is relatively inconsequential in vision\footnote{Marr's theory can be quite easily extended to include analog computation performed by the periphery. However, this would demonopolize the brain (and neurons in general) from being the sole information processor of the animal---something that is not currently discussed in biology and neuroscience, to the best knowledge of the author.}, but it can be misleading in hearing, because if an auditory image exists anywhere, then it is most likely concealed within the auditory pathways in the brain and not in the periphery (\cref{RigorousAnalogies}). Alternatively, the eye can be recast as an analog computer, whose goal is to create an image, which can be understood as a solution to the imaging equations for light waves, using the mechano-optical periphery of the eye\footnote{The following passage about analog computation captures this idea almost fully \citep[p. 27]{Maclennan2007}: ``\textit{...we may define computation as a physical process the purpose of which is the abstract manipulation of abstract objects (i.e., information processing); this definition applies to analog, digital, and hybrid computation... Therefore, to determine if a natural system is computational we need to look to its purpose or function within the context of the living system of which it is a part. One test of whether its function is the abstract manipulation of abstract objects is to ask whether it could still fulfill its function if realized by different physical processes...}''}. 

In vision, ``object'' is an overloaded term. It refers to the optical object that is positioned in front of the lens and is projected as an upside-down image on the retina. It also refers to the visual object that is experienced in perception as a result of the image sensation. According to the philosopher Thomas Reid \citep[quoted in][]{Duggan1960}: ``\textit{Sensation is a name given by philosophers to an act of the mind which may be distinguished from all others by this, that it hath no object distinct from the act itself.}'' For Reid, it is necessary to attend to the sensation in order for its output to turn into an object\footnote{Other common definitions for sensation and perception tend to be somewhat circular. For example, in \citet[p. 415]{Goldstein2014}, sensations are defined as: ``\textit{Elementary elements that, according to the structuralists, combine to create perceptions,}'' whereas perception is ``conscious sensory experience'' (p. 412). Definitions in \citet[pp. 140--141]{Mather2011} are slightly more helpful---sensation is ``\textit{An elementary experience evoked by stimulation of a sense organ, such as brightness, loudness, or saltiness,}'', whereas perception is ``\textit{A complex, meaningful experience of an external event or object, created from a combination of many different sensations.}'' According to Merriam-Webster dictionary, sensation is ``\textit{a mental process (such as seeing, hearing, or smelling) resulting from the immediate external stimulation of a sense organ often as distinguished from a conscious awareness of the sensory process.}'' And perception is ``\textit{awareness of the elements of environment through physical sensation.}'' Perhaps it is not a coincidence that the two terms are often used interchangeably in literature.}. But optics does not really care about attention or intent---the image exists by virtue of the illuminated optical object, the lens, and the screen. Thus, the image of the eye is primarily a product of sensation. Why should the auditory image be any different from vision? Why should the auditory object be dependent on the intent of the listener? We would like to break the circular logic of indirect perception theory by physically relating to the acoustic object and the auditory image as components of sensation, even if they are formed neurally---well within the brainstem or midbrain. This is the key for the present analysis.

\subsection{A note about subsequent auditory imaging terminology}
Given the above review and discussion, a note about the terminology that is going to be used in this work would be in place. In the literature, different concepts are encountered that variably relate to objects and images such as acoustic source, acoustic object, sound source, sound object, auditory object, auditory image, and sound image. Sometimes the object is material, whereas in other cases it is strictly visual and does not apply to hearing, and in modern usage it is often perceptual. It is a similar case with the image, which is sometimes taken to exist outside of the listener, or inside as a neural or mental representation. In this work, unless referring to specific jargon of another work, the adjective \term{auditory} is reserved for signals, stimuli, or entities that are observable within the animal's hearing system, especially within its neural pathways. External sources and objects are invariably considered \term{acoustic}. Similarly, in reference to vision, external objects and the retinal image are \term{optical} and all internal representations are \term{visual}. Although the word ``physical'' is often useful in this context too---to describe the acoustic or optical objects---perceptual images or objects are physical in the sense that all information has to be manifested physically, even if in neurally encoded form. Thus, the term \term{material object} is preferred to describe the physical object that is sensed by the animal. 

\section{Rigorous analogies between hearing and vision}
\label{RigorousAnalogies}
In the previous sections, two weak aspects of auditory theory were highlighted. One aspect was the relative opaque role of some of the main auditory areas in the brain---mainly the brainstem. While it is known to be critical in extracting all sorts of low-level auditory cues such as used for localization, it does not have a well-defined function that can be communicated in simple terms, as part of a modular system. The second aspect was the unsatisfactory importation of the object and image concepts into hearing. The main concern in this work is the latter aspect, but through the development of the idea of hearing as an imaging system, several ideas will be explored concerning new hypothesized roles of the auditory brainstem and midbrain as well. 

There are at least five different perspectives that motivate the temporal imaging theory that is at the heart of this work: the prominence of direct versus reflected radiation, anatomy and physiology, imaging mathematics, information and communication, and coherence. These perspectives overlap and complement one another and should not be taken as independent. Except for the anatomical perspective that is argued for in more depth, the other perspectives are presented qualitatively and briefly, and more rigorous derivations will be left for their respective chapters. 

\subsection{The prominence of direct versus reflected radiation}
\label{DirectProminence}
Among the senses, vision, hearing, and touch are specifically geared to deal with wave stimuli. In the case of touch, low-frequency vibrations ($<$ 500 Hz) from an object require direct contact with the skin (at least at typical amplitudes)  \citep{Bolanowski1988}, which suggests conduction rather than radiation. Hearing and vision are unique in that the radiated waves propagate in the three-dimensional space around the animal, where reflections tend to be rampant---mainly from large objects and boundaries (which are also objects). Perceptually, however, the two senses treat the direct and reflected energy of the radiating source in completely different manners. The visual image is largely based on reflections of the light source, which is not particularly interesting as an object in its own right. In contradistinction, hearing is primarily interested in the source itself and much less in the reflections. And yet, just as the visual image reveals something about the source of light, which was not imaged directly (e.g., the location of the source, its color temperature and identity, its uniformity, its power), so does the auditory image contains some information about the reflecting enclosure that is not directly imaged (e.g., indoor or outdoor environment, room volume, the boundary materials, proximity to walls). In both hearing and vision, the supplementary information might be useful for the animal, but is considerably less accessible in perception than the primary object of interest. Why is this the case? 

\begin{figure} 
		\centering
		\includegraphics[width=.8\linewidth]{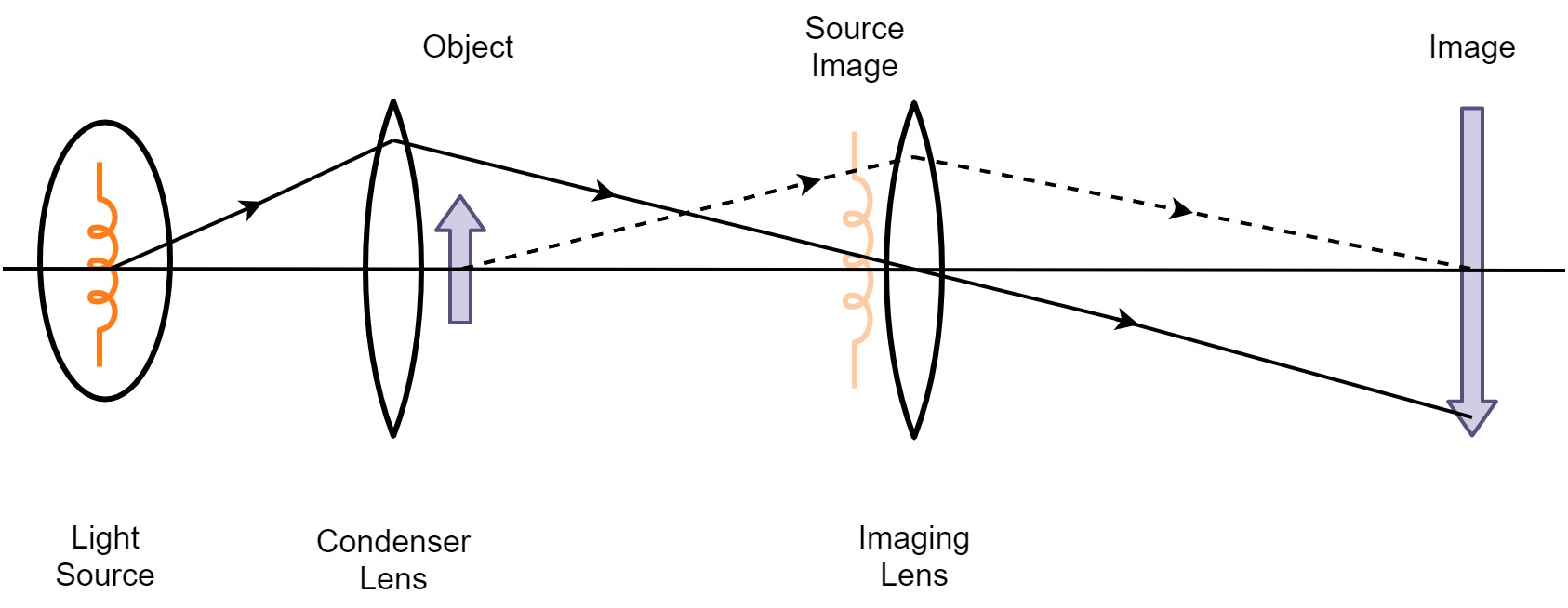}	
		\caption{A simplified diagram of K\"ohler illumination. The light source is placed behind a condenser lens that distributes the light uniformly on the object. The object is in sharp focus of the imaging lens on the image plane, whereas the light source is completely defocused at this point. The illustration is based on Figure 32 in \citet[p. 153]{Goodman1988Malacara}.}
		\label{KohlerIllum}
\end{figure}

To illustrate the difference between vision and hearing with respect to their preferential treatment to reflected or emitted radiation, it can be telling to compare these situations to the K\"ohler illumination system, which is the most prevalent illumination technique used in microscopy and other optical systems \citep{Kohler1893,Goodman1988Malacara}. The problem that this technique solves is how to use an external source of light to illuminate an object and produce its image on a screen or in an eyepiece, but without getting the features of the lamp geometry (the shadow of its filament) and light distribution mixed with the features of the object itself (see Figure \ref{KohlerIllum}). The way in which K\"ohler solved it was to defocus the lamp's filament, so that the light that illuminates the object arrives to it as a uniform beam. Effectively, this amounts to producing a very blurry image of the lamp, which does not reveal any of its undesirable features\footnote{In modern implementations, the first lens is usually supplemented with a diffuser that further decoheres the beam.}. Then, the uniform light is used to illuminate the object that is imaged in sharp focus by a second lens, which in turn projects the image on a screen (or inside the eyepiece). Now, if we compare this system to daylight vision, we see that the normally diffuse sunlight anyway does not disclose any easily observable features of the surface of the sun, nor its exact shape. Therefore, in daylight vision, the first lens of K\"ohler illumination is superfluous, whereas the lens of the eye assumes the role of the second lens in this system, which produces the sharp image of the object, but not of the sun.

Contrast this with hearing an object inside a room. For the sake of this presentation, we can assume that a source that is clearly heard may be considered to be in sharp focus. But the reflections from the surrounding surfaces, while they go on simultaneously with the source radiation, are generally blurry. Even if we want to discern them, we generally cannot, as numerous reflections mix together and lose their individual character. Not being able to hear these reflections means that we do not directly ``hear the walls'', although information about them is contained in the acoustic field. Why is it so different from the visual objects? How does hearing achieve this?

Part of the explanation lies in the different wavelength and frequency ranges that are associated with light and sound. Audible frequencies are low enough to be amenable to direct neural processing that has small time constants, which are suitable for analysis of the acoustic signal phase. As it turns out, the phase functions that are associated with the acoustic source and reflections are qualitatively different, in a way that will be quantified later using coherence theory (\cref{CoherenceTheory}). The auditory system can take advantage of this difference and accentuate it using transformations that are analogous to the optical ones from imaging theory. Effectively, the ear further defocuses partially coherent reflections so they do not come at the expense of the coherent object itself (\cref{AudImageFun}). The combined partially coherent image contains information about both the source and its environment. Once again, this is in contradistinction to vision, which is based exclusively on incoherent imaging, as natural light sources and most artificial lighting are incoherent sources too.

\subsection{Anatomy and physiology}
\label{AnaPhysioComp}
A fair comparison between hearing and vision should be anchored to anatomically or physiologically homologous structures of both sense organs. Let us try to identify what these structures are.

In the cochlea, each inner hair cell (IHC) is innervated by about 10 nerve fibers (in humans), which then project to the cochlear nucleus (CN) and on to the superior olivary complex (SOC), lateral lemniscus (LL), inferior colliculus (IC), medial geniculate body (MGB), and the primary auditory cortex (A1). Different signal pathways downstream from the CN exist, including via subnuclei that are not mentioned here, although nearly all of them synapse at the IC (see Figure \ref{BrainstemFig}). 

\begin{figure} 
		\centering
		\includegraphics[width=.7\linewidth]{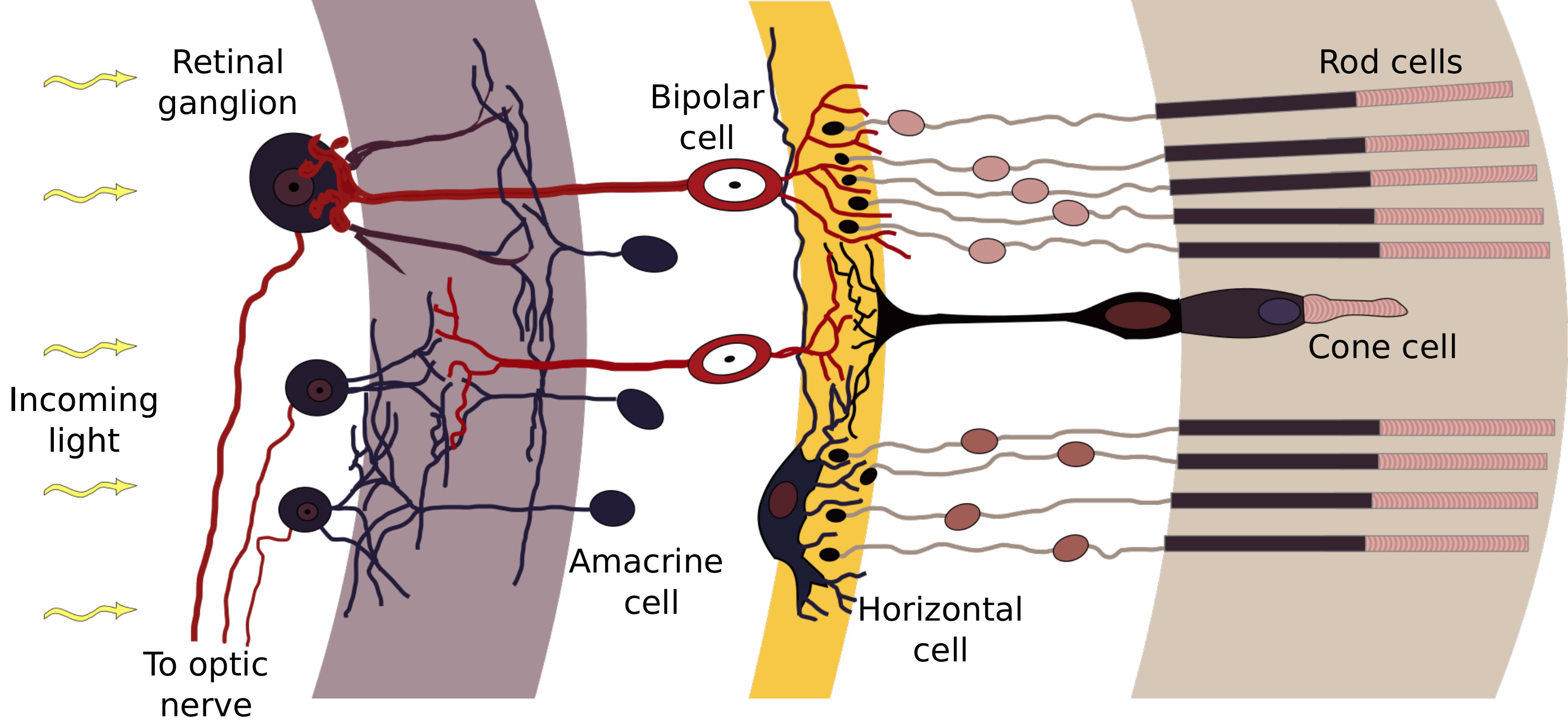}	
		\caption{A section of the convergent network of the human retina. Light arrives from the left and is detected by the photoreceptor layer on the right after traversing through the intermediate transparent cell layers. Additional processing then goes from right to left. Labels added to illustration by Anka Friedrich and Chris, whose original illustration was derived from Ramón y Cajal (1911), \url{https://en.wikipedia.org/wiki/Retina\#/media/File:Retina-diagram.svg}.}
		\label{RetinalLayers}
\end{figure}

In contrast to the ear, the photoreceptors of the retina are innervated by a network of neurons both vertically (leading to the retinal ganglion cells) and horizontally with interneurons \citep{Sterling2003Chalupa}. Depending on lighting conditions, photoreceptor type (cones or rods), and place on the retina, there are usually 1--3 neurons in the direct path between the photoreceptor and the ganglion cells (a combination of horizontal interneuron cells, bipolar cells, and amacrine interneuron cells; see Figure \ref{RetinalLayers}). The ganglion cells project primarily to the lateral geniculate body (LGB), and from there to the primary visual cortex (V1). In low-light conditions, rod photodetection is dominant, where many rods (120 on average) converge to a single ganglion cell. Considerably less cones (6 on average) converge to each ganglion cell in daylight conditions, with the least convergence taking place around the fovea at the center of the retinal field. 

\begin{figure} 
		\centering
		\includegraphics[width=1\linewidth]{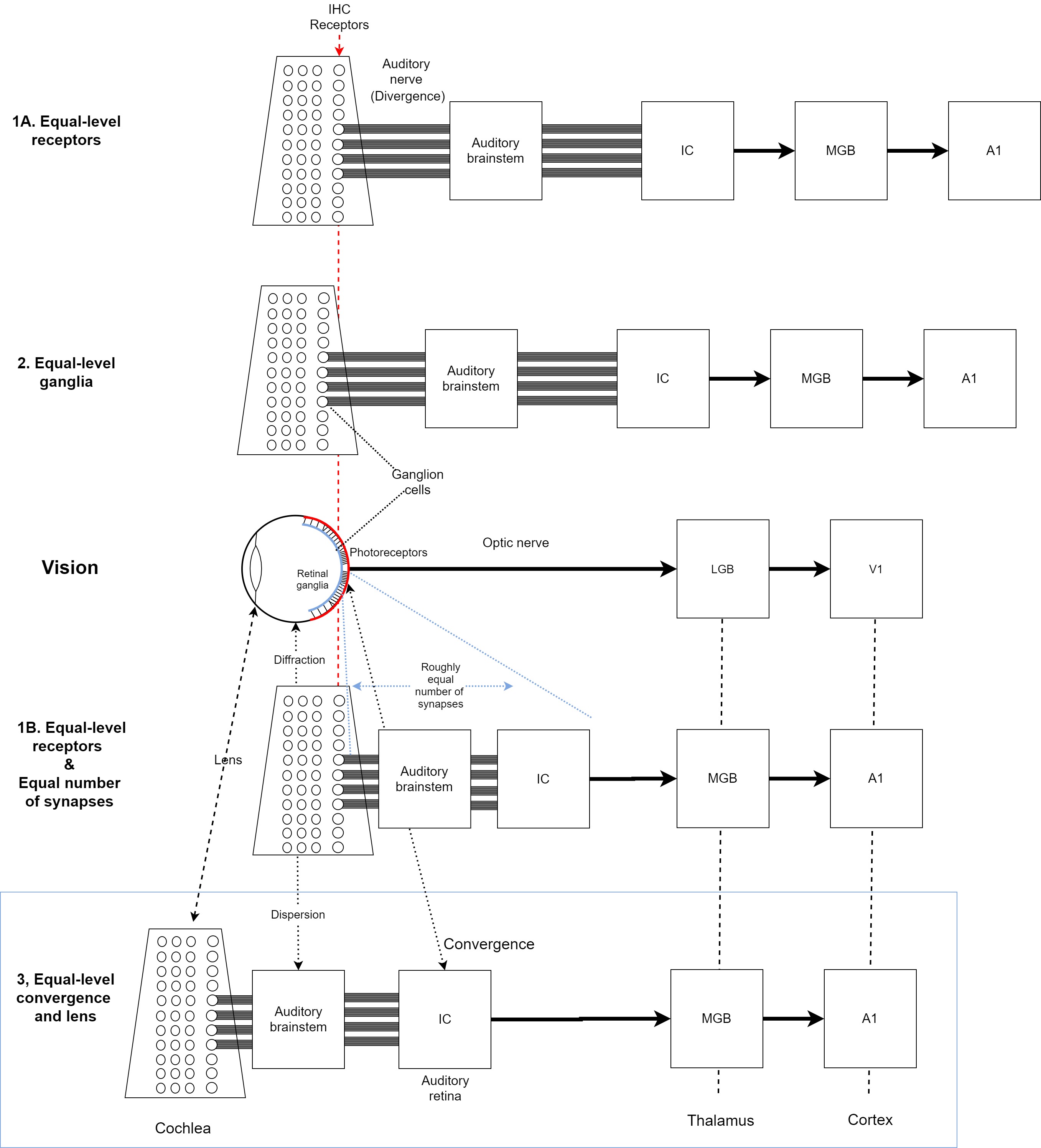}	
		\caption{Illustration of different anatomical and physiological analogies that are possible between the eye and ear. Four alignments of the auditory system are considered relative to the visual system, which is sketched in the middle row of the plot. If brain areas of the two systems are vertically aligned, it indicates that they are on equal-processing level. In areas that are too crowded in the image, dashed lines indicate alignment. Note that the photoreceptors are located on the right of the retina (drawn in red), while the retinal ganglions are on the left (in blue). See text for further details.}
		\label{EyeEarComps}
\end{figure}

Four possible comparisons between the eye and the ear are considered, which are illustrated in Figure \ref{EyeEarComps} and are explained below. 

\subsubsection{Equal-level receptors}
The naive way to compare the two systems is by setting the sensory receptors of the ear---the IHCs---and the photoreceptors of the eye---cones or rods---on comparable levels (Alignment 1A in Figure \ref{EyeEarComps}).

According to this comparison, the auditory brainstem should be at a comparable processing level to the retinal neurons \citep[e.g.,][]{Sitko2021}. However, the brainstem contains more synapses and has a completely different architecture than the different neural networks possible in the retina. One interpretation for this disparity is that the auditory signal has gone through more processing than the visual one by the time they reach their respective cortices \citep{Nelken2003, King2009}. This account is unattractive from a system-design perspective, because cortical theory would be more parsimonious if the functions of the cortex and thalamus are relatively consistent regardless of modality, unlike what is suggested by this comparison. For example, it has become increasingly clear that the cortex is highly plastic and areas that were once thought to be dedicated to one function can be repurposed by another, given the right circumstances. So, the areas associated with the auditory cortex in deaf people (and animals) may be used for visual processing \citep{Kral2007}. Therefore, on some level of processing abstraction, different modalities may be processed as equals, either due to the stimuli or due to the circuitries that process them \citep[e.g.,][]{Handel2006,Sievers2021}, which makes this comparison unconvincing.

A variation of this comparison (Alignment 1B in Figure \ref{EyeEarComps}) is that the retina and auditory brainstem contain a comparable number of synapses, which is just enough to have A1 and V1 on analogous processing levels \citep{Rauschecker2015}. However, the architectural differences between the retina and the brainstem are vast and their functions and complexity do not obviously overlap. Probably the most damning difference between the two networks is that the photoreceptors \textbf{converge} to fewer ganglion cells, whereas the IHCs \textbf{diverge} to many more auditory nerve fibers than there are IHCs. It suggests that the retina is constructed so to reduce the amount of information that reaches the eye before relaying it to the brain\footnote{In low-light conditions, the convergence of rod inputs optimizes for low signal-to-noise ratio conditions, so that photon-activated rods are enhanced, while the noise from adjacent rods is inhibited. In daylight conditions, the cones are configured to avoid saturation, which also entails information overload \citep{Sterling2003Chalupa}.}, while the auditory periphery is designed to conserve as much information as possible before central processing commences. 

Therefore, comparing the hearing and vision by setting their receptors at equal levels seems misguided. 

\subsubsection{Equal-level ganglion cells}
Another variation of the previous comparison is that the peripheral nerve cells of the ear (the spiral ganglion cells) and of the eye (the retinal ganglion cells) are set on comparable levels (Alignment 2 in Figure \ref{EyeEarComps}). This alignment suffers from unequal processing levels of A1 and V1 and it cannot deal with the convergence/divergence asymmetry, as it places the auditory and optic nerves on equal levels. The former is still composed of many more fibers than IHCs that feed into the auditory brainstem. The optic nerve contains less fibers than photoreceptors and projects primarily to the LGB of the thalamus. It is usually noted that although both are counted as cranial nerves, the optic nerve is, in fact, part of the central nervous system, whereas the auditory nerve is part of the peripheral nervous system. Thus, this comparison suggests that by the time that the two signals reach the cortex, considerably more neural processing of the auditory signal has taken place than of the visual one, which once again implies a processing disparity. Therefore, this comparison seems misguided as well.

\subsubsection{Equal-level cortex and thalamus}
The last possible comparison aligns A1 and V1 of the cortex and the LGB and MGB of the thalamus (Alignment 3 in Figure \ref{EyeEarComps}). Aligning A1 and V1 makes sense due to the many perceptual analogies that were found between hearing and vision. This, in turn, automatically aligns the auditory and visual thalamic levels as well. The photoreceptors and the retinal neurons should be then aligned against the IHCs, the auditory nerve, and the auditory brainstem and midbrain nuclei. The major point of connection to the auditory thalamus in the MGB is the IC, which should then be comparable to the retinal ganglion cells\footnote{An anatomical equivalence between the IC and the retina was noted by \citet{Carney2018}. A functional equivalence was indirectly implied as well, although instead of an optical image, the IC deals with ``fluctuation profiles''.}. However, to avoid the convergence/divergence asymmetry, rather than comparing individual neuron layers, it is more sensible to place the retina as a whole against the IC, which is the only auditory nucleus with considerable and unmistakable convergence (Figure \ref{BrainstemFig}). 

This solution sets the most peripheral parts of the eye---the lens and the vitreous humor (between the lens and the retina; Figure \ref{TheEye})---on an equal level to the auditory periphery and brainstem. It suggests that part of the early hearing function is better achieved in the neural domain than in the analog domain. But once processing moves to the neural domain, it is subjected to the same neural reality that other parts of the brain are subjected to, as various auditory circuits can be excited, inhibited, or neuromodulated by other circuits. These diverse functions may be contrasted with visual accommodation, which provides a neuro-mechanical modulation of the peripheral visual function---a kind of preprocessing that is applied before the optical signal reaches the retina. 

This anatomical comparison between the eye and the ear blurs the traditional distinction between peripheral and central processing. Or rather, if hearing requires neural processing to achieve a function that is achieved in the ``analog'' visual domain, then the border between auditory sensation and perception arguably becomes less obvious.

~\\

A cartoon comparison of the anatomical levels of the auditory and visual systems up to the image level is given in Figure \ref{AudVisSystems}. A more thorough overview of the auditory anatomy is given in \cref{EarAnatomy} and arguments for why the IC is the most adequate organ for the auditory image are provided in \cref{paramestimate}.

\subsection{Imaging theory}
The eye is an optical imaging system. As imaging theory in optics has been thoroughly studied at different levels of abstraction and given that it has several parallels in hearing, it can aid us in refining the concept of auditory imaging. 

A perfect optical image is a linearly scaled light pattern of (the projection of) an object (\cref{spatialsimaging}). When the object is illuminated and placed in front of a lens and a screen is placed behind it, its image is obtained from a simple geometrical law that relates the curvature of the lens to the distances between the lens and the object and the screen. A critical condition for the image to remain linearly scaled (i.e., uniformly magnified or demagnified) is that the object should subtend only small angles from the imaginary axis that connects the object and the lens centers. 

Applying more rigorous wave physics, spatial imaging can be also shown to be a combination of three processes---a diffraction, a lens curvature operation, and another diffraction. Mathematically, the three can be expressed as  quadratic phase transformations, which cancel out when the imaging system is in sharp focus. The image itself is then understood as an intensity pattern of a spatially modulated light source of a much higher carrier frequency.  Typically in the optical analysis, light is taken to be monochromatic---a constant frequency, effectively like a pure tone in acoustics, or a fixed carrier in communication---so the changes in intensity relate only to brightness, but not to color, perceptually. In human color vision, three narrowband channels of the cone photoreceptors are normally available, which can be mapped to three monochromatic images. Each image is a demodulated version of the light intensity pattern, in which the high-frequency carrier is discarded. A polychromatic (color) image may then be expressed as the combination (i.e., an incoherent sum) of different monochromatic images within its frequency range. 

The mathematical analogy here requires us to use the space-time duality, which was the perceptual analogy found between the spatial dimensions of vision and the temporal dimension of hearing (\cref{HearingTheoryVision}). Originally, the mathematical space-time duality was discovered in nonlinear optics by Akhmanov \citep{Akhmanov1968, Akhmanov1969}, who obtained a formulation to the wave equation that is in every way analogous to the wave equation used in imaging optics, only with interchanged time and space coordinates. To use this analogy, we retain the monochromatic carrier, but apply the modulations in the time domain as instantaneous frequency variations around the carrier, instead of spatial frequency modulations as in standard optics. Mathematically, we lose the two spatial dimensions of the image, which were previously applied to the spatial frequency, leaving intact only the propagation coordinate of the plane wave. Thus, the quadratic phase transformations no longer describe changes in the spatial object, but rather temporal changes to a pulse, which correspond to amplitude and frequency modulation and to the effect of group-velocity dispersion\footnote{When the velocity of the wave depends on frequency, then the medium is considered dispersive (\cref{PhysicalWaves}). If the group velocity is also dependent on frequency, then the effect is of group-velocity dispersion (\cref{ParatonalApprox}). It is equivalent to talk about group-delay dispersion instead, which relates to the same physics.}, such as exists in the cochlea itself. 

\begin{figure} 
		\centering
		\includegraphics[width=0.8\linewidth]{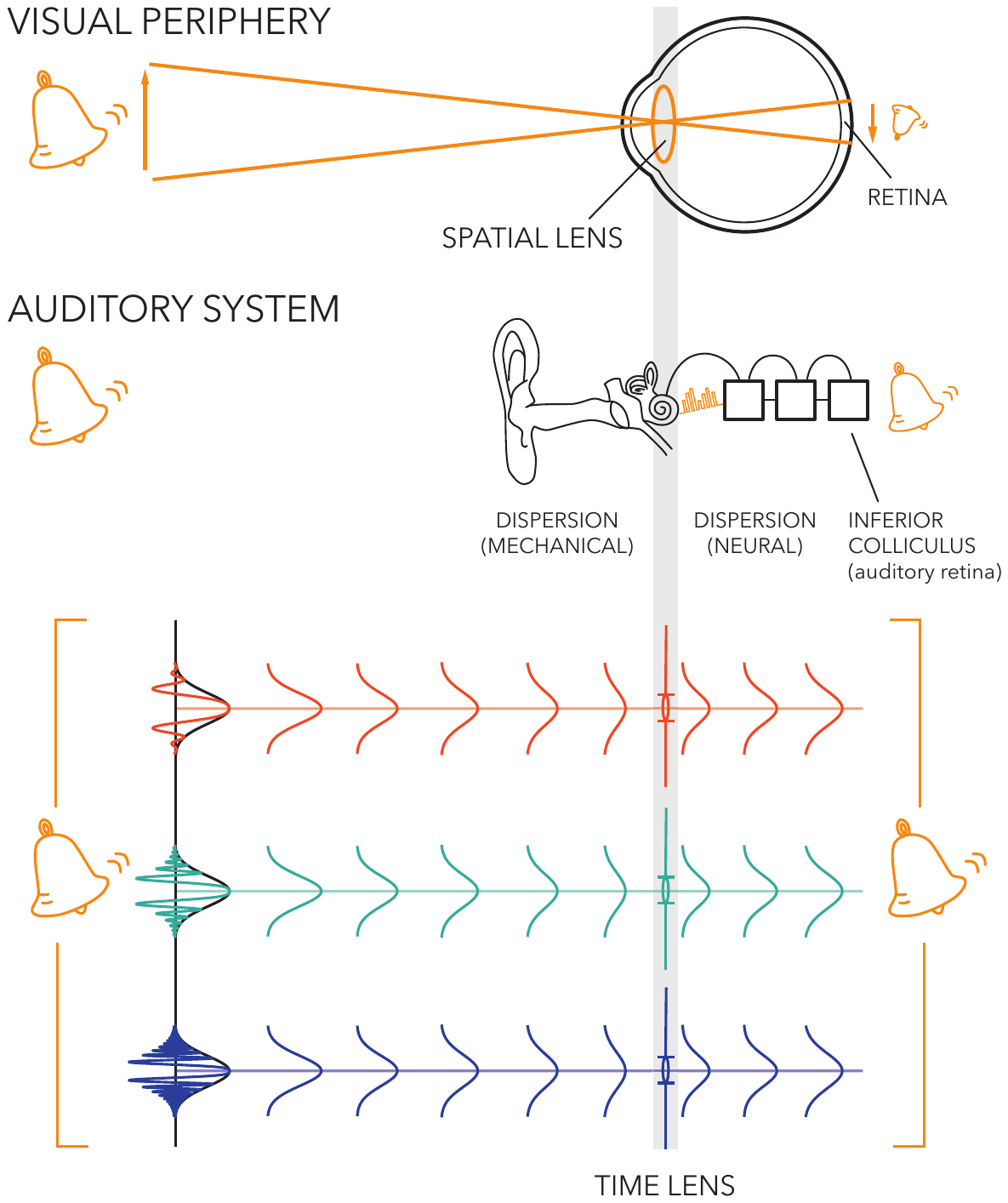}	
		\caption{Cartoon comparison between auditory temporal and visual spatial imaging. A time lens resides in the cochlea and is analogous in processing level to the crystalline lens of the eye. The auditory image appears in the inferior colliculus, which is analogous in level to the retina. Unlike vision, information from the carrier phase may be conserved in addition to the envelope information that is used in both senses. In both cases, the image is a combination of the monochromatic images from different frequency channels---either tone or color. Original illustration by Jody Ghani (2020).}
		\label{AudVisSystems}
\end{figure}

In temporal imaging we also resort to narrowband channels that are associated with a monochromatic carrier, which  in itself is not used directly in the imaging calculation. The temporal object is the envelope of a pulse input, which relates to the finite range of spectral and temporal variations that can be captured by each narrowband filter. Using the auditory filters of the cochlea as narrowband channels, each one can linearly handle small temporal modulations relatively to the pulse center, which is analogous to the small-angle condition in spatial imaging. The existence of a lens in the auditory system will be explored in depth in \cref{lenscurve}. The image of the pulse is obtained after additional processing beyond the lens, where demodulation may take place as well. It should be mentioned that unlike the eye, which produces a demagnified image, we know of no scaling that takes place in the auditory system. It suggests that the system magnification may be very close to unity. An illustration of the auditory and the visual system analogous parts and functions is given in Figure \ref{AudVisSystems}.

Dealing with sound pulses as images means that the input is composed of short samples that have to be integrated to be perceived continuously. This is in analogy to the photoreceptors that spatially sample the entire area of the retina, whose exact distribution sets the limits on the maximum spatial frequency that can be detected by the eye. 

The analogy between spatial modulation in optics and temporal modulation in psychological acoustics motivated \citet{Houtgast1973} to adopt the modulation transfer function in acoustics, which directly impacts speech reception in reverberation \citep{Houtgast1985}. More directly related to hearing, \citet[pp. 544 and 565]{Joris2004} suggested, in passing, that the spectral contents of the temporal envelope of sound may be analogous to the spatial frequencies in vision. 

While the mathematical space-time duality is very robust, fully motivating it may seem far-fetched without additional theory that establishes the concepts of group-delay dispersion, time lens, aperture, and coherence, in a way that is  applicable to acoustics and hearing. The relevant theory will be developed over the next chapters and the components of the auditory imaging system itself will be explored from \cref{temporaltheory} onwards. 

\subsection{Information and communication}
It is common to ascribe information processing to the brain function, especially with respect to perception of information-bearing signals and stimuli that are detected through sensory channels \citep[e.g.,][]{Sterling2015}. Such analyses tend to examine the receiver's side of the communication chain (comprising a source, a channel, and a receiver), which neglects the physical transfer of information from the source, through the environment, and into the sensory receptors of the animal. However, actual transmission of information depends on engineering a communication system that is shared between the source and receiver over a physical channel\footnote{This communication framework is found in Claude Shannon's seminal work that was originally named ``A mathematical theory of communication'' \citep{Shannon1948}, but presents a system that is completely abstracted from the nitty-gritty mechanics of generating the communication signals in reality. The universal principles of his theory, which is now referred to as \term{information theory}, are integrated into the somewhat more mundane communication engineering, which deals with the mathematical principles that are embedded in the electronic system designs that actually deliver the information. Both sciences are tightly related and both contain universal principles that are not limited to a specific implementation in hardware, software, or the choice of physical medium and transmission energy. See \cref{InfoTransfer} for further details.}. 

Any kind of communication depends on the ability to physically transmit and receive waves of an arbitrary type that are manipulated to carry messages. Communication signals can be formulated, without loss of generality, as a product of a high-frequency carrier and a slow-varying complex envelope, which modulates the carrier. The message is generally taken to be contained in the complex envelope of the signal rather than in the carrier itself, and is then referred to as baseband signal. In communication, the goal of the receiver is to recover the message from the received signal with minimum loss of information or distortion. This process is done by demodulation, which entails the separation of the envelope and discarding of the carrier. All long-range communication is physically realized by modulating well-defined channels with known frequency bandwidth. If the channels are narrowband (their bandwidth is much smaller than the carrier frequency), then it is possible to unambiguously recover the message\footnote{It should be understood that modulation and demodulation happen below the coding level that is familiar from information theory, even though some complex digital modulation techniques appear like codes in and of themselves. Therefore, possible message encoding can only take place before modulation, while decoding must take place after demodulation (see \cref{InfoTransfer}).}. 

There are cases in which the amount of information in the message requires a large bandwidth that cannot be fitted in the narrowband channel without breaching it, which would result in distortion and loss of information. The simplest solution is to employ a higher-frequency carrier and use a channel that has a relatively narrow but absolutely wide bandwidth. In practice, this is not always possible, if such high frequencies are not available for different reasons, such as prohibitive energetic cost of generating the carrier, interference from other communication or background radiation that occupies the same bandwidth, high absorption by the medium, strong interaction with objects in the environment, etc. If this solution is ruled out, communication cannot be considered narrowband. The main difficulty arises because wideband signals cannot be expressed in a mathematically unique form, so the concept of slowly-varying envelope cannot be well-defined for them. Therefore, receiving wideband signals may give rise to ambiguity in the demodulation and message recovery, unless a system is devised to disambiguate the received signals and, hence, the messages. Technically, it means that the received message may suffer some distortion. Hearing overcomes this problem in part by dividing its bandwidth to narrowband channels, which overlap to such an extent that there can easily be some redundancy in their detection that may be used to minimize ambiguity. While the broadband signal that is recovered from such a configuration may not be a faithful reconstruction of the original input (should this be its goal), it is possible that its informational content may remain largely intact. 

There are two general categories of signal modulation detection. In noncoherent detection, only amplitude information is demodulated and the phase is ignored. In coherent detection, both the phase and the amplitude of the complex envelope are demodulated, which is essential in frequency- and phase-modulation techniques. In order to coherently detect the modulation phase, it is necessary to track the signal phase, since its carrier frequency tends to drift in transmission. The most common electronic circuit used to facilitate coherent detection in various engineering applications is the phase-locked loop (PLL), whose output is literally locked on to the carrier phase, so that no information is lost. In general, modulation detection is a well-studied feature of hearing, which behaves as if both coherent and noncoherent detections may be employed in different situations, as will be discussed throughout this work. Interestingly, phase locking is a robust feature of the auditory system, which is nevertheless limited to low-frequency carriers.

Evolutionarily, the hearing system long preceded auditory communication, at least in its verbal, human form. How did the hearing and communication functions converge? Acoustic objects and the cavities they are often coupled to are characterized by resonance frequencies that depend on their geometrical and physical properties. When objects oscillate, it is a result of them being forced into motion, using internal or external sources of energy. The forcing pattern itself can be thought of as modulation that shapes the vibration patterns of the object. For example, listening to a plucked guitar string, the pitch of the string is determined by its resonant frequencies, whereas the plucking is determined by the modulation, which has onset and offset temporal patterns, as well as input power. If, in addition to plucking, the guitar player also bends the string, it results in frequency modulation---a time-dependent change in pitch around its natural tuning. Both plucking and bending may be expressed as slow changes to the complex envelope of the string resonant frequencies. The constant-level frequency resonance is mathematically no different from a carrier. Each frequency may relate to its own channel, as long as it is analyzed independently of other resonant frequencies. Therefore, the mathematical basis for communication and acoustic source hearing has natural commonalities. Treating the guitar sound as a communication signal gives the listener the access to information both about the string itself, as well as to how it was forced into vibration. The two constitute two separate dimensions of information, which nevertheless tend to overlap in frequency and to interact in complex ways.

While the present work has been written with optical imaging theory as its main source of inspiration, communication theory became an indispensable element in many of its chapters. The most pertinent aspects to hearing of communication, information, and imaging theories are presented and contrasted in \cref{InfoTransfer}. The analytic signal and the envelope domain in the context of hearing are reviewed in \cref{PhysicalSignals}. The auditory phase locked loop is in \cref{PLLChapter}. Realistic acoustic sources are reviewed in \cref{InfoSourceChannel}. Aspects relating to the balance between noncoherent and coherent detection are discussed throughout Chapters \crefrange{accommodation}{GeneralDisc}.

\subsection{Coherence}
The concept of coherence was used above in three different contexts: as a feature of communication detection that preserves the signal phase, as a necessary quality of the complex auditory stimulus that endows it with objecthood, and as a property of the wave field that determines how certain information about the source propagates. These definitions and additional ones that are relevant to hearing are reviewed in depth in \cref{IntroCoh}, where a unifying definition is sought, and in \cref{CoherenceTheory} where the relevant coherence theory from physical optics is introduced in consistent manner. Source coherence and its acoustic and then physiological propagation are shown to be key to understanding auditory imaging. However, coherence goes beyond that, because the key to conserve coherence is in phase locking (\cref{PLLChapter}), which is a feature of synchronization in the brain that seems to have correlates that are more universally important than just in hearing. Brain synchronization to a stimulus is an indicator for attention and thus for the engagement of the animal with certain input channels and actions. While it is perhaps a speculation of a sort, on a high level of abstraction, coherence seems to be the currency of important signals in the brain, regardless of their modality. This is speculated to be the key to auditory accommodation in \cref{accommodation} and an overarching processing design motivation in the auditory system as a whole (\cref{GeneralDisc}).

\section{Conclusion}
We began the chapter with an overview of some of the milestones of hearing science that have attracted the most attention in research, but also emphasized some gaps in what appears to be a rather loose and fragmented theory. Then, we dwelt on how some commonalities and differences of hearing and vision have repeatedly attracted scholars to formulate hearing models using visual concepts. These ideas have also been used to elucidate how each modality is specialized within perception as a whole, with the recurrent observation that hearing is predominantly temporal whereas vision is predominantly spatial. Perhaps the most common of all visual concepts that have found their way to hearing are that of the object and the image. However, these terms have been inconsistently applied in hearing and have failed to retain the insight that they carry for vision. Subsequently, we presented five different perspectives that can be developed to obtain more insight into hearing theory, using concepts from vision, optics, communication, and wave physics.

Table \ref{tab:acousticvision} compares some basic parameters of hearing and vision in humans, which also relate to conclusions from the five specific comparisons above. Some parameters are well-known from literature, whereas others are based on the present work and are explained throughout. A similar (and shorter) parametric comparison is found in \citep[p. 24, Table 1.1]{Handel2006}.

\begin{figure}
		\centering
		\includegraphics[width=1\linewidth]{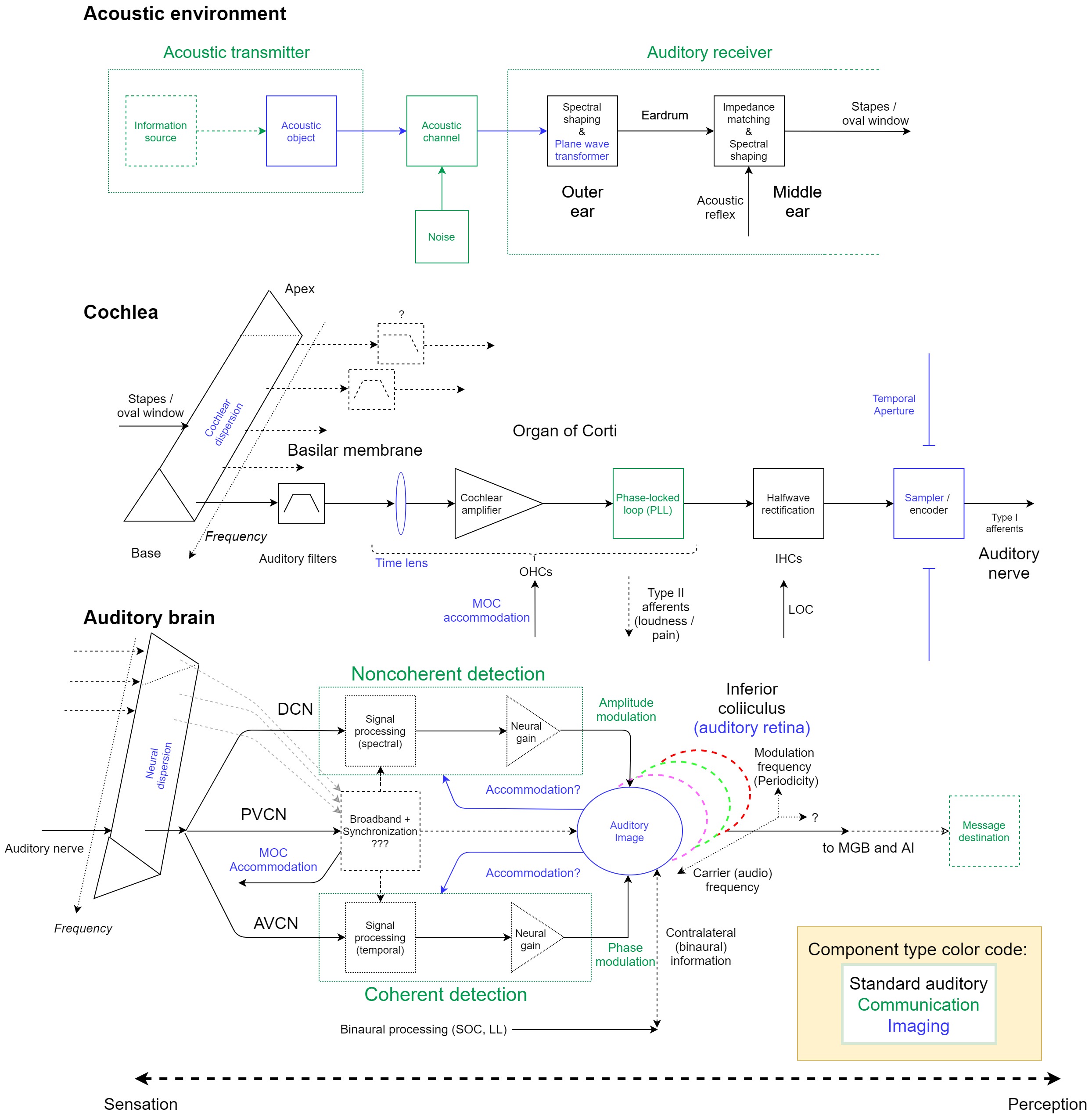}	
		\caption{A functional diagram of the monaural auditory system as theorized in this work. The model contains standard auditory elements (in black). Novel elements to the standard auditory system are shown in green (inspired by communication) or in blue (imaging). The new components will be considered throughout the text and the full model is revisited in \cref{GeneralDisc}.}
		\label{AuditoryCoherence0}
\end{figure}

The complete auditory system that will be explored in this work in parts is displayed in Figure \ref{AuditoryCoherence0} for reference, but will be explained in detail only in \cref{CompleteModel}.

\begin{table}
\footnotesize\sf\centering
\begin{tabular}{P{3.9cm}P{6.4cm}P{6.4cm}}
\hline
&\textbf{Hearing}&\textbf{Vision}\\
\hline\hline
\multicolumn{3}{c}{\textbf{Carrier}}\\
\hline
Energy & Acoustic (pressure) & Light (electromagnetic)\\
Typical speed & 343 m/s (in air at $20^\circ$C) & 299,792 km/s (vacuum)\\
Frequency range & 20--20000 Hz & 400--750 THz (nominal); 360--830 THz (maximum range)\\
Wavelengths in air & 17.15 m--1.71 cm  & 400--700 nm (nominal); 360--830 nm (maximum range)\\ 
Period & 50 ms--50 $\mu$s & $\approx 10^{-15}$ s\\
System bandwidth & 10 octaves & 0.9 octave \\
Channel bandwidth$^1$ & 5--30\% & 15--20\% \\
Modulation bandwidth & $\le$ 2000 Hz (broadband noise) & $\le$ 90 cycles / degree\\ 
Dynamic range & $>$ 120 dB & $10^{-6}-10^{8}$ cd/m$^2$ (140 dB)\\ 
\hline
\multicolumn{3}{c}{\textbf{Physical image}}\\
\hline
Type & Temporal & Spatial \\
Governing equation & Paratonal (dispersion) & Paraxial (diffraction) \\
Image field & Complex temporal amplitude envelope, $a(t)$ & Spatial intensity envelope, $I(x,y)$\\
Inverse domain & Complex spectral modulation envelope, $A(\omega)$ & Spatial frequency intensity envelope, $I(k_x,k_y)$\\
Main assumption & Paratonal---slowly varying envelope, narrowband, short aperture (time window) & Paraxial---small angles, monochromatic\\  
Spatial assumption & Plane waves & Paraxial / Gaussian beams\\
Typical invariance (still image) & Spatial & Temporal\\
Linearity & Complex amplitude (within channel), Intensity (across channel) & Intensity\\
Typical mode of imaging & Partially coherent & Incoherent \\
Primary objects imaged & Direct acoustic sources & Light reflecting objects\\
Sensory sampling rate & 50--2500 Hz & 25--75 Hz\\
\hline
\multicolumn{3}{c}{\textbf{Detector}}\\
\hline
Number of detectors & 2 & 2 \\
Channel frequencies & $\approx$ 3000--3500 & 3 colors (cones) + 1 intensity (rods) \\ 
Ability to move & No & Yes\\
Lensing & Cochlear time lens & Crystalline lens and cornea of the eye\\
Type of aperture stop & Temporal (neural, cochlear at low frequencies) & Spatial (pupil)\\
Image information medium & Neural & Optical \\
Image locus & Inferior colliculus & Retina \\
Sensitivity & quantum-limited (basilar membrane motion of the order of $10^{-12}$--$10^{-11}$ m) & quantum-limited $<1$--$7$ photons \\
Near point& N/A &25 cm\\
Far point& N/A &6 m\\
Means of accommodation & Lens curvature; phase locked loop gain; coherent to incoherent weighting (likely); level gain & Lens curvature (and pupil size and vergence)\\
Accommodation time & ? & Minimum reaction time 0.4 s and response time 0.6 s \citep{CharmanBass3}\\
\hline
\hline
\end{tabular}
\caption{Comparison of properties and attributes of human vision and hearing. $^1$Equivalent rectangular bandwidth of the auditory depends on the method used to measure frequency selectivity. The range in the table is based on minimum and maximum values from \citet{Glasberg1990,Shera2002}. A rough estimate of the full-width half maximum (FWHM) of the photoreceptor sensitivity curves was obtained from \citet[Figure 2.2a]{Hunt2004}. Relative channel bandwidth in both vision and hearing decreases with frequency.}
\label{tab:acousticvision}
\end{table}

\chapter{The anatomy and physiology of the mammalian ear}
\label{EarAnatomy}
\section{Introduction}
The ear structure is similar across all mammals and interspecies differences appear to be primarily morphological, relating to the size, shape, and layout of the different parts. The mammalian auditory periphery is subdivided into outer (or external) ear, middle ear, and inner (or internal) ear, which includes both the cochlea and the auditory nerve. Subsequently, the auditory nerve projects to the central nervous system, which comprises a number of auditory nuclei in the brainstem, midbrain, thalamus, and cortex. 

Much of the experimental and theoretical work in physiological hearing research has revolved around the cochlea, where the most significant signal transformations take place. Whatever kind of transformation or processing occurs in the cochlea, its effects can usually be measured downstream, in the auditory brain. However, since the cochlea is remarkably complex, concealed, and vulnerable, its exact function and micromechanics are not well understood. Other elements of the central part of the auditory system are even less well understood.

As the anatomy and physiology of the ear has been covered in numerous texts in great detail, it will not be covered here in depth. Instead, the emphasis in the following is on coarse-grained components and functions of the human auditory system that are relatively uncontroversial, up-to-date, and informative for a high-level analysis. Therefore, the typical description of the micromechanics of the cochlea or the cell types in the brain with their specific responses is usually avoided. Unless referenced otherwise, the review in \cref{Periphery} and \cref{CentralNeuroanatomy} is largely based on \citet{Pickles}. Additional presentations of the human ear's structure are found also in \citet{ReesPalmer2010a} and \citet{ReesPalmer2010b} and shorter introductions in \citet{Moller2012} and \citet{Gelfand2018}. 

When discussing the neurophysiology of the auditory brain, the lack of a coherent theory for it, and for hearing in general, becomes a stumbling block for the understanding of what this complex system does. Therefore, before reviewing the central auditory system, a discussion is provided in \cref{OrgPrinciples} that provides a few general guiding principles and theoretical ideas that have been commonly employed in auditory neuroscience research, as well as some of their critical shortcomings. The few theories about specific auditory nuclei are then briefly mentioned, where available, throughout the review in \cref{CentralNeuroanatomy}.

Since much of the knowledge about the human ear's anatomy has been gathered from animal data, it is essential to know how the auditory systems of these animal species differ. The final section (\cref{ComparativeHearing}) deals specifically with the comparative anatomy between mammals. It too is a selective and coarse-grained review of some of the notable differences between animals, given the commonalities covered in \cref{Periphery} and \cref{CentralNeuroanatomy}. Therefore, it is somewhat unlike typical comparative hearing texts that tend to focus on specific organs within a subset of animals and to review their similarities as well. This part of the review is based primarily on \citet{Rosowski1994Fay} for the outer and middle ears, \citet{Echteler1994Fay} for the cochlea, and \citet{Glendenning1998} for the central pathways. In-depth reviews of various topics in comparative hearing of mammals (and other vertebrates) can be found in volumes edited by \citet{Popper1980,Fay1994,Manley2000} and \citet{Koppl2014}.

\section{The peripheral ear}
\label{Periphery}
Many parts of the peripheral ear have at least two anatomical terms that are in regular use. Where available, we present both terms with the lesser used term in the present text given in parentheses. 

\subsection{The outer ear}
\begin{figure} 
		\centering
		\includegraphics[width=0.6\linewidth]{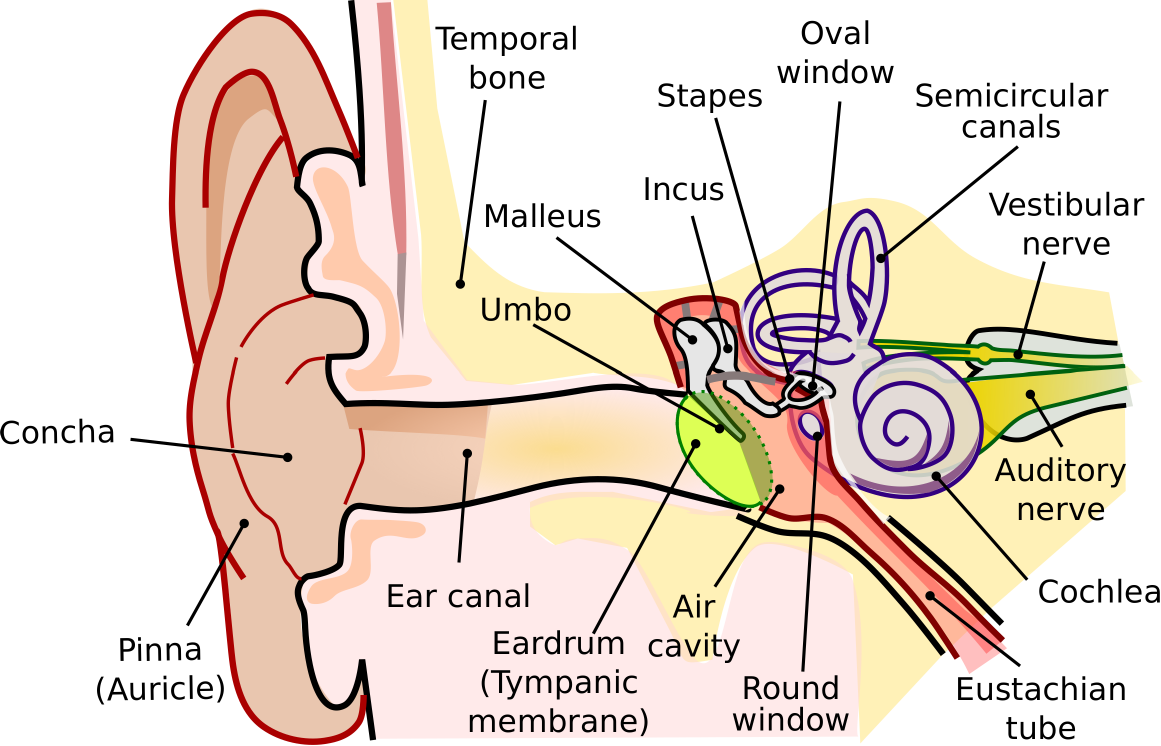}	
		\caption{The main parts of the peripheral ear (not in scale). The drawing is adapted from \url{https://commons.wikimedia.org/wiki/File:Anatomy_of_the_Human_Ear.svg}, which was redrawn after \citet{Chittka2005}.}
		\label{ExternalEar}
\end{figure}

Acoustic waves in the environment propagate from their source through a medium---either air or water---before entering the external ear, whereupon they begin a sequence of transformations. Sound is diffracted by the \term{pinna} (or \term{auricle})---the cartilaginous external part of the ear, which is connected to the \term{concha} that is shaped like an irregular horn (Figure \ref{ExternalEar}). The pinna and concha collect the impinging sound to efficiently radiate it into the ear.

The sound then passes through a hard-walled \term{ear canal} (or \term{auditory meatus}) that terminates with the \term{eardrum} (\term{tympanic membrane}), which is a thin membrane that vibrates with the sound pressure. The ear canal has a few resonances that emphasize certain frequencies (most prominently at around 2.5--4 kHz for humans), which are determined mainly by its length \citep{Wiener1946,Shaw1974,Mehrgardt1977}. The sound field that reaches the eardrum is one-dimensional, in very good approximation, especially at low and midrange frequencies, since the ear-canal diameter is much smaller than the wavelength at this range \citep{Rabbitt1988}. The pinna spectrally shapes the sound as well, but in a direction-dependent manner that has a role in localizing the elevation angle of the sound source (whereas the azimuth is detected using information both ears). This feature, along with its efficient sound power delivery, are thought to be the two main acoustic functions of the outer ear \citep{Rosowski1994Fay}.

\subsection{The middle ear}
\label{MiddleEar}
On the internal side of the eardrum, the vibrations in the air or water medium push a mechanical system made of three small bones, the \term{ossicles}---the \term{malleus} (hammer), the \term{incus} (anvil), and the \term{stapes} (stirrup)---which constitute together the \term{ossicular chain} of the middle ear (Figure \ref{ExternalEar}). The ossicles are positioned within an \term{air cavity}, whose pressure is regulated through the \term{eustachian tube}, which is connected to the pharynx at the back of the mouth. This anatomy was already described in great detail by \citetalias{Helmholtz} (pp. 129--135) along with a hypothetical model of operation that has been partially confirmed since. The malleus is attached to the stretched eardrum in the \term{umbo} at its center, whereas the stapes is attached via the \term{annular ligament} to another membrane---the \term{oval window}, which is part of the cochlear wall. The eardrum, umbo, and ossicles are coupled rigidly at low frequencies, but higher modes of vibration (e.g., bending of the ossicles) and other degrees of freedom in their movement (compression of the ossicular joint or the ligaments between the ossicles, or multimodal motion of the ear drum) may diminish the effectiveness of the coupling at higher frequencies. 

The outer and middle ears work together as a non-ideal (frequency-dependent) transformer between the acoustic impedance (the ratio between the pressure and volume velocity) of the air and that of the incompressible fluid of the cochlea. The middle ear also functions as a mechanical lever that transforms the acoustical power exerted on the eardrum to much larger vibrational power on the oval window, which is of a smaller area than the eardrum. In humans, the gain it provides has a bandpass filter characteristic with a maximum at 1200 Hz and approximately -6 and -7 dB per octave below and above, respectively \citep{Aibara2001}. However, the impedance is not necessarily resistive and complex relations between the pressure and velocity may exist. The vibration of the middle ear is linear with input level below 96 dB SPL and has negligible nonlinear distortion up to 130 dB SPL---well above the ecological sound range \citep{Guinan1967,Aerts2010}. The middle ear output is the displacement of the stapes that is coupled to the cochlea, which moves one-dimensionally like a piston below 6.7 kHz, but has more complex rocking motion at higher frequencies \citep{Aibara2001}. 

Two muscles are connected to the ossicles that control the \term{acoustic reflex}---the \term{tensor tympani} is connected to the malleus and the \term{stapedius} is connected to the stapes. Both appear to regulate the transmission gain of the middle ear, particularly below 1 kHz, by stiffening the ossicular movement, which causes an increase of acoustic impedance. The result may provide some protection against loud sounds and may act as automatic gain control. It is also thought to reduce self-generated sounds by the listeners and to selectively reduce low frequency sounds that can cause undesirable masking. However, the function of these reflexes is not well understood. 

The middle ear is a critical stage in hearing that provides necessary amplification in hearing. An impaired middle ear can lead to a \term{conductive hearing loss}. 

\subsection{The inner ear}
\label{TheInnerEar}
\subsubsection{The cochlea}
\begin{figure} 
		\centering
		\includegraphics[width=0.5\linewidth]{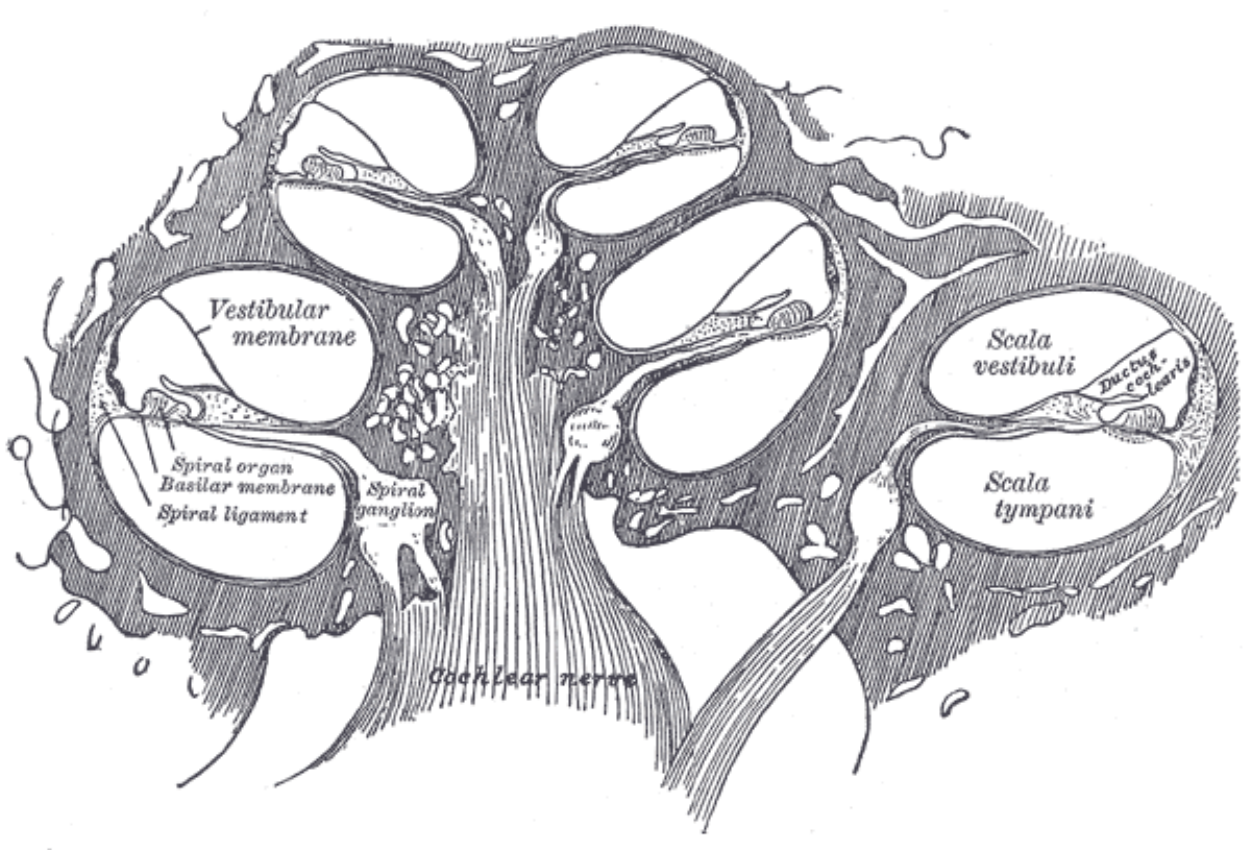}	
		\caption{A cross section of the in the human cochlea. The narrow sections are apical and the wide ones are basal, showing the typical 2.5--3 turns of the human cochlea \citep{Pietsch2017}. The helicotrema is found at the top of the cochlea (not shown). Image by Henry Gray from the ``Anatomy of the Human Body'' (1918), taken from \url{https://www.bartleby.com/107/illus928.html}.}
		\label{CochlearTurns}
\end{figure}

The cochlea is a spirally shaped structure that is located inside the anterior portion of the petrous region of the temporal bone, along with the semicircular canals of the vestibular system that occupy its posterior region \citep{Echteler1994Fay}. The coiled tube of the cochlea is longitudinally divided into three fluid-filled compartments, whose cross section is illustrated in Figure \ref{CochlearTurns}. The first compartment, the \term{scala vestibuli}, is where the oval window is located. The motion of the stapes presses the elastic oval window and thus the cochlear fluid, the \term{perilymph}, which is largely incompressible (with approximate fluid properties of water and composition similar to extracellular fluid). Scala vestibuli terminates in an opening called the \term{helicotrema}, where the compartment connects to \term{scala tympani}, which is another compartment that leads back through the entire length of the cochlea. The oval window displacement causes a corresponding displacement of the \term{round window}, at the other end of scala tympani. 

\begin{figure} 
		\centering
		\includegraphics[width=1\linewidth]{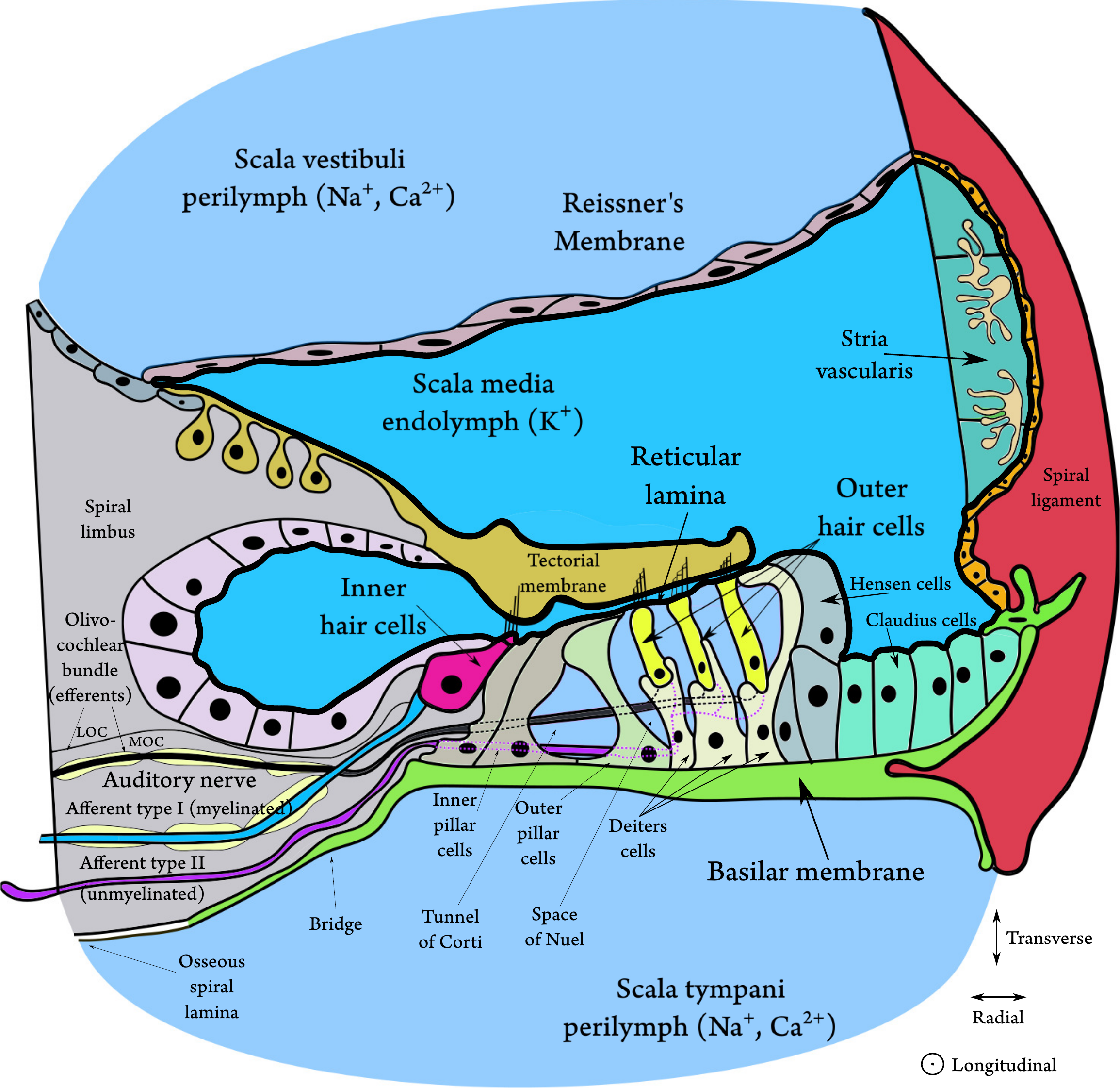}	
		\caption{A cross section of the organ of Corti, showing its most important elements. The figure is based on Figures 2.2, 2.3, 2.13, and 2.21 in \citet{SlepeckyDallos1996} and on Figure 3.5 in \citet{Pickles}. The thick black line surrounding the scala media and the endolymph represents the tight junctions that electrically separate it from the perilymph and mark the boundaries of the cochlear duct. New evidence has shown that the stereocilia of both hair cell types are deeply embedded in the tectorial membrane (TM) and that there is no gap between it and the reticular lamina without acoustic stimulation \citep{Hakizimana2021}. There is a gap of endolymph illustrated, as a halfway reminder of older observations that showed free standing IHCs under the TM. The bridge tissue is after \citet{Raufer2019}.}
		\label{OrganCorti}
\end{figure}

The two scalae are separated by elastic membranes, which tightly enclose on the \term{scala media} (or \term{cochlear duct})---another compartment that is not directly connected to the other two and is filled with \term{endolymph}, or \term{endocochlear fluid} (see Figure \ref{OrganCorti}). The endolymph is similar in properties to intracellular fluid that is rich with potassium and low in sodium ions that make the scala positively charged in reference to the perilymph by about about 80--100 mV. The supply of ions originates in the \term{stria vascularis}, on the wall of scala media, which has a capillary supply. The membrane below scala vestibuli is called \term{Reissner's membrane} and it does not have a direct role in the sound transduction chain, although it has been proposed that it takes part in the reverse transmission the otoacoustic response from the ear \citep{Reichenbach2012}. The fibrous membrane above scala tympani is called the \term{basilar membrane} (BM) and it is attached to the \term{organ of Corti}---a complex structure of cells that transduces the mechanical movement to electric discharges in the auditory nerve. The elastic BM moves in response to a pressure gradient between the scala vestibuli and scala tympani. This movement manifests as a slow \term{traveling wave} that propagates from the oval window in the \term{base} of the cochlea toward the helicotrema in its \term{apex}. 

The cochlea has unique frequency analysis properties, whose mechanics was largely elucidated by \citet{Bekesy1960}. High frequencies conducted through vibrations in the stapes cause the BM to vibrate at the base, while low frequencies vibrate at the apex, in terms of the traveling wave peak response. For a given stimulus frequency, the wavelength and phase of the traveling wave change quickly and slows down just before the frequency-dependent peak, whereupon it dies out almost immediately after the peak with little to no energy reflected back to the base. The peak sharpness is reduced at high intensities. The frequency analysis property is achieved by the particular geometry and mechanics of the cochlea, which is tapered in two opposite directions: the ducts are wide at the base and narrow at the apex, whereas the BM is wide at the apex (about 0.5 mm) and narrow at the base (0.1 mm). Critically, the forward direction of the movement is largely determined by the stiffness of the BM that is high at the base and gradually diminishes towards the apex, where the BM movement is mass-controlled by the perilymph. The peak sharpness is limited by damping of the cochlear partition (the organ of Corti including the BM).

The organ of Corti is attached to the BM by supporting cells that embed the \term{inner hair cells} (IHCs) and \term{outer hair cells} (OHCs). Both hair-cell types comprise of a \term{soma} (cell body) and a \term{hair bundle}, or \term{stereocilia}. These stiff hair-like structures protrude from the apex of the cell, which is flushed with the \term{reticular lamina}---a thin membrane inside the organ of Corti, which chemically isolates the perilymph from the endolymph. The IHCs are organized in a single row and the OHCs in three or more rows along the cochlea. The hair bundle is graduated in height and it forms straight (IHCs) or V-shaped (OHCs) patterns that are apparent on the reticular lamina. The stereocilia within one hair cell are connected to one another via small \term{horizontal top connectors} and \term{tip links}, so that the entire hair bundle rigidly deflects upon mechanical force and it bends around its base. The \term{side links} of the OHCs provide mechanical reinforcement against loud sounds \citep{Han2020}.

The traveling wave in the BM causes a shearing movement of the organ of Corti, which leads to deflection of the stereocilia of the hair cells. In turn, this deflection of the IHCs causes the tiny tip links between the hairs to open and close mechanotransducer channels in the apical end of the stereocilia, close to the tip links. This leads to ionic current flow from the positively-charged endolymph to the negatively-charged cell and then to neurotransmitter release (likely, glutamate, although not exclusively; \citealp{Eybalin1993}) to the synapse of the auditory nerve. The opening of the channel is asymmetrical with respect to the direction of deflection of the hairs, which results in a mixed AC and DC response potential. The hair cell response reflects the mechanical response of the BM around its peak, whose corresponding frequency is referred to as the \term{characteristic frequency} (CF)\footnote{The majority of the classical studies of the cochlear mechanics in vivo targeted basal CF sites, whose responses were later extrapolated to apical (low-frequency) sites. Recent studies have begun challenging the validity of this extrapolation, as they consistently revealed lower frequency selectivity at the apex, which does not correspond with a bandpass model. The interaction between the filtering of the cochlea and the auditory nerve is likely to be different there as well. See \cref{LowFreqCorr} for a short review and relevant references.}. 

The \term{tectorial membrane} (TM) is an elastic gelatinous membrane that is flapped over the hair bundle rows and the reticular lamina. Unlike the BM, the TM is found in all tetrapods, but its role in the cochlear micromechanics remains uncertain. The tallest OHC sterocilia are connected to one another and to the TM by \term{TM-attachment crown}, which is essential for the normal OHC function \citep{Han2020}. It has been recently found in guinea pigs that both the IHCs and OHCs are deeply embedded inside the TM \citep{Hakizimana2021}. It was also found that all hair cells are connected to the TM with filamentous tubes called \term{Ca$^{2+}$ ducts}, which directly supply calcium ions to the hair cells that are required for the OHC motility. In the same study it was shown that the TM rests on the RL with no gap between them in quiet, and that their movement remains tight also at large amplitudes, only with momentary gaps when the system returns to quiet. During stimulation, the OHCs and the IHCs appear to be phase locked. These results contradict past findings that were done using coarser methods that found that the only Ca$^{2+}$ supply to the bundle is from the endolymph, that only the tallest OHC tips are embedded in the TM, and that the IHCs are not directly connected to it, although their tips fit into a groove on the bottom of the TM (\term{Hensen's stripe}). 

It was demonstrated in human cadavers and in mice that the TM sustains a traveling wave as well, which in comparison with the BM has sharper tuning, apically shifted frequency mapping, and larger dynamic range \citep{Lee2015,Farrahi2016}. Shearing motion of the RL translates to TM motion that can move the IHC hair bundle through fluid forces \citep{Zwislocki1980FiveDecades,Cheatham1999,Robles2001}. Other recent findings in mice suggest that the role of the TM may be in stabilizing the cochlear mechanical amplification (see below), as detached TMs result in high incidence of spontaneous otoacoustic emissions \citep{Cheatham2016}. 

The passive system in the cochlea cannot achieve on its own the tuning sharpness of the BM that is observed empirically, so there is little doubt today that the OHCs provide the compressive amplification necessary for the healthy cochlea to hear sounds at adequate levels \citep{Robles2001}. The major candidate mechanism that can drive the amplification is somatic motility, but another active mechanism exists in the OHCs---the bundle motility---whose potential role in amplification has not be clearly elucidated. Both mechanisms are induced by the shear forces on the organ of Corti, which moves with the vibrations of the traveling wave in the BM. This cochlear motion causes deflection of the stereocilia, which gates mechanotransducer ion channels that set the current flow into the cells. 

The main active mechanism proposed for amplification---and the most touted one---is the \term{somatic motility}, or \term{electromotility}, of the hair cells. It involves length changes of the hairs when intracellular electric current passes through the cells and causes somatic lengthening with hyperpolarization and shortening with depolarization \citep{Brownell1985}. In vitro, this movement can take place at ultrasonic frequencies (shown at least up to 79 kHz, \citealp{Frank1999}). However, the exact dynamics of this mechanism---how it actually delivers power that impacts the signal picked up by the IHCs---is not entirely clear. One influential class of amplification models requires that the motile changes happen in phase with the traveling wave, on a cycle-by-cycle basis \citep{Dallos2008,Guinan2020}. However, this view has been challenged because electromotility may be limited to much lower frequencies by the capacitive cell membrane in-vivo and more realistic in-vitro conditions \citep{Dallos1995,Ashmore2008,Vavakou2019, Santos2019Review, SantosTan2019}. Electromotility is associated with the protein prestin \citep{Zheng2000}, and in mutated mice that lack prestin, there is no amplification or compression \citep{Liberman2002}. Counterexamples exist for the presence of prestin with no amplification in birds (e.g., in chickens, \citealp{Xia2016}), and electromotility on its own does not guarantee amplification also in mammals (gerbil in \citealp{Strimbu2020}). Also, amplification with no prestin was observed in prestin-knock-out mice at ultrahigh frequencies ($>$ 40 kHz) \citep{Li2022}.  Alternative models to electromotility-based amplification work by locally modifying the impedance of the basilar membrane, thereby creating a sharper resonance with less dampening \citep[e.g.,][]{Kolston2000,Vavakou2019}. These length changes, as well as other biophysical changes in the cell, can result in substantial OHC stiffness modulation \citep{Ashmore2008}. 

The OHCs have an additional active part, which is the hair bundle (the stereocilia), whose movement is called \term{bundle motility}. The hair bundle protrudes from the apex of every hair cell, at the reticular lamina, and is most likely mechanically connected to the TM. It has been proposed that this motion that happens during depolarization of the OHCs can impart power to the motion of the reticular lamina and, therefore, may provide amplification as well \citep{Kennedy2006}. However, much evidence to support this mechanism is still lacking. 

It seems that whichever mechanism is responsible for the cochlear amplification, it requires a certain feedback between the BM, the OHC soma, the hair bundle, and perhaps other elements such as the TM (see \cref{PuttingPLL} for feedback model references). It was also hypothesized that the two motile mechanisms may be both necessary to achieve the OHC amplification in mammals \citep{Peng2011}.

The OHCs are implicated in a host of nonlinear phenomena in the cochlea. First and foremost, they provide amplification around the characteristic frequency for low-level inputs, which results in effective lowering of the hearing threshold. Consequently, it also sharpens and shapes the filter response around the CF. Additionally, the OHCs cause compression of the dynamic range of the input at medium levels (approximately 30--90 dB SPL), they are implicated in intermodulation distortion generation, in two-tone suppression, and in otoacoustic emissions. The OHCs require metabolic supply of energy, which means that their associated functions cease to work after death. 

Recent evidence has suggested that, the amplification and the other active effects of the OHCs take place inside the organ of Corti, so they are measurable on the reticular lamina, but not necessarily in the BM \citep[e.g.,][]{Ren2016Reverse,Cooper2018,Nuttall2018,HeRen2021,Lin2024}. Furthermore, the stiffness and ionic regulation that are observed in the response of the OHCs are regulated by the supporting cells, whose role is gradually being uncovered \citep{Lukashkina2022,Zhou2022}. It was suggested that the force generated by the OHCs may in fact be directed toward the TM and the reticular lamina, and hence the entire Organ of Corti, which then indirectly applies pressure that amplifies the traveling wave, rather than by forcing the BM directly (\citealp{Altoe2022}; see also \citealp{Guinan2022}).

\subsubsection{The auditory nerve}
The vestibulocochlear nerve, which is the eighth cranial nerve, is shared between the vestibular and auditory (cochlear) nerves \citep[pp. 81--93]{Rea2014}. The human cochlea is innervated by approximately 30000 afferents of the \term{spiral ganglion} neuron type and about 1400 efferents, which form the \term{auditory nerve}. Most afferent fibers are \term{Type I}, which are fast (both fibers and soma are myelinated) bipolar cells, whose dendrites synapse to IHCs and axons synapse to the auditory brainstem nuclei. Each IHC is connected to 10--30 Type I fibers with a single \term{ribbon synapse} per fiber, which are characterized by their reliable, temporally precise, and sustained responses. \term{Type II} fibers are slow (unmyelinated) unipolar cells that innervate up to 50 OHCs each. The OHCs are synapsed to multiple Type II fibers. Both types project to the cochlear nucleus in the brainstem, which is the first nucleus in the central auditory pathway.

It is common to refer to the discharge patterns in the auditory nerve as the earliest part of an auditory \term{neural code}. It has been extensively studied in single-unit recordings in many species, mostly for relatively simple signals \citep[e.g.,][]{Eggermont2001,Rutherford2021}. The code varies in its rate and inter-spike patterns, with respect to the input stimulus level, temporal, and spectral characteristics. Afferent fibers (Type I) are characterized by their spontaneous discharge rate, which reflects their inherent noise level and may be low, medium, or high. High-spontaneous-rate fibers are most sensitive to low-level inputs and are the most prevalent (90\% of all fibers), but have a relatively limited dynamic range of about 20--30 dB. The low-spontaneous-rate fibers are sensitive to high levels and have wider dynamic range of up to 60 dB. The frequency tuning of the fibers reflects the bandpass characteristics of the cochlea, which is often illustrated using \term{tuning curves} of the effective filtered response. At the CF, the spiking rate is maximal (for a given input level). Another important feature of the auditory nerve spiking is \term{phase locking}---a precise temporal correspondence between the (tonal) stimulus wave phase and the neural spiking pattern in the auditory nerve and most other nuclei. Phase locking is limited to low frequencies (estimated to be below 4 kHz for humans). The auditory nerve can also track the low-frequency stimulus envelope for all carrier frequencies (also above 4 kHz), which is also conserved throughout all auditory nuclei further downstream to the auditory cortex \citep{Souffi2023}.

The function of Type II fibers has been recently elucidated in experiments on mice in vivo, where a non-acoustic nociceptive (pain) reaction to very loud sounds (120 dB SPL) has been established, which leads to avoidance behavior \citep{Flores2015}. In-vitro tests found that these fibers react to OHC damage, as the kind that occurs with the traumatic exposure to loud sounds and is irreparable in mammals \citep{Liu2015}. 

Descending efferent innervation to the hair cells exists in the cochlea as well, which is called the \term{olivocochlear bundle}. It arises in the superior olivary complex (SOC) of the brainstem and has two divisions. The medial olivocochlear (MOC) efferents, are myelinated and they project both to the contralateral and ipsilateral OHCs with cholinergic synaptic contacts to the cell body. The lateral olivocochlear (LOC) fibers are unmyelinated and they project mainly to the ipsilateral afferent dendrites of the IHCs, but not directly to the hair cell body. The MOC receives afferent input from the auditory nerve via the poteroventral cochlear nucleus (PVCN), which projects to the contralateral SOC. Activation of the MOC causes a bilateral change of threshold through the \term{medial olivocochlear reflex}, which has two pathways---an ipsilateral reflex and a contralateral reflex \citep{Liberman1986,Brown1989}. Fibers that respond to monaural sounds are the majority and a much smaller number of binaural fibers respond to stimulation from both ears. Both ipsilateral and contralateral fibers synapse to the OHCs in the same way, although there are about double the number of ipsilateral as there are contralateral . Accordingly, the ipsilateral reflex effect has been consistently stronger than the contralateral effect in humans \citep{Salloom2021}. The MOC efferents are tuned to the same frequencies as the OHCs they innervate, although their activation is proportional to the bandwidth of the stimulus, where broadband noise is typically used. While uncertain, it has been proposed that the MOC may provide some protection against acoustic trauma including own voice attenuation, regulate the operating point of the amplification, optimize detection of signal in noise (i.e., reduce masked tone threshold), and reduce the sensitivity to unattended stimuli (see also \cref{MOCR}). The role of the LOC remains unknown at present. See \citet{Romero2021} for a more detailed neurophysiological account of the olivocochlear system. 

\section{Organizing principles and common threads in the\\central auditory system}
\label{OrgPrinciples}
Unlike many parts of the peripheral ear, the functions of the different auditory nuclei have so far eluded a plain explanation. The auditory neuroanatomy has been thoroughly charted, but the complexity of the parts forming the system as a whole is prohibitive for the formation of a simple functional account. In order to facilitate the description of the system, several general and particular aspects of the auditory brain are discussed in brief. Some of these principles are not unique to hearing.

The central auditory system comprises several nuclei and multiple pathways for the signal to travel downstream to the cortex (the major pathways are illustrated in Figure \ref{BrainstemFig}). Each nucleus has several neuron types, which can be characterized by different morphologies and typical responses to different stimuli, usually using single-unit (e.g, one neuron) measurements. The auditory nuclei further project to other auditory and occasionally non-auditory nuclei, which may be either excitatory or inhibitory. The connections and responses of the different neuron types have been studied in depth over the last decades, and functions of the respective areas were sometimes inferred from these responses. However, the complexity and diversity of the complete system makes it very challenging to attribute a ``closed-form'' function to most, if not all, of the auditory nuclei. 

\subsection{Tonotopy}
\label{Tonotopy}
Perhaps the most characteristic organizing feature of the auditory system is that the cochlear frequency axis, its \term{tonotopy}, is conserved throughout the auditory brain. Thus, the same monotonic arrangement of frequencies from high to low is found in all of the main auditory nuclei. For example, in the cat, tonotopy was recorded in the cochlear nucleus, lateral lemniscus, inferior colliculus (IC), medial geniculate body (MGB), and auditory cortex \citep{Bourk1981, Aitkin1970, Merzenich1974, Aitkin1972, Reale1980}. Importantly, even though the early brainstem nuclei have their own tonotopic maps, they all converge to a single map in the inferior colliculus \citep{Casseday1996}. Additionally, in the human primary auditory cortex, tonotopy appears to be more complex than in the more peripheral nuclei, as it maps to pitch rather than frequency, as was initially demonstrated measured in the magnetoencephalogram (MEG) response to the missing fundamental \citep{Pantev1989}. The most recent measurements using functional magnetic resonance imaging (fMRI) suggest that frequency and pitch are both topographically mapped in the auditory cortex by different neural populations that have only limited overlap \citep{Allen2022}. 

Starting from the IC and extending to the MGB and auditory cortex, a useful classification of the different auditory stations is into \term{lemniscal pathways} that are more sharply tuned and are organized tonotopically. They are the primary ascending conduit of auditory sound information from the periphery \citep{Carbajal2018}. In contrast, \term{non-lemniscal pathways} constitute the ``belt'' areas that are less responsive to auditory stimuli, are not organized tonotopically, while they send ascending pathways to the next nucleus, and receive descending connections from the cortex. However, it has been recently suggested that the non-lemniscal corticocortical circuits may be involved in complex processing of speech that is parallel to that of the core and its associated lemniscal pathway \citep{Hamilton2021}. We will mostly consider the lemniscal nuclei in the review below. 

\subsection{Synchronized responses}
Another important property that is used to characterize different neuron types is their degree of phase locking, or synchronization---how well they track the acoustic waveform of the stimulus (see \cref{Phaselocking}). Some nuciei and particular cell types seem to excel in synchronizing either to the carrier or to the envelope. The degree of synchronization to the stimulus can be enhanced through an increase in the number of dendrites that synapse the synchronizing cell. This is the case since multipolar cells generally have high threshold for firing, which means that several input fibers have to fire simultaneously in order for the cell to fire. When all the inputs are excitatory, such a configuration is sometimes referred to as a \term{coincidence detector}---a useful circuit in the enhancement of the temporal response of the neuron (see \cref{InterauralCoherence}). Another aspect in which cells vary is whether they synchronize to stimulus onsets or to sustained stimuli, as some cells synchronize to one feature and not to the other. Either way, the response usually shows spiking adaptation over time---a reduction in the average spiking rate---in response to an unvarying stimulus.


\subsection{Generalizations from single unit recordings}
\label{SingleUnitGeneral}
Because of the complexity of the auditory system, as well as neural systems in general, it is difficult to attribute specific functions for the brainstem circuitry. Hence, the understanding of the roles of many auditory nuclei remains relatively vague. \citet[pp. 155--157]{Pickles} categorizes the central auditory system research findings into overarching themes, all of which are also encountered in non-auditory sensory research\footnote{\citet{Pickles} presented three themes, but we merged the first two into a single theme, as they are difficult to distinguish in the original text.}:
\begin{enumerate}
	\item Feature detection and extraction---Relevant auditory features from the stimulus (e.g, spectrum, temporal fluctuations, sound location) may be observable through recordings of a single unit. It may also be associated with a population of neurons that is localized in a certain brain area and has a modular role in the complete signal processing chain. 
	\item Hierarchical analysis---It is sometimes possible to demonstrate progression in complex auditory processing that emerges in one auditory area and culminates in another, usually more central in the ascending pathways (e.g., the auditory cortex as the destination area of auditory scene analysis). This theme introduces continuity and causality into the information-processing logic of different auditory areas in the brain. 
\end{enumerate}

These themes are interdependent and find wide use in auditory research. However, they require external theories to direct the experimental explorations and inspire plausible interpretations. A lack of theory makes the interpretation much more challenging due to the complexity of the system. This was illustrated in a critical paper called ``\textit{Could a neuroscientist understand a microprocessor}?''\footnote{It was inspired by an earlier paper called ``\textit{Can a biologist fix a radio?---Or, what I learned while studying apoptosis}'' \citep{Lazebnik2002}, which may have some relevance to complex structures in the peripheral ear, such as the organ of Corti.} \citep{Jonas2017}. The premise of the paper is that reverse engineering of even a relatively simple computational circuit may be downright impossible using many of the standard methods in neuroscience. Their example was a (simulation of a) popular microprocessor from 1975 that contains 3510 transistors in total, which in today's standards would be considered primitive. It was made to run a number of simple video games as known ``behaviors''. Transistors in the circuit were taken as rough analogs of neurons. Employing classical neuroscientific methods produced a wealth of data and many distinct patterns, which were nevertheless useless in explaining what the microprocessor actually does on a level that can be generalized beyond the specific game being played. 

While these conclusions from \citet{Jonas2017} may appear dispiriting (and perhaps controversial), they are invoked to highlight the nontrivial implications of a lack of a coherent theory of the auditory brain pathways. Local simplicity of brain functions, including low-level ones as observed using simple stimuli in single unit recordings, might be misleading, especially if tested with a narrow range of stimuli. As illustration, a recent comparison of several phenomenological and biologically-inspired auditory signal processing models demonstrated that the simplest model has the highest correspondence to single-unit recordings from the ferret primary auditory cortex, using several simple stimuli (clicks, pure tones, white noise, speech snippets; \citealp{Rahman2020}). The authors suggested that this surprising result can entail that the total effect of the auditory system is much simpler than its complexity may imply. However, their interpretation entails (with some exaggeration, admittedly) that for the tested stimuli the brainstem processing may as well be replaced with direct connections to the cortex, as long as the necessary frequency weighting and nonlinear compression is reproduced. While not unique to this model or study, such oversimplification leads to tagging entire networks with the infamous ``relay neuron'' role (often designated to the thalamus). To this \citet[p. 46]{Winer2005Winer} commented: ``\textit{The concept of a relay nucleus requires critical scrutiny as all central nuclei transform and modify the information that passes through them...}'' Another problem with this interpretation is that it implies that the single-unit primary auditory cortical response is tantamount to a perceptual and behavioral output. It was recently argued that spiking patterns cannot be simply replaced in analysis with more informative signals or inputs, as these patterns do not generalize to real-world stimuli with arbitrary context \citep{Brette2019}. 

Another point of view to consider continues the microprocessor analogy, where the auditory brain is thought to perform some computational task. Hypothetically, any computation requires not only the transmission of the data, but also the transmission of various control signals to modulate its processing, according to the situation. The auditory system processes signals that are on the same temporal scale as the neural processing speed limit. Thus, neuron spiking patterns may correlate to acoustic signals with relatively few transformations throughout the brain, which reduce its complexity and maintains clear correspondence between stimulus and response. However, internal control signals that may be internally used for computational purposes may be correlated with the stimulus as well. This may result in a cacophony of control and sensory signals in the brain that are correlated on a population level, but have fundamentally different roles in the system as a whole. 

In summary, getting a handle on the brain function requires a theory for guidance, which is especially pertinent in the auditory system due to its high complexity. Developing a theory from the reduced low-level components may be hypothetically possible, but it will require a way about compressing the amount of low-level details to a compact description, which can be formulated using high-level concepts. There is no guarantee that this is the case, though. Quoting \citet[p. 45]{Winer2005Winer} again: ``\textit{It may not be possible or even appropriate to delegate functions to nuclei, as functions are global constructs, while neurons and nuclei and circuits are restricted to local operations.}'' 


\subsection{Dual-stream models}
We shall mention one organizational theory that has been influential in recent hearing research that may apply to the entire auditory system---\term{the dual stream model}. It was originally proposed for vision by \citet{Trevarthen1968} and \citet{Schneider1969} and has been revised several times since. According to this model, the processing of visual objects bifurcates in the cortex. The \term{ventral stream}, or the \term{what stream}, processes the pattern (shape) and the identity of the object, whereas the \term{dorsal stream}, or the \term{where stream}, processes its position in space \citep{Mishkin,Wilson1993}. Analogous functions were identified in two distinct anatomical streams going out of the auditory cortex that pertain to identification and localization \citep{Rauschecker1998,Kaas1999,Romanski}.

The neat labor-division offered by the dual-stream model turned out to be more complex both in vision and in hearing. In vision, object localization processing is tied to motor control functions that may be required for actions based on the visual input \citep{Goodale1992}. Thus, a revised dual-stream model in vision distinguishes vision-for-perception (ventral) and vision-for-action (dorsal) streams \citep{Goodale2011}. A more recent version of the model identifies a third visual processing stream that specializes in motion or socially relevant objects (e.g., faces, body movements) \citep{Pitcher2020}. A similar refinement has been applied in hearing, as it became clear that the dorsal stream does not process localization exclusively, but also has some roles in speech and even music processing that require tracking acoustic changes over time \citep{Belin, Hickok2004, Hickok2007}. The ventral stream in speech may be suitable for recognition of words, which are sometimes classified as auditory objects. Thus, a unified model is emerging that posits that language may require processing that relies on both streams, which include sensorimotor connections with articulators that may be used to produce speech \citep{Rauschecker2017,Rauschecker2018}. This interpretation would be a special feature of the human cortex, with unknown applicability to other mammals.

The dual-stream model has also influenced the interpretation of subcortical processing that is known to bifurcate early in processing (in the cochlear nucleus) into several parallel streams that converge again only in the midbrain (inferior colliculus; \citealp[Chapter 6]{Pickles}). The early ventral stream does seem to be dedicated mostly to binaural localization, but the roles of the other two streams are not obvious. Another version of the dual-stream model for speech suggested that the brainstem may be separately processing the fundamental frequency of speech and spectral changes in time---at least as early as the lateral lemniscus and inferior colliculus \citep{Kraus2005}. However, as has turned out in the cortical stream modeling, the subcortical streams do not readily lend themselves to neat classification as the original what/where model suggested, or to other obvious signal processing, so their explicit functions are not necessarily clarified by this model.


\section{Central auditory neuroanatomy}
\label{CentralNeuroanatomy}
This section provides a simplified overview of the complex circuitry in the auditory brainstem. Only high-level details that are deemed to have greater functional significance for the complete system are mentioned, while most of the fine-grained details are omitted. 

\begin{figure} 
		\centering
		\includegraphics[width=0.85\linewidth]{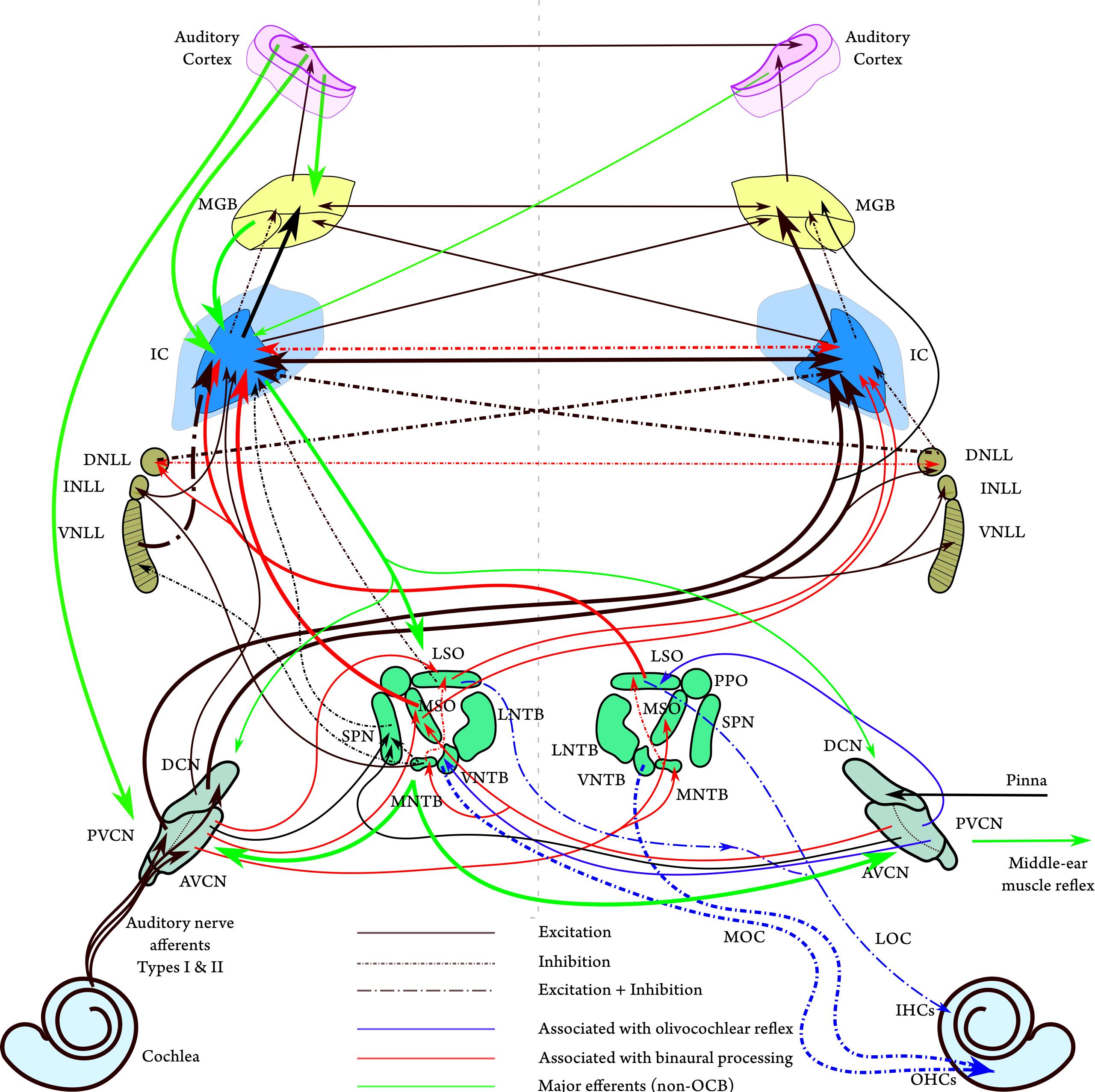}	
		\caption{A schematic diagram of the main connections in the auditory brain, with emphasis on major afferents (in black) up to the central nucleus of the inferior colliculus (IC). Note that some projections are united to avoid graphical clutter, which may not reflect the actual anatomy. The most prominent projections are plotted with thicker lines. The most important efferent connections without specific subnucleus destination are plotted in green, based on Figures 3.2--3.5 in \citet{Schofield2010}. The olivococohlear bundle (OCB) is displayed in blue along with the afferent projections that inform it according to Figure 1 in \citet{Lopez2018}. Excitatory projections are in solid lines, inhibitory in short-dash-dot, and mixed inhibitory-excitatory in long-dash-dot (see \citealp{Larsen2010} for a specific discussion of the lateral olivocochlear, LOC). The main structure is plotted after Figure 6.12 in \citet{Pickles}, Figure 1A in \citet{Felix2018}, and Figure 2.10 along with additional information from \citet{Malmierca}. The main projections that are associated with binaural processing are plotted in red after Figures 6A and 8D in \citet{Grothe2010}. The contour of the cochlear nuclei (CN) in humans was plotted after Figure 7 in \citet{Moore1979}. The superior olivary complex (SOC) contour of humans is plotted after the micrograph of Figure 2A in \citet{Weinrich2018}. The nuclei of the lateral lemniscus (DNLL, INLL, and VNLL) are plotted after that of the cat in Figure 2 in \citet{Glendenning1981} and Figure 11.11 in \citet{Langner2015}. The human inferior colliculus (IC) contour was plotted after \citet[Figure 3]{Mansour2019}. The human medial geniculate body (MGB) was plotted after \citet[Figure 1]{Winer1984}. The auditory cortex was plotted after \citet[Figure 2B]{Kaas2000}. Connections to and from the IC, MGB, and auditory cortex do not imply specific target subnuclei. Note that the direct projection from the CN to the MGB is plotted as if coming from the VCN, but some studies pointed to the DCN as the origin of this pathway \citep{Schofield2014}. For a complete map of known projections from the CN see \citet{Cant2003}. A diagram of the interneuron connections within the CN is provided in \citet{Young2018Shepherd}. Nonauditory connections to the IC are reviewed in \citet{Gruters2012}. Connections between the CN, IC, A1, and A2 and the cerebellum are discussed in \citet{Mennink2020}.}
		\label{BrainstemFig}
\end{figure}

\subsection{Medulla and pons}
Auditory nerve fibers innervate the cochlear nucleus, where they branch to three \term{cochlear nucleus} (CN) areas. The axons from these nuclei run in three different tracts, or \term{acoustic strias}, which lead to further nuclei \citep{Cant2003}. The rostral branch begins from the \term{anteroventral cochlear nucleus} (AVCN), which is one of the two divisions of the \term{ventral cochlear nucleus} (VCN)---the other being the \term{posteroventral cochlear nucleus} (PVCN). The AVCN receives its input from the \term{endbulbs of Held} --- giant synaptic terminals with multiple synapses. The endbulbs of Held are characterized by reliable spiking in response to auditory nerve spikes, as well as very short delays as the synapses are positioned on the soma itself. The AVCN appears to have little or no inhibitory inputs, so its response largely reflects that of the auditory nerve. 

The AVCN projects contralaterally to the \term{medial superior olivary} (MSO) and ipsilaterally to the \term{lateral superior olivary} (LSO)---both are nuclei of the \term{superior olivary complex} (SOC). The LSO also receives inhibitory input from the contralateral VCN via the \term{medial nucleus of the trapezoid body} (MNTB), which contains the largest synaptic terminals in the brain---the \term{calyx of Held}---that enable temporally precise spiking. The SOC is critical in binaural functions, where the LSO primarily extracts interaural loudness difference (ILD) cues at high frequencies, while the medial superior olive (MSO) primarily extracts interaural time difference (ITD) cues at low frequencies. The MSO has additional capabilities in extracting fine temporal cues that are monaural (e.g., echo suppression), at least in small mammals that do not exploit interaural localization cues, such as bats and rats \citep{Grothe2000}. The SOC is also where the olivocochlear bundle arises, which projects contralaterally to the OHCs (the MOC), ipsilaterally to the IHCs (the LOC), as well as collaterally to the CN. 

The caudal branch of the CN innervates the PVCN and the \term{dorsal cochlear nucleus} (DCN). The PVCN contains four different cell types with unique responses. Among them are onset detection by \term{octopus cells} of broadband sound (e.g., clicks), which are temporally precise. Another interesting cell type is the \term{chopper} (\term{T-stellate}) multipolar neuron, which is sharply tuned in frequency and fires with a sustained period that is independent of the input frequency but closely tracks its envelope. The T-stellate cells project to the MOC cells in the SOC, among others. The PVCN cells project ipsilaterally and contralaterally to most neighboring nuclei, including the VCN, the ventral nucleus of the lateral lemniscus (VNLL), and the \term{inferior colliculus} (IC). Additionally, they project to the \term{superior paraolivary nucleus} (SPN), which is a part of the SOC, but has a monaural function in precise offset and (much less in) onset detection in complex sounds \citep{Felix2017,Kopp2018}. Therefore the PVCN plays a role in all auditory functions, yet a single specific role cannot be easily pinned down. 

The DCN is composed of a complex network of cells with diverse responses and numerous connections to other nuclei. In addition to auditory inputs, it receives and projects to somatosensory cells around the pinna. This is thought to reflexively affect localization optimization in animals that can move their external ears, which can have a role in detecting elevation localization cues. It may also have a role in suppression of the input of own vocalizations. Like other areas in the CN, the DCN receives excitatory and inhibitory inputs from most other auditory nuclei, which can change the tuning curve bandwidth of the different cells, or modify their dynamic range by shifting the threshold for firing. Its main projections are to the inferior colliculi on both sides. The DCN is also implicated in sound offset detection and as such it may constitute a part of a specialized ``offset pathway'' along with the SPN, the IC and the medial geniculate body, which appears to be distinct from an ``onset pathway'' in the auditory brain \citep{Kopp2018}.

The CN receives centrifugal (descending) efferent innervation from the IC and from the auditory cortex, which is implicated in modulation of the critical bandwidth of the auditory filters and changing of the masked threshold for tone in noise. The centrifugal inputs may also affect the time constants associated with masking, as well as spike timing modulation by somatosensory inputs to DCN neurons. 

The next stage after the CN and SOC is the \term{lateral lemniscus tract}, which has three nuclei---the \term{ventral}, \term{intermediate}, and \term{dorsal nuclei of the lateral lemniscus} (VNLL, INLL, and DNLL, respectively). The DNLL is part of the localization stream as it receives inputs from the LSO, MSO and CN and projects to the IC on both sides. It has been suggested that the DNLL is key in inhibiting the reflection (the lag) in the (binaural) precedence effect, which then gives precedence to the direct (the lead) sound at the level of the IC \citep{Brown2015Review}. 

The VNLL receives projections from the ipsilateral MNTB and the contralateral VCN, but is completely bypassed by projections from the DCN and SOC. The VNLL projects mainly to the ipsilateral central nucleus of the IC (ICC). As it receives its input mainly from the octopus cells, it exhibits temporally precise but complex responses, which do not disclose an obvious function. An unusual feature of the VNLL is that its tonotopic axis is folded in a three-dimensional helicoidal topology, so that its peripheral laminae are mapped to low frequencies of the IC and the central laminae are mapped to high frequencies \citep{Merchan1996}. This structure was compared with the musical pitch helix and associated with auditory sensitivity to periodicity, and hence, was implicated with the basis for harmonicity detection (\citealp{Langner2015}; but see \citealp{Regev2019}). 

The function of the INLL is even less clear than the VNLL and DNLL. The INLL primarily projects to the ipsilateral IC and receives its major projections from the ipsilateral MNTB (inhibitory) and the contralateral AVCN and PVCN (excitatory), but with lighter projections from other ipsilateral CN and SOC nuclei \citep{Kelly2009,Yavuzoglu2010}. It was shown in bats that high-frequency INLL units can be inhibited by low-frequency non-tonotopic sounds---spectral shaping that most likely carries over to the IC and maybe to the auditory cortex \citep{Yavuzoglu2010}.

~\\

There is no generally accepted theory for the function of the lower brainstem nuclei, except perhaps for the localization done by the SOC \citep[e.g.,][]{Glendenning1998}. The early bifurcation in the auditory pathways at the CN suggests that its different subnuclei have different functions in processing the stimulus. Studies in the avian analogs of the CN, which are simpler than the mammalian CN\footnote{The avian auditory brainstem pathway splits to two well-separated streams, instead of three. In the barn owl, this separation is retained also in the processing of interaural time and level differences until they converge at the inferior colliculus \citep{Takahashi1984}.}, suggest a simpler role division in the barn owl \citep{Sullivan1984} and chicken \citep{Warchol1990}. The \term{cochlear magnocellularis} (the avian homolog to the AVCN) is particularly sensitive to temporal changes, judging by its enhanced phase locking to stimuli. At the same time, it is relatively insensitive to intensity changes and has a small dynamic range. The opposite is the case in the \term{cochlear angularis} (the avian homolog to the PVCN and the DCN), which is sensitive to intensity changes, but shows negligible phase locking. Somewhat similar role division in the brainstem was hypothesized to be separately processing envelope and temporal fine-structure cues that may underlie the ``where'' and ``what'' streams, respectively \citep{Smith2002}. Such streams would be analogous to intensity and phase cues from the avian brainstem. However, this dual stream interpretation has been challenged \citep{Zeng2004}, and the separation to these two domains turned out to be nearly impossible to accomplish in practice (see \cref{AnalyticChallenges}), so if this functional framing has merit, it has to be refined. 

A recent review of subcortical auditory processes by \citet{Felix2018} makes a case for the hierarchical processing scheme mentioned in \cref{SingleUnitGeneral}. According to their model, several features that are useful in auditory scene analysis---the segregation and grouping of sound streams in complex acoustic environments---emerge in the brainstem and become more salient in pathways downstream. These processes include fundamental frequency extraction and harmonicity detection, gap detection, forward masking, improvement of signal in noise and reverberation, spatial segregation, as well as early selectivity to species-specific vocalizations. Brainstem involvement in speech processing goes counter to the traditional approach that associates it with the cortex alone and attributes only general-purpose processing to subcortical areas \citep{Scott2003}. Findings that consider speech specialization in humans generally begin in the IC rather than in the CN nuclei, although they may be constrained by methods that do not allow for recordings in humans further upstream \citep{Krishnan2009}. In general, the brainstem appears to be involved in early extraction of features that are meaningful in the description of sound sources, or acoustic objects, rather than in the extraction of mathematical features per se \citep{Masterton1992}. It has been noted that the brainstem is where the auditory system stabilizes the signal in face of level fluctuations and noise and generally provides the fidelity that is needed to temporally resolve complex sequences of sound \citep{Young2018Shepherd}. 

\subsection{Midbrain}
The \term{inferior colliculus} (IC) is the primary auditory nucleus in the midbrain and is considered an ``obligatory'' pathway, as nearly all major and secondary pathways cross it between the CN and the cortex \citep{Aitkin1984} (see Figure \ref{BrainstemFig}). In the IC, the different streams that split at the CN converge while retaining their tonotopy. The topographic layout of the \term{central nucleus of the IC} (ICC) achieves this with thin layers of neurons that form \term{isofrequency laminae}, which are sharply tuned (they cover about 0.3 octaves, on average), also at high input levels\footnote{Isofrequency laminae are sometimes ascribed to the CN topography as well \citep{Young2018Shepherd}.}. Each lamina maps to a discrete CF, which is thought to mirror the critical-band concept from psychoacoustics \citep{Schreiner1997, Malmierca2008}. However, there are ``fine structure'' changes in the frequency tuning around the CF along the extent of the two dimensions of the lamina. 

Various cells in the IC were found to have specialized responses that are tuned to features such as the direction of sound, the range of amplitude modulation frequencies, or frequency modulated sweeps (in rats and bats). The two-dimensional shape of the isofrequency laminae suggests that they map additional features of sound that are orthogonal to frequency. One influential observation (in the cat) is that periodicity is one such mapped feature, as cells tuned to different modulation frequency ranges were found to be distributed on the lamina, on an axis orthogonal to frequency \citep{Langner1988,Schreiner1988}. A spatial map for monaural front-back localization was found in the \term{external nucleus of the inferior colliculus} (ICX) of the guinea pig and a homologous region in the barn owl's brain. The ICX and the \term{dorsal cortex of the inferior colliculus} (ICD) also receive projections from somatosensory and trigeminal inputs and show broadly tuned responses that are not necessarily auditory in function. It has been suggested that in these peripheral nuclei of the IC, the auditory system begins its engagement with predictive coding (stimulus deviance detection), which becomes more extensive in the thalamus and cortex \citep{Carbajal2018, Carbajal2024}. 

The IC also receives centrifugal efferent projections from the auditory cortex and projects to the CN. These projections can have a localized effect on tuning and sensitivity to specific stimuli. Some efferents from the IC (and the VCN) specifically target the MOC neurons in the ventral nucleus of the trapezoid body (VNTB) and neuromodulate the dynamic range of these cells, which is likely reflected in the MOC effect in the cochlea \citep{Romero2021}. 

The unique structure of the IC has led to relatively mature research efforts to pin down its strategic involvement in different hearing functions \citep{Winer2005}. A theory about the function of the IC was proposed by \citet{Casseday1996} (see also \citealp{Casseday2002Oertel}), which was inspired by bat echolocation, but is generalizable to mammals and other vertebrates. The theory has two main hypotheses: ``\textit{(1) Tuning processes in the IC are related to the biological importance of sounds. (2) The change in timing properties at the IC, from rapid input to slowed output, is related to the timing of specific behavioral responses}'' \citep[p. 312]{Casseday1996}. The authors used five lines of evidence to substantiate these hypotheses: the IC has homologs in all vertebrates, it receives inputs from all auditory brainstem nuclei, it has rife connections to the motor system, it has many neurons that are tuned to highly specific sounds that are behaviorally relevant (like parts of species-specific vocalizations or echolocation, such as frequency sweeps), and there is a change to a much slower neural coding. The role of the slowing down at the IC was suggested to be necessary to match the motor responses (e.g., speech production) and to set a slow pace for the cortex to act upon. Alternatively, it was suggested that the slower processing introduces the necessary delay into the system to allow for adequate processing of sound. Overall, this theory implicates the IC with rather general-purpose roles that are neither completely automatic and low level, nor are they particularly complex or in anyway conscious.

\subsection{Thalamus and cortex}
\label{MGBA1}
From the IC, the auditory signal continues to the \term{medial geniculate body} (MGB) in the thalamus. The MGB receives projections from different modalities and is also implicated in fear response to auditory stimuli due to direct projections to the amygdala. The responses that have been recorded in the MGB are complex and diverse and, somewhat like the IC, are relevant to all auditory signal processing aspects. Although it has traditionally been considered to be an auditory relay layer before the cortex (see \cref{OrgPrinciples}), this view is gradually shifting as it has been revealed that the MGB is tightly integrated with the auditory cortex through multiple diverging and converging (thalamocortical) pathways of great diversity in terms of their functions and connections \citep{Winer2010Winer2}. Aside from providing the main link between the IC and the cortex, information that passes through the MGB can be shaped and optimized using feedback loops from the cortex (corticothalamic projections). Additionally, after the auditory signal convergence in the IC, the MGB diverges, so that several parallel streams project to different places in the auditory cortex. For example, the MGB features significant convergence of receptive fields\footnote{The \term{receptive field} refers to the range of input parameters that triggers a response in a neuron. For example, in a central auditory neuron, this may relate to the range of frequencies and interaural timing differences. In general, the more central the neuron is, its receptive field tends to be more specialized to a particular combination of input parameters.} that are more acoustically complex than in the brainstem and IC and culminate in spectrotemporal maps in the auditory cortex \citep{Miller2001}. 

The \term{auditory cortex} is found in the upper surface of the temporal lobe and is subdivided to \term{core}, \term{belt}, and \term{parabelt} areas that are also called the primary, secondary, and tertiary auditory cortex (A1, A2, and A3), respectively. The core projects to the obligatory belt area. The parabelt is connected to the frontal lobe and eye movement control areas. All three areas receive inputs from different areas of the MGB and from one another. Tonotopy occurs independently in several fields of the core and belt where iso-frequencies are organized in strips. Yet other peripheral areas do not show tonotopy, or rather, they exhibit fragmented frequency maps. This may be indicative of parallel processing that is going on in the auditory cortex. Tonotopic areas have centrifugal projections to other tonotopic areas in the MGB, which potentially gives rise to feedback loops. Other features that have corresponding organized areas in A1 include spatial distribution of monaural or binaural stimuli, areas that show narrow or broad tuning, and cell groups that respond to frequency modulation. Single units also vary in tuning characteristics (narrow or broad), may have multiple peaks, and be influenced by inhibitory or excitatory inputs from other areas. Cortical cells sometimes respond to very specific types of stimuli of growing complexity in areas farther away from A1. Both A1 and A2 also contain cells that are tuned to simultaneous temporal and spectral (\term{spectrotemporal}) envelope modulations of low frequencies \citep{Schonwiesner2009}, which may first appear at the level of the IC, at least in some animals \citep[e.g.,][]{Poon2000,Qiu2003}. A1 cells are shown to specifically respond and adapt to stimuli of different time scales of many orders of magnitude ($10^{-1} - 10^2$ s)---something which does not arise at the processing level of the MGB \citep{Ulanovsky2004}.

In general, compared to the core areas, the belt seems to have more complex processing that may be related to meaningful signals, such as communication, and is likely engaged in parallel processing (i.e., bypassing A1; cf. \citealp{Wang2024}) of complex signals like speech, along with other cortical areas \citep{Hamilton2021, Whalen2024}. Although the auditory system is quite capable without an auditory cortex, it is required for sound localization, sound detection and frequency discrimination tasks---sometimes in a manner that depends on the input to one ear only. Being part of the cortex, the auditory areas seem to exhibit a considerable degree of plasticity, which is predicated on the animal's individual developmental experience and conditions found related to other sensory pathways. Information from the auditory cortex may be used for further decision making and action. 

Two complementary approaches for the role of the auditory cortex are that it is either the culmination of hierarchical processing in preceding auditory pathways, or that it serves as a control center that modulates and controls the lower-level input processing via the descending efferent network \citep{Cariani2012Poepel}. The auditory cortex is hypothesized to be where auditory objects are formed and localized in space and are brought closer to conscious awareness, which itself may be distributed and not confined to one area within the auditory cortex \citep{Dykstra2017}. 

Several hypotheses for the lateralization of the human cortex have been made over the years. For example, that the left hemisphere excels in fast temporal processing \citep{Schwartz1980}, that the left hemisphere of the auditory cortical areas specializes in temporal processing such as speech, whereas the right hemisphere excels in spectral processing, such as music \citep{Zatorre2002}, that the left hemisphere specializes in temporal modulations and the right hemisphere in spectral modulation processing \citep{Flinker2019}, and that the processing of the left auditory cortex is critical for communication (both language and vocalizations; \citealp{Ruthig2022}). However, the auditory cortex response has not lent itself to straightforward interpretation as the visual cortex, due to highly specialized and plastic effects, as well as complex subcortical preprocessed input, which is dynamically controlled by the descending centrifugal network \citep{King2009}.

Although the concept of auditory objects is of prime theoretical interest in this work (\cref{ObjImg}), we will not be dealing directly with the auditory thalamic or the cortical areas. We rather dwell on the concepts of the acoustic object and its corresponding auditory image at the level of the IC, which will likely have some implications on the ideas behind auditory objects. 

Formulating a theoretical account of the functions of the auditory cortex is the intersection of a complete hearing theory and a theory of the sensory cortex in general. These theories are in flux and are nowhere near consensual within the research community. A complete review beyond the above sections and the relevant sections in \cref{HearingTheory} is outside the scope of this treatise. See \citet{Cariani2012Poepel} and \citet{Heilbron2018} for relevant models and further literature. 

\section{Hearing in humans and other mammals}
\label{ComparativeHearing}
\begin{figure} 
		\centering
		\includegraphics[width=0.9\linewidth]{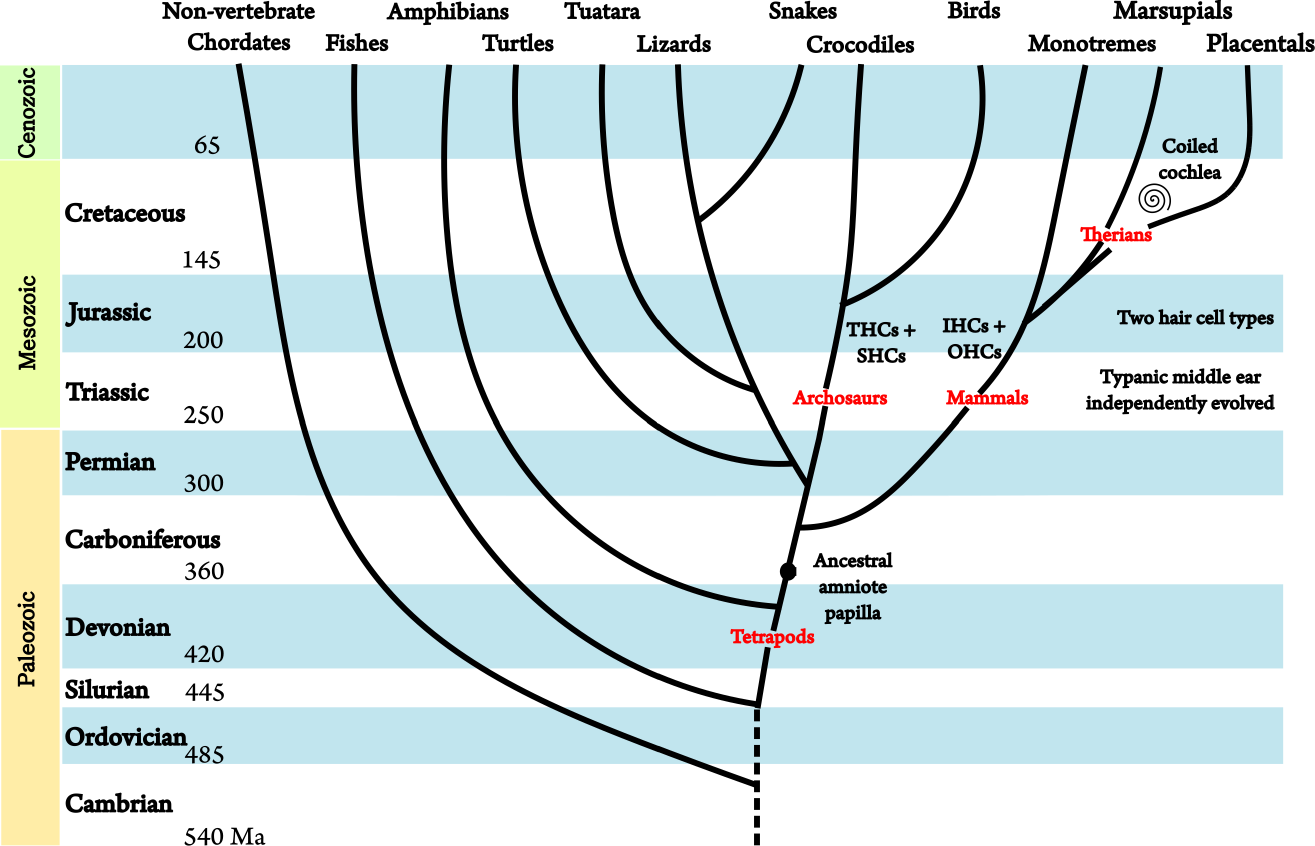}	
		\caption{A coarse-grained evolutionary tree of the main extant animal clades. The ancestral amniote ear contained a papilla that later evolved into the cochlea in mammals. The tympanic middle ear evolved independently in different animal classes. Similarly, a specialization of hair cells to two types also evolved independently. In birds, these are \term{tall hair cells} (THCs) and \term{short hair cells} (SHCs). The figures are adapted from \citet{Manley2017} and \citet{Manley1998}. The additional split in mammals is based on \citet{Luo2011}. The numbers represent the estimated start dates of the different geological periods in million years (megaannum, Ma).}
		\label{EarEvo}
\end{figure}

An overarching goal in much of the auditory research is to better understand human hearing. Detailed knowledge about the biology that underlies the hearing system owes much to various animal species, whose auditory systems are remarkably similar to that of humans. In the case of mammals, there is a great similarity between animals, but also differences in morphological details, which may have effects on physiology that are not always well understood. The most popular species used as animal models are cats, followed (in no particular order) by mice, rats, Mongolian gerbils, chinchillas, guinea pigs, squirrel monkeys (New World monkeys), rhesus macaque monkeys (Old World monkeys), ferrets, marmosets, rabbits, and bats. The latter are often studied as an entirely different specialty within bioacoustics and hearing. The auditory systems of other vertebrates also have significant similarities to humans, so they have been occasionally studied for the same purpose with focus on birds, lizards, and frogs. In this work, notable bird data are brought from the European starling, barn owl, budgerigar, and chicken, as well as bullfrog, tree frog, and bobtail skink lizard data. However, it is useful to remember that the earliest ancestors to the mammalia lineage split from the amniotes (the egg-laying tetrapods that adapted to terrestrial life) before the other vertebrate taxa did---like birds, lizards, or amphibians---so their auditory evolution has been independent of one another throughout the last 320 million years \citep{Manley2017}. See Figure \ref{EarEvo} for a coarse time-line of the ear evolution in amniotes.

Invasive research on the living human hearing organ is out of the question, except for very rare cases in which other medical procedures call for surgical intervention in the ear's area. Thus, invasive experimentation on humans is restricted mainly to studies of cadavers, which allow for anatomical examinations, but only limited physiological and no behavioral observations, to be collected. All other data about human hearing are gathered using indirect methods that may be either objective (e.g., electrophysiology, brain imaging), or psychophysical, which is the only method that can benefit from verbal accounts of the perceptual experience by the listeners. Ideally, invasive and noninvasive methods should all converge to the same results as predicted by theory, should one exist.

On the whole, the auditory systems of all mammals are qualitatively similar but they quantitatively differ: in dimensions, geometries, and sensitivities. However, there are neither extra nor missing auditory organs in any known mammalian order. The differences tend to be striking in the external ears and become more nuanced in the central pathways. One interesting difference is between hearing generalists and hearing specialists, which appears to relate to their cochlear structure---its geometry and organ of Corti configuration. The main focus of the review below is on the  differences in auditory systems between mammals, whereas commonalities should be taken from the general description of the hearing organs in \cref{Periphery} and \cref{CentralNeuroanatomy}. 

The subsection about the external and middle ears is heavily based on \citet{Rosowski1994Fay} and that about the inner ear is based on \citet{Echteler1994Fay}, in the same volume. 


\subsection{Outer and middle ear differences}
\label{OuterMiddleDiffs}
The most prominent auditory differences between mammalian species are visible in their peripheral ears, which to a large extent determine their audible frequency range.  Generally, the dimensions, shapes, and structural complexity of all ear parts vary between mammals, often to a great extent. This applies to the shapes of the pinna and the concha, the ear canal length, the diameter (cross section), the canal bends, the eardrum shape, the middle ear air cavity, and the ossicles. The maximum audible frequencies are inversely proportional to the body and head sizes of the animal. As a rule of thumb, small and stiff ears are most effective for high frequencies, whereas large and compliant ears are better for low-frequency hearing. Where a species is able to hear both low and high frequencies well (e.g., cats, gerbils), it is indicative that its middle ear had specialized by evolving its geometry to overcome the conductive limitations imposed by its size. The degree of coupling and type of movement between the eardrum and the ossicles also vary between species. The eardrum and stapes footplate size are approximately scaled like the fourth root of the entire animal body mass, and are linearly correlated with one another. Finally, the effectiveness of the acoustic reflex muscles varies---it is most effective in humans and cats below 1 kHz, and in species with stiff middle ear like bats it is effective over a much broader bandwidth, up to 80 kHz. In several mammalian orders either one of the middle-ear muscles was either lost or degenerate \citep{Mason2013}. For example, the stapedius is the only active reflex muscle in the guinea pig (and perhaps in the chinchilla), but appears to be degenerate and have little-to-no effect. Several subterranean species have independently lost their tensor tympani muscles, although the advantage in that remains unclear. 

The morphological variations have a significant effect on the impedance matching and hence on the power transmission efficiency between the external, middle, and inner ears. External ears collect acoustic power in a frequency-dependent manner, whose efficiency peaks between 10\% and 100\% somewhere above 1.5 kHz, depending on the particular mismatch between the external ear radiation impedance and the middle ear input impedance of the animal. The impedance matching between the outer and the middle ears is generally poor below 1 kHz and its most effective frequency is usually around 2 kHz, depending on the species.  The larger is the area of the eardrum and the air cavity in the middle ear, the smaller is the total middle ear stiffness that dominates its impedance. However, the interdependence of the various parts of the middle ear may be too complex to follow simple laws and the individual anatomy of the ear may be required to determine its sound conduction properties. So, in some cases the stiffness of the eardrum dominates, whereas in other cases the air cavity stiffness dominates. The total transfer of power from the diffuse acoustic field to the oval window, which has the cochlear impedance as its load, is far from ideal and it peaks only at high frequencies, depending on the species. For example, for the cat it is highest at 2--10 kHz and for humans it is much lower and peaks at 1--4 kHz. In the case of gerbils, they excel in low-frequency ($<$1 kHz) sound collection (which tends to be poor for most animals) that is attributed to their hypertrophied eardrum and middle ear cavity.

The geometry of the pinna and the relative positioning of the ears also affects the directional dependence of incoming sound. For example, the human ears have small pinnae relative to the head size and they are located at its sides, whereas the cat pinnae protrude above it while the ears are still at the sides. Other animals have much larger ear flange sizes relative to their body size, such as the wallaby that has an exceptionally large pinna size to its body. Some mammals can move their external ears and further optimize them for directional localization. In cats, the asymmetrical pinna movement was shown to be coordinated with their eye movements, and the dynamic pinna movement was suggested to improve sound processing \citep{Populin1998}. Scattering of low frequencies is affected by the body shape as well, which in the case of human is caused by the torso below about 600--800 Hz. The external ear resonance frequency also varies between species, and is generally higher with shorter ear 9canals, but depends on incoming sound direction and on additional geometrical features of the ear. In general, large ears collect higher sound power at all frequencies and are more directional at low frequencies than small ears. 

Marine mammals evolved independently from terrestrial mammalian orders serveral times, so specializations for underwater hearing in the different marine taxa do not necessarily have much in common. However, the challenges of hearing in water had to be overcome by all marine mammals, which evolutionarily converged to similar phenotypic solutions, arising from similar genomic expression \citep{Foote2015}. Theoretically, the aqueous medium has almost identical impedance to the cochlear perilymph, so the impedance matching function may no longer be as necessary underwater as it is overground, but at least in cetaceans the middle ear cavity is filled with air, which allows for pressure gradient separation of the oval and round windows \citep{Ketten1998}. Thus, the middle ear function is more ambiguous in these animals. Some of them are active terrestrially as well (primarily pinnipeds---seals, sea lions, and walrus), so their ears still have to function both in air and in water, which makes the middle ear necessary. To this effect, pinnipeds are able to regulate the contact with the acoustic medium, as they are equipped with an epithelial layer, which is controlled by vascular activity, that can open and close the ear canal. In general, marine mammals do not have pinnae, apart from sea lions and otters, although some vestigial remnants may be present. In contrast, the ear canal openings of cetaceans is either filled with wax, or is altogether absent. Both cetaceans and pinnipeds have special lining (cavernous mucosa) inside their middle ears to regulate its pressure while diving. Due to the large external pressures, cetaceans evolved massive ossicles in comparison with terrestrial animals. Underwater, it is likely that sound arriving to the inner ear has a dual pathway both through the bones and the tissue. In odontocetes (toothed whales, dolphins, and porpoises), one pathway (through the jaw) may be more suitable for ultrasonic echolocation, whereas the other (through the blocked ear canal) for low-frequency communication \citep{Ketten1992,Popov2008}.

\subsection{Cochlear and auditory nerve differences}
\label{ComparativeCochlea}

Mammals are unique among vertebrates in their high frequency ($>$ 10 kHz) hearing capabilities (but see \citealp{Nothwang2016}, for counterexamples). The specific frequency map between the CF to the relative position of the BM can be described using this power law \citep{Greenwood1990}:
\begin{equation}
	f = A(10^{ax} - k)
	\label{GreenwoodFunction}
\end{equation}
with $f$ being the frequency, and $A$, $a$, and $k$ are constants specific to the animal. For humans, $A=165.4$, $k=0.88$, and $a=0.06$ when $x$ is in millimeters. This power law is common to many mammals, for which it only differs in the parameters that scale the specific audio bandwidth they respond to according to the BM length. It entails a logarithmic frequency map, as more BM length is dedicated per unit frequency to low frequencies than to high frequencies. So using this formula, it is always possible to express the distance $x$ in relative units to obtain scale-free cochlea that only differs in the audio range. In this case, $x$ would be the proportion distance between 0 and 1, whereas $a=2.1$ in all animals. 

The resonance itself is most conveniently described as a bandpass filtering operation on the incoming sound. The human BM length is taken to be 35 mm on average, although a recent meta-analysis determined it is $33.088 \pm 0.452$ mm for the highly variable population average \citep{Atalay2020}. However. studies contributing to this figure may have not taken into account the difference between the BM and the cochlear duct lengths due to the helicotrema, which is 1.6 mm on average \citep{Helpard2020}. Furthermore, it was found that the interindividual morphological differences in the spiral shape of the human cochlea are substantial \citep{Pietsch2017}. 

Mammals can be roughly categorized in two groups, in terms of their hearing. \term{Hearing generalists} are those whose hearing threshold is smooth and they do not concentrate on particular spectral regions. Humans, cats, mice, and guinea pigs belong to this category. In contrast, \term{hearing specialists} have spectral regions of greater fidelity that stand out with respect to their entire audible spectrum. Hearing specialists include bats, rats, dolphins, and gerbils. These preferential spectral regions appear to be behaviorally motivated at least in some of these species. When they exist, these regions are characterized by a relatively constant stiffness in the BM (achieved by unvarying width and thickness gradient of the BM), along with a discontinuous geometry where the stiffness changes quickly over the cochlear length. Therefore, in hearing specialists, relatively narrow spectral regions occupy large portions of the cochlear length. In bats, these narrowband regions correspond to the second harmonic of their constant-frequency echolocation, where this region is referred to as the \term{auditory fovea} and it is tonotopically conserved also in the auditory pathways \citep{Covey2005}. As a result, cochlear power-law scaling applies much better to the generalists than to the specialists. 

The cochlea of different mammalian species can be visually distinguished by the geometry of the bony structure---the number of turns in the spiral (2--4), and in the length, width, and thickness of the basilar membrane. None of these parameters, however, correlates well with the basic hearing variables, such as the audible range, or its limits, if the specialists and generalists are mixed, or if large terrestrial animals are included. If only the generalists are included, then longer BMs tend to correlate with lower high- and low-frequency cutoffs. There are indications, though, that the low-frequency cutoff is determined by the size of the opening of the helicotrema that connects the vestibular and tympanic scalae---the larger it is, the higher the cutoff is. High frequency in ten primates (including humans) was robustly shown to be inversely correlated with cochlear volume, independently of body mass, which is itself highly correlated with the BM length \citep{Kirk2009}. Another detailed morphological and audiological survey of the cochleas of 33 different mammalian (therian) species---primarily rodents and primates---suggests that rodents may have evolved to have extended low-frequency audible range through a distinctive ``tower-shaped'' cochlea (i.e., more elongated in volume) that is achieved through extra coiling \citep{delRio2023}. In contrast, for similar cochlear lengths, primates have extended high-frequency audible range by having wider cochleas.

Another difference between mammals that has been recently highlighted is the area of the cochlear partition that supports the traveling wave. Traditionally, it has been considered to be only the BM, but it turns out that the structures that connect the BM to the cochlear wall---a soft \term{bridge} that connects the BM to the hard plate-like \term{osseous spiral lamina} (see Figure \ref{OrganCorti})---also vibrate and may even occupy a larger relative surface vibrating area than the BM \citep{Raufer2019}. In humans, the resonant peak of the transverse movement is located underneath the inner pillar cells---in the interface between the bridge and BM. In other mammals, the relative proportion of the vibrating parts and the location of the bridge and the peak relatively to the hair cells can be dramatically different than in humans \citep[Supplementary information]{Raufer2019}. The effect of these differences on cochlear models is unknown, but may be significant\footnote{In this work we shall refer to the BM as the substrate of the traveling wave. As we will not be concerned with detailed mechanical modeling, this should be understood more generally, as the BM and any connecting structure that vibrate with the traveling wave.}.

In various cetaceans, the BM is supported by the osseous spiral lamina in the basal turn of the cochlea, which (along with other morphological differences) gives rise to ultrasonic hearing range that violates the generalists' scaling law as well \citep{Ketten1992,Ketten1998}. The basal stiffness itself seems to be correlated with the high-frequency limit of the animals, whereas the apical stiffness is not correlated with the low-frequency limit. Correlations were observed between the thickness of the TM and the pronounced frequency regions in bats and rats. 

The length and the quantity of the hair cell bodies and stereocilia vary between different animals. The length of hair cells of all types increases toward the apex. The variation is larger in OHCs (compared to IHCs), whose maximum length in the apex is correlated with the low-frequency limit of the animal. There are on average 3000--3500 IHCs in humans, 2600 in cats, 960 in rats, and about 1100 in gerbils \citep{Ulehlova1987,Nadol1988,Hutson2021}. Another prominent variation between animals is in the number of OHC rows, which is at least three (as in humans), but can be up to six in some rat species in certain cochlear regions. While there may be 11000--16000 OHCs in total in humans, there are less than 10000 in the cat, about 4600 in the gerbil, 3500 in the rat, and 2400 in the guinea pig \citep{Hutson2021,Nadol1988}. The number of cilia in each cell is also widely different and is highest for humans and monkeys (up to 150 in the base and 46 in the apex) and less in rats and cats. In the guinea pig, there is almost no variation in cilia count between the base and apex. Several additional morphological differences were found inside hearing specialists' organ of Corti in the form of hypertrophied supporting cells, different TM shapes, a second spiral lamina in some species, and other finer differences \citep[pp. 158--162]{Echteler1994Fay}. It is interesting to note that while all vertebrates appear to have some types of hair cells with bundle motility, electromotility is uniquely found in mammalian OHCs \citep{Peng2011}. Another interesting point that has been recently found is that the OHCs and their bundle motility in mice, and perhaps in other small mammals, is required for hearing ultrasound above 16 kHz \citep{Li2021}.

The innervation of the IHCs is determined by the morphology of the spiral ganglion cells and their synaptic terminals. In humans, each IHC is innervated by about half the number of fibers (9--11) than in the cat (20--26), but each human nerve terminal has multiple synapses (15--16) and only one synapse in the cat \citep{Nadol1988}. The differences in the numbers are reflected also in the total number of spiral ganglion cells in the cochlea, which is 25000--30000 in humans, 45000--58000 in cats, and 15800 in rats. Furthermore, the number of fibers is about 31000 in humans and rhesus monkeys, 52000 in cats, but only 24000 in guinea pigs. It is probably the highest in dolphins, whose exceptional hearing may be superior to humans, where the bottlenose dolphins have approximately 105000 ganglion cells in their cochlea \citep[Table 35.1]{Ketten1992}. Significant structural differences in the bony compartment that houses the spiral ganglia (Rosenthal's canal) of echolocating bats of the suborder Yangochiroptera have enabled larger and more numerous ganglia to evolve, which may have been key in the diversity of echolocation displays that are found in this suborder compared to Yinpterochiroptera \citep{Sulser2022}. The last factoid we shall mention is that only 5\% of the cell bodies in humans are myelinated, whereas in cats it is about 95\% \citep[Table VI]{Nadol1988}. See \citet{Nayagam2011} for additional species-specific data. 

\subsection{Central differences}
Few studies systematically compared the central auditory pathways between mammals. Because of the relative obscure function of most auditory nuclei, it is not obvious what the most informative level of comparison may be, once we factor out the differences in the various receptive fields of the different cells types that may be learned and determined by the animal's environment. \citet{Glendenning1998} compared the absolute and relative sizes of the ten most prominent ascending auditory nuclei in a sample of 53 mammalian species\footnote{This set does not constitute an unbiased sample of all mammals. For example, the marsupial class is over-represented compared to placentals, while monotremes are altogether absent. Within the placentals, the orders of rodents and chiroptera (bats) are under-represented, while primates are over-represented. However, this sample covers many of commonly used animal models and reveals unmistakable patterns among them, which are more than satisfactory in the present context.}. This study proposed that, in first approximation, the size is indicative of the relative importance of the particular nuclei for the animal. Perhaps the most striking example is that the MSO appeared absent in both mice and hedgehogs, although at least in the case of mice it probably has to do with the method, since the MSO does exist \citep{Fischl2016}. Due to their small head size, ITD cues are unavailable to mice (and other small mammals), and they have to rely on high-frequency ILD cues for localization. The relative size of other nuclei follows a relatively robust mean, in which the IC is by far the largest nucleus and the MGB is the second largest. In some species this is reversed, as the MGB is slightly larger or similar in size to the IC. It is interesting to note that both in the albino rat and the rhesus monkey, it was found that the IC has the highest glucose consumption of all auditory system nuclei, followed by the auditory cortex, and then the MGB \citep{Sokoloff1977,Kennedy1978}. Most animals have uniform CN size relative to the entire auditory system, with the feathertailed glider (a small marsupial) having an unusually large DCN and bats having a large AVCN---measured relative to the total CN, or total auditory system size. However, the interspecies variations of the DCN, PVCN, and AVCN seem to be large. The AVCN and PVCN division itself is not as well-distinguished in humans as it is in other mammals \citep{Moore1979}. The MSO of cats, llamas, and foxes is untypically larger than the LSO, which in most species is larger than the MSO. 

Although echolocation has been documented in several animal species that evolutionarily converged to a similar orientation principle---notably, bats and toothed-whales, but also some types of rodents and birds \citep{Shen2012,Parker2013,He2021}---it has been studied most intensively in bats. Echolocating bats have highly specialized hearing that makes use of the same auditory nuclei, but sometimes in a different way than all other mammals \citep{Covey2005}. Some of their auditory organs are hypertrophied---much larger than would be suggested by their brain size---like the cochlea, the IC, the auditory cortex, the VNLL, and the INLL. Notably, the bat's MSO seems to be used for monaural tasks that require temporal precision, rather than for interaural time difference detection, as is the case in other mammals that can hear low frequencies. In some bat species, the VNLL features cells that respond only to particular patterns of modulated sound, with high temporal precision. The IC in echolocating bats is thought to relate their emitted frequency-modulated pulse to the delayed echo from the environment using specialized \term{delay-tuned} (or ``\term{FM-FM}'') neurons (that are also found in the INLL, VNLL, MGB, and A1; \citealp{Wenstrup2011}). This information is then used to derive the distance to the target that is passed on to the thalamus and to the auditory cortex, where pulse-echo maps are formed (organized by the delay time and the FM harmonic that is analyzed) that can inform decision making. Furthermore, dense projections from the IC to premotor areas (pretectal and pontine nuclei) can rapidly stir motor action in flight and dynamically adjust subsequent vocalizations with respect to the target. 

A more detailed comparison of the human and cat brainstem nuclei revealed local morphological differences in dimensions (larger in human) and several underdeveloped auditory nuclei in humans---the LSO, the MNTB, and the VNLL \citep{Moore1987Comp}. The latter is by far the most poorly developed nucleus in humans, in line with other primates and new-world monkeys, and in opposition to bats and porpoises that have highly-developed VNLL. Nevertheless, a double helical structure of the VNLL was identified in humans too that has 7--8 turns---each turn is thought to correspond to one octave \citep[pp. 174--176]{Langner2015}. The human MNTB is also notoriously difficult to identify, but it was argued to positively exist in \citet[Appendix A]{Grothe2010}. The implications of these differences and others on the morphogenic and cytoarchitectural levels are unknown. 

At the level of the auditory cortex there is great morphological variability with different tonotopic maps abound, whose frequency axes are similar down to a scaling factor, despite different boundary shapes \citep{Goldstein1980Fay,Merzenich1992Webster}. The exception is, yet again, echolocating bats, whose auditory fovea gives rise to tonotopic maps with magnified areas of the foveal narrowband range. Even mammals with relatively primitive cortex like hedgehogs and possums have an auditory cortex with similar responses to more developed and fully laminated cortices \citep{Gates1982,Batzri1990}. More generally, it is probably relevant to note that the corpus callosum, which connects the left and right hemispheres, is found only in placental mammals, but not in marsupials and monotremes, or in other vertebrates \citep{Kaas2013}---something that undoubtedly has to have some effect on auditory perception and processing as well. 

Finally, the scaling of the auditory system relative to the size of the brain is also compressed, as its absolute size is largest in humans, while its relative size to the brain is smallest. It is the opposite in bats that have the largest auditory system relative to their brain size.

\chapter{The acoustic source and environment}
\label{InfoSourceChannel}

\section{Introduction}
Our sense of hearing is concerned much more with acoustic sources than with their reflections from the environment. This is in stark contrast to vision that primarily relies on optical images that are formed using reflected light from objects. Therefore, understanding the acoustics of typical sources in our living environment can potentially inform us about the type of signals that hearing has to deal with. However, much of hearing science was established using observations and insights obtained from a number of mathematically idealized and primitive stimuli that rarely (or never) occur in nature: pure tones, clicks, white noise, complex (harmonic) tones, and to a lesser extent, amplitude- and frequency-modulated tones. Judging from the prevalence of these stimuli in experiments in mammals (perhaps except for bats), it may be naively concluded that pure tones and harmonicity are common, that modulation is a relatively special feature in natural sounds, and that white noise is a common type of noise. The reality is not so clear-cut, though. Of course, nobody has made explicit claims that these stimuli are literally representative of realistic acoustics, and in fact, there has been a greater push in recent years for employing more complex stimuli and creating heightened realism in the laboratory. But the legacy of the mathematically simple stimuli still dominates the field.

The potential misrepresentation of the acoustic world in auditory experimentation is not the most problematic implication that these stimuli have on our understanding of the hearing system. Rather, it is the idea that signals can contain no modulation information. Real sounds have a beginning and an end, which means that they are modulated, even if the modulation appears extremely slow or aperiodic. Furthermore, in many signals, amplitude modulation and frequency modulation happen concurrently. Therefore, an acoustic source can become an object of hearing only through modulation---only by forcing it into vibration. Acoustic objects can become meaningful only if we consider these two necessary constituents of sound: carriers and modulators. Information transfer as sound requires both domains to be present. 

The goal of this chapter is to provide a counter-narrative to the classical textbook approach of the ideal acoustic stimuli. At least a subset of the facts that are included in this overview are going to be familiar to readers with background in acoustics---only not in the way that they are brought together here. We will generally attempt to show that frequency is seldom constant, harmonicity and periodicity are rare, dispersion is common, and modulation of all kinds is ubiquitous. This will be illustrated using available examples from literature. After a short introduction that provides universally applicable tools to mathematically represent waves, the chapter proceeds to cover aspects of the acoustical sources themselves, as well as their acoustical environment. The conclusion is that the most general representation of a broadband acoustical sound is also the most suitable one: carrier waves modulated by complex envelopes. The changes incurred by the environment are most generally understood as changes to the complex envelope, but they may also lead to the stochastic broadening of the carrier. 

\section{Physical waves}
\label{PhysicalWaves}
Intuition into many fundamental problems in acoustics and hearing comes from linear wave theory and with it, Fourier analysis. Due to the equivalence between the spectral and the temporal domain representations, the linear perspective tends to be heavily reliant on the spectral nature of the solutions, which is most suitable for stationary signals. In reality, though, hearing deals with nonstationary signals. Using Fourier analysis, signals such as frequency-modulated tones that elicit pitch change over time, require broadband representations that do not correspond well to perceptual insight, even if they are mathematically correct (e.g., \citealp[pp. 383--395]{Zverev}; see Figures \ref{defocused} and \ref{OffFreqFM} for examples of the Fourier spectra of frequency-modulated signals). This divide has required analytical tools that allow the spectrum to change over time, which were sometimes imported from time-frequency analysis or communication signal processing, and have been gradually incorporated and standardized in hearing theory. While this development enabled more freedom in accounting for the sensation of signals that vary both in frequency and in time, it has widened the gulf between classical acoustical theory, where much of the intuition lies, and the physical and perceptual reality. 

This section relies heavily on an approach that was crystallized by \citet{Whitham}, but has antecedents in \citet{Havelock1914} and \citet{Lighthill1955I}. We will use this approach to create a firm link between the signal representation that is more appropriate for temporal analysis and the acoustic waves in the physical world. 

\subsection{Linear analysis}
\label{LinAnaDisp}
Waves describe a very broad class of physical phenomena, which include acoustic, elastic, and electromagnetic fields, among many others. Incidentally, the simplest problems of these three fields are also described by the same hyperbolic differential equation---the homogeneous wave equation, which in three dimensions is
\begin{equation}
	\nabla^2 \psi - \frac{1}{c^2}\frac{\partial^2\psi}{\partial t^2} = 0
\label{WaveEq}
\end{equation}
where $\nabla^2$ is the Laplacian operator, which in Cartesian coordinates and three dimensions is $\nabla^2 = \frac{\partial^2}{\partial x^2} + \frac{\partial^2}{\partial y^2} + \frac{\partial^2}{\partial z^2}$. Then, $\psi$ is some field function---e.g., pressure or velocity potential in acoustic waves, displacement in elastic waves, and electric and magnetic fields in electromagnetic waves. $c$ is the propagation speed of the wave in the medium. A simple change of variables leads to the general solution of the wave equation 
\begin{equation}
	\psi(x,t) = f(x - ct) + g(x + ct)
\label{WaveSolutions}
\end{equation}
we use the scalar one-dimensional equation for simplicity, but the results are easily generalized to three dimensions. The solution is therefore a superposition of two waves going in opposite directions. In fact, these two waves are the solutions to simpler differential equations that can be obtained by factoring Eq. \ref{WaveEq} into
\begin{equation}
	\left(\frac{\partial}{\partial x} - \frac{1}{c}\frac{\partial}{\partial t}\right)\left(\frac{\partial}{\partial x} + \frac{1}{c}\frac{\partial}{\partial t} \right)\psi = 0
\label{WaveEq2}
\end{equation}
The forward-propagating wave is therefore represented by the second of these two first-order differential equations
\begin{equation}
	\frac{\partial\psi }{\partial x} + \frac{1}{c}\frac{\partial \psi}{\partial t} = 0
\label{WaveEq3}
\end{equation}
A solution for this linear equation is 
\begin{equation}
	\psi(x,t) = a e^{i(\omega t - kx)}
\label{wavesol}
\end{equation}
where the angular frequency $\omega$ and the wavenumber $k$ are real constants and the complex amplitude $a$ are all determined by the initial and boundary conditions. It is implied that only the real part of the solution is used in physical problems, although the imaginary may be used just as well. From the solution, we have the speed of sound, or the \term{phase velocity}, which is defined as the ratio between the temporal (radial) and the spatial frequencies 
\begin{equation}
	c = v_p = \frac{\omega}{k}
\label{phasevel}
\end{equation}
However, since the realistic physical medium in which the wave propagates is generally nonuniform, the phase velocity depends on the frequency. Then, as the spatial and temporal frequencies are interdependent, their relation can be expressed using either one of the two complementary forms of the \term{dispersion relations}\footnote{More precisely, Eqs. should be considered to be the \term{dispersion formula}, rather than the more general integral transformations that are implied by the dispersion relations and are due to causality constraints \citep[footnote 13, p. 46]{Nussenzveig}.}
\begin{equation}
	\omega = \omega(k) \,\,\,\,\,\,\,\,\,\,\,\,\,\,\,\,\, k = k(\omega)
\label{disprel}
\end{equation}
General solutions to the linear wave equation and related problems can then be obtained in the frequency domain using the Fourier integral, so that
\begin{equation}
	\psi(x,t) = \int_{-\infty}^\infty F(k) e^{i\left[\omega(k)t - kx \right]} dk
\label{FourSol}
\end{equation}
where $F(k)$ is a function that can be determined from the boundary and initial conditions. This approach often results in series of solutions, or \term{modes}---each of which is associated with a specific combination of $\omega$ and $k$. The superposition of all the modes gives rise to the (full-spectrum) wave shape in the time-domain. Many of the famous problems in acoustics have been solved using this and related methods, which result in series of modes---sometimes with harmonic dependence.

When the propagation is composed of several waves of different frequency pairs $(\omega_n,k_n)$, it becomes meaningful to divide it into a fast-moving carrier and a slowly-varying envelope (or modulation). The simplest illustration of this procedure is given by the superposition of two waves of equal amplitudes and proximate frequencies, so that $\omega_1 = \omega_c + \Delta \omega$, $\omega_2 = \omega_c - \Delta \omega$, $k_1 = k_c + \Delta k$, and $k_2 = k_c - \Delta k$. The two frequencies beat as \citep[\S 191]{Rayleigh1945}
\begin{equation}
	\psi(x,t) = a \cos(k_1 x - \omega_1 t) + a \cos(k_2 x - \omega_2 t) = 2a\cos(\Delta \omega t-\Delta kx )\cos(\omega_c t - k_c x)
\label{FourSol}
\end{equation}
The high-frequency part of the wave, the carrier $(\omega_c,k_c)$, moves at phase velocity, $v_p=\omega_c/k_c$, whereas the low-frequency envelope moves at a velocity
\begin{equation}
	v_g(k) = \frac{\Delta \omega}{\Delta k} 
\label{groupvel}
\end{equation}
where $v_g$ is called the \term{group velocity}.  In the limit of small frequency and wavenumber differences, $v_g$ can be replaced with derivative
\begin{equation}
	\underset{\Delta\omega,\Delta k = 0}{\lim} v_g(k) = \frac{d\omega}{dk}
\label{groupvel2}
\end{equation}
As it turns out, this definition holds in general and can be derived in a number of different ways and not necessarily through beating \citep{Brillouin1960,Lighthill,Whitham}. 

\begin{figure} 
		\centering
		\includegraphics[width=1\linewidth]{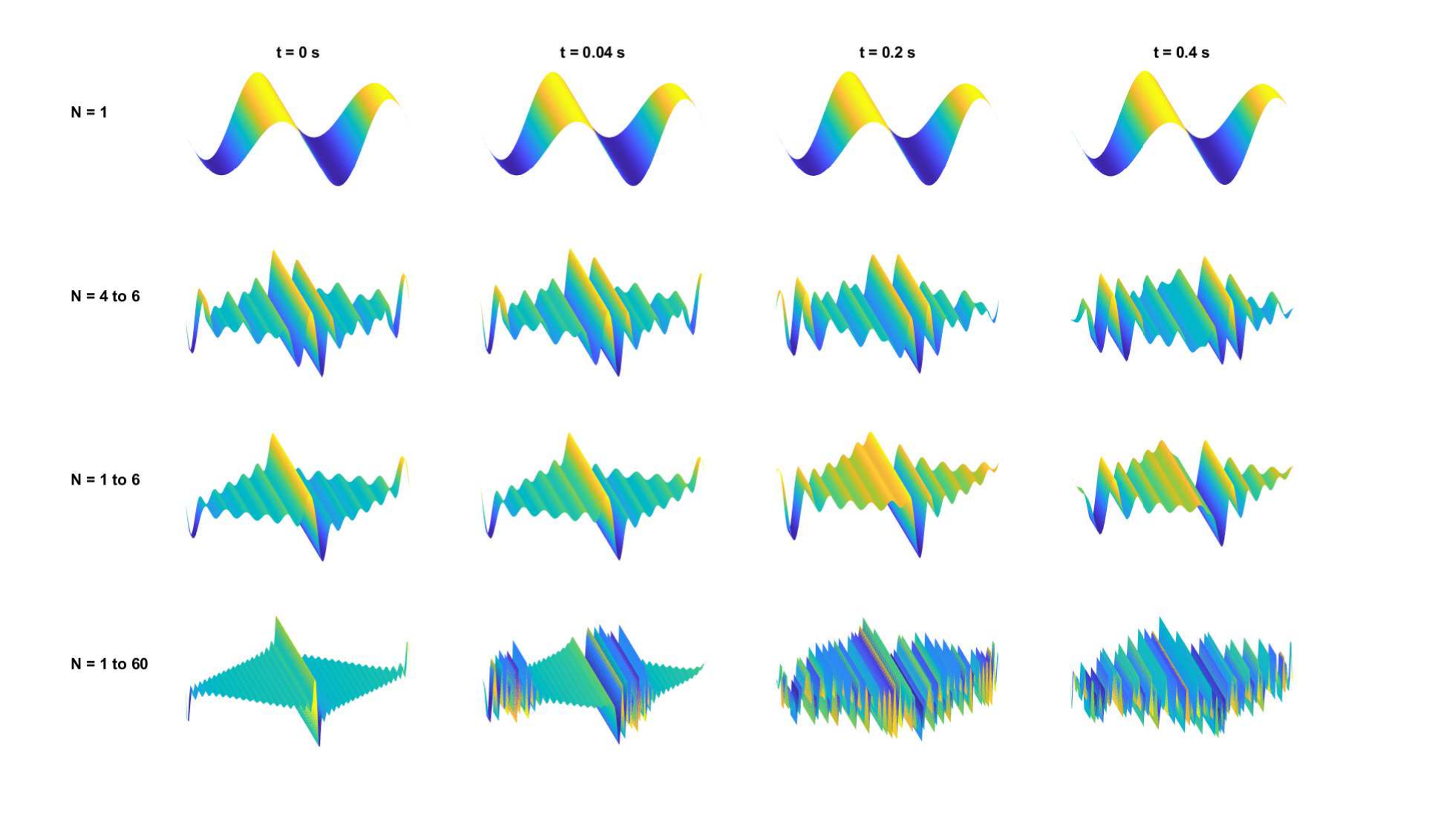}	
		\caption{The effects of dispersion on the spatial acoustic field on (from top to bottom) pure tone, amplitude-modulated tone (200\% depth), complex tone with six components, and a complex tone with 60 components. All components (marked with $N$) were summed with zero initial phase. Four conditions were computed corresponding to different time points measured in the same area, from left to right, at 0, 40, 200, and 400 ms. The dispersion relation is of the form $\omega(k) \sim k^2$, which describes a thin plate \citep[pp. 76--77]{FletcherRossing} set with the approximate properties of steel of 1 mm thickness, bulk modulus of 100 GPa, and density 8000 kg/$\m^3$. The fundamental frequency is 100 Hz. All waveforms were normalized to maximum amplitude of 1, for clarity.}
		\label{dispXY}
\end{figure}

\begin{figure} 
		\centering
		\includegraphics[width=1\linewidth]{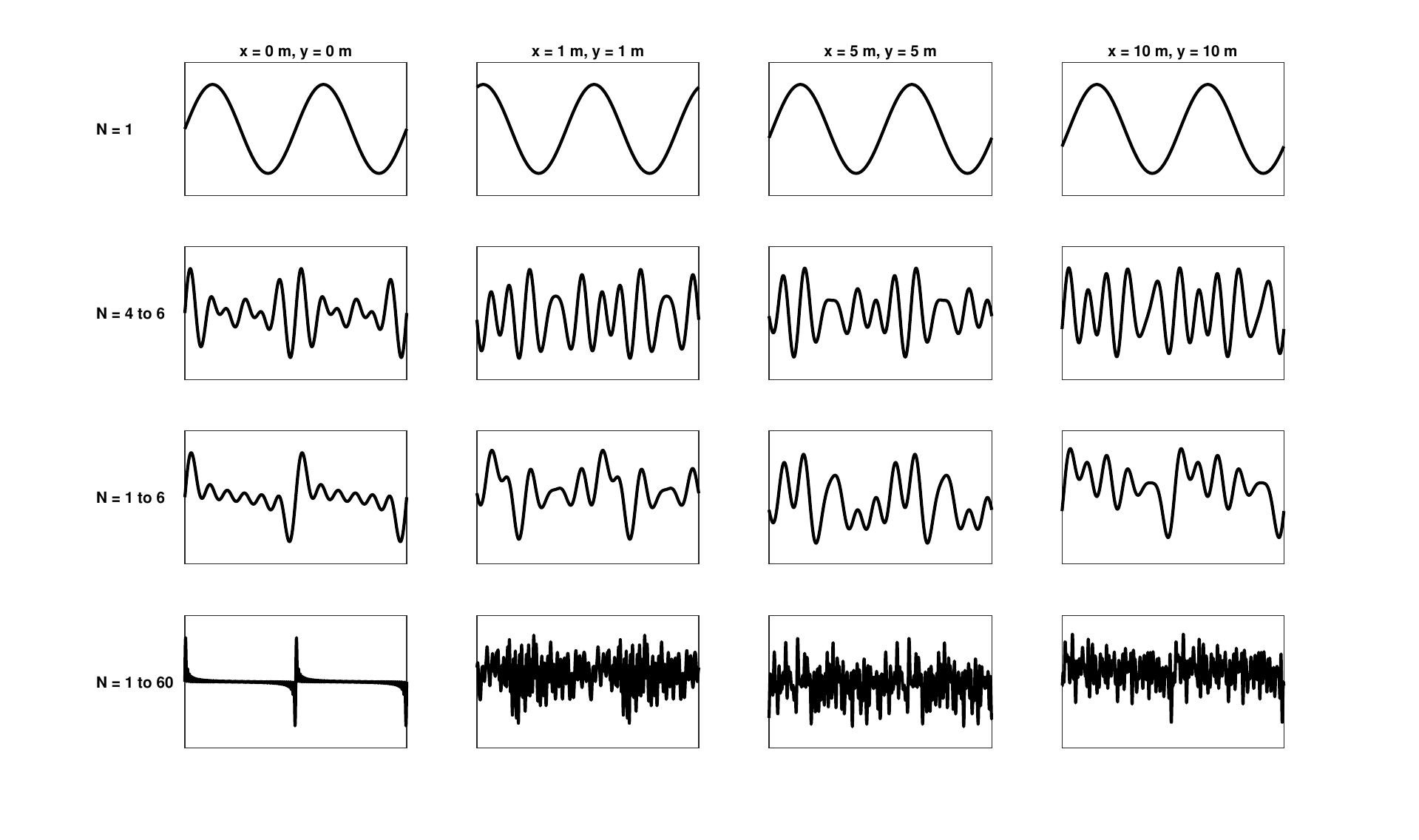}	
		\caption{Similar to Figure \ref{dispXY}, but displaying the time signal measured at different points on the X-Y plane. The dispersion relation of Figure  \ref{dispXY} was inverted so that $k(\omega) \sim \sqrt{\omega}$.}
		\label{dispT}
\end{figure}

In uniform, isotropic and linear systems, $v_g = v_p$ and the system is \term{dispersionless}. More generally, though, all physical media are dispersive, so $v_g \neq v_p$. As the shape of the wave propagation is determined by its envelope, it becomes distorted through propagation in dispersive media, as the different phases that give rise to the envelope shape become misaligned far away from the source of oscillation. Dispersive spatial and temporal effects are illustrated in Figures \ref{dispXY} and \ref{dispT} for dispersion relations of the form $\omega(k) \sim k^2$ and $k(\omega) \sim \sqrt{\omega}$ that characterize vibrations in thin plates.

Dispersion and nonlinearities of the field give rise to more complex wave equations even in the simplest problems. For example, Eq. \ref{WaveEq3} becomes quasilinear as the velocity $c$ is indirectly dependent on the field itself
\begin{equation}
	\frac{\partial \psi}{\partial t} + c(\psi)\frac{\partial\psi }{\partial x} = 0
\label{WaveEq4}
\end{equation}
This equation applies to a wide range of wave problems that are not necessarily linear (or even directly relevant in acoustics). However, the concept of dispersion holds for all wave problems, including those that are represented by other differential equations. It can be shown that the dispersion relation of a problem contains the same information as in the differential equation itself. 


\subsection{Dispersion analysis}
\label{DispersiveAn}
While Fourier analysis is limited to linear systems, the concepts of dispersion and group velocity are applicable in nonlinear problems as well. A universal solution form, which is suitable for dispersion problems and for general nonlinear systems, allows both $\omega$ and $k$ to locally vary in space and time. Consider a wave that has a well-defined amplitude $a(x,t)$ and phase $\varphi(x,t)$ that can be expressed using
\begin{equation}
	\psi(x,t) = \Re \left[ a(x,t) e^{ i\varphi(x,t)}\right]
\label{generalwave}
\end{equation}
We consider solutions with a constant amplitude and a phase function that varies in time and space
\begin{equation}
	\varphi(x,t) =  \omega t - kx
	\label{generalphase}
\end{equation}
Differentiating the phase, we have
\begin{equation}
	k(x,t) = -\frac{\partial \varphi}{\partial x}  \,\,\,\,\,\,\,\,\,\,\, \omega(x,t) = \frac{\partial \varphi}{\partial t}
\label{KandOmega}
\end{equation}
Differentiating these expressions again yields
\begin{equation}
	\frac{\partial k(x,t)}{\partial t} = -\frac{\partial^2 \varphi}{\partial x \partial t}  \,\,\,\,\,\,\,\,\,\,\, \frac{\partial \omega(x,t)}{\partial x} = \frac{\partial^2 \varphi}{\partial t \partial x}
\label{KandOmega2}
\end{equation}
Summing the two equations then results in
\begin{equation}
	\frac{\partial k(x,t)}{\partial t} + \frac{\partial \omega(x,t)}{\partial x} = 0
\label{WaveConservation}
\end{equation}
Using the first dispersion relation in \ref{disprel} with Eq. \ref{WaveConservation} gives
\begin{equation}
	\frac{\partial k(x,t)}{\partial t} + v_g(k) \frac{\partial k(x,t)}{\partial x} = 0
\label{WaveConservation2}
\end{equation}
where the dependence on $k$ of the group velocity is now made explicit in $v_g(k)$. We can also use the second dispersion relation in Eq. \ref{disprel} to get a similar equation to \ref{WaveConservation2} using $\omega$ instead of $k$
\begin{equation}
	v_g(k) \frac{\partial \omega(x,t)}{\partial x} + \frac{\partial \omega(x,t)}{\partial t} = 0
\label{WaveConservation3}
\end{equation}
where the second term is the local frequency, frequency velocity, or frequency slope, which is introduced in Eq. \ref{GeneralizedPhase} and is characteristic of frequency modulation. Equations \ref{WaveConservation2} and \ref{WaveConservation3} have the same form of the first-order hyperbolic equation as \ref{WaveEq4}, although they were obtained in a different way. These equations take the universal form of conservation laws, which in this case was referred to as \term{wave conservation} by \citet{Whitham}. In analogy to other conservation laws, $k$ can be thought of as the flow of the wave and $\omega$ as its flux. 

In dispersionless systems, the wave propagation is uniform and both $k$ and $\omega$ are space-invariant and time-invariant, so their partial derivatives in Eqs. \ref{WaveConservation2} and \ref{WaveConservation3} are zero and the equations become trivial. However, in dispersive media and nonlinear systems they convey information about the time dependence of the waves in the system, which may not be accessible using constant frequencies. 

As the ears work largely as point receivers that detect acoustic waves that are one-dimensional throughout most of the audio range (see \cref{outerear}), signal processing of sound can factor the wavenumber $k$ as a constant phase term, $e^{ikx_0}$ at a point $x_0$. In dispersionless systems, this phase is linear in frequency and produces a delay $e^{i\omega x_0/c}$. Drawing from filter theory \citep[pp. 66--67]{Zverev}, we refer to this phase as \term{phase delay}, which is generally given by
\begin{equation}
		\tau_p = -\frac{\varphi}{\omega} = \frac{k(\omega)x}{\omega}
\label{GroupDelay1}
\end{equation}
for phase of the general form of Eq. \ref{generalphase} that excludes the usual $\omega t$ term. In signal processing of time signals, the wavenumber $k$ is not considered directly---only the frequency. In the present treatment, considering the spatial dependence too, results in $\tau_p = x/v_p = t$ for a linear dispersionless system. 

Similarly, it is also useful to quantify the deviation between the linear phase delay and any higher-order phase dependence on $\omega$, which can be indicative of dispersion. Most generally, we define the \term{group delay} with respect to the phase as\footnote{For various derivations of this important formula, see \citet{Ville1948} and \citet{Boashash1992}.}
\begin{equation}
		\tau_g = -\frac{d\varphi}{d\omega}
\label{GroupDelay1}
\end{equation}
Using the same general phase and the dispersion relation of $k(\omega)$ we get
\begin{equation}
		\tau_g = x\frac{dk(\omega)}{d\omega} = \frac{x}{v_g} 
		\label{GroupDelay2}
\end{equation}
which can provide more insight if written as
\begin{equation}
		\tau_g =  \frac{v_p}{v_g} t
\label{GroupDelay3}
\end{equation}
Therefore, in the dispersionless case where $v_g = v_p$, the group delay is equal to the phase delay, which is another way to say that the modulation that is being carried by the wave retains its shape---it is invariant to spatial and temporal shifts. In all other cases, the group delay quantifies the amount of dispersion, which grows the farther away the wave is from the origin and when the two velocities are markedly different. In normal dispersion (as opposed to anomalous dispersion) the group velocity is lower than the phase velocity, and we obtain a positive group delay \citep{Brillouin1960}. 

The group delay is especially useful in a narrowband range of frequencies, for which the carrier and modulation frequencies are well-separated. With sufficiently narrow bandwidth, we are able to treat the group delay as approximately constant, even in dispersive systems. In broadband, dispersive media, and nonlinear systems, the group delay generally varies with frequency and becomes less linear the farther away it is from the carrier. This reasoning can be inverted and provide a useful operational definition for the rather vague ``narrowband condition'' that we highlight throughout \cref{PhysicalSignals}: a narrowband range of frequencies is taken such that the group delay changes only a little from its mean value at the center frequency of the band (the carrier). 

Note that if we let the phase be complex, this general wave analysis can be expanded to encompass the variable amplitude $a(x,t)$ as well, which can be insightful in absorptive systems \citep{Vakman1997a}. However, as most of this work is concerned with phase and dispersion, the explicit consideration of absorption will be relatively secondary. 

\subsection{Conclusion}
This very brief introduction presents a universal approach to the representation of almost any wave, using a generalized phase function. Essentially, we have two approaches that hinge on the fundamentally different definitions for frequency that are discussed in \cref{FreqInstFreq}---the Fourier frequency and the instantaneous frequency. The two are the same only in the simplest of cases, which can be considered approximately linear. Even if they yield mathematically identical solutions, each carries a different insight with it that may be incongruent with the other. When the acoustic system is nonlinear, but is still spectrally analyzed using Fourier analysis, we risk not only losing insight of the physics of the problem, but also downright misrepresenting its nature. Usually, the acoustic conditions are close enough to linear that we can retain a certain flexibility in switching back and forth between the Fourier and the instantaneous representations, depending on the problem. This will enable us to reexamine some of the familiar acoustic problems that are most relevant to our hearing world. 


\section{Acoustic sources}
\label{AcousticSources}
Three general mechanisms can cause acoustic wave generation in a closed region of fluid: solid body forces that create pressure gradients in the fluid, injection or removal of material from the region, and flow within the fluid that creates turbulence \citep[pp. 140--142]{Kinsler}. Complex acoustical sources such as human speech may contain elements of all three. The few examples that are mentioned below are taken from the simplest and best studied sources, which illustrate that even simple acoustics is susceptible to ``ill-behaved'' phenomena that may be naively associated only with much more complex acoustics. 

\subsection{Primitive sound sources}
\label{PrimitiveSources}
Harmonic intervals are characterized by integer ratios of their constituent fundamental frequencies\footnote{Note that the word ``harmonic'' is used in two related meanings in the wave physical literature. Harmonic dependence entails that the temporal dependence of the solution goes as $e^{i\omega t}$. Harmonic intervals or sounds assumes that a few such solutions have natural frequencies that are related by integer ratios. In the present work, unless we refer specifically to a harmonic solution, harmonic should always be understood in the second meaning.}. The acoustic source itself may be considered harmonic if it produces an overtone series that is inherently harmonic (and hence periodic)---something that is captured in the popular complex tone stimulus. Such are the overtone series of the ideal string and the resonance series of the air column (pipe). However, arbitrarily shaped structures do not generally produce harmonic overtones when they vibrate. As harmony is a pillar of music that underlies consonance, it had to be progressively engineered into musical instrument design over generations. Therefore, the comprehensive analysis of acoustic sources in musical acoustics may be ideally-suited to identify ``well-behaved'' sources that have mathematically convenient properties such as harmonicity.

This section is heavily based on \citet{FletcherRossing}, who systematically reviewed the physical acoustics of musical instruments. Many of the basic problems have been compiled by \citetalias{Rayleigh1945}, but modern observations and modeling have considerably supplemented the classical (and sometimes idealized) solutions with a degree of realism.  

\subsubsection{Solid objects}
Vibrating objects that are positioned in a fluid medium generate an acoustic field, which is determined by the object structure and its mechanical properties. The vibrations of solid objects can be shown to consist of \term{normal modes} of oscillation (also called \term{eigenmodes}), which are mathematically orthogonal (independent). Each mode is spectrally characterized by a \term{natural frequency} (also \term{eigenfrequency}) and, spatially, by a geometrical pattern of vibration (\term{eigenfunction}). Temporal characterization is divided into transient response (in terms of damping constant and decay time), and a steady-state time dependence that is generally taken to be sinusoidal. Depending on how and where the object is excited, combinations of the normal modes with different weights can be observed. Therefore, without loss of generality, the vibration of the object in the coordinate system of the object itself can be represented by
\begin{equation}
	\eta_{nml}(x,y,z,t) = \sum_{n=0}^{\infty}\sum_{m=0}^{\infty}\sum_{l=0}^{\infty}{a_{nml}\Psi_{nml}(x,y,z) e^{i\omega_{nml} t}}
	\label{NormalModes}
\end{equation}
where the displacement from equilibrium $\eta$ varies in time and space, here in Cartesian coordinates. The most general case is given, which is three-dimensional with no special symmetry. Three degrees of freedom correspond to the three dimensions (e.g., of a thick plate), where each one is associated with an integer $n$, $m$, and $l$. The eigenfunction is given by $\Psi_{nml}(x,y,z)$ with its respective eigenfrequency $\omega_{nml}$, which is factored into the steady-state harmonic dependence. The weight of each mode in the sum is given by the complex amplitude $a_{nml}$.

Only systems with very simple geometries can be studied in closed-form---strings, bars, membranes, and plates are the primary ones among them. Typically, the normal modes of acoustic sources are studied spectrally---by noting prominent peaks in their spectrum---and mapping them to corresponding eigenfunctions. Most research characterizes the sources based on their steady-state response. Some instruments, such as the grand piano, have an idiosyncratic transient response that has been studied in depth, which showed the temporal envelope of each mode, as well as a distinct attack (onset) noise from the hammer action \citep[pp. 390--396]{FletcherRossing}.

The simplest vibrating solid object is the string. The ideal string is one-dimensional and its normal modes of vibration are exactly those studied by Pythagoras and found in the Fourier series solution of the string equation (Ibid., pp. 39--44). This string requires perfectly rigid support at its ends and zero bending stiffness. If these requirements are relaxed, the string overtones tend to deviate from harmonicity, as they are no longer integer multiples of the fundamental. So, if the end supports of the string can move, then the overtone series ratios become more compressed than the ideal integer overtone series (Ibid., pp. 52--53). Or, when the bending stiffness of the string (its two-dimensional cross-sectional elasticity) is taken into account, the overtone series become stretched (Ibid. 64--66). 

The string can be mathematically extended into a two-dimensional thin bar or membrane. Like strings, bars vibrate harmonically in their longitudinal modes (Ibid., pp. 56-57), yet they vibrate inharmonically in their much more important transverse modes (Ibid., pp. 58--63). The ideal rectangular thin membrane also contains harmonic overtones, but they co-occur with inharmonic ones as well. If air loading, bending stiffness, or stiffness to shear is introduced to the membrane, or if the geometry is non-square (e.g., circular membranes) the overtone series can become completely inharmonic. Other shapes that have been modeled such as plates (thick membranes) and shells generally have inharmonic modes, even when their geometry is relatively symmetrical. 

\subsubsection{Modal dispersion}
The dispersion of the normal modes is well-studied in many vibrating objects, which entails that the speed of vibrations in the different modes within the object depends on their respective eigenfrequency. Simple dispersive sources are the stiff string, the plate (Figure \ref{dispXY}), and transverse waves in bars \citep[pp. 59--60, 65--66, and 77]{FletcherRossing}. Therefore, an envelope traveling within the vibrating object does not retain its shape in the object area and over time. The simplest case study for envelope dispersion is a single pulse, as its shape is dependent on the relative timing of the superimposed modes. This means that when an object is impacted---forced by an impulse somewhere on its surface, rubbed, broken, deformed---its response is not only affected by its unique vibrational behavior, but also by how and where it was impacted. As an impulse excites all modes in the structure, the transient pulse shape will necessarily depend on where it is measured in the field surrounding the object in the medium\footnote{Interestingly, as noted by \citet{Brillouin1960}, dispersion in bars was analyzed first by Lord Rayleigh, which led him to formally talk about group velocity \citep[\S 191]{Rayleigh1945}, although the concept was originally introduced by \citetalias{Hamilton} without calling it by name.}. 

\subsubsection{Resonators}
Complex vibrating objects are often systems that consist of simple solid components, which are coupled to additional structures that vibrate and resonate. This coupling may accentuate and sometimes shift the modes of the standalone oscillator \citep[pp. 102--132]{FletcherRossing}. Many musical instruments employ resonant tubes and cavities, which are also commonly found in animal vocalization systems, and are of particular importance for modeling the vocal tracts of mammals and birds. 

Pipes, like strings, are in good approximation one-dimensional at low frequencies where plane waves are the only normal mode that can propagate and carry sound in the pipe. An ideal pipe has a nearly harmonic series of resonances that depend on its length and termination (open or closed). The resonance frequencies tend to be slightly stretched due to reflections from the pipe ends that increase with diameter (Ibid., pp. 196--205). Other realistic deviations from harmonicity can take place when the walls of the pipe are not hard but yielding, which means that they have a finite (reactive) impedance (Ibid., pp. 202--205). Additionally, air has a finite viscosity near the walls, which causes viscous losses (mostly during flow) that affect the modes as well (Ibid., pp. 193--196). Also, thermal losses may be caused at the boundaries through compression, as the thermal conductance of the boundary material tends to be much higher than air \citep[pp. 290--292]{Morse}. Pipes shaped as conical horns can also exhibit nearly harmonic modes---in particular conical and compound (Bessel) shaped horns \citep[216--218 and 461--464]{FletcherRossing}. At high frequencies, additional propagation modes can exist in the resonators \citep[pp. 492--498]{Morse}. Each mode has a cutoff frequency below which energy cannot propagate in the mode, as well as a characteristic spatial distribution and phase velocity associated with it, making the pipe dispersive. 

Pipes have been mostly analyzed in the spectral domain, which provides steady-state solutions for harmonic inputs. The computation and measurement of transient signals is much more challenging and reports have been scarce. For example, a model of the attack transient of a flute (simple air column with holes) showed that it takes the instantaneous frequency about 5 ms to settle to a steady-state value to produce notes at 1000--1500 Hz \citep{Keefe1990}.  

\subsubsection{Mode-Locking}
Despite the inherent inharmonicity of many of the vibrational systems mentioned, inharmonic modes can sometimes phase-lock to produce precise harmonic oscillations \citep{Fletcher1978}. There are several conditions that have to be met for this to take place. First, the modes have to be nearly harmonically related with simple ratios of small integers $n/m$ (for $n+m < 4$). Additionally, the modes should be strongly coupled and driven in large amplitudes by a nonlinear force. These conditions are relatively restrictive and the prevalence of mode-locking in natural acoustic sources is unknown. Outside of the realm of musical instruments, mode-locking was measured in parts of the zebra-finch song, which suggests that the dynamics of their syrinx can be directly responsible for mode-locking \citep{Fee1998}.

\subsubsection{Airflow generators}
There are two primary airflow sound generation mechanisms that are employed in various musical instruments. The first is buzzing a valve (such as a reed or lips), as in saxophone, harmonica, and trumpet, as well as in the human vocal folds and the bird syrinx. As these generators are often coupled to an air column in a pipe, the combined system can give rise to harmonic sounds, which are excited by the nonlinear dynamics of the reed that modulates the airflow in the pipe \citep[pp. 418--424]{FletcherRossing}. The second mechanism generates jet airflow over a sharp edge as in whistles, flutes, and organs (Ibid., pp. 418--424). This mechanism is also nonlinear and is responsible for creating broadband noise when the jet flow is strong enough to cause turbulence, which is then shaped (filtered) by the resonances of the pipe (Ibid., pp. 528--529).

\subsubsection{Stochastic sources}
Completely fluid sound sources (e.g., water waves, waterfall, wind blowing) or those generated by large ensembles of similar units (e.g., walking on gravel, audience hum, rain) are very common and may also generate unique sounds \citep[pp. 158--160]{Schafer}. These sounds are broadband and stochastic and do not have any stable vibrational modes that are associated with specific objects. Therefore, the analysis of turbulent and other complex sounds is generally done using stochastic tools \citep[pp. 768--772]{Morse}. Friction between objects is another common sound source that comes across as noise, although it may be a result of complex excitation of different harmonic modes that are generally not harmonic \citep{Serafin2004}. A qualitative taxonomy of everyday acoustic sources, based on their physical state and the type of force or function that excites them can be found in \citet{Gaver}.

\subsection{Speech and other animal sounds}
\label{SpeechAnimals}
Human speech and animal vocalizations constitute complex acoustic sources that can be modeled as several basic components coupled together to generate sound. Two influential theories were developed with relation to human speech (vowel production, in particular). The most influential speech model is the \term{source-filter theory} (originally introduced by \citealp{Chiba}, according to \citealp{Arai2004}; \citealp{Fant1970}). It states that the harmonically-rich vocal-cord oscillations are independently shaped by the vocal tract resonances, which act as filters that endow the various vowels with their characteristic timbre. According to the \term{myoelastic-aerodynamic theory}, which complements the source-filter theory, sound is generated by aerodynamic energy that produces a jet airflow from the lungs and through the larynx, where it is converted to acoustic oscillations by the vibrating vocal cords \citep{denBerg1958}. 

Detailed mechanical and acoustical models have been further introduced to describe speech production acoustics. \citet[pp. 55--126]{Stevens} listed four specific mechanisms for sound production in speech---two of them were mentioned with respect to musical instruments---vocal-cord periodic modulations (similar to musical reeds), turbulence generation variations close to constrictions, sudden release of air from a pressurized cavity to the vocal tract (a short transient that is part of stop consonant production), and inward air suction using the tongue for constriction. The vocal cavities themselves are typically modeled as one-dimensional pipes with hard walls, where only plane waves can propagate. The associated filtering by the cavities is modeled as time-invariant linear functions, effective per specific articulation configuration. This configuration is determined by the size of the compartments (e.g, pharynx and the mouth cavity) and how they are connected among each other and to the nasal cavities. The specific resonances of the vocal cavities (\term{formants}) shape the rich harmonic spectrum coming from the larynx. The acoustic coupling between the different cavities, the effect of end-corrections, realistic wall properties, and perturbations to the cross-section of the pipes---all shift the resonances away from the natural harmonic series, although often not dramatically \citep[pp. 127--202]{Stevens}. 

Vocalizations of most mammals and birds are largely based on organs similar to humans, to the extent that the myoelastic-aerodynamic and source-filter theories can be applied just as well as they are for humans \citep{Suthers}. There are also many exceptions, such as songbirds and toothed whales, which have a dual noise generation capability in their larynx/syrinx, and other animals that use highly nonlinear and even chaotic generation of sounds using the same mechanics \citep{Suthers,Herbst}. Simpler animals like some anthropods can mechanically rub body parts to produce sounds (\term{stridulation}), whereas others like certain species of fish may not even be capable of producing sound. 

The standard source-filter theoretical view of speech production has been challenged over the last decades, with the growing understanding that the transient nature of real speech signals cannot be satisfactorily accounted for by time-invariant linear filters \citep{Teager1990}. Nonlinear jet airflow speeds that were measured in the vocal tract suggested that vortices (turbulence) significantly contribute to the total speech energy, which entails a mechanism that is governed by the fluid-mechanics and not only by the acoustical dynamics \citep{Teager1990, Barney, Shadle, Sharma2017}. 


\subsection{Complex source modulation}
\label{ComplexSourceMod}
The understanding of how the sound produced by acoustic sources is shaped would be incomplete without considering the role of modulation, which is known to be rife in natural sound sources, both as amplitude and as frequency modulations \citep{Attias1997,Singh2003}. 

An important example of these effects has come to recent attention in speech modeling. In practical modeling of recorded speech signals, the inherent transience of real-world vocalizations is not adequately captured by any of the models mentioned in \cref{SpeechAnimals} \citep{Sharma2017}. From the signal perspective, animal (including human) vocalizations can be reduced to fundamental building blocks that vary dynamically: amplitude modulation, downward or upward glides (frequency modulation), broadband noise, constant (periodic) frequencies, and (near) pure tones \citep{Klug2010}. \citet{Moody1989} noted that when animals frequency-modulate their calls, it is always done gradually as a sweep and never in discrete steps. Species-specific vocalization systems have been shown to be matched by auditory systems that can be specifically tuned to receive the acoustic building blocks relevant to this particular vocalization, and thereby achieve processing efficiency \citep{Casseday1996,Klug2010, Theunissen2014}. These sonic elements, at least in speech, are most effectively analyzed using time-domain methods that can extract the instantaneous envelope and frequency of the signals and avoid smearing effects that are involved in strictly frequency-domain methods \citep{Huang2009, Sharma2017}. Therefore, practical analysis of realistic speech applies somewhat generic solutions of time-frequency techniques, which decompose broadband signals to different modes that change in time---modes that are dynamically amplitude- and frequency-modulated (AM-FM). The superposition of all modes yields a representation of the complete broadband signal $s(t)$ \citep{Sharma2017}
\begin{equation}
	\hat{s}(t) = \sum_{n=1}^N a_n(t)\cos\left[ \omega_n t + \int_0 ^t m_n(\tau)d\tau + \varphi_n \right] 
	\label{SpeechAMFM}
\end{equation}
Here the speech signal is estimated as a multi-component signal of $N$ modes, each of which has its own time-varying envelope $a(t)$ and frequency $m(t)$ and initial phase $\varphi_n$. Ideally, the modes correspond to the harmonics and the formants of speech that are seen also in the static models (e.g., source-filter). In reality, determining the center frequencies $\omega_n$ is the biggest challenge and different algorithms have been proposed to achieve it in the time-domain without resorting to Fourier analysis. Depending on the precise algorithm and signal, there may be a residual signal $e(t) = s(t)-\hat{s}(t)$ that is not fully captured by the finite number of modes. However, in theory, the decomposition in Eq. \ref{SpeechAMFM} overcomes the situation caused by harmonic analysis that generates an infinite series of frequencies for every abrupt discontinuity in the signal\footnote{The abrupt discontinuity can be modeled using a step function that modulates the amplitude of the signal. Using the standard Fourier-transform analysis, the step function $u(t)$ has a hyperbolic continuous spectrum with an infinite support: ${\cal F}\left[ u(t) \right]= \frac{1}{i\omega} + \pi\delta(\omega)$. }. 

We recognize Eq. \ref{SpeechAMFM} as a summation of waves of the form of Eq. \ref{generalwave}, which has time- and space- dependent amplitude and phase functions. While we did not deal directly with time-dependent amplitude in the discussion about dispersion above, it can be incorporated into the same framework if necessary by allowing the wavenumber $k(\omega)$ to be complex. A case will be made later (\cref{RealandComplex}) for employing a complex envelope as a catch-all modulation domain part of an arbitrary narrowband signal. This is a convenience that bundles the slowly varying AM and FM together along with a high-frequency and constant carrier. It is clear that the AM-FM modeling of speech can be applied using complex envelopes for the modes. In fact, the same procedure can be made to correspond to the generic sum of normal modes of Eq. \ref{NormalModes}, as long as the amplitude $a_{nml}$ is made time-dependent in addition to being complex, $a_{nml}(t)$. Of course, at the signal level, the factors that include spatial dependencies in Eq. \ref{NormalModes} can be replaced by constants. The same goes for Eq. \ref{SpeechAMFM}, which can be expressed using a complex envelope that includes both AM and FM. 

There are several mechanisms that can constitute complex modulation in acoustic systems. The inclusion of these mechanisms in the complex envelope is a mathematical convenience, which does not change the physical signal. Three general categories of complex modulation are considered below. Note that in the hearing literature the category of spectral modulation is often invoked as one-half of the important spectrotemporal modulation. Spectral modulations can be the result of reflections that cause interference between the incident and reflected sounds. They are only mentioned in passing in \cref{Reflections}.

\subsubsection{Explicit forced modulation}
The most obvious modulation is also unequivocally presented as such in literature. It is caused by setting the oscillator into vibration using a periodic external force. For example, according to the source-filter theory, the vocal folds modulate the airflow from the lungs at the fundamental frequency \citep{Stevens}. Additionally, the outgoing sound is temporally modulated by the tongue and lips, which also spectrally modulate the sound along with the other cavities in the vocal tract \citep{Plomp1983}. At much lower frequencies, the speech modulation spectrum is widely used in research, mostly referring to the natural amplitude modulation it has, which peaks at around 3--4 Hz \citep{Steeneken1983} and is considerably diminished above 64 Hz \citep{Drullman1994a, Drullman1994b,Singh2003}. A vibrato (frequency modulation) effect in singing is caused by a periodic modulation of the fundamental frequency \citep{Sundberg1995}, similarly to vibrato in string instruments \citep[pp. 317--318]{FletcherRossing}. Also, string instruments (especially the viola and cello) are susceptible to the dreaded ``wolf tone'', which is amplitude modulation caused by periodic low-impedance coupling between the string note and the normal mode of the instrument body (\citealp[pp. 312--313]{FletcherRossing} and \citealp[p. 629]{Chaigne}). Vibraphone notes can be amplitude-modulated (tremolo) upon periodically opening and closing its resonators using an electric motor \citep[pp. 638--639]{FletcherRossing}. As modulation that is not necessarily periodic, siren and guitar string pitch bending were given as realistic examples for frequency modulation in \citet[pp. 22 and 31]{Schnupp}, as are a few of the characteristic sound effects produced by the floating bridge and drum-head assembly of the banjo \citep{Politzer2015}. Finally, echolocating bats use frequency-modulated chirps as their main targeting signal. 

\subsubsection{Implicit modulation}
Several types of signals that are sometimes seen as stationary are amenable to reformulation as modulated sounds. 

The case of mode beating---the interference between two normal modes that are close in frequency---is somewhat problematic. Beating unmistakably behaves as amplitude modulation, but it lacks a physical carrier \citep[pp. 10 and 105]{FletcherRossing}. The standard mathematical solution is to set the carrier at the average frequency of the two mode frequencies with the difference frequency as the modulator (Eq. \ref{FourSol}), although this description may not correspond to how the wave is generated at the source.

Constant tones are another common case. By making the carrier constant and assigning all the amplitude changes to the envelope, we obtain a very broad category of implicit modulations. They are typically overlooked or implicitly attributed to the carrier, despite the inherent ambiguity in this, as they can just as well belong to the envelope (see \cref{RealandComplex}). A pure carrier is a pure tone---it has no beginning and no end---a precondition for keeping its frequency fixed. Therefore, sound onset and offset should count as a segment of a slow and long amplitude-modulated envelope, which includes the particular ramps associated with the onset and offset. 

Nonperiodic changes in amplitude or frequency are also modulations, but with more complex envelope spectra that are not strictly sinusoidal or linear. In his landmark book about the source-filter theory, \citet[p. 18]{Fant1970} commented (emphasis in original): ``\textit{The terms \textbf{harmonic} or \textbf{periodic} are not adequate, from a strictly physical standpoint. Because of the variations always present it would be more appropriate to speak of voiced sound as \textbf{quasi-periodic}}.''

By including phase and frequency modulation effects, this category can be further expanded to all forces that cause some change to the otherwise static sound. For example, realistic vowel sounds contain fundamental frequency modulations (glides) that affect its formants before the vowel becomes static \citep{Hillenbrand2013Morrisson}. 

\subsubsection{Transient response}
Oscillators require a source of energy in order to vibrate. Sometimes it is provided from the outside by a distinctly separate object, and sometimes from within---in complex sources that have additional moving parts or internal sources of energy (e.g., a pendulum clock, a vocalizing animal, a fizzing chemical reaction). The mass, momentum, and material, as well as the area and duration of coupling between the force and the sounding mechanism affect the timbre of sound, often dramatically. Specifically when animal vocalizations (and human speech) are produced, they require coordinated muscle action, which may be periodic, impulsive, gradual, etc. and can instantaneously depend on numerous mechanical and aerodynamic factors in the system. 

The solution to the inhomogeneous wave equation, which describes the forced source motion, contains two parts---a transient solution and a steady-state solution. The transient solution exhibits the normal modes of the oscillator, which inevitably decay. Each mode can be mathematically expressed as the product of a constant carrier and a decaying exponential envelope. The steady-state solution is caused by the external force. An external impulse reveals only the transient response, but all other forces produce more complex responses, which may involve dynamic frequency changes that can be incorporated into the complex envelope.  

Another indirect modulation can take place if the source is moving while producing sound, which can create subtle frequency shifts through the \term{Doppler effect} \citep[pp. 699--700]{Morse}:
\begin{equation}
	f' = \frac{c+v}{\lambda}
\end{equation}
where $f'$ is the Doppler-shifted frequency for an object moving at speed $v$ producing sound with wavelength $\lambda$. Therefore, relative acceleration of the source or receiver can cause effective frequency modulation to a listener at rest. For example, an observer at rest, listening to a vocalization centered at 1000 Hz of an animal moving at an instantaneous velocity of 5 km/h will experience a maximum frequency shift of about 4 Hz. The modulatory effect on the spectrum and other room acoustic parameters that were produced in playing orchestral musical instruments, expressively (in motion), is documented by \citet{Ackermann2024}.

\section{The acoustic environment}
\label{acoustenv}
The acoustic environment contains the medium for the radiated waves from the source. Given an arbitrary acoustic source, we would like to know how its sound is transformed by the time it reaches to a listener at a distance. As in the previous section, this review is highly selective and emphasizes acoustical effects that have received less attention in the context of hearing. The main focus is transient effects and how they distort the acoustic envelope through propagation in air, reflections, and various room acoustic effects\footnote{A few audio demos that demonstrate the aggregate effect of some of the phenomena that are discussed in this section are provided in \textsc{/Section 3.4 - Radiation, dispersion, reflection, and reverberation/}. The demos bring five complex scenes that were recorded in-situ from a far distance (10--1400 m), and thus demonstrate the effects of dispersion and reflections in complex environments and how they interact with the source type. Unfortunately, the recordings were all made using a mobile phone and are therefore of poor quality and in two far-field recordings of relatively low signal-to-noise ratio. The low-frequency content below 100-300 Hz was low-pass filtered to eliminate wind, vehicle, and handling noises. Nevertheless, the recordings may still serve as examples for how sound becomes distorted and decohered over distance---something listeners are all familiar with, but is rarely dealt with directly as stimuli in hearing research.}.

\subsection{Radiation patterns}
Vibrating objects of arbitrary geometrical shapes can produce complex radiation patterns in the medium, which determine the spatial distribution of the acoustic field (both pressure and velocity) away from the object \citep[pp. 306--394]{Morse}. In near-field, the radiation pattern may vary significantly as a function of position, especially if the source shape is not simple (i.e., if it does not possess any symmetry). Solid sources produce normal modes that vibrate with different amplitudes in different parts of the object geometry. Additionally, if the object is modulated by an external force, then their point of contact generally causes some normal modes to vibrate more than others. As most sources are inherently dispersive (\cref{PrimitiveSources}), different elements of their external surface area tend to vibrate with slightly different phase, which varies as a function of frequency. Therefore, sources with multiple oscillating modes tend to sound different from different positions around them.

In farfield, the complexity of the acoustical radiation pattern may be reduced by approximating the sources to point size, or to other theoretically useful types of sources (e.g., dipole, spherical, line). This approximation dispenses with the precise object shape and replaces it with simpler and more symmetrical radiation patterns, compared to  realistic conditions, in which arbitrary three-dimensional objects radiate asymmetrically. There is a practical limit to the usefulness of such approximations, as they tend to get too complex to derive much intuition from when higher-order terms are introduced (i.e., quadrapoles and higher-order terms than dipole). When this happens, the power spectrum of the source can be used alongside statistical tools to model the source radiation \citep[pp. 329--332]{Morse}.

Regardless of the specific geometry and modeling approach, many complex sound fields exhibit approximate plane wave behavior far from the source (i.e., when $kr \gg 1$; see Table \ref{soundvslight}).

\begin{table}
\footnotesize\sf\centering
\begin{tabular}{P{3cm}P{6cm}P{6.5cm}}
\hline
&\textbf{Sound}&\textbf{Light}\\
\hline
Field variables & pressure $p$, velocity $\bm{u}$ & electric field $\bm{E}$, magnetic field $\bm{H}$ \Tstrut\Bstrut\\
Wave speed & $c = \frac{1}{\sqrt{\rho \kappa}}$ & $c = \frac{1}{\sqrt{\mu \epsilon}}$ \Tstrut\Bstrut\\
Characteristic impedance & $Z_0 = \rho c = \sqrt{\frac{\rho}{\kappa}} $  (air) & $ Z_0 = \sqrt{\frac{\mu_0}{\epsilon_0}}$  (vacuum) \Tstrut\Bstrut\\
Scalar wave equation$^\dagger$ (homogenous) & $\nabla^2 p = \frac{1}{c^2}\frac{\partial^2p}{\partial^2t}$ & $\nabla^2 E= \frac{1}{c^2}\frac{\partial^2 E}{\partial^2t}$, $\nabla^2 H= \frac{1}{c^2}\frac{\partial^2 H}{\partial^2t}$ \Tstrut\Bstrut \\ 
\hline
\multicolumn{3}{c}{Plane waves with solutions of the form $\bm{F} = F(\bm{k} \cdot \bm{r}  - ct) + F(\bm{k} \cdot \bm{r}  + ct)$}\Tstrut\Bstrut\\
\hline
Scalar field relations & $p = \rho c u$ & $ E = \sqrt{\frac{\mu}{\epsilon}} H$ \Tstrut\Bstrut\\
Wave energy density & $w = \frac{1}{2}\rho|u|^2 + \frac{1}{2}\kappa|u|^2 = \frac{|p|^2}{\rho c^2}$ & $ w = \frac{\epsilon}{2}E^2= \frac{\mu}{2}H^2$  \Tstrut\Bstrut \\
Intensity (energy flux) & $I = cw = \frac{1}{2}(p^*u + pu^*) = \frac{|p|^2}{\rho c} $ & $|S| = \sqrt{\frac{\epsilon}{\mu}}c w = \frac{\epsilon}{2\mu}E^2 = \frac{\mu}{2\epsilon}H^2 = \frac{1}{2}|E||H|$ \Tstrut\Bstrut \\
\hline
\multicolumn{3}{c}{Far-field monopole (point source)}\Tstrut\Bstrut\\ 
\hline
Scalar field & $p = -\frac{i\rho ck}{4\pi r}S_\omega e^{i(\omega t - kr)}$ & N/A \Tstrut\Bstrut\\
Intensity (energy flux) & $I_r = \frac{|p|^2}{\rho c} = \frac{\rho c k S_\omega}{4\pi} \frac{1}{r^2} \,\,\,\,\,\,\, I_\theta=I_\phi = 0 $ & N/A \Tstrut\Bstrut \\
\hline
\multicolumn{3}{c}{Far-field dipole}\Tstrut\Bstrut\\ 
\hline
Scalar field & $p = -\rho c \frac{k^2 D_a \cos\theta}{4\pi r} e^{i(\omega t - kr)}$ & $E = - \sqrt{\frac{\mu}{\epsilon}} c \frac{k^2 D_e \sin\theta }{4\pi r}e^{i(\omega t - kr)}$ \Tstrut\Bstrut\\
Intensity (energy flux) & $I_r = \rho c \left( \frac{k^2 |D_a|\cos\theta}{4\pi } \right)^2  \frac{1}{r^2} \,\,\,\,\,\,\, I_\theta=I_\phi = 0 $   & $I_r =  \sqrt{\frac{\mu}{\epsilon}} \frac{c^2}{2} \left(\frac{ k^2|D_e|\sin\theta }{4\pi }\right)^2 \frac{1}{r^2} $ \Tstrut\Bstrut \\
\hline
\end{tabular}
\caption{Comparison between the acoustic and electromagnetic analogous wave relations that are most relevant to hearing and vision, due to scalar fields and plane waves. Vector variables are printed in boldface. $\bm{E}$ is the electric field, $\bm{H}$ is the magnetic field, and $D_e$ is the electromagnetic dipole strength. $S_\omega$ is the acoustic point source strength and $D_a$ is the acoustic dipole strength. The medium constants used are the fluid density $\rho$, the adiabatic compressibility of the fluid $\kappa$, the dielectric constant or permittivity $\epsilon$, and the magnetic permeability $\mu$, where in vacuum they are designated as $\epsilon_0$ and $\mu_0$. The intensity vector in electromagnetic theory is referred to as Poynting vector $S$, but is called irradiance in radiometry. The standard medium is air for both sound and light, but in the case of light the values of vacuum are usually used instead as they are close enough to air. Because electromagnetic monopoles cannot be time-dependent (and there are no known magnetic monopoles), the simplest system that is directly comparable is a dipole in the far-field approximation. Naturally, its expressions are close to the acoustic dipole, which is less commonly used than the point source. The important thing to notice is that all source types have the same inverse-law dependence on distance in far-field (spherical divergence), up to scale constants. The acoustic expressions are taken from \citet[pp. 243, 258, 311--312]{Morse} and the electromagnetic expressions from \citet[pp. 15, 24--25]{Born} and \citet[pp. 410--413]{Jackson}.  Note that electromagnetic theory has multiple standardized normalizations, which means that similar equations often appear with slightly different coefficients, depending on the choice of units. See \citet[pp. 775--784]{Jackson} for further details. $^\dagger$In the context of acoustics, the homogenous wave equation applies both to the linearized elastic wave equation in solids and to the linearized acoustic wave equation in fluids, which are correct only for small amplitudes \citep[pp. 4--5]{Whitham}. For a rigorous comparison of scalar and vector electromagnetic and acoustic potentials and fields (particularly geared for quantum mechanics), see \citet{Burns2020}.}
\label{soundvslight}
\end{table}

\subsection{Acoustic information propagation in air}
\label{airtravel}
All material media exhibit acoustic dispersion and absorption, which affect the radiated waves in the medium and accumulates over distance \citep{Brillouin1960}. Absorption is responsible for dissipating acoustic energy through three primary mechanisms: viscosity, thermal conduction, and relaxation phenomena (causing molecular vibrations, rotations, ionization, or short-range ordering; \citealp[pp. 210--241]{Kinsler}). At low frequencies, most mechanisms produce absorption that is quadratic in frequency, but this dependence can change as a function of the relaxation frequency of the gas, which in turn depends on temperature (and humidity, in the case of air). Absorption is associated with broadening of pulses on top of the spherical divergence attenuation of the wave (i.e., the drop in intensity as $1/r^2$; see Table \ref{soundvslight}). In realistic conditions, though, the exact power-law of the frequency dependence may have to be determined empirically, as it can be a fractional power that is smaller than two \citep{Treeby}.

Absorption is always accompanied by dispersion, which causes pulse broadening as well as phase distortion that can lead to chirping. Unlike absorption, dispersion does not result in energy loss. The combined effect of dispersion and absorption is modeled using a complex wavenumber \citep[e.g.,][]{Markham}
\begin{equation}
	k(\omega) = k_r(\omega) + ik_i(\omega)
\end{equation}
where the real part $k_r(\omega)$ corresponds to dispersion and the imaginary part $k_i(\omega)$ to absorption. Importantly, the absorption and dispersion of $k$ are not independent and the two form a Hilbert-transform pair. This is a result of the \term{Kramers-Kronig relations}, which connect the real and imaginary parts of functions in causal systems \citep{Kramers, Kronig, Toll, Nussenzveig}\footnote{The Kramers-Kronig relations were validated for a range of problems in acoustics \citep[e.g.,][]{Ginzberg, ODonnel1978, Waters, Alvarez}. In their polar representation formulation, these relations can be shown to be analogous to the concept of \term{minimum-phase} filters in signal processing. Any linear, time-invariant filter that is stable and causal with zeros and poles on the left half of the s-plane is minimum-phase, which entails the minimization of its group delay. It also means that its phase response is uniquely determined by its magnitude response and vice-versa, up to a constant (\citealp[pp. 54--61]{Manolakis}; \citealp[p. 533]{HartmannRossing}; \citealp{Toll}). Both responses can be derived from one another using the Hilbert transform. Non-minimum-phase filters can be expressed as the combination of a minimum-phase and an \term{all-pass} filter---a filter that affects only the phase. Thus, the effect of absorption is roughly analogous to the magnitude response and dispersion to the phase response in filter theory. In audio filters, however, signals and operations are generally expressed as time and frequency functions, while keeping the spatial dependency implicit, whereas the corresponding wave functions involve both spatial and temporal coordinates in physical systems.}. 

The magnitude of the associated absorption and dispersion of the audio range in normal atmospheric conditions is small and is typically neglected in laboratory settings, except for large spaces or distances and high frequencies \citep[pp. 122--124]{Vigran}. The most common analyses examine the effects of propagation of pure tones or broadband noise. Both types are stationary signals that do not provide much insight about the possible effects of the atmosphere on temporal modulation, which is a necessary element in acoustic communication\footnote{It can be seen in Figures \ref{dispXY} and \ref{dispT} how pure tones are not affected by dispersion directly.}. A more informative signal for this purpose is the pulse, which is transient and can be designed as narrowband (long duration) or broadband (short duration). A narrowband pulse moves in group velocity centered around its carrier frequency, whereas a broadband pulse (like a delta function) reflects more clearly the uniformity of the group velocity function in the medium, or the lack thereof. In any case, pulse propagation depends on the extent of the variations in the group velocity function, which are encapsulated in the dispersive properties of the medium. As the pulse shape is determined by the alignment of the phases of its spectral components, dispersion generally causes pulse deformation in propagation, to a degree that is proportional to the distance traversed (\citealp{Vainshtein1976} and \cref{DispersiveAn}). 

Natural weather conditions entail fluctuations in temperature, density, medium velocity and scattering properties that make the atmosphere inhomogeneous and further affects the sound propagation. For example, it has been long known that sound propagation in fog causes attenuation in the outdoors \citep{Tyndall1874,Cole1970}\footnote{The earliest published observations about this topic were by R. Derham in 1708, according to \citet{Tyndall1874}.} and also faster decay in closed rooms filled with smoke \citep{Knudsen1948}. Another important example is turbulent atmosphere that can be formed in windy conditions, which causes both pulse (temporal) and spectral broadening \citep{Havelock1998}. It is possible to get a handle on the magnitude of absorptive effects by evaluating the \textbf{excess attenuation} of the atmospheric absorption, which is estimated after subtracting the effect of spherical divergence loss of the signal \citep{Wiener1959, Morton, Lengagne}. The extent of this effect can vary significantly, as can be gathered from measurements of frog calls in a forest after rain \citep{Penna2012}, which revealed significant excess attenuation, while penguin calls in the dry and cold plains of Antarctica followed spherical divergence with no excess attenuation \citep{Robisson1991} .

\begin{figure} 
		\centering
		\includegraphics[width=.85\linewidth]{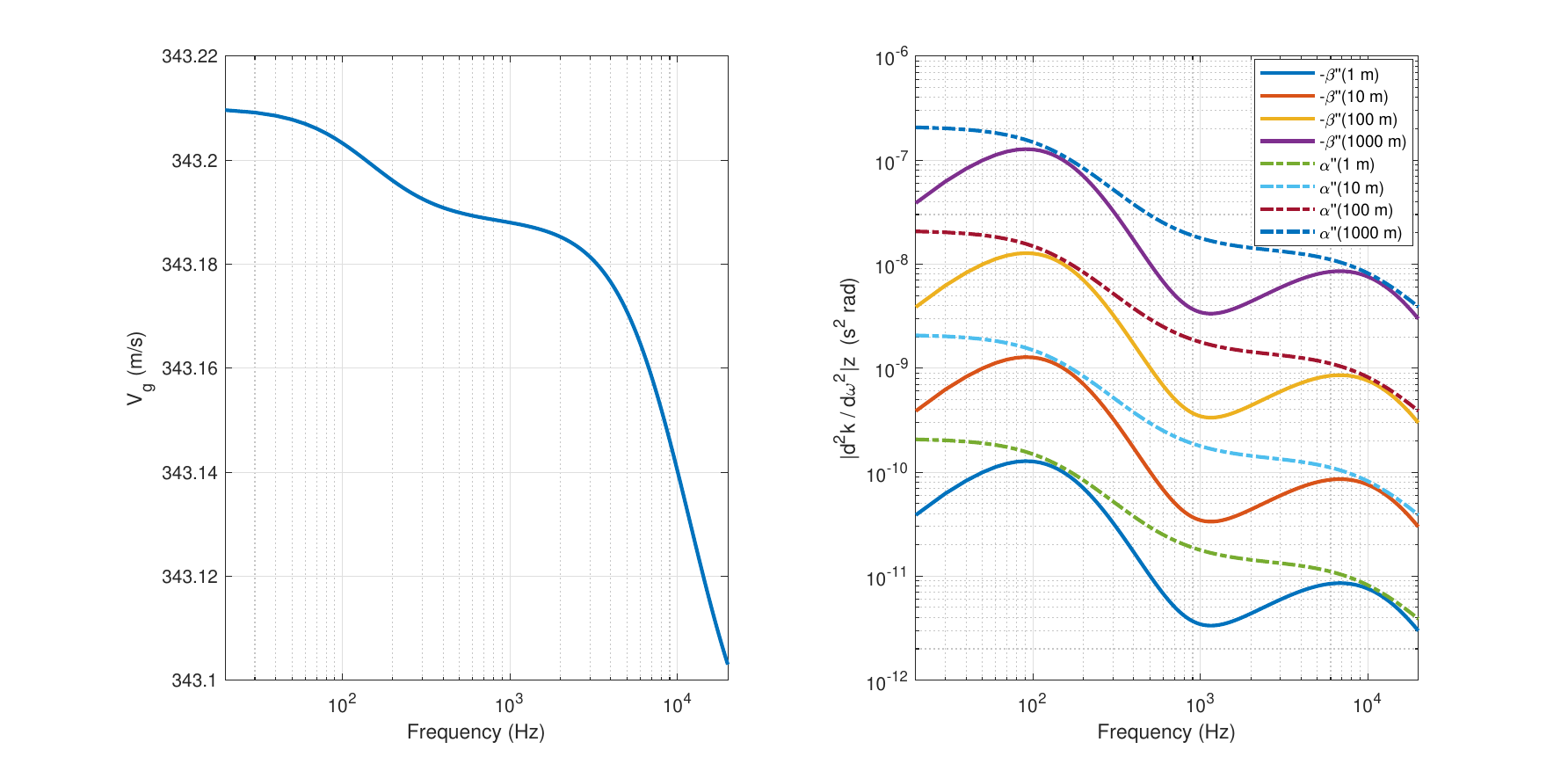}	
		\caption{Sound propagation in air at $20^\circ C$ and 50 \% humidity, based on \citet{Alvarez}. \textbf{Left:} Approximate group velocity is based on phase velocity of 343.21 m/s. \textbf{Right:} The absolute value of the (negative) dispersion and (positive) absorption curvatures (i.e., their quadratic frequency dependence). The dispersion was derived numerically from Eq. 9 in \citet{Alvarez}. However, the same could not be done for the absorption of their Eq. 8 due to multiple sign reversals, but the trend was closely matched by dividing Eq. 8 by $\omega^2$ to obtain the approximate absorption curvature.}
		\label{atmo}
\end{figure}

In this work we will be primarily interested in the quadratic frequency dependence (the curvature) of the dispersion (the group-velocity dispersion; see \cref{temporaltheory}), whose small values are shown in Figure \ref{atmo} for different distances in normal atmospheric conditions. Absorption effects are not studied directly in this work, but their potential role will be hypothesized at some points.

\subsection{Reflections}
\label{Reflections}
Apart from birds and bats in flight far from the ground, or animals in the depth of the ocean far from the ocean floor and water surface, all animal communication sounds are inevitably reflected from nearby surfaces---soil, rock, vegetation, or water. Theoretical reflection analysis ranges from simple to highly complex, depending on the amount of assumptions made about the reflecting boundary. As in propagation problems, much of the literature is concerned with pure tones and broadband noise, which model the effects of reflection on static amplitude or intensity, phase, and reflected angle. However, we continue to emphasize the effect of reflections on pulses and other transient signals that are more relevant for communication. 

Several theoretical treatments of the reflection effects on sound pulses are found in literature. Pulses of arbitrary shape can be computed by convolving the impulse response function with a pulse of arbitrary shape \citep[pp. 259--270]{Morse}. The total field is a superposition of the incident pulse and a reflected pulse, which forms a wake of negative pressure. Depending on the delay between the incident and reflected waves, the two may interfere at the point in space and time of measurement and produce a deformed pulse. Another treatment of acoustic pulse reflection (and transmission) between two homogenous media was provided in \citet[pp. 113-125]{BrekhovskikhLayerI}, where the reflected pulse is shown to be composed of a superposition of the incident pulse of and another pulse that is proportional to its Hilbert transform. In general, the reflection of a pulse by a realistic surface (of a finite acoustic impedance) causes deformation of the pulse shape. Hence, reflection can distort both spectral and temporal envelope, in a similar manner to dispersion \citep[p. 123]{BrekhovskikhLayerI}. Most environments also contain objects of dimensions of the same order of magnitude as the sound wavelength that cause wave scattering, which is even more complex than simple reflection from large surfaces \citep[pp. 400--466]{Morse}. For a geometrical acoustical treatment of sound pulse reflection, see also \citet{Friedlander1958}.

Pulse deformation has been demonstrated in several measurements in-situ. For soil reflection, most measurements and applications rely on steady-state signals to evaluate the reflection properties, but in some cases pulses were used instead, which revealed severe pulse shape deformations following reflection \citep{Don, Cramond}. Reflection characteristics are sometimes known to be affected by surface waves at low heights over the ground \citep{Daigle}. In outdoor sound propagation, the acoustic impedance, porosity, and multilayeredness of different soils (e.g, grassland, sand, forest floor, porous asphalt) have been evaluated in several studies, and results were reproduced using different models of varying complexity \citep{Attenborough2011}. In underwater measurements, \citet{Cron} showed that Gaussian and rectangular pulse-modulated pure tones become deformed as a result of reflections at angles greater than the critical angle (defined by the ratio of speeds of sound in the two media, water and ocean floor), when the pulse was relatively broad. 

\subsection{Room acoustics}
\label{RoomAc}
An acoustic source radiating in a closed space produces numerous reflections between its boundaries. Other objects in the enclosure give rise to additional scattering that can be substantial. Also, in very large spaces, the atmospheric effects on propagation may be observable, especially as such effects accumulate with successive reflections. Therefore, depending on the complexity of the space, its dimensions, and its materials, we can expect that reflected pulses and other temporally modulated sounds may become severely deformed the farther they are in time and space from the sound at the source. Taking the sound field as a whole, it becomes gradually more diffuse the farther it propagates from the source and the longer it takes it to die out due to absorption. 

\subsubsection{Steady-state response}
Two prominent approaches for the analysis of sound propagation in rooms exist in the literature---explicit wave equation solution for particular boundary conditions in one extreme, and a geometrical statistical approach in the other.

The first approach involves obtaining a closed-form solution of the wave equation with the boundary conditions of the enclosure, which yields a series of normal modes in three-dimensions \citep[pp. 554--576]{Morse}. The energy of any acoustic wave that propagates in the room has to be carried by its normal modes. This approach is insightful for large wavelengths (low frequencies) that are comparable with the dimensions of the enclosure. Solutions are generally interpreted in the frequency domain and relate to steady-state resonances, where each mode has its characteristic time constant for delay. At high frequencies, the increasingly high number of modes per unit frequency accounts for the erratic frequency response that rooms typically have \citep{Lubman1968}. 

\term{Geometrical acoustics}, the second approach, employs idealized ``sound rays'' of negligible wavelength compared to the surrounding enclosure dimensions (\citealp[pp. 576--599]{Morse}; \citealp[pp. 81--102]{Kuttruff}). A necessary condition for this to work is that the normal-mode density\footnote{Normal-mode density is a statistical measure that quantifies the number of normal modes in a structure per unit frequency. In rooms, the lowest normal mode is determined by its largest dimension. With higher frequencies, there are increasingly more modes per unit frequency, so at very high frequencies, there are no ``holes'' left as arbitrary frequencies are carried by numerous modes.} is sufficiently high, so that an arbitrary frequency component can be carried by several normal modes simultaneously. This condition is generally fulfilled for sufficiently high frequencies (see \cref{CoherenceReverb}). 

The ability to use statistical methods opens the door for a practical definition of \term{reverberation}---the sound made by the ensemble of all the reflections. It is quantified by the \term{reverberation time}, which is defined as the duration it takes for a steady-state sound source that is switched off to decay by 60 dB (see \cref{CoherenceReverb}). Its value depends on the room volume, the surface area of its boundaries, and their corresponding frequency-dependent absorption. While reverberation time is a temporal measure, it does not give sufficient information to infer the transient properties of ongoing nonstationary sounds. 

It is often more practical to measure the \term{room impulse response} and extract different parameters from it, without having to make too many assumptions about the analytical problem \citep[pp. 193--218]{Kuttruff}. This facilitates the separation of the direct and reverberant regions of the sound field. In the \term{direct field} region (measured by its distance from the source), the signal from the source is relatively intact and suffers the least deformation, as its level is higher than the field generated by the reflections---the \term{reverberant field}. The reverberant portion of the impulse response is typically separated to early reflections, which may be individually distinguishable as echoes, and to late reflections, which asymptotically behave as a statistical ensemble that is diffuse---it has a random phase function \citep[pp. 86--90]{Kuttruff}. 

\subsubsection{Transient response}
There are relatively few clear examples of temporal effects in room acoustics, beyond the obvious reverberant decay or distinct echoes in large spaces. \term{Flutter echo} may be the most familiar example---periodic reflections between parallel walls of long structures (like corridors) that are slow enough (longer period than about 25 ms) to be heard as temporal modulations---usually of low-frequency sounds \citep[pp. 90 and 167]{Kuttruff}. More obscure effects include the chirping echo caused by diffraction from the stairs of the Mayan pyramid at Chich\'en Itz\'a in Mexico \citep{Lubman1998,Declercq2004}, or the sweeping echo following an impulse in some rectangular rooms \citep{Kiyohara}. 

Note that shorter reflection periods than 20--25 ms can produce sound \term{coloration} \citep{Atal1962}, which is perceived spectrally rather than temporally \citep[e.g.,][]{Rubak2004}. In general, the interference between incident and delayed (reflected) wavefronts gives rise to spectral modulation, which can be measured with broadband sounds, and was demonstrated in a handful of natural sources in unspecified acoustic conditions \citep{Singh2003}, as well as for read speech \citep{Elliott2009}. The resultant interference, however, is never completely destructive in realistic conditions, due to the partial coherence between the incident and reflected waves (\cref{CoherenceTheory}). Sinusoidal spectral modulation of the broadband sound are usually referred to as \term{ripples} in the auditory literature. 

More mundane transient room acoustic effects exist as well. At low frequencies, the normal-mode density in rooms is relatively small, which can make specific modes stand out. For example, \citet{Knudsen} demonstrated that even with a pure tone source in a small lightly-damped rectangular room of about 17 m$^3$, low-frequency tones (around 100 Hz) changed their frequency during the decay, when they did not coincide with the frequencies of the normal modes of the room. Additionally, if the tone frequency fell between two modes, it exhibited noticeable beating, as its energy was shared between the modes. Similar phase and amplitude modulations obtained in tone decay with two, three, and four-wall structures \citep{Berman}. 

In general, exact solutions of the room boundary condition problem lose their appeal at high frequencies, where numerous normal modes exist. \citet[p. 393]{Morse1948} suggested that for a pulse transmitted in a room to retain its shape, a large number of modes ($>$10) has to overlap with a pulse carrier. In contrast, an overlap of three modes only is considered sufficient to ensure a smooth response of steady-state sounds \citep{Schroeder1996}. As signals are generally not steady-state and can be highly variable, the realistic effect of reflections and reverberation can be more complex when it comes to transient signals. 

In the reverberant field, the phase response of the room appears to be inconsequential, since listeners are able to detect phase differences only very close to the source and where the fundamental frequency is low, as measured with various broadband steady-state signals (\citealp{Kuttruff1991}; see also \citealp{Traer2016}). Nevertheless, as is seen in \cref{SpeechAnimals}, phase information is also necessary to appropriately reconstruct speech and other sound sources. Reverberation by its nature randomizes the signal phase when it is sufficiently far away from the source in time and space. The effect it has on transient sound may be appreciated from the modeling of the response to short Gabor pulses in two room geometries, which was found to be significantly closer to measurement when the complex acoustic impedance of the walls was included in the model \citep{Suh1999}. While the basis of that model was geometrical, the complex impedance could propagate the (non-geometrical) effect of interference in successive reflections. In another study, it was demonstrated that the instantaneous frequency and amplitude of linear and sinusoidal FM signals become distorted in the room, when the involved modulation is fast relative to the inverse of the reverberation time \citep{Rutkowski1997} . 

With increasing reverberation, the sound envelope decays more slowly and can energetically mask subsequent sounds, if new sounds from the source are emitted before the decaying sound subsides. This effect is captured by the modulation transfer function (MTF) concept, which was imported into acoustics from optics, and has been used as a proxy to estimate envelope smearing effects \citep{Houtgast1973,Houtgast1985}. It is measured by applying sinusoidal amplitude modulation to bandlimited continuous noise bursts. The relative difference between the smeared output and the clean input can be averaged over all center frequencies for each modulation frequency band, typically of the range 0.25--16 Hz. In general, longer reverberation times decrease the received modulation depth \citep{Schroeder1981}, which entails a decrease in audible contrast between the high and low points of the envelope (\cref{AudSenseEnv}). 

An analogous measurement to the MTF in the transmission of sinusoidally frequency-modulated narrow noise bands (i.e., the center frequency of the noise band was frequency-modulated) was demonstrated by \citet{Rutkowski1996Comp}. It was found that the frequency deviation of the modulation---analogous to modulation depth in AM---tends to decrease with higher modulation frequency and reverberation time, just like the MTF. However, in some carrier bands, the FM MTF was not monotonically decreasing and showed enhanced transmission. 

It should be mentioned that the MTF, reverberation, and other room-acoustic principles do not apply only to closed spaces. For instance, sound propagation in a flat deciduous forest introduced significant amplitude modulation, attenuation, and reverberation in both pure and amplitude modulated tones (\citealp{Richards}; see also, \citealp{Padgham2004}). It was hypothesized to have an impact on animal communication in these habitats, as vocalizations may have to be adapted to spectral windows where the information-distorting acoustics is minimal. 

\section{Transitioning from waves to stimuli}
\label{WavesStimuli}
The previous sections made the case for high complexity of realistic acoustic sources and the various transformations that can befall them on the way to the receiver. In terms of analysis, it paints a somewhat bleak picture of acoustic waves that are not only complex right at their origin, but may also become hopelessly deformed in propagation. Such sources are a far cry from the popular and mathematically convenient pure and complex tones, and their environments are nothing like the safe acoustic spaces offered by the anechoic chamber or the audiometric booth. However, it is also evident that the direct sound path from the source to the receiver suffers the least deformation compared to longer acoustic paths due to large distances or multiple reflections. In closed spaces, where the direct and reflected fields are superimposed, there are additional issues of interference and signal-to-reverberation ratio, which make the direct field even more precarious than in the outdoors, where there are fewer reflections, typically. 

As was argued in \cref{HearingTheoryVision}, hearing is primarily temporal, whereas many of the complications of realistic sources and fields manifest spatially. This means that the spatial dependence between two points can be generally reduced to a transfer function, while leaving any temporal effects explicit, as is customary in auditory signal processing\footnote{In this sense, an oscillator with a single degree of freedom is a more suitable physical model for the acoustic (or audio) signal than the acoustic wave, which is continuous and comprises multiple degrees of freedom. The former is generally modeled using ordinary differential equations (e.g., the harmonic oscillator), whereas the latter with partial differential equations. However, the generic solutions of these equations are almost the same, except for the specific physical constants that enter the phase terms in both cases and exclude the wavenumber in the oscillator case.}. The difference is that we advocate for using instantaneous quantities (phase, frequency, and amplitude) throughout the signal representation, in order to be able to factor in all kinds of modulations, either inherent to the sound or as a result of its propagation. 

We saw that a practical method---maybe the most practical method---to represent a complex signal like speech is by decomposing it to a sum of carriers with slowly-varying envelopes, which account for instantaneous changes both in amplitude and in frequency (if the envelope is allowed to be complex). We also know that the room acoustics and reverberation interact with the modulation domain, which should entail modulation deformation of some sort as well (usually a low-pass filtering of modulation frequencies). At the same time, the carrier domain becomes broader with accumulating dispersion and reflections, which leads to an overall loss of phase structure of the signal. However, even after transmission, the signal may still be representable as a component of a sum with slowly-varying complex envelope. In the direct field, the phase function and the exact frequency matter. In the reverberant field, the phase does not matter, and the sinusoidal carrier may have to be replaced with a stochastic carrier. In reality, neither the direct nor the reverberant fields accurately describe the acoustic field, which may be better understood as a mixture of both field types. 

The distinction between the direct and reverberant fields has far-reaching implications for sound detection as is performed by hearing. Generally, the direct field provides a deterministic phase transfer function. To be able to make full use of the phase information, communication theory requires precise determination of the carrier frequency before demodulation (\cref{CommunicationTheory}). Typically, it is achieved using phase-locking that synchronizes to the carrier. In reverberant fields and in the direct fields of random sources, the phase is random and demodulation can be applied much more simply to the intensity envelope only---no longer requiring phase-locking (which is anyway impossible when the phase is truly random). Both detection methods are useful for somewhat different purposes. We will later refer to the direct field as coherent, the reverberant as incoherent, and their corresponding detections as coherent and noncoherent. The process in which the phase of the direct field becomes randomized through reflections and reverberation will be called decoherence. These concepts are central in this work and will be explored from different perspectives along the subsequent chapters.

\chapter{Optical imaging}
\label{ChapterImaging}
\section{Introduction}
Visual sensation critically depends on the optics of the eye that constitutes an imaging system. This connection implies that different concepts pertaining to the optics of image formation and its quality are deeply ingrained within our visual perception. In turn, it suggests that the study of imaging can be both intuitive and palpable. It was argued in \cref{ObjImg} that the various auditory image models in literature provide an incomplete analogy to the visual image. These models are mute with respect to some of the most powerful concepts of the optical image such as focus, sharpness, blur, depth of field, and aberrations. In most cases, it is not even clear what the object of the putative auditory image is. This chapter provides a general and selective overview of optical imaging theory, in order to be able to establish a rigorous analogy to the auditory image later in this work. The proposed auditory image will be amenable to the derivation of mathematically and conceptually analogous candidates for almost every aspect of imaging that vision has\footnote{The exception are three-dimensional aberrations such as astigmatism and curvature of field, which will not have an obvious analog in hearing.}. Ideally, some of the intuition from the optical image formation theory can then be translated to hearing as well.

A very brief introduction to spatial optics is given in the next section, where different but complementary analytic frameworks to understand the imaging process are highlighted based on geometrical, wave, and Fourier optics. Then, a simple account of the eye as an optical imaging system is presented. Finally, some links between imaging optics and acoustics and noted. 

Readers who are well-versed in imaging optics may comfortably skip this chapter, which is mainly intended for acoustics and hearing specialists with no background in optics. There are numerous introductory texts that cover the topics mentioned below in great depth. The overview is primarily based on general introductory texts on physical optics by \citet{Lipson} and \citet{Hecht2017}, Fourier optics by \citet{Goodman}, lasers by \citet{Siegman}, geometrical optics by \citet{Ray1988} and \citet{Katz2002}, and on the comprehensive text by \citet{Born}. 

\section{Optical imaging theory}
\label{spatialsimaging}
Light is the narrow portion of the electromagnetic spectrum that can be detected by human vision in the wavelength range of 400--700 nm and frequency range of 430--750 (THz). This spectral range constitutes a significant portion of the electromagnetic radiation from the sun. It is exploited by many animals as it is biologically stable and detecting it does not suffer from high internal physiological noise \citep{Soffer1999}. Optics is the physical science that deals with light and visible phenomena, but many of its methods are applicable to electromagnetic and scalar wave fields in general. 

In daylight, broadband light from the sun is reflected in all directions by objects in the environment. In every point in space, we can place an optical imaging system and obtain an image of a part of the environment, by manipulating the reflected light waves that impinge on it. However, without meeting certain necessary conditions, all that is obtained of other objects are patterns of shadows and light: no image of a tree will spontaneously appear on an adjacent wall---only its shadow. 

Let us first define the ideal image $I_1$, which appears on a plane at a certain distance $z$ away from the object $I_0$, as measured along the z-axis, which is referred to as the \term{optical axis}. The object and image are referenced to their own parallel rectangular axes $(x,y,z)$ and $(x',y',z')$, respectively. We examine a three-dimensional object that is positioned perpendicular to the optical axis and is described as a light intensity pattern in the X-Y plane with depth along $z$, $I_0(x,y,z)$. The ideal image is then \citep[pp. 43--73]{Goodman}
\begin{equation}
\label{Idealimage}
	I_1(x',y',z') = \frac{1}{\left|M\right|}I_0\left(\frac{x}{M},\frac{y}{M} ,\frac{z}{M} \right)
\end{equation}
where $M$ is a constant scaling factor, which entails \term{magnification} for $M>1$ and \term{demagnification} for $M<1$. Thus, an image is a linearly-scaled version of the object in space, also measured as a light intensity pattern. In other words, the image is geometrically similar to the object (see also \citealp[pp. 152--157]{Born}). Often, the projection of the object is all that we care about, so both object and image are treated as two-dimensional and perpendicular to the optical axis. 

Different theoretical frameworks exist that can account for image formation. We introduce below the two that will be most relevant to the temporal auditory imaging theory.

\subsection{Geometrical optics}
\label{GeometricalOptics}
Geometrical optics provides a relatively intuitive framework for analyzing optical systems. It is valid in the limit of infinitesimally small wavelengths, which is often adequate to fully account for the behavior of many important optical instruments \citep[pp. 116--285]{Born}. 

In geometrical optics, light propagates in ``\term{rays}'', which are normal to the geometrical wavefront of the light arriving from the object. Normally, both spherical and plane waves are considered in the analysis, although local propagation is approximated as plane waves. The space is completely described by the \term{refractive index} $n$, which encapsulates the speed of light at the medium of $c/n$, where in vacuum and air $n=1$ and $c$ is the speed of light. Most geometrical systems of interest describe isotropic and homogenous regions of the medium, in which $n$ is constant, except for abrupt boundaries where the ray changes direction due to \term{refraction}. Critically, we are interested in \term{Gaussian optics}, which considers only small angles subtended by the system---the angles that the rays form with respect to the optical axis. This is the \term{paraxial approximation}, which leads to simplifying conditions of $\sin\theta \approx \tan\theta \approx \theta$ and $\cos \theta \approx 1$. 

\begin{figure} 
		\centering
		\includegraphics[width=.8\linewidth]{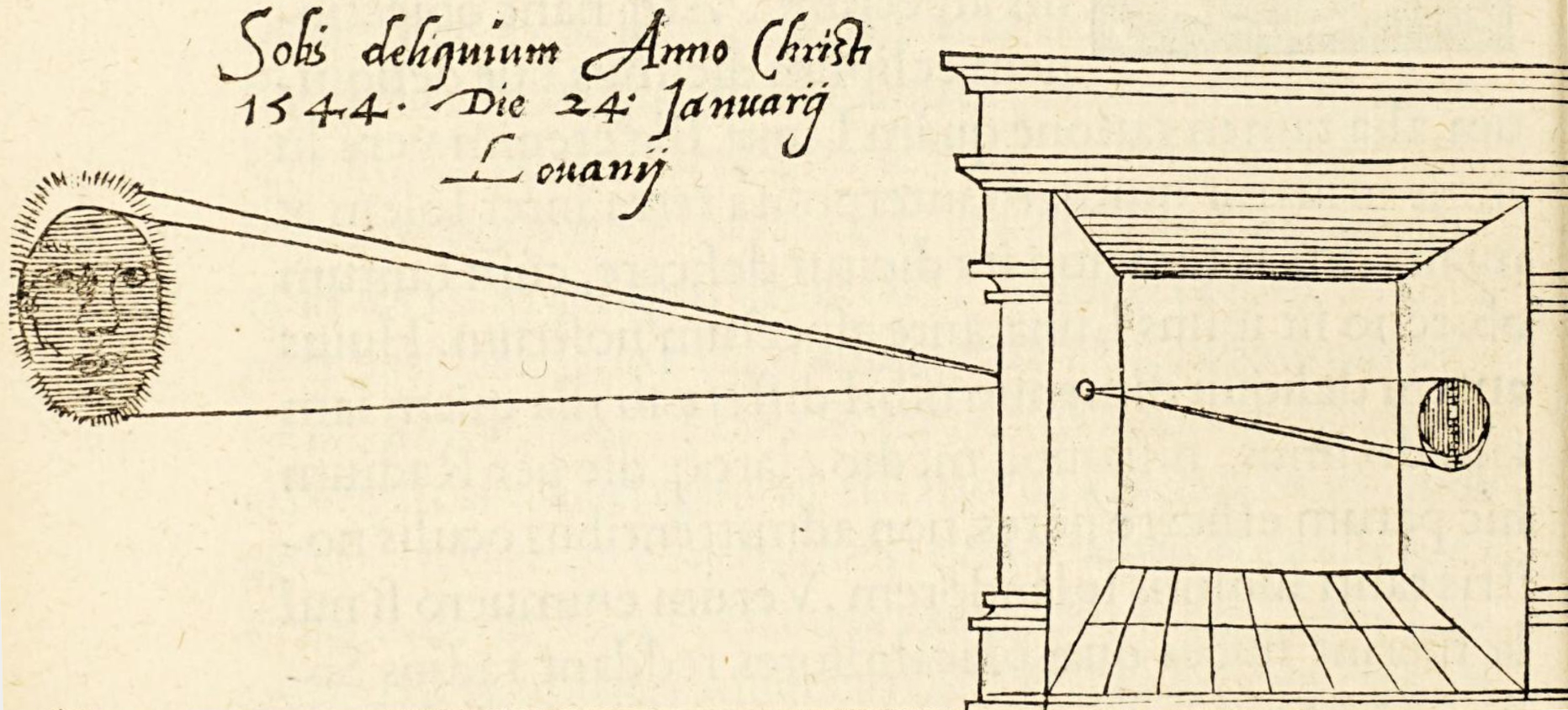}	
		\caption{The simplest optical imaging system is the camera obscura. The illustration is taken from \citet[Cap. XVIII, p. 31]{Frisius}.}
		\label{CameraObs}
\end{figure}

The simplest system that can produce an image is called a \term{pinhole camera} (\citealp[p. 229]{Hecht2017}), or a \term{camera obscura}---referring to its original incarnation as a light-tight room with a small hole in one of its walls that allows for sunlight to enter \citep{Renner2009}. It comprises three elements that are common in all imaging systems: a pinhole aperture, a darkened compartment, and a screen (see Figure  \ref{CameraObs}). The role of the pinhole aperture is to limit the angular extent of light that arrives directly from the object. For an infinitesimally small aperture, every point of the object is mapped to a point-sized sharp region on the screen, which is positioned on the imaged plane. But a very small aperture may reduce the amount of energy too much, so that the image brightness suffers. A large aperture produces brighter images, but allows multiple paths between object points to the screen. The effect is to (geometrically) blur the resultant image, so that fine details of the object cannot be discerned. The darkened compartment is necessary to block all ambient light that can wash out the image. The image that is formed on the screen is scaled by a magnification factor that is determined by the ratio of the distances from the object to the aperture and from there to the screen.

\begin{figure} 
		\centering
		\includegraphics[width=.8\linewidth]{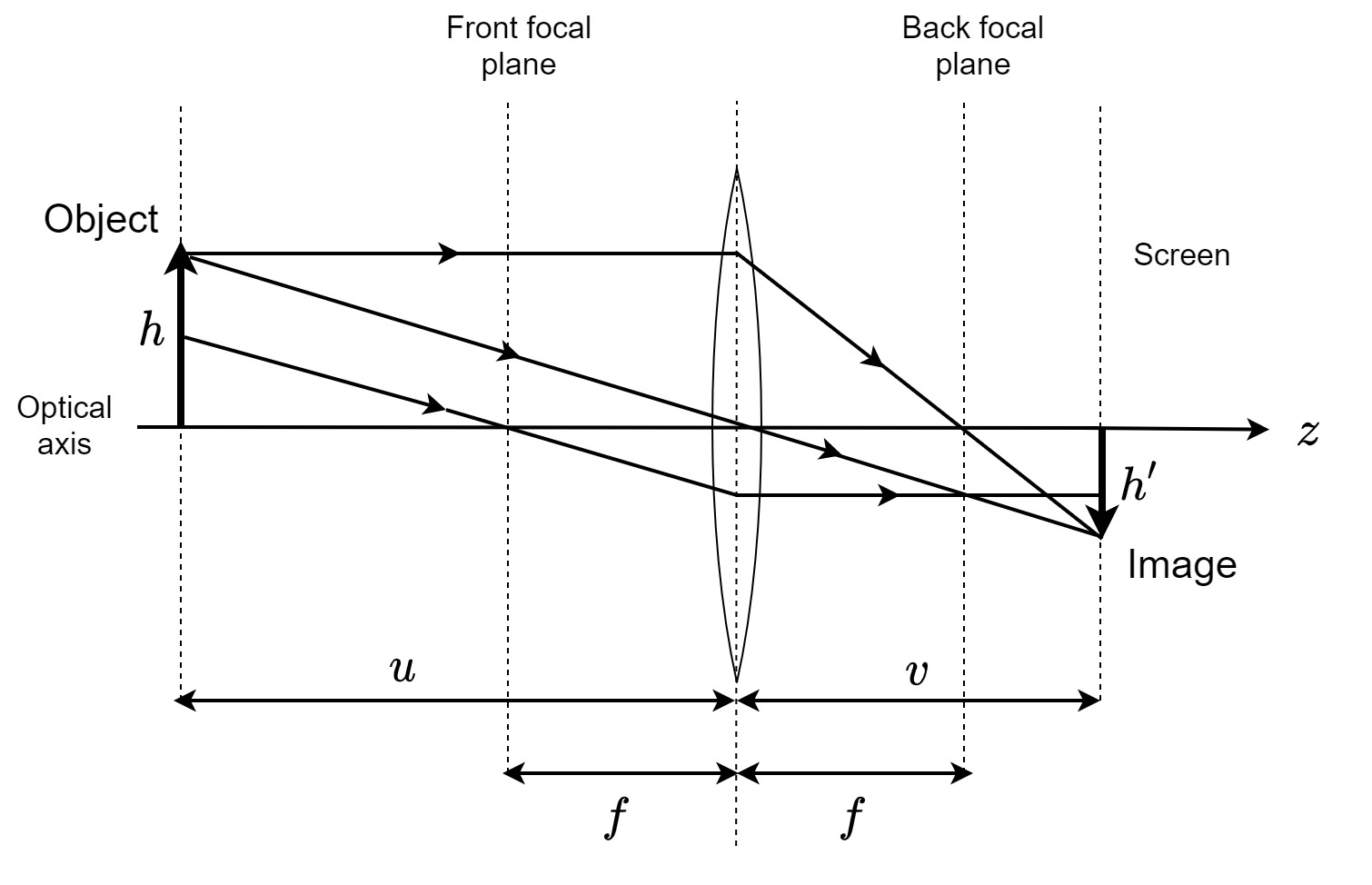}	
		\caption{A single-lens imaging system. The image of an object of height $h$ perpendicular to the optical axis is  imaged by a thin lens with focus $f$. The image height $h'$ is scaled by a magnification factor $M = -v/u$, where $u$ is the distance from the object to the lens and $v$ is the distance from the lens to the image.}
		\label{NormalImage}
\end{figure}

An improved design over the pinhole camera is the single-lens camera, which places a \term{thin lens} in front of the aperture \citep[pp. 223--228]{Katz2002} (Figure \ref{NormalImage}). The lens is made from a dense optical material, which has a higher refractive index $n$ than the surrounding medium. An ideal thin spherical lens is defined by its radii of curvature, $R_1$ and $R_2$, and its refractive index, according to the lens-makers' formula (assuming air to be the surrounding medium)
\begin{equation}
\frac{1}{f}  = (n - 1)\left( {\frac{1}{{R_1 }} - \frac{1}{{R_2 }}}\right)
\end{equation}
where $f$ is the focal length of the lens, measured along the z-axis. The main property of the lens is its ability to transform a beam of light that is parallel to the axis (a plane-wave front that corresponds to an infinitely distant object) to a spherical wavefront that converges at the focus of the lens. Also, parallel rays that are oblique with respect to the optical axis converge to the same point in the focal plane of the lens. The converging property of the lens (for positive lenses with $f>0$) makes it suitable to use with larger apertures than are admissible in the pinhole camera, by correcting for the geometrical blur that would have been formed without it. This is expressed through the \term{imaging condition}, which ties the object and lens distance $u$ and the lens and image distance $v$ to the focal length
\begin{equation}
\frac{1}{u} + \frac{1}{v} = \frac{1}{f}
\label{imaging_condition}
\end{equation}
The ratio between the object size (or height) and the image is the magnification, which is given by the distance ratio $M = -v/u$ (when the distances are both positive), as in the pinhole camera. A negative value of $M$ implies that the image is inverted.

This simple imaging description breaks down under several conditions. First, real lenses are thick---they have a non-negligible width---which has to be taken into account in the analysis \citep[pp. 171--175]{Born}. Second, light rays at large angles that violate the paraxial approximation would not converge in the same point as is predicted by the spherical thin lens \citep[pp. 178--180]{Born}. This deviation from paraxial conditions can increase the likelihood of different distortions and blur in the image, which are variably referred to as \term{aberrations}. For example, if a lens has a \term{spherical aberration}, then parallel rays arriving from different regions of the lens do not all meet in one focus, but are distributed around it on the optical axis, which makes the image appear blurry. However, the aberrations may be tolerable for the resolution or degree of spatial detail that is required by the system. When using more complex systems with several lenses, it is possible to optimize the image by compensating for some of the aberrations, or trading them off with one another \citep{Mahajan2011}. 

In some cases, the broadband nature of light can become apparent due to the dispersion of light in different media. Dispersion is dependence of the phase velocity on frequency (see \cref{PhysicalWaves}), which is expressed as measurable differences in the refractive index in the entire spectrum covered. If left uncorrected, dispersion may lead to noticeable \term{chromatic aberration}, whereby images of different wavelengths (colors) appear at somewhat different distances from the lens \citep[pp. 186--189]{Born}. In the extreme, a glass prism introduces enough dispersion to spectrally decompose white (or any broadband) light into all of its constituent wavelengths, so that each one refracts in a slightly different direction. 

The smallest constriction in the system is called the \term{aperture stop} and it controls the amount of light power that is available for imaging. Together with the focal length of the lens, the aperture stop determines the light power per unit area on the imaging plane. This is expressed with the \term{f-number} of the system, which is defined as
\begin{equation}
f^{\#} = \frac{f}{D}
\label{eq:fnumber}
\end{equation}
Where $D$ is the aperture size (i.e., its diameter). The f-number is the reciprocal of the \term{relative aperture}. It expresses the available light power as a unitless number, which enables comparison across different lenses and optical systems, also with more complex designs than the single-lens camera.

\begin{figure} 
		\centering
		\includegraphics[width=1\linewidth]{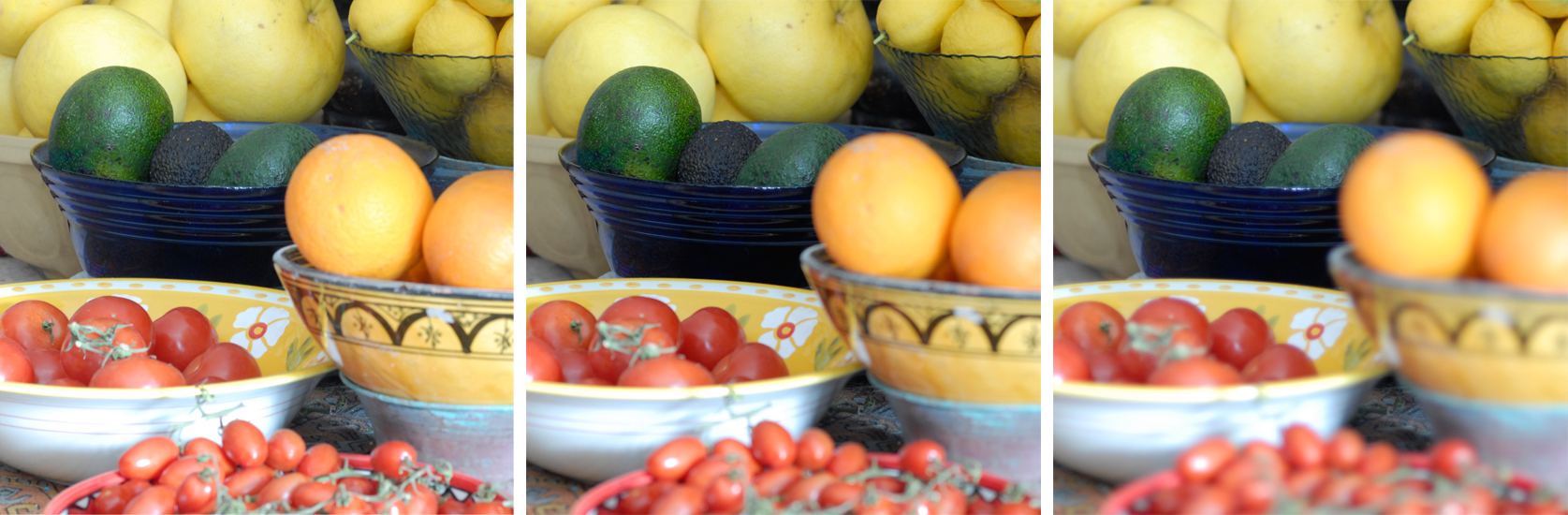}	
		\caption{A demonstration of depth-of-field in still photography. The avocados are in sharp focus in all three photos, whereas the fruit that are nearest and farthest from the lens become gradually more blurry between the photo on the left through to the right, as their depth-of-field decreases. This effect is achieved by fixing the focal length of the camera while increasing the aperture size, as is implied in Eq. \ref{eq:fnumber}, with f-number values of f/8, f/4, and f/1.4 (left to right) and a lens with focal length of 50 mm.}
		\label{DoF}
\end{figure}

It is sometimes useful to define a range of object distances for which the image obtained can be considered sharp. This range is called the \term{depth of field} of the system (see Figure \ref{DoF}) \citep[pp. 180--193]{Ray1988}. For a system set to sharp focus, the depth of field can be approximated geometrically with
\begin{equation}
	\Depth\,\of\,\Field \approx \frac{2u^2 f^2 f^{\#} C}{f^4 - {f^{\#}}^2 u^2 C^2 } \approx \frac{2u^2 f^{\#} C}{f^2}
\label{eq:DoF}
\end{equation}
which is measured in units of length, and $C$ is called the \term{circle of confusion}---a limit to the size of the finest detail that is of practical interest in the image. As can be seen, the depth of field is strongly dominated by the distance of the object from the lens $u$ and by the focal length $f$. The farther the object is and the smaller the focal length is, the larger the depth of field will be. The depth of field also increases with smaller aperture sizes. The effect of the aperture size can be explained rather intuitively. A system with a large $f^{\#}$, where $f \gg D$ implies that the angular extent of light rays that are let into the lens is kept relatively small. This means that there is less light power forming the image, but also that the image is sharper, as information limited to small angles produces less aberrations. If we fix the other parameters in Eq. \ref{eq:DoF}, then a high $f^{\#}$ corresponds to larger depth of field (e.g., Figure \ref{DoF}, left). In contrast, when $D \gg f$ the image is formed by rays arriving at larger angles that make the image blurry quickly away from the focal plane. Images here are much brighter, but they suffer from a shallow depth of field (e.g., Figure \ref{DoF}, right). For the effect of the focal length see Figure \ref{SpatialDOFLens}.

A complementary concept is the \term{depth of focus}, which expresses the tolerance in positioning the screen, on which the image of an object at a given distance would appear sharp. It is approximated as 
\begin{equation}
	\Depth\,\of\,\Focus \approx \frac{2vf^{\#} C}{ f} \approx 2Cf^{\#}
\label{eq:DoF2}
\end{equation}
where the second approximation is valid for small magnification systems. Thus, the depth of focus is determined almost exclusively by the relative aperture, with larger aperture causing a reduction in depth of focus. 

When an object distance violates the imaging condition (Eq. \ref{imaging_condition}) and is well outside of the depth of field of the system, its image suffers a form of blur that is called \term{defocus}, which is considered a kind of aberration as well.

~\\

In reality, the spatial resolution of the image is finite due to diffraction, so that even an infinitesimally small light point on the object would map to a finite disc of light of the image. When the aperture size is very small, the wavelength of the light is no longer negligible as the assumptions of geometrical optics require, in which case the image has to be analyzed using wave theory. 

\subsection{Diffraction and Fourier optics}
\label{DiffractionFourier}
According to geometrical optics, the pinhole camera should have given an ideal image for an infinitesimally small aperture pinhole size, as long as the illuminating light is bright enough. In reality, the small aperture causes image blur due to diffraction \citep{Young1971}. In diffraction, wavefronts of different paths sum in amplitude rather than in intensity, as is assumed in geometrical optics. This leads to local constructive and destructive interference patterns that are visible on the image and can severely reduce its quality. Diffraction may be noticeable especially along the edges of the image, which may not appear as sharp as predicted by geometrical optics, but appear as dark and bright fringes (spatial oscillations). In effect, diffraction sets a bound on the achievable image resolution.

Diffraction theory is also based on scalar wave fields, just like geometrical optics, where polarization and electromagnetic effects are neglected\footnote{A rigorous analysis that takes into account electromagnetic effects requires much more complex analysis, which yields measurable differences to the scalar theory only close to the edges of the aperture \citep[pp. 633--673]{Born}.} (\citealp[pp. 31--61]{Goodman}; \citealp[pp. 412--516]{Born}). The solution is based on the Huygens-Fresnel principle. It is an extension to Huygens' theorem, which states that\footnote{Quoted in \citet[p. 141]{Born}, from Christiaan Huygens' ``Treatise on light'' (1690), as translated by S. P. Thompson (1912).} ``\textit{Each element of a wave-front may be regarded as the centre of a secondary disturbance which gives rise to spherical wavelets}.'' Also, ``\textit{that the position of the wave-front at any later time is the envelope of all such wavelets}.'' Augustin-Jean Fresnel extended this principle by adding that the secondary wavelets interfere with one another, which he used in his solution for the diffraction problem. 

A full solution for the diffraction problem was obtained by Gustav Kirchhoff, who continued from Fresnel, based the homogenous (source-free) Helmholtz wave equation
\begin{equation}
	(\nabla^2+k^2)E = 0
\end{equation}
where $E$ is the complex amplitude of the field variable and $k=2\pi/\lambda$ is its wavenumber. Harmonic time dependence is assumed and factored out. The Kirchhoff-Helmholtz integral is then utilized, which gives a closed expression for the field at an arbitrary point inside a closed boundary on which the field is known. The derivation of the diffraction formula is obtained by summing the amplitudes of all the different optical paths between two points, which have an aperture positioned somewhere between them \citep[pp. 412--430]{Born}. 

\begin{figure} 
		\centering
		\includegraphics[width=.6\linewidth]{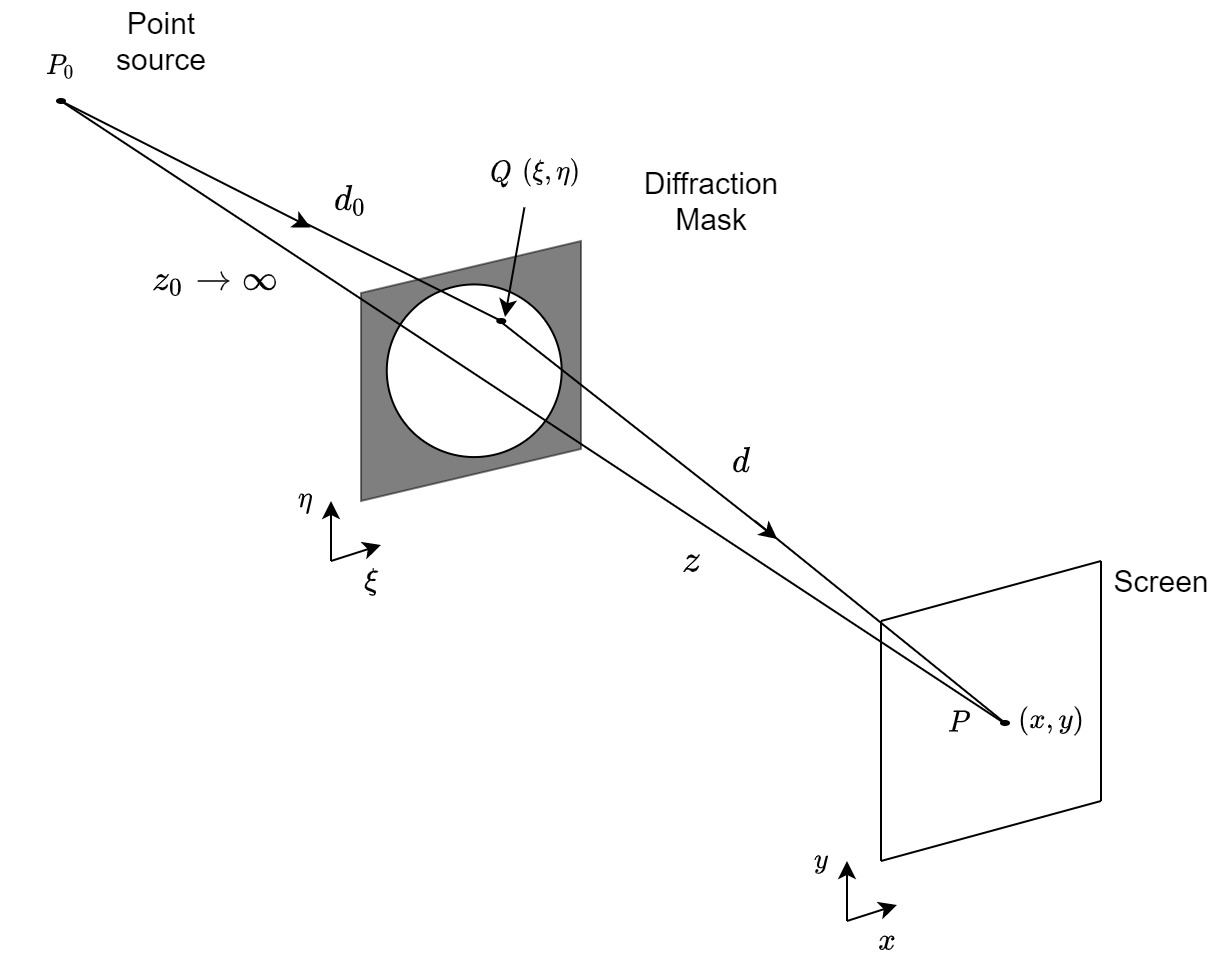}	
		\caption{The geometry of the Fresnel integral (Eq. \ref{FresenlDiff}). The source is taken to be very far from the mask, so the light is considered parallel to the optical axis. The illustration is drawn after Figure 7.3 in \citet{Lipson}.}.
		\label{FresnelIntegral}
\end{figure}

The resultant Fresnel-Kirchhoff diffraction formula requires further approximations in order to be usable. We specifically consider the paraxial approximation \citep[pp. 75--114]{Goodman}. It coincides with the Fresnel approximation that neglects higher-order terms than quadratic, which nevertheless yields precise results for small angles. In the solution, a planar \term{mask} is illuminated by a plane wave field, whose distribution in front of the mask is known. The mask---essentially a geometrical pattern of transparent openings that in imaging can refer to the aperture---is perpendicular to the optical axis (see Figure \ref{FresnelIntegral}). Two solution regimes are obtained---\term{Fresnel diffraction} (near-field) and \term{Fraunhofer diffraction} (far-field). There are different formulations to the Fresnel diffraction integral, but we refer to the one from \citet[p. 79]{Goodman}, which highlights the resemblance between the two regimes
\begin{equation}
 E(x,y) = \frac{e^{ikz}}{i\lambda z} e^{\frac{ik}{2z}(x^2+y^2)} \int_{-\infty}^\infty\int_{-\infty}^\infty E(\xi,\eta) e^{\frac{ik}{2z}(\xi^2+\eta^2)} e^{-\frac{2i\pi}{\lambda z}(x\xi + y\eta)} d\xi d\eta
\label{FresenlDiff}
\end{equation}
where  $E(\xi,\eta)$ is the complex amplitude of the field on the mask plane, $\lambda$ is the wavelength of the monochromatic wave, and $z$ is the distance between the mask and the observer. The double integral sums all the contributions from the mask plane  $(\xi,\eta)$ that have optical paths reaching the observer at $(x,y)$. The diffraction is observed as an intensity pattern $I = EE^* = |E|^2$, which means that the initial quadratic phase factor in Eq. \ref{FresenlDiff} cancels out. The presence of the quadratic phase term in the integrand of the Fresnel diffraction integral entails that a closed solution can only be obtained for relatively few problems. 

In the limit of an infinitely far source and observer $z \gg k(\xi^2+\eta^2)_{max}/2$, so the quadratic phase term in the Fresnel diffraction integral is approximately unity and it simplifies to a special case---the Fraunhofer diffraction
\begin{equation}
 E(x,y) = \frac{e^{ikz}}{i\lambda z} e^{\frac{ik}{2z}(x^2+y^2)} \int_{-\infty}^\infty\int_{-\infty}^\infty E(\xi,\eta) e^{-\frac{2i\pi}{\lambda z}(x\xi + y\eta)} d\xi d\eta
\end{equation}
This integral assumes the form of a scaled two-dimensional Fourier transform of the field $E(\xi,\eta)$. This opens the door for Fourier analysis and linear system theory, in which the inverse domain corresponds to \term{spatial frequencies} that can be used to decompose the field functions\footnote{Fourier optics was pioneered in the 1940s by \citetalias{Duffieux1983}, but took a couple of decades to properly catch.}. The spatial frequencies are derived from the propagation vector of the field $\bm{k} = \frac{2\pi}{\lambda}(\alpha \hat{x} + \beta \hat{y} + \gamma \hat{z} )$, where $\alpha$, $\beta$, and $\gamma$ are the direction cosines of $\bm{k}$, so that $k_x = \alpha/\lambda$ and $k_y = \beta/\lambda$ are component spatial frequencies. This means that, in this case, the Fourier decomposition is made into component wavefronts that propagate in specific directions with respect to the optical axis. As the wavefront propagates, so does the corresponding angular spectrum in $k$, which accumulates a component-dependent phase. 

The scope of Fresnel diffraction can be generalized for light beam propagation even when no aperture or mask are involved. It is particularly useful in the analysis of lasers beams, which have a Gaussian amplitude profile that can be propagated in closed form. This analysis expresses the field as a traveling wave along the optical axis $E(x,y,z) = a(x,y,z)e^{-ikz}$, so that $a(x,y,z)$ is its slowly changing spatial envelope, or essentially, it is the transverse amplitude and phase variation of the beam that is carried over the spatial frequency $k$. This field can be applied to the homogenous Helmholtz equation by assuming that the change in the beam profile is slow in comparison to the wavelength. The resultant equation is the \term{paraxial Helmholtz equation} (\citealp[pp. 276--279]{Siegman}; \citealp[p. 74]{Goodman})
\begin{equation}
	\nabla^2 a + 2ik\frac{{\partial a}}{{\partial z}} = 0
\label{eq:spatialdiff}
\end{equation}
The solution to this equation is a Fresnel integral, which for a Gaussian beam input results in a Gaussian profile as well. This identity between general field propagation and Fresnel diffraction underlines the generality of the concept of diffraction, which was defined by Sommerfeld (quoted in \citealp[p. 44]{Goodman}) as ``\textit{any deviation of light rays from rectilinear paths which cannot be interpreted as reflection or refraction}.''

The thin lens can be analyzed in a manner similar to diffraction as well, if the relative phase delay of the field is computed in different paths \citep[pp. 155-167]{Goodman}. The ideal lens converges a parallel wavefront to the a single point in the focal plane. Mathematically, each such wavefront may be mapped to a single spatial component (angle). The thin lens is a phase transformation of the form
\begin{equation}
	E_1(x,y) = E_0(x,y)\exp\left[-ik\frac{(x^2+y^2)}{2f} \right]
\label{lenstransfer}
\end{equation}
for an input field $E_0(x,y)$, whose angular extent is smaller than the extent of the lens, so no additional aperture is required. If the lens is smaller than the angular extent of the field, it is necessary to constrain it using a \term{pupil function}, which is the aperture associated with the finite extent of the lens. In small angles, the transformation approximately converts plane waves to spherical waves. The spherical wavefront is converging for a positive focus and diverging if it is negative.

In the paraxial approximation of the wave equation, the lens transformation is the (mathematical) dual operation of Fresnel diffraction. The single-lens imaging system can be analyzed by placing a lens between two propagation/diffraction sections. When the imaging condition (Eq. \ref{imaging_condition}) is satisfied, all the quadratic phase terms in the integration cancel out, and we are left with a double Fourier transform, which produces the ideal image as in Eq. \ref{Idealimage} \citep[pp. 168--174]{Goodman}. Unlike geometrical optics, the effects of diffraction can now be accounted for when the aperture is small. When it is large, wave theory asymptotically produces the same results as geometrical optics\footnote{It should be emphasized that the results of geometrical optics can be rigorously derived from Maxwell's equations and also directly from the scalar wave equation, either for the electric or for the magnetic field, assuming there are no currents or charges in the region. Additionally, this derivation and many others in optics assume harmonic time dependence of the carrier wave---a monochromatic field whose explicit time dependence is usually factored out. Obtaining results for a broader spectrum requires summation or integration of the intensity, as interactions between components are assumed to not take place.}. 

\begin{figure} 
		\centering
		\includegraphics[width=.4\linewidth]{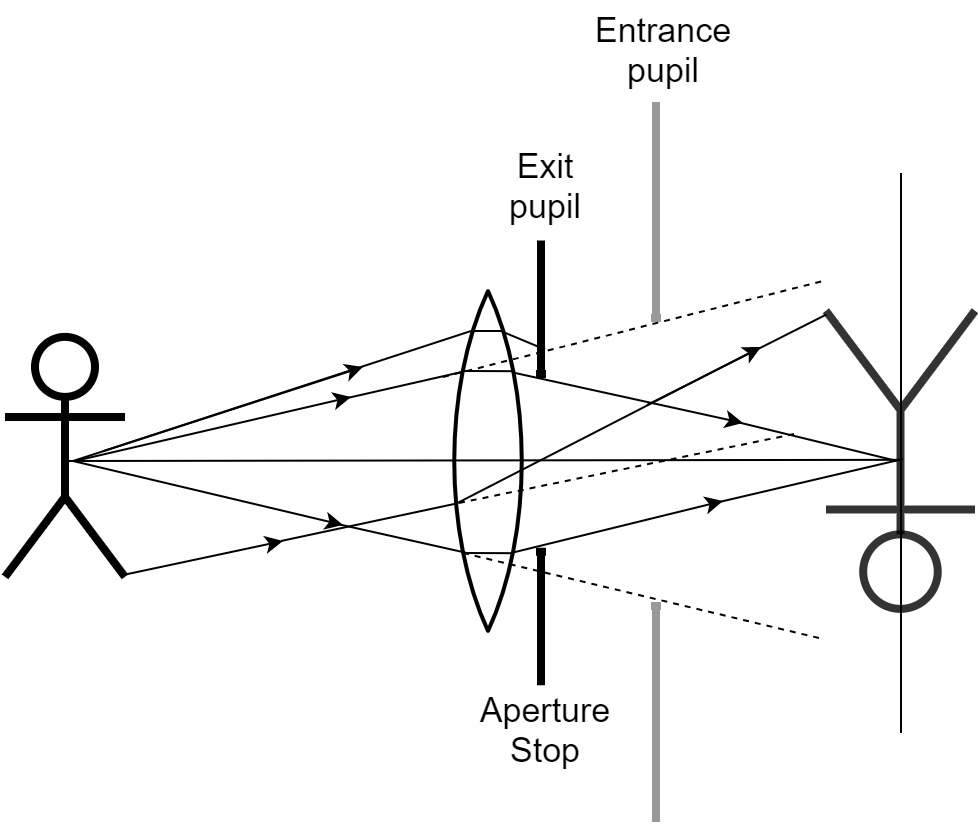}	
		\caption{Entrance and exit pupils in a simple imaging system with a rear aperture stop that is placed behind the lens. The entrance pupil is the apparent size of the aperture as seen from the object point of view. Therefore, it is the image of the aperture on the side of the lens. The exit pupil is the image of the aperture seen from the image point of view, which in this case is equal to the aperture stop itself. This illustration is a variation on Figure 5.44 in \citet{Hecht2017}.}
		\label{EntExitPupils}
\end{figure}

\begin{figure} 
		\centering
		\includegraphics[width=0.7\linewidth]{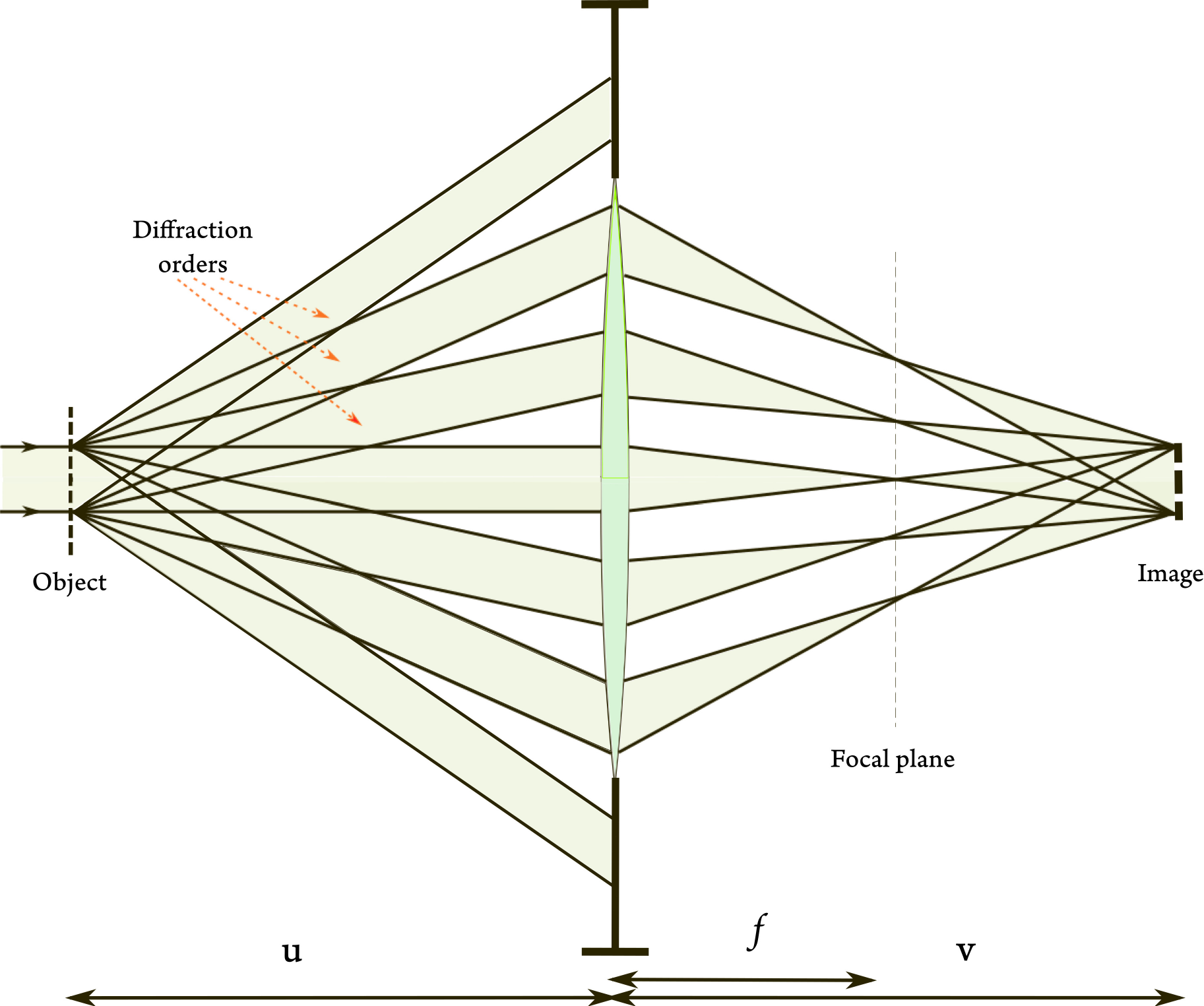}	
		\caption{Abbe's imaging theory. Light is diffracted from the object, so that different diffraction orders refer to different angles subtended from the optical axis. The lens performs a Fourier-transform-like operation on the light field envelope, so that different angles are mapped to points on the focal plane. From there they continue to the image plane and form the image as an interference pattern between the available diffraction orders. The highest positive and negative orders that are transformed by the lens are cut off by the aperture stop. The illustration is based on Figure 12.3 in \citet{Lipson} and on Figure 7.2 in \citet{Goodman}.}.
		\label{AbbeModel}
\end{figure}

The significance of the aperture in imaging cannot be overestimated. Perhaps the most revealing result from Fourier optics relates the ideal image to the aperture \citep[pp. 185--189]{Goodman}. The classical way to think of this problem goes back to Ernst Abbe and later to Lord Rayleigh, who introduced the concepts of \term{entrance pupil} and \term{exit pupil}---the images of the aperture, as seen from the object and image points of view, respectively (Figure \ref{EntExitPupils}). Using this approach, the image amplitude can be shown to be the convolution between the ideal image (of geometrical optics, Eq. \ref{Idealimage}) and the \term{point-spread function} (the optics jargon for the two-dimensional impulse response) of the aperture, which is the Fraunhofer diffraction pattern of the exit pupil. For a simple pinhole aperture, the pupil effect can be thought of as a filter that removes high spatial-frequency components that subtend the largest angles to the axis and are more likely to contribute to image aberrations (see Figure \ref{AbbeModel}). Therefore, designs based on Fourier optics can harness the aperture to process the image in the inverse spatial-frequency domain. 

The Fourier representation of imaging as in Figure \ref{AbbeModel} is important in illustrating how every spatial frequency component---one that corresponds to a particular direction of light from the object---is mapped to a particular spatial frequency in the image. This one-to-one mapping takes place at the focal plane when the image is sharp. A blurry image as is caused by defocus or other aberrations is composed from spatial frequencies that are not mapped uniquely to one frequency each, but spread some of their energy to other frequencies at their vicinity. This effectively distorts the spatial representation by not exactly retaining the same spatial pattern, as contours, edges, and other fine details become less well-defined. This is the basis for blur in spatial imaging. 

An important example of how the point spread function can be used is to compute the finest details that can be resolved in a \term{diffraction-limited} image---an image that does not suffer from any geometrical blurring aberrations. For a circular aperture of diameter $D$, an object point appears in the image as concentric rings with a fuzzy disc in the image called \term{Airy disc}. The \term{Rayleigh criterion} sets the minimum resolvable angular detail $\theta_{min}$ at the first zero of the Airy disc, where a second disc of a second point may be placed to be distinct, and hence resolvable \citep[pp. 413--414]{Lipson}: 
\begin{equation}
	\theta_{min} = \frac{1.22\lambda}{D}
\label{RayleighCriteriom}
\end{equation}
for a given wavelength $\lambda$.

\section{The human eye}
\label{TheHumanEye}
\begin{figure} 
		\centering
		\includegraphics[angle=90,origin=c,width=.5\linewidth]{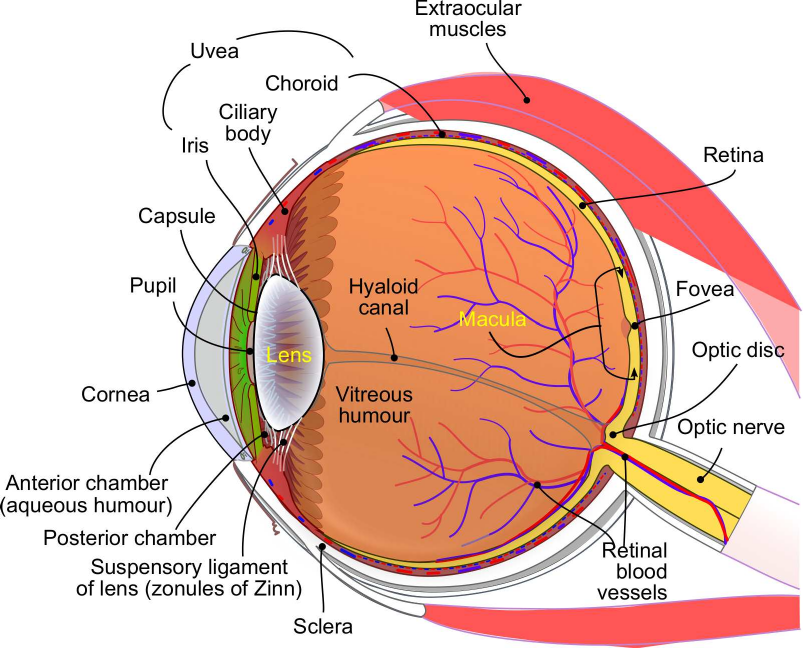}	
		\caption{The main parts of the human eye (adopted from artwork by Rhcastilhos and Jmarchn, \url{https://commons.wikimedia.org/wiki/File:Schematic_diagram_of_the_human_eye_en.svg}).}
		\label{TheEye}
\end{figure}

Explaining the function of the visual periphery, the eye, becomes trivial, once the basic optical theory of imaging is understood---much unlike the ear. Although the eye does not strictly follow the paraxial approximation, its function is analyzed most regularly using geometrical optics, while allowing for aberrations to account for the realistic image imperfections \citep{LeGrand1980}. The human eye (Figure \ref{TheEye}) is a wide-angle single-lens imaging system and is similar to a simple camera \citep[pp. 261--263]{Born}. Objects in front of the eye emit (or rather reflect) light that is focused by the \term{cornea} and the \term{crystalline lens} of the eye, both of which have a refractive index larger than unity. The lens itself is convergent and \term{aspheric}---it has a graduated refractive index between its center and its periphery---which counteracts some of the natural aberrations of the eye. From there, the light refracts through the \term{pupil}, which serves as an aperture stop, and into the \term{vitreous humor} that has refractive index similar to water. At the focal plane, on the back of the retina, a real upside-down image is formed\footnote{Images may be classified as real or virtual. \term{Real images} appear on the screen behind the lens, whereas \term{virtual images} appear to be in front of the lens to a viewer observing from the behind the lens \citep[p. 32]{Ray1988}. The magnifying glass is the simplest example for a virtual imaging system.}. Additional light scattering from the boundaries between the ocular elements within the eye adds relatively little blur to the image \citep{Williams2003Chalupa}. In daylight conditions and in young healthy eyes, a pupil of about 3 mm in diameter can achieve nearly diffraction-limited imaging, which is almost free of aberrations \citep{CharmanBass3}. \term{Refractive errors} that cause the focal plane to appear in front (\term{myopia}) or behind (\term{hyperopia}) the retina are prevalent in the population, but can be optically corrected in many cases using glasses, contact lenses, or a corrective surgery. 

The retina itself is spherical and is studded with photoreceptors that are directed to the exit pupil and sample the image plane \citep{CharmanBass3}. The \term{cones} detect narrowband light at moderate and high intensity, which gives sensation to color. The \term{rods} are broadband and become dominant only at low light levels. A high density of cone photoreceptors is found at the center of the retina\footnote{The visual axis of the eye does not overlap with the nominal optical axis, which means that the fovea is not located exactly in front of the lens \citep{CharmanBass3}.} in an area called the \term{fovea}, which is responsible for the central vision. In the peripheral regions of the retina, the photoreceptors become sparser and more dominated by rods. A network of neurons processes the information from the cones and rods in a convergent fashion (see Figure \ref{RetinalLayers}), which delivers the neurally coded response to the brain via the optic nerve \citep{Sterling2003Chalupa}. 

The simple design of the eye is complemented by adaptive mechanisms that are critical for image quality. Three mechanisms enable the eye to control the image quality, which work as a triad, or a reflex---sometimes called the \term{accommodation reflex} \citep{Charman2008}. The first mechanism controls the aperture of the eye, which is the pupil. Its size is controlled by the \term{iris} that closes and opens the pupil using annual sphincter and dilator muscles, respectively. In bright illumination conditions, pupil contraction cuts off the high spatial frequencies and increases the apparent depth of field. The second mechanism directly relates to the image sharpness. Ciliary muscles that connect to the lens through tiny ligaments (the \term{zonules of Zinn}) control the shape of the eye itself, which modifies its focus and improves the image of objects at different distances in a complex process called accommodation \citep{Toates,Charman2008}. The third mechanism is called \term{vergence} and is responsible to rotate the eyes inwards and outwards according to the distance of the object. This mechanism is important for the unified binocular perception of the two eyes. The three mechanisms usually work involuntarily, in concert. 

Another reflexive feature of the eyes has to do with their movement, which can be either voluntary or involuntary. The eye movement inside its orbit is accomplished by three pairs of muscles. The rapid muscular movements, \term{saccades}, scan the scene and are necessary to center the required image on the fovea. During the saccades, vision is suppressed and is activated when the eye is at rest. Further \term{miniature eye movements} or \term{microsaccades} are continuously present in the eye and, although their role is not entirely clear \citep{Collewijn2008}, visual perception of the image may disappear completely in their absence \citep{CharmanBass3}. 

The main measure to quantify the visual image quality is called \term{visual acuity} and it relates to sharpness---the finest detectable spatial details that can be sensed by the observer \citep{WestheimerBass3}. It is a psychophysical measure, which, to a large degree, is constrained by the periphery. It is maximal in small pupil openings (smaller than 3 mm), which eliminate most aberrations that blur the image and is dictated by the diffraction limit. The best performance that is optically possible in the spatial domain has a spread of 1.5 arcmin or 90 cycles/degree in the spatial frequency domain. These numbers are further constrained by the limited spatial sampling of the cones in the retina, whose average spacing is 0.6 arcmin in the fovea, where they are densest and provide the best resolution.

\section{Some links between imaging optics and acoustics}
The fact that much of the theory in optics relates to scalar effects, where the electromagnetic nature of the wave field is irrelevant, makes many of the results from optics applicable in other wave theories as well. Almost every basic concept from optics was at some point adopted in acoustics, where the scalar pressure field is of highest relevance (see Table \ref{soundvslight}). Geometrical acoustics applies the concepts of sound rays and bases its calculations on intensity only, as phase effects are neglected \citep[pp. 135--140]{Morse1944,Kinsler}. Imaging theory is used in ultrasound technologies, as are commonly encountered in medicine \citep{Azhari2010}. The Kirchhoff-Helmholtz integral that was mentioned as a stepping stone for the diffraction problem solution has found use in wavefield synthesis technology for spatial sound reproduction \citep{Berkhout1993}. Diffraction problems, however, are relatively uncommon in acoustics, which tends to emphasize scattering problems, in which the boundary condition objects are not significantly larger than the wavelength \citep[pp. 449--463]{Morse}. The results tend to be very complex and are not nearly as influential as those in optics, where the simple effective equivalence between the Fraunhofer diffraction and Fourier transform led to numerous applications.

While both Fourier optics and Fourier acoustics (as applied to auditory signals) are founded on linear system theory and harmonic analysis, they are different in one important respect, which will become clearer over the next two chapters. Acoustic theory generally applies the various transforms as baseband signals, which are real and are calculated from 0 Hz. In contrast, Fourier optics treats its images as bandpass signals. As the optical signals are represented as phasors with slow spatial envelopes carried on a fast traveling wave, the exact carrier frequency is often inconsequential. Thus, the modulation domain in Fourier optics is always complex. Spatial envelopes in auditory signals are of relatively little interest, but temporal envelopes are key. However, they are considered to be real, rather than complex, and are rarely processed using Fourier transforms. We will challenge this convention in acoustics in the chapters to come.

\chapter{Three approaches to information transfer}
\label{InfoTransfer}
\section{Introduction}
Information theory, communication theory, and imaging optics all deal with information transfer using different but complementary paradigms. Information theory deals with abstract and probability-based information transfer between a source and a receiver through a noisy communication channel, and how coding can be used to optimize the communication. Communication theory offers mathematical tools that prescribe how information transfer can be achieved in practice, by considering real physical signals, which can be applied in hardware and in software. Optical imaging specifically deals with the transfer of the information contained in spatial objects, most conventionally by means of quadratic transformations that represent how light waves propagate in space. All three theories complement one another in different ways, and while they are occasionally packaged into common applications, they are usually studied separately. 

While by no means foreign to hearing science, the three theories had arguably little direct influence on its progress and are nowhere considered ``staple'' disciplines for hearing. However, several concepts from each field were imported into hearing and are occasionally used without disclosing their parent discipline, which runs the risk of losing the grounding and intuition that can be garnered by having the full context. 

The goal of this chapter is to show how some of the basic paradigms of information, communication, and imaging theories overlap. Some of the most useful concepts in information theory are presented in a short overview and several historical connections with auditory research are highlighted. Then, the general communication system  is presented as a practical realization of the ideal one described in information theory. It will be argued in broad strokes that hearing can be readily fitted into a communication system paradigm. We also discuss the various similarities and differences between the general communication system and a generic single-lens imaging system. The perspectives offered in the following sections are intended to bolster our confidence in borrowing from the rich methods that have been developed in communication and imaging theories. Still, the knowledge about these theories that will be required later in this work is relatively limited. Hence, the review of these theories is brief and qualitative and is mainly geared to familiarize the reader with basic concepts. Showing that hearing can be also recast as an imaging system will be the subject of later chapters in this work.

As in other introductory chapters, the material presented in the following sections may appear trivial to communication or electronic engineers and to people with similar backgrounds. However, the connections to hearing and the interrelationship between the three approaches to information transfer have not been previously presented in the context of hearing or acoustics, to the best knowledge of the author.

\begin{figure} 
		\centering
		\includegraphics[width=1\linewidth]{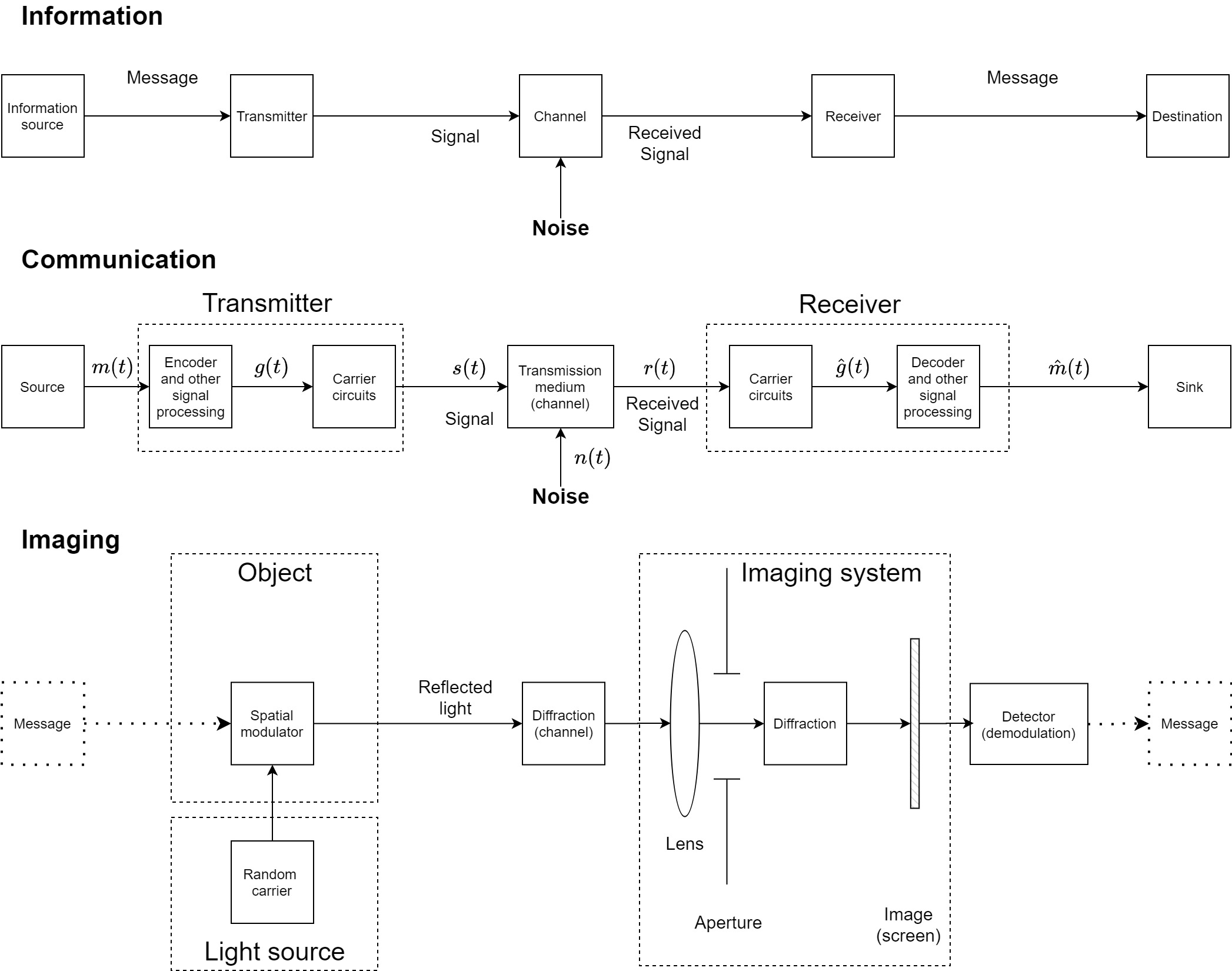}	
		\caption{The three basic paradigms of information theory, communication theory, and optical imaging. The upper diagram is a reproduction of Figure 1 in \citet{Shannon1948}. The middle diagram is a reproduction of Figure 1-1 in \citet{Couch}. The dotted lines in the imaging system around the message on the source and destination ends imply that information transfer in imaging is optional.}
		\label{ThreeParadigms}
\end{figure}

\section{Information}
The term \term{information} is used differently in lay language and in the branch of mathematics known as information theory. In common usage, it is typical to attribute information to facts and data that convey meaning to people, whereas in information theory it is none of these things. Even when used in other branches of science, the term is often applied quite intuitively and non-technically, which tends to make the initial encounter with the formal and abstract nature of information within information theory quite overwhelming for newcomers. Nevertheless, the impact of information theory on modern science and technology cannot be overstated. Still, there have been several ongoing important controversies that are more philosophical in nature, regarding the true nature of information, such as its applicability to systems that are not probabilistically and symbolically closed, and to its complete disallowance of meaning to have anything to do with the theory. 

While several theories of information have been proposed over the past century, Claude Shannon's information theory rules supreme \citep{Shannon1948}. His seminal article, originally targeting the problem of communication, is self-contained and provides most of the insight needed for familiarization with the basic concepts. Despite its highly mathematical nature, information theory has several general results that are qualitatively useful for this work, although usually indirectly, so the overview below is very concise. Interested readers can find out more from Shannon's original paper (also in the edition with Weaver's introduction, \citealp{ShannonWeaver}) and many introductory texts at different levels of abstraction and rigor (e.g., \citealp{Pierce1980,Cover,BenNaim2008}).

\subsection{Information theory in a nutshell}
\label{InfoNutshell}
\subsubsection{Discrete communication}
Shannon analyzed the transfer of information within a communication system that consists of an \term{information source} that sends a \term{message} through a \term{transmitter}, a \term{channel}, and a \term{receiver}, to the information \term{destination} (top diagram in Figure \ref{ThreeParadigms}). The content of the message is immaterial for the analysis---only the relative probability in which it appears out of the ensemble of all possible messages that can be sent over the channel. Thus, the messages are defined with respect to a closed-set of \term{symbols} that are commonly shared by the transmitter and receiver (sometimes called an \term{alphabet}). Each symbol can appear in communication at a certain probability and be part of a sequence of symbols that forms the message. The lower the likelihood of a symbol to appear, the more information it carries. In contrast, the more predictable a symbol is, the less information it carries. Therefore, according to this definition the amount of information relates to the ability of a given message to remove uncertainty from the communication. 

The simplest communication system consists of only two symbols, such as ``yes'' and ``no'', or 0 and 1. More complex messages can always be expressed as sequences of 0s and 1s (or answers to yes/no questions), without the loss of generality. By stating a few very basic requirements, Shannon was able to arrive at a unique measure of information, which he called \term{entropy} (borrowing from statistical mechanics, see \cref{PhysicalInfo})
\begin{equation}
	H = -\sum_{n=1}^N p_i \log_2{p_i}
\end{equation}
where $H$ is the entropy (or Shannon's entropy) that is computed over the probability mass function $p$, which is defined over a set of $N$ symbols (symbols with probability 0 are defined to have a corresponding zero entropy). For a two-symbol communication, if the two symbols appear at equal probability, then their entropy is $H=1$, so that each message is said to carry on average one \term{bit} of information\footnote{The requirements for a unique entropy function are: 1. The measure should be continuous with respect to probability. 2. When all symbols appear with equal probability, the amount of information carried by each symbol is inversely proportional to the number of symbols in the set. 3. The total information expressed by a certain choice (represented by a symbol) should remain invariant to branching to several sub-choices---each of which contains a smaller amount of information. In other words, if the symbol is replaced with a number of symbols, they would together carry the same amount of information as that one symbol. 4. The measure should be normalized to yield 1 bit of information when there are two symbols with equal probabilities. Implicit to these requirements is that probability theory---itself can be derived from three postulates---is valid. It is, however, possible to invert the logic and assume that information is the more primitive concept and derive probability theory from it instead \citep{Ingarden1962}. This unusual approach is important to attest to the primacy of the information as was defined by Shannon.}. 

It should be underlined that because messages are modeled as if drawn from a specific probability distribution, the entropy carried by a single message can only be understood as an element of an ensemble. Therefore, it is not meaningful to think of a standalone message that is being communicated if it is not embedded in a statistical relationship that exists within the full communication system. 

The channel in information theory is also defined probabilistically. In noise-free communication, the message that is transmitted is received unchanged. In the presence of noise, the likelihood for communication errors increases, which means that the wrong symbol can be received compared to what was sent. Errors create overall ambiguity in the message reception. Because of noise, channels are physically limited in their \term{channel capacity}, which is the maximum amount of information (or number of symbols) that can be transmitted in the channel per unit time and cannot exceed the information rate of the source\footnote{Strictly speaking, the channel is taken to be ``memoryless'', which means that the present output is determined by the present input, independently of past inputs. The various bounds on channel capacity are unaffected by the presence of feedback in the channel (but see \citealp{Massey1990}).}. If a higher rate than the channel capacity is transmitted, then it is certain that there will be errors in reception. However, it is theoretically possible to communicate information with an arbitrarily low error rate within the channel capacity. Achieving a low error rate requires \term{coding} of the messages. 

It is possible to reframe the coupling between any two physical systems as a channel, if we retain the abstract mathematical point of view. Hence, channels are not realized only in the form of cables or wireless transmission. Accordingly, the start and end points of communication may be somewhat arbitrary and can depend on the desired analysis. However, according to the \term{data-processing inequality}, information that passes through a cascade of channels can remain at most equal to the amount of information in the original message, or get gradually lost between channels---a gain of information down the communication chain is impossible \citep[pp. 34--35]{Cover}. 

Coding entails a transformation to the message representation that does not change its contents and its source entropy, but impacts its length. If the coded message contains predictable elements---for example, when a symbol identity can be confidently guessed by the identity of the previous two symbols---then it is said to have a \term{redundancy}. For example, in the statement ``the integer between 1 and 3 is 2'', the ``is 2'' can be considered redundant. There is flexibility in designing codes that serve different purposes---either to enhance redundancies, or eliminate them using (\term{lossless}) \term{compression}. Adding redundancies serves to optimize the communication robustness to errors (noise), as they can facilitate \term{error correction}, whereby the redundant information can be used to ensure the veracity of the message. In contrast, compression increases the transmission economy by employing a minimal number of symbols to communicate a particular message. In practice, a minimal error rate has to be tolerated, which means that some of the information from the source becomes lost in transmission. Alternatively, limitations on the maximum rate may exist, which force the communication system designer to exclude some information from the coded messages through \term{lossy compression}. 

\subsubsection{Continuous communication}
The description above applies to discrete messaging systems, in which the symbols are fixed. Information theory can be applied to continuous systems as well---something which introduces additional challenges to communication and to its mathematical representation. The same relations apply using probability density functions and integrals instead of probability mass functions and sums. Strictly speaking, continuous signals contain an infinite amount of information because it is always possible to represent a real continuous quantity with increased precision that requires more bits of information (e.g., to represent a real number with more digits after the decimal point). Thus, continuous entropy is defined as a relative measure and requires a reference level to enable its calculation. 

In practice, continuous information is often manipulated by discretization---by \term{sampling} the signal in time and by \term{quantizing} its level. The number of bits allocated to each quantized sample depends on the available dynamic range in the system. The larger the dynamic range, the higher is the attainable fidelity of the quantized signal, which can be measured in number of bits per sample. The channel capacity is a function of the signal-to-noise ratio (SNR) 
\begin{equation}
	C = B\log_2 \left( 1+ \frac{S}{N} \right)
\end{equation}
in which $B$ is the bandwidth, $S$ is the power in the signal and $N$ is the power of the noise. Higher channel capacity, measured in bits per second, can be obtained with higher SNR.

A complete equivalence can be drawn between discrete and continuous communication through sampling the analog signal. The sampling process transforms a bandlimited signal of bandwidth $B$ to a sequence of discrete samples. These samples can be used to mathematically reconstruct the original signal perfectly, as long as the signal is sampled with at least twice the bandwidth of the highest frequency component in the bandlimited signal, according to the sampling theorem \citep{Nyquist1928,Shannon1948}\footnote{The sampling theorem has been discovered several times before it was popularized by Shannon \citep{Luke1999}.}
\begin{equation}
	f_s \geq 2B
\end{equation}
where $f_s$ is the \term{sampling rate} or \term{sampling frequency}. The bound that is achieved when $f_s=2B$ is called the \term{Nyquist rate} (or \term{Nyquist frequency}). When a signal is regularly sampled below the Nyquist rate, the reconstructed signal may be distorted due to \term{aliasing}. This is caused when frequency components in the passband $f>f_s/2$ are wrongly reproduced as folded components within the new passband range (e.g., at $f_s/2-f$) (see also \cref{SamplingBasics} and Figure \ref{AliasDemo}).

The source entropy is maximal when all symbols can appear at equal probabilities. This is equivalent to saying that the message symbols are completely random and no additional information is available that can reduce the length of the message that has to be communicated. The continuous analog distribution that maximizes the entropy of the channel is the Gaussian distribution, or white noise.

\subsection{The physicality of information}
\label{PhysicalInfo}
The abstract nature of information as was defined by Shannon---a mathematical entity that depends on the probability distributions of arbitrary symbols---should not be interpreted as though information can exist without a physical substrate. Regardless of how the information bits manifest physically in transmission, the physical parts of the communication system should be able to resolve the differences between different levels or different combinations of bits. The manifested difference between symbols may be mapped to any measurable quantity that is being transmitted in the channel such as power, frequency, or periodicity pattern, whereas higher-level symbolic representations may relate to shape, color, pitch, duration, etc. The simplest symbol set contains only two distinct states, 1 and 0, so a physical system that can represent them must have at least two stable states, which can be mapped to the two symbols \citep{Landauer1961}. It should be possible to change the output of the transmitter, at will, to either one of these two states. Similarly, the receiver must be capable of resolving the two states of its input stage, so they can be mapped to two different symbols.

Integrating information into theoretical physics has been fraught with controversy ever since Shannon's work and possibly even before \citep{Szilard1929}. One major point of contention has been Shannon's choice to name the information measure ``entropy'' after a quantity from statistical mechanics that has the same mathematical form. As the two quantities are derived from probability distributions, some scholars have argued that they are in fact the same, whereas others have argued that the overlap is a mere coincidence that produces incoherent interpretation. However, sidestepping the controversy, a unified concept of entropy and information has been successfully treated as a de-facto physical quantity that plays a key role in modern scientific fields such as astrophysics and quantum computing. The most embracing take was probably the one expressed by the physicist John Archibald Wheeler, who has argued that information is a fundamental property of the universe \citep{Wheeler1990}. Others have proven that information may be defined axiomatically, so probability can be derived from information instead of the other way round \citep{Ingarden1962,Jumarie2000}. 

While the notion that ``\textit{information is physical}'' \citep{Landauer1996} has been one of the inspirations for the present work, the intricacies of this topic are not directly relevant to its main thread. The interested reader may consult \citet{BenNaim2008} for an engrossing treatment of the relation between physics and information theory. 

\subsection{Information theory and hearing}
Because of its highly general formulation, information theory seems to have been applied in almost every domain of science, but at varying degrees of rigor. Information theory has indirectly had the most impact on hearing science through its significant role in digital signal processing (by bridging analog and digital signal representations) and in audio compression technology (combining perceptual coding and data compression). Additionally, various effects in hearing were modeled with reference to information theory, but with different degrees of adherence to the mathematical theory of information. Undoubtedly, cognitive psychology and neuroscience are where it has resonated the most as it seems that ``information processing'' is a given in brain circuits that relate to cognition. 

An early and influential adoption of key concepts such as information bit and channel capacity can be seen in early cognitive psychology (along with computation theory), which tended to apply it in a more metaphorical way that did not always cite Shannon's work directly \citep[e.g.,][]{Miller1956,Chomsky1956,Broadbent1958,Kahneman1973}. Side by side, information theory conceptually inspired several seminal psychoacoustical papers that introduced ideas such as the redundancy in speech \citep{Miller1950}, confusion matrices in speech reception \citep{Miller1955}, the cocktail party effect \citep[p. 420]{James1890} representing limited attentional and auditory channel capacity \citep{Cherry1953}, and lipreading as a parallel information channel to acoustic speech \citep{Sumby1954}. Typically, these studies selected the parts of speech that should be treated as the information to be quantified, which is convenient symbolically, but cannot be easily generalized to other signals. An early estimate of a general auditory channel capacity in bits per second was attempted by \citet{Jacobson1950,Jacobson1951b}. 

In neuroscience, Edgar Adrian used terminology that borrowed from early communication theory that predated information theory \citep{Garson2015}. Even much later, neuroscience still held on to a conceptualization of neural coding and information processing that is independent of the information theoretical concepts that bear the same names \citep{Perkel1968}. However, some ideas about neural coding have clearly been influenced by information theory. Perhaps the most famous example is the \term{efficient coding hypothesis}, which states that as sensory information is processed in more central areas, the brain gradually eliminates redundant information that is a characteristic of natural signals \citep{Attneave1954,Barlow1961}. More rigorous usage of information theory has become more prevalent in neuroscience in the last decades \citep{Rieke1999} and, gradually, in auditory neuroscience \citep[e.g.,][]{Chechik2006,Nelken2007} and speech perception \citep{Gwilliams2022}. 

There has been an ongoing controversy about the usefulness and validity of implementing information theoretic concepts both in cognitive psychology and in neuroscience, as well as in other sciences, which Shannon himself warned against \citep{Shannon1956}. For example, to \citet[pp. 7--8]{Neisser2014}, information processing was the essential variable of cognitive psychology---``\textit{Information is what is transformed, and the structured pattern of its transformations is what we want to understand}''---yet he considered the quantitative measures of information theory overall unfruitful in psychology (unlike computation theory that has led to more insights). Another criticism about the use of information theory in psychology is the indiscriminate reliance on statistics, which suggests that sequences of events (e.g., sensory inputs) exist only as probabilistic entities, whereas in reality they are rarely random and can be significant to the person \citep{Luce2003}. Similarly, in neuroscience, \citet{Brette2019} contended that the popular concept of neural coding contains the conditions and context that are specific to the experimental setting. He argued that coding should be instead taken only metaphorically, if neural patterns are to be modeled with information theoretical tools. Information theory itself may be deficient in explanatory power, as it does not provide meaning to the coded sequences, which have to be sought externally. Another recent critique has pointed to that the neuroscientific literature often neglects to identify the different parts of the communication system, like source, channel, and receiver within the brain or the external environment, which results in incoherent modeling \citep{Nizami2019}. 

\subsection{Auditory information}
\label{AudInfoConserve}
In the context of the present work, it is the acoustically transmitted information that is being tracked from the object to the brain. It is explicitly assumed that however the information is physically expressed in the signal domain, it is largely conserved throughout the various transformations that the signal undergoes: from acoustical waves in air or water, through to the outer ear waveguide, elastic vibrations of the eardrum, mechanical motion of the stapes, compression waves in the cochlea, elastic traveling wave of the basilar membrane, hair-cell deflection, mechanoelectric transduction, and neural spikes. The most critical stage is the final transduction between the fluid motion in the inner ear and the neural domain, in which the carrier energy and form are markedly changed from the mechanical waves, and where information processing conventionally begins (i.e., according to neuroscience and cognitive psychology). An implicit information conservation assumption has been repeatedly made in numerous other models of the auditory system, which apply a continuous signal processing, energetic, or phenomenological analysis between the cochlear and neural parts of the auditory system. Arguably, such an assumption is necessary to meaningfully interpret the function of any sensory systems---hearing being no exception. A reinterpretation of various auditory mechanisms was presented in \citet[pp. 123--162]{Weisser2019,WeisserPhD}, where it was contended that cumulative information loss is a unifying principle of auditory perception, which is necessary to avoid perceptual and cognitive overload.

As will be seen below, communication theory introduces additional layers of signal processing that bring Shannon's stripped-down communication system a few steps closer to physical realization. In visual and auditory sensation, the communication system parts are relatively unambiguous, which either obviates or defers the discussion about meaning and coding and pushes it further downstream in the brain processing. In a sense, we will be taking \citet{Luce2003} up on his (somewhat overstated) observation: ``\textit{The elements of choice in information theory are absolutely neutral and lack any internal structure; the probabilities are on a pure, unstructured set whose elements are functionally interchangeable. That is fine for a communication engineer who is totally unconcerned with the signals communicated over a transmission link; interchanging the encoding matters not at all.}'' Therefore, we shall shift our attention to communication theory in the following discussion. 

\section{Communication}
Communication theory and engineering historically preceded information theory, but are a logical and technical elaboration of its principles. Here, the communication process is physical and no longer abstract. The reach of communication theory within hearing science is found in the methods and jargon that have entered the field. Any mention of amplitude-, phase-, or frequency-modulation, envelope, carrier, and instantaneous amplitude, phase, and frequency owes something to communication theory. Half-wave rectification that is commonly attributed to the inner hair cell transduction operation is also a common building block that is used in communication demodulation. While these terms are routinely found in other disciplines as well, they are most instrumental in communication technology. It is worth noting that acoustic devices that rely on the same communication-theoretic principles as have been established for electromagnetic waves has found increasing use in underwater applications, most notably in acoustic modems and telemetry \citep[e.g.,][]{Stojanovic1993,Sendra2015}.

There is a wealth of literature on communication theory. The brief overview below is mostly based on \citet{Couch} with some input from \citet{Middleton}, \citet{Proakis2013}, and \citet{Ling2017}. 

\subsection{Communication theory basics}
\label{CommunicationTheory}
All communication systems consist of a transmitter, a channel, and a receiver (middle diagram in Figure \ref{ThreeParadigms}). The system is capable of transmitting low-frequency messages from an information source to a remote destination. The information may be either analog or digital, in which case several steps of encoding are included in the process. The information is fed into the transmitter, which processes the messages and \term{up-converts} the resultant signal to a suitable frequency for transmission (always as an analog signal). Mathematically, it is done by modulating the low-frequency \term{baseband} signal (that corresponds to the information) onto a high-frequency carrier, which forms the signal for \term{bandpass} communication\footnote{While uncommon over large distances, it is also possible to communicate the message directly in baseband, without a carrier.}. The modulated carrier is amplified and transmitted into a channel---a physical medium---where the signal generally becomes attenuated and corrupted by noise through propagation. A remote receiver then \term{down-converts} the signal through \term{demodulation} to low frequency and usually performs additional signal processing to extract the baseband message, which can be read or further processed by the operator. 

A plethora of signaling techniques have been developed that are widely applied in modern electronic hardware and in software. However, the basic analysis leading up to these techniques is done without considering the electronic or computational implementation, and is therefore relevant to any technology that can assume the same mathematics.

A general bandpass signal $s(t)$ can take either one of three canonical forms. The first one is
\begin{equation}
	s(t) = a(t)\cos[\omega_c t + \varphi(t)]
	\label{CanonicalReal}
\end{equation}
where $a(t)$ and $\varphi(t)$ are the real, time-dependent, non-negative envelope and phase functions, respectively. Both modulate a sinusoid carrier of frequency $\omega_c$. The signal is sometimes more conveniently expressed as a sum of two orthogonal channels of the same carrier, but $90^\circ$ shifted
\begin{equation}
	s(t) = x(t)\cos(\omega_c t) - y(t)\sin(\omega_c t)
	\label{CanonicalQuad}
\end{equation}
with the two real functions being $x(t)$, the \term{in-phase} modulation, and $y(t)$, the \term{quadrature} modulation associated with $s(t)$. Another equivalent representation of the bandpass signal uses the real part of the complex signal
\begin{equation}
	s(t) = \Re \left[ g(t)e^{i\omega_c t}\right]
	\label{CanonicalComplexComm}
\end{equation}
where $g(t)$ is the \term{complex envelope} that is related to the previous expressions through $g(t) = a(t)e^{i\varphi(t)}$ and also $g(t) = x(t) + iy(t)$. The three expressions are completely general for any time signals (see \cref{NarrowbandApp} and \citealp[pp. 238--241]{Couch}). In the context of communication, the complex envelope is also referred to as a \term{mapping operation} of the message $m(t)$, so that $g(t) = g[m(t)]$. 

Modulation can be applied to any of the real functions that are used in the canonical signal representations. The simplest type is \term{amplitude modulation} (AM) on $a(t)$, but the term is also used to specifically refer to modulation of the form $a(t) = 1 + m\cos(\omega_m t)$. Another basic type is \term{angle modulation}, which can refer directly to the phase in phase modulation (PM), or to its derivative in frequency modulation (FM). If $x(t)$ and $y(t)$ are modulated, then it is called \term{quadrature modulation} (QM). 

\begin{figure} 
		\centering
		\includegraphics[width=.8\linewidth]{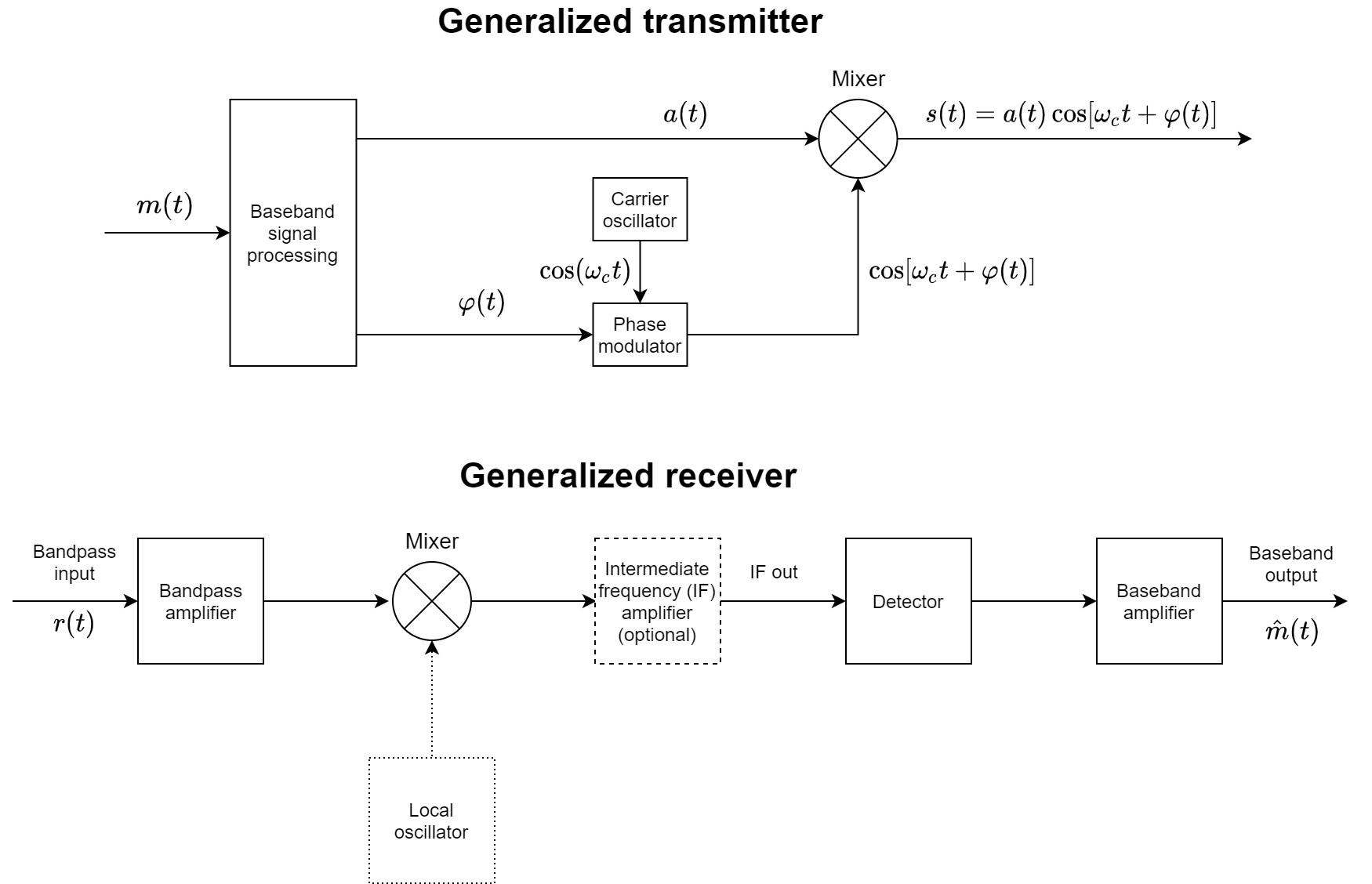}	
		\caption{Generalized transmitter and receiver diagrams. The receiver contains an optional intermediate-frequency stage that facilitates baseband signal processing. The local oscillator is used only in coherent detection. The diagrams are redrawn after Figures 4-27 and 4-29 in \citet{Couch}}
		\label{TransRec}
\end{figure}

According to the modulation type produced by the transmitter, the receiver has to perform an inverse demodulation operation in order to recover the baseband message. This constrains the possible modulation to relatively simple mathematical operations that are one-to-one in the frequency range of interest. There are two types of \term{generalized transmitters}, which can produce any type of signal modulation. One type is capable of producing AM and PM (top diagram of Figure \ref{TransRec}) that is ideal for Eqs. \ref{CanonicalReal} and \ref{CanonicalComplexComm}. The second type (not shown) produces QM, which is particularly handy to use with Eq. \ref{CanonicalQuad}. In this work we will focus on the complex canonical signal form (Eq. \ref{CanonicalComplexComm}), and hence adopt the generalized transmitter of the first type. 

The modulated carrier is transmitted into a communication channel---a physical medium that connects the transmitter and the receiver. In wireless communication it is usually the atmosphere, and in wired communication it is a cable or an optical fiber. When propagating in the atmosphere, the signal is attenuated and is subject to distortion from dispersion and variable atmospheric conditions. Moreover, the received signal may be the sum of several reflections (a \term{multipath} channel) that cause pulse broadening at the receiver. Additional detrimental effects of the channel are interference from other transmissions and noise (usually thermal). In general, for a linear and time-invariant channel with impulse response $h(t)$ and noise $n(t)$, the received signal $r(t)$ is of the form:
\begin{equation}
	r(t) = s(t)*h(t) + n(t)
	\label{receivedsig}
\end{equation}
where $*$ designates the convolution operation. The ideal communication system provides immunity to the forms of degradation encompassed by $h(t)$, so that the receiver is able to recover the message without any errors caused by the transmission. This means that the ideal channel can provide the receiver with a distortionless signal that is as close to the output of the transmitter as possible and differs from it only by a constant gain factor and a constant delay. Using the third canonical signal of Eq. \ref{CanonicalComplexComm}, the received distortionless signal becomes
\begin{equation}
	r(t) = \Re\left\{ Kg(t-\tau_g) e^{i\left[\omega_c t + \varphi(\omega_c)\right]} + n(t) \right\}
	\label{receivedsig2}
\end{equation}
where the complex envelope is attenuated by a constant factor $K$, delayed by group delay $\tau_g$, and suffers a carrier-dependent phase shift $\varphi(\omega_c)$. Ideally, these changes are negligible and/or are compensated for by the receiver. 

A central design consideration in communication engineering is choosing the type of signaling, which entails a specific kind of modulation at the transmitter and a corresponding demodulation at the receiver. The particular choice may have different advantages in terms of the attainable signal-to-noise ratio for a given bandwidth (which is proportional to the maximum information rate in the channel), robustness to noise and distortion, and ease and cost of technical implementation. In turn, these considerations may be constrained by the available channels in the electromagnetic spectrum (at least in wireless communication). The choice of channel depends on the power that is required and available for transmission, the absorption characteristics of the medium in the particular frequency range, dispersion profile with respect to bandwidth, the distance that the signal should travel, interference from other communication in overlapping channels, susceptibility to eavesdropping and jamming, effects of finite wavelength on transmission, and more. 

At the receiver, two different kinds of detection methods are distinguished. \term{Noncoherent detection} uses the signal alone for demodulation with no additional inputs. Commonly, it allows for neglecting the carrier phase, which makes it attractive for AM reception. However, it is possible to demodulate FM and PM noncoherently as well. \term{Coherent detection} involves an additional input from a local oscillator that is used to eliminate the carrier through destructive interference, and therefore retain the phase information. Coherent detection is generally (but not universally) able to achieve better signal-to-noise ratio, remove phase distortion, and follow potential carrier frequency drift. However, it tends to be more complicated and costly to implement. Some modulation types, like QM, for example, can only be demodulated using coherent detection. Others, like FM, can be demodulated in both ways, but tend to benefit significantly from coherent detection.

A \term{generalized receiver} that can demodulate an arbitrary signal is shown in Figure \ref{TransRec} (bottom diagram). The receiver has a local oscillator that can constitute the coherent source, if necessary. In some receivers, before fully modulating the signal to obtain the baseband, a modulated \term{intermediate frequency} is obtained, which can be advantageous for removing interference and for filtering in the carrier band. This stage is also optional. 

In modern communication technology that transmits digitally coded information, coherent detection has another critical feature---it enables the receiver to fully synchronize with the transmitter. This can have different advantages, depending on the application. For example, the clock of a synchronized receiver can be synchronized to the transmitter, so that signal processing can be made much more precise and flexible, without additional layers of sampling and resampling that can exacerbate potential phase distortion and errors. 

Advanced communication techniques sometimes incorporate multiple carriers or a wideband spectrum as carrier. By increasing the bandwidth, they can have advantages in achieving better signal-to-noise ratio and lower error rate, or even reduce the likelihood of eavesdropping or interference. Such techniques may be more complex and expensive to implement. However, the same general building blocks and principles guide the design of wideband systems as narrowband and single-carrier communication.

\subsection{Acoustic and auditory communication}
\label{AcouAudComm}
The analysis of the auditory system and its acoustic environment poses an inverse problem to what is paradigmatically solved in communication engineering. Instead of designing a signaling system based on general requirements, we would like to see how communication can arise from naturally occurring acoustic signals and from the auditory system existing design. As is discussed in \cref{WavesStimuli}, the majority of acoustic signals can be represented most generally in the form of a sum of AM-FM narrowband modes (Eq. \ref{SpeechAMFM}). This representation can be justified based on the physical acoustics of typical sources, on the realistic necessity to represent their transient properties, and on the effects of the environment. Thus, the typical broadband acoustic signal is a superposition of narrowband signals that are suitable for communication, as in Eq. \ref{CanonicalComplexComm}. This matches with the canonical transmitter design in communication, only with the option for multiple carriers. Therefore, many acoustic sources are natural candidates for being a communication transmitter of acoustic signals. Acoustic sources that are stochastic and do not have a fixed carrier can still be modulated and are amenable to noncoherent reception methods. 

Acoustic signals that propagate in the environment are susceptible to various distortions such as dispersion, reflections, and effects of variable weather conditions, as was discussed in \cref{acoustenv}. As a rule, the first wavefront to arrive to the ear is the least distorted one and has the highest likelihood to retain a form---and in particular a phase function---that is closest to the original source. Additionally, the acoustic receiver picks up noise from the environment and interference from competing sources that occupy the audio range, which is also considered to be noise.

It is worth dwelling on the concept of \term{noise}, which has several related meanings that have been used somewhat interchangeably in traditional hearing research. One meaning is ``unwanted sound'' \citep[e.g.,][p. 273]{Schafer}, which in research can be any out-of-context sound source, according to how it is defined by the experimenter who has also designed the hearing task. In signal processing and much of classical psychoacoustics, noise has been modeled as white noise, or a spectrally weighted version of it. In communication theory, white Gaussian noise is appropriate, because it exactly models (random) thermal noise, which is indeed the most conspicuous noise type in electronic circuitry. However, other types of unwanted signals according to the communication jargon would be considered interference, but not noise. The acoustic equivalent would be competing speech and other non-random sources from the environment.  As will be seen in \cref{PLLNoise}, there is an aspect of noise that is relative and is determined by the ability of a system to track the incoming signal. When the signal is too fast and too unpredictable to track, it can be considered noise. Therefore, constraining the range in which the ``noise-signal'' can vary, e.g., by a filter, removes some of the unpredictability and makes it somewhat less noise-like, as would be in the unfiltered version. Because the audio range has such low frequencies involved compared to electromagnetic communication, this has some implications on low-frequency auditory processing. 

These acoustic communication challenges are qualitatively identical to those experienced in communication across electromagnetic channels. Whether the transmission is in radio, microwave, or light frequencies, it is susceptible to atmospheric dispersion and absorption effects, to reflections, and to variable weather conditions\footnote{The electromagnetic signal may be also susceptible to effects that are strictly electromagnetic arising from conducting surfaces, charged objects, magnetization, polarization, etc.}. Multipath propagation is the analogous concept in communication engineering to reverberation in acoustics, but is more general. Its effects may be detrimental to reception \citep[e.g.,][]{Saleh1987}. Both phenomena are characterized by pulse broadening.  

As a receiver, the auditory system has mechanisms that are suitable for both coherent and noncoherent detection. In demodulating AM, noncoherent detection is the most straightforward detection as it requires envelope extraction and it discards the carrier phase through squaring. One of the simplest envelope detector designs is a half-wave rectifier, which coincides with the mechanoelectric transduction input-output response, as no spiking occurs during the hyperpolarizing phase of the inner hair cell receptor potential \citep{Brugge1969,Russell1978, Joris1992}. Depending on the particular low-pass filtering that exists in the transduction stage, some of the phase information is retained after rectification \citep{Heil2013,Sanderson2003}. At the same time, the auditory system exhibits neural phase locking (estimated to be effective below about 4 kHz in humans), which means that the phase of the incoming signals can be conserved in transduction from mechanical signals. This is a necessary condition for coherent detection, which is ideally suited for FM, but can also improve the performance of most other demodulation types. See \cref{AuditoryEnvPhase} for a further review of the auditory sensitivity to both AM and FM.

All in all, the auditory system has the same features as a generic communication system, as long as the acoustic object is treated as a transmitter. Then, the mathematical form of the signal, the effect of the channel, and the basic operations performed at the auditory receiver, are all standard parts of the communication signal processing. The only missing component is the messaging intent from the side of the transmitter, which is anyway not modeled mathematically. In communication, it is assumed that there is an agent behind the transmitter that tries to send an informative message to the receiver. Behind the receiver there is also an agent who accepts the recovered message.  Modulation of acoustic sounds, however, is not always intentional and can be caused by the oscillator itself (e.g., by beating modes or nonlinear transients), or by transformations imposed in propagation through the medium (\cref{ComplexSourceMod}). This means that not all modulation in sound necessarily stand for intentionally sent information. Putting it all together, we can reframe the acoustic-auditory signal processing chain as a \textit{potential} communication system, which can become de-facto communication if the roles of information source and destination are engaged. Given that the potential and the de-facto communications are mathematically indistinguishable, they are both amenable to the same analytical concepts and tools of communication theory. Conceptually, this logic transforms the acoustic wave to an acoustic signal. 

That hearing can be formally recast as a communication system is hardly a surprising conclusion, since its role in communication is ingrained in much of human and other animal life. Thus, robust ``coupling'' between active cortical brain areas of talker and listener is to be expected and has indeed been demonstrated, representing the message origin and destination \citep{Stephens2010}. Still, while the communication engineering jargon and background has been in some use in auditory research, the analogy has not been pursued to a great length (e.g., \citealp[p. 11]{Truax2001}; \citealp{Brumm2005}, \citealp{Blauert2005Blauert}) and it has never been formally integrated into the auditory theory. 

The novelty of the present approach, then, is that the link between communication and hearing is made openly and is based on more intricate mathematical and functional similarities. Borrowing from communication here is not done in a metaphorical way, but rather as a well-justified analytical step. Nevertheless, we shall use a rather limited set of qualitative results from communication, namely, the classification to detection types, and the theory of phase-locked loops for coherent detection (\cref{PLLChapter}). While this leaves much room for borrowing more ideas from communication, it will be sufficient in combination with our main concern of the imaging nature of the system. With this in mind, let us turn to optical imaging as see how it relates to communication.

\section{Imaging and communication}
\label{ImagingCommunication}
The basic principles of imaging systems were presented briefly in \cref{ChapterImaging}. Imaging and communication are in many respects complementary \citep{Rhodes1953}, but present different ways of thinking, which occasionally overlap in optical communication applications. However, there are several parallels and differences between the imaging and communication perspectives, which limit the extent of the analogy and have to be clarified first. In this section we will show how an imaging system can be interpreted as a communication system and highlight some of the similarities and differences between the two disciplines. 

It is going to take much more work to prove that the auditory system can be rigorously interpreted as an imaging system than it has taken us to prove that it is a communication system, so it will be deferred to \crefrange{temporaltheory}{ImagingEqs}. 

\subsection{Imaging as communication}
\label{ImageCommunication}
A block diagram of the most basic single-lens imaging system is given in the bottom of Figure \ref{ThreeParadigms}. Unlike the general communication system, message transmission is optional. Furthermore, the modulation is two-dimensional spatial rather than one-dimensional temporal. Instead of a transmitter, we have an object that modulates a light source carrier. Thus, the order of the carrier and modulation operations is inverted compared to a standard transmitter, but the resultant signal is mathematically identical and can take the canonical communication signal form of Eq. \ref{CanonicalComplexComm}, as long as the modulation and carrier domains are separated. However, a random carrier is a better model for sunlight and most artificial light sources than a sinusoidal (monochromatic) carrier, as was considered in \cref{CommunicationTheory}. 

The light propagation in the medium is a form of diffraction---a quadratic phase transformation that varies along the cross-section of the wavefront, normal to the optical axis (see \cref{DiffractionFourier}). Over long distances, absorption, dispersion, and atmospheric disturbance may have a significant effect on the image quality and visibility, just like in other radiation types. Noise is something that is less of an issue in normal daylight conditions, but can become significant in low-light imaging. For example, stargazing is highly sensitive to light pollution, and it is impossible in daylight with the naked eye and difficult even with a telescope.

The light signal enters the lens and its extent is limited by the aperture (which is sometimes the lens itself). The lens is optional when the aperture is very small (a pinhole camera imaging system). The light propagates further to the screen, where an image is formed that can be demodulated by a suitable detector and further processed from there. The detection that is applied in vision is noncoherent---the carrier phase plays no role in the image formation on the retina, which results in incoherent intensity imaging.

If we recast the lens, aperture, and internal diffraction as a signal-processing stage, then the analogy to the generic communication receiver becomes clearer. Hence, the transmitter is a combination of the object information source, while the light source generates the carrier. Then, the optical medium is the channel, and the receiver comprises the imaging system elements---the lens, aperture, second diffraction, screen, and detector. 

While imaging is clearly similar to a generic communication system, there is no mandatory messaging intent that is associated with the transmission, just as was the case in the auditory-communication analysis above. But functionally there is no difference, as the modulation and subsequent transmission and demodulation take place anyway. Nevertheless, sending information is once again true in \textit{potential} using the imaging system, which is mathematically set up as a spatial communication system. It conceptually transforms the optical wave into a signal. 

\subsection{Similarities and differences between imaging and\\communication}
Despite the clear high-level similarities between imaging and communication, it is worth drawing a more nuanced comparison that can sharpen the uniqueness and strength of each approach. 

The standard goal of imaging is expressed as an ideal image---a geometrical replica of the original spatial pattern that only differs by a constant factor (Eq. \ref{Idealimage}). This is somewhat analogous to the ideal transmission through a distortionless channel, which only differs by a constant gain factor and a constant delay (Eq. \ref{receivedsig2}). The temporal delay of communication is factored in the different coordinate systems of the spatial object and image, which are separated by their distance. Imaging allows for magnification on the temporal domain, whereas gain is scaling of level only, so for the two ideals to be the same, the magnification must be set to unity. 

The most obvious difference between imaging and communication is the number of dimensions that are actively involved. Imaging is spatial and is usually taken as two-dimensional, but can be also one- or three-dimensional, or even four-dimensional when motion is considered. The signal processing of the image is spatial and not temporal, so it is a function of the position. The temporal factor is implicit in the high-frequency carrier (often omitted due to harmonic time dependence). However, it may be explicitly included in the imaging by temporally modulating the object features, or through relative motion of the object and the imaging system. Communication is primarily temporal and its spatial extent is irrelevant for most signal processing. Spatial considerations in communication enter the design only in the antenna, or when the channel or electronic components have to be modeled as transmission lines, for relatively short carrier wavelengths compared to the dimensions of the electronics. 

The basic components in the signal processing of the image are based on all-pass quadratic phase transformations, which do not have counterparts in communication. Additionally, single-lens imaging is modeled as a double Fourier transform in the modulation domain as a result of the lens curvature and the diffraction. But this highlights a deeper difference in the approaches of the two perspectives. The signal processing of imaging, including the Fourier transforms, is calculated in the modulation domain, which is taken as independent from the carrier. In communication, it is done primarily in the carrier domain, even if the ultimate concern is the baseband (demodulated) domain. However, convenient transformations exist between bandpass to low-pass filter transfer functions, which essentially relate to the modulation domain as well and can be used to draw additional parallels \citep[pp. 248--250]{Couch}. 

In communication, the type of modulation has to be designed or programmed in the transmitter and receiver. In standard imaging theory as applied in vision, complex modulation fully describes the object and is demodulated as (spatial) AM, although spatial FM effects can be visually observed as well \citep{Stromeyer1975}. Therefore, an intensity image is obtained based on the spatial modulation pattern. With coherent imaging that is limited to a monochromatic light source, it is possible to conserve the phase and obtain an amplitude image. However, such an image would typically appear only as an intermediate stage, as the eye and standard optical equipment cannot detect the fast amplitude variations and require conversion to intensity images as the final format. 

Communication emphasizes the minimization of errors in the recovered message that are caused by distortion and noise. Imaging attempts to minimize aberrations of any kind, which are well-defined forms of distortion in one, two, or three dimensions. Aberrations are not usually presented as imaging errors per se, but optical designs usually try to correct them just as well \citep{Mahajan2011}. Aberrations are determined by the imaging system and the object properties, which make them deterministic. Communication errors are random, whereas distortion may be random if it arises in the channel, or deterministic if it arises in the receiver signal processing. In the latter case, carefully designed systems try to minimize these distortions. 

In addition to aberrations, imaging has several conceptual tools that are intuitive and do not have clear analogs in communication. First and foremost, the image can be focused or blurry. The focused image relates to the condition in which the different quadratic phase transformations exactly cancel out. It is somewhat analogous to the receiver signal processing that should precisely invert the signal processing done by the transmitter. However, a markedly blurry communication would be deemed distorted and possibly ridden with errors, with no special value for the receiving agent. In contrast, in vision, the blurry image provides information about the relationship between foreground and background, and sometimes about the distance from the object, the available light, the object colors, or the state of the optical system (in cameras). Therefore, there are more degrees of freedom in image interpretation than there are in standard communication, which is tailored to more specific requirements. 

Finally, unless it is also temporally modulated, the spatial image exists ``in parallel'', or simultaneously, whereas the received communication is obtained sequentially. Image processing, however, can still be sequential (e.g., by scanning the image), although the spatial order of processing needs not be linear.

\chapter{Physical signals}
\label{PhysicalSignals}

``\textit{In time, I came to the conclusion that the dehydrated cats and the application of  Fourier analysis to hearing problems became more and more a handicap for research in hearing.}'' \citep{Bekesy1974}.

\section{Introduction}
Sound propagation is most generally represented as a wave that is distributed in space and time. But in all common audio applications, it is customary to work with signals rather than waves, which are expressed using a single dimension of time. The spatial dependence is then factored into a frequency-dependent phase and amplitude terms that are fixed as long as the source and receiver are stationary. Mathematically, it simplifies all calculations a great deal. Using the term ``signal'', although sometimes synonymous with wave, also confers some intentionality that may be lacking in the normal acoustic wave and implies that it carries information. Typically, signals within the auditory system too are strictly explained in the time domain, which matches the perception of hearing as a temporal sense (\cref{HearingTheoryVision}). Nevertheless, we would like to keep sight of the auditory signal and be cognizant that its source is a physical acoustic wave, which may not always be as mathematically idealized as the temporal signal representation implies. 

The analytic signal is a powerful signal analytic tool that goes a long way to simplify the treatment of physical signals. It was introduced by Dennis Gabor\footnote{The term ``analytic signal'' was coined by \citetalias{Ville1948}.}, when he tried to identify the minimal unit of sound in hearing---the \term{logon}---using the uncertainty principle dictated by the Fourier transform \citep{Gabor}. This technique has been widely employed in different fields including hearing research, primarily to decompose signals to their real envelope and temporal fine structure. Unlike the bulk of hearing research, we shall make use of another variation of the analytic signal with a fixed carrier and a complex envelope, which includes all sources of modulation. The complex envelope is commonly used in optics and coherent communication theories, which also provide the theoretical foundation for the present work---in coherence, phase-locked loop, and temporal imaging theories. In these theories, the complex envelope is particularly handy, because it exactly coincides with mathematical solutions to the physical wave problems, which are based on complex amplitudes, or phasors. Therefore, the introduction of the complex envelope will allow us to align the physical solutions in acoustics and optics with the intuition obtained from communication theory regarding modulation.

In this chapter, the analytic signal is derived along with the complex envelope. Emphasis is placed on the narrowband channel condition that is required for the analytic signal to make sense. Then, we take a detour and present an overview of the auditory perception of envelope and phase, or temporal fine structure, as it is commonly referred to. Challenges to the classical and contemporary views are highlighted and a formalism using the complex envelope is contrasted with the standard usage of real envelope. Finally, the status and the implications of the idea that the auditory system performs real demodulation is discussed in light of all of the above, where sound is understood to be best represented by a dual spectrum of carrier and modulation domains.

\section{The analytic signal}
\label{AnalyticSignals}
The following overview of the analytic signal compiles a few standard results and draws on derivations from \citet[pp. 27--43]{Cohen1995}, \citet[pp. 92--102]{Mandel1995}, and \citet[pp. 557--562]{Born}. 

We are generally interested in real-valued signals, which are measurable over time regardless of the specific physical system that is being analyzed. Let us consider a time signal $x(t)$ that has finite energy in the range $-\infty < t < \infty$, which implies that it is square-integrable
\begin{equation}
	\int_{-\infty}^\infty x^2(t) dt < \infty
\end{equation}
Thus, the Fourier transform $X(\omega)$ of this signal exists
\begin{equation}
	X(\omega) = \int_{-\infty}^\infty x(t)e^{-i\omega t} dt 
	\label{invFTX}
\end{equation}
where $\omega = 2\pi f$ is the angular frequency. The inverse Fourier transform is similarly defined as
\begin{equation}
	x(t) = \frac{1}{2\pi}\int_{-\infty}^\infty X(\omega)e^{i\omega t} d\omega
	\label{xXFT}
\end{equation}
Real signals have the convenient property that their Fourier transform is a Hermitian symmetrical function,
\begin{equation}
	X(\omega) = X^*(-\omega)
	\label{Hermiticity}
\end{equation}
where the $^*$ is the complex conjugate operation. This property indicates that all the information in $X(\omega)$ is contained in half the complex plane (including $\omega = 0$). Intuitively, positive frequencies are more physically meaningful. As $X(\omega)$ is generally a complex function, we can redefine the time signal $x(t)$ as the real part of a complex function $z(t)$, which is the inverse Fourier transform of the complex function 
\begin{equation}
	Z(\omega) = \left\{ \begin{array}{l}
	X(\omega) \,\,\,\,\,\,\,\,  \omega \geq 0\\
	0 \,\,\,\,\,\,\,\,\,\,\,\,\, \omega < 0
	\end{array} \right.\ 
\end{equation} 
so the spectrum $Z(\omega)$ is non-zero only for non-negative frequencies, as physical intuition would have. The corresponding time signal is then
\begin{equation}
	z(t) = \frac{1}{2}[x(t) + iy(t)] = \frac{1}{2\pi} \int_{-\infty}^\infty Z(\omega)e^{i\omega t} d\omega = \frac{1}{2\pi} \int_0^\infty Z(\omega)e^{i\omega t} d\omega = \frac{1}{\pi} \int_{0}^\infty X(\omega)e^{i\omega t} d\omega 
	\label{ztX}
\end{equation} 
where the last equality used Eq. \ref{Hermiticity} and introduced a factor of 2 to balance the energy of $z(t)$ and $x(t)$. We can reintroduce the definition of $X(\omega)$ from Eq. \ref{invFTX} into \ref{ztX}
\begin{equation}
	z(t) = \frac{1}{\pi} \int_{0}^\infty \int_{-\infty}^\infty  x(t')e^{-i\omega t'} e^{i\omega t} dt' d\omega = 
	\frac{1}{\pi} \int_{0}^\infty \int_{-\infty}^\infty  x(t')e^{i\omega (t-t')} dt' d\omega
	\label{ztX2}
\end{equation} 
This integral can be solved by using the known transform of the complex exponential\footnote{See \citet[pp. 389--390]{Nussenzveig} for a rigorous derivation of this expression.}
\begin{equation}
	\int_{0}^\infty e^{i\omega t} d\omega = \pi \delta(t) + \frac{i}{t}
	\label{expFT}
\end{equation} 
which then yields
\begin{equation}
	z(t) = \frac{1}{\pi} \int_{-\infty}^\infty  x(t') \left[\pi \delta(t-t') + \frac{i}{t-t'}\right] dt' d\omega = x(t) + \frac{i}{\pi} \int_{-\infty}^\infty \frac{x(t')}{t-t'} dt' 
	\label{ztX3}
\end{equation} 
The final integral is the \term{Hilbert transform} of $x(t)$, so
\begin{equation}
 y(t) = {\cal H} [ x(t) ] \equiv \frac{1}{\pi} {\cal P} \int_{-\infty}^\infty \frac{x(t')}{t-t'} dt' 
	\label{HT1}
\end{equation}
in which at $t' = t$ the integral is evaluated using the Cauchy principal value denoted by ${\cal P}$ (see \citealp[pp. 92--97]{Mandel1995}, for a more rigorous derivation). Similarly, the real part of $z(t)$ can be obtained from its imaginary part using the inverse Hilbert transform
\begin{equation}
	x(t) = {\cal H}^{-1} [ y(t) ] \equiv -{\cal H} [ y(t) ]  = -\frac{1}{\pi} {\cal P} \int_{-\infty}^\infty \frac{y(t')}{t-t'} dt' 
	\label{HT2}
\end{equation}
$x(t)$ and $y(t)$ therefore make a Hilbert-transform pair. $z(t)$ is called the \term{analytic signal} of $x(t)$ and it can be expressed as
\begin{equation}
	z(t) = x(t) + i{\cal H} [ x(t) ] 
\end{equation}
Using this definition, the following equalities follow from Eqs. \ref{xXFT} and \ref{ztX}, and from Plancharel theorem of the equality of energy in the time and frequency representations of the signal
\begin{equation}
 \int_{-\infty}^\infty x^2(t) dt = \int_{-\infty}^\infty y^2(t) dt = \frac{1}{2}\int_{-\infty}^\infty z(t)z^*(t) dt = \frac{1}{2}\int_{-\infty}^\infty |Z(\omega)|^2d \omega = 2\int_{0}^\infty |X(\omega)|^2d \omega
\end{equation}
Additionally, the real and imaginary parts of the analytic signal satisfy
\begin{equation}
 \int_{-\infty}^\infty x(t)y(t) dt = 0
\label{propriety}
\end{equation}
This property is used in communication theory to obtain two locally-independent components in the same channel: the real part $x(t)$ is referred to as the in-phase component and the imaginary part $y(t)$ is the quadrature-phase component (\cref{CommunicationTheory}). 

\section{The narrowband approximation and the\\complex envelope}
\label{NarrowbandApp}
The analytic function is most useful for \term{narrowband} or \term{quasi-monochromatic} signals---different terms that imply that the spectrum is concentrated in a relatively narrow frequency band $\Delta \omega$ around its center frequency $\omega_c$
\begin{equation}
 \frac{\Delta \omega}{\omega_c} \ll 1
	\label{narrowbandcondition}
\end{equation}
Throughout this work, we shall refer to this inequality as the \term{narrowband condition} and to its corollaries as resulting from the \term{narrowband approximation}. In practice, values as large as $\Delta \omega / \omega_c \approx 0.2$ may sometime count as narrowband\footnote{Another definition was proposed in \cref{DispersiveAn} based on dispersion, which may be more intuitive when frequency-dependent group delay is taken into consideration: A narrowband range of frequencies is taken such that the group delay changes only a little from its mean value at the center frequency of the band (the carrier).}. For these signals it is implied that all the energy is contained in the range
\begin{equation}
 \omega_c - \frac{\Delta \omega}{2} \leq |\omega| \leq \omega_c + \frac{\Delta \omega}{2} \,\,\,\,\,\,\,\,\,\,\,\, \Delta\omega , \omega_c > 0
	\label{narrowbandrange}
\end{equation}
and outside this range the spectrum is zero, $X(\omega) = 0$. More relaxed bounds may be implied by the terms \term{bandpass} or \term{bandlimited} signals, which are not necessarily narrowband. 

The analytic signal is a complex function and as such, it can be represented in polar form
\begin{equation}
 z(t) = a(t)e^{i\varphi(t)}
\end{equation}
with $a(t)$ and $\varphi(t)$ the amplitude and phase functions, respectively. In the limiting case of a \term{pure tone} or \term{strictly monochromatic} signal, only a single component exists in the spectrum of $z(t)$ at $\omega_c$, as the bandwidth is infinitesimally small, $\Delta \omega \rightarrow 0$. In this case, $\varphi(t) = \omega_c t$, $a(t) = a$ is a complex constant, and the real signal can be written as,
\begin{equation}
 x(t) = 2\Re \left[ z(t) \right] = 2\Re ( a e^{i\omega_c t} ) = a_0 \cos(\omega_c t + \varphi_0)
\end{equation}
where $a = \Re \left(\frac{1}{2}a_0 \right)$. 

\begin{figure} 
		\centering
		\includegraphics[width=.5\linewidth]{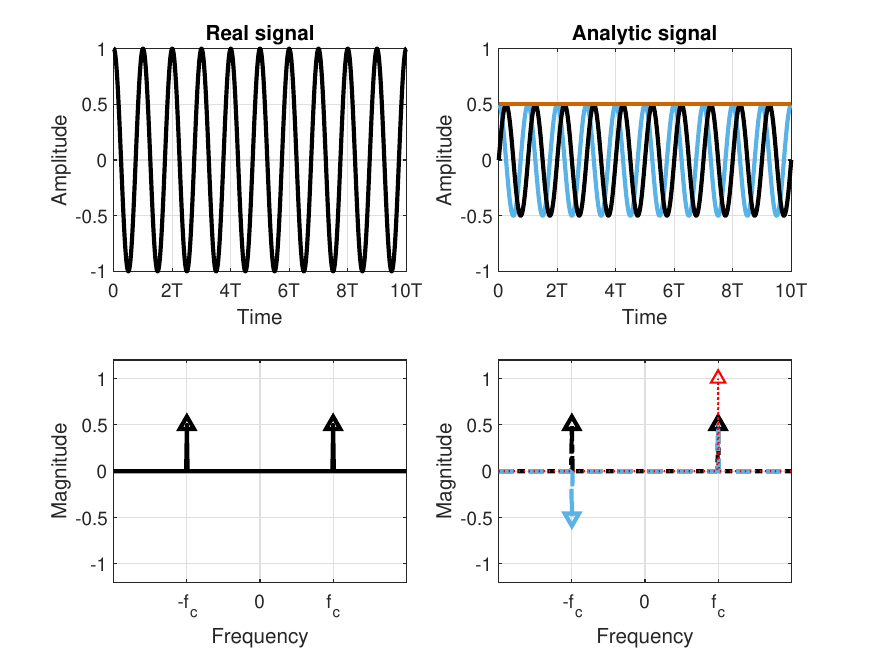}	
		\caption{The real (standard) time representation of a pure tone / monochromatic signal (top left), its spectrum (bottom left), and their analytic signal counterparts on the right. The real part of the analytic signal is in black and its imaginary part is in blue. On the top right, the constant envelope of the tone is marked in red. On the bottom right, the even and odd components (plotted in black and blue, respectively) are summed to yield the single-sided spectrum component (in red).}
		\label{Analytic1}
\end{figure}

This procedure may be extended for narrowband signals that are \term{quasi-monochromatic}, in which both amplitude and phase functions vary in time,
\begin{equation}
 x(t) = a(t) \cos[\omega_c t + \varphi(t)]
\end{equation}
The choice of $a(t)$ and $\varphi(t)$, however, is not unique for $x(t)$ \citep{Voelcker1966I}. It is made unique by using the corresponding analytic signal definition when $\omega_c$ is known
\begin{equation}
 z(t) = \frac{1}{2} a(t) e^{i\omega_c t + i\varphi(t)}
\label{ztfull}
\end{equation}
where both $a(t)$ and $\varphi(t)$ are real functions, and $0 \leq \varphi(t) < 2\pi$. From $z(t)$, the quadrature component can be readily computed 
\begin{equation}
 y(t) = a(t) \sin[\omega_c t + \varphi(t)]
\end{equation}
(see Figure \ref{Analytic1}). If we dissociate the analytic signal from its linear carrier phase $\omega_c t$, we obtain
\begin{equation}
	g(t) = a(t) e^{i\varphi(t)} = 2z(t) e^{-i\omega_c t}
\label{complexenv}
\end{equation}
This expression is referred to as the \term{complex envelope} of the real quasi-monochromatic signal $x(t)$. It is manipulated just like a phasor or a general complex amplitude, which have the same functional forms \citep{Wheeler1941}. This important concept allows us to examine the envelope independently of the specific carrier frequency, as long as the narrowband approximation holds. The narrowband range of Eq. \ref{narrowbandrange} means that with the frequency shift of Eq. \ref{complexenv}, the complex envelope is non-zero for $-\Delta \omega \leq \omega \leq \Delta \omega$, which is well-separated from the carrier at $\omega_c$. This requirement effectively entails that the envelope functions vary much more slowly than the carrier, as $T \sim 1/\Delta \omega \gg 1/\omega_c \sim T_c$, where $T$ is the period of the envelope and $T_c$ is the period of the carrier. When the separation between the envelope and carrier bands is incomplete, the signal is no longer narrowband, and the complex envelope (or any modulation function for that matter) no longer has a useful interpretation, even though the equations still hold \citep{Rihaczek1966}.

In fact, the complex envelope leads to a very general result in communication theory, which shows that any bandpass waveform may be expressed as the real part of the product of the complex envelope and a carrier
\begin{equation}
	x(t) = \Re[g(t)e^{i\omega_c t}] 
\end{equation}
following Eq. \ref{complexenv} (we referred to this result as the third canonical signal form in \cref{CommunicationTheory} and Eq. \ref{CanonicalComplexComm}). It may be drawn directly from the analytic signal definition, and can also be proven relatively straightforwardly in an alternative way, by expanding the bandpass signal $x(t)$ to its Fourier series \citep[pp. 239--240]{Couch}
\begin{equation}
	x(t) = \sum_{n = -\infty}^{n=\infty} c_n e^{ i n \omega t} \,\,\,\,\,\,\,\,\,\,\, \omega = \frac{2\pi}{T}
\end{equation}
For a general non-periodic signal, we examine the series in the limit of $T \rightarrow \infty$. Once again, we require that the signal is real, so the negative frequency coefficients are Hermitian symmetric to the positive ones: $c_{-n} = c_n^*$. Therefore, we can rewrite the series as a combination of positive frequencies only
\begin{equation}
	x(t) = \Re \left(c_0 + 2\sum_{n = 1}^{n=\infty} c_n e^{ i n \omega t} \right) 
\end{equation}
where the DC constant $c_0$ was taken out of the summation, which was multiplied by a factor of two to compensate for the energy in the negative frequencies. Now, a bandpass signal does not have a DC component, by definition, and it is in fact concentrated around the center frequency of the band $\omega_c$. This can be expressed by rewriting the summation
\begin{equation}
	x(t) = \Re \left\{\left[ 2\sum_{n = 1}^{n=\infty} c_n e^{i n (\omega-\omega_c) t}\right]e^{i\omega_c t} \right\}
\end{equation}
By comparing this expression to Eq. \ref{complexenv}, we immediately see that the complex envelope is equal to the shifted Fourier series
\begin{equation}
	g(t) =  2\sum_{n = 1}^{n=\infty} c_n e^{ i n (\omega-\omega_c) t}
\end{equation}
which means that its spectrum is distributed around $\omega = 0$ and it has the bandwidth of the bandpass signal.

As was mentioned in the introduction, the complex envelope is not standard in the interpretation of the analytic signal, which is normally decomposed to a real low-frequency envelope and a fluctuating, high-frequency carrier \citep{Dugundji1958}. This standard definition works well only when the frequency ranges of the envelope and carrier do not breach one another \citep{Bedrosian1963}, although it results in some ambiguity and practical challenges notwithstanding. It appears that a formulation of the analytic signal that accepts the complex envelope may solve some of the challenges that riddle it, as is progressively being concluded in a few recent studies and will be discussed in \cref{AnalyticChallenges}. The complex envelope formulation, which effectively treats the carrier frequency as a constant, has been used independently of the concepts of envelope and analytic signal throughout the development of modern optics theory (e.g., \citealp{Born,Goodman}). In coherence theory, it appears to have been formally connected to the ideas of analytic signal and envelope only in hindsight \citep{Mandel1967}. Practically the same formulation---of a slowly varying complex amplitude---also arose in the derivation of the paraxial dispersion equation by \citet{Akhmanov1968,Akhmanov1969}, which is used in the latter half of this work (\cref{temporaltheory}).

\begin{figure} 
		\centering
		\includegraphics[width=.5\linewidth]{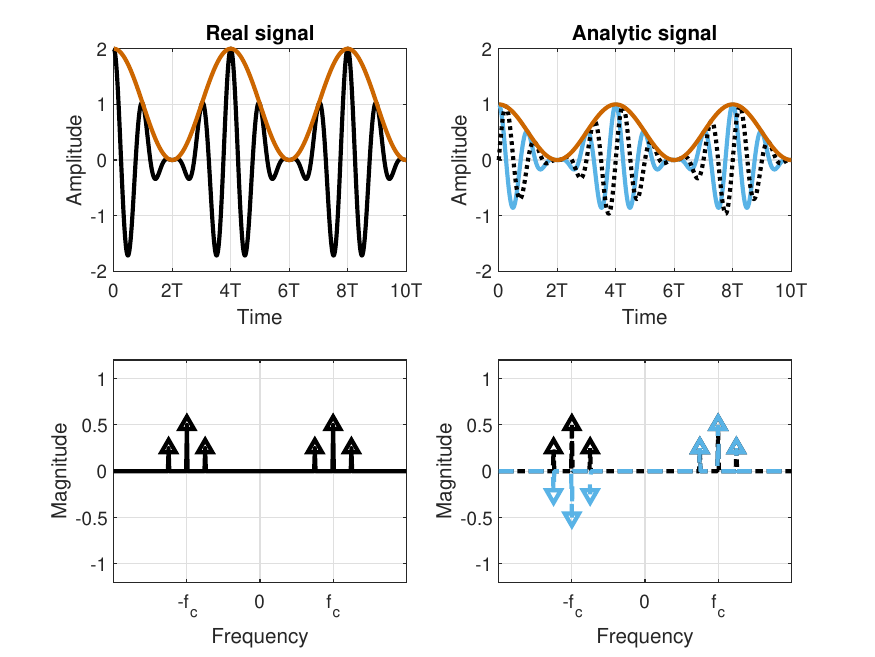}	
		\caption{This figure is similar to Figure \ref{Analytic1}, only with an amplitude-modulated tone. The temporal envelopes of both real and analytic signals are plotted in red in the top plots. The spectrum contains side bands around the tone carrier. The real signal (bottom right) has the usual negative frequencies, whereas they cancel out in the analytic signal (bottom right), because of the contributions of the even (solid black) and odd spectra (dash blue).}
		\label{Analytic2}
\end{figure}

The flexibility in constructing real signals by distinguishing the complex envelope and carrier domains enables us to construct high-frequency signals that carry low-frequency information using modulation\footnote{Of course, modulated signals can be formed by real envelopes and fluctuating carriers, as there is ambiguity in the definition. But we shall usually refer to the complex envelope formulations.}. The information is assumed to vary much more slowly than the carrier and can be fully contained in a low-frequency complex envelope. The low-frequency information is then referred to as the baseband or the \term{modulating} signal, and its transformed high-frequency version is the bandpass or the \term{modulated} signal \citep[p. 238]{Couch}. It can be made to modulate either $a(t)$---amplitude modulation (AM) (see Figure \ref{Analytic2}), $\varphi(t)$---phase modulation, or a combination of both (AM-FM, see Figure \ref{Analytic3}). In this context, $a(t)$ is called the \term{instantaneous amplitude}, and $\varphi(t)$ is the \term{instantaneous phase}. Its derivative is the \term{instantaneous frequency} \citep{Carson1922,Carson1937}
\begin{equation}
 \omega = \frac{d\varphi(t)}{dt}
\label{InstantFreq}
\end{equation}
which can be made to carry the information using frequency modulation (FM). 

Therefore, without loss of generality, physical communication takes place when a modulated signal of carrier frequency $\omega_c$ and bandwidth $\Delta \omega$ is transmitted to a receiver over a channel. The convenient properties of the bandpass system make it particularly attractive for communication, whereas wideband or even ultra-wideband may raise many technical difficulties in practice (see \cref{CommunicationTheory}), partly because of their more ambiguous mathematical and physical nature (see \cref{AnalyticChallenges}).

The instantaneous quantities are readily computed from the analytic signal, as these following relations hold
\begin{equation}
 a(t) = \sqrt{x^2(t) + y^2(t)} = \sqrt{z(t)z^*(t)} = |z(t)| 
	\label{AmplitudeDef}
\end{equation}
\begin{equation}
 \varphi(t) = \omega_c t + \tan^{-1}\frac{y(t)}{x(t)} = \omega_c t + \tan^{-1}\left(\frac{z^*(t)-z(t)}{z^*(t)+z(t)}\right)
	\label{PhaseDef}
\end{equation}

It will be useful to generically expand the instantaneous phase function to different terms of a power series around $t=0$ with\footnote{This useful nomenclature was presented by Mark Wickert in \url{http://ece.uccs.edu/~mwickert/ece5675/} without reference, but these terms occasionally appear in literature.}
\begin{equation}
	\varphi(t) = \varphi_0 + \Delta \omega t + \frac{1}{2}\Delta \dot{\omega} t^2 + \frac{1}{6}\Delta \ddot{\omega} t^3 + \cdots 
	\label{GeneralizedPhase}
\end{equation}
We refer to the terms from left to right as phase, frequency (also, \term{phase ramp} or \term{phase velocity}), \term{frequency velocity}, \term{frequency ramp}, or \term{phase acceleration}, and the last term is \term{frequency acceleration} or \term{phase jerk}---borrowing from the naming convention in mechanics \citep{Thompson2011}. 

\section{Auditory envelope and phase}
\label{AuditoryEnvPhase}
If the physical time signal is allowed to be completely arbitrary, then it can change equally likely in its instantaneous envelope and phase functions. Nevertheless, auditory effects of phase and envelope have been traditionally studied in isolation (e.g., measuring either AM or FM stimuli). An overarching framework for the two functions has been more prevalent over the last three decades \citep{Rosen1992}, although they are still largely viewed as separate entities. However, the separability of phase and envelope is not obvious if we employ the complex envelope, which requires a combination of both. In the remainder of this chapter, some of the milestones in auditory research of envelope and phase sensation are reviewed with emphasis on current trends and challenges, which will help bolster the case for using a complex envelope in hearing.

\subsection{Auditory sensitivity to temporal envelope}
\label{AudSenseEnv}
The following is a brief review of a few key findings in the study of amplitude-modulated sound perception. Historical studies of envelope perception are reviewed in \citet{Kay1982} and a modern synthesis is found in \citet{Joris2004} with emphasis on physiological findings.

\begin{figure} 
		\centering
		\includegraphics[width=.5\linewidth]{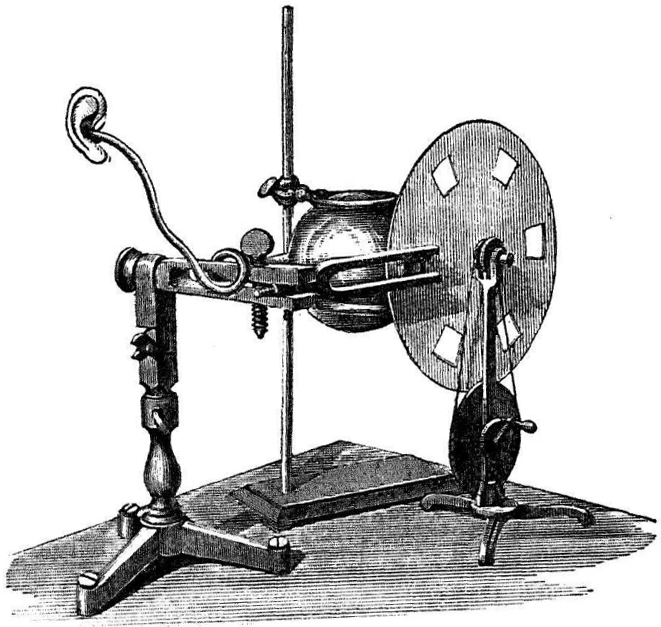}	
		\caption{The original setup used by \citet{Mayer1874} to test the continuity of amplitude-modulated tones that were produced by tuning forks and were interrupted by the rotating perforated disc.}
		\label{MayerFlicker}
\end{figure}

While AM has been long been used as a musical effect---\term{tremolo} (\textit{tremble}, in Italian)---at least since 1617 \citep{Carter1991}, AM as a test stimulus had been tested very sporadically until relatively recently. The first report appears to be by \citetalias{Mayer1874}, who used a perforated disc in front of a vibrating tuning fork to test at what frequency of interruptions the sound would be perceived as continuous (Figure \ref{MayerFlicker}). However, with more careful methods, it turned out that the interrupted tone never quite appears continuous, even at high modulation frequencies \citep[pp. 408--412]{Wever1949}. 

Perhaps the most interesting early AM research related the significance of the speech envelope to intelligibility. AM was used in a simple apparatus for speech synthesis, the \term{vocoder}, which partially modeled the speech signal by using the level information from a bank of bandpass filters to modulate narrowband noise generators \citep{Dudley1939}, a principle that was simplified later and was shown to produce intelligible speech \citep{Shannon1995}. A similar principle of bandpass filter bank analysis was also used in the design of the speech spectrograph, which displayed the running level fluctuations in the speech signals in different bands \citep{Potter1945,Koenig1946, Potter1950}. The importance of hearing the speech envelope properly was demonstrated with the aid of the modulation transfer function (MTF). Initially introduced into acoustics as a tool to measure the interaction between the temporal envelope and the room acoustics, it showed how reverberation tends to reduce the modulation depth of received speech, which corresponds to a loss of intelligibility (\citealp{Houtgast1973,Houtgast1985}; see also \cref{RoomAc}). Using the analytic signal, it was later found that severe loss of intelligibility can be the result of low-pass and high-pass modulation frequency filtering of running speech signals---especially when frequencies in the range of 4--16 Hz are removed \citep{Drullman1994b,Drullman1994a}. 

Modern psychoacoustic research of the perception of electronically-controlled AM may have begun with \citet{Riesz1928}. In this experiment, the audibility of beating between two sinusoidal tones was manipulated by varying the amplitudes of one of the tones, which resulted in maximum sensitivity among listeners for beating frequency of 3--4 Hz. Much of the subsequent AM research focused on the most common stimulus type, sinusoidal AM, which has been the standard throughout the literature
\begin{equation}
 p(t) = [1 + m\sin(\omega_m t)]\sin(\omega_c t)
	\label{SineAM}
\end{equation}
The carrier frequency is set by $\omega_c$ and the real envelope is defined by two parameters: the modulation frequency $\omega_m$ and the \term{modulation depth}, $0\leq m \leq1 $. It can be shown that the modulation depth can be also expressed by the minimum and maximum fluctuating intensity, $I_{min}$ and $I_{max}$, respectively:
\begin{equation}
 m = \frac{I_{max}-I_{min}}{I_{max} + I_{min}} 
	\label{DepthAudibility}
\end{equation}
this expression describes the audible contrast of the modulated sound. 

The perceptual sensitivity to AM was formalized with the introduction of the \term{temporal modulation transfer function} (TMTF) that directly quantifies the minimum audible modulation depth at a given modulation frequency \citep{Viemeister1973,Viemeister1979}. The sinusoidal carrier is sometimes replaced with broadband or narrowband noise, which yield different responses. The effect of the auditory filter bandwidth is clearly noticeable in sinusoidal and narrowband-noise modulation, which is not the case in broadband stimuli. The AM signal can be ``resolved'' to its sinusoidal components (e.g., bottom plots in Figure \ref{Analytic2}) when it is detected using spectral cues. It takes place when the signal is analyzed using adjacent narrow bandpass filters that are separated by less than the AM bandwidth, which results in a much improved threshold. These results will be discussed in depth in \cref{TheTMTF}.

As may be expected from the psychoacoustic observations, envelope processing can also be traced along all the main auditory nuclei with different specificity in different cell types. Instead of varying the modulation depth as was done in the earliest studies \citep[from][]{Nelson1966}, modern animal experiments often test the synchronization to a fully modulated ($m=1$) stimulus instead (beginning in \citealp{Palmer1982}), as it decreases with lower modulation depths \citep{Joris2004}. Functionally, the nonlinearity in the transduction of sound to neural discharges provides the necessary mechanism for demodulation \citep{Khanna1989AM, Nuttall2018}. 

The neurally encoded envelope is qualitatively different centrally, as a \term{rate code}---spiking patterns at the average AM frequency where instantaneous changes are no longer informative---becomes much more prevalent downstream after the inferior colliculus (IC) than the temporal code, which responds to instantaneous changes in the modulation frequency that is observable upstream\citep{Joris2004}\footnote{The distinction between temporal and rate encoding is discussed in \citet{Theunissen1995}. Note that even if rate coding is more prevalent downstream, variations of spiking patterns around the mean can still carry information \citep{Qasim2021}.}. Additionally, the maximum modulation frequency that can be gradually encoded decreases downstream from the auditory nerve, by up to an order of magnitude. It is found that some cells in the brainstem are tuned in the modulation domain and can have either a low-pass or a bandpass frequency response. While some areas are not highly specified as they synchronize to both carrier and envelope (e.g., the auditory nerve), others seem to be exceptionally geared to track the envelope information and sometimes even improve on the auditory nerve (e.g., by sharpening the response, or improving the signal-to-noise ratio). 

The AM stimulus is often  treated as magnitude only, without specifically considering the relative phase of the modulation. However, listeners are known to be sensitive also to modulation phase across different channels \citep[e.g.,][]{Bregman1985,Yost1989Across,Strickland1989,Moore1990,Lentz2015}, an effect that was also recorded in the auditory cortex of awake marmoset monkeys \citep{Barbour2002}. It appears that not all tasks are as sensitive to modulation phase changes \citep{Lentz2015}, though, and not all experiments establish phase effects \citep{MooreGlasberg1991, Furukawa1997, MooreSek2000}.

An influential theory suggests that not only is the IC organized tonotopically according to characteristic frequencies, but also according to modulation frequency, which may explain the salience of periodicity pitch \citep{Schreiner1988,Langner1997} (see \cref{CentralNeuroanatomy}). Another influential psychoacoustic signal processing model suggests a modulation filter bank that leads to modulation channels, for which periodicity would be only a special case \citep{Kay1982,Dau1997a,Jepsen2008}.

\subsection{Auditory sensitivity to phase}
\label{AuditoryPhaseReview}

\subsubsection{``Phase deafness'' and its discontents}
The role of phase in hearing has been much more contentious than that of the envelope. Especially, models that considered it to be immaterial for the ear seemed to have had considerable influence on the course of research and the general understanding of the ear. 

In his monumental work about hearing, Helmholtz concluded the following \citep[p. 127]{Helmholtz}: \textit{``...differences in musical quality of tone depend solely upon the presence and strength of partial tones, and in no respect on the differences in phase under which these partials enter into combination.''} With this conclusion he had vindicated the theory of \citetalias{Ohm}, whose original proposal that the ear performs Fourier-series analysis on incoming complex tones was heavily criticized at the time by August Seebeck \citep{Turner1977}. Helmholtz spelled out \term{Ohm's law} \citep[p. 33]{Helmholtz}: \textit{``Every motion of the air, then, which corresponds to a composite mass of musical tones, is, according to Ohm's law, capable of being analysed into a sum of simple pendular vibrations, and to each such single simple vibration corresponds a simple tone, sensible to the ear, and having a pitch determined by the periodic time of the corresponding motion of the air.''} This is usually summarized by the idea that the ear is ``phase deaf''. However, Helmholtz was almost exclusively concerned with compound \textit{``musical tones''}, which lend themselves to neat interpretation as the sum of \textit{``simple tones''}, limited only by the frequency resolution of the ear. All other nonperiodic and transient sounds (e.g., \textit{``jarring, scratching, soughing, whizzing, hissing''}) were classified as \textit{``noise''} \citep[pp. 119 and 127]{Helmholtz}. Incidentally, Ohm himself never discussed the role of phase in his original paper, which makes Ohm's law a misnomer \citep[Appendix A]{Goldstein1967phase}. Helmholtz had planted the seeds of doubt notwithstanding, by carefully avoiding to make his conclusion about phase too sweeping. Referring to transient noisy sounds, he added \citep[p. 127]{Helmholtz}: \textit{``we must leave it for the present doubtful whether in such dissonating tones difference of phase is an element of importance.''}

Nevertheless, Helmholtz's experimental results could not be replicated and evidence for the audibility of phase had accumulated over the twentieth century (for historical reviews and results see \citealp{Craig1962, Goldstein1967phase, Plomp1969,Patterson1987phase, Moore2002, Laitinen2013}). Typically, phase-detection experiments compared signals that have identical magnitude but different phase spectrum and documented noticeable shifts in timbre of complex tones or in masking threshold, as a consequence of their different phase spectrum. In his \term{pulse ribbon model} for phase detection, \citet{Patterson1987phase} distinguished local phase effects, which stem from phase differences between adjacent unresolved frequency components that pass through the same auditory filter, from global phase effects, which correspond to timing differences across different filters (but see \citealp{Laitinen2013}). As all of these studies concluded, the sensitivity to local phase changes may be attributed to changes to the within-channel signal envelope rather than to the broadband spectrum associated with a pure place model. Therefore, accounting for phase effects requires temporal processing---likely based on phase locking.

Despite its incorrectness, the ``phase deafness'' adage has effectively transmigrated into various formulations of the \term{power spectrum model} of hearing---the necessary conclusion out of a strict place model of the cochlea, as Helmholtz theorized. In its simplest version, the power spectrum model is based on the critical-band concept discovered by \citet{Fletcher1940}, who demonstrated psychoacoustically that sound is analyzed in the ear by a bank of bandpass filters that together cover the entire audio range. Different subjective correlates have been found that correspond to the amount of energy that goes into each filter. Important applications include the pure tone audiogram \citep{Fletcher1922}, the speech spectrogram \citep{Potter1945,Koenig1946, Potter1950}, the articulation index of speech \citep{French1947}, loudness models \citep{Zwicker1957,Moore1987Loud}, masking models \citep[e.g.,][]{Moore2013}, reverberation time and other room-impulse-response indices \citep{Kuttruff}, and the modulation transfer function \citep{Houtgast1985}. Except for the speech spectrogram that is based on short-term samples, these applications rely on long-term averages of narrowband spectra of the full broadband signals, which eliminate any contribution of phase terms (see also \citealp[p. 20]{Fant1970}). Perhaps it may seem surprising, then, that it is possible to reconstruct speech signals from their phase spectrum without its corresponding magnitude information, as long as some relatively general conditions hold \citep{Oppenheim1981}. In general, the phase spectrum contains more information than the magnitude spectrum, so that arbitrary signals suffer from smaller distortion when reconstructed from the phase spectrum alone, compared to the magnitude spectrum alone \citep{Ni2007}\footnote{It is worth noting that there are serious practical difficulties in obtaining the phase spectrum of arbitrary signals using standard methods for acoustic signal processing (e.g., using the phase obtained by short-time Fourier transforms), primarily due to phase wrapping and truncation effects of signal windowing. Modern techniques for overcoming these challenges have been an active field of research in the signal processing of speech \citep{Mowlaee2016}.}.

\subsubsection{Frequency modulation}
A much more obvious type of phase change that listeners can detect is frequency modulation (FM), which entails the modulation of the phase derivative. Perhaps the most common FM stimulus in hearing research is sinusoidal modulation (sinusoidal FM) of the phase around the carrier
\begin{equation}
	p(t) = a \sin\left[ \omega_c t + \frac{\Delta \omega}{\omega_m} \sin(\omega_m t)  \right]
	\label{SFM}
\end{equation}
where the modulation term is defined by the \term{modulation index} $\Delta \omega / \omega_m$ and the modulation frequency $\omega_m$. The peak \term{frequency deviation} from the carrier is $\Delta \omega$. While periodic, this signal has a rather complicated harmonic series, whose relative amplitudes are determined by Bessel functions of the first kind, $J_n\left( \Delta \omega/\omega_m \right)$ for harmonic $n$ \citep{Carson1937}. The larger the modulation index is, the more harmonics there are with non-negligible amplitudes. 

Sinusoidal FM has been used in different paradigms in hearing and only a handful are mentioned here. The musical counterpart to it is called \term{vibrato} (in analogy to tremolo for AM) and its associated perceived pitch approximately corresponds to the carrier frequency \citep{Tiffin1931, Iwamiya1984}. If the frequency deviation is very small, then the pitch of the carrier may sound like a pure tone, which was used to estimate the frequency discrimination associated with pure tones \citep{Shower1931}. 

Different psychoacoustic results of FM recognition that depend on the modulation frequency, modulation index, duration, and occasionally on its AM envelope, suggest that different auditory signal processing may be dominant at low and high modulation frequencies. For example, some sensitivity to FM may be explainable by conversion from FM to AM, as the spiking rate in the channel that analyzes the stimulus depends on its instantaneous frequency, which corresponds to a level change that is indistinguishable from that caused by an AM signal in some conditions \citep{Zwicker1952,Saberi1995}\footnote{The opposite conversion---between AM and PM---has been recently demonstrated in humans, where the level-dependent phase response of amplitude-modulated otoacoustic emissions was shown to correlate with the sensitivity to AM, at frequencies where phase locking was available \citep{Otsuka2021}.}. The span of FM signals affects several filters simultaneously, whose outputs are thought to combine to an \term{excitation pattern} that is only dependent on cochlear place---in the spirit of Helmholtz and Ohm's law \citep{Zwicker1956,MooreSek1994Disc,MooreSek1994Mixed}. However, this view has been challenged, as the responses to some low carrier-frequency ($<4$ kHz) and low modulation-frequency ($<5$ Hz) stimuli appear to rely on temporal (phase) information as well (\citealp{Edwards1994,MooreSek1995,MooreSek1996,MooreSk2002,He2007}, but see \citealp{King2019Mod}). More recent modeling of AM and FM thresholds suggests that auditory processing of FM may be altogether distinct than the assumed processing for AM \citep{Attia2021}. A further possibility is that the auditory sensitivity to FM is determined centrally (perhaps cortically) by the sensitivity to the fundamental frequency ($f_0$) of the sound, which is primarily a function of precise place coding in the cochlea \citep{Whiteford2023}.

Physiological measurements of the auditory nerve of the cat also suggest that while the instantaneous frequency of sinusoidal FM is phase-locked to the signal, it is, in fact, converted to AM by the auditory filter \citep{Khanna1989FM}. Different cell types of the ventral cochlear nucleus (VCN) of the guinea-pig were shown to either phase lock to the FM or to synchronize to its envelope, in a manner that depends on the characteristic frequency (CF) and the bandwidth of the cell's receptive field \citep{Paraouty2018}. In general, low-frequency VCN cells of larger bandwidth tend to phase lock to the carrier with little effect of intensity, whereas above 4 kHz, where cells also tend to be narrowband, they are synchronized primarily to the envelope. However, chopper and onset cells excel in envelope coding and do poorly in phase locking to the carrier even at low frequencies. In the different subnuclei of the cochlear nucleus of the bat, single units synchronize to maximum modulation frequency of 400--800 Hz of sinusoidal FM \citep{Vater1982}. In general, the degree of specialization of neurons to FM patterns increases the closer the auditory signal is to the IC \citep{Koch1998,Yue2007}. Despite these findings, the low-level physiological availability of phase cues may not necessarily translate to high-level perception \citep{Kale2014}.

Linear FM is a mathematically simpler signal, but has not been studied as much as sinusoidal FM. It is defined by a single parameter---the frequency slope (or velocity) $\Delta\dot{\omega}$
\begin{equation}
	s(t) = a \sin\left( \omega_c t + \frac{\Delta \dot{\omega}}{2} t^2  \right)
	\label{SimpleLinearFM}
\end{equation}
where the carrier frequency $\omega_c$ is better understood as a center frequency of a chirped pulse around $t=0$, whose frequency slope is $\Delta\dot{\omega} = 2\pi B /T$, where $B$ is the bandwidth, and $T$ is the pulse width. It can be seen that this stimulus requires band-limitation in order to contain its spectral range, which means that a specific envelope must be implemented. Thus, linear FM stimuli must involve some AM as well. 

One of the most interesting features of these stimuli is that sensitivity to them is direction-dependent, as listeners exhibit lower masking thresholds to upward ramps than to downward ramps, which emphasizes the role of their phase spectrum \citep{Nabelek1978,Collins1978}. Moreover, it has been found in mammals (beginning in \citealp{Whitfield1957} and \citealp{Whitfield1965}) that some cells in the IC and the auditory cortex are specialized in detecting particular patterns of FM (e.g., tuned to a certain bandwidth and to the direction of sweep---up or down). While these results are particularly telling about echolocating bats, they are found in different species and are thought to reflect the important role that linear FM has in vocalizations \citep[e.g.][]{Casseday1992,Klug2010}. Only one study appears to have tested the linear FM response at the auditory nerve level. In the auditory nerve of the cat, phase locking to the instantaneous frequency was demonstrated---both in upward and downward linear chirps \citep{Sinex1981}. 

Linear FM is of particular importance in this work, because of its relation to group-delay dispersion, its simplicity as a mathematical basis for arbitrary chirps, and the fact that it often leads to tractable expressions. The jargon associated with it includes many nearly-synonymous terms from different fields, which may seem obfuscating for the unacquainted reader. These terms are summarized in Table \ref{LinearFMterms} with clarifying, informal definitions, along with near synonyms that do not necessarily imply linearity.

\begin{table}
\footnotesize\sf\flushleft
\begin{tabular}{P{6cm}P{10cm}}
\hline
\textbf{Term} & \textbf{Definition}\\
\hline
\multicolumn{2}{c}{\textbf{Frequency}}\\
\hline
\textbf{Linear frequency modulation (FM)}& Refers to the mathematical signal or modulation technique (Eq. \ref{SimpleLinearFM}) that entails a linear change in the instantaneous frequency.\\
\textbf{Glide} & The term used most often in the hearing and speech/phonological literature for signals that rise or fall in frequency. It does not necessarily imply linearity, although often it is linear.\\
(\textbf{Frequency}) \textbf{ramp} & A linear FM signal, but with an emphasis of the upward or downward direction of the instantaneous frequency.\\
(\textbf{Linear}) \textbf{sweep} & Similar to a ramp, but emphasizes the coverage (either continuous or discrete) of a range of frequencies, as is often required in measurements.\\
(\textbf{Linear}) \textbf{chirp} & Similar to a glide, but used more often in technical literature such as signal processing and radar communication, and in the context of bat echolocation and birdsong.\\
\textbf{Frequency velocity} & The coefficient $\Delta \dot{\omega}$ of the instantaneous frequency, or of the quadratic term of the phase (Eq. \ref{GeneralizedPhase}).\\
(\textbf{Instantaneous}) \textbf{frequency slope} & Similar to frequency velocity, but linearity is implied, at least locally.\\
\hline
\multicolumn{2}{c}{\textbf{Phase}}\\
\hline
\textbf{Phase curvature} & The same as frequency slope, with emphasis on the quadratic (or higher-order) term in the phase function. The term may also relate to the second-order frequency dependence of the (arbitrary) phase spectrum. \\
\textbf{Quadratic phase} & The general phase function that contains a quadratic term in frequency or time. This term is most common in Fourier optics, where the linear phase term is either omitted or treated separately.\\
\textbf{Phase acceleration} & The quadratic coefficient of the time-dependent phase function.\\
\hline
\multicolumn{2}{c}{\textbf{Group delay}}\\
\hline
\textbf{Frequency-dependent group delay} & The negative derivative of the phase function with respect to frequency is the group delay of the system, which is non-zero if the system is dispersive. If the second-derivative is non-zero, then the group delay is frequency dependent, which implies frequency modulation for signals that go through the system.\\
\textbf{Group-delay dispersion (GDD)} & When the group delay is frequency dependent, then it is itself dispersive. The group-delay dispersion represents this characteristic of a medium or a structure in which the wave propagates.\\
\textbf{Group-velocity dispersion (GVD)} & The group velocity and group-delay dispersions contain the same information, but GVD refers more specifically to the effect on group velocity.\\
\hline
\multicolumn{2}{c}{\textbf{Oscillators}}\\
\hline
(\textbf{Long-term}) \textbf{frequency drift} & A term that indicates a slow change in frequency that is sometimes used in oscillator modeling. A functional form is not implied, as it can be affected by random processes.\\
\hline
\multicolumn{2}{c}{\textbf{Music}}\\
\hline
\textbf{Glissando} & Discrete and usually rapid playing of the intermediate notes contained in the interval between two notes.\\
\textbf{Portamento} & The continuous changing (or bending) of pitch of the interval between two notes.\\
\textbf{Vibrato} & The periodic changing of the pitch around its tuning.\\
\hline
\end{tabular}
\caption{A comprehensive list of jargon related to linear frequency-modulation used in different disciplines. Many terms do not require the FM to be strictly linear and can be synonymous in some contexts. The phase- and group-delay terms imply (linear) FM indirectly. Most terms do not have a closed definition, so their definitions here are provided according to the best understanding of the author.}
\label{LinearFMterms}
\end{table}

\subsection{The envelope and the temporal fine structure}
\label{EnvelopeTFS}
Despite the evident link between the temporal envelope and the phase spectrum, until recently they have been framed in research largely as separate entities. Using the unified framework of the analytic signal, a more recent trend in research has started juxtaposing the slow-varying envelope and the rapid phase changes---the \term{temporal fine structure} (TFS). \citet{Rosen1992} defined the TFS as \textit{``...variations of wave shape within single periods of periodic sounds, or over short time intervals of aperiodic ones as fine-structure information.''} The envelope was defined there as: \textit{``fluctuations in overall amplitude at rates between about 2 and 50 Hz as envelope information''}\footnote{In a footnote (p. 74), Rosen noted that this definition is not the same as the envelope of the analytic signal, although the two are related. This assertion is attributed to findings by Seggie (1986), but it is not clear why this is the case. At present, the community appears to treat the envelope as in the analytic signal \citep[e.g.][]{Moore2008}.}. This approach is consistent with hearing research that has not adopted the idea of complex envelope, but instead has applied the standard analytic-function real envelope with positive frequencies\footnote{It is curious to note that the conceptual origins of the analytic signal and the temporal fine structure are both rooted in quantum mechanics. The analytic signal was introduced by \citet{Gabor}, who used the complex quantum wave function formalism to try and establish the uncertainty relations in hearing, similarly to the ones from quantum mechanics \citep{Heisenberg1927}. The ``fine-structure constant'' was introduced in quantum mechanics by Arnold Sommerfeld in 1915--1916, who used it to account for the spectral line splitting of the hydrogen atom, due to relativistic and spin effects \citep{Kragh2003}. It was probably imported to hearing science by \citet{Licklider1952}, who was also trained as a physicist. Many of the famous results in quantum mechanics are based on stationary wavefunctions that are time-independent, as harmonic solutions to Schr\"odinger's equation are assumed. In these cases, the power spectrum is the only measurable property, and the phase cancels out. Gabor too suggested (in a footnote) that Ohm's law for hearing holds, yet his analytic signal accounted for the phase notwithstanding and therefore became a much more general tool than what it was originally developed for.}. 

The analytic signal hypothetically enables the separation of the envelope and the TFS from an arbitrary signal. This idea has been applied several times to illustrate how different aspects of the auditory signal processing may be dominated by one or the other. For example, speech intelligibility appears to be dominated by the temporal envelope \citep{Drullman1994a,Drullman1994b}, even when the TFS is removed \citep{Shannon1995,Smith2002,Zeng2005}. In contrast, the TFS appears to be important in sound localization and pitch (melody and tonal vowels) perception \citep{Smith2002,Xu2003}, voice identification and intelligibility with competing speech \citep{Zeng2005}, and listening in the dips \citep{Lorenzi2006}. However, the underlying methods have been challenged for being mathematically unsound (see \cref{AnalyticChallenges}), as the information in the discarded part (the envelope or the TFS) is in fact preserved in the remaining part. Indeed, speech intelligibility in quiet is also rich with TFS-only cues \citep{Lorenzi2006}. 


\subsection{A final remark}
The partial reviews above reveal a complex interplay between the envelope and phase functions with respect to the auditory perceptual roles they realize. However, it is not intuitively obvious why some situations and stimuli can be processed, ostensibly, with no phase information, whereas in other cases the phase either contains either an equal amount or most of the information. In order to be able to answer this question, a coherence theory is required that is suitable for auditory processing, which makes a distinction between coherent, partially coherent, and incoherent signals. This distinction delineates the signals for which processing the phase matters and those that do not. Essentially, these provide the basic conditions for phase locking (processing of TFS) to be altogether possible and, beyond that, useful. The foundations of such a theory are presented in \cref{IntroCoh}, \cref{CoherenceTheory}, and \cref{PLLChapter}. 

\section{Challenges to the analytic signal formulation}
\label{AnalyticChallenges}
Even though the analytic signal decomposition is unique, applying it universally to arbitrary signals is not without complications. This section reviews some of these challenges from different interrelated perspectives: mathematical, auditory, and conceptual. While the importance of the challenges to the interpretation of various results is yet unclear, they reflect a real engineering problem that the hearing system has to routinely solve: to construct unambiguous perceptual representations of broadband signals that are not mathematically unique. 


\subsection{Auditory challenges}
\label{AudChallenges}
\subsubsection{Empirical methods}
As was noted in \cref{EnvelopeTFS}, the standard analytic signal is often employed to obtain an estimate of the real envelope and TFS. So, for example, one common method has been to resynthesize signals using decompositions to envelope and TFS, which can be manipulated independently \citep[e.g.,][]{Drullman1994b,Drullman1994a,Smith2002,Zeng2005,Lorenzi2006}. However, this procedure has been criticized on several grounds. \citet{Ghitza2001} demonstrated how the envelope information is regenerated from a Hilbert-transformed signal, which was supposed to retain only TFS information. These results were confirmed physiologically from auditory nerve measurements in the chinchilla \citep{Heinz2009}. \citet{Schimmel2005} showed how the sub-band (i.e., narrowband filter bank) decomposition of arbitrary signals to a carrier and a real envelope is both contrived and incomplete, as it requires the modulated signals to be symmetrical around the filter center frequency---an unrealistic assumption\footnote{Another way to frame it is to say that the carrier frequency must be precisely estimated at all times. This requires some kind of phase locking mechanism \citep{Clark2012phd}.}. Instead, they suggested a two-dimensional bi-frequency decomposition (carrier and modulation frequencies) which includes complex envelope modulation (see also \citealp{Atlas2003} for a discussion about the bi-frequency decomposition relevant to audio signals). \citet{Apoux2011} built on these findings and emphasized how envelope and TFS manipulation that is suitable for synthetically produced analytic signals cannot be generalized to arbitrary signals, such as natural speech. The authors demonstrated how when the assumption of real non-negative envelope is used to decompose signals with a negative envelope, wideband TFS-only signals that span several critical bands become contaminated with envelope information in multiple sub-bands. Importantly, these findings were not specific for Hilbert decomposition and were shown for another envelope extraction technique (squared, low-pass filtered, half-wave rectification)\footnote{It is interesting to note that a similar discussion is ongoing in neuroscience, where cortical synchrony may be quantified purely temporally (by measuring phase locking), or both temporally and amplitudinally (using coherence) \citep{Lachaux1999}. However, in order for this separation to be meaningful, the frequency range has to be narrowband and even then it is not obvious that the two measures are truly independent \citep{Srinath2014,Lepage2017}. These concepts will be discussed with relation to hearing in \cref{IntroCoh}, \cref{CoherenceTheory}, and \cref{PLLChapter}.}.

According to the definition of the analytic signal, the low-frequency amplitude and phase---the components of the complex envelope---are not meant to be independent. Rather, it is the real and imaginary parts of the signal that are separable and even they constrain one another (at least long term). The polar representation of complex functions is neither unique nor separable and both amplitude and phase are dependent on the real and imaginary values of the function, as is seen from Eqs. \ref{AmplitudeDef} and \ref{PhaseDef} (see also \citealp{Boashash1992} and \citealp{Picinbono1997}). While it casts a doubt on some past claims in studies that found high speech intelligibility after removing the envelope or TFS information \citep[e.g.,][]{Smith2002,Lorenzi2006}, it is noteworthy that they produced meaningful results that appear to be relatively robust to variations and consistent with research based on other methods \citep{Gilbert2006,Sheft2008,Moore2008,Moore2019}.

\subsubsection{Bandwidth}
As was emphasized throughout this chapter, the analytic signal relies on the narrowband approximation, as are the majority of communication systems. But, while individual auditory channels are narrowband, the overall audio bandwidth is ultra-wideband (see \cref{HearingModDemod} below). Additionally, due to the low frequencies involved, there is a large overlap between the overall low-frequency audio and modulation spectra (see Table \ref{tab:acousticvision} and Figure 2 in \citealp{Joris2004})\footnote{Depending on the spectral resolution, noise, and complexity of the signals and detector involved, the modulation can appear as sidebands in the spectrum around the main lobe, or broaden the spectral bandwidth of the center frequency. Modulation frequencies, though, constitute an additional dimension to the signal and are sometimes thought of as hidden, as they cannot be straightforwardly detected using normal spectral analysis. Instead, detecting them requires so-called \term{cyclostationary} techniques, such as autocorrelation \citep{Gardner1988,Gardner1991,Antoni2009}. Demodulation can be seen as a rather specific cyclostationary application that targets a specific channel and does not necessarily aims to measure the specific spectral modulation content at its output.}. When signals are modulated within individual filters, the two spectra are independent and they are perceived in qualitatively different manner, so they are unambiguous. However, if the modulation frequency (in AM) or maximum frequency deviation (in sinusoidal FM) are large enough to be resolved by adjacent filters, then the modulation components disappear from the perceived modulation spectrum and appear in the audio spectrum. Conversely, when spectral components are too close to be resolved, they may beat---appear as modulation instead of two individual pitches. Therefore, this interdependence between the two domains is dictated by the auditory filter bandwidths throughout the spectrum. It means that the physical information that produces the acoustic vibrations is conserved either in the audio or in the modulation domain, but may not necessarily correspond to the physical mechanism that produced it, which may be either a mode of vibration or modulation, but not both (see \cref{InfoSourceChannel}). It also means that if signals are dramatically scaled either spectrally or temporally, they may undergo qualitative perceptual changes that render the message they carry less recognizable, as different components move between the two spectra.

This challenge is not unique to the analytic signal framework, but it is made especially pertinent because of its predication on narrowband signals. 

\subsection{Mathematical challenges}
As was noted earlier, the analytic signal prescribes a unique combination of real and imaginary signals, which do not necessarily translate to unique amplitude and phase, once converted to polar representation. As the amplitude function determines the degree of AM and the phase function determines the FM, complex AM-FM signals may not be uniquely decomposable, which makes their decomposition an ill-posed problem. \citet{Loughlin1996} suggested a method to uniquely determine the AM-FM while satisfying these four conditions (cf. \citealp{Vakman1996}): signal level boundedness must stem from the AM part, bandwidth limitation must apply to the FM, pure tones entail constant AM and FM, and level scaling applies to AM but not to FM. These conditions are very reasonable, as long as the system is linear. The solution Loughlin and Tacer proposed is based on the complex envelope, effectively (see also \citealp{Atlas2004}). The carrier is obtained from the time-frequency distribution, which can be used to coherently demodulate the signal (e.g., by subtracting its carrier; see \cref{CommunicationTheory}).  \citet{Cohen1999} further explored the problem and proposed that it may be resolved either with a non-negative amplitude and occasionally a discontinuous phase, or by forcing continuous phase and accepting the existence of negative amplitudes. The latter solution can be made to work as a complex envelope. However, the conclusion that the analytic signal produced ambiguous instantaneous quantities was criticized by \citet{Hahn2003}, who showed that by using complex phase and frequency functions the ambiguity disappears. While mathematically correct, it requires us to embrace these complex functions that are not particularly intuitive. Additionally, it does not solve the auditory challenges (\cref{AudChallenges}) that were observed after the analytic signal was applied to separate the TFS and envelope. 

Broadband signals pose an even more challenging problem than narrowband signals, as they do not have a unique representation that is based on specific decomposition \citep[e.g.,][]{Boashash1992}. The perceived signal must be the end-result of a combination of outputs from all the active auditory filters, which gives rise to the auditory experience. This has led to the development of different time-frequency analysis methods that are capable of extracting the narrowband frequencies from broadband signals---often still resorting to the analytic signal along the process \citep{Huang2009}.

An additional challenge for analytic signals has to do with nonstationarity and is briefly mentioned in \cref{Nonstationarytheory}.

\subsection{Frequency and instantaneous frequency}
\label{FreqInstFreq}
The concept of instantaneous frequency was presented earlier, in passing, where it was defined as the derivative of the phase with respect to time. For a pure tone that has a linear phase function, the frequency is a constant and is exactly equal to the instantaneous frequency. However, standard frequency, as is obtained from Fourier analysis of time signals, measures repeating patterns over the entire duration of the signal from minus to plus infinity, but is not time dependent in itself. For example, an arbitrary FM signal can be expressed in the form 
\begin{equation}
	x(t) = a \exp \left[ i\left(\omega_c t + \gamma \int_{-\infty}^t m(\tau) d\tau \right)\right]
\end{equation}
the instantaneous frequency being $\omega_c + \gamma m(t)$, which describes the frequency deviation from the fixed carrier frequency. For any $m(t)$ that is non-constant, $x(t)$ is probably not going to be periodic. 

How is one to understand the meaning of an instantaneous frequency function that describes a fluctuation that need not be periodic? The conceptual discrepancy between the two frequency definitions was noted already by \citet{Carson1922} in his early work on FM and has been debated ever since \citep[pp. 39--41]{Boashash1992,Cohen1995}. \citet{Gabor} echoed this question when he introduced the analytic signal and noted that our auditory perceptual intuition does not necessarily coincide with the traditional Fourier view of frequency, which is much more rigid than the common perceptual experience of time-varying frequencies. Rather, he noted, our hearing experience lies somewhere in between the steady-state Fourier frequency and the instantaneous one. An ad-hoc solution in numerous auditory models and other acoustic applications has been to use some variation of the \term{short-time Fourier transform}, which is integrated over short (overlapping) time windows. This retains short periods of constant spectrum, which can be merged together into a continuous signal, when the time windows are partially overlapping.

As long as the narrowband approximation is carefully maintained, the instantaneous frequency appears to have a plausible physical meaning as the deviation from the carrier (plus the carrier) \citep{Boashash1992}. Other useful definitions exist, then, which may provide further intuition. For example, the instantaneous frequency is the first frequency moment of the time-frequency distribution of the signal, such as the Wigner-Ville distribution. Or, for a monotonic instantaneous frequency and large time-bandwidth product, it also determines the group delay of the signal as a function of frequency (also defined locally for a narrowband signal; see \cref{PhysicalWaves}).

\begin{figure} 
		\centering
		\includegraphics[width=.5\linewidth]{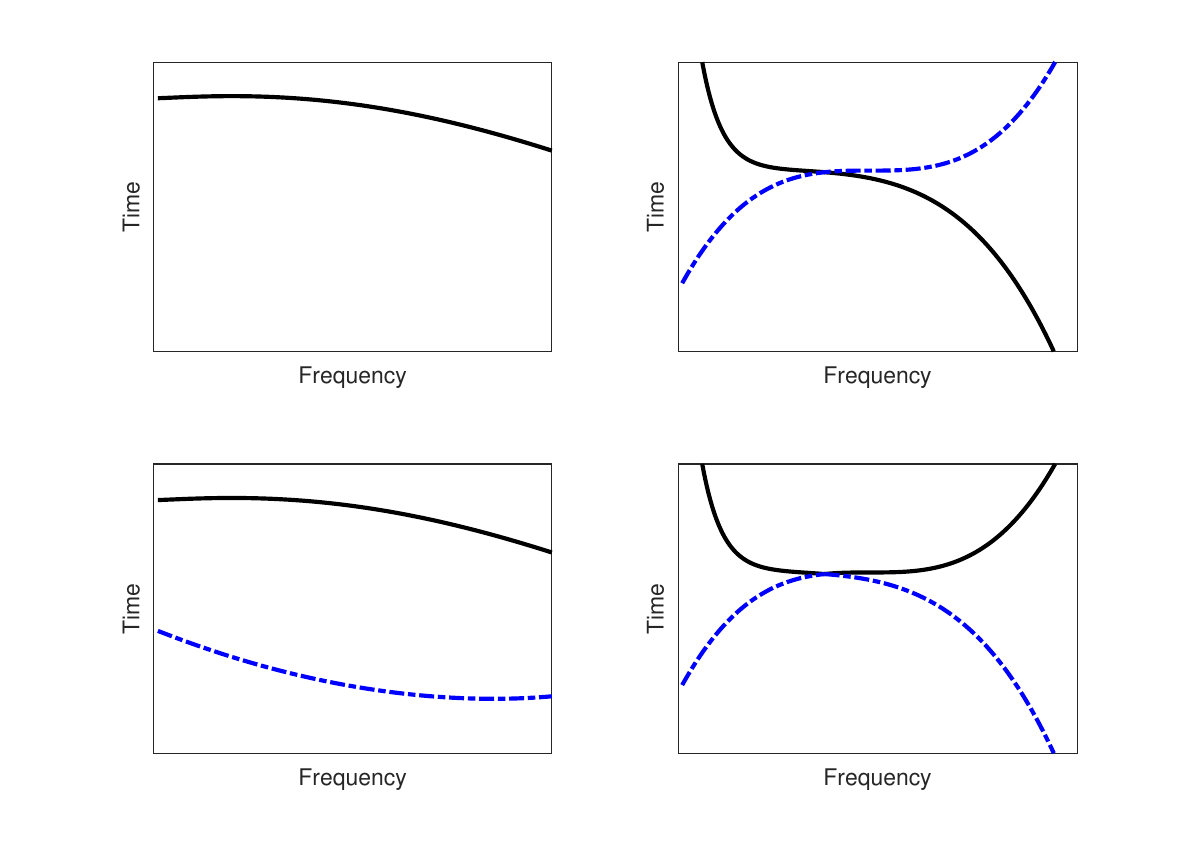}	
		\caption{Different types of signals qualitatively plotted on the time-frequency phase plane. \textbf{Top left}: A monochromatic signal can be unambiguously described in terms of its time-frequency distribution, although it has to be narrowband in order for the instantaneous frequency to be meaningful. \textbf{Bottom left}: A multicomponent signal with two components that can be described unambiguously, since their trajectories do not intersect. \textbf{Top and bottom right}: A multicomponent signal with two components, whose trajectories intersect, is ambiguous in terms of which trajectory belongs to which component. This figure is based on Figures 3 and 13 from \citet{Boashash1992}.}
		\label{multcomp}
\end{figure}

Importantly, the instantaneous frequency loses any coherent meaning for broadband and multicomponent signals. This situation is particularly delicate, because multicomponent signals do not have a unique representation in terms of variable components, if they intersect on the time-frequency plane. This is the reason why, in general, broadband signals do not have a unique representation. An example from \citet{Boashash1992} that illustrates this situation for two components is reproduced in Figure \ref{multcomp}. Nevertheless, long and broad sweeps, for example, do not sound ambiguous, so the instantaneous frequency has a role there that extends beyond a single narrowband filter. The instantaneous frequency is also indispensable in nonlinear and nonstationary systems, where the Fourier transform is generally invalid (\citealp{Huang1998,Huang2009}; see also \cref{PhysicalWaves}). 

\section{Hearing, modulation, and demodulation}
\label{HearingModDemod}

\subsection{Real and complex envelopes}
\label{RealandComplex}
\begin{figure} 
		\centering
		\includegraphics[width=.5\linewidth]{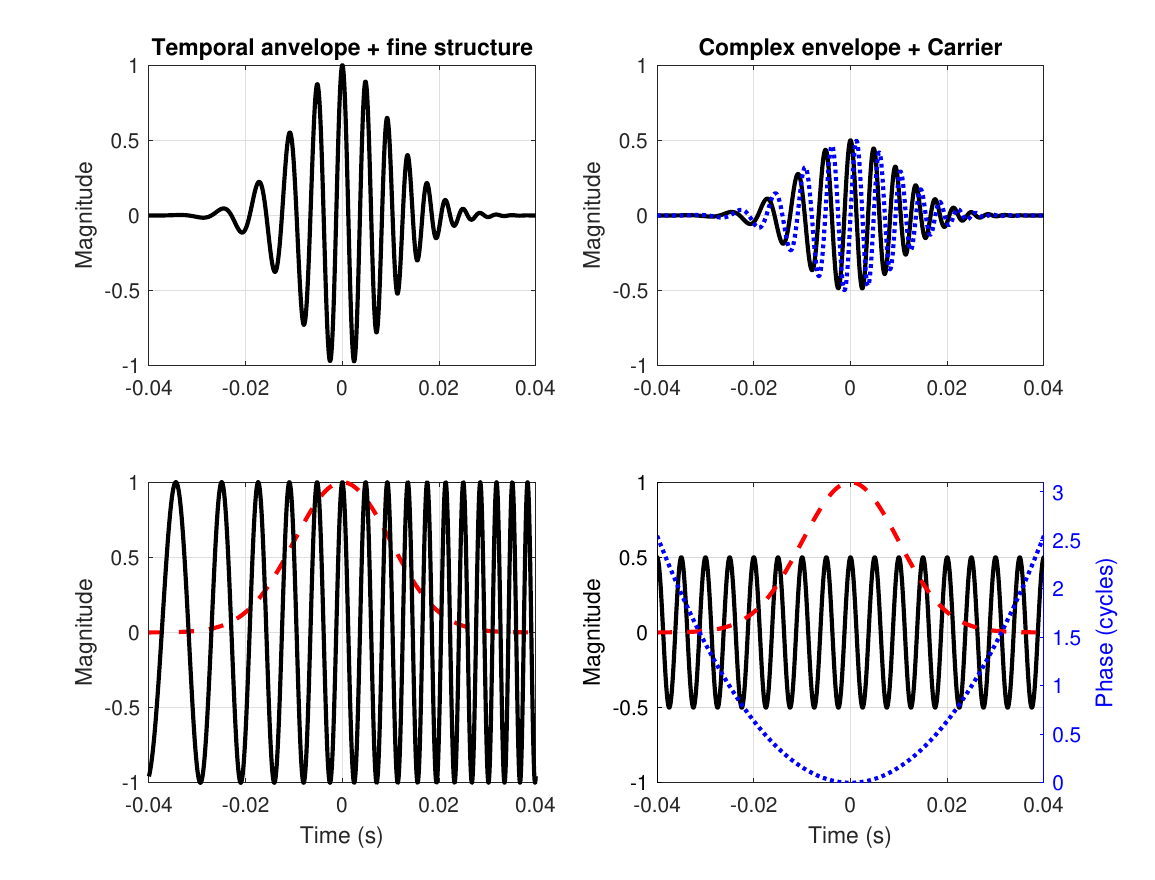}	
		\caption{The real time-signal of a linear chirp (top left) and its analytic signal representation---real and imaginary parts (top right). In the bottom figures, two decompositions of the signals are displayed. \textbf{Bottom left}: The standard auditory decomposition to a real envelope and temporal fine structure. \textbf{Bottom right}: The complex envelope and carrier of the analytic signal. The complex envelope is displayed as magnitude (dash red) and unwrapped phase in cycles (dotted black). The constant carrier is in solid black.}
		\label{Analytic3}
\end{figure}

The difference between real (standard auditory interpretation) and complex (nonstandard) envelopes is in the phase term. The complex envelope includes the frequency deviation around the carrier, which is treated as a constant that is subtracted from the total frequency. Thus, the spectrum of the complex envelope is double-sided and generally includes negative frequencies. A real envelope function is obtained only when the spectrum is Hermitian-symmetrical, which requires precise subtraction of the carrier. So, when the real envelope representation is employed, the carrier is taken to be part of the TFS, which leaves the phase modulation in the bandpass domain. Mathematically, this difference can be thought of as two different ways to associate the signal components. Using the complex envelope, the signal is $\Re\left\{[a(t)e^{i\varphi(t)}] \cdot [e^{-i\omega_c t}]\right\}$, whereas the real-envelope signal is $\Re\left\{|a(t)| \cdot [e^{-i\omega_c t+i\varphi(t)}]\right\}$ (see Figure \ref{Analytic3} for an illustration) (cf., \citealp{Shekel1953}). In other words, according to this standard view of hearing science, the auditory system does not necessarily demodulate the incoming signal phase. To the extent that any demodulation takes place, it occurs only in amplitude, which can theoretically happen noncoherently (without the phase information, or precise knowledge of the carrier; see \cref{CommunicationTheory}). 

There is another important difference that the complex envelope and the demodulation process emphasize, which the real envelope approach tends to overlook. Modulation is about changes from the mean---the mean being the static carrier amplitude and frequency. Thus, expressing the changes from the mean is much more efficiently done with low modulation frequencies rather than high frequencies, as it requires a smaller amount of information to be communicated between the transmitter and the receiver. The low-frequency deviation of FM, which the complex envelope isolates from the carrier along with AM frequencies, may be sampled at a low rate, effectively achieving lossless compression. It is impossible to take advantage of this economy when the fast high-frequency changes in the TFS are those that are being communicated. 

Whether the TFS is demodulated or not, a complex envelope representation is more general, as it retains the option for noncoherent detection. Indeed, some hearing research trends suggest that a conceptual transition from real to complex envelope may be timely. The significance of the TFS information (as currently defined) in normal hearing is gradually understood to be critical to many of its functions \citep{Moore2008,Moore2014,Moore2019}. At the same time, there may be a warming up to the idea of auditory demodulation, especially when the baseband components can be directly identified in the system \citep[e.g.,][]{Khanna1989AM,Khanna1989FM,Teager1990,Feth1992,Khanna2001,Khanna2002,Cooper2006,Nuttall2018}. The hypothesis that AM and FM may be centrally coded in the same channel, which depends only on the modulation frequency of either AM or in FM, is also notionally closer to the demodulation idea \citep{MooreGlasberg1991}. In animals such as bats, whose audible bandwidth can reach up to 100 kHz, modulation of very high-frequency carriers is standard so that synchronization can only happen in baseband---effectively after the signal has been noncoherently demodulated \citep[e.g.,][]{Vater1982,Bodenhamer1983}. As was implied in \cref{AnalyticChallenges}, the analysis of these cases may benefit from the intuitive appeal of the complex envelope, which appears to have the potential to solve at least some of the mathematical difficulties associated with real envelopes. A complex envelope (even when this term is not used explicitly) seems to better represent practical problems in hearing, such as estimating the fundamental frequency of speech through frequency-following response (FFR; \citealp{Aiken2006}), or accounting for the effects on speech intelligibility  of differential envelope delay or time reversal, which are applied to different bands of the broadband speech spectrum \citep{Greenberg1998,Greenberg2001}. Importantly, the complex envelope as a mathematical tool also coincides with the common complex amplitude or phasor formalism that is used throughout wave physics, including most of those that are used later in this work. This may enable a more intuitive interpretation of some solutions, which are readily understood to be separable to carrier and modulation domain information\footnote{While approached from a numerical perspective without reference to the mathematically complexity of the envelope, a recent computational model of cochlear processing using the envelope as its primary object has shown an impressive predictive power for several key auditory phenomena, including an effective extraction of the auditory filter responses \citep{Thoret2023}. It relies on a variation of the empirical mode decomposition algorithm (limited to a single iteration), which separates the upper and lower envelopes and inherits the periodicity contained in both \citep{Huang1998}. While not referred to as complex envelope, this method retains all the information in the signal---notably any asymmetry that is contained in its envelope---which is equivalent to using the complex envelope of the signal minus the static carrier signal. This procedure is indeed expected to be most suitable for dealing with nonstationary ``improper'' real-world signals (\cref{Nonstationarytheory}).}. 


It must be remembered that this discussion merely relates to different ways to represent signals mathematically, which ultimately all map the very same real physical signal. However, different representations may better correspond to different perceptual counterparts produced by the hearing system. 

\subsection{Two spectra}
\label{TwoSpectra}
Aside from phase deafness, Helmholtz's legacy reflects an additional important cause that may be implicated in the predominance of a real envelope interpretation that is divorced from true demodulation---as is the norm in hearing science. Ever since ancient Greece, knowledge in acoustics was intertwined with musical understanding---of intervals, tuning, instruments, consonance, dissonance, etc. \citep{Hunt1992}. The scientific and mathematical advents of the 19th century---mainly Fourier's theory---provided appealing solutions to simple vibrating systems and, as Ohm proposed, had a high explanatory power for a more general range of problems. However, without fail, the emphasis throughout has been tones, both simple and complex---how they interact, or how musical they sound. Pitch, according to this view, is the ultimate acoustical percept. Because pure-tone pitch (rather than periodicity, residue, or binaural pitches) is determined primarily by frequency, it is perhaps reasonable that precise spectrum analysis, as implied by the pure place model, is the most important aspect of hearing. It is natural then that frequency resolution becomes key to understand auditory perception and design. For example, \citet{ZweigLipes1976} discussed the ``\textit{cochlear compromise}'' of the cochlear mechanical design, which entails a trade-off between minimal reflection of sound and maximal spectral resolution. Extrapolating from this perspective, envelope detection becomes a secondary feature of the system, which can use it to extract additional information from the signal that relates the varying level of specific spectral components. Furthermore, if the signal information is mainly in its magnitude spectrum, then modulating it is an extra layer of information that is optional. This logic entails that the signal does not have to be demodulated, because there may be no underlying message to be received, aside from the spectrum itself. However, if the modulation spectrum is equally important as the audio spectrum, then it is not at all obvious that a very sharp frequency resolution is the correct design goal for the system. 

Let us compare the auditory filter design problem with that of generic communication and optical imaging systems. In standard communication (of a single carrier), it is the low-frequency information that is of the ultimate importance (e.g., the message). It modulates a somewhat arbitrary carrier, which is selected according to its physical properties and can address different requirements of information transmission: What is the required channel bandwidth? How much energy is required to generate the carrier? How far can it travel? How noisy does it become? How much interference is it subjected to by competing transmissions? How much error does it typically accumulate? How complicated is it to modulate and demodulate these frequencies? And so forth. However, once the carrier is received and demodulated, its role is over and it does not hold any additional information but its very own channel identity\footnote{Incidentally, a general and modality-independent neural model suggested that communication between neurons in the thalamus and cortex does exactly what we would expect it to do according to the communication theory: it rate-codes only the channel identity, while it temporally-codes the actual message \citep{Hoppensteadt1998}.}.

In optical imaging, as in the eye, the image \textbf{is} the modulation pattern---the complex envelope that is manifest in colors, which are analogous to pitch. Light energy is carried at frequencies that cannot be biologically tracked and are immediately demodulated in the retina. The color---really, the visual channel---obviously carries some information (similar to cochlear place information). But a colorless (black and white) image still contains much of the optical information. Either way, just as in other communication systems, the message is contained in the (complex) envelope patterns, which modulate the light. Therefore, visual imaging can be understood as a form of communication, as was argued in \cref{ImagingCommunication}. 

Given the finite bandwidth of the auditory system---its very low ($f_L = 20$ Hz) and not very high ($f_H = 20$ kHz) cutoff frequencies---its inputs and outputs are often treated as baseband signals that can be directly sampled without demodulation, e.g., at a sampling rate of 44.1 kHz, which samples all frequencies from 0 Hz. For example, \citet[p. 158]{Fastl} conveniently assume that the lowest critical band runs from 0 Hz, or \citet{Oppenheim1981} directly compares processing the speech audio phase spectrum to a visual image modulation phase spectrum. This switch from bandpass to baseband signaling can only be done because of the relatively low frequencies involved in hearing, but it would be out of the question if they were much higher. Informationally speaking, however, it is wrong, since hearing is a true bandpass system.

One may ask, then, why hearing is any different from standard communication systems? After all, it is certainly used for communication, so is there any particular reason why it should not be configured as a communication system? One answer may have to do with the fact that the ear contains many more channels than the eye, or most typical radio receivers. It allows for very rich spectral coding, for which the modulation information is relatively secondary. However, even as a classical spectrum analyzer or pitch interpreter, it still has to acquire the onsets and offsets of the different tones, as well as their varying levels. These can all be thought of as AM envelope functions, without any loss of generality \citep[e.g.,][]{Kay1982}. 

A related argument for why hearing is different from other communication systems is that it clearly violates the narrowband approximation, if all the channels are considered en masse rather than individually. Thus, the human ear may be more usefully classified as an \textbf{ultra-wideband} (UWB)  system, which can be defined somewhat differently according to the application. In radar systems, for example, an UBW system is said to have a relative bandwidth that satisfies $(f_H-f_L)/(f_H+f_L) \geq 0.25$ \citep[p. 2]{Taylor1995}, which for human hearing is close to unity(!). In communication systems, UBW systems are defined by having bandwidth larger than 20\% of the center frequency \citep[p. 11]{Arslan2006}---a limit that is breached in hearing no matter how the center frequency is defined. Similarly engineered systems to hearing may exist in \term{impulse radio} or \term{multiband UBW} communication systems---two UBW techniques which distribute the large bandwidth in pulse trains, or divide the signal into several carriers, respectively \citep[pp. 11-12]{Arslan2006}. The same bandwidth criteria for UWB systems are also applied in optical communication \citep{Yao2007}. Thus, as a communication system, hearing appears to have a peculiar design that may well qualify as UBW.

Because the carrier-centered and envelope-centered perspectives on hearing are complementary points of view, deciding which one is more faithful to the system's internal logic may seem like a moot exercise. Nevertheless, the envelope-centered perspective---that of true modulation---would imply that the specific identity of the carrier---the exact pitch---is of secondary importance. If the same message can modulate different carrier(s) and it can be decoded equally well, then it would strongly support an envelope-centered perspective. This ``spectrum-invariance'' is indeed the case often---within some constraints---as a verbal message can be equally well received if spoken by two different voices of different fundamental and formant frequencies, or a musical piece can be similarly well-received if it is played for different arrangements and in different keys. This does not diminish from the importance of spectral range and across-channel periodicity effects that are uniquely auditory. And indeed, music and speech can be equally well perceived also at different tempi, which entails scaling of the low-frequency envelope, but not of the carrier spectrum. Therefore, there is a broad category of messages that appears to be relatively invariant to both spectral and temporal changes, which suggests that the auditory system simultaneously extracts information from at least two dimensions. It also suggests that the system may not be constrained to the spectrum nor to the complex envelope, as long as the acoustic information can be extracted from one of them, or from the combination of both.




\chapter{Toward a unified view of coherence}
\label{IntroCoh}
 
''\textit{It seems however that the best way to reach a conclusion in the near or far future is to try and apply to optics  the central problem of communication theory. In order to do this one should not indulge too much in dealing with coherent or semi-coherent  illumination, frequency analysis and all straightforward translations of well-known radio communication results into optical terminology. Such topics have been perhaps a little overstressed in optics. The eye is not the ear and any interpretation of vision in terms of frequencies is really too far-fetched.}'' \citepalias{diFrancia1955}

\section{Introduction}
Hearing science is manifestly interdisciplinary and as such it has absorbed ideas and methods from numerous fields in science and engineering. As every hearing researcher brings their own expertise and perspective into the science, the collective knowledge about hearing has a distinct richness to it. This has meant that the different inputs from all these fields have often been introduced independently of one another, so they do not always readily coalesce into a coherent whole. A key example is the concept of coherence in hearing (no pun intended), which has been imported to auditory science from several different disciplines---optics, communication, and neuroscience are the primary ones---each at a different period, and often without consideration of one another. Therefore, coherence has become an ambiguous term that is not well-defined  within hearing, despite its widespread use. 

On an intuitive level, \term{coherence} quantifies how closely related two (or more) signals are. This can be applied to observations of the same signal at different coordinates, which correspond to different signal evolution or processing. It can also apply to different signals that originate from the same source, or to different signals that are subjected to common modulation. Essentially, coherence provides information about the identity of different signals or measurements---an identity that may be difficult to ascertain using other measures. Coherence theory has been developed mainly with respect to optics and is critical in imaging theory, as imaging is usually classified as either coherent or incoherent. It has also been used in a more ad-hoc way in communication engineering with the design of receivers that employ either coherent or noncoherent detection, which depend on whether they contain a local oscillator that can be made to track the carrier phase. In acoustics and hearing, coherence theory has been adopted in a rather sporadic manner in physical and room acoustics, signal processing, and hearing. 

\term{Synchronization} is the effect of binding together the outputs of independent oscillating systems, in a way that confers the temporal pattern of one oscillator to the other. Synchronization effects have been used extensively in engineering, including in coherent detection in communication systems. Just like coherence, synchronization has also become more prominent in neuroscience as a likely universal feature of the brain operation. In parallel, synchronization effects have been identified throughout the auditory pathways as a hallmark of hearing and are gradually being recognized for their significance. From the definition of coherence above, we can see that when an output is synchronized to the input, then the two are effectively coherent. Indeed, when referring to neural transmission  it is nearly synonymous to talk about it as being coherent or synchronized.

In order to track the coherence properties of the acoustic source all the way to the brain, it is necessary to have a unified view of what it is exactly that is being tracked. However, none of the available coherence theories can continuously map the entire auditory processing chain. Classical coherence theory in optics, for example, deals mostly with stochastic and stationary processes that do not represent very well the dynamic nature of the kind of signals that are regularly encountered in acoustics and hearing. Acoustic theory has a few tools that are useful in room acoustics and reverberation, but not elsewhere. Communication engineering methods may provide an adequate classification for the detection and signal processing that is desirable, at least at the interface of the auditory system with its environment, but requires committing to a specific modulation and detection method. Finally, synchronized brain activity models (auditory and others) have been completely divorced from these disciplines and they tend to remain confined to the neural domain, which makes their connection with realistic acoustic environments weak. 

\begin{table}
\footnotesize\sf\centering
\begin{tabular}{P{7cm}P{7cm}}
\hline
\textbf{Optics} & \textbf{Acoustics}\\
\hline
Temporal coherence & Temporal correlation\\
Spatial coherence & Spatial correlation\\
Cross-spectral density / Mutual spectral density & Coherence\\
Coherence length & Correlation length\\
Coherence time & Correlation time\\
\hline
\end{tabular}
\caption{A jargon ``thesaurus'' for coherence functions used in optics and their most common counterparts in acoustics, where the optical terms were not employed. The terms from optics are adopted in this work and are explained in \cref{CoherenceTheory}. For a few additional terms in optics see \citet[Table 5.1, p. 185]{Goodman2015}.}
\label{coherenceterms}
\end{table}

The following contains short quasi-historical and conceptual reviews of coherence in the different domains that are relevant to hearing, which have also guided this work, to a large degree. The reviews enable the synthesis of the various perspectives, which can then form a basis for a unified understanding of the topic in general. In \cref{CoherenceTheory} and \cref{PLLChapter} we will provide more detailed accounts of coherence theory and synchronization, with the phase-locked loop as a primary component that is at the heart of the mammalian auditory system.

\section{Perspectives on coherence}
Because the concept of coherence has been used in somewhat different contexts within communication, optics, physical acoustics, hearing, and neuroscience, it is instructive to trace back a few of the milestones in its development in these disciplines. The short introductions below are not intended to be thorough reviews, but rather provide vignettes on different needs and systems that motivated the development of this important concept. It will turn out that hearing theory requires a hybrid approach, with elements borrowed from all the particular coherence theories. The development of optical coherence theory is reviewed in \citet[pp. 554--557]{Born}, but no such historical accounts were found for any of the other fields. 

The first technical use of the term coherence appeared in one of the earliest inventions of the radio days---\term{the coherer}. It was invented by Edouard Branly in 1890, and was perfected by Oliver Lodge who patented it in 1898 \citep{Dilhac2009}. Lodge called it ``Branly's coherer'' (from Latin: cohaere---\textit{to stick}). The coherer was capable of remotely detecting a spark discharge by a change in resistance of a tube filled with granules of oxide metal, as a result of an electric charge, which made metal surfaces in the vicinity of the charge  momentarily ``fuse together'' \citep[p. 64]{Garratt1994}. This device was widely used by Guglielmo Marconi in his early attempts for radio transmission, until more suitable inventions became available that enabled continuous wave transmission and reception at practical power levels. 

\subsection{Optics}
\label{OpticalCoh}
\begin{figure} 
		\centering
		\includegraphics[width=0.6\linewidth]{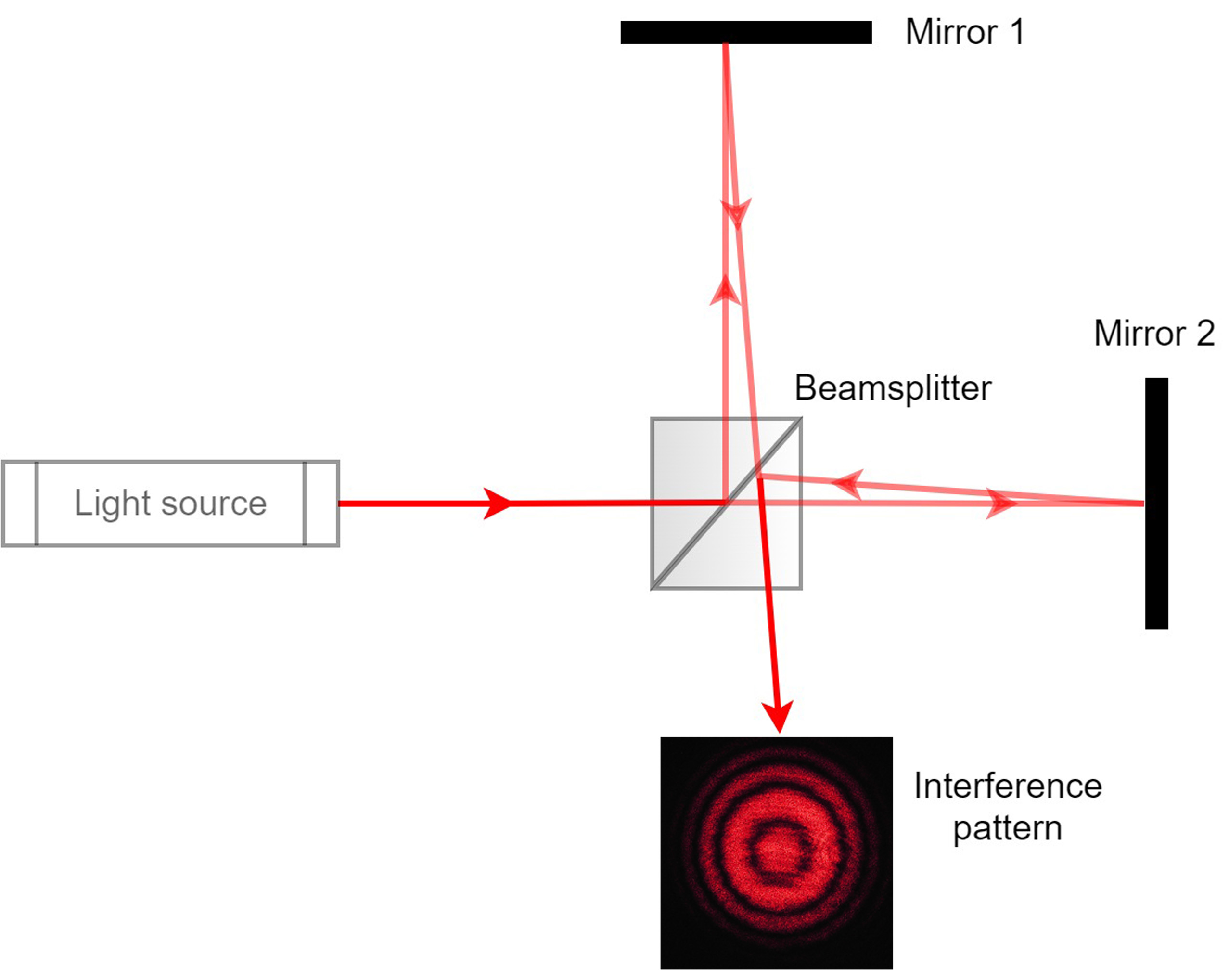}	
		\caption{Michelson interferometer. A beam of light from a coherent source is split evenly in the beamsplitter, with one beam going toward mirror 1 and the other continuing to mirror 2. The returning beams are united at the beamsplitter and are measured at the detector, where they form an interference pattern. The interference pattern photo at the detector was obtained from a He-Ne Laser (633 nm), by FL0, \url{https://en.wikipedia.org/wiki/File:Interferenz-michelson.jpg}.}
		\label{Michael2}
\end{figure}

\begin{figure} 
		\centering
		\includegraphics[width=0.7\linewidth]{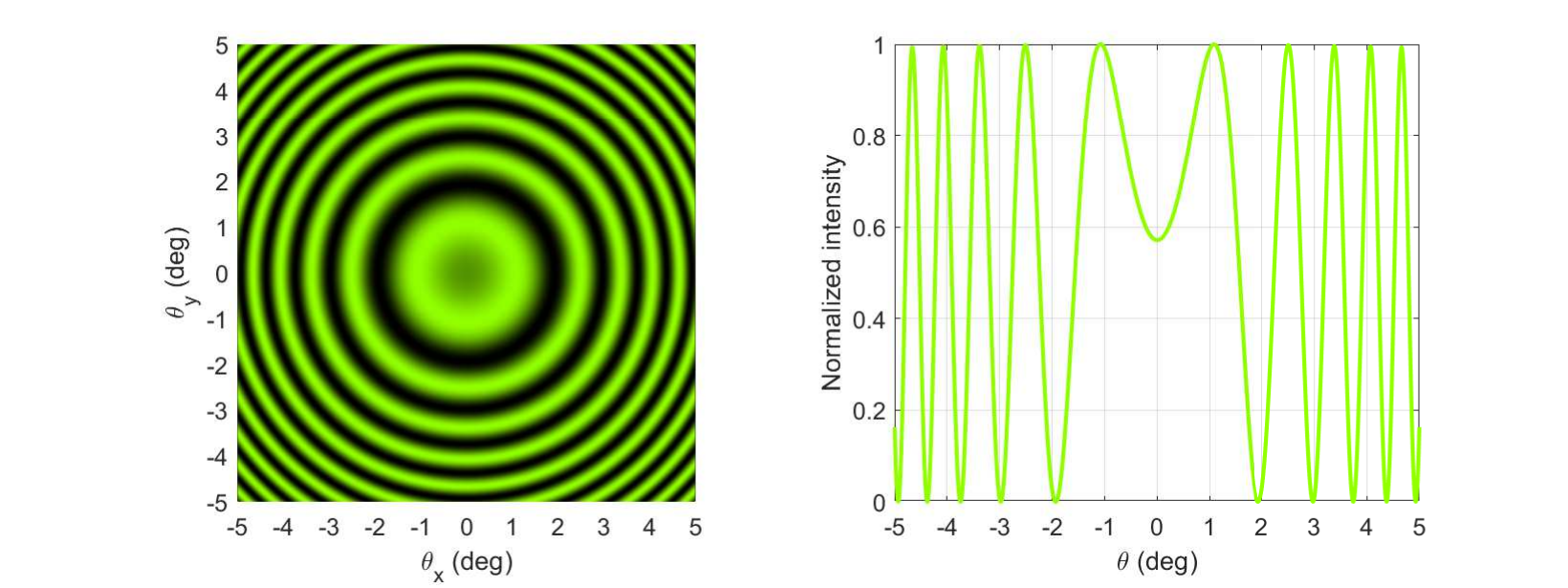}	
		\caption{A simulated example of an interference pattern in Michelson interferometer. A monochromatic and coherent green laser beam is split in two and the beams meet at the same point on the screen (see Figure \ref{Michael2}). The interference of the two fields produces the visible fringes of light intensity. The projection of the interference pattern on a flat screen is shown as a function of the angles from the optical axis (left). The one-dimensional profile of the intensity is shown on the right. The simulation was produced using a Matlab code by Ian Cooper (2019).}
		\label{Michael}
\end{figure}

Interference of waves was widely known since Thomas Young's famous slit experiments, and the observation of a limited region of coherence with polarized light modes has been recognized already by \citet{Verdet1865}. But coherence as a dedicated term describing waves that possess the ability to interfere may have appeared much later \citep[p. 60]{Schuster}: ``\textit{Two centres of radiation emitting vibrations which are related in phase owing to their having originated at the same ultimate source are said to be ``coherent''.}'' Schuster also defined the contrary, which he did not name, but was later referred to as incoherent: ``\textit{Independent sources of light even when emitting quasi-homogeneous light do not give rise to interference effects.}'' The coherence between two point sources that belong to an extended light source at a distance was explored by Pieter Hendrik van Cittert using statistical correlation and ideal imaging conditions, which yielded a theorem that has been used extensively in astronomy \citep{VanCittert1934,VanCittert1938}. The most formal introduction of coherence into optics was done by \citetalias{Zernike1938}, who tied it to a quantity introduced by \citetalias{Michelson1890} earlier to analyze interference patterns, using the intensities of two beams of light---visibility\footnote{This expression is identical to the definition of modulation depth in amplitude modulation that is often used in auditory stimuli, Eq. \ref{DepthAudibility}.}
\begin{equation}
	V = \frac{I_{max}-I_{min}}{I_{max}+I_{min}}
	\label{Visibility}
\end{equation}
where $I_{max}$ and $I_{min}$ refer to the maximum and minimum intensity of the fringes in the interference pattern measurement (see example in Figures \ref{Michael2} and \ref{Michael}). Zernike referred to visibility as the degree of coherence, or partial coherence, which is the important state between complete coherence and complete incoherence. He then generalized the theory and defined the complex degree of coherence and mutual intensity, and showed how they propagate to point images from a common distant source, validating earlier results by van Cittert. The theory was further simplified and was formulated using more deterministic (less statistical) principles by \citetalias{Duffieux1983} and independently by \citetalias{Hopkins1951}, who introduced very useful connections between the degree of coherence and image formation theory. Importantly, they showed how an amplitude image is obtained with fully coherent objects, whereas in incoherent imaging, it is an intensity image that is independent of the phase. Finally, over several publications, Emil Wolf further generalized these tools to polychromatic waves and introduced more rigor to the theory, as in proving that coherence propagates according to the wave equation, amongst other contributions \citep{Wolf1954,Wolf1955}. An alternative version of the theory based on the cross-spectral density instead of the cross-correlation function was also developed by \citet{Wolf1982,Wolf1986} and provided tools to express partially coherent fields as incoherent sums of coherent modes.


As the optical applications were primarily concerned with static spatial images and interference patterns, and given that light frequency is very high, coherence theory was safely formulated using stationary signals and corresponding statistical tools\footnote{These conditions entail that at light frequencies (of the order of $10^{14}$ Hz) the electromagnetic wave would require umpteen periods to lose its degree of coherence.}. This renders the theory ideal for time-invariant signals and systems, but is inadequate to deal with non-stationary signals and dynamic systems. A rigorous adaptation of optical coherence theory to nonstationary signals was introduced much more recently \citep{Bertolotti1995,Sereda1998,Lajunen2005}.

\subsection{Communication}
\label{CommunicationCoherence}
From the communication engineering side, a fundamental distinction is made between coherent and noncoherent modulation detection methods, which influences the complexity of the design and many of its characteristics (\cref{CommunicationTheory}). The principles were discussed in some depth in \citet{Lawson1950}, but this terminology appears to have been introduced earlier---perhaps by A. G. Emslie in 1944 \citep[p. 331]{Lawson1950}. This distinction exists because there are inevitable fluctuations in the carrier frequency and phase when it is generated using equipment that can drift around the channel frequency, and is transmitted over long distances in variable atmospheric conditions, and likely undergoes multipath propagation. These fluctuations amount to noise on the receiver's end, whose impact should be minimized, ideally.

In noncoherent detection, the modulation band is recovered without using the carrier phase, which is assumed to be random and uniformly distributed between $0$ and $2\pi$ (\citealp[pp. 103--107]{Viterbi1979}, \citealp[p. 277]{Couch}). Noncoherent detection implies that the envelope is squared to remove all phase effects, and its most basic implementation is completely passive. The simplest example is an envelope detector that tracks the real envelope of the an amplitude-modulated carrier. 

In contrast, coherent detection always requires harnessing a local oscillator in the detector that ensures that the carrier phase is tracked on a cycle-by-cycle basis, so that any phase errors are minimal. Different schemes have been devised to detect the phase, depending on the modulation method used, which often implies the use of a phase-locked loop (see \cref{PLLs}). 

In general, not all modulation techniques are suitable for both types of detection. The choice between coherent and noncoherent detection technique is not trivial and it depends on the specifications of the system being designed, such as the type of signaling (modulation technique) applied and the target error rate of the demodulated message.

\begin{figure} 
		\centering
		\includegraphics[width=0.6\linewidth]{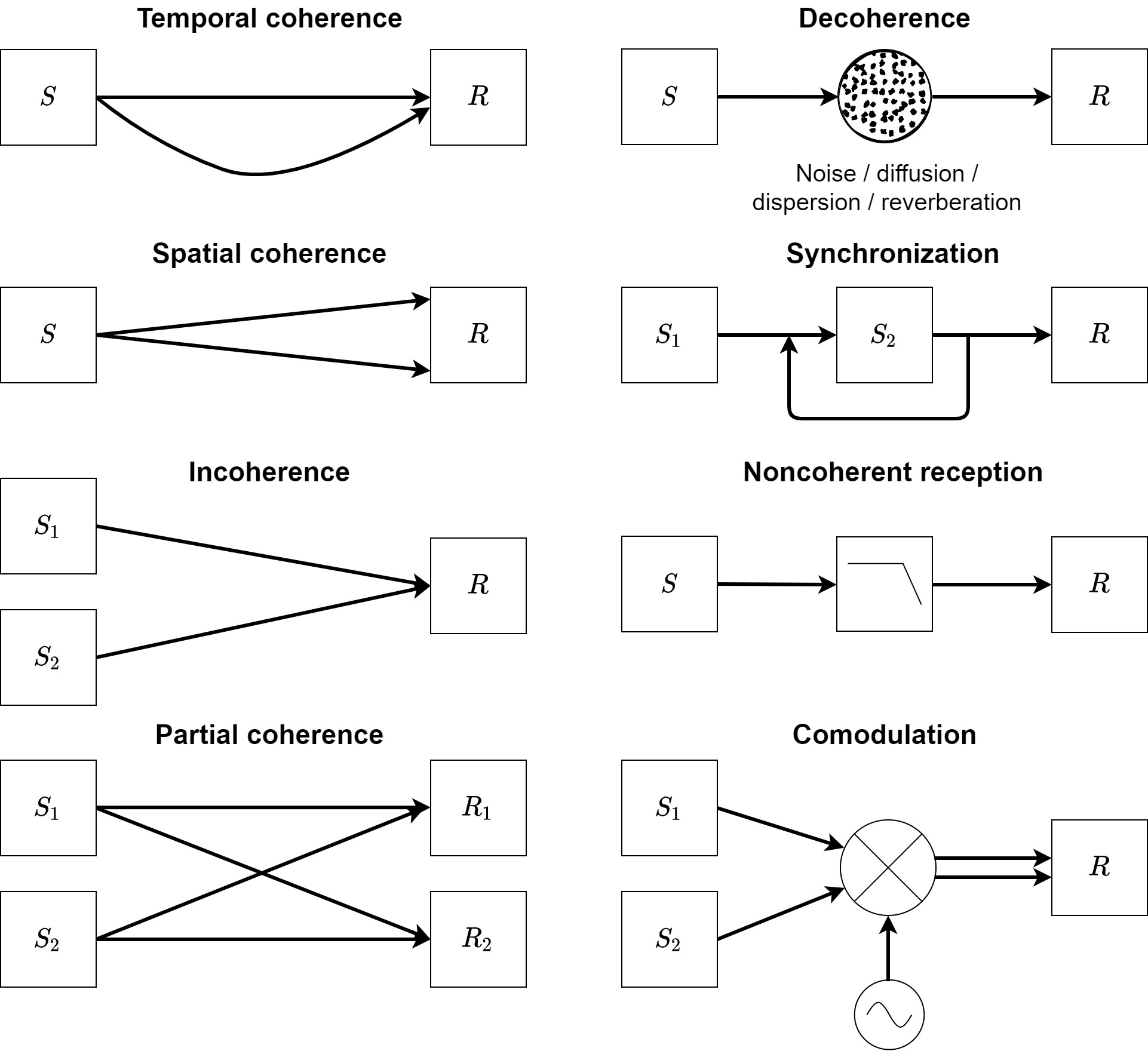}	
		\caption{Simplified cartoon representation of key types of coherence using arbitrary source(s) $S$ and receiver $R$.}
		\label{CoherenceTypes}
\end{figure}

\subsection{Acoustics}
\label{CoherenceinAcoustics}
Despite being applicable to all wave phenomena, application of coherence theory has been somewhat sporadic in acoustics and hearing research. Moreover, the terminology used in acoustics usually draws on that used in stochastic processes instead of wave theory. This is inconsistent with the optical and communication terminology and may have created some confusion (Table \ref{coherenceterms}). 

The first usage of the term coherence for sound waves may have been in the context of room acoustics. \citet{Morse1944} analyzed the steady-state response of point sources in rooms and drew the distinction between coherent and incoherent waves. Coherence here relates to the direct sound from the source and (and sometimes to the first reflection), which has a deterministic direction. Incoherence is the residual sound that is reflected from the walls. It does not have a deterministic direction and has to be analyzed in terms of its power---neglecting the phase. This distinction underlies the treatment of wave acoustics (coherent) versus statistical and geometrical acoustics (incoherent), which is often not clear-cut. Typically, in the statistical approximation rooms are taken to be large, highly reflective, and irregular in shape. 

Several additional studies about coherence appeared somewhat later. \citet{Cook1955} investigated the effect of a reverberation chamber acoustics on random sound fields using cross-correlation\footnote{\citet{Cook1955} cited  S. G. Hershman, Zhur. Tekh. Fiz. 21, 1492 (1951), who originated the idea to use cross-correlation measures in acoustics.}. The authors defined a random sound field as one that has uniform probability to propagate in any direction and with any phase around a frequency band over sufficiently long time. They derived the theoretical correlation coefficients as a function of distance for fields in two and three dimensions. These expressions were then compared to measurements of a broadband source in a reverberation chamber, which had to be averaged over a duration of 12 s in order to match the theoretical prediction. Save from the finite integration time, the correlation coefficient of \citet[Eq. 1]{Cook1955} is identical to the real part of the degree of coherence in optics (\citealp[e.g.,][Eq. 10.3.10]{Born}, and Eq. \ref{Degreeofcoherence}). Similarly, \citet{Schroeder1962} obtained an expression for the autocorrelation of the frequency response functions of large rooms (with high modal density), as a function of spectral distance. \citet[pp. 329--323]{Morse} introduced expressions for acoustic sources that fluctuate too randomly to be modeled using anything but their autocorrelation function. They provided an estimate for the source correlation length and correlation time---quantities that indicate how far and long the spatial and temporal correlations remain high, respectively. While these definitions convey the same information as the coherence length and time in optics \citep[e.g.,][p. 554]{Born}, they have considerably fewer applications. Coherence of acoustic fields, however, finds a greater role in the analysis of propagation and scattering in random media (e.g., turbulence), where acoustic and electromagnetic waves have been often treated together \citep{Tatarskii1971,Ishimaru1978I,Ishimaru1978II}.

The standard definition of coherence adopted in acoustics is taken from signal processing theory. It follows Norbert Wiener's coherency matrix theory, which was defined using the linear combination of the cross term between a set of complex time signals \citep[pp. 182--195]{Wiener1928,Wiener1930}. When the functions do not interfere, the coherency matrix is diagonal, which can then represent incoherent waves that do not interact. If the functions are not independent, their coherency matrix can be diagonalized. Similarly, the squared coherency function is defined in the context of spectral analysis methods in order to deal with stochastic signals and linear systems \citep{Jenkins1968}. The coherency function gives a dimensionless measure of the correspondence between the output and the input (between 0 and 1, for completely uncorrelated and completely correlated, respectively). Along with the phase spectrum, coherency gives a complete description of the system. Coherency, which is also referred to as the normalized cross-spectral density function, is related to the standard coherence measure in acoustics today simply by squaring (e.g., \citealp[pp. 284--287]{Shin2008}). For systems with additional inputs, partial coherence designates the contribution of each input to the total coherence function of the output \citep[pp. 364--370]{Shin2008}, which is unrelated to the definition of partial coherence in optics. See also \citet{Paez2006} for a historical review of stochastic signal processing concepts. 

Notably, \citet{Derode1994} suggested a correction to the standard optical definition of coherence that makes it suitable to be adopted in acoustics, by generalizing it to nonstationary signals (e.g., pulses). This correction mainly targets applications in ultrasound. 

\subsection{Hearing: Psychoacoustics}
\label{PsychoacousticsCoh}
In hearing, the coherence terminology has been used in two somewhat different contexts. Initially, it was applied to binaural processing, which is thought to cross-correlate between the left- and the right-ear signals, corresponding to the interaural phase difference \citep{Licklider1948}. Listeners are remarkably sensitive to changes in the interaural coherence, which also affects the perceived source size (the apparent source width) \citep{Jeffress1962}. 

The second context in which coherence is used in hearing emphasizes the simultaneity of sound components or events, regardless of whether they are monaural or binaural. The term ``coherence'' was used by \citet{Cherry1956} for discussing dichotic signals that were played with variable delay between the two ears. The stimulus coherence (perceptual \term{fusion} or integration) was determined according to how its perceptual location moved on the horizontal plane (instead of being perceived as separated). Temporal coherence was later defined as ``\textit{the perceived relation between successive tones of a sequence, characterized by the fact that the observer has the impression that the sequence in question forms a whole which is ordered in time}'' \citep{Noorden}. \citet[p .181]{McAdams1984} further defined, ``\textit{It is possible to focus one's attention on a given stream and follow it through time; this means that a stream, by definition, exhibits temporal coherence.}'' 

Unlike optics and communication, coherent events in the psychoacoustic parlance entail that different frequency components of a broadband sound have exactly the same envelope---that they are comodulated (Figure \ref{CoherenceTypes})---even if the carrier phases are unrelated (inharmonic). This definition of temporal coherence found extensive use in auditory scene analysis \citep{Bregman} and is conceptually closer to the modern use of the term in auditory research, which is itself influenced by similar ideas in neuroscience\footnote{\citet{Handel2006} dedicated an entire book to ``perceptual coherence'' in visual and auditory phenomena, where commonalities in the processing of the two are contrasted. Unfortunately, the term is not explicitly defined in that work and seems to be used in more than one meaning throughout the book, which makes it difficult to exactly state what it means. It appears to measure how clearly mental objects are perceived---how sharp, differentiated from their surroundings, or stable they are.}. 

\subsection{Neuroscience}
An influential brain function theory, the \term{temporal correlation hypothesis}, argues that synchronized activity in networks of neurons (cell assemblies) can explain how the brain recognizes and attends to visual objects, by enhancing their features using neural networks with feedback \citep{Milner1974}. In his independent formulation of this theory,  \citet{Malsburg1981} generalized the hypothesis to arbitrary modalities and included processes between perception and thought. According to his theory, the dynamics of the network activity is determined by the synaptic conductivity that may be modulated on very short time scales ($<1$ s). Such dynamics may also be suitable for forming mental objects and separating them from the background. The main motivation was to explain how the brain produces specific responses to a nearly infinite set of possible objects in the external world. Initially, Malsburg used the term ``correlation'', which was not well-defined, but later became more concrete with the requirement that cross-correlated cortical neural activity should be synchronized on a millisecond time scale \citep{Singer1995}. The temporal correlation hypothesis has been developed mainly with respect to cortical processing and to the binding problem\footnote{The \term{binding problem} refers to the computational task of producing unified perceptual objects from various features that are recognized using different circuits (e.g., color, shape, location). The task often takes place in several modalities simultaneously---each with its own characteristic processing time. Binding is generally attributed to the cortex.} of different features within one modality that are thought to be processed in localized areas of the visual cortex. However, coherence at different frequencies over large distances between different brain areas has been demonstrated \citep{Varela2001}. These ideas were further generalized to account for binding of multisensory features \citep{Senkowski2008}. In quantifying neural synchrony between different cortical assemblies, a distinction has been made between phase-locking or temporal synchronization and amplitude correlation \citep{Lachaux1999}. According to Lachaux et al, coherence captures both phase and amplitude similarities, but temporal synchrony is a much more sensitive and critical quantity in the brain. 

\subsection{Hearing: Neurophysiology}
The amalgamation of the second coherence meaning in psychoacoustics and ideas from neuroscience have heralded a refined usage of the term in auditory neurophysiology. Here, the perspective of neural temporal correlation has been applied to auditory object-formation models as well, which, incidentally, may have started with an early attempt to account for the cocktail party problem \citep{Malsburg1986}. What may be unique in auditory temporal coherence, unlike spatial images in vision (and most other modalities), is the highly dynamic nature of sound streams that allows direct correlation at the scale of temporal unfolding of stimuli and their corresponding synchronized brain response. Synchronized cortical processing is thought to manifest through selective attention to specific sound events, which exhibits synchronization across different channels following tonotopy \citep{Elhilali2009}. This is expected to be representative of many spectrally-rich natural stimuli, whose spectral components are comodulated \citep{Nelken1999} and are preferentially processed together in the auditory cortex \citep{Barbour2002}. This was referred to, yet again, as ``temporal coherence'', which was redefined---``\textit{temporal coherence between two channels ... denotes the average similarity or coincidence of their responses measured over a given time-window}'' \citep{Shamma2011}. The latter definition specifically refers to cortical (i.e., slow) neural synchronicity between common features of acoustic events, which is perceived on scales of 50--500 ms ($<$ 20 Hz) and can be made more coherent through attention. It is distinguished from incoherent responses that may be part of the same auditory stream, but are not attended to. Although the cited research mainly discussed processing in the cortex or thalamus, we will be more interested in the upstream temporal activity that precedes these high-level effects---something which has been highlighted in the context of temporal coherence in recent studies as well \citep{Viswanathan2022}. See \citet[pp. 115--122]{Elhilali2017} for a recent review of auditory temporal coherence and alternative object formation models.


\section{Synthesis}
The quote by \citet{diFrancia1955} that was cited in the preamble provides an ironic historical vignette that is triply wrong. The author suggested to not dwell on the question of coherence and frequencies in optics as would be more relevant for communication and for the ear. Since then, Fourier optics has been vindicated and become much more mainstream, so that the question regarding the relevance of coherence to imaging is no longer in doubt\footnote{This point was made by the unnamed translator in the ``Translator's Preface'' of \citet{Duffieux1983}.}. But the reader of this quote may also get the false impression that acoustics and hearing science had been invested in the study of coherence back in the 1950s, which was barely the case, save for a handful of studies that were cited above. Moreover, the author compared the spatial frequencies of optics to the audio frequencies of hearing, which are two separate domains that should not be compared (although they often are). However, the reviews above should enable us to pick up from the point where di Francia left us, and motivate us to strengthen some of the connections between optics, acoustics and communication.

~\\

The six perspectives on coherence that were outlined above are at least partially incongruent. The definitions in optics and acoustics are easily merged, as long as we are careful to use a common terminology. In most (but not all) texts in acoustics, acoustical correlation is the same as optical coherence, whereas acoustic coherence is the cross-spectral density in optics (see Table \ref{coherenceterms}). Similarly, the interaural cross-correlation is functionally identical to spatial coherence in optics (\cref{InterauralCoherence}). In this work, we shall adhere to the optical terminology, which will be elaborated in \cref{CoherenceTheory} with emphasis on points of intersection with acoustic theory.

In communication, the concept of coherence has a narrower scope, even though the wave theoretical coherence from optics applies just as well. Coherent detection implies that there is a local oscillator that is capable of synchronizing (phase locking) to an external carrier. Once the internal and external oscillations are synchronized, they are effectively coherent as well, just as in optical coherence theory. We shall retain the communication jargon of ``noncoherent'' only in the context of communication reception and modulation detection, but refer to ``incoherent'' signals in other, more general contexts. 

The psychoacoustic and neurophysiological definitions of coherence are looser, but appear to be synonymous with comodulation of the envelope domain of different bands of a broadband input. In order for it to coincide with the optical theory, the stimulus has to be demodulated first as a multi-carrier signal, so that each channel is demodulated independently, and the outputs from the different channels are used as inputs to the coherence function. In the brain, the latter operation is thought to be accomplished using coincidence detectors, which perform an instantaneous cross-correlation operation in neurons with multiple inputs (see \cref{InterauralCoherence}). In this text, we generally refrain from using the term ``coherence'' if it refers to the envelope domain only, unless it is made explicit. 

Similarly, neural synchrony can be recorded in different time scales. Synchronization (phase locking) to carrier frequencies takes place at low frequencies (estimated to be $<$ 4--5 kHz in humans, \cref{Phaselocking}). In higher-frequency channels, only slow-varying modulation information can be synchronized to. Low-frequency channels can be tracked both in the carrier and in the modulation domains. However, in the strict communication theory usage of the term, only the carrier tracking is real synchronization. We will nevertheless resort to talking about ``envelope synchronization'', if only to to distinguish it from carrier phase locking.

What the brain theory and its auditory variations give us, which the wave theories do not, is the insight that synchronization can be linked to object formation and to selective attention. In other words, if these theories are correct, then synchronization to the stimulus on different levels in the brain may imply that it is being actively perceived, or is in attentional focus.

Cascading all these coherences into a continuous chain of processes, we see how coherence may propagate from the physical source of waves, through the medium, into the auditory system, and culminating in the conscious brain. As the brain processing speed is limited to low frequencies, it may not be able to fully synchronize to arbitrary stimuli, but there can still be low-frequency components that are tracked in the brain. Some of these points will be revisited in the chapter about the auditory phase locked loop (\cref{PLLChapter}).

\chapter{Acoustic coherence theory suitable for hearing}
\label{CoherenceTheory}
\section{Introduction}
\label{CoherenceIntro}
In the previous chapter several perspectives on coherence were reviewed that have found their way into acoustics and auditory research in one way or another. In synthesizing the different perspectives, we aim to be able to track the coherence of a sound signal all the way from the source to the brain. For this purpose, the scalar coherence theory from optics seems to be almost ready-made to import into acoustics. The only adaptations that have to be made are in emphases and in the connections to existing research in acoustics, which have often employed a different jargon (see Table \ref{coherenceterms}). The optical theory provides a rigorous basis, upon which it will be easier to introduce the topic of synchronization. This should then enable us to apply it to the coherence perspectives of communication theory and to a large extent---to the auditory brain. This chapter is therefore dedicated to the presentation of the scalar-wave coherence theory along with examples that are relevant to hearing. As experimental data about the coherence properties of acoustical sources are relatively scarce, the chapter is supplemented also by data presented in \cref{ExCohere}, which demonstrate some of the concepts using realistic sources. 

It should be emphasized that this chapter does not present any new science, as most ideas from coherence theory have appeared in acoustics at one point or another. However, to the best knowledge of the author, there has been no systematic effort to formally integrate coherence and acoustic theories in a consistent and rigorous way. Nor has there been an effort to intuitively integrate the fundamental concepts of coherence into hearing theory, even though some interpretations of it have found wide use. Therefore, the presentations in this chapter and in the appendix (\cref{ExCohere}) attempt to bridge this gap. The intuition and concepts that are obtained here will be useful throughout the remainder of this work.

In adopting the scalar coherence theory, we are knowingly going to neglect the velocity field, which has been often analyzed in the context of acoustic coherence (e.g., in sound intensity, \citealp{Jacobsen1979,Jacobsen1989}, or in modeling of microphone array responses, \citealp{Kuster2008}). Because the ear is sensitive only to acoustic pressure, coherence expressions involving velocity are of relatively little appeal. Therefore, we shall limit our attention to plane waves only, where the pressure and velocity are in phase and the intensity can be expressed using the pressure alone. Where it applies to source functions that are non-planar, the far-field approximation to plane waves should be assumed (see Table \ref{soundvslight}). Alternatively, secondary sources can be used (\cref{DiffractionFourier}). 

The terminology used throughout this chapter adheres to the physical optics theory of coherence, rather than to the acoustical one. The main reason for this is that it is more consistent and is embedded more deeply in wave physics, whereas acoustical correlation has not been equally elaborated and integrated in standard acoustical theory and practice. Additionally, by using the term coherence rather than correlation, we distinguish between wave-physical similarity, as can be manifested in interference phenomena, and the mathematical and statistical operation that represents it in theory, which is largely based on the correlation function. Refer to Table \ref{coherenceterms} for coherence related terms in optics and acoustics. 

Unless otherwise noted, the presentation of coherence theory is based on the texts by \citet{Mandel1995}, \citet[pp. 554--632]{Born}, \citet{Wolf2007}, and to a lesser extent on \citet{Goodman2015}.

\section{Coherence and interference}
The standard way of introducing coherence in wave optics is by analyzing the interference pattern that is observed from a source of light that radiates through two pinholes (see Figure \ref{CohSetup}). According to Huygens principle (\cref{DiffractionFourier}), when the pinhole diameter is small with respect to the wavelength of the light, it can be taken as a (secondary) point source. As the theory deals with a high-frequency electromagnetic field, it is assumed that only average intensity patterns are measurable and, importantly, that the interference pattern is static, due to stationarity. The latter assumption stems from the statistical regularity of the source, which goes through the order of $10^{14}$ cycles per second for light in the visible spectrum. None of these assumptions is particularly relevant to the acoustic signals sensed by the auditory system, but it will be easier to develop insight for the basic coherence concepts through this standard theory, and then modify it as necessary. The basic theory deals with classical scalar electromagnetic fields, which entails that polarization effects that stem from differences in the electric and magnetic components are neglected\footnote{More advanced theory deals with polarization effects as well, which are of no relevance in acoustics. The classical domain is sufficient to explain most imaging effects, but an extensive quantum coherence theory has also been developed. See \citet{Mandel1995}.}. This means that this theory can hold for the scalar acoustic pressure field as well. While there has been much work done on the acoustic velocity-pressure field coherence in acoustics \citep[e.g.,][]{Jacobsen1989}, this topic seems to be of little direct relevance to the pressure-based hearing and will not be explored here. Hence, sound intensity is used here in its scalar form, which is proportional to sound power and the pressure squared only in the plane wave approximation (Table \ref{soundvslight}). Thus, intensity will be treated as a scalar and its directionality will be neglected. 

\subsection{The coherence function}
\label{BasicCoherenceDeriv}
\begin{figure} 
		\centering
		\includegraphics[width=0.5\linewidth]{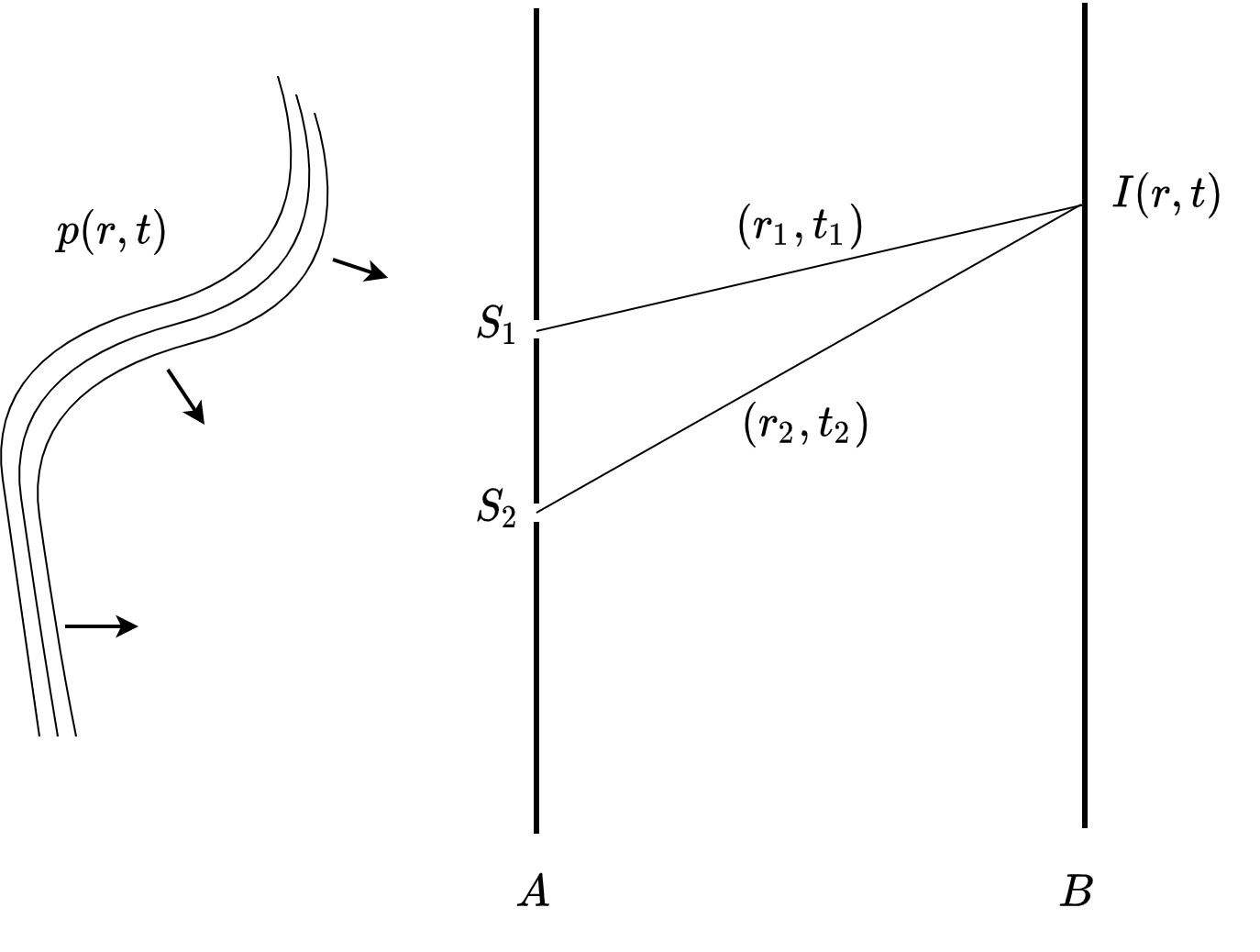}	
		\caption{The basic setup of the coherence problem. The pressure field $p(r,t)$ is represented by the wavefront arriving from the left that impinges on two pinholes on plane $A$. The radiated field from the pinholes is modeled as secondary point sources, neglecting any effects of diffraction from the finite pinhole size. The sources interfere on a screen $B$, where the intensity $I(\bm{r},t)$ is recorded.}
		\label{CohSetup}
\end{figure}

Let $p(\bm{r_1},t)$ and $p(\bm{r_2},t)$ be the sound pressures at time $t$ and points $\bm{r_1}$ and $\bm{r_2}$, respectively (Figure \ref{CohSetup}). The measurement is set up in a way that the pressure field at point $\bm{r}$ is completely determined by the contributions arriving from $\bm{r_1}$ and $\bm{r_2}$. This is a given in a free-field measurement where the sound pressures represent point sources. It can also be ensured far from the source by encircling the points with pinholes, which approximate the pressure arriving from them to point receivers that become secondary point sources, according to Huygens principle. We would like to calculate the total intensity that is measured at point $\bm{r}$, as a function of the contributions from $p(\bm{r_1},t)$ and $p(\bm{r_2},t)$. We consider the intensity to be a \term{random process}\footnote{A random or \term{stochastic process} relates to a variable or variables that are drawn from a certain probability distribution. Without getting into any mathematical formalities, whenever we discuss arbitrary signals or waves, they can be modeled as a random process, which facilitates certain types of analyses. Stochastic processes are distinguished from \term{deterministic processes}. See the cited literature in \cref{CoherenceIntro} for more formal short introductions. For an extensive introduction, see, for example, \citet{Bendat2010}. For a shorter introduction, see \citet{Middleton}.}, which has to be averaged over time and sampled multiple times in order to be estimated properly. The intensity in $\bm{r}$ is given by the \term{ensemble average} of the pressure square in $\bm{r}$
\begin{equation}
 \langle I(\bm{r},t)\rangle_t = \langle p^*(\bm{r},t)p(\bm{r},t)\rangle_t
\label{IntensityO}
\end{equation}
The angle brackets $\langle \cdots \rangle_t$ represent the ensemble averaging operation with respect to samples in time, which are indicated by the subscript $t$. The asterisk denotes the conjugate value operation. We further assume that the pressure field is stationary\footnote{Two general types of stationary processes are distinguished. A \term{strict-sense stationary process} is defined to be independent of time. Thus, such a process can be sampled at any time point without affecting the average. White noise is the simplest such process. A \term{wide-} or \term{weak-sense stationary process} has a constant average, and a time-invariant cross-correlation function, which depends only on the time difference and not on the absolute time points of the process. Wide-sense stationarity requires that the second moments of the process exist. In the text, whenever stationarity is invoked, it is meant in the wide sense.}, so the intensity is independent of time, by definition
\begin{equation}
 \langle I(\bm{r},t)\rangle_t \equiv I(\bm{r})
\label{IntensityO2}
\end{equation}
In general, the expectation value (i.e., the average) for a function $g(t)$ is defined as
\begin{equation}
\langle g(t) \rangle_t = \lim_{T \rightarrow \infty} \frac{1}{2T} \int_{-T}^T g(\tau)d\tau
\end{equation}
If $g(t)$ represents a stationary process which is also \term{ergodic}, then its long-term average value as defined by this integral is equal to the ensemble average, as any sample from its probability distribution at any particular moment is representative of the entire ensemble. 

\begin{figure} 
		\centering
		\includegraphics[width=1\linewidth]{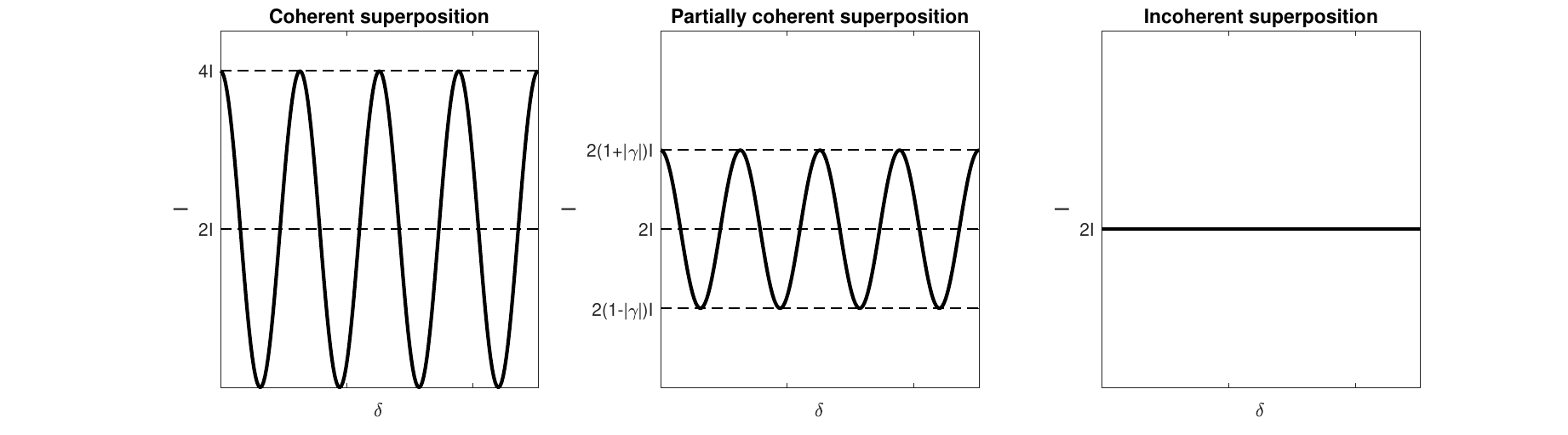}	
		\caption{Three kinds of coherence in the simple interference experiment. The plot on the left describes complete coherence, which shows constructive interference using a coherent sum (linear in amplitude). The plot on the right shows an incoherent sum (linear in intensity) that exhibits no interference. The middle plot describes the general situation in which interference exists, but is not complete, giving rise to partial coherence. The curves can represent the visibility pattern of the fringes in an interference experiment. The plots are a reproduction of Figure 3.2 from \citet[p. 35]{Wolf2007}.}
		\label{partialdemo}
\end{figure}

For a pressure wave propagating from $\bm{r_1}$ and $\bm{r_2}$ to $\bm{r}$, the associated propagation time delays to the screen are
\begin{equation}
	t_1 = \frac{|\bm{r}-\bm{r_1}|}{c} \,\,\,\,\,\,\,\,\,\,\,  t_2 = \frac{|\bm{r}-\bm{r_2}|}{c}
\end{equation}
where $c$ is the speed of sound in the medium. For a small source (or pinhole) size, the pressure amplitude in $\bm{r}$ is inversely proportional to the distance. We can introduce complex amplitudes $a_1$ and $a_2$ to designate the respective complex amplitudes in $\bm{r}$, which include also the effect of the distance. The intensity in $\bm{r}$ is therefore the ensemble average of the superposition of the pressure contributions arriving from $\bm{r_1}$ and $\bm{r_2}$, according to Eq. \ref{IntensityO}
\begin{multline}
	I(\bm{r}) = |a_1|^2\langle p^*(\bm{r_1},t-t_1)p(\bm{r_1},t-t_1)\rangle_t + |a_2|^2\langle p^*(\bm{r_2},t-t_2)p(\bm{r_2},t-t_2)\rangle_t + \\
	2\Re \left[ a_1^*a_2\langle p^*(\bm{r_1},t-t_1)p(\bm{r_2},t-t_2)\rangle_t \right]
 \label{Interference1}
\end{multline}
Using the stationarity property again, we define the \term{mutual coherence function} from the last term of Eq. \ref{Interference1}
\begin{equation}
	\Gamma(\bm{r_1},\bm{r_2},\tau) = \langle p^*(\bm{r_1},t)p(\bm{r_2},t + \tau)\rangle_t
\end{equation}
with a time difference variable $\tau = t_1-t_2$, which can be measured from any time $t$ because of stationarity. Specifically, at $\tau=0$, the mutual coherence function $\Gamma{\bm{r_1},\bm{r_2},0}$ is referred to as the \term{mutual intensity}. This is the cross-correlation function of the two pressure functions. Additionally, the average intensities right at the two pinholes are given by
\begin{equation}
	I_1(\bm{r}) = |a_1|^2\langle p^*(\bm{r_1},t)p(\bm{r_1},t)\rangle_t \,\,\,\,\,\,\,\,\,\,\, I_2(\bm{r}) = |a_2|^2\langle p^*(\bm{r_2},t)p(\bm{r_2},t)\rangle_t
\end{equation}
where it is implied that they are shifted to $t_1=t_2=0$ due to stationarity. Therefore, Eq. \ref{Interference1} can be rewritten as
\begin{equation}
	I(\bm{r}) =  I_1(\bm{r}) + I_2(\bm{r}) + 2\Re\left[ |a_1| |a_2| \Gamma(\bm{r_1},\bm{r_2},\tau) \right]
	\label{interference2}
\end{equation}
The first two terms represent the intensity contribution of each pinhole when the other one is blocked. We can now define the \term{complex degree of coherence}
\begin{equation}
	\gamma(\bm{r_1},\bm{r_2},\tau) = \frac{\Gamma(\bm{r_1},\bm{r_2},\tau)}{\sqrt{I_1(\bm{r})}\sqrt{I_2(\bm{r})}}
	= \frac{\Gamma(\bm{r_1},\bm{r_2},\tau)}{\sqrt{\Gamma(\bm{r_1},\bm{r_1},0)}\sqrt{\Gamma(\bm{r_2},\bm{r_2},0)}}
	\label{Degreeofcoherence}
\end{equation}
whose absolute value is bounded between 0 and 1 because of the normalization. We will often refer in the text to the complex degree of coherence simply as ``coherence''. Using all of these, we can write the \term{law of interference for stationary pressure fields} (of plane waves), which is naturally identical to the one for scalar optical fields,
\begin{equation}
	I(\bm{r},\tau) = I_1(\bm{r}) + I_2(\bm{r}) + 2\Re \left[ \sqrt{I_1(\bm{r})} \sqrt{I_2(\bm{r})} \gamma(\bm{r_1},\bm{r_2},\tau) \right]
	\label{InterferenceLaw1}
\end{equation}
In the narrowband approximation, the pressure wave is monochromatic and $\gamma$ can be meaningfully expressed using a complex envelope and the following identity instead
\begin{equation}
	\gamma(\bm{r_1},\bm{r_2},\tau) \equiv |\gamma(\bm{r_1},\bm{r_2},\tau)| e^{i\left[ \alpha(\bm{r_1},\bm{r_2},\tau) -\overline{\omega} t \right]}
	\label{ComplexCoherenceDef}
\end{equation}
with $\overline{\omega}$ being the average angular frequency and 
\begin{equation}
	\alpha(\bm{r_1},\bm{r_2},\tau) = \overline{\omega} \tau + \arg \gamma(\bm{r_1},\bm{r_2},\tau)
\end{equation}
The angle $\alpha$ changes very slowly relatively to the period of the mean frequency, $1/\overline{\omega}$. We define 
\begin{equation}
	\delta = \overline{\omega} \tau = \overline{\omega} (t_1 - t_2) = \frac{2\pi}{\overline{\lambda}}(|\bm{r_1} - \bm{r_2}|)
	\label{deltaangle}
\end{equation}
where $\overline{\lambda} = 2\pi c/\overline{\omega} $ is the mean wavelength. Thus, the interference law of Eq. \ref{InterferenceLaw1} can be rewritten 
\begin{equation}
	I(\bm{r},\tau) = I_1(\bm{r}) + I_2(\bm{r}) + 2\sqrt{I_1(\bm{r})} \sqrt{I_2(\bm{r})} |\gamma(\bm{r_1},\bm{r_2},\tau)| \cos\left[ \alpha(\bm{r_1},\bm{r_2},\tau) - \delta \right] 
	\label{Interference3}
\end{equation}
In case that the two pressure source contributions at $\bm{r}$ have equal levels, $I = I_1(\bm{r}) = I_2(\bm{r})$, then Eq. \ref{Interference3} simplifies to
\begin{equation}
	I(\bm{r},\tau) = 2I\left\{ 1 + |\gamma(\bm{r_1},\bm{r_2},\tau)| \cos\left[ \alpha(\bm{r_1},\bm{r_2},\tau) - \delta \right] \right\}
	\label{GeneralInterference}
\end{equation}
which describes the intensity fringes that form as a result of interference (for example, see Figures \ref{Michael2} and \ref{Michael}). When $|\gamma| = 1$, the two waves are said to be \term{completely coherent} and the fringes exhibit maximum contrast with areas of no light between the peaks. When $|\gamma| = 0$, no interference takes place and the two waves are \term{completely incoherent}, yielding a constant intensity pattern with no fringes. Everything in between, $0 < |\gamma| < 1$, is \term{partially coherent}, which entails reduced fringe contrast with decreasing degree of coherence, until they are no longer visible (see Figure \ref{partialdemo}). It is possible to quantify the fringe contrast using the maximum and minimum intensities obtained when the cosine is maximum
\begin{equation}
	I_{max}(\bm{r}) = 2I(1 + |\gamma(\bm{r_1},\bm{r_2},\tau)|)
\end{equation}
and minimum
\begin{equation}
	I_{min}(\bm{r}) = 2I(1 - |\gamma(\bm{r_1},\bm{r_2},\tau)|)
\end{equation}
Accordingly, the \term{visibility} of the interference pattern is defined as
\begin{equation}
	V = \frac{I_{max}(\bm{r})-I_{min}(\bm{r})}{I_{max}(\bm{r}) + I_{min}(\bm{r})} 
	\label{Visibility}
\end{equation}
which is $V = |\gamma(\bm{r_1},\bm{r_2},\tau)|$ when $I = I_1(\bm{r}) = I_2(\bm{r})$. In this view, the argument of $\gamma$ determines the relative positioning of the fringes\footnote{When applied to pressure waves, the term \term{audibility} will be more appropriate than visibility, or the more neutral term contrast, as audibility is often used in hearing to designate that a stimulus or a manipulation thereof is above the individual's threshold. However, visibility is the more standard term in optics and we will used it below.}. For arbitrary intensities, $I_1(\bm{r})  \neq I_2(\bm{r})$, the visibility can be computed in general from Eqs. \ref{GeneralInterference} and \ref{Visibility},
\begin{equation}
	V = \frac{2}{\sqrt{\frac{I_1(\bm{r})}{I_2(\bm{r})}} + \sqrt{\frac{I_2(\bm{r})}{I_1(\bm{r})}}} |\gamma(\bm{r_1},\bm{r_2},\tau)|
	\label{GeneralVisibility}
\end{equation}
The interference law of Eq. \ref{Interference3} can be rewritten in yet another way, which provides additional insight into partial coherence, 
\begin{multline}
	I(\bm{r},\tau) =  |\gamma(\bm{r_1},\bm{r_2},\tau)| \left\{I_1(\bm{r}) + I_2(\bm{r}) + 2\sqrt{I_1(\bm{r})} \sqrt{I_2(\bm{r})} \cos\left[ \alpha(\bm{r_1},\bm{r_2},\tau) - \delta \right] \right\}\\
	+ \left[1 - |\gamma(\bm{r_1},\bm{r_2},\tau)|\right] \left[ I_1(\bm{r}) + I_2(\bm{r}) \right]
	\label{Interference4}
\end{multline}
This formulation splits the interference pattern to a coherent part with intensities proportional to $|\gamma(\bm{r_1},\bm{r_2},\tau)|$ and relative phase difference of $\alpha(\bm{r_1},\bm{r_2},\tau) - \delta $, and an incoherent part with intensities proportional to $1 - |\gamma(\bm{r_1},\bm{r_2},\tau)|$. Thus \textbf{the total intensity} $I_{tot}$ \textbf{is a sum of coherent and incoherent contributions}, $I_{coh}$ and $I_{incoh}$, respectively
\begin{equation}
	I_{tot} = I_{coh} + I_{incoh}
	\label{totalcoherence}
\end{equation}
Following Eq. \ref{Interference4} and neglecting the cosine term
\begin{equation}
	\frac{I_{coh}}{I_{incoh}} = \frac{|\gamma(\bm{r_1},\bm{r_2},\tau)|}{1-|\gamma(\bm{r_1},\bm{r_2},\tau)|}
	\label{cohtoincohratio}
\end{equation}
and therefore
\begin{equation}
	\frac{I_{coh}}{I_{tot}} = |\gamma(\bm{r_1},\bm{r_2},\tau)|
\end{equation}
This insightful formulation indicates that any partially coherent intensity pattern can be decomposed into completely coherent and completely incoherent intensity contributions. 

\begin{figure} 
		\centering
		\includegraphics[width=0.9\linewidth]{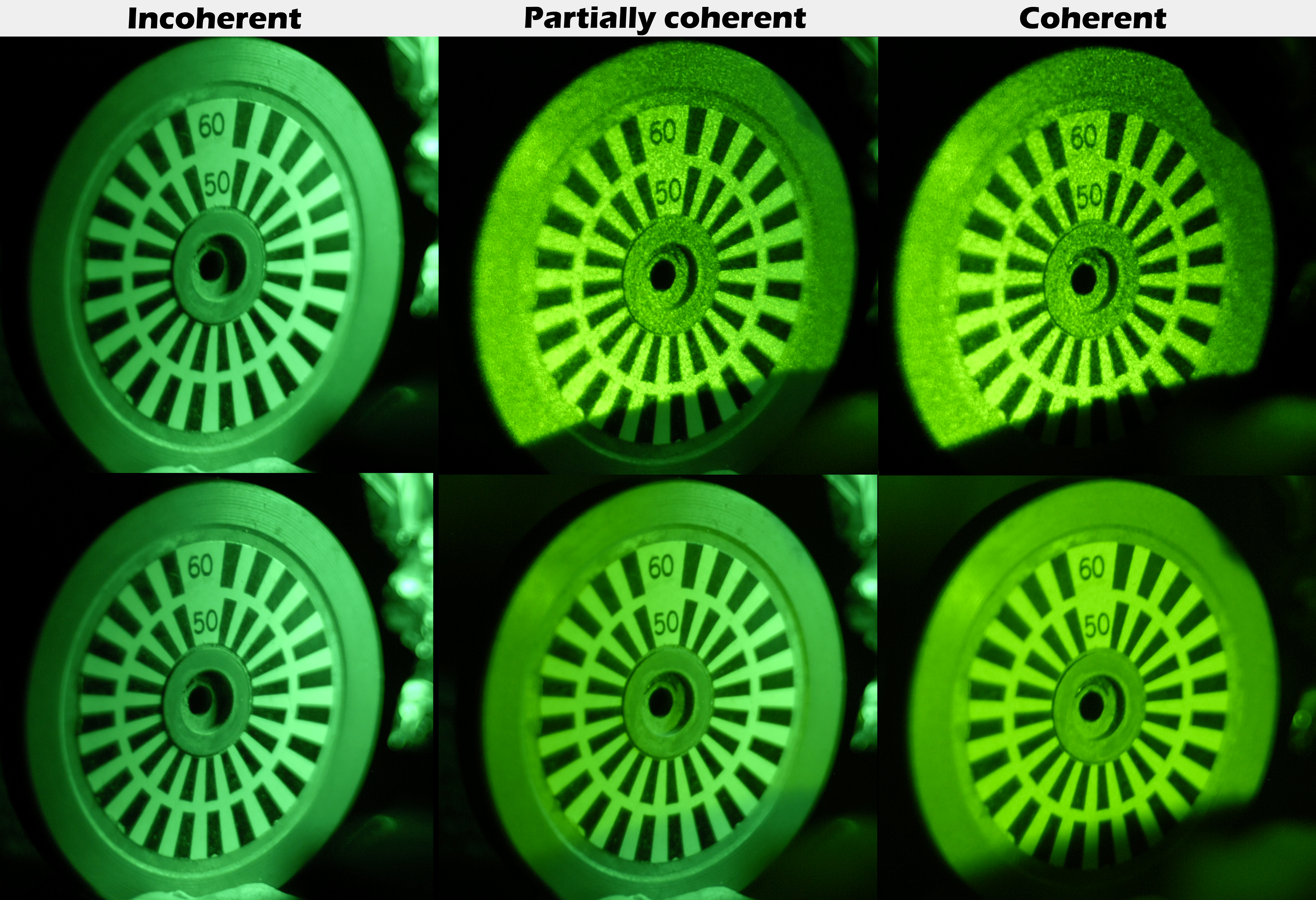}	
		\caption{Two sets of images done under incoherent (left), partially coherent (middle), and coherent (right) illumination. The images were obtained using two different light sources. The coherent source was a green Nd:YAG laser ($\lambda = 532$ nm), whose beam was spread with a diverging lens and a one (top) or two (bottom) different types of diffusers. The incoherent images employed a green light-emitting diode (LED) light, embedded in a bulb, of a slightly different wavelength than the laser, which is almost completely incoherent and diffuse from the source (see, for example, \citealp{Mehta2010}). The two light sources were illuminating the object simultaneously in the partially coherent image. However, as the laser light was significantly more intense than the LED, it was employed with different amounts of diffusion, to get different degrees of partial coherence, according to Eq. \ref{totalcoherence}. Note that the two light sources arrive from different angles: the coherent source was almost frontal, whereas the incoherent source came from the top right of the object.}
		\label{partialimages}
\end{figure}

Examples of coherent, partially coherent, and incoherent illuminations are given in Figure \ref{partialimages}. The partially coherent objects are obtained by combining a coherent source and an incoherent source in different amounts, according to Eq. \ref{totalcoherence}. 

\subsection{Temporal coherence and spatial coherence}
\label{TempSpatCoh}
Two idealized types of coherence are distinguished based on the expressions above. The first one is \term{spatial coherence}, $\gamma(\bm{r_1},\bm{r_2},\tau_0)$, which is estimated at two different points in space, but employs a fixed time difference $\tau_0$, usually $\tau_0 = 0$, for convenience. The second type is \term{temporal coherence}, $\gamma(\bm{r},\bm{r},\tau)$, which is determined solely by the time difference $\tau$ assuming that the measurement positions of the two disturbances are the same, $\bm{r_1} = \bm{r_2}$. In real fields, the temporal and spatial coherence are not necessarily independent. Cartoon illustrations of temporal and spatial coherence are given in Figure \ref{CoherenceTypes}. 

For stationary acoustic signals, the temporal coherence is, in fact, the normalized autocorrelation function of $p(\tau)$
\begin{equation}
	\gamma(\bm{r_1},\bm{r_2},\tau) = \gamma(\bm{r},\tau) = \frac{\Gamma(\bm{r},\tau)}{I(\bm{r})} = \frac{\int_{-\infty}^\infty p^*(\bm{r},t)p(\bm{r},t+\tau)dt}{\int_{-\infty}^\infty p^*(\bm{r},t)p(\bm{r},t) dt} = R_{pp}(\bm{r})
\end{equation}
using Eq. \ref{Degreeofcoherence}. The notation $R_{pp}$ is standard for autocorrelation in signal processing, but will not be used here. 

The distinction between temporal and spatial coherence provides an informative tool for analysis on different levels. It is also regularly employed in acoustics and hearing, only not by this name. For example, one may think of interaural acoustic cross-correlation (IACC) as a spatial coherence function for the two fixed locations of the left and right ears (\cref{InterauralCoherence}). Temporal coherence, in contrast, is used in some models that probe monaural hearing, which is sensitive to the time course of signals and specifically to periodicity, which appears as peaks in the autocorrelation function. However, the input in these models is often broadband, which violates the narrowband assumption and therefore behaves somewhat differently.

\subsection{The cross-spectral density and spectrum}
\label{CrossSpect}
According to the \term{Wiener-Khintchin theorem}, the power spectral density of a wide-sense stationary random process with zero mean is the Fourier transform of its autocorrelation function. In our case, it can be applied to the mutual coherence function, so
\begin{equation}
	S(\bm{r},\omega) = \int_{-\infty}^\infty \Gamma(\bm{r},\tau) e^{-i\omega \tau} d\tau 
\end{equation}
with $S$ being the power spectral density of the process. The inverse transform applies as well
\begin{equation}
	\Gamma(\bm{r},\tau) = \frac{1}{2\pi}\int_{0}^\infty S(\bm{r},\omega) e^{i\omega \tau} d\omega
\end{equation}
Only positive frequencies are used in the integral with the assumption that the signal is taken to be analytic (see \cref{AnalyticSignals}).

Similarly, according to the \term{generalized Wiener-Khintchin theorem}, the mutual coherence function itself---a cross-correlation function---is the Fourier transform pair of the \term{cross-spectral density} $W$
\begin{equation}
	W(\bm{r_1},\bm{r_2},\omega) = \int_{-\infty}^\infty \Gamma(\bm{r_1},\bm{r_2},\tau) e^{-i\omega \tau} d\tau
	\label{CrossSpec}
\end{equation}
and the inverse applies again,
\begin{equation}
	\Gamma(\bm{r_1},\bm{r_2},\tau) = \frac{1}{2\pi}\int_{0}^\infty W(\bm{r_1},\bm{r_2},\omega)  e^{i\omega \tau} d\omega
	\label{CrossSpec2}
\end{equation}

It is important to emphasize that the cross-spectral density function spectrally depends on $\omega$ alone only in the case of stationary signals, but it is not true in the more general case of nonstationary signals that are broadband. It can be seen by inspecting the correlation function of the pressure field in spectral-spatial coordinates, assuming that the pressure function has a Fourier transform \citep{Mandel1976}
\begin{multline}
	\langle  P^*(\bm{r_1}, \omega_1) P(\bm{r_2}, \omega_2) \rangle_t = \int_{-\infty} ^\infty\int^\infty_{-\infty} \langle  p^*(\bm{r_1}, t_1) p(\bm{r_2}, t_2) \rangle_t e^{i\omega_1 t_1}e^{-i\omega_2 t_2}dt_1 dt_2 \\
	= \int_{-\infty} ^\infty\int^\infty_{-\infty} \langle  p^*(\bm{r_1}, t_1) p(\bm{r_2}, t_1+\tau) \rangle_t e^{i(\omega_1-\omega_2) t_1}e^{-i\omega_2 \tau}dt_1 d\tau 
	\label{SpectralCorrelation1}
\end{multline}
The ensemble average in the integrand is simply the mutual coherence function, which is independent of $t_1$ for stationary signals. Therefore, the first transform gives a delta function with the frequency difference in the argument, whereas the Fourier transform of the mutual coherence is simply the cross-spectral density
\begin{equation}
	\langle  P^*(\bm{r_1}, \omega_1) P(\bm{r_2}, \omega_2) \rangle_t = W(\bm{r_1},\bm{r_2},\omega_1) \delta(\omega_1-\omega_2)   
	\label{SpectralCorrelation1}
\end{equation}
Hence, for $\omega_1 \neq \omega_2$, the cross-spectral density is 0, due to stationarity, which means that waves at different frequencies are completely incoherent to one another (see also \citealp[pp. 442--448]{Bendat2010}). This is not true for nonstationary processes and coherence, as will be discussed in \cref{Nonstationarytheory}.

The spectral density expressions are known in acoustics simply as ``coherence'', adopted from signal processing of random processes \citep[e.g.,][]{Shin2008}. They are much more commonly used in acoustics than the coherence representation (using temporal coordinates), which is sometimes referred to as (cross-)correlation and sometimes as coherence\footnote{\citet{Jacobsen1987} made the distinction that the acoustic correlation function is bandlimited, whereas the coherence is defined in the frequency domain.}. A complete coherence theory using spectral coherence was derived by \citet{Wolf1982,Wolf1986} and will be reviewed in \cref{SpectralCoherence}.

\subsection{Coherence time and coherence length}
\label{CoherenceTimeLength}
It has long been known that interference effects in light cannot be empirically observed if the disturbances are too far from the (secondary) source (e.g., obtaining visible interference fringes with sunlight is limited to very short distances from the pinholes; \citealp[pp. 72--124]{Verdet1869}). This is true for sound waves too---when the pressure disturbances propagate they gradually acquire phase distortion, which eventually makes them too dissimilar to be capable of interfering. It manifests as spectral broadening of the original source output, and misalignment of the phases in the superposed field functions. Two quantities are particularly handy in quantifying the reach of coherence, inasmuch as interference effects can be measured (i.e., the fringes are visible, or $V>0$ in Eq. \ref{Visibility}). The first one is \term{coherence time}, which is the relative delay that can be applied in the autocorrelation function before the fringes of the interference pattern disappear, so it can be considered effectively incoherent. The coherence time $\Delta\tau$ is inversely proportional to the spectral bandwidth of the source $\Delta \omega$, so that 
\begin{equation}
	\Delta\tau \sim \frac{2\pi}{\Delta \omega}
	\label{CoherenceTimeEst}
\end{equation}
There are different ways to prove this expression, such as by considering an ideal detector with finite integration time that would measure a different level if the input interferes with itself during integration \citep[pp. 352--359]{Born}. A more rigorous way to define the coherence time is based on the self-coherence of the source
\begin{equation}
	\Gamma(\tau) = \langle p^*(\bm{r},t) p(\bm{r},t+\tau) \rangle_t 
\end{equation}
which is essentially its autocorrelation function. The coherence time can then be defined using the second moment of the squared modulus of the self-coherence function (the first moment is zero due to stationarity) \citep[pp. 176--177]{Mandel1995}
\begin{equation}
	\Delta\tau^2 = \frac{\int_{-\infty}^\infty \tau^2 |\Gamma(\tau)|^2 d\tau}{\int_{-\infty}^\infty |\Gamma(\tau)|^2 d\tau }
	\label{FancyCohTime}
\end{equation}
Similarly, the effective spectral width of the source can be defined as the second moment of the spectral density function, centered around the mean frequency $\bar{\omega}$
\begin{equation}
	\Delta \omega^2 = \frac{\int_{0}^\infty (\omega-\bar{\omega})^2 S(\omega)^2 d\omega}{\int_{0}^\infty S(\omega)^2 d\omega }
\end{equation}
where the mean frequency is the first moment
\begin{equation}
	 \bar{\omega} = \frac{\int_{0}^\infty \omega S(\omega)^2 d\omega}{\int_{0}^\infty S(\omega)^2 d\omega }
\end{equation}
which, for narrowband modulated sources, we normally equate with the carrier $\bar{\omega} = \omega_c$. With some effort, it may be shown that $\Delta \tau$ and $\Delta \omega$ are related by the inequality \citep[pp. 176--180]{Mandel1995}
\begin{equation}
	 \Delta \tau \Delta \omega \geq \frac{1}{2}
	\label{CoherenceUncertain}
\end{equation}
This expression is reminiscent of the uncertainty principle for time signals and their frequency representation. The similarity is perhaps unsurprising, given the Fourier-transform pair that the autocorrelation and the spectral density form (although in this case it was established for wide-sense stationary random processes according to the Wiener-Khintchin theorem, unlike regular time signals). Just as for time signals, the inequality \ref{CoherenceUncertain} becomes an equality only for a Gaussian source distribution---both its spectrum and its autocorrelation \citep[pp. 158--162]{Gabor, Goodman2015}. 

Even with the new definition of the coherence time of Eq. \ref{FancyCohTime}, it is somewhat arbitrary and may be used primarily as an approximate measure. In reality, the visibility may oscillate around $\Delta \tau$, depending on the exact distribution of the spectrum around the mean frequency. Furthermore, obtaining a meaningful interpretation of the coherence time when the source is not monochromatic or narrowband is not straightforward, as both spectrum and self-coherence functions have multiple peaks. This is the case in most realistic acoustic sources, which requires more ad-hoc estimates, as is demonstrated in \cref{ExCohere}. 

Given a finite coherence time for the source, it is also possible to define a corresponding \term{coherence length} $\Delta l$
\begin{equation}
	\Delta l = c \Delta\tau \sim \frac{2\pi c}{\Delta \omega}
	\label{CoherenceLength}
\end{equation}
The coherence length is more intuitive in spatial coherence propagation problems and may therefore be handier in binaural hearing. In contrast, the coherence time is more immediately relevant in monaural listening, as will be explored below and throughout this work. 

It will be useful to refer to the coherence time (or length) as a figure of merit for the degree of coherence of a source or a signal. So a signal with relatively high coherence time and length has a high degree of coherence and vice versa.

\subsection{The wave equation for the coherence functions}
\label{CoherenceWaveEquation}
A profound property of the mutual coherence function is that it satisfies the wave equation \citep{Wolf1955}. This can be established relatively easily. Starting from the scalar wave equation for the pressure field
\begin{equation}
	\nabla^2 p(\bm{r},t)= \frac{1}{c^2}\frac{\partial^2p(\bm{r},t)}{\partial t^2}
\end{equation}
We can take the complex conjugate of $p$ on both sides of the equation, switch to a local coordinate system with $\bm{r_1}$ and $t_1$, and multiply both sides of the equation by $p(\bm{r_2},t_2)$
\begin{equation}
	\nabla_1^2 p^*(\bm{r_1},t_1) p(\bm{r_2},t_2)= \frac{1}{c^2}\frac{\partial^2p^*(\bm{r_1},t_1)}{\partial t_1^2}p(\bm{r_2},t_2)
\end{equation}
where $\nabla_1$ is the Laplacian in local coordinates. Now if we take the ensemble average of both sides and interchange the order of integration and differentiation
\begin{equation}
	\nabla_1^2 \langle  p^*(\bm{r_1},t_1) p(\bm{r_2},t_2) \rangle_t= \frac{1}{c^2}\frac{\partial^2}{\partial t_1^2}\langle p^*(\bm{r_1},t_1)p(\bm{r_2},t_2) \rangle_t
\end{equation}
Applying the wide-sense stationarity property, we can use the fact that $\tau = t_1-t_2$ and $\partial^2/\partial t_1^2 = \partial^2/\partial \tau^2$ and replace the ensemble average with the mutual coherence function
\begin{equation}
	\nabla_1^2  \Gamma(\bm{r_1},\bm{r_2},\tau) = \frac{1}{c^2}\frac{\partial^2 \Gamma(\bm{r_1},\bm{r_2},\tau)}{\partial \tau^2}
	\label{CohWaveEq1}
\end{equation}
Similarly, 
\begin{equation}
	\nabla_2^2  \Gamma(\bm{r_1},\bm{r_2},\tau) = \frac{1}{c^2}\frac{\partial^2 \Gamma(\bm{r_1},\bm{r_2},\tau)}{\partial \tau^2}
	\label{CohWaveEq2}
\end{equation}
Therefore, the homogenous wave equation is satisfied for $\Gamma(\bm{r_1},\bm{r_2},\tau)$, which indicates that the coherence function, while being an average quantity, propagates in space deterministically and is an inherent property of the scalar pressure field. Coherence propagation according to the wave equation is derived for different types of fields, including inhomogeneous ones with a primary radiating source \citep[pp. 181--196]{Mandel1995}. To the best knowledge of the author, the above equations have not been discussed with respect to acoustic fields, except in specific underwater acoustic problems \citep{McCoy1976,Berman1986}. 

Finally, it is possible to Fourier-transform both sides of Eqs. \ref{CohWaveEq1} and \ref{CohWaveEq2} using the Wiener-Khintchin theorem (Eq. \ref{CrossSpec}) and obtain the corresponding Helmholtz wave equations for the cross-spectral density $W(\bm{r_1},\bm{r_2},\omega)$
\begin{equation}
	\nabla_1^2  W(\bm{r_1},\bm{r_2},\omega) + k^2 W(\bm{r_1},\bm{r_2},\omega) = 0
	\label{SpecCohWaveEq1}
\end{equation}
\begin{equation}
	\nabla_2^2  W(\bm{r_1},\bm{r_2},\omega) + k^2 W(\bm{r_1},\bm{r_2},\omega) = 0
	\label{SpecCohWaveEq2}
\end{equation}
where we changed the order of differentiation and Fourier-transform integration in the two equations and used the wavenumber definition $k = \omega/c$.

\subsection{Spectral coherence}
\label{SpectralCoherence}
A relatively late development in optical coherence theory has been the introduction of a rigorous derivation of coherence in the spatial-spectral domain \citep{Wolf1982, Wolf1986}. It bridges the gap with the coherence functions used in signal processing and acoustics. One important insight that this theory is able to provide is in accounting for the effect of passive narrowband filters on coherence. Importantly, this theory is the basis for what appears to be the most rigorous extension to date of optical coherence theory to nonstationary processes \citep{Lajunen2005}, which is most relevant to acoustics and hearing. Dispersive propagation also requires nonstationary coherence theory \citep{Lancis2005}, as does non-periodic frequency modulation in general. 

The following is a non-rigorous sketch of the main steps of derivation of the spectral coherence expressions, primarily based on \citet[pp. 60--69]{Wolf2007}. The initial steps are particularly technical, but they lead to familiar and intuitive results. The full derivation is found in \citet{Wolf1982,Wolf1986} and \citet[pp. 213--223]{Mandel1995}, which is more rigorously approached than an initial derivation of the theory that appeared in \citet{Mandel1976}. An alternative way of derive spectral coherence was outlined by \citet{Bastiaans1977}, but is not reviewed here.

The main obstacle in formulating a spectral-coherence theory in the first place is that stationary random processes do not have a valid Fourier representation of the time signal $p(t)$\footnote{We have dispensed with the formal discussion of \term{harmonizable random processes} due to Lo\`eve that is usually invoked here and may be relevant to the discussion of the analytic signal (\cref{PhysicalSignals}). In general, it deals with nonstationary random processes with infinite energy that may also be discontinuous, for which the standard Fourier transform does not exist---only the power spectrum via the autocorrelation function (see \cref{CrossSpect}). Instead, it then applies the Riemann-Stieltjes integral for the spectral density, $x'(t) = \frac{1}{2\pi}\int dX(\omega) e^{i\omega t}$. For further details, see \citet[pp. 173--178]{Clark2012phd} for a friendly and brief introduction, or for a more complete treatment \citet{Napolitano2012}.}. Thus,
\begin{equation}
	P(\omega) = \int_{-\infty}^\infty p(t) e^{-i\omega t}dt
\end{equation}
does not exist for stationary random processes, because $p(t)$ is not integrable as it does not converge to zero when $t \rightarrow \pm \infty$. This means that the spectrum in its ordinary definition as a function of the ensemble average of the squared modulus of the Fourier transform of $p(t)$ does not formally exist either. The solution to this problem harnesses the existence of a more general definition of the spectrum found in generalized harmonic analysis \citep{Wiener1930}, which is based on the autocorrelation function of $p(t)$. In order to apply it, \citet{Wolf1982} used very general assumptions about the source field---that $\Gamma(r_1,r_2,\tau)$ is absolutely integrable with respect to $\tau$ and that it is continuous and bounded in the domain $\Omega$ that contains the source and the relevant points, for all $\tau$. He showed that these assumptions can lead to certain basic conditions on the cross-spectral density that can make it suitable to be a kernel of the Fredholm integral equation\footnote{The conditions are that the cross-spectral matrix $W(\bm{r_1},\bm{r_2},\omega)$ is non-negative, definite, and Hermitian. These conditions make it suitable for expansion into eigenfunctions (Eq. \ref{CrossSpecSum2}) according to Mercer's theorem. In this case, the modes satisfy the Fredholm equation (\ref{CrossSpecSum}) for which $W$ is the kernel.}
\begin{equation}
	\int_\Omega W(\bm{r_1},\bm{r_2},\omega)\Psi_n (\bm{r_1},\omega) d^3r_1 = \lambda_n(\omega) \Psi_n (\bm{r_2},\omega) 
	\label{CrossSpecSum}
\end{equation}
where $W(\bm{r_1},\bm{r_2},\omega)$ is a matrix of the cross-spectral density, $\Psi_n (r,\omega)$ are eigenfunctions with corresponding positive eigenvalues $\lambda_n(\omega) \geq 0$, which are solutions to the Fredholm integral equation. A complete solution for the equation can be expressed as a sum of all the eigenfunctions that solve the equation
\begin{equation}
	W(\bm{r_1},\bm{r_2},\omega) = \sum_n \lambda_n(\omega) \Psi_n^* (\bm{r_1},\omega) \Psi_n (\bm{r_2},\omega)
	\label{CrossSpecSum2}
\end{equation}
This sum is called the \term{coherent-mode representation} of the cross-spectral density function, as each mode can be shown to be completely spatially coherent. It is a linear combination of orthonormal coherent modes at frequency $\omega$, whose orthonormality condition is given by
\begin{equation}
	\int_\Omega \Psi_n^* (\bm{r},\omega) \Psi_m (\bm{r},\omega) d^3r = \delta_{nm}
	\label{Orthonormal}
\end{equation}
where $\delta_{nm}$ is the Kronecker delta that is 1 when $n=m$ and 0 otherwise. Using this condition it can be readily shown that each individual mode $\Psi_n$ also satisfies the wave equations \ref{SpecCohWaveEq1} and \ref{SpecCohWaveEq2}.

Let us now continue to construct functions that can form an ensemble from which a correlation function of the cross-spectral density function can be derived. We construct a pressure-field function using the superposition of the modes $\Psi_n$
\begin{equation}
	P(\bm{r},\omega) = \sum_n b_n(\omega) \Psi_n(\bm{r},\omega)
	\label{USumofModes}
\end{equation}
where the random finite coefficients $b_n$ are related to the eigenvalues $\lambda_n(\omega)$ through
\begin{equation}
	\langle b_n^*(\omega) b_m(\omega) \rangle_\omega = \lambda_n(\omega) \delta_{nm}
	\label{CoeffDel}
\end{equation}
Then, we can solve for the corresponding cross-correlation function
\begin{multline}
	\langle P^*(\bm{r_1},\omega)P(\bm{r_2},\omega) \rangle_\omega = \sum_n\sum_m \langle b_n^*(\omega) b_m(\omega)\rangle_\omega \Psi^*_n(\bm{r_1},\omega)\Psi_m(\bm{r_2},\omega) = \sum_n \lambda_n(\omega) \Psi^*_n(\bm{r_1},\omega)\Psi_n(\bm{r_2},\omega) \\ = W(\bm{r_1},\bm{r_2},\omega)
\end{multline}
where the order of ensemble-averaging and summation was interchanged in the first equality, Eq. \ref{CoeffDel} was used in the second equality, and Eq. \ref{CrossSpecSum2} in the last equality that established equivalence with the cross-spectral density. With these relations at hand it is now possible to derive a corresponding expression for the spectrum
\begin{equation}
	S(\bm{r},\omega) = \langle P^*(\bm{r},\omega)P(\bm{r},\omega) \rangle_\omega 
	\label{SpectrumRig}
\end{equation}
This formula is intuitively appealing because it has the same form of the naive interpretation of the spectrum as the squared modulus of the Fourier transform components of the field function. Instead, it is the ensemble average of its monochromatic eigenfunction representation, rather than of the direct (forbidden) Fourier transform of the stationary field. 

Now, since $P(\bm{r},\omega)$ is a linear combination of the eigenfunctions $\Psi_m (\bm{r},\omega)$ (Eq. \ref{USumofModes}), each of which satisfies the wave equation, then $P(\bm{r},\omega)$ itself satisfies it too,
\begin{equation}
	\nabla^2 P(\bm{r},\omega) + k^2 P(\bm{r},\omega) = 0
\end{equation}
The interpretation of this version of the wave equation is that $P(\bm{r},\omega)$ is the spatial part of the pressure field that has a simple harmonic (monochromatic) dependence $p(\bm{r},t) = P(\bm{r},\omega)e^{i\omega t}$. It emphasizes the fact that all eigenfunctions in $P(\bm{r},\omega)$ are at the same frequency and each one may be coherent in its own right. However, if the sum of $P(\bm{r},\omega)$ (Eq. \ref{USumofModes}) contains more than a single mode, then the ensemble may be only partially coherent. 

\subsection{Broadband interference with spectral coherence}
\label{SpectralCoherenceTheory}
It is now possible to derive analogous expressions for the interference setup described in \cref{BasicCoherenceDeriv} and Figure \ref{CohSetup}, but in terms of the cross-spectral density rather than intensity. The derivation follows \citet[pp. 63--66]{Wolf2007}. For an alternative proof, see also \citet[pp. 585--588]{Born}.

The frequency-dependent pressure field at point $\bm{r}$ is the sum of the contributions of the pressure from points $\bm{r_1}$ and $\bm{r_2}$,
\begin{equation}
	P(\bm{r},\omega) = a_1 P(\bm{r_1},\omega)e^{ikr_1} + a_2 P(\bm{r_2},\omega)e^{ikr_2}
\end{equation}
with $a_1$ and $a_2$ being the complex amplitudes of the waves traveling from $\bm{r_1}$ and $\bm{r_2}$ to $\bm{r}$, respectively. Using the expression for $S(\bm{r},\omega)$ in Eq. \ref{SpectrumRig}, let us write the spectral density in $\bm{r}$
\begin{equation}
	S(\bm{r},\omega) = |a_1|^2 S(\bm{r_1},\omega) + |a_2|^2 S(\bm{r_2},\omega) + 2\Re\left[ a_1^*a_2 W(\bm{r_1},\bm{r_2},\omega) e^{-i\delta}\right]
	\label{SpectralInterference1}
\end{equation}
where $\delta = \omega |\bm{r_1}-\bm{r_2}|/\lambda$ is once again the phase associated with the path difference, $\lambda$ is the wavelength, and the interference is now expressed by the cross-spectral density $W(\bm{r_1},\bm{r_2},\omega)$. The first two terms in Eq. \ref{SpectralInterference1} are the contribution to the spectral density in $\bm{r}$ when one of the sources is switched off, so 
\begin{equation}
	|a_1|^2 S(\bm{r_1},\omega) = S_1(\bm{r},\omega)  \,\,\,\,\,\,\,\,\, |a_2|^2 S(\bm{r_2},\omega) = S_2(\bm{r},\omega)
\end{equation}
The \term{spectral degree of coherence} $\mu(\bm{r_1},\bm{r_2},\omega)$ is defined as
\begin{equation}
	\mu(\bm{r_1},\bm{r_2},\omega) = \frac{W(\bm{r_1},\bm{r_2},\omega)}{\sqrt{W(\bm{r_1},\bm{r_1},\omega) W(\bm{r_2},\bm{r_2},\omega)}} \\
=	\frac{W(\bm{r_1},\bm{r_2},\omega)}{\sqrt{S(\bm{r_1},\omega) S(\bm{r_2},\omega)}} 
\end{equation}
We also set
\begin{equation}
	\mu(\bm{r_1},\bm{r_2},\omega) = |\mu(\bm{r_1},\bm{r_2},\omega)|e^{i\alpha(\bm{r_1},\bm{r_2},\omega)} 
\end{equation}
with $\alpha(\bm{r_1},\bm{r_2},\omega) = \arg \mu(\bm{r_1},\bm{r_2},\omega)$. Hence, we can rewrite Eq. \ref{SpectralInterference1} as
\begin{equation}
	S(\bm{r},\omega) = S_1(\bm{r},\omega) + S_2(\bm{r},\omega) + 2 \sqrt{S_1(\bm{r},\omega) S_2(\bm{r},\omega)} |\mu(\bm{r_1},\bm{r_2},\omega)| \cos\left[\alpha(\bm{r_1},\bm{r_2},\omega)-\delta \right]
	\label{SpectralInterference2}
\end{equation}
This is the \term{spectral interference law}. Just as with the complex degree of coherence in the temporal-spatial coherence treatment earlier, it can be shown that the spectral degree of coherence $\mu$ is bounded $|\mu(\bm{r_1},\bm{r_2},\omega)| \leq 1$, where 1 indicates complete coherence, and 0 complete incoherence \citep[p. 911]{Born}. Unlike the temporal degree of coherence, the spectral degree of coherence is defined for a single frequency component. In general, this quantity is suitable for broadband measurements and can reveal spectral effects that have relatively small intensity changes per frequency. In contrast, temporal-spatial coherence is applicable for narrowband signals, where intensity effects are visible, with few spectral effects. 

Interestingly, $\mu(\bm{r_1},\bm{r_2},\omega)$ is none other that the coherence function that is commonly used in acoustics and signal processing \citep[e.g.,][pp. 284--285]{Shin2008}, only backed by physical conditions that correspond to the familiar interference experiment. 

When the power contributions from the two points are equal, $S(\bm{r_1},\omega) = S(\bm{r_2},\omega)$, the spectral interference law simplifies to 
\begin{equation}
	S(\bm{r},\omega) = 2 S_1(\bm{r},\omega) \left\{ 1 +  |\mu(\bm{r_1},\bm{r_2},\omega)| \cos\left[\alpha(\bm{r_1},\bm{r_2},\omega)-\delta \right]\right\}
	\label{SpectralInterference3}
\end{equation}
which assumes the form of sinusoidal modulation envelope as a function of position when two sources interfere. It also implies that, in general, the superposed spectrum of the two contributions to the field is different than the spectrum of a single one. Unlike the interference with narrowband sounds that was analyzed using temporal coherence, spectral modulation is not sensitive to the distance from the source in the same way, and fluctuations in the spectrum may be observed well beyond the coherence time and length of the source. In some conditions, this can potentially give rise to one-half of spectrotemporal modulation that has been studied with broadband acoustic stimuli \citep{Aertsen1980I,Aertsen1980II,Aertsen1981}, primarily with respect to their cortical responses (see also \cref{MGBA1} and \cref{RoomAc}). 

\subsection{Narrowband filtering and coherence}
\label{CoherentFiltering}
Before leaving the realm of stationary coherence, let us also explore the effect of linear narrowband filtering on the broadband spectral coherence function (\citealp{Wolf1983}, \citealp[pp. 73--76]{Wolf2007} and \citealp[pp. 174--176]{Mandel1995}). This problem is particularly relevant in hearing, because of the cochlear bandpass filtering property---albeit nonlinear in its transient response---which affects the received coherence of the auditory signal.

The output cross-spectral density function $W_o$ of the input pressure field $P$, which passes through a bandpass filter that has a transfer function $H(\omega)$, is
\begin{multline}
	W_o(\bm{r_1},\bm{r_2},\omega) = \langle H^*(\omega)P^*(\bm{r_1},\omega) H(\omega)P(\bm{r_2},\omega) \rangle_\omega = |H(\omega)|^2 \langle P^*(\bm{r_1},\omega) P(\bm{r_2},\omega) \rangle_t \\
	= |H(\omega)|^2 W_i(\bm{r_1},\bm{r_2},\omega)
\end{multline}
where $W_i$ is the cross-spectral density function before the filter, at the input. In the second equality, $H(\omega)$ was taken out of the ensemble average because it is deterministic. It results in the familiar input-output relation of the linear filter, with the cross-spectral density function assuming the role of the signal. It is then straightforward to show that the spectral degree of coherence is unaffected by the filter. At the input it is
\begin{equation}
	\mu_i(\bm{r_1},\bm{r_2},\omega) = \frac{W_i(\bm{r_1},\bm{r_2},\omega)}{\sqrt{W_i(\bm{r_1},\bm{r_1},\omega) W_i(\bm{r_2},\bm{r_2},\omega)}}
\end{equation}
And at the output
\begin{multline}	
	\mu_o(\bm{r_1},\bm{r_2},\omega) = \frac{ W_o(\bm{r_1},\bm{r_2},\omega)}{\sqrt{W_o(\bm{r_1},\bm{r_1},\omega) W_o(\bm{r_2},\bm{r_2},\omega)}} \\
	= \frac{|H(\omega)|^2 W_i(\bm{r_1},\bm{r_2},\omega)}{\sqrt{|H(\omega)|^2 W_i(\bm{r_1},\bm{r_1},\omega) |H(\omega)|^2 W_i(\bm{r_2},\bm{r_2},\omega)}} =	\mu_i(\bm{r_1},\bm{r_2},\omega) 
\end{multline}
Moving now to the spatial-temporal complex degree of coherence $\gamma(\bm{r_1},\bm{r_2},\tau)$, we repeat Eq. \ref{Degreeofcoherence} using the self-coherence functions in the denominator:
\begin{equation}
	\gamma(\bm{r_1},\bm{r_2},\tau) = \frac{\Gamma(\bm{r_1},\bm{r_2},\tau)}{\sqrt{\Gamma(\bm{r_2},\bm{r_2},0)}\sqrt{\Gamma(\bm{r_2},\bm{r_2},0)}}
		\label{CoherentFiltering1}
\end{equation}
We would like to find out whether $\gamma(\bm{r_1},\bm{r_2},\tau)$ is affected by the filter, unlike $\mu(\bm{r_1},\bm{r_2},\omega)$. It is possible to use the Wiener-Khintchin theorem and express the cross-correlations as the inverse Fourier transform of the cross-spectral density (Eq. \ref{CrossSpec2}), so at the input to the filter we get
\begin{equation}
	\Gamma_i(\bm{r_1},\bm{r_2},\tau) = \frac{1}{2\pi} \int_{0}^\infty W_i(\bm{r_1},\bm{r_2},\omega) e^{i\omega \tau} d\omega
\end{equation}
whereas at the output it is
\begin{multline}
	\Gamma_o(\bm{r_1},\bm{r_2},\tau) = \frac{1}{2\pi} \int_{0}^\infty W_o(\bm{r_1},\bm{r_2},\omega) e^{i\omega \tau} d\omega =  \frac{1}{2\pi} \int_{0}^\infty |H(\omega)|^2 W_i(\bm{r_1},\bm{r_2},\omega) e^{i\omega \tau} d\omega \\
	\approx  \frac{1}{2\pi} W_i(\bm{r_1},\bm{r_2},\omega_c) \int_{0}^\infty |H(\omega)|^2  e^{i\omega \tau} d\omega 
		\label{CoherentFiltering2}
\end{multline}
where the final approximation used was done by assuming that the cross-spectral density is about constant across the narrow filter bandwidth, so it could be taken out of the integral. Generalizing across the two positions of $\bm{r}$
\begin{equation}
	\Gamma_o(\bm{r_m},\bm{r_n},\tau) \approx \frac{1}{2\pi} W_i(\bm{r_m},\bm{r_n},\omega_c) \int_{0}^\infty |H(\omega)|^2  e^{i\omega \tau} d\omega  \,\,\,\,\,\,\,\,\,\, m,n = 1,2
		\label{CoherentFiltering3}
\end{equation}
Therefore, the complex degree of coherence is
\begin{equation}
	\gamma_o(\bm{r_1},\bm{r_2},\tau) \approx \frac{W_i(\bm{r_1},\bm{r_2},\omega_c) \int_{0}^\infty |H(\omega)|^2  e^{i\omega \tau} d\omega}{ \sqrt{W_1(\bm{r_1},\bm{r_1},\omega_c) W_2(\bm{r_2},\bm{r_2},\omega_c)} \int_{0}^\infty |H(\omega)|^2 d\omega  } = \mu_i(\bm{r_1},\bm{r_2},\omega_c) \frac{\int_{0}^\infty |H(\omega)|^2 e^{i\omega \tau} d\omega}{\int_{0}^\infty |H(\omega)|^2 d\omega} 
	\label{CoherentFiltering4}
\end{equation}
where we used Eqs. \ref{CoherentFiltering1}, \ref{CoherentFiltering2}, and \ref{CoherentFiltering3}. The integral of Eq. \ref{CoherentFiltering4} can be simplified by inspecting the value of $\gamma_o(\bm{r_1},\bm{r_2},\tau)$ around $\tau = 0$. The quotient is the normalized Fourier transform of $|H(\omega)|^2$---essentially the normalized impulse response function of the filter, measured for intensity, and assuming that the filter is linear. This gives us a maximum at $\tau = \tau_0 \geq 0$, for a causal filter, so that 
\begin{equation}
	\gamma_o(\bm{r_1},\bm{r_2},\tau) \leq \gamma_o(\bm{r_1},\bm{r_2},\tau_0) = \mu_i(\bm{r_1},\bm{r_2},\omega_c) 
	\label{mugamma0}
\end{equation}
This result mirrors the narrowband nature of the coherence function $\gamma_o(\bm{r_1},\bm{r_2},\tau)$. It predicts that the temporal degree of coherence is not guaranteed to remain the same given the narrowband filtering. Even for very narrow bandwidths it may well be less than unity. But we can obtain further insight than that, by testing the effect of the bandwidth of an ideal rectangular bandpass filter on the coherence function. If we set 
\begin{equation}
	|H_r(\omega)|^2 = 1 \,\,\,\,\,\,\,\, \omega_c - \frac{\Delta\omega}{2} \leq \omega \leq \omega_c + \frac{\Delta\omega}{2}
\end{equation}
for a filter with bandwidth $\Delta\omega$. Then we can directly compute Eq. \ref{CoherentFiltering4}, which gives
\begin{equation}
	\gamma_o(\bm{r_1},\bm{r_2},\tau) \approx \mu_i(\bm{r_1},\bm{r_2},\omega_c) \Delta\omega e^{i\omega_c\tau}\sinc\left( \Delta \omega \tau \right)
	\label{CoheringFilter}
\end{equation}
where $\mu_i(\bm{r_1},\bm{r_2},\omega_c)$ has the dimensions of spectral density, so it cancels out the scaling by $\Delta \omega$. The main lobe of the sinc function becomes narrower with increasing filter bandwidth. We remember that the coherence time of a narrowband source is inversely proportional to its bandwidth (Eq. \ref{CoherenceTimeEst}). Here we showed how the narrowband filtering can effectively turn a broadband source into a narrowband one when the bandwidth is narrow enough, which results in an increase of the coherence time. This would make a broadband source appear more coherent than it is at the origin. This is an important result that we will revisit throughout this work. See also \citet{Jacobsen1987} for an alternative derivation. 

\subsection{Nonstationary coherence}
\label{Nonstationarytheory}
Coherence theory as has been formulated until recently is largely based on the theory of random stationary processes. Even though it has proven to have a tremendous explanatory power in optics and is just as applicable in other scalar wave phenomena, it is inadequate to deal with realistic acoustic signals that are, by and large, nonstationary. The importance of nonstationarity was expressed by \citet{Antoni2009} in the context of cyclostationary acoustic signal analysis, which reveals hidden periodicities (modulations) in the standard power spectrum: ``\textit{Let us insist on the assertion that nonstationarity---as evidenced by the presence of transients---is intimately related to the concept of information. This is completely analogous to speech or music signals that can carry a message or a melody only because they consist of a succession of nonstationarities.}'' 

Formal work on nonstationary optical coherence theory started only relatively recently with \citet{Bertolotti1995}. We shall adopt a similar but more rigorous theory by \citet{Lajunen2005}, which is a close variation on the spectral coherence theory presented in \cref{SpectralCoherence}. Therefore, it is not going to be necessary to consider all the steps in the derivation, but rather to discuss the main differences and implications. 


Nonstationary coherence theory is defined for a pulse train, in which the pulses are well-separated in time from one another, so they do not mix. Pulses here should be seen as a very general way to express a finite wave that has a beginning and an ending, although its duration is not constrained by the analysis. The pulses can be different from one another in level, spectrum, position, and duration, which means that the ensemble is certainly not ergodic. Hence, nonstationary ensemble averages that are employed here are not time averages as before, but have to be computed on a pulse-by-pulse basis, where every pulse is defined with its own time reference---independently of the other pulses. Therefore, the expression for the mutual coherence function takes the form
\begin{equation}
	\Gamma(\bm{r_1},\bm{r_2},t_1,t_2) = \langle p^*(\bm{r_1},t_1)p(\bm{r_2},t_2) \rangle = \frac{1}{N}\sum_{n=1}^N p_n^*(\bm{r_1},t_1)p_n(\bm{r_2},t_2)
\label{nonstationGamma}
\end{equation}
where the pressure field  $p$ of each pulse is marked with a subscript $n$. Because of the nonstationarity of the pulse train, the cross-correlation is not time-invariant, in general, which is underscored by the ensemble averaging operation $\langle \cdots \rangle$ not carrying the subscript $t$ this time. The nonstationary complex degree of coherence is \citep{Bertolotti1995}
\begin{equation}
	\gamma(\bm{r_1},\bm{r_2},t_1,t_2) = \frac{\Gamma(\bm{r_1},\bm{r_2},t_1,t_2)}{\sqrt{I(\bm{r_1},t_1)}\sqrt{I(\bm{r_2},t_2)}}
\end{equation}

Similarly, the nonstationary cross-spectral density is 
\begin{equation}
	W(\bm{r_1},\bm{r_2},\omega_1,\omega_2) = \langle P^*(\bm{r_1},\omega_1)P(\bm{r_2},\omega_2) \rangle = \frac{1}{N}\sum_{n=1}^N P_n^*(\bm{r_1},\omega_1)P_n(\bm{r_2},\omega_2)
\end{equation}
Here the most striking feature of the nonstationary theory is revealed, as disturbances of different frequencies can interact---unlike in the stationary case where different frequencies are completely incoherent. The nonstationary spectral degree of coherence is then
\begin{equation}
	\mu(\bm{r_1},\bm{r_2},\omega_1,\omega_2) = \frac{W(\bm{r_1},\bm{r_2},\omega_1,\omega_2)}{\sqrt{S(\bm{r_1},\omega_1)}\sqrt{S(\bm{r_2},\omega_2)}}
\end{equation}
All steps and expressions in the derivation are completely analogous to the stationary case as described in \ref{SpectralCoherence}, but with the frequency variables differentiated $\omega_1 \neq \omega_2$ everywhere where cross-correlation is applied. Notably, the nonstationary mutual coherence and cross-spectral density both satisfy the wave equation, just like Eqs. \ref{CohWaveEq1}, \ref{CohWaveEq2}, \ref{SpecCohWaveEq1}, and \ref{SpecCohWaveEq2}.

The option of using time-dependent ensemble averages instead of mutual coherence was mentioned in \citet{Mandel1965}, where it was referred to as the \term{ensemble correlation function}. It was not further pursued there, because of the greater usefulness of stationarity. \citet{Derode1994} suggested, independently, to use this alternative averaging in acoustical nonstationary signals, anticipating the nonstationary optical theory. However, their acoustic coherence theory did not consider frequency interaction in its spectral coherence functions. Nonstationary acoustic signal theory was also derived by \citet{Mark1970}, but without considering an interaction between frequencies either, so $\omega_1=\omega_2$, just as in stationary processes. The coherence between of two frequencies that propagate in turbulent conditions was investigated by \citet{Havelock1998}, but the analysis referred to an envelope-like fluctuation term that is common to the two frequencies and becomes gradually decohered with increasing turbulence strength and range, for fixed points in time and space. This usage is more similar to the definition of coherence in hearing science (\cref{PsychoacousticsCoh}).

In general, nonstationary processes are much more complicated to work with than stationary processes. As a rule, since the frequency components are not independent in nonstationary processes, it is possible that the usual self-coherence function does not fully capture its second-order statistical properties as in stationary processes (e.g., the spectrum and the coherence time). This can result in analytic signals that are not ``\term{proper}'', which means that their real and imaginary parts are not independent and their cross-correlation does not cancel out as Eq. \ref{propriety} predicts \citep{Neeser1993,Picinbono1994}. It has been suggested that realistic signals such as speech are \term{improper}, which may require different applied tools to deal with them \citep{PicinbonoBondon1997,Schreier2003,Clark2012phd}. It was recently suggested by \citet{Clark2012phd} that phase-locked coherent tracking (using the complex envelope formalism) may go a long way to accurately demodulate improper signals (i.e., obtain accurate audio and modulation spectra), which coincides with what the ear appears to be doing, at least in part (\cref{PLLChapter}). While this topic has some clear relevance to the present theory, as both deal with the borderline between constant and instantaneous quantities, this approach is outside the scope of this work. 

\subsubsection{Beating}
\label{CohereBeating}
An early example of nonstationary interference has been the demonstration of the beating between two laser sources that have nearly identical line frequencies \citep{Magyar1963}. This would have not been considered a particularly interesting achievement in acoustics, but the analysis provides a useful formulation to beating using coherence theory. In this case, the ensemble average was not employed, because it would have effectively randomized the initial phase, which in the long run eliminates the contrast of any measurable beating. The spatial coordinate is assumed constant in all expressions, so it is incorporated into the phase
\begin{equation}
	p_1(t) = \sqrt{I_1(t)} \exp ( i\omega_1 t + i\varphi_1 ) \,\,\,\,\,\,\, p_2(t) = \sqrt{I_2(t)} \exp ( i\omega_2 t + i\varphi_2)
\end{equation}
with the initial phase terms $\varphi_1$ and $\varphi_2$. In general, there is a different initial phase, as well as a possible path difference $c\tau$ between the wavefronts that corresponds to a delay $\tau$. We can write the interference law for the fields $p_1(t-\tau/2)$ and $p_2(t+\tau/2)$ using all the phase contributions
\begin{equation}
	I(t) = I_1(t) + I_2(t) + 2\sqrt{I_1(t)I_2(t)} \cos\left[ (\omega_2-\omega_1)t + \frac{1}{2}(\omega_2+\omega_1)\tau + \varphi_2 - \varphi_1\right]
\end{equation}
The intensity will vary in time as long as $\omega_1 \neq \omega_2$. This beating can be thought of as temporal fringes given that mathematically there is no difference with spatial interference except for the changed space-time coordinate. Assuming a detector that has a rectangular input window $T$, we can integrate the intensity and obtain the detected beating
\begin{multline}
	 \hat{I}(t,T) = \frac{1}{T}\int_t^{t+T} I(t')dt' \\
	= I_1(t) + I_2(t) + 2\sqrt{I_1(t)I_2(t)} \sinc \left[\frac{(\omega_1-\omega_2)T}{2}\right] \cos\left[ (\omega_2-\omega_1)\left(t+\frac{T}{2}\right) + \frac{1}{2}(\omega_2+\omega_1)\tau + \varphi_2 - \varphi_1\right]
\end{multline}
Based on Eq. \ref{GeneralVisibility} we obtain the visibility pattern, which is stationary in comparison with the time-dependent cosine term
\begin{equation}
	|\gamma(t)| = \frac{2}{ \sqrt{\frac{I_1(t)}{I_2(t)}} + \sqrt{\frac{I_2(t)}{I_1(t)}} }\sinc (\pi \Delta f T ) 
	\label{visibilityrect}
\end{equation}
for spacing between the frequencies of $\Delta f = |f_2-f_1|$. By exchanging audibility for visibility, this equation provides a condition for beating detection. However, the rectangular integration window causes the audibility to fluctuate with the $\Delta f$ due to the side lobes of the sinc function, which is psychoacoustically not the case. Rather, the beating audibility should diminish monotonically with $\Delta f$. Therefore, the rectangular window may be replaced with a Gaussian window that has the same equivalent rectangular bandwidth $T$ (see \cref{PulseCalc})
\begin{multline}
	 \hat{I}(t,T) =  \frac{1}{T} \int_{-\infty}^{\infty} I(t')\exp \left( -4\ln2 \,\, \frac{t^{'2}}{T^2} \right) dt' \\
	= I_1(t) + I_2(t) + \sqrt{\frac{\pi}{\ln 2}}\sqrt{I_1(t)I_2(t)} \exp\left[ -\frac{(\pi T \Delta f)^2}{4\ln2 \,\,}\right] \cos\left[ \frac{1}{2}(\omega_2+\omega_1)\tau + \varphi_2 - \varphi_1\right]
\end{multline}
with the respective visibility/audibility
\begin{equation}
	|\gamma(t)| = \frac{\sqrt{\frac{\pi}{\ln 2}}}{ \sqrt{\frac{I_1(t)}{I_2(t)}} + \sqrt{\frac{I_1(t)}{I_2(t)}} }\exp\left[ -\frac{(\pi T \Delta f)^2}{4\ln2 \,\,}\right] 
	\label{visibilitygauss}
\end{equation}
For a given $T$ of the detector and for $I_1 = I_2$, the audibility with the Gaussian window will drop to half for $\Delta f = \frac{2\ln2 \,\,}{\pi T} \approx 0.441/T$, whereas with the rectangular window, it is $\Delta f = 0.5/T$ (the first zero of the sinc function). We shall revisit this solution in \cref{BeatingCurves}, once the values of the auditory $T$ will be estimated. 

The phase of the degree of coherence depends on the values of $\varphi_1$, $\varphi_2$ and $\tau$, which are not necessarily controlled. It means that had we taken the ensemble average, the temporal interference pattern---the beating---may no longer be audible, as the relative position of the temporal fringes would move with the phase. Therefore, in order to measure beating, we have to look at the specific instance in time, as nonstationarity requires. 

\subsection{Discussion}
The classical theory of coherence was reviewed with emphasis on the most relevant aspects to auditory processing of acoustic signals. Although many of the theoretical results have been undoubtedly known and used by acousticians, there is an obvious lack of systematic treatments or reviews of this important topic within the field. It is appreciated that some of the results are imprecise in the near-field approximation, where the sound intensity is directional and evanescent modes can be dominant. Nevertheless, the squared pressure is the most useful proxy for acoustic intensity and power in the majority of practical cases. It is also the most immediate measurement when using pressure detectors such as the ears and standard pressure microphones. This makes the classical coherence theory highly relevant for hearing research as well. 

~\\

The remainder of this chapter is therefore dedicated to the exploration of how coherence theory has been applied in hearing-relevant acoustics, sometimes using different terminology than in optics. In turn, the insight garnered in these sections will provide the necessary basis for understanding the coherent-incoherent distinction that is key to the imaging auditory system.

\section{Coherence of typical acoustic sources}
\label{TypicalSourceCoh}
\subsection{Mathematical sounds and realistic light sources}
It can be easy to forget that natural acoustic signals are not pure in the mathematical sense. In the context of coherence, the infinitesimal spectral width of the pure tone entails that, asymptotically, it has infinite coherence time and length. From the coherence time definitions (Eq. \ref{CoherenceTimeEst}), the coherence time for a mathematical pure tone is
\begin{equation}
\Delta \tau_{pure\,tone}\rightarrow \infty
\end{equation}
whereas for white noise---the other mathematically idealized sound stimulus---it is
\begin{equation}
\Delta \tau_{white\,noise} \rightarrow 0
\end{equation}
While somewhat trivial, these extreme cases are important because they map to many of the experiments done in both acoustic and auditory research since the advent of electronics. In the present work, these extreme coherence properties inform the interpretation of published results that were obtained using such signals. Unlike signals of finite bandwidth, broadening of pure tones (i.e., through absorption, dispersion, reflections, and the Doppler effect) accumulates very slowly, to the point of being negligible in normal settings (see Figures \ref{dispXY} and \ref{dispT} for illustrations using dispersion). This means that effects that have to do with partial coherence may not be readily encountered in laboratory-based experiments and calculations based on pure tones. 

In contrast to both pure tones and white noise, realistic optical sources have finite bandwidths, so they exhibit a broader range of coherence phenomena. Only with the invention of the laser (and maser, both in the 1950s) did experimental optics obtain stable quasi-monochromatic sources, which have much narrower spectral bandwidths than natural sources (yet still finite)\footnote{Before the advent of laser technology, the alternative in optics has been to use a \term{monochromator}, which produces a narrowband spectrum from a broadband light source. This is usually achieved by spatially selecting the diffracted or dispersed product of the broadband light. This technology is still widely used in applied optics.}. For example, \citet[p. 5]{Wolf2007} compared the coherence time and length of a gas discharge lamp of $\Delta\tau \approx 0.01\,\,\mu\s$ and $\Delta l \approx 19$ m to a stabilized laser of $\Delta\tau \approx 100\,\,\mu\s$ and $\Delta l \approx 190$ km(!). Note, however, that these sources still have finite bandwidths of $\Delta f \sim 10^{8}$ Hz for the lamp and $\Delta f \sim 10^{4}$ Hz for the laser. While the exact center frequency was not provided by Wolf, we can roughly assume it to be $ 10^{14}$ Hz in both cases, which would make the relative bandwidth of the two sources $10^{-6}$ and $10^{-10}$, respectively. While in optics these sources are still not considered pure tones---their phase cannot be controlled as in sound generation---equivalent precision in audio would be very impressive. For a tone at 1000 Hz, it would entail approximate precision of 0.001 Hz and 0.0000001 Hz, respectively---well below the resolution of any standard acoustic instrumentation.

In the following, we will try to understand what kind of coherence properties may be expected from realistic acoustic sources. Due to the relative scarcity of data in literature, this question can be answered only in part. 


\subsection{Effective duration of realistic acoustic sources}
\label{CoherentSources}
The coherence time of acoustic sources may be estimated directly from the autocorrelation function of the radiated output in free field (Eq. \ref{FancyCohTime}). There is relatively little information available in literature about the coherence properties of typical acoustic sources and most work has focused on (broadband) music and the interaction it has with room acoustics \citep[e.g.,][]{Ando1985}. As most natural acoustic sources are nonstationary, some kind of running autocorrelation is required to be able to estimate their temporal coherence. In general, these sources are not at all as well-behaved as the optical sources used in stationary coherence theory, and their coherence time can only be estimated approximately. If the stationary autocorrelation function is computed using the limit
\begin{equation}
	\Gamma(\tau) = \lim_{T \rightarrow \infty} \frac{1}{2T} \int_{-T}^T p^*(t)p(t+\tau)dt
	\label{AutoCorrRep}
\end{equation}
Then, in the nonstationary case, the integration constant $T$ is finite and its value affects the measurement. A longer $T$ can be used to detect slower periodicities and features in the coherence function. Examples of the effect of the choices of $T$ on the autocorrelation function of several musical pieces are shown in Figures 3.4--3.6 in \citet[pp. 14--18]{Ando1998} and for other sounds in Figure \ref{SaxT}. These figures demonstrate how signals differ in their sensitivity to $T$, which may or may not capture their characteristic coherence time. Therefore, $T$ has to be selected with consideration to the degree of coherence of the sound that is being analyzed.

In several acoustic studies that estimated the running autocorrelation, a measure similar to coherence time---the \term{effective duration} (usually denoted with $\tau_e$)---was defined as the -10 dB drop from the autocorrelation peak at $\tau=0$ (i.e., when $|\gamma|=0.1$), but extrapolated from the initial 5 dB drop in the response peak. This definition produces an exaggerated effect compared to the coherence time in the optical definition that entails only a 3 dB drop from the peak (half-width maximum power). As the procedure to obtain the 10 dB point normally involves a linear fit to the autocorrelation function of the A-weighted pressure \citep[e.g.,][pp. 12--13]{Ando1998}, the conversion entails that $\Delta\tau \approx 0.3 \tau_c$, if the autocorrelation function is well-behaved. Different methods exist to estimate the running autocorrelation and the slope of its main peak, so the estimates of the effective durations of acoustic sources tend to be somewhat inconsistent between reports and should be taken only as approximate ranges \citep{Dorazio2011}. This is also apparent from the supplementary analyses presented in \cref{ExCohere}.

While the criterion for the effective duration definition seems to have been arbitrarily set, it has been associated with some behavioral measures. Notably, identification of monosyllables by moderately hearing-impaired elderly listeners was highly correlated ($r=0.87$) with the effective duration of the syllables, but not with the broadband coherence time analog\footnote{In the paper the coherence time is referred to simply as the width of the autocorrelation peak and is notated with $W_{\phi(0)}$.} ($r=0.19$) \citep[Figure 5d--5e]{Shimokura2017}. In another study, \citet[pp. 88--119]{Ando1998} found several correlations between the preference of ideal reverberation time for specific music types and the effective duration of the music. A ``temporal factor'' was also defined that is equivalent to coherence time of the broadband autocorrelation, set at half the power of the main lobe, and was found to correlate with the timbre perception of spectral tilt of distortion guitar \citep[pp. 11 and 120--124]{Ando2009}.

The effective duration---and more generally, the degree of coherence---of speech signals has not been systematically surveyed in the literature, even though autocorrelation is frequently used in signal processing of speech (e.g., for extracting the instantaneous fundamental frequency; \citealp{Kawahara1999} and \citealp{Cheveigne2002}). While speech contains periodic or quasi-periodic components---especially in vowels (see \cref{SpeechAnimals})---it is highly nonstationary and dynamic, and therefore cannot be expected to have very long effective duration on average, especially when it is calculated for the broadband signal. According to \citet{DOrazio2017}, speech has an effective duration of somewhere between 8 and 16 ms. This estimate was probably based on \citet[Figure 4.2]{Ando1998}, who found a range of effective durations with an average of 12 ms (measured with a Japanese poem spoken by a female reader). That even speech vowels are not completely periodic can be seen in the autocorrelation curves in \citet[Figure 2]{Hillenbrand1988}, which reveal large, but nowhere near infinite effective duration for a sustained /a/ vowel---about 10--20 pitch periods for both real and synthesized male and female voices. Converted to time units, it can correspond to a range of 50--200 ms for a male with fundamental frequency of $f_0=100$ Hz at the low end, or to $f_0=200$ Hz for a female voice at the high end. A more detailed study of female Japanese monosyllables found that the effective duration is 30--70 ms \citep{Shimokura2017}. An older study with no specific details about the exact signal used reported male speech to have effective duration of about 100 ms, but the autocorrelation curve never tapered off below $|\gamma| \approx 0.15$ \citep[Figure 2 and table]{Fourdouiev1965}. Interpolation of the same curve gives a coherence time of about 31 ms for speech.  

More data has been collected about the coherence of musical sources than about speech sources. For example, \citet[pp. 13--19]{Ando1998} compared the running autocorrelation of different piano performances and found a wide range of effective durations with a minimum of 20--30 ms (e.g., Ando's Figure 3.5d), along with instances of much longer durations ($>$ 200 ms, Figure 3.4d), depending on the integration time used in the calculation. 

Effective duration data of non-musical acoustic sources have been reported sporadically. For example, autocorrelation measurements of narrowband noise centered around 250, 1000, and 4000 Hz with $\Delta f = $ 213, 554, and 1556 Hz, respectively, found coherence time (for $|\gamma|=0.5$) that is scaled by the noise bandwidth of about $\Delta \tau = $ 3, 1.5, and 0.4 ms, respectively \citep{Ando1982}, as expected from the coherence time definition. In another study, the temporal characteristics of ground-level broadband noise from a landing airplane at 1 km altitude were estimated and found to become incoherent very quickly (1--2 ms) \citep{Fujii2001}. However, there were additional peaks in the autocorrelation function after the initial drop that represent spurious coherence due to periodic components of the aircraft noise.  

\begin{figure} 
		\centering
		\includegraphics[width=0.5\linewidth]{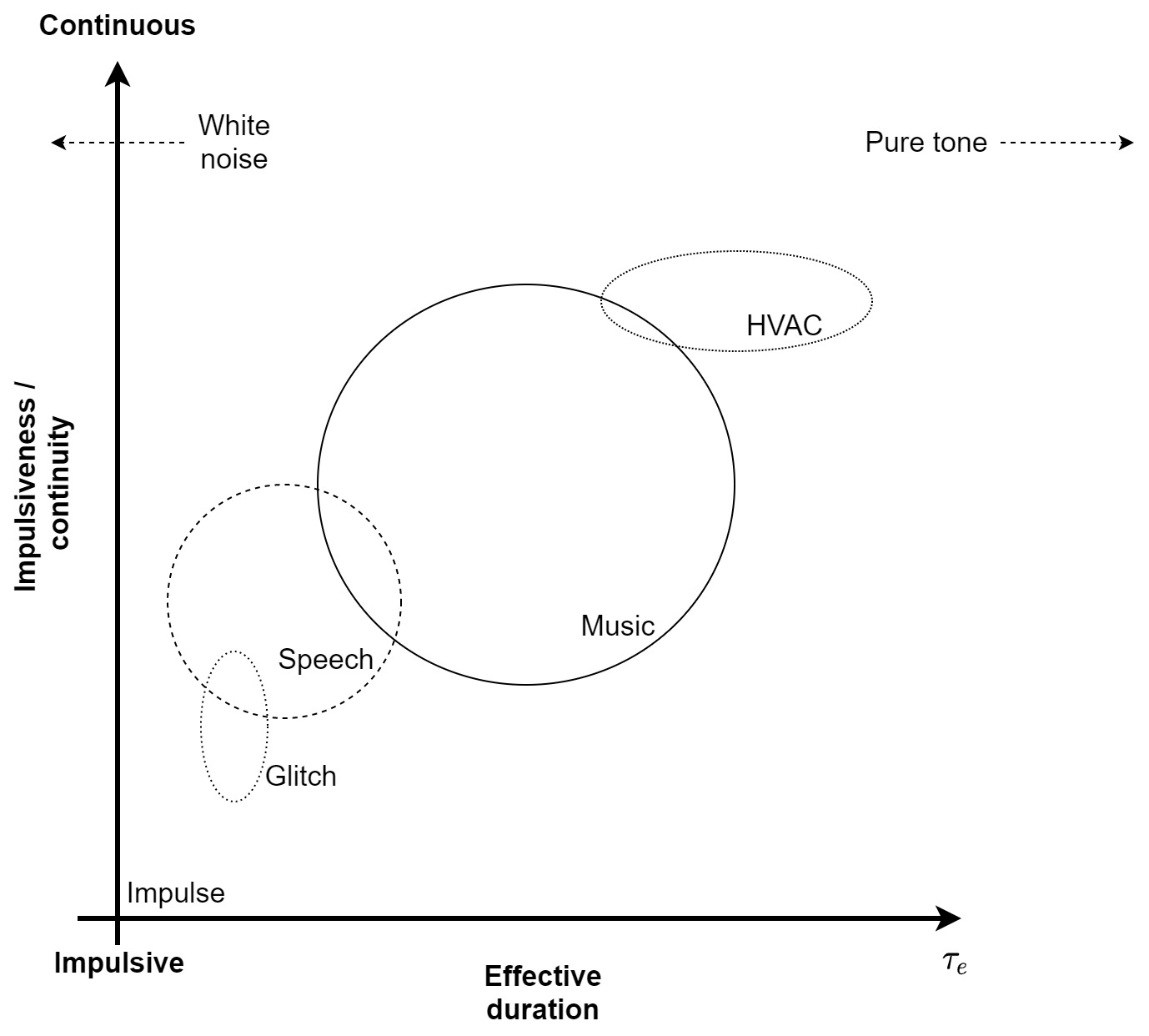}	
		\caption{The rough relative range of the effective duration of different types of acoustic signals and their impulsiveness. This is drawn according to \citet[Figures 4]{DOrazio2017}, with slight modifications.}
		\label{TypCoh}
\end{figure}

A qualitative breakdown of the coherence of various signal families was provided recently by \citet[Figures 4 and 6]{DOrazio2017}, in which the relative effective duration of speech, music, pure tones, which noise, impulses, ventilation sound, and ``glitch'' sounds were all plotted against an impulsiveness scale (ranging between completely impulsive and completely continuous). As many musical pieces contain sounds that are both continuous and periodic (e.g., harmony of a string section) as well as impulsive sounds with very slow rhythmic periods (e.g, drum beat), its effective durations occupies a large range of times (16--512 ms), depending on the musical content and instrumentation. The figure is roughly reproduced in Figure \ref{TypCoh} with slight modifications. 

\subsection{Coherence time of realistic sources}
Given the relative lack of narrowband coherence data in literature, several additional acoustic sources were analyzed in \cref{ExCohere}. The results show the interactions between coherence and the spectral features in the analyzed channel, the instantaneous tonality, and the type of source. Sources that have sustained and resonant normal modes are more coherent than transient sources, including speech. However, the variability is very large, and becomes even larger when it interacts with the room acoustics. 

One immediate observation is that coherence time values are long at lower frequencies and are almost always very short at high frequencies (i.e., above 2000 Hz). This can be due to the narrowband filtering and the sources themselves. The analysis filters impose some coherence on the signals, which is proportional to the absolute bandwidth of the filter, rather to its relative bandwidth (Eqs. \ref{CoherenceTimeEst} and \ref{CoheringFilter}). Since filter banks (including the cochlear one) tend to have a relatively small variation in the relative bandwidth as a function of frequency, then high frequencies that are analyzed by them will always be less coherent, by the definition of coherence time. Therefore, inasmuch as the filters themselves have a cohering effect on the signals, it decreases with frequency. Perhaps more importantly, the acoustic sources surveyed were musical and human sounds, which tend to be well-tuned only at low frequencies. At high frequencies, their high modal density may preclude stable narrowband sounds with the filters used. As the survey of sources was anything but exhaustive, more data will have to be analyzed in the future to enable a better grasp of these issues. 

Audio demos are available for some of the samples used in the experiment, which may be informative for the reader to get a handle of the different coherence regimes---especially when heard along with the various plots in \cref{ExCohere}.

\subsection{Discussion}
The combined data from literature and from \cref{ExCohere} are sufficient to give an idea about the ranges of coherence times associated with different sources. Importantly, they enable us to relate to a continuum of coherence times, which may be indicative of the subjective perception of coherent, partially coherent, and incoherent sources. As a rule, natural sources are rarely completely coherent or completely incoherent, but sources with coherence time on the order of a couple of milliseconds or less will be assumed incoherent, whereas those in the hundreds of milliseconds will be assumed coherent. These bounds will demarcate the relevance of acoustic and auditory phenomena that can be explained by interference. 

A common theme in this chapter and the rest of this work is that realistic acoustic sources and fields are partially coherent. Now that we have the extreme bounds of coherent and incoherent sources---what sounds could count as partially coherent? As was shown earlier (Eq. \ref{totalcoherence}), partial coherence can be expressed as a sum of completely coherent and completely incoherent contributions that are weighted by $|\gamma|$ and $1-|\gamma|$, respectively. Furthermore, according to Eq. \ref{cohtoincohratio}, $|\gamma|=0.5$ when the coherent and incoherent contributions are equal. For a purely temporal coherence measurement, this equality is satisfied exactly at the coherence time $\tau = \Delta \tau$, namely, at the full-width half-maximum of the coherence distribution. 

As a rule of thumb we shall consider the completely coherent regime of a coherence function to be contained within the coherence time interval $|\tau| \geq \Delta\tau$. Similarly, the completely incoherent regime will be outside of the effective duration $|\tau| \leq \tau_e$. The partially coherent regime is going to be everything in the middle $\Delta\tau \leq |\tau| \leq \tau_e$, which leaves some room for ambiguity that may depend on the individual responses and their interaction with the particular source. 

Note that even though we have discussed the results based on autocorrelation (self-coherence), it does not imply that such an operation is necessarily being performed within the auditory system---something that has been questioned in literature \citep{Loeb1983,Kaernbach1998}. Rather, autocorrelation here is the mathematical tool that reveals what kind of acoustic source is being radiated. This information affects the receiver whether or not it has a physiological autocorrelator. See \cref{BrainCoherence} for further discussion about autocorrelation in the auditory system.

\section{Decoherence through reflections and reverberation}
\label{DecoherenceRev}
Out of all the branches of acoustics, room acoustics may have made the most extensive use of stochastic tools that have parallels in scalar electromagnetic coherence theory (\cref{CoherenceinAcoustics}). This is because reverberation is most effectively described as an ensemble of reflections with random phase, whereas a complete solution of the wave equation for all room modes becomes untenable at high frequencies, where the modal density of the room is prohibitively high. The fundamental effect of reverberation on coherence is to gradually \term{decohere} the sounds through reflections, spectral broadening, and mixing with numerous other reflections in the field. Therefore, the reverberant sound field is usually divided into two limiting cases---the coherent/direct sound and the incoherent/reflected sound \citep{Morse1944}. The details matter, though, as the source type, boundary (room) structure, and construction materials determine the extent of decoherence that is ultimately detected by the ears. While the topic of room acoustics is too large to fully reframe in light of coherence theory, several basic results will be reviewed below that will provide the foundation for later analysis of the auditory system response to signals in their everyday environments. 


\subsection{Coherence and reflections}
\label{CohereReflections}
A complete description of the pressure waves reflected from a surface can be prohibitively complex and only relatively simple cases can be solved analytically (see \cref{Reflections}). The coherence approach avoids this difficulty by looking only at the relative point-to-point transformation of the wave function and at its potential to cause interference, rather than at a complete description of the entire field. This is particularly useful because any one-dimensional field function---fully describable as point source and point receiver (as is the effective pressure field arriving in each ear)---may be fully characterized by its instantaneous amplitude and phase (see \cref{WavesStimuli}).

Following the propagated field from a source with a known degree of coherence, we would like to know how its radiated pressure changes as a result of reflections. We can make a few inferences based on \cref{Reflections}. In the simplest case, a hard wall of very large dimensions (with respect to the longest wavelength of the sound field) acts as a perfect mirror. The pressure at any point in front of the wall is a superposition of the direct sound and its reflection from the wall. The time delay of the reflection that corresponds to $\tau$ in the coherence function $\gamma(\tau)$ is determined by the distance from the wall, the angle of incidence, and the wall impedance. If $\tau$ is smaller than the coherence time of the source, then the direct and reflected sounds interfere and level variations will be measurable around the point, embodying the interference fringes in the acoustic field. When the time delay is not much larger than the coherence time, partial coherence leads to partial interference that exhibits poor fringe visibility (measured in space). Finally, when the wall is not hard, then the reflected wave is generally attenuated and exhibits a frequency-dependent phase response, which can also lead to a loss of coherence (if the source is not monochromatic), even when the associated time delay is well below the coherence time. Partially coherent broadband sound that is mixed with its reflection will exhibit spectral ripples, or spectral modulation. Such ripples were observed in various geometrical configurations, where the source was placed between two, three, or four reflecting walls in \citet{Berman}. Even though the source was coherent (pure tones), the spectrum was affected by the modes that existed between the walls, which gave rise to a complex interference pattern that is seen in the spectrum. 

\subsection{Coherence and reverberation}
\label{CoherenceReverb}
Reverberation is caused by the ensemble of reflections between an acoustic source and its environment, which arrive to a receiver asynchronously. The reflected waves are further reflected between surfaces until the reflected energy eventually becomes fully dissipated and the sound dies out. The  delay associated with each discrete reflection is determined by the distance from the source, angle of incidence, and the surface impedance---summed over multiple reflections. Reverberation can be statistically characterized using its \term{reverberation time}, which is proportional to the duration of the sound decay in the reverberant space. In its simplest formulation\footnote{For a brief review of alternatives and limitations see \citet{Xiang2020}.}, the reverberation time is a function of the total surface area $\mathit{S}$, its relative absorption $0 \leq \mathit{\alpha} \leq 1$ (0 is completely absorbent, 1 completely reflective), and the volume $V$ of the space---through Sabine's formula \citep{Sabine}:
\begin{equation}
T_{60} = 6\ln 10 \cdot \frac{4V}{c\mathit{S \alpha}}
\end{equation}
Technically, it is defined as the time it takes for stationary sound to drop by 60 dB (to one millionth of its original level) from when it is switched off. Nowadays, it is usually evaluated directly from the impulse response function of the space. Although reverberation has been explored mainly inside built rooms, its applicability is completely general to other spaces with enough reflective surfaces to elicit statistical behavior (see \cref{RoomAc}). Therefore, the reference to rooms provides a convenient system that is also relevant to much of humanity, but many of the conclusions most certainly go beyond built enclosures. 

The validity of the statistical approach to reverberation largely depends on how random the sound field is---a property which is captured by the \term{diffuse field} model. The diffuse field has been a highly useful theoretical model for investigating the effects of reverberation. In a perfectly diffuse field, the number of reflections, their phase, and their direction are completely random (the quantities are average, so the randomness is evaluated over time) and behave ergodically \citep{Morse1944}. \citet[p. 12, definition II with interpretation B]{Jacobsen1979} explored several related definitions for a diffuse field in a reverberation chamber and zeroed in on: ``\textit{A diffuse sound field comprises an infinite number of plane propagating waves with random phase relations, arriving from uniformly distributed directions.}'' And additionally, ``\textit{At an arbitrary position in a pure-tone field the phase relations of the waves constitute a fixed set of random variables.}'' These somewhat technical definitions have been mainly used with spectral coherence, although it was developed independently from that reviewed in \cref{SpectralCoherenceTheory}.

A room that approximates a diffuse field is called a \term{reverberation chamber}. It is notionally antonymous to the \term{anechoic chamber}, whose design goal is to approximate free-field behavior that renders exclusive the direct sound from the acoustical source by minimizing reflections through heavy absorption \citep[pp. 151--154 and 221--225]{Kuttruff}. In interpreting experimental acoustic data, it is perhaps handy to consider the following analogies. \textbf{The diffuse field is to spatial coherence what white noise is to temporal coherence}---both approach the limit of complete incoherence with infinitesimally small spatial or temporal shifts in their respective self-coherence functions. Similarly, \textbf{the free field is to spatial coherence what the pure tone is to temporal coherence}---both approach the limit of complete coherence with infinitesimally large spatial and temporal shifts in their respective self-coherence functions. As always, reality lies somewhere in between these idealized model fields, as \citet{Morse1944} admitted that ``\textit{...room acoustics finds itself in the difficult intermediate region,}'' referring to the gulf between statistical modeling of rooms using geometrical acoustics and accurate modeling with wave acoustics that is often impractical. In a similar vein, \citet[p. 140]{Kinsler} commented on the difference between coherent and incoherent summation of a monochromatic source: ``\textit{A typical case of continuous wave propagation may lie somewhere between these two idealizations. Coherence is favored by short-range, low-frequency, smooth boundaries, few boundary reflections, and a stable and smooth speed of sound profile. Random phasing is favored by the converse conditions.}''

It becomes valid to model an acoustic field as diffuse where the enclosure supports a high modal density with respect to frequency. This is the case for large rooms, as is expressed by the \term{Schroeder cross-over frequency} \citep{Schroeder1996}
\begin{equation}
	f \approx 2000 \sqrt{\frac{T_{60}}{V}} \,\,\, \Hz
	\label{SchroederCrossover}
\end{equation}
for $T_{60}$ measured in s and $V$ in $\m^3$. This formula guarantees that frequencies above $f$ resonate due to at least three normal room modes, on average---so no normal mode ``isolates'' are allowed. Below the cross-over frequency, single modes may dominate the response, which is no longer well-described by statistical considerations. By inverting Eq. \ref{SchroederCrossover} and turning it into an inequality, we can relate to ``large rooms'' with an approximate diffuse field if they have a volume $V > T_{60}(2000/f)^2$ \citep[p. 68]{Kuttruff}. In contrast, ``small rooms'' may not have a meaningful $T_{60}$ associated with them, because individual normal modes occupy significant bandwidths. However, $T_{60}$ is generally dependent on frequency, so its blanket application in all these formulas requires caution.

Despite its power as a theoretical tool, the diffuse field approximation is often violated in real rooms and should only serve as a limiting case \citep{Morse1944,Waterhouse1968}. There are correlations in the diffuse sound field that demonstrate the limits of its randomness, which therefore induce extended coherence volumes than theory predicts\footnote{The \term{coherence volume} is defined as an extension to coherence length in three dimensions, as the volume around a point in which interference may be observed \citep[pp. 8--10]{Wolf2007}. The concept is not used in acoustics, but seems appropriate in the context of three-dimensional fields.}. For example, \citet{Waterhouse1955} showed both theoretically and empirically that interference is observable along the room boundaries even in a highly reverberant chamber. The interference is constructive close to the boundary with the strongest effect for a corner, followed by an edge, and finally a wall that has the weakest effect. Naturally, the interference is most pronounced for coherent sources, as pure tone fringes are measurable up to a distance of a couple of wavelengths from the walls. With broadband noise, the effect decreases with increasing bandwidth, as its coherence length drops (see Eq. \ref{CoherenceLength}; \citealp[Figure 7]{Waterhouse1955}). Even far away from the room boundaries, the spatial coherence of the diffuse broadband pressure field at two points is finite. Thus, an interference pattern can be observed, especially at low frequencies, given by the spatial coherence 
\begin{equation}
	\mu(\bm{r_1},\bm{r_2},\omega) = \sinc \left( \frac{\omega |\bm{r_1}-\bm{r_2}|}{c} \right)
	\label{roomdeocherence}
\end{equation}
where $\bm{r_1}$ and $\bm{r_2}$ are the two measurement points \citep{Cook1955}. This formula was found to be correct also in rectangular rooms as long as the reverberant field is undamped, modal eigenfrequencies are unique, and the modal density is high enough to be approximated as continuous \citep{Morrow1971}. 

Just as correlations exist in the temporal and spatial domains, they also characterize the spectral domain in the diffuse field. Sound components in the spectral domain are not entirely incoherent in large reverberant rooms. \citet{Schroeder1962} showed that if the frequency response between two points is modeled as a random process, then the autocorrelation of the power spectrum (defined as a function of spectral distance $|\omega_1-\omega_2|$), goes as
\begin{equation}
	\mu(\bm{r},\omega_1,\omega_2) = \frac{1}{1 + \left(\frac{|\omega_1-\omega_2| T_{60}}{13.8}\right)^2}
	\label{DiffDiffF1F2}
\end{equation}
which we associated with the spectral nonstationary self-coherence function $\mu$. The factor $T_{60}/13.8$ represents the time it takes for an impulse to decay to a value of $1/e$ from its initial value. The autocorrelation drops by about 10 dB for $|f_1-f_2| = 6.6/T_{60}$, so that it takes the frequency spacing of a few Hertz for two components to become completely incoherent. A very similar result was obtained for the pressure amplitude autocorrelation function. Eq. \ref{DiffDiffF1F2} is a neat illustration of the decohering effect of reverberation---the longer the reverberation time is, the more decorrelated adjacent spectral components become, as the field becomes closer to a theoretically stationary field in which frequencies are incoherent (Eq. \ref{SpectralCorrelation1}). However, caution must be taken when the spectral resolution of the measurement is higher than the spacing condition for incoherence, because the measurement filter can make the coherence appear higher than it is in reality \citep{Jacobsen1987, Jacobsen2000}. Some examples of spectral coherence at different measurement points in a room are given in \cref{Crossspectral}, where it is shown that realistic sources can be much more spectrally correlated in standard (non-diffuse) rooms than is predicted by Eq. \ref{DiffDiffF1F2}.

At the sound receiver in the reverberant space, three coherence regimes may be noticeable---two distinct and one that is more elusive. First, the direct sound from the source arrives uninterrupted to the receiver before the first reflection does. When the source is coherent, then its mutual coherence function is largely retained in this duration, as it propagates according to the wave equation (Eqs. \ref{CohWaveEq1} and \ref{CohWaveEq2}) and does not suffer from the decohering effects of reflection (\ref{CohereReflections}). Second, if a certain surface is much closer to the receiver than other surfaces, then a single reflection may interfere with the direct sound, if the direct sound and reflection are both coherent. Additional ``early reflections'' are possible over a short time window that produce some interference, depending on the surrounding geometry\footnote{\label{Diffusers}It is common to distinguish between specular and diffuse reflectors in this context. Specular reflections tend to be coherent, as they may give rise to interference between the incident and reflected sounds. The reflectors are usually hard and flat surfaces with no discontinuities. In contrast, a diffuse reflector (a \term{diffuser}) is designed to emulate a more randomized phase response as much as possible and thereby decohere the reflected sound and minimize interference effects. It therefore attempts to maximize the surface geometrical randomness and discontinuities in order to decohere the impinging field. \citet[p. 30--31]{Cox2005} noted that strictly speaking all passive reflectors are coherent, because they are time invariant and deterministic. They additionally noted that the effect of diffusers is then better thought of as temporal and spatial dispersion of the field and not of decoherence.}. Third, late reflections are no longer distinct and they arrive with random phases, which add incoherently in intensity without interfering. The first and third regimes can be distinguished by noting that the power of the reverberant field is independent of position and constitutes a constant fraction of the source power. Thus, a room may be characterized by the distance in which the direct and reverberant powers are equal---the \term{critical distance} or the \term{diffuse-field distance} \citep[p. 118]{Kuttruff}
\begin{equation}
	r_c \approx 0.1 \sqrt{\frac{V}{\pi T_{60}}}
\end{equation}
The separation between the early and late reflection regimes may not be clear-cut. Rarely, a distant and high-level reflection can stick out as an echo that is a distinct coherent reflection within the incoherent reverberant tail. A more common coherent effect may be a series of regular reflections that are caused by structural geometrical regularities such as parallel walls or spherical domes, which can give rise to flutter echo and/or coloration. These are modulation effects, which are imposed as peaks in the coherence function (either spectral or temporal) of the sound pressure in the room (see \cref{RoomAc}). 

The various effects of reverberation on radiated sound are of major importance in hearing, as they represent every possible degree of coherence that has to be dealt with by the hearing system. Stationary signals are of rather limited usefulness in everyday listening, especially when information transfer is considered, which contains modulations that can be on the temporal scale of the temporal fluctuations of the field itself. When the sound source is continuous but nonstationary, ``fresh'' direct sound is continuously mixed with the early and late reflections of previously emitted sound. In this case, if the source is highly coherent, the early reflections are dominant, and the source variation is rapid, then new and old information from the source can interfere and make any message communication difficult. Also, when the reverberation time is long---its energy decays slowly---it produces an approximately constant intensity level that ``washes out'' or reduces the contrast of fresh acoustic information (or the modulation depth in amplitude modulation; \citealp{Houtgast1973}). 

A few examples for some of the above effects are provided in \cref{ExCohere}. In brief, they illustrate how realistic room acoustics produces a partially coherent field that varies in time, space, and frequency, and depends on the acoustic content of the source and on the reverberant space. The analysis also demonstrates how typical signals are generally nonstationary and it is often not straightforward to predict how a signal is affected by reverberation in a field that is not ideally diffuse. Nevertheless, statistical effects that are predicted for stationary sounds in diffuse fields are occasionally observed, locally. 

\section{Interaural coherence}
\label{InterauralCoherence}
We will be remiss without mentioning the binaural function, which can be readily included in the general framework of coherence theory. Interaural cross-correlation between the two ear signals has been originally proposed by \citet{Licklider1948} as a means to separate speech from broadband noise that the hearing system employs. Diotic speech intelligibility significantly improved when either the noise or the speech was out of phase between the two ears, while the other was in phase. The advantage was retained for partial coherence of the noise $\gamma(R, 0) \leq -0.75$ for in-phase speech or $\gamma(R, 0) \geq 0.75$ when it was out of phase, where we set $R = |\bm{r_1} - \bm{r_2}|$ to be the fixed distance between the two ears of the listener. Another effect is of the perceived size of the acoustic source---its \term{apparent source width} in the listener's ``phenomenal space''---turns out to also be a function of the interaural coherence \citep{Licklider1948}. It was shown to be a quantity that listeners are very sensitive to (down to the $|\gamma| = 0.1$ level; \citealp{Jeffress1962}). Other effects can be at least partially explained by interaural correlation, such as perceived binaural loudness, binaural beats, and the binaural ``Huggins'' pitch. 

The spatial coherence function of the head interacts with the room acoustics. \citet{Lindevald1986} presented coherence data of a person walking in a reverberant field ($T_{60} \approx 1$ s) of a large lecture hall ($V=1000$ $\m^3$) that had a loudspeaker playing sinusoidal tones. The interaural coherence was clearly affected by the head shadow and the two ear signals became incoherent at a frequency as low as 500 Hz (the first zero of a sinc function). When the measurement was repeated for two microphones spaced as the two ears ($R = 15$ cm), the first zero was reached at 1100 Hz, so the response was more coherent. Thus, the head has a measurable spatially decohering effect on the sound field.

Binaural research has not formally engaged with coherence theory beyond the interaural-correlation. The one (informal) exception to this is \citet{Cohen1988}, who proposed that the interaural function is, in fact, an interferometer (see also \citealp{Dietz2021Litovsky}). This idea can be readily developed using the theory we presented earlier in the chapter. In principle, the classical duplex cues for binaural localization---\term{interaural time difference} (ITD) and \term{interaural level difference} (ILD) \citep{Rayleigh1907Duplex}---may be recast as elements within the interaural coherence function.  The ILD relates to the level difference, which leads to partial incoherence and audibility of less than unity (Eq. \ref{GeneralVisibility}). 

The ITD relates to delay---a constant phase term in the cross-term of the coherence function (i.e., $\delta$ in Eqs. \ref{deltaangle} and \ref{Interference3}). Substantial efforts have been dedicated to modeling physiological mechanisms that fulfill the role of \term{coincidence detectors}---neurons with two excitatory inputs that fire only when the binaural inputs are synchronized and perform an instantaneous cross-correlation operation. It has been proposed that coincidence detectors are accompanied by matching signal processing that can extract the necessary information \citep{Jeffress1948,Licklider1951,Colburn1973,Colburn1977,Loeb1983,Shamma1989}. While in reality the interactions and deviations from such a clean interpretation of the binaural operation are messy (e.g., \citealp{Buchholz2018}), there is recent physiological evidence from simultaneous juxtacellular recordings that the simple coincidence model may be sufficient to explain ITDs \citep{Plauvska2016}. Regardless of the precise mechanism for computing the correlation, the effectiveness of using it to predict binaural unmasking has been found to be very high, for a host of listening stimuli and conditions \citep{Encke2022}.

\section{Coherence processing in the brain}
\label{BrainCoherence}
Several models of pitch and harmonicity perception hypothesized brainstem circuitry that contains coincidence detectors that are not necessarily used in binaural processing. The important aspect about their operation is that they more-or-less instantaneously respond to input coherence, which means that their operation is adapted to nonstationary coherence by definition. 

For example, \citet{Loeb1983} posited a cross-correlation function, explicitly within the brainstem---in particular in the medial superior olive (MSO)---that can work either monaurally or binaurally. According to this model, the spectral resolution of the ear can be explained using coincidence detectors that receive inputs from channels whose spectral distance corresponds to 0.3--0.4 of the respective wavelength on the basilar membrane. This model does not resort to delay lines---a signal processing component that was hypothesized by \citet{Jeffress1948} that may be valid in birds, but probably not in mammals \citep{Grothe2010}. \citet{Shamma2000} hypothesized that a matrix of coincidence detectors may be responsible for harmonicity perception and various types of pitch (periodicity and residue), which are not easily explained without a cross-correlation setup that includes delay lines (for which there is no clear evidence). The model requires phase locking (see \cref{Phaselocking}) and distortion products (obtained through squaring) to have the information necessary for synchronization within the channel. It also hypothesizes spectral and temporal sharpening and that the inherent dispersion between the channels is necessary for eliciting their synchronization later in processing, regardless of the signal type. A related model by \citet{Carney2002} proposed a coincidence detector network that can detect pitch in noise. The network comprises all the adjacent channels that are affected by the stimulus and individually phase lock. Roughly, the idea here is that when sound elements coincide, they add up coherently and produce  a sharp frequency response, whereas the noise is broad and does not add cumulatively to the sharpness of its spectrum. The model requires for specific phase relations to exist between adjacent channels. As a final example, the model by \citet{Langner2015} identified periodicity mapping (periodopy) at the central nucleus of inferior colliculus (ICC), which is orthogonal to its tonotopy. In this model, the unique mesh-like topology of ICC neurons enables them to work as coincidence detectors for spikes that arrive from the dorsal and ventral cochlear nuclei (delayed), and the ventral nucleus of the lateral lemniscus (inhibition) \citep[see Figures in][pp. 129 and 168]{Langner2015}.

\section{Discussion}
Basic principles of coherence theory that are germane to acoustics and hearing science were reviewed above. While perhaps none of the ideas presented in this chapter are new to the field, they have never been presented en masse in this context in a comprehensive and rigorous manner. Nevertheless, acoustic coherence is intimately related to the identity of sound sources, which the auditory system can detect and react to. Later in this work, it will become apparent that the auditory system is inherently set up to differentially process coherent and incoherent stimuli, so the basic distinction between them is critical. Because every partially-coherent stimulus can be expressed as a sum of coherent and incoherent components, such a decomposition potentially endows the system with a unique handle to identify and process different sound sources.

The review dwelt on room acoustic effects, which were originally described using stationary signals in diffuse sound fields. These are generally not representative of most common acoustic stimuli, as was demonstrated in \cref{CoherentSources} and \cref{ExCohere} for a handful of human and musical sources. As these signals arrive to the outer ear and propagate to the cochlea via the middle ear, it can be expected that their coherence function will vary within the ear as well, albeit only slightly due to the very small distances involved. Things are more complex inside the cochlea because of dispersion, phase locking, and discretization, as will be discussed in subsequent chapters. 

It is important to make a distinction between signal processing that requires coherence detection as part of the computation and coherence that is propagated within the system, which represents an inherent property of the signal. Coincidence detectors are of the first kind, whereas other references to coherence in the brain (e.g., temporal self-coherence or autocorrelation) may be a propagated property of the second kind. The latter is the most important type of coherence in the present context, because it is an inherent feature of the signal that directly corresponds to its wave nature. Removing coherence from the signal is tantamount to removing some information that it carries about its source. Therefore, for coherence to \textbf{not} propagate into the auditory brain may be just as nontrivial a processing step as the opposite case. In the next chapter we will focus mainly on phase locking as a feature that facilitates coherence propagation.

\chapter{Synchronization and phase-locked loops}
\label{PLLChapter}
\section{Introduction}
In the previous chapter we laid out the basics of coherence theory as applies primarily to waves. Once inside the brain, the transduced sound no longer retains its wave characteristics, with the exception of its temporal properties that are synchronized to the external wave, either in carrier or in envelope. This enables a temporal continuity that applies to coherence, as is used in neurophysiological modeling. Specifically in hearing, carrier synchronization is mediated through the phase-locking process, which is possible at frequencies that are low enough for the neurons to track. What causes phase locking to appear in the auditory nerve pattern? This is usually not accounted for in cochlear mechanical models and is taken for granted in phenomenological models of the auditory nerve, whose input originates right at the synaptic interface between the inner hair cells (IHCs) and the auditory nerve. We would like to fill in this gap by showing that the organ of Corti and the outer hair cells (OHCs) can work as a phase-locked loop (PLL)---a circuit that synchronizes a local oscillator to an external source. As it turns out, different elements of the OHCs can assume the necessary components of a PLL---a phase detector that generates quadratic distortion components ($f_2-f_1$), a low-pass filter, and an oscillator that feeds back to the phase detector and serves as an output. While this model does not necessarily contradict the presently known or assumed functions of the OHCs (notably, amplification), it indicates that an important function of the OHCs may have been overlooked and that the cochlea and the auditory brain work in concert. In \cref{MOCR} we will additionally explore the possibility that efferent-mediated auditory accommodation has something to do with actively setting the PLL. In a narrower sense, though, the purpose of this chapter is to prove that external signal coherence can be conserved by the system, but not in an unconditional way.

In the next sections, we will review the general features of nonlinear synchronization in dynamical systems and specifically focus on PLLs. Then, a short proof of coherence conservation is sketched, using tools from the previous chapter. We will then quickly review the equivalence between a PLL and a general nonlinear oscillator, and draw parallels to known characteristics of the OHCs. We will then associate the different PLL components with known mechanisms in the OHCs, including some that are not so well understood that have not been well-accounted for in current models of the auditory system. We will finally consider potential evidence that can attest to some qualitative predictions that are made by the PLL model. 

\section{Background}
\label{SyncBackground}
\term{Synchronization} is a universal nonlinear phenomenon that applies to a very broad class of systems in the natural world and in engineering. It occurs when the oscillations or rhythms of two or more independent systems adjust to one another due to (weak) coupling between them \citep{Pikovsky2001}. It is critically distinguished from systems that contain a resonant component that would not oscillate on its own (i.e., it does not have an internal energy source). It is also distinguished from systems that are coupled so rigidly that they effectively become one, which means that the output phase of the forced system is tightly determined by the input. Hence, the response of two synchronized oscillators does not behave like a resonant filter, even if the two outputs appear to be the same under certain conditions. 


\begin{table}
\footnotesize\sf\centering
\label{syncterms}
\begin{tabular}{P{4cm}P{4cm}}
\hline
\textbf{Noun} & \textbf{Adjective}\\
\hline
synchronization & synchronized\\
(phase) synchrony/synchronicity & synchronous\\
entrainment & entrained\\
phase locking, phaselock & phase-locked, locked\\
frequency locking & frequency-locked\\
mode locking & mode-locked\\
coherence, coherency & coherent\\
correlation & correlated\\
coincidence & coincident\\
\hline
\end{tabular}
\caption{A list of near-synonyms that relate to the concept of synchronization. The differences in meaning are generally subtle and not always consistent between different texts and authors. Specific meanings that are employed in this work are defined in the text.}
\end{table}

The first documented synchronized systems were originally discovered in mechanics and later in acoustics. The discovery of mode locking is attributed to Huygens, who patented the pendulum clock in 1656 and found out that the pendulum motion of two clocks hanging from the same wooden beam became tightly synchronized, only in anti-phase, due to imperceptible coupling through the beam \citep{Pikovsky2001}. More than two centuries later, Rayleigh observed that when the ends of two organ pipes that are slightly mistuned are brought close together, they tend to resonate in unison, instead of beating slowly together (\citealp[p. 322c]{Rayleigh1879b,Rayleigh1945}; \citealp[see also,][]{Gripon1874} for an even earlier account; additionally, see \citealp{Abel2006} for modern measurements and model). Effectively, the fundamental modes of the two pipes become locked when the difference in their resonance frequencies is small and when the pipes are closely positioned. Using more precise measurement tools, \citet{Rayleigh1907} also found a very similar effect in two vibrating tuning forks that are slightly detuned, even when their coupling was very weak---through the air, or with a thin cotton thread. 

Further notable discoveries of synchronization effects were mainly in electronic circuits with coupled oscillators \citep{Vincent1919,Eccles1920,Appleton1922,VanderPol1926}. The associated electronic principles were critical in applied communication engineering and therefore received much theoretical attention subsequently. Many other phenomena in different domains (e.g., chemical, ecological) have been identified since (see \citealp{Pikovsky2001} for a review and further bibliography).

A more general type of synchronization can take place within a single oscillator that has multiple normal modes or coupled oscillators with different frequencies. Such \term{higher-order synchronization} between different modes is relatively general and can occur when the natural frequencies of two coupled systems are nearly harmonically related \citep[p. 104]{Pikovsky2001}:
\begin{equation}
	n\omega_1 \approx m\omega_2
	\label{modelock}
\end{equation}
where $n$ and $m$ are two integers and $\omega_1$ and $\omega_2$ are angular frequencies. Thus, even though few real-world vibrating objects are precisely harmonic, it is theoretically possible to induce nearly-harmonic modes to vibrate harmonically by coupling the objects together. By virtue of nonlinear mode-locking, musical instruments that produce sustained notes often end up producing harmonic sound under certain conditions \citep[pp. 143--144]{FletcherRossing}. It is difficult to estimate how prevalent these conditions may be encountered in real life, though, as they are rather restrictive (see \cref{PrimitiveSources} for more details and examples). 

Synchronization phenomena consist of numerous systems and a very rich set of effects and methods that appear in diverse contexts. We would like to focus on one particularly influential system, the \term{phase-locked loop} (PLL), which encapsulates some of these effects in a way that can be applied to hearing, without loss of generality. In addition to being a powerful model with high explanatory power, the PLL also provides a biomechanical and neurophysiological link to the signal coherence arriving from the environment, as was described in the previous chapter. 

\section{The phase-locked loop (PLL)}
\label{PLLs}
In the following, the principles of operation of the PLL, its main applications, and some of its key specifications and limitations are going to be briefly reviewed, based on texts by \citet{Wolaver1991,Stephens2001,Margaris2004,Gardner2005} as well as a short introduction in \citet[pp. 282--290]{Couch}.

One of the most common requirements in communication systems and in numerous other electronic systems is to force a local oscillator to track the instantaneous phase variations of an external (reference) signal. The PLL is perhaps the most common device that achieves this function. It has had a century-long history in modern electronic engineering and control theory \citep[pp. 1--9]{Stephens2001}. It is not an overstatement to say that the fidelity of modern communication technology would have been impossible without the invention and perfection of the PLL, which enables the synchronization of receivers to transmitters at arbitrary distances, frequency channels, and modulation techniques, often in prohibitive noise conditions.  

\begin{figure} 
		\centering
		\includegraphics[width=0.4\linewidth]{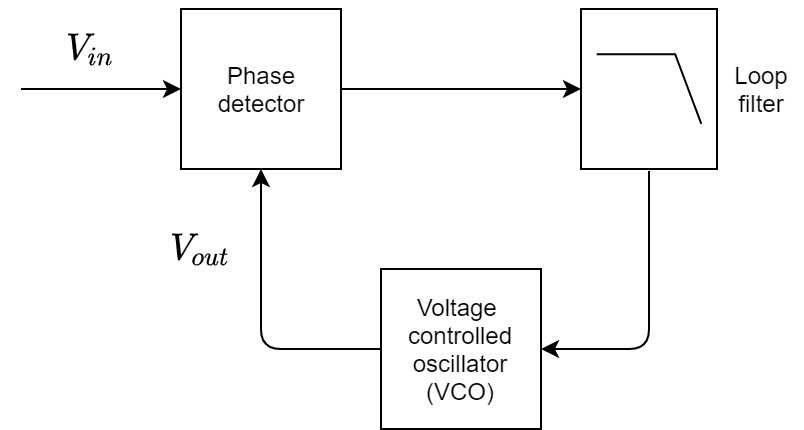}	
		\caption{The basic elements of a phase locked loop (PLL).}
		\label{PLLfig}
\end{figure}

The most basic PLL architecture can appear deceptively simple---it consists of an external signal fed into a phase detector that is connected to a local oscillator, which is itself connected by a feedback loop back to the phase detector (Figure \ref{PLLfig}). The fed back signal ensures that the phase detector output works as a correction control signal that always keeps the oscillator locked to the input. The phase detector is by definition a nonlinear device, where the difference between the input and the fed-back output is one of its distortion products. The instantaneous difference between the two signals corresponds to their phase difference. The PLL typically contains an additional low-pass filter that removes any high-frequency distortion products that are not needed by the oscillator. More importantly, the filter can determine the dynamical properties of the PLL---its stability and frequency response. This basic architecture can be complexified to any degree necessary and there are numerous ways to implement it in practice, both in analog and in digital electronic systems. 

The PLL is considered ``locked'' when the phase difference is zero (or constant). In reality, even in the locked state there is a finite steady-state phase error, which fluctuates around zero. This can often make the locked PLL appear as a narrowband filter that is capable of rejecting much of the noise, especially if the signal is well-behaved \citep[e.g.,][p. 2]{Gardner2005}. Since the frequency is the derivative of the phase, phaselock entails (instantaneous) frequency-lock as well.

\begin{figure} 
		\centering
		\includegraphics[width=1\linewidth]{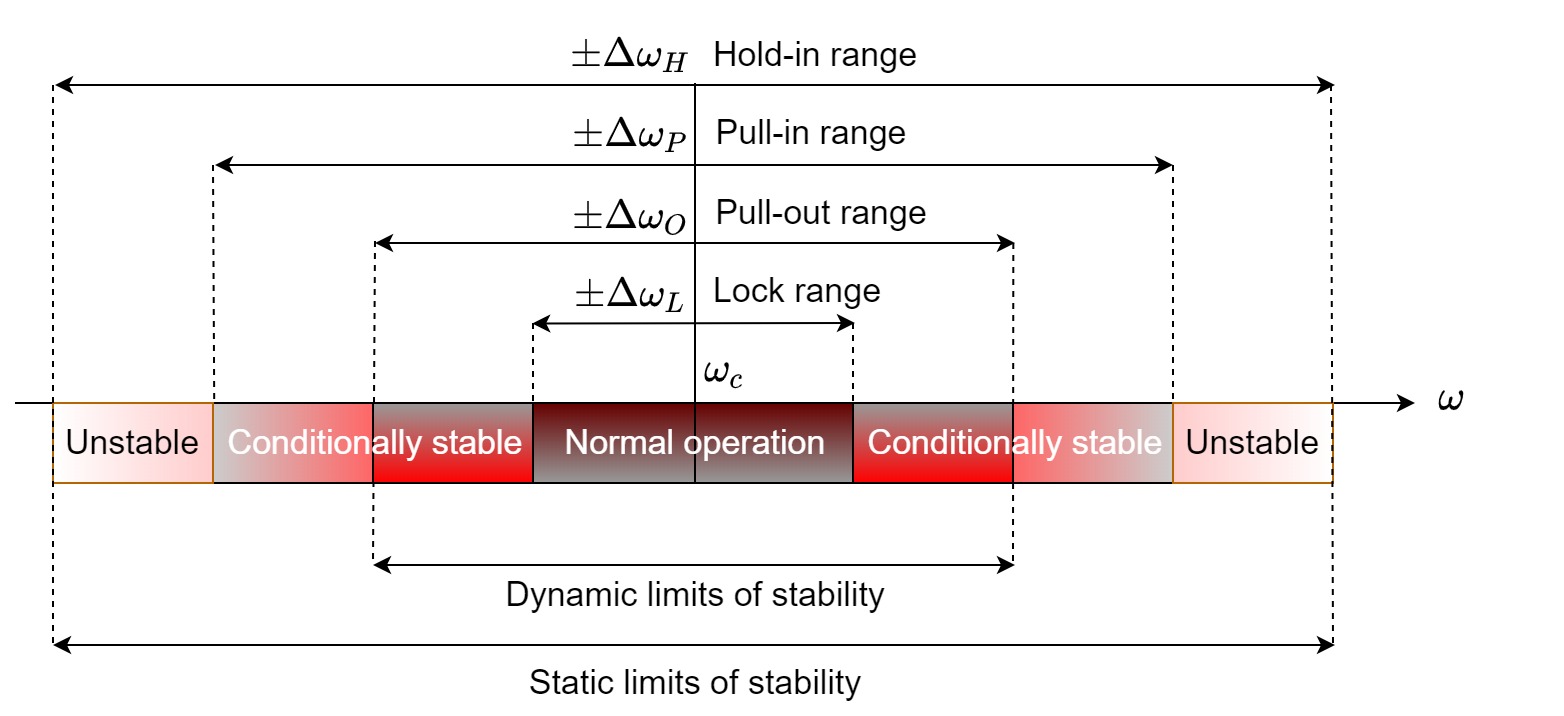}	
		\caption{The four basic frequency ranges of the PLL, which characterize its operating range and typical dynamic stability. This illustration is based on Figure 2.29 in \citet{Best2003}.}
		\label{PLLrange}
\end{figure}

Different frequency ranges and characteristic time constants are associated with the PLL operation. A PLL in lock can only remain so within the \term{hold-in range}---the maximum bandwidth around the carrier frequency in which the PLL works. If, however, the signal frequency changes quickly yet gradually, the PLL may not be able to maintain lock and its phase error will increase. However, as long as the change is within the \term{pull-in range}, which is narrower than the hold-in range, then the PLL will be able to reacquire its lock slowly through a \term{pull-in process}. Similarly, if the input signal frequency is changed abruptly (i.e., with a frequency step), the PLL will not be able to retain its lock if the step breaches the \term{pull-out range}, which is narrower than the pull-in range. Finally, the \term{lock range} (also, the \term{capture range}) is the narrowest of the PLL ranges and is where it can quickly achieve lock to input signal changes---without a single beat note---in what is called the \term{lock-in process}. The different frequency ranges associated with the PLL are illustrated in Figure \ref{PLLrange}. Care must be taken in the design of the PLL to not make its lock range too narrow, because it can result in excessive increase of the pull-in time, as tracking a fluctuating signal becomes more difficult. 

The \term{phase detector} is a nonlinear device that produces a distortion product that is proportional to the phase difference of its two input signals. In simple analog designs, it is modeled as an ideal mixer, or a multiplier, where it is made clear that it produces the sum and the difference of the two input signal frequencies. The sum is of no use and is removed by the loop filter to avoid adding noise to the control signal of the oscillator. Real phase detectors are also characterized by a ripple response that is superimposed on their phase difference output function. 

The PLL oscillator has to have a tunable frequency, typically achieved using a DC voltage control-signal, giving it the name \term{voltage-controlled oscillator} (VCO). The control signal is biased along with the error signal from the phase detector, which results in the dynamic tuning of the VCO frequency. The oscillator is inherently a nonlinear device as well, although in linear treatments of PLL theory it is modeled as an ideal integrator with stable tuning. This alludes to one of the main motivations for having a PLL in the first place---all oscillators (and clocks) tend to drift, so if left uncontrolled, they lose their tuning over time in an unpredictable way \citep[e.g.,][]{Eccles1920}. In coherent demodulation tasks such a drift is particularly detrimental, because it leads to an accumulated phase noise in the demodulated product, which amounts to distortion and reduced signal-to-noise ratio (SNR) at the output. Another nonlinear property of oscillators is their tendency to phase-lock to directly injected signals (i.e., in the case of a PLL, not via the phase detector and filter). This tendency can disrupt the controlled operation of the full PLL and is therefore kept to a minimum by design. 

The loop filter, generally low-pass, is placed between the phase detector and oscillator, but might be altogether absent in the most primitive designs. The loop filter is critical in determining the stability characteristics of the PLL---the susceptibility for self-oscillation or inability to pull out of the locked position, as the feedback loop becomes positive. The order of the filter determines the order of the PLL itself plus one. For example, a second-order low-pass filter makes for a third-order PLL\footnote{This should not be confused with the PLL type number, which is determined according to the number of integrators in the design---the oscillator is counted as one, and any additional filter pole at $\omega=0$ counts as an additional integrator. Therefore, a type 1 PLL contains only one integrator---the oscillator itself.}. In simple first-order PLLs, there is a single parameter that can be optimized (gain). However, first-order PLLs are inadequate in most applications, as they cannot lock to arbitrary signal types. Increasing the filter order provides the circuit designer with additional parameters that can be better tailored to achieve certain tasks. For example, a third-order PLL is better-suited to maintain lock to the received signal despite relative motion of the transmitter, which induces a Doppler shift on its spectrum. Similarly, only third-order PLLs can track a linear frequency modulation ramp. Third-order is also typically used for frequency synthesis, which is another common application for PLLs. However, in second- and third-order PLLs there is a risk of instability as the poles may be sensitive to gain and input level changes. PLLs of higher order than three are uncommon in typical applications. 

\section{The linearized PLL model}
\label{LinearizedPLL}
The PLL equations are not going to be used explicitly in this work, but they are presented because of the valuable insight that they provide for the understanding of the PLL principles of operation.

Several different models have been developed that describe the PLL operation. Of them, the linearized model of the PLL is the simplest to derive---as it requires only one approximation in the nonlinear equations. It preserves sufficient complexity that captures the central aspects of the PLL operation, which is completely satisfactory in steady-state conditions. It is the transient behavior---mainly pulling in (acquiring lock) and pulling out---that is manifestly nonlinear and sometimes requires other analytic methods. The linearized PLL model derivation is reproduced below, originally developed by \citet{Jaffe1955} and elaborated by others. A nonlinear approach that accounts for the transient behavior of the PLL is given in \citet{Margaris2004}. 

For a system that is described in Figure \ref{PLLfig}, let us assume a reference input signal around a carrier $\omega_c$ with an arbitrary phase function $\varphi_i(t)$ and amplitude $\sqrt{2} A_i$
\begin{equation}
	V_{i}(t) = \sqrt{2} A_i \sin \left[ \omega_c t + \varphi_i(t) \right]
\end{equation}
The output from the VCO is chosen to be in quadrature, with an output phase function $\varphi_o(t)$ and amplitude $\sqrt{2} A_o$
\begin{equation}
	V_{o}(t) = \sqrt{2} A_o \cos \left[ \omega_c t + \varphi_o(t) \right]
\end{equation}
where the instantaneous phase function of the VCO is determined by its voltage input $V_o(t)$
\begin{equation}
	\varphi_o(t) = K_v\int _{-\infty}^t V_o(\tau)d\tau
\end{equation}
with $K_v$ being the gain of the VCO. We can express the same relation in differential form
\begin{equation}
	\frac{d\varphi_o(t)}{dt} = K_v V_o(t)
	\label{OutPhaseKv}
\end{equation}
which directly quantifies the sensitivity of the VCO output voltage to the change in the instantaneous frequency of the input. 

Both $V_{i}$ and $V_o$ are inputs to the phase detector, whose operation is to multiply them with sensitivity $K_m$. Because we selected the reference and VCO to be in quadrature, their sum and difference terms are both sine functions
\begin{multline}
	V_d(t) = 2K_m A_o A_i \sin \left[ \omega_c t + \varphi_i(t) \right]\cos \left[ \omega_c t + \varphi_o(t) \right]\\
	= K_m A_o A_i \left\{ \sin \varphi_e (t)+ \sin \left[ 2\omega_c t +\varphi_o(t) + \varphi_i(t) \right] \right\}
	\label{PDout}
\end{multline}
where we defined the difference phase to be the phase error,
\begin{equation}
\varphi_e(t) \equiv \varphi_i(t) -\varphi_o(t)
\label{PhaseErrorDef}
\end{equation}
which is zero when the PLL is locked in, by definition. 
Next, we assume that the low-pass loop filter removes the sum term on the right of Eq. \ref{PDout}, and that the remaining difference term is convolved with the impulse response of the filter $h_f(t)$
\begin{equation}
	V_c(t) = K_m A_o A_i \sin \varphi_e(t) * h_f(t)
\end{equation}
that yields the output signal $V_c(t)$ from the filter, which is the input to the VCO that scales it by a gain $K_v$ (Eq. \ref{OutPhaseKv}), as by definition, the output phase of the VCO is proportional to its input. Thus, we can now construct an integro-differential equation for $\varphi_e(t)$ by differentiating Eq. \ref{PhaseErrorDef}
\begin{equation}
	\frac{\varphi_e(t)}{dt} = \frac{\varphi_i(t)}{dt} - K_v K_m A_o A_i \int_0^t \sin\varphi_e(\tau)h_f(t-\tau)d\tau
	\label{IDPLL}
\end{equation}
This is the basic nonlinear equation that describes the PLL, but it does not have a closed-form solution without some approximations. The linearized solution approach has the sine function approximated for small angles with 
\begin{equation}
\sin\varphi_e(t) \approx \varphi_e(t)
\end{equation}
just like in the paraxial optics approximation of geometrical optics (\cref{GeometricalOptics}). 

Without explicitly solving it, we can find the theoretical hold-in frequency range around $\omega_c$ for which the PLL can maintain its lock. If the input frequency $d\varphi_i(t)/dt$ is changed infinitesimally slowly, then the low-pass filter response can be approximated to its DC gain, $K_f$. Using the definition of $\varphi_e(t)$, Eq. \ref{IDPLL} can be rewritten as 
\begin{equation}
	\frac{d\varphi_o(t)}{dt} = K_v K_m K_f A_o A_i \sin\varphi_e(t)
	\label{PLLTheta}
\end{equation}
Eq. \ref{PLLTheta} is maximized for $\sin\varphi_e(t) = \pm 1$, and we can obtain the maximum frequency range,
\begin{equation}
	\Delta \omega_h = K A_o A_i  
	\label{PLLbandwidth}
\end{equation}
where we compacted all the gain constants into a single symbol $K$ 
\begin{equation}
	K = K_v K_m K_f
\end{equation}
which is called the \term{loop gain} of the PLL. $K$ is also called the \term{PLL bandwidth}. In the first-order PLL, the bandwidth is equal to the hold-in, pull-in, pull-out, and lock ranges, so $K = \omega_h$. All PLL orders fundamentally depend on the  loop gain $K$ \citep[pp. 20--22]{Gardner2005}. 

The simplest time-domain solution to Eq. \ref{IDPLL} can be obtained for an ideal filter with pure attenuation (or gain) and no zeros or poles, so that $h(t) = K_f\delta(t)$. Using the linear approximation, this leaves us with a first-order PLL
\begin{equation}
	\frac{\varphi_e(t)}{dt} = \frac{\varphi_i(t)}{dt} - K A_o A_i \varphi_e(t)
\end{equation}
which has closed-form solutions for specific input types \citep[e.g.,][pp. 16--19]{Stephens2001}. 

In the linear approximation, the transfer function of the PLL exists, and it is possible to analyze the closed-loop circuit in the frequency domain using the Laplace transform. The most useful \term{closed-loop transfer function} is a phase function ratio
\begin{equation}
	H(s) = \frac{\Phi_o(s)}{\Phi_i(s)} = \frac{K_v K_m A_o A_i H(s)}{s + K_v K_m A_o A_i H(s)}
	\label{ClosedLoop}
\end{equation}
where $s$ is the Laplace-transform complex frequency variable, which can be set to $s = i\omega$ to convert it to standard frequency. For the first-order PLL, this expression simplifies to 
\begin{equation}
		H(\omega) = \frac{K A_o A_i}{i\omega + K A_o A_i}
\end{equation}
which is a low-pass response and is unconditionally stable. In the majority of situations, this design is impractical, as it requires high gain in order to achieve lock, which directly increases the bandwidth of the PLL (Eq. \ref{PLLbandwidth}) and hence the susceptibility to noise. Also, it is often the case that a spurious pole (e.g., a delay) appears in the loop as a result of the phase detector, for example, which would then make it unstable when combined with high gain. With low gain this PLL would be unable to maintain lock for most variable signals, so a more sophisticated filter is required in almost all applications, namely, a second- or third-order PLL. It is important to note that the PLL response varies with the level of the reference input (and its SNR), which makes the PLL highly nonlinear. For this reason, some designs try to stabilize or compress the input level, or design an appropriate filter to minimize this dependence \citep{Jaffe1955}.

\begin{figure}
		\centering
		\includegraphics[width=1\linewidth]{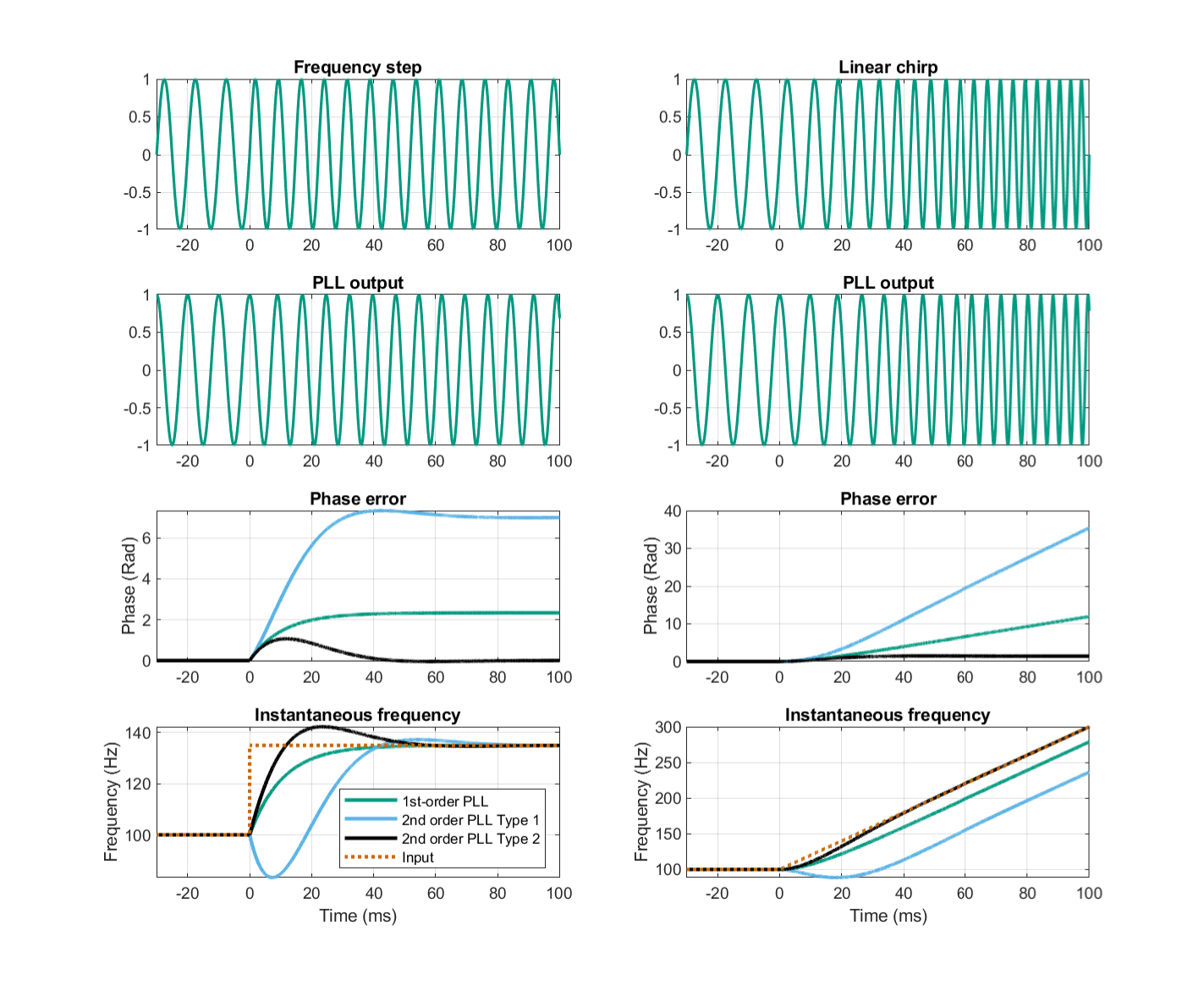}	
		\caption{Examples of linearized phase-detector ($\sin\varphi_e(t) \approx \varphi_e(t)$) PLL responses to frequency step of +35 Hz (left) and frequency linear chirp of 4000 Hz/s (right) starting at $t=0$, from a constant frequency at $t<0$ of 100 Hz. The output responses (second row) are from the first-order PLL as follows. The responses of three basic PLL topologies are shown in the bottom two rows: first-order PLL with $K = 15$ Hz (green curves); second-order PLL type 1 with $f_n = 35$ Hz, $\zeta = 0.707$, $K_v = 5$, low-pass with lead compensation filter, $F(s) = K_v\frac{1+s\tau_2}{1+s\tau_1}$ with $\tau_1 = \frac{K_v}{(2\pi f_n)^2}$, $\tau_2 = \frac{2\zeta}{2\pi f_n}\left( 1 - \frac{2\pi f_n}{2K\zeta} \right)$ (blue curves); and second-order PLL type 2, $f_n = 35$ Hz, $\zeta = 0.707$, an integrator and lead compensation $F(s) = \frac{1+s\tau_2}{s\tau_1}$ with $K = 4\pi\zeta f_n$, $\tau_2 = \frac{\zeta}{\pi f_n}$ (black curves). The simulation is based on a Matlab code by Mark Wickert, 2014, \url{http://ece.uccs.edu/~mwickert/ece5675/lecture_notes/PLL_simulation.pdf}.}
		\label{PLLdemos}
\end{figure}

Basic examples for the responses of three simple PLL topologies are displayed in Figure \ref{PLLdemos}. The numerical values were selected to exaggerate the duration of the transient responses, so it should not be implied that these values are representative of real-world designs. The damping factor in the filters of the second-order PLLs was chosen to be 0.707 as an ideal trade-off between filter ringing and damping. All PLL topologies would acquire lock more quickly with higher bandwidth and/or gain, as long as they are not overdamped or underdamped. Note how all PLLs are able to track the frequency step successfully, but only the second-order PLL with an integrator can (almost) track the linear chirp. Also, before locking onto the frequency step, the second-order type 1 PLL (bottom left of Figure \ref{PLLdemos} in blue) \term{phase-slips} and misses about one cycle---a common PLL characteristic during lock, which is tantamount to an error.

The PLL operation is susceptible to poor SNR, which translates to phase noise in the phase detector that increases the error, increases the pull-in time, and decreases the hold-in, pull-in and other frequency ranges. The phase error dependence on SNR is generally nonlinear---it is almost unaffected at positive SNRs, but deteriorates quickly at around 0 SNR. In a narrowband channel with bandlimited Gaussian noise, the effect of noise is well-modeled as equivalent to being added from within the loop rather than through the phase detector input. This means that the noise translates to additive phase noise inside the PLL. However, the internal SNR in the loop is actually 3 dB better than the external one in the input. Additionally, just as the loop performance depends on the signal level, so does the phase detector gain may depend on the SNR. It is possible to mitigate the noise effects by correctly designing the loop filter, so that the phase error is minimal for given SNR and input level, e.g., by decreasing the bandwidth when the SNR is low. It should be noted that internal noise generated by any of the PLL components may be significant and has to be considered as part of the total noise model. It is sometimes useful to refer to the noise-equivalent bandwidth $B_n$, which can be computed from the closed-loop transfer function (Eq. \ref{ClosedLoop}) with
\begin{equation}
	B_n = \frac{1}{2\pi} \int_0^{\infty}|H(\omega)|^2 d\omega = \frac{K A_o A_i}{4} \,\,\,\ (\Hz)
\end{equation}
for a first-order PLL. This expression and the bandwidth $\Delta \omega_h$ become more complicated for higher-order PLLs. 


\section{PLL coherence}
\label{PLLCoherence}
It is a convention in communication theory that employing a PLL in the circuit enables coherent detection (\cref{CommunicationTheory}). Therefore, as a corollary to the PLL basic function, we would also like to find out under what conditions a PLL can conserve the arbitrary coherence properties of an input signal. The phase detector continuously compares the input and output signals by multiplying them (Eq. \ref{PDout}), so it can be thought of as a correlator, which produces the instantaneous cross-term of the coherence function. The coherence function is therefore expected to be dependent on the phase error between the input and the output. 

Let us inspect a narrowband input of the form
\begin{equation}
	p_i(t) = A(t)\exp \left\{ i[\omega_c t +\varphi(t)] \right\}
\end{equation}
The input can be described by its (nonstationary) self-coherence function, $\gamma_{ii}(t,\tau)$. Without loss of generality, the output signal is the same but is amplified with gain $B(t)$, and differs from $p_i(t)$ by the instantaneous phase error $\varphi_e(t)$
\begin{equation}
	p_o(t) = B(t)A(t)\exp \left\{ i[\omega_c t +\varphi(t) + \varphi_e(t)]\right\}
\end{equation}
The corresponding self-coherence function of the output is then $\gamma_{oo}(t,\tau)$. 

Assuming that the gain varies very slowly in time compared to the phase tracking operation, we assume that $B(t) \approx \const$, so we can use normalized amplitudes and the normalized coherence function $\gamma_{io}$ instead of $\Gamma_{io}$ (see \cref{BasicCoherenceDeriv} for definitions). According to Eq. \ref{nonstationGamma}, the nonstationary coherence function is given by
\begin{equation}
	\gamma_{io}(t,\tau) = \frac{1}{N}\sum_{n=1}^N p_i^*(t_n)p_o(t_n + \tau)
\end{equation}
where we substituted, for simplicity, $t=t_1$ and $\tau=t_2-t_1$, in Eq. \ref{nonstationGamma}, which corresponds to the update (processing) time of the feedback loop of the PLL\footnote{This concept is more relevant in discrete PLLs, which update every clock cycle, but is somewhat more abstract with analog continuous PLLs.}. Each time instance $t_n$ may relate to a period in which the input self-coherence function is more or less stable. We examine only the low-frequency terms of the coherence function and neglect the fast-varying terms in the sum
\begin{multline}
	\gamma_{io}(t , \tau) = \frac{1}{N}\sum_{n=1}^N \exp\left\{ i[\omega_c \tau + \varphi(t_n +\tau) - \varphi(t_n) + \varphi_e(t_n + \tau)] \right\}\\
	=  \frac{\exp( i\omega_c \tau)}{N}\sum_{n=1}^N \exp\left\{ i[\varphi(t_n +\tau) - \varphi(t_n) + \varphi_e(t_n + \tau)] \right\}
	\label{PLLgamma}
\end{multline}
where the constant difference in the carrier phase was taken out of the sum and has no effect on the magnitude of the coherence function. 

There are three phase terms in the argument of the exponential in the sum of \ref{PLLgamma}. The first two, $\exp\left[\varphi(t_n + \tau) - \varphi(t_n)\right]$, are equivalent to the autocorrelation function of the signal $p_i(t)$, $\gamma_{ii}^n(\tau)$, evaluated for a delay $\tau$. The faster the PLL is, the closer the phase difference is to 0 and the autocorrelation function is closer to its peak value of 1. However, the less coherent the signal is and the shorter is its coherence time, the more erratic is its phase function, and a smaller $\tau$ will be required to stay close to the autocorrelation peak (see \cref{WhiteCoherence}). For signals that are asymptotically completely incoherent (white noise), the coherence time $\Delta \tau \ll \tau$, i.e., no $\tau$ will be short enough to minimize $\varphi(t_n + \tau) - \varphi(t_n)$, which will assume a different value between time instances $t_n$ and will average to zero for the entire ensemble. In contrast, for signals that are asymptotically completely coherent along with small $\tau$ from the PLL, or $\Delta \tau \gg \tau$, then $\varphi(t_n + \tau) - \varphi(t_n) \approx \const$---a constant phase term that may be also taken out of the sum and does not affect the coherence magnitude. 

Now, the operation of the PLL is also tied to the third term in Eq. \ref{PLLgamma}, which is the instantaneous value of the phase error term $\varphi_e(t_n + \tau)$. When the PLL is locked to the signal, the phase error is small and fluctuates around 0 (or another constant) $|\varphi_e(t)| < \epsilon \approx 0$. Once again, this is a constant term that can be taken out of the sum, effectively making the $\gamma_{io}(t,\tau) \approx \gamma_{ii}(t,\tau)$ a self-coherence (autocorrelation) function of $p_i(t)$, evaluated at $\tau$. When the PLL is unlocked---either because it is in the transient lock-acquisition stage, or because the signal cannot be locked to---then $\varphi_e(t)$ is unbounded and its effect is identical to dealing with an incoherent signal, whose ensemble-average coherence is $\gamma_{ii}(t,\tau) \ll 1$. 

We should remember, though, that both the coherence function and the PLL generally operate in narrowband. If the signal is passed through a bandpass filter in addition to the PLL, then it exaggerates its apparent coherence, which increases with narrower filters (\cref{CoherentFiltering} and \cref{WhiteCoherence}). 

~\\

In summary, when the PLL receives a coherent input and locks on to it, the output remains coherent, except for the initial transient locking in stage. The output coherence and the input coherence would be asymptotically equal, or 
\begin{equation}
	\lim_{\tau,\varphi_e(t) \rightarrow 0} \gamma_{oo} = \gamma_{ii}
\end{equation}
as long as the finite time of the PLL update speed is virtually instantaneous ($\tau \rightarrow 0$) and there are no fluctuations in the phase error ($\varphi_e(t) \rightarrow 0$). As the PLL cannot lock in to an incoherent input, its output will remain incoherent over time (subject to the cohering properties of the channel filter).

The coherence of partially coherent signals may be conserved as well, but will be generally somewhere in between the two extremes. For example, some PLL designs cannot lock onto fast linear frequency sweeps. In such a case, $\varphi_e(t)$ may be unbounded, but deterministic. So depending on the specific signal evolution, coherence may be gradually lost, but not altogether absent such that it acquires a completely random phase error. 

We also assumed independence of the channel gain, which may not be true in general. If the gain fluctuates rapidly, the coherence $\gamma_{io}(t,\tau)$ is expected to decrease. 

In conclusion, the above reasoning proves that the PLL is able to conserve the coherence of the input signal under some conditions that are not particularly limiting. 

\section{Motivation for an auditory PLL}

While phase locking is a well-studied hallmark feature throughout the auditory pathways (but primarily in the auditory nerve and brainstem), it is usually discussed as a de-facto property of the system \citep[e.g.,][]{Heil2015,Verschooten2019} rather than as an intended result of a mechanism that acquires locking. The author is unaware of any mention in the auditory research literature of a PLL as a module that is integral to the system---at least not at low-level processing. There has been, however, sporadic jargon borrowed from PLL theory in the context of synchronized spontaneous or evoked otoacoustic emissions (e.g., ``pulled in'' in \citealp{Wilson1981Tin}; ``frequency locking'' in \citealp{Probst1991}; ``capture'' in \citealp{Miller1997}). 

Given the complexity of the PLL circuit and operation, why should we invest in applying such a model to the auditory system? 

Tying phase locking to an actual PLL module is not pursued merely out of academic interest. There are a few strong reasons that make it especially pertinent to identify a PLL in the auditory system:
\begin{enumerate}
	\item Phase locking entails conservation of coherence of the input stimulus in the receiver. As will turn out later in this work, coherence and the lack thereof inform much of the intuition of what the auditory system does, also in the context of achieving sharp temporal images. Establishing an earlier source of phase locking in the auditory system is critical in understanding how it deals with different types of signals and what kind of responses can be expected.
	\item Tying different organs or circuits in the auditory system with the PLL function may help us demystify their function and potential impairments that are associated with them. A stronger version of this argument is that given that the PLL requires a closed feedback loop to work, studying its components in isolation, as though they can function as part of an open-loop circuit, is mistaken. We will see a clear example for this in the loop filter of the putative PLL. 
	\item Both PLL theory and practice are vast, so they can undoubtedly provide a conceptual framework and added insight in the analysis of various auditory phenomena that may have resisted treatment with other tools. For example, it may apply to the distinction between transient and steady-state effects, along with a distinction between trackable and untrackable signals.
\end{enumerate}

Because the main focus of the present work is to apply imaging theory concepts to hearing, it is mainly the first reason that has motivated the development of this model. It is impossible to meaningfully understand auditory imaging without a notion of coherence. And it is impossible to understand how coherence is conserved or eroded without a basic notion of phase locking, and arguably for that matter, of PLLs.

\section{Nonlinear synchronization recast as a PLL}
\label{RecastingPLL}
In this section, we attempt to connect a few loose ends between the standard accounts of the cochlear nonlinear dynamics and the phase locking in the auditory-nerve. This will be a stepping stone in identifying a PLL module in the auditory system.

\subsection{Nonlinear oscillators and PLLs}
Despite many commonalities, nonlinear synchronization phenomena and PLL design are generally studied independently. This is not entirely surprising, because opposing aspects of these systems are of particular interest in different disciplines. In synchronization processes, they include nonlinear and chaotic phenomena such as the underlying physical mechanisms of phase locking, Hopf bifurcation, instability (through the Lyapunov exponents), limit cycles and attractors, and synchronization dynamics of multiple oscillators. So a recent interest in PLLs from the nonlinear dynamics perspective has been to uncover how various second- and third-order PLL topologies may be susceptible to chaotic dynamics under some conditions \citep[e.g.,][]{Endo1988,Chu1990,Harb,Piqueira}. Chaotic dynamics can be invoked with the right choice of parameters, by modulating the input signal just around the pull-in range, which throws the PLL in and out of lock \citep{Endo1988}, and have a response that strongly depends on the initial conditions \citep{Chu1990}. Additionally, period doubling and chaos can be observed when a key parameter is set above a certain threshold (Hopf bifurcation) \citep{Harb,Piqueira}. In stark contrast, in PLL engineering, instability is carefully studied in order to be avoided like the plague in all applications. A PLL that self-oscillates, or becomes chaotic, is useless as a module within a larger system. 

As it turns out, several generic nonlinear systems that contain a free-running (uncontrolled) oscillator that exhibits synchronization to weakly coupled inputs (i.e., through injection locking) may be remodeled as PLLs without loss of function \citep{Couch1971,Schmackers2005}. This has been shown in specific cases for the Van der Pol oscillator, whose equations can be brought to the same form as either first- or second-order PLLs. Additionally, there exists a transformation that can map between the PLL and the other nonlinear system phase-space representations (both system types are generally modeled in different phase-space coordinates) \citep{Schmackers2005}. 

Therefore, it is perhaps unsurprising to see how the different nonlinear effects that have been observed in OHC models are also found in PLLs---Hopf bifurcations with limit cycle regime, synchronization, phase slips, possible instability, suppression (referred to as \term{oscillation quenching} or \term{death} in nonlinear dynamics), and dependence on the input level \citep[e.g.,][p. 229]{Roongthumskul2013,Hudspeth2014,Chakraborty2016,Roongthumskul2021,Pikovsky2001}. This equivalence enables us to treat the OHC(s) either as a synchronized oscillator or as a PLL, interchangeably. However, as \citet[pp. 40--41]{Pikovsky2001} commented, it is frequently very difficult to identify the feedback loop in natural systems with synchronization. Therefore, a practical transformation from the physical nonlinear oscillator to a PLL may not be obvious. 

\subsection{Auditory neural phase locking}
\label{Phaselocking}
In the absence of acoustical input to the auditory system, the auditory nerve discharges spontaneously, in a stochastic manner, at rates that correspond to the fiber sensitivity to sound level---low rates correspond to high threshold units, medium rates to medium thresholds, and high rates to low threshold ones \citep{Kiang1965,Liberman1978}. In the presence of pure-tone input from the cochlea, spikes become synchronized to the carrier phase, so their overall temporal pattern is no longer random. This phase locking between the pure tone stimulus and its neurally encoded version is a characteristic of the mammalian auditory nerve \citep{Galambos1943,Tasaki1954,Kiang1965,Rose1967}. See \citet{HeilPeterson2017} for a review of phase locking in the auditory nerve.

Phase locking has a high-frequency cutoff that varies between mammals and other vertebrates. In humans this limit has been often associated with degraded perception of pitch and melody that is observed above 4--5 kHz \citep{Moore2019}. In all other mammals tested it is lower than in humans \citep{Koppl1997, Palmer1986}, whereas in the barn owl it is much higher (9--10 kHz) \citep{Sullivan1984, Koppl1997}. While a 4--5 kHz phase-locking cutoff in humans is taken as a standard figure, there has been an ongoing controversy regarding its precise value, and more generally, of its significance in hearing, especially given its limited bandwidth \citep{Verschooten2019}. 

In the context of auditory phase locking, a distinction is sometimes made between synchronization and \term{entrainment}. The former relates to the temporal precision of the spikes, whereas the latter to the number of spikes per stimulus cycle \citep{Rhode1986ventral,Joris1994}. The two factors are independent dimensions of auditory temporal coding and were shown to be improved by the existence of the refractory period in the auditory nerve \citep{Avissar2013}. 

It is also common to refer to ``envelope phase locking'' or to ``envelope synchronization'', but for reasons that will be discussed in \cref{PLLNoise}, we will avoid the former expression\footnote{Phase locking to a stimulus with a known frequency is typically quantified in either one of two ways, based on spike recordings from a single unit. Most commonly, the \term{synchronization index} or \term{synchronization strength} is estimated by testing for the regularity of the spike timing, as a function of stimulus phase \citep{Goldberg1969}, $R = \sqrt{\left(\sum\limits_{i}^N\sin \varphi_i\right)^2 + \left(\sum\limits_{i}^N\cos \varphi_i\right)^2}/N$, where the spike phase $\varphi_i$ is determined by segmenting the recordings to windows corresponding to the period of the stimulus tone. If the spiking is completely randomly distributed, then $R$ approaches zero (often, 0.1--0.2 is considered completely random). The other method of estimating phase locking is based on the post-stimulus time histograms, which provide a time series of spikes that can disclose possible periodic regularity. The histograms may be Fourier-transformed to obtain estimates of potential non-random spectral components \citep{Joris2004}. \label{SyncIndex}}. Envelope synchronization is distinguished from phase locking to the carrier---both of which are observable in different stimulus conditions \citep{Javel1980}. 

Phase locking is also found throughout the central auditory system \citep{Joris2004}. The high-frequency cutoff within the system progressively deteriorates from the auditory nerve through the brainstem, and midbrain, as the temporal coding is replaced with an average rate coding that dominates the spiking patterns in the thalamus and cortex \citep{Joris2004}. For instance, the ventral cochlear nucleus (VCN) is exceptionally precise in coding the temporal fine structure (even more than the auditory nerve) \citep{Rhode1986ventral}. In contrast, cells of the dorsal cochlear nucleus (DCN) show almost no phase locking to the carrier, despite being sharply tuned spectrally \citep{Rhode1986dorsal}, but they show enhanced synchronization to the envelope in comparison with the auditory nerve \citep{Kim1990,Rhode1994Encoding,Joris1994}. 

A variation of the more general kind of synchronization---mode-locking---was also demonstrated in the auditory system. Mode-locking here refers to exact harmonics of modulation-band frequencies, which were detected as multimodal distributions in the interstimulus intervals of the spiking pattern recordings. The detected peaks had approximately integer ratios, which is indicative of harmonicity \citep{Laudanski2010}. Steady-state mode-locking response was demonstrated in the guinea-pig VCN onset and chopper units using sinusoidal AM tones, synthesized vowels, and harmonic complexes. In a more recent study, \citet{Lerud2014} presented a mode-locking model that could account for 68\% of the frequency-following response (FFR) data variance of two musical intervals in the brainstem by \citet{Lee2009}. The model relies on dynamic nonlinearities in the brainstem, which generate distortion products that can be used for mode-locking between independent channels. Critically, these studies relate a meaning to mode-locking that is different from the one that is used in nonlinear dynamical systems (\citealp{Fletcher1978}; see also \cref{PrimitiveSources} and \cref{SyncBackground}), which locks several nearly-harmonic modes to a harmonic oscillation. In contrast, in the two studies mentioned, the modes refer to frequencies that are already part of the modulation spectrum within the channel, or to difference tones between channels. The absence of the standard mode-locking effect might be gathered from measurements of mistuned complex tones in the chinchilla's cochlear nucleus, where primary-like units followed the carrier and were not reported to mode-lock to a nearby harmonic \citep[e.g.,][]{Sinex2008}. For further treatment of mode-locked neural synchronization patterns involving sensorimotor cortical areas see \citet{Tass1998}.

\subsection{The origin of auditory phase locking}
\label{PLLSource}
It has been occasionally acknowledged that neural phase locking may well originate in the inner ear. For example, \citet{Rose1967} wrote: ``\textit{It seems thus reasonable to assume that events which determine the effectiveness of the cycle take place peripheral to the fiber, possibly in the nerve endings or other structures of the inner ear.}'' \citet{Russell1978} found that the intracellular receptor potential of guinea-pig IHCs consisted of AC and DC components, but that above 4 kHz, the AC component disappeared and the response was dominated by DC. Given that this frequency is also the phase-locking limit, the authors conjectured, in passing, that the phase locking reflects the receptor AC potential of the IHC. In another introduction, \citet{Miller1997} stated that ``\textit{Both the spread of synchrony and the capture phenomenon reflect nonlinear signal processing by the cochlea, in that they are not predictable from a fiber’s tuning properties.}'' In their modeling of auditory nerve responses, \citet{Peterson2020} relate the fact that phase locking does not clip as a function of stimulus level to the mechanoelectrical transduction (MET) nonlinear characteristic response of the IHCs, as well as to their additional low-pass filtering property. Phase locking in the vestibular system has been similarly documented and directly related to the hair bundle deflections \citep{Curthoys2021}: ``\textit{For phase locking of the primary vestibular afferent to occur, the hair bundle of the receptor(s) must be deflected and activated once per cycle...}'' See also Figure 1B in \citet{Felix2018}.

Several studies suggested that the spontaneous discharges in the auditory nerve depend upon the input from the cochlea. In chinchillas whose IHCs were ototoxically lesioned, the spontaneous rates dropped significantly in affected fibers---something that could be explained in more than one way, including direct damage to the IHCs \citep{Wang1997}. Additionally, the spontaneous rates in the auditory nerves of cats were shown to logarithmically depend on the endocochlear potential in the scala media, which modulated the rate of action potentials from the IHCs \citep{Sewell1984}. Interestingly, in direct electric stimulation of the auditory nerve, which bypasses the cochlea and the hair cells, the cat's synchronization index to tones remains significant at least up to 8--10 kHz \citep{Dynes1992}. Indeed, a correspondence between the inner hair cell synchronized intracellular receptor potential and the auditory nerve cells was pointed to in \citet{Weiss1988}. It was suggested that the high-frequency temporal synchrony degradation is a result of a cascade of three low-pass filters between the hair cell receptor and the auditory nerve. See also \citet{Rutherford2021}.

These findings suggest that the spontaneous activity in the auditory nerve is driven by the IHC activity, which is itself irregular, in the absence of any acoustic input. However, spontaneous rate decrease was also observed after de-efferentation of the lateral and medial olivocochlear (LOC and MOC) nerves in the cat \citep{Liberman1990}. While not considered in the original paper, a mechanical effect of the normal OHC spontaneous activity may drive the IHCs, so its absence after the loss of the MOC may have caused the IHCs to move less. A similar idea was explored in a model by \citet{Camalet2000}, where it was proposed that the auditory hearing sensitivity was accomplished by keeping the OHC hair bundle near level-dependent self oscillation, which can explain spontaneous motion as well. Therefore, synchronized activity in the auditory nerve represents a coherent motion of the IHC stereocilia, which may depend to some extent on the OHC oscillation, as weak coupling would entail (see \cref{SyncBackground}). 

An additional review of the effects of OHC impairment and consequent hearing loss on phase locking function is provided in \cref{OHCimpair}. The results are somewhat inconsistent and are not always easy to interpret, although association between normal OHC function and phase locking has been documented in several studies, including some very recent ones. In the following, we explore the possibility that phase locking in the IHCs and auditory nerve is impacted by the OHCs.

\section{The organ of Corti as a PLL}
\label{CortiPLL}
The outer hair cells (OHCs) in the organ of Corti have been a long-standing conundrum in hearing science. Because of their concealed nature and high sensitivity to mechanical insults, direct measurements in the live cochlea are impossible using traditional methods. Additionally, modeling of the OHC dynamics using data gathered in other methods is complicated because the OHCs are deeply embedded in the organ of Corti and are coupled to other structures within it---the reticular lamina, the tectorial membrane, the endocochlear fluid (scala media), connections between the stereocilia via tip links and side connectors, and indirectly to the basilar membrane (BM) through Deiters and pillar cells. Thus, the OHC importance has only started to be unraveled in the last four decades using indirect measures---ever since the discovery of otoacoustic emissions by \citet{Kemp1978}. Some of the most characteristic features of the mammalian hearing have been associated with the OHCs: the amplification of low-level signals, the nonlinear compression of high-level inputs, the sharpening of the auditory filters, the generation of intermodulation distortion products, and two-tone suppression. The OHCs themselves have several unique biomechanical features that may account for these effects in some configurations, although few if any models are unanimously accepted among specialists \citep{Ashmore2010}. 

Despite the slowly accumulating data, there is still a lack of firm association between the organ of Corti and phase locking, which makes the integration of the mechanical and neural auditory segments somewhat conjectural. Nevertheless, we argue in this section that even with the current state of knowledge, the necessary and sufficient components to make a complete PLL circuit are all in place within the organ of Corti and the OHCs, which permits us to treat them as a system.

\subsection{Identifying the PLL components}
\label{PLLcomponents}
A PLL should have at least two elements---a phase detector and an oscillator---and preferably three---including a loop filter between the other two. These elements must be connected with a feedback loop in order to work as a PLL. The forward gain of the entire module may be larger than unity with no loss of generality, as long as it is stable. Below, we attempt to associate these functions with the known physiology of the organ of Corti, and in particular, the OHCs. 

\subsubsection{The phase detector}
A phase detector is a nonlinear device that produces an output that is proportional to the difference between two inputs. In audio it is referred to as intermodulation distortion, which translates to difference tones, in the case of two pure-tone inputs. In general, the nonlinearity also produces summation tones, which should be removed by the low-pass filter. It is well-known that the ear produces combination tones of frequencies $mf_1 \pm nf_2 >0$, for two closely-spaced pure tones $f_1$ and $f_2$ and positive integers $m$ and $n$, with notable difference components that are related to cubic distortion (i.e., $2f_2-f_1$) \citep{Goldstein1967}. These tones are psychoacoustically audible and are associated with \term{distortion-product otoacoustic emission} (DPOAE) that is measurable in the ear canal \citep{Kim1980}, which appears to cover the entire audio range, as it has been shown to exist in bats at least up to 95 kHz \citep{Kossl1992}. The quadratic component $f_2 \pm f_1$ is present too, but it is less dominant in the DPOAE and BM spectra (outside of the organ of Corti) than in the reticular lamina (inside the organ of Corti) \citep{Ren2020}. Further, it is more dominant than the cubic component at the level of the IC than in the cochlea \citep{Arnold1998}. The essential role of intermodulation distortion (mainly the quadratic component) in hearing was recently proposed by \citet{Nuttall2018}, as a processing stage that extracts the amplitude envelope of the signal. 

While there are several nonlinearities associated with the ear, most of them do not contribute to its intermodulation distortion \citep{Avan2013}. With high confidence, the source for these distortion products is the transduction process of the OHC hair bundle deflections to ionic current through the MET channels. There are two interdependent causes for this nonlinearity, referred to as \term{gating compliance}. First, the elasticity of the hair bundles is asymmetrical with respect to the resting position and, additionally, at small displacement amplitudes its stiffness is not constant \citep{Jaramillo1993}. Second, the potassium ion (K$^+$) current is nonlinearly dependent on the deflection amplitude and has a sigmoid-like characteristic transfer function \citep{Avan2013}. These distorting nonlinearities depend on the joint deflection of adjacent stereocilia that is achieved by horizontal connectors. The connectors are positioned within and across rows of stereocilia and they also connect the tallest row to the tectorial membrane\footnote{This degree of embedding should be seen as the minimum, as likely it is much higher. It is based on the classically established interfacing of the OHC and the tectorial membrane. As was noted in \cref{TheInnerEar}, a recent study in guinea pigs demonstrated that the OHCs are fully embedded in the tectorial membrane in the intact organ of Corti \citep{Hakizimana2021}.}. When these connectors are missing, as was the case in a special strain of mutant mice, most distortion products disappear \citep{Verpy2008}. These features ensure that the distortion is prominent also at low levels---something that would be unattainable with a static nonlinearity \citep[supporting information]{Barral2012}. As the locus of nonlinear behavior is in the hair bundle, the distortion products are considerably stronger in the reticular lamina than in the BM \citep{Ren2020,HeRen2021}. 

The $f_2-f_1$ (quadratic) and $2f_2-f_1$ (cubic) distortion components have different properties, somewhat depending on where they are measured within the auditory system. Their relative intensity depends on the symmetry of the operating point of the nonlinear MET transfer function, which can be biased with DC current, low-frequency tones, or OHC motility blockers \citep{Frank1996,Brown2009}. Additionally, the DPOAE level of the quadratic but not cubic distortion products changes over time in response to ipsilateral and contralateral tones and broadband noise, which suggests a modulatory role of the olivocochlear efferent system (\citealp{Brown1988,Kirk1993}; but see \citealp{Kujawa1995}). Furthermore, only the quadratic distortion product significantly interacts with low-frequency amplitude modulation, and its strength and phase are affected by the activation of the contralateral reflex \citep{Abel2009}. Another difference was noted when the distortion products were measured using electrodes in the IC of the awake chinchilla---the quadratic component was up to an order of magnitude stronger than the cubic component \citep{Arnold1998}. Also, it was shown to depend only on the difference between the primaries, to have a low-pass response for small difference products ($f_2-f_1 < 100$ Hz), and to monotonically increase with level. 

It can be concluded that the hair bundle produces the required distortion that can make the necessary output of a phase detector. However, most studies tested the case where two or more external inputs produced a measurable distortion either in the reticular lamina or the basilar membrane. In the context of a PLL, one of the inputs should be the internal oscillator signal. This may be difficult to measure with steady-state pure-tone measurements that are likely to cause the PLL to lock almost instantly, which entails a negligible or DC phase error signal---the very output that is expected from a phase detector. 

\subsubsection{The loop filter}
The role of the loop filter of the PLL is to set the dynamics of the feedback loop, in addition to the removal of high-frequency components from the output of the phase detector. While the filter is not essential for the operation of a (first-order) PLL, an unfiltered design is rather limited in function and not used much in practice. 

If we consider the MET channels of the hair bundle to be the ``epicenter'' of the phase detector, it is natural to look for a filter within the soma of the OHC, where the ionic current flows, as a function of the hair bundle deflection. In fact, the low-pass response of the cell membrane has been a notorious stumbling block in cochlear amplification modeling---the so-called ``\term{RC time constant problem}'', which is caused by the low-pass characteristics (resistance-capacitance, RC) of the receptor potential of the cell membrane  \citep{Ashmore1987,Housley1992,Santos1992}. The low-pass filter characteristic is consistently found in vitro (see \citealp{Santos2019Iwasa} for a summary of relevant studies) and was recently shown to be the case in vivo as well (in gerbils), with cutoff frequencies lower than 3 kHz for basal and lower for apical CFs \citep{Vavakou2019}. Even in apical channels, the cutoff frequency is well below the characteristic frequency (CF) \citep[Figure 4D]{Vavakou2019}. In contrast, in a more recent in-vivo study in mice, the low-pass filtering effect was confirmed, but appeared relatively small and insufficient to counter the high-frequency cycle-by-cycle amplification generated by somatic motility, as measured in the organ of Corti (most strongly in the displacement of the tectorial membrane; \citealp{Dewey2021}). Moreover, its effect in the guinea-pig OHCs in vitro did not produce significant reduction in amplification up to 80 kHz \citep{Santos2023}. 

This filter was studied in the context of cochlear amplification, as it may be seen as a hindrance in several cochlear amplifier models, since it prevents them from using high-frequency somatic motility in a straightforward manner as can be observed in vitro \citep{Frank1999,Ashmore2008,Santos2019Review}. Without high-frequency response, it is unlikely that the electromotile force can be efficiently used in cochlear amplification over a broad frequency range. Models often try to explain this low-pass filtering away by including alternative electrical pathways, such as extracellular potential or endolymphatic ionic currents in the OHCs \citep[e.g.,][]{Dallos1995,Johnson2011}, which also received some support in recent in-vivo measurements in mutant mice \citep{Levic2022}. An alternative solution to the low-pass problem was recently proposed by \citet{Rabbitt2020}, who showed that, because of various nonlinear effects, the membrane capacitance of the OHCs is both highly nonlinear and complex. These properties were shown, in vitro, to account for both the low-pass behavior and the full-bandwidth, which is thought to be responding to the stimulus on a cycle-by-cycle basis, as in \citet{Frank1999}. Another alternative analysis by \citet{Iwasa2017} suggested that under some mechanical load conditions, effective negative capacitance can mitigate some of the low-pass characteristics of the membrane. Finally, a simpler analysis using a linearized analog circuit model of the OHC amplifier convincingly argued that given the known estimates of the RC filter values, it poses no serious problem of amplification at high frequencies, as it still produces the correct OHC elongation that is known to be required for adequate gain \citep{Altoe2023}. At present, one available in-vivo study suggests that the filter does have a detrimental effect on high-frequency amplification \citep{Vavakou2019}, another study suggests that it does not under some conditions \citep{Dewey2021}, and a third one suggests that different mechanisms may be responsible for the low and high frequency responses, so that the effect of the filter itself is limited \citep{Levic2022}.

In the case of a PLL, the low-pass filter is a desirable feature, since it can serve as the loop filter. Some measurements established low-pass characteristics that were conveniently fitted with a two-pole low-pass filter \citep{Ashmore1987}, or rearranged as a four-pole low-pass transfer function with all but one pole well above the effective passband \citep{Santos2019Iwasa}. These would imply, respectively, a second-order PLL with three spurious poles, or an inconvenient fifth-order PLL if all poles count. Either way, the existence of this filter implies that the low-frequency distortion products can be used for the PLL function and may have good tracking capability, but may be relatively susceptible to noise, due to its relatively wide bandwidth.

Note that beneficial roles of the RC filter were argued recently. \citet{Altoe2023} showed that such a filter can improve the signal-to-noise ratio at the output, in the condition of incoherent noise and coherent signal. Along with model data from \citet{Peterson2020}, it additionally supports the idea that such a filter is beneficial in removing undesirable high-frequency harmonic distortion components. 

\subsubsection{The local oscillator}
\label{LocalOsc}
The next step is to identify an oscillator embedded in the organ of Corti that can be synchronized with an external stimulus. The oscillator has to have two important features. The first feature is that it should be free-running, so to produce periodic output even in the absence of reference input (i.e., a stimulus). The second feature is that it must be tunable using slow error-correction inputs from the phase detector. 

The most obvious candidate for the oscillator resides in the OHCs, which are typically identified as the source that generates the spontaneous otoacoustic emission (SOAE)---a feature of an active component in the system, perhaps an amplifier. Thomas Gold attempted to measure such emissions, yet was unsuccessful, probably due to inadequate measurement technology at the time \citep{KempManley2008}. \citet{Glanville1971} may have been the first to report auditory sound production from a whole family that had emitted sounds from their ears that were externally audible, interpreted as ``\term{objective tinnitus}'', whereas \citet{Wilson1980KempEchos} recorded ``\term{subjective tinnitus}'' emitted from the ear. SOAE was more formally discovered by \citet{Kemp1979} and confirmed shortly after by \citet{Zurek1981}. When it is measurable, the emitted tones vary in level, number of components, and frequency between individuals and animals \citep{LonsburyManley2008}. Tests using ototoxic drugs, as well as the absence of SOAE in hearing-impaired people suggest that SOAE must come from the cochlea \citep{Probst1991}. It was phenomenologically shown early on that the SOAE can be the result of active filtering that contains an oscillator and a positive feedback loop, which can also exhibit the typical sharp response and phase locking of the ear \citep{Bialek1984}. Various tests localized the source of the oscillations at the OHCs, as contralateral stimuli were shown to affect the frequency and amplitude of the SOAE, likely as a result of the olivocochlear efferent bundle \citep{Mott1989}. The emissions were also shown to stem from a nonlinear amplifying mechanism as reflectance and impedance measurements in the ear canals of people with strong SOAEs showed net gain in the reflected energy compared to the incident energy \citep{Burns1998}\footnote{An alternative model to SOAE formation, which has been quite successful in accounting for empirical data, hypothesizes that the cochlea works as a cavity that supports standing waves. The resonant modes of the cavity are determined by irregularities in the geometrical arrangement and mechanical parameters of the OHCs in the organ of Corti \citep{Shera1999,Shera2003SOAE}. Such a system is able to sustain its modes because it has an amplifying medium with active elements, which give rise to what may appear as local oscillators, but is in fact an emergent global oscillatory behavior with very sharp modes. Even this model, however, accepts that there are active cellular elements that perform the amplification and may function as local oscillators, though they cannot be directly tapped by external measurements. See also a recent review of SOAE oscillator models by \citet{Wit2024}. The present theory is agnostic to the precise external manifestation of the oscillator, although such a global resonance system is somewhat more awkward to integrate in light of some of the empirical data that is brought below to substantiate the PLL model.}. 

Animal data provide some evidence for the specific source of emissions from within the OHCs. In a series of in-vitro experiments on the bullfrog's hair cell, the hair bundle movements were identified as a source of spontaneous action \citep{Martin1999,Martin2001,Martin2003}. The bullfrog's hair bundles in vitro were also found to synchronize with one another, when they were coupled together with an elastic thread, simulating the otolithic membrane that is homologous to the tectorial membrane \citep{Zhang2015}. The combined power of the synchronized cells could potentially give rise to SOAE, if the same mechanism is at play in the mammalian cochlea, which seems to be the case. 

Additionally, the hair bundle in the bobtail skink lizard was identified to be the source of amplification, which is thought to be also the source of SOAE \citep{ManleyKirk2001}. This lizard has a unique cochlear anatomy with hair cells lined up on both sides of the auditory papilla---the analog organ of the basilar membrane. In this in-vivo study, electrically evoked OAE (EEOAE)---known in mammals to produce acoustic emissions that can be detected externally \citep{Zheng2006}---were produced by injecting a low AC electric current to the scala media, which was then acoustically modulated at a lower frequency. A mechanical geometrical argument was invoked to predict whether hair bundle motility or lateral membrane motor movement produce either one of two fundamentally different emission patterns. If the modulation causes radial motion of the hair bundles that lie opposite to one another, it should produce beating due to destructive interference between adjacent bundles. Alternatively, vertical motion due to lateral membrane movement of the hair cells should produce in-phase motion due to constructive interference. The EEOAEs measured at the ear canal showed unmistakable beating, leading to the conclusion that the hair bundle is the source of motility (and hence amplification, according to the authors). A direct comparison to mammalian OHCs is impossible, unfortunately, but \citet{Manley2000b} suggested that since all nonlinear effects (including amplification) in mammals are found also in other vertebrates, then the cochlear amplification in the various vertebrates may relate to a common ancestor. This may further suggest that the hair-bundle mechanism for the SOAE has not changed significantly between animals, as long as the connection between the SOAE and amplification is also stable between all species. Nevertheless, a definitive proof for the connection between the hair bundle and SOAE in mammals is still lacking.

The second feature of the PLL oscillator---of responding to slow error signals---is the one that enables control and error correction of the PLL output. In the PLL, the phase detector produces a signal that is proportional to the instantaneous phase difference between the oscillator and the reference. When the PLL is locked in, the phase difference is zero, or more realistically, it fluctuates around zero. If it is out of lock, then the difference is a low-frequency error signal (of the order of the difference between the local and input instantaneous frequencies). Therefore, we would like to find out whether the spontaneous oscillations can be tuned or biased using a low-frequency signal. There are a few studies that suggest that this is indeed the case, based on the original observation that SOAEs can be entrained to external tones---usually for emissions between 1 and 2 kHz \citep{Wilson1980Evidence,Wilson1981Tin}. For example, \citet{Bian2008} found that in the presence of low-frequency tones (25--100 Hz) the frequency of prominent SOAE peaks increased, although by a smaller frequency interval than the tones. The effect depends on the tone level as well and is generally nonlinear, as at very high levels it can cause suppression of the SOAE. The effect on the SOAE may be prolonged after the offset of a long-term exposure to the tone \citep{Drexl2016}. Similarly, low frequency tones also amplitude-modulate the DPOAEs, where the modulation depth is higher for quadratic than for cubic distortion products, despite their lower absolute levels \citep{Drexl2012}. This effect is interpreted as a modulation of the operating point of the MET nonlinear transfer function. 

The SOAE response to low-frequency tones can be directly attributed to the unusual response that the OHCs have to infrasound that is orders of magnitude more sensitive than that of the IHCs, which ultimately determines the audio threshold \citep{Hensel2007,Salt2010}. There are three primary reasons for this peculiar low-frequency response (starting below 1000 Hz and reaching 150 dB attenuation at 1 Hz for the IHC input; \citealp[Figure 3B]{Salt2010}): strong attenuation by the middle ear (6 dB/oct, below 1000 Hz), shunting by the helicotrema (6 dB/oct, below 100 Hz), and the fact that at low frequencies the OHCs respond to displacement, whereas the IHCs respond to velocity changes (6 dB/oct below 500 Hz) \citep[e.g.,][]{Dallos1972,Russell1983}. The strong attenuation of infrasound by the middle ear and helicotrema suggests that external tones have to be very loud in order to cause any discernible changes in local oscillations. However, this may not be the case for low tones produced internally through nonlinear distortion. For example, the same 30 Hz from \citet{Drexl2018VanDijk} could have been presented inside the cochlea at 60 dB SPL instead of 120 dB SPL outside of it to obtain the same effect. Presumably, spatially localized low-frequency distortion products can be of much lower level and have a more localized effect if they are generated directly at the level of the reticular lamina. The special propensity of infrasound to elicit auditory synchronization has been recently demonstrated using infrasound tones (8 Hz) at threshold level and above, as can be learned from its frequency-following response (FFR) even in the absence of any audible sound \citet{Jurado2023}.

~\\

All these effects point to a sensitivity of the OHCs to low-frequency biasing and modulation, which interacts with both the putative oscillator and the phase detector. However, injecting the system with such low-frequency inputs is unlikely to be exactly equivalent to a generation of a similar distortion product (phase error signal) within the organ of Corti in the presence of an external stimulus. Therefore, establishing a phase-correction effect within an auditory PLL is likely to require more delicate experimentation in which the complete loop is in operation. If this system indeed works as a PLL, then it is not far-fetched to hypothesize that it has evolved to isolate against external low-frequency noise as much as possible, in order to eliminate spurious biasing of the PLL feedback loop, as well as to maintain a biologically useful crossover between vestibular and acoustic frequencies \citep{Lewis1992Webster}. 

\subsection{Putting together the auditory PLL}
\label{PuttingPLL}
Now that the three PLL components have been identified, we can put them together to make a full circuit. What was omitted from the account above, though, is the somatic motility of the OHCs, which receives the low-frequency error signal as a voltage input from the phase detector and should move accordingly \citep{Evans1993}. If the somatic motility amplifies the error signal, then it effectively injects additional power to the open-loop path, which can be taken as gain along with the low-pass filtering. This interpretation may be supported by a recent in-vivo measurement of quadratic distortion products in the gerbil, which showed that the maximum vibration amplitude is not sharply tuned and is confined to a relatively small ``hotspot'' that includes the OHCs and the Deiters' cells \citep{Cooper2018}. If it is accepted that the source of the distortion product is the MET channels, then these findings suggest that the phase error is amplified between the hair bundle and the Deiters' cells---along the OHC soma\footnote{Note that this is not the interpretation given by \citet{Cooper2018}, who did not discuss the actual source of distortion. The authors considered the gradient of the distortion product level to be aligned with the forward sound path, from the basilar membrane to the reticular lamina, instead of a reverse effect that we propose here. However, the reticular lamina response always showed a phase lead relative to the basilar membrane, which supports our interpretation.}. The idea that somatic motility amplifies the distortion product may also be deduced from findings by \citet{Jia2005}, who showed in cochlear preparations of gerbils and prestin-knock-out mice that OHC electromotily is necessary for voltage-induced hair bundle motility at medium or large amplitudes. Only small amplitudes could be supported by direct electrical stimulation of the hair bundle when electromotility was inactive. However, ultra-high-frequency ($>$ 40 kHz) hearing was nevertheless present in prestin-knock-out mice \citep{Li2022}.

\begin{figure}
		\centering
		\includegraphics[width=0.8\linewidth]{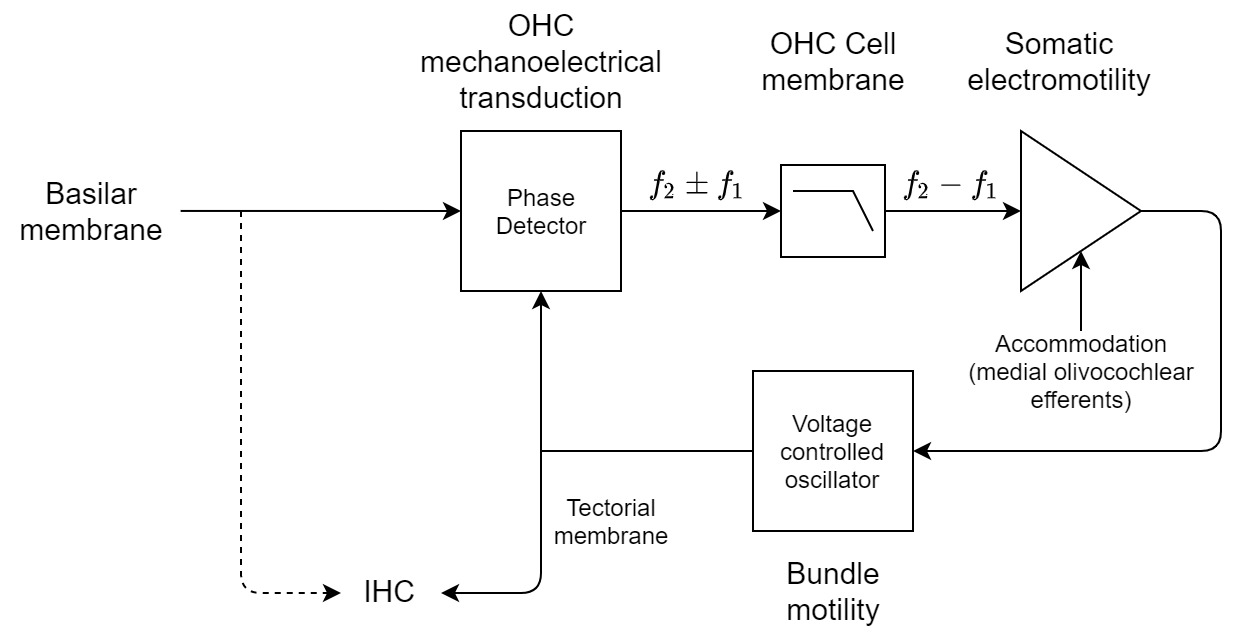}	
		\caption{The auditory PLL with the three usual components of phase detector, low-pass filter, and oscillator, along with an amplification that probably functions as loop gain. The dashed arrow can function as the passive path for the signal, which goes directly to the IHCs and can be dominant at very low or very high levels, when the PLL does not lock, or when there is damage to the OHCs. Otherwise, the output from the PLL and the OHCs is coupled to the IHCs, most likely through the tectorial membrane. An additional control input was added to the somatic motor due to the olivocochlear efferent bundle to the OHCs, which is discussed in \cref{MOCR}. }
		\label{auditoryPLL}
\end{figure}

The feedback loop of the cochlear PLL may work as follows. The hair bundle moves with spontaneous frequency in the vicinity of the resonance of the BM, which is not necessarily detected in otoacoustic recordings. At the CF resonance, the signal carried with the traveling wave causes shear forces to deflect the OHC hair bundles, which mix with the local spontaneous oscillation. The spontaneous and incoming signals are multiplied as a result of the MET channel nonlinearity, but only the low-frequency products pass through as ionic current into the OHC soma. The current causes respective electromotile contractions and elongations of the soma, which in turn feed the local oscillation of the hair bundle. As the local oscillation becomes synchronized to the external signal, the distortion product gets closer to zero, and the somatic motility becomes minimal. The oscillator itself is coupled to the IHCs (probably through the tectorial membrane; \citealp{Hakizimana2021}) which transduce the phase-locked output to neural action potentials. See Figure \ref{auditoryPLL} for a corresponding diagram of the PLL. While all the components are found within one OHC, it is fully embedded in the organ of Corti, which provides the right amount of coupling to the BM vibrations, as well as isolation between the different parts of the system---mainly below and above the reticular lamina. Note that the entire system is bandlimited by virtue of the passive cochlear filter. 

A positive feedback design in the organ of Corti has been hypothesized many times starting from \citet{Gold1948}, but with amplification as its foremost goal (e.g., \citealp{Kemp1980Towards,Davis1983,Bialek1984,Kemp1986,Patuzzi1989,Dallos1992,Yates1992,Robles2001,Lu2006,Bell2007,Ashmore2011,Avan2013} and Fettiplace in \citealp{Ashmore2010}). The PLL uses a negative feedback loop, which is nevertheless susceptible to instability because of the gain in the low-pass filter stage and the delay that the loop incurs, which can turn the feedback to positive under some conditions. The feedback loop described above has the same route as the one described in \citet[Figure 4]{Avan2013}, although with somewhat different details that need not be contradictory. The other feedback mechanisms cited above involve returning energy back to the traveling wave in the BM as part of the loop, sometimes with the explicit requirement for a cycle-by-cycle amplification \citep[e.g.][]{Dallos1992}, which is yet to be demonstrated in vivo \citep{VanHeijden2015,Ren2016Reticular,He2018}. Instead, our PLL model makes this particular design requirement obsolete at phase locking frequencies. 

\subsection{Interim discussion}
While the above arguments do not constitute a direct empirical proof for the existence of a PLL in the cochlea, they entail a reorganization of known mechanisms of the OHCs that suggests a novel solution to several puzzles. These include the difficulty to establish a valid model for the cochlear amplifier, partly due to the time-constant problem, as well as a clearer delineation of the roles of the hair bundle and somatic electromotility (for the extent of these controversies, see for example the transcripts of discussions in \citealp[pp. 468--482]{Cooper2009}; see also \citealp{Ashmore2010}). However, the regenerative amplifying nature of the system makes the ear a nonstandard design as far as PLL theory goes. Here we have a set of oscillators that are running at very low levels, may provide more gain than typical PLLs in lock, and possibly retain some inertial oscillations after the input signal terminates. Such a circuit may require a more refined modeling effort if proven correct. Most likely, many of the insights obtained by nonlinear dynamic models that have been applied to the OHCs \citep{Hudspeth2014} can be reframed as part of a PLL design, although it is not certain that approaching the problem from its nonlinear dynamics is the simplest approach to the problem. 

There are additional hints that suggest that the PLL design is probably incomplete in its own right. For example, the various extracellular and perilymphatic currents may still have roles in auditory signal processing that are not captured in the PLL model, or that provide access for dynamic biasing of some of the PLL parameters. There are additional operating point changes and efferent connections to the OHCs that suggest a complex function \citep[e.g.,][]{Fettiplace2017}. Finally, as is theorized in \cref{lenscurve}, the OHCs produce modulatory changes on the basilar membrane stiffness, which impacts the velocity of the traveling wave that seems to result in a phase modulation (time-lens). These stiffness changes are dependent on the same somatic motility as is thought to be oscillating with the phase detector output, which may not correspond to the modulation required for stiffness modulation through the Deiters cells. Therefore, at present it is not clear how the PLL and time lensing functions coalesce and whether they interact (but see \cref{MOCR}).

Other critical issues about the putative oscillator remain unanswered. Are different oscillators shared between channels? Can every inner hair cell have its own local oscillator? Or are there several dominant oscillators that serve multiple channels in their vicinity? Is there any substantive difference in oscillation capability between apical and basal sites? Is there any function for basal high-frequency circuits where they cannot provide phase locking to the carrier frequencies? Are the multiple OHCs in every row perfectly synchronized? 

\section{Corollaries to the PLL}
\label{Corollaries}
In this section we return to what has originally motivated us to search for an auditory PLL: the conservation of coherence between the mechanical and neural domains. However, coherence conservation cannot be unconditional and is dependent on the specifics of the PLL operation and its ability to lock in to arbitrary signals. While we do not have quantitative details about the actual operation of the putative PLL, it will be insightful to consider some of its properties that are central to its operation. Definitive answers to many of the questions that are raised through the analysis below will have to wait for careful testing of the theory in the future. Nevertheless, a few attempts will be made to provide qualitative bounds on the PLL behavior that follow from theory and from diverse empirical findings. 

\subsection{What signals can acquire lock?} 
PLLs require coherent inputs to lock onto signals with deterministic phase. Pure tones are obviously coherent, as are different kinds of modulated pure-tone carriers. In contrast, true random Gaussian noise is incoherent by definition and does not contain any predictable phase structure to lock onto (but see \cref{PLLNoise} for some qualifications). Most realistic signals fall somewhere in between---they are partially coherent---so they can be locked onto with some error that corresponds to their incoherent part (e.g., Eq. \ref{totalcoherence}). 

Synchronization to more complex signals than pure tones has been demonstrated as well. Phase locking in the cat was demonstrated with upward and downward linear FM ramp with slopes of up to 15.7 kHz/s, as long as the instantaneous frequency was within the channel's bandwidth \citep{Sinex1981}. Also, sinusoidally frequency-modulated tones are phase-locked at low carrier frequencies ($<$1--2 kHz) and low modulation frequencies in the ventral cochlear nucleus of the guinea pig \citep{Paraouty2018}. Synchronization to synthetic speech vowels has also been demonstrated a few times. For example, in cats, formant components dominated the synchronized response, along with distortion components, as some other frequencies were suppressed \citep{Young1979}. However, the synchronization strength varied significantly across the spectrum and did not necessarily achieve the full strength as pure tones would, maybe because of suppression. The synchronization strength was relatively robust down to a SNR of 9 dB. Additional evidence for the negative effect of noise on auditory nerve (steady-state) phase locking was shown to be stronger with poorer SNR and broader noise bandwidth in squirrel monkeys \citep{Rhode1978}. Similar findings were shown in the tree frog \citep{Narins1989} and to a lesser degree in the bullfrog \citep{Freedman1988}. Phase noise ``cleaning'' and high robustness to noise are not uncommon in electronic PLL designs---a feature that makes them appear as sharp narrowband filters. 

Lock acquisition is more difficult to establish based on FM psychoacoustic data, which suggest that temporal processing may be effective for very slow modulation rates, but probably not so much for fast rates (\cref{AuditoryPhaseReview}). In a study that tested discrimination between different glide shapes that connected two frequencies over a fixed duration (i.e., with different curves of instantaneous frequency), the average performance of listeners was not very sensitive and detection could be explained mainly using a place of excitation model \citep{Thyer2006}. These results, as well as those reviewed in \cref{OutofLockChanges} and the example provided in \cref{FurtherPsycho} may suggest that the auditory system capability to lock onto frequency glides is limited. 

\subsection{Is there any spurious or residual phase locking even if the PLL receives incoherent signals?} 
\label{PLLNoise}
Synchronization to an input can take place when it has a well-defined phase function over a narrow frequency band, so that any ambiguity about the resultant output can be avoided. The narrowband constrains the range of instantaneous phase changes that are permitted around the linear phase progression of the center frequency. Thus, the input signal provides some predictability for a PLL that reacts to the input within a fraction of a period without getting out of lock. Broadband noise signals are, by definition, random, which means that their phase is also random and cannot be predicted. Still, even a random carrier may carry low-frequency information in the envelope domain that is slowly varying and nonrandom. In hearing research, synchronization to the envelope is also referred to as ``phase locking'', occasionally, which can be confusing, because it fails to imply that different detection processes may be employed to lock to the carrier and to the envelope. Synchronization to the envelope can be achieved using noncoherent detection, without resorting to carrier phase locking, which occurs on a much shorter time scale. 

There has been significant research about auditory phase-locking to broadband noise---both continuous and clicks (for an exhaustive bibliography of broadband studies see \citealp{HeilPeterson2017}). At the heart of the studies that tested phase locking to noise is the assumption that stationary noise can reveal the inner workings of the linear and static nonlinear auditory processing, as the effects of any dynamic nonlinearities (e.g., spike generation, refractory period) are relatively minor, or can be circumvented in analysis \citep{deBoer1978,Eggermont1993}. A consistent finding is that the auditory nerve locks onto the characteristic frequency of the channel, on average, which reflects the auditory filter resonance \citep[e.g,][]{Ruggero1973,Louage2004}. The main explanation for this apparent contradiction (of locking to noise) is that the phase-locked response to broadband noise is dominated by low-frequency components that represent slow modulations to the characteristic frequency carrier of the auditory channel \citep[e.g.,][]{HeilPeterson2017}. It can be seen, though, that relatively high frequencies are not tracked and the phase locking better represents the envelope of the filtered signal \citep[e.g.,][Figures 9C-D, 10C-D]{Moller1983}. The extent to which the channel coheres the random noise can be gathered from measurements that display quantities that are similar to coherence time, such as the cross-correlograms in \citet[Figures 1 and 3]{Moller1983} and the ``Difcor\footnote{The Difcor is derived from the shuffled correlogram---itself a variation of cross-correlation that was adopted to spike trains \citep{Joris2003}. The Difcor is suitable to test synchronization to the temporal fine structure of the signal \citep{Louage2004}.}'' half-width in \citet[Figure 7]{Louage2004}. Typically, both measures are less than 1 ms long and get gradually shorter as the characteristic frequency increases. These values can be compared with simple calculations of bandpass-filtered white noise in Figure \ref{filtertype}, which produce somewhat longer coherence times, depending on the filter type, order, and center frequency. 

In light of the present auditory PLL theory, the underlying assumptions and interpretation of the broadband phase-locking studies are unsatisfactory, for two reasons. First, the assumption that the auditory channel can be modeled using a static nonlinearity is wrong, given an active phase-locking device, be it a PLL or other. A PLL would appear linear in steady-state only, but its transient response is dynamically nonlinear and not static---it depends on the initial conditions of the input and on the previous state of the system (i.e., whether it is already in lock)---it is not memoryless. Furthermore, the transient response to coherent signals is degraded by random noise, which is the very signal that is being tested. 

The second reason is that phase locking fundamentally depends on the degree of coherence of the input signal, which is used as reference phase. As the input is initially bandpass-filtered by the passive cochlear mechanics, the input to the PLL is more coherent than the original stimulus. Therefore, even random broadband noise that is completely incoherent at the input can acquire some residual coherence once it goes through a bandpass filter, depending on its bandwidth (\citealp{Reddy1979,Jacobsen1987}, \cref{WhiteCoherence} and \cref{CoherentFiltering}). This is the general effect we derived of the increased temporal coherence between the output and the input of a bandpass filter, which depends on its bandwidth only. While the auditory channels are relatively sharper at high frequencies than they are at low frequencies (e.g., \citealp{Glasberg1990,Shera2002}; Figure \ref{Q10}, right), they are much narrower at low frequencies in terms of their absolute bandwidth---the primary parameter of coherence time (\cref{CoherenceTimeLength})---and hence appear to be more coherent. The low-frequency components that are extracted from the spectrum \citep{HeilPeterson2017} exhibit spurious phase structure primarily as a result of the broadband input becoming partially coherent. To properly track broadband noise, a PLL must have instantaneous lock-in time and large hold-in range that can accommodate arbitrary phase modulation within its bandpass---unphysical requirements. Even if spurious low-frequency components in the filtered noise can be captured by the PLL's lock, they tend to vanish quickly in real stochastic noise, so the PLL would get in and out of lock intermittently. We will revisit this topic in the context of the psychoacoustic response to narrowband noise modulation input (\cref{NarrowTMTF}).

In conclusion, in experiments that employ noise-only stimuli, the noise serves as an energy source that activates the channel---including any oscillators in its range. The actual phase of the noise is immaterial, except for instantaneous low-frequency phase trajectories that are formed by the cohering passive bandpass filter. In the communication engineering framework, this situation maps to noncoherent detection, as amplitude modulation data can still be extracted from such an input signal without its carrier phase. Broadband synchronization data do not represent true phase locking as is understood to be possible below 4 kHz, but rather it relates to very short-term tracking that is at most of the order of the coherence time of the filtered signal and directly depends on the filter bandwidth. It is therefore not clear whether the PLL module has any role in achieving this slow synchronization. 

Despite the reservations about using broadband noise in synchronization measurements, a great deal of information can be gathered from such data, as long as care is taken in interpreting them. Several such studies will be cited throughout the text.

\subsection{What is the approximate pull-in time?} 
\label{LockInPLL}
Phase locking is achieved within a finite duration. Since the knowledge about phase locking is based on discrete neural spike patterns, usually measured in steady-state, it may be difficult to estimate the pull-in time of the PLL, especially since it may co-occur with neural transient phenomena, such as refractoriness and adaptation. Note that we do not distinguish here between the pull-in and the lock-in times, since we miss the necessary details about the PLL architecture to tell them apart. However, the pull-in time is longer than the lock-in time and may be easier to observe in the right conditions.

Auditory nerve adaptation to 300 ms tone bursts in the gerbil was effectively modeled using two frequency-dependent time constants: one that describes the \textbf{rapid} changes after onset and is on the order of 0.5--4 ms, and another \textbf{short-term} time constant that is an order of magnitude longer and characterizes the discharge pattern before the \textbf{steady-state} response \citep{Westerman1984}. During rapid adaptation, the spike rate is much higher than in the short and steady-state stages and is sensitive to intensity changes. The time constants decrease with frequency. \citet{Rhode1985} measured the auditory nerve responses to 25 ms tone pips in cats, which enabled better resolution of the rapid stage of the adaptation that was largely in agreement with the gerbil data. Phase synchronization was not reported quantitatively, but from the peristimulus time histograms, it appears almost instantly with relatively little jitter. 

In the chicken, refractoriness has been shown to improve the synchronization and entrainment of 40 ms low-frequency tone bursts in the auditory nerve, whenever the period was larger than the refractory period \citep{Avissar2013}. Additionally, the jitter of single-unit responses to the same tone bursts was about equal during the 10 ms onset and offset duration of the stimulus, which suggests uniform phaselock precision throughout \citep{Avissar2007}. However, the first 2.5 ms of the signal onset (a ramp) was excluded from the analysis, so the full transient effect may have been lost. Finally, measurements of the cat's auditory nerve show decreased synchronization and entrainment during the first 5--8 ms from the onset of the low-frequency stimulus---a duration that can count as part of the adaptation period \citep[Figures 8A and 8B]{HeilPeterson2017}. 

In addition to direct physiological effects, it is plausible that the locking-in stage has perceptual correlates as well. For example, a tone burst sounds like a click without tonality if it is extremely short, below the click-pitch threshold.  \citet{Doughty1947} found that for tone bursts presented at 110 dB SPL this threshold is shorter than 4.5 ms for frequencies above 1000 Hz and decreased slightly up to 8000 Hz (4 ms)\footnote{The first such experiments were reported by \citet[pp. 266--268]{Mach}, who found much longer thresholds for pitch (20--30 ms at 128 Hz), using a rather crude method.}. Thresholds were higher for lower levels and frequencies below 1000 Hz. Somewhat longer tonal threshold were found as a function of the number of periods at 60 dB by \citet{Mark1990}. Shorter thresholds for tonality were uncovered by \citet{Mohlin2011}, who used much improved methods to control for loudness and introduced proper attack and decay to the envelopes to avoid extra peaks in the spectrum. While the definition used was not exactly identical to the click-pitch threshold by \citet{Doughty1947}, the just-audible tonality threshold values were as low as 2.6 ms for 8000 Hz bursts and were often below 3 ms for frequencies above 3000 Hz and up to 20--23 ms at 150 Hz. Despite these results, we do not know if the pitch perception corresponds to a locking in period in the periphery, or if it depends on a critical time constant that is more central. 

The above evidence may be indirectly indicative of the existence of finite but short PLL lock-in time for tones---on the order of 2.5--10 ms, or longer at low frequencies and levels. This duration overlaps with the rapid adaptation stage, although the exact interplay between it and psychoacoustic correlates has to be further explored. 

\subsection{What is the pull-in (capture) frequency range for a given\\auditory channel?}
In principle, because of the presumable multiplicity of OHC oscillators, the totality of the cochlea may achieve simultaneous phaselock in multiple channels, subject to the effects of suppression. Within a single channel, the input to the PLL is bandlimited by the cochlear filter before any phase locking takes place. When the synchronization of a single auditory-nerve unit is measured as a function of frequency, it overlaps with the range of its tuning curve only for low-frequency CFs, but at high frequencies (above 1 kHz) phase locking diminishes quickly and the response no longer reflects the filter. For example, in the guinea-pig, it was shown that for a fiber with CF of 600 Hz, the synchronization is nearly constant for a range of 200--1000 Hz with no clear peak, while for CFs of 2 and 3.5 kHz, the tuning curves were not predictive of synchronization, which was limited to low frequencies only \citep[Figure 2]{Palmer1986}. Similarly, in measurements of the chinchilla's auditory nerve of both low and high CFs, apical (average driven rate was maximum at CF = 413 Hz) synchronization strength was nearly independent of frequency at 200--2000 Hz \citep[Figure 6]{Temchin2010}. A more basal measurement (CF = 2477 Hz) revealed a bandpass synchronization response that had a sharp peak below 1000 Hz. For all CFs, maximum synchronization strength as a function of CF was highest below 1000 Hz and then gradually fell to negligible phase locking for fibers with $\CF>4$ kHz and had somewhat better phase locking for fibers with $\CF<4$ kHz \citep[Figure 7]{Temchin2010}. The authors concluded that the differences found between apical and basal sites of the chinchilla are suggestive of other mammals as well. Data from the rat suggest that at very high CFs, synchronization in the auditory nerve is negligible also at low frequencies unless they are intense ($> 60$ dB SPL), although it progressively improves in the ventral cochlear nucleus and trapezoid body of the olivary complex \citep{Paolini2001}. 

These results suggest that the bandwidth of the PLLs is relatively broad, as long as the carriers are low frequency. The effect of the cochlear bandpass filter is not apparent in many cases---synchronization bandwidth is broader at low frequencies and is anyway limited to them, while at high frequencies it drops. What cannot be gathered from these and similar measurements is what the cause of the phaselock is: weak synchronized within-channel OHCs that correspond to the IHC and its fiber, or perhaps OHCs of a remote channel that resonate with the tone and are weakly coupled to the channel under test. Either way, the low-frequency range may be suitable for coherent detection of only low-frequency carriers, or perhaps for coherent detection of low amplitude-modulated frequencies across the audible spectrum. This can be justified in light of the findings of amplitude-demodulated baseband frequencies caused by the intermodulation distortion in the organ of Corti, which were identified by \citet{Nuttall2018}. Theoretically, the baseband components can be coded by the auditory nerve independently of the carrier, given that the auditory nerve in mammals is not tuned. 

A large PLL bandwidth suggests a proportionally large loop gain (which usually goes as the square root of the bandwidth in higher-order PLLs). However, the exact relation depends on the PLL order and on other parameters, so no further statements can be made based on the available data.

\subsection{What changes in the input signal can pull the PLL out of lock?}
\label{OutofLockChanges}
Changes in the input reference stimulus are potentially capable of pulling the PLL out of steady-state lock. The higher order the PLL is, the more resilient it is to higher-order abrupt changes in the phase function derivatives. Three types of changes are customarily examined: phase step, frequency step, and frequency ramp. A well-designed PLL should be able to track the changes while in lock if they are within its pull-in range (slow) or hold range (fast). A lower-order PLL may struggle to maintain lock and can accumulate phase error in the process, mainly outside of its hold range. 

\subsubsection{Phase step}
The phase step is the most basic signal available to test the PLL transient response. It was tested using a modified slot pattern in a mechanical siren, which produced a phase step of half a cycle and sounded as an interruption in the tone \citep{Hartridge1936}. It was explained using the cochlear resonator theory of the time, but without any formal proof or quantitative data to support it. Informal tests by the author confirm that the phase step creates an interruption and in some combinations of short durations and carriers, it sounds like the tone effectively starts afresh and becomes double. However, if the phase step is small (around $10^\circ$ at 500 Hz and increasing to about $25^\circ$ at 6000 Hz), it becomes imperceptible. When the duration is very short, it sounds like the tone is superimposed with a harshness at the step, but its tonality is not broken. It can be shown that the spectrum of a phase-stepped tone shows a slight spectral broadening of the spectral line, which may not be sufficient to provide a spectral cue for detection. Audio demos for different combinations of carriers and phase steps can be found in \textsc{/Section 9.9.5 - Frequency and phase steps/Phase steps/}.

An indirect effect of phase step may have caused erratic thresholds in a tonal gap detection test by \citet{Shailer1987}. A 200 ms tone (200, 400, 1000, and 2000 Hz) in notched-noise masker was tested, with the gap duration applied at its middle. The gap threshold as a function of gap duration was monotonic only when the phase of the tone was made to be continuous, so it continued from the same phase as though the signal continued uninterrupted. When the phase started anew or was reversed after the gap, the threshold oscillated until it settled at longer gaps (4--10 ms, with longer times for short frequencies). This response could be only partially explained by the ringing of the filters that decays over a few milliseconds after the sound subsides. However, no combination of time constants was able to fully account for the results, so the authors suggested that the effect could be also related to phase locking that has to settle before and after the gap. This explanation also coincides with the PLL model, where the local oscillator retains its lock, at least for a few milliseconds after the sound has ceased, and requires a finite duration to obtain lock to the new phase, after a phase step. 

\subsubsection{Frequency step}
The second basic test for the PLL transient response is a frequency step, which was explicitly reported only once as well. A frequency step is perceived as a click if larger than a semitone, unless it is compensated with an adequate amplitude change (\citealp{Neustadt1965}, abstract only). Once again, an informal listening test by the author, suggested that the effect of the step is to produce a subtler interruption than the phase step, which is less noticeable than a click. A large step (say, 10--20\% of the carrier) sounds less harsh than a small step. When the duration of the tone is long enough, then even a small step (e.g., 0.1\%) is noticeable. Placing the step in the zero crossing has no perceptible effect. Another interesting effect is the apparent continuity of the pitch change, which for small steps may sound more like a ramp than a discrete step---something that may reflect a transient locking-in duration. Audio demos of different carrier and step combinations are available in 
\textsc{/Section 9.9.5 - Frequency and phase steps/Frequency steps/}. As in the phase step, the putative PLL seems to track this change well, for the most part, although large steps may not tracked per se, but are taken over by other channels, maybe with their own PLLs, which avoids any click-like change. 

\subsubsection{Frequency ramp}
A single study directly tested phase locking in the cat of upward and downward frequency-modulated ramps---the third basic signal that is commonly considered for the PLL. The authors used the inter-stimulus spike intervals as a proxy for instantaneous frequency and showed that it follows closely the instantaneous period of the ramps, or its subharmonics, for slopes up to 15.7 kHz/s, as long as the corresponding instantaneous frequencies were within the passband \citep[Figure 10]{Sinex1981}. Unfortunately, the temporal resolution of the original plots does not allow for close inspection of the phase tracking precision---especially in the largest slope, which occurs over about 30 ms, but appears almost like a vertical line in the plots. In another study that tested the phase locking to sinusoidal FM in ventral cochlear nucleus (VCN) units, no degradation of synchronization was reported as a function of the maximum frequency velocity, which depended on the modulation frequency and index \citep{Paraouty2018}\footnote{In the sinusoidal FM of Eq. \ref{SFM}, the frequency velocity is zero, but we can replace the sine with a cosine in the phase term to obtain a maximum velocity of $\Delta \omega$ around the carrier. From Eq. \ref{GeneralizedPhase}, $\Delta \dot{\omega} = \omega_m\Delta\omega$. In \citet{Paraouty2018}, the maximum $f_m$ was 10 Hz, and the maximum $\Delta\omega=0.32\omega_c$. Taking $f_c=2000$ Hz, we obtain $\Delta \dot{f} \approx 1020$ Hz/s.}. 

In a psychoacoustic pitch matching test in humans, tones that were simultaneously frequency and amplitude modulated (triangular modulation at 6 Hz) did not elicit equal pitch as did pure tones of equal carriers (440--1500 Hz) \citep{Iwamiya1984}. Therefore, the AM and FM interacted and the pitch was either overestimated or underestimated as a function of modulation depth. When either the AM or the FM was switched off, the pitch matching was still imperfect. This suggests that insofar as pitch reflects frequency tracking, then it is not instantaneous in humans at these frequencies. 

In a series of psychoacoustic experiments, Demany and colleagues showed how the pitch of slow and wide frequency-modulated sounds is perceived asymmetrically with respect to instantaneous frequency trajectories, as peaks are always perceived more distinctly than troughs \citep{Demany1994,Demany1995a,Demany1995b,Demany1997}. So, for example, musicians perceived fewer notes in a frequency-modulated melody than in a discrete-tone version of the same melody, which had its notes placed in the peaks and troughs of the continuous melody \citep{Demany1994}. In another experiment, it was found that the asymmetry was smaller for a center frequency of 4000 Hz than 250 and 1000 Hz, which suggested that the effect depends on phase locking or on the lack thereof \citep{Demany1995a}. While a satisfactory explanation for this effect is lacking, it was hypothesized that the perception of the peaks and troughs relies, at least in part, on memory, which suggest that the effect is retroactive and central in nature \citep{Demany1997}. These results may be reinforced by those from \citet{dAlessandro1998}, who tested the perception of the start and end frequencies of synthetic vowel glissandos of different durations and extents. Subjects' judgment could be modeled using a weighted time average, in which the extremities received more weight than intermediate frequencies. In another study that tested the pitch matching between asymmetrical (but periodic) FM patterns and a pure tone, subjects tended to assign a higher or lower pitch than the geometric mean (either 1 or 2 kHz) \citep{Etchemendy2014}. The direction of the difference was determined by where the asymmetry was and its relative degree. Modeling had only a limited success and could produce approximately correct predictions using a weighted average of the instantaneous frequency across several channels. 

Other studies focused on the discrimination of the rate of linear frequency ramps and found that subjects were able to match it across different frequency ranges \citep{Divenyi2004}, as well as to discriminate between different slopes \citep{Pollack1968,Nabelek1969,Dooley1988}. The slope of frequency acceleration could be discriminated as well, although not as well as the frequency velocity \citep{Divenyi2004}. 

A few studies tested the continuity illusion of frequency glides that are interrupted either by silence or a white noise burst, to find out how well listeners match the frequency trajectory before and after the interruption. Depending on the task, this requires the listeners to perform an interpolation of the trajectories they hear before or after, or an extrapolation of the sound before the interruption. \citet{Ciocca1987} employed logarithmic frequency glides that were interrupted by a loud white noise burst before it was continued again from a different frequency. It was found that subjects were able to match the initial trajectory of the glide (slope and initial frequency), as though the noise was not there, effectively interpolating the sound, often with a correct starting point and slope, on average. \citet{Kluender1992} found that with bark-scale (logarithmic) glides disrupted by white noise, listeners tended to underestimate the required post-noise glide. The steeper the glide was, the larger was the underestimation. This corresponded with the underestimation of linear frequency slopes composed of discrete tonal pulses interrupted by noise that was observed by \citet{Pollack1977}. However, his results are difficult to interpret because of the unspecified pulses that elicit constant pitch, ostensibly. Because the PLL should lock also to short tones, the results from these studies may indicate that, at least in part, it is higher-level processes that are responsible for the incorrect matches, rather than to the poor tracking of the PLL. In any case, it is a given that a correct answer in any of these tests is possible only if the listener is able to track the initial glide well enough to be able to have the right frequency end point from which to interpolate/extrapolate. Once this is achieved, higher-level cognitive processing may be involved that performs the actual interpolation or extrapolation. Unfortunately, it makes the interpretation of incorrect results in these studies ambiguous, as they can be attributed either to low-level tracking error or to an incorrect cognitive process. 

\subsubsection{Conclusion}
In summary, both physiological and psychoacoustic data that can be indicative of phaselock retention are currently lacking and do not allow for a definitive verdict, especially with regards to linear frequency ramps. The scant data about phase and frequency steps suggest that the PLL remains locked, despite the audible brief interruption associated with the steps. Data from the cat show a clear lock to linear ramps, but it is unknown how well the ramps were tracked, because of poor display resolution. The various listening tests, which spanned across multiple channels (and perhaps multiple PLLs), suggest that in some cases phase tracking lags and is not achieved in real time. However, there could be alternative interpretations for the data and tasks, which would make this conclusion uncertain. 

This incomplete analysis can entail that the human PLL is second-order, which accounts for the poor tracking of frequency ramps. This would match the loop filter response of the OHCs, which according to one model could be first order, yielding a second-order PLL. If frequency-ramp tracking turns out to be robust after all, then a third-order PLL may be a better model. Given that some animals rely much more heavily on chirps, it is curious to know if their loop filter properties are more clearly second-order. This applies mainly to bats, as some bat species compensate for Doppler shifts in motion, which theoretically requires a third-order PLL. In reality, this may be unnecessary, since bats appear to achieve their Doppler compensation noncoherently using combined amplitude modulation synchronization and place information, rather than coherent carrier processing that would require very high frequency phase locking \citep[e.g.,][]{PollakGrinnell1995}.

\subsection{Can the PLL become unstable?}
A well-designed PLL should be unconditionally stable in the range of operation. Otherwise, it can oscillate with no relation to the input signal. Therefore, a healthy cochlea should have no issue with PLL stability. However, SOAEs that are uncontrolled, a form of objective tinnitus, may be a result of unstable oscillators and PLLs, but is a relatively small fraction of tinnitus suffers---between 1.1\% and 9.05\% \citep[95\% confidence interval,][]{Penner1990}. Loss of cochlear function is also often assumed to be involved in subjective tinnitus cases, perhaps where dead regions are implicated \citep[e.g.,][]{Henry2014Tin}. However, the fact that it is not generally accompanied with SOAEs may indicate that the problem is of different nature and has a central cause.

\subsection{What is the effect of the input level on the PLL response?}
It was seen earlier that the PLL loop gain grows with the level of the input signal, which is one of its characteristic nonlinear features. The effect of gain in any PLL order is to increase its bandwidth and therefore its ability to track more variable inputs. However, in second-order PLLs and higher, increasing the gain may cause instability and change the various phase locking parameters (e.g., lock-in time, hold-in range). Therefore, maintaining relatively constant level to the PLL can make its operation more predictable and stable. 

Intensity effects in neural phase locking are well-documented, so this is only a brief overview of the main ones. The auditory nerve spiking rate to phase-locked tones increases with intensity below 70 dB SPL in the squirrel monkey, and it remains saturated at higher levels \citep{Rose1967}. However, the synchronization strength itself also peaks at 60 dB SPL, but is nearly saturated already at 35 dB SPL and grows only marginally from there. At high levels there is a slight deterioration in synchronization. Somewhat different values were measured for the cat in terms of the peak intensities of the rate and synchronization functions, although the general trends are the same \citep[Figure 8G-H]{HeilPeterson2017}. In the chinchilla, synchronization saturates at low levels in low-frequency fibers, whereas in high-frequency fibers synchronization keeps increasing with level, also in terms of its bandwidth \citep[Figure 6]{Temchin2010}. In the guinea pig, it was shown how in high-frequency channels ($>$ 1 kHz)---where no phase locking is measurable---intensity does not have any effect and synchronization remains poor \citep{Palmer1986}. 

The effect of increased bandwidth with higher intensity can be predicted from the PLL theoretical point of view, as the PLL bandwidth and transfer general is dependent on the input level (Eq. \ref{PLLbandwidth}). However, the saturation of synchronization with input levels is nonstandard and may be related to other factors in the cochlear PLL design and the compressed dynamic range of the neural signal. There could be a design benefit to maintain a more or less stable input level that achieves a good phase locking, despite input fluctuations. A similar function was hypothesized by \citet{Carney2018} with emphasis on envelope synchronization rather than on phase locking to the temporal fine structure. 

\subsection{Conclusion}
The above analysis revealed a few of the main features of the auditory PLL in a qualitative manner, albeit patchy. Thus, any conclusions made here will remain at least somewhat uncertain until the system can be analyzed in a more rigorous manner. Nevertheless, this level of analysis is sufficient for the present work, which requires the PLL function primarily in order to be able to account for coherence conservation between the acoustic signal and its neural representation. 

The PLL appears to be second-order---or perhaps third-order allowing for efficient linear FM tracking---with relatively large bandwidth that is also level-dependent but dynamically limited. It has a pull-in time of less than 7--10 ms for most signals and probably much less (1--3 ms) in the case of simple changes (perhaps tapping to the hold-in time instead). Lock is maintained in adverse SNR conditions, but broadband noise itself appears to have some residual coherence that makes it not entirely random, as PLL theory normally assumes. The PLL is limited to low-frequency carriers, which vary between species. In humans, the lock is strongest for carriers below 1 kHz and appears to become negligible at 4--5 kHz.

There are several aspects that make the auditory PLL unusual within standard PLL theory. The main one is that it contains a chain of tuned PLLs that roughly correspond to channels, which are activated simultaneously and are weakly coupled. The individual PLL is also level-limited and amplificative, and its output is neurally and nonuniformly sampled. Furthermore, there is a duality between the mechanical OHC oscillator and the neural spike generator that involves spontaneous discharging and internal synchronization between the OHCs, IHCs, and auditory nerve. Differences between the biological and electronic systems are to be expected, although the fundamental principles and related concepts from the electronic and control PLL theory appear to hold for the biological one as well. 

\section{Further PLLs downstream}
\label{NPLLs}
This chapter has dealt primarily with the most well-documented kind of phase locking in the auditory system, whose origin was traced to the cochlea, but is normally studied by observing neural spiking patterns. However, the advantages of using a PLL are not limited to high-frequency and low-level auditory processing. As we saw earlier, the PLL can be synchronized to an external signal and dispense with the need for a local clock to process it coherently. Phaselock is also robust against noise and various changes in the signal that are not predictable otherwise, which can make it appear like a narrowband filter. Thus, the possibility that the brain employs neural PLLs (NPLLs) may be attractive in the realization of higher-level synchronization functions. 

NPLLs have been hypothesized in different contexts than hearing \citep{Hoppensteadt1997,Ahissar1998}, which were recently extended to hearing as well \citep{Ahissar2023}. Simple NPLL models are based on a VCO neuron model that is embedded in various neuron networks, which are studied with respect to their nonlinear dynamics, amongst other aspects \citep{Hoppensteadt1997}. In another class of models, it was proposed that NPLLs may be common thalamocortical modules that are instrumental in the transformation between a temporal code to a rate code \citep{Ahissar1998, Ahissar2001}. Examples were provided primarily from tactile and visual data, but the principles are general enough to be applied to any modality. The involved frequencies that are synchronized to in rate coding in the NPLL scheme are relatively low compared to those in hearing (smaller than 14 Hz; \citealp{Ahissar2001}). It is hypothesized the NPLL thalamocortical loops may be concentrated in the non-lemniscal areas of the auditory thalamus, where it can point the auditory cortex to home in on the syllabic onset of, for example, at typical running speech rates \citep{Ahissar2023}. However, auditory encoding shifts to a rate code mainly around the IC, where the maximum frequencies are in the hundreds of Hertz, with only few instances of rate coding documented in the cochlear nucleus for the carrier but not the envelope \citep{Joris2004}. Feedback loops of different sorts are found in the auditory system, given its widespread efferent system. This includes thalamocortical loops that have been hypothesized in the context of tinnitus \citep[pp. 201--202]{Eggermont2012a}, or in attention switching between auditory streams \citep{Kondo2009}. Different neural oscillators exist in the subcortex as well (e.g., the chopper cells in the CN and IC), which may hypothetically function as local oscillators with a feedback loop. There is no particular reason why an NPLL cannot be conceived further upstream from the thalamus, where it can be used to improve or maintain phaselock, or synchronize together (bind) different events and control processes. 

One hypothetical use of NPLLs is to facilitate synchronization to undetected modulation-band modes, which are not available on the level of the individual cochlear channel and require inputs across channels. Direct physiological evidence for different across-channel integration of coherent information has not been framed with respect to NPLLs, though. Different mechanisms may exist that can achieve coherence or synchronization manipulation. The example of coincidence detection was briefly mentioned in \cref{BrainCoherence}, as a popular neural model for nonstationary coherence detection between different inputs, mode-locking was mentioned in \cref{Phaselocking}, and broadband information contribution in processing was suggested by \citet{Pressnitzer2001}. 

\section{Detection schemes with and without phase locking}
\label{DetectionSchemes}
Our discussion has revolved around the function of one major phase-locking site, which we identified as the auditory PLL, whereupon it was briefly considered that additional sites could exist that produce a similar function in the brain. Hence, auditory synchronization phenomena do not stop in the cochlea or in the auditory nerve and can take additional forms that may not be independent of one another: within-channel coherence manipulation, across-channel synchronization, and modulation phase-locking within and across channels. Advantages exist to increasing or decreasing the synchronization in all types of processing, depending on the stimulus and context. Theoretically, such an accommodation in synchronization may be instigated either locally or through top-down control. These ideas will be explored in some depth after we discuss auditory sharpness and blur, which we argue are ultimately determined by the degree of coherence of the stimulus (\cref{AudImageFun}). According to this logic, an increase in synchronization of any kind of object enhances its sharpness over a background that is less coherent, or that became desynchronized through processing (or dysfunction) in what we shall later refer to as auditory accommodation (\cref{accommodation}). These functions might be helped by the availability of NPLLs throughout the auditory brain. 

\begin{figure} 
		\centering
		\includegraphics[width=0.5\linewidth]{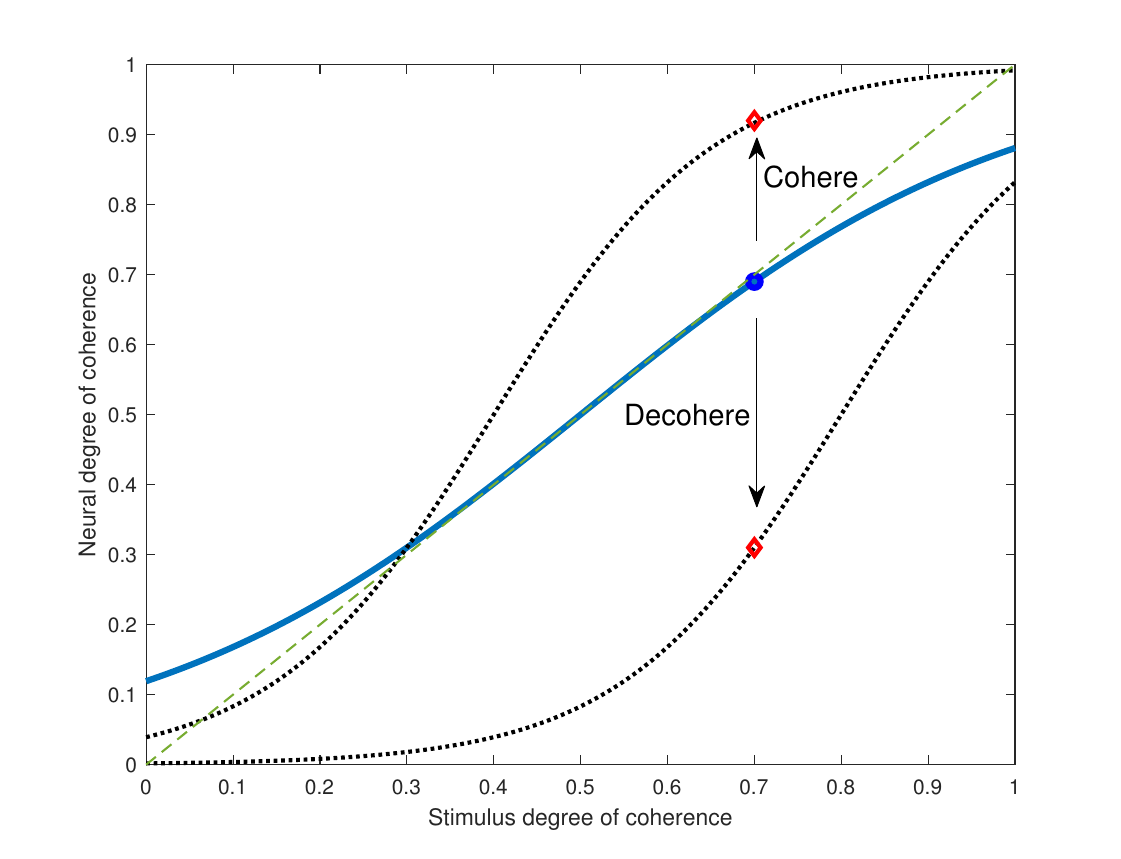}	
		\caption{A cartoon illustration of the hypothetical operating point of the coherence in the auditory system. The abscissa designates the degree of coherence of the stimulus. The auditory degree of coherence is designated by the ordinate, which relates to the phase-locked neural signal somewhere in the system after the cochlear PLL. The blue circle on the middle curve marks a hypothetical operating point which may be further cohered (i.e., its synchronization increased and be more than the input degree of coherence) by pushing the curve upwards, or decohered with the absence of phase locking or the addition of phase noise by pushing the curve downwards, as marked with the red diamonds. There may be advantages in both types of processing in different situations.}
		\label{operatingpoint}
\end{figure}

How do these additional synchronization manipulations square with the idea that the auditory PLL conserves the input coherence? The PLL and the ensuing neural synchronization endow the system with an initial degree of coherence that closely corresponds to the acoustic input (within some limitations, see \cref{PLLCoherence}). However, by not maximizing it right at the beginning of the processing chain, the system may be benefiting from an operating point from which it may be simpler to decohere/desynchronize the signal or to further cohere/synchronize it (Figure \ref{operatingpoint}). Which way the processing should go then likely depends to some extent on top-down control that can bias the processing in order for the auditory scene analysis to go as desired, within physical limitations. Therefore, the auditory PLL may provide the system with the \textbf{potential} to conserve the coherence, rather than relay a fixed coherence associated with the input. The PLL feeds into several pathways that can be processed as coherence-conserving, coherence-enhancing (sharpening), or decohering (blurring). Some of it is determined by virtue of the input signal itself (e.g., if it is completely coherent or incoherent). As the majority of realistic stimuli are partially coherent, they may benefit from dual processing: coherent and noncoherent. 

Acknowledgment of a dual-processing strategy, beyond the traditional duplex pitch \citep{Licklider1951,Licklider1959} and binaural models \citep{Rayleigh1907Duplex}, has begun to emerge in literature. It embraces the distinction between the slow temporal envelope and the temporal fine structure (TFS---an indicator of phase locking to the carrier; see \cref{AuditoryEnvPhase}), rather than choose one over the other. Examples can be found in discussions about the avian brainstem processing \citep{Sullivan1984,Warchol1990}, in the attempt to associate the TFS and the envelope with the where/what streams \citep{Smith2002}, an extension of dorsal and ventral stream model to the brainstem in \citep{Pickles}, normal speech spectrograms are a combination of both envelope and TFS processing that are about equally balanced---almost irrespective of vocoding \citep{Shamma2013}, normal and impaired balance of TFS and envelope processing \citep{Henry2013,Anderson2013,Hao2018,Parida2021b}, binding of TFS and envelope responses to binaural stimuli \citep{Luo2017,Wang2018}, differential processing of AM and FM tones as inferred from inconsistent internal noise levels required in modeling \citep{Attia2021}, and mixed place/time pitch models \citep{Cheveigne,Moore2013,Oxenham2022}\footnote{A somewhat different dual-processing strategy was proposed more openly by \citet{Coffey2016}, regarding the pitch perception of different listeners. Perceived pitch was correlated with frequency-following response (FFR) recordings, which related either to the spectral contents of the stimulus (physical spectrum) or the periodicity pitch, as in the case of the missing fundamental, which requires additional processing. The latter strategy was more dominant among musicians.}. As some studies demonstrated how one of these two processing types prevails using different stimuli and conditions, different models selectively indicated which type of processing (envelope or TFS) is required for a particular task (e.g., speech reception, pitch discrimination, localization, musical listening). We would like to reframe this discussion from a communication standpoint. 

\begin{figure} 
		\centering
		\includegraphics[width=0.85\linewidth]{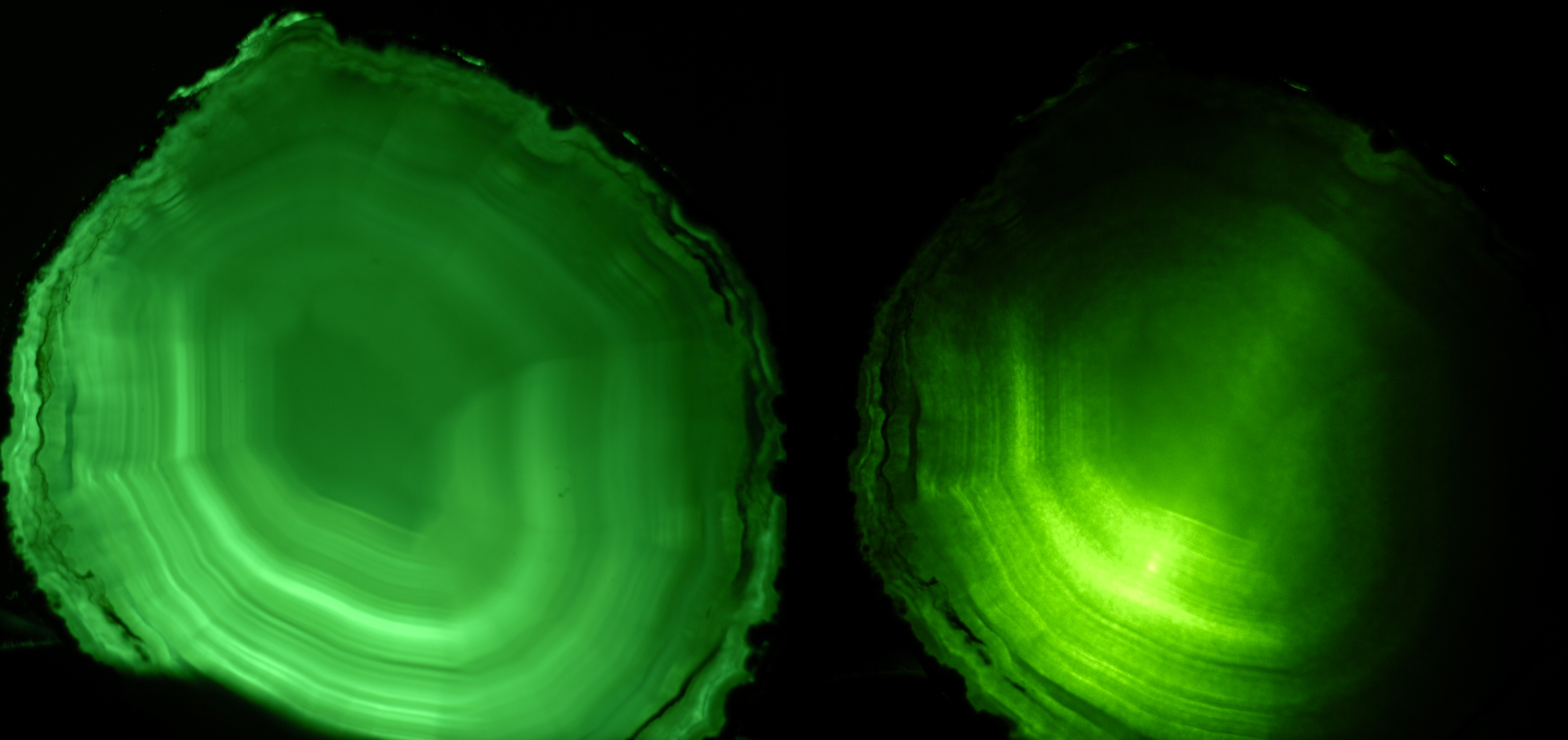}	
		\caption{Two images of a polished slice of agate rock back-lit by an incoherent LED source (left) and a coherent laser source (right). The laser beam was spread with a diverging lens and a glass diffuser and its center can be observed in the brightest point from behind the agate. More technical details about the sources are in Figure \ref{partialimages}. These images illustrate how the information about the object can be observed using both types of illumination, which nevertheless produce different effects. The incoherent image is smooth and clear, whereas the coherent image appears grainy, as the light scatters from the various imperfections in its path and in the agate.}
		\label{CohIncoh}
\end{figure}

As was argued in \cref{AcouAudComm}, being a communication system, the auditory system has to detect, or demodulate, arbitrary signals that were modulated in arbitrary ways at the source. Each of the two general detection types excels in the demodulation of some signal types and not in others. The PLL necessarily belongs to a coherent detection scheme, but as was seen in \cref{Corollaries}, it is not unconditionally effective and can sometimes be out of lock even within the phase-locking limit. This can happen, for instance, with incoherent signals, in transience (while locking), or because the acoustic signal propagates toward the listener and picks up noise and becomes distorted, so its initial degree of coherence changes. In such cases, it is advantageous to have noncoherent detection that is more general and can intercept the signals that for whatever reason cannot be coherently detected. Parallel coherent and noncoherent detection can therefore provide a fail-proof strategy for processing arbitrary sounds in arbitrary conditions. Hypothetically, the doubly-detected product---the demodulated message---can then be optimally weighted and mixed to produce a superior output. Optical images were interpreted as two-dimensional communication in \cref{ImageCommunication} and as such they can visually provide a demonstration of the difference between coherent and noncoherent detection. Examples are given in Figures \ref{partialimages} for a gradation of coherent, partially coherent, and incoherent images, and in Figure \ref{CohIncoh} for extreme differences between incoherent and coherent imaging. 

Out of the different available experimental methods that have been used to demonstrate envelope or TFS processing in different conditions, the frequency following response (FFR) is probably the only one that can reveal concurrent information of both processing types. The FFR is a brain potential measurement technique, whose response to sustained and transient sounds corresponds to the low-frequency range (typically $<$ 2 kHz; up to 1.4 kHz phase locked response has been observed by \citealp{Bidelman2018}) of broadband signals at a short latency from the input \citep{Moushegian1973}. The potential is the summation of multiple areas in the brain and is not necessarily well-localized, although the inferior colliculus (IC) has been usually implicated as the strongest source compared to other subcortical and cortical areas \citep{Chandrasekaran2010,Tichko2017,Coffey2019, Bidelman2021}. The FFR generally relies on synchronization to different parts of the signal that include a mixture of envelope and TFS. It is possible to estimate the phase locking to TFS by repeating the FFR measurement with inverted polarity (in anti-phase) \citep{Aiken2008}. Because the auditory nerve codes only half the wave \citep{Brugge1969}, the polarity inversion forces it to code the half-wave that was rejected in the noninverted run. Subtracting the responses of the two runs eliminates the common factor, which is taken to be the slow-varying temporal envelope, so the remainder is then the response to the fast-varying TFS, or $\FFR_{TFS} = (\FFR_+ - \FFR_-)/2$, where the sum yields the envelope, $\FFR_{ENV} = (\FFR_+ + \FFR_-)/2$. 

It should be emphasized that the relative magnitude of the FFR measurements necessarily depends on the degree of coherence of the signal that is being heard. As was seen in \cref{CommunicationTheory} and throughout the present chapter, processing the carrier phase requires coherent detection, whereas the envelope can be detected both coherently and noncoherently. Therefore, we can roughly consider $\FFR_{ENV}$ data to be telling of noncoherent detection and $\FFR_{TFS}$ of coherent detection. Partially coherent signals can always be represented as a weighted sum of the incoherent and coherent parts of the signal (Eq. \ref{totalcoherence}), so in principle, this role division can be universally suitable for arbitrary signal coherence. However, this method cannot tell us with confidence that the temporal envelope was not processed by coherent detection mechanisms---only what was certainly not detected noncoherently. 

The FFR sensitivity to the degree of coherence is useful in quantifying the effects of broadband noise and reverberation on hearing. As was mentioned in \cref{CommunicationTheory} and \cref{PLLs}, an effective PLL should be able to provide more robust detection of coherent signals in low broadband noise than noncoherent detection---especially at a poor SNR level. A good illustration of this was shown in normal-hearing chinchillas, where the $\FFR_{TFS}$ for speech in pink-noise-masker (high-pass filtered at 600 Hz) had discernible low-frequency ($< 750$ Hz displayed) peaks at 0 and -20 dB SNR, whereas $\FFR_{ENV}$ hardly had any in that range \citep[Figures 4--5]{Parida2021}. Additionally, at 10 dB SNR or less, the $\FFR_{TFS}$ power was larger than that of the $\FFR_{ENV}$. 

It is also interesting to consider the effect of reverberation that decoheres the signal, so that the direct sound may remain coherent (if it is coherent to begin with), whereas cumulative reflections gradually decohere it. Thus, we would expect to observe effective coherent detection only at the onset of the stimuli, as long as the reverberation time is not too long. In normal-hearing listeners, it was found that the fundamental frequency of a vowel is almost immune to the effects of reverberation, while the power of the formant harmonics decreases with increasing reverberation time \citep{Bidelman2010}. While these results are based only on the standard envelope FFR (with no polarity inversion), they are consistent with the idea that reverberation decoheres the parts of the signal that are best detected coherently. In contrast, the fundamental pitch can be detected noncoherently and is therefore more robust to decoherence.

As a final note, it should be mentioned that a pitch extraction algorithm  has been recently introduced, which is based on a digital PLL-like principle---the Period-Modulated Harmonic Locked Loop (PM-HLL) \citep{Hohmann2021}. The PM-HLL is based on variable delays that track the period of the signal (narrowband, or broadband with multiple PM-HLL implementations) and does not employ an oscillator per se. It has been shown to be highly efficient and precise in tracking the pitch of complex signals, including frequency sweeps and complex tones. The processing has additionally implemented qunatization and stabilization, as is prescribed by the auditory image model of (\citealp{Patterson1992}; see \cref{TheAuditoryImage}). While the algorithm does not attempt to adhere to the auditory physiology, its advantage in accounting for TFS processing has been briefly discussed. 

\section{Conclusion}
This chapter introduced an auditory PLL model that tailors aspects of operation of the mechanical cochlea and neural phase locking---two auditory subsystems that are known to work in concert, but are not usually modeled together. Additionally, the model provides a continuous link, if only conceptually, between the stimulus coherence and the degree of synchronization that is found in the auditory nerve and beyond. While it is undoubtedly speculative in spirit, much of the evidence that is required to make such a system work is already in place. A degree of speculation, however, may be inevitable in the analysis of any system that works as a loop. Breaking the loop into individual subcomponents makes little sense in explaining the function of such a system as a whole. If accepted, the PLL model has the potential to explain a range of physiological and psychoacoustical phenomena, as all sound that is transduced has to pass through it. 

As part of the entire work presented here, the role of the PLL model is relatively minor, perhaps with the exception of the analysis of likely mechanisms to drive auditory accommodation (\cref{HypoAcco}). It has been developed in order to shed light on the continuity of the signal coherence between the external environment and the brain. Certain aspects of temporal imaging depend on the degree of coherence that the signal or the system have. Therefore, without a clear account of how the coherence function propagates up to the level of perception, much of that discussion will stay murky. This has also been the main reason for not delving into quantitative modeling of the auditory PLL, which is undoubtedly necessary in order to provide the full picture about this system. However, with the amount of unknowns and the general difficulty to simulate PLLs (let alone a chain of coupled PLLs), this likely deserves a lengthy treatment in its own right.

Even if the PLL model will be eventually rejected, a PLL-like modeling should still be valuable in providing deterministic answers to the standard parameters that are generally employed to characterize phaselock. Some of these parameters were explored in \cref{Corollaries}, but data to confidently estimate them have been generally lacking. 

In the second half of this work, the ideas of separate coherent and incoherent processing will be discussed using an imaging optics theoretical framework, which considers the two types of processing as two extremes that are key in the understanding of partially coherent processing. 


\chapter{The paratonal equation}
\label{temporaltheory}
\section{Background}
\label{ParaxialBackground}
Up until now, various topics from a range of fields have been introduced that are deemed critical for the understanding of the temporal imaging theory, which we shall start presenting in this chapter. The concepts that were presented earlier will be used throughout this work---some more than others---and to a large degree serve to motivate it.

The basis for the temporal imaging theory is a solution to the scalar Helmholtz wave equation that was developed in the late 1960s, but has never been used in acoustics. There are two features of this solution that make it particularly attractive in hearing. First, it deals with the complex envelope of the signal and not with the carrier, which is immediately useful in problems of modulation, where nonstationarity is key. It therefore dodges the static nature of classical Fourier analysis that has us infer modulation indirectly from its spectrum, or apply time-windows on otherwise infinitely long transforms \citep[pp. 383--395]{Zverev}. Second, it is mathematically analogous to the paraxial equation for light. This means that it should be possible to devise an analogous imaging system in sound by combining dispersive elements, in analogy to spatial optics, where diffraction is used instead. Both these aspects have far-reaching implications to hearing theory---some of which will be explored in the final chapters of this work. 

The temporal paraxial equation was derived by the nonlinear-optics physicist Serge\v{i} Aleksandrovich Akhmanov and colleagues in \citet{Akhmanov1968, Akhmanov1969}\footnote{See \citealp{Drabovich} for a short professional biography of Akhmanov.}. It was done in the context of a comprehensive space-time equivalence theory, which exploits mathematical analogies between time-modulated to narrow bounded-beam waves to account for the propagation of \term{ultrashort pulses} of high-energy laser in dispersive media. In fact, using different considerations, a completely independent treatment of dispersive waves got close to a complete optical imaging system analogy was presented earlier by \citetalias{Tournois}. He also proposed to apply this solution to acoustic waves \citep{Tournois67}, but that work has never been followed up. Related concepts have been applied even earlier, in chirp radar technology, using a technique called \term{pulse compression} \citep{Klauder1960}. Pulse compression in radars employs frequency modulation (FM) to obtain high-power signals that have a large time-bandwidth product. It requires matched filtering at the receiver, which inverts the FM and undoes the pulse compression. In applied optics, dispersion has been successfully employed to stretch, amplify, and recompress an ultrashort low-energy laser pulse into a powerful one \citep{Strickland1985}\footnote{This seminal experiment won its authors a Nobel Prize in Physics in 2018.}. While we adhere to the formalism developed in optics, it should be noted that ideas from radar technology have been influential in the context of bat echolocation for a long time \citep[pp. 342--346]{Griffin}, including pulse compression signal processing. 

The basis for the space-time analogy is valid for carrier waves of visible light, such as those carried by thin, axially uniform, single-mode optic fibers, which can be modulated by low-frequency microwaves, in a source-free space \citep[pp. 178--187]{Haus}. The description of this propagation is scalar, because electromagnetic polarization that would have made the description vectorial, may be neglected. It means that with careful inspection of the assumptions, it is straightforward to adapt the plane-wave scalar equations of light to the plane waves of sound. Therefore, in the following we will derive the ``paraxial'' dispersion equation and follow with a presentation of the time lens. 

\section{The ``paraxial'' approximation of the dispersion\\equation}
\label{ParatonalApprox}
We consider the propagation of a plane wave at $z \geq 0$ with known spectrum at $z=0$. This refers to a so-called secondary source, according to Huygens principle, which treats any point on the wavefront away from the source as a point source (\cref{DiffractionFourier}). 

Starting from the homogeneous three dimensional acoustic wave equation \citep[e.g.,][p. 282]{Morse},
\begin{equation}
	\nabla^2 p= \frac{1}{c^2}\frac{\partial^2p}{\partial^2t}
	\label{eq:wave}
\end{equation}
Where $p$ is the pressure, and $c$ is the speed of sound. In the plane-wave approximation, the direction of propagation is arbitrarily set parallel to $z$, so that the wavenumber components $k_x=k_y=0$ and the profile of the propagating wave may be ignored (i.e., an infinite plane wavefront). This results in a one-dimensional equation
\begin{equation}
	\frac{\partial^2 p}{\partial z^2}= \frac{1}{c^2}\frac{\partial^2p}{\partial^2t}
	\label{eq:1Dwave}
\end{equation}
A monochromatic solution that satisfies Eq. \ref{eq:1Dwave} can then take the general form (\cref{LinAnaDisp})
\begin{equation}
\label{eq:carriermod}
	p(x,y,z,t) = p_0e^{i\left[\omega_c t-k(\omega)z\right]}
\end{equation}
Where $p_0$ is the pressure wave amplitude, and $k(\omega)$ is the frequency-dependent wavenumber in the medium, which may entail dispersion, if $k(\omega) \neq \const$.

The derivation of the dispersion equation below follows \citet[pp. 179--181]{Haus} and \citet[pp. 128--129]{New2011} for scalar electromagnetic fields. The central assumption in this derivation is that the pressure envelope varies slowly in space, so that the modulation wavelength is much larger than the carrier, $\lambda_m \gg \lambda_c$, and the corresponding period $T_m\gg T_c$ \citep{Akhmanov1968}.

Let us posit a known frequency dependence of the medium dispersion $k(\omega)$ and a corresponding dependence on $z$ of the spectrum $P(z,\omega)$, assuming that $p(z,t)$ is square-integrable, so it has a Fourier representation
\begin{equation}
P(z,\omega) = A(\omega,0)e^{-ik(\omega)z}
\label{eq:narrowband}
\end{equation}
Where $A(\omega,0)$ is the initial complex spectral envelope of $P(z,\omega)$. Then, along the z-axis, the following differential equation is satisfied
\begin{equation}
	\frac{{\partial P(z,\omega)}}{{\partial z}} =  - ik (\omega )P(z,\omega)
	\label{eq:diff1}
\end{equation}
For a slowly-varying complex envelope, the dispersion relation of $k(\omega)$ can be approximated using Taylor series around $\omega_c$
\begin{equation}
k (\omega) = k_c + \frac{{dk }}{{d\omega }}(\omega- \omega_c)  + \frac{1}{2}\frac{{d^2 k }}{{d\omega ^2 }}(\omega- \omega_c)  ^2  + ...
\label{beta_taylor}
\end{equation}
$k(\omega)$ is a complex function, the real part of which is the dispersion and the imaginary part is the absorption (\cref{airtravel}). $k_c$ is the plane wave phase $k_c=\omega_c/\lambda$. In a small range around $\omega_c$, we consider the second-order approximation to be accurate enough, given the above slow-varying modulation condition (though, higher-order terms are frequently used in nonlinear optics). In hearing, the shifted frequency $\omega-\omega_c$ is referred to as the \term{envelope frequency} or \term{modulation frequency} and in communication theory the terms \term{frequency deviation} and \term{baseband frequency} are used. 

We can plug Eq. \ref{beta_taylor} in Eq. \ref{eq:diff1} up to the second derivative of $k(\omega)$
\begin{equation}
\frac{{\partial P(z,\omega)}}{{\partial z}} =  - i\left[k_c + \frac{{dk }}{{d\omega }}(\omega  - \omega_c ) + \frac{1}{2}\frac{{d^2 k }}{{d\omega ^2 }}(\omega  - \omega_c )^2 \right]P(z,\omega)
\label{eq:diff_series}
\end{equation}
If the spatial dependence of the envelope is shifted to be around $\omega_c$, then $P(z,\omega)$ can be reformulated to factor out the constant phase using
\begin{equation}
P(z,\omega) = A(z,\omega-\omega_c)e^{-ik_c z}
\label{eq:narrowbandA}
\end{equation}
Where the shifted complex spectral envelope around $\omega_c$, $A(z,\omega-\omega_c)$, was introduced and now contains the spatial dependence in $z$ through the higher order terms of $k$. Now Eq. \ref{eq:diff_series} in $P$ can be reformulated as an equation in $A$, eliminating the mean spatial frequency $k_c$
\begin{equation}
\frac{{\partial A(z,\omega-\omega_c)}}{{\partial z}} =  - i\left[\frac{{dk }}{{d\omega }}(\omega  - \omega_c ) + \frac{1}{2}\frac{{d^2 k }}{{d\omega ^2 }}(\omega  - \omega_c )^2 \right]A(z,\omega-\omega_c)
\label{eq:diff_series2}
\end{equation}
In order to obtain a time-domain expression of Eq. \ref{eq:diff_series2}, it will be necessary to have the inverse Fourier transform of the shifted envelope, which can be obtained from the full signal and Eq. \ref{eq:narrowbandA}
\begin{multline}
p(z,t) \equiv \frac{1}{2\pi}\int _{-\infty}^\infty P(z,\omega) e^{i\omega t} d\omega = \frac{1}{2\pi}\int _{-\infty}^\infty A(z,\omega-\omega_c) e^{-ik_c z}e^{i\omega t} d\omega \\
=\frac{1}{2\pi} e^{i\omega_c t - ik_c z } \int _{-\infty}^\infty A(z,\omega-\omega_c) e^{i(\omega -\omega_c) t} d(\omega -\omega_c) = a(z,t) e^{i\omega_c t - ik_c z}
\end{multline}
which means that the temporal envelope $a(z,t)$ is simply the inverse Fourier transform of the complex spectral envelope, in the modulation frequency coordinate
\begin{equation}
\frac{1}{2\pi}\int _{-\infty}^\infty A(z,\omega-\omega_c) e^{i(\omega-\omega_c) t} d(\omega -\omega_c) \equiv {\cal F}^{-1} \left[ A(z,\omega-\omega_c) \right]= a(z,t)
\label{temptospectenv}
\end{equation}
This identity enables us to manipulate the complex envelope independently of the carrier, as a function of modulation frequency, in the narrowband approximation. Note this additional relation for the inverse Fourier transform
\begin{equation}
{\cal F}^{-1}\left\{ \left[ i(\omega-\omega_c) \right]^n A(z,\omega-\omega_c) \right\}=\frac{\partial ^n}{\partial t^n}a(z,t)
\label{eq:FourDer}
\end{equation}
The inverse Fourier transform can now be used to convert Eq. \ref{eq:diff_series2} to the time domain, using Eq. \ref{eq:FourDer}, which yields the following differential parabolic equation for the temporal pressure envelope
\begin{equation}
\label{eq:dispersion_p}
\left( {{\partial  \over {\partial z}} + {\frac{dk}{d\omega}}{\partial  \over {\partial t}}} \right)a(z,t) = i\left( {\frac{1}{2}\frac{{d^2 k }}{{d\omega ^2 }}{{\partial ^2} \over {\partial {t^2}}}} \right)a(z,t)
\end{equation}
In most physical conditions, frequency-dependent absorption is usually even with dependence on $\omega^2$ (see \cref{airtravel}) and not on $\omega$, so it is therefore assumed that $dk/d\omega$ is real. Then the (real) group velocity is defined as usual (Eq. \ref{groupvel2})
\begin{equation}
\frac{1}{v_g}  = \frac{dk }{d\omega }
\label{eq:vg}
\end{equation}
Additionally, it is convenient to separate the real and imaginary parts of the second derivative of $k$. In general, $k(\omega) = \beta(\omega) + i\alpha(\omega)$, and the real part of its second derivative is
\begin{equation}
\beta'' = \Re \left(\frac{{d^2 k }}{{d\omega ^2 }}\right)
\end{equation}
which gives a measure of the \term{group-velocity dispersion} (GVD) of the medium, whereas the imaginary part relates to the absorption, or \term{gain dispersion} \citep[p. 335]{Siegman}
\begin{equation}
\alpha'' = \Im \left(\frac{{d^2 k }}{{d\omega ^2 }}\right) 
\end{equation}
Finally, Eq. \ref{eq:dispersion_p} may be further tidied up with change of variables to a traveling coordinate system, using
\begin{equation}
 \tau  =  (t - t_0) - \frac{(z - z_0) }{v_g}
\label{traveltau}
\end{equation}
\begin{equation}
 \zeta  = z  - z _0  \\ 
\end{equation}
This change of variables means that the frequency dependence of $a$ is always centered around the group velocity at $\omega_c$, when it is situated at a distance $\zeta$ from the origin. The new time coordinate is in fact the difference between the phase time coordinate and the group delay of the pulse, measured against its reference at $(t_0,z_0)$ (see \cref{PhysicalWaves} and Eq. \ref{GroupDelay2}). Using the chain rule on Eq. \ref{eq:dispersion_p},
\begin{equation}
\frac{\partial a(\zeta ,\tau )}{\partial \zeta } \frac{\partial \zeta}{\partial z }  + \frac{\partial a(\zeta ,\tau )}{\partial \tau } \frac{\partial \tau }{\partial z }  +\frac{1}{v_g} \frac{\partial a(\zeta ,\tau )}{\partial \tau } \frac{\partial \tau }{\partial t } = \frac{i\beta''-\alpha''}{2} \left[ \frac{\partial a^2(\zeta ,\tau )}{\partial \tau^2 }\left(\frac{\partial \tau}{\partial t} \right)^2 + \frac{\partial a(\zeta ,\tau )}{\partial \tau }\left(\frac{\partial^2 \tau}{\partial t^2} \right)  \right]
\label{eq:dispersionabs}
\end{equation}
where we neglected the term with $\partial \zeta /\partial t = 0$. Then, after using the definitions of $\zeta$ and $\tau$, the equation reduces to
\begin{equation}
\frac{{\partial a(\zeta ,\tau )}}{{\partial \zeta }} = \frac{i\beta''-\alpha''}{2}\frac{{\partial ^2 a(\zeta ,\tau )}}{{\partial \tau ^2 }}
\label{eq:dispersionabs}
\end{equation}
As Eq. \ref{eq:dispersionabs} incorporates complex dispersion, it combines effects from the diffusion equation (for absorption) and Schr\"odinger's equation (for dispersion). This coupling may significantly complicate the treatment of the imaging system, so in optical treatments of this equation the absorption term is generally neglected. As it turns out in this work, there is a broad range of useful results that can be obtained for hearing without resorting to absorption. However, a discussion about the significance of this term will be revisited in several places (mainly in \cref{PsychoEstimation}). Therefore, in the remainder of this work, we will set $\alpha=0$ and assume a ``classical imaging system''. Therefore, 
\begin{equation}
\frac{{\partial a(\zeta ,\tau )}}{{\partial \zeta }} = \frac{i\beta''}{2}\frac{{\partial ^2 a(\zeta ,\tau )}}{{\partial \tau ^2 }}
\label{eq:dispersion}
\end{equation}
This is the parabolic dispersion equation, which belongs to the heat equation family and has a very similar mathematical form to the spatial paraxial equation (\ref{eq:spatialdiff}). A solution can be given using an inverse Fourier transform on the initial spectrum at $z=0$ \citep[pp. 349--354]{Haberman}\footnote{Note that the standard diffusion equation has interchanged time and space variables compared to the dispersion equation (\ref{eq:dispersion}). As a result, the initial spectrum condition for the diffusion equation is at $t=0$, whereas here it as at $\zeta=0$.}, multiplied by a quadratic complex kernel
\begin{equation}
a(\zeta ,\tau ) = \frac{1}{{2\pi }}\int_{ - \infty }^\infty  {A(0,\omega )\exp \left( - i\frac{\beta'' \zeta}{2} \omega ^2 \right) e^{i\omega \tau} d\omega } 
\label{eq:Fsolution}
\end{equation}
Eq. \ref{eq:dispersion} is the basic propagation transformation in a dispersive medium. Note the similarity to the Fresnel diffraction integral from spatial optics (Eq. \ref{FresenlDiff}). This solution, along with the paraxial-like differential equation \ref{eq:dispersionabs}, constitute for us the basic space-time analogy. 

The term ``paraxial'' in geometrical optics refers to the sound rays that are considered in the analysis, which are in the vicinity of the optical axis. In wave optics, the same idea is expressed using the spatial frequency $k$, whose range expresses the limited angular variation around the optical axis. As the dispersion equation strictly deals with uniaxial plane waves, referring to this approximation as ``paraxial'' makes little sense here, and we shall instead rename Eq. \ref{eq:dispersion} (and Eq. \ref{eq:dispersionabs}) to be the \term{Helmholtz paratonal equation}, in analogy to Eq. \ref{eq:spatialdiff}. Here, \term{paratonal} conveys the intention of the main approximation that we employ and is applicable to the auditory system---that of narrowband channels that are centered around a particular carrier, a characteristic frequency, or a tone. Arguably, the term ``paratemporal'' may be just as valid as a name, because temporal imaging is also based on the limited extent of the temporal aperture---a physical time window that effectively turns a continuous wave into a finite received pulse---in analogy to the finite extent of the spatial image \cref{ImagingEqs}.

It is important to dwell on the $\beta'' \zeta/2$ factor in the exponent, of \ref{eq:Fsolution} which is what characterizes the group-velocity dispersion of the medium. Its units are $\s^2/\rad$, and the units of $\beta''$ are $s^2/\rad\m$. As the group-velocity dispersion grows with increasing length of the dispersive path $\zeta$, the total dispersion between the measurement points is of interest, rather than the specific magnitude of $\zeta$. This may be avoided by using the group delay definition instead. For spectral phase dependnce around $\omega_c$ of the form
\begin{equation}
 \varphi(\omega) = - k(\omega) \zeta
\label{phase1}
\end{equation}
the group delay is (Eq. \ref{GroupDelay1})
\begin{equation}
 \tau_g = -\frac{d\varphi(\omega)}{d\omega}
\end{equation}
so differentiating this definition produces an alternative expression for $\beta''z$ as well 
\begin{equation}
\frac{d\tau_g}{d\omega} = -\frac{d^2\varphi}{d\omega^2} = \zeta \frac{d^2 k}{d\omega ^2} = \beta''\zeta
\label{GDDder}
\end{equation}
Therefore, the group-velocity dispersion parameter $\beta''z$ (that typically appears with the factor of $1/2$ from the Taylor expansion) expresses also the curvature of the frequency-dependent phase function. This equation also shows that the same information contained in the group-velocity dispersion is available in the group delay derivative. Therefore, we will always prefer to relate to {group-delay dispersion} (GDD) instead of group-velocity dispersion, because it better reflects the method of calculation, when the distance $\zeta$ is unknown. 

Basic examples of explicit solutions to the paratonal equation and a review of useful related expressions are found in \cref{AppWaves}.

\section{The time lens}
\label{TheTimeLens}
The dispersion integral of Eq. \ref{eq:Fsolution} can be thought of as an all-pass filter in the frequency domain, which has a similar operation to the diffraction integral in the spatial frequency domain. Thus, it is curious to look for a propagation medium that can produce a multiplicative quadratic phase function that is analogous to the normal lens, but in the time coordinate instead of the spatial coordinate. 

A \term{time lens} was defined in an analogous way to the spatial lens (Eq. \ref{lenstransfer}), using the following phase function \citep{KolnerNazarathy}
\begin{equation}
\varphi (\tau ) = \frac{{\omega _c \tau ^2 }}{{2f_T }}
\label{eq:phase1}
\end{equation}
where the time-dependent phase (in the traveling pulse coordinate system) $\varphi(\tau)$ is a quadratic function of $\tau$ and is also dependent on the \textbf{focal time} $f_{T}$, the temporal equivalent to the focal length of the spatial lens. The phase function too can be generically expressed using Taylor series around $\tau _0 $
\begin{equation}
\varphi (\tau ) = \varphi _0 (\tau _0 ) + {\frac{{d\varphi }}{{d\tau }}} (\tau  - \tau _0 )  + \frac{1}{2} {\frac{{d^2 \varphi }}{{d\tau ^2 }}} (\tau  - \tau _0 )^2  + ...
\label{eq:phase2}
\end{equation}
Comparison with Eq. (\ref{eq:phase1}) at $\tau_0=0$ suggests that the focal time should be related to the phase with
\begin{equation}
\label{eq:ft}
f_{T}  = \frac{{\omega _c }}{{\frac{d^2 \varphi}{ d\tau ^2 }}}
\end{equation}
The respective response for a time lens with this phase function is
\begin{equation}
h_L (\tau ) =  \exp \left( {i\frac{{\omega _c \tau ^2 }}{{2f_T }}} \right)  
\label{lens_function}
\end{equation}
This is the transfer function of a time-domain all-pass filter, which is mathematically analogous to the spatial lens. It will sometimes be more useful in its frequency-domain form, which can be obtained by Fourier transforming $h_L(\tau)$,
\begin{equation}
H_L (\omega ) =  \int_{ - \infty }^\infty \exp \left( {i\frac{{\omega _c \tau ^2 }}{{2f_T }}} \right) e^{-i\omega\tau}d\tau = \sqrt{\frac{2\pi i f_T}{\omega_c} } \exp \left( -\frac{if_T \omega ^2}{2\omega_c} \right)
\label{TimeLensOmega}
\end{equation}
Where $H_L(\omega )$ is the frequency-domain impulse response of the filter. This kind of integral repeats in several places along the text and is most easily solved using ``Siegman's Lemma'' \citep[p. 337]{Siegman}, which is given by 
\begin{equation}
	\int_{-\infty}^\infty e^{-Bx^2-2Cx}dx = \sqrt{\frac{\pi}{B}} e^{C^2/B} \,\,\,\,\,\,\,\, \Re(B)>0
	\label{SiegmansLemma}
\end{equation}
for any complex constants $B$ and $C$. It can be readily derived by completing the square of the exponent. 

Note that the focus $f_T$ has the units of time, so the quantity $f_T/2\omega_c$ has the same units as the group dispersion factor $\beta''\zeta$ ($\s^2 / \rad$). Thus, we define the lens curvature $s$,
\begin{equation}
s = \frac{f_T}{2\omega_c}
\label{TimeLenss}
\end{equation}
And Eq. \ref{TimeLensOmega} can be brought to the form \citep{Kolner}
\begin{equation}
H_L (\omega ) = \sqrt{4\pi i s } \exp \left( - is\omega ^2 \right)
\label{TimeLensOmega2}
\end{equation}

The time-lens processing redistributes the power of the various spectral components around the carrier---an operation that requires an active nonlinear device \citep[pp. 278--279]{Yariv}. In optics, such a lens is realized using a \term{phase modulator}, which is an active component unlike the passive dispersive medium. \citet{Kolner} emphasized that the ideal (electro-optic) phase modulator has to have a linear phase response---independent of the incoming wave amplitude. In general, the focal time is frequency-dependent, just as in spatial lenses that have a refraction index that depends on the wavelength of light and affects the lens curvature. 

Different principles of phase modulation have been developed in optics, but the one that was favored by \citet{Kolner} is a traveling-wave modulator. It harnesses a microwave oscillation that is much lower frequency than the carrier and modulates the traversing wave phase as it passes through the device. Even if a reflected backward propagating wave inside the modulator exists, a good modulator should be coupled only to the forward-propagating mode. Optical modulators generally rely on the anisotropic index of refraction and two polarization modes of the electromagnetic radiation in the medium. Optical traveling-wave phase modulators work by slowly electro-optically modulating the index of refraction in a crystal, along which the light carrier propagates \citep[pp. 429--431]{Yariv}. The modulation frequency may be smaller than the carrier, or equal to it for maximum effect. However, analogous properties and related phenomena (e.g., the electro-optic and acousto-optic effects, which can slowly modulate light) do not exist in pure acoustic fields and other physical effects may have to be harnessed in order to obtain phase modulation. Therefore, more specific details about the optical phase modulators are not central to this analysis, as this is a point of divergence for the electromagnetic and acoustic scalar wave theories. It will be sufficient to know the mathematical principle, when we attempt to identify the relevant organ within the auditory system, even if other mechanisms may be conceived to realize these functions in acoustics. To the best knowledge of the author, phase modulators have not been systematically discussed outside of the photonics/optics literature.    

\section{Summary of assumptions}
\label{SummaryofAssumptions}
Throughout the above derivation, a number of assumptions were used beyond those of classical linear acoustics, which enabled the solution. The first three assumptions are synonymous with one another and can fall under the paratonal approximation definition:
\begin{enumerate}
	\item Source: Narrowband signal
	\item Source: Slow envelope in comparison with the carrier
	\item Source/Medium: $k(\omega)$ changes slowly around the carrier; higher order terms than quadratic in the Taylor series are negligible
	\item Medium: Source-free (secondary source wavefront)
	\item Medium: Plane waves---one-dimensional propagation with no higher modes
	\item Medium: Constant absorption around the carrier, but the derivatives of $k(\omega)$ are real, or much greater than the imaginary part
	\item Time lens: phase function is level-independent (linear)
	\item Time lens: phase function is quadratic
 \end{enumerate}

Unlike diffraction and scattering problems, non-planar spatial modes---functions that vary in the $x$ and $y$ dimensions---are neglected. Therefore, the paratonal equation is particularly attractive for communication, because limiting the carrier to a single (planar) mode eliminates any interaction between envelopes of different modes, which corrupts their shape. When light propagates in an optical fiber in different fiber modes that have different dispersions associated with them, the mode can interact (beat) and give rise to \term{dispersion distortion}\footnote{This is \textbf{not} the same as intermodulation disortion, where different carrier frequencies interact. Dispersion distortion relates to the same frequencies carried in different modes with different group velocities associated with them. For a rare example in underwater acoustics, see \citet{Zhang2019}.}. In fiber optics, this is the prime reason for using single-mode fibers, whose dispersion is well-behaved, rather than multi-mode fibers that are prone to exhibit dispersion distortion, especially over long distances \citep[pp. 1--17]{Haus,Agrawal}. Single-mode optical fibers are employed almost exclusively in the communication industry, where high channel capacity is required \citep[p. 8]{Mitschke}. In the ideal design of single-mode optical fibers, the dispersive and absorptive effects of the fiber for particular carriers are minimal. 

\section{Conclusion}
The paratonal equation was introduced along with the simplest dispersive medium transformation and its active dual---the time lens. While these equations have not been used in acoustics or hearing before, they naturally fit them, given that real sources and their signals can be universally described as modulated carriers with complex envelopes. Not only do these equations provide a convenient mathematical analogy to the familiar spatial imaging theory from optics, but they also tackle the problem of envelope propagation, which has not received any rigorous closed-form treatment in acoustics. 

At this point, all the basic components are available for constructing a complete temporal imaging system, based on a cascade of dispersion, time-lens, and another dispersion, in analogy to normal imaging with spatial lens that is sandwiched by diffraction. Before continuing to develop the theory for such a system in \cref{ImagingEqs}, we would first like to identify the different dispersive auditory elements and estimate their magnitudes in \cref{paramestimate}. This will ground the discussion about the complete imaging system to the relevant parameters of the ear.

\chapter{Estimating the auditory imaging parameters}
\label{paramestimate}
\section{Introduction}
All physical systems are dispersive to a certain extent \citep{Brillouin1960} and various dispersive phenomena have been reviewed in \cref{InfoSourceChannel} to illustrate it in acoustical systems. Therefore, it should not be surprising to encounter dispersive effects in the hearing organs as well. While it has been commonly accepted that the cochlea is dispersive, other parts of the system are not explicitly considered to be so. Most importantly, second-order dispersive effects---group velocity (or group delay) dispersion---are even more rarely considered and are not given any special significance, and are often assumed to have negligible effects on hearing. 

\begin{figure} 
		\centering
		\includegraphics[width=0.9\linewidth]{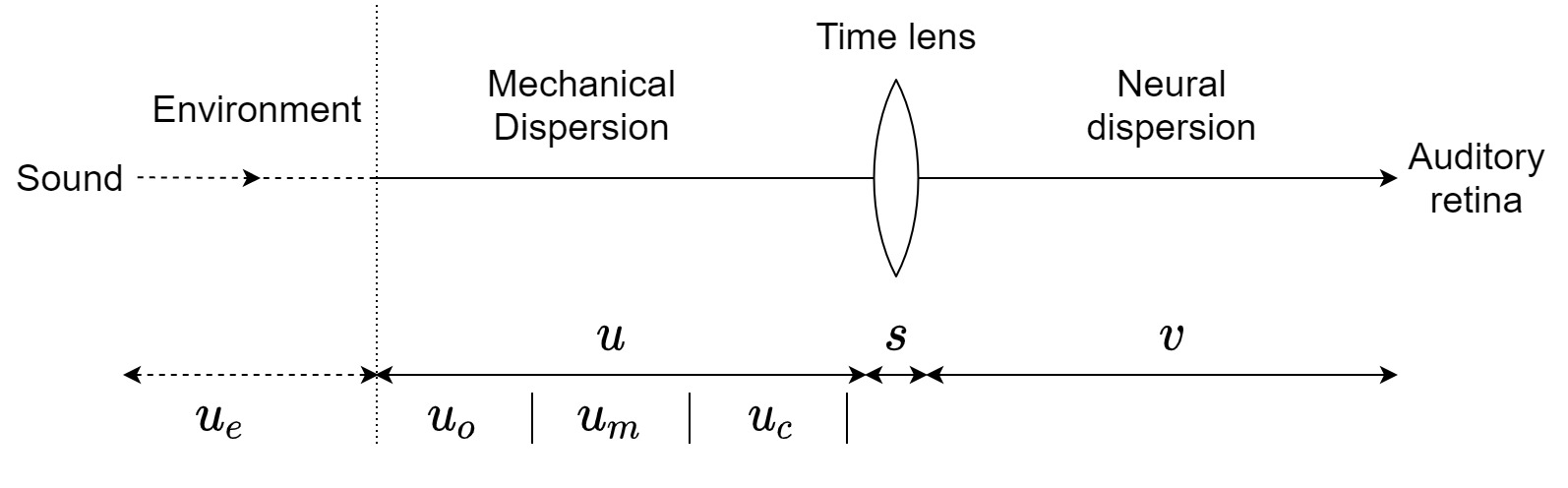}	
		\caption{The rough division of the auditory system to dispersive elements. The mechanical parts of the ear are combined into one group-delay-dispersive parameter, $u$, with contributions from the outer ear $u_o$, the middle ear $u_m$, and the cochlea $u_c$. The time lens with curvature $s$ is hypothesized to be mechanically incorporated in the organ of Corti and neurally controlled (accommodated; \cref{MOCR}). The neural group-delay dispersion $v$ begins as late as the auditory nerve, but may also comprise the inner hair cells and the passive transmission in the organ of Corti. The external group-delay dispersion in the environment $u_e$ is not considered directly in the analysis, but is assumed to be relatively low compared to $u$ in normal atmospheric conditions and over short distances.}
		\label{DispStages}
\end{figure}

Based on available findings from literature, in this chapter we attempt to estimate the human group-velocity dispersion along the auditory system---the outer ear, middle ear, inner ear, and brainstem. As will be seen, dispersion is always frequency-dependent, which results in non-negligible group-delay dispersion. Conveniently, consecutive dispersive paths are additive (being phase arguments) and may be factored into single parameters (\cref{PulseCalc}). The main segments that will be identified are the cochlear dispersion up to the organ of Corti, the time lens of the organ of Corti, and the neural dispersion from the inner hair cells (IHCs) to the inferior colliculus (IC). These segments will inform the subsequent temporal imaging analysis. See Figure \ref{DispStages} for the rough segmentation of the system considered here.

Throughout this chapter, the effects of bone conduction hearing are neglected. While it is well-known that the outer and middle ear stages of the ear may be bypassed through bone conduction \citep{Bekesy1948}, the effect is not dominant in normal conditions, with the exception of some sea mammals (\cref{OuterMiddleDiffs}) and in listeners with severe conductive losses that must rely on bone conduction for hearing.

All data from published figures in this chapter and throughout this work were digitized using WebPlotDigitizer\footnote{\url{https://apps.automeris.io/wpd/} by Ankit Rohatgi.} and analyzed in Matlab (the Mathworks Inc.).

\section{The outer ear}
\label{outerear}
The outer ear is the first organ to receive the information carried by acoustic waves in air or in water, neglecting bone conduction. By the time the sound reaches the outer ears, it has accumulated a degree of dispersion that is proportional to the distance it has traversed in the medium, often including additional reflections. As was shown in \cref{airtravel}, the atmospheric dispersion is generally negligible over small distances, but it may be susceptible to unpredictable weather conditions and to other environmental factors, which likely affect the group-velocity dispersion as well. Another uncertainty is the combination of acoustic modes that carry the information at the point of entrance to the ear. As was noted in the previous chapter, the temporal imaging equations are well-defined only for plane waves, where higher-order modes are absent. In this light, the outer ear seems to play an important role, as it imposes a unimodal, plane-wave-only, transmission over a significant portion of the audio spectrum. 

\subsection{The waveguide approximation}
To a first approximation, the outer ear is an acoustic waveguide, shaped as a pipe that is closed in one end. Wave propagation in pipes is typically analyzed in terms of normal modes, which can be spatially distributed in different ways, according to the geometry of the pipe (for illustration, see \citealp[p. 193]{FletcherRossing}). Ideal waveguides act as transmission lines and allow acoustic energy to be carried only in the plane wave mode, as long as the wavelength of the sound is much longer than the diameter of the tube, $\lambda \gg D$ \citep[pp. 471--472]{Morse}. Higher modes do not exist below a certain cutoff frequency, where their phase velocity is infinite. Above this cutoff, the phase velocity decreases quickly. If the tube walls are yielding---if they are not entirely rigid, but have a finite compliance---then they locally react to the pressure gradient of the plane wave\footnote{If the point of the surface that is impacted by the wave does not interact with its vicinity, then the surface is said to be \textbf{locally reactive}. If the surface reacts as a whole (like a membrane), then its impedance is of \textbf{extended reaction}.} \citep[pp. 475-477]{Morse}. This results in frequency-dependent phase velocity as a function of the wall material and resonances for the particular pipe geometry. It means that the low-frequency waves travel faster than the high-frequency ones---dispersion. When the one-dimensional plane-wave approximation breaks down, additional modes appear as some of the sound waves begin to propagate along the waveguide circumference rather than in its center \citep[pp. 688--689]{Morse}. Applying the simplistic rigid-wall waveguide limit to the human ear canal geometry, and using a typical ear canal diameter of 0.7 cm \citep[cited in \citealp{Rabbitt1988}]{Goode}, or from cross-sectional data 0.75 cm \citep{Rosowski1994Fay}, indicates a strict plane-wave propagation of up to about 24.5 kHz of sound in air, at normal room temperature, for an open-ended waveguide. \citet{Keefe1993} reported growing diameters with age with adult diameter of 1.04 cm, corresponding to a 16.5 kHz cutoff. However, these cutoff values are unrealistic, as is shown next. 

\subsection{Higher-order modes}
In a real outer ear, strict plane-wave propagation breaks down at much lower frequencies than predicted by the waveguide approximation due to the complex geometry of the ear canal. When sound arrives from the environment to the outer ear, it is scattered by the concha, which creates various non-planar modes, mainly at high frequencies \citep{Rabbitt1988, Rabbitt1991}. However, these modes quickly vanish once inside the ear canal, as the plane-wave mode becomes dominant within a few millimeters, even at high frequencies \citep{Rabbitt1991, Hudde}. Three-dimensional simulations of sound waves in the bent human ear canal showed that the ear canal has additional non-planar modes that are trapped around its bends, but these modes also vanish very quickly and do not interfere with the plane wave propagation \citep{Hudde}. 

Analytic approximation to the solution of the wave equation of the ear canal found that higher vibrational modes start to be present from about 4 kHz \citep{Rabbitt1988}. These modes are formed by the eardrum (the pars tensa) itself due to its elasticity and geometry that is detached from the ear canal walls. With increasing frequency the eardrum modes tend to extend spatially to the interior of the ear canal, and these modes are reflected back to the canal and interact with its trapped modes. This may explain an effect of probe microphone response variance as a function of distance from the eardrum above 4 kHz \citep{Caldwell2006}. Holographic measurements of the eardrum revealed modes above 1 kHz, which grow in dominance at higher frequencies \citep{Cheng}. The effect extends even to the middle ear, as impedance measurements of the cat's middle ear were best modeled by including standing waves of the eardrum above 3 kHz, which produced a measurable transmission delay \citep{Puria1998}. Therefore, the dominant non-planar, and thus dispersive, mechanism in the ear canal is not a result of yielding walls, but rather of the tube coupling to the compliant, oddly shaped eardrum, which is itself yielding.

All together, the pressure wave that arrives to the middle ear is the sum of all the modes that make it to the eardrum. So, at frequencies below 5 kHz the relative coupling of the non-planar higher-order modes in the ear canal to the movement of the eardrum is about 10\% for children and close to 30\% for adults, and about 25\% at 4 kHz \citep{Rabbitt1988}. \citet{Hudde} found that the eardrum minimally disturbs the plane-wave mode below 4 kHz, despite its compliance and middle ear resonances above 1 kHz. 

One question remains unanswered regarding the high-frequency domain above 4 kHz, where non-planar modes carry more energy: is there any dispersion distortion (\cref{SummaryofAssumptions}) that affects the information entering the middle ear? The topic has not been considered at all in the acoustic literature. However, indirect data from the cat suggest that dispersion distortion may be a real problem. Probe microphone measurements in the cat's ear canal show that above 10 kHz the variation of pressure over the eardrum surface makes it impossible to have one reference or mean level that is confidently conducted to the middle ear, due to anomalous high-frequency response \citep{Khanna}. In another perspective, it was demonstrated through simulations that the multitude of normal modes at high frequencies is advantageous in terms of energy distribution, and hence, power transmission to the middle ear \citep{Fay2006}. 

In conclusion, treating the ear canal as one-dimensional plane wave conduit is a justified assumption below 4 kHz in humans, in line with the dispersion equation assumptions. At higher frequencies, the validity of this assumption is expected to progressively drop, but to an unknown degree. The specific cutoff is most certaintly different in other animals with different ear geometries. 

\subsection{The group-velocity dispersion of the outer ear}
The group-velocity dispersion will be estimated using published phase or group delay data.

\begin{figure} 
		\centering
		\includegraphics[width=0.9\linewidth]{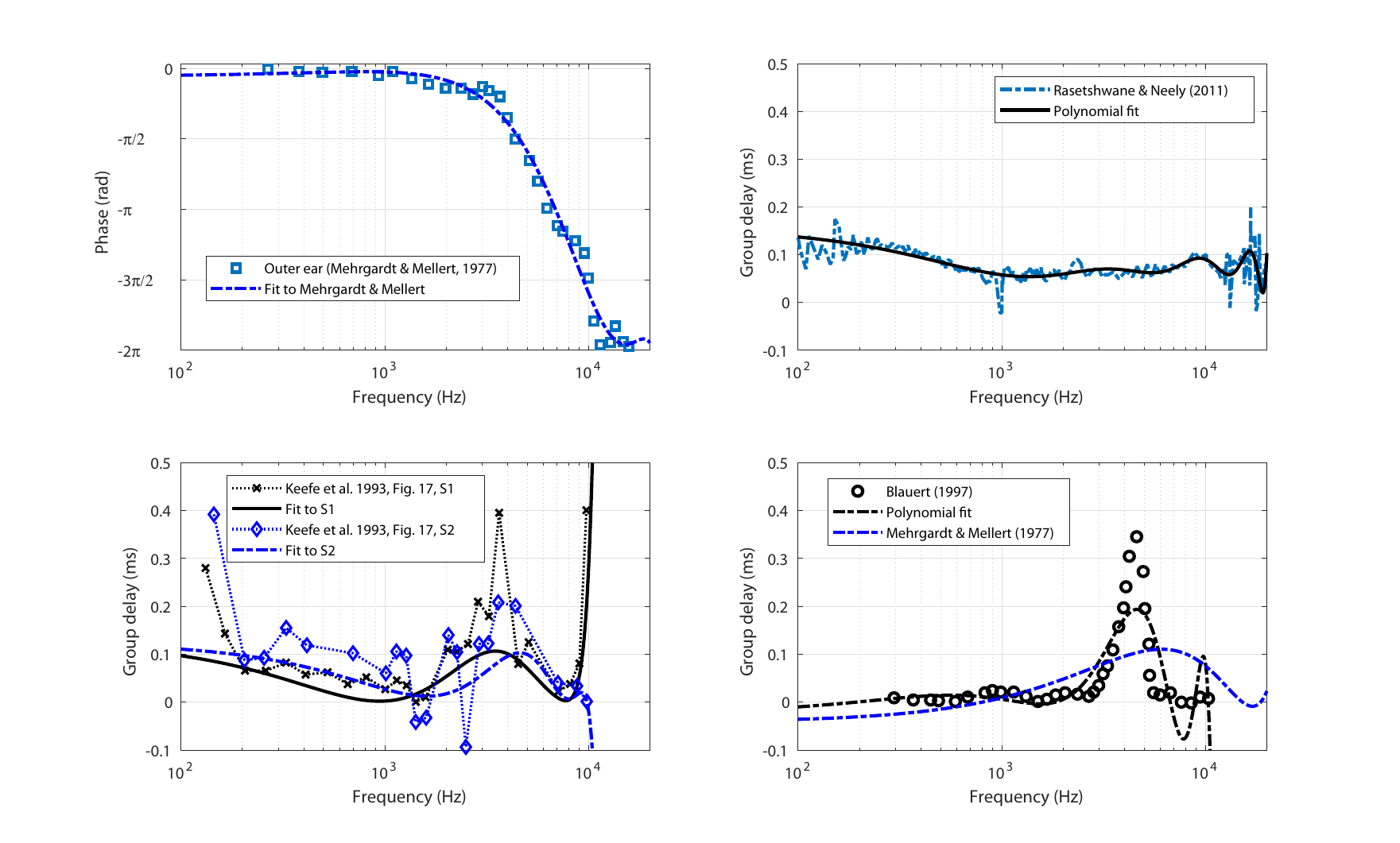}	
		\caption{Extracted and fitted phase and group delay of ear canal data from literature. \textbf{Top left:} Phase response of a single subject from \citet[Figure 6, bottom]{Mehrgardt1977}, using a probe microphone at the eardrum referenced to the ear canal entrance. \textbf{Top right:} Group delay data based on reflectance measurements on 24 subjects \citep[Figure 4, bottom]{Rasetshwane2011}. \textbf{Bottom left:} Ear canal reflectance group delay data of two subjects \citep[Figure 17, right]{Keefe1993}. \textbf{Bottom right:} Direct ear canal group delay measurements of 11 subjects, source at 4 mm inside the ear canal, and probe microphone by the eardrum \citep[Figure 18, bottom, $0^\circ$]{Blauert1997}. The bottom polynomial fits are 6th order, the top left is 4th order, and top right is 8th order.}
		\label{earcanalGDphase}
\end{figure}

In measuring the phase response of the outer ear, the results may be susceptible to large errors due to small variations in measurement positions at frequencies above 4 kHz, the small dimensions involved, the resonances of the ear canal, finite dimensions of the microphone, and access to the eardrum \citep{Brass1997,Caldwell2006}. Figure \ref{earcanalGDphase} reproduces ear-canal phase and group delay data compiled from various studies, which employed different techniques to obtain phase measurements, all at slightly different measurement positions. Ear canal phase data was obtained from three subjects by \citet{Mehrgardt1977} for a free-field source by subtracting the response of a free-field microphone positioned at the ear canal entrance from the response of a probe microphone near the eardrum, 20 cm from the entrance (top left). Data from \citet{Rasetshwane2011} (top right) are full-spectrum reflectance group delay measurements averaged from 24 individual subjects. Similar data from two additional subjects were reported by \citet{Keefe1993} for a narrower spectrum (bottom left). These measurements were obtained by sealing the ear canal, and flush-mounting a miniature sound source on the seal, while a probe microphone was positioned right outside the eardrum. The group delay of these measurements accounts for the round trip of the pressure wave, so they were divided by two \citep{Voss}. Finally, direct measurements of the group delay in the ear canal of 11 subjects were also provided by \citet{Blauert1997}, for a sound source positioned 4 mm inside the ear canal, and a probe microphone close to the eardrum (bottom right). All datasets were polynomially fitted in order to obtain smooth functions of group delay from phase, and group-delay dispersion from the group delay. The polynomial fits are displayed in Figure \ref{earcanalGDphase} as well.  

\begin{figure} 
		\centering
		\includegraphics[width=0.5\linewidth]{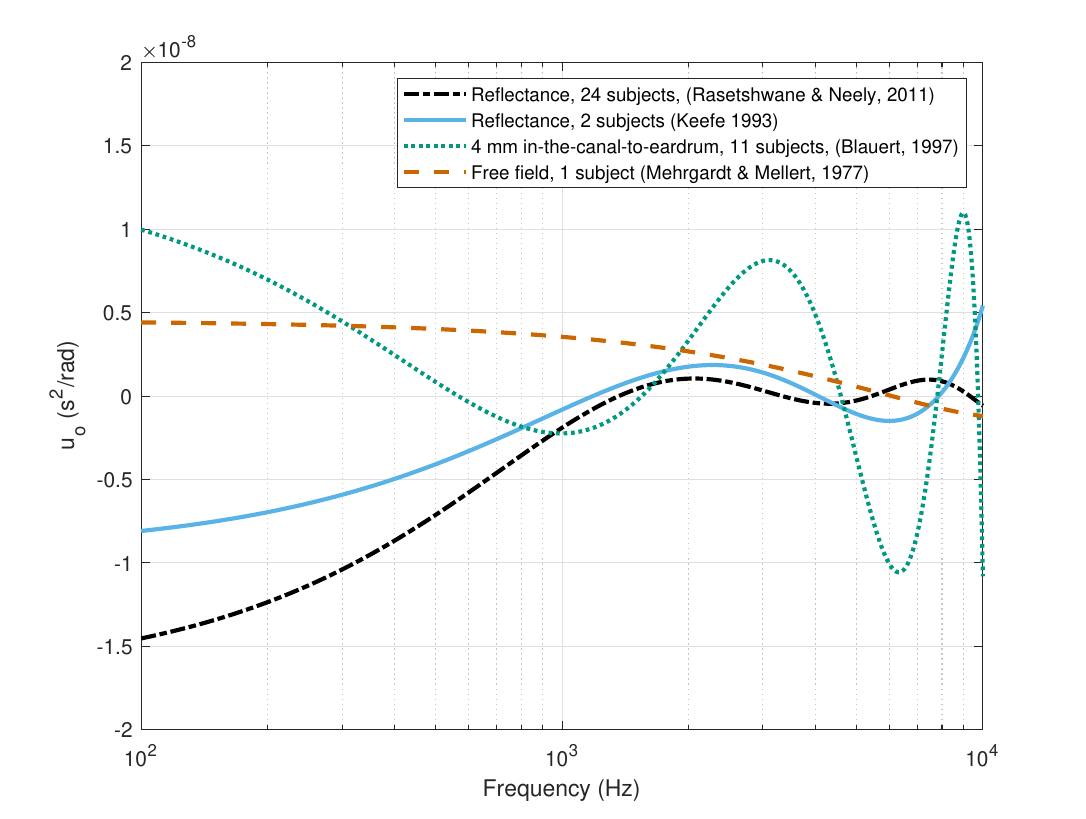}	
		\caption{The group-velocity dispersion of the group delay data plotted in Figure \ref{earcanalGDphase}, according to Eq. \ref{GDvsGVD}.}
		\label{earcanalGDphase2}
\end{figure}

Using the fitted phase and group delay functions, the outer ear dispersion coefficient $u_o$ can be readily computed with
\begin{equation}
u_o = \frac{1}{2}\frac{d\tau_g}{d\omega} = -\frac{1}{2}\frac{d^2\phi}{d\omega^2} = \frac{\beta^{''}_o\zeta_o}{2}
\label{GDvsGVD}
\end{equation}
according to Eq. \ref{GDDder}. The resultant group-delay dispersion from all datasets is plotted in Figure \ref{earcanalGDphase2}. The data exhibit large variability that reflects the relative microphone and source positions and, perhaps, the measurement methods themselves. The estimates fluctuate between negative and positive values at different spectral regions, but is bounded for $|u| \le 1.5 \cdot 10^{-8}$ $\s^2/\rad$. Below 100 Hz and above 10 kHz the estimates are not displayed because of insufficient data and, hence, poor fits. 

The estimated values of the ear canal dispersion indicate that unless a larger and stabler group-velocity dispersion segment follows the outer ear, auditory imaging may suffer as a result of the frequent sign changes as a function of frequency. 

\section{The middle ear}
The middle ear appears to have relatively simple vibrational dynamics in comparison with both outer and inner ears, as its movement is essentially uniaxial and linear. While plane-wave movement is irrelevant here, the paratonal conditions can equivalently apply as long as the vibration is unimodal and one-dimensional.

\subsection{The middle ear vibrational modes}
The ossicular chain of the middle ear receives vibrational energy from the eardrum movements, which reflect the summed modes at the output of the outer ear. Measurements on human cadavers show a rather linear response in amplitude and phase of the middle ear, but they also indicate that there are several resonant modes between 1.2 kHz and 2 kHz \citep{Homma}. \citet{Voss} found that the middle ear ossicular movement is dominated by translational movement of the bones up to 1 kHz, which may be combined with additional higher-order modes at higher frequencies. High order modes are dominant at high frequencies in different animals (above 3--4 kHz in humans), where they are thought to improve sound transmission in mammals despite the ossicular mass and may even be a factor in their extended hearing range, in comparison with other vertebrates \citep{Puria2010,Rosowski2020}. Thus, we may also expect some level of dispersion distortion from the middle ear, which increases with frequency. Nevertheless, the frequency and phase responses generally show a linear, well-behaved transfer function of a low-Q bandpass filter centered at around 1 kHz \citep{Voss, Aibara2001, Sun, Homma}. Therefore, if the middle ear has any impact on the single-mode transmission, then it should appear only above 1--2 kHz and is not expected to be particularly strong. Thus, the middle ear dynamics appears to be effectively aligned with the plane-wave single mode assumption required for the temporal imaging theory. Any anomalous behavior in its response likely reflects the higher-order modes (and possibly the dispersion distortion) of the outer ear. 

Note that this analysis neglects the middle ear reflex, which acts as an automatic gain control at medium-high sound pressure levels (\cref{MiddleEar}). However, during fast transitions, the reflex may have a transient dispersive effect as well.

\subsection{Middle ear group-velocity dispersion} 
\begin{figure} 
		\centering
		\includegraphics[width=1\linewidth]{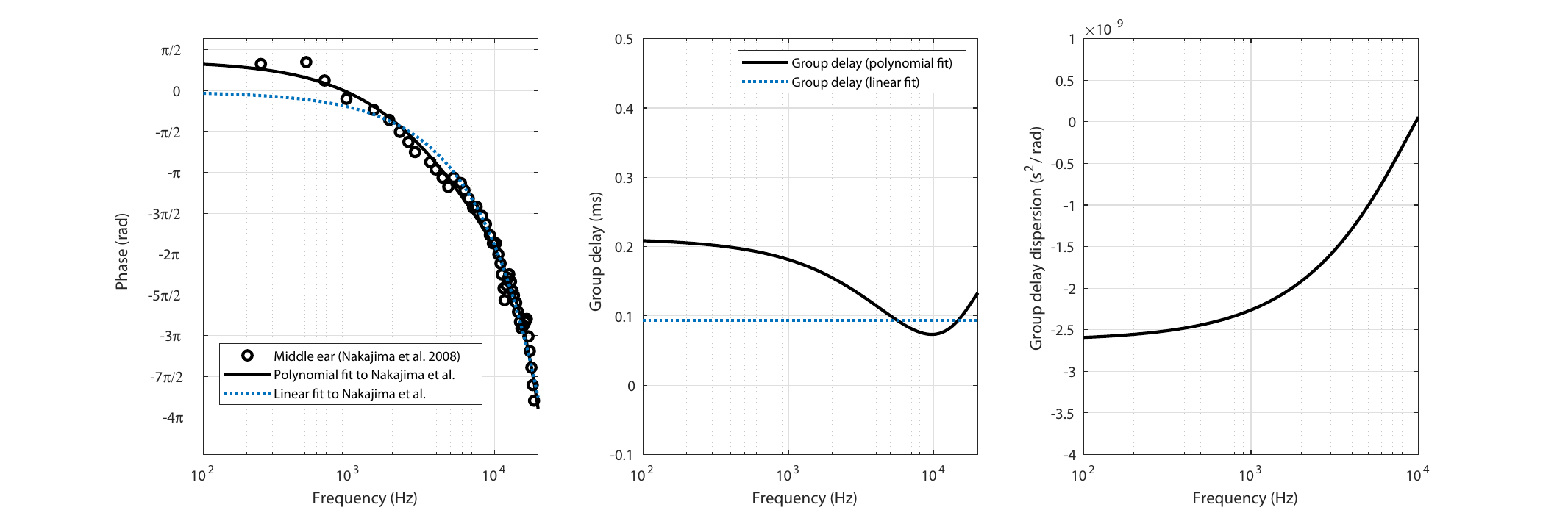}	
		\caption{Middle ear phase response data from six temporal bones of human cadavers, extracted from \citet[Figure 5, bottom]{Nakajima}. \textbf{Left:} Phase response data fitted with linear and fourth-order polynomial functions. \textbf{Middle:} The derived group delay data, which is constant for the linear fit. \textbf{Right:} The group-delay dispersion is very small for the polynomial fit and is identically zero for the linear fit (not shown).}
		\label{middleearphase}
\end{figure}

The middle ear phase response was measured in six temporal bones of human cadavers by \citet{Nakajima}, from which the group delay and group-delay dispersion could be calculated (Figure \ref{middleearphase}, left). The phase response is very close to being linear, which means that it has about constant group delay, and almost negligible group-delay dispersion. Nevertheless, a fourth-order polynomial better modeled the data than a linear fit. The polynomial fit was used to calculate the group delay (middle) and the group-delay dispersion (right), which has a smaller magnitude than the outer ear with $|u_m|<3\cdot 10^{-9}$ $s^2$ / rad. The linear-phase alternative produces $u_m = 0$ throughout the spectrum, which is an unphysical result.

\section{The inner ear: oval window to the outer hair cells}
As the complexity of the cochlear anatomy and mechanics is much greater than both the outer and middle ears, there are several ways to segment the wave propagation before it reaches the auditory nerve. Unlike the other parts of the ear, cochlear dispersion is relatively well-documented and is sometimes considered a defining feature of the cochlear structure\footnote{For example, see references to the cochlea as an ``acoustic prism'' in \citet{Shera2002}, \citet{Oxenham2014}, and \citet{Altoe2020}.}. However, the presence of the outer hair cells (OHCs) does not square with the conditions for a source-free propagation, since they generate sound through their motility \citep{Kemp1978, Ashmore2008}. Additionally, they do not constitute a passive medium for propagation, as their nonlinear amplificative nature may be taken as negative absorption, whereas the traveling wave of the passive basilar membrane itself is highly dampened within the cochlea. Therefore, the cochlear region of dispersion is defined here to include the passive path only up to the OHCs, which will be dealt with separately in \cref{lenscurve}. 

\subsection{Single-mode traveling wave}
The cochlear fluid is forced by the oval window movement that is driven by the one-dimensional movement of the stapes footplate---the last bone of the ossicular chain. According to one of the simplest and most influential one-dimensional models of the cochlear dynamics, fast pressure waves in the incompressible cochlear fluid propagate from the oval window to the round window---first through scala vestibula via the helicotrema and into scala tympani \citep{Peterson1950}. The pressure difference between the two chambers produces a much slower differential pressure wave that produces the transverse traveling wave along the cochlear partition, and specifically the basilar membrane (BM), which separates the two scalae. The fast and slow waves can be viewed as independent modes of transmission that can be expressed as plane waves. In this sense, both modes contain the acoustic information from the outside world. While the slow traveling wave theory has received most of the attention in modeling over the years \citep{Bekesy1960}, there is still some controversy as for the exact energy balance between the two modes and the exact role of the fast wave \citep{Robles2001}. For example, there are various documented conductive loss cases where information reaches the auditory nerve, despite of a lack of a traveling waves \citep{Sohmer2015}, or hearing in the absence of traveling wave in the ears of lizards and frogs, which have close but somewhat different auditory anatomy and mechanisms to mammals \citep{Bell2012b}. The role and relative effect of the fast wave are controversial, but they appear to not be altogether  negligible \citep[e.g.,][]{Lighthill1981, HeRen2008, Bell2012, Recio2015}.

Once it is transformed to a traveling wave inside the cochlea, the movement is often considered one-dimensional, although more realistic models of the cochlea are two- or three-dimensional. In the basal region before the characteristic frequency (CF) peak, most analytical models, including nonlinear ones, assume a one-dimensional wave motion with no additional modes \citep{Zweig2015}. This assumption is often generalized to the peak area itself, where the fluid is said to maintain laminar flow \citep[pp. 58, 109--110]{Duifhuis}. In the class of solutions referred to as the ``long-wave approximation'' models, the geometrical distribution of the peak resonance over the width of the BM is neglected, and the velocity of the fluid around the peak is redistributed to reduce the problem to a single dimension---still giving good agreement with observations---even though the latter assumption is wrong \citep{deBoer1996}. While increased dimensionality in the modeling is physically essential to produce the resonance of the BM, the modeling advantage of going to three over two dimensions may be marginal \citep{Zweig1991,deBoer1996}. Higher-dimensional models sometimes treat the fluid as three-dimensional, but still assume a one-dimensional array of resonators \citep{Zweig2015, Zweig2016}, or a transmission line \citep{Peterson1950,Verhulst2012}. 

The traveling wave itself is usually modeled as unimodal as well, but there are indications that it may not be the case throughout the cochlea. A second mode was suspected as contributing to the nonlinear dynamics unraveled by \citet{Rhode1971}. Higher-order vibrational modes that were found useful in early modeling attempts of the cochlear partition were also considered to be a necessary ingredient of cochlear models that should account for anomalous click glides \citep{Lin2004}. A finite-element method (FEM) simulation of a simplified passive cochlea (a straight box model with a single partition as the BM) decomposed the traveling wave to orthogonal modes \citep[Figure 7]{Elliott}. It was found that the fundamental mode at 1 kHz is 25--30 dB stronger than the second strongest mode. Evanescent modes became more significant only more apically than the CF (after the peak), but they decayed relatively quickly farther away. These findings are similar to \citet{Watts2000}, where some cochlear modeling inconsistencies were resolved by adding a second mode after the peak, which was also hypothesized to account for Rhode's observations. 

Another assumption that is important to keep in check is the lack of dominant reflections that affect the forward-propagating waves in the cochlea. According to some evoked otoacoustic emission (OAE) models, the emitted spectrum is the result of multiple reflections from the basal end of the cochlea or from the helicotrema \citep{Kemp1978}. However, the existence and exact nature of such reflections are not settled matters \citep{KempManley2008}. For example, reflections from irregularities in the cochlear walls may interfere with the propagating wave in the BM and there is some evidence from interferometric and OAE measurements of the chinchilla that it creates ripples (micro-structure) in the BM spectrum, phase, and multiple-lobe envelope response to clicks (\citealp{Shera2013}, but see \citealp{He2013, Wit2015, Shera2015}). While these small ripples seem to occur in many click measurements, some argue that the contribution of reflections to the overall cochlear response may be safely neglected \citep{deBoer1984}. A reverse traveling wave was inferred to be present from the measurements of ex-vivo gerbil cochleas both at basal and apical positions relative to the characteristic frequency in the first and second cochlear turns \citep{Zosuls2021}.

In summary, the assumption of the single-mode transmission appears to be good only in first approximation, as it may be violated more apically than the CF peak. The exact effect of the higher-level modes or internal reflections on the neural coding and eventual perception, however, is not at all clear, especially since much of their analysis has been done in simulations, simplified theoretical models, or animals. Nevertheless, we shall assume that these effects are small enough to be negligible, while focusing on the qualitative first-order response of the cochlea in their absence. This approach is going to be surprisingly effective, despite the mitigating approximations.

\subsection{Cochlear dispersion and group-velocity dispersion}
\label{sec:curvature}
It was \citet{Bekesy1949} who first observed that different pure-tone frequencies appear with different delay between the stapes and their corresponding CF resonance on the basilar membrane. In the most immediate interpretation, the differential delay reflects the different paths that the traveling wave information takes to arrive to the peak region. The basal end of the BM, close to the oval window input, responds to high frequencies faster (it peaks earlier) than the apical end responds to low frequencies, due to the frequency-dependent impedance of the BM. A more physically rigorous explanation was provided by \citet{Ramamoorthy2010}, who showed that even a simplified system with a uniform plate (modeling the BM) coupled to a fluid-filled duct exhibits dispersion. The mechanical dispersion translates to dispersion in the neural encoding of different frequencies.

As it turns out, the group delay itself is also frequency dependent as was first observed neurally in rats, where it was found that the input frequency slopes of FM tones were not conserved at the output \citep{Moller1974}. The change in instantaneous frequency is a characteristic of the impulse response of the basilar membrane and is referred to as a frequency \term{glide} (\citealp{deBoer1997}; see Table \ref{LinearFMterms}). Further direct observations were obtained in different animals, although the glide direction may vary between species and CFs \citep[e.g., ][]{Recio1998,Carney1999,Recio2005, Wagner, Recio2015}. Pyschoacoustic confirmation for dispersion has been obtained several times as well \citep[e.g.,][]{Smith1986, Kohlrausch,OxenhamDau, OxenhamDau2, Summers, Shen2009}, where the curvature has been found to be negative and to increase with frequency, contrary to findings in certain animals. \citet{OxenhamDau2} noted that the phase behavior cannot be predicted by simple auditory filter models. Indeed, inconsistent estimates of group delay as a function of frequency were computed using seven different cochlear models \citep[Figure 6A]{Saremi}. Simulating clicks of 1 kHz carriers, the modeled group-delay slopes around 1 kHz were found to be inconsistent in sign and in their functional form (linear or curved). However, many of these studies do not make a clear distinction between dispersion arising in the cochlea itself and other dispersive contributions from the rest of the auditory system (but see \cref{NeuDispExist}).

While there is some inconsistency regarding the exact mechanism behind the frequency glides, as well as their exact frequency dependence in humans, there is no doubt that they exist. Although the glide slopes are not always straight, none of the cited studies advocated for phase terms that are higher than quadratic. Thus, in the vicinity of the CF, a linear curvature seems to be an acceptable assumption. This assumption will be challenged in \cref{PhysioOffFreq}.

\subsection{Estimating the cochlear group-delay dispersion}
Several attempts at estimating the group delay of the cochlea have been published that employed different methods, all of which contain rather strong assumptions that make the estimates uncertain to some extent. For example, both evoked auditory brainstem response (ABR) and evoked transient OAE (TOAE) have been used as indirect methods to estimate the cochlear group delay. For this to be the case, their output must contain exactly the same dispersive contribution from the cochlea and it should be ensured that neural group-delay dispersion is negligible. This was the conclusion of an early attempt to compare the estimates from the two methods in \citet{Neely}, where data from separate TOAE and ABR studies were similar enough, so that the contribution of the neural pathways to the responses was considered to be a constant delay (i.e., that results in zero group-delay dispersion). However, a more recent study that repeated the comparison using simultaneous measurements of ABR and TOAE, using the same stimuli and subjects, could not establish an identical group delay of the two measures, regardless of the specific parameters used for the stimuli \citep{Rasetshwane}\footnote{The ABR and TOAE methods will be revisited in the section about neural dispersion \cref{NeuralDisp}, where these methods appear to have no alternatives, at present.}.

A somewhat more transparent cochlear group delay estimation method was therefore favored, based on a group delay map measured in the chinchilla and transformed to human \citep{Temchin2005,Ruggero2007}. In-vivo cochlear group delay was measured between the eardrum and the auditory nerve of the chinchilla using the Wiener-kernel method for obtaining the nonlinear impulse response from white noise \citep{Temchin2005}. Additional post-mortem group delay measurements of the chinchilla allowed \citet{Ruggero2007} to form a re-tuned map for the cochlea that could be used to transform between the live and post-mortem measurements. It was also corrected for frequency-dependent phase shifts as a result of death, which reflect the active effect of amplification in the live cochlea. Then, due to scaling similarities between all mammals and in particular the similarity between the chinchilla and human hearing ranges, the authors were able to transform human cadaver data to a live map of group delay \citep[Figure 7]{Ruggero2007}. The group delay was corrected also for the middle ear and constant synaptic and neural conduction delays \citep[see their Figure 8 caption]{Ruggero2007}. The group delay data were shown to agree with a large pool of animal data, including non-mammalian vertebrates, despite widely different morphologies.

It is arguable whether the post-mortem or the live group delay data should be used in the computation of the cochlear group-delay dispersion. The post-mortem data entails OHC inactivity that removes any amplificative phase effects from the total dispersion, which are present especially at low levels. But it also broadens the cochlear filter significantly, which has an effect that extends apically from the best frequency site and may distort the phase response. The live data, in contrast, has a normal filter response, but ostensibly includes the active OHC effect\footnote{The live data refer to the active status of the OHCs, which provide frequency selectivity and compressive gain in normal listening conditions. In this work, the OHCs also have a role in time lensing, which is phase modulation in the time domain (\cref{lenscurve}). Hence, the effect of phase modulation can be thought to affect the cochlear group delay and group-delay dispersion figures from \citet{Ruggero2007}. However, the nonlinear analysis in \citet{Temchin2005} and \citet{Ruggero2007} is based on Wiener-kernel method, which requires white noise as input to the nonlinear system. Typically, the system nonlinearity is considered static (time-invariant) \citep{deBoer1978,vanDijk1994}. However, there is nothing about the operation of the OHCs that suggests that it is static. While we do not know if the two functions are related, the amplificative OHC function could theoretically be almost instantaneous \citep{Altoe2017}, whereas the time lens operation may require buildup time to arrive to the operation point of its modulated stiffness (\cref{lenscurve}; see also \cref{LockInPLL}). For such a system to work, the input has to be (partially) coherent, rather than totally incoherent (white noise). This means that the Wiener-kernel method may fail to engage the time-lens functionality of the OHCs and therefore may not disclose any phase modulation curvature. The exception may be at low frequencies, where the auditory narrowband filters can significantly cohere the input (\cref{PLLNoise}). Of course, bypassing the time-lens is exactly what we would like to achieve in order to get a clean estimate of the cochlear dispersion.\label{WienerKernel}}. Both responses include a mechanical dispersive path associated with the IHCs, which cannot be subtracted using the available data. As it turns out, the difference between the two datasets is relatively small, although the live data seem to produce stabler results in some of the calculations throughout this work. 

\begin{figure} 
		\centering
		\includegraphics[width=1\linewidth]{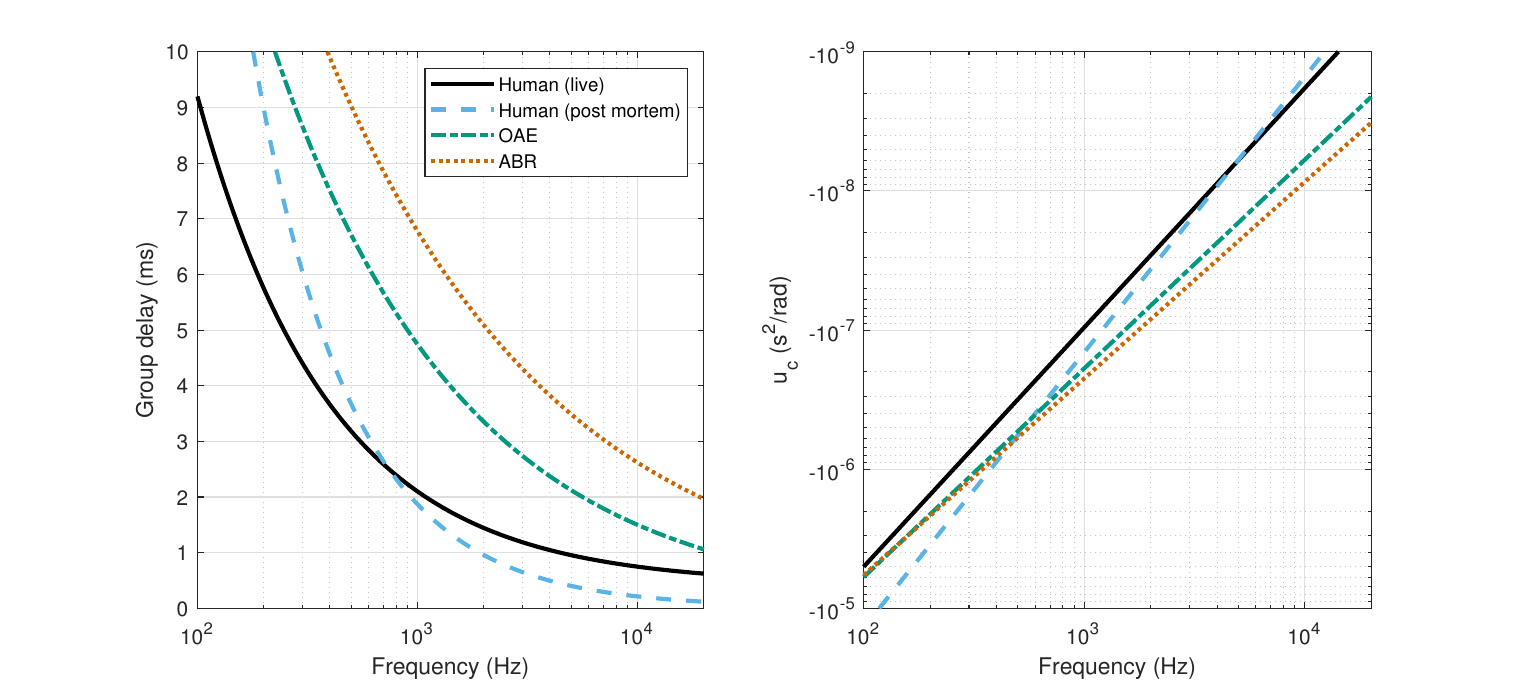}	
		\caption{\textbf{Left:} Live (black solid) and post-mortem (blue dash) group delay of the human cochlea, based on human cadaver data compiled by \citet[Figure 7]{Ruggero2007}, which were corrected for the effects of death using a chinchilla group delay cochlear map from \citet{Temchin2005}. Additional estimates based on closed-form functional fits are based on OAE measurements (green dash dot) \citep{SheraGuinan2000, Fobel} and ABR (red dot) \citep{Neely}. \textbf{Right:} Group-delay dispersion derived from the group delay curves on the left.}
		\label{inner}
\end{figure}

The group delay functions are plotted in Figure \ref{inner}, left, for the live and post-mortem human responses. The functions are affine power-law fits, reproduced from the functions in \citet[Figure 13]{Temchin2005}. The live data fit (solid black) is
\begin{equation}
	\tau_{g,live} = 0.43 + 1.67f_{kHz}^{-0.72} \,\,\,\, \ms
\end{equation}
where the group delay $\tau_{g,live}$ is given in ms and the frequency $f$ in kHz. Similarly, the post-mortem group delay function is given by
\begin{equation}
	\tau_{g,dead} = 0.02 + 1.85f_{kHz}^{-0.98} \,\,\,\, \ms
\end{equation}
The cochlear group-delay dispersion $u_c$ can be directly obtained by differentiating these expressions with respect to $\omega$ and dividing by 2, according to Eq. \ref{GDvsGVD} (Figure \ref{inner}, right). Additionally, for comparison, some of the above-mentioned evoked ABR and OAE group delay data are plotted as well. The OAE is from \citet{SheraGuinan2000} and \citet{Fobel} and the ABR is from \citet{Neely}.

\section{Total group-delay dispersion of the inner ear}
\label{Total_u}
\begin{figure} 
		\centering
		\includegraphics[width=.5\linewidth]{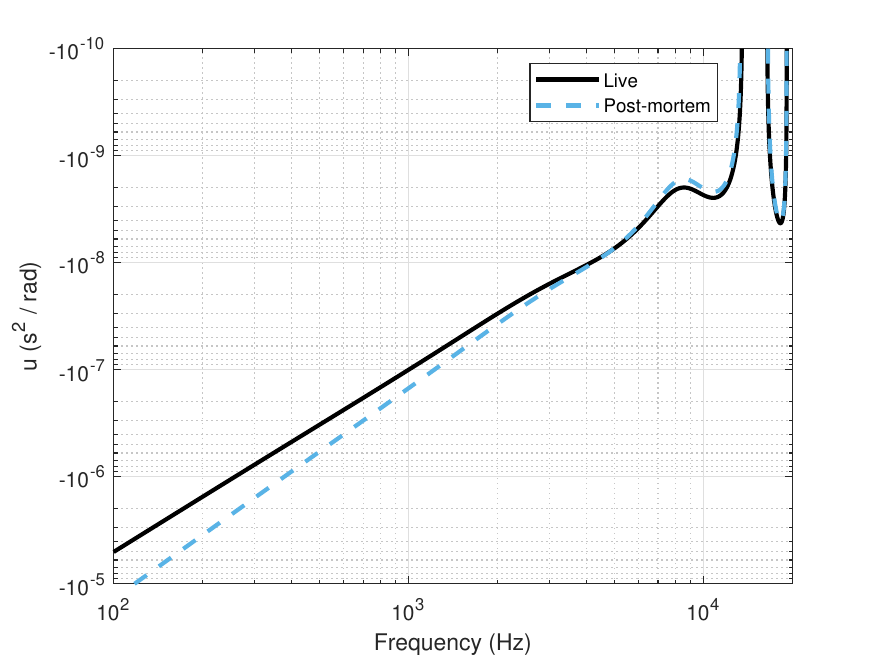}	
		\caption{Total group-delay dispersion of the outer ear \citep{Rasetshwane2011}, middle ear \citep{Nakajima}, and cochlea \citep{Ruggero2007}. The discontinuity above 10 kHz represents a sign change, where the cochlear dispersion no longer dominates the input dispersion and all three estimates are unreliable.}
		\label{Reu}
\end{figure}

Combining the dispersions of the outer ear ($u_o$), middle ear ($u_m$), and cochlea ($u_c$), we can obtain an estimate for the total input dispersion of the human auditory system
\begin{equation}
	u = u_o + u_m + u_c
\label{utotal}
\end{equation}
The total group-delay dispersion is plotted in Figure \ref{Reu} both for the live and for post-mortem responses, which merge above 3 kHz. The outer ear was taken as the relatively ``well-behaved'' average response from \citet{Rasetshwane2011}, which was based on many more subjects than the other datasets (Figure \ref{earcanalGDphase}). The middle ear data were based on the only dataset that was analyzed here from \citet{Nakajima}. 

The most important feature of the total input group-delay dispersion is that it is mostly dominated by the cochlear group-delay dispersion. It means that it is not subjected to fluctuations in frequency caused by the outer ear acoustics. Otherwise, more sign-change ``holes'' in the group-delay dispersion curve could have dominated the total group-delay dispersion, as is seen at around 16 kHz in Figure \ref{Reu}. The same logic applies to the atmospheric dispersion that can be dominant in extreme weather conditions or very long distances (Figure \ref{atmo}). These effects may be absorbed by the relatively large cochlear dispersion (Figure \ref{DispStages}).  

Because of its dominance, we will often refer to the total input group-delay dispersion $u$ simply as cochlear dispersion. 

\section{The inner ear: time lensing by the outer hair cells}
\label{lenscurve}
The time lens is the second new function of the OHCs that is hypothesized in this work. The first one was of a phase-locked loop (PLL; \cref{PLLChapter}). Despite their differences, the two may not be altogether independent as will be suggested in \cref{MOCR}. However, this section, while presenting five separate lines of evidence and a hypothetical mechanism, may be rightly considered speculative---even more than the PLL---especially given that it relies on an acoustic phenomenon that has not been previously modeled (phase modulation)\footnote{But see a discussion about an apparent phase modulation in the chinchilla's BM basal response to clicks in \citet{Recio2000}, where it was suggested that the modulation is a result of nonlinear and compressive processes and is likely not a mere artifact of BM motion.}. Nevertheless, the utility of this proposal will be essential for the temporal imaging theory, and hence for the rest of this work. As will turn out, the empirical evidence we have is sufficient to demonstrate that phase modulation does exist, but extrapolating its magnitude to humans will prove challenging due to several unknowns in the process. We will therefore aim to estimate the upper and lower bounds for the phase modulation in humans and later discuss how the different bounds can relate to different known responses of the ear. 

\subsection{Stiffness-dependent traveling-wave phase modulation}
\label{StiffnessPM}
In the following, a general formulation of acoustic phase modulation will be proposed, which depends on stiffness variation of the medium. Specifically, a corresponding mechanism will be proposed for how phase modulation of the traveling wave can emerge as a result of the unique features of the OHCs, and by proxy, the organ of Corti. Because of the paucity of direct empirical data, it is kept largely qualitative and, arguably, oversimplified. 

Let us examine the phase velocity of a narrowband disturbance, as it propagates from the base of the cochlea to its apex as a traveling wave, through the site of the CF resonance. The speed of propagation depends on the local density of the BM and its Young's modulus (or its stiffness, if it is modeled as a one-dimensional oscillator array). It is well-established that the BM stiffness (and the stiffness of other supporting cells in the organ of Corti) varies continuously and monotonically along the BM due to geometrical changes \citep[pp. 466--469]{Naidu1998,Emadi2004,Teudt2014,Bekesy1960}. Additionally, around the resonance, the stiffness of the BM changes with the electromotile actuation of the OHCs \citep{He1999, Zheng2007}, which are embedded in the organ of Corti that is attached to the BM with the supporting Deiters cells \citep{SlepeckyDallos1996}. When the traveling wave moves from the base toward the site of resonance, its movement gradually causes more vigorous hair bundle deflections, which in turn gate a stronger current in the OHCs and raises their intracellular potential. Apical to the resonance, the mechanoelectric activity decreases. Therefore, the somatic stiffness of the OHC is effectively modulated with the OHC potential, which in turn modulates the speed of propagation and the phase of the traveling wave in the BM. While there is some controversy about the voltage dependence of the OHC stiffness in vivo \citep{Hallworth2007,Dallos2008,Liu2009}, even a small effect can produce the phase modulation needed in a way that does not violate the observations by \citet{Hallworth2007}, who did not find significant stiffness-voltage dependence in vitro. 

Let us look at a forward traveling wave around $\omega_c$, 
\begin{equation}
	p(z,t) = a \exp\left[i(\omega_c t - kz)\right] 
\end{equation}
Using the phase velocity definition $c = \omega/k$, we would like to find the phase of the wave at point $z$, which is within the region of the OHC modulation that is associated with the CF
\begin{equation}
	p(z,t) = a \exp\left[i\left(\omega_c t - \frac{\omega_c z_0}{c} - \varphi(z,t)\right)\right] 
\end{equation}
The instantaneous phase $\varphi(z,t)$ is determined by the traveling wave path between $z_0$ and $z(t)$. The speed of sound in a fluid is defined as
\begin{equation}
	c = \frac{1}{\sqrt{\rho \kappa}}
\end{equation}
where $\rho$ is the fluid density, and $\kappa$ is its adiabatic compressibility \citep[p. 229]{Morse}. In the case of a one-dimensional oscillator array, $\rho$ is instead the mass per unit length, and $\kappa$ is longitudinal compressibility---the reciprocal of stiffness per unit length $K$ \citep[p. 84]{Morse}. It is convenient to adapt an \term{acoustic index of refraction}, which enables using a relative stiffness measure. The index of refraction $n$ is generally defined with reference to the speed of light in vacuum \citep[e.g.,][p. 10]{Yariv}, but in the acoustic case with reference to the speed of sound in air \citep[p. 136]{Kinsler}
\begin{equation}
	v_p = \frac{c}{n}
\end{equation}
Where $v_p$ is the phase velocity in the medium. The speed in vacuum has no analog here, so let us instead define the index of refraction relative to the speed of the traveling wave in the passive BM
\begin{equation}
	n = \sqrt{\frac{\rho K_{BM}}{\rho_{BM} K}}
\end{equation}
Realistically, it may be much easier to modulate the compressibility than the density of the medium \citep[cf., ][p. 37]{Azhari2010}. In this case, the index of refraction simplifies to $n = \sqrt{K_{BM}/K}$. We assume that the phase velocity $v_p$ is a function of position, because of the BM-width and voltage-dependent stiffness. Putting it all together, the instantaneous phase is
\begin{equation}
	\varphi(z,t) = \int_{z_0}^{z(t)} k(\omega) dz = \frac{\omega_c}{c}\int_{z_0}^{z(t)} \Delta n(z,t) dz = \frac{\omega_c}{c}\int_{z_0}^{z(t)} \sqrt{\frac{K_{BM}}{K\left[z(t),V(t)\right]}}dz
	\label{AudPhaseMod}
\end{equation}
where $\Delta n$ is the change in index of refraction along the acoustical path, which is calculated in analogy to optics, and is equal to 0 at $z_0$. The end point of $z(t)$ may be on the BM, inside the organ of Corti, or on top of it---on the reticular lamina. In this case it is determined by the voltage- and place-dependent stiffness $K(z,V)$. Note that to obtain the most relevant results, the coordinates must be of the traveling wave system, $\zeta$ and $\tau$ \citep{Kolner}. Note also that this expression is valid in a linear medium, but within a strong negative damping medium the conditions may change and make the phase level-dependent as well (see indications for a ``null-frequency'' point where the cochlear phase is level-independent; \citealp{Geisler1982,Ruggero1997} and \citealp{Palmer2009}). 

One implicit condition for this system to be efficient is that the voltage signal must precede the traveling wave in order to instantaneously modulate the stiffness, before it reaches the CF site. This can happen in either one of two ways. One option is for the potential to build up over time (say, within several periods) after it has been triggered by the electromotile response of the OHC from the BM---effectively sustaining a feedback loop. This option is relatively unfavorable because it requires the signal to be spectrally narrow and periodic and it prevents the system from reacting instantly. Rather, it ``sacrifices'' the onset of the signal, before stiffness can become modulated. Nevertheless, inasmuch as this mechanism is related to the compressive nonlinearity of the OHCs, there are indications that the compression onset is not instantaneous (\citealp{CopperVanDijk2018}; see also \citealp{Altoe2017}). Another phenomenon that suggests it may be the case is that pitch perception from very short sinusoidal stimuli builds up over a few milliseconds, as was reviewed in \cref{LockInPLL}. It was interpreted as part of the PLL pulling in time, but it may have a parallel effect also on activating the time lens. 

The second option is that the electromotile response is triggered by a faster wave that deflects the hair bundle beforehand. This may happen if the bundle is sensitive to the compression wave in the fluid. Alternatively, it can happen if the traveling wave of the TM, which is connected to the tips of the stereocilia, is simultaneous but a bit faster than the traveling wave of the BM. Current data suggest that the velocities of the traveling waves in the BM and TM are comparable \citep{Stenfelt2003, Farrahi2016}, although they do not allow for conclusively determining which one leads over the other in the live cochlea. 

A completely different and passive alternative cause for the production of phase modulation is if the stiffness function is frequency-dependent in a manner that is tuned according to distance from the base (i.e., according to the CF). Such a condition would effectively mean that every frequency component can be subjected to a somewhat different impedance, which changes according to the channel in which it is being analyzed. So, for example, 950 Hz component would be subjected to a somewhat different stiffness when it traverses the 900 Hz and the 1000 Hz channels. As stiffness is usually measured statically and not dynamically, there is only scant evidence for frequency-dependent stiffness in the cochlea \citep{Scherer2004,Rochefoucauld2007}. This stiffness function may additionally interact with the stiffness gradient that has been observed between the different supporting cells and the hair cells within the organ of Corti \citep{Babahosseini2022}. Passive stiffness modulation may seem mathematically indistinguishable from the voltage-modulated medium that was proposed as the primary mechanism. However, this possibility seems relatively tenuous at present, if only because of the limited evidence to support it, and will not be explored further. 

In conclusion, we identified a general mechanism by which the traveling wave may be phase-modulated by the electromotility of the OHCs that causes stiffness modulation. Since we do not know the actual stiffness function of the BM and the organ of Corti, this expression will provide a theoretical anchor for the underlying cause for the modulation, rather than be used analytically. Instead, we will resort to empirical data that suggest a slow modulatory effect in the cochlea that can provide the evidence for a quadratic time-lens operation. 

It should be mentioned that research of stiffness modulation in non-biological systems is a topic that has received some attention, but is still relatively nascent \citep{Trainiti}.

\subsection{Phase modulation evidence}
\label{PhaseModEvidence}
Five different studies were identified in the hearing literature that can be directly interpreted as showing phase modulation in the cochleas of gerbils and guinea-pigs. Four of them are amenable to numerical phase curvature estimation \citep{Guinan2008,Dong2013,Zosuls2021,Meenderink2022}, whereas the fifth one will only be treated qualitatively \citep{Cooper2018}. As is discussed below, a degree of uncertainty about the precise values will accompany us for the rest of this work, which is compounded by a high likelihood that the lens curvature is variable due to auditory accommodation. Therefore, throughout this work, we may occasionally consider particular bounds of time lens values rather than a fixed value. 

\subsubsection{Negative resistance due to outer hair cell activity}
What appears to be an explicit demonstration of a cochlear phase response that can qualify as a time lens was shown in the Mongolian gerbil by \citet{Dong2013}. Using a spatially-coincident voltage and pressure dual-sensor to track the BM dynamics, it was possible to estimate the temporal response of the OHCs in vivo with high precision. In particular, the phase responses of the extracellular voltage, the BM displacement, and the pressure were measured around the resonance site of 24 kHz \citep[Figure 4]{Dong2013}. The extracellular voltage was measured in the scala typmani close to the BM (a \term{cochlear microphonic} potential), which implies that it is proportional to the intracellular voltage of the OHCs \citep{Davis1965}. This was indirectly confirmed in \citet[Figure 3]{Dong2013}, where both evoked pressure and voltage are displayed and show a peak around the CF in the live cochlea, whereas the voltage vanished post portem while the pressure remained unchanged. It was found that below and above the CF, the displacement phase leads the pressure phase, which entails that negative resistance is in effect. Critically, the voltage phase led the displacement by about 0.4 cycles at the CF, but that lead decreased both below and above the CF (in forced oscillators, the displacement lags the force and is at quarter-cycle lag at resonance; \citealp[pp. 46--49]{Morse}). This is indicative that the OHCs impart power to the traveling wave, which then produces the nonlinear amplification of low-level inputs \citep[Figure 4D]{Dong2013}. But the fact that the phase drops above CF is unlike a classical oscillator (where the voltage phase lead is expected to go to $\pi$ at $f \rightarrow \infty$) and appears rather like symmetrical phase modulation that co-occurs with the forced amplification. 

\begin{figure} 
		\centering
		\includegraphics[width=.8\linewidth]{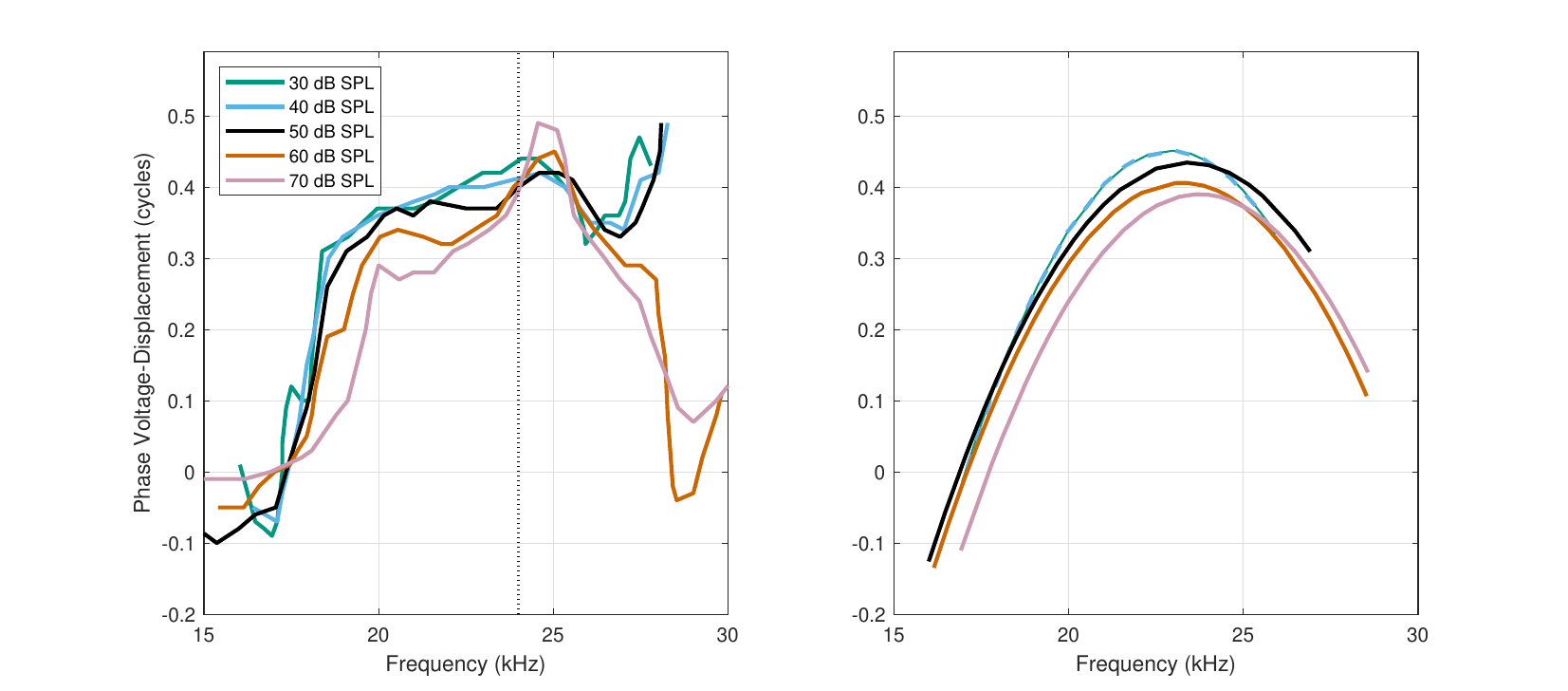}	
		\caption{\textbf{Left:} Simultaneous relative voltage-to-displacement phase data of the gerbil's basilar membrane around 24 kHz at 30--70 dB SPL, as was measured by \citet[Figure 4B]{Dong2013}. The extracellular voltage reflects the intracellular voltage of the outer hair cells. The displacement captures the movement of the traveling wave. Around the characteristic frequency, the voltage phase leads by about 0.4 of a cycle over the displacement. Two additional measurements at 80 and 90 dB SPL were likely contaminated by effects of the fast pressure wave modes rather than the traveling wave, which violate the measurement assumptions and are therefore not displayed. \textbf{Right:} Quadratic phase functions fitted to the measurements on the left. The parabola peaks were constrained to the CF in all cases.}
		\label{DongPhase}
\end{figure}

Figure \ref{DongPhase} reproduces Figure 4B of \citet{Dong2013}. It shows the relative phase between the voltage and the displacement around a CF of 24 kHz. Similar phase data for the same frequency in another animal were obtained between the voltage and the pressure, which is itself in phase with displacement, although with varying levels of smoothness and symmetry \citep[Figures 5E and 6]{Dong2013}. As is seen in Figure \ref{DongPhase}, around the CF the voltage leads by almost half a cycle, but is approximately in phase with the displacement below and above the CF region. 

The nonlinear phase shift appears as would be expected from a phase modulator: it has an apparent symmetrical form, which suggests that the frequency-dependent phase function may contain a quadratic component. If, as Eq. \ref{AudPhaseMod} requires, the stiffness of the OHCs is indeed voltage dependent, then there has to be a modulatory effect on the traveling wave speed at the CF, or in its propagation inside the organ of Corti. There are no direct estimates of either the stiffness or the velocity in \citet{Dong2013}, but a peak in the BM velocity can be derived from the displacement peak at the CF, as is also commonly observed elsewhere \citep[e.g.,][]{Ren2002,Zheng2007}. Additionally, a slowing down of the group velocity of the traveling wave at places basal to the CF was observed in vivo in the gerbil, as well as in other mammals \citep{Heijden2015}. Finally, the phase variations in the BM motion just underneath the OHCs coincide with the constant phase difference observed at the reticular lamina \citep{Chen2011, Ren2016Reticular}, although it is seen below that phase modulation may occur around ``hotspots'' inside the organ of Corti itself \citep{Cooper2018}\footnote{The phase difference for a pure tone between the BM and the reticular lamina was recently measured in vivo in the gerbil ($N=8$) and presented in \citet{HeRen2021} in their Figure 4g and Supplementary Data 4. The two responses do not match, though, but both can be shown to have a small quadratic component once the linear phase component is removed from the CF region. In the case of the plotted response, the peak is above the CF and the curvature is an order of magnitude larger than in the spreadsheet data, which produce an almost negligible curvature that is part of faster broadband oscillation. Neither case is obviously consistent or inconsistent with the data from \citet{Dong2013}.}. 
Therefore, it can be deduced that any phase modulation---manifest as the difference between the extracellular voltage and displacement in the BM---should be reflected in the output of the cochlea at the IHCs and then encoded in the auditory nerve.

Indeed, auditory nerve phase measurements at low frequencies show a distinct curvature around the CF once their linear component (e.g., their mean constant group delay) is removed (or ``detrended'', \citealp{Temchin2010,Palmer2009})\footnote{Because the linear term is usually very dominant, typical phase responses may appear completely linear, similar to a simple resonance of a bandpass filter. So curved components were removed from low-frequency auditory nerve responses in \citet{Allen1983}, which appear completely linear. In another more recent example, \citet[Figure 6]{Lewis2002} used the Wiener-kernel technique to nonlinearly estimate the phase response from spike timing patterns in the auditory nerve, using a white noise input. However, no curvature information could be seen there, maybe due to the incoherent nature of the signal and the inability of the OHCs to phase lock to it.}. Additionally, the curvature is often not centered around the CF \citep{Palmer2009}, and is not always symmetrical, or quadratic looking, probably depending on its cochlear position \citep{Temchin2010}. Whatever curvature was measured in \citet{Dong2013}, it incorporated also effects of adjacent dispersive paths before and after the CF. For the time being, the asymmetries that are also noticeable in the data from \citet{Dong2013} will be ignored.

\subsubsection{Olivocochlear efferent bundle effects}
Using a displacement-sensitive interferometer to measure the vibrations of the BM, \citet{Guinan2008} found that the phase response of a click depended on whether the medial-olivocochlear (MOC) efferent was activated (i.e., if it caused inhibition to the OHCs). A slow phase lag was observed between the onset and the first minimum of the envelope response to the click when the MOC was inhibiting compared to when it was not (no inhibition was observed in the click amplitude during the first half period). The slow change took place over several carrier cycles, so it had little effect on the instantaneous frequency of the click. We may expect that the MOC reflex (MOCR) has some effect on the time-lens curvature, perhaps in analogy to the ocular accommodation that controls the curvature of the crystalline lens. While this possibility will be explored only in \cref{MOCR}, we shall accept it as correct, at present, and obtain estimates for the phase modulation value changes that were observed before and after efferent stimulation. 

\begin{figure} 
		\centering
		\includegraphics[width=1\linewidth]{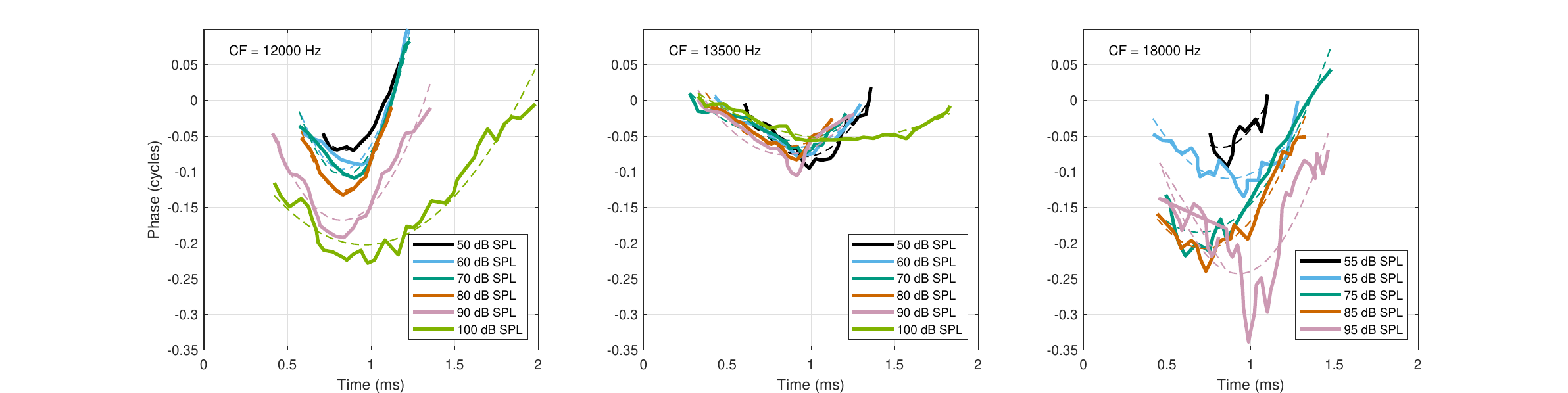}	
		\caption{Phase response change in the guinea pig as a result of the medial olivocochlear efferent excitation at three characteristic frequencies: 12 kHz (\textbf{left}), 13.5 kHz (\textbf{middle}), and 18 kHz (\textbf{right}). The data are taken from Figures 3E, S1G, and S2G in \citet{Guinan2008}. The dashed lines are quadratic fits to the measurements that are shown in solid lines.}
		\label{MOCphasediff}
\end{figure}

Figures 3E and 6 in \citet{Guinan2008} display the phase difference and the zero-crossing values, respectively, of the two efferent modes for one CF in the guinea pig first (basal) turn, which allows for direct estimation of the temporal phase curvature, using Eq. \ref{eq:phase2}. Supplementary Figures S1G and S2G of \citet{Guinan2008} provide similar data from two other guinea pigs and CFs. The authors also stated that similar responses were obtained for the chinchilla. The apparent phase curvature, which is reproduced in Figure \ref{MOCphasediff}, covers about 60 dB of input dynamic range and seems to be level dependent. At low levels, a curvature change as a function of the MOC inhibition is hardly visible. While these measurements provide a relatively extensive dataset in the present context, it is not obvious how to extract a relevant baseline phase from it, so it relates only to changes induced by the MOC, which we assume represent the entire curvature. 

It should be also noted that \citet{Guinan2008} ruled out that OHC stiffness change can be a likely cause of the click responses they obtained, which revealed fast inhibition (of the amplitude) after the first half cycle, whereas the stiffness changes slowly. However, the slow phase-modulation effect that we saw was predicted regardless of amplitude inhibition that may or may not appear within a few cycles. What more, the very slow phase modulation has exactly the effect we expect to have from such a nonlinear system.

\subsubsection{Radial displacement of inner hair cell stereocilia}
\begin{figure} 
		\centering
		\includegraphics[width=1\linewidth]{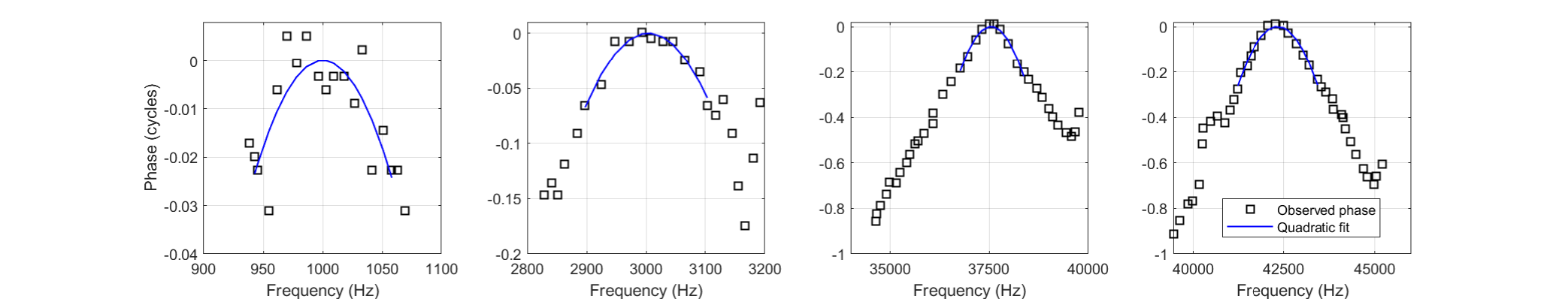}	
		\caption{Phase data (squares) from ex-vivo gerbil cochleas, reproduced from Figures 4F (1 kHz), 4G (3 kHz), 3G (37.5 kHz), and 3H (42.5 kHz) in \citet{Zosuls2021}. The original abscissas were a function of distance from the CF with a range of $\pm 200$ $\mu m$, but are here converted to frequency (and hence linearized), using cochlear scaling parameters from \citet{Greenwood1990} (see Eq. \ref{GreenwoodFunction} and  \cref{ComparativeCochlea}). The quadratic phase fits for the points around the center frequency are plotted with solid lines. Note the different ordinate ranges of the different subplots.}
		\label{ZosulsPhase}
\end{figure}

Traditional methods of measuring the response of the organ of Corti to external stimuli have focused on the transverse movement of the of the BM (see Figure \ref{OrganCorti}). Using the mechanical properties of the cochlear partition, it is then possible to deduce the shear force that acts on the IHCs, which causes their movement in the radial direction. In a study by \citet{Zosuls2021}, ex-vivo samples of gerbil cochlea were used to directly measure the radial motion of the IHCs, which were stimulated by mechanically actuating the BM using a probe that was placed under the outer pillar cells, and whose longitudinal position could be adjusted in increments of 2 micrometers. An inverted microscope with stroboscopic imaging and custom digital image processing were used to record the fine motion of the stereocilia in resolution of 8 nanometers. While the measurement was done on a small subsection of the organ of Corti at a time, its mechanical and biophysical properties were shown to be close enough to live animal and intact conditions, which would yield data that is sufficiently valid. We assume that the OHC section of the organ of Corti around the CFs was intact in all cases. Four measurements are presented in \citet{Zosuls2021}, which provide the spatial response function of the IHC displacement, including the phase as a function of (longitudinal) distance from the CF along the BM. At four frequencies, 1 kHz, 3 kHz, 37.5 kHz, and 42.5 kHz, the phase function is presented and in all cases it shows a maximum at the CF position, in what could be well approximated using quadratic phase modulation. The relevant data is reproduced in Figure \ref{ZosulsPhase}. Note that the equivalent sound pressure level that would have produced the mechanical actuation here is unknown.

\subsubsection{Vibration ``hotspots'' in the organ of Corti}
Using high-speed optical coherence tomography imaging of the gerbil's organ of Corti, \citet{Cooper2018} found that the vibrations between the BM and reticular lamina exhibit ``hotspots'' in the region between the Deiters cells and the OHCs. In phase measurements along the path between the two surfaces (the BM and reticular lamina, see Figure \ref{OrganCorti}), the spatially and spectrally dependent phase function (relative to the BM) clearly oscillated around the hotspot, before it returned to about zero---amounting to a symmetrical phase modulation that may have a quadratic component. The degree of modulation depended on frequency and on the exact path that was imaged in the organ of Corti, which in turn determined the modes of vibration that were imaged. In one case in which a transverse path was tracked, the modulation was positive (about 0.1 cycle), tuned to the CF (23 kHz), and decreased symmetrically at lower frequencies \citep[Figure 7c]{Cooper2018}. At CF of 40 kHz and a slightly different path with a longitudinal cross-section, the modulation was negative (minimum -0.15 cycles) at low frequencies, but rather shallow and mistuned at the CF \citep[Figure 8f]{Cooper2018}. If these results can be generalized, then a traveling wave propagating from the BM to the reticular lamina is subjected to an internal phase modulation. Furthermore, in some cases the modulation may appear to have never happened if measured at the BM or reticular lamina alone. Similar phase patterns were also recorded in mice using related methods, only that the phase does not return to its initial value between the BM and the reticular lamina / tectorial membrane \citep[See, ][Figures 1G, 1H, and 3A]{Dewey2021}.

\subsubsection{Angle-dependence phase measurements of the organ of Corti}
\begin{figure} 
		\centering
		\includegraphics[width=0.6\linewidth]{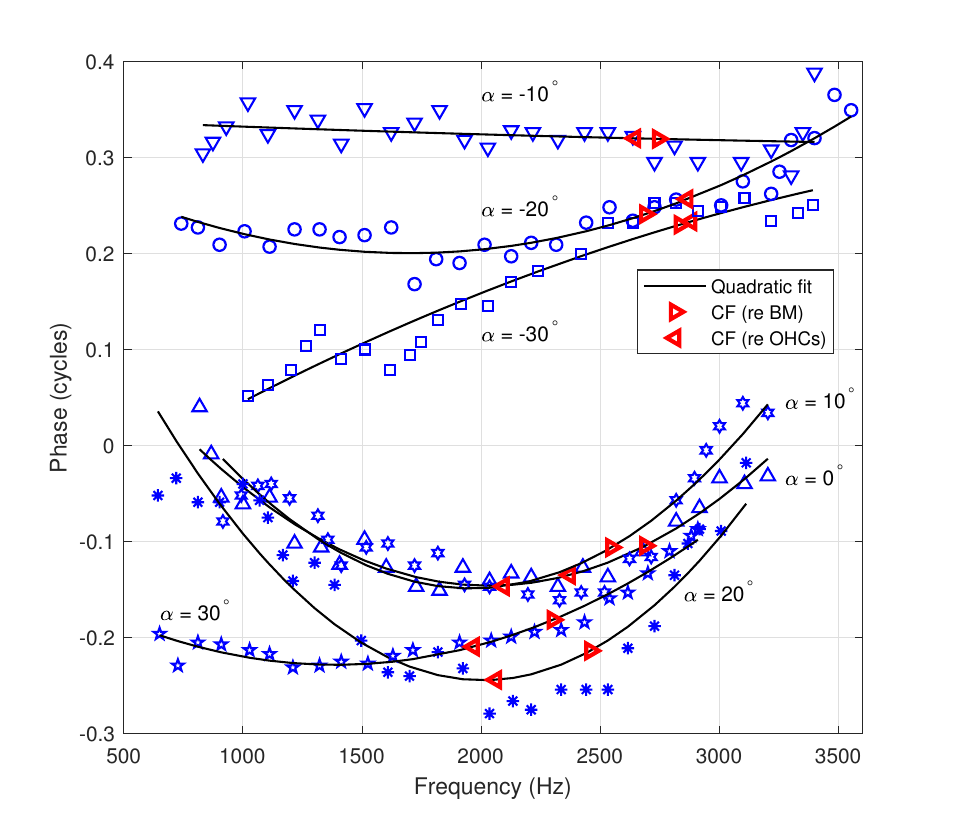}	
		\caption{Displacement phase difference between the BM and OHC motion in the gerbil, as a function of measurement angle, and hence of longitudinal part of the trajectory in the organ of Corti. The phase data is replotted after Figure 2e in \citet{Meenderink2022}. Best frequencies for each plot was different for the BM and for the OHC site and they are marked with left pointing and right pointing red triangles, respectively, after Figures 2c and 2d in \citet{Meenderink2022}. Angles varied between $-30^\circ$ and $+30^\circ$, as are marked beside each quadratic plot fitted. The input level of the stimulus was 30 dB SPL.}
		\label{MeerDong2022fig}
\end{figure}

In a study by \citet{Meenderink2022}, the phase of the motion of the organ of Corti was measured in vivo using optical coherence tomography as a function of the angle between the laser beam and the longitudinal direction of the BM. This angle relates to different acoustic paths within the organ of Corti, whose angular dependence suggests that the OHC motion have a non-negligible longitudinal component. The phase was measured along different points between the BM and the OHCs in the second turn of the gerbil's cochlea. The angle was varied between $-30^\circ$ and $+30^\circ$ and produced a different frequency dependence of the phase $\phi_{OHC}-\phi_{BM}$ in every angle, similarly to what was found in \citet{Cooper2018} and reviewed above. The phase has a clear peak, also at $0^\circ$, which may be therefore taken to have a quadratic component, as is seen in Figure \ref{MeerDong2022fig}. However, as is shown in the next subsection, the curvature of the $0^\circ$ measurement is in opposite sign to those extracted from other studies, and only at angles of $-30^\circ$ appears to change the sign, whereas at $-10^\circ$ the curvature becomes negligible. Furthermore, the discrepancy between the two CFs given (for both BM and OHC positions) and the phase curvature center frequency, as exists in most other measurements reviewed above, is relatively large and it is not clear which value should be used. 

\subsection{Estimation of the auditory time-lens curvature}
\label{OHCtimelens}
From all the studies reviewed in \cref{PhaseModEvidence} that may be suggestive of a time-lensing function, only the phase data in \citet{Cooper2018} is directly given in the time domain. Insofar as they can be interpreted as a time lensing operation, both time- and frequency-domain representations have almost the same mathematical form (complex Gaussians, but with different signs of the argument; Eqs. \ref{TimeLensOmega2} and \ref{lens_function}, respectively) and thus the procedures to extract their curvatures are about the same in all cases.

Wherever reported, the phase modulatory effect is dependent on level, although the spread is small in the gerbil \citep{Dong2013} and large in the guinea pig \citep{Guinan2008}. To constrain the spread and match it to the levels we are working with, the average curvatures were calculated from data points at 75 dB SPL or lower. Additionally, the quadratic fit was performed as a rough approximation to the curves (that were converted from cycles to radians) that change monotonically around the peak and were truncated where additional oscillations and phase shifts became visible. The resultant fits are displayed in Figures \ref{DongPhase}--\ref{MeerDong2022fig}. The linear and constant terms in the fits are immaterial and were dropped in the subsequent analyses. The quadratic coefficient was readily applied in the time lens expressions to obtain the curvature and focal time in the time domain using Eqs. \ref{lens_function} and \ref{TimeLenss} and in the frequency domain using Eqs. \ref{TimeLensOmega2} and \ref{TimeLenss}. 

\begin{figure} 
		\centering
		\includegraphics[width=1\linewidth]{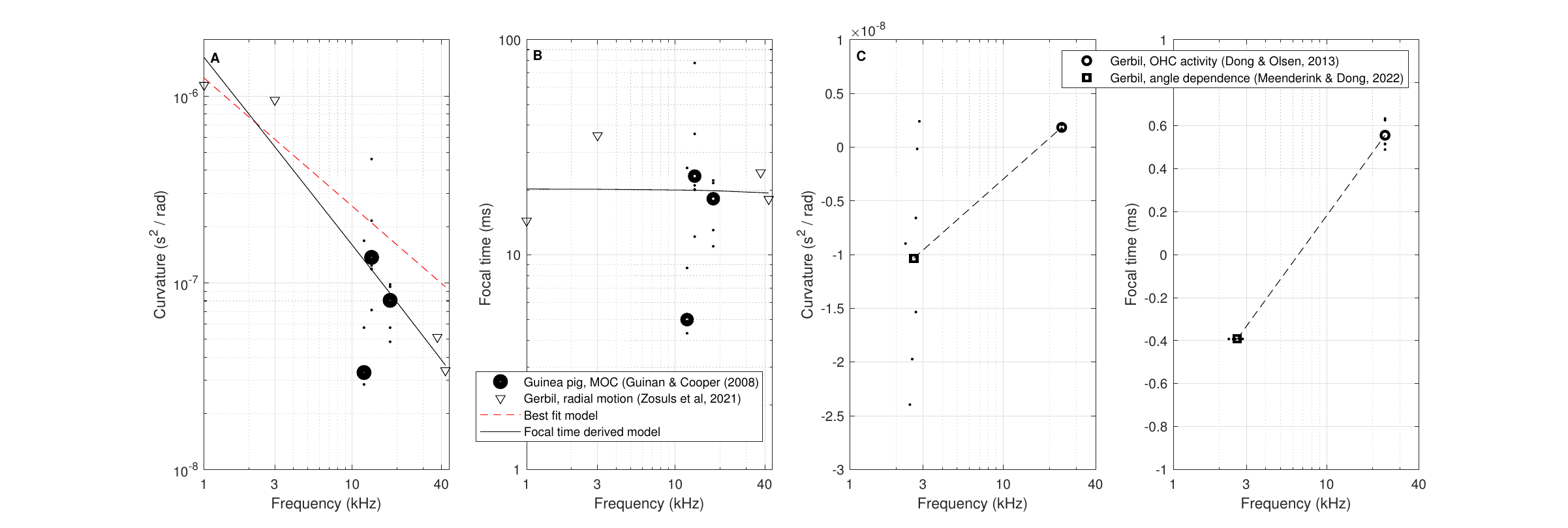}	
		\caption{Estimated time-lens curvature (\textbf{A},\textbf{C}) and focal time (\textbf{B}, \textbf{D}) in the cochlea of the gerbil and guinea pig, based on four independent measurements \citep{Guinan2008,Dong2013,Zosuls2021,Meenderink2022}. The data are clustered into two groups: large-curvature observations in panels \textbf{A} and \textbf{B} and small-curvature in \textbf{C} and \textbf{D}. Each dot marker relates to a single level/curve that appears in Figures \ref{DongPhase}--\ref{MeerDong2022fig} and whose means are marked with circle. Mean values were used to generate the power-law fit (red dotted line) for the large curvature (\textbf{A}) and a linear fit was used for the focal time (\textbf{B}). For consistency between models, an additional fit to the curvature was generated from the linear fit of the focal time and is plotted in solid black in \textbf{A} and is the one that is used throughout the text. Linear fits were used in \textbf{C} and \textbf{D} for the small-curvature data.}
		\label{gerbilphase}
\end{figure}

All phase-curvature data and derived focal times are shown in Figure \ref{gerbilphase}. The data can be readily clustered into two groups. High and positive curvature values $s>3 \cdot 10^{-8}$ $s^2/\rad$, with corresponding focal times $f_T> 4$ ms from \citet{Guinan2008} and \citet{Zosuls2021}, and small-curvature (both positive or negative) data $|s|<3 \cdot 10^{-9}$ $s^2/\rad$ and corresponding focal times $|f_T|<0.7$ ms in \citet{Dong2013} and \citet{Meenderink2022}. While the data point at 24 kHz from \citet{Dong2013} may be considered a mere outlier of the large-curvature group, the sign changes and very low magnitude of the rest of the data points at 2-3 kHz are completely distinct from the other low-frequency data. The low-frequency clustering may be further supported by the fact that all of these data points came from the gerbil, which otherwise yielded large-curvature values. 

The large-curvature data were well fitted with a power-law model, whereas the focal time data points were nearly constant ($f_T \approx 20$ ms) and were fitted with a linear function. However, the independent modeling of these two linearly dependent variables are inconsistent, as the curvature does not yield a constant focal point function. The other direction---of deriving the curvature from the modeled focal time---does indeed yield a satisfactory fit (if only graphically) so that this fit will be used throughout this section. The focal time for the gerbil and guinea pig is
\begin{equation}
f_{T,gg}(f) = -2.06 \cdot 10^{-8} f + 0.0202 \,\,\,\,\Hz
\label{GGfocalTime}
\end{equation}
for frequency in Hz and focal time in s.  To obtain the curvature, we simply divide this expression by $2\omega_c$ (a power law with exponent -1)
\begin{equation}
s_{gg}(f) = \frac{f_T}{2\omega_c} = \frac{0.0016}{f} -1.639 \cdot 10^{-9} \,\,\,\,\, \s^2/\rad
\label{ggsconst}
\end{equation}
The small-curvature data suffer from a dearth of frequency points, which may or may not be fitted with a linear function. We note that while the two animals have comparable hearing ranges \citep{Fallah2021}, it is possible that the phase measurement methods do not quantify exactly the same process or anatomy, although this seems rather unlikely. While the two clusters seem to complicate the analysis and make the data appear inconsistent, variable curvature is going to be perfectly consistent with an accommodating hearing system, in analogy to the eye. This will be reviewed in \ref{accommodation}. 

\subsection{Extrapolation of time-lens curvature to human hearing}
\label{TimeLensExtrapolation}
Short of carrying out direct measurements of the human time-lens curvature values, additional assumptions must be made in order to transform the animal data obtained to values that are valid for humans. There are several approaches that can be taken based on the available data. For example, the focal time curve appears to be approximately constant at 20 ms (large curvature). This constant may apply to all mammals, or be unique to the rather similar gerbil and guinea pig (and likely other rodents), whose data coincided. A similar option is that the focal time of 20 ms in these animals should map to the same area in the auditory brain as in humans and remain a constant. Yet another option is that the phase curvature could be scaled just like other cochlear parameters. For example, it may be scaled in accordance with the cochlear filter bandwidths that might also apply to the phase modulation function (in the previous versions of this manuscript, the latter option yielded plausible values, despite limited data). A final option is that the curvature we obtained depends primarily on the transverse cochlear geometry rather than on the longitudinal place alone (i.e., on the tissue between the BM and the reticular lamina rather than on CF alone), so it should be scaled accordingly. If a mechanism along the lines hypothesized in \cref{StiffnessPM} turns out to be correct, then this last option may be the most precise. However, it depends on unknown parameter values such as the stiffness distribution in the organ of Corti, but its histological complexity \citep{Naidu1998} defies simple scaling and detailed cross-species values are not available. Therefore, this approach will not be further pursued. The three remaining approaches to derive the human curvature entail rather strong assumptions, so none of them will be completely satisfactory before they can be cross validated with other methods and data. 

\subsubsection{Constant focal time}
The large-curvature data in both gerbil and guinea pig yielded a nearly flat focal time as a function of frequency, with only slight decrease at high frequencies (19.4 ms at 44 kHz), and unknown response at frequencies lower than 1 kHz (20.2 ms). This relative constancy ($\pm 2\%$) may be a desirable feature for the auditory system, so achieving it may be a design goal that applies to all mammals. In this case we can take the same focal time curve and apply it to humans, but using the scaling property between the gerbil and human cochleas, remap it to human frequencies and find the new curvature that would produce it. 

The focal time in Figure \ref{gerbilphase} B was fitted by the linear function in Eq. \ref{GGfocalTime}. We use the very same function, but now express the frequency as a function of relative cochlear place, as is shown in Figure \ref{Q10} D. The human focal time is then shown in Figure \ref{Focaltimefromphase} B and follows the linear function
\begin{equation}
	f_{T,h}(f) = -5.15 \cdot 10^{-8} f + 0.0202
\end{equation}
The corresponding curvature is then obtained using Eq. \ref{ggsconst} and is displayed in Figure \ref{Focaltimefromphase} A
\begin{equation}
	s_{h}(f) = \frac{-5.15 \cdot 10^{-8} f + 0.0202}{4\pi f}
	\label{ConstantCurvatureModel}
\end{equation}

\subsubsection{Analogous focal time target}
A stronger assumption that may be invoked to derive the focal time is that it points to a region in the auditory system that should be analogous in the gerbil, guinea pig, and human. In evoked potential auditory electrophysiology, the 20 ms value is considered a \term{middle latency response} (MLR) potential, whose latency lies between the brainstem (ABR) and cortical potentials \citep{Picton1974}. The morphology of the MLR varies between animals, as it also depends, among others, on the individual animal, the stimulus used to obtain it, its intensity, how the recordings are filtered, and how the electrodes are placed, which itself is suggestive of multiple generators that produce some of the peaks in the MLR \citep[e.g.,][]{McGee1991,Musiek2018}. The generators are thought to lie in the thalamocortical pathways, but there is also strong evidence that the inferior colliculus plays a role in the early MLR peaks \citep{McGee1991}. In the adult gerbil, three positive peaks are distinguished around the 20 ms time frame, which measured at the temporal lobe: positive peaks at 11 ms (wave A) and at 25 ms (wave C), and a negative peak at 16 ms (wave B) \citep{Kraus1987}. However, these values vary between studies, so it is not uncommon to find wave B peaking at around 20 ms and wave C at 35 ms. When measured at the midline, the morphology changes and there is a negative peak $M-$ at -10.5 ms and a positive peak $M+$ at 19.2 ms. The human MLR morphology is less complex and it involves a first negative wave $Na$ with a peak at about 12-21 ms and a first positive wave $Pa$ at about 21-38 ms, followed by second wave with $Nb$ and $Pb$. The exact human generators are also in doubt, but the $Na$ is sometimes thought to arise in the midbrain (IC) \citep{Hashimoto1982,McGee1991}, or in the thalamocortical pathways, in which case it may be centered in the medial geniculate body (MGB) of the thalamus, as well as other subcortical regions such as the reticular formation \citep{Musiek2018}. 

It seems that the human $Na$ potential is closest to the $M-$ potential in the gerbil and guinea pig, both in generator site and in latency \citep{McGee1991}, which may suggest that $M+$ and $Pa$ are also analogous. Given the variance in the latencies that appear in literature for all waveforms, it will be difficult to precisely determine which latency in human would be most correctly mapped to 20 ms in gerbil, but anything between that same value and, say, 25-30 ms, may be adequate to bracket the actual focal time. This means that the above solution may be adapted as is to humans, but a range of focal times around that value may be useful to look at. As can be seen in Figure \ref{Focaltimefromphase}, the constant difference leads to a relatively modest change in the curvature itself.

\subsubsection{Filter bandwidth scaling}
Normally, we would like to take advantage of the scaling property of the cochlea, which enables the transformation of quantities according to their relative distance along the basilar membrane or their characteristic frequency (\cref{ComparativeCochlea}). While the CFs associated with the time lens can be transformed easily, we do not know if and how the curvature scales in the normal cochlea. However, the model that was obtained in Eq. \ref{ggsconst} is a function of frequency, as are all scalable cochlear parameters. To tie the animal curvature data, different proxy variables for scaling the curvature can be conceived aside from frequency. A plausible scaling can be conjectured that is tied to the bandwidth of the auditory channel that is related to the time-lens CF, even though the time lens itself functions as an all-pass filter that needs not obey the same scaling rule as the bandpass filters. Indeed, it has been recently shown that the bandwidth does not change significantly along the active path between the BM and OHCs in both gerbils and guinea pigs---the same area that corresponds to the vibration hotspot where phase modulation seems to take place \citep[Figure 9F and 9G]{Fallah2021}. As we apply this particular scaling, we are confronted by additional uncertainties regarding the correct bandwidth values that should be used for animals and human. 

\subsubsection{Guinea pig and gerbil $Q_{10}$ spread}
\begin{figure} 
		\centering
		\includegraphics[width=1\linewidth]{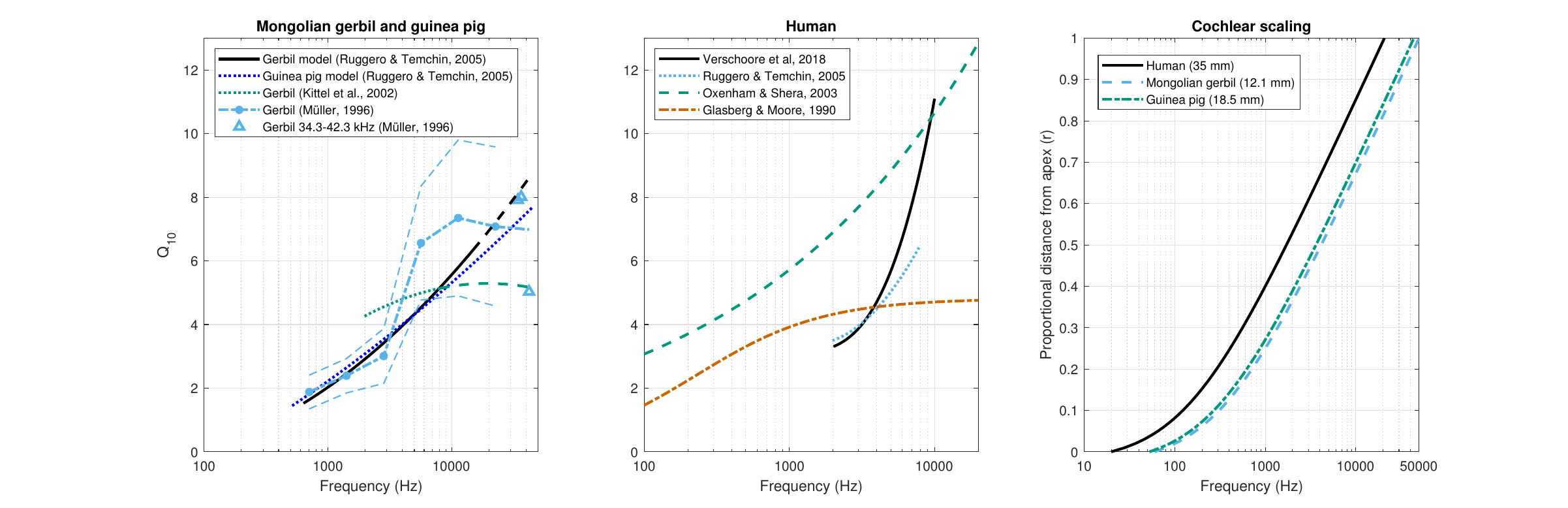}	
		\caption{The auditory filter bandwidths, expressed as $Q_{10}$, of the Mongolian gerbil and guinea pig, and human, as well as their cochlear scaling functions. \textbf{Left:} Animal $Q_{10}$ data were collected from several sources, namely from gerbil and guinea pig models by \citet[Figure 6A]{Ruggero2005} that are based on several auditory nerve tuning function datasets and allow for easy extrapolation across the audible range, on modeled gerbil data by \citet[Figure 4]{Kittel2002}, and on extensive gerbil dataset in \citet[Figure 4 and Table 1]{Muller1996}, including confidence intervals that are marked with dashed lines at the $\pm 1$ standard deviation. \textbf{Middle:} In humans, the sharpest filters are based on \citet{OxenhamShera2003}, who fitted psychoacoustic data with the power law $Q_{ERB} = 11f^{0.27}$ ($f$ in kHz), which can be multiplied by a factor of 0.52, to convert to $Q_{10}$ (\citealp{Verschooten2018}; this factor can be also directly computed from the filter models in \citealp{OxenhamShera2003}). The broadest filter estimates are derived from psychoacoustic estimates by \citet{Glasberg1990} of the equivalent rectangular bandwidth (ERB). Medium-sharp estimates are based on \citet[Figure 1]{Verschooten2018} and on \citet[Figure 6]{Ruggero2005}. \textbf{Right}: The frequency to relative cochlear functions of the three animals based on the scaling law (Eq. \ref{GreenwoodFunction}) by \citet{Greenwood1990}.}
		\label{Q10}
\end{figure}

We can break down the uncertainty in the filter bandwidth into that related specifically to the guinea pig and gerbil and that related to humans. Conveniently, the gerbil and guinea pig have audible frequency range that appears to be close enough to one another (Figure \ref{Q10} C; \citealp{Greenwood1990}), so combining their few available data points together was preferred here, for simplicity. Similar logic applies to the channel bandwidth, as is seen below.

Most animal frequency selectivity data are based on neural tuning curves, which directly relays the effect of cochlear processing. They are usually characterized using the 10 dB bandwidth ($Q_{10}$), as is plotted in Figure \ref{Q10} A and B for the three species. For the gerbil, the most detailed auditory-nerve tuning curves data are available from \citet{Muller1996}, which reveal a substantial spread that reflect the broad sample of tuning curves and is marked on Figure \ref{Q10} A. The confidence intervals ($\pm$ 1 standard deviation in the plot) are nevertheless consistent with estimates based on data modeled by \citet{Kittel2002} and \citet{Ruggero2005}, as can be seen in the figure. Unfortunately, there are almost no $Q_{10}$ data available directly for the 37.5-42.5 kHz basal frequency range that was targeted in \citet{Zosuls2021}. Therefore, while the distribution provided by \citet{Muller1996} in tabular form  was cut off at 32 kHz, the additional three data points of higher frequencies were available in his measurements and are used to form a rough estimate of the bandwidth at these frequencies. From Muller's measurements the bandwidth dependence on frequency in gerbil decreases rather than increases above 20 kHz. In similar measurements of the mouse $Q_{10}$ a similar kink in the bandwidth curve is observed at around 30 kHz, but increases again by 50 kHz \citep{Taberner2005}. \citet{Ruggero2005} have also provided a trend line for the guinea pig, which is quite similar to that of the gerbil---a similarity that repeats in many species regardless of their cochlear dimensions \citep{Ruggero2005}. These trend lines are still contained within the confidence intervals by \citep{Muller1996}. Therefore, the guinea pig model will be implemented as a first-order approximation for a usable bandwidth scaling as a function of frequency (Figure \ref{Q10} A). 

\subsubsection{Frequency selectivity in humans}
While the auditory nerve tuning curves are frequently considered to be the gold standard for the peripheral filtering estimation, accessing them in humans is possible only post-mortem---after the cochlear nonlinearity disappears. Thus, the live neural tuning curves are unknown in humans and therefore require an animal reference, whose bandwidth can be compared and extrapolated between methods and species. However, the human bandwidth that should be paired with the gerbil's and guinea pig's $Q_{10}$ is uncertain, as there has been an ongoing controversy in literature with regards to the relative sharpness of the human auditory filters. This is a twofold controversy, in fact, which relates to the absolute filter bandwidth in humans, as well as to the relative bandwidth compared to other mammals. 

According to several studies, human hearing has a superior frequency resolution compared to other mammals, perhaps except for other primates \citep[e.g.,][]{Shera2002,OxenhamShera2003,Shera2010,Joris2011,Verschooten2018,Sumner2018, Burton2018, Walker2019}. Other studies found that the filters are equally sharp for all mammals \citep{Shofner2005,Ruggero2005,Siegel,Ruggero2007,Martin2011,LopezMartin2013, Manley2016} and that they share other common features with vertebrates in general \citep{Manley2015}. These conclusions depend on the specific methods employed in each study, as well as on the theories or models used to interpret them, which in themselves are often not in consensus. For example, some results rely on stimulated-frequency OAE data \citep{Shera2002} or spontaneous OAE \citep{Manley2016}, which require a theory to interpret them. Another example is from studies that involve either simultaneous masking or forward masking in notched-noise data, which requires control of nonlinear suppression and level dependence, as well as a clear understanding of the operation of central processing \citep{OxenhamShera2003,Ruggero2007,Martin2011,LopezMartin2013,Verschooten2018}. Yet other experiments relied on pitch discrimination tasks that require some involvement of central processing as well \citep{Shofner2005,Walker2019}. 

Some human $Q_{10}$ modeled data from different sources are presented in Figure \ref{Q10} (middle), which highlights the two extremes of sharp and broad filtering. Both the bandwidth and the frequency dependence are markedly different between studies. Sharp filter responses were found in \citet{OxenhamShera2003} using forward-masking notched-noise psychoacoustic experiments, which contrast with broad filters based on simultaneous-masking notched noise by \citet{Glasberg1990}. Additional compound action potential data from \citet[Figure 1]{Verschooten2018} are plotted, exhibiting sharp filters at high frequencies and broad filters at low frequencies. Modeled tuning curve $Q_{10}$ by \citet[Figure 6]{Ruggero2005} are displayed as well, showing broad filters, along with extrapolation to 20 kHz\footnote{While it is not attempted here to resolve it, a few provocative remarks should be made regarding the human frequency selectivity controversy. First, the very notion of a constant bandwidth that is fundamental in classical linear filters can be elusive in systems that exhibit suppression, emissions, feedback, level dependence, compression, and other nonlinear effects. The ongoing attempt to remove the confounding effects of these phenomena and obtain a reduced linear filter ``kernel'' may belie the generality and hence the usefulness of the concept of bandwidth, which drives this exploration in the first place, as each bandwidth applies only to a limited set of stimuli (see also, \citealp{Thoret2023}). Inasmuch as the PLL theory put forth in \cref{PLLChapter} may turn out to be correct, it will most certainty change the interpretation of some of the involved models (e.g, of OAEs and suppression) and their corresponding results. This is so because the PLL appears linear only when it is in lock and it has several bandwidths associated with different modes of operation (Figure \ref{PLLrange}). If we additionally consider a passive linear filter that precedes and ``contains'' the PLL, then some quasi-linear effects may be obtained that produce broad filtering, whereas under locked conditions, the filtering appear narrower. Combined with the filter sharpness controversy, the discussion about the correct place of filtering is reminiscent of the dreaded \term{second filter} problem, which suggested that there can be two stages of bandpass filtering in the cochlea. The problem was originally framed by \citet{Evans1972} and \citet{Evans1973}, who noted that the neural and mechanical data available at that time did not match. It has been considered more or less resolved ever since modern methods converged on very similar neural and mechanical results \citep{Sellick1982, Khanna1982}. However, recent in-vivo measurements of vibrations within the organ of Corti have shown that the BM tuning is not as sharp as that recorded on the reticular lamina \citep{Ren2016Reverse}, which may be interpreted to show the existence of a second filter after all. See \citet{Cooper2008Manley} for a historical review and \citet{Bell2005} for an alternative point of view. In order to keep the temporal imaging and PLL theories independent, we will leave the bandwidth interpretation question unanswered, at present.}.

\subsubsection{Human time-lens curvature estimation}
The short reviews above indicate that there are several possible combinations of animal-to-human scalings that can be invoked to derive the human time-lens curvature, but none that is clearly more correct than the others. We will therefore aim to bracket the time-lens curvature in human and then explore the curvature-space, as needed, throughout this work. To simplify this procedure, we will use a single curve for guinea pig and gerbil filter sharpness and apply the broad human tuning according to \citet{Glasberg1990} and the sharp tuning from \citet{OxenhamShera2003}. The following first applies the scaling only the large-curvature estimates (Figure \ref{gerbilphase} A and B). The small-curvature estimates (Figure \ref{gerbilphase} C and D) for humans rely on less data and are simply down-scaled in frequency from the animal data, as is presented below.

\paragraph{Large-curvature scaling}
In order to perform scaling of curvature, we will apply the following procedure in all cases. The quality factor definition of a bandpass filter is $Q_{10} = f_c/\Delta f$, where $f_c$ is the center frequency and $\Delta f$ is the bandwidth of the filter at 10 dB down from its peak. If the animal's bandwidth is $\Delta f$, for a given $Q$, then we can compute the bandwidth-phase pair $(\Delta f, \phi_{\Delta f/2})$ from the quadratic phase function fits around $f_c$ (Figure \ref{gerbilphase}). The argument of the time-lens transfer function (Eq. \ref{TimeLensOmega2}) in the (one-sided) bandwidth corner frequency around $\omega_c$ is: $\omega^2 s = 4\pi^2 (\Delta f /2)^2 s = (\pi\Delta f)^2s = \phi_{\Delta f/2}$, where $s$ is the animal's lens curvature. Therefore,
\begin{equation}
		\phi_{\Delta f/2} = (\pi\Delta f)^2 s =  s \left(\frac{\pi f_c}{Q}\right)^2  
	\label{CurvePhiTrans}
\end{equation}
Next, we would like to use the obtained animal's phase values $\phi_{\Delta f/2}$ for the human's equivalent CF and $Q_{10}$ and extrapolate it from there for the entire spectrum using scaling. This can be done by using one of the modeled animal bandwidths from Figure \ref{Q10}. We shall use the \citet{Muller1996} $Q_{10}$ data at frequencies above 1200 Hz, which extend all the way to 42.5 kHz (albeit with very few data points), but use the \citet{Ruggero2005} function at lower frequencies in order to obtain smooth extrapolation. 

The calculation is repeated twice for the two different human filter types. First, the broad human filter bandwidth are, using the ERB approximation by \citet{Glasberg1990}
\begin{equation}
	\Delta f_{h,broad} = \frac{\ERB}{0.52} = \frac{0.108f + 24.7}{0.52} = 0.208 f + 47.5
	\label{ERB10}
\end{equation}
with the subscript $_h$ designating human values, and the conversion factor 0.52 was introduced to obtain the equivalent 10 dB bandwidth \citep{Verschooten2018} and can be gathered directly from the filter models in \citet{OxenhamShera2003}. Second, for the sharp filter relations, using the approximation from \citet{OxenhamShera2003} and the same correction factor
\begin{equation}
	\Delta f_{h,sharp} = \frac{f}{Q_{10}} = \frac{f}{0.52 Q_{ERB}}= 1.129 f^{-0.73}
	\label{Q10ERB}
\end{equation}
Finally, using these relations, we can plug in the argument of the time lens again in 
\begin{equation}
	s_h = \frac{\phi_{\Delta f/2}}{(\pi\Delta f_h)^2}
\end{equation}
This curvature applies now to a new CF that has the same proportionate distance on the human's cochlea as the one on the gerbil's, according to their respective Greenwood function (Eq. \ref{GreenwoodFunction}) \citep{Greenwood1990}. It allows us to compute the human's focal time using $f_T = 2\omega_c s_h$. In all cases we have to carry over the sign of the curvature from the animals to humans---a positive curvature both in the gerbil and in the guinea pig cases (the arguments in the frequency and time domain time lens representations have opposite signs).

\begin{figure} 
		\centering
		\includegraphics[width=0.9\linewidth]{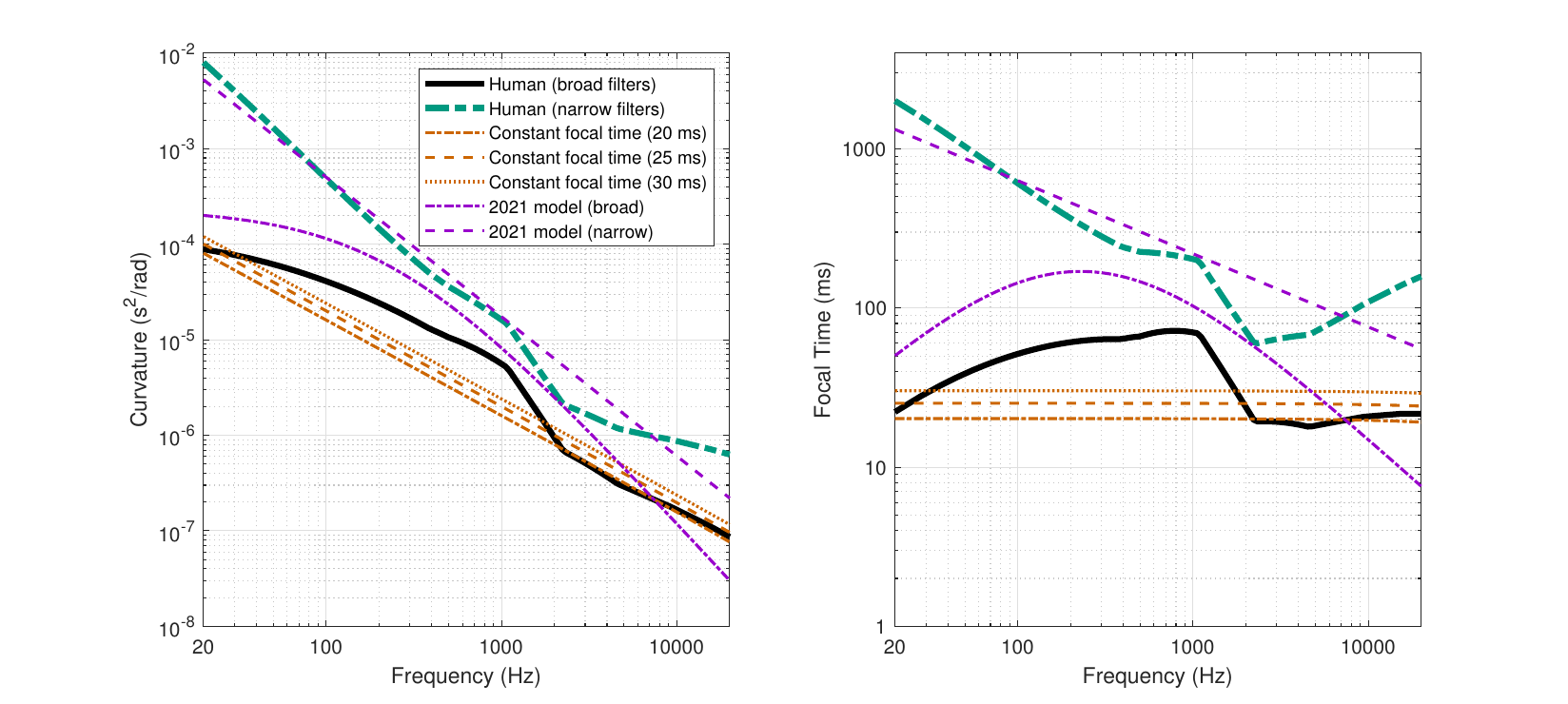}	
		\caption{Human large-curvature mode time-lens estimates (left) and their respective focal time (right) based on different combinations of animal and human filter bandwidths and different modeling assumptions. The first two curves are calculated based on gerbil to human data scaling of the $Q_{10}$, derived from data by \citet{Muller1996} at high frequencies ($>1200$ Hz), and an extrapolated model by \citet{Ruggero2005} at low frequencies ($<1200$ Hz). The solid black curve applied the human broad auditory filters based on \citet{Glasberg1990} and the narrow filters in dash-blue on \citet{OxenhamShera2003}. The next three curves in red are based on the assumption that the nearly constant focal time of the gerbil applies to humans, as is with $f_T = 20 ms$ (dash-dot), $f_T = 25 ms$ (dash), and $f_T = 30 ms$ (dot). The last two curves are based on an earlier estimate of the curvature, when fewer data points were available. They were also based on scaling according to $Q_{10}$ and are plotted in purple dash-dot for the broad filters in humans and dash for the narrow filters.}
		\label{Focaltimefromphase}
\end{figure}

The large-curvature results are displayed in Figure \ref{Focaltimefromphase}. They show a large spread of possible values of curvatures and focal times that are easily one order of magnitude apart between the constant-focal-time model and the various scaled models. However, the spread tends to be largest the lowest and highest frequency ranges, where the estimates rely on extrapolation and are much less reliable. Interestingly, the constant focal-time curvature estimate of 20 ms nearly coincides with the broad-filter curvature that was computed based on scaling at frequencies above 2000 Hz. The choice we made about the animal filter data made the bandwidth function jagged and the scaled curves non-monotonic. 

\begin{figure} 
		\centering
		\includegraphics[width=0.9\linewidth]{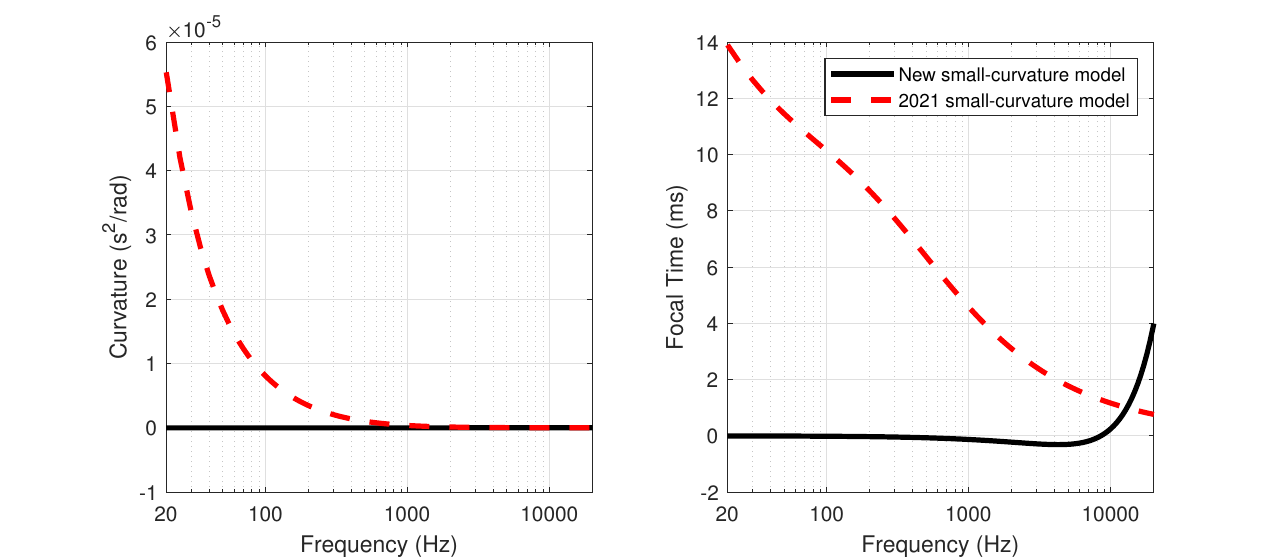}	
		\caption{Human small-curvature mode time-lens estimate (left) and its respective focal time (right). As bandwidth scaling does not seem to apply to this limited dataset, the animal data was simply scaled (translated) to human frequencies with curvature (left) and focal time (right). The old small-curvature model from the previous version of this that applied to the one data point from the gerbil at 24 kHz is shown in dashed red for comparison and shows stark difference in magnitude and sign.}
		\label{FocaltimefromphaseLow}
\end{figure}

\paragraph{Small-curvature scaling}
The small-curvature estimates in Figure \ref{FocaltimefromphaseLow} are based on simple scaling of the frequencies between the animals and humans of the two-point data obtained above, using the respective gerbil and human scaling functions from \citet{Greenwood1990}. Here the focal time and curvature are usually very close to zero and negative ($ -0.1 < f_T < -0.5 $ ms), which is 2-3 orders of magnitude lower than the large-curvature mode estimates.

We will subsequently refer to these two estimates as the \term{small-curvature time lens} the \term{large-curvature time lens}. 

\subsection{Discussion}
While we were able to obtain estimates for the time-lens curvature that will turn out to be plausible later in this work, this section may have been one of the most speculative part in the entire work (not a small feat, admittedly). This is despite the relatively simple physical model it alluded to, given the variable stiffness of the BM. 

According to the above analysis, the time lens is effectively a result of an active and nonlinear function in the organ of Corti, which is naturally associated with the electromotility of the OHCs. This explanation may coincide with recent physiological and psychoacoustic findings by \citet{Nuttall2018}, which conclusively determined that the place of envelope information generation is found within the organ of Corti---between the BM and the reticular lamina---as a consequence of its nonlinear distorting properties. This is exactly the effect we would expect to see following a time lens---effectively a nonlinear phase modulator. 

Inasmuch as phase modulation depends on the OHCs, their number in the healthy organ of Corti of the different animals is substantially different: 11000--16000 in humans, 4600 in gerbils, and 2400 in guinea pigs (\cref{ComparativeCochlea}). While we do not know the exact effect of these large differences, they may impact the stiffness and its variability, and hence the degree of phase modulation that the OHCs can generate.

Identifying a basic time lens mechanism and obtaining estimates of its magnitude will turn out useful throughout this work, despite the inevitable lack of certainty in the estimation. It should be recognized that a very simple imaging system may be designed without a lens altogether---a pinhole camera that has only a small aperture instead of a lens (\cref{GeometricalOptics})---so the possibility that the auditory system might be lens-less after all will be considered at several points later on. 

Another challenge to the time lens is that its effect is not apparent in the auditory nerve and other measurements. It may be because it is too small, too short, too slow, or too localized. The choice of stimulus may also be critical in observing the time lens effect anywhere beyond the organ of Corti. The effect of the phase modulation is subtle in the time domain when measured in the auditory nerve because of the limited duration of the temporal aperture---a feature of the system that is coupled to the filter bandwidth and will be examined in later chapters in detail. 

At least three things were neglected in the derivation of the lens that should be eventually corrected in more refined analysis: compressive nonlinear dependence on level, asymmetry of the curvature with respect to the carrier (or CF), and phase function components that are higher-order than quadratic---perhaps contributing to the asymmetry. These will be briefly explored in \cref{HigherOrderAb}.

The values obtained for the focal time of the lens can be interpreted as the additional group delay that would be required to cancel out the effect of the lens after the wave left it \citep{Kolner1994b}. Given that the input curvature is relatively small and the distances between the auditory nuclei are short, the highest values obtained in the focal-time range of some models ($>$ 100 ms) appear to be grossly overblown to be compensated for by dispersion, as would be expected from an imaging system in sharp focus (\cref{ImagingEqs}). However, the interpretation of the extreme values is going to turn out to be nontrivial, as it does not follow the same design logic as the eye. 

It will be argued much later in this work that the lens curvature is variable by design through accommodation in both visual and auditory systems. Accepting that the time-lens curvature (and associated effects) may be variable poses a complicating factor to modeling, though, because we do not know what is the ``relaxed'' or ``normal'' position for the time lens. In the spatial lens of the eye, this position is referred to as \term{emmetropic} whereby the lens focus is set to ``infinity''. Effectively, objects that are 6 m or farther from the eyes are at infinity \citep[e.g.,][]{CharmanBass3}. 

While we do not know what was the root cause for the difference between the animals that exhibited small-curvature vs. large-curvature time lens estimates, it is not impossible that the state of the accommodation at the time of measurement had that effect, although with unknown experimental conditions that have led to it. Hypothetically, this is supported by the study by \citet{Guinan2008}, whose results were used to derive large-curvature values in \cref{PhaseModEvidence}. The only obtainable data from that study was relative, but we used it as absolute curvature for lack of an absolute reference. The corresponding underlying assumption in doing so is that the uninhibited MOC produces nearly zero curvature. This is not too different from the small-curvature values we got, which fluctuated around zero.

Finally, the physiological mechanism of achieving phase modulation that was explored here is yet another function that is stacked on the organ of Corti, in addition to the PLL that was examined earlier. The two functions do not necessarily have to interact and they may be realized by different parts of the organ of Corti or the OHCs. Specifically, the phase modulation may be a result of the bottom part of the basilar membrane, whereas the PLL is dependent on the hair bundle and a specific feedback path through the OHC soma. Both functions assume a role for the somatic motility of the OHCs, which supplies power either to the PLL loop gain, and to the stiffness modulation. Independently, the operation of the OHC is thought to utilize the very same mechanisms to achieve its amplification function. 


\section{Neural dispersion}
\label{NeuralDisp}
Having estimated the dispersive properties of the BM and the OHCs, we are left with the final part of the periphery---the IHCs and the auditory nerve---before entering the central nervous system at the brainstem. In a single-lens imaging system, as the auditory system likely is, this segment behind the lens is most conveniently modeled as a single dispersive unit. This is true even if it combines several media with different dispersive properties (as is the cochlear group-delay dispersion $u$), for the reason that dispersion is mathematically additive (see \cref{PulseCalc}). However, the transduction of the sound wave to neural action potentials represents a fundamental departure from the more explicitly-physical mechanical waveforms. 

Different paradigmatic approaches are common in the modeling of the acoustic-to-neural transduction. In signal-processing-oriented models, the hair bundle motion and neural transduction coupling are usually accounted for by signal rectification and low-pass filtering, while still treating the signal as continuous. These two operations entail amplitude demodulation, or (real) envelope extraction (\cref{AcouAudComm}). As the paratonal equation is already framed in the modulation domain, these operations may be neglected, as long as the carrier informs the envelope. In other words, tonotopy dictates that even a demodulated response would always be associated with its high-frequency carrier. The approach in neuroscience is usually to conceptualize neural transduction as a coding operation, which emphasizes the representational transformation that the physical referent (e.g., the mechanical wave) undergoes \citep{Perkel1968}. Instead, in the present work, we would like to employ a more primitive operation that conceptually precedes coding (in the informational theoretic sense)---sampling. This ensures that information is conserved in the process through discretization, which can also provide several insights into the system processing for later (see \cref{TemporalSampling} for a more in-depth discussion). It simplifies the discussion by avoiding coding intricacies such as the spontaneous rate of the different auditory fiber types (high, medium, or low; \citealp{Liberman1978}).

\subsection{The inferior colliculus is the candidate auditory retina}
Having delineated its reach and contents, the analysis of the final segment of the dispersive path of the sound signal boils down to a single question: what is the destination of the signal? Or alternatively---if a complete imaging system view is adopted (as will be shown in \cref{ImagingEqs})---what is the ``screen'' on which the final image is ``projected''? Candidate areas can be argued for given their key roles in auditory perception and processing: the auditory nerve, the cochlear nucleus (CN), the inferior colliculus (IC), and the primary auditory cortex (A1). We would like to argue that the IC is the destination and its role can be likened to an ``\term{auditory retina}''\footnote{This term has been used once in literature to analogize the function of the fish ear, but with no particular reasoning for why that is so \citep{Yoda2002}.}. There are several arguments that can be made to support this claim, each from a different standpoint. The first two arguments complement those that were made in \cref{AnaPhysioComp}:
\begin{enumerate}
	\item Anatomical analogy---The retina is where an optical image is formed, which the visual system can then process, whereupon it culminates in visual perception. The retinal connection to the brain is unique among the peripheral senses, because the second cranial nerve (the optic nerve), which connects to the retinal ganglion cells, is in fact part of the central nervous system that projects from the forebrain \citep[pp. 7-10]{Rea2014}. Specifically, the optic nerve is projected from the lateral geniculate body (LGB) in the thalamus and from there to the visual cortex in the occipital lobe. Only a small fraction of the optic fibers bypass the LGB and lead to the pretectal nucleus and to the superior colliculus (SC)---two midbrain structures that are responsible for various reflexive visual functions. In analogy, the primary projection from from the IC is also to the thalamus---to the medial geniculate body (MGB), which is considered the main nucleus between the IC to A1 \citep{Malmierca}. Incidentally, some of the IC subnuclei project to the SC and the pretectal nucleus as well \citep{Kudo1980}. More intricate analogies between the IC and the retina exist, based on function and processing \citep{Kvale2004}.
	
	\item System physiology---All information from the CN, the superior olivary complex (SOC), and the lateral leminiscus (LL) converges in the IC \citep[demonstrated on the cat]{Aitkin1984}, with very few exceptions of fibers that directly project from the CN to the contralateral MGB (see Figure \ref{BrainstemFig}). Its importance is also manifested in the number of neuron cells it has compared to other subcortical auditory structures---an average of 373,000 in the rat, which is one or two orders of magnitude more than in the CN, the SOC, the LL and the MGB \citep{Kulesza2002}. Also, the IC appears in the auditory system of all mammals and in birds \citep{Casseday1996}, and has a homologous structure (the torus semicircularis) in the midbrain of amphibians, reptiles, and fish \citep{BassWiner2005}. Finally, out of all the brain structures (including all auditory nuclei), the glucose metabolized by the IC is the highest---about twice as high as the superior colliculus in rhesus monkey \citep{Kennedy1978} and in albino rats \citep{Sokoloff1977}.
	
	\item Function---Another unique feature of the IC is that different signal processing pathways converge to organized maps that share the same tonotopy. In the IC, tonotopic maps are organized in characteristic iso-frequency laminae, which are thought to be orthogonal to a further map of periodicity that is then propagated to A1 \citep{Langner1997}. This property of the IC suggests that the information necessary for cortical processing is complete at that stage, even if certain dimensions (e.g., spectral, temporal, and spatial) of the stimulus are separately processed downstream. 
	
  \item Information---The coding of the modulation transfer function in the brain distinctly shifts in the IC to a rate code from a temporal code that is more characteristic in the CN and SOC \citep{Casseday1996,Joris2004}. It suggests that a maturer degree of signal processing may be possible at this stage, which was unavailable in the brainstem nuclei. It may be deduced, for example, from studies in bats who could still echolocate in part after the ablation of their A1 \citep{Suga1969AI}, but not at all after ablation of their ventral IC \citep{Suga1969IC}. From a system design point of view, it seems efficient that there should not be a change in coding before the imaging process is complete.
\end{enumerate}

\subsection{The existence of the neural dispersion}
\label{NeuDispExist}
We would like to estimate the dispersion of the auditory signal path from the back of the time lens---presumably in the organ of Corti at the reticular lamina before being coded in the IHCs---all the way to the IC. This should include the effects of the auditory nerve and the IC, as well as the intermediate pathways in the brainstem. Unfortunately, it will be impossible to isolate the response of the IHCs from the previously calculated live cochlear dispersion. The synaptic delay of the auditory nerve is often taken to be constant \citep[e.g.,][]{Palmer1986,Ruggero1987}, which would mean that its group delay and group-delay dispersion are both zero. This leaves us with the dispersive contribution of the auditory nerve fibers and the central nervous system as the dominant component of this dispersion. Hence, we refer to it as neural dispersion, even though it begins in the cochlea.

Neural dispersion has been hypothesized a number of times in the past \citep[e.g.,][]{Neely, Fobel, Harte2009}, but was ruled out more often than not. For example, \citet{Neely} estimated the difference between the latency of wave V in ABR and OAE measurements of two evoked-response datasets with similar tone-burst stimuli \citep{Gorga, Norton}. In theory, the ABR includes both the mechanical and neural pathways, whereas the OAE response includes only (approximately double the) mechanical path (more about it below). The group delay was estimated according to the two measurements and a good match was obtained to within $\pm2$ ms \citep[Figure 3]{Neely}. The two estimates were assumed to differ mainly due to the neural pathway that has a constant delay. The authors postulated that any significant neural effect is practically eliminated at low stimulus levels, as the two measurements differ only by a frequency-independent delay. Indeed, the ABR and OAE group delay measurements appear to have converged. In another example, the group delays of waves I, III, and V were compared using derived-band ABR and were found to vary by a constant, which implied that they are determined by the auditory nerve alone and any frequency-dependent group delay propagates downstream from there, unchanged \citep{Don1978}.

Even though auditory neural pathways may appear to be dispersionless, they are physical transmission paths and as such must have a finite dispersion. To the best knowledge of the author, the only data that explicitly demonstrate it is from a study by \citet{Morimoto2019}. The objective of that study was to maximize the peak response of either wave I or wave V of a chirp-evoked ABR measurement, which was designed to compensate for the cochlear dispersion and concentrate as much energy as possible at the peak of wave V \citep{Elberling2010b}\footnote{The chirp-evoked electrophysiological measurement was originally introduced by \citet{Shore1985}, in an attempt to counter the asynchronous auditory channel activation due to dispersion in click-evoked compound action potential measurement. The underlying principle here is essentially the same as chirp radars and ultrashort pulse generation, where inverse operations are used to compress otherwise long pulses whose power is too dispersed (\cref{ParaxialBackground}).}. Using data from 25 normal-hearing subjects, different chirp slopes were found that maximized the two wave peaks, albeit with large individual variation. It suggests that the path between the areas that corresponds to wave I (the auditory nerve) and wave V (the contralateral LL or the IC) is (group-delay) dispersive\footnote{Note that the IC itself does not produce strong enough electric field that can be detected with ABR due to its disorganized layout. Hence, wave V corresponds to an earlier timing than would characterize the central nucleus of the IC---usually attributed to the area between the LL and the IC \citep[p. 45--46]{Hall2007}.}. Note that these frequency-dependent differences between wave I and wave V delays were not reliably reproduced, as observed using the same chirps and other click stimuli in derived-ABR measurements by \citep{deBoer2022}. If neural dispersion difference between waves I and V appeared at all, it was relatively small (especially on the group level) and its effect was not monotonic in frequency. 

More indirect evidence for neural dispersion can be gathered from octopus-cell recordings in the mouse by \citet{McGinley2012}. The octopus cells in the posteroventral cochlear nucleus (PVCN) work as broadband coincidence detectors (\cref{InterauralCoherence}), where dendrites from different cochlear locations converge. The different dendrite lengths compensate for the across-channel delay that is caused by the cochlear dispersion and thereby allow for the temporally precise detection to take place, effectively time-compressing the broadband output from the cochlea\footnote{Recent recordings of the octopus cells in gerbils uncovered high temporal precision that is also sensitive to the direction of linear frequency sweeps around frequency ``hot spots'' (i.e., tuned input fibers from the auditory nerve)---a cellular detection mechanism that appears to be independent of the coincidence detection mechanism \citep{LuSmith2022}.}. As these findings apply only to one specific cell type and function, it is unknown at this stage if and how they should be generalized to the other brainstem nuclei. This may be reinforced by findings from the big brown bat, which showed that tuned units had a range of latencies that grew from the CN (smallest) to the LL, and through to the IC (largest) \citep{Haplea1994}. These differential delay lines inevitably create neural dispersion, although with patterns that may be difficult to pin down using a single parameter. 

Simultaneous measurements of evoked ABR and TOAE may be also used to show the existence of neural dispersion, as there is generally a small but consistent difference between the slopes of the two. An example of this difference was displayed in Figure \ref{inner}, where the unused OAE and ABR estimates of cochlear group delay and group-delay dispersion are plotted. If the two represented only cochlear dispersion, as standard theory has it \citep[e.g.,][]{Neely}, then they would only differ by a constant delay that does not affect group-delay dispersion. However, their slopes are different, which means that their group-delay dispersions are frequency dependent and different from one another. This subtle difference opens up the possibility of computing the neural dispersion from the difference between these ostensibly identical estimates of the ear's group delay. Hence, differentiating the group delay difference between ABR and OAE measurements (using Eq. \ref{GDDder}) should give us the neural group-delay dispersion $v$
\begin{equation}
v = \frac{\beta_2''\zeta_2}{2}= \frac{1}{2}\frac{d}{d\omega}(\tau_{gABR} - \tau_{gOAE})
\label{neuralD} 
\end{equation}
where the frequency-dependent group delay of the ABR is $\tau_{gABR}$ and of the OAE is $\tau_{gOAE}$. 

To make things more complicated, though, the interpretation of the various types of OAEs, the TOAE amongst them, requires a model that identifies both the generator and/or the source of reflection in the cochlea that accounts for the return travel time of the emission. As in other cochlear research questions, there is no universal agreement about these issues. The main point of controversy is determining whether different types of evoked OAEs occur due to reflections from irregularities in the cochlear geometry and mechanics, or rather from active nonlinear mechanisms that act as sources, or some combination of both \citep[e.g.,][]{Probst1991,Shera2008Manley}. It appears that both reflection and nonlinear models identify the OHCs as the site of (re-)emission. Even then, there remains an additional uncertainty regarding the return path of the reverse wave, which is factored into the group delay estimation. If a reverse wave returns through the BM, then we can expect that the group delay is double the forward path, $\tau_{OAE}(\omega) \approx 2\tau_{BM}(\omega)$ \citep{Shera2003}. However, in species where the group delay could be measured directly from the mechanics, it turned out smaller than this prediction (e.g., for the chinchilla the factor is 1.86 instead of 2; \citealp{Cooper2004}). It suggests that at least some of the energy may be returned to the middle ear in another (faster) path different from the forward path. In the present context, the important distinction is between an OAE measurement that includes or excludes the dispersive contribution of the phase-modulating organ of Corti, both in the forward and reverse paths. While resolution of this controversy is beyond the scope of this work, using the OAE in Eq. \ref{neuralD} as though it includes the time-lens curvature seems to work reasonably well and does not require a correction at this stage. However, we note that there is an unknown error expected using the general method of the neural dispersion estimation based on Eq. \ref{neuralD}. 


\subsection{Neural dispersion estimation}
\label{NeuralDispEst}
The neural dispersion will be estimated here based on Eq. \ref{neuralD} and on the small but persistent difference in group delay of ABR and OAE measurements found in literature. Comparable measurements of OAE and ABR were reported several times, but despite the qualitative similarity of the results, they are numerically inconsistent. This inconsistency is exacerbated upon differentiation, which is where group-delay dispersion arises. 

\begin{figure} 
		\centering
		\includegraphics[width=1\linewidth]{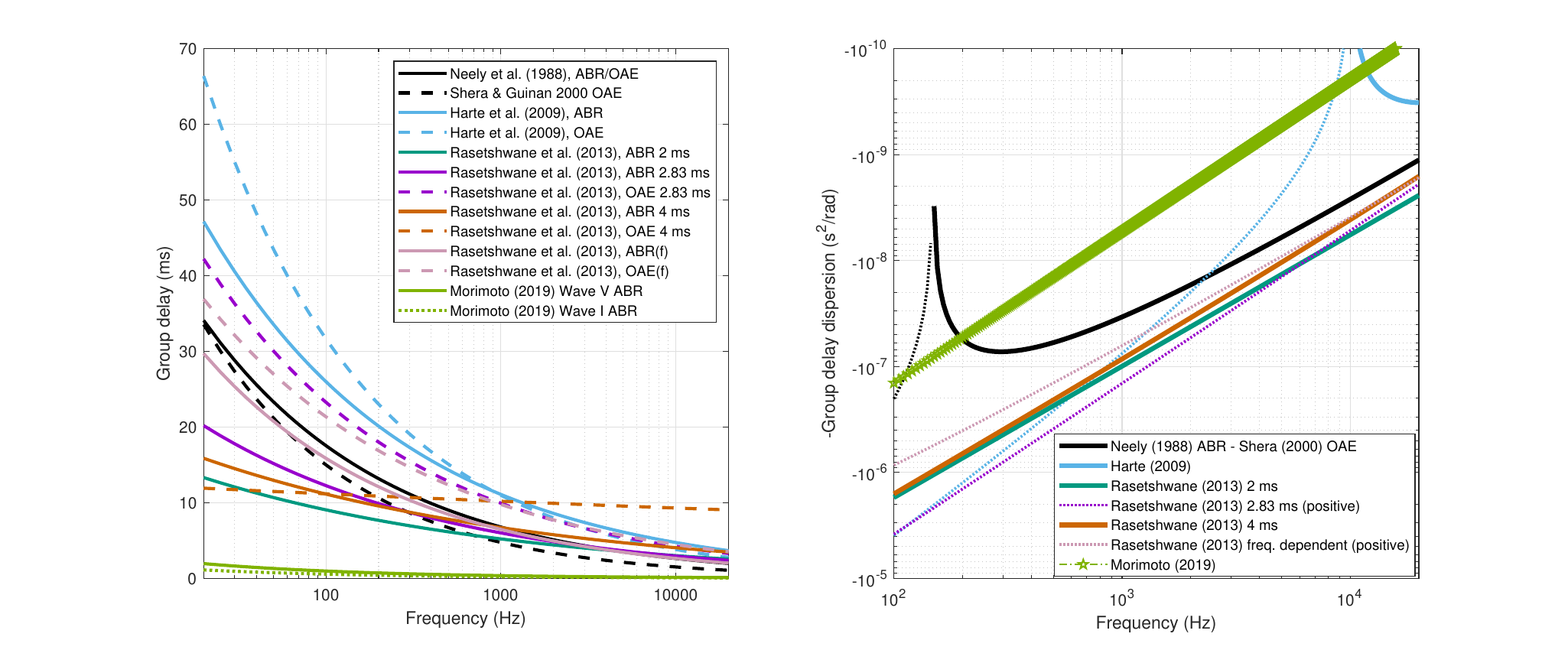}	
		\caption{Neural group-delay dispersion estimates based on power-law fits from literature. \textbf{Left:} Cochlear (and neural) group delay fitted according to the power-law functions summarized in Table \ref{OAEABRdata} (ABR---solid curves, OAE---dashed curves), omitting constant delays. Note that the curve for the 2 ms OAE condition of \citet{Rasetshwane} is identically 0 and does not appear on the plot. Two curves of frequency-dependent tone bursts are marked in the legend with (f). \textbf{Right:} The (negative) neural group-delay dispersion based on differences between the paired ABR and the OAE curves, which were taken as the neural group delay (Eq. \ref{neuralD}). The dash-dot-star green curve marks the $\tau_{gV} - \tau_{gI}$ dispersion according to \citet{Morimoto2019}, which sets a lower bound for the complete neural dispersion path. Solid curves mark the (desired) negative dispersion, whereas dotted curves are positive.}
		\label{neuraldisp}
\end{figure}

The OAE and ABR measurements by \citet{Neely} were originally fitted to a level-dependent power law that has also been adopted in several other studies later (see also \citealp{Anderson1971})
\begin{equation}
\tau_g = a + bc^{-i}f^{-d}
\label{eqNeely}
\end{equation}
where $i$ is the ratio of the input in dB SPL to 100 dB SPL, and $f$ is given in kHz. The intercept $a$ was added to account for the constant neural delay, typically set to 5 ms. The constants $b$, $c$, and $d$ are provided in Table \ref{OAEABRdata}\footnote{Note that the group delay level dependence itself appears to be frequency dependent, as was recently shown for ABR measurements by \citet{Huang2022}.}. The corresponding group delay curves by \citet{Neely} as well as additional measurements that used the same power law form as Eq. \ref{eqNeely} are all plotted in Figure \ref{neuraldisp} (left) and are summarized in Table \ref{OAEABRdata}. Studies were generally preferred if their ABR and OAE responses were recorded simultaneously, or at least had individual-subject-matched data\footnote{Other ABR and OAE estimates that were not necessarily fitted by power laws were compiled in \citet{Moleti2008}, but were not explored here.}. The resultant group-delay dispersions of these fits (Eq. \ref{neuralD}) are shown on the right plot of Figure \ref{neuraldisp}. 

\begin{table}
\centering
\scriptsize
\begin{tabular}{p{2.5cm}|p{1.5cm}|p{1cm}|p{2.5cm}|p{1cm}|p{1cm}|p{3.5cm}}
\textbf{Study}&\textbf{b}&\textbf{c}&\textbf{i}&\textbf{d}&Level (dB SPL)&\textbf{comments}\\
\hline
\multicolumn{7}{c}{\textbf{Auditory brainstem response (ABR)}}\\
\hline
\citet{Neely}&12.9&5&L dB / 100 dB SPL&0.413&10--100 &non-simultaneous ABR and OAE\\
\citet{Harte2009}&11.09&1&1&0.37& 66 &Tone burst ABR, 0.5--8 kHz\\
\citet[Table III]{Rasetshwane}&9.99&5.1&L dB /100 dB& 0.24&20--90&ABR, 2 ms tone bursts\\
&11.47&5.05&L dB /100 dB&0.31&20--90&ABR, 2.83 ms tone bursts\\
&13.89&6.17&L dB /100 dB&0.22 &20--90&ABR, 4 ms tone bursts\\
&12.63&5.34&L dB /100 dB&0.39&20--90&ABR, frequency-dependent tone bursts\\
\citet{Morimoto2019}&$0.00920n_k/9$&1&1&0.4356& 60 dB HL, or 104 dB SPL (peak) &$0\le n_k \le 9$, $f$ in Hz and $t_g$ in s; chirp ABR to find maximum wave-I response\\
\hline
\multicolumn{7}{c}{\textbf{Otoacoustic emissions (OAE)}}\\
\hline
\citet{SheraGuinan2000}& 0.15 & 1 & 1 & 0.5&40 &OAE model was provided by \citet{Fobel}\\
\citet{Harte2009}&10.98&1&1&0.46& 66 & Tone burst OAE, 0.5--8 kHz\\
\citet[Table V]{Rasetshwane}&16.40&9.32&L dB /100 dB& 0.00&20-90&OAE, 2 ms tone bursts\\
&20.41&6.06&L dB /100 dB&0.37&20--90&ABR, 2.83 ms tone bursts\\
&19.00&4.75 &L dB /100 dB&0.04&20--90&ABR, 4 ms tone bursts\\
&20.56&6.44&L dB /100 dB&0.34&20--90&ABR, frequency-dependent tone bursts\\
\end{tabular}
\caption{Summary of various power-law fits found in literature for evoked otoacoustic emissions or auditory brainstem response data according to the power law prescribed by \citet{Neely}, Eq. \ref{eqNeely}, where the frequency $f$ is in kHz, and the group delay $\tau_g$ in milliseconds. Wherever $i \neq 1$, the fraction of the dB level $L$ over 100 dB is used. Constant delays (e.g., 5 ms neural delay) are omitted. ABR level dependence of group delay that was observed to also be frequency dependent was modeled using a somewhat different power law in \citep{Huang2022}.}
\label{OAEABRdata}
\end{table}

Since neural dispersion is currently a hypothetical property of the auditory system, the results from the study by \citet{Morimoto2019} were used to cross-validate it. The group delay of the chirps that were used as stimuli compensated for the neural group delay of the form given by \citet{Elberling2008} (``CE-chirp''):
\begin{equation}
\tau_{g} = 0.00920\frac{n_k}{9} f^{-0.4356}
\end{equation}
with the frequency $f$ in Hz, and the integer parameter $n_k$ varying between 0 and 9. By using chirps with the corresponding negative group delay, it was found that $n_k=4$ maximized the wave-I response and $n_k=7$ maximized the wave-V response. Thus, the same method as above can be used to estimate the group delay associated only with the wave-I to wave-V path, and differentiate it once to obtain the respective neural group dispersion:
\begin{equation}
v_{V-I} =  \frac{1}{2}\frac{d}{d\omega}(\tau_{gV} - \tau_{gI}) =  \frac{-0.4356}{4\pi}0.00920\frac{7-4}{9} f^{-1.4356} = -0.0001063f^{-1.4356}\,\,\s^2 /\rad
\end{equation}
Ideally, this partial group-delay dispersion (thick green curve in Figure \ref{neuraldisp}, right) is smaller than the total path, $|v_{V-I}|<|v|$, and has the same sign, which implies that the growth of the group-delay dispersion is a monotonic function of neural distance. This reasonable (yet unproven) assumption can be used as a key to select the negative group-delay dispersion results (solid curves in Figure \ref{neuraldisp}), over the positive ones (dotted curves in Figure \ref{neuraldisp}), which rules out the positive, frequency-dependent tone-burst measurements obtained by \citet{Harte2009} and \citet{Rasetshwane}. Incidentally, the method in the latter responses was deemed invalid due to stimulus duration dependence \citep{Ruggero2007}. Furthermore, the curve derived from \citet{Neely} and \citet{SheraGuinan2000} is also positive at low frequencies and becomes smaller than the wave-V to wave-I data. This leaves the 2 and 4 ms fixed-rise-time responses by \citet{Rasetshwane} as the most favorable candidates from which to obtain the neural group-delay dispersion. However, for an unknown reason, the 2.83 ms fixed rise-time response \citet{Rasetshwane} did not produce the desired curve, which was expected to lie between the 2 and 4 ms curves. This discrepancy produces some uncertainty as for how confident we can be in the data using this method. However, as will turn out, the remaining 2 and 4 ms neural group dispersion data from \citet{Rasetshwane} both produced plausible values that could be employed in the rest of this work. Specifically, the 4 ms dataset produces slightly better results and was used throughout. 

\section{Discussion}
This chapter systematically analyzed the dispersive properties of the human auditory path from the outer ear to the inferior colliculus (IC) in order to have plausible estimates of its group-delay dispersion. While the possibility of dispersion should not be controversial anywhere in the system, the segmentation process that leads to associating various elements in the system (summarized in Figure \ref{DispStages}) with different measurements is not free of assumptions that may turn out to be inaccurate. Nowhere has it been more conspicuous than in and around the organ of Corti, wherein we hypothesized the time lens resides. As another layer of complexity, this work hypothesizes a mechanism for phase modulation that can neatly function as a time lens, but has not been explicitly measured to date. This adds up to the earlier theory that established a PLL function as yet another role for the OHCs. While the effect on the present data is inconsequential in many of the results obtained in the next chapters, at least some of these controversies will eventually have to be empirically settled in order to be able to get better estimates of the dispersive system parameters. 

Several simplifying assumptions have been made to be able to parse the system more efficiently, which will have to be relaxed when higher certainty is obtained. The two main ones are the neglecting of all level considerations (data were obtained for low levels or 40 dB SPL, whenever possible) and the treatment of the entire audio range as scalable, with no regard for anomalies of low or high frequencies. Indeed, a correction will be required for frequencies below 500 Hz in some of the results obtained later. Another simplifying and necessary assumption has been to treat the different auditory pathways between the brainstem and the IC as a single dispersive path. We do not know whether the parallel processing of the brainstem is precisely timed, so that outputs from the two or three branches (VCN, DCN, PVCN) simultaneously converge in the IC, but two studies were mentioned that indicate that this may not be universally the case \citep{Haplea1994,McGinley2012}. 


Neural transmission codes the information carried by the mechanical waves from the cochlea to the brain. As such, it was taken here as a proxy of a real physical process that has non-zero dispersion, by definition (see \cref{airtravel}). Whether the dispersion itself is a feature of the auditory code or an epiphenomenon of its transmission is subject to future exploration. In this work, only the latter alternative is directly explored---dispersion that is evident from brainstem potentials that are not decoded, but are treated as aggregate activity that can be neurophysiologically localized.

An alternative derivation of the dispersion parameters will become possible at a much later stage of this work---once the full theory is developed---using a battery of four psychoacoustic tests (\cref{PsychoEstimation}). With a limited pool of available data, most of the general trends observed above can be tentatively cross-validated, as long as the parameters are allowed to be complex, so absorption becomes more dominant. They suggest that at the lowest frequencies, one of the dispersion parameters---most likely the time-lens curvature---changes sign. However, this psychoacoustic solution is not going to be used in the text, as the majority of the studied effects can be studied qualitatively with the parameters obtained above, and without the troubling inclusion of absorption in the theory. 

All in all, we obtained estimates for the cochlear and neural group-delay dispersions that turned out both negative. The time-lens curvature, in contrast, is positive. The combination of these three frequency-dependent parameters will be used to explore the temporal auditory imaging equations in the next chapters.

\chapter{The temporal imaging equations}
\label{ImagingEqs}
\section{Introduction}
Having obtained a paraxial-equation analog as well as two distinct dispersive auditory segments separated by a time lens, we are now in a position to put them together and see how they work as a full imaging system. This chapter is therefore dedicated to the presentation of the temporal imaging equations. The equations will be then applied to psychoacoustic data of Schroeder-phase complex thresholds that specifically targeted the cochlear phase curvature, although we reinterpret them as applying to the entire auditory dispersive path. Derived aperture time durations will be compared with direct measurements of temporal window of the chinchilla along with a few more human psychoacoustic temporal window data points.

Using the paratonal equation (\ref{eq:dispersion}) (originally, the ``paraxial'' dispersion equation) of \citet{Akhmanov1968,Akhmanov1969} as a foundation, the analogy to a full spatial imaging system in the temporal domain was made complete by \citet{KolnerNazarathy}. \citet{Kolner} then perfected the solution and elaborated the theory \citep{Kolner1994b,Kolner1997}, and together with colleagues included treatments for aberrations, additional imaging applications, and some alternative configurations to the basic setup \citep{Bennett1995,Bennett1999a,Bennett1999b,Bennett2000,Bennett2000b,Bennett2001}. A very similar analogy to spatial imaging that anticipated their work was suggested earlier by \citet{Tournois1968}. Another similar technique using electric signals was also introduced independently by \citet{Caputi}, where the time lens was replaced with a standard mixer, but did not receive much attention at the time. Aspects of this theory and other applications of the dispersion equation have been successfully applied in many high-end optical systems from the 1990s. Most notably, it is applied in real-time fast optical spectroscopy that can image microscopic events at the femtosecond range. For reviews of principles and applications of the space-time analogy in optics, see \citet{Salem}, \citet{Goda}, \citet{Torres2011} and \citet{Kolner2011Chen}.

\section{Temporal imaging with a single time lens}
\label{ImagingDerivation}
The derivation of the temporal imaging equations is based on \citet[Section VII]{Kolner}, which itself is completely analogous to the standard Fourier optics approach to spatial imaging, as is derived in \citet[pp. 155-177]{Goodman}. Kolner's derivation will be then supplemented with equations that include the effect of defocusing. 

The analysis follows the complex envelope of a modulated signal (a pulse) through a first dispersion, into a time lens, and out through a second dispersion, where an image is formed. The modulation is assumed to be narrowband around a constant carrier that will be implicit throughout the analysis---a condition that in the context of hearing we called paratonal. Therefore, while the carrier changes between acoustic, vibrational, compression, mechanical traveling wave, shear motion, ciliary movements, electrical, and neural forms, the information that it carries is assumed to be conserved (\cref{AudInfoConserve}). As in \cref{temporaltheory}, the medium absorption (or gain) is taken to be constant in frequency with negligible higher-order terms. 

At the origin, an envelope $a(0,\tau)$ with spectrum $A(0,\omega)$ is carried by a plane wave carrier through a dispersive medium. Let us define the group-velocity dispersion transformation $D$ based on the kernel of the solution of the paratonal equation \ref{eq:Fsolution},
\begin{equation}
D (\zeta  ,\omega ) = \exp \left( { \frac{{- i \beta''}\zeta \omega ^2}{2} } \right) 
\end{equation}
The input dispersion in the auditory system was defined by the combined effect of the outer, middle, and inner ears, up to the time lens in the organ of Corti (Eqs. \ref{GDvsGVD} and \ref{utotal}). Therefore, any reference to $\beta''$ and $\zeta$ can be made implicit
\begin{equation}
u \equiv \frac{{\beta''_1}\zeta _1}{2}
\end{equation}
The first dispersion $D_1$ is therefore defined as
\begin{equation}
D_1 (\zeta _1 ,\omega ) = \exp \left(  -iu\omega ^2 \right) 
\label{disp_stage}
\end{equation}
Using $D_1$, the general inverse-Fourier transform solution of Eq. \ref{eq:Fsolution} is rewritten to obtain the time-domain envelope at the output of $D_1$, in the traveling-wave coordinate system
\begin{equation}
a(\zeta _1 ,\tau ) = {\cal F}^{ - 1} \left[ {A(0,\omega )D_1 (\zeta _1 ,\omega )} \right]
\end{equation}
Following the first dispersion, we would like to add the time lens to the signal path. The effect of the time lens is multiplicative in the time domain (Eq. \ref{lens_function}) and is
\begin{equation}
h_L (\tau) =  \exp \left( {\frac{{is\tau ^2 }}{{4}}} \right)  
\label{lens_function_s}
\end{equation}
for a lens with curvature $s = f_T/2\omega_c$, $f_T$ being the focal time of the time lens, as was identified in the organ of Corti in \cref{OHCtimelens}. The time lens size is not factored in directly in its transfer function, so we take it to be a ``thin time lens'', that extends infinitesimally to $\zeta = \zeta_1+\varepsilon$. The envelope is then multiplied by $h_L(\tau)$ and becomes
\begin{equation}
a(\zeta _1  + \varepsilon ,\tau ) = {\cal F}^{ - 1} \left[ {A(0,\omega )D_1 (\zeta _1 ,\omega )} \right]h_L(\tau ) 
\label{radareq}
\end{equation}
and
\begin{equation}
A(\zeta _1  + \varepsilon ,\omega ) = \frac{1}{{2\pi }}\left[ {A(0,\omega )D_1 (\zeta _1 ,\omega )} \right] * H_L(\omega )
\label{radareqw}
\end{equation}
where the convolution operation is marked by $*$ and is applied according to the convolution theorem in the second equality. Behind the lens (where $\zeta > \zeta_1+\epsilon$) the wave propagates through second dispersion $D_2$, which is defined just like in Eq. \ref{disp_stage}, but this time for the neural dispersion, which includes, roughly, the contributions from the inner hair cells, auditory nerve, brainstem, all the way to the inferior colliculus (\cref{NeuralDispEst})
\begin{equation}
D_2 (\zeta _2 ,\omega ) = \exp \left(  -iv\omega ^2 \right) 
\label{disp_stage2}
\end{equation}
This dispersion then multiplies Eq. \ref{radareqw}, to obtain the frequency-domain transform of the complete path.  Applying the inverse Fourier transform on it, the final envelope is then obtained
\begin{equation}
a_n(\zeta _2,\tau ) = {\cal F}^{ - 1} \left\{ \left[ \left( \frac{1}{2\pi} A(0,\omega )D_1 (\zeta _1 ,\omega ) \right) * H_L(\omega )\right] D_2 (\zeta _2 ,\omega ) \right\}
\label{fullpathfreq}
\end{equation}  
where the subscript $n$ was added to emphasize that the envelope is sampled (or encoded) in the neural domain. This expression can be solved by writing the convolution integral and inverse Fourier transform explicitly, and by changing the order of integration
\begin{equation}
a_n(\zeta _2 ,\tau ) = \frac{1}{{4\pi ^2 }}\int_{ - \infty }^\infty  {A(0,\omega ')D_1 (\zeta _1 ,\omega ')d\omega '} \int_{ - \infty }^\infty  {e^{i\omega \tau }  D_2 (\zeta _2 ,\omega )H_L(\omega  - \omega ') d\omega}
\label{Fullpath}
\end{equation}
The lens transfer function $h(\tau)$ (Eq. \ref{lens_function_s}) was Fourier transformed earlier (Eq. \ref{TimeLensOmega2}) and is repeated with the time shifted argument of the convolution integral
\begin{equation}
H_L(\omega  - \omega ') = \sqrt{4\pi is} \exp \left[-is (\omega  - \omega ')^2\right]
\label{timelensF}
\end{equation}
The second integral in (\ref{Fullpath}) can be solved first
\begin{multline}
 \frac{1}{{2\pi }}\int_{ - \infty }^\infty  {D_2 (\zeta _2 ,\omega )H_L(\omega  - \omega ')e^{i\omega \tau } d\omega }  =  \sqrt {\frac{{is}}{\pi }} \int_{ - \infty }^\infty  {\exp \left( { - iv\omega ^2 } \right)\exp \left[ { - is(\omega  - \omega ')^2 } \right]} e^{i\omega \tau } d\omega  \\ 
  =  \sqrt {\frac{{is}}{\pi }} \exp ( - is\omega '^2 )\int_{ - \infty }^\infty  {\exp \left( { - i(v + s)\omega ^2 } \right)\exp \left[ {i\omega (\tau  + 2s\omega ')} \right]} d\omega \\
  = \sqrt {\frac{s}{{v + s}}} \exp ( - is\omega '^2 )\exp \left[ {\frac{{i(\tau  + 2s\omega ')^2 }}{{4(v + s)}}} \right] \\ 
\end{multline}
Using this expression back in (\ref{Fullpath}), we obtain a closed-form formula for the output temporal envelope, as an integral transform of the input envelope spectrum \citep{Kolner}:
\begin{equation}
a_n(\zeta _2 ,\tau ) = \frac{1}{{2\pi }}\sqrt {\frac{s}{{v + s}}} \exp \left[ {\frac{{i\tau ^2 }}{{4(v + s)}}} \right]\int_{ - \infty }^\infty A(0,\omega ')\exp \left[ { - i\left( {u + s - \frac{{s^2 }}{{v + s}}} \right)\omega '^2  + \frac{is\tau \omega '}{{v + s}}} \right] {d\omega '}
\label{totalpath}
\end{equation}
This is essentially a Fourier transform of the input envelope spectrum in a scaled coordinate system after it has been modulated by a quadratic phase term that depends on all three parameters $u$, $v$, and $s$. An additional quadratic phase term modulates the integral and depends only on $v$ and $s$. Eq. \ref{totalpath} is analogous to the full solution of the spatial imaging system \citep[p. 169, Eq. 6-33]{Goodman}, and it belongs to the important family of linear canonical transforms (LCT)---a generalization of Fourier and other integral transforms (see \cref{AppLCT}). 

\subsection{Ideal temporal imaging}
\label{imagingequations}
In order to recover the original envelope after the dispersive effect, it is necessary to look for a condition under which the quadratic phase term in the integrand of Eq. \ref{totalpath} vanishes, namely,
\begin{equation}
u + s - \frac{{s^2 }}{{v + s}} = 0 
\end{equation}
which is satisfied when 
\begin{equation}
\frac{1}{u} + \frac{1}{v} = -\frac{1}{s}
\label{temporal_imaging_condition}
\end{equation}
This \term{imaging condition} is the temporal analog to the spatial lens law (Eq. \ref{imaging_condition}) and is nearly identical functionally, only with opposite signs. 

\begin{figure} 
		\centering
		\includegraphics[width=1\linewidth]{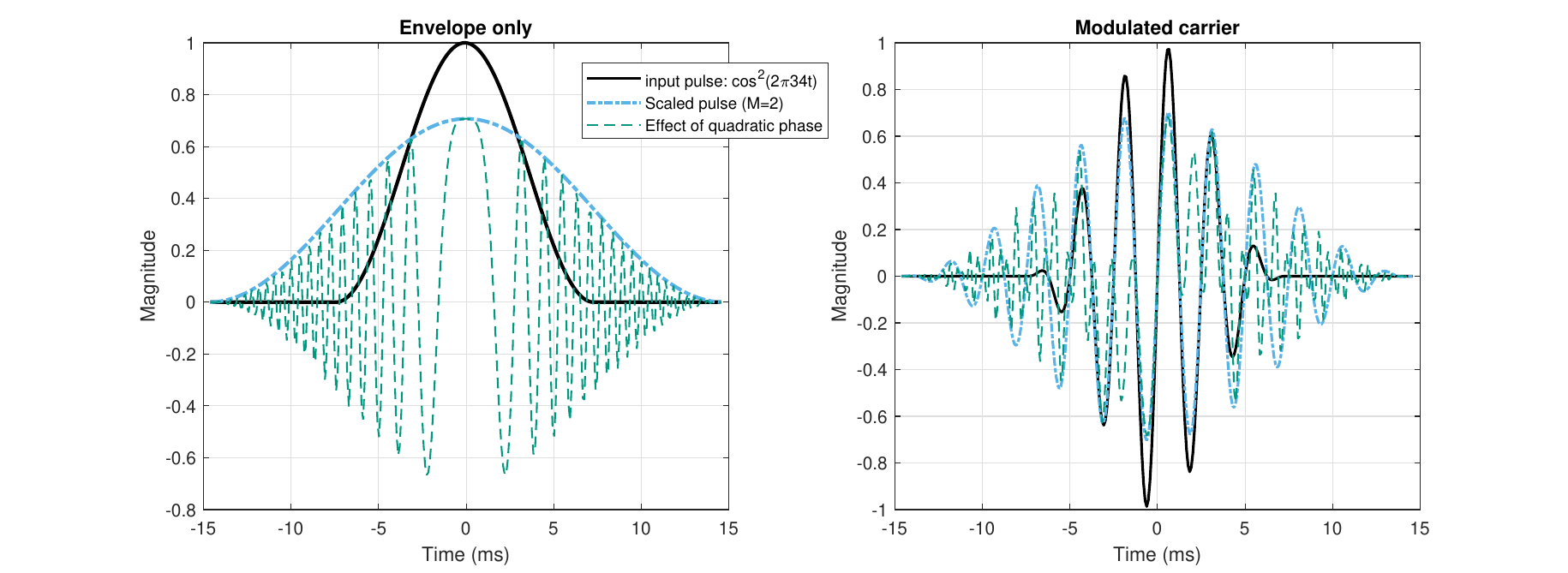}	
		\caption{An illustration of temporal imaging (Eq. \ref{eq:image}) for a cosine-squared pulse with: $f_c=400$ Hz, $f_{AM}=34$ Hz, $f_{T}=0.001\,\,\s$, and $M=2$ (magnification). \textbf{Left:} The initial envelope (narrow pulse), the magnified pulse envelope (wide), and the effect of the quadratic phase on the amplitude (real part only; green dash). \textbf{Right:} The modulated (real) signals before and after imaging, showing the chirping effect of the quadratic phase (in green) and the perfectly imaged unchirped amplitude image (in blue). Only the center-most portion of the pulse is unaffected by the quadratic phase, while the side lobes are corrupted by the global quadratic phase.}
		\label{figdispimage}
\end{figure}

Let us redefine the scaling factor in the inverse Fourier transform of Eq. \ref{totalpath} using the magnification $M$, also in analogy to spatial imaging,
\begin{equation}
M\equiv\frac{v+s}{s}
\label{MagnificationDef}
\end{equation}
When the temporal imaging condition (Eq. \ref{temporal_imaging_condition}) is satisfied then the magnification is additionally equal to $M_0$,
\begin{equation}
M_0 = -\frac{v}{u} = M
\label{MagnificationDef2}
\end{equation}
which is sometimes more convenient to use. Plugging the imaging condition of Eq. \ref{temporal_imaging_condition} and the magnification definition in Eq. \ref{totalpath}, and using the definition of $s$ (Eq. \ref{TimeLenss}) the output transform simplifies to
\begin{equation}
 a_n (\zeta _2 ,\tau ) = \frac{1}{{2\pi \sqrt M }}\exp \left[ {\frac{{i\omega _0 \tau ^2 }}{{2Mf_T }}} \right]\int_{ - \infty }^\infty  {A(0,\omega )\exp \left( {\frac{i\tau \omega}{M}} \right)d\omega }  
\label{eq:image0}
\end{equation}
which is identically equal to
\begin{equation}
 a_n (\zeta _2 ,\tau ) = \frac{1}{{2\pi \sqrt M }}\exp \left[ {\frac{{i\omega _0 \tau ^2 }}{{2Mf_T }}} \right]a \left( 0,\frac{\tau }{M} \right) 
\label{eq:image}
\end{equation}
This remarkable result indicates that at the output of the dispersive system, when the imaging condition of Eq. \ref{temporal_imaging_condition} is satisfied, a scaled version of the original temporal envelope is obtained. Except for the quadratic phase, this is an ideal image according to the general definition of an image (Eq. \ref{Idealimage}). It becomes exact for intensity imaging for which this global quadratic phase term vanishes. The conditions for which the quadratic phase can be neglected in amplitude imaging will be investigated later in the chapter (\cref{GlobalQuadPhase}). Once again, this temporal imaging result is completely analogous to the known two-dimensional spatial images \citep[pp. 169--172]{Goodman}. An example of this transform on an arbitrary pulse that is scaled by a factor of $M=2$ is displayed in Figure \ref{figdispimage} (left). The effect of the quadratic phase can be seen only on the amplitude waveform and not on the envelope. When the narrowband modulated waveform with the carrier is displayed, then the chirping effect of the quadratic phase is made visible (right). 

\subsection{Nonideal imaging of a Gaussian pulse}
\label{GaussImaging}
In general, we cannot assume that the imaging condition is satisfied and that the image is in sharp focus. Thus, the full transform of Eq. \ref{totalpath} has to be evaluated to obtain the defocused image. A general solution for arbitrary envelopes cannot be obtained for this transform, so the simplest illustration to its effect would be to use a particular envelope---a real (unchirped) Gaussian pulse of width $t_0$
\begin{equation}
a(0,\tau) = a_0\exp \left(-\frac{\tau^2}{2t_0^2} \right)
\label{inputpulse}
\end{equation}
The envelope spectrum is given by
\begin{equation}
A(0,\omega) = \int_{ - \infty }^\infty a(0,\tau) e^{-i\omega t}dt = \sqrt{2 \pi} a_0 t_0 \exp\left(-\frac{t_0^2 \omega^2}{2} \right)
\label{inputenvspec}
\end{equation}
Plugging it in Eq. \ref{totalpath} gives
\begin{equation}
a_n(\zeta _2 ,\tau ) = \frac{a_0 t_0}{\sqrt{2 \pi}} \sqrt {\frac{s}{{v + s}}} \exp \left[ {\frac{{i\tau ^2 }}{{4(v + s)}}} \right]\int_{ - \infty }^\infty  {d\omega '} \exp \left\{ \left[  \frac{- t_0^2}{2} - i\left( {u + s - \frac{{s^2 }}{{v + s}}} \right) \right]\omega '^2 + {\frac{i s \tau\omega '}{{v + s}} } \right]
\label{totalpathabs}
\end{equation}
Using Siegman's Lemma, the solution to Eq. \ref{totalpathabs} is given by
\begin{equation}
a_n(\zeta _2 ,\tau ) = a_0 t_0 \sqrt{ \frac{\frac{s}{v + s}}{2 \left[  \frac{t_0^2}{2} + i\left( {u + \frac{{vs }}{{v + s}}} \right) \right]}} \exp \left[ {\frac{{i\tau ^2 }}{{4(v + s)}}} \right]  \exp \left\{  \frac{-\left(\frac{s}{{v + s}}\tau\right)^2}{ 4\left[  \frac{t_0^2}{2} + i\left( {u + \frac{{vs }}{{v + s}}} \right) \right]} \right\} 
\label{absonlyfull}
\end{equation}
which reduces to the form of Eq. \ref{eq:image}, if the imaging condition is satisfied, so the imaginary term in the denominator of the second exponent and in the square root becomes zero. However, if the imaging condition is not satisfied, then the magnification factor cannot be completely separated from the pulse width itself, which results in an effective pulse broadening and chirping. We can define a new complex width $t'_1$ for the output
\begin{equation}
t'_1 = \frac{\sqrt{ t_0^2 + 2i\left( u + \frac{vs }{v + s} \right) }}{\frac{s}{v+s}} = t_0 M\sqrt{ 1 + \frac{2i}{t_0^2}\left( u + \frac{vs }{v + s} \right) } = M't_0
\label{altt0}
\end{equation}
where we also defined a new magnification factor $M'$, which is a complex function of the input width and the parameters of the system, including $M$,
\begin{equation}
	M' = M\sqrt{ 1 + \frac{2i}{t_0^2}\left( u + \frac{vs }{v + s} \right) }
\label{altM}
\end{equation}
The imaginary part of $M'$ results in chirping that may be undesirable in imaging. It can be reduced by minimizing the imaginary term under the radical in Eq. \ref{altM}, which can be rewritten 
\begin{equation}
2i\left(u + \frac{vs }{v + s} \right) = \frac{2i}{v}\left(\frac{1}{M} - \frac{1}{M_0} \right) \,\,\,\,\,\,\,\,\, v \neq 0
\end{equation}
using the definitions of $M$ and $M_0$ (Eqs. \ref{MagnificationDef} and \ref{MagnificationDef2}). Therefore, $M=M_0$ only in ideal imaging, which also does not chirp.

Finally, the output pulse of Eq. \ref{absonlyfull} can be written more compactly using the definition of $M'$, in a similar form to the ideal image
\begin{equation}
a_n(\zeta _2 ,\tau ) = \frac{a_0}{\sqrt{M'}} \exp \left[ {\frac{{i\tau ^2 }}{{4(v + s)}}} \right] \exp\left[-\frac{1}{2}\left(\frac{\tau}{M't_0}\right)^2\right] 
\label{absonlyfull2}
\end{equation}
Although specific to a Gaussian input, this expression is almost of the same form as Eq. \ref{eq:image}. As will turn out in the next section, this solution is more relevant to the auditory system than the ideal imaging transform. The implications of having a defocused imaging will be explored in depth later in this work. 

Although realistic signals are not limited to pulses, the system (mainly the time lens) can process only a limited extent of duration of every object. This is analogous to the spatial angle that an object occupies with respect to a camera lens, so it may still be fitted on a single image. In audio signal processing this is achieved with a \term{window function}, which is called an aperture in optics. The existence of an aperture in the auditory system will arise organically through the analysis of psychoacoustic data in the remainder of this chapter. A more detailed analysis of the aperture and its critical role on imaging will be deferred to the next chapter, where adequate inverse-domain tools will be developed. 

\section{The imaging condition and the auditory system}
\label{ImagingDefocusFound}
At this point, we can put to test the imaging equations using the frequency-dependent parameters of the human auditory system obtained in the previous chapter---the cochlear dispersion $u$, the neural dispersion $v$, and the various estimates of the time-lens curvature $s$ (\cref{lenscurve}). This will give us an indication of what kind of imaging may be possible in hearing. In the forthcoming analysis. the three parameters will provide surprisingly effective predictions, despite the manifest uncertainty in their values. 

\begin{figure} 
		\centering
		\includegraphics[width=0.9\linewidth]{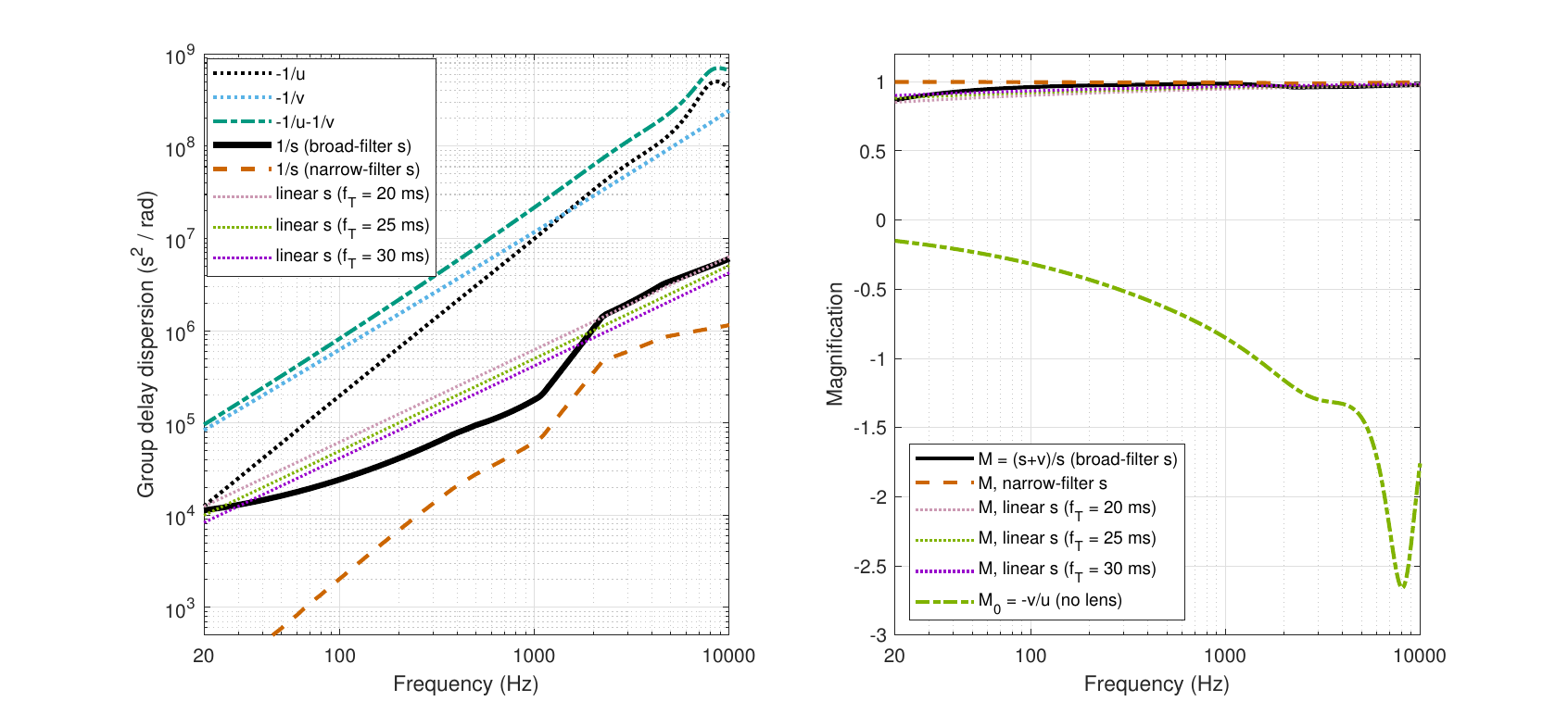}	
		\caption{Two aspects of the built-in defocus in the human auditory system. \textbf{Left:} Different terms in the temporal dispersive imaging condition, Eq. \ref{temporal_imaging_condition}. The reciprocal of the negative input dispersion ($-1/u$) and negative neural dispersion ($-1/v$) are in dotted black and blue lines, respectively. Their sum (dash-dot green) should be compared to the reciprocal of the time-lens curvature(s) ($1/s$), which is between one and two order2 of magnitude smaller for the broad-filter large-curvature lens estimate (solid black) and more than two orders of magnitude smaller for the large-curvature narrow-filter estimate (dash red). The intermediate constant focal-time estimates are displayed as well, as the three parallel dotted lines. The differences between the different terms of the equation illustrate how far the system is from sharp focus that would satisfy the imaging condition. \textbf{Right:} The alternative expressions for magnification are displayed. According to $M = (s+v)/s$, the magnification is relatively flat and asymptotically unity for all the large-curvature cases with the narrow-filter curvature being the closest to 1 (dash red). The other expression for magnification $M_0 = -v/u$ is unequal to $M$ when the system is not in focus, but is applicable to lens-less systems. It comes out negative and highly frequency dependent for the parameters we have (dash-dot green). The small-curvature estimates produced curves that were off the chart in all cases and are not displayed here (see text).}
		\label{imagingcond}
\end{figure}

The first aspect to test is whether the imaging condition (Eq. \ref{temporal_imaging_condition}) is satisfied with the calculated parameters. In short, the answer is a resounding ``no''. That the system is naturally defocused, given the parameters found above, can be readily seen from Figure \ref{imagingcond} (left), where the two sides of the imaging condition are displayed---the different large-curvature time lens estimates, $1/s$ (solid black), against $-1/u - 1/v$ (dashed dot green). In principle, the group-delay dispersion signs match, as both $u$ and $v$ are negative, as is the large-curvature $-s$, but not the small-curvature (not displayed). Depending on the particular estimate of the curvature---based on scaling, or constant focal time---the curvature is 1--2.5 orders of magnitude off-target to be in sharp focus. In order to bring the two curves together to sharp focus, other combinations of much larger input and neural dispersions would be needed, e.g., if both $u$ and $v$ were 10-20 times larger. Therefore, despite the uncertainty in the estimation process of the different dispersion and curvature values we obtained earlier, the defocus in the human auditory system seems relatively robust, as it is unlikely that the estimates are so off. 

The image magnification can be plotted now as well (Figure \ref{imagingcond}, right). We introduced two alternative expressions for magnification (Eqs. \ref{MagnificationDef} and \ref{MagnificationDef2}) that converged when the imaging condition is satisfied, but not when the system is defocused, as appears to be the case here. Once again, we explore the space of time-lens curvatures based on the different estimates. The large-curvature lenses all produce $M=(s+v)/s > 0.9$ throughout the frequency range, with the narrow-filter curvature achieving near unity magnification $M \approx 0.995$ in large portions of the spectrum, whereas the constant focal-time estimates obtaining lower magnification that slowly decrease at lower frequencies. To have a unity magnification may seem like a desirable quality in imaging, but the flat curve may be a result of overestimated time-lens curvatures in this case. Either way, when the imaging condition is satisfied, $M$ is also equal to $M_0 = -v/u$. However, this is clearly not the case here (the green dashed-dot curve on Figure \ref{imagingcond}, right), as $M \neq M_0$ across the audible frequency range. 

Once again, the small-curvature estimates are off the chart and are not displayed. They produce magnification of $M>3.5$ at 7 kHz that increases rapidly to $M\approx 9$ at 1 kHz and then to $M>>10$ at lower frequencies. These large and variable values are likely excessive and may entail unrealistic imaging. They additionally underscore the uncertainty of the small-curvature estimates. 

It is interesting to consider the effects of $M_0$, in case it turns out to be the defining magnification constant in the system. First of all, it is negative, which means that it inverts all the images. Second, it also varies considerably throughout the audio spectrum, which implies that similar envelopes that modulate different carriers can be significantly incongruent in duration with respect to one another\footnote{This form of chromatic aberration will be considered in \cref{TransChromAb}.}, which does not correspond to any known observation about the auditory system at present.

Aside from the fact that the inequality of $M$ and $M_0$ underlines the unfocused state of the system, it indirectly affirms the necessity of a lens in the auditory system. The simplest imaging system can be constructed without a lens, only with a pinhole---the pinhole camera. For such a design in temporal imaging, the magnification is exactly $M_0$ \citep{Kolner1997}. This suggests that the auditory system must have a lens, even of arbitrary curvature, in order to constrain the magnification and keep it positive. More about this later in the chapter. 

Assuming that these results are more or less correct, it is puzzling that the system may be defocused by design for a large range of time-lens curvatures. This question will be tackled directly in \cref{AudDefocus}, but we will build towards the answer throughout the subsequent sections. 

The defocused nature of the system suggests that it is chirping. There are ample pyschoacoustic data that indeed suggest that the auditory system exhibits natural chirping, which usually shows in asymmetrical sensitivity to up- and down-chirps \citep[e.g.,][]{Collins1978, Nabelek1978,Cullen1982,Schouten1985,Gordon2002}. However most of the relevant literature has been concerned either with sinusoidal frequency modulation (FM), or with broadband linear FM (linear chirps), whereas the temporal imaging equations are formulated for narrowband linear FM. A better stimulus paradigm that can be tested against the finding of defocus may be psychoacoustic data based on the Schroeder-phase complex, which emulate linear FM as a periodic signal. 

It was briefly mentioned in \cref{ParaxialBackground} that echolocating animals may use pulse compression signal processing, similar to chirp radars that undo the initial time compression after receiving the reflected pulse. This idea was originally proposed as a model for bat echolocation \citep{Strother} and was also considered in beluga whales \citep{Johnson1992}. However, a recurrent problem with this model is that it relies on cochlear dispersion that has to undo the chirp \citep{Altes1975}. While a large range of delays was measured in different units sampled in the big brown bat's brainstem and inferior colliculus \citep{Haplea1994}, a universal de-chirping of arbitrary calls is unlikely to be found in general, especially given the diversity of call types \citep{Boonman2005}. The existence of chirp units in the inferior colliculus is a further evidence that any pulse expansion in bats is partial at best, as they may not have the necessary neural delay lines to produce the matching filters to their returning chirps \citep{Simmons1971,Suga1973}. It is possible that this problem is a manifestation of defocusing of the auditory system in bats, only that it is exploited in different ways in these animals. 

\section{Psychoacoustic glides in temporal imaging}
\label{CurvatureModeling}
The human auditory dispersion appears to form an imaging system that is out of focus---perhaps by design (Figure \ref{imagingcond}). This entails that for the constant-frequency pulse objects, a chirped image is received and maybe even heard under particular conditions. If this is true, then it may be the underlying cause of psychoacoustic observations that quantify the cochlear phase curvature, only that according to the temporal imaging theory, these observations should be attributed to the entire dispersive system and not only to the cochlea. This interpretation is therefore tested in the remainder of this chapter. Using the temporal imaging equations for defocused images, the human dispersive curvature can be neatly accounted for down to 500 Hz. Frequencies below 500 Hz require a correction that is based on the modulation domain and on irregularities in the cochlear filtering at these frequencies. The most revealing outcome of this exercise, however, is the extraction of what will turn out to be the frequency-dependent aperture time durations. In turn, these values will be shown to closely match direct measurements from the chinchillas, as well as several other measurements done in humans.

\subsection{Modeling of Schroeder-phase complex curvature measurements}
\label{ModelSchr}
\begin{figure} 
		\centering
		\includegraphics[width=0.65\linewidth]{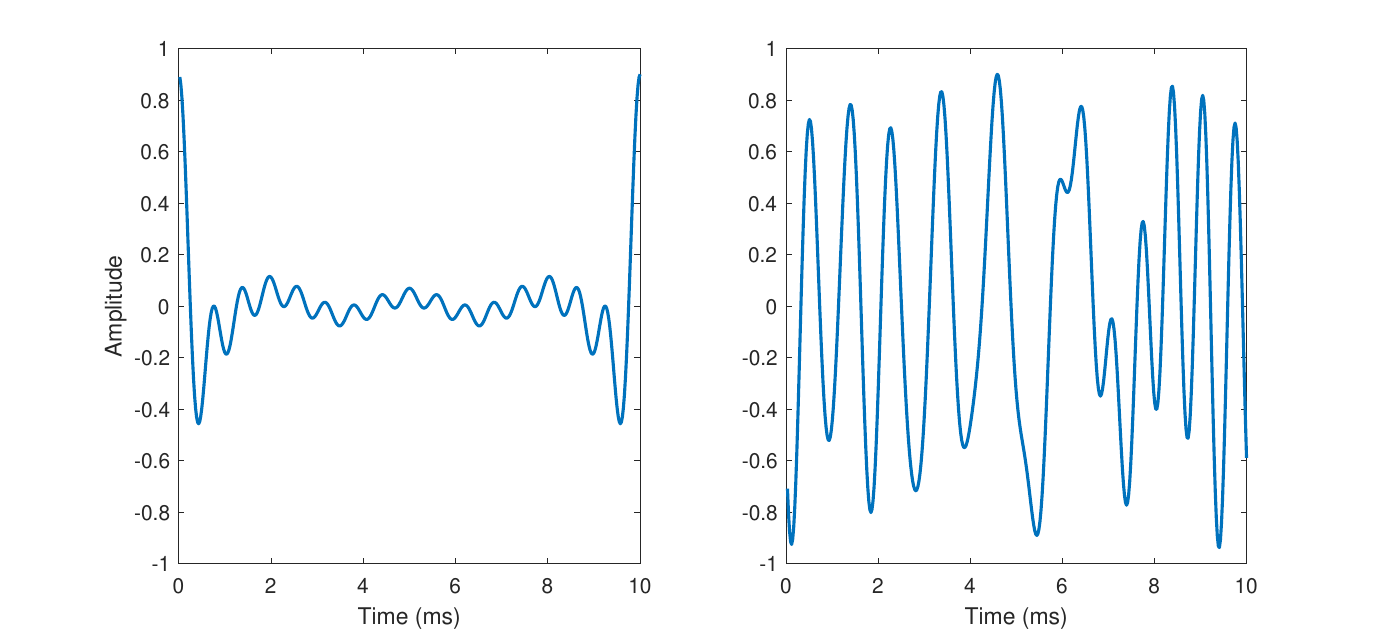}	
		\caption{Schroeder phase complex with fundamental frequency $f_0=100$ Hz, $N=13$, with components between 400 Hz and 1600 Hz (Eqs. \ref{schphase0} and \ref{schphase}), which is used to mask a 1000 Hz signal with random initial phase. \citep[e.g.,][]{OxenhamDau}. \textbf{Left:} With $C=0$ the masker is highly modulated. \textbf{Right:} With $C=1$ the masker envelope is nearly flat.}
		\label{Schrphase}
\end{figure}

Perhaps the most comprehensive data about the inherent chirping of the human auditory system was measured by \citet{OxenhamDau}, who obtained estimates for the auditory phase curvature by generating a periodic quasi-linear chirp. This was produced by the broadband Schroeder phase complex at 75 dB SPL (64 dB SPL for each harmonic component; see Figure \ref{Schrphase}), which is thought to approximately cancel out the internal auditory chirp of the system. By adding a pure tone that is one of the harmonics in the Schroeder phase complex, the masking caused by it could be psychoacoustically estimated \citep{Schroeder1970,Kohlrausch}. The curvature estimates were consistent with other measurements and were summarized in \citet[Figure 8]{OxenhamDau}, which is also reproduced in Figure \ref{PhaseCurveOxen}. 

In the Schroeder phase paradigm, a periodically rising or falling glide stimulus is constructed using a sum of $N$ harmonics of a fundamental frequency $f_0$,
\begin{equation}
p(t) = \sum_{n = n_1}^{n_1+N} \sin(2\pi n f_0 t + \phi_n)
\label{schphase0}
\end{equation}
where the phase $ \phi_n$ is given by
\begin{equation}
\phi_n = C\frac{\pi n(n-1)}{N} \,\,\,\,\,\, -1 \le C \le 1
\label{schphase}
\end{equation}
where $C$ is a curvature parameter and $n$ is the harmonic number. When $C=0$ the signal is a normal sine complex tone, but for other $C$ values, it has a phase curvature that is given by
\begin{equation}
\frac{d^2\phi_n}{df^2} = C\frac{2\pi}{Nf_0^2} \,\,\,\,\,\, -1 \le C \le 1
\label{schrcurv}
\end{equation}
This formula first appeared in \citet{Kohlrausch} without derivation. The stimulus in the test comprises an additional target stimulus---a pure tone at a frequency that coincides with one of the harmonics (typically $n=10$). In \citet{OxenhamDau}, each target frequency was tested with different $C$ values in steps of 0.25. The target is minimally masked (lowest threshold) at different $C$ values that depend on frequency, presumably because this is when the Schroeder complex cancels out the inherent curvature of the auditory system. The net effect is a stimulus that is highly modulated, and thus enables masking release of a sort, by ``listening in the dips'' between the envelope peaks. 

\begin{figure} 
		\centering
		\includegraphics[width=1\linewidth]{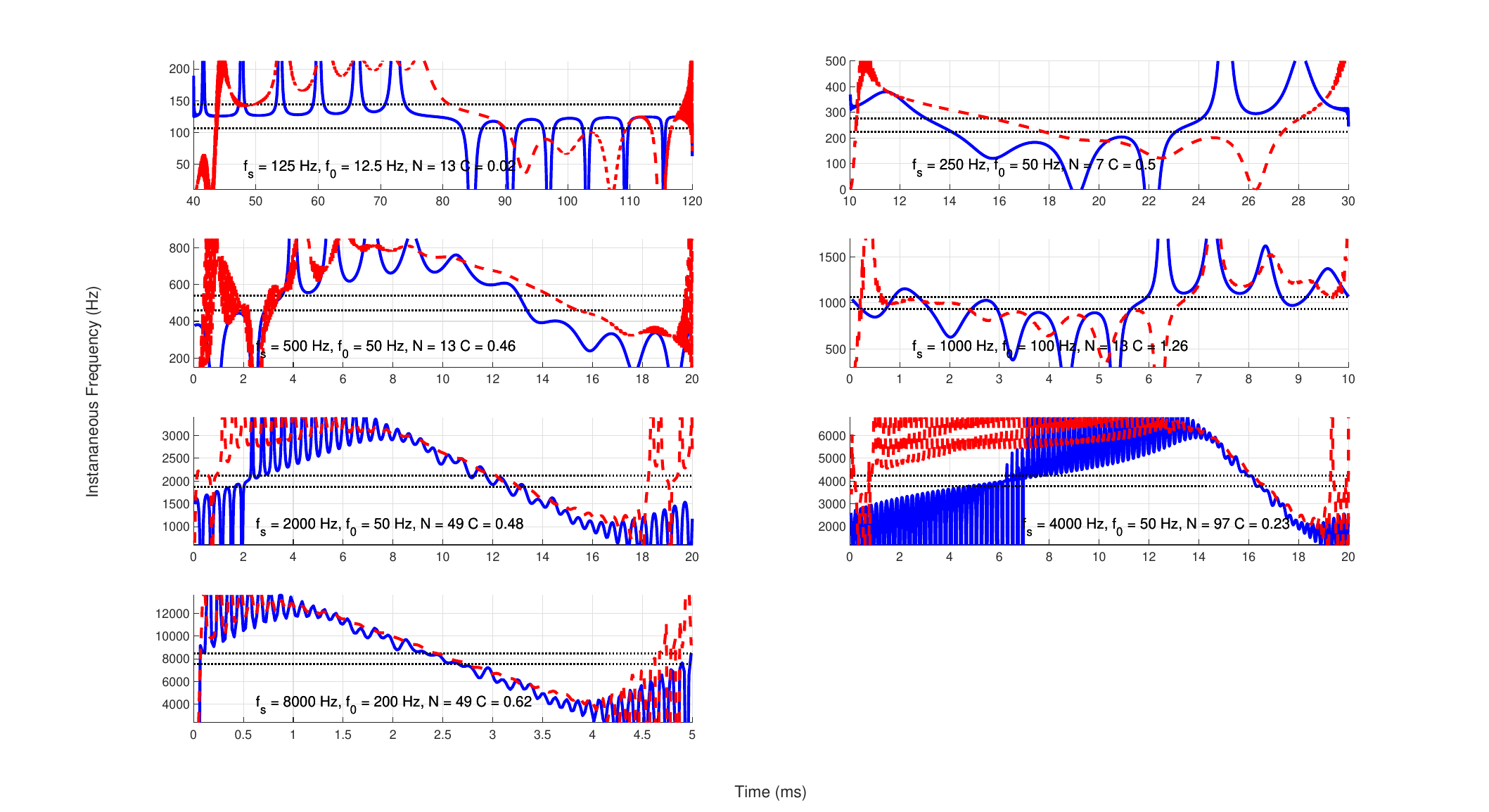}	
		\caption{The instantaneous frequency of a single period of the seven Schroeder complex stimuli, designed to mask a pure tone frequency $f_s$ \citep{OxenhamDau}, which elicited the lowest thresholds averaged over four listeners. Each masker contained $N$ harmonics of fundamental frequency $f_0$ with curvature value $C$, as printed on each plot, and computed according to Eq. \ref{schphase}. The instantaneous frequencies were computed using the Hilbert transform of the broadband signal (solid blue) and narrowband signal (dash red), which were sixth-order Butterworth bandpass filtered around $f_s$ (between $0.6f_s$ and $1.4f_s$). In all plots, the limited range of the linear frequency modulation is visible, except for the 125 Hz, where it is smooth but not linear and the curvature is nearly zero. The auditory filter bandwidth around $f_s$ (one ERB, according to \citealp{Glasberg1990}) is marked with the dotted lines. Note the different time and frequency scales of the different plots.}
		\label{SchrIF}
\end{figure}

\begin{figure} 
		\centering
		\includegraphics[width=1\linewidth]{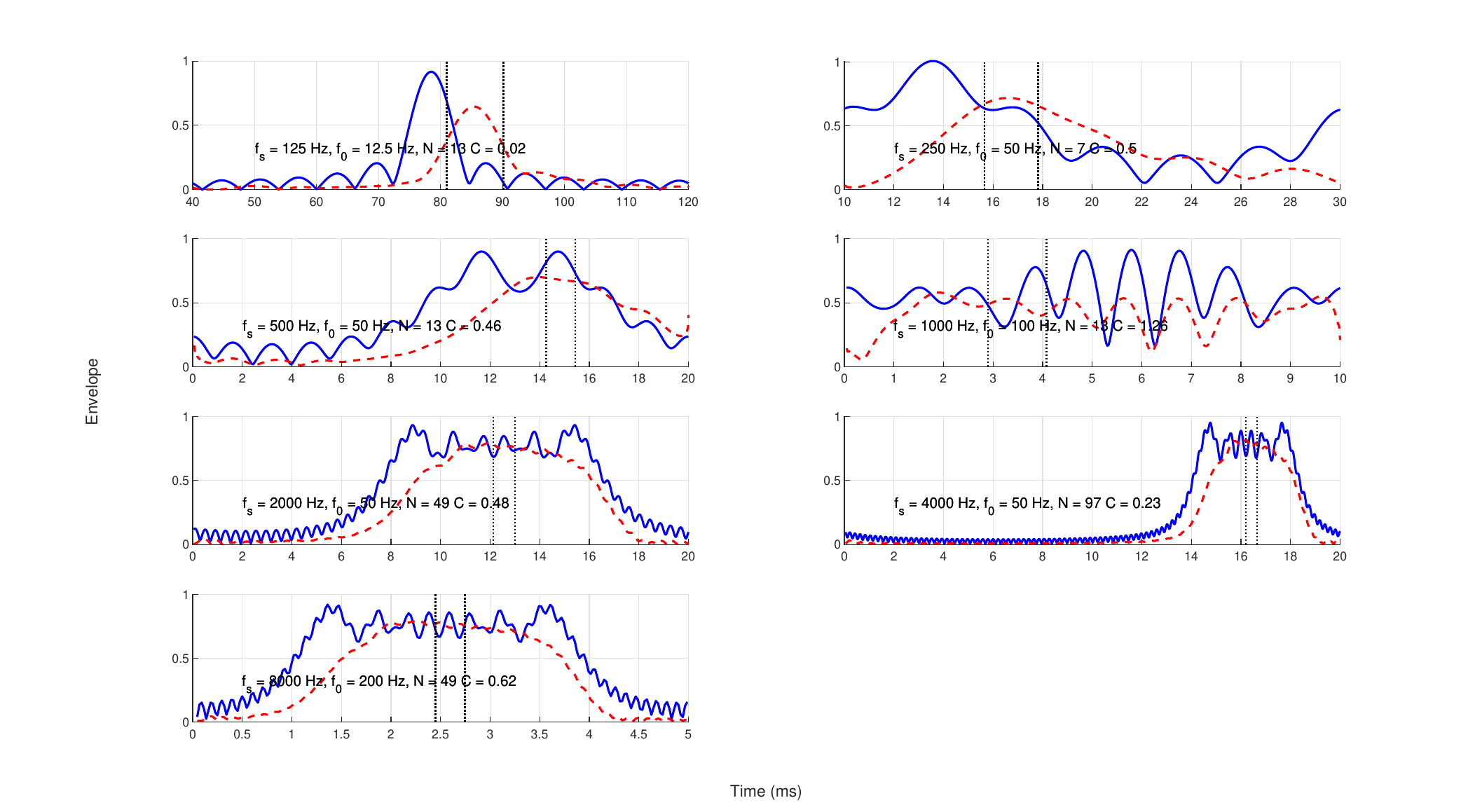}	
		\caption{The (magnitude) envelopes of a single period of the same seven Schroeder complex stimuli of Figure \ref{SchrIF}. In all cases, the signals had about flat envelope where they were also approximately linearly chirped. Some of the borderline cases in terms of chirping linearity also exhibit fluctuating envelopes. The solid blue curves are broadband envelopes, whereas the red dashed curves are bandpass filtered using the sixth-order Butterworth filters discussed in the text. The dashed vertical lines mark the equivalent duration of crossing one ERB around the signal frequency according to Figure \ref{SchrIF}. Note the different time scales of the different plots.}
		\label{SchrEnv}
\end{figure}

It is instructive to examine the stimuli used in \citet{OxenhamDau} in detail. Figures \ref{SchrIF} and \ref{SchrEnv} show the instantaneous frequencies and envelopes (using the Hilbert transform of Eqs. \ref{InstantFreq}, \ref{AmplitudeDef}, and \ref{PhaseDef}), respectively, of the seven main stimuli used to obtain Figure 8 in \citet{OxenhamDau}. Because of the ambiguous nature of the instantaneous frequency concept (see \cref{FreqInstFreq}), the linear chirping effect is most visible when the stimulus is bandpass filtered. Therefore, the stimuli are drawn twice---both broadband and bandpass filtered (sixth-order Butterworth), with bandwidth according to the equivalent rectangular bandwidth (ERB) from \citet{Glasberg1990}
\begin{equation}
	\ERB = 0.108f + 24.7
	\label{ERB}
\end{equation}
Even after filtering (the auditory filter bandwidth is marked in the figures at $f_s \pm ERB/2$) the Schroeder complex is linear only around a narrowband region, whereas outside of it the instantaneous frequency often oscillates rapidly and cannot be meaningfully approximated by a linear slope. A quasi-linear portion is visible in all stimuli but the lowest one, designed to mask 125 Hz pure tones, which has two inflection points within the desired bandwidth. These quasi-linear portions are also visible as being relatively constant in terms of their Hilbert envelopes in Figure \ref{SchrEnv}. It can be seen that the 125 and 250 Hz stimuli clearly exhibit spurious periodic chirps within the same ERB that are not the intended downward linear chirp, unless they are bandpass filtered (but note that the curvature is nearly zero for 125 Hz). These ambiguous regions become shorter at higher frequencies, but their effect on the listening tests are unknown.


While \citet{OxenhamDau} treated the curvature data as resulting from the cochlear dispersion alone, the temporal imaging theory would require it to be the combination of cochlear, outer-hair cells, and neural dispersive effects. In the simplest case, the imaging condition (Eq. \ref{temporal_imaging_condition}) is not satisfied, and then the neat scaling expected from imaging (Eq. \ref{eq:image}) is also not obtained. Instead, using a Gaussian pulse as input, we obtained a closed-form solution that has the same form as the ideal image, but with a complex pulse width (or complex magnification). This was given in Eq. \ref{altt0} and is repeated
\begin{equation}
t_1 =  M\sqrt{ t_0^2 + 2i\left( u + \frac{vs }{v + s} \right) } 
\label{altMrep}
\end{equation}
where $t_1$ is the effective Gaussian pulse width. The additional global phase term in the closed-form solution of Eq. \ref{absonlyfull2} is ignored at the moment and will be revisited in \cref{GlobalQuadPhase}. According to Eq. \ref{altMrep}, when the imaging condition is not satisfied, the expression in parenthesis does not cancel out, and a real pulse with a linear chirp will result in a complex Gaussian of width $t_1$. By squaring and inverting the equation, 
\begin{equation}
\frac{1}{t_1^2} = \frac{1}{M^2} \frac{t_0^2 - 2ix}{t_0^4+4x^2}
\label{t1M}
\end{equation}
with the substitution 
\begin{equation}
x = u + \frac{vs }{v + s}
\label{xdefocus}
\end{equation}
It is straightforward to calculate the linear chirp slope $m_1$ at the output by isolating the imaginary part of this equation, using the Gaussian relations of Eq. \ref{unknownpulseparams}, $m_1 = -\Im\left(1/t_1^2\right)$
\begin{equation}
m_1 = \frac{2x}{M^2(t_0^4+4x^2)}
\end{equation}
This chirp is measured at the output---it is an image that may be perceivable to the listener. Note that we also ignore the effects of neural sampling (\cref{TemporalSampling}), as the assumption here is that the chirping information is contained in a single sample---essentially a complete pulse---within the linear duration that sweeps over the pure tone. The stimulus duration, therefore, must be long enough to allow for at least one spike to be fired at the auditory nerve and everywhere downstream. 

The complex envelope of the Schroeder complex $A(0,\omega)$ is locally equivalent to a linear chirp, as its phase curvature is (approximately) constant (Eq. \ref{schrcurv}), with the convenient feature of being periodic. To eliminate the auditory curvature, the Schroeder phase curvature must exactly cancel out the quadratic phase term in the full transform of Eq. \ref{absonlyfull}. Therefore, the phase curvature of the input complex envelope $d^2\angle A(t)/dt^2 = m_0$ undergoes transformation and only then matches $m_1$, but with an opposite sign. This may be expressed using a complex (AM-FM) pulse (\cref{PulseCalc}) with $t_0^{'2} = t_0^2/(1-im_0t_0^2)$ that has an initial width of $t_0$. Placing it back in Eq. \ref{altMrep},
\begin{equation}
t_1 =  M\sqrt{ t_0^{'2} + 2ix} =  M\sqrt{ \frac{t_0^2}{1-im_0t_0^2} + 2ix} = M\sqrt{ \frac{t_0^2 + i(2xm_0^2t_0^4 + m_0t_0^4 + 2x)}{1+m_0^2t_0^4}}
\label{t1complex}
\end{equation}
We impose the condition that the imaginary term under the radical is zero, so to ensure that the image at the output is real and has no chirp. After some manipulation, a quadratic equation is obtained
\begin{equation}
m_0^2 + \frac{1}{2x}m_0 + \frac{1}{t_0^4}=0 \,\,\,\,\,\,\,\, x,t_0 \ne 0
\label{quad1}
\end{equation}
with the roots
\begin{equation}
m_0 = - \frac{1}{4x} \pm \frac{1}{2}\sqrt{\frac{1}{4x^2} - \frac{4}{t_0^4}} 
\label{msolution}
\end{equation}
and with the constraint
\begin{equation}
t_0 \geq 2\sqrt{|x|}
\label{msolutionINEQ}
\end{equation}
this inequality ensures that $m_0$ is real (i.e., $m_0$ and $t_0$ can be set independently). Otherwise, when $m_0$ is complex, the output pulse $t_1$ will anyway contain a chirp that cannot be canceled out. 

It appears that determining the precise value of the pulse width $t_0$ is critical to get a correct estimate of $m_0$. While the stimulus is continuous, the system can ``see'' only a fragment of it through a temporal aperture or window, which must not be too short. In the case of equality, $t_0=2\sqrt{|x|}$, there is only one solution that cancels out the internal dispersion of the system. For larger $t_0$ values, two perceived curvature minima should have been obtained per subject per condition in the Schroeder's phase experiments. In intermediate cases, when the actual value of $t_0$ is close to equality, the two curvature minima may merge to a single broad minimum. There is currently no published data to support this effect, although some individual datasets show rather broad and indistinct curvature minima (e.g., \citealp[Figures 1, 5 and 6]{OxenhamDau} and \citealp[Figures 1 and 2]{Shen2009}). Therefore, we shall start from equality and attempt to minimally perturb the solution from there.

The chirp slope $m_0$ is defined with respect to the time-dependent phase function, whereas the curvature that appears in Eq. \ref{schrcurv} was expressed as a frequency derivative. The phase term of the Fourier transform of a linear chirp gives a curvature that is simply $-1/m_0$\footnote{This can be seen by Fourier transforming a complex Gaussian. More generally, it applies to other pulse shapes, such as a rectangular pulse chirp \citep[e.g.,][p. 63]{Levanon}.}. Additionally, $m_0$ was divided by $4\pi^2$ to convert it to the same units as the frequency curvature in the original paper\footnote{The units of the data from \citet{OxenhamDau} are given in $\left[\rad \right] \left[ s \right]^{2}$ (Eq. \ref{schrcurv}), which is suitable for ordinary frequency $f$, where the quadratic phase argument of the signal can then be expressed as $\frac{1}{2}\frac{d^2\phi_n}{df^2} f^2$. However, the units of $m_0$ are based on angular frequency. As the time-domain phase argument appears as $\frac{m_0t^2}{2}$ (Eq. \ref{BasicAMFM}), $m_0$ is given in $\left[\rad\right] \left[s\right]^{-2}$. Therefore, upon inversion, $1/m_0 = \left[s\right]^2 \left[\rad\right]^{-1}$, so $-\frac{1}{m_0} = \frac{1}{4\pi^2} \frac{d^2\phi_n}{df^2}$.}. 

\subsection{Initial estimates of the phase curvature and $t_0$}
\label{InitialCurveEstimates}
\begin{figure} 
		\centering
		\includegraphics[width=0.65\linewidth]{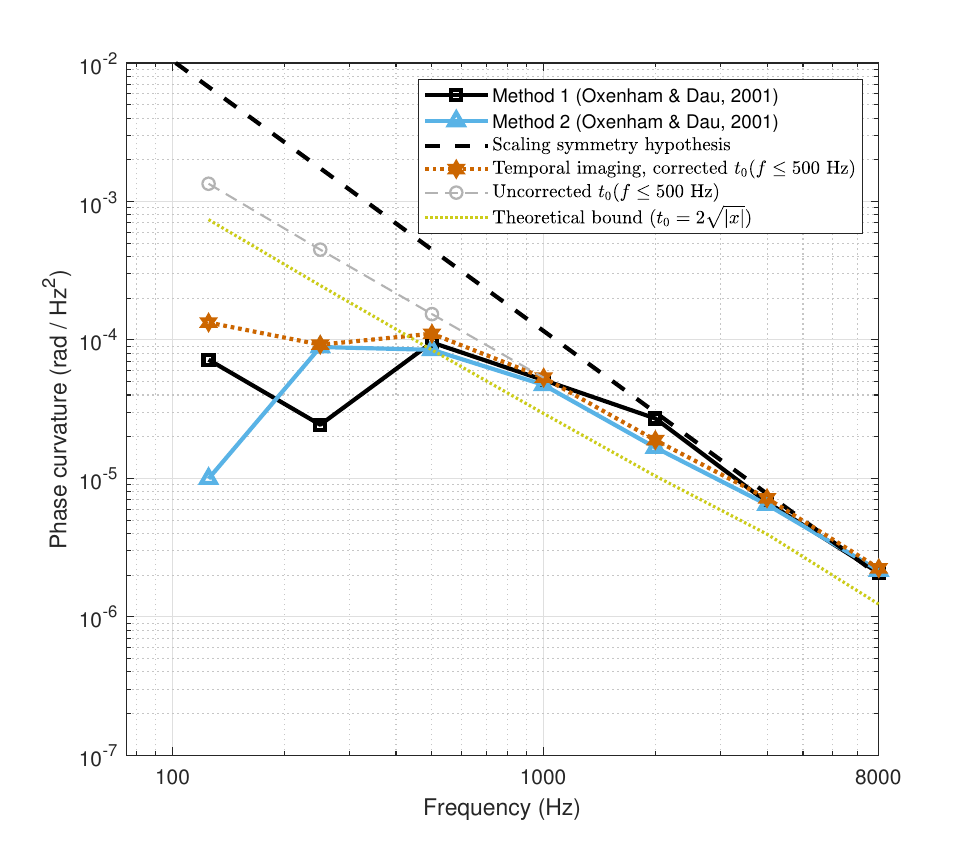}	
		\caption{The measured and fitted masker frequency slope needed to cancel out the auditory dispersion, based on Schroeder phase complex stimuli \citep{OxenhamDau}. Lines corresponding to Method 1 (solid, black squares) and Method 2 (solid, blue triangles) are reproduced from the data points in \citet[Figure 8]{OxenhamDau}, respectively. The scaling symmetry hypothesis from \citet[Figure 8]{OxenhamDau} is reproduced as well (dash black). According to the temporal imaging solution (large-curvature time lens), the duration of $t_0$---the effective pulse width at the input--- is plotted with red asterisks-dot. Theoretically, these values cannot be smaller than the limit in Eq. \ref{msolutionINEQ}, based on the amount of defocus in the system (dot orange). However, at low frequencies ($\leq$ 500 Hz), the predictions had to be corrected, as a result of the modulation filter bandwidth limitation that will be discussed in \cref{TheTMTF}. The uncorrected prediction is marked in dash gray circles.}
		\label{PhaseCurveOxen}
\end{figure}

The original observations by \citet[Figure 8]{OxenhamDau} are reproduced in Figure \ref{PhaseCurveOxen} (solid curves with black squares and blue triangles) and are used as targets, although their Method 2 was considered more reliable by the authors. The lowest two frequencies measured (125 and 250 Hz) were reportedly less certain than others due to the near-zero curvature involved\footnote{Another technical reason for why the lowest frequencies are off may be related to the type of headphones used in the original experiment (Sennheiser HD-580), which likely had a comparable group-delay dispersion to that of the ear. This can be gathered from headphone group delay data about a newer but very similar model of the headphones (Sennheiser HD-650), whose group delay was reported in \citet[Figure 3]{Laitinen2013}. The data translate to about $-1.9\cdot 10^{-6}$ $\s^2$/rad, compared to about $-3.4\cdot 10^{-6}$ $\s^2$/rad computed for the ear at 125 Hz. This headphone group-delay dispersion, however, increases more for lower frequencies than 125 Hz that are required to produce the masking Schroeder phase harmonic components correctly.}, and are also the two that deviate the most from scaling symmetry (dashed black line)\footnote{\citet{Shera2001a} modeled the auditory-nerve instantaneous-frequency data from \citet{Carney1999} and hypothesized that the slope is invariant to frequencies above 1500 Hz, if the auditory filters are assumed to be symmetric. This is obtained by normalizing the glide slope by the square of the characteristic frequency of the auditory nerve fiber. \citet{OxenhamDau} could not establish a scaling symmetry at 1000 Hz and below, and deviations from its predictions are already visible at 2000 and 4000 Hz, as can be seen in Figure \ref{PhaseCurveOxen}.}. 

The estimate for $m_0$ that is based on the $t_0$ model is plotted as well, using the negative-sign solution of Eq. \ref{msolution}. The first solution was based on mathematically satisfying the condition of Eq. \ref{msolutionINEQ} that ensures a real solution using the known system group-delay dispersion parameters, while achieving an optimal fit. This was done by slightly perturbing the equality 
\begin{equation}
	t_0 = 2 \sigma \sqrt{|x|} \,\,\, \s
	\label{t0xindp}
\end{equation}
where the multiplicative factor $\sigma = 1.058$ was introduced. $x$ was computed using Eq. \ref{xdefocus} based on the estimated parameters: the cochlear dispersion $u$ (\cref{Total_u}), the broad-filter large-curvature time lens $s$ (\cref{OHCtimelens}), and the neural dispersion $v$ (\cref{NeuralDispEst}). The obtained $t_0$ is 6\% larger than the theoretical bound. All other large-curvature time-lens models produced nearly identical predictions with $1.058 \leq \sigma \leq 1.061$, whereas the small-curvature model produced a somewhat poorer fit with a relatively large $\sigma = 1.19$ (not displayed).

The result is shown as a red dotted line above 500 Hz and dashed gray line below 500 Hz in Figure \ref{PhaseCurveOxen}. It reveals a close match at frequencies down to 500 Hz, but with poor fits of the 125 and 250 Hz frequencies (continued as gray dashed line and circles in the plot). A much better fit for the two frequencies can be achieved by taking into consideration an external constraint that will be introduced in \cref{LowFreqCorr}, which stems from the modulation transfer function analysis of \cref{TransFun} and will be discussed in the next subsection. The pulse durations derived from the estimates are summarized in Table \ref{twidths} using the ``uncorrected'' $t_0$, which relates to the computation done without the external constraint from the modulation transfer function. 

Large individual variations have been reported in curvature estimations \citep[e.g.,][]{OxenhamDau, Shen2009}. For example, under different masker conditions, individual differences in curvature of up to five times were sometimes recorded \citep[Figure 3]{Shen2009}, while differences of about two times were common across most conditions \citep[Figures 3 and 5]{Shen2009}. Nevertheless, above 500 Hz, the match between the computed $t_0$ and the data is excellent and only at low frequencies the differences become significant (gray curve in Figure \ref{PhaseCurveOxen}). Once again, using the large-curvature time-lens values produces a very small change to the values in the table, which are therefore not shown.

\begin{table} 
\centering
\scriptsize
\begin{tabular}{|l|c|c|c|c|c|c|c|}
\hline
\textbf{Duration (ms)}&\textbf{125 Hz}&\textbf{250 Hz}&\textbf{500 Hz}&\textbf{1000 Hz}&\textbf{2000 Hz}&\textbf{4000 Hz}&\textbf{8000 Hz}\\\hline
Uncorrected $t_0 \propto \sqrt{|x|}$ & \textit{4.53} & \textit{2.64} & \textit{1.56} & 0.94 & 0.57 & 0.36 & 0.22 \\
Uncorrected $t_{0,RECT} \propto \sqrt{|x|}$ & \textit{10.67} & \textit{6.21} & \textit{3.67} & 2.21 & 1.34 & 0.86 & 0.52 \\
\hline 
Corrected $t_0 \propto \sqrt{|x|}$ & 2.12 & 1.33 & 1.32 & 0.94 & 0.57 & 0.36 & 0.22 \\
Corrected $t_{0,RECT} \propto \sqrt{|x|}$ & 4.98 & 3.14 & 3.10 & 2.21 & 1.34 & 0.86 & 0.52 \\
\hline
$1/f_s$ & 8.33 & 4.17 & 2 & 1 & 0.5 & 0.25 & 0.125 \\
\hline
\end{tabular}
  \caption{The different frequency-dependent durations that were fitted to the cochlear curvature data from \citet{OxenhamDau}. The $t_0$ durations that are proportional to the internal dispersion $\sqrt{|x|}$ are given in the first row, and their rectangular equivalent in second row (first row times 2.355). The corrected values that consider the modulation bandwidth at low frequencies are given in the third and fourth rows. The period of the target tone $1/f_s$ is given in the last row for reference. The three values in italics indicate that they have to be corrected. See text for further details.}
	\label{twidths}
\end{table}

\section{The temporal aperture based on psychoacoustic glide data}
\label{TempAperture}
\subsection{The entrance pupil, temporal aperture, and exit pupil}
\label{SpecTempApertureCalc}
The most striking aspect of the dispersion-dependent fit to the data above 500 Hz is that the pulse width $t_0$ could be derived independently of the stimulus duration, only as a function of the combined dispersion parameter $x = u + \frac{sv}{s+v}$, at very close to the minimum theoretical value allowed according to Eq. \ref{msolutionINEQ}. The signal is continuous and much longer than the system is able to process at once. We obtained this signal duration limit $t_0$, which turned out to be constrained by a function of the internal group-delay dispersions through $x$ that was estimated independently of the auditory channel response. It can be therefore deduced that $t_0$ is none other than the \term{temporal aperture} of the system---the time-limited window for processing the incoming acoustic signals. More precisely, $t_0$ is the \term{entrance pupil}---the image of the aperture as viewed from the position of the object (see \cref{DiffractionFourier} and Figure \ref{EntExitPupils}).

While the imaging magnification is smaller than unity (Figure \ref{imagingcond}), there are small differences in the temporal aperture between the object and the image plane, and right at the lens, where the aperture is assumed to be located. At the image plane, the temporal aperture is imaged and appears as the \term{exit pupil}. Its duration can be readily computed by using the real part of Eq. \ref{t1M} and Eq. \ref{unknownpulseparams}
\begin{equation}
t_1 = M \frac{\sqrt{t_0^4+4x^2}}{t_0}
\label{exitpupilt1}
\end{equation}
The same equation may be used to find the actual temporal aperture after the lens, $T_a$, by setting $v=0$ and $M=1$ (see Eq. \ref{pressureclear}, for an example, where the real Gaussian term has the same form, if we set $x=u$). 

\subsection{The low-frequency correction}
\label{LowFreqCorr}
As will turn out in \cref{TheTMTF}, using the available values for $v$ and $T_a$, as prescribed above, leads to unphysical modulation bandwidths at 125, 250 and 500 Hz for the coherent modulation transfer function\footnote{The aperture time can be thought of as a time constant in the low-pass filtering that characterizes the modulation impulse response of the channel.} (Eq. \ref{wcoh}). In other words, if the 3 dB cutoff frequencies of the low-pass filters that are associated with the modulation transfer functions are calculated using the very same parameter values that are used to obtain the off-fitted 125, 250 and 500 Hz curvatures, then these filters would have modulation bandwidths that are larger than their carriers, which is impossible. Therefore, the current parameter estimates below 500 Hz according to the inequality \ref{msolutionINEQ} must be wrong. Thus, the cutoff frequencies have to be corrected to only allow modulation bandwidths that are smaller than the carrier. This can be done either by tweaking $t_0$, $v$, or both. The effect of $v$ on the curvature is negligible, whereas reducing $t_0$ so that the cutoff frequencies at 125--500 Hz are equal to the carrier immediately leads to a much better fit of the off 125 and 250 Hz curvature points, and a little improvement to the off 500 Hz point. In order to obtain the corrected values for $t_0$, the new aperture was used in Eq. \ref{wcoh}
\begin{equation}
T_a = \frac{ 4\sqrt{2}|v| }{\FWHM} 0.9\omega_c 
\label{newTa}
\end{equation}
where 0.9 is the arbitrary fraction of the carrier $\omega_c$ bandwidth that replaces the previous bandwidth\footnote{The $0.9f_c$ bandwidth factor is arguably still too large, as the modulation and carrier frequencies begin beating when they are spectrally close, so the modulation effect is no long as intended from the standard AM operation. However, this provisional choice has provided reasonable results for this work, given that the full correction is currently unknown (see text below).\label{ModCarrierFraction}}. We are interested in non-rectangular (Gaussian) aperture duration, which is why the expression is divided by $\FWHM \approx 2.355$\footnote{The full-width half maximum (FWHM) is a numerical factor that converts the standard Gaussian width $t_0$ to a rectangular pulse equivalent that has the same bandwidth. For Gaussian, $\FWHM = 2\sqrt{2\ln2 \,\,} \approx 2.355$. See \cref{PulseCalc} for derivation.}. Finally, the new aperture duration of Eq. \ref{newTa} has to be back-propagated to the entrance of the system, to obtain the entrance pupil $t_0$. This is done with 
\begin{equation}
t_0 = \sqrt{\frac{T_a^2 \pm \sqrt{T_a^4 - 16u^2}}{2}}
\label{newt0}
\end{equation}
This expression was derived from Eq. \ref{pressureclear} by comparing the real Gaussian width to $T_a$ and solving the biquadratic equation in $t_0$. Note that with the corrected values of the dispersion term $x$, $m_0$ has to be complex to satisfy Eq. \ref{quad1}, which means that there will be a glide at the output anyway---potentially an audible one. Therefore, $x$ must be corrected as well to satisfy the inequality \ref{msolutionINEQ}. It was done by using the new $t_0$ from Eq. \ref{newt0} with the empirically obtained inequality \ref{t0xindp}. However, both solutions in Eq. \ref{newt0} are complex, so we shall discard the imaginary part of the obtained $t_0$ and note that this solution is provisional.

Results for the three tweaked points are displayed in Figure \ref{PhaseCurveOxen} (dotted red curve). In addition to the corrected $t_0$, the corresponding $x$ values were reiterated using Eq. \ref{t0xindp} to force zero chirping at the output---something that made the fit a little worse. The required changes in $x$ are significant, but there are no other data to guide this correction and pinpoint which parameter(s) within $x$ should be specifically modified (i.e., $u$, $v$, or $s$). 

The final entrance and exit pupil and the aperture time values are summarized in Table \ref{t0opts}. As can be seen from the table, the differences between the three measures tend to be small, but are more pronounced at low frequencies. Note that the term \term{temporal aperture} is used to to refer to the device or feature that imposes the temporal constriction in the signal path, whereas \term{aperture time} is the particular duration that is associated with it. 

\begin{table} 
\centering
\scriptsize
\begin{tabular}{|c|c|c|c|c|c|c|c|}
\hline
\textbf{Duration (ms)}&\textbf{125 Hz}&\textbf{250 Hz}&\textbf{500 Hz}&\textbf{1000 Hz}&\textbf{2000 Hz}&\textbf{4000 Hz}&\textbf{8000 Hz}\\\hline
$t_0$ (entrance pupil) & 4.98$^*$ & 3.14$^*$ & 3.01$^*$ & 2.15 & 1.28 & 0.79 & 0.44 \\
$T_a$ (temporal aperture) & 9.13$^*$ & 4.87$^*$ & 3.31$^*$ & 2.21 & 1.30 & 0.80 & 0.44 \\
$t_1$ (exit pupil) & 5.27$^*$ & 3.35$^*$ & 3.33$^*$ & 2.32 & 1.35 & 0.83 & 0.47 \\
\hline
$\Delta t_{opt}$ & 3.18 & 1.96 & 1.19 & 0.72 & 0.43 & 0.26 & 0.13 \\
$T_a/\Delta t_{opt}$ & 2.87 & 2.49 & 2.78 & 3.09 & 3.07 & 3.08 & 3.40 \\
\hline\end{tabular}
  \caption{The entrance and exit pupils and aperture times of the full auditory system, based on Eqs. \ref{t0xindp} and \ref{exitpupilt1}. The entrance pupils are the corrected $t_{0,RECT}$ values from Table \ref{twidths}, where for the 125, 250, and 500 Hz values (marked by an asterisk) were corrected using modulation filter considerations (see text). The $T_a$ values were computed by setting $x=u$ in Eq. \ref{exitpupilt1} (the effect of $s$ is multiplicative chirping and thus does not matter here). The exit pupil values were calculated using Eq. \ref{exitpupilt1}. Additionally, the theoretical values that exactly balance the geometrical and dispersive blurs are given with $\Delta t_{opt}$, which was computed using Eq. \ref{minaperture}. The ratio between the actual aperture stop and the optimal values are given in the last row, showing that geometrical blur is three times larger than dispersive blur, on average. All values are equivalent rectangular bandwidths in milliseconds.}
	\label{t0opts}
\end{table}

Several causes may account for the anomalous low-frequency results. As was discussed when $u$, $s$, and $v$ were initially estimated, there are some uncertainties associated with all of them, which can be substantial at low frequencies, where they were sometimes extrapolated. Another possible cause for the discrepancy in $x$ is that even the allowed modulation bandwidth results in over-modulation, since the narrowband approximation breaks down and the governing equations no longer hold, as higher-order dispersion terms become dominant \citep{Bennett2000}. 

Another likely cause for the anomalous behavior we observed below 500 Hz and the unusually broadband channels is that the cochlear filters are not truly bandpass toward the apex, but rather low-pass. This has been observed in several studies in the last years that specifically targeted the apical cochlear mechanics, using various novel imaging techniques that do not damage the delicate cochlear structures, where noticeable differences have been observed in comparison with the better studied basal mechanics. For example, the guinea pig apex (characteristic frequencies, CF $<$ 2 kHz) was tested in vivo, where it was found that the cochlear response is low-pass, while only neural filtering imposed the bandpass response \citep{Recio2017}. Using the scaling property of the cochlea to transform this frequency range from the guinea pig to humans, it maps to CFs below about 900 Hz \citep[Figures 1 and 4]{Greenwood1990}. The physical modulation limitation that resulted in the above correction holds below 660 Hz, according to Figure \ref{OTAcutoff}. If a similar response characterizes the low frequencies in humans as in the guinea pig, then its effect is to change the aperture duration and shape, and possibly tamper with the neural dispersion computation. In any event, a low-pass component in the dispersive system would challenge the narrowband approximation and may have to be replaced with a more accurate model. The guinea-pig findings are compounded by a followup study that additionally found that, unlike the group delay in basal locations, in apical locations the group delay is almost independent of level \citep{Recio2018}. Another study of the guinea-pig found little to no variation in best frequencies between apical and middle cochlear sites (different from basal sites), which also exhibited very little group delay dispersion in comparison with the standard place-model predictions \citep{Burwood2022}. It was also demonstrated that the outer hair cells are responsible for distributing the mechanical response between apical sites and among auditory nerve channels. Low-pass responses were also shown in the gerbil, where CFs in the apical first turn also exhibited a different compressive nonlinearity than either the second or the third (basal) turns \citep{Dong2018}. Finally, \citet{Warren2016} found that the displacement amplitude of the basilar membrane in the apical turn of the guinea-pig was considerably smaller than in more basal sites, and that the motion was significantly larger on the reticular lamina---that is, after the outer hair cell bodies. It was suggested that these differences are fundamental, so much so that straightforward scaling of the cochlear properties from basal data to apical sites may be invalid \citep{Dong2018}. 

In humans, in a study that compared stimulus-frequency and distortion otoacoustic emissions (OAEs), it was found that OAE recordings returning from the cochlea have three regions that are separated by two bends in their phase response: basal above 2500 Hz and apical below 900 Hz \citep{Christensen2020}. This matches the remapped frequency range estimate from the guinea pig data by \citet{Recio2017} above.

\subsection{Comparison with temporal windows from psychoacoustic\\literature}
\label{AperturePsych}
The entrance pupil values we computed, which refer directly to the duration of the input stimulus, can now be compared with some psychoacoustic data from the literature.

\begin{figure} 
		\centering
		\includegraphics[width=0.5\linewidth]{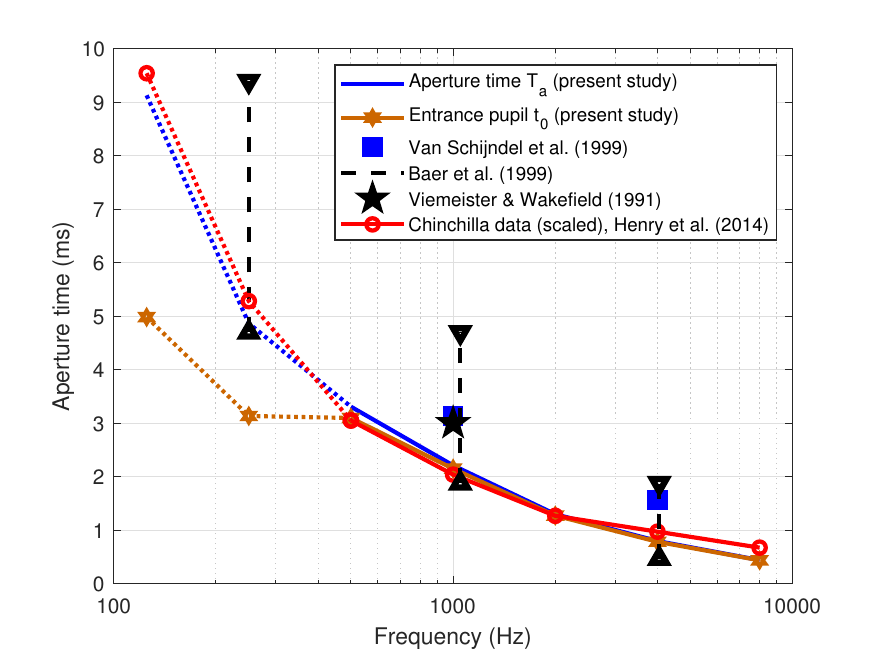}	
		\caption{The aperture time and entrance pupil estimates from the present study (second and first rows in Table \ref{t0opts}, respectively) and comparable estimates from literature. All values are given in equivalent rectangular durations. Psychoacoustic data are taken from \citet{vanSchijndel1999,Baer1999} and \citet{Viemeister1991}. The physiological data from \citep{HenryKale2014} were scaled for the slight difference in the cochlear mapping between the chinchilla and human \citep{Greenwood1990}. The present data are uncertain at 125, 250, and 500 Hz because of the exact relationship between the carrier and modulation bandwidths (see text), but they provide a reasonable fit to the curvature data from \citet{OxenhamDau}. Data for these frequencies from \citet{HenryKale2014} were linearly extrapolated from the reported 500--8000 Hz data.}
		\label{AllApertures}
\end{figure}

Values of $t_0$ may be tested against independent psychoacoustic data measured by \citet{vanSchijndel1999} of intensity discrimination tests of Gabor pulses of different durations (``shape factors''), but equal power. The masking thresholds of three subjects were measured at various levels using a pink-noise masker. Similar to Gabor's logons \citep{Gabor}, the authors hypothesized that the auditory stimuli may be perceived in multiple time-frequency grid windows, but that with the right choice of time-bandwidth product the number of active windows can be minimized. For a Gaussian envelope $\exp\left[-\pi (\alpha f_c t)^2\right]$, the shape factor $\alpha$ produced the worst just-noticeable intensity difference for $\alpha = 0.3$ with carrier $f_c = 1000$ Hz and $\alpha = 0.15$ with $f_c = 4000$ kHz carrier \citep[Figure 3]{vanSchijndel1999}. By equating these coefficients to the standard Gaussian used throughout this chapter (Eq. \ref{inputenvspec}), which explicitly contains the entrance pupil, we obtain $t_{0,RECT} = \FWHM /(\sqrt{2\pi} \alpha f_c)$, which results in 1.6 ms at 4 kHz and 3.1 ms at 1 kHz. These values are longer than the values obtained in the present study of 0.8 ms at 4 kHz and 2.2 ms at 1 kHz (see Table \ref{t0opts}). The data points are shown alongside the temporal aperture and entrance pupil values from the present study in Figure \ref{AllApertures}. 

The \citet{vanSchijndel1999} experiment was replicated with three other subjects and extended to 250 Hz and additional conditions by \citet{Baer1999}, whose results were added to Figure \ref{AllApertures}. The mean results for 1 and 4 kHz of \citet{vanSchijndel1999} were replicated, but higher individual variability of the pulse was observed in different intensity conditions. The test included a quiet condition that in absolute level was closest to 40 dB SPL \citep[Figure 2]{Baer1999}, as was used to derive our temporal aperture. At 1 kHz, the peaks are distributed between 1.9 and 4.7 ms---longer in the quiet condition, and shorter but less peaky at the 40 dB masked condition. Similar ranges were obtained for the other frequencies, which always contain the estimates from our theoretical prediction, as can be seen in Figure \ref{AllApertures} (dash lines, triangles).

The value we obtained of $t_{0,RECT}$ at 1 kHz (2.2 ms) is also smaller than the equivalent rectangular width of a single ``look''  of the order of 3 ms at 1 kHz, as was estimated in \citet{Viemeister1991}\footnote{\citet{Viemeister1991} introduced the \term{multiple-look model}, which relates to discrete samples that the hearing system may be hypothetically generating to acquire sound input (see also \cref{AliasIntro}).}. 

The discrepancy between the values may be attributed to individual variations, possible level-dependent differences in all the various studies, and essentially different methods that may not exactly tap into the same quantity. 

\subsection{Comparison with physiological chinchilla data}
\label{ChinAperture}
A much more precise comparison to our aperture time can be made using physiological data of the temporal windows of anesthetized chinchillas from \citet{HenryKale2014}. The data are based on single-unit measurements of auditory nerve fibers, whose response to Gaussian white noise input was used to derive the second-order Wiener kernel of the system. The temporal windows of normal-hearing controls (10 animals, 143 fibers) were obtained at different frequencies by computing the first eigenvector of the second-order Wiener kernel and applying Hilbert transform on it. The Hilbert envelope facilitated a direct estimate of the temporal window and it was calculated at different levels relatively to the peak. Since the relative cochlear frequency maps of humans and chinchillas are almost identical \citep{Greenwood1990}, it is possible to directly compare the human and chinchilla data with minimal error. Nonetheless, a scaling correction was applied by using the chinchilla's cochlear map to derive the relative cochlear place for the reported data, and then obtain the equivalent place-frequency for the human cochlea \citep{Greenwood1990}. From these values, slightly modified temporal window estimates were interpolated from the 50\% (FWHM) temporal window values from \citet[Figure 7, top]{HenryKale2014} for characteristic frequencies of 500--8000 Hz, which are reproduced in Figure \ref{AllApertures}. Once converted to equivalent rectangular widths, these values are directly comparable to the aperture time values of the second row in Table \ref{t0opts}. 

The linear regression trend line from \citet[Figure 7, top]{HenryKale2014} was linearly extrapolated to have estimates also for 125 Hz---9.54 ms and 250 Hz---5.28 ms. Note that the two extrapolated values are expected to be a little longer, because the chinchilla's frequency map deviates from the power law below 500 Hz \citep{Greenwood1990}. The differences between the present aperture time estimates and the transformed chinchilla's values are 4.3--8.4\% for frequencies below 2000 Hz, but at 4 kHz it is 18\% and at 8 kHz it is 34.6\%. This gives an average error of 12\% for the seven frequencies. Regardless, these aperture time estimates are in excellent agreement with the present study, given the wildly different methods and conditions used to obtain them, as well as the uncertainty in the dispersion parameters and the possible low-frequency anomaly.

\begin{figure} 
		\centering
		\includegraphics[width=0.5\linewidth]{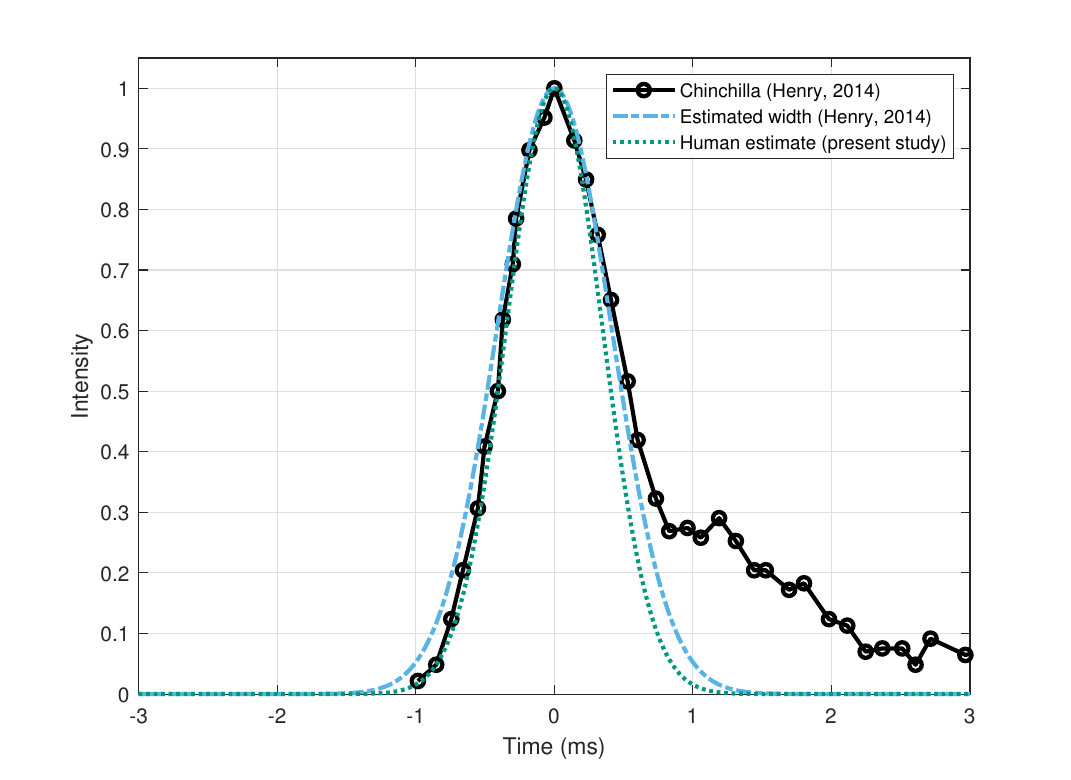}	
		\caption{An example of a temporal window of the chinchilla measured at about 3.9 kHz (black solid circles), obtained from data by \citet[Figure 1E]{HenryKale2014}. The authors provided the 50\% window duration, which is used in a Gaussian pupil function (blue dash-dot line). The present-study estimate from human data for the same frequency has a slightly narrower duration, which is plotted as another Gaussian pupil in green dotted line.}
		\label{ChinWin}
\end{figure}

It is remarkable that the Gaussian pupil shape gave such close results to the physiological data of the chinchilla. In addition to the specific temporal window values, the Henry et al's paper provided one instance of an actual window function that was estimated using the Wiener kernel method \citep[Figure 1E]{HenryKale2014}. Using a zero crossing estimate from the figure, the carrier is about 3.9 kHz, for which the authors provided a 50\% temporal window duration of 0.93 ms. The envelope is replotted in Figure \ref{ChinWin}, after centering and normalizing the peak. The window has a long tail, which extends for another 2--3 ms after the peak. Using the duration from the paper and the present estimate of the temporal aperture at 3.9 kHz (0.87 ms), it is possible to compare the measurements to a theoretical Gaussian pupil. Gaussian pupils of these two durations are plotted in Figure \ref{ChinWin} as well. Excluding the tail, the measured and theoretical pupils are nearly identical. This suggests that the Gaussian pupil function may be a valid approximation for the aperture for most of the stimulus\footnote{We should also allow for a completely different explanation for the remarkable match between the measured and theoretical apertures, aside from the long tail. The proximity to a mathematical Gaussian might also suggest that it reflects an artifact in the measurement and averaging method that is based on Wiener kernel that was used in \citet{HenryKale2014}. However, this hypothesis is unsubstantiated at present.}. However, the Gaussian pupil may not account for slow and low-energy signal decay and possible aberrations induced by the asymmetrical tail (\cref{HigherOrderAb}). 

\subsection{Comparison with psychoacoustic beating data}
\label{BeatingCurves}
\begin{figure} 
		\centering
		\includegraphics[width=0.5\linewidth]{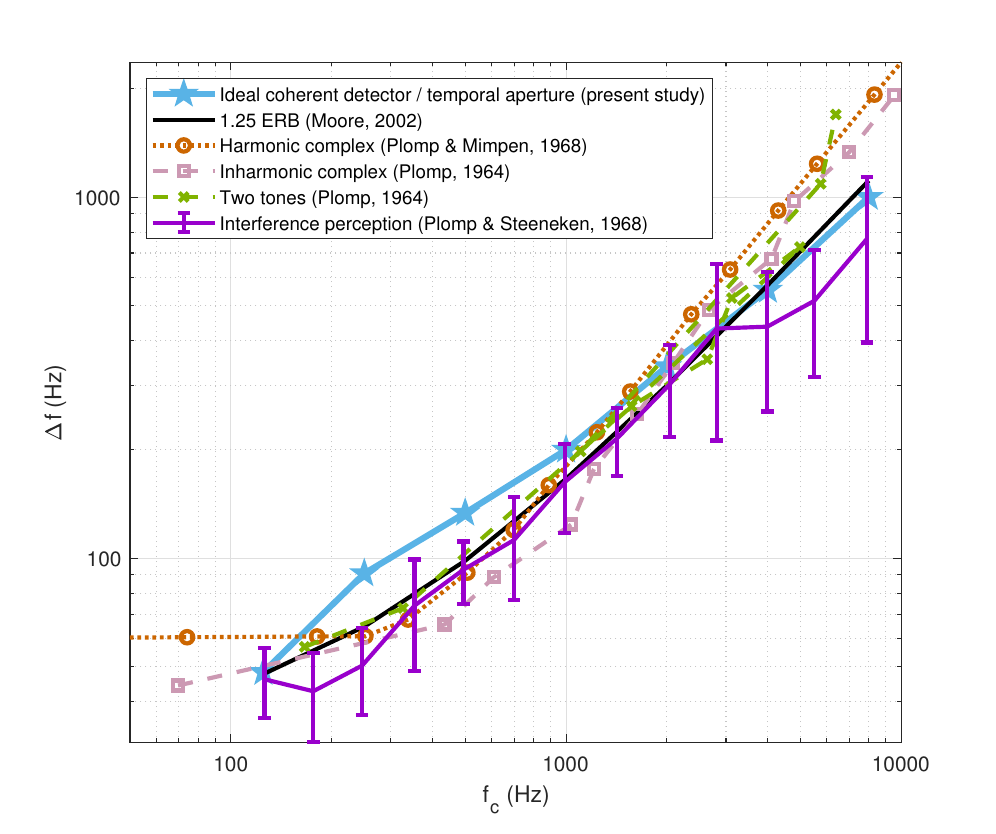}	
		\caption{Auditory beating perception of two tones expressed through their frequency spacing $\Delta f$, as a function of average frequency $f_c$. Predictions according to Eq. \ref{visibilitygauss} and the aperture time values from Table \ref{t0opts}, row 2, are plotted in solid blue stars. Different data from literature are plotted in comparison, which are based either on harmonic complexes \citep{Plomp1964,PlompMimpen1968}, inharmonic complexes \citep{Plomp1964}, or two-tone interference \citep{Plomp1964,PlompSteen1968}. The additional rule of thumb of $\Delta f \approx 1.25 \ERB$ by \citet{Moore2002} is plotted for reference too. }
		\label{bearingfig}
\end{figure}

As a final test of the temporal aperture values and shape, we can compare the present values to psychoacoustic results from literature that quantify the beating threshold between two tones. Acoustic interference has been localized to be taking place within the auditory filter and the periphery and anyway not centrally \citep{Krumbholz1998}. In \cref{Nonstationarytheory}, we derived an expression for the frequency range of beating between two tones with frequency difference $\Delta f$, based on nonstationary coherence theory, which depends only on the temporal integration time of the hypothetical intensity detector with a Gaussian window. The FWHM of the detector was set as a frequency spacing criterion, as beating can be distinctly observed only for $\Delta f < 0.44/T$. It is interesting to examine how the temporal aperture compares with published estimates for this threshold, by setting $T=T_a$. Results from literature are compiled from \citet{Plomp1964,PlompMimpen1968} and \citet{PlompSteen1968}, and to a rule of thumb provided by \citet{Moore2002}, which sets the limit of beating at maximum spacing of $\Delta f \approx 1.25 \ERB$\footnote{Interestingly, \citet{Moore2002} provided an additional rule of thumb for dissonance or roughness between two pure tones that is maximal for frequencies separated by approximately $0.44\ERB$ for center frequencies up to 2000 Hz. This rule was referenced to several psychoacoustic studies, but was not explicitly derived in any of them.}. The results are all plotted in Figure \ref{bearingfig}. As the curves show, all frequencies but 250 and 500 Hz are within the spread of the data from literature. This discrepancy reinforces the suspicion that the low-frequency data may be misestimated. Nevertheless, the beating measurement may provide an alternative and more direct way to estimate $T_a$ that is independent of all other dispersion parameters. Thus, if we use Moore's $1.25\ERB$ law as approximating the literature average, we would obtain $T_a=6.8$ ms at 250 Hz and $T_a = 4.5$ ms at 500 Hz. These values are significantly longer (by 29\% and 26\%, respectively) than the values from Table \ref{t0opts}. The value found at 1 kHz is also longer (17\%) than the dispersion-derived value. All other beating duration are less than 11\% different from the estimates, and the seven frequency average difference is 14\%.

\section{The global quadratic phase term}
\label{GlobalQuadPhase}
The role of a finite aperture is critical in suppressing the oscillations of the global quadratic phase term that appears in the imaging transforms (e.g., Eqs. \ref{eq:image}, \ref{totalpathabs}, and \ref{impresponse5}) and has been neglected earlier. The inclusion of the global phase term is of no consequence in intensity images (because of squaring), but can severely distort the amplitude image if it is not truncated by the aperture.

\begin{figure} 
		\centering
		\includegraphics[width=0.9\linewidth]{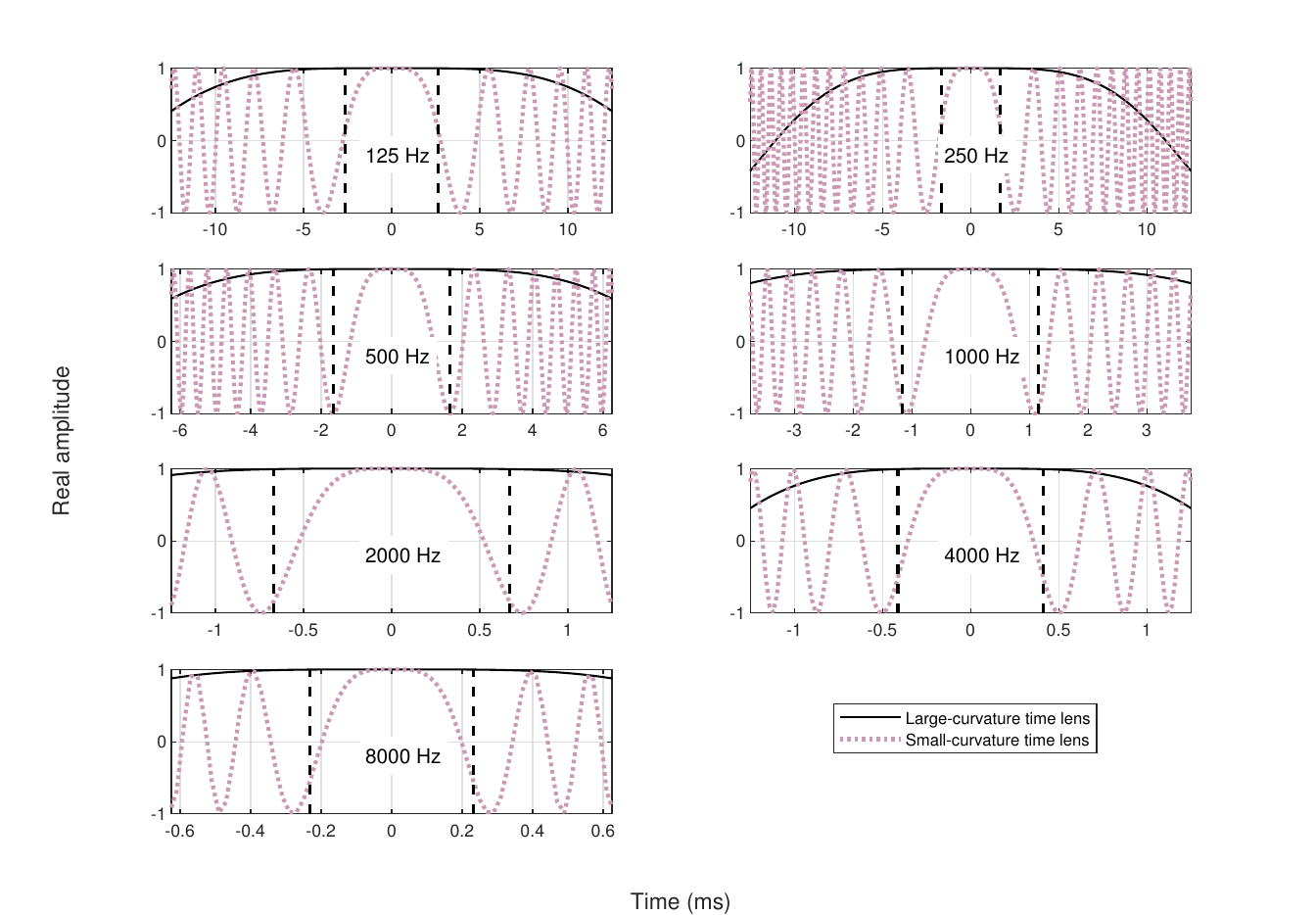}
		\caption{The global quadratic phase term in the imaging transform, $\exp\left[\frac{i\tau^2}{4(v+s)} \right]$, at the seven frequencies tested above. This phase does not matter in intensity imaging, but may distort the image in amplitude imaging if left un-windowed. Therefore, it is important that the temporal aperture truncates the quadratic phase function while it is still in the main lobe, where the amplitude magnitude is as close to 1 as possible. With the small-curvature time lens (red dot), it is never the case, as the phase completes up to half a cycle within the window, as can be seen by the vertical dashed lines that mark the (rectangular) $\pm t_1$ times, obtained from Table \ref{t0opts}. If we use the large-curvature time-lens data (solid black), then the problem disappears completely, as the phase becomes effectively constant within the aperture time. Only the real part of the complex envelope amplitude is displayed.} 
		\label{quadT}
\end{figure}

The quadratic phase functions at the seven frequencies are shown in Figure \ref{quadT} along with where the (rectangular equivalent) exit pupil (i.e., the image of the aperture) truncates them. In the case of the large-curvature time lens (all models), there is no effect of the quadratic phase within the limits of the exit pupil, as the amplitude hardly drops from its maximum value. Therefore, the quadratic phase is effectively constant within the temporal window and no plausible risk of distortion is present (solid black curves in Figure \ref{quadT}). In contrast, the small-curvature time-lens exhibits half a cycle of the quadratic phase within the limits of the exit aperture. While it does not oscillate within the aperture, the effect of such a lens may still be distorting. It  might also cause a loss of perceived loudness, as there is a smaller amount of power per image in comparison to the large-curvature image. 

For additional validation in the context of the Schroeder-phase complex data, the system including the phase term can be analyzed in closed form. Starting from the full solution of Eq. \ref{absonlyfull}
\begin{equation}
a_n(\zeta _2 ,\tau ) = a_0 t_0 \sqrt{ \frac{\frac{s}{v + s}}{2 \left[  \frac{t_0^2}{2} + i\left( {u + \frac{{vs }}{{v + s}}} \right) \right]}} \exp \left[ {\frac{{i\tau ^2 }}{{4(v + s)}}} \right]  \exp \left\{  \frac{-\left(\frac{s}{{v + s}}\tau\right)^2}{ 4\left[  \frac{t_0^2}{2} + i\left( {u + \frac{{vs }}{{v + s}}} \right) \right]} \right\} 
\label{GlobalPhaseFull}
\end{equation}
According to the formalism of \cref{PulseCalc}, the frequency slope of the first quadratic phase term is $m_g = \frac{1}{2(v+s)}$. Using the complex output pulse solution of Eq. \ref{t1complex}, the frequency slopes of the two quadratic phase terms in Eq. \ref{GlobalPhaseFull} can be summed, according to Eq. \ref{GaussProduct}. Using $m_g$, then once again the imaginary part of the output signal should cancel out when applying an external chirp with slope $m_0$. This can be written as
\begin{equation}
	m_g + \frac{m_0t_0^4 + 2m_0^2t_0^4x +2x}{M^2\left[t_0^4(1+2xm_0)^2 + 4x^2\right]} = 0
\end{equation}
which translates into this quadratic equation
\begin{equation}
	2t_0^4x(2m_g xM^2+1)m_0^2 + t_0^4(1+4x m_g M^2)m_0 + 2x+ m_g M^2(t_0^4+4x^2) = 0
\end{equation}
Given the values obtained of $t_0$, $x$, and $M$, all the terms involving $m_g$ are negligible in comparison with the terms that do not contain it, which asymptotically reduces this quadratic equation to the original one in Eq. \ref{quad1}. Therefore, this configuration is indistinguishable from the full system, at least as far as the psychoacoustic phase curvature is concerned.

The importance of amplitude imaging at this stage is not entirely well-understood. In the case of intensity imaging, the entire transformation is squared, so that the quadratic phase is anyway canceled out.

\section{Discussion}
We presented the derivation of the basic temporal imaging theory according to \citet{KolnerNazarathy} and \citet{Kolner} and applied it to the auditory system of humans. The system appears to be out of focus by design, which can be used to account for its chirping property (glides). Several important observations could be made based on the successful modeling of the psychoacoustic curvature data based on the theory. Most central to them is the identification and estimation of the temporal aperture. 

\subsection{The imaging equations}
The basic equations of the temporal imaging theory were adopted from \citet{Kolner}, almost verbatim, only with occasional changes in notation. The temporal imaging equations are exactly analogous to those of Fourier optics \citep{Goodman}, as long as we exchange the spatial modulation for temporal modulation, and with two less spatial dimensions. Temporal imaging is conceptually more complex because it overloads the time domain with both modulation and carrier evolution, whereas in light frequencies of spatial optics the two are distributed in different dimensions and domains that have close to no interaction. This mixture between the carrier and modulation domains is propagated inside the hearing system, as the two frequency ranges largely overlap, sometimes within the same pathways. 

\subsection{Inherent auditory defocus}
Unlike Kolner's applications and most others' in the burgeoning field of ultrashort pulse imaging, we were led to concentrate on a defocused system configuration that appears to be relevant to the human auditory system, and thus has a very different parameter space than is typical in optics. Many attempts were made by the author to perturb the obtained system parameters (not presented here, except for two time-lens curvature data sets) using alternative datasets from literature that may be used for dispersive parameter derivation, but all led to the inevitable conclusion that the defocus term is too large to be an estimation error (see Figure \ref{imagingcond}). It begs the question: Why is the auditory system configured to be defocused? In optical imaging systems, defocusing is sometimes used to achieve selective sharp focusing on one object, while blurring undesirable parts of the scene. We will tie this principle of operation to coherence over the next chapters and specifically in \cref{AudDefocus}.

Perception of the auditory phase curvature was hitherto associated with the cochlear dispersion alone, but was modeled here successfully using a single pulse image, whose imaginary part that normally produces a linear chirp was set to zero. The internal chirp that had to be canceled out can be understood as a defocusing side-effect (formally, an aberration)---which we represented by $x$---the defocusing parameter. It may be taken as an additional confirmation that the auditory system is inherently defocused. 

It is possible to obtain another bit of intuition about the defocus, by computing how the auditory system balances with geometrical and diffraction blur (see \cref{DiffractionFourier}). In good optical designs, the two can be traded off to optimize the image sharpness, so that the defocus is minimal, but also not much object detail is lost to diffraction, so a diffraction-limited image can be realized. We will obtain the formulas to compute this optimum in \cref{ImagingNotSatisfied}, but the results according to Eq. \ref{minaperture} are given in the bottom two rows of Table \ref{t0opts}. As can be seen, the computed durations of the auditory temporal aperture are suboptimal, in that they are about 2.4--3.5 times longer than the optimum. This means that the auditory system works above its dispersion limit, within its geometrical range. In other words, the images are not significantly corrupted by dispersive effects, which would have required much shorter aperture times to be distinctly noticeable. Thus, the auditory system is not dispersion-limited, unlike the eye that is close to being diffraction-limited, at least for average pupil sizes (\cref{TheHumanEye}). 

\subsection{The temporal aperture}
\label{TheTemporalAperture}
The temporal aperture is a mandatory component in any system that dynamically processes signals over time and is inevitable in any imaging system of finite extent. However, the temporal aperture mathematics does not indicate where it is exactly localized in the auditory system. Various psychoacoustic models have attempted to approximate the temporal response of the auditory system, (see \cref{AliasIntro} for a short review), but the temporal imaging theory may only agree with something like the multiple-look model \citep{Viemeister1991} or variations thereof, which assume a discretized representation of sound without committing to a specific physiology. Our aperture analysis became better grounded with the ability to closely match the obtained psychoacoustic results to direct physiological measurements of the temporal window of the chinchilla by \citet{HenryKale2014}. This surprising finding serves two important purposes. First, it localizes the aperture in the cochlea and auditory nerve---probably behind (after) the time lens. The neural firing itself is the most natural candidate to constitute the aperture, given its finite sampling period. However, at low frequencies, where the cochlear mechanics appears to behave differently, there may be a temporal effect on the aperture of the specific mechanical filtering as well. Second, it provides a very strong indication that the Gaussian function can well describe the aperture shape, at least around its center. This will be of utmost importance in the chapters dealing with modulation transfer functions. 

Another surprising fact is that the apparent general validity of the inequality \ref{t0xindp}, from which we derived the temporal aperture. This expression was motivated through the particular problem of accounting for the stimuli in Schroeder phase complex experiments and it appeared as a mathematical-physical constraint to the actual solution. It is likely that this inequality can be motivated and derived in a more general way that is independent of this particular problem. Once such an approach is found, it may become easier to develop further insight about temporal imaging in the auditory system, and in defocused systems in particular.

\subsection{The auditory dispersion parameters}
The convergence of the numerical values obtained by wildly different methods is an important indicator that applying the temporal imaging theory to the auditory system is warranted. The temporal aperture function constitutes the final parameter that is needed to describe the basic imaging system of hearing, along with $u$, $v$, and $s$. Interestingly, we saw how the temporal aperture duration above 500 Hz can be very closely approximated by a combination of the other three parameters, which serve as a mathematical lower bound that enabled the cancellation of the internal chirp (Eq. \ref{msolutionINEQ}). The bound called for a factor of 2, whereas we obtained 2.116 as a preferred factor to optimize the fit to the data ($\sigma = 1.058$ in \ref{t0xindp}). It is likely that a more precise estimate of the other parameters will lead to it being even closer to 2. This will also ensure a unitary minimum for the curvature masking of the Schroeder phase complex, per frequency band. 

There was much uncertainty with respect to values that time-lens curvature should take, according to the different auditory filter models \cref{OHCtimelens}. While all the models produced a very close fit to experiment with the right $\sigma$, the broad-filter time lens curvature minimized it, while achieving a somewhat closer fit. In a later analysis we shall explore the possibility that the system can accommodate the time-lens curvature, somewhat similarly to vision (\cref{MOCR}). 

Throughout this analysis, we have excluded one key component from the system. To account for the continuous nature of the signal, the aperture has to be triggered at a certain rate. In the moving image camera it is achieved with a \term{shutter} (e.g., a rotary disc shutter in analog cameras). The solution provided above was derived based on a single sample, so the shutter function was implicit. If the auditory nerve firing represents the temporal aperture and the sample, then the neural spiking rate may be thought of as a (nonuniform) trigger that works as a shutter. This will be explored in more depth in \cref{TemporalSampling}.

\subsection{Pinhole camera design}
\label{PinholeArch}
The auditory imaging system appears to combine a short temporal aperture and a time lens with uncertain, and possibly variable, curvature. This architecture closely resembles the pinhole camera design. As the time-lens curvature was based on relatively few data points and its effect on the perceptual data appears to be limited (because of its small curvature), its role and very existence remain somewhat uncertain at this stage of the analysis. If the lens would be altogether missing---in line with standard cochlear models that do not attribute any phase-modulatory role to the organ of Corti / outer hair cells---then the system would assume the ``classic'' temporal pinhole camera design \citep{Kolner1997}. In the pinhole camera, the magnification depends on the input and neural dispersion alone ($M=-v/u$, \citealp{Kolner1997}), which would yield an inverted (time-reversed) image, as both $u$ and $v$ are negative, according to our analysis. In contrast, the scaling involved in the full system with time lens depends on the other expression of the magnification $M = v/(s+v)$, which is positive and close to unity (see Figure \ref{imagingcond}, right). Keeping the image non-inverted may be one significant role of the time lens. A counter argument, then, may be that such time-reversal has not been documented in the auditory literature because it can only be detected on a very short time scale in narrowband. However, the values of the $-v/u$ magnification are also extremely variable in frequency, which does not seem like a desirable design characteristic for the auditory system.

Despite its unlikeliness, one of the most important features of the pinhole camera is that it has a theoretical infinite depth of field. This also implies an atypically large depth of focus. Similar properties can be expected from the single-lens system with very small aperture. This aspect of the system will be explored in \cref{AudImageFun}.


\subsection{Absorption and high-order dispersive aberrations}
In the entire analysis, there was no obvious need to invoke the group-delay absorption (imaginary parts of the $u$, $v$, and $s$ parameters) to model the data with a reasonably convincing fit and, indeed, such attempts have not been presented. It may provide limited confirmation for the validity of neglecting the absorption in the dispersion equation (\cref{temporaltheory}), at least at high frequencies. Thus, the auditory system may form images in a dispersion-dominant fashion, just like the original temporal imaging theory by \citet{KolnerNazarathy}. Nevertheless, in modeling the perceptual curvature response there were some uncertainties involved at low frequencies, which may be accounted for by high-order dispersive aberrations or by absorption, as well as by an altogether different filter topology (e.g., low-pass instead of bandpass). Theoretically, the effects of such an absorption could be image deformation and chirping that are unaccounted for by the input chirp $m_0$ or the simple defocus term $x$. Except for possible over-modulation at low frequency carriers, there is no data to support these claims at present. 

The need to elucidate the significance of absorption receives more currency in \cref{PsychoEstimation}, where the entire dispersive parameter set ($u$, $v$, $s$, and $T_a$) is re-derived using strictly psychoacoustic effects and available data, including the phase curvature and beating effects explored earlier. This alternative solution is largely consistent with other results in this work based on our earlier physiological estimates of the dispersion parameters, but only as long as the parameters are allowed to take complex values. Thus, absorption becomes dominant notwithstanding. However, the validity of these results cannot be ascertained at this stage, because it is not clear that the equations that were used to obtain them are all valid for absorption. Interestingly, this psychoacoustic solution suggests that in the 125 Hz band at least one of the parameters flips sign. If this will be shown to extend to higher frequencies, then it may explain the anomalous responses found by \citet{OxenhamDau} and match the cat data from \citet{Carney1999}.

\chapter{The impulse response and its associated modulation transfer functions}
\label{ImpFun}
\section{Introduction}
At the end of the previous chapter, we saw how the finite duration of the auditory temporal aperture creeps into the auditory response to the Schroeder phase complex stimuli. We also saw how the dispersion parameters of the system bound the aperture duration from below and that a Gaussian function approximates its physiological shape very closely, except for a long one-sided tail. The aperture shape, also called the pupil function, is of special significance in imaging as it can be used to fully describe the system's imaging resolution and its various imperfections. This treatment requires the impulse response of the system (its point spread function in the optics jargon), which can be then used along with the pupil function to obtain the modulation transfer function of the system. Things get more complicated as a distinction has to be made between classes of imaging and signals---coherent and incoherent---which requires having an appropriate coherence theory. Additionally, in the hearing system, it is impossible to avoid the effect of nonuniform sampling on the image, which has to be considered on top of the transfer functions. 

To get a handle on the various functions associated with the imaging system response, we begin from the existing impulse response function from \citet{Kolner}, who derived it in complete analogy to how it is done in spatial imaging  \citep[pp. 169--174]{Goodman}. The validity of the impulse response depends on a local time-invariance property, concentrated around the carrier of the traveling wave system. This is analogous to the space invariance that characterizes spatial imaging systems about the optical axis, if the image magnification and inversion are factored out. When space is divided to small isoplanatic patches, each patch is approximately space-invariant \citep[pp. 27 and 173]{Goodman}. It should be underscored, though, that the auditory system is not exactly time-invariant because of the stochastic nature of its neural sampling, which will have to be taken into consideration in more advanced analyses. Nevertheless, we will neglect this complication at present and derive the amplitude transfer function (ATF), the optical transfer function (OTF), and the modulation transfer function (MTF) of the system---for focused and defocused, coherent and incoherent cases---using a Gaussian pupil. Additional expressions for rectangular pupil will be derived as well for later comparison. These derivations follow \citet[pp. 195--211]{Goodman}, and to the best knowledge of the author have not been introduced previously within temporal imaging theory in optics. 

In the final section, we will review the temporal modulation transfer function in the hearing literature and see how its various findings can be connected to the functions we derive here. We shall argue that the nonuniform sampling of the system may no longer be ignored. We will also point to how partial coherence has to be taken into consideration more rigorously within hearing.

\begin{table}
\footnotesize\sf\centering
\begin{tabular}{P{4cm}P{8.5cm}P{4cm}}
\hline
\textbf{Term}&\textbf{Definition}&\textbf{Acoustic analog}\\
\hline
\textbf{Aperture} & An opening that limits the amount of light that enters the system. Every element in the imaging system may function as an aperture & \\
\textbf{Aperture stop} & The smallest aperture in the system & \\
\textbf{Pupil} & The image of the aperture stop & \\
\textbf{Entrance pupil} & The image of the aperture on the object plane & \\
\textbf{Exit pupil} & The image of the aperture on the image plane & \\
\textbf{Pupil function} & The functional form of the aperture, which weights the transfer function of the system (e.g., of the lens) & \\
\textbf{Apodization} & A mask or pupil that is designed with a particular function, which includes graduated intensity filtering & Windowing (in time)\\
\textbf{Transmittance} & A transparent object that spatially modulates incident light & \\
\textbf{Power} (of a lens) & A measure of the reciprocal of the focal length of a system; measured in [diopters] or [D], equivalent to [$m^{-1}$] &\\
\hline
(Coherent) \textbf{Point Spread Function (PSF/cPSF)} & The amplitude impulse response function of an object point to its image. The term \term{Line spread function} is sometimes used for the one-dimensional response of a two-dimensional PSF \citep[p. 225]{Goodman} & Impulse response function $h(t)$ \\
\textbf{Amplitude transfer function (ATF)} & The Fourier transform of the cPSF (a function of spatial frequency) & Transfer function, frequency response\\ 
(incoherent) \textbf{Point Spread Function (PSF/iPSF)} & The intensity point spread function, which is the modulus of the squared cPSF & Impulse response function $|h(t)|^2$\\
\hline
\textbf{Optical transfer function (OTF)} & The Fourier transform of the iPSF in frequency coordinates. It is also the normalized autocorrelation function of the ATF. & Complex modulation transfer function \\
\textbf{Modulation transfer function (MTF)} (incoherent) & The modulus of the complex OTF & Modulation transfer function\\
\textbf{Phase transfer function (PTF)} (incoherent) & The phase of the complex OTF &\\
\textbf{Contrast sensitivity function (CSF)} & The combined modulation transfer function of the periphery and the neural pathways in vision & Temporal modulation transfer function (TMTF)\\
\hline
\textbf{Radiometry} & Objective measurement of light (and other electromagnetic) radiation & \\
\textbf{Radiant energy} & The energy that propagates onto, through, or from a given surface area and time duration [J] &  \\
\textbf{Radiant flux} (radiant power) & Radiant energy per unit time [W] & Acoustic source power \\
\textbf{Irradiance} & Received radiant flux per area, coming from all directions [W/m$^2$]. It is specified for a given point on the surface & Sound intensity\\
\textbf{Radiant intensity} & Direction-dependent radiant flux density, measured per unit of solid angle [W/st] & \\
\textbf{Radiance} & The direction- and position-dependent radiant flux per unit of planar area and unit solid angle [W/st m$^2$] & \\

\hline
\textbf{Photometry} & Radiometry that is adapted to human vision, where the energy is weighted by its relative visible sensitivity per wavelength & \\
\textbf{Luminous flux / power} & Photometric equivalent to radiant flux [lm] & \\
\textbf{Illuminance} & Photometric equivalent to irradiance [lm/m$^2$] & \\
\textbf{Luminous intensity} & Photometric equivalent to radian intensity [candela] = [lm/st] & \\
\textbf{Luminance} & Photometric equivalent to radiance---close to subjective brightness [candela/$m^2$] & Loudness [phon] \\

\hline
\end{tabular}
\caption{A jargon glossary for common functions used in optics with occasional analogs in acoustics. The analogies are usually associative as the optical terms are used in the X-Y plane, whereas in acoustics they are used for in time-frequency plane, which means that the space-time analogy has to be invoked. The radiometry and photometry definitions are from \citet{McCluney1994}. [J] is Joule, the energy unit. [st] is steradian, the unit of solid angles. [lm] is lumen, the unit of luminous flux.}
\label{acousticsvsoptics}
\end{table}

\section{The impulse response of the imaging system}
\label{AudImpulseRes}
The goal in the following is to find a time-invariant impulse response of the complete imaging system as was presented in \cref{ImagingEqs}, so that it does not depend on the absolute time $\tau_0$ of the input pulse, but rather on the time difference between the output and the input $\tau-\tau_0$. This should enable the standard convolution integral computation
\begin{equation}
a_2(\tau) = \int_{-\infty}^\infty h(\tau-\tau_0) a_0(\tau_0) d\tau_0
\label{generalconv}
\end{equation}
for the envelope input $a_0(\tau)$ and imaging output $a_2(\tau)$. We shall omit from here on the spatial coordinate $\zeta$, which is implicit in the dispersion parameters based on group-delay dispersion rather than on group-velocity dispersion. The response for an arbitrary input envelope that propagates from $\zeta=0$ is
\begin{equation}
a_2(\tau)= \left\{\left[a_0(\tau)*d_1(\zeta_1,\tau)\right]h_L(\tau)P(\tau) \right\}*d_2(\zeta_2,\tau) 
\end{equation}
where the same dispersive stages were followed as before (see Eq. \ref{fullpathfreq})---dispersion ($d_1$), time lens ($h_L$), and another dispersion ($d_2$)---only that all transformations are represented in the time domain. Additionally, right after the lens we added a pupil function $P(\tau)$, whose role is to apply the aperture by constraining the temporal extent of the pulse. Plugging in the input envelope $a_0(0,\tau)=\delta(\tau_0)$, it becomes an impulse response
\begin{equation}
h(\tau;\tau_0) = \left[d_1(\zeta_1,\tau-\tau_0)h_L(\tau)P(\tau) \right]*d_2(\zeta_2,\tau)
\label{imp0}
\end{equation}
The dispersive stages have the time domain transfer functions that are the Fourier transform of Eq. \ref{disp_stage}
\begin{equation}
d_1(\tau) = {\cal F}^{ - 1} \left[ D_1 (\zeta _1 ,\omega ) \right] = \frac{1}{\sqrt{4\pi iu}} \exp\left(\frac{i\tau^2}{4u} \right)
\end{equation}
and similarly for $d_2$
\begin{equation}
d_2(\tau) = {\cal F}^{ - 1} \left[ D_2 (\zeta _2 ,\omega ) \right] = \frac{1}{\sqrt{4\pi iv}} \exp\left(\frac{i\tau^2}{4v} \right)
\end{equation}
Now the convolution integral Eq. \ref{imp0} can be solved explicitly, by using also the time lens relations of Eqs. \ref{lens_function} and \ref{TimeLenss} 
\begin{multline}
h(\tau;\tau_0) = \int_{-\infty}^\infty d_1(\zeta_1,T-\tau_0) h(T) P(T) d_2(\zeta_2,\tau-T) dT\\
=\frac{1}{4\pi i \sqrt{uv}} \int_{-\infty}^\infty \exp\left[\frac{i(T-\tau_0)^2}{4u} \right] \exp \left( \frac{iT^2}{4s } \right)  P(T) \exp\left[ \frac{i(\tau-T)^2}{4v} \right] dT \\
= \frac{1}{4\pi i \sqrt{uv}} \exp\left[\frac{i}{4}\left(\frac{\tau_0^2}{u} + \frac{\tau^2}{v}\right) \right]\int_{-\infty}^\infty  P(T) \exp \left[ \frac{i}{4}\left( \frac{1}{u} + \frac{1}{v} + \frac{1}{s}\right) T^{2}\right] \exp \left[-\frac{iT}{2}\left(\frac{\tau_0}{u}+ \frac{\tau}{v} \right) \right] dT
\label{impresponse3}
\end{multline}
where the time is designated by $\tau_0$ at the object coordinate system, and by $\tau$ in the image coordinate system. 

\subsection{Imaging condition satisfied}
\label{ImpulseResponseDer}
The quadratic phase in the first term of the integrand of Eq. \ref{impresponse3} contains the familiar imaging condition from Eq. \ref{temporal_imaging_condition}, which can be eliminated if it is satisfied, as is done in \citet{Kolner} and in the analogous spatial case \citep[pp. 169--170]{Goodman}. We will do the same as an intermediate step, before solving for the more general case when it is not satisfied. 

The scaled Fourier exponential of Eq. \ref{impresponse3} can be simplified using the magnification $M_0 = -v/u$
\begin{equation}
\exp \left[-\frac{iT}{2}\left(\frac{\tau_0}{u}+ \frac{\tau}{v} \right) \right] = \exp \left[-\frac{iT}{2v}\left(\tau-M_0\tau_0 \right) \right]
\label{changevar1}
\end{equation}
The final simplification step concerns the initial quadratic phase term in Eq. \ref{impresponse3}, which contains two terms that depend on the object and image time coordinates, but whose effects are altogether undesirable as they may distort the image \citep[pp. 169--172]{Goodman} (see also \cref{GlobalQuadPhase}). The term belonging to the image coordinate $\exp(i\tau^2/4v)$ becomes negligible in intensity imaging---when the final image is detected as an intensity pattern rather than amplitude. Note that unless the imaging condition is satisfied, then $M_0 \neq M = s/(v+s)$. The term belonging to the object coordinate $\exp(i\tau_0^2/4u)$ will have a negligible effect if every interval in the object envelope $\delta \tau_0$ is mapped only to a small region in the output $\delta \tau$. In other words, the effect of an infinitesimal unit of time from the object affects only a limited duration in the image time. The latter condition is approximated by replacing this instance of $\tau_0$ with $\tau/M$, which can be thought of as an extension of the coordinate transformation between $t$ and $\tau$ to include the quadratic phase term that is not subjected to the full imaging transformation. Placing it back in the quadratic term makes it dependent only on the image coordinates
\begin{equation}
\exp\left[\frac{i}{4}\left(\frac{\tau_0^2}{u} + \frac{\tau^2}{v}\right) \right] \approx \exp\left[\frac{i\tau^2 }{4}\left(\frac{1}{uM^2} + \frac{1}{v} \right) \right] = \exp\left( \frac{i\omega_c\tau^2}{2Mf_T} \right)
\label{quadphaseaprrox}
\end{equation}
where the equation on the right is true only if the imaging condition is satisfied, so that $M = M_0$ and the term depends only on the image time coordinate.

Using the results of Eqs. \ref{temporal_imaging_condition}, \ref{changevar1} and \ref{quadphaseaprrox} in Eq. \ref{impresponse3}, we obtain the (time-variant) impulse response
\begin{equation}
h(\tau;\tau_0) \approx \frac{1}{4\pi i \sqrt{uv}} \exp\left( \frac{i\omega_c\tau^2}{2Mf_T} \right) \int_{-\infty}^\infty  P(T) \exp \left[-\frac{iT}{2v}\left(\tau-M_0\tau_0 \right) \right] dT
\label{impss}
\end{equation}
Thus, up to the quadratic factor, the impulse response is a scaled and shifted Fourier transform of the pupil function in ideal imaging conditions. Finally, two more variable changes will be applied to Eq. \ref{impss}: a change of integration variable $\tilde T  = T/2v$, and a change to the so-called \term{reduced coordinate} of the object time $\tilde\tau_0  = M_0\tau _0$
\begin{equation}
h(\tau-\tilde\tau_0) = \frac{\sqrt{M}}{2\pi} \exp\left( \frac{i\omega_c\tau^2}{2Mf_T} \right) \int_{-\infty}^\infty  P(2v\tilde T) \exp \left[-i\tilde T\left(\tau-\tilde\tau_0 \right) \right] d\tilde T
\label{impss2}
\end{equation}
This makes the Fourier integral dependent only on the time difference $\tau - \tilde\tau_0$ in the traveling wave coordinate system, and the impulse response is then time invariant in these coordinates. Convolving the envelope object with this expression as in Eq. \ref{generalconv}, we obtain
\begin{equation}
a_2(\tau) = \int_{-\infty}^\infty \frac{h(\tau-\tilde\tau_0)}{M} a_0\left(\frac{\tilde\tau_0}{M}\right) d\tilde\tau_0
\label{convintegral1}
\end{equation}
Redefining the impulse function according to
\begin{equation}
\tilde h (\tau-\tilde\tau_0) = \frac{1}{\sqrt{M}} h(\tau-\tilde\tau_0)
\label{Mimpss}
\end{equation}
We obtain this convolution integral
\begin{equation}
a_2(\tau) = \int_{-\infty}^\infty \tilde h(\tau-\tilde\tau_0)\frac{1}{\sqrt{M}} a_0\left(\frac{\tilde\tau_0}{M}\right) d\tilde\tau_0
\label{convintegral2}
\end{equation}
The envelope on the right-hand side is the familiar ideal one-dimensional image prediction according to geometrical optics (Eq. \ref{imaging_condition})
\begin{equation}
a_g(\tau) = \frac{1}{\sqrt{M}} a_0\left(\frac{\tau}{M}\right)
\label{geomimage}
\end{equation}
which can be summarized by the convolution of the ideal image with the impulse response of the pupil, which is in itself a scaled Fourier transform of the pupil function
\begin{equation}
a_2(\tau) = \tilde h(\tau) * a_g(\tau)
\label{geomconv}
\end{equation}
This is completely analogous to the important result from spatial Fourier optics (cf. \citealp[pp. 172--174]{Goodman}), which is based on Abbe's imaging theory (\cref{DiffractionFourier}). In spatial imaging, it designates the effects of geometrical projection and of diffraction, which are completely determined by the aperture and its own image as the exit pupil. This can be seen if the aperture is completely open, $P(\tau) = 1$, then its impulse response (or the \term{point spread function}, or PSF) is a delta function and the image obtained is the ideal geometrical image. In the context of temporal imaging, the analogous result enables us to refer to \textbf{dispersion-limited imaging}---imaging that does not suffer from any aberrations except for dispersion. 

\subsection{Imaging condition not satisfied}
\label{ImagingNotSatisfied}
When the imaging condition is not satisfied, it is convenient to define a \term{generalized pupil function} \citep[p. 205]{Goodman}, which includes the quadratic phase term, whose effect is a defocus aberration
\begin{equation}
{\cal P}(\tau) =  P(\tau) \exp \left[ \frac{i}{4}\left( \frac{1}{u} + \frac{1}{v} + \frac{1}{s}\right)\tau^{2}\right]
\label{generalizedpupil}
\end{equation}
As the defocus term is going to repeat throughout this work, we set 
\begin{equation}
W_d = \frac{1}{u} + \frac{1}{v} + \frac{1}{s}
\label{generalizedpupil}
\end{equation}
Using this generalized pupil and Eqs. \ref{impss} and \ref{Mimpss}, the impulse response integral is
\begin{equation}
\tilde h_d(\tau-\tilde\tau_0) = \frac{1}{2\pi} \exp\left( \frac{i\omega_c\tau^2}{2Mf_T} \right) \int_{-\infty}^\infty  P(2v\tilde T) \exp ( iv^2 W_d \tilde T^2) \exp \left[ { - i\tilde T (\tau  - \tilde\tau_0 )} \right] d\tilde T
\label{impresponse4}
\end{equation}
where the subscript $d$ was added to designate the defocused impulse response. In order to make the integral analytically solvable and the subsequent solutions well-behaved, it is convenient to assume a particular Gaussian-shaped pupil function \citep{Kolner1997}
\begin{equation}
P_g(\tau) = \exp\left[ -4\ln2 \,\,\left(\frac{\tau^2}{T_a^2} \right) \right]
\label{GaussPupil}
\end{equation}
where the aperture full width at half maximum is $T_a$ (see \cref{PulseCalc}). This is essentially the impulse response of a Gaussian low-pass filter for the modulation band with a time constant $T_a$. The full impulse response including the pupil can be solved and is therefore
\begin{multline}
\tilde h_d(\tau-\tilde\tau_0) = \frac{1}{2v} \sqrt{\frac{1}{\pi\left(\frac{16\ln2 \,\,}{T_a^2} -iW_d\right)}} \exp\left( \frac{i\omega_c\tau^2}{2Mf_T}\right) \exp \left[ -\frac{(\tau  - \tilde\tau _0)^2}{4v^2\left(\frac{16\ln2 \,\,}{T_a^2} -iW_d\right)}\right] \\
= \frac{1}{2v} \sqrt{\frac{1}{\pi\left(\frac{16\ln2 \,\,}{T_a^2} -iW_d\right)}} \exp\left( \frac{i\omega_c\tau^2}{2Mf_T} \right) \exp \left[ - \frac{\frac{16\ln2 \,\,}{T_a^2} + iW_d }{\left(\frac{16\ln2 \,\,}{T_a^2}\right)^2 + W_d^2} \frac{(\tau  - \tilde\tau_0)^2}{4v^2} \right]
\label{impresponse5}
\end{multline}
Where the final term is a product of a real Gaussian and a linear chirp. When the aperture duration is too long it gives rise to geometrical blur, as the aperture's own image is superimposed on the object image. 
This will cause temporal smearing of the image---loss of detail and contrast. When the aperture is too short, then the response will be dominated by dispersion effects that distort the image and also produce blur. Therefore, the choice of the aperture time is a tradeoff between geometrical blur and dispersion, which can be determined by minimizing the real part of the denominator in the complex Gaussian exponential in Eq. \ref{impresponse5}, similarly to \citet{Kolner1997}
\begin{equation}
\frac{16\ln2 \,\,}{T_a^2} =  |W_d|
\label{minaperture} 
\end{equation}
Or
\begin{equation}
T_a =  4 \sqrt{\frac{\ln2 \,\,} {\left|W_d\right|}}
\label{minaperture} 
\end{equation}
This expression was used along with the aperture time values that were computed in \cref{SpecTempApertureCalc}, The frequency-dependent values appear as $\Delta t_{opt}$ in Table \ref{t0opts}, as well as the ratios between $T_a$ and $\Delta t_{opt}$. The large ratios ($\Delta t_{opt} \approx 3$) indicate that the auditory system is heavily skewed toward geometrical blur and is therefore not dispersion-limited. 

It will be informative throughout the text to consider a different pupil function---the rectangular window (a slit, spatially), which is not necessarily more realistic than the perfect Gaussian pupil. While we have already seen that the Gaussian pupil may be a good model for the auditory aperture (\cref{ChinAperture}), sometimes the rectangular window provides more intuition. However, given the results of \cref{paramestimate} (Figure \ref{imagingcond}), the constant $W_d$ is negative in the entire audible spectrum, which makes it impossible to obtain closed-form solution in all of the cases under study (see \cref{RectImp}). 

\section{The Modulation transfer functions}
\label{TransFun}
To complete the analogy with Fourier optics, we shall derive the various modulation frequency-domain transfer functions based on the impulse response functions we derived above. Namely, we would like to derive the amplitude transfer function, the optical transfer function, and the modulation transfer function (see Table \ref{acousticsvsoptics}). All three should be derived for both the focused and the defocused cases and for both coherent and incoherent objects. Once available, these transfer functions will be primarily employed for qualitative analysis, because they do not take into account the neural sampling process that is taking place halfway through the signal propagation inside the auditory system. As was shown several times before for spatial systems that employ sampling (e.g., charge-coupled device cameras, CCD, that have grids of discrete detectors as pixels), the assumption of time-invariance (spatial invariance) is no longer correct for arbitrary signals with respect to an independent sampling grid \citep[pp. 35--50]{Wittenstein, Park, deLuca, Boreman}. Some implications of this will be discussed in \cref{TemporalSampling} and throughout the work, but for now will be neglected in order to obtain tractable expressions. They will be qualitatively reiterated later, once sampling is considered. 

The \term{amplitude transfer function} (ATF) is defined as the Fourier transform of the point spread function, $h(\tau)$, which can be thought of as the modulation-domain transfer function in acoustics. This relationship is particularly straightforward to obtain, since the impulse response in itself is already a scaled Fourier transform of the (generalized) pupil function (Eqs. \ref{impss} and \ref{impresponse4}; see \citealp[pp. 194--195]{Goodman}). Therefore, the ATF, $H(\omega)$, can be obtained from the double Fourier transform
\begin{multline}
H(\omega) = \int_{-\infty}^\infty \tilde h(\tau) \exp\left(-i\omega \tau \right) d\tau = \frac{1}{4\pi v} \int \int_{-\infty}^\infty  P(T) \exp \left(-\frac{iT\tau }{2v} \right) \exp\left(-i\omega \tau \right) dT d\tau \\ 
=\frac{1}{2\pi} P(- 2v \omega) 
\label{ATF} 
\end{multline}
where the global quadratic phase term was neglected as is customary for intensity imaging (see also \cref{GlobalQuadPhase}). Additionally, as the pupil function is generally symmetrical, the negative sign may be dropped, so $P(-2v \omega) =  P(2v \omega)$.

\subsection{Gaussian pupil}
In the Gaussian pupil case (Eq. \ref{GaussPupil}), the ATF is
\begin{equation}
H(\omega) = \exp\left[ -\frac{(16\ln2) v^2}{T_a^2}\omega^2 \right]
\label{GaussPupil2}
\end{equation}
where we dropped the $1/2\pi$ factor. Similarly, the ATF of the generalized Gaussian pupil is 
\begin{equation}
H_d(\omega) = H(\omega) \exp \left( iv^2W_d\omega^{2}\right) =  \exp\left[ -\frac{(16\ln2) v^2}{T_a^2}\omega^2 \right]\exp \left( iv^2W_d\omega^{2}\right)
\label{GenGaussPupil}
\end{equation}
This function will be referred to as the \term{defocused ATF} and is marked by the subscript $d$. 

The two ATFs are suitable for working with coherent objects---sounds that have a well-defined phase function. It is now possible to obtain the corresponding \term{optical transfer function} (OTF), which is defined as the normalized Fourier transform of the squared impulse response. It is the appropriate transfer function when working with incoherent objects that have stochastic and functionally undefined phase function. It can be computed from the normalized autocorrelation function of the ATF, using the Wiener-Khintchin theorem (\cref{CrossSpect}; \citealp[pp. 197--199]{Goodman}; see also \citealp{Schroeder1981})
\begin{equation}
{\cal H}(\omega) = \frac{\int_{-\infty}^\infty H(\omega'-\frac{\omega}{2})H^*(\omega'+\frac{\omega}{2})d\omega'}{\int_{-\infty}^\infty |H(\omega')|^2 d\omega'}
\label{OTFdef}
\end{equation}
Obtaining the OTF of the Gaussian pupil will be done in stages\footnote{See the appendix in \citet{Jiang} for a derivation of a similar two-dimensional OTF.}. The normalization in the denominator of Eq. \ref{OTFdef} is the same for both focused and defocused pupils
\begin{equation}
\int_{-\infty}^\infty |H(\omega')|^2 d\omega' = \int_{-\infty}^\infty |H_d(\omega')|^2 d\omega' = \int_{-\infty}^\infty \exp\left[ -\frac{\omega^{'2}}{2\left(\frac{T_a}{8\sqrt{\ln2}v}\right)^2}\right] d \omega' = \frac{\sqrt{\pi}}{4\sqrt{2\ln2}} \frac{T_a}{v}
\label{OTFnormal}
\end{equation}
For the standard pupil, the pupil and hence and numerator of Eq. \ref{OTFdef} are real
\begin{multline}
\int_{-\infty}^\infty H(\omega'-\frac{\omega}{2})H^*(\omega'+\frac{\omega}{2})d\omega' = v^2 \int_{-\infty}^\infty \exp\left\{ -\frac{(16\ln2) v^2}{T_a^2} \left[ \left(\omega'- \frac{\omega}{2}\right)^2 + \left(\omega'+ \frac{\omega}{2}\right)^2  \right] \right\} d\omega'\\
= v^2  \exp\left[ -\frac{(8\ln2) v^2}{T_a^2} \omega^2 \right] \int_{-\infty}^\infty \exp\left[ -\frac{\omega^{'2}}{2\left(\frac{T_a}{8\sqrt{\ln2}v}\right)^2}\right] d\omega' =  \frac{\sqrt{\pi}}{4\sqrt{2\ln2} } \frac{T_a}{v}\exp\left[ -\frac{(8\ln2) v^2}{T_a^2} \omega^2 \right]
\label{stanOTFnum}
\end{multline}
Putting the last three equations together we obtain the standard focused pupil OTF
\begin{equation}
{\cal H}(\omega) = \exp \left[-\frac{(8\ln2) v^2}{T_a^2} \omega^2 \right]
\label{ACGaussPupil}
\end{equation}
Moving on to the generalized defocused pupil, it is possible to recycle part of the standard OTF solution of Eq. \ref{stanOTFnum}
\begin{multline}
\int_{-\infty}^\infty H_d(\omega'-\frac{\omega}{2})H_d^*(\omega'+\frac{\omega}{2})d\omega'\\
= \exp\left[ -\frac{(8\ln2) v^2}{T_a^2} \omega^2 \right] \int_{-\infty}^\infty \exp\left[ -\frac{\omega^{'2}}{2\left(\frac{T_a}{8\sqrt{\ln2}v}\right)^2}\right] \exp \left\{ iv^2W_d\left[ \left(\omega'- \frac{\omega}{2}\right)^2 - \left(\omega'+ \frac{\omega}{2}\right)^2 \right] \right\} d\omega' \\
= \exp\left[ -\frac{(8\ln2) v^2}{T_a^2} \omega^2 \right] \int_{-\infty}^\infty \exp\left[ -\frac{\omega^{'2}}{2\left(\frac{T_a}{8\sqrt{\ln2}v}\right)^2}\right] \exp \left(- 2iv^2\omega W_d\omega' \right) d\omega' \\ 
= \frac{\sqrt{\pi}}{4\sqrt{2\ln2} } \frac{T_a}{v} \exp \left[-\frac{(8\ln2) v^2}{T_a^2} \omega^2 \right] \exp{\left(-\frac{v^2T_a^2W_d^2 }{32\ln2 }\omega^2 \right)}
\label{HdGaussUnnormal}
\end{multline}
where Siegman's lemma (Eq. \ref{SiegmansLemma}) was used again to solve the integral. 

Therefore, due to the generalized pupil, the OTF contains an additional Gaussian term compared to the standard OTF, which accounts for the defocusing phase term of the defocused ATF
\begin{equation}
{\cal H}_d(\omega) = {\cal H}(\omega) \exp{\left(-\frac{v^2T_a^2W_d^2 }{32\ln2 }\omega^2 \right)} = \exp \left[-\left(\frac{8\ln2 }{T_a^2} + \frac{T_a^2W_d^2 }{32\ln2 } \right) v^2\omega^2\right]
\label{ACGenGaussPupil}
\end{equation}
As the OTF is always positive due to the choice of pupil function, it is also identical to the modulation transfer function (MTF), which is defined as the modulus of the OTF. However, this is not the case in general and a \term{phase transfer function} (PTF) may have to be obtained as well.

\subsection{Rectangular pupil}
\label{RectPupil}
The Gaussian function is clearly a theoretical shape for the aperture, which is useful because of its convenient mathematical properties, as well as the intuition it can provide for some problems. It is going to be instructive to have the ATF and OTF of all varieties available for another theoretical aperture form---the rectangular pupil function. Let us define a rectangular pupil $P_r$ of width $T_a$. 
\begin{equation}
P_r(\tau ) = \rect\left(\frac{\tau}{T_a}\right)  = \left\{ \begin{array}{l}
1\,\,\,\,\,\,\,\,|\tau | < T_a/2\\
\frac{1}{2}\,\,\,\,\,\,\,|\tau | = T_a/2\\
0\,\,\,\,\,\,\,\,|\tau | > T_a/2
\end{array} \right.\
\label{RectPupil}
\end{equation}
The ATF readily follows from Eq. \ref{ATF} and the symmetry of the rect function
\begin{equation}
H_r(\omega) =  P_r(- 2v \omega) = \rect\left(\frac{2v\omega}{T_a}\right)
\label{ATFrect} 
\end{equation}
the OTF can be obtained from the autocorrelation of the ATF (Eq. \ref{OTFdef})
\begin{equation}
\int_{-\infty}^\infty H_r(\omega'-\frac{\omega}{2})H_r^*(\omega'+\frac{\omega}{2})d\omega' = \int_{-\infty}^\infty \rect\left[\frac{2v\left(\omega'-\frac{\omega}{2}\right)}{T_a}\right]\rect\left[\frac{2v\left(\omega'+\frac{\omega}{2}\right)}{T_a}\right] d\omega'
\end{equation}
This can be solved separately for the negative and positive ranges of overlap of the two rect functions
\begin{equation}
{\cal H}_r(\omega) = \left\{\begin{array}{l}
 \int_{-\frac{T_a}{4v} - \frac{\omega}{2}}^0 d\omega' = \frac{T_a}{4v} + \frac{\omega}{2}\,\,\,\,\,\,\,\,\, -\frac{T_a}{2v} \leq \omega  \leq  0 \\
\int_{0}^{\frac{T_a}{4v} - \frac{\omega}{2}} d\omega' = \frac{T_a}{4v} - \frac{\omega}{2}\,\,\,\,\,\,\,\,\,\,\, 0< \omega  \leq \frac{T_a}{2v} \\
\end{array} \right.\
\label{OTFRect}
\end{equation}
This is the triangle function ($\Lambda$) at double the support of the coherent ATF, which is obtained after normalization
\begin{equation}
{\cal H}_{r}(\omega) = \Lambda \left( \frac{2v\omega}{T_a} \right)  \,\,\,\,\,\,\,\,\, |\omega| \leq \frac{T_a}{2v}
\label{TriangleOTF}
\end{equation}
The triangle function is non-zero when the absolute value of its argument is smaller than 1, which is by definition double than the rect function support. The generalized rectangular pupil is
\begin{equation}
H_{dr}(\omega) = H_r(\omega) \exp( iW_dv^2\omega^{2}) = \rect \left( \frac{2v\omega}{T_a} \right) \exp ( iW_dv^2\omega^{2})
\label{GenRectPupil}
\end{equation}
Finally, calculating the defocused OTF requires a bit more work
\begin{equation}
{\cal H}_{dr}(\omega) =  \int_{-\infty}^\infty \rect\left[\frac{2v\left(\omega'-\frac{\omega}{2}\right)}{T_a}\right]\rect\left[\frac{2v\left(\omega'+\frac{\omega}{2}\right)}{T_a}\right] \exp (- 2iv^2\omega W_d \omega' ) d\omega' 
\end{equation}
The integral contains the same complex exponential as in Eq. \ref{HdGaussUnnormal}, but with the integral limits of Eq. \ref{OTFRect}
\begin{equation}
{\cal H}_{dr}(\omega) =  \left\{ \begin{array}{l}
\int_{-\frac{T_a}{4v} - \frac{\omega}{2}}^0 \exp (- 2iv^2\omega W_d \omega' ) d\omega' \,\,\,\,\,\,\,\,\, -\frac{T_a}{2v} \leq \omega  \leq  0 \\
\int_{0}^{\frac{T_a}{4v} - \frac{\omega}{2}} \exp (- 2iv^2\omega  W_d \omega' ) d\omega' \,\,\,\,\,\,\,\,\,\,\,\,\,\,\,\, 0 \leq \omega  \leq \frac{T_a}{2v} \\
\end{array} \right.\
\end{equation}
Solving for both intervals yields
\begin{equation}
{\cal H}_{dr}(\omega) =  \left\{ \begin{array}{l}
-\frac{1}{2iW_dv^2 \omega}\left\{ 1 - \exp\left[ -2iW_dv^2\omega \left( -\frac{T_a}{4v} - \frac{\omega}{2} \right) \right] \right\} \,\,\,\,\,\,\,\,\, -\frac{T_a}{2v} \leq \omega  \leq  0 \\
-\frac{1}{2iW_dv^2 \omega}\left\{ \exp\left[ -2iW_dv^2\omega \left( \frac{T_a}{4v} - \frac{\omega}{2} \right) \right] - 1 \right\} \,\,\,\,\,\,\,\,\,\,\,\,\,\,\,\, 0 \leq \omega  \leq \frac{T_a}{2v} \\
\end{array} \right.\
\end{equation}
Inside the parentheses in the arguments of the exponents of both parts of the integral, the same interval is covered as a function of $\omega$ of $[-\frac{T_a}{4v},\frac{T_a}{4v}]$, so they can be both united using the absolute value of $\omega$ in the argument and dividing their sum by 2
\begin{equation}
{\cal H}_{dr}(\omega) =  \frac{1}{4iW_dv^2 \omega}\left\{ \exp\left[ -2iW_dv^2\omega \left(- \frac{T_a}{4v} + \frac{ |\omega|}{2} \right) \right] - \exp\left[ -2iW_dv^2\omega \left( \frac{T_a}{4v} - \frac{|\omega|}{2} \right)  \right] \right\} \,\,\,\,\,\,\,\,\,\,\,\,\,\, |\omega|  \leq \frac{T_a}{2v}
\end{equation}
The two exponentials can now be replaced with a sine function and then with a sinc function
\begin{multline}
{\cal H}_{dr}(\omega) =  \frac{1}{2W_dv^2 \omega} \sin \left[  W_dv^2\omega \left( \frac{T_a}{2v} - |\omega| \right) \right] = \left( \frac{T_a}{2v} - |\omega| \right) \sinc  \left[  W_dv^2\omega \left(  \frac{T_a}{2v} - |\omega| \right) \right] \\
 =  \frac{T_a}{2v\omega}\Lambda \left(\frac{2v\omega}{T_a} \right) \sinc  \left[  W_dv^2\omega \left(  \frac{T_a}{2v} - |\omega|\right) \right] 
\end{multline}
Thus, the effect of the defocus results in a sinc function term, a remnant of the rectangular window, which multiplies the focused OTF, ${\cal H}_{r}(\omega)$. After normalization we obtain the final OTF
\begin{equation}
{\cal H}_{dr}(\omega) = \Lambda \left(\frac{2v\omega}{T_a} \right) \sinc  \left[  W_dv^2\omega \left(  \frac{T_a}{2v} - |\omega|\right) \right] 
\label{increctH} 
\end{equation}
Note that unlike the Gaussian pupil, the rectangular pupil is not non-negative, which means that its MTF and OTF are not identical, as the PTF changes between -1 and 1 for every zero crossing of the sinc function.

\subsection{Power modulation spectra and bandwidths}
\label{PowerMods}
We conclude with the general solution for the spectrum of a sinusoidal amplitude modulation component using the ATF and the OTF (see \citealp[pp. 215--217]{Goodman}). The ATF determines the modulation spectrum of coherent signals, whereas the OTF determines the incoherent sound modulation spectrum. Assume an amplitude modulation signal envelope
\begin{equation}
a(\tau) = \cos (\omega_m \tau) \,\,\,\,\,\,\,\, I(\tau) = \cos^2(\omega_m \tau)
\end{equation}
where $a(\tau)$ and $I(\tau)$ represent the amplitude and intensity, respectively, that are modulated at frequency $\omega_m$. Accordingly, the amplitude and intensity spectra $S$ are given with
\begin{equation}
S_a(\omega) = \frac{1}{2}\left[\delta(\omega-\omega_m) + \delta(\omega+\omega_m)\right]  \,\,\,\,\,\,\,\, S_I(\omega) = \frac{1}{2} \delta(\omega) +  \frac{1}{4}\left[\delta(\omega-2\omega_m) + \delta(\omega+2\omega_m)\right] 
\end{equation}
Given $H(\omega)$, the intensity and intensity spectrum of the coherent image output can be computed directly from the ATF by
\begin{equation}
I_{coh}(\tau) = |h(\tau) \ast a(\tau)|^2 
\label{cohconv}
\end{equation}
\begin{equation}
S_{coh}(\omega) = \left[H(\omega) A(\omega)\right]\star \left[H(\omega) A(\omega) \right]
\end{equation}
Where the $\star$ symbol designates the autocorrelation operation. Similarly, in the incoherent case
\begin{equation}
I_{inc}(\tau) = |h(\tau)|^2 \ast |a(\tau)|^2  = |h(\tau)|^2 \ast I(\tau)
\label{incconv}
\end{equation}
\begin{equation}
S_{inc}(\omega) = \left[H(\omega)\star H(\omega)\right] \cdot \left[A(\omega)\star A(\omega)\right] = {\cal H}(\omega)\cdot \left[A(\omega)\star A(\omega)\right] = {\cal H}(\omega) S_I(\omega)
\end{equation}
 Specifically for the input modulation
\begin{equation}
S_{coh}(\omega) =  \frac{1}{2} |H(0)|^2\delta(\omega) + \frac{1}{4} |H(2\omega_m)|^2\left[\delta(\omega-2\omega_m) + \delta(\omega+2\omega_m)\right] 
\end{equation}
\begin{equation}
S_{inc}(\omega) = \frac{1}{2} {\cal H}(0)\delta(\omega) + \frac{1}{4} {\cal H}(2\omega_m)\left[\delta(\omega-2\omega_m) + \delta(\omega+2\omega_m)\right] 
\end{equation}
It is important to remember that the power spectrum of a modulation frequency component $\omega_m$ is associated with a transfer function of double the frequency $H(2\omega_m)$\footnote{In hearing, this envelope is represented by beating, where we do not hear any difference between the negative and positive modulation half cycles, which amounts to an effective period doubling. In contrast, classical amplitude modulation (with envelope of the form $1+ m \cos(\omega_m t)$) has a linear component as well that is directly associated with $H(\omega_m)$. The simple trigonometric identity underlying it may be also explained more intuitively using sampling theory and is revisited in \cref{Aliasing}.}. 

The spectral relations above enable us to directly compare the effect of the ATF for coherent sounds versus the OTF for incoherent sounds. Both functions behave as low-pass filters, whose cutoff frequencies can be readily calculated and compared as well. In the coherent case, half the intensity level can be obtained from the square of the modulus of Eq. \ref{GaussPupil2} for the Gaussian pupil
\begin{equation}
\omega_{coh} =  \frac{T_a}{4\sqrt{2}|v|}
\label{wcoh}
\end{equation}
Similarly, the incoherent cutoff is obtained from \ref{ACGenGaussPupil}
\begin{equation}
\omega_{inc} = \frac{32(\ln2) \omega_{coh}}{\sqrt{256(\ln2)^2 + W_d^2T_a^4}}
\label{winc}
\end{equation}
Thus, when the system is in sharp focus, the second term in the denominator cancels out and we are left with an incoherent cutoff frequency that is double as large as the coherent cutoff---in line with results from spatial optics \citep[p. 203]{Goodman}. In general, though, because of the defocus, it is the opposite, or $\omega_{inc} \leq \omega_{coh}$, so the comparison will turn out to be much subtler, exactly because of the defocus term in the denominator. The analogous expression for the rectangular aperture that has an identical $T_a$ is obtained from Eq. \ref{TriangleOTF}
\begin{equation}
\omega_{coh,r} =  \frac{T_a}{4|v|}
\label{wcohrect}
\end{equation}
It can be seen that the modulation band support is larger in the rectangular aperture, which has a larger time-bandwidth product than the Gaussian aperture. 

The incoherent rectangular cutoff may be obtained numerically for specific parameter values from \ref{increctH}.

The range between these two extreme conditions of coherent and incoherent sounds represents the partially coherent domain. In linear systems, it can be represented as a combination of coherent and incoherent effects (\ref{totalcoherence}). A direct treatment of partial coherence that is comparable to the one above of these two extremes requires more advanced tools and is beyond the scope of this work (but see, for example, \citealp[pp. 599--606]{Born}). However, an intuitive understanding of partial coherence as an intermediate coherence regime, which combines weighted coherent and incoherent images, will be at the heart of explaining the range of operation of the auditory system as whole.

\section{The modulation transfer function in hearing}
\label{TheTMTF}
We are now in a position to test the predictions of the OTF against the temporal modulation transfer function (TMTF), using the dispersion parameters we obtained in \cref{paramestimate} and the temporal aperture from \cref{TempAperture}. TMTFs are commonly used to estimate the sensitivity to amplitude modulation in human hearing (see \cref{AudSenseEnv}). The visual analog of the TMTF is the contrast sensitivity function (CSF), which is defined as the combined MTF of the eye and the visual neural system, so whatever part of the threshold that cannot be explained by peripheral optics is typically attributed to the neural pathways \citep{VanNes1967,Bour1996}. They have been measured using sinusoidally amplitude modulated broadband, narrowband, and tonal carriers. For broadband carriers, the TMTF threshold generally has a low-pass filter response, as the sensitivity drops with increasing modulation frequency, and has a nominal range in broadband carrier of 4 kHz \citep{Viemeister1979} or 2 kHz \citep{Forrest1987}. Narrowband-noise and sinusoidal thresholds yield morphologically different responses for the same modulation depths, depending on the bandwidth of the carriers used \citep{Fleischer1983, Dau1997a, Kohlrausch2000}. Intuitively, we would like to be able to relate the theoretical coherent and incoherent modulation transfer functions from \cref{TransFun} to the tonal and broadband TMTFs, respectively. As it turns out, the comparison of the various transfer functions is not straightforward and will be mostly qualitative in first approximation, apart from extreme cases. However, this comparison will provide important insight about the temporal imaging in the auditory system, and initiate a discussion about partial coherence and the effects of sampling.

\subsection{Low-frequency modulation bandwidth correction}
\label{LowFreqCorr}
As was seen in the psychoacoustic curvature data analysis, aperture time based on a strictly dispersive criterion (Eq. \ref{t0xindp}) yielded wrong predictions at low frequencies (below 500 Hz). The correction to the temporal aperture durations that was offered is based on the following observations about the auditory MTFs. 

\begin{figure} 
		\centering
		\includegraphics[width=0.8\linewidth]{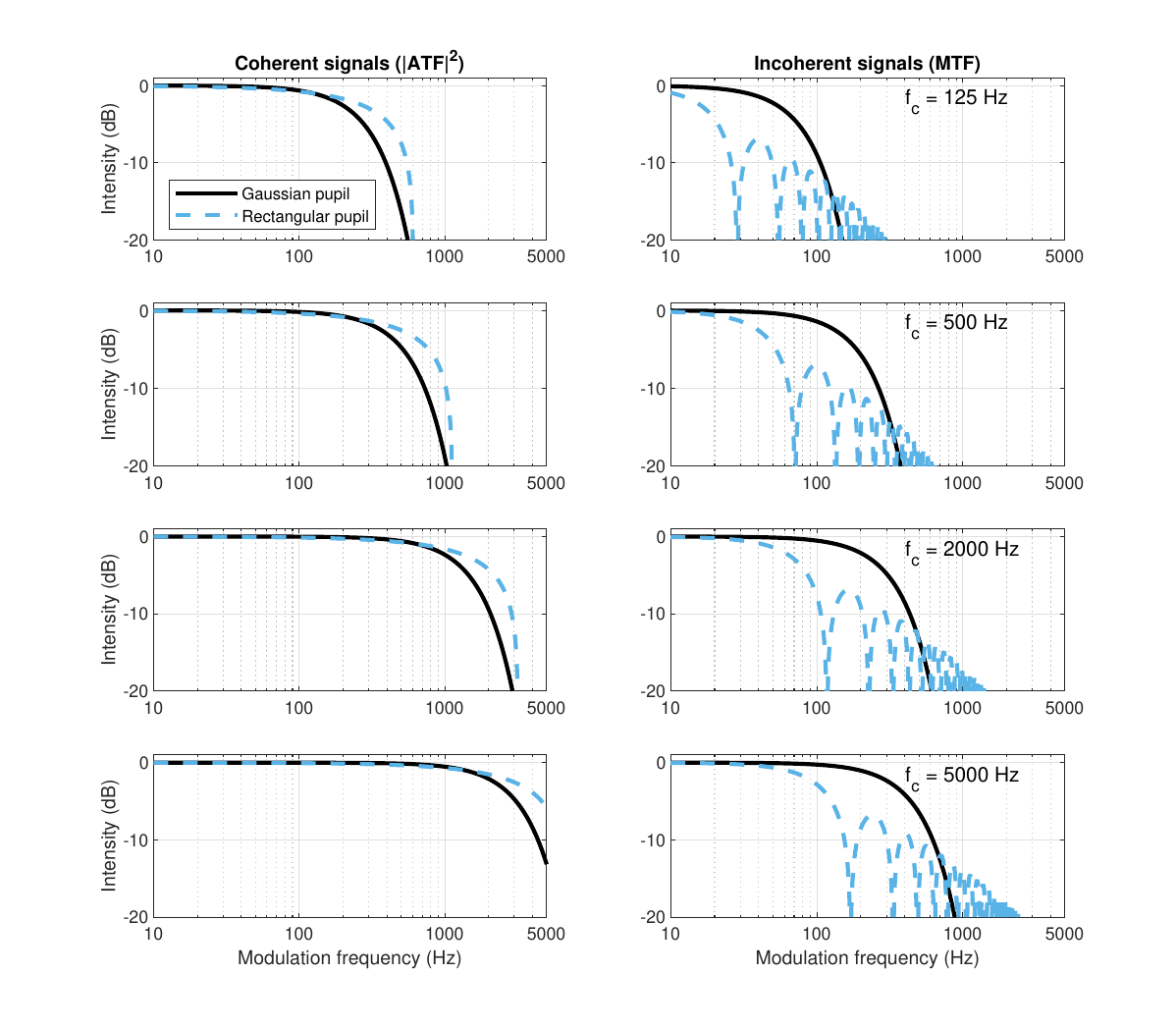}
		\caption{Estimated amplitude and modulation transfer functions of the human auditory system, with Gaussian and rectangular pupils and with ideal sampling, at 125, 500, 2000, and 5000 Hz carriers. For coherent inputs (plotted on the left), the modulus square of the ATF is computed, to obtain an intensity measure that is comparable to the incoherent MTF (right). The rectangular-pupil response (dashed blue) is broader than the Gaussian-pupil (solid black) for coherent inputs, but it is much narrower than the Gaussian for incoherent inputs. Additionally, the sinc function (Eq. \ref{increctH}) makes the incoherent rectangular-pupil response oscillate many times before it completely decays.} 
		\label{ATFvsOTF}
\end{figure}

The coherent and incoherent modulation transfer functions for perfectly-sampled inputs\footnote{The perfect-sampling condition allows us to interpret the continuous signal expressions at face value. This condition will be relaxed later, as evidence will emerge that can be interpreted as suboptimal sampling that degrades the various MTFs.} were given in Eqs. \ref{GenGaussPupil} and \ref{ACGenGaussPupil} for a Gaussian pupil function and in \ref{ATFrect} and \ref{increctH} for a rectangular pupil function. For coherent signals, the amplitude transfer function (ATF) is in effect, which entails linearity in amplitude. Its normalized autocorrelation forms the MTF\footnote{As the Gaussian pupil function is real, its OTF is always positive, which makes it identical to the MTF. Therefore, we will refer to it as MTF, to suggest similarity to the TMTF and simplify the terminology going forward. However, in the case of the rectangular pupil, the OTF changes signs, so a distinction will be made between the OTF and MTF.}, which linearly weights the intensity spectrum of the signal and is valid only for incoherent inputs. The functions are plotted for several carrier frequencies in Figure \ref{ATFvsOTF}. All transfer functions exhibit a low-pass behavior for modulation frequencies, but the cutoff frequency is about five times lower for incoherent signals with the Gaussian pupil, due to the inherent defocus aberration in the system. This difference is much more pronounced with the rectangular pupil with about 20 times the difference between coherent and incoherent in cutoff frequency for some of the carriers. 

\begin{figure} 
		\centering
		\includegraphics[width=0.7\linewidth]{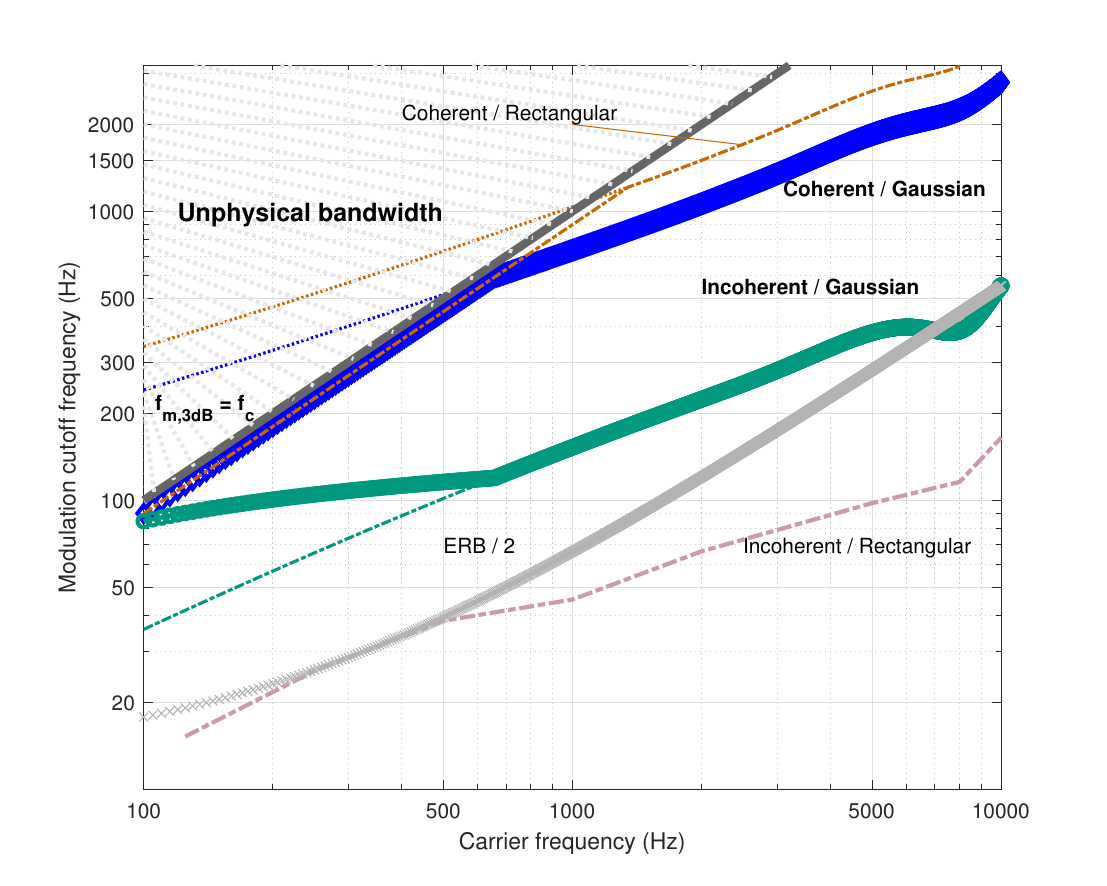}
		\caption{The 3 dB cutoff frequencies of the ideal modulation transfer functions with Gaussian and rectangular pupils. The coherent and incoherent Gaussian modulation bandwidths are plotted in thick blue and green lines, respectively. However, if left uncorrected, the original cutoff frequencies would be broader than the carrier, which is physically impossible (the limit $f_m=f_c$ is in dash gray and the corresponding unphysical area is hatched above it). Using a maximum bandwidth that is somewhat arbitrarily set to $0.9f_c$, the coherent correction takes place below 660 Hz. For comparison, the responses of rectangular apertures are plotted as well. The coherent rectangular bandwidth (dash-dot red) is $\sqrt{2}$ larger than the Gaussian, based on Eqs. \ref{wcoh} and \ref{wcohrect}. For incoherent modulation, the rectangular bandwidth is 3.37 times narrower than the Gaussian bandwidth (purple dash-dot). The cutoff was computed numerically for the rectangular pupil and according to Eq. \ref{winc} for the Gaussian. For comparison, the growth of half the auditory filter bandwidth (equivalent rectangular bandwidth, ERB) is plotted in gray crosses (Eq. \ref{ERB}), based on \citet{Glasberg1990}.} 
		\label{OTAcutoff}
\end{figure}

The modulation bandwidth generally increases with the carrier frequency, as is displayed in Figure \ref{OTAcutoff}. At low frequencies, the theoretical coherent MTF has a modulation bandwidth that is larger than the carrier, which is physically meaningless. While these cutoff frequencies are undoubtedly excessive, surprisingly high cutoff frequencies were measured for low tonal carriers in the cat's auditory nerve fibers \citep[Figure 13]{Rhode1994Encoding}, where exceptionally broad TMTFs can be seen of carrier $f_c \approx 350$ Hz, and cutoff frequency $f_m \approx 295$ Hz $=0.84 f_c$, as well as for $f_c \approx f_m \approx 500$ Hz\footnote{The modulation filters were characterized by the cutoff frequency, which is where the synchronization coefficient drops to 0.1 and is therefore higher in frequency than the 3 dB cutoff \citep[Figure 1]{Rhode1994Encoding}.}. Additionally, recent in-vivo measurements of the intact guinea-pig apical channels found that the (mechanical) cochlear response at frequencies lower than 2 kHz (equivalent to 900 Hz in humans; see \cref{LowFreqCorr}) is low-pass and not bandpass \citep{Recio2017, Recio2018}. Of course, this comparison is problematic not only because human, cat, and guinea pig all have different neural group-delay dispersion magnitude, but also because the value we obtained for $v$ applies to signals whose destination is the inferior colliculus and not the auditory nerve. However, auditory nerve dispersion alone is most likely smaller than $v$ (\cref{NeuralDisp}), which would entail even broader TMTFs than with our estimation using the human $v$ (Eq. \ref{wcoh}). 

Therefore, we artificially force the modulation bandwidth for low-frequency carriers ($\leq 660$ Hz for a Gaussian aperture; $\leq 1350$ Hz for rectangular) to be equal to 0.9 of the carrier, somewhat arbitrarily pushing the limit of the filter bandwidth (see \cref{ModCarrierFraction}). Additionally, even with this conservative correction made to the cutoff frequency, it appears that the effect of defocus diminishes as the two transfer functions are brought closer together at very low carrier frequencies, depending on the pupil function. Interestingly, because the coherent cutoff frequency should depend only on $v$ and $T_a$ (Eq. \ref{wcoh}), this correction generates an important constraint for these two parameters that could not be justified otherwise. By tweaking the temporal aperture $T_a$, the prediction of the psychoacoustic curvature data at 125 and 250 Hz readily fits the experimental data from \citet{OxenhamDau}. However, an additional correction to the defocus term is probably required as well, in order to cancel out any chirping at the output image, although this was not pursued further due to lack of sufficient information (see \cref{LowFreqCorr} for more details). 

Using the corrected values for the low-frequency carrier aperture times, we can now revisit our auditory MTF predictions (figure \ref{ATFvsOTF}). For example, using a 5 kHz carrier, the Gaussian pupil 3 dB cutoff frequency is about 340 Hz in the incoherent case, whereas it is 2360 Hz for the coherent case. For the rectangular pupil, the cutoff frequencies are 3400 and 100 Hz, for the coherent and incoherent cases, respectively. The incoherent rectangular MTF oscillates many times before dying out completely. If the rectangular shape had some resemblance to the physiological window, then we would expect to have non-monotonic incoherent TMTF due to oscillations. As will be seen below using data from literature, this is certainly not the case, so we shall stick to the Gaussian pupil function, in line with our earlier analysis (\cref{TempAperture}). 

\subsection{Empirical TMTF data from literature}
With the theoretical MTFs now available, we would like to compare their predictions to empirical data. The dependent variable in behavioral data of the TMTF is the hearing threshold needed for detection of the standard amplitude modulation (AM) signal,
\begin{equation}
	a(t) = \left[ 1 + m \cos(\omega t) \right] \sin(\omega_c t) \,\,\,\,\,\,\,\,\, 0 \leq m \leq 1
	\label{StandardAM}
\end{equation}
as a function of the modulation frequency $\omega$ and carrier frequency $\omega_c$, when the carrier is tonal. The modulation depth $m$, which is equivalent to contrast, is expressed in dB, where $m=1$ is considered 100\% modulation depth. In behavioral studies, the TMTF is tested as a detection threshold---sensitivity to any modulation. Thus, it does not give information about whether the listener hears the particular modulation frequency, or something else. We will return to this subtle point later. In physiological studies of the brainstem and midbrain, the stimulus is usually in full modulation, and the sensitivity is quantified using the synchronization strength to the envelope, producing a TMTF as a function of modulation frequency \citep{Joris2004}. 

\begin{figure} 
		\centering
		\includegraphics[width=1\linewidth]{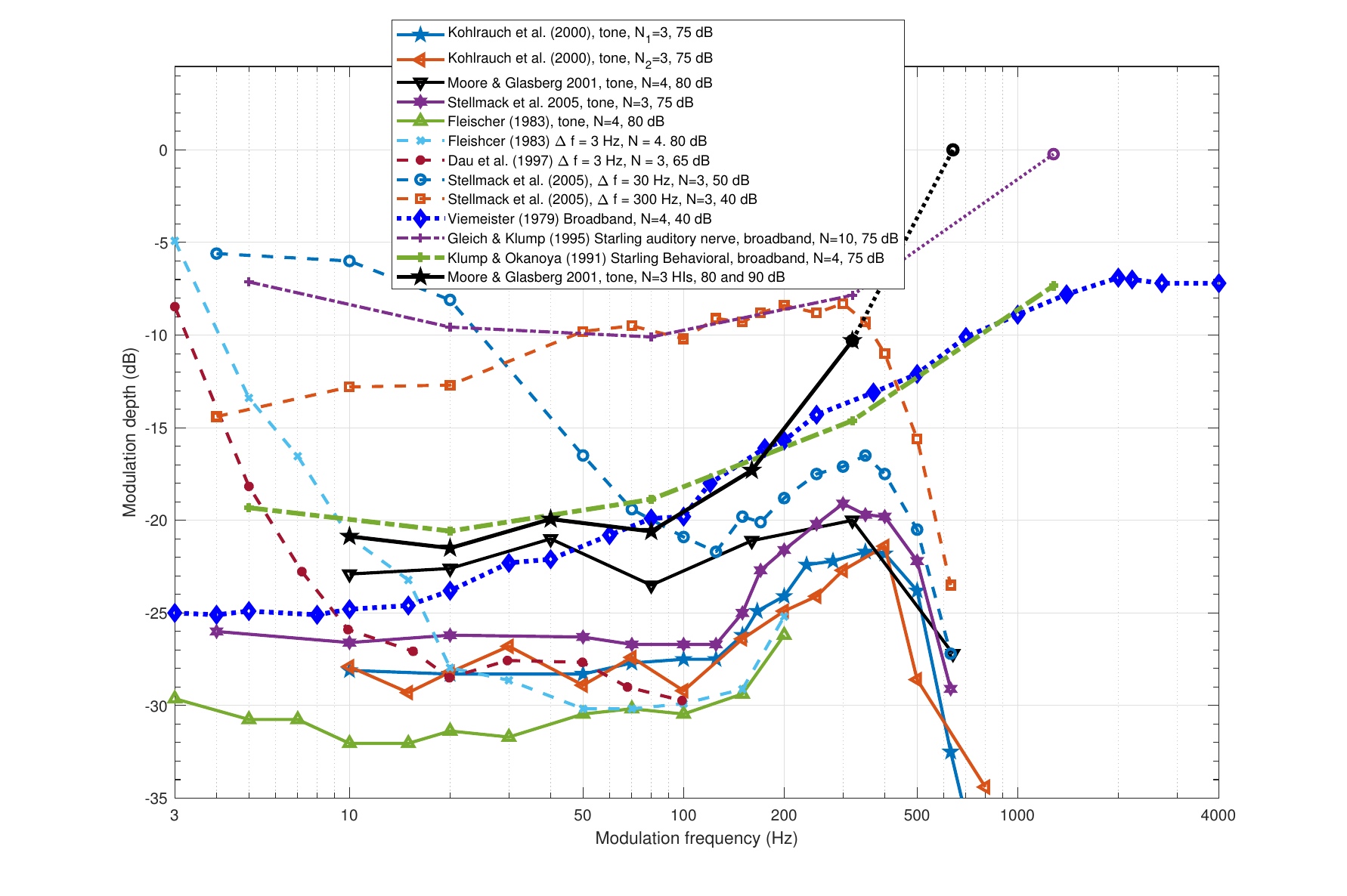}
		\caption{Various temporal modulation transfer functions (TMTFs) from literature. Pure tone and narrowband data have all been measured with a 5 kHz carrier, which is therefore the most readily comparable frequency and the only one displayed for tonal and narrowband carriers. The threshold is given in amplitude modulation depth dB ($20\log m$), where 0 dB designates 100\% modulation ($m=1$) in Eq. \ref{StandardAM}. The data were usually collected using a small number of subjects ($N$), which is noted in the corresponding legend, along with the type of signal. The curves include two datasets from \citet[Figures 2 and 3]{Kohlrausch2000}, which were measured with different modulation frequencies and were separated to two groups of three subjects; normal hearing and hearing-impaired tonal data from \citet[Figures 2 and 3]{Moore2001}; tonal and narrowband data ($\Delta f = 30,\,\, 300$ Hz) from \citet[Figure 3, top right]{Stellmack2005}, which reproduced very closely the main observations (but over an extended frequency range) of \citet[Figures 4--5]{Dau1997a} with $\Delta f = 31,\,\, 314$ Hz that are therefore not plotted here; tonal and narrowband data from \citet[Figure 1]{Fleischer1983}; narrowband data from \citet{Dau1997a} of $\Delta f = 3$ Hz; broadband data from \citet[Figure 2]{Viemeister1979}. Finally, two animal broadband datasets are presented of the European starling as measured directly from the auditory nerve \citep{Gleich1995} and behaviorally \citep{Klump1991a}---both curves were extracted from Figure 9A in \citet{Gleich1995}. Note that at 1280 Hz, there was no measurable physiological response in the starling even at $m=0$, which is therefore plotted in dotted line and a circle instead of a cross.} 
		\label{TMTFsAll}
\end{figure}

Several published TMTF curves of different types are compiled in Figure \ref{TMTFsAll} and will serve as a reference in the subsequent analysis with several subsets of these curves presented later. As the curves represent thresholds and not sensitivities, they are plotted upside-down compared to the MTFs. The lowest threshold (most sensitive) is rarely lower than about -30 dB. The lowest value should be compared to the 0 dB passband level of the theoretical MTFs, which do not account for internal noise in the auditory system. Therefore, the most, or perhaps the only, relevant parameter of the TMTF that can be compared with the MTF prediction is the cutoff frequency.

At first glance, the theoretical values we obtained for the cutoff frequencies for both pupil functions (Figure \ref{OTAcutoff}) appear completely at odds with much of the human and animal behavioral TMTF data of Figure \ref{TMTFsAll}, which show a much narrower modulation bandwidth. This is most readily seen in data using the 5 kHz carrier, which has probably been the most tested frequency of the human TMTF. The 3 dB cutoff frequency was estimated to be anywhere from 144 to 229 Hz for 5 kHz tonal carriers \citep{Stellmack2005}. Additionally, narrowband data exhibit altogether different TMTF morphology, as it has higher threshold at low modulation frequencies---a departure from pure tone TMTFs, which behave like a typical low-pass filter. At high frequencies, the threshold drops again, as the input is spectrally resolved and the modulation can be also detected by adjacent auditory channels, producing a sensitivity that cannot be temporally achieved by a single channel. 

All in all, there is a substantial discrepancy between the idealized predictions and the empirical data. While not displayed, it is observed that it is impossible to straightforwardly tweak the imaging parameters---mainly $v$ and $T_a$---to retain consistent results of the temporal resolution and phase curvature data of the previous sections, along with the empirical TMTF data. Therefore, any discrepancy with the empirical data is not only due to misestimation of the dispersion parameters. However, in order to account for this discrepancy, a more detailed analysis of the empirical data is first provided for tonal (coherent) and broadband (incoherent) in the next two subsections. The narrowband data will be used to usher in the discussion about partial coherence in the next section. All together, these analyses will provide some of the necessary insight for the understanding of the role of the auditory defocus.

\subsection{Tonal TMTFs}
\label{TonalTMTFs}
Pure tones are the ultimate coherent carrier and as such should theoretically reflect the low-pass behavior of the coherent auditory ATF. When the pure tone is amplitude-modulated at high enough a frequency, its sidebands are resolved by the adjacent filters. Therefore, with low- and mid-frequency carriers, the low-pass characteristics of the TMTF may not be easily observed, because the modulation is spectrally detected by other channels before the threshold drops, which results in a flat response. At 5 kHz, the equivalent rectangular bandwidth (ERB; Eq. \ref{ERB}) is 565 Hz \citep{Glasberg1990}, so usually some threshold increase can be observed before it drops again when the sidebands are resolved, as can be seen in Figure \ref{TMTFsAll}.

\begin{figure} 
		\centering
		\includegraphics[width=0.5\linewidth]{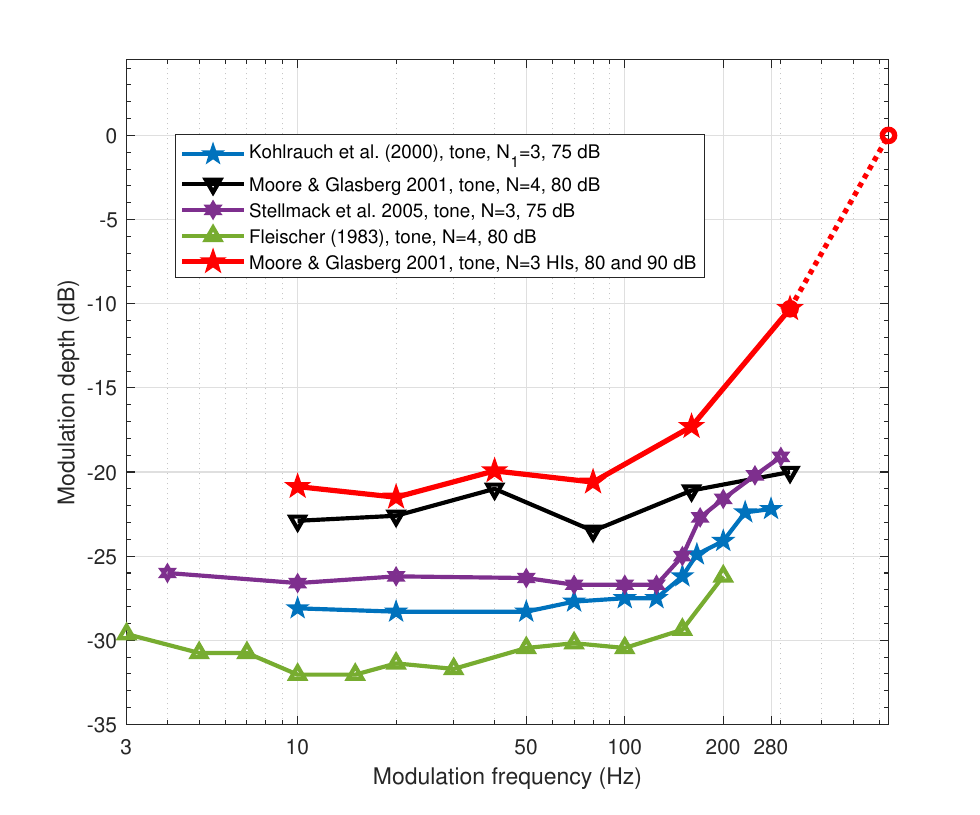}
		\caption{A subset of tonal TMTFs at a carrier of 5 kHz. Refer to Figure \ref{TMTFsAll} for details. Note that no response could be measured at 640 Hz with the hearing-impaired subjects, even at 100\% modulation depth \citep{Moore2001}. The curves were truncated before the sidebands are spectrally resolved by the adjacent filters, in order to show only the monotonically decreasing responses associated with a single channel.}
		\label{TMTFsTone}
\end{figure}

The tonal TMTF curves from Figure \ref{TMTFsAll} are replotted in Figure \ref{TMTFsTone}, where the responses are truncated at $\ERB/2 \approx 283$ Hz, for clarity, just when the threshold is at its highest. All datasets except for those from \citet{Moore2001} show a sharp bend at around 150 Hz, and a cutoff at slightly higher frequency. The normal hearing data from \citet{Moore2001} is inconsistent with these trends, showing a nearly flat response below the half-ERB frequency. However, additional data of three hearing-impaired subjects---likely with broadened auditory filters---revealed a distinct low-pass filter response with a similar cutoff (160 Hz), and no response at 640 Hz modulation. Two subjects had similar responses at 2000 Hz as well, which are not displayed\footnote{Few other studies are also inconsistent with the low-pass response and tend to show bandpass behavior at very low frequencies \citep[e.g.,][]{YostSheft1997}. These effects may be related to signal presentation methods and to longer temporal integration effects that are beyond the scope of this analysis. See \citet{Kohlrausch2000} for further discussion.}. 

Regardless of the specific cutoff frequency of the low-pass modulation filter, it is almost an order of magnitude lower than the ideally-sampled coherent ATF that was obtained earlier (about 2 kHz for a 5 kHz carrier, Figure \ref{ATFvsOTF}, bottom left). Therefore, comparison to within-channel auditory nerve measurements may be more revealing than behavioral measurements. For instance, the bandwidth of the cat's auditory nerve TMTF increases to a maximum of about 1500 Hz \citep[Figures 11 and 14]{Joris1992}. For carriers below 10 kHz, the modulation frequency scales with the carrier and the cutoff frequency reaches a maximum of 1300 Hz. At high tonal carriers it levels off and reaches the absolute maximum at a carrier of 27 kHz. However, several instances of maximum modulation cutoff frequencies that are even higher (1500--2500 Hz) were recorded in the auditory nerve of the cat for carriers between 10 and 30 kHz \citep[Figure 13]{Rhode1994Encoding}. Given the high variability in the data and that it may be constrained by physiological limitations on the firing rates and nonexistent phase locking at these carrier frequencies, it is difficult to directly compare these cat data to our predictions. Nevertheless, our very high cutoff prediction (2000 Hz for carrier of 4000 Hz and 3000 Hz for a carrier of 10000 Hz, Figure \ref{OTAcutoff}) may be more relevant to the specific channel, which carries ``pre-perceptual'' information that is partially lost on the way downstream \citep{Weisser2019}. 

\subsection{Broadband TMTFs}
\label{EmpiricalBB}
Broadband carriers in the audible spectrum represent incoherent signaling and is thus suitable to be tested vis-\`a-vis the ideal MTF, which predicts a lower cutoff frequency than the coherent ATF, due to defocus. However, the fact that full-spectrum white noise simultaneously excites multiple channels complicates the analysis. It can been dealt with either by filtering out part of the signal, or by combining several auditory-channel-wide stimuli that together produce the broadband stimulus bandwidth. There is an important caveat to this, though---it is not obvious that an auditory-channel-wide white noise signal may be considered a true incoherent carrier, in a sense that is equivalent to complete incoherence in optics, which has vanishingly small coherence time (Eq. \ref{CoherenceTimeEst}; see also \cref{PLLNoise}). The frequencies associated with electromagnetic coherence theory are many orders of magnitude larger than the audible range, so that the normal bandwidth of quasi-monochromatic light can have a true randomized phase that covers its entire modulation bandwidth and accrues over numerous periods over short physical distances \citep[originally commented about ultrasound frequencies, which are less affected by this problem than audio frequencies]{Tarnoczy}. This means that true incoherent light need not interact with the carrier bandwidth or violate the narrowband assumption. It does not appear to be exactly the case in narrowband sound. While seemingly a technical point, it is nevertheless of great importance, as will be argued below.

\begin{figure} 
		\centering
		\includegraphics[width=0.5\linewidth]{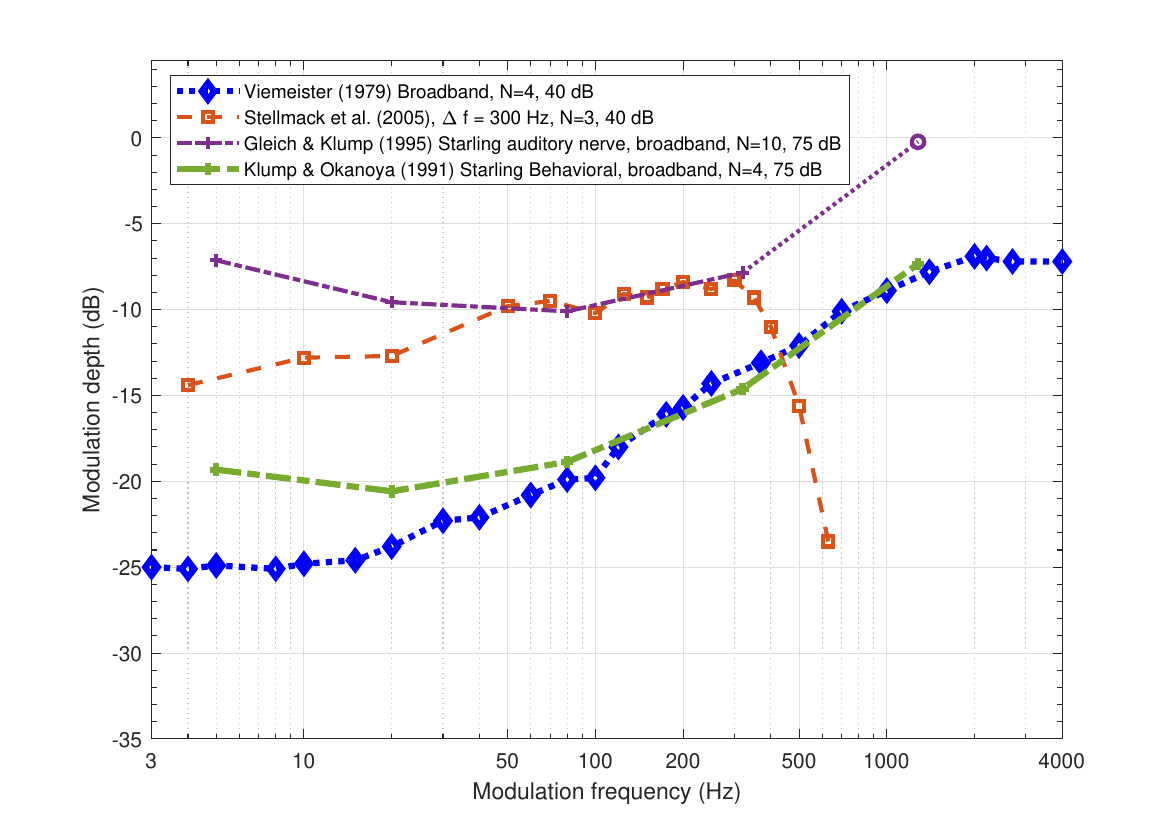}
		\caption{Broadband TMTFs. Refer to Figure \ref{TMTFsAll} for details.}
		\label{TMTFsBB}
\end{figure}

Four broadband curves are plotted in Figure \ref{TMTFsBB}. The most well-known TMTF of sinusoidally modulated white noise was measured by \citet{Viemeister1979} and unlike the tonal TMTFs, it is monotonically rising with a low-frequency cutoff of approximately 40 Hz (compared with about 150--200 Hz in the tonal case at 5 kHz). While a lower cutoff frequency is expected from the general relation between the defocused coherent and incoherent transfer functions (Figure \ref{ATFvsOTF}), it is not obvious how the auditory filter outputs are combined to yield this broadband threshold. 

Some insight may be garnered from animals that were tested using comparable stimuli. Observations of the European starling are particularly revealing, because they were obtained both behaviorally \citep{Klump1991a} and physiologically \citep{Gleich1995}. The behavioral starling data follow the human data very closely for modulation frequencies above 80 Hz (green dash-dot curve in Figure \ref{TMTFsBB}). In contrast, the response to the same stimulus measured in single units of the auditory nerve has a very similar shape, but at a 7--10 dB reduced sensitivity (purple dash-dot curve). Now, returning to human behavioral data, a narrowband carrier of 300 Hz bandwidth produced a threshold that is identical to the starling's auditory nerve at (roughly) 80--320 Hz. Above 320 Hz, the two thresholds diverge, as the human auditory system appears to resolve the narrowband noise, or at least it has access to spectral cues from adjacent filters that reduce the threshold. Hence, it seems that pooling modulation information across fibers can reduce the TMTF threshold in both humans and starlings. Note that additional behavioral data from other animals suggest that the overall threshold (its most sensitive portion) can vary between species (\citealp[Figure 4]{Dent2002} and \citealp[Figure 5]{Klump1991a}).

Assuming that the single-unit starling response to broadband sound does indeed qualify as completely incoherent, it is possible to test it against the ideal-sampling MTF. As we do not have the imaging parameters of the starling system, we can make use of the incoherent temporal acuity expression of Eq. \ref{TemporalRes} that will be introduced in \cref{GapDetect}. Combining it with the expressions for the incoherent cutoff frequency of Eq. \ref{winc} and the coherent Gaussian-pupil cutoff of \ref{wcoh}, the following expression is obtained after some algebraic manipulation
\begin{equation}
f_{inc} = \frac{4 \ln2}{\sqrt{2}\pi} \cdot \frac{1}{d} \approx \frac{0.624}{d}
\end{equation}
that ties together the temporal acuity $d$ and the incoherent cutoff frequency $f_{inc}$ of the MTF. Interpolation of the starling neural TMTF curve in Figure \ref{TMTFsBB}, sets the 3 dB cutoff at 365 Hz \citep[the curves were averaged over 30 units with CFs between 200 and 3500 kHz,][Figures 6 and 7a]{Gleich1995}. Single-unit auditory nerve broadband gap detection measurements of the starling identified an absolute minimum of $d=1.6$ ms \citep[at CF = 1.3 kHz,][Figure 4a]{Klump1991b}. Assuming that the minimum gap is achieved with ideal sampling allows us to estimate that $f_{inc} = 390$ Hz---a fairly close result to $f_{inc} \approx 365$ Hz (re 80 Hz) from the auditory nerve TMTF measurement from \citet{Gleich1995} (dash-dot purple in Figure \ref{TMTFsBB}). In contrast, the behavioral broadband starling cutoff is only 150--180 Hz \citep{Klump1991a}---about half of the physiological bandwidth. However, linking these physiological and behavioral TMTFs requires consideration of the additional processing stages following the auditory nerve, as well as adequate translation to human hearing. 


Two studies directly compared the behavioral and the physiological tonal and broadband TMTFs in the same animals and recording sites. In both studies, the TMTFs were obtained from the inferior colliculus (IC), where the maximum cutoff frequencies observed are expected to be lower than those found the brainstem. Multi-unit recordings of amplitude modulated noise and tonal carriers were compared to behavioral responses of the awake budgerigar (an Australian parrot), whose hearing and vocalization capabilities resemble those of humans \citep{Henry2016}, and in rabbits that have higher thresholds than humans \citep{Carney2014}. In the budgerigar, lower noise-carrier TMTF cutoffs were shown in IC neurons that exhibited higher peak synchrony and higher best modulation frequencies for tonal carriers \citep[Figure 5]{Henry2016}. Additionally, for the noise carrier, both the across-channel pooled rate and neural synchrony thresholds were less sensitive to high modulation frequencies (256--512 Hz) and showed no best modulation frequency neurons that are tuned to 512 Hz---the only frequency band measured above 256 Hz \citep[Figure 7]{Henry2016}. Very similar results were shown in the rabbit---only less sensitive overall \citep{Carney2014}. In comparison with the budgerigar's data, the cutoff of the ideal Gaussian-pupil incoherent MTF of Figure \ref{OTAcutoff} increases slowly with carrier frequency, but also never goes beyond 400 Hz, and even this bound is likely to be reduced by the time it reaches the IC, as was seen in the starling data in \cref{EmpiricalBB}. Notably, in both animal studies, both rate and envelope synchrony information was recorded separately and was associated with behavioral performance in the rabbit (spiking rate) and budgerigar and human (synchrony). This observation may suggest that different animals may employ different detection methods to process the modulated sound. While this separation to two coding regimes is commonly employed, they should be both understood as two complementary and indispensable aspects of sampling. If sampling is precise, it must be synchronized to the envelope at an appropriate rate. Otherwise, it necessarily generates sampling errors.

The incoherent-broadband prediction seems to be much more realistic than the coherent predictions for the tonal data, in terms of the modulation bandwidth. Nevertheless, even for the incoherent TMTF, the drop in rate between the physiological and behavioral animal data is substantial. Possible causes for these differences are discussed in \cref{MTFandSampling}.

\subsection{Narrowband TMTFs}
\label{NarrowTMTF}
Narrowband stimuli reveal a range of TMTF responses that are a hybrid of the tonal and the broadband responses, but are neither. Typically, in order to measure the narrowband TMTF, full-spectrum white noise is bandpass-filtered around the carrier and then sinusoidally amplitude-modulated\footnote{The order of the bandpass filtering and modulation operations was not the same in different studies, but was shown to elicit very similar TMTFs \citep{Dau1997a, Stellmack2005}. It appears that the two orders produce responses that are close enough both qualitatively and quantitatively, so whatever difference exists between the two is ignored in the analysis.}. Because of the relative and absolute proximity of the modulation and the carrier frequency bands, the stochastic carrier bleeds into the lowest frequencies in the modulation band, in proportion to the bandwidth of the carrier\footnote{\citet{Dau1997a} highlighted that according to a proof by \citealp{Lawson1950}, a rectangular carrier envelope in the spectral domain results in a triangular envelope in the spectral modulation domain of the same bandwidth. We saw in \cref{TransFun} that this is a general result of the autocorrelation of rectangular windows, only that the resultant bandwidth is doubled. See \citet{Dau1999} for further investigations.}. This results in considerable modulation masking energy at low modulation frequencies, which steers the TMTF away from its typical low-pass filter behavior.

\begin{figure} 
		\centering
		\includegraphics[width=0.5\linewidth]{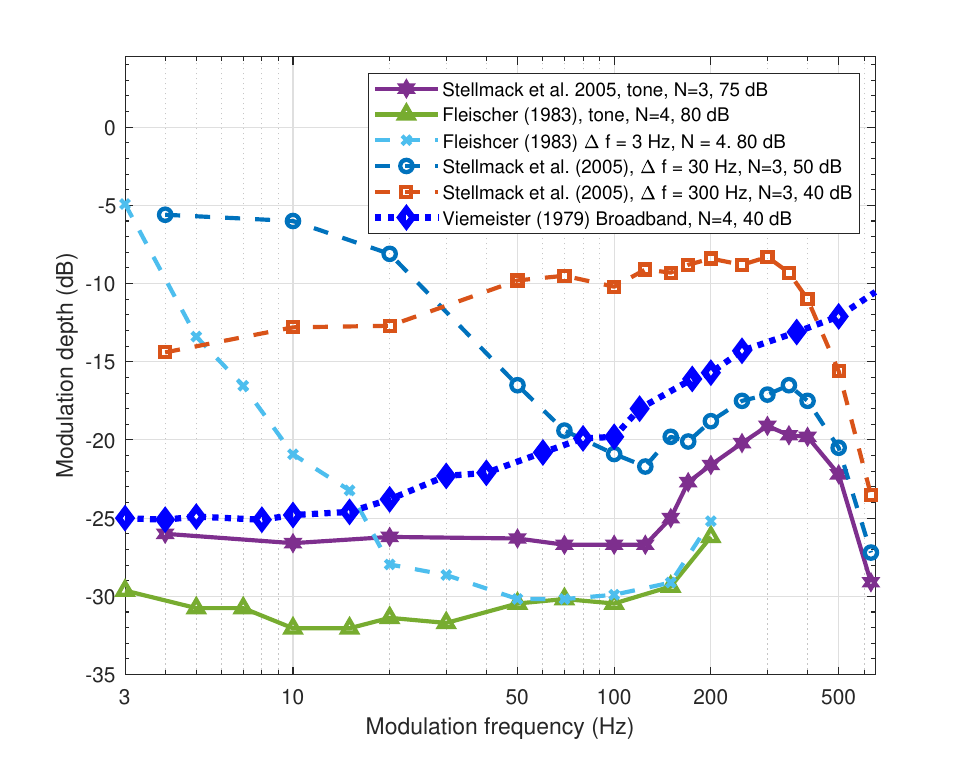}
		\caption{Narrowband TMTFs. Refer to Figure \ref{TMTFsAll} for details. }
		\label{TMTFsNB}
\end{figure}

Consider the five TMTF curves that are replotted in Figure \ref{TMTFsNB}. Pure tone modulations elicit the flattest and most sensitive responses. The tonal data by \citet{Stellmack2005} provide the widest frequency range that includes the usual low-pass response before the sidebands are resolved and the threshold drops again beyond the half-ERB frequency. The tonal data by \citet{Fleischer1983} are a bit more sensitive, but are identical in shape. This tonal TMTF was measured in the same setup along with a 3 Hz narrowband carrier TMTF. The tonal and narrowband responses diverge below 50 Hz, but then merge and become indistinguishable (very similar data are given in \citealp[Figure 3]{Dau1997a}, where the responses merge already at 15 Hz). As the narrowband bandwidth broadens (measured for 30 and 300 Hz bandwidths), the responses diverge from the tonal TMTF by becoming less sensitive over a wider modulation bandwidth. The 300 Hz bandwidth curve is much flatter than the 3 and 30 Hz curves, presumably due to the triangular distribution of the modulation spectrum that gets flatter with increased bandwidth \citep{Dau1997a}. As was noted in the broadband TMTF analysis, in the limit of broadband stimuli \citep{Viemeister1979}, the response shape is about the same as the 300 Hz narrowband, but the sensitivity improves by about 10 dB and continuously decreases, instead of being resolved by adjacent filters.   

It is apparent that the responses generated by the narrowband stimuli are not well-described by either coherent or incoherent MTFs. This type of signals requires the much more general framework of partial coherence, which better taps realistic signals as well. It is discussed below.

\section{Modulation and partially coherent sound}
\label{PartialModu}
The partial overlap between the narrowband and tonal TMTFs reveals a great deal about the auditory system sensitivity to carrier types. A narrowband carrier---even if stochastic by design---cannot be considered truly incoherent if it contains discernible modulation components that compete with the target modulation. Moreover, as it becomes indistinguishable from a fully coherent carrier at higher modulation frequencies, it may have effective qualities of a coherent carrier. If this is correct, then narrowband carriers such as the 3 Hz and the 30 Hz bandwidths from Figure \ref{TMTFsNB} should be able to interfere and beat together, even though they are not tonal. This should not be surprising, as both coherence time and length---the measures of the degree of coherence over time and space, respectively---are inversely proportional to the bandwidth of the carrier (\cref{CoherenceTimeLength}). So, by definition, a pure tone has a degree of coherence of 1, white noise 0, and narrowband signal somewhere in between. It was also shown how the filter bandwidth in which sounds are analyzed can increase the apparent coherence of otherwise incoherent signals, even if they never become exactly the same (\cref{CoherentFiltering}). Either way, the compiled data of tonal, broadband, and narrowband TMTFs strongly suggest that narrowband sounds may be best understood as partially coherent, rather than completely incoherent. 

\begin{figure} 
		\centering
		\includegraphics[width=0.75\linewidth]{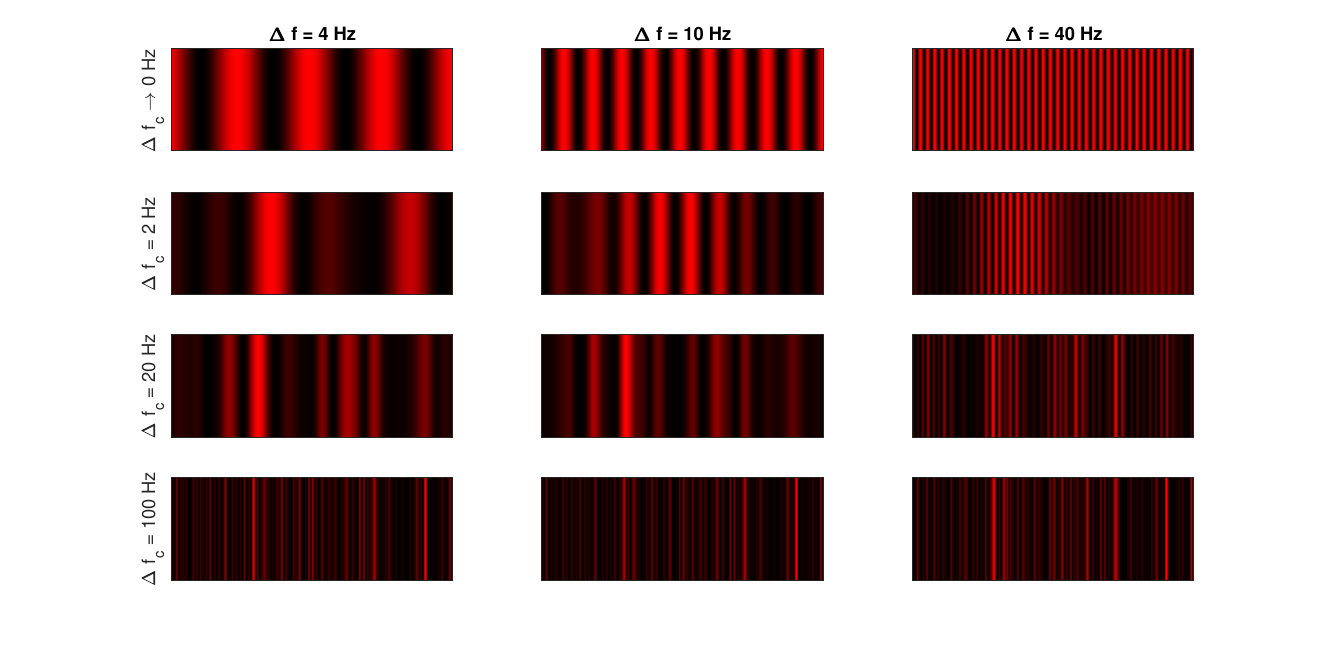}
		\caption{The interference caused by beating of two tones (top row) or narrowband sounds (bottom three rows). The patterns were generated by summing two time signals centered at $f_c = 5$ kHz of bandwidth $\Delta f_c$ that are separated by $\Delta f$, squaring them, and low-pass filtering the output using a fourth-order Butterworth filter with 400 Hz cutoff. The top row is the interference pattern of two pure tones, with no random components. The duration is always 300 ms. Narrowband carriers were generated by zeroing the out-of-band frequencies from the Fourier transform of broadband Gaussian white noise signals, as in \citet{Dau1997a}. The output is mapped to 256-level intensity color map.}
		\label{InterfereTMTFbeat}
\end{figure}

\begin{figure} 
		\centering
		\includegraphics[width=0.75\linewidth]{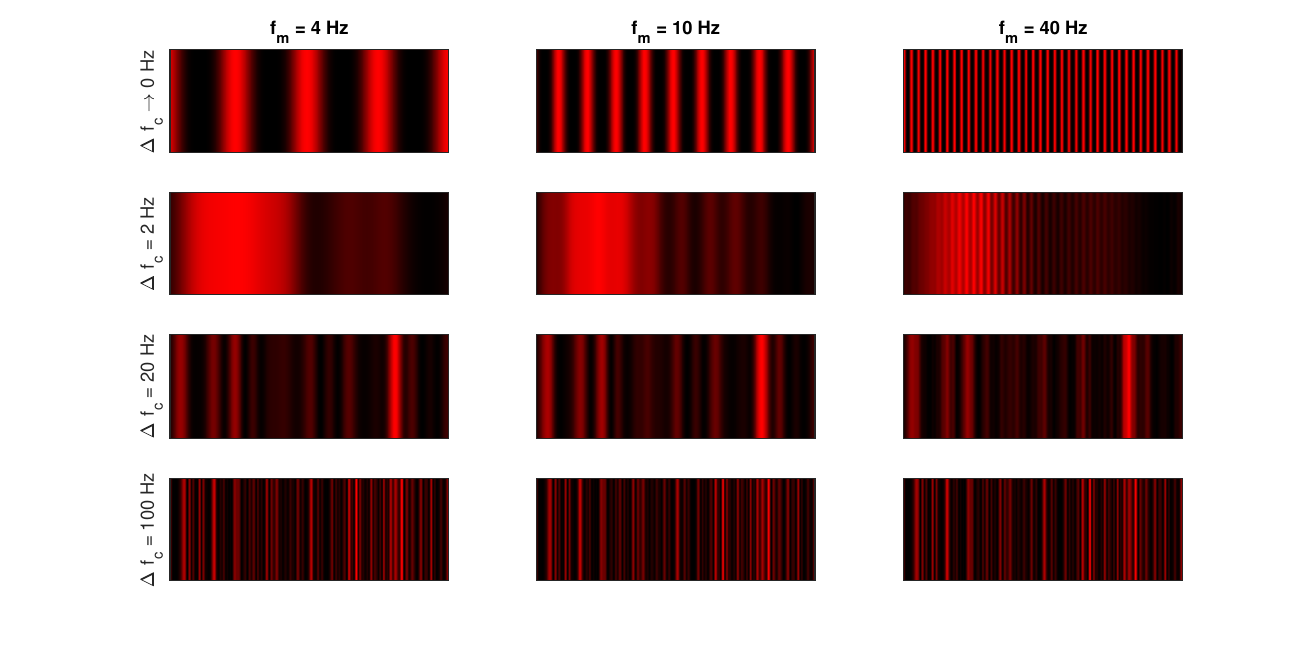}
		\caption{The interference caused by amplitude modulation of pure tones (top row) or narrowband sounds (bottom three rows). The patterns were generated by summing three time signals centered at $f_c = 5$ kHz and $f_c \pm f_m$ at half the carrier level, with additional details identical to Figure \ref{InterfereTMTFbeat}.}
		\label{InterfereTMTFam}
\end{figure}

To illustrate the partial coherence of narrowband sounds, the beating of two narrowband sounds with different bandwidths is visualized in Figure \ref{InterfereTMTFbeat} using one-dimensional and monochromatic interference patterns, or \term{interferograms}. Additionally, the effective amplitude modulation of three narrowband sounds (carrier plus two sidebands) is shown in Figure \ref{InterfereTMTFam}. The interferograms are particularly descriptive in illustrating the basic measure of interference that is fundamental in imaging as well---visibility or contrast (i.e., modulation depth)---which is not as vividly represented in normal time signal plots. Unlike optical interferograms, the x-axis represents time instead of a static spatial dimension. The plot height contains no information and is used only for visualization. Audio demos corresponding to the two figures are found in \textsc{/Section 13.5 - Modulation and partially coherent sound/}.

In Figure \ref{InterfereTMTFbeat}, the interference of two narrowband sounds of bandwidth $\Delta f_c$ and frequency spacing $\Delta f$ is shown. After demodulation, the beating between the components at frequency $f_c \pm  \Delta f/2$ translates to $\Delta f$ in the intensity pattern, which was obtained after low-pass filtering the carrier and squaring. In the top row of the figure, $\Delta f_c = 0$, which is the classical beating of two pure tones for three different frequency separations $\Delta f$ of 4, 10 and 40 Hz, over 1 s duration. The interference patterns are fully periodic and exhibit high contrast between the peaks and the troughs, showing as completely black background. In the next three rows, three narrowband carriers with $\Delta f_c$ of 2, 20 and 100 Hz are interfered for the same spacing. As the bandwidth of the carrier increases, the image becomes gradually irregular, the periodicity is disrupted, and the contrast is lost, so the interference pattern turns more uniform, on average. 

The loss of contrast is more apparent in the second set of interferograms in Figure \ref{InterfereTMTFam}, which illustrates AM interferences between a narrowband carrier and two narrowband sidebands. The loss of contrast is most apparent in the second row. As three sounds with random phase interfere here, the coherent-looking patterns are visible only for the carrier with 2 Hz bandwidth, where the modulation frequency and carrier frequency appear superimposed. For larger bandwidths, the patterns no longer show visible interference and the modulation rate sounds almost unrecognizable. Thus, the degree of coherence for such an AM stimulus is smaller than the beating source. Note that if the sidebands would have been produced directly using a sine function instead of narrowband sidebands, then the interference would have been much stronger and closer to the results from literature.  

In these examples, the effect of the auditory filter itself was not taken into account, but the corresponding perceptual effects can be heard in the supplementary sound demos. The interference between narrowband noise maskers and narrowband noise or pure tone targets and the likely effect of beating has been occasionally discussed in literature. See for example, \citet{Egan1950} and \citet{Moore1998}. 


\section{Discussion and conclusion}
In this chapter we derived the various impulse response functions for the human auditory system, as well as its modulation domain transfer functions. These functions bring to light the significance of the defocus in the system, as it differentiates the coherent and incoherent signal regimes that enter the system. 

We compared the predictions from this theory in terms of MTF bandwidth to human data of broadband, tonal, and narrowband TMTFs. That there are significant differences between the functions that could be predicted from theory, but the actual bandwidths were grossly overestimated when compared to behavioral thresholds and only roughly corresponded to single-channel measurements from the auditory nerve of animals. This discrepancy is puzzling given the robust prediction we obtained in the time-domain analysis in \cref{CurvatureModeling} using the very same parameters. As will be shown in the next chapter, this can be explained by incorporating sampling into the system, which is known to degrade the spatial MTF of light in optics. 

Despite the discrepancy in the TMTF and predicted MTF, the intuition that is garnered by the idealized frequency-domain analysis will remain valuable in the chapters to come. Namely, coherent signals have broader modulation bandwidth than incoherent signals, due to the inherent defocus of the auditory system that differentiates the ATF and the MTF. For this conclusion to be meaningfully used, the concept of partial coherence must be considered in the analysis of typical auditory stimuli. It should be emphasized, though, that the sensed coherence is not necessarily equal to the stimulus coherence. The sensed degree of coherence of the source is directly tied to the filter bandwidth that processes the signal and on the ensuing phase locking provided by the system. These factors will be discussed in \cref{HypoAcco} in the context of auditory accommodation.

Using the modulation spectrum analysis, we obtained two further insights about the temporal aperture of the system. First, its pupil function is indeed closer to a Gaussian than to a rectangular window, in line with animal data from the chinchilla, and earlier results in \cref{TempAperture}. Second, at low frequencies, the aperture stop is likely be the result of the cochlear filters, perhaps combined with other elements, rather than the neural spiking that dominates the high frequencies. This justifies the correction to the aperture time values that was required to account for the low-frequency cochlear phase curvature prediction data in \cref{TempAperture}.

\chapter{The use of sampling for imaging continuous signals}
\label{TemporalSampling}

\section{Introduction}
The visual image consists of relatively static frames that are perpetuated in time, which is then perceived as a degenerate dimension of the object. Such images can be thought of as time-independent, as they represent spatial patterns that appear to be frozen in time. Visible temporal changes in the spatial image often require macroscopic movement to take place. It is mapped to an extra dimension and requires sequential imaging that is continually refreshed. The same holds also in the case of still and moving-image cameras. Both devices contain almost the same optics, but only the latter has an explicit added dimension of time and, hence, movement. 

In contrast to light that is produced by quantum processes on the atomic and molecular levels, sound production requires continuous macroscopic movement (vibration), whose temporal dependence is much closer to relevant biological time constants than light, by many orders of magnitude. The way that the auditory system is designed to receive the signals confines them to being spatially one-dimensional, which means that all temporal and spatial dependencies are mathematically projected on a one-dimensional audible time signal. This means---to complete the analogy with vision---that a static image of a pulse, or part thereof, must be taken as a stepping stone to a continuous image of moving objects. Unlike photography though, sound recording is only interesting inasmuch as it can capture ``moving'' sound, and not just a single pulse. Therefore, the concept of an ``acoustic still shot'' is of little use outside of theoretical research. 

Acoustic signals are generally continuous, whereas the various expressions developed in the temporal imaging theory relate to pulses. In fact, nowhere in the temporal imaging equations is it made explicit that the envelopes must be shaped as pulses to satisfy the imaging operation. By the physical nature of the system, it has a finite temporal aperture, which represents what the system can ``see'' at any one time. Through modeling the psychoacoustic phase curvature data using the continuous temporal imaging equations, it became abundantly clear in \cref{CurvatureModeling} that the system has a finite aperture. A finite aperture was then included in the impulse response and modulation transfer functions, as the defining feature of the pupil function (\cref{AudImpulseRes}). Effectively, it is this aperture that imposes a pulse shape on the incoming signal---it is not necessary for the signal itself to enter the system in a pulse form. 

In this chapter we will integrate some of the most elementary principles of sampling theory into the auditory temporal imaging theory. Auditory sampling, which happens in the transduction of the auditory nerve and in all subsequent synapses, is nonuniform and extends to a finite duration per sample, unlike classical sampling techniques. The implications will be discussed, mainly with respect to the narrowing of the temporal modulation transfer function between the periphery and higher auditory areas. An additional discussion about the implications of discrete processing and how it compares with continuous temporal models of hearing is provided in \cref{Aliasing}.

\section{Sampling in hearing theory}
\label{SamplingInTheory}
While sampling is an indispensable operation in modern signal processing that is based on discretized representations of all analog signals, it is not an integral concept in hearing theory itself, although it arguably deals with neurally discretized signal representations. In this brief section, the exceptions to this trend in literature are mentioned---nearly all of which were modeled independently of one another. 

In psychoacoustics, two models exist that employ either ``looks'' \citep{Viemeister1991} or ``strobes'' (\citealp{Patterson1992}; adopted also in \citealp{Lyon2018})---discrete samples in the auditory signal processing. In some conditions, these models provide better predictions to experimental data than continuous ``leaky integration'' or ``sliding window'' models. These models are not given precise physiological correlates, or detailed technical parameters about the sampling. See \cref{Aliasing} for a further discussion.

Sampling is somewhat more common in physiological models of hearing. Several auditory signal processing models exist that were inspired by nonuniform or irregular sampling of wavelet frames, whose exact physiological correlate is not made explicit \citep{Yang1992,Benedetto1993,Benedetto2001}.

Most sampling models relate directly to the auditory nerve. \citet{Lewis1995} and \citet{Yamada1999} referred to the noise from the high spontaneous rate auditory nerve fibers as performing dithering\footnote{\term{Dithering} is smoothing of sampling fluctuations, which are caused by the minimum quantization level (its finite resolution), through the addition of random low-level noise.}---a term that is normally used only in the context of sampling and conversion between digital and analog signal representations. A more specific mechanism of sampling was considered by \citet{HeilIrvine1997} and \citet{Heil2003}, where the auditory nerve coding of the onset of temporal envelopes was modeled as equivalent to point-by-point sampling of the envelope function, which tracks it at high resolution, limited by the spike/sampling rate. Another neural processing model makes use of the concept of stochastic undersampling to show how deafferentation of the auditory nerve is analogous to noise \citep{Poveda2013, Poveda2014}. This model has some parallels to the classical volley principle, whereby the acoustic input is adequately sampled (or even oversampled) by a population of neural fibers, each of which by itself undersamples the signal (\citealp{Wever1930Present}; \cref{ElementsHearing}).

Similar ideas were sometimes attributed to higher-level nuclei such as the brainstem. \citet{Warchol1990} suggested that high spontaneous rates in the avian auditory cochlear nucleus enable better sampling of the stimulus. Additionally, \citet{Yang1992} noted that the anteroventral cochlear nucleus (AVCN) receives inputs from the auditory nerve, which could be instantaneously mismatched and then lead to effective lateral inhibition. This perspective may be interpreted as another form of nonuniformity in the sampling that exists beyond the stochastic auditory nerve spiking pattern. Further downstream, \citet{Poeppel2003} suggested that the two auditory cortices work by asymmetrically sampling the incoming sound---the left hemisphere samples the auditory cortex at around 40 Hz, and the right hemisphere at 4--10 Hz. This sampling strategy is thought to be particularly advantageous for speech processing, where information on different time scales can be extracted simultaneously. 

Finally, sampling in the spectral domain of the spectral envelope was also considered in the context of a model for vowel identification, which can be degraded when the harmonic content is rich and the fundamental frequency is high, because of spectral undersampling and resultant aliasing distortion \citep{deCheveigne1999}. The model was also formulated in the temporal domain using autocorrelation, which may have a physiological correlate. More generally, the model was applied for pitch perception as well \citep{Cheveigne}.

\section{Basic aspects of ideal sampling}
\label{SamplingBasics}
The temporal imaging theory is based on envelope objects that are shaped as finite pulses. However, realistic signals are long and continuous, so the theory must be extended to include signals of arbitrary durations. A natural solution is to internally construct the signal as a sequence of pulses. Each pulse is an image in its own right, but can also be thought of as a sample. It is possible in general to use the system linearity and time-invariance (only approximate features in our case) to impose the pulse internally and relax the condition that the input is shaped as a pulse. This was demonstrated naturally in our model of the psychoacoustic phase curvature responses, which assumed that the continuous signal is shaped according to the aperture stop of the system. In theory, if the signal is sampled frequently enough, an accurate reconstruction can be made that recovers the complete, arbitrary signal, based on sampling theory. If this happens on multiple channels in parallel, then the complete auditory image can be reconstructed from the individual channel images\footnote{The existence of an actual reconstruction operation will remain, at this stage, completely hypothetical insofar as it pertains to auditory perception.}. 

\begin{figure} 
		\centering
		\includegraphics[width=0.6\linewidth]{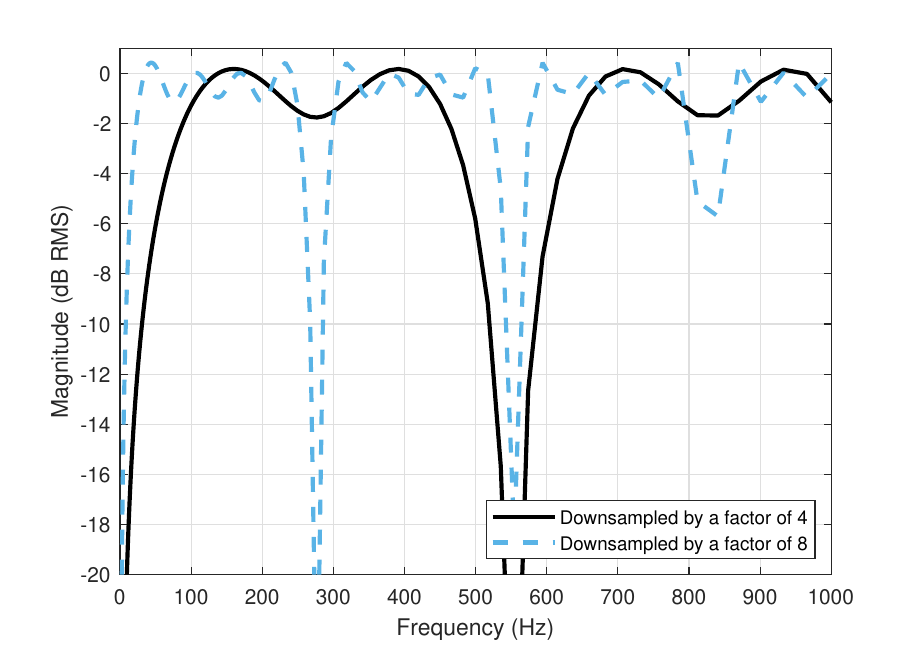}
		\caption{The apparent frequency response caused by downsampling a signal when it contains frequencies higher than the Nyquist limit and no anti-aliasing filter is employed. In the examples, sine tones generated at a sampling rate of $f_s = 4410$ Hz were resampled using delta impulse functions at rates of 1102 Hz (solid black) and 551 Hz (dash blue). All frequencies above half the downsampling rates are aliased products caused by folded tones of lower frequencies. This aliasing gives rise to distinct notches in the frequency response.}
		\label{aliasexample}
\end{figure}

Ideal sampling is represented as a periodic array of delta functions (also called a \term{comb function}):
\begin{equation}
	s(\tau) = \sum_{n=-\infty}^{\infty} \delta(\tau-nT_s)
\end{equation}
with period $T_s$ that is determined by the sampling rate $T_s = 1/f_s$. The sampled version of an input $a(t)$ is then:
\begin{equation}
	a_s(\tau) = \sum_{n=-\infty}^{\infty} a(nT_s)
\end{equation}
Using Fourier series with period $T_s$, it is possible to obtain the spectrum of the sampler (see, for example \citealp[pp. 133--136]{Pelgrom}):
\begin{equation}
	S(\omega) = \frac{1}{T_s}\sum_{m=-\infty}^{\infty} \delta(\omega - m \omega_s)
\end{equation}
And the corresponding spectrum of the sampled signal is 
\begin{equation}
	A_s(\omega) = \frac{1}{T_s}\sum_{m=-\infty}^{\infty} A(\omega - m\omega_s)
\end{equation}
Therefore, the spectrum of the sampled signal is periodic as well and repeats in multiples of $f_s = \omega_s/2\pi$. As long as the sampled signal is bandlimited and satisfies the Nyquist criterion of $B < 2f_s$, the periods of the spectrum stay clear of one another and perfect reconstruction of the signal is possible, according to Shannon's sampling theorem (\cref{InfoNutshell}). But if this bound is breached, then the reconstruction will include spectral components that are not in the original input and give rise to aliasing (see an example in Figure \ref{aliasexample}). Therefore, an anti-aliasing (low-pass) filter is commonly employed to ensure that the input to the sampler is bandlimited. The ideally reconstructed signal is then given by
\begin{equation}
	a_s(\tau) = \sum_{n=-\infty}^{\infty} a(nT_s) \sinc\left(\frac{\tau - nT_s}{T_s}\right)
\end{equation}
In practical engineering designs, a sinc filter is never used, but there are many approximations that can achieve almost as good a reconstruction as this theoretical one \citep{Unser2000}. 

\section{Auditory sampling and imaging}
\label{temporalmodels}
In spatial imaging, the single samples are bits of intensity at a certain wavelength channel, corresponding to color pixels. In vision, photoreceptor activation was shown to occur from single photons \citep{Bialek1987}. Is the correct way to understand hearing equivalent to vision---point-by-point sampling of the acoustic signal? We shall go over the canonical sampling types to try and shed some light over this question.

\subsection{Impulse sampling}
\label{ImpulseSampling}
\begin{figure} 
		\centering
		\includegraphics[width=1\linewidth]{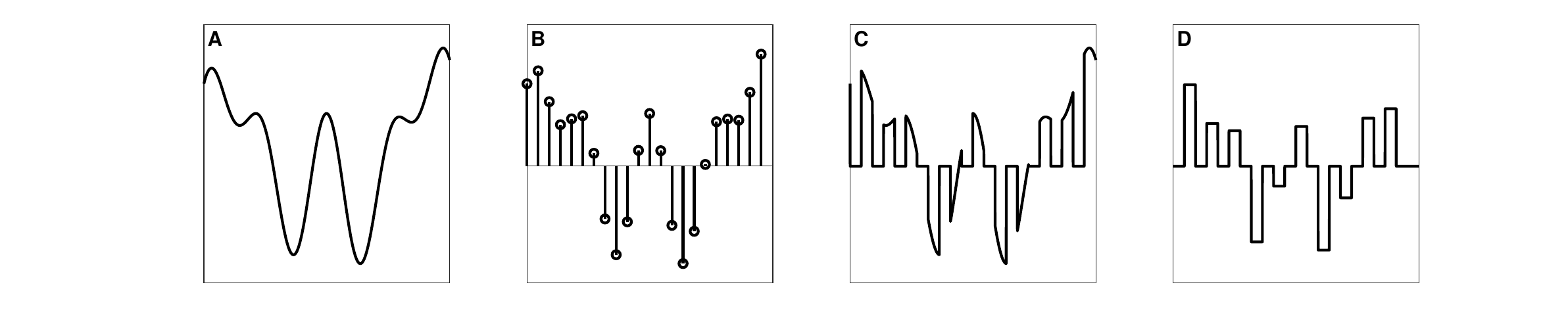}
		\caption{Different sampling types. \textbf{A}: The original signal. \textbf{B}: Impulse sampling. \textbf{C}: Natural sampling. \textbf{D}: Flat-top sampling.}
		\label{samptypes}
\end{figure}

\term{Impulse sampling} is achieved when the sampling windows are infinitesimally short, as each one is well-approximated by a delta function (\citealp[pp. 93--95]{Couch}; see A and B in Figure \ref{samptypes}). Obviously, the sampling window always has a finite width, even if it is approximated mathematically as a delta function. Insofar as the entrance pupil that was estimated in \cref{TempAperture} reflects the sample duration, an infinitesimally short impulse sampling is not supported by the data. In practice, however, the effective pulse must have been represented by a population of neurons in the auditory channel that fired more or less simultaneously. It is possible that each neuron in itself samples the image that we considered as a pulse in the temporal imaging equations. The sampled and coded pulse then propagates downstream, where the impulsive nature of the initial sampling may or may not be perceived. At present, this sampling model cannot be confirmed or rejected. 

\subsection{Natural sampling and images}
\label{NaturalSamp}
A realistic sampler cannot have infinitesimally short samples, so the delta function should be replaced with a window function, or in our case, with the image of the aperture as it appears on the input---what we referred to earlier as the entrance pupil, $P_e(\tau)$. Each delta function in the grid of samples, can be therefore replaced with the entrance pupil, which yields
\begin{multline}
s(\tau) = P_e(\tau) \ast  \sum_{n=-\infty}^{\infty} \delta(\tau-nT_s) = \int_{-\infty}^{\infty} P_e(\tau') \sum_{n=-\infty}^{\infty} \delta(\tau -nT_s -\tau')d\tau'\\
 =\sum_{n=-\infty}^{\infty} \int_{-\infty}^{\infty}  P_e(\tau')  \delta(\tau -nT_s -\tau')d\tau'= \sum_{n=-\infty}^{\infty} P_e(\tau-nT_s) 
\end{multline}
And accordingly,
\begin{equation}
	a_s(\tau) = a(\tau) \sum_{n=-\infty}^{\infty} P_e(\tau - nT_s) 
\end{equation}
This turns the continuous object into a periodic sequence of pulses in a shape that is constrained by the duration of the entrance pupil (see \cref{TempAperture}). As is familiar from temporal windowing in signal processing, the shape of the entrance pupil interacts with the output spectrum of the input as well. 

When the sample has a finite duration, but it faithfully tracks the signal amplitude while it is on, it is referred to as \term{natural sampling} (\citealp[pp. 133--137]{Couch}; see A and C in Figure \ref{samptypes}). Once again, the original signal can be reconstructed using the same principles as impulse sampling, but the spectrum depends also on the duty cycle of the sampler---the relative signal duration that is being sampled---in addition to the original spectrum. In our case, we successfully used a Gaussian pupil function, which suggests that the sample is natural, in the sense that it is possibly weighted in amplitude before it is being coded. However, as spikes do not convey amplitude directly (only through their aggregate spiking rate), the concept of weighting may not be necessarily valid on the individual-sample basis. 

In two static spatial dimensions, natural sampling is really a complete image, where the sample duration is analogous to the aperture stop. The interesting thing about it is that a natural sample contains much information in itself, unlike the impulse sample that only encodes the amplitude of the signal at one time point. In monaural human hearing, standalone signals that are less than, say, 2--3 ms, do not appear to convey any useful information, so they may not count as anything more than a single data point. This is not the case in echolocating bats (Eptesicus fuscus), though, which were shown to be able to extract information from single reflected pulses that are about 0.4 ms long (the maximum resolution of single pulses), possibly by comparing them to stored signal data. The information contained in that pulse was shown to be discernible by bats with variations of tens of nanoseconds or longer (!) \citep{Simmons1990, Simmons1996}. While this information may be extracted in different ways \citep{Sanderson2003}, it may attest to that, at least for some mammals, a single pulse might not be just a single data point. In contrast, in humans, all major percepts (e.g., pitch and loudness) are integrated over tens of milliseconds, or longer. A single impulse has no distinct pitch \citep{Doughty1947}. At 1 ms or less, only binaural cues are used to determine localization through interaural-time difference, which does not require the detailed information of the sample, but only its onset data or relative phase. Nevertheless, the curvature data extracted in \cref{CurvatureModeling} was mathematically based on a single sample---a ``snapshot'' of the periodic chirping masker as it repeatedly crossed the narrowband pure-tone target. 

There is another way to look at natural sampling that is rather counterintuitive, if one is primarily used to think of spectral resolution as the main limitation that underlies the auditory system's performance. When high frequency-sensitivity is a critical goal, then the uncertainty principle dictates that the temporal window should be as long as possible to achieve high spectral precision and allow for the output of the slow high-Q bandpass filter to build up. According to this perspective, the identity of the tuned filter discloses the signal frequency, as its only measurable output is signal power at that frequency. However, the temporal aperture itself---which is determined by the bandpass filter of the channel only if it is also the aperture stop---can be also seen as a modulation-band low-pass filter, whose cutoff frequency increases with the window duration (Eqs. \ref{wcoh} and \ref{winc}). The auditory temporal aperture used for sampling fulfills an opposite target in the case of impulse sampling, where it has a zero modulation bandwidth that completely excludes any fluctuations, whereas an infinitely long sample would contain all the variations of the object modulation band. Here, the longer the natural sample is, the wider is its passband and more modulation information is preserved in the image. Then, if the passband is too broad, the modulation information may be smoothed if it is carried by too many frequencies that are all averaged together \citep[cf.,][]{OxenhamKreft2014}. These reciprocal relations suggest that if natural sampling is relevant in hearing, there may be a trade-off in setting the ideal temporal aperture, modulation bandwidth, spectral bandwidth, and sampling rate of the channel---parameters that are not all independent of one another. See also the discussion in \cref{TwoSpectra} for a complementary view about the information conveyed by the spectral and modulation spectra. 

While the natural sample view is mathematically sound, it is challenging in terms of the underlying neuronal mechanisms that are supposed to sample or communicate such information, as they may have to carry more information than may appear to be borne by a single discharge. As was mentioned in \cref{ImpulseSampling}, it could be the result of firing by a few sampling neurons that together combine into a higher-level sample that contain more information. However, given the example of the hyperacute temporal perception of the bat, we should remain open to the conceptually challenging option that a single spike contains more information than a digital sample does in engineered applications. This point of view is currently not supported by neuroscientific theory.

\subsection{Flat-top sampling}
\label{FlatTop}
\term{Flat-top sampling} is a combination of impulse and natural sampling in that it samples the input signal over a finite duration, in which one level of the signal is recorded (\citealp[pp. 137--141]{Couch}; see A and D in Figure \ref{samptypes}). In electronics, this is typically the onset or peak level, using a \term{sample-and-hold} circuitry, which makes this popular technique simple to implement. The original spectrum can be recovered as in impulse and natural sampling. However, individual samples do not contain extra information that can be interpreted as own images. Instead, the optical analogy would be that of a pixel---a point of finite width and constant amplitude that comprises a single datum of a larger image, when viewed from the distance. A pixel, however, is finite in spatial size and is usually some kind of an average of the intensity pattern of light that is distributed on the area of the detector. This model is more attractive in terms of the underlying neuroscience, as it carries less information than in natural sampling. However, there is no evidence at present to support this kind of sampling in hearing.

\section{Auditory coding or sampling?}
Because hearing is very much a time-based sense (see \cref{HearingTheoryVision}), the distinction between coding and sampling matters a great deal. Both arise in the context of information theory \citep{Shannon1948}, where uniform sampling was presented as a necessary step in the conversion between arbitrary continuous signals of finite bandwidth to discrete sequences that can be directly quantified in bits. Early mentions of neural coding predated Shannon's information theory and are found in the reports of the pioneering experiments by Edgar Adrian in the late 1920s \citep{Garson2015}. With the introduction of information theory, coding received a theoretical boost that was embraced in neuroscience. Shannon introduced coding as an additional layer to communication in which raw information is transformed to combat noise, minimize errors, and compress message length (\cref{InfoNutshell}). However, Shannon himself warned against the indiscriminate employment of information theory outside of communication engineering \citep{Shannon1956}. Indeed, even to this day, the usage of the coding concept in neuroscience, despite its ubiquity, is controversial \citep[e.g.,][]{Brette2019}. 

\citet[p. 232]{Perkel1968} proposed a working definition of neural coding in their seminal report: ``\textit{The representation and transformation of information in the nervous system.}'' This general and somewhat vague definition was intentional, given a few competing meanings of the term ``code'', and the fact that the kinds of information that are carried by the neural system are too diverse to be confined to just one narrow definition. Nevertheless, the observation that at the locations of sensory transduction the physical signal is mathematically sampled seems either to be taken for granted, or to be encapsulated in other features of the neural code (e.g., the ``transformation'' of the ``referent'' of a physical signal---a measurable property such as intensity or brightness; \citealp[p. 233]{Perkel1968}). 

Samples of analog bandlimited signals are also expressed in bits---bits of digital data, though, that constitute the required information for signal reconstruction. This implies that sampling is a lower-level or ``dumber'' operation than coding (in the information theoretic sense), which does not involve direct decision making in selecting which elements of the signal to process or suppress, or how to deal with signal redundancies, to give just two examples. The sampler itself should be ``hard-coded'' to deliver a certain fidelity in terms of quantization noise, dynamic range, etc. At the level of the sample---a sequence of bits on a clocked grid---the additional level of coding, as in Shannon's theory, either does not exist or is superfluous. 

It may be argued that sampling and coding are intertwined, because a sequence of samples forms a primitive code in its own right, as it is based on a rule-based transformation of physical signals. However, there are several important differences between the coded and sampled sequences. The goal of sampling is to represent a physical signal to the degree that it can be potentially reconstructed---up to a certain bandwidth---at an arbitrary degree of precision. Sampling errors have the potential to cause reconstruction errors, which are usually detectable as various forms of distortion and noise at the output. In contrast, coding does not necessarily entail reconstruction. Coding errors can manifest in different ways, depending on the application of the decoder, which has to process the received message using a program (that is also referred to as a code, confusingly). For example, coding errors can cause a misidentification of a symbol, result in inefficient processing speed that may cause a processing bottleneck, mislearn patterns and misestimate averages (e.g., pitch), or give rise to false predictions or illusory percepts, which may even lead to misguided decision making. These are all high-level effects on time scales larger than the individual sample. A sophisticated coding scheme may be designed to be robust despite random errors (e.g., error correction through redundancy), which is a primary strength of digital computation that is very difficult (maybe even downright impossible) to attain with analog computation \citep{Landauer1996}. 

In the case of the ear, the two neural operations---sampling and coding---may not be amenable to a clear-cut distinction, due to the evident complexity of the hard-coded apparatus of the auditory nerve. Also, this operation does not generate obvious symbols (\cref{InfoNutshell}). More fundamentally, both sampling and coding share a common receiver as the message destination---the conscious listener. 

\section{``It from bit\footnote{\citet{Wheeler1990}.}''}
Three different lines of evidence are presented below that directly link (or strongly correlate) the perception of distinct sound events and the firing of a single neural spike, or the lack thereof. This is intended to bolster the low-level auditory sampling operation perspective, rather than the traditional one of auditory coding.

The first line of evidence is based on a series of physiological and behavioral gap detection experiments conducted on the European starling \citep{Klump1989,Buchfellner,Klump1991b}. In these studies, a gap in continuous broadband noise could be observed in the bird's auditory nerve \citep{Klump1991b}, forebrain \citep{Buchfellner}, and behavioral \citep{Klump1989} thresholds. A gap in the stimuli was observable in the spike train of single auditory nerve fibers. The authors referred to coding of the gap in two different ways---a decrement in the spike response as seen in the peristimulus time histogram for gaps of 12.8 ms or less, or through the ensuing neural onset excitations for gaps of 25.6 ms and longer \citep{Klump1991b}. The median minimum gap detected in this method was 1.6 ms, whereas it was 1.8 ms in the behavioral test \citep{Klump1989}. Interestingly, a subset of forebrain neurons were sensitive to even smaller gaps---as small as 0.4 ms. Despite the (small) threshold differences, these results strongly suggest that the information about a gap that is coded in the auditory nerve propagates to the central brain and thus enables appropriate decision making. Given that the neural firing is stochastic, then if the gap is registered in the auditory nerve---effectively sampled with adequate resolution---then it can be further processed by the system downstream---encoded, recoded, or simply, coded.

Another line of evidence comes from results that are presented in \cref{Aliasing}, which may be interpreted in a similar way to that from the starling: information that is not yet encoded in more elaborate patterns can elicit responses that are well explained using a sampling model. The main premise of this series of experiments is that instantaneous undersampling may cause momentary aliasing, as listeners can confuse the number of pulses in very fast sequences of one, two, or three pulses. Using this aliasing model, it was possible to derive bounds for the instantaneous sampling frequency that could underlie the responses. This interpretation enables direct comparison with typical spiking patterns that are known from animal studies. For example, onset and steady-state responses are commonly distinguished through their increased spiking rate that decays very quickly---\term{neural adaptation} \citep{AdrianZott,Galambos1943,Kiang1965}. \citet{Westerman1984} obtained high-resolution spike data of Mongolian gerbils that exhibited very rapid adaptation to tone burst within a few milliseconds, which has a time constant in the order of 2 ms for inputs at 40 dB SPL, and somewhat longer for lower inputs. The corresponding rate during this onset was 358--927 spikes per second, with a mean of 642 spikes per second. A further short-term adaptation period was observed before the units dropped to the steady-state rate, which had a corresponding range of 20--89 ms time constant (mean 57 ms), and rates of 35--261 spikes per second (mean 122 spikes per second)\footnote{Higher onset discharge rates were recorded in the auditory nerve of the chinchilla of about 2700 spikes per second (!) \citep[Figure 2.6]{Ruggero1992Popper}.}. These mean rates are comparable to the instantaneous sampling frequencies that were estimated in Appendix \cref{Aliasing}, which indicates that they may represent the short duration in which adaptation was minimal. Confusion threshold range was 313--660 Hz using a non-adaptive test method (Experiment 1), and 191--352 Hz using an adaptive test method (Experiment 2). When a rapid pulse was hidden within a longer slow pulse train that is likely long enough to evoke a steady-state response, the effective sampling frequency range dropped to 41--77 Hz, which corresponds well to the slower time constant range after onset from the gerbil\footnote{Note that in an attempt to model psychoacoustical forward masking patterns using either neural adaptation or temporal integration models, the former was disfavored and produced somehow less good fit than the latter \citep{Oxenham2001}. In that analysis Oxenham suggested that neural adaptation effects may be altogether ignored at the level of the periphery. For a higher-level review of auditory adaptation, see \citet{Willmore2023}.}. Finally, while not statistically significant, the estimated mean sampling rate for 20 dB louder clicks was 132 Hz higher in \cref{Aliasing}, which is to be expected from higher discharge rates observed for stimuli of high intensity \citep{Galambos1943,Kiang1965}. These figures suggest a circumstantial link between the behavioral results and physiological measurements, which can be interpreted as stemming from different sampling effects of the acoustic stimulus. 

\begin{figure} 
		\centering
		\includegraphics[width=0.6\linewidth]{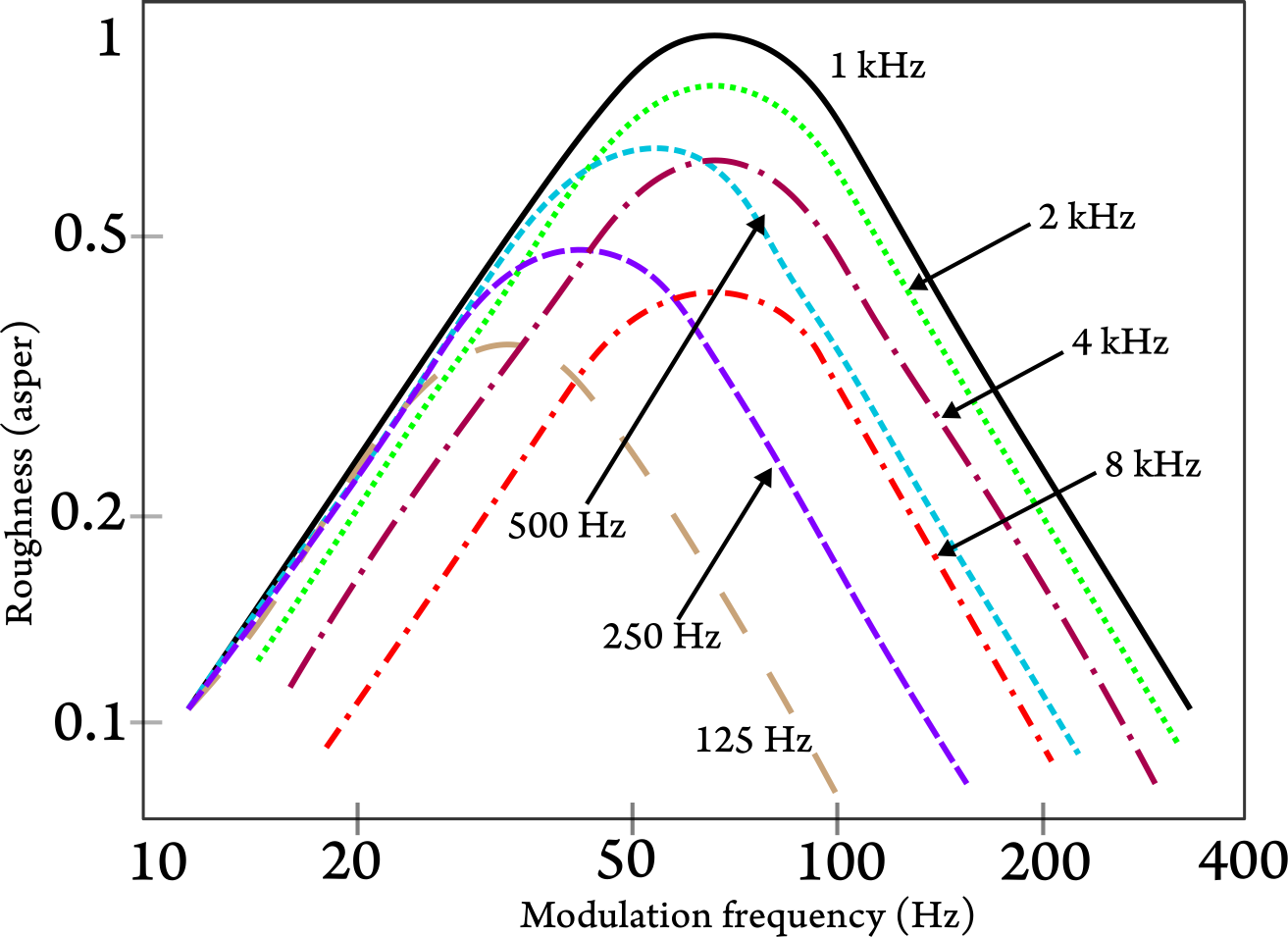}
		\caption{Psychoacoustic sensitivity to roughness at different carrier frequencies, as a function of the amplitude modulation frequency of the tones. The modulation depth is 1 in all cases. The figure was redrawn after \citet[Figure 11.2, p. 259]{Fastl}.}
		\label{RoughFastl}
\end{figure}

A related psychoacoustic measure that can be neatly fitted into the sampling model is roughness perception, which is a sensation that is heard with amplitude- or frequency-modulated sounds at frequencies of 15--300 Hz \citep[pp. 257--264]{Fastl}. It is a distinct sensation in comparison with the more general TMTF, which relies on discrimination thresholds that can be a result of temporal resolution, spectral resolution, or even intensity cues at very slow rates. Roughness taps to the temporal nature of the signal, which completely disappears only if the signal is fully resolved (without any audible or residual input to the auditory filter that corresponds to the carrier). However, if sampling is taken into account, then there must be a modulation frequency threshold above which discrete fluctuations in level cannot be distinguished anymore, due to undersampling. This is exactly the case, as can be seen in responses to high carrier and modulation frequencies. For example, at 8 kHz, where the equivalent rectangular bandwidth (ERB) according to Eq. \ref{ERB} \citep{Glasberg1990} is 888 Hz, one would expect spectral resolution to fully take over at modulation rates above 444 Hz, approximately, or even higher (see \cref{TempAperture}). Instead, as Figure \ref{RoughFastl} that reproduces \citet[Figure 11.2, p. 259]{Fastl} shows, roughness sensation almost disappears at a modulation rate of about 250 Hz for an 8 kHz carrier\footnote{The 8 kHz frequency band can be made to work with the sharper filters reported in \citet{OxenhamShera2003} (see \cref{OHCtimelens}), which produce about half the bandwidth at this frequency (212 Hz), but predicts better spectral resolution at 250 Hz and less roughness. Figure \ref{RoughFastl} shows no roughness at 250 Hz. However, this explanation does not work for lower carrier frequencies, which have a smaller absolute bandwidth, but high maximum modulation frequency for roughness sensation. For example, at 2 kHz carrier, the same calculation produces a limit of 75 Hz, while the data show that some roughness exists at least up to 300 Hz.}. \citet{Fastl} noted that the maximum roughness at low frequency bands is limited by the frequency selectivity, whereas at high frequencies it is limited by the temporal resolution of the system. These results may tap to the sensation of auditory flicker \citep[pp. 408--416]{Wever1949}, whose visual counterpart is the perception of flicker from a moving image. In cinema, 24 frames per second (each projected twice, so effectively 48 frames per second) have been the golden rate for decades. With the advent of computer monitors, refresh rates had to be raised to 70--120 Hz to eliminate the annoying flicker perception they would otherwise have. 

A related and informal example of the effect of sampling on the TMTF can be seen in the incoherent carrier data from \citet{Viemeister1979}, who measured thresholds of modulation frequencies of up to 4000 Hz and showed that they are still audible at a -5 dB modulation depth (see Figure \ref{TMTFsBB}). However, listening to such fast modulation, it is clear that the modulation is no longer perceived as temporal at this modulation frequency, but they rather give a timbral change to the carrier, without being resolved. Thus, roughness does not apply here directly, but a similar effect happens at very high frequency, in which the temporal fluctuations are no longer audible above a certain frequency. 

In summary, the empirical data presented can be used to draw a causal link between neural spikes and perceptible events. The simple consideration that has been used in all cases is whether the instantaneous sampling rate is fast enough to capture fast events in the temporal domain---a short gap, a fast pulse train, or very high modulation rates. Wheeler's ``\textit{it from bit}'' was coined to emphasize a fundamental relation between measurable physical quantities and their informational manifestation. We borrow this expression to also include perceptual events that are carried by neural spikes. 

These arguments do not preclude the existence or validity of coding of the auditory events that were described. But in the above examples, coding does not provide a parsimonious description of the measurements, because the type of errors reported (e.g., failure of detection, or drop in sensitivity) are straightforward and do not seem to merit any high-level transformations. However, hardwired aspects of coding that define the sampling patterns, such as the adaptation patterns to stimulus onset, may be considered a low-level amalgamation of sampling and coding that may be conceptually useful in all these cases. 

\section{Nonuniform undersampling}
\label{nonuniformunder}
The imaging system has a finite temporal window, which must be ``refreshed'' in order to generate multiple images during a continuous signal, just as moving-image cameras employ shutters to periodically generate discrete images. Electronic samplers employ clocks that trigger the precisely timed samples of the analog input. Neural spikes in themselves act as irregular triggers, due to the underlying stochastic biophysical process. More centrally, the chopper cells in the cochlear nucleus might also serve the purpose of ad-hoc clocks that oscillate regularly for a brief period (\cref{CentralNeuroanatomy}). Sampling may also be the aggregate of all neural firing along the different auditory nuclei on the way downstream.

If the sampling perspective is embraced, then neural spiking represents a nonuniform-rate sampling pattern, which is conceptually different from the familiar fixed-rate sampling that was presented in \cref{SamplingBasics}. It is not the intention here to review the theory of nonuniform sampling in any depth (for example, see \citealp{Marvasti}), but a few key points will be highlighted, which may be intuitively understood from the vantage point of the more familiar uniform sampling.

In nonuniform sampling we generally adhere to a temporal approach that dynamically compares the instantaneous sampling rate to the Nyquist rate that is dictated by the signal spectral content. If the instantaneous frequency of the signal is above the Nyquist rate, then the signal is going to be instantaneously undersampled. The other extreme---instantaneous oversampling---is also interesting to analyze using information load considerations, rather than signal processing per se. 

As real-world acoustic sources are not necessarily bandlimited, it raises a concern for aliasing that may be caused by undersampling. This is relevant here only in the context of the signal modulation spectrum bandwidth in a given auditory channel, when the modulation is not resolved in adjacent channels. High modulation frequency components that are contained in the passband have to be sampled at an appropriate rate, to avoid aliasing, unless they can be removed with an anti-aliasing filter. However, there is an interesting twist to this rule. As was first proven by \citet{Shapiro}, randomizing the timing of the samples can prevent aliasing from happening in undersampling systems. Therefore, if the nonuniformity is truly random (jittering the sampling around the period does not work), then \textbf{aliasing may be traded off with noise}. \citet{Shapiro} also showed that if the samples are generated according to a Poisson distribution, then the reconstructed signal will be essentially alias-free\footnote{A modern signal-processing approach to undersampling that is called \term{compressed sensing} is based on different mathematical principles than those of \citet{Shapiro}. Compressed sensing has been successfully employed in signal processing applications to beat the aliasing limit imposed by the Nyquist rate \citep{Candes}. It makes use of a known but general functional basis (e.g., wavelets) that is taken as background information, which reduces the need for many samples to be generated with regular clocking. Neural correlates of this techniques have been theorized \citep{Ganguli}, but none were studied in the auditory system to date, although the earlier model by \citet{Benedetto1993} and \citet{Benedetto2001} is also based on nonuniform sampling with a wavelet basis and might be conceptually related. Compressed sensing has found much use in image compression \citep{Patel2013}, for example, but will not be dealt with here, as it is not clear that it reflects the physiological auditory signal processing.}. 

Another aspect of nonuniform sampling is related to the interplay between stationary (or periodic) and nonstationary elements within continuous acoustic signals. Much of the information about the sounding objects is contained in the transient part of the signal that arrives first, before the steady-state part ensues \citep[e.g.,][]{Kluender2003}. Specifically, transient sounds can be rich in harmonics both in the carrier and in the modulation domains---depending on how they are analyzed. Therefore, it makes for an economical system design to invest more resources in sampling the transient sounds more finely than the steady-state parts of the signal that are less informative and can be sampled more slowly. This is an intuitive justification for neural adaptation in hearing, which potentially budgets the available spikes where they are most needed.

These two nonuniform sampling aspects---random sampling and denser sampling of information-rich regions of the signal---were shown to work in concert in the rhesus monkey's retina, which can be modeled as a two-dimensional spatial sampling grid \citep{Yellott1983}. Photoreceptor (cones) arrays on the retina are distributed in decreasing density from the highst density in the fovea, then the parafovea, the periphery, and far periphery, where the cones are sparse. One may expect that the visual image would be plagued by aliasing, if high spatial frequencies that are well-sampled by the fovea at above the Nyquist rate, are imaged by the low-density sampling periphery. However, this does not happen because the retinal sampling grid is quasi-random (not exactly Poisson-like, because of distance constraints between cones), and higher frequencies than the Nyquist rate are converted to noise in the periphery. An insightful primer about the relationship between nonuniform sampling and aliasing and how it plays out in vision was given by \citet[pp. 245--263]{Resnikoff}. 

The adaptive nature of auditory processing---its nonuniform sampling---suggests that aliasing, if observable, evokes only a fleeting sensation, as it is either replaced with internal noise, or resolved to adjacent frequency bands. In the auditory system, aliasing was conjectured to take place only during very short sequences that were tested in \cref{Aliasing}. Physiological evidence is more scant and has not invoked the terms aliasing or undersampling (with the exception of the spectral sampling model of \citealp{deCheveigne1999}). A clear evidence for aliasing in auditory-nerve fiber synchronized patterns is found in \citet[Figures 9 and 10]{Sinex1981}, where instantaneous frequency was estimated by the interstimulus intervals of the spiking pattern. The fiber tracked the frequency well, but when the frequency was larger than could be sampled by the fiber, the instantaneously tracked frequencies were subharmonics of the actual instantaneous frequency---$f/2$, $f/3$,... $f/6$. In other examples, occasional neural recordings of temporal modulation transfer functions seem to suggest that aliasing can be invoked artificially. See for example, Figures 2D, 5D and 5F in \citet{Kale2012}, which show inexplicable above-cutoff-frequency responses that look like aliasing. Typically, these tuned units are taken to exhibit a band-stop (or band-reject) characteristic response \citep{Krishna2000,Nelson2007,Carney2014}, which is nevertheless difficult to distinguish from an aliased response as is seen in Figure \ref{aliasexample}. This ambiguity is perhaps most conspicuous in \citet{Krishna2000}, where inferior colliculus (IC) units were (over-)modulated at frequencies close to the characteristic frequency and were considered to exhibit suppressive regions caused by inhibition in the reject band, but with no clear role in sound processing (e.g., Figures 2, 4, 6, and 16).

Nonuniform sampling is presented here as an underlying reality of neural spiking, which is in large part due to its stochastic nature. However, this is not an entirely accurate representation, since the system is able to phase-lock to the carrier and synchronize to the envelope. We modeled phase locking earlier as a result of a dedicated phase-locked loop (PLL) module in the cochlea (\cref{PLLChapter}). One of the most important usages of PLLs in communication circuits is synchronization of the internal signal to the external one, which was produced remotely using a local oscillator, which works as a clock. The PLL therefore obviates the need for a local clock at the receiver, as long as ad-hoc samples can be generated that are synchronized to the signal\footnote{Clocks have appeared rather rarely in auditory models, although usually only with short-term regularity and without a clear physiological correlate \citep[e.g.,][]{Whitfield1970,Suga1973,Yang2018}.}. If the PLL is out of lock, then the randomness of its samples is expected to increase. 

\section{The effect of sampling on the modulation transfer\\function}
\label{MTFandSampling}

Having clarified some of the main features of the sampling auditory system, we would like to account for the discrepancy in the bandwidth of the predicted and the empirical TMTFs we analyzed in \cref{TheTMTF}. From the analysis of the various auditory TMTFs, it appears that the cutoff frequency predictions made using the ideal ATF (Figure \ref{OTAcutoff}) grossly overestimate the behavioral coherent response values compared to findings from literature (Figure \ref{TMTFsTone}). The same is true for the incoherent TMTF, although the general property of a smaller incoherent than coherent TMTF bandwidth was predicted correctly. How can this discrepancy be explained, given that the very same parameters have been used fairly successfully to predict the behavioral curvature and temporal aperture data in \cref{CurvatureModeling}?

One possible explanation for the MTF and TMTF discrepancy may be related to the fact that some auditory brainstem units are better described by bandpass  rather than by low-pass modulation response. For example, in the guinea-pig's ventral cochlear nucleus (VCN), several tuned unit types (primary, sustained choppers, and onset chopper cells) have characteristic responses to sinusoidal amplitude modulation that are less low-pass and more bandpass in their response, as they display a clear peak at their best modulation frequency \citep[e.g.,][]{Sayles2013}. At full modulation depth most units appear low-pass, but become bandpass with lower modulation depth values---mainly the chopper units that are most sensitive to AM. However, the high-frequency response of these bandpass units overlaps that of the low-pass units, so even if the characteristic filtering of these cells is relatively dominant in the VCN overall response (and eventual perception), it still does not explain the complete lack of sensitivity to high frequencies in behavioral tests. Rather, it appears to be shaping the low-frequency irregularities that are sometimes observed in empirical data (Figure \ref{TMTFsAll}), which were also discussed in the context of narrowband responses (\cref{NarrowTMTF}).

An alternative explanation for the discrepancy in the MTF prediction is due to sampling limitations. There are at least three known causes for MTF degradations and bandwidth reduction in optics that are caused by finite spatial sampling of the image \citep[pp. 35--50]{Wittenstein, Park, deLuca, Boreman}. First, the assumption of time-invariance\footnote{Shift-invariance in spatial imaging.} is generally not true for a sampler that has a finite sampling frequency $f_s$. This means that small delays of the object---of the order of one sampling period---can heighten the sensitivity to slightly different features that are captured by its finite-sized samples. This is compounded by the second cause, which is that real-world objects are generally not bandlimited, so that arbitrary inputs that contain modulation frequencies that are higher than its instantaneous Nyquist rate $f_m>f_s/2$ can be aliased.  The third cause is the finite footprint of the optical detector, which in our case is equivalent to sampling that is not impulse-like (\cref{temporalmodels}). In optical detectors, the basic unit of detection is at minimum one pixel, which assumes a constant intensity that effectively produces flat-top sampling (\cref{FlatTop}). While we do not expect it to be exactly the case in hearing, we know that the finite size and response of each sample has an effect on the image. In spatial optics, the compounded effect of sampling on MTF is captured by the \term{sampling MTF} following the detector, which is located at a distance behind the lens. In the hearing system, the first sampling location is right after the lens, before the neural group-delay dispersion. We will nevertheless consider the effect to be qualitatively the same as in spatial imaging. 

The main difference between the psychoacoustic phase curvature (\cref{CurvatureModeling}) and the MTF predictions is that the former was based on the time-domain imaging transform and the latter on frequency-domain transfer functions. The time-domain model we developed for the phase curvature data relied on a single image of the stimulus---a snapshot of a pulse---which contains all the information necessary for solving that problem. Using frequency-domain methods for long stimuli, though, we are required to integrate over many pulses that are sampled across fibers over a long time. For this to work, sampling has to be perfect, so that the sampling MTF can be exactly equal to the MTF, over several synapses---each of which resamples the previously sampled MTF. As perfect sampling is unlikely to take place, we can expect the signal to be gradually undersampled by the time it reaches the auditory retina at the IC. 

The role of undersampling in the received TMTF is supported by animal data---as long as it is reinterpreted according to a sampling framework. On the whole, TMTFs that are measured along the auditory pathways, between the auditory nerve and the midrain and auditory cortex, gradually become less sensitive to high modulation frequencies and rate coding becomes more prevalent beyond the IC (for a comprehensive summary, see \citealp{Joris2004}, Figure 9). For example, despite broadened cochlear filters due to noise-induced hearing loss, broadening had no effect on the modulation bandwidth in the auditory nerve of chinchillas. It was suggested that the TMTF is constrained by non-cochlear factors, such as neural adaptation and refractoriness (\citealp{Kale2010,Kale2012,Kale2014}; See \cref{GroupDelayDisp} for further discussion). Incidentally, it fits our earlier conclusion that the aperture stop of the system is determined by the neural transduction and not by the cochlear filters. 

In another recent study, data from the Mongolian gerbil, which were based on realistic stimuli from its natural environment, showed that some parts of the stimulus are postsynaptically inhibited between the auditory nerve and the spherical bushy cells in the anteroventral cochlear nucleus (AVCN) \citep{Keine2017}. Sampling considerations can suggest that many samples are lost between the auditory nerve and the AVCN, which effectively results in a substantial drop in the transmitted information and, hence, in a drop in the effective sampling rate. Notably, though, the recorded rate in these cells was comparable to their spontaneous activity rate. Additionally, the reduction in the instantaneous firing rate resulted in a complementary increase in the temporal precision of the samples. Similar results were reported by \citet{Dehmel2010} and \citet{Kuenzel2011}.

The decrease in sampling rates in the gerbil is reminiscent of the starling data we reviewed earlier by \citet{Klump1991b} and \citet{Gleich1995}, who found that the behavioral modulation bandwidth was about half of the physiological one (\cref{EmpiricalBB}). Our prediction was only 7\% higher than the physiological value that was recorded from the auditory nerve of the starling. We can try to supplement these findings by estimating how critical the effect of undersampling is, by comparing the predictions we obtained for the perfectly sampled MTF in humans to the fastest known synchronized modulation rates that were reported in physiological studies. As was reviewed in \cref{TonalTMTFs}, the highest modulation frequencies that were measured in the auditory nerve of any species are from the cat and are in the range of 1500--2500 Hz for carriers of 10--30 kHz \citep[Figure 13]{Rhode1994Encoding}. Therefore, our predictions still overestimate the frequency cutoff of the MTFs, as it we see up to 3000 Hz modulation cutoff at 10 kHz (Figure \ref{OTAcutoff})\footnote{Unfortunately, the dispersion parameters above 10 kHz are extrapolated and are very unstable and therefore could not be reliably used to estimate higher MTF cutoff frequencies.}. 

Our predictions of human compared to both cat and starling maximum cutoff measurements suggest that in any case they are overestimated. Accounting for undersampling definitely appears to contribute to the discrepancy, especially if it begins early on---as soon as the mechanical signal is transduced. But these results are also confounded by the uncertainty we have about the dispersion parameters, especially at high-frequency carriers, which likely skew the results. Furthermore, the simple modeling offered by the MTF does not take into account the evident physiological limitation for synchronizing to high modulation frequencies, perhaps due to refractoriness. 

A final mechanism for degradation of the coherent TMTF should be mentioned, which is not independent from the undersampling mechanism. Both undersampling and nonuniformity effectively contribute to the decoherence of the coherently detected signal. This is encountered in the auditory nerve, but we also mentioned examples for AVCN units that retain their low modulation frequency precision in spite of undersampling, although these observations did not apply for all cells \citep{Dehmel2010,Kuenzel2011,Keine2017}. This may mean that other subpopulations of cells in AVCN and perhaps in other nuclei may not conserve the coherence of the incoming signal, which depends on the phase-locking precision, on top of the MTF bandwidth. It is therefore possible that as the signal becomes partially coherent, its MTF becomes closer to that of the incoherent MTF, which has a more limited bandwidth.

\section{Discussion}
The role of sampling in auditory processing was analyzed with special attention given to the type of sampling window, the effect of nonuniform sampling, and potential aliasing. It was then used to account for the bandwidth narrowing in the measured TMTF compared to the predicted MTF, which was based on perfectly sampled frequency-domain transforms. We identified the auditory nerve as the primary site of sampling, but noted that repeated resamping throughout the brain is a likely source of further degradation in the perceived MTF. 

The sample, which is the sound image of a pulse, may represent the minimal unit of auditory perception that can be experimentally observed both physiologically and psychoacoustically. We produced some evidence to attest to that the image might contain more information than an optical pixel, which is defined by its fixed place and intensity reading. Further work will have to be done to find out what the most appropriate relationship is between the image and the sample, or how much information is contained in a single spike. Also, given that information is pooled over populations of fibers in every channel and usually over several channels, then the perceptual significance of a single image-sample as the atom of hearing (similar to Gabor's logon, \citealp{Gabor}) has to be clarified. 

Neural sampling of continuous signals poses a few conceptual challenges that have been either neglected or treated only in passing in various hearing models. We would like to flag some of these challenges in advance, as they can guide the discussion that follows. These challenges are not going to be necessarily resolved in this work, though, as our main goal is to see how sampling implicate temporal imaging. 

Embracing the notion of neural sampling comes close to suggesting that the auditory brain operates discretely and not continuously. This is a highly contentious topic in neuroscience that has been avoided in auditory research almost completely (but see \citealp{VanRullen2014}). Although the experience of our auditory perception (and other modalities too) feels continuous and not discrete, there are more than a handful of results that indicate that it is a finely sampled sequence of discrete images, which only appear continuous. If the IC is assigned the role of the auditory retina (\cref{NeuralDisp}), then beyond it the discrete/continuous representation problem is reduced to that of visual processing downstream from the retina, as long as thalamic and cortical processing is directly comparable between the two modalities (see \cref{AnaPhysioComp}). Additional discussion is found in \cref{GenDisAlias}.

A major requirement in sampling, which was implicit throughout the chapter, is that it should lead to reconstruction of the signal at the output. However, we have no access to the notion of signal reconstruction in perception, so we cannot estimate what the relevance of this mathematical concept is to the brain. Nevertheless, it has been insightful to discuss sampling as a goal-oriented process, which can be evaluated by comparing it to an ideal process from which the signal can be faithfully recovered. 

We discussed the implications of nonuniform sampling with emphasis on undersampled inputs that cause aliasing, but we did not touch the concept of oversampling. In theory, oversampling provides a fail-proof solution to distortion from aliasing, so it may appear like a good strategy. However, it is also wasteful in that it produces larger quantities of data than may be needed and can load the system informationally, cognitively, and also in terms of the incurred energy costs. Therefore, it can be argued that unnecessary information is readily discarded in hearing, using only a minimal number of samples that are needed to just represent the stimulus correctly and maintain a sense of continuity \citep{Weisser2019}. See also \cref{AliasingExp3}.

\chapter{Auditory image fundamentals}
\label{AudImageFun}
\section{Introduction}
In the previous chapters, we established a theoretical and quantitative analytical basis for temporal imaging in the auditory system, along with a firm basis for an auditory-relevant notion of coherence. The original temporal imaging theory applies directly to individual samples or pulses, but we qualitatively extended it to include nonuniform sampling, in order to make better use of the modulation transfer functions that can be applied over longer durations than single samples. It highlighted the distinction between coherent and incoherent sound, which can be traced to the inherent defocus of the auditory system (and its preferential phase locking), and is supported by empirical data about the temporal modulation transfer functions. This distinction indicates that there is an increased sensitivity to high frequencies that modulate coherent carriers in comparison to incoherent carriers. 

There appears to be a built-in system that may be more optimal for coherent than for incoherent sound detection, insofar as the within-channel imaging is analyzed. This is so because of the relative broader bandwidth of the coherent modulation transfer function (MTF), which is completely flat in the effective passband---at modulation frequencies that do not get resolved in adjacent filters. In contrast, incoherent carriers carry random modulation noise at low frequencies and tend to have poorer sensitivity, unless information from several channels is pooled together. These observations are not trivially integrated when it comes to the design of the complete auditory imaging system. Analogy to spatial imaging in vision is also not helpful, as vision is strictly incoherent. In fact, incoherent-illumination imaging normally produces superior images to coherent illumination, which produces images that are much more sensitive to diffraction effects and speckle noise (e.g., dust particles on the objects or on the system elements that become visible in coherent illumination). Additionally, incoherent imaging is strictly intensity-based (i.e., linear in intensity), whereas coherent imaging is amplitude-based (linear in amplitude), although its final product is usually an intensity image as well. Direct comparison between visual and auditory imaging can only be made based on intensity images, but because of the longstanding confusion about the role of phase in hearing (\cref{AuditoryPhaseReview}), it is not immediately clear that amplitude imaging does not have a role in hearing and that the comparison is valid. 

Many natural sound sources are coherent, but the acoustic environment and medium tend to decohere their radiated sound through reflections, dispersion, and absorption (\cref{acoustenv}). Other sounds of interest are generated incoherently at the source, but their received coherence depends on the bandwidth of the sound and the filters that analyze it. Therefore, we would like to formulate the action of auditory imaging, so that it can differentially respond to arbitrary levels of signal coherence. The effects of complete coherence or incoherence are well known (mainly for stationary signals), but much of the intuition in them may be lost because they are usually not framed as part of an auditory-relevant coherence theory. This means that for the majority of signals that are neither coherent nor incoherent there are no presently available heuristics that can be used to analyze them. We would like to provide some conceptual tools that bridge this gap in intuitive understanding of realistic acoustic object imaging. 

This chapter follows a broad arc that encompasses several key topics in auditory imaging. It begins with discussions about sharp and blurry auditory images, which enables us to make sense out of the substantial defocus that is built into the auditory system. Suprathreshold imaging is then discussed, based on an extrapolation of threshold-level masking responses. The notion of polychromatic images is applied to hearing by way of analogy with a number of known phenomena that are reframed appropriately to support it. The special case of acoustic objects that elicit pitch is briefly reviewed and is also reframed with imaging in mind. Then, several sections deal with various polychromatic and monochromatic aberrations, as well as an interpretation of the depth of focus of the system. Finally, we provide a few rules of thumb that aid the intuition of how images are produced in the system. 

\section{Sharpness and blur in the hearing literature}
Sharpness and blur are central concepts when discussing the optics of the eye \citep[e.g.,][pp. 52--61]{LeGrand1980,PackerShevell} and imaging systems in general. If a visual system does not produce sufficiently sharp images due to blur, it can cause various levels of disability if not corrected. Therefore, identifying the auditory analogs of sharpness and blur---should they exist---may be a powerful stepping stone in understanding how the ear works and where things can go wrong.  

Currently, there is no analogous concept in psychoacoustics for sharpness that resembles the optical one and references to it in the auditory literature are scarce. Sharpness was introduced in psychoacoustics as a consistently large factor of timbre \citep{Bismarck1974}. Ranging on a subjective scale between dull and sharp, sharpness was modeled using the first moment of the sum of the loudness function in all critical bands, where the high frequency content above 16 Bark (3.4 kHz) affected the rated sharpness of the stimuli tested---typically, noise and complex tones \citep[pp. 239--243]{Fastl}. These conceptualizations of psychoacoustic sharpness appear to be irrelevant to the present discussion. 

The antonymous notion of sharpness, \textbf{blur}, is more frequently encountered in the hearing literature, and is closer to how it is used in imaging. It is perhaps because of the association of the convolution with blurring operations that makes this term somewhat more commonplace in hearing research \citep[e.g.,][]{Stockham1975}. For example, blur is occasionally invoked in the context of modulation transfer function fidelity that is impacted by room acoustics \citep{Houtgast1985}, or through manipulation of the speech envelope through modulation filtering \citep{Drullman1994a,Drullman1994b}. Similarly, in bird vocalizations, the in-situ degradation of envelope patterns over time and distance were quantified and referred to as blur \citep{Dabelsteen}. Another typical usage was exemplified by \citet{Simmons1996}, who referred to the blurring effects of the long integration window on the perception of minute features in the echoes perceived by bats over durations shorter than 0.5 ms. Similar references to temporal or spectral blur occasionally appear in the hearing literature. For example, \citet{Carney2018} raised the question of whether there is an auditory analog to the visual accommodation system that reacts and corrects blur to achieve (attentional) focus. 

A single study tested the focus of sound sources directly. A subjective rating of the perceived source focus of anechoic speech superposed either with a specular or with a diffuse reflection was obtained in \citet{Visentin2020}. Subjects were instructed that focus ``\textit{should be considered as the distinction between a ``clear'' or ``well-defined'' sound source and a ``blurred'' sound image.}'' Two clear patterns were observed. First, when the angle of diffuse reflection was increased (from $34^\circ$ to $79^\circ$), the rated focus dropped to the point that it became equal to the specular reflection. Likely, at large angles the reflection was coherent-like and interfered with the source definition. Similarly, the rated focus was also correlated with the interaural time difference (ITD) averaged over 500, 1000, and 2000 Hz octave bands. So the focus was rated highest for ITD when it was about 0. As was argued in \cref{InterauralCoherence}, the ITD directly quantifies spatial coherence. So maximum coherence correlated with the maximum perceived focus, as long as the source direction is unambiguous. Incidentally, the focus highly correlated ($r=0.84$) with speech intelligibility, which was also highly correlated with rated loudness.

\textbf{Clarity} is an altogether different concept, which is probably associated with sharpness to some extent, and is more common in different audio-quality and audiometric evaluations. It was designated as a fundamental component of hearing-aid performance that is hampered by noise and distortion \citep[p. 61]{Katz2015}---two factors that affect the imaging quality independently \citep[pp. 8--9]{Blackledge}. In room acoustics, clarity ($C_{80}$) is often used to estimate the power of the earliest portion of the room impulse response in which single reflections are still relatively prominent, in comparison with the late portion (after 80 ms) \citep[pp. 169--170]{Kuttruff}. This quantity has been used to estimate the sound transparency in concert hall acoustics. Clarity is also used more informally in audio quality evaluations \citep{Bech2006,Toole}, and was defined by \citet{Bech2006} as: ``\textit{Clarity---This attribute describes if the sound sample appears clear or muffled, for example, if the sound source is perceived as covered by something.}''. Muffled sounds often suggest high frequency content roll-off due to absorption---perhaps the opposite quality to Bismarck's sharpness. High modulation frequency roll-off is also a common feature of blur in spatial imaging, as the removal of high spatial frequencies causes the blur of sharp edges.

\section{Sharpness, blur, and aberrations of auditory images}
\label{SharpBlurAud}
In vision, sharpness characterizes static images and can be extended to moving images without much difficulty. In hearing, even static images (with constant temporal envelopes, as pure tones) unfold over time, which is physically, perceptually, and conceptually unlike images of still visual objects, despite the mathematical parallels garnered by the space-time duality. Although we now have the imaging transform of a single pulse or sample, the short duration of the aperture does not truly allow for any appreciation of the image sharpness. Therefore, it is only through the concatenation of samples over time that auditory sharpness can be sensibly established. Still, it is much simpler to analyze the conditions for the loss of sharpness---the creation of blur---than those that give rise to sharpness. If the sources of auditory blur are negligible or altogether absent, relative sharpness can be argued for and established. In other words, we can define auditory sharpness by negation: \textbf{The auditory image is sharp when different sources of image blur are either negligible or imperceptible.} Therefore, the remainder of this chapter is dedicated to elucidating the different forms of blur and related aberrations that can be found in human hearing. 

\subsection{The two limits of optical blur}
In both spatial and temporal imaging, the information about blur is fully contained in the impulse response function (or the point spread function of the two spatial dimensions), which relates a point in the object plane to a region in the image plane. In two dimensions, the effect of blur is to transform a point into a disc. As we saw in \cref{ImpFun}, the point spread function is fully determined by the pupil function of the imaging system and, specifically, it depends on the neural group-delay dispersion (analogous to the distance between the lens and the screen in spatial optics). 

There are two limits that characterize the possible blur in the image. When the aperture is large compared to the light wavelength, the image is susceptible to geometrical blur. It can be explained by considering the different paths that exist between a point of the object to the image, which do not all meet in one mathematical point. Thus, in geometrical blur, multiple non-overlapping copies of the image are overlaid in the image plane. Consequently, the fine details and sharp edges of the object are smeared and the imaged point appears blurry. Geometric blur can be further specified according to the exact transformation that causes an object point to assume a distorted shape on the image plane. These distortions are called aberrations and among them, defocus is the simplest one that causes blur.

In the other extreme, when the aperture size is comparable or smaller than the wavelength, the image becomes susceptible to effects of diffraction. A point then turns into a disc with oscillating bands of light and shadow (fringes), which makes fine details less well-defined, and hence, blurry. 

The aperture size for an ideal imaging system should be designed to produce blur between these two limits. An image that does not have any aberrations is referred to as diffraction-limited. 

\subsection{Contrast and blur}
It is worthwhile to dwell on the two image fidelity characteristics that often appear together---contrast and blur---and elucidate their differences. Contrast quantifies the differences in intensity between the brightest and darkest points in the image, or a part thereof. Hence, it is a measure of the dynamic range of the image, which ideally maps the dynamic range of the object, so that intensity information is not lost in the imaging process. As the image is made of spatial modulations, contrast is quantified with visibility (Eq. \ref{Visibility}), which we also referred to as modulation depth (\ref{AudSenseEnv}). 

Blur refers to the transformation that the image undergoes that makes it different from a simple scaling transformation of the object. The effects of geometrical and diffractive blurs are not the same here, though. In geometrical blur, the envelope broadens, as spectral components share energy with neighboring components and overall distort the image, while retaining its general shape. In diffractive blur, new spectral components can emerge in the envelope that are not part of the original object, but appear due to wave interaction with small features in the system or object that have similar dimensions to the wavelength of light carrier. In both cases, both the envelope and its spectrum change due to blur. 

While blur and contrast can and often interact, they represent two different dimensions of the imaging system and we will occasionally emphasize one and not the other. Contrast does not interact directly with the spectral content of the image, whereas blur does. Note that when the envelope spectrum is imaged, it is scaled with magnification, which is a transformation that is integral to the imaging operation and does not entail blur or contrast effects.

\subsection{Auditory blur and aberrations}
\label{BlurandAbbs}
The temporal auditory image blur can be understood in analogous terms to spatial blur by substituting the aperture size with aperture time and diffraction with dispersion. Thus, the temporal image can be geometrically blurred if the aperture time is much longer than the period of the sound. If it has no aberrations, then it can be considered to be dispersion-limited. A very short aperture with respect to the period may produce audible dispersion effects, at least when fine sound details are considered. Thus, a good temporal imaging design should strike a balance between geometrical blur and blur from dispersion. In fact, what we saw earlier is that the system is set far from this optimum, since it is significantly skewed toward a geometrical blur, which is seen in the significant defocus we obtained---an aberration (\cref{ImagingDefocusFound}, \cref{TempAperture}, and \cref{ImagingNotSatisfied}).

However, things get more complicated with sound in ways that do not have analogs in vision. The most significant difference, as was repeatedly implied throughout the text, is that unlike visual imaging that is completely incoherent, sound is partially coherent, but some of the most important sounds to humans have strong coherent components in them. As was shown in \cref{TheTMTF} and will be discussed in \cref{AudDefocus}, the defocus blurs incoherent objects more than it does coherent ones. Thus, it can be used to differentiate types of coherence by design, instead of achieving nominally uniform blur across arbitrary degrees of input coherence. 

The second complication in sound is due to the nonuniform temporal sampling by the auditory nerve that replaces the fixed (yet still nonuniform) spatial sampling in the retina. The loss of high modulation frequency information from the image because of insufficient (slow) sampling rate (and possibly other factors) can be a source of blur as well, which is neither dispersive nor geometrical per se. This effect will not be considered here beyond the earlier discussion about its effects on the modulation transfer function in \cref{MTFandSampling}.

Another nonstandard form of geometrical blur may occur if the sampling rate and the aperture time are mismatched, so that consecutive samples overlap (i.e., the duty cycle is larger than 100\%). Effectively, this kind of blur is produced outside of the auditory system in the environment through reverberation, where multiple reflections of the object are superimposed in an irregular manner that decoheres the signal (\cref{RoomAc}).   

A combined form of geometrical and dispersive blur is caused by chromatic aberration---when the monochromatic images from the different channels are not exactly overlapping or synchronized, which gives rise to the blurring of onsets and other details. Virtually all polychromatic (broadband) optical systems have some degree of chromatic aberration and the possibility of encountering it in hearing will be explored in \cref{ChromaticA}.

Image blur may be caused by other aberrations, as a result of the time lens imperfect quadratic curvature, when its phase function contains higher-order terms in its Taylor expansion (see Eq. \ref{eq:phase2}). Similarly, phase distortion may be a problem in the group dispersive segments of the cochlea or the neural pathways, if their Taylor expansion around the carrier (Eq. \ref{beta_taylor}) has higher terms \citep{Bennett2001}. While this large family of aberrations is very well-studied in optics and vision, the existence of their temporal aberrations in hearing is difficult to identify at the present state of knowledge. They will be explored in \cref{HigherOrderAb}.

In general, blurring effects can be also produced externally to the temporal imaging system. Outside of the system, it can accrue over large propagation distances, and likely occurs in turbulent atmospheric conditions (\cref{airtravel}). Phase distortion can also be the result of individual reflections from surfaces (\cref{Reflections}). As was reviewed in \cref{outerear}, the plane-wave approximation gradually breaks down in the ear canal above 4 kHz, which means that modulation information may be carried by different modes with different group velocities. This situation causes a so-called dispersion distortion in optics (\cref{SummaryofAssumptions}), and may theoretically cause distortion and blur also in hearing at high frequencies. 

\section{Suprathreshold masking, contrast, and blur}
\label{SupraMasking}
Few auditory phenomena have received more attention in psychoacoustic research than masking. Beyond the curious nature of its effects, interest has stemmed from the usefulness of masking in indirectly estimating many hearing parameters and thereby inferring various aspects regarding the auditory system signal processing. Additionally, it has been implied that masking effects can be generalized to everyday hearing and can significantly impact its outcomes---especially with hearing impairments. 

The definition of masking normally refers to an increase in the threshold of a stimulus in the presence of another sound \citep{Oxenham2014}. However, the change in threshold can be caused by more than one process and several peripheral and neural mechanisms have been considered in literature in different contexts \citep{Oxenham2001,Moore2013}. But since the discussion of masking is strictly framed around the change of threshold, it leaves out a no-less important discussion about how audible, or suprathreshold, signals sound in the presence of masking. Put differently, a complete knowledge of the masking threshold does not necessarily mean that the suprathreshold signal combined with the masker is going to sound identical to a lower-level version of the original signal in the absence of masking. 

We can think of four general classes of interactions between masker and signal, or even more generally, between any two acoustic objects. 

The first class of masking relates to masking that only causes the signal to sound less intense, while it is otherwise unchanged when it is presented above its masking threshold. This effect can be analogized to the apparent dimming of one object in the presence of another (e.g., viewing a remote star in broad daylight). In a perfectly linear system, amplification of the dim target leads to its perfect recovery with no distortion or loss of information. When the system exhibits (nonlinear) dynamic range compression, perfect recovery of the envelope requires variable amplification (i.e., expansion) and may be impossible to realize in practice. When the sound in question is modulated, this is akin to loss of contrast---the difference between the envelope maximum and minimum. Loss of contrast can also happen if only part of the modulated sound is masked, whereas the rest is above threshold. Or, it can take place if the loudest parts of the sound saturate and do not allow for a linear mapping of intensity. In any case, this type of masking is strictly incoherent, as the signal and masker only interact by virtue of intensity superposition. 

In the second class of masking, the suprathreshold signal interacts with the masker and its fine details change as a result, even if they are still recognizable as the target sound. This can obviously happen only if the two sounds interfere, which is possible when they are mutually coherent or partially coherent within the aperture stop of the relevant channel(s). As this phenomenon involves interference, the corresponding optical analogy here is diffraction blur. Note that this analogy does not specify the conditions for an interference-like response, which is usually referred to as \term{suppression} phenomena in hearing. Suppression is thought to involve nonlinearities, which extend beyond the normal bandwidth of the auditory channel. In this case, the response may not be obviously interpreted as interference, since it can also be caused by inhibition in the central pathways, which may produce a similar perceptual effect, but have a different underlying mechanism. In his seminal paper of two-tone suppression masking, \citet{Houtgast1972} compared it to \term{Mach bands} in vision, which appear as change in contrast around the object edge, although it is not caused by interference, but rather by inhibition. And yet, it is now known that suppression is cochlear in origin and has indeed been shown to be caused by the nonlinear cochlear dispersion \citep{Charaziak2020}, as we should expect from the space-time analogy between diffraction (interference) and dispersion. Unlike the loss of contrast, the resultant image here does not necessarily involve loss of information, only that some information may be difficult to recover after interference.

The third class of masking involves ``phantom'' sounds, whose response persists in the system even after the acoustic masker has terminated. This nonsimultaneous (forward or backward) masking is measurable in the auditory pathways and is not exclusively a result of cochlear processing that ``recovers'' from the masker (e.g., when the compression is being released, or after adaptation is being reset due to replenishment of the synapses with vesicles; \citealp{Spassova2004}). The existence of nonsimultaneous suppression effects appears to be much shorter than the decay time of forward masking \citep{Arthur1971}, so that even if it is measurable over a short duration after the masker offset, it is unlikely to be in effect much later. However, short-term forward entrainment effects in the envelope domain have been sometimes demonstrated in cortical measurements \citep{Saberi2021}. This means that suprathreshold sounds playing during the perceived masker decay may be comparable to those under weak (decayed) masking of the first class of incoherent sounds, since the sounds do not directly interact and may only result in loss of contrast. Although physically and perceptually it is nothing of the sort, the forward masking decay effectively produces a similar interaction effect that would be experienced in sound reverberation. The reverberation decay is incoherent (\cref{CoherenceReverb}) and itself produces an effect that resembles geometrical blur (\cref{BlurandAbbs}). However, since the effect is internal to the auditory system, perhaps the adjective ``fuzzy'' might describe the masking objects better than blurry. 

A fourth class of maskers does not belong to any of the above, which are referred to as \term{energetic masking} effects. \term{Informational masking} has been a notable effect, where sounds are masked in a way that cannot be explained using energetic considerations only. It appears to have a central origin that is physiologically measurable as late as the inferior colliculus \citep{Dollevzal2020}. The analogous effect here is of deletion: elements from the original objects do not make it to the image and are effectively eliminated from it. The type of maskers and stimuli involved in these experiments are usually tonal---multiple short tone bursts scattered in frequency \citep{Kidd2008}. Each tone burst that is simultaneously played with such a masker may be comparable to a type of coherent noise that is called \term{speckle noise}---distinct points that appear on the image because of dust on the imaging elements, for example, but do not belong to the object of interest. Speckle noise can be effectively removed through incoherent or partially coherent imaging, which averages light that arrives from random directions so that small details like dust do not get imaged (for example, compare the coherent and incoherent images in Figures \ref{partialimages} and \ref{CohIncoh}). In hearing, the removal of details like tones in informational masking tests may stem from dominant incoherent imaging. If this is so, then suprathreshold sounds under informational masking may suffer some geometrical blur. Interestingly, not all listeners exhibit informational masking, so some studies preselect their subjects accordingly \citep{Neff1993}. Note that the time scales involved here may be longer than those that relate to a single pulse-image that is coherent or incoherent, which may require integration over longer time scales. Nevertheless, the logic for all types of images should be the same. 

In realistic acoustic environments, we should in all likelihood expect to continuously encounter all types of masking in different amounts, but using much more complex stimuli. If the analogies above have any merit, then they entail that masking does not only dictate the instantaneous thresholds of different sound components in the image, but it also determines their suprathreshold contrast (available dynamic range) and relative blur. It may strike the reader as sleight of hand to be appropriating this host of well-established masking effects into the domain of imaging. However, it should be underlined that the emphasis is on signals \textbf{above} the masking threshold, which are what is being perceived, not at and below the thresholds which usually receive much of the attention in research. This intermingling of imaging and masking terminologies will enable us to make occasional use of the vast trove of masking literature that has been accumulating over a century of investigations. 

\section{The auditory defocus}
\label{AudDefocus}
The inherent auditory defocus may have been the most unexpected feature that was uncovered in this work, since one of its original goals was to show how normal hearing achieves focus, in close analogy to vision. But the unmistakable presence of a substantial defocus term is a divergence from vision theory. Interpreting its meaning requires further input from both Fourier optics and coherence theories. 

Valuable insight may be gathered from two optical systems, where defocus is employed intentionally, either to blur unwanted objects, or to achieve good focus with an arbitrary depth of field. The first case, which was already presented in \cref{DirectProminence} and Figure \ref{KohlerIllum}, is the most common form of illumination in microscopy (K\"ohler illumination), where a small specimen is the object that is mounted on a transparent slide between the light source and the observer \citep{Kohler1893}. The function of the microscope requires that the image of the object on the observer's retina is in sharp focus, whereas the image of the light source (usually a thin filament) should be nonexistent. This is achieved by placing the object beyond the focal point of the condenser lens that converts the point source radiated light to parallel plane waves. Because the source of light is spatially coherent, if its image is not sufficiently diffused and defocused at the object plane, then an image of the filament would be projected on the object (the specimen) and give rise to a distorted image. In hearing, the design logic is diametrically opposite, as the acoustic sources are of prime interest, whereas the passive objects that only reflect sound are of much less importance, relative to the sources. 

A second system that employs a similar principle is an optical head-mounted display for producing visible text that is overlaid on whatever visual scene the wearer is looking at \citep{Waldkirch2004}. By applying defocus and using partially coherent light to produce the text, it is possible to create a sharp image of the letters regardless of the accommodation of the eye (\cref{OcularAcco}), as accommodation reacts only to incoherent light. If the text is lit incoherently as well, then it becomes extremely difficult to keep it sharp with accommodation continuously modifying the focal length of the lens. In contrast, in hearing, coherence itself is a parameter of the sound source and  its environment. Using defocus, the hearing system may be able to differentiate the amount of blur assigned to different signals types (or signal components) according to their degree of coherence. It is possible also that a partially-coherent image is obtained from coherent and incoherent imaging pathways, which are combined to produce an optimal quality with an appropriate amount of blur. 

The effect of defocus was indirectly examined in \cref{TheTMTF} through its effects on the modulation transfer function (MTF) for incoherent and coherent inputs\footnote{In fact, the amplitude transfer function squared, $|ATF|^2$, was used in the coherent case.}. Because of the relatively high cutoff frequencies associated and the effect of the nonuniform neural downsampling in the system, it was difficult to identify many interesting test cases that differentiate the two coherence extremes. This is largely because the auditory filters overlap and normally do not run into the modulation bandwidths associated with these MTFs. Nevertheless, the analysis enabled us to predict that differentiation between coherent states is dependent on the input bandwidth, as the degree of coherence is inversely proportional to the spectral bandwidth of the signal. 

\begin{figure} 
		\centering
		\includegraphics[width=0.85\linewidth]{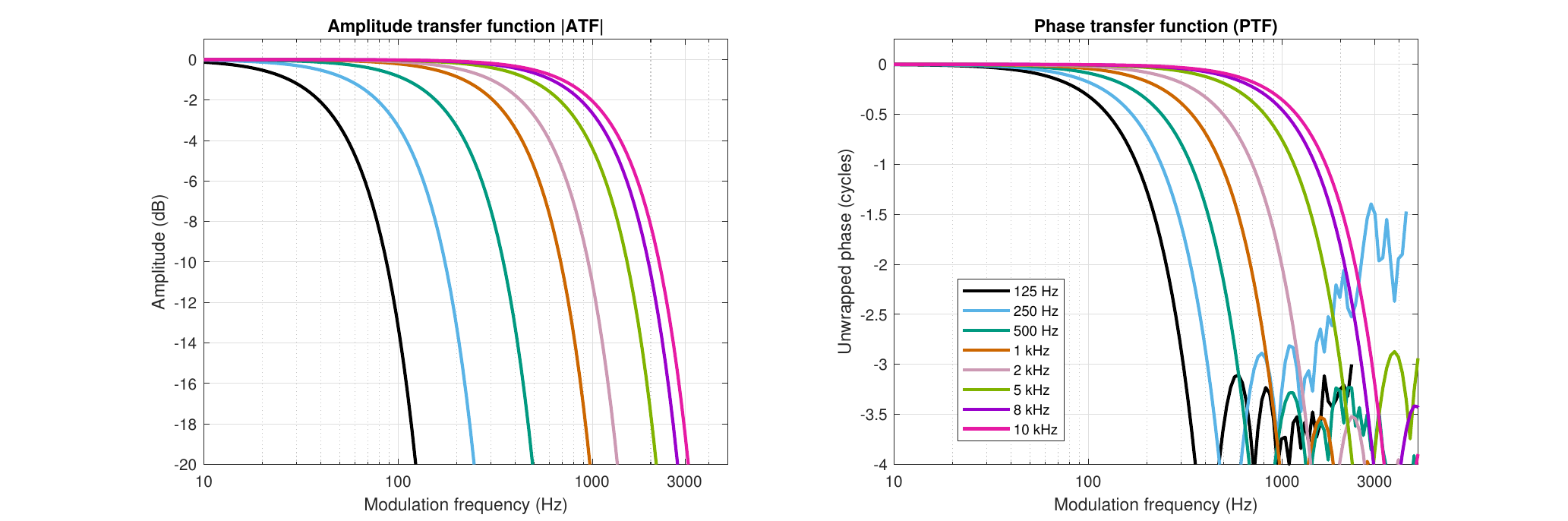}
		\caption{The estimated amplitude transfer function (ATF, left) and (unwrapped) phase transfer function (PTF, right) of the human auditory system (Eq. \ref{GenGaussPupil}), using the parameters found in \cref{paramestimate} and low-frequency corrections from \cref{TempAperture}.}
		\label{ATFPTF}
\end{figure}

The (coherent) amplitude transfer function (ATF) and phase transfer function (PTF) are plotted in Figure \ref{ATFPTF}. Remembering that partially coherent signal intensity can be expressed as the sum of the coherent and incoherent intensities (Eq. \ref{totalcoherence}), these functions apply also to the coherent component in partially coherent signals. Wherever the signals become truly incoherent, the phase response of the function becomes meaningless, and it is reduced to the familiar optical transfer function (OTF) or MTF. In pure tones too, which are completely coherent, the PTF is inconsequential as it phase-shifts the tone by a constant factor, which cannot be detected by the ear without additional interfering carriers. Therefore, the best test cases for the defocus may be signals of finite bandwidth that are sensitive to phase, where the power spectrum model breaks down. 

\begin{figure} 
		\centering
		\includegraphics[width=0.85\linewidth]{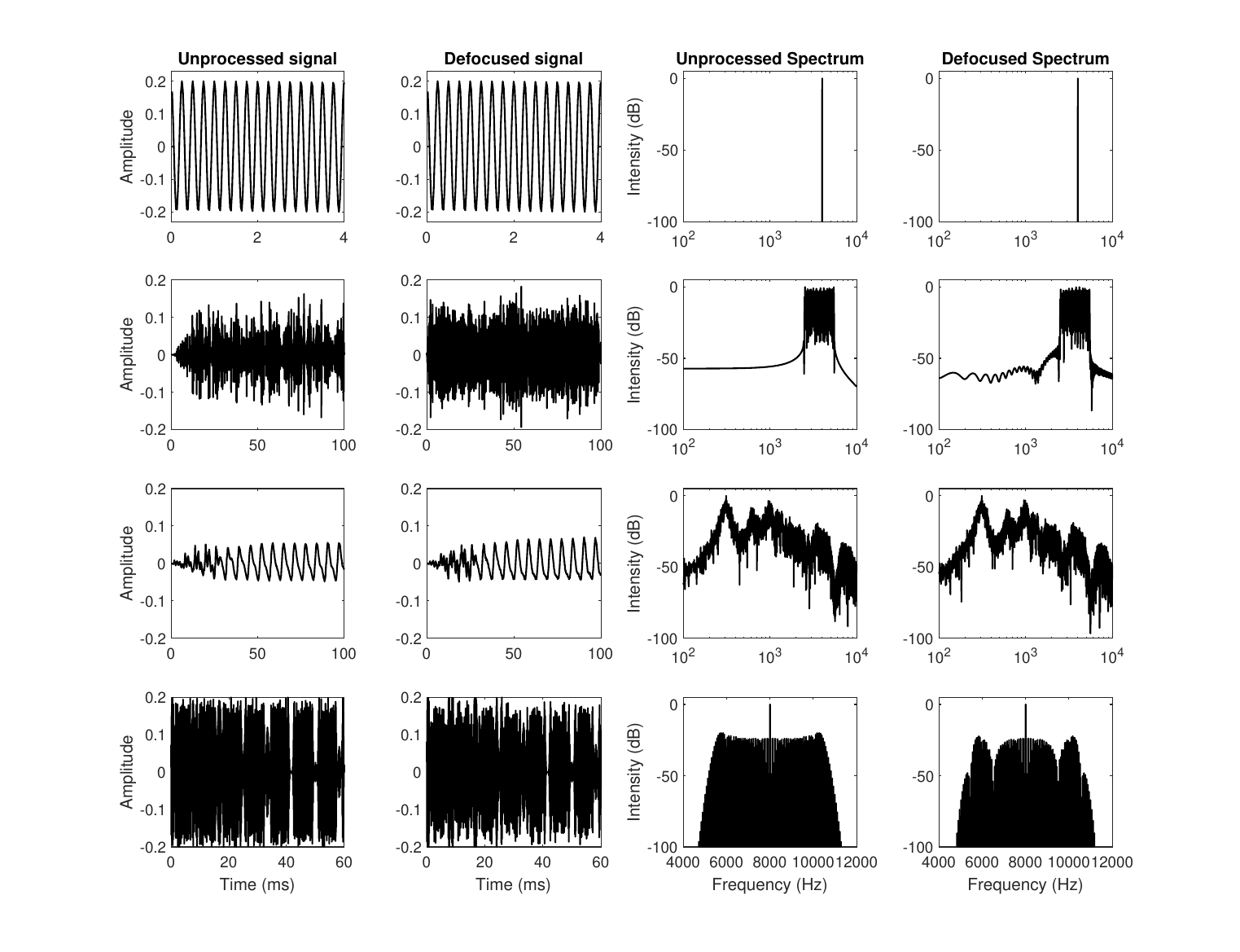}
		\caption{The effect of the auditory defocus on four signals of different modulation spectra and levels of coherence. Each row corresponds to a signal type, where the leftmost plot is the unprocessed time signal, the second from the left is the defocused time signal obtained by multiplying its modulation spectrum with the complex ATF (see Figure \ref{ATFPTF}), the third is the power spectrum of the unprocessed signal and the rightmost plot is the power spectrum of the defocused signal. \textbf{The first row} is a 4 kHz pure tone, which is unaffected by the modulation filter. \textbf{The second row} is a rectangular-shaped narrowband noise centered at 4 kHz with 3 kHz total bandwidth. The processed and unprocessed sounds were RMS equalized to make their loudness comparable. The audible effect of the defocus filter is subtle, as it slightly lowers the pitch of the noise, and it is caused by both the aperture and quadratic phase. \textbf{The third row} is taken from a 2 s long male speech recording in an audiometric booth (only the first 100 ms are displayed) that was bandpass-filtered around seven octave bands (125--4000 Hz, 4-order Butterworth, bandwidth equal to the equivalent rectangular bandwidth (ERB) (Eq. \ref{ERB}) \citep{Glasberg1990}. The unprocessed version was obtained by using Hilbert envelope and phase as a complex envelope to modulate a pure tone carrier in the respective octave band and the total signal is the summation of all seven bands. The defocused version was the same, but the complex envelope was filtered in the modulation frequency domain by the ATF before modulating the carrier. The audible effect was not dominated by the quadratic phase itself, but rather by the aperture, which results in a thinner sound. \textbf{The last row} is of a strongly modulated 8 kHz carrier, whose AM-FM envelope is: $1+\sin\left[2\pi 50t + 40\cos(2\pi 60t)\right]$. Once again, the audible effect is subtle and makes the defocused sound slightly muffled and lower-pitched, although this time it was caused mainly by the quadratic phase. Refer to the audio demo directory \textsc{/Figure 15.2 - auditory defocus/} to hear the corresponding examples.}
		\label{defocused}
\end{figure}

Several signal types are displayed in Figure \ref{defocused} with and without the effect of the estimated auditory ATF\footnote{Refer to the audio demos in the directory \textsc{/Figure 15.2 - auditory defocus/} to listen to the corresponding stimuli.}. The responses were obtained by generating a complex temporal envelope and multiplying it in the frequency domain with the ATF\footnote{Mathematically speaking, this procedure is not strictly valid, as stochastic signals do not have a proper time-domain representation. This is regularly circumvented in optics using measurable autocorrelation and intensity functions, as was discussed in \cref{TransFun}. Because of the low-frequencies involved, obtaining an ad-hoc time-domain representations for white noise is technically straightforward in acoustics and hearing. However, a more mathematically rigorous procedure may have to be developed for partially coherent signals---something that should be relevant in hearing as well.}. Critically, the negative frequencies of the modulation domain were retained in sound processing. The most coherent and narrowest signals are also the ones that are unaffected by the defocus and by the MTF on the whole, as its low-pass cutoff is higher than half the signal bandwidth (for signals whose carrier is centered in the auditory filter). The effect of the aperture low-pass filtering is illustrated using the different signals---notably their amplitude-modulation (AM). The defocus (i.e., its quadratic phase) directly affects the signal phase and any frequency-modulated (FM) parts (see caption in Figure \ref{defocused} for further details). Both the narrowband noise and the AM-FM signal were deliberately designed to have relatively broadband spectra, so to emphasize the dispersive effect. In all cases but the pure tone, it is the author's impression that the defocused version may be considered less sharp than the unprocessed versions, although the effect is not the same in the different cases. The narrowband noise sounds narrower and with less low-frequency energy, which raises the perceived pitch of the filtered noise. The speech sample sounds duller, but the effect is subtle with the neural group-delay dispersion value obtained. The FM sound is made duller after filtering, and its pitch is pushed both lower and higher than the pitch of the unfiltered version. However, to simulate a more correct auditory defocusing blur may require tweaking of the parameters (neural dispersion, temporal aperture, or filter bandwidth) and more importantly, to apply nonuniform sampling that resembles adaptive neural spiking and that captures the low-pass filtering and decohering effects that it may have after repeated resampling and downsampling (\cref{MTFandSampling}). 

It appears that the auditory defocus tends to interact with sounds that are clearly broadband. When they are not resolved to narrowband filters, these sounds potentially contain high modulation frequencies that may be affected either by the low-pass amplitude response or by the phase of the ATF. The narrowband noise In Figure \ref{defocused} is an example for such a sound, whose random variations cause instantaneous phase changes that map to very high nonstationary FM rates.

The prominence of defocus in listening to realistic sound environments is unknown at present. The most useful portion of the (real) modulation spectrum of (anechoic) speech is well-contained below 16 Hz \citep{Drullman1994a} mostly for consonants, whereas temporal envelope information for vowels was mostly contained below 50 Hz \citep{Shannon1995}. Using the estimated auditory dispersion parameters from \cref{paramestimate}, such a modulation spectrum is unlikely to be strongly affected by group-velocity dispersion alone, which may have a larger effect on the FM parts of speech in case they can be associated with higher instantaneous frequencies. This conclusion is somewhat supported by the crudely processed speech example of Figure \ref{defocused}, which disclosed only a subtle blur compared to the unprocessed version. Nevertheless, the analysis in the final chapters of this work will suggest that the role of defocus is greater than initially may appear from the examples that were given above.

\section{The temporal resolution of the auditory system}
\label{GapDetect}
\subsection{Temporal acuity}
One of the most basic aspects of any imaging system is its resolving power, which quantifies the smallest detail that can be imaged distinctly from adjacent details. In spatial imaging, it is intuitively conveyed by the point spread function, which converts a point in the object plane to a disc with a blurred circumference in the image plane. The size of the disc is minimal when the system is in sharp focus and with no aberrations, as it is only constrained by the blurring effect of diffraction. An analogous effect is obtained using the impulse response function in temporal imaging, only that the limiting effect is caused by group-velocity dispersion instead of diffraction. Using the estimated system parameters, it is possible to use the impulse response function of Eq. \ref{impresponse5} to compute the theoretical temporal resolution values at different frequencies. 

There are several established criteria in optics for imaging resolution based on two-dimensional patterns (usually assuming circular or rectangular apertures). The most famous one is the \term{Rayleigh criterion}, which is based on the image of two object points. When the center of one imaged point falls on the first zero of the diffraction pattern (an Airy disc) of the second point, the two points are just about resolved (\citealp[pp. 216--219]{Rayleigh1879, Goodman}; \cref{DiffractionFourier}). There is no standard criterion in temporal imaging, although \citet{Kolner} suggested one for a system with a rectangular aperture in sharp focus. In analogy to spatial imaging, he equated the location of first zero of the corresponding impulse response sinc function to the spacing required for resolving two impulses. 

\begin{figure} 
		\centering
		\includegraphics[width=0.9\linewidth]{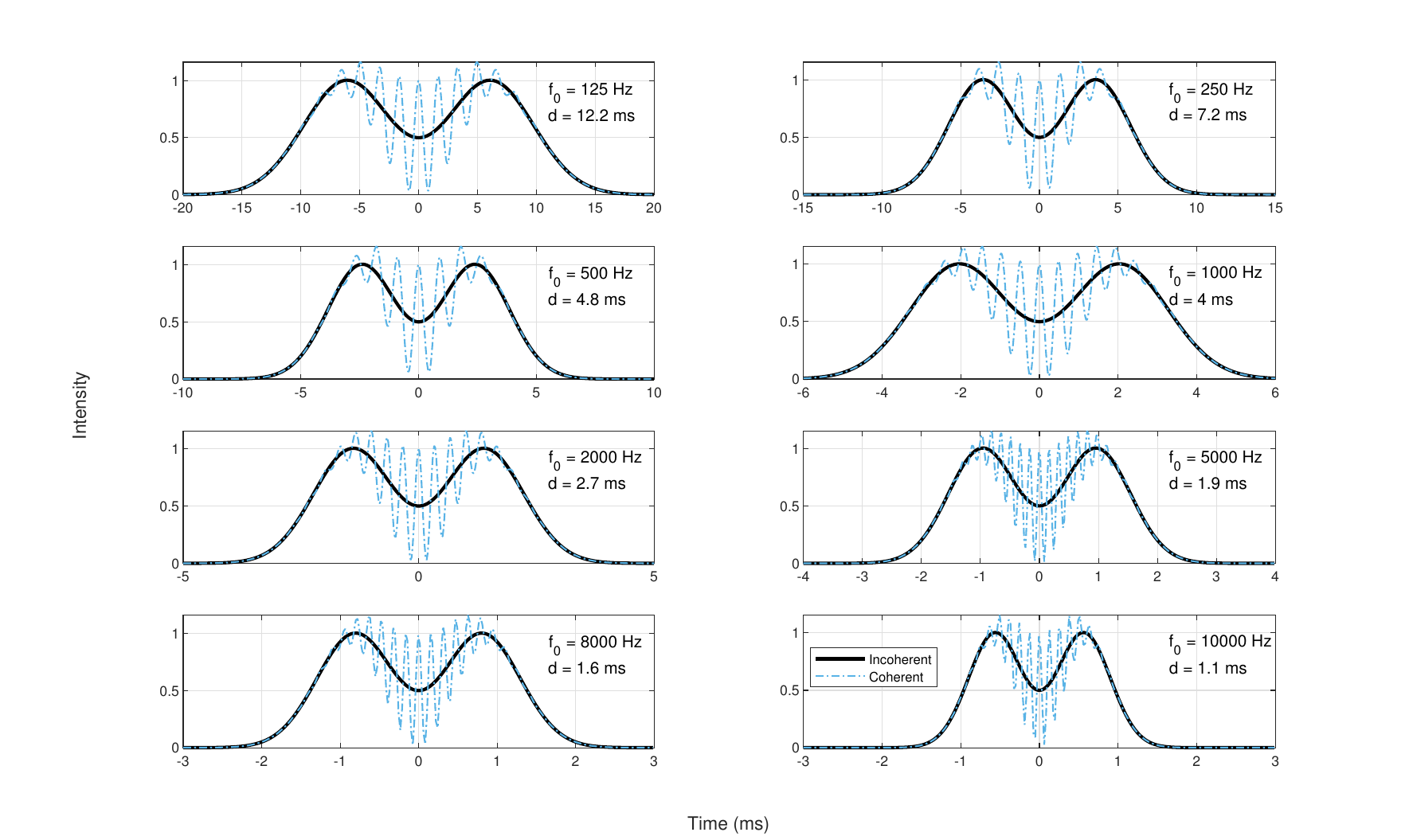}
		\caption{The temporal resolution of two consecutive impulses spaced by $d$ milliseconds, according to the impulse response function of Eqs. \ref{impresponse5} and \ref{TemporalRes} and the parameters from \cref{paramestimate}. The spacing was chosen so that pulses are considered resolved when their (summed) intensity image is exactly at half of the peak value. The black solid curve shows the incoherent pulse response. The blue dash-dot curves show the ``coherent impulse responses'' as a demonstration of the effect of the phase term in the defocused impulse response function. The oscillations are a result of interference, which subsides when the gap is increased.} 
		\label{PulseRes}
\end{figure}

As the present work employs a Gaussian aperture shape---an unphysical shape that has no zeros, but is convenient to work with analytically and appears to correspond to the auditory pupil shape---the criterion here will be somewhat more arbitrary. We shall consider a sequence of two impulses as resolved if their responses intersect at the half-maximum intensity. Using the auditory system parameters found in \cref{paramestimate} and the generalized (defocused) impulse response for a Gaussian aperture (Eq. \ref{impresponse5}), the resolution at different spectral bands can be readily obtained by using an input of the form $\delta(-d/2) + \delta(d/2)$, for two pulses separated by a gap of $d$ seconds, measured between their peaks. The gap duration can be calculated by convolving the impulse response with a delta function and finding the time $d/2$ at which the respective intensity drops to a quarter of the maximum. The two incoherent pulses then intersect at half the maximum level. After some algebraic manipulation, we obtain the gap duration:
\begin{equation}
d = v T_a\sqrt{\left(\frac{16\ln2}{T_a^2}\right)^2 + \left(\frac{1}{u} + \frac{1}{v}+ \frac{1}{s} \right)^2 }
\label{TemporalRes}
\end{equation}
In order to obtain the intensity response for the incoherent case, the response to each pulse was squared independently (Eq. \ref{incconv}): $I = I(-d/2) + I(d/2)$ (black curves in Figure \ref{PulseRes}). In the coherent case (Eq. \ref{cohconv}), the summation preceded the squaring: $I = [a(-d/2) + a(d/2)]^2$, which is displayed in blue dashed lines of Figure \ref{PulseRes}. 

As delta impulses are incoherent by definition, the coherent responses of Figure \ref{PulseRes} are shown more for academic interest, although short Gabor pulses produce very similar results (see \cref{PsychoEstimation}). The coherent pulses give rise to visible interference (fast oscillations) in the final intensity pattern, which does not match known auditory responses. These oscillations disappear upon increasing the gap $d$, so that $1.1d$ produces a more visible gap in the interference pattern, which almost vanishes completely at $1.4d$ (some oscillations are visible in the trough). The incoherent pulses produce smooth responses that are more easily resolved using the chosen criterion, which is also how the resolution limits were determined. The response was extended up to 10 kHz, but above this frequency the estimates of the cochlear dispersion could not be trusted (see Figure \ref{Reu}). 

The results in Figure \ref{PulseRes} represent the output of a single auditory channel and should be compared to narrowband temporal acuity data from literature, if available. In auditory research, temporal acuity assessment and definition is challenged by the fact that multiple cues (e.g., spectral, intensity) can account for just-noticeable differences between stimuli. Therefore, different methods have been devised to obtain specific types of acuity, which are not always consistent with each other \citep[pp. 169--202]{Moore2013}. As a rule of thumb, the auditory system is able to resolve 2--3 ms \citep[p. 200]{Moore2013}, which corresponds well only to the obtained resolution at 2 kHz (2.7 ms) and maybe at 4 and 8 kHz (1.8 and 1.6 ms, respectively) and 1 kHz (4 ms). At low frequencies, the resolution drops (12.2 ms at 125 Hz, 7.2 ms at 250 Hz, and 4.8 ms at 500 Hz), but less so than it would have dropped had the temporal aperture remained uncorrected (i.e., if we let it be unphysically long; see \cref{LowFreqCorr}). The 4 and 8 kHz predictions are a bit shorter than the data we obtained using Gabor pulses (see \cref{Experiment2}), which had a median acuity of 2.3--3.4 ms at 6 kHz, and 2.8--4.2 ms at 8 kHz, with the two best performing subjects having 1.8--2.7 ms and 2.1--4 ms, respectively. The two lowest frequencies as well as frequencies above 8 kHz are particularly susceptible to errors, because of greater uncertainty in the dispersion parameters and temporal aperture\footnote{The gap detection predictions are not very sensitive to the time-lens curvature, if the large curvature values are used.}. Such frequency dependence of the temporal resolution has usually not been observed in past studies that employed narrowband noise \citep[e.g.,][]{Green1973,Eddins1992}, but values as short as obtained here at high frequencies are typical with broadband stimuli \citep[e.g.,][]{Ronken1971}. However, gap detection ($>$ 50\% threshold) using sinusoidal tones in \citep[Figures 2--3]{Shailer1987} was 2--5.5 ms at 400 Hz, 2.5--3.5 ms at 1000 Hz, compared to predicted gap thresholds of 4.1 ms (not plotted) and 3.6 ms, respectively.

Implicit to this discussion is that the built-in neural sampling of the auditory system can deal with arbitrarily proximate pulses. This is probably true for broadband sounds that stimulate multiple channels along the cochlea, but may be a stretch for narrowband sounds, whose response depends on fewer fibers\footnote{\citet{Kiang1990} discussed whether all the fibers that synapse to a single inner hair cell have correlated responses. While no direct multi-unit data were available that could be confidently be associated with the same hair cell, he surmised that the correlation between fibers is partial, at best. Nevertheless, we maintain that whatever correlation exists between fibers, it must depend on the stimulus coherence at the point of transduction.}. At least in the bandwidth for which data are available, this assumption appears to be met. 

This gap detection computation was based on a Gaussian-shaped pupil function, which appears to be approximately valid (\cref{TempAperture}). However, we do not know the actual pupil function shape in humans, which may eventually alter these estimates to some extent. 

\subsection{Envelope acuity}
While the temporal acuity as quantified above defines the shortest sound feature that can be resolved within a channel, such a fine resolution of 2--3 ms is not generally found in continuous sound, where the acuity tends to significantly degrade. This was measured in Experiments 4 and 5 in \cref{Aliasing} with click trains that included 8 or 9 clicks, to reduce the onset effect and induce some pattern predictability in the listener. When these short events are interpreted as modulations, they suggest a drop in the instantaneous modulation rate from 300--600 Hz to 60--100 Hz. This suggests that the bandwidth of the various psychoacoustic TMTFs may not necessarily reflect the fidelity as it relates to the specific contents of the temporal and spectral envelopes. 

Several studies can attest to this assertion. For example, the discrimination between amplitude-modulation frequencies has been tested a few times with tonal, narrowband noise, and broadband noise carriers \citep[e.g.,][]{Miller1948,Buus1983, Formby1985, Hanna1992, Lee1994, Lemanska2002}. Roughly, these measurements reveal that the discrimination is fairly good (about 1--2 Hz) for low modulation frequencies (below 20 Hz) and gets gradually worse at high modulation frequencies (approximately 20 Hz for 150 Hz modulation, for unresolved modulations). This repeats for different carrier types and frequencies and does not vary much between listeners. Such findings support the model of an auditory modulation filter bank that has broadly tuned bandpass filters, which get broader at higher modulation frequencies, similar to the critical bands in the audio domain \citep{Dau1997a,Dau1997b,Ewert2000,Moore2009}. A more recent study demonstrated how listeners are not particularly sensitive to irregularities in the temporal envelope, which were superimposed at lower frequencies than the regular AM frequency under test \citep{Moore2019b}. The results of this experiment suggest that fluctuations in the instantaneous modulation frequency can go unnoticed by many listeners, who perceive a relatively coarse-grained version of the envelope. However, the individual variation in this experiment was large. 

Since we defined blur earlier as a change in the contents of the temporal, and hence in the spectral envelope, then these studies suggest that the system may not be particularly sensitive to blur in the modulation domain, at least in conditions of continuous sounds, as opposed to onsets or impulsive sound. In a way, this perception may represent a form of internal blur that corresponds to a tolerance that the auditory system has to different complex stimuli.

\section{Polychromatic images}
\label{polychromatic}
\begin{figure} 
		\centering
		\includegraphics[width=0.7\linewidth]{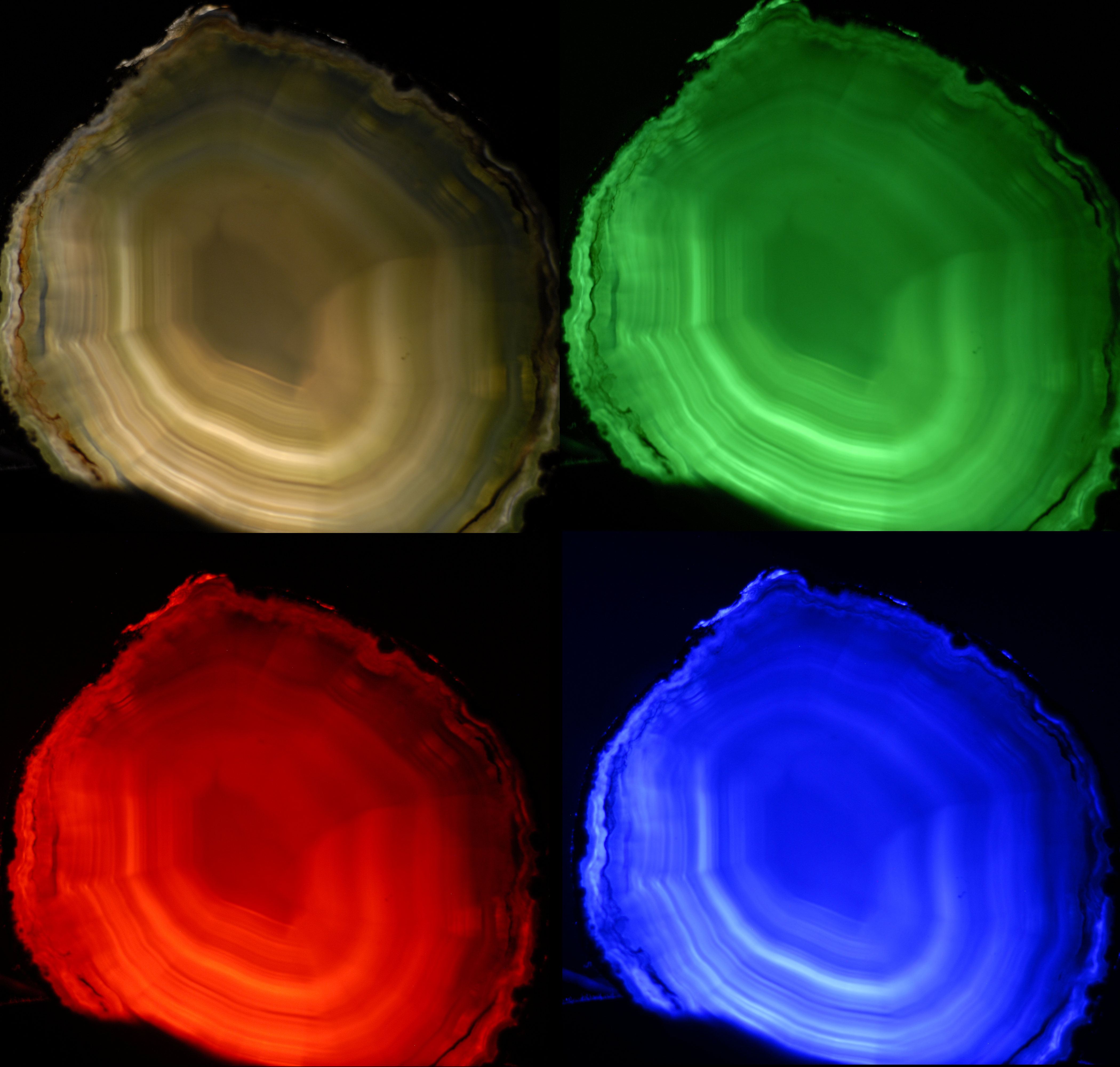}	
		\caption{Four images of a polished agate rock that is back-lit by an LED incoherent light source with variable colors. The top-left image is illuminated with white light (full spectrum) that produces the polychromatic image. The other three images are all monochromatic. Different details of the agate surface are emphasized under each light. More technical details about the LED sources are in Figure \ref{partialimages}.}
		\label{PolyCrystal}
\end{figure}

\begin{figure} 
		\centering
		\includegraphics[width=0.7\linewidth]{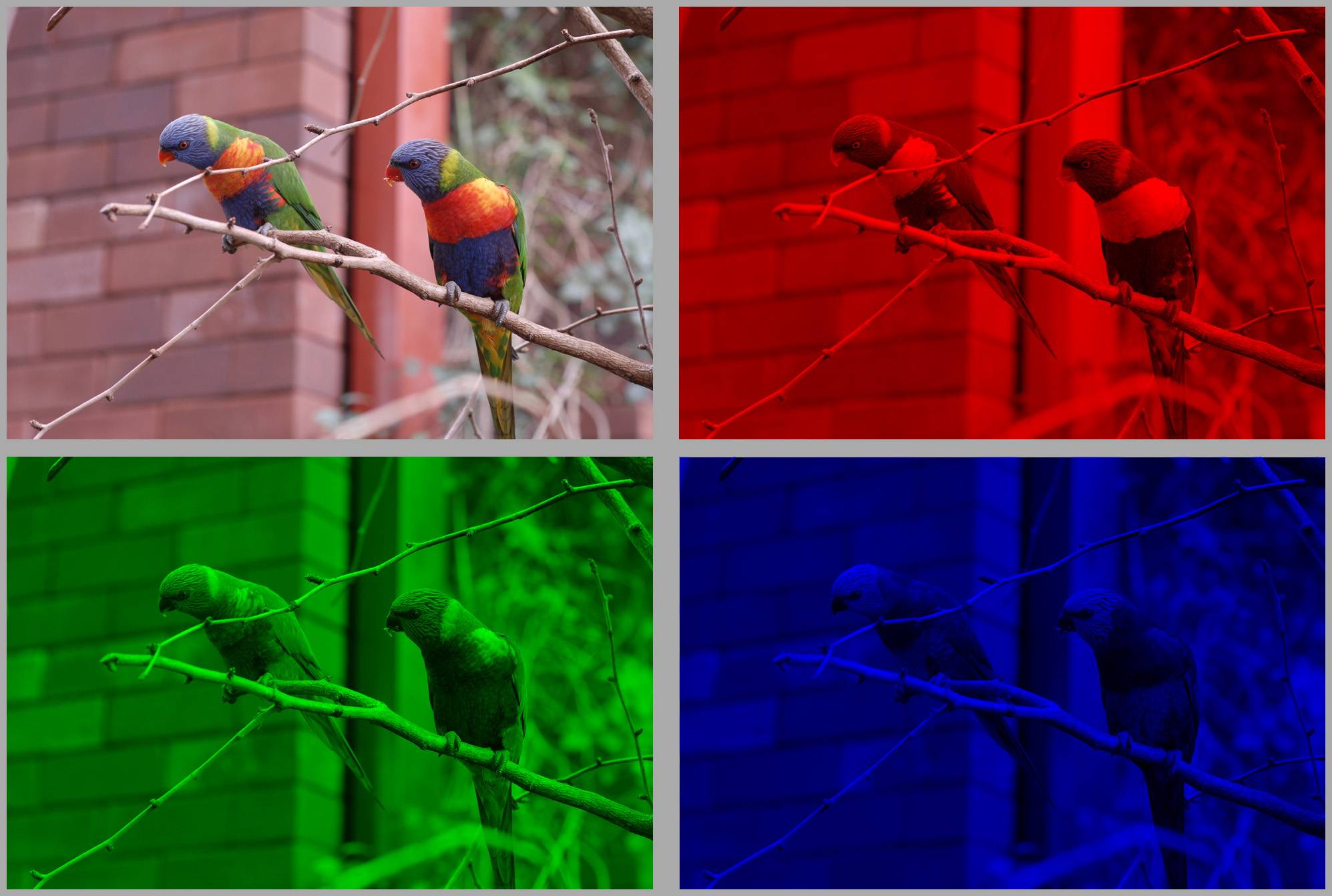}
		\caption{An illustration of a polychromatic image decomposed into three monochromatic color channels. The image is a spatial modulation pattern carried by incoherent broadband light, which is detected in three narrowband channels in the retina by red, green, and blue photoreceptors  (long, medium, and short wavelength cones, respectively). The three monochromatic images are very similar, but some object details are not observable in all of them. For example, the birds' eyes appear to be almost uniform in blue light, whereas the existence of the iris and pupil can be seen most clearly in red and much less clearly in green.}
		\label{polybirds}
\end{figure}

So far, the discussion and analysis of the auditory temporal images were done within a single narrowband channel. However, the modulation in real acoustic objects tends to affect multiple vibrational modes and is rarely limited to narrowband vibrations (single modes). For example, during a drum roll, the modes of vibrations of the drum get excited together---they are \term{comodulated}---in a unique pattern of sound. More complex sounds such as speech also show high correlation across the spectrum in different modulation frequency bands---especially below 4--6 Hz \citep{Crouzet2001}. The analogous optical situation is of an object that is lit by white light, whose spatial modulation spectra---one spectrum per color channel---are highly correlated as a result of being reflected from the same object, whose geometry affects all wavelengths nearly equally. The imaged object is then sensed as a superposition of very similar images of different primary colors, which can be processed by the corresponding cone photoreceptors (Figures \ref{PolyCrystal} and \ref{polybirds}). Importantly, the images created by the photoreceptor channels spatially overlap in a way that facilitates the perceptual reconstruction of the original object. In analogy, we would expect that the broadband auditory image would stem from a perceptual re-synthesis of temporally modulated narrowband images in different channels, which represent the acoustic object response as a whole. This across-channel broadband image is referred to as \term{polychromatic}---an adjective that distinguishes it from a mere broadband sound, for which the identity of the constituent monochromatic channels may be inconsequential . 

\begin{figure} 
		\centering
		\includegraphics[width=0.9\linewidth]{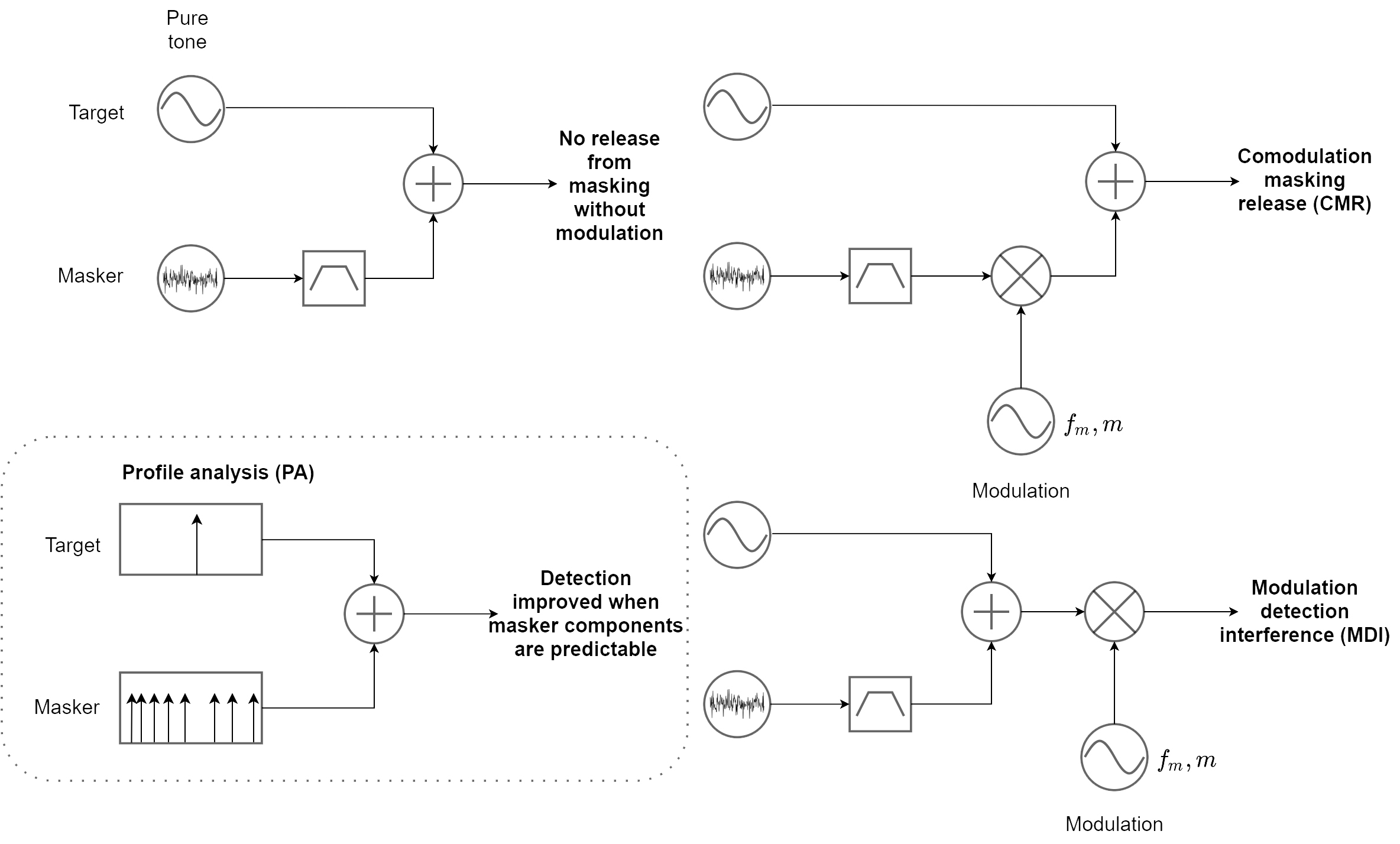}
		\caption{Three psychoacoustic paradigms that entail broadband information integration, beyond a single auditory channel. On the top left, the standard paradigm is shown of a target tone in masking noise. The bandpass filter symbol indicates that the masker bandwidth is a parameter in these experiments. Modulation is indicated with a sinusoidal source of frequency $f_m$ and depth $m$, which multiplies the noise or signal and noise. Multiplication is indicated by the mixer (cross sign).}
		\label{CMRMDI}
\end{figure}

That the auditory system indeed combines across-channel modulation information as generated by common events has been repeatedly demonstrated in several effects---most famously, through the phenomenon of \term{comodulation masking release} (CMR; \citealp{Hall1984}). Normally, when a target tone is embedded in unmodulated broadband noise, its detection threshold increases as the noise occupies more of their shared bandwidth within a single auditory channel. The strength of this masking effect depends on the noise bandwidth as long as it is within the channel bandwidth, but is unaffected by noise components that are outside of the channel. However, if the noise is extended beyond the channel bandwidth and is also amplitude-modulated (not necessarily sinusoidally), then the masking effect decreases---information from adjacent and remote noise bands is used by the auditory system to release the target from the masking effect of the noise (see top plots in Figure \ref{CMRMDI}). The effect is robust and has been shown to yield up to 10--20 dB in release from masking, depending on the specific variation \citep[pp. 102--108]{Verhey2003, Moore2013}. The effect is also more or less frequency-independent, as long as the bandwidth of the noise is scaled with reference to the auditory filter bandwidth of the target \citep{Haggard1990}. 

CMR has been interpreted as a form of pattern recognition and comparison across different bands of the signal, which is representative of real-world regularities of sounds \citep[e.g.,][]{Hall1984,Nelken1999}. As such, it is also considered an important grouping cue in auditory scene analysis \citep[pp. 320--325]{Bregman}, which is effective as long as the different comodulated bands are temporally synchronized (or ``coherent'', in the standard psychoacoustic jargon, \cref{PsychoacousticsCoh}) \citep{Christiansen2015}. In stream formation, temporal coherence (of the envelopes) can act as a strong grouping cue that binds across-frequency synchronized tones, but not asynchronous tones whose onsets do not coincide \citep{Elhilali2009}. 

There has been only one physiological demonstration of CMR at early processing stages \citep{Pressnitzer2001}. Spiking pattern correlates of CMR were found in the guinea pig anteroventral cochlear nucleus (AVCN) units---mainly of the primary-like and chopper-T types. In order for this effect to work, low-level integration is required that relies on the modulated masker in different channels to be in phase. The authors modeled the results using a multipolar broadband unit that receives excitatory off-frequency inputs, which in turn inhibits a narrowband in-channel unit---inhibition that results in masking release. The existence of a broadband processing stage is supported also by psychoacoustic data, which ruled out a model of across-channel comparison of the different narrowband envelopes \citep{Doleschal2020}. Furthermore, the CMR model of \citet{Pressnitzer2001} was successfully used to demonstrate how across-channel information may be advantageous in consonant identification under different conditions (i.e., in noise or when the temporal fine structure was severely degraded; \citealp{Viswanathan2022}).

Other effects exist that demonstrate the polychromatic auditory image primacy over spectral mechanisms, as the system prioritizes temporal cues of multiband signals at the apparent expense of unmodulated narrowband target signals. For example, in \term{modulation discrimination interference} (MDI), an amplitude- or frequency-modulated masker causes the decrease in detection sensitivity of a similarly modulated target at a distant channel (Figure \ref{CMRMDI}, bottom right) \citep{Yost1989Mod, Wilson1990,Cohen1992}. Thus, the modulated target cannot be easily heard as being separate from the masker. However, FM elicits a more limited MDI effect that does not always provide sufficient resolution across channels and modulation patterns, at least at high modulation frequencies \citep[e.g.,][]{Lyzenga2005}. \term{Profile analysis} is another phenomenon, whereby the detection of a level change of one of the components of a multicomponent masker depends on the entire across-frequency profile of the masker (Figure \ref{CMRMDI}, bottom left) \citep{Spiegel1981}. The detection of a single-component target improves when the masker has known frequencies compared to when there is some uncertainty in its component frequencies. The spectral profile is often modeled as spectral modulation (e.g., \citealp{Chi1999}).

It may be argued that in all three effects mentioned---CMR, MDI, and profile analysis---the experimenters' designation of signal and noise (or target and masker) is incongruent with what the auditory system determines. Once the system identifies a potential across-frequency image, it attempts to optimize it as a whole, rather than as a loose collection of monochromatic images.

The importance of the polychromatic representation in an ensemble of monochromatic channels may be gleaned from a study in cochlear implant processing by \citet{OxenhamKreft2014}. The authors showed how, unlike normal-hearing listeners, the speech-in-noise performance was identical for the cochlear-implant users regardless of the type of masker used: broadband Gaussian noise with random fluctuations, broadband tone complex with the same spectral envelope as the broadband noise, and the same tone complex with the superimposed modulation of the Gaussian noise. It was found that this undifferentiated pattern is not caused by insensitivity to temporal fluctuations, but rather by spectral smoothing, as the cross talk between the implant electrodes within the cochlea causes the different channel envelopes to mix across the spectrum. This was shown from speech intelligibility scores of normal-hearing listeners, who could no longer take advantage of the fluctuation difference between masker types, after the temporal envelopes extracted from 16 channels were summed and identically imposed on all 16 channels.

These phenomena and others may all attest to the image dominance, where modulation is involved, compared to unmodulated isolated sounds\footnote{For a review of the above phenomena, see \citet{HallMoore1995}. Related examples of multichannel images and speech segregation are discussed in \citet{Patterson1992}, who had already recognized that CMR is an interesting test case in their Auditory Image Model. However, no interpretation was offered there as for the underlying cause of this effect.}. The CMR effect originally appeared to violate the critical band theory, which predicts that only information within auditory filters should be fused. However, by analogy to vision, this effect may be predicable if the auditory system ``overlays'' multiple monochromatic images to produce a single polychromatic image, or a fused or coherent stream, according to auditory scene analysis. MDI too is consistent with the system trying to form auditory images (``objects'' in \citealp{Yost1989Mod}) from a common acoustic source. Unlike vision that has its three monochromatic detectors interwoven in the same spatial array on the retina, overlaying the auditory imaging has to be done in time. Such a mechanism has been discussed at length with regards to periodicity in the inferior colliculus (\term{periodotopy}; \citealp{Langner2015}), which is nevertheless more restricted than general modulation patterns that are not necessarily periodic.  

As a final note, it should be emphasized that some broadband sounds may not be amenable to representation as polychromatic images, if they vary across channels and time in an independent manner across channels. 

\section{Pitch as an image}
It is worth dwelling briefly on tones and pitch, which have captured the spotlight of auditory research throughout most of its history in one form or another. This section is not concerned directly with how the pitch is determined within the auditory system as a function of the stimulus properties, or in which frequency range each of the pitch types exists. It does, however, attempt to illustrate how the idealized auditory image is related to this standard stimulus family, in order to elucidate interrelated aspects in the pitch and imaging theories. Different pitch types are analyzed using three spectra---of the carrier (through a filter bank), the envelope (or baseband), and the power spectrum that contains the same information as the broadband autocorrelation. The last two spectra require nonlinearity to either demodulate the signal, or generate harmonic distortion that produces the square term that explicitly reveals the fundamental. 

\begin{figure} 
		\centering
		\includegraphics[width=1\linewidth]{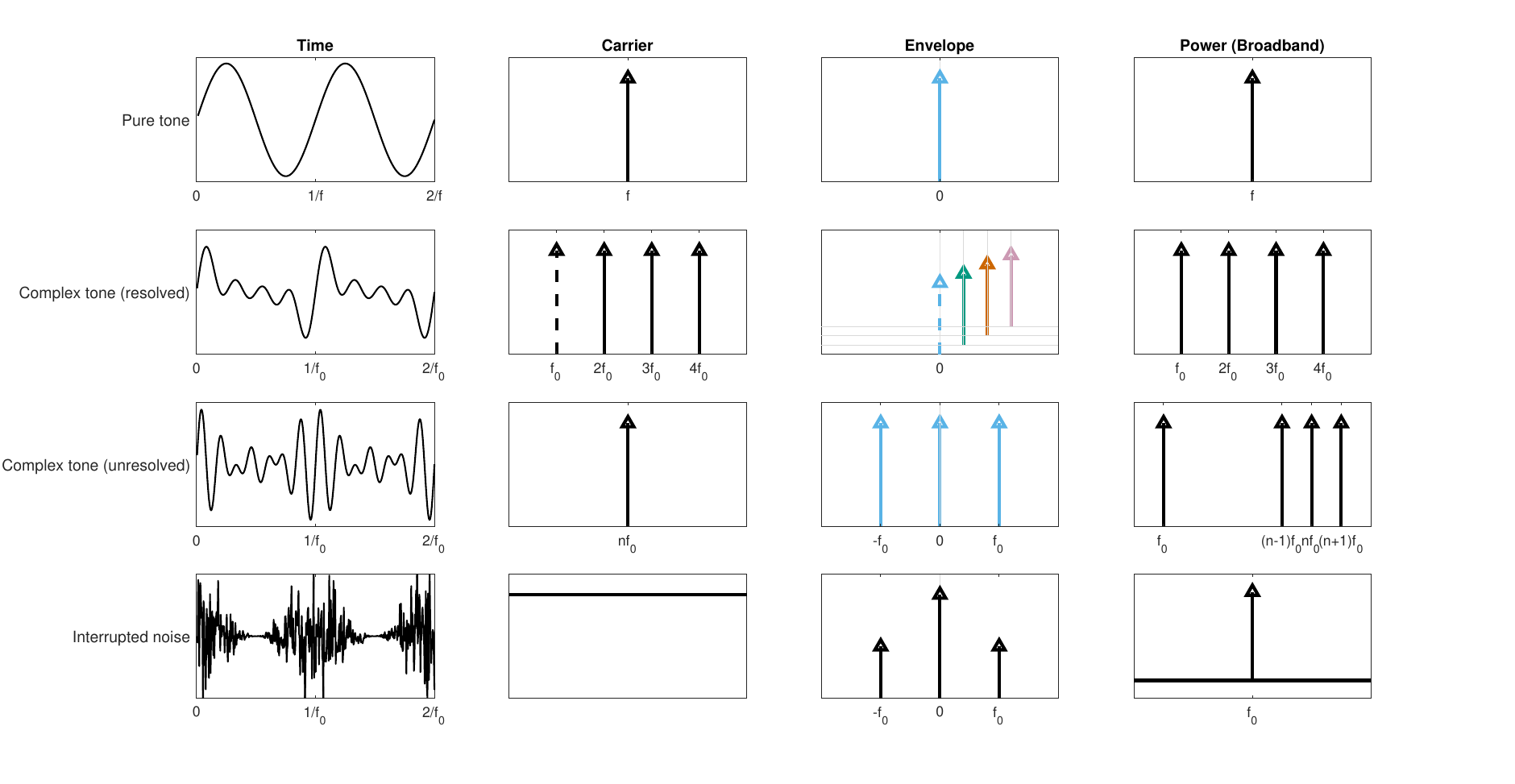}	
		\caption{Cartoon spectra of four stimuli that elicit four common types of monaural pitch. Two periods of each time signal appear on the left along with three different spectra on the right. On the second column is the filter-bank spectrum that can identify the carrier. On the third column is the modulation spectrum, which is the baseband spectrum of the filter output after demodulation. This is where the monochromatic object and image reside. Finally, on the rightmost column is the power spectrum, which has to follow some nonlinearity (e.g., half-wave rectification and squaring), although here it excludes additional harmonic distortion products, for clarity. \textbf{On the top row} is a pure tone, which has a single component in all three spectra that makes its image degenerate. \textbf{On the second row} is a complex tone, whose components are individually resolved in one filter each, with its fundamental frequency $f_0$ intact, or missing (dashed spectral lines). Its envelope spectrum contains only the DC component in each channel, which is associated with a harmonic, and is thus analogous to a polychromatic image. The power spectrum contains $f_0$, whether it appears in the stimulus or not. \textbf{On the third row} is an unresolved complex tone, whose components are analyzed by a single filter, where it appears like a single component. Its ideal modulation image contains all three components, whereas the power spectrum also contains the missing fundamental (although it is heard more faintly). \textbf{On the bottom row}, a form of periodicity pitch---interrupted noise---is produced by amplitude-modulating white noise, which assumes some pitch if it is fast enough. The power spectrum can reveal the modulation period that is seen in the envelope spectrum, on top of the spectral distribution of the noise itself.}
		\label{Pitches}
\end{figure}

It should be noted that while the information required to give rise to monaural pitch is available already at the level of the inferior colliculus (IC), pitch perception is most likely generated in the auditory cortex after appropriate coding transformation \citep{Plack2014}. The early availability appears to hold also more generally to nonstationary pitch, such as the fundamental frequency in speech, which is tracked at the level of the IC \citep{Forte2017}. Inharmonic complex tone pitch and binaural pitch, though, may require information that is derived after some processing that may take place at a higher level than the IC as well \citep{Gockel2011}. Therefore, these general results suggest that an image that appears at the IC can be used to infer some properties of pitch, even if it is produced and perceived further downstream. We shall use employ our own auditory image concept loosely here to relate directly to the physical stimulus, but also link it to the psychoacoustic percept, which is assumed to be directly derived from the image, at least for the simple examples used below. Cartoon spectra of four of the most common monaural pitch types are displayed in Figure \ref{Pitches} and are described below in some detail.

The pure tone is probably the most widely used and abused stimulus in all of acoustics.  But what does it entail within the temporal imaging framework? As the pure tone has a real and constant envelope, its ideal image also has a constant envelope with an arbitrary magnification (gain). We hear the constant envelope along with a uniform pitch percept without perceiving the tonal oscillations. Therefore, this is an intensity image, rather than an amplitude image. Such an image is completely static (or ``stable'' according to \citealp{Patterson1992}), because it is time-invariant. It is contrasted with arbitrary acoustic objects that have time-dependent amplitude and phase functions that give rise to complex envelopes. Therefore, a pure tone is also coherent in two senses. In the classical sense, a pure tone is obtained from a perfectly (temporally) coherent signal that can always interfere with itself\footnote{Any phase difference or time delay between a pure tone and its replica only causes the degree of coherence to become complex, but its absolute magnitude does not change, $|\gamma|=1$ (see \cref{BasicCoherenceDeriv}).}. In the auditory jargon usage of coherence, as the pure tone has no beginning and no end, it is always coherent in the envelope domain as well, whose spectrum contains only a single line at zero (DC). Therefore, we can think of the pure tone as a \term{degenerate image}\footnote{Borrowing from the concept of degenerate frequencies in physics (e.g., \citealp[pp. 464--466]{Goldstein2014M}).}. In analogy to vision, such an image would correspond to a monochromatic dot object that is fixed in space right on the optical axis and is projected as a still (spread) point at the center of the fovea.

Complex tones refer to series of pure tones with fixed spacing between their frequency components and common onsets and offsets. The series is usually harmonic, which means that the spacing between the component frequencies follows an integer ratio. It is possible to generate harmonic series so that each tone is resolved in its own dedicated auditory channel, which prevents audible beating between components from taking place. This gives rise to a series of degenerate images. But, the auditory system also extracts the periodicity of the harmonic series, which in this case corresponds to the lowest-frequency spectral line in its broadband power spectrum. Thus, while more complex, this image is still static and has a distinct pitch that corresponds to the fundamental frequency---the spacing between the components. Additionally, because of the integer ratios between the harmonics and the frequency spacing, it is also a degenerate image, but in a different sense: the fundamental frequency of the harmonic series coincides with the periodicity from the power spectrum. This gives rise to the famous missing fundamental effect, when the harmonic series excludes the fundamental---the perception that the pitch corresponds to the fundamental even when it is absent from the stimulus. There is no complete analog to this image in vision, but the periodicity spectrum is analogous to a grating, whereas each component corresponds to a color. However, we cannot represent their harmonic relations visually\footnote{\citet{Julesz1972} considered the possibility to have the horizontal spatial dimension in vision analogous to auditory time and the vertical spatial dimension analogous to frequency. If this were the case, then harmonic relations could be presented geometrically using shapes with integer ratios. However, this is an arbitrary analogy that bears little physical resemblance to the complexity that is offered by real harmonicity.}$^,$\footnote{\citet{Shamma2001} argued that periodicity pitch perception is analogous to bilateral symmetry in vision, as both provide grouping and segregation mechanisms that can be used in the perception. This analogy relies on coincidence detectors that perform the comparison between inputs to different spectral/spatial channels. However, it is not clear why this analogy should be more appropriate than directly comparing the (bilateral) spatial dimensions of both senses, as hearing can detect spatial symmetry as well. Additionally, this analogy does not address the special status that integer ratios between multiple channels have in hearing, but not in vision that is only trichromatic.}. 

For limited harmonic series with small frequency spacing and only few components, the harmonics may not be resolvable, so they are analyzed mainly within a single channel. In general, such series can be mathematically represented as interference or beating patterns that modulate a carrier. It means that the envelope spectrum contains at least two components (i.e., a DC component at zero and at the beating frequency), whereas the carrier domain has only one. The broadband power spectrum again shows the fundamental that corresponds to a residual pitch. However, the image in itself---the temporal envelope---is monochromatic. This pitch usually produces a more faint pitch sensation than other pitch types.

Interrupted pitch is another interesting type of pitch that is produced strictly by modulating broadband noise that does not contain any tonal information. Thus, it has no distinct components in its carrier spectrum. It has nontrivial components only in its envelope and power spectra.

The objects of more realistic sounds are generally not time invariant, so they give rise to nondegenerate images. These images may have variable carriers that are more intuitively expressed using frequency modulation than using  a stationary carrier spectrum with multiple components (for example, the two bottom-right plots in Figure \ref{defocused}). Realistic objects can also have nonuniform frequency spacing, which eliminates periodicity and makes the within-channel broadband envelope spectrum nonstationary as well. Complex tone objects may also be frequency-shifted---an operation that retains time-invariant carriers and envelopes in resolved channels, but produces aperiodic broadband spectra in unresolved channels (with more than two components). 

Each type of pitch, therefore, reveals a different feature of the auditory system. When the corresponding spectrum or feature of the stimulus is degenerate, it elicits a stable sensation that we call pitch. Part of the multiplicity of pitch types may go back to the fact that, in general, there is no unique mathematical representation for broadband signals, so the auditory system may have had to develop ways to ``corner'' the signal analysis and make it sound unique by comparing information from different spectra. 

Three questions can be raised following this high-level characterization of pitch. First, must all images be perceived with pitch? Second, do ``pitchiness'' and sharpness refer to the same underlying quality in hearing? Third, does the perception of pitch always require periodicity? According to our image definition---the scaled replica of a temporal envelope pulse---the answer is ``no'' to the first question, since the image appears at a more primitive processing stage than the pitch. As for the second question, the temporal auditory image refers to an arbitrary complex envelope of a single mode of the acoustic object, but the imaging condition says nothing about its duration, which is essential to elicit pitch. This situation is complex, because the perception of pitch entails sharpness. Also, increase in blur can erode the pitch of a sound, but not eliminate it. Therefore, there is a strong association between pitch and sharpness, but there are examples for pitched sounds that are not sharp (Huggins pitch is a clear one), and for sharp sounds that are not pitched (perhaps like a snappy impulse, such as a snare drum). Therefore, while it seems that the answer to this question is a cautious ``no'', a more exhaustive answer probably demands further research. As for the third question, linearly frequency-modulated tones (glides) do not have a fixed pitch and they are not periodic. However, they certainly elicit a perception of pitch, albeit a dynamic one \citep[pp. 206--207]{Cheveigne}. Therefore, the answer here is negative as well. 

Interestingly, the property of pitchiness, or \term{pitch strength}, is inversely proportional to the bandwidth of the signal \citep[pp. 135--148]{Fastl}, or maybe to its coherence time that is directly dependent on the bandwidth (Eq. \ref{CoherenceTimeEst}). So, tones have a very long coherence time, whereas broadband noise have a negligible one, and narrowband noise somewhere in between. It also indicates that the auditory system is configured to have complete incoherence only for full bandwidth inputs, which include several critical bands. The corollary is that a single channel may have residual coherence even with random narrowband noise, by virtue of its limited bandwidth. This is in line with the conclusions from literature review about apparent phase locking to broadband noise (\cref{PLLNoise}) and the discussion about temporal modulation transfer functions of partially coherent signals (\cref{NarrowTMTF}).

\section{Higher-order monochromatic auditory aberrations}
\label{HigherOrderAb}
\subsection{General considerations}
Basic spatial imaging harnesses the paraxial approximation, which requires light propagation in small angles about the optical axis and perfectly spherical or planar wavefronts. In realistic optical systems, these approximations are increasingly violated the larger the light angle is and when the various optical elements are imperfect---imperfections that are collectively called aberrations. The nonideal image exhibits various aberrations that can be studied as departures from ideal imaging. It can be done through wavefront and ray analysis, by comparing the path difference of different points along the same wavefront, as it propagates in space, which should have equal optical length in aberration-free imaging. In general, all imaging systems have a certain degree of primary higher-order aberrations and the eye is no exception---something that was already recognized by \citet[pp. 353--376]{Helmholtz1867}. In the design of optical systems, aberrations are eliminated or mitigated by balancing them with other aberrations in specific conditions, although this process results in the generation of yet higher-order aberrations \citep{Mahajan2011}. 

In wave optics, the geometrical optical wavefront analysis is elaborated by the inclusion of higher-order phase effects, beyond the quadratic phase terms that characterize the diffraction integral and the lens curvature. Similarly, in deriving expressions for the time lens and group-velocity dispersive medium, the phase functions used in the theory were expanded only up to second order that implied quadratic curvature (\cref{temporaltheory}), which can account for defocus and chromatic aberrations. The existence of additional phase terms that depend on higher powers of frequency or time would drive the channel response away from its ideal imaging \citep{Bennett2001}. However, because the dimensionality in temporal imaging is lower than in spatial imaging, not all of the known spatial aberrations have relevant temporal analogs. 

Because the auditory channels are relatively narrow and the aperture stop is very short, higher-order aberrations may be difficult to observe, or they may appear completely absent---something that reflects the paratonal approximation that is analogous to the paraxial approximation. It is not impossible that the normal functioning hearing system circumvents higher-order aberrations by having a dense spectral coverage with dedicated fine-tuned filters along the cochlea---each of which has diminishingly low aberration around its center frequency. In other words, the various auditory filters are optimal around their characteristic frequency but are overtaken by other filters away from it, off-frequency. An additional mitigating factor is that higher-order aberrations are severer for magnification values that are much different than unity $|M| \gg 1$ or $|M| \ll 1$ \citep{Bennett2001}, whereas our system is much closer to $M\approx1$. Finally, the defocus itself, which is a second-order aberration, may mask the smaller effects of the higher-order (third and above) aberrations. So for example, \term{spherical aberration} is symmetrical around the channel center frequency and generally results in increased blur away from the (spatial or temporal) image center, which may be masked in hearing. 

To the extent that higher-order aberrations are a real concern, they may also be difficult to identify using our pupil function and, hence, the point spread function analysis \citep[pp. 77--137]{Mahajan2011}, which we estimated only up to second order. While it was assumed for convenience that the auditory aperture is Gaussian, the single measurement that determined its shape directly, had an asymmetrical tail attached to the Gaussian from the right (the forward-time direction; \cref{TempAperture}). It can be expected to cause the point spread function of the system to have asymmetrical (odd) higher-order phase terms. 

The implications of having higher-order phase terms can be made more concrete by closely examining the dispersive elements of the system. If the phase curvature in the filter skirts is not exactly quadratic, then various asymmetrical dispersive aberrations analogous to spatial optics may exist---\term{coma} (third-order phase term in the Taylor expansion, Eq. \ref{eq:phase2}), and spherical aberration (fourth-order term) \citep{Bennett2001}. These terms may be detrimental to perceived sound quality when the excited channel is either over-modulated (reaching high instantaneous frequencies that should be normally resolved into multiple filters), or is more simply excited off-frequency---away from its center frequency\footnote{It can be argued that over-modulation and off-frequency excitation are essentially the same thing, only that over-modulation is typically symmetrical around the carrier (characteristic frequency), whereas off-frequency is generally asymmetrical.}. These situations entail that the auditory channel works beyond its paratonal approximation---well-beyond its center frequency, whose role is analogous to the spatial optical axis\footnote{Rays that cross the optical axis at the position of the aperture stop (\textbf{chief rays}) are aberration-free by definition. In analogy, the center frequency of the temporal channel is aberration-free.}. 

There is little physiological evidence that directly indicates that the phase curvature of the cochlear filters is asymmetrical away from the characteristic frequency, or even not perfectly quadratic. In contrast, a few psychoacoustical studies may be interpreted as showing such an asymmetry. We briefly mention evidence to the former and then focus on evidence to the latter and offer another example of our own to demonstrate this effect. 

\subsection{Physiological evidence}
\label{PhysioOffFreq}
There is mixed physiological evidence for an ideal quadratic phase response of the auditory system. In physiological recordings of frequency glides in auditory nerve fibers of the cat, the instantaneous frequency chirps were best fitted by linear functions that indicated a quadratic phase term only \citep{Carney1999}. In several instances the glides were not linear, but better fits could not be obtained using higher order regression, including those made with log frequency. Linearity in the instantaneous frequency slopes was generally observed in the barn owl as well \citep{Wagner}. In contrast, in impulse responses measured using different methods in the guinea pig, chinchilla, and barn owl, the slopes of the instantaneous frequency or the phase were usually linear only in part of the response, or they changed only well below the characteristic frequency \citep{deBoer1997,Recio1997,Heijden2003,Fontaine2014}. This may be indicative of some asymmetry in the phase response of these auditory channels. 

It should be noted that these measurements consider the auditory system to be dispersive only within the cochlea. This necessarily includes one segment of the neural dispersion (i.e., the path between the inner hair cells and the auditory nerve), which may have a complex phase function in itself. Additionally, these measurements tend to treat the phase response (and the filtering in general) as time-invariant, which is questionable if the outer hair cells produce active phase modulation. Therefore, we shall look for more qualitative psychoacoustic evidence of higher-order aberrations, which includes the complete dispersive path. 
 
\subsection{Psychoacoustic evidence}
If the normal auditory system has a temporal coma aberration, then it might be possible to observe its effect when a filter is excited asymmetrically (off-frequency) with an adequate modulation spectrum. With coma, point images are smeared asymmetrically to one side, which is also the reason for the name coma---it refers to the distinct comet-like smear of the affected image points in two dimensions. If a coma-free dispersive channel is completely uniform over its entire bandwidth, then, to a first approximation, its demodulated temporal image is invariant to how the envelope object is oriented about its center. \term{Distortion} is another type of aberration, which has not been considered in the temporal imaging literature, that is sensitive to the orientation about the center frequency. In spatial imaging it appears as uneven magnification of the image, so that its circumference is deformed in relation to its center, endowing the image with the familiar barrel or pin-cushion deformations. Unlike transverse chromatic aberration that also exhibits nonuniform magnification in the context of polychromatic imaging, distortion does not cause any blur within the monochromatic image. Just like the auditory spherical aberration, distortion aberration is presently impossible to estimate. However, the combined effect of these hypothetical temporal aberrations may be tested using off-frequency stimuli. Three examples were found in literature that support this idea are described below.

Direct psychoacoustic evidence that the phase curvature is not constant across the auditory channel has been provided by \citet{Oxenham2005} and \citet{Wojtczak2009}. These studies are near replications of the curvature estimation study by \citet{OxenhamDau} that was described in \cref{ModelSchr}. For Schroeder phase complex maskers centered on the 1, 2 and 6 kHz target tones, the threshold was minimized for a particular masking curvature, as shown previously. However, for maskers that were centered off-frequency, below the target frequency, there was no optimal curvature at 1 kHz and 2 kHz, and zero curvature at 6 kHz \citep{Wojtczak2009}, and also zero or not well-defined optimal curvature at 2 kHz in \citet{Oxenham2005} with a slightly different stimulus. In these studies, upward spread of masking was harnessed to set the maskers below the target frequency. Along with the physiological data mentioned above in \cref{PhysioOffFreq}, these studies suggest that the auditory channels suffer from coma aberration, or from another aberration or a combination of aberrations that can account for the phase curvature differences between the on- and off-frequency excitation. 

Another direct test case for off-frequency effects is found in hearing-impaired listeners with \term{dead regions}. These refer to either inner hair cells or auditory nerves that are completely dysfunctional and no longer conduct any information to the brain \citep{Moore2004}. Therefore, frequency bands that are normally transduced at the center of healthy channels are instead transduced by channels located at the edge of the region that is still capable of hearing. The response of the edge channels can thus provide a direct indication of how off-frequency inputs are perceived by the listener. The only caveat is that these channels are generally impaired as well---they have elevated thresholds and broader filters than unimpaired channels. In general, the farther the pure tone frequency is from the edge of the dead region, the more noise-like its sound seems, with less clear and less distinct pitch \citep{Huss2005a}. Different descriptions have been given by hearing-impaired listeners to describe their sensations for tones well within their dead regions: ``\textit{noise-like, distorted, hollow, without a body, very-soft, screechy}'' \citep[including descriptions that were quoted from previous studies]{Huss2005a}. This was contrasted with normal-hearing listeners who rated extreme frequencies below 125 Hz and above 12 kHz as noise-like and described the low frequencies as ``\textit{buzzing}''---especially at high levels. Otherwise, the normal-hearing listeners had clear perception of tones, especially at 500--4000 Hz, with slight degradation at higher and lower frequencies and high levels at these frequencies. The most compelling explanation for this impaired perception is that it corresponds to a discrepancy between temporal and place information received by the ear, since the place is determined by the edge frequency, but the temporal (periodic) information corresponds to the dead region's place. Irrespective of the coding difficulties, these channels were obviously driven well outside of their paratonal approximation, where the signal-to-noise ratio is poor and the phase linearity is in question. In a related study, the same coding difficulties also translated to difficulties in pitch matching between better and worse ears, or within the same ear \citep{Huss2005b}. The farther into the dead regions the tones went, the more erratic the pitch matching was. Once again, it cannot be determined from these experiments what kind of aberration(s) drive the responses to sound the way that they do. 

The final evidence for off-frequency aberrations may be deduced indirectly from the responses to frequency-shifted complex tones. \citet[Experiment 1]{Moore2003} presented a complex tone with unresolved harmonics that were shifted by a constant factor, to produce various degrees of inharmonicity. Unlike similar complex tone studies, the excitation pattern that was produced by the tone was fixed by controlling the amplitudes of the various harmonics (based on an excitation pattern model), so it remained about constant despite the spectral shift. Normal-hearing subjects had to match the pitch of the shifted complex tones to harmonic references of the same $f_0$ at a lower spectral region (lower harmonics), but almost identical envelope spectrum. Interestingly, the matched pitch of complex tones with $f_0$ of 100, 200, and 400 Hz was invariant to shifts of up to $0.24f_0$ in frequency (the UNRES conditions in Figures 2 and 3 and Tables II and III in \citealp{Moore2003}). The results were about the same with and without noise that was supposed to mask any combination tones that may have been used as cues. This means that in the vicinity of the complex tone and auditory filter center (specifically, the 16th harmonic in this condition), the perceived pitch was a function of the envelope spectrum alone and not of the carrier. However, away from the center frequency, for inharmonicity shifts larger than $0.24f_0$, subjective matches were deemed ``\textit{very difficult}'' in pilot studies and were not pursued further. This is reflected even in the results of the resolved conditions (centered around the fifth harmonic) that show performance leveling off and variance increasing between $0.16f_0$ and $0.24f_0$ with $f_0 = 400$ Hz \citep[Figure 2 and 3]{Moore2003}. Since these conditions contain both fine-structure and envelope cues but are still challenging for listeners, this brings about the possibility of a dispersive aberration---either symmetrical or asymmetrical.  


\subsection{Further psychoacoustic evidence}
\label{FurtherPsycho}
\begin{figure} 
		\centering
		\includegraphics[width=1\linewidth]{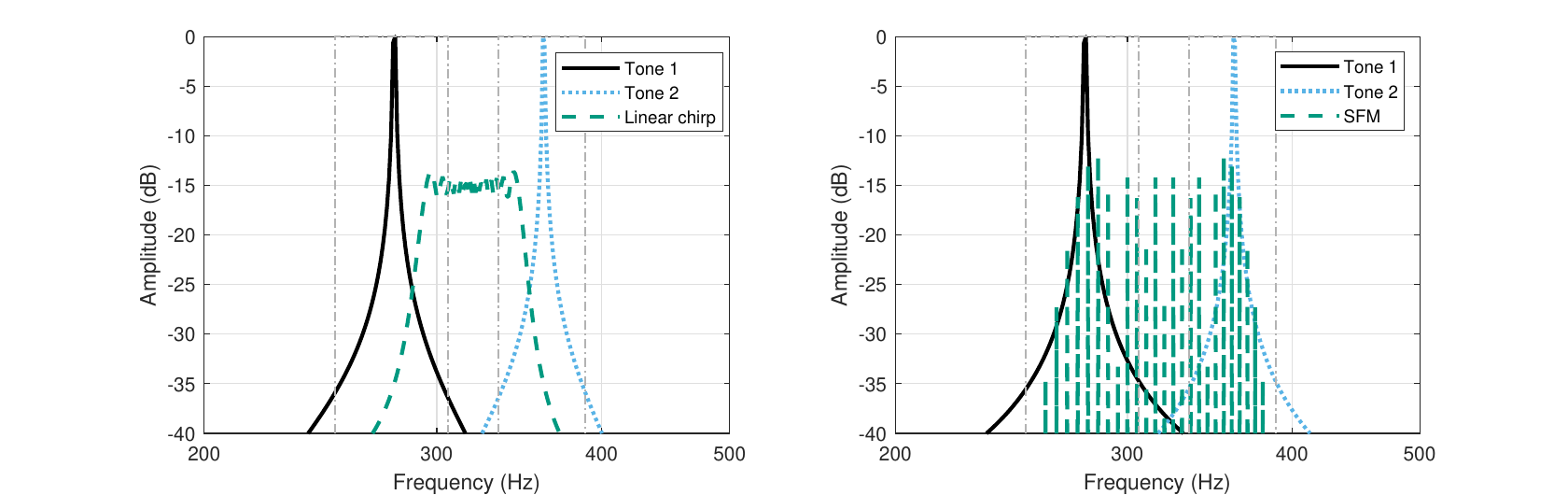}	
		\caption{Examples for the combined spectra of two tones and a linear frequency modulated chirp with slope $\Delta \dot{f} = 60$ Hz/s (\textbf{left}) or a sinusoidal frequency modulated (SFM) tone with $f_m = 5$ Hz, $\Delta f = 44$ Hz (\textbf{right}). In both cases, when played together, the FM sounds severely distorted, which may suggest a strong off-frequency phase distortion---a high-order aberration. The equivalent rectangular bandwidths of the filters centered at the tones are marked with dash-dot lines.}
		\label{OffFreqFM}
\end{figure}

The reviewed off-frequency psychoacoustic data strongly suggest that higher-order aberrations exist in human hearing. However, they are insufficient to confidently determine whether a particular aberration can account for the pitch encoding difficulties, especially since pitch is ultimately a higher-level percept. We propose a novel and simple stimulus that is based on the above analysis, which can be used to further probe the existence of off-frequency higher-order aberrations. A variation of the off-frequency paradigm \citep{Patterson1980} is used to test the effect of a linear or sinusoidal FM that is processed off-frequency. In the original paradigm, sounds are made to be processed off-frequency by noise that engages the neighboring auditory channels. Thus, the signal can be positioned anywhere in the notched part of the spectrum, which may or may not correspond to the characteristic frequency. This is based on the power-spectrum model of hearing and on the assumption that the filter center is located where the maximum signal-to-noise ratio is produced. Here, instead of using a notched broadband noise, which is a poor masker for FM, we shall engage two adjacent filters with pure tones, separated just enough to avoid any beating between them (about 1.5 ERB; see \citealp{Moore2002}). In between the tones, at a lower level, an FM signal is produced that occupies the off-frequency bandwidth between the engaged filters (see Figure \ref{OffFreqFM}). If the auditory filters indeed exhibit significant off-frequency aberration, then a distorting effect should be audible, the farther the FM carrier is from the center frequency of the passband. If in addition the stimulus does not sound identical (to the extent that it can be compared) below and above the filter center, then an asymmetrical aberration may be inferred. 

Several audio examples\footnote{The examples are found in audio demo directory \textsc{/Figure 15.20 - Higher-order aberration/}.} are provided of clear cases where FM signals and tones are distorted. It is the impression of the author that for a relative level that is just about lower than the tones, most FM signals are severely distorted in the presence of the tones with high-frequency carriers, whereas at low frequencies the distortion to linear FM is more obvious than sinusoidal FM. In general, at low frequencies, the auditory filters are relatively broad (in terms of their bandwidth to center-frequency ratio), but occupy a small absolute frequency range, whereas at high frequencies they are relatively narrow, but occupy a larger frequency range in absolute terms. Thus, high-frequency channels encompass more phase cycles, which make phase aberrations there more likely. Several alternative explanations may be brought up, such as a masking effect by difference tones, strong instantaneous interference between the FM and the pure tone sounds, or a complex suppression pattern between the tones and the FM. As can be gathered from \cref{SupraMasking}, these explanations are not necessarily in contradiction to the aberration one proposed here. It will be left for future experiments to determine---and for the reader to judge---which of the explanations is the most plausible.

At present, we shall retain the hypothesis that off-frequency higher-order aberration causes the loss of FM details. This is likely exacerbated by a mismatch between the phase aberration in the overlapping flanks of the two adjacent filters. The overall effect is blurring of the FM of the signal. It is possible that in typical acoustic objects and normal listening such extreme off-frequency responses are largely avoided due to the narrow bandwidth of the filters combined with lateral inhibition. Implications for the perception of temporal fine-structure in hearing-impaired listeners are discussed in \cref{HigherAbbImpair}.

\section{Chromatic aberration}
\label{ChromaticA}
Polychromatic images may be subject to \term{chromatic aberration} that can give rise to a distinct type of blur. In spatial imaging, it occurs as a consequence of dispersion (also called in this context \term{chromatic dispersion})---the dependence of the speed of light on wavelength in the optical medium. There are two main types of chromatic aberration. In \term{axial} or \term{longitudinal chromatic aberration} the focal length depends on the wavelength, so the different monochromatic images do not align in the same plane and are therefore not all equally sharp. In \term{transverse} or \term{lateral chromatic aberration}, magnification itself is wavelength dependent, so that off-axis images in different wavelengths do not exactly overlap (\citealp[pp. 24.17--24.19]{CharmanBass1}; \citealp[pp. 15.22--15.23]{MillerBass3}). Because of the relatively large bandwidth of visible light, axial chromatic aberration in the normal eye causes a shift in focus of 2.25 diopters between the short and long wavelength range of light \citep[p. 58]{PackerShevell}, whereas lateral chromatic aberration is 36 arcsec \citep[p. 24.18]{CharmanBass1} or $\pm3$ arcmin \citep[p. 15.23]{MillerBass3}, depending on how it is measured. While the ocular chromatic aberration probably does not have a perceptible blurring effect in normal vision \citep[pp. 59--60]{PackerShevell}, axial chromatic aberration must be compensated for in the design of virtually all polychromatic optical instruments. 

\begin{figure} 
		\centering
		\includegraphics[width=0.85\linewidth]{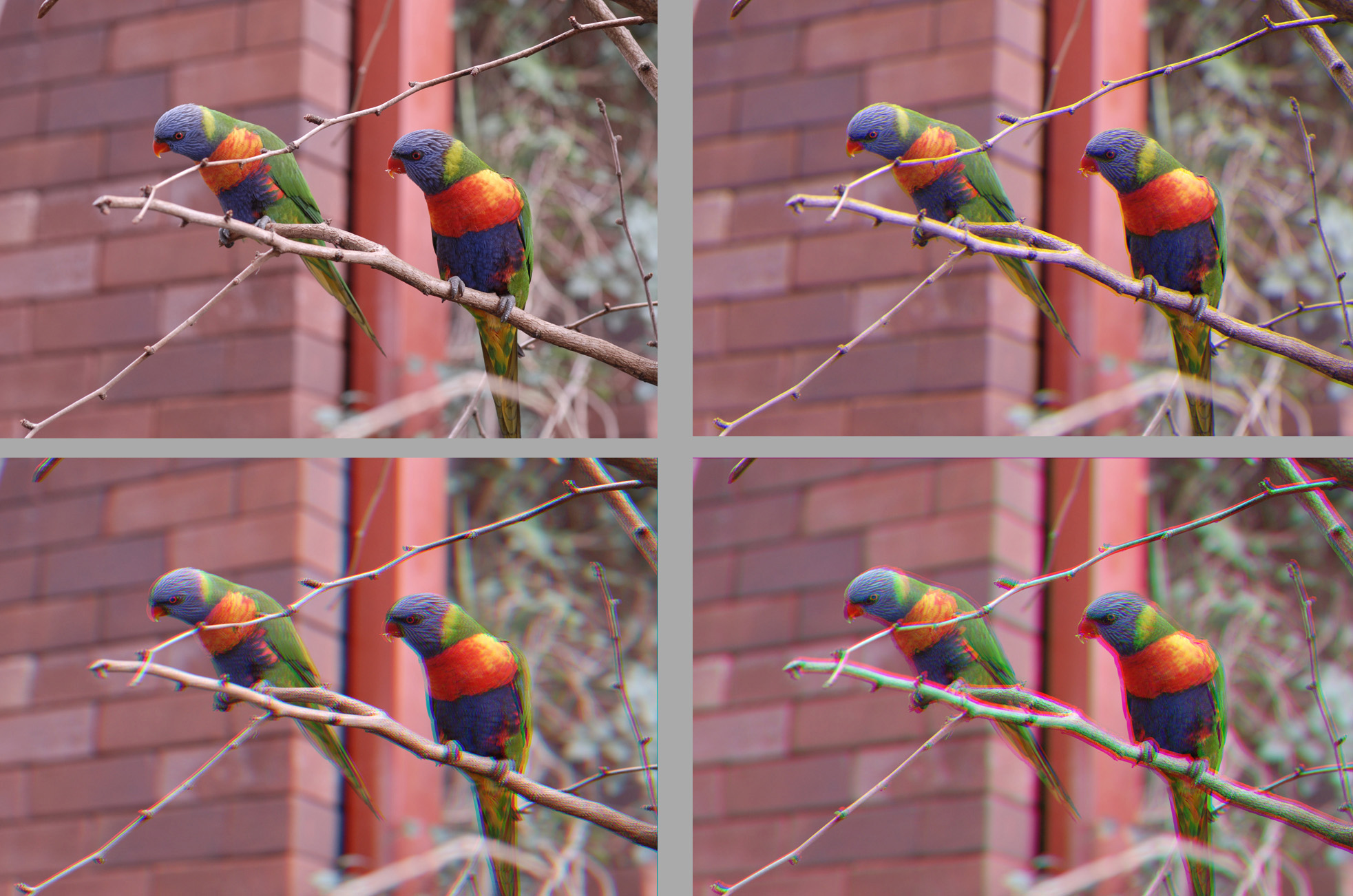}
		\caption{Blur caused by chromatic aberration. The polychromatic image on the \textbf{top left} is obtained by exactly aligning the three color channels (shown in Figure \ref{polybirds}). By translating the monochromatic images by a few pixels relatively to each other, a gradually increasing blur may be observed, where the \textbf{top right} is slight blur (green 1 pixel to the right, 4 pixels up; red 2 right, 4 up), the \textbf{bottom left} medium blur (green 1 up, 4 left; red 1 up, 9 left) and \textbf{bottom right} most blur (green 6 left, 2 down; red 1 right, 4 up). Note how the effect of blur is strongest for the objects that are already in focus, whereas the effect on the out-of-focus background is much subtler. This figure is a spatial analogy to the temporal chromatic aberration. However, its blurring effect also resembles transverse chromatic aberration, which produces a distinct colorful halo around objects in the image. For example, on the bottom right photo, the twig on which the birds are sitting has a pink halo from above.}
		\label{polybirdsaber}
\end{figure}

In temporal imaging, the dispersive effect causes envelopes that are carried by different frequency bands to be delayed in a frequency-dependent way. Thus, in broadband signals that are subjected to frequency-dependent dispersion, the envelope contents of different components do not arrive together. The corresponding form of chromatic aberration may therefore take place when the image of the polychromatic complex envelope is not temporally synchronized across channels. We shall refer to it as \textbf{temporal chromatic aberration} (cf. this term in ultrashort pulse optics, \citealp{Andreev2008}, and in motion vision, \citealp{Mullen2003}). An analogous visual effect is illustrated in Figure \ref{polybirdsaber}, where gradually increasing amounts of blur are produced by translating the monochromatic images (from Figure \ref{polybirds}) by a few pixels, relatively to each other.

In most optical systems that employ ultrashort-pulse temporal imaging, it is unusual to have an ultra-wideband spectrum as is common in hearing (see \cref{TwoSpectra}), so chromatic aberration has not been formally analyzed. In spatial imaging, the dispersive and diffractive effects operate on different dimensions, so there are more degrees of freedom than in temporal imaging, also with regards to aberrations. In temporal imaging, frequency-dependent dispersion is the cause of group-delay dispersion, so some level of chromatic aberration is almost inescapable. Particularly, the very existence of defocus is indicative of its temporal chromatic aberration. This is because only a sharply focused system has a total group-velocity dispersion that is zero, which entails constant group delay. In contrast, the defocused system has a non-zero group-delay dispersion, which entails a non-constant group delay (Eq. \ref{GDvsGVD}). Because of the inherent defocus in the system, the role of axial chromatic aberration is not expected to be critical insofar as the individual channels are not sharply focused\footnote{Interestingly, a beneficial role in cephalopods (octopus, cuttlefish, and squid) of inherent axial chromatic aberration has been hypothesized and demonstrated in simulation, where the amount of blur along the optical axis can provide sensitivity to color in an otherwise monochromatic-photoreceptor visual system \citep{Stubbs2016}.}. However, the two other types of chromatic aberration will be explored below. Chromatic aberration will be a necessary building block in the analysis of hypothetical dispersive hearing impairments, where more results from literature will be analyzed (\cref{impairments}). 

\subsection{``Transverse'' chromatic aberration}
\label{TransChromAb}
\begin{figure} 
		\centering
		\includegraphics[width=1\linewidth]{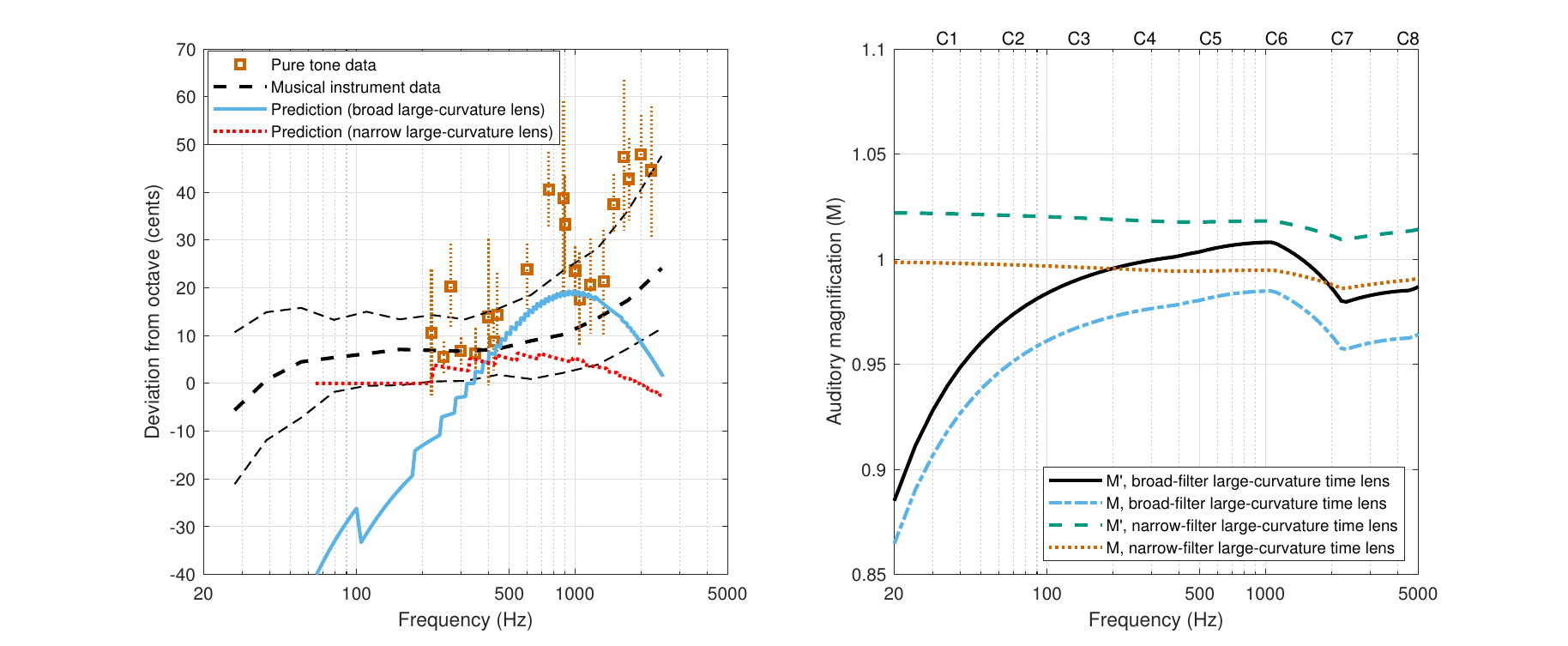}
		\caption{Magnification and stretched octave data. \textbf{Left:} Stretched octave psychoacoustic data and prediction reproduced from \citet[Figure 1]{Jaatinen}. The red squares and confidence intervals represent pure tone data compiled from literature by \citet{Jaatinen}. The thick black dashed line is the threshold to their simulated orchestral instrument (complex tone) sounds, including its confidence intervals in thin dash black. The prediction according to auditory magnification is plotted twice, once for the broad-filter (large-curvature) time lens (solid blue) and another for the narrow-filter in dotted red, both according to Eq. \ref{MagStretch1}. \textbf{Right:}  The auditory magnification curve replotted from \ref{lenscurve}, for both filter types in the large-curvature time lens. Musical notes are marked on the top abscissa between C1 and C8 for reference.}
		\label{FigStretched}
\end{figure}

Up until this point, auditory magnification has not been given any explicit significance. However, its numerical proximity to unity may make its effect elusive. While the variation in magnification in most of the audio spectrum is not large---it varies by about 14\% or less over three decades according to our estimates (reproduced in Figure \ref{FigStretched}, right), it can be expected to influence across-channel timing. This is because the local time variable is scaled by the channel magnification (even for a stationary pure tone) according to $\tau \rightarrow \tau/M$, as dictated by both the focused and defocused imaging transforms of Eqs. \ref{eq:image} and \ref{absonlyfull2}, respectively. If pitch detection is a temporal process or is anyway conveyed by a local time variable, then even slight changes in magnification may lead to a discrepancy between the perceived pitch and the physical frequency prediction, which is scaled as the reciprocal of the time, $f \rightarrow Mf$ (see also \citealp{Bennett2001}, Eq. 27). This discrepancy enables us to tap into the effects of transverse chromatic aberration. 

The magnification variation in frequency may be applicable in the analysis of the \term{stretched octave} phenomenon. In order to obtain a subjective octave perception of double the pitch of a two-note sequence\footnote{The two-tone sequence refers to a \term{melodic octave}, as opposed to a simultaneous \term{harmonic octave}, which is generally perceived more closely to the physical 2:1 ratio \citep{Demany1990,Bonnard}.}, it is necessary to stretch the octave tuning to a slightly larger ratio than 2:1 \citep{Ward1954}. More specifically, the stretching effect levels off at 200--400 Hz of the reference tone (i.e., the lower note) and is stronger at high frequencies, but seems to be reduced with complex tones \citep{Jaatinen}. These relations are summarized in the left plot of Figure \ref{FigStretched}, which reproduces pure-tone data compiled from literature by \citet[Figure 1]{Jaatinen}, as well as from simulated musical instrument octaves across the musical range from the same article. Stretched intervals are so ingrained in musicians' hearing that \citet{Jaatinen} advocated for adapting the stretched tuning curve as a new standard that can replace the usual equal-tempered tuning\footnote{The effect is considered to be a likely cause for the standard practice in piano tuning to have octaves stretched, although another potential cause may be an attempt to reduce unwanted beating from the inharmonic partials of the piano \citep[pp. 388--390]{Railsback1938,FletcherRossing}.}. 

Several hypotheses have been proposed that account for the stretched octave effect. They include an excitation pattern place model that is centrally learned just like speech \citep[e.g.,][]{Terhardt1974}, an auditory-nerve neural firing temporal model attributed to a frequency-dependent refractory period \citep{Ohgushi1983, McKinney}, and a cochlear model based on the geometry of outer hair cell row arrangement \citep{Bell2019}. A study by \citet{Hartmann1993} of binaural Huggins pitch revealed the same octave stretched response as in diotic pure-tone tests, which led him to propose a modified neural timing model, based on autocorrelation. However, this model could not be corroborated using cat data \citep{McKinney}. This, as well as findings by \citet{Bonnard} and \citet{Demany1990}, disfavor the learned-template place model of \citet{Terhardt1974}, but do not explicitly rule out a central autocorrelation model. Additionally, \citet{McKinney} reexamined Ohgushi's model and emphasized that the timing deviations that were attributed to refraction by \citet{Ohgushi1978} may stem from an earlier (cochlear) cause.

A stretched octave theory should be also able to explain the difference between the pure tone and the complex tone values. It may be similar in nature to the difference between the melodic and harmonic octaves: as multiple partials vibrate together, there is a clear subjective tendency to prefer the ``simple'' over the stretched frequency ratio \citep{Bonnard}. This may represent a trade-off between the two pitch dimensions---\term{tone chroma} that is relevant for harmonic relations and \term{tone height} that is relevant to pure tones across the audible spectrum \citep{Bachem1950,Warren2003}. 

A prediction based on the magnification ratio between the reference frequency and its octave, which are functions of local time variables, can be easily generated by looking at an auditory $f_1$:$f_2$ magnified interval, such that the two frequencies $f_2 \approx 2 f_1$, which satisfies the relation:
\begin{equation}
	2f_1 M(f_1) = f_2M(f_2)
	\label{basicinterval}
\end{equation}
where the magnification $M(f)$ explicitly depends on frequency. This equation assumes that the locally perceived (magnified) frequencies have to be mathematically doubled in value in their internal auditory representation, in order for the interval to be perceived as having a true octave relation. Ideally, the magnification is exactly unity and $f_2 = 2f_1$. In reality, there is a discrepancy, because $M(f_1) \neq M(2f_1)$, so a different frequency $f_2>2f_1$ has to be found to satisfy Eq. \ref{basicinterval}. For convenience, we can define the deviation between the ideal and the real frequency using $\Delta f = f_2 - 2f_1$. Typically, the results in the cited studies are expressed in cents\footnote{One octave is logarithmically divided into 1200 cents, so a semitone is 100 cents.}. Therefore, we take the logarithm base 2 of Eq. \ref{basicinterval}, multiply by 1200, and after some rearranging obtain
\begin{equation}
	1200\left(\log_2 \frac{f_1}{f_2} + 1 \right) = 1200 \log_2 \frac{M(f_2)}{M(f_1)}
\end{equation}
Using $\Delta f$, we can obtain an expression that depends only on $f_1$, albeit indirectly through $M$
\begin{equation}
1200 \log_2 \frac{M(f_1)}{M(2f_1 + \Delta f)} = 1200\left[\log_2 \left( \frac{\Delta f}{f_1} + 2 \right) -1 \right]
	\label{MagStretch1}
\end{equation}
A graphical solution may be found by solving both sides of the equation for different $\Delta f$ around $2f_1$. Note that only when the magnification is identical at the two frequencies, then $\Delta f = 0$. Note also that the magnification used may be either the general expression $M = (v+s)/s$ (Eq. \ref{MagnificationDef}) or the effective magnification for the defocused pulse $M'$ (Eq. \ref{altM}), which also depends on the aperture size $t_0$ and cochlear group delay dispersion $u$, but produces the same predictions. 

\citet{Ward1954} and \citet{Jaatinen} emphasized and verified the requirement of interval additivity, which entails that the sum of the tuning factors for two consecutive intervals $f_1$:$f_2$ and $f_2$:$f_3$ must be equal to $f_1$:$f_3$. Using the magnification rule above, this requirement is automatically satisfied by virtue of having the frequency scaled by a multiplicative factor. If the first two magnified intervals are set according to
\begin{equation}
	I_1 = \frac{M_2f_2}{M_1f_1} \,\,\,\,\, I_2 = \frac{M_3f_3}{M_2f_2} 
\end{equation}
then their combined interval is
\begin{equation}
	I_3 = \frac{M_3f_3}{M_1f_1} = \frac{M_3f_3}{\frac{M_2f_2}{I_1}} = I_1 I_2
\end{equation}
which is a sum in the log cent scale, as interval additivity requires.

To the extent that the prediction correctly captures the stretch effect, it is limited in frequency to 200--500 Hz for the narrow-filter large-curvature time lens (red dotted curve, Figure \ref{FigStretched}, left), where the predicted effect is at most 5 cents. At higher frequencies, this curves remains flat and then decreases, opposite to empirical data where the stretch shoots up above 500 Hz. Similarly, the broad-filter large-curvature time lens roughly captures this stretch size increase at 500--1000 Hz, but completely misses the target outside this range. As the predicted octave stretching effect depends on the magnification $M$, it is effectively dependent on the time-lens curvature $s$ and on the neural dispersion $v$ (but not on the cochlear dispersion $u$). The uncertainty we have in the estimates of $s$, which has hitherto been of negligible importance, becomes particularly noticeable in modeling this phenomenon, as two different predictions in Figure \ref{FigStretched} show. Note that at frequencies lower than 220 Hz, no empirical data is available (for pure tones), and the magnification values are also likely to be off, being based on untested assumptions and extrapolations (\cref{OHCtimelens}). The source of discrepancy above 1000 Hz is less certain and may be the result of mis-estimated $s$, $v$, or both. The other time lens estimates (the constant focal-time and small-curvature) all produced predictions that are completely off target and are not displayed here. 

The temporal imaging theory considers the dispersive function of elements that are both peripheral and central, and requires both place and temporal theories to interact, in similarity to modern pitch theories \citep[e.g.,][]{Cheveigne,Moore2013,Oxenham2022}. Place information determines the rough identity of the narrowband channel, whereas temporal information conveys the dynamics of the modulation variables. The magnitude of the magnification itself is dependent on the time lens and neural group-delay dispersion, but is negligibly dependent on the cochlear group-delay dispersion. Inasmuch as the magnification implicates the perception of pitch height, then both peripheral and central modeling have some merit, as the effect of the fine details of the dispersive elements may be factored into their group dispersive properties (e.g., a frequency-dependent neural refractory period). As temporal imaging is a coarse-grained approach to the auditory system, it is agnostic to the fine-grained explanations of the stretched octave that were mentioned above. 

The smaller stretch factor required in complex tones may be explained using mode locking, rather than with dispersion. Mode locking causes an inharmonic overtone series to synchronize into a harmonic series. If a similar mechanism occurs in the auditory brain, then it may contribute to a significantly smaller stretch factor in complex tones. (See \cref{Phaselocking} for discussion\footnote{Note that we distinguished this type of mode locking from other harmonic synchronization effects that were measured in the auditory brainstem and were also referred to as mode locking in the hearing literature.}).

Incidentally, the connection between pitch height and magnification may indirectly support the existence of a time lens in the system---something that was questioned in \cref{PinholeArch}, given the similarity of the system to a pinhole camera design. This is important because the effect of the lens appears to be almost negligible in most other contexts (e.g., curvature and modulation perception), while it dominates magnification and keeps it at near unity level (Fig. \ref{imagingcond}). The magnification analysis may be a useful tool for calculating the individual human time lens properties indirectly, once the individual neural group-delay dispersion is known. For example, the stretched octave effect and data are used in \cref{PsychoEstimation} along with three additional psychoacoustic effects to derive a strictly psychoacoustic estimation of all the dispersion parameters in the auditory temporal imaging system. That calculation was designed to perfectly match the empirical data. It produces a magnification that is much closer to 1 below 1000 Hz ($M>0.966$), which decreases more rapidly above. However, no estimate is available below 125 Hz.

The magnification analysis in conjunction with pitch perception lends itself to another interesting hypothesis, which is more general and is likely relevant even if the precise values of the magnification are still uncertain and do not produce satisfactory predictions. As is seen in Figure \ref{FigStretched}, at least one magnification estimated (based on the broad-filter curvature) is relatively flat in a narrow frequency range (200-1000 Hz), but it decreases more sharply in the bass and treble ranges. Comparing the frequency range in which the magnification is flat and close to unity with the melodic range in music (roughly C2--C6, where C4 is Middle C is set at 261.6 Hz), begs the question of whether the melodic range itself exists in the region of the flattest magnification. This range roughly contains the vast majority of melodic and harmonic instruments and spans the conventional ranges of bass, tenor, alto, and most of the soprano voice types. 

Other side-effects of the auditory transverse chromatic aberration may exist that are not necessarily related to pitch, but depend on the magnification distortion of the local time variable.  

A challenge that can be leveled against the magnified frequency explanation to the stretched octave effect is that it is not reflected in physiological measurements, which directly quantify the period in the auditory nerve and other pathways. These responses are locked to the input over the (usually long) time frames in which the neural activity recording takes place. In order for the magnification stretch and the long-term frequency synchronization to coexist, it entails that any scaling takes place on the image (sample) level that should correspond to perceived pitch, and/or is limited to shorter time frames. However, there is no evidence at present to support these ideas and more research will be needed to uncover the exact relations between these levels of operation.

\subsection{Temporal chromatic aberration}
\begin{figure} 
		\centering
		\includegraphics[width=.7\linewidth]{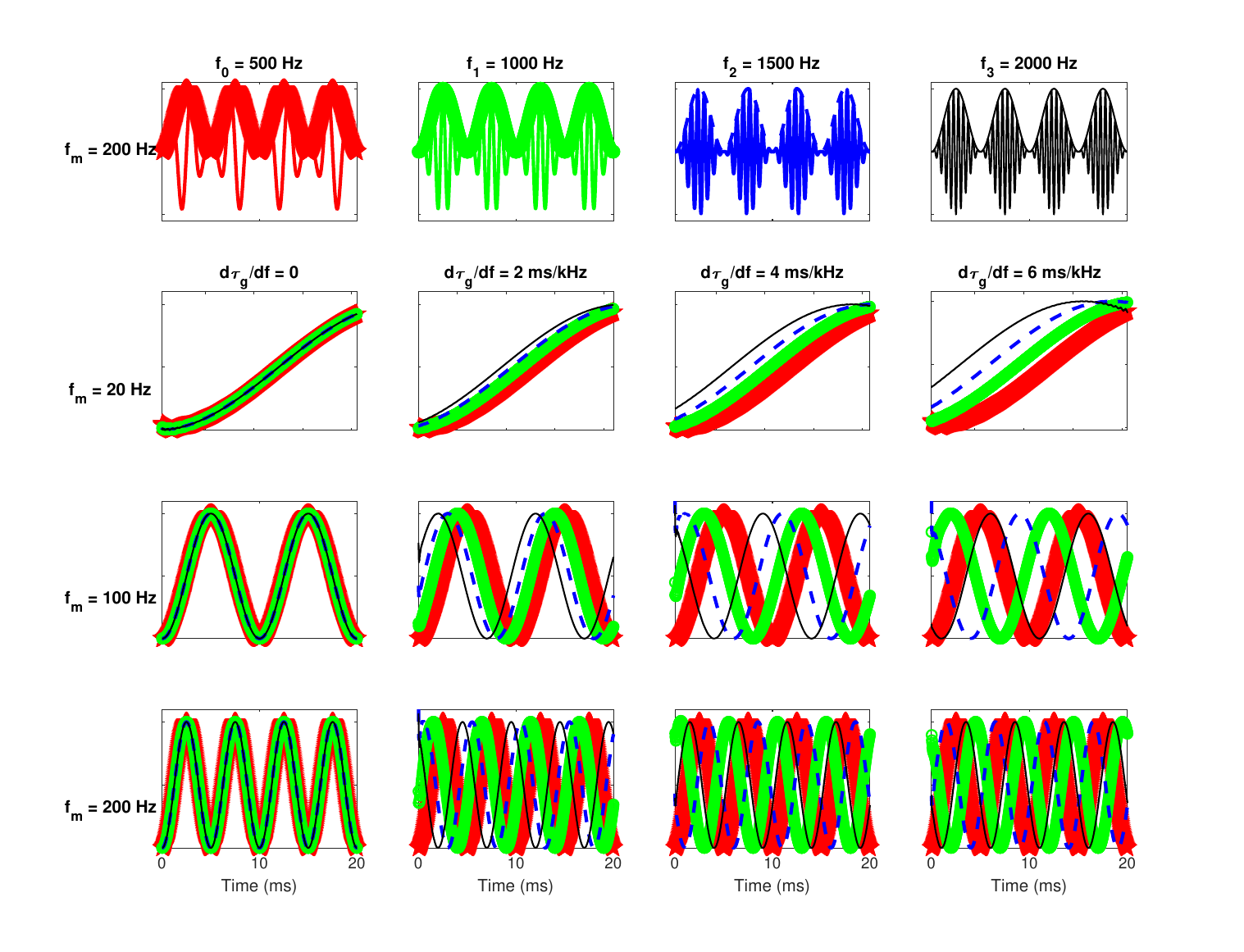} 
		\caption{The effect of temporal chromatic aberration on an amplitude-modulated complex tone with $f_0 = 500$ Hz and its first three harmonics. The four components (carriers and envelopes) are plotted separately in the top row of the figure, each with its own color and line-style to emphasize the analogy with spatial chromatic aberration. The image is the intensity envelope of the individual channels, indicated only with colors, but no carriers, which are assumed to have been removed after demodulation. When the group delay is identical in all channels (leftmost column), the polychromatic envelope shape is well-defined. As the group delay is increased (left to right), the blurring effect becomes more apparent, in an increasing manner with higher modulation frequencies (from top to bottom).}
		\label{chromatic}
\end{figure}

The effect of temporal chromatic aberration on pulses is illustrated in Figure \ref{chromatic}, as a simplified analogy to the spatial example of Figure \ref{polybirdsaber}. The figure shows the components of a modulated harmonic polychromatic tone, whose intensity envelopes are carried in four channels. The synchronized (coherent) version where the tone overlaps in all frequencies is the sharpest polychromatic image, whereas the aberrated version shows a blurring effect. Temporal chromatic aberration is nearly synonymous with group-delay distortion, which is known from audio engineering and is considered undesirable \citep[e.g.,][]{Blauert1978, Flanagan2005}, and in some cases is directly redressed by dedicated corrective filters (e.g., for the mixing sweet spot in a control room) that boast improved stereo imaging. However, group-delay distortion is rarely discussed with respect to its blurring effect (but see \citealp[p. 6-27]{TalbotSmith}). 

Just as in vision, it is necessary to distinguish between the intrinsic chromatic aberration of the auditory system, and the system's sensitivity to chromatic aberration induced by external factors. Evidence for the internal level of chromatic aberration may be gathered, for example, from click auditory brainstem response (ABR) dispersion, which is typically attributed to cochlear dispersion only, but actually accounts for the dispersion along the entire system. \citet{Don1978} and \citet{Don1998,Don2005} isolated the contributions of different frequency bands to the broadband click ABR wave-V response and found 1--5 ms latency differences between high and low frequencies, depending on level and subject. Similar values may be gathered from the group delay measurements that were summarized in Figure \ref{neuraldisp} (left) based on several studies (Table \ref{OAEABRdata}), which were used for the derivation of the cochlear and neural group-delay dispersion.

The internal chromatic aberration pattern that exists in normal hearing under normal acoustic conditions may be assumed to be perceptually well-tolerated by design. This is supported by several psychoacoustic studies that show how clicks rather than chirps are perceived as the most compact stimuli \citep{Uppenkamp}, perhaps through central compensation of otherwise asynchronous channels \citep[e.g.,][]{McGinley2012}. A similar conclusion may be drawn from a study that found how the synchrony of two tones was perceived to be maximal when they were gated simultaneously, independently of the spectral distance between the two tones \citep{Wojtczak2012}. According to the pulse ribbon model, global (across-channel) phase misalignment is corrected centrally as long as the misalignment between channels is relatively small (4--5 ms across the bandwidth; \citealp{Patterson1987phase} and \cref{AuditoryPhaseReview}). According to this model, only relatively large global phase effects could be heard. 

The sensitivity to external chromatic aberration may be deduced in part from available psychoacoustic data. As was mentioned above, group-delay distortion threshold measurements essentially quantify the same function, but without an explicit reference to the temporal imaging quality and the effect on modulation. Relevant studies typically employed various all-pass filter designs, which were used to process the phase of broadband impulses, but only over a narrow spectral range \citep[see review in][]{Moller2007}. The results varied depending on the specific methods, but showed group delay detection thresholds between 0.5 and 2 ms, with 1.5 ms being a typical average value that is relatively frequency-independent. The effect on the impulse sound was described as increased ``pitchiness'' or extended ringing. However, when applied to music or speech, the effect is often inaudible, probably due to masking \citep{Moller2007}. Nevertheless, an across-spectrum gradual delay was applied to speech in another line of studies, where tolerance in terms of intelligibility was assessed by delaying the large spectral ranges in a frequency-dependent manner. Performance is much more robust than may be inferred from the impulse response studies, as intelligibility dropped only with delays longer than 60 ms across the spectrum \citep{Arai1998}. High frequency bands, however, deteriorated more quickly with increasing delays with performance flooring at 240 ms. In another experiment where only two narrow bands of speech were presented, the performance was shown to depend much more critically on timing information and delays as short as 12.5 ms already had a strong effect \citep{Healy2007}. 

Audio examples that demonstrate the temporal chromatic aberration are provided in the audio demo directory \textsc{/Section 15.10.2 - Chromatic aberration/}, where a male speech excerpt was bandpass-filtered into seven bands, which were then summed back together, compensating only for the internal group delay of the center frequency of the filters. Other versions progressively applied differential group delay to the different bands before summing them up, in a way that sounds more and more objectionable, when the group delay is long enough. Very short group delays, however, are almost inaudible, in line with psychoacoustic studies. At relatively long group delays, the ringing effect \citep{Moller2007} can be heard more as a ``chirpiness'', because it is not confined to a single frequency. Intermediate group delays tend to sound more metallic. These group delays, however, do not seem sufficient to significantly degrade intelligibility of anechoic speech in quiet. It requires about 50--100 ms of delay between consecutive channels to render the speech unintelligible, albeit in a very artificial way. 

\section{Auditory depth of field?}
\begin{figure} 
		\centering
		\includegraphics[width=1\linewidth]{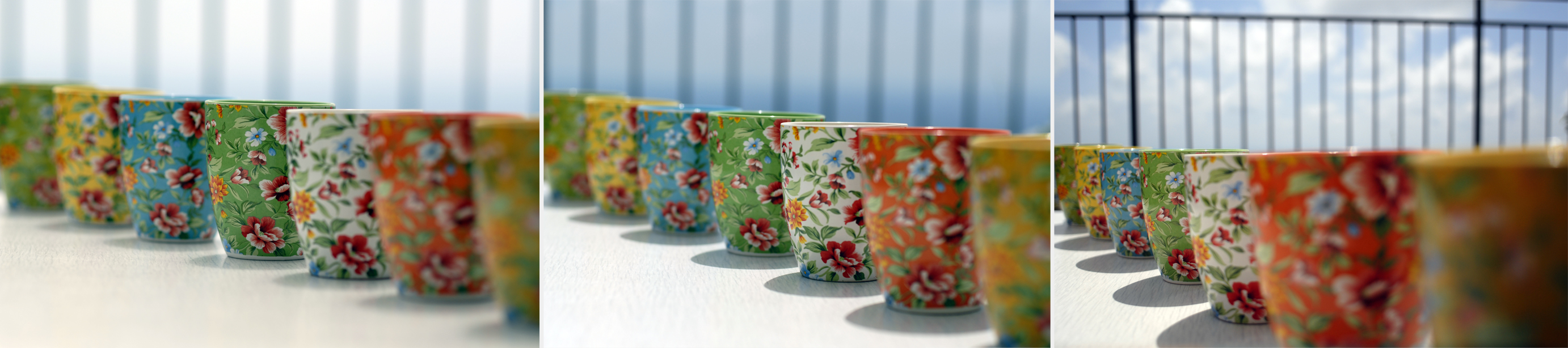}	
		\caption{Depth of field as a function of focal length of the lens. In the photos, the sharp focus is set to be on the middle (green) cup or on the white cup in front of it. As the focal length is \textbf{decreased} (from left to right), the depth of field \textbf{increases}, which can be seen in the cups that are most distant from the lens, which are less blurry and in the background---the sky and the metal railing. Note how the visual attention is automatically drawn to the middle cups in the left photo, which are in sharp focus. This attentional effect is visible but weaker in the middle photo, but not so much in the right photo, which is perceived more as an aggregate and is also dominated by the magnified and blurry cups in the front. The region of sharpest focus and its surrounding blur can be clearly seen in the marks on the tabletop, which demonstrate how the depth of field increases for smaller focal length (on the right). The photos were taken with constant (nominal) f-number of f/2.8 and lenses of 85 mm, 50 mm, and 24mm focal length (from left to right, respectively). Additionally, in order to conserve the angular extent of the image, the camera was brought closer to the objects for smaller focal lengths, to obtain approximately the same image.}
		\label{SpatialDOFLens}
\end{figure}

The visual depth of field quantifies the distance range (from the lens) within which an object can be positioned  in order to produce a sharp image (see examples in Figures \ref{DoF} and \ref{SpatialDOFLens}). Blurred objects in a defocused spatial scene encroach into neighboring objects, as their contours and fine details blend in and the overall contrast in the corresponding area of the image is lost---depending on the amount of blur. For example, a close inspection of the tabletop texture in Figure \ref{SpatialDOFLens} reveals how when the texture lines become blurry close to the lens they also become broader and fainter, until they completely disperse into a featureless surface. In contrast, in the focused area of the image, points of the object do not visibly encroach into the space of neighboring points, which remain well-resolved (by definition). 

According to the space-time duality, if spatial depth of field manifests in space as a function of the spatial envelope that represents the object features, then the temporal depth of field should manifest temporally through changes to the temporal envelope. Similarly, the encroachment in space beyond the ideal-image locus due to spatial blur can be analogized to temporal features that encroach in time beyond the local time of their own coordinate system. 

The spatial depth of field is determined by the focal length of the lens, the distance from the object to the lens, and by the relative aperture size---the f-number, $f^{\#} = f/D$, where $f$ is the focal length, and $D$ is the aperture size (\cref{GeometricalOptics} and Eq. \ref{eq:DoF}). Two parameters have an immediate counterpart in temporal imaging---focal time and f-number (see \cref{AudFNum}). A distance counterpart is more complicated, because it is the total amount of dispersion that we care about, irrespective of the distance over which it builds up. Also, the range of allowable group-delay dispersion is unintuitive and is not mapped well using the space-time analogy. Instead of a range of distances, we would ideally prefer to obtain a time interval. A simple manipulation of the imaging conditions can provide us with a suitable quantity (\cref{AudDepth}). Finally, we have to consider coherence as a parameter that is crucial in the auditory depth of field, but has no practical significance in vision (at least not in daily circumstances and natural light). We will consider this factor in \cref{NonSimultDepth}. We will return to the question of focal time in \cref{MOCR}.

\subsection{The auditory f-number}
\label{AudFNum}
An analogous temporal-imaging expression to the f-number that was proposed by \citet{Kolner} is simply
\begin{equation}
	f_T^{\#} = \frac{f_T}{T_a}
	\label{AudDepthEq}
\end{equation}
where $f_T$ is the focal time and $T_a$ is the aperture time. This number quantifies the inverse of the relative aperture time---how much of the signal energy enters the system for its particular configuration. 

\begin{figure} 
		\centering
		\includegraphics[width=0.4\linewidth]{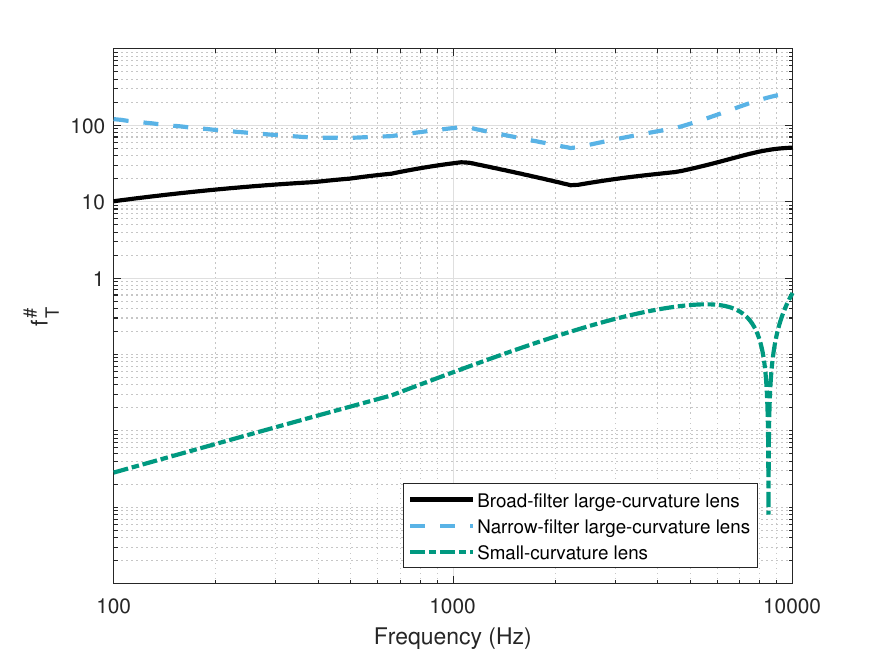}	
		\caption{Estimated auditory f-number, based on three different time-lens curvatures: the broad-filter and the narrow-filter large-curvatures that have been used throughout the text, and the small-curvature time lens (the absolute value f-number displayed) that has been largely omitted from analyses due to its implausible values (\cref{lenscurve}).}
		\label{AuditoryDOFx}
\end{figure}

The auditory f-number according to Eq. \ref{AudDepthEq} is plotted in Figure \ref{AuditoryDOFx} for three lens curvature estimates we obtained in \cref{lenscurve} and the aperture time from Table \ref{t0opts}. In spatial imaging, the f-number represents the relative irradiance (intensity per unit area) that is available for the image and it should have a similar logic in hearing too. In the case of the auditory system, the combination with neural sampling and the fact that we have no reference for this number in other acoustic systems makes it somewhat difficult to interpret.  More importantly, since the auditory system is defocused and it is governed by the signal coherence, the usefulness of the f-number concept is not exactly clear at this stage.

\subsection{Auditory depth of field (time)}
\label{AudDepth}
The spatial depth of field is expressed as a range of distances for the imaging system, when it is in sharp focus. While we do not have a sharply focused system, we can use it as a starting point from which it may be possible to obtain a temporal quantity instead of group-delay dispersion. Starting from the imaging condition:
\begin{equation}
\frac{1}{u} + \frac{1}{v} = -\frac{1}{s}
\end{equation}
we can express $s$ with the focal time (Eq. \ref{TimeLenss}) and $u$ and $v$ with their group delay derivatives (Eq. \ref{GDDder}) to get
\begin{equation}
	\frac{2}{\frac{d\tau_{g,u}}{d\omega}} + \frac{2}{\frac{d\tau_{g,v}}{d\omega}} = -\frac{2\omega_c }{f_T}
\end{equation}
where $\frac{d\tau_{g,u}}{d\omega} = 2u$ and $\frac{d\tau_{g,v}}{d\omega} = 2v$. This expression entails
\begin{equation}
	\frac{1}{\omega_c \frac{d\tau_{g,u}}{d\omega}} + \frac{1}{\omega_c\frac{d\tau_{g,v}}{d\omega}} = -\frac{1}{f_T}
\end{equation}
All terms of this expression have the dimensions of time and are determined by the center frequency of the time lens. As long as we stick to the narrowband approximation, this expression may be approximately correct also for off-frequency conditions. In the cochlea, however, we have assumed throughout the text (without proving) that time lensing is available on-frequency throughout the audible range. If this is so, then $\omega_c$ can continuously vary with center frequency. Otherwise, it should be fixed to the channel carrier that is determined by the time lens. In analogy to spatial optics, in a sharply focused system that satisfies this condition, the depth of field should be defined for an interval of $\omega_c \frac{d\tau_{g,u}}{d\omega}$ for which the image appears sharp. While this expression can endow us with the confidence that the depth of field can be expressed as a temporal range, it is not necessarily straightforward to apply in practice.

\begin{figure} 
		\centering
		\includegraphics[width=1\linewidth]{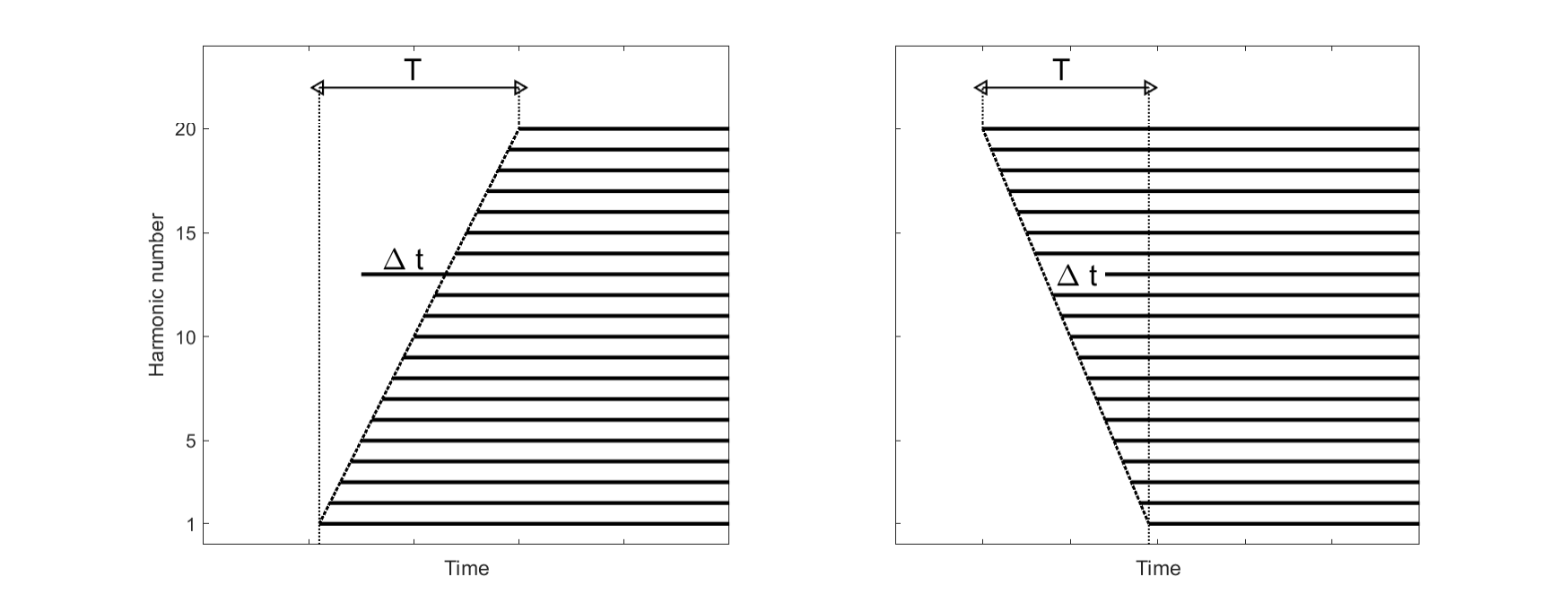}	
		\caption{An illustration of the stimuli used in \citet{Zera1993a}. Complex tones with 20 components were varied with respect to their relative onset time, whose asynchrony slope is determined by the standard $T$, which may be either positive (left) or negative (right). An additional parameter is the deviation of a single component from the standard, $\Delta t$, which can also be either positive or negative. Inharmonic versions of this stimuli were tested as well using the same number of components that were logarithmically spaced.}
		\label{zerastan}
\end{figure}

Empirical temporal ranges that can represent the depth of field may be derived from data of several psychoacoustics studies. To this end, we will reinterpret observations of asynchrony perception of different components of a polychromatic object that temporally stick out. \citet{Zera1993a} measured the effect of differentially gating the onset time of a single component of complex tone with a fundamental of $f_0=200$ Hz and 20 components between 200 and 4000 Hz (Figure \ref{zerastan}). Complex tones were defined by a ``standard'' asynchrony $T$---the delay time between the lowest and highest components of the tone, which could be either negative or positive. The other components were delayed accordingly, except for a single component, which started with an extra delay $\Delta t$ that deviated from that standard. Subjects had to discriminate between the standard tone and the one that contained a delayed component. Measurements repeated for harmonic tones and logarithmically-spaced inharmonic tones in several conditions. In the harmonic condition, the in-phase components fused and formed an acoustic object whose highest harmonics were unresolved. In contrast, the inharmonic tone components were almost always resolved into separate auditory filters, but did not fuse to create unified perceptual objects. As the experiment manipulated the onset time of the components, it directly affected the (spectrotemporal) envelope, and hence the auditory image of the complex tones\footnote{Responses for offset gating were tested as well, but are not treated here. They were usually worse (less sensitive) than onset responses.}.   

\begin{figure} 
		\centering
		\includegraphics[width=1\linewidth]{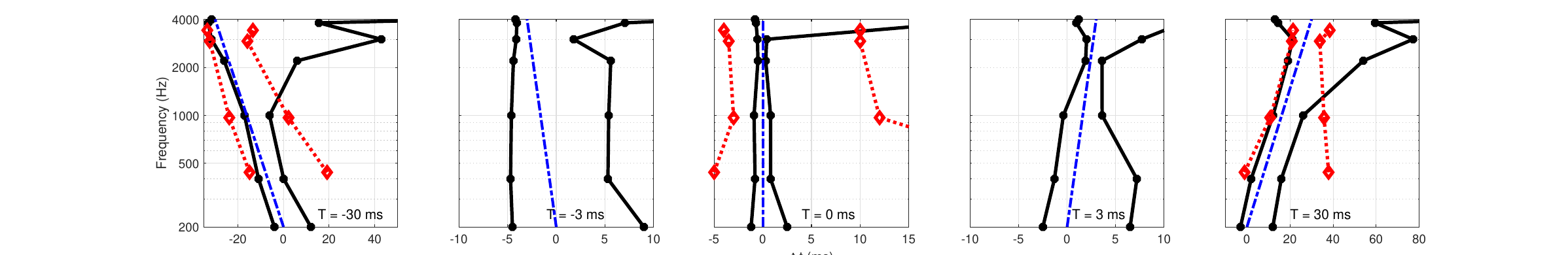}	
		\caption{Replotted data from \citet[Figures 2, 4, and 5]{Zera1993a}, which show the discrimination threshold for a manipulated single component of harmonic and inharmonic complex tones, whose onset is delayed by $\Delta t$ relatively to the tone onset standard, as shown in Figure \ref{zerastan}. The complex tone standard is defined by the onset asynchrony of the components ($T$), which is defined as the delay between the lowest and the highest components. The standards are plotted with dash-dot blue lines and are, from left to right: -30, -3, 0, 3 and 30 ms. The negative values indicate that the high-frequency component onsets come first, whereas low-frequency component onsets come first when they are positive. Responses to harmonic tones ($f_0=200$ Hz, 20 harmonics, zero initial phase for all components) are plotted in solid black and are based on three subjects, except for the $\pm3$ ms conditions that are based on one subject. Inharmonic tones had the same fundamental frequency and number of components, but were logarithmically spaced at a 1.17 ratio between components. The responses to the $T=$ -30, 0, and 30 ms inharmonic standards are plotted in dotted red and are based on two subjects. It is hypothesized here that the area bounded between the lines on both sides of the standard slope corresponds to the auditory depth of field of the ear for these stimuli.}
		\label{zerachrom}
\end{figure}

The results of \citet{Zera1993a} are rearranged and replotted in Figure \ref{zerachrom}. There are at least two alternative ways to associate the depth of field with the results. Resolving the component from the complex tone is possible when both are sharp (large depth of field), or when one is sharp and the other is blurred (small depth of field). In general, the smaller the absolute value of the asynchrony is, the smaller is the effective depth of field, which is represented in the figure by the black contours for the harmonic tones and by red contours for the inharmonic tones. So, the highest sensitivity to single-component delays shows when all the tone components are in phase in the synchronous condition ($T=0$, middle plot) with some specific components being as sensitive as $\Delta t \le 0.6$ ms. Additionally, there is higher sensitivity to components starting before the standard than to those starting after (i.e., the solid black lines to the left of each standard in blue are closer to it than the solid black lines to its right). Importantly, the sensitivity to these early-onset components is relatively frequency-independent, with the synchronous condition varying by less than 1 ms throughout the entire spectrum (at least below 4000 Hz), in line with group-delay distortion detection studies \citep{Moller2007}. The exception are the two highest (unresolved) components tested of 3800 and 4000 Hz that produced anomalous thresholds of 40--100 ms in other conditions. The response to inharmonic tones (shown in red in three plots) is relatively frequency independent for $T = \pm 30$ ms, but is more dispersed for $T = 0$ ms and $\Delta t < 0$. It should be noted that the synchronous results are similar to those obtained in \citet[Figure 9]{Zera1993b}, where it was found that increasing the number of delayed components in the harmonic condition further enhanced the sensitivity to changes (reduced $\Delta t$). The opposite occurred for the inharmonic tones \citep[Figure 10]{Zera1993b}. Discrimination thresholds also depended on the onset phase of the delayed component \citep{Zera1995}.

These results from Zera et al. are telling about the system sensitivity to chromatic irregularities that affect its image. As they were tested specifically with complex tones, it is unknown to what extent they can be generalized. For example, does the in-phase standard $T = 0$ elicit the minimum possible depth of field? Does it depend on the bandwidth of the complex tone? Is it highest for completely coherent stimuli? Also, these observations comprised data from three subjects or less, which is insufficient for more sweeping generalizations. Nevertheless, the data are in accord with related observations by \citet{Wojtczak2012} of a better-controlled, yet simpler across-frequency asynchrony detection task\footnote{Using the common jargon in hearing research, when the depth of field is applied to polychromatic images, then they are considered ``coherent'', as their temporal modulation is in phase. It is commonly used to describe stimuli rather than images. Combining the two terminologies, the depth of field relates to the set of objects that sound the same and may be considered effectively coherent for the subject. It should be stressed, though, as was argued in \cref{IntroCoh}, this terminology is ambiguous, because it relates to different things in the carrier and modulation domains.}

\subsection{Auditory depth of field (coherence)}
\label{NonSimultDepth}
The depth of field is generally defined with respect to an object in focus and is given as a range in dimensions of length. If we draw direct analogy between spectral and temporal imaging, then the analogous depth of field in temporal imaging systems should specifically relate to the sensitivity of the image to dispersive path from the object to the lens \citep{Shateri2020}---a quantity that can be combined with the cochlear dispersion $u$. As was discussed in \cref{acoustenv}, the farther the object is, the more dispersed it tends to be, subject to conditions in the environment. However, a much more dominant effect of the environment that interacts with dispersion is a general loss of signal coherence between the source and the receiver due to weather conditions, reflections, and noise---both coherent and incoherent. These effects also strongly depend on the acoustic source coherence, which dictates the kind of transformations that the waves may be most sensitive to (\cref{CoherenceWaveEquation}). Here, dispersion is only one factor that leads to signal decoherence, according to the signal bandwidth. Therefore, dispersion alone is probably not the most informative parameter that should be employed to characterize the auditory depth of field. But, if dispersion is considered along with source randomness, its effect can be approximated to decoherence, for a certain degree of detail or temporal resolution (see \cref{Diffusers}). 

Given its similarity to the pinhole camera, we may expect the auditory system to have a relatively high depth of field for signals that are anyway in focus---namely, completely coherent narrowband stimuli, which are largely insensitive to defocusing, but are instead sensitive to interference and decoherence. In contrast, incoherent stimuli are sensitive to defocus, but not to interference and decoherence. Hence, partially coherent objects are sensitive to dispersion, interference, and decoherence in different amounts. This suggests a departure from the straight analogy to the spatial depth of field, as external coherence constitutes the most dynamic parameter of the environment and objects can vary in more than a single dimension (i.e., distance). Additionally, we should consider that the extensive signal processing in the auditory brain may be set to exaggerate the depth of field, as may be expected from a standard imaging system that is completely passive. Let us explore how the effective auditory depth of field may arise in common situations.

According to the space-time duality applied to the auditory depth of field, defocused temporal objects extend in time beyond the temporal boundaries of the original signal (from the acoustic source), which manifests somehow on the time-frequency plane\footnote{Forward masking in the temporal domain does exist in vision, where the perceived object size that is followed by a flash of light changes over 300 ms after the object disappears \citep{Wilding1982}. A much smaller effect exists if the object is preceded by the flash (backward masking condition). The effect is monoptic and may be a result of retinal processes only. Arguably, this type of masking does not satisfy the space-time analogy, as was defined earlier.}. If this is the case, then it should be possible to measure the effect of this extension, as the boundaries of the image encroach into the images of adjacent objects---both in time and in frequency. In other words, a finite auditory depth of field anticipates the existence of \term{nonsimultaneous masking}---the change in the response of one sound by another non-overlapping sound. 

\begin{figure} 
		\centering
		\includegraphics[width=1\linewidth]{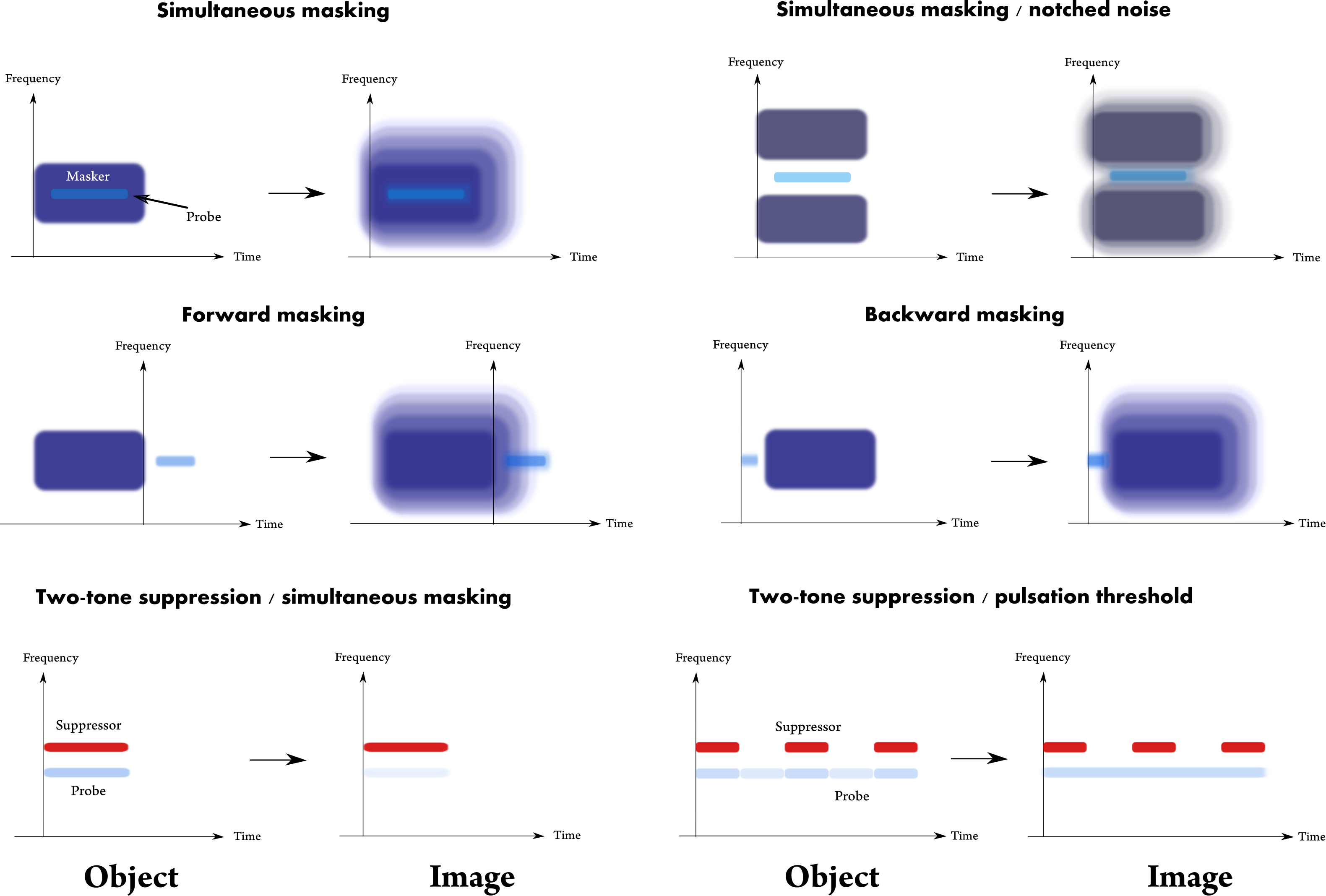}	
		\caption{Cartoon summary of common types of masking paradigms reframed as objects and images on the time-frequency plane. The stimulus is considered the object and is illustrated on the left. It is perceived as an image, which is illustrated on the right of each pane. The rectangle designates the masker, which is usually a narrowband or broadband noise, although it is sometimes a sinusoid. The probe is typically a sinusoid whose frequency coincides with the center of the masker in on-frequency conditions. If it is presented with the masker, then it is referred to as simultaneous masking (\textbf{top left}). A variation on this type of masking is the notched-noise with tone, which has been extensively used to estimate the auditory filter bandwidth by varying the spectral gap between the two bands of noise (\textbf{top right}). In the example, the probe is ``on-frequency'', as it is centered right in the middle of the notch. Non-simultaneous masking may be either forward masking, in which the probe is presented after the masker has been switched off (\textbf{middle left}), or the much weaker and shorter backward masking, in which a short probe is presented before the masker (\textbf{middle right}). Upon imaging, the masker gets a ``halo'', so it extends beyond its original temporal and spectral limits in a diminishing way. When the masker image collides with that of the probe, it makes it more difficult to hear, which elevates its detection threshold. In the text, we refer to this extended blur between two objects as their relative depth of field. Objects are easier to tell apart over the background when they are sharp and the background is blurry. The bottom two plots describe two different effects of two-tone suppression. On the \textbf{bottom left} is the increase in the threshold of a tone as a result of another tone that is either higher or lower in frequency. On the \textbf{bottom right} is the pulsation threshold, which is the effect of a tone that is modulated, which sounds continuous at low thresholds as a result of another suppressing tone. Note that in the top four masking types, the image of the masker always extends asymmetrically toward high frequencies, in what is called \term{upward spread of masking}.}
		\label{MaskingCartoons}
\end{figure}

\subsubsection{Nonsimultaneous masking}
Typically, the stimuli used in nonsimultaneous masking experiments are classified as either \term{maskers} or \term{probes} according to the specific experimental paradigm that is being used (Figure \ref{MaskingCartoons}). When the probe follows the masker, it is called \term{forward masking} (or \term{postmasking}), and the effect tends to be robust and extend up to 100--200 ms from the offset of the masker. When the probe precedes the masker, it is called \term{backward masking} (or \term{premasking}), and there the effect tends to be less robust and it operates on considerably shorter time scales (0--10 ms from the probe onset to the masker onset). Extensive literature exists about masking in general and about forward masking specifically, where it has been typically found that the exact change to probe threshold depends on all the stimulus parameters: the masker and probe duration, the gap between them, their absolute and relative levels, their absolute and relative spectra, the possible addition of more masking components, or the addition of other conditions (e.g., contralateral stimulation). The topic of nonsimultaneous masking is too vast to comprehensively treat here, in part because it is tightly related to the issues of frequency selectivity, temporal integration, compression, and suppression. We only review a handful of results that are amenable to the logic of temporal imaging and coherence, albeit in a more empirical way. Reviews of masking effects can be found in \citet[pp. 61--110]{Fastl} and \citet[pp. 67--132]{Moore2013}. Early literature was reviewed by \citet{Duifhuis1973}.

There are two reference thresholds against which forward masking can be tested. When the masker and probe overlap, simultaneous masking may take place, which can be largely (but definitely not always) accounted for using energetic measures when the two stimuli are analyzed within the same auditory channel. Then, the signal-to-noise ratio (SNR) of the input (i.e., the probe-to-masker ratio) can be employed as the main parameter that explains the degree of masking \citep[e.g.,][]{Fletcher1940}. On the other extreme is the response to the probe in the absence of any masking. Forward masking is then the time-dependent residual masking between the simultaneous masking threshold and the threshold in quiet.

A central point in research about forward masking has been to account for the long time constants that are associated with the decay of masking after the external stimulus has long been switched off. At low frequencies, forward masking depends on the frequency of the stimuli, which suggests that the cochlear filter ringing may have some effect. But two more general mechanisms have been proposed to account for forward masking: neural adaptation in the auditory nerve and persistence of the stimulus in the central pathways \citep{Oxenham2001}. Adaptation may have a masking-like effect due to a reduction in spiking rate in the auditory nerve following the onset of the masker and subsequent slow recovery \citep{Smith1977,Harris1979}. However, the physiological effect characteristics are inconsistent with psychoacoustic data, which have to consider detectability of the probe that is based on both the response and its variance \citep{Relkin1988}. Furthermore, listeners with cochlear implants also experience forward masking, despite the fact that their auditory nerve is directly stimulated without emulating adaptation \citep{Dent1987,Shannon1990}. Also, it was found that centrifugal efferents to ventral cochlear nucleus (VCN) units in the guinea pig can either facilitate or delay the recovery from forward masking, beyond the response that is measurable in the auditory nerve \citep{Shore1998}. It has led to the conclusion that forward masking is caused by persistence that has a central origin, which may be modeled through temporal integration after preprocessing that occurs in the cochlea \citep{Oxenham2001,Digiovanni2018}. This explanation defers the mechanism to yet unidentified brain circuits, although correlates have been found in the ventral and dorsal cochlear nuclei \citep{Frisina2001}, the IC \citep{Nelson2009}, and the auditory cortex \citep[e.g.,][]{Calford1995,Brosch1997,Wehr2005}. The forward masking effect seems to be largely complete at the level of the IC, where it was suggested that the neural persistence is the result of adaptation or offset inhibition \citep{Nelson2009,Middlebrooks2013}. An analogous forward masking effect has been also found in amplitude and frequency modulated tones \citep{Wojtczak2005,Byrne2012,Fullgrabe2021}, but this type of masking (at least the AM) does not appear to be reflected in subcortical processing \citep{Wojtczak2011}. 

Arguably, the conclusions from current research suggest that the auditory system is actively geared to produce and even exaggerate forward masking, rather than try to minimize it, as may be naively surmised from standard signal processing considerations. Indeed, it has been suggested that forward masking may be useful in object formation, as part of early scene analysis \citep{Fishman2001,Pressnitzer2008}. More germane to the present work, it further suggests that forward masking cannot be fully captured by the time-invariant imaging transfer functions. However, if the exaggerated function relies on a time-invariant imaging effect that is too short to be useful on its own (perhaps, similar in duration to backward masking), then new intuition may be garnered from qualitative depth-of-field predictions and observations. Specifically, if forward masking can be indeed reframed as a depth-of-field effect, then it should be possible to relate the coherence of both signal and masker to their measured response patterns. The coherence-dependent differences may be noticeable both in the magnitude and in the duration of the masking effects of stimuli of different kinds. 

\subsubsection{Nonsimultaneous masking as depth of field}
Throughout this section, we have treated the masker and the probe as two temporally distinct objects in the external environment, whose corresponding images may be difficult to resolve because of masking that is generated internally, within the auditory system (Figure \ref{MaskingCartoons}). We know from vision, by way of analogy, that \textbf{the auditory depth of field should endow the listener with a percept to distinguish between foreground and background, but not between background and background or foreground and foreground.} But, what is foreground and what is background between two acoustic objects? Since coherent signals are theoretically always in focus in the defocused system, then the relative degree of coherence between the objects may be a natural cue to distinguish between foreground and background. Thus, we would like to find out whether coherence can be used as a cue for masking release between these two images, which may then be reinterpreted as an auditory depth-of-field effect.

Two examples are presented that illustrate the effect of coherence on the depth of field. The first example is a simplified model of data presented by \citet{Moore1985}, which demonstrate a release from forward masking effect when a reference masker was mixed with another type of masker. We would like to show how the relative amount---or rather, the rank order of the relative amount---of masking release can be predicted from the coherence function. The reference masker was a 400 ms narrowband noise with 100 Hz bandwidth centered at 1 kHz, at 60 dB SPL, that included 5 ms cosine squared onset and offset ramps. The probe was a 20 ms 1 kHz tone burst at 60 dB SPL that had 5 ms onset and offset ramps and a 10 ms steady-state middle. The probe was presented immediately after the masker, at 0 ms offset-onset delay time. The authors mixed the reference masker with 75 dB SPL tones of 1.15, 1.4, 2, and 4 kHz, as well as 1.4 kHz tones at 50 and 60 dB SPL, and low-pass filtered (4 kHz cutoff) white noise at three different levels\footnote{An additional contralateral condition was not modeled and is omitted here. However, a recent model by \citet{Encke2021,Encke2022} quantitatively accounts for the release from masking in a broad range of binaural conditions using interaural coherence, at a higher level of precision than was obtained (or attempted) here. The complex-valued coherence was derived from the spectral coherence (cross-spectral density function) using gammatone filter transfer functions to model the auditory filters and signal detection theory to model the subjective threshold.}. All masker combinations resulted in some release from masking---sometimes very substantial---that was presented as individual data for three subjects. The authors argued that the release from masking is unlikely to be caused by suppression for the majority of the masker types. Instead, they suggested it is caused by \term{perceptual cueing}, whereby the additional masker components disambiguate the masker and the signal where they are too similar to tell apart. The cueing relates to a nonspecific change in quality such as a level change, or any other temporal or spectral distinction that signals to the listener that the probe is distinct from the masker \citep{Terry1977,Weber1981}. 

\begin{figure} 
		\centering
		\includegraphics[width=1\linewidth]{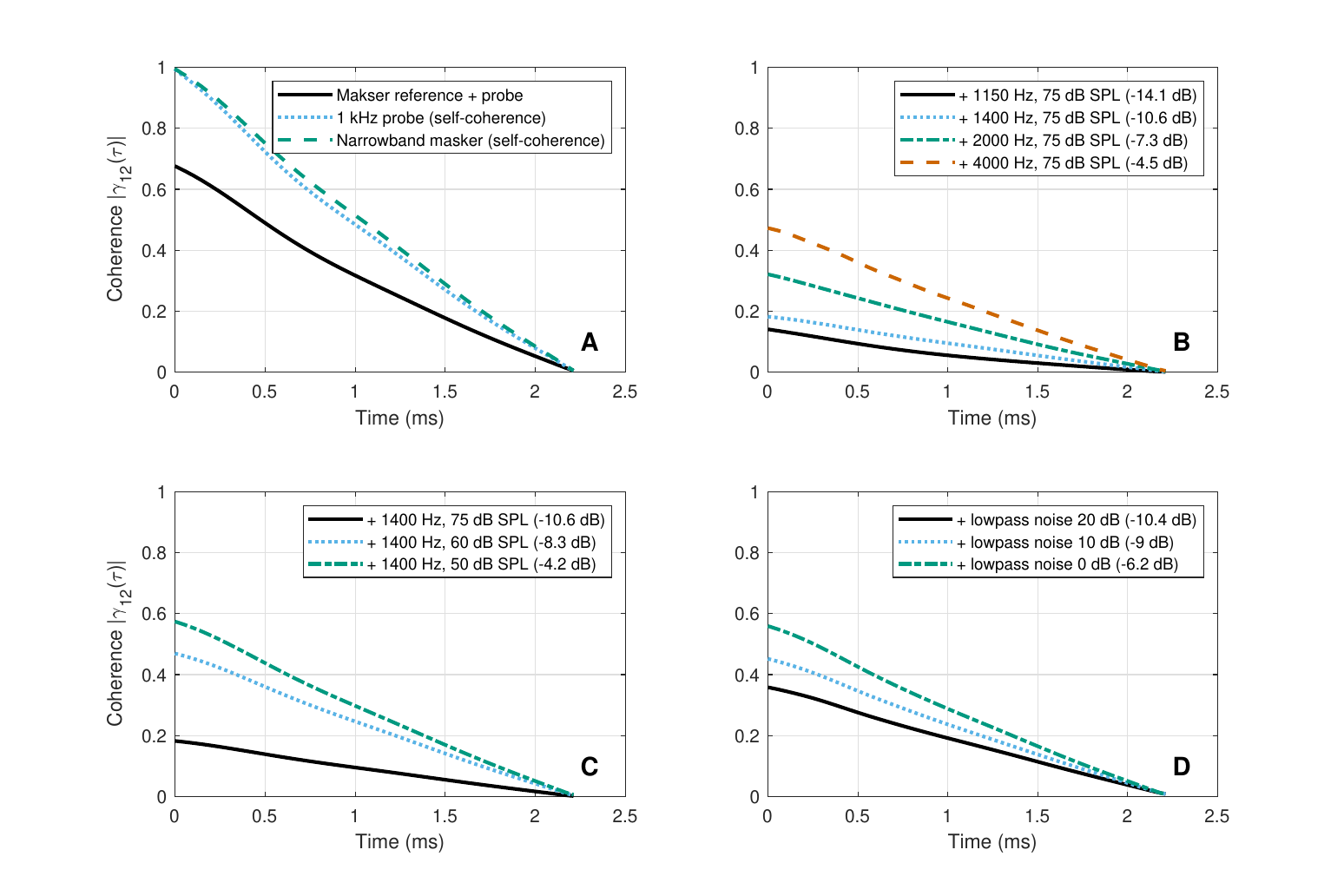}	
		\caption{The coherence functions of the different maskers and probe from \citet{Moore1985}. The Hilbert envelope of the nonstationary coherence (Eq. \ref{nonstationGamma}) was obtained using the estimated aperture time at 1 kHz of 2.26 ms and 50\% overlap between cross correlated segments (see \cref{FourSources}). All stimuli are bandpass filtered at 1 kHz, using a second-order Butterworth filter. As the coherence function is symmetrical, it is displayed only for positive time. The results were averaged over 100 runs with randomized masker and noise. \textbf{A:} The self-coherence functions of the narrowband masker (centered at 1 kHz with 100 Hz bandwidth) and the probe (1 kHz tone), and their mutual coherence function. \textbf{B--D:} The mutual coherence of the narrowband masker and additional components with the tonal probe. The mean masking release compared to the reference masker is given in parentheses in the legend of all plots based on the three subjects in Table I of \citet{Moore1985}. \textbf{B:} Coherence of pure-tone maskers added to the reference. The closer the added component is to the 1 kHz probe, the lower is their degree of coherence, which results in higher release from masking. \textbf{C:} The same for 1.4 kHz tones at different levels. The higher the tone level is, the less coherent it is with the probe and the more release from masking is obtained. \textbf{D:} The same for low-pass filtered white noise at three different levels. The higher the noise level is, the lower is the coherence function which achieves increased release from masking.}
		\label{MooreData1985}
\end{figure}

Perceptual cueing is a lower-level explanation of the more abstract depth-of-field effects that separate different types of objects according to their degree of coherence, which we argue is the most salient auditory analog to distance in vision. The coherence functions of the probe and the various maskers are plotted in Figure \ref{MooreData1985}, for bandpass filtered stimuli. It can be seen from plot A in Figure \ref{MooreData1985} that the choice of narrowband masker and short tone burst produces coherence functions that are very similar, which means that the masker and stimulus are partially coherent to a similar degree. The fact that both have the same center frequency as well results in significant mutual coherence (cross-correlation) for the short temporal aperture duration that was applied (2.26 ms for 1 kHz; see Table \ref{t0opts}). Mixing the maskers with additional components always results in a further decrease of mutual coherence (plots B--D), which effectively segregates the various probe-masker pairs. Within each plot, the smaller the slopes and maxima of the obtained coherence functions are relative to the masker-probe coherence in plot A, the larger is the release from masking. This can be compared to the measurements by \citet{Moore1985} (shown in the legends of Figure \ref{MooreData1985}), which correspond to the same rank order of coherence curve within each plot. However, the global rank order of the different stimuli does not correspond exactly to the measured one, possibly due to contributions from off-frequency channels \citep[e.g.,][]{Moore1981}, which were not modeled here.

\begin{figure} 
		\centering
		\includegraphics[width=1\linewidth]{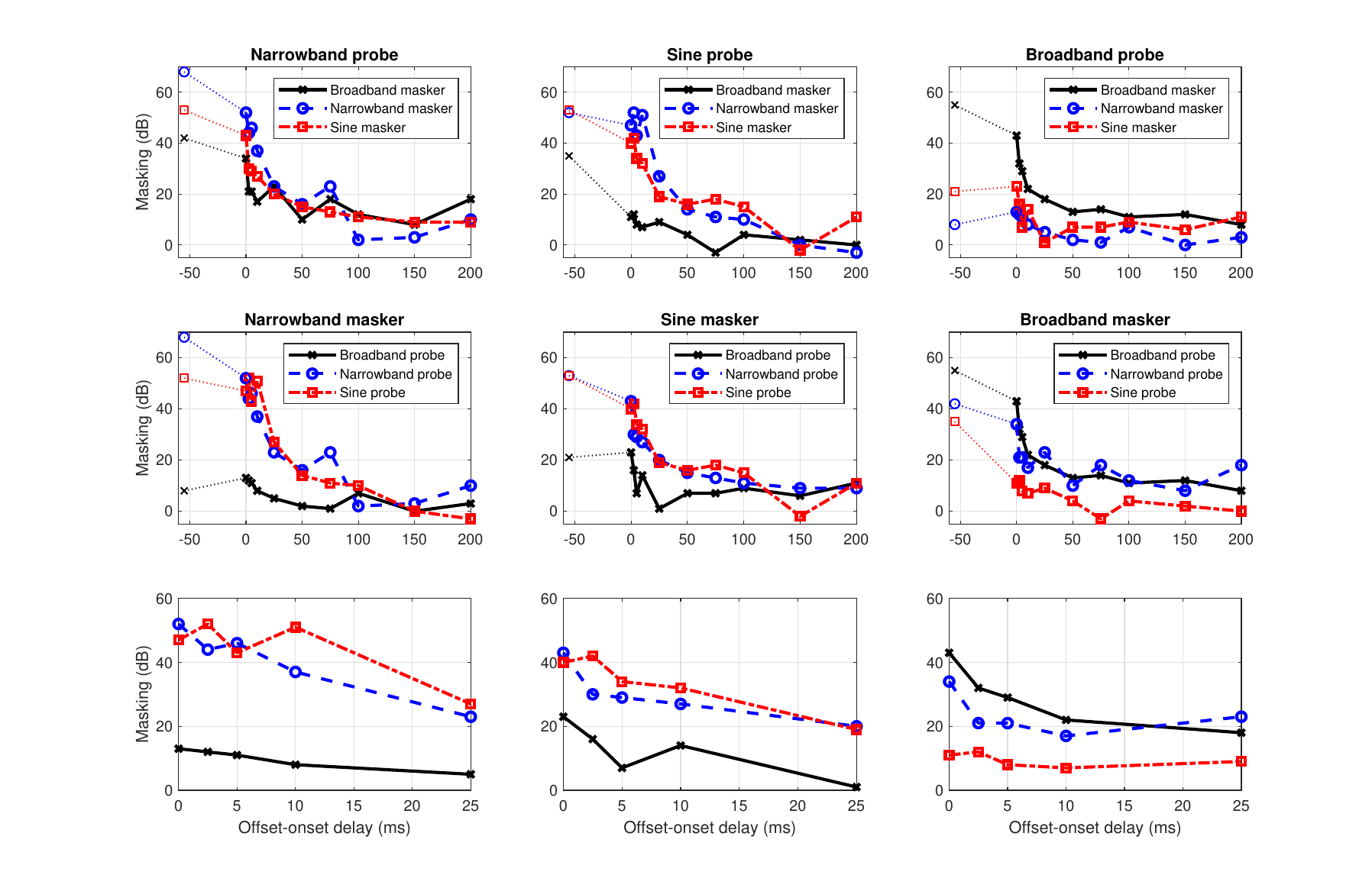}	
		\caption{Forward masking decay curves for three masker and probe combinations of a single normal-hearing listener. The masker was always 100 ms and the probe 10 ms, both with 5 ms cosine squared fades. The sine masker and probe are at 1000 Hz, and the narrowband is centered at 1000 Hz with 100 Hz bandwidth. The broadband masker and probe were full bandwidth. Offset-onset delay times tested were 0, 2.5, 5, 10, 25, 50, 75, 100, 150, 200, and 400 ms. The amount of masking was referenced to the threshold measured with an onset-offset delay of 400 ms (not shown). Simultaneous masking thresholds were also measured for probes that were exactly centered within the maskers and are shown at -55 ms with dotted lines. The two top rows contain the same data organized according to common masker or common probe. The third row contains the same information as the second row, but zoomed in on offset-onset delays of 0--25 ms. The root mean square (RMS) level of the sine and narrowband maskers were equalized and the broadband masker was set to -10 dB from that value to be approximately the same loudness. The absolute presentation level could not be determined, but it was fixed at a comfortable level. The testing equipment was identical to that described in \cref{DoublePulseGap}.}
		\label{ForMaskSelf}
\end{figure}

The second example that illustrates the application of depth of field to forward masking presents the author's own data using all nine combinations of broadband, narrowband, and tonal maskers and probes. The three maskers were 100 ms in duration, where their onset and offset were ramped with cosine square functions over 5 ms. The narrowband masker was centered at 1000 Hz and was 100 Hz wide. The sine masker was a 1000 Hz tone and 10 ms with the same 5 ms ramps at onset and offset with no steady-state portion. The sine and narrowband maskers were normalized to have the same RMS level, but the broadband masker was set to 10 dB lower (RMS) to have approximately equal loudness for the three maskers, at a comfortable listening level. The forward masking thresholds for different offset-onset delay values are plotted in Figure \ref{ForMaskSelf} grouped for probes and maskers, referenced to the threshold at 400 ms delay. The simultaneous masking thresholds were measured as well, as an additional reference. The results follow the familiar forward masking patterns that were sometimes modeled as exponential decay \citep[e.g.,][]{Plomp1964b} with much of the decay taking place within the first 10 ms and then more slowly up to 200 ms and beyond. 

The most striking aspect in the results of Figure \ref{ForMaskSelf} is that the masker is always highly effective for a probe of the same type: broadband for broadband, narrowband for narrowband, and sine for sine. In contrast, masking is least effective for the most dissimilar types: narrowband and sine maskers for broadband probes, and broadband masker for sine probe. In the other cases, the sine and narrowband probes tend to behave similarly and elicit about the same thresholds in the case of narrowband and sine maskers, but somewhat more masking in the case of broadband masker and narrowband probe. 

\begin{figure} 
		\centering
		\includegraphics[width=0.5\linewidth]{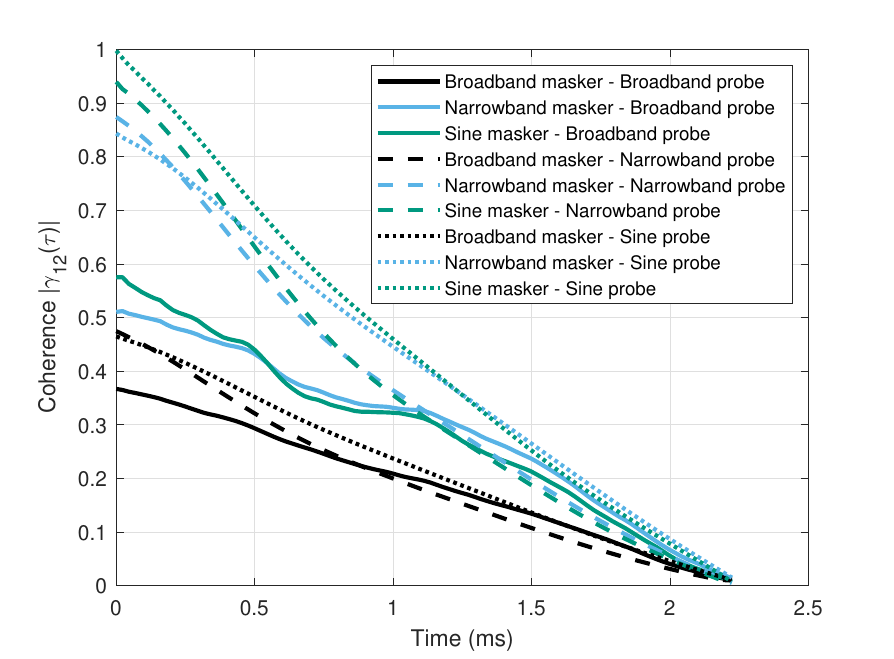}	
		\caption{Mutual coherence plots of the stimuli described in Figure \ref{ForMaskSelf}, computed as were detailed in Figure \ref{MooreData1985}. Masker-probe pairs that are have higher mutual coherence are expected to be less easily distinguishable and hence elicit higher forward masking compared to those with low mutual coherence function. However, the predictions are not always correct here, as can be seen in the case of the broadband masker-probe pair. This may be explained by considering that the coherence modeling here was done on a single channel, whereas the broadband stimuli are multichannel.}
		\label{ForMaskSelfCoh}
\end{figure}

As before, these results are intuitively explained, in part, by appealing to coherence. Roughly speaking, the broadband stimuli are incoherent and the narrowband are partially coherent. The sine masker is close to being fully coherent, whereas the sine probe is also partially coherent due to its short duration---perhaps similar to the narrowband probe. This is captured in part in Figure \ref{ForMaskSelfCoh}, which again brings the mutual coherence curves of all masker-probe combinations. If we look at the rank order of forward masking at 0 ms, as in the first example, then most of the main trends are well predicted by the curves with the exception of the broadband masker and probe case. In this case the coherence calculation predicts to have the least masking of all pairs, which was not the case. Off-frequency modeling of coherence using additional channels can easily achieve the correct rank order of these pairs, but it has not been pursued here.
 
Regardless of the modeling details, the main result stands: signals of different degrees of coherence elicit less masking the more dissimilar they are. This is measurable both in the threshold themselves, beginning in the simultaneous masking thresholds, and continues to the decay function, which determines the rate of release from masking. This was also seen in the comparisons between sine and narrowband masker-probe pairs in \citet{Weber1981}, where differences were attributed to both stimulus energy and perceptual cueing. These findings underline how the system is geared to differentially process objects according to their individual signal properties, such as their degree of coherence. 

\subsection{Discussion}
The auditory depth of field is not entirely analogous to the visual depth of field, since there is no naturally-occurring coherence variation in optical objects that produce images whose sharpness is not affected by their distance from the eyes\footnote{We did see an example in \cref{AudDefocus} for an artificial visual system that exploits coherence to maintain sharp text for any focal length that the lens assumes \citep{Waldkirch2004}.}. Even pure tones, which we had considered to be in sharp focus in standard temporal imaging processing (\cref{AudDefocus}), produce substantial forward masking when paired as both masker and probe, unless they are sufficiently well-separated in frequency \citep[e.g.,][]{Miyazaki1984}. In contrast, if a sharply-focused optical object is visually imaged, then an adjacent object at the same distance is not going to differ in relative sharpness as a result of a depth-of-field effect. The simple (and likely, simplistic) examples we explored revealed that objects may be differentiated according to their relative degree of coherence, but they do not exhibit preferential sharpness of a specific type of object as a function of its coherence. Therefore, the degree of coherence is not exactly analogous to distance in spatial optics and vision. This subtle point highlights the importance of partial coherence as the default mode of operation of the auditory system (see also \cref{DetectionSchemes}, \cref{StreamMixing},  \cref{CoherenceAndDefous}, \cref{CoherentNonCoherent}, and throughout \cref{accommodation}).

Nevertheless, more realistic and complex stimuli may shed light on the idea of auditory depth of field. We explored rather narrowly the effect of coherence, but there are clearly additional parameters that determine the effective auditory depth of field such as dispersion, focal time, and f-number. Moreover, given that the auditory system clearly exaggerates the forward masking, it is not impossible that it can also control masking release under some conditions that are more dynamic in nature. We will touch upon this idea in \cref{accommodation} in the context of accommodation. Future studies will ultimately enable us to determine how useful this concept actually is within hearing science. 

\begin{figure} 
		\centering
		\includegraphics[width=0.75\linewidth]{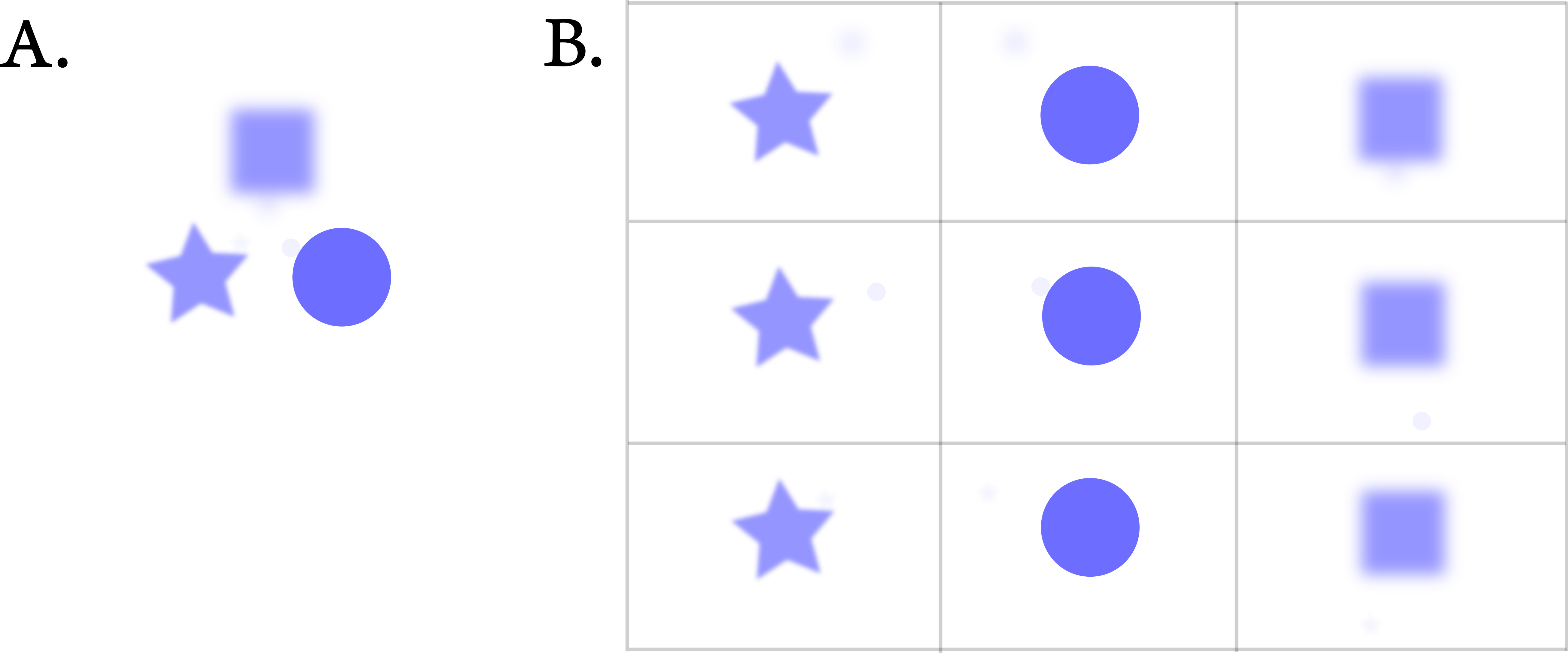}	
		\caption{A visual analogy of auditory depth of field. Different shapes relate to different auditory objects. They vary in size, brightness, position, and blur, but not in color. A large and bright shape corresponds to a loud masker. A small dim shape corresponds to a probe. In \textbf{A}, all the maskers are displayed with the probes just next to them. In \textbf{B}, the nine probe-masker (small-large shape) pair combinations are displayed. Some probes are more readily detectable than others, based on their relative position to the masker and their degree of blur, which determines the sharpness of their contour. Note that this analogy does not map easily to the three masker and probe types used above (i.e., broadband, narrowband, sine). Still, it illustrates how all the parameters in the simple ``scenes'' are relevant to the detection.}
		\label{DoFAnalogy}
\end{figure}

A final cartoon analogy of visual objects is displayed in Figure \ref{DoFAnalogy}. It illustrates how different relations between shape parameters (brightness, size, shape, relative distance, and blur) can determine the relative visibility of a small and dim ``probe'' next to a large and bright ``masker''. These images do not appear as real spatial depth-of-field effects of a real complex scene, but rather underscore how the complex interplay between all object parameters affects perception that gives rise to the sense of foreground and background among the images. This figure is a reminder of how all factors may matter in the scene analysis and not all of them are necessarily acoustic or peripheral, when it comes to auditory processing, since attention, context, and other high-level cognitive factors all provide input to perception. 

\section{Aberrations not found in hearing}
Because of the lower-dimensional space of the temporal image compared to the spatial image, some of the most important aberrations in optics are irrelevant, as they require at least two spatial dimensions to manifest. Specifically, these are \term{astigmatism}, which causes blur because of asymmetrical focus about the optical axis that causes the loss of rotational symmetry, and \term{curvature of field}, which happens when the image is projected on a flat plane but is in fact focused on a curved surface \citep[pp. 240--244]{Born}. It may be argued, though, that curvature of field is possible in temporal imaging as well, if the sharp image is formed on a mathematically curved surface in the time domain. However, such an effect in a single dimension may then be expressed as a combination of spherical aberration and distortion. In hearing, analogous aberrations to these two---if they exist---may manifest in nontrivial ways, such as in the binaural integration to a unitary image from the individual channels. These aspects may be more pertinent to bat echolocation fidelity, for example, or in applications where high-precision, spatially-faithful, three-dimensional audio reproduction is of primary interest. 

\section{Aberrations due to nonlinearities}
Ideal temporal imaging, as does any type of imaging, requires linearity of its lens \citep{Kolner}, so there is a design advantage in obtaining auditory components that are not dependent on the signal amplitude. However, the question of whether the human auditory curvature is amplitude-dependent has not been settled. There is psychoacoustic evidence for linearity \citep{Summers2000,OxenhamDau,Tabuchi2016} and against it \citep{Shen2009}. Similarly, there is physiological data from animals that suggest that the instantaneous frequency is amplitude-dependent \citep{Wagner}, and amplitude-independent \citep{Carney1999,deBoer1997,Recio1997}, at least up to input levels of 80 dB SPL. Other more well-studied aspects of the auditory system are definitely nonlinear whenever the outer hair cells are involved. This includes the time-lens curvature phase modulation that appears to depend on level (\cref{OHCtimelens}), as well as amplitude dependence of the complete dispersive pathway in the auditory system that was seen in auditory brainstem response and otoacoustic emission measurements (\cref{NeuralDispEst}). Whether, and if so how, any of these amplitude dependencies affect the final image quality is presently unknown and is beyond the scope of this work.

\section{Rules of thumb for auditory imaging}
Seven rules of thumb are given below that encapsulate some of the intuition about the nature of the auditory image. They are all corollaries of having a partially coherent intensity imaging system, which has the object coherence as its currency. The first three are more-or-less synonymous with one another and are readily supported through empirical findings from literature, so they merely rephrase known system behavior that describe different aspects of auditory interference. The fourth one is a general corollary of the previous rules and much of the acoustic source analysis in \cref{InfoSourceChannel}. The last three are more directly drawn from analogies with optics and vision.

Note that throughout this work, the term ``image'' is used in three different meanings. First, it refers to a single image of the pulse object---the sample. This is equivalent to a ``sound pixel''. Second, it is the time sequence of samples within a channel, which corresponds to a monochromatic image. Finally, it is the combined sequences of all channels, which give rise to a polychromatic image. It also includes any cross-channel effects that have to do with harmonicity. We have mainly referred to the second image type, although explicit calculations were done using the first image type only. 

\begin{enumerate}
	\item \textbf{Coherent and partially coherent signals interfere within the auditory filter}---Within a duration that is determined by the aperture stop of a single channel, signals may interfere through the superposition of their complex amplitudes, according to their degree of coherence. This is the case in the simple beating effect (see \cref{Nonstationarytheory} and \cref{TempAperture}). More complex interference patterns are heard with the addition of tonal components within the same channel. For example, the timbre produced by complex tones or vowels that contain high-frequency harmonics that are not resolved can exhibit sensitivity to phase (see review of related phenomena in \citealp{Moore2002}). Other narrowband sounds also interfere intermittently (\cref{PartialModu}) or cause suppression partly because of cochlear dispersion. which can also be taken as a form of interference (\cref{BlurandAbbs}). Incoherent signals or signals that are resolved in separate channels do not interfere, by definition. The image of an individual channel is approximately monochromatic.

	\item \textbf{An intensity-image is perceived at the auditory retina and more centrally}---Despite the interference effects throughout the early auditory stages, the final image is perceived as an intensity pattern. Therefore, there is no perceptual positive or negative amplitude, but rather a strictly scalar sensation of level. So, in the simple beating example, the modulation frequency heard in beating is double than is implied by the amplitudinal interference ($\Delta f$ instead of $\Delta f/2$)
	\begin{equation}
	I(t) \propto  p^2(t) \propto \cos^2(\pi\Delta f t) = \frac{1 + 2\cos (2\pi\Delta f t)}{2} 
	\end{equation}
This holds also for frequency-modulated sounds, which are represented as intensity images that are manifested in dynamic pitch.
	
	\item \textbf{Polychromatic (broadband) images may mask, but they do not interfere}---The superposition of monochromatic images (i.e., in different carriers with overlapping modulation spectra) does not lead to destructive interference. However, images in different channels are not independent either, since the information across channels can be pooled in different ways if they are coherent in envelope (\cref{polychromatic}). In the case of simultaneous but dissimilar intensity envelopes in different channels, listening in the dips may be possible, as quiet instances of one envelope overlap high-level informative instances of the other, just enough to enable speech recognition \citep{Miller1950, Cooke2006, Edraki2022}. In other conditions, dissimilar images may mask one another or compete for attention. 
	
	\item \textbf{Real-world signals that are partially coherent produce mixed behavior}---Unlike visual objects, acoustic objects of auditory significance are very often nearly coherent over behaviorally-meaningful durations (\cref{TypicalSourceCoh}). However, this coherence tends to be lost over distance, with reflections, and with reverberation (\cref{DecoherenceRev}). Therefore, the most general type of auditory stimulus is partially coherent. It means that the response to arbitrary sounds is somewhere between the coherent and incoherent modes---interference tends to be partial as well. The response to the superposition of a sound and its own reflection also obeys their relative coherence time\footnote{Some of these effects are treated as part of the (monaural) precedence effect by distinguishing between signals that are fused (usually within an early 1--5 ms window), or perceived as separate for longer intervals (that become distinct echoes if very long). In the binaural effect, a distinction is also made for a shift in localization that characterizes signals with very short delays ($<1$ ms) \citep{Litovsky,Brown2015Review}. In our framework, it would be natural to expect three regimes that correspond to fusion (coherence) and echo (incoherence), but also including a long intermediate one of coloration (\cref{RoomAc}), which corresponds to partial coherence between signals. In binaural hearing, which has received most of the attention in this line of research, there is an additional dimension of spatial coherence that is not directly relevant in monaural or diotic hearing.}.
	
	\item \textbf{Pitch is the closest percept in hearing to color in vision}---While there are numerous examples of stimuli that produce either pitch or color in nontrivial ways, in their most mathematically straightforward elicitation (pure tones / primary colors), they both refer to carrier channels of sound or light. Therefore, these sensory channels may be treated as communication channels, irrespective of their modality. Each channel is itself monochromatic, but the availability of multiple channels, which are also ordered on a continuous scale (cyclical or not), makes color and pitch comparable. The specific combination of light wavelengths can give rise to the perception of secondary or nonbasic colors that are elicited by a simple mapping to the three primary colors. In sound, these chromatic combinations may reflect some aspects of timbre (including temporal variations of timbre that are analogous to spatial variations of color). The analogy clearly breaks down when harmonicity is included, at least as a spectral domain phenomenon. Human vision is effectively less than one octave wide (Table \ref{tab:acousticvision}), so the very possibility of harmonic effects is excluded by its limited bandwidth. However, temporal pitch or periodicity pitch can be directly gathered from the envelope, so analogies to optical periodicity in the spatial envelope domain may have some merit. 
	
	\item \textbf{The characteristic frequency of the auditory channel corresponds to the optical axis of the eye}---An object point on the optical axis should suffer no aberrations and obtain an ideal image. In hearing, this limit can be achieved only by the pure tone. This represents the fundamental assumption we have made---the paratonal approximation, which is analogous to the paraxial approximation in optics. This assertion should be qualified, though, since the visual and the optical axes of the eye are not exactly aligned, and there is a blind spot on the retina, which does not have an auditory analog. 

	\item \textbf{Depth of field is measured over time, between temporally close objects}---The blurring effect is exaggerated by the auditory system in the form of forward masking. Objects that sound similar and share similar coherence properties tend to mask one another more effectively than objects that are acoustically dissimilar and do not correlate well with each other. This may aid the listener in the formation of perceptual objects that are distinct from a competing background. 
\end{enumerate}

\section{Discussion}
This chapter began with an open question about the possible meaning of sharp auditory images. Instead of answering it directly, it led to an extensive analysis of the possible sources of aberration and blur in the system---some of which are hypothetical at present. This knowledge allows us to answer what sharp images are like, mainly by understanding what they are not: they do not suffer from any significant blurring aberrations. Namely, sharp auditory images are minimally defocused, exhibit negligible temporal and transverse chromatic aberrations, and inasmuch as they exist, they do not suffer from spherical and coma aberrations. Several extrinsic factors were mentioned as well that can blur acoustic objects and make them fuzzy upon arrival to the ears: high reverberation and multiple reflections, large distances, and turbulent atmospheric conditions. However, the acoustic source itself may be unstable, if it intrinsically produces aperiodic sound that varies too quickly for the auditory system to settle, which elicits a fuzzy auditory sensation. By definition, these are incoherent sources, which have random attributes that dominate their sound generation. 

The relationship between sharpness and defocus in the hearing system is less obvious than in vision, due to considerations of coherence and phase perception, which do not exist in vision. Sharp images are unlikely to be audibly defocused, which means that the corresponding objects should have a high degree of coherence. Specifically, images of narrowband objects (either as standalone components, or as polychromatic combinations of sharp components) may have a ``well-behaved'' phase-function that is insensitive to defocus. In parallel, avoidance of chromatic aberration implies high across-channel coherence, which is both spectral and temporal.  But, as the system appears to generally detect signals both coherently and noncoherently, the ideally combined image may be partially coherent, in reality. This means that a combination of defocus and decoherence may come into play within the imaging system and contribute to the perceived image. These considerations also permeate the logic of auditory depth of field, which provides a handle on temporal distinction between acoustic objects of different degrees of coherence. Some of these ideas will be explored in the next chapter, where different accommodating mechanisms are hypothesized that all require different capabilities of dynamic changes in the system signal processing. 

In practice, no spatial imaging system can be designed to be aberration-free and different types of aberrations may have to be traded off \citep[pp. 243--244]{Mahajan2011, Born}. An equivalent statement has not been proven for temporal imaging, but if it is understood mathematically as a one-dimensional analog of spatial imaging, there is no reason to expect it to be any different. If the existence of aberrations in auditory images can be established, then it opens the door for a novel assessment of sound perception that is much more rigorous than is in use today, whether in normal audio perception, or in hearing impairments. In any case, much work will have to be done to establish baseline performance levels and methods to measure them before such a feat may be possible. 

Another potential use for the auditory aberration concept may be the elimination of higher-order aberrations from normal hearing. In vision, it has been shown that visual acuity can be improved beyond its normal function by optically correcting for coma and spherical aberrations that characterize the normal eye \citep{Liang,Yoon}. While a similar target seems to be far from our current understanding of the auditory system, it is not unthinkable that similar manipulations may be possible there too. Hearing systems that have evolved to be exceptional in some animals like bats and dolphins may have evolved to have relatively aberration-free hearing that is closer to be dispersion-limited than in humans. 

Aberrations will provide an important stepping stone with respect to the analysis of hearing impairments, which will be attempted in \cref{impairments}.

\chapter{Auditory accommodation}
\label{accommodation}
\section{Introduction}
Equipped with a rudimentary understanding of what distinguishes sharp and blurry auditory images, we are now able to explore the final prominent functional analogy between vision and hearing: accommodation. This adaptive feature is indispensable in vision and provides the basis for the sharp optical imaging when objects are positioned at arbitrary distances from the eye. In a broader context, the accommodation of the lens is part of a reflex that includes the adaptation of the pupil size and the convergence of the eyes that enables a proper combination of the two images in binocular vision. While there is much evidence of various adaptations that the auditory system makes over different time scales, none has been framed in analogy to ocular accommodation. 

Because of the uncertainty about the exact architecture of the auditory imaging system, potential accommodating elements will be considered in terms of their physiological plausibility, supporting empirical evidence, and hypothetical advantage for hearing. Thus, should auditory accommodation exist, it may manifest as variable cochlear group-delay dispersion, time-lens focal time (curvature), neural group-delay dispersion, duration and/or shape of the aperture, auditory filter bandwidth, instantaneous sampling rate, synchronization regularity, coherent/incoherent product weighting in the inferior colliculus, or any combination thereof. Dynamic range compression, middle-ear acoustic reflexes, and other mechanisms that dynamically vary the gain applied in the signal chain---``\term{gain accommodation}'', perhaps---are outside the scope of this work and will be mentioned only tangentially in relation to the medial olivocochlear reflex.

This chapter briefly reviews the main features of ocular accommodation. It then proceeds to hypothesize an imaging-centered analogous function of auditory accommodation along with some evidence that can be used to support this idea. The different parameters that can be hypothetically accommodated are then considered in terms of their usefulness and likelihood to exist. Finally, based on the analysis, we further hypothesize that so-called ``listening effort'' is an emergent response to difficulties in auditory accommodation.

\section{Ocular accommodation}
\label{OcularAcco}
Accommodation is the automatic process that dynamically adjusts the optical power of the eye to maintain sharp focus on a visual target, with respect to its distance. The following brief summary---largely based on a review by \citet{Charman2008}---attempts to communicate only aspects that may have relevance for the hypothetical analogous mechanism in hearing. 

The biomechanical process in which the crystalline lens in the eye is stretched and flattened (accommodated) for distant vision is a complex coordinated action of ciliary muscles, which are connected to the capsule that contains the lens (see Figure \ref{TheEye}). A fine interplay of elastic forces is then responsible for the changes in shape and power of the lens. The accommodation of the lens is always accompanied with synchronized horizontal rotation of the eyes---\term{convergence} (also called \term{vergence}), which ensures that the two images lie in the central field of both eyes, so the visual cortex can produce a fused binocular image. Additionally, accommodation is generally accompanied with changes in the pupil constriction. These three processes are sometimes referred to as the \textbf{accommodation reflex} (also, the \term{near reflex} or the \term{near triad}), but only the change of the lens focal length is considered as accommodation proper. Normal accommodation is binocular and symmetrical, just like the pupil constriction, as the two eyes assume closely matched muscular movements \citep{Campbell1960,Flitcroft1992}. When asymmetrical targets are presented to the two eyes, the accommodation sets on a lens power that better matches the farthest object \citep{Koh1998}.

Accommodation is controlled automatically by the Edinger-Westphal nucleus in the midbrain that is located at the level of the superior colliculus. Control is done primarily via the (parasympathetic/cholinergic) oculomotor nerve, which also controls convergence and pupil constriction, although some minor inhibitory sympathetic innervation exists as well \citep{Gilmartin}. Despite its automaticity, it is possible to trigger accommodation by consciously attending to the object distance, or in the case of some individuals, to voluntarily control accommodation. The accommodative change of the lens power takes about 300 ms---a duration that most likely increases with age---with an additional 1000 ms that are needed for stabilization. However, even then, the focusing precision is imperfect and can fluctuate and may cause blur that is comparable to that cause by some of the higher-order aberrations of the eye. 

The optical stimulus that cues the visual system to accommodate its focus has been debated extensively and several mechanisms may be at play \citep{Toates}. In recent studies, optical wavefront-vergence (not to be confused with the convergence reflex mentioned above) has been consistently shown to cue accommodation even in monocular monochromatic vision \citep{MarinFranch, DelAguila}, in line with a classical hypothesis by \citet{Fincham1951}. This means that the eye is sensitive to the direction of retinal defocus, which corresponds to whether the blurred wavefront is convergent or divergent with respect to the focus. The response to these cues, however, is facilitated exclusively through foveal cones, before the control information is neurally fed back to the ciliary muscles \citep{Toates}.  

\section{What may auditory accommodation be like?}
\label{PsychAcco}
As was argued in the last chapters, auditory imaging is different from visual imaging, primarily because the acoustic phase information is slow enough that it can be processed directly by the neural system---at least at low carrier frequencies. This means that the intrinsic defocus of the system has the potential to differentiate sounds according to their degree of coherence. In particular, defocus differentiates the incoherent and coherent temporal modulation transfer functions (TMTF) according to their cutoff frequencies. In analogy, we would like to get a better understanding of the hypothetical auditory accommodation function, but also distinguish it from  accommodation in vision\footnote{The word accommodation has been occasionally used in auditory research in reference to a host of adaptive behaviors that involve listening over time \citep[e.g.,][]{Holt2000,Carlile2014}. However, these word usages appear to have no direct relevance to auditory accommodation as is defined here.}.

Visual objects are critically characterized by their distance from the lens, which determines the focal length that the lens must assume for the eye to achieve sharp focus. Additionally, when the light intensity is high, the pupil (the spatial aperture) closes and lets less light into the system by eliminating high spatial frequencies, which causes an increase in depth of field and, as a result, sharper imaging. When this happens, the tolerance of precise focal accommodation can be somewhat relaxed, depending on the degree of sharpness afforded by the instantaneous depth of field. The situation is quite different in hearing, where depth of field manifests temporally as nonsimultaneous masking and appears to be exaggerated by the system for some types of acoustic objects. Therefore, unlike vision, the most relevant independent variable of the acoustic object is the degree of coherence of the source in its environment, rather than the distance from the object, which is factored into the propagated coherence function only indirectly. 

\subsection{Relevant empirical evidence}
Several studies have demonstrated an apparent adaptation that the auditory system makes for reverberation, which results in improved reverberant speech intelligibility scores when the system is given sufficient time to adapt. The interpretation of this effect has been challenged in that it can be elicited without reverberation, only through the manipulation of the modulation depth and spectrum of the stimulus. After reviewing these results we analyze them with a coherence framework instead. Subsequently, it is maintained that auditory accommodation may be understood rather organically by considering the function of a dual coherent-noncoherent detection system. 

The initial series of studies that may offer a window into accommodation-relevant hearing was pioneered by Watkins \citep{Watkins2005a,Watkins2005b}. In these studies, listeners were presented with a word token, ``\textit{sir}'' or ``\textit{stir}'', whose phonological category boundary was discretely varied between these two words using interpolation. The words were embedded in a carrier sentence and subjects were required to determine which of the two tokens they heard. But, the subjective categorical judgment between the two words can be impaired by reverberation, which reduces the temporal envelope cues necessary to hear the /t/ in ``\textit{stir}''. \citet{Watkins2005a} found that if the listeners are given a congruent acoustical context before the target---matching reverberation of both carrier and token---they are able to compensate for the blurring effect of reverberation, as the boundary between their categories is unaffected by the reverberation. This was unlike presentations with incongruent context, for which the token and context acoustics did not match, which caused a shift in the perceived categorical boundary. This effect was stronger for rapid speech than it was for slow speech presentations. Further, diotic presentations produced a larger effect than dichotic ones, which suggests a monaural mechanism. Also, single reflections enabled a similar but smaller compensation, maybe because they lack the long tail of reverberation decay that can inform the subjects about the appropriate acoustic context \citep{Watkins2005b}. Tests with narrow- and wide-band noises that were temporally comodulated as the speech context, showed that compensation becomes more effective the wider-band the context is and is almost completely ineffective with single auditory-channel-wide contexts \citep{Watkins2007}. 

The interpretation of the auditory effect as a mechanism that specifically adapts to reverberation was challenged by \citet{Nielsen2010}. They replicated the experiment done by Watkins et al. using carriers with no reverberation: white noise, a silent interval, speech-shaped noise, and that same noise with random amplitude modulation at 4--8 Hz. In all cases but one, the category boundary was preserved as in the congruent case, despite the absence of any reverberation cues. The modulated speech-shaped noise showed a slight change in the category boundary, but was still close to congruent results. The authors suggested that the observed effect is the result of forward masking in the modulation domain, because the context in the near condition has a larger modulation depth, so it tends to cause larger modulation forward masking than the other signals that have smaller modulation depth. Thus, if adaptation takes place here it may relate to the modulation and not the reverberation. This alternative interpretation was partially addressed by \citet{Beeston2014} using additional data that included the silent interval. They found that the adaptation was fine-tuned enough to have to rely at least in part on the reverberation information within the token word and not only on the preceding carrier context. 

Perhaps a more compelling indication for adaptation to reverberation may be inferred from a related series of studies, which generalized the findings by Watkins et al. through testing of speech-in-noise thresholds in a room acoustical context. In several experiments it was found that when subjects were pre-exposed to the room acoustics they had an average of 2.7 dB improvement in performance, compared with the unexposed conditions \citep{Brandewie2010}. However, this improvement did not translate well to anechoic conditions \citep{Brandewie2010, Zahorik2016}. It was optimal (up to 3 dB improvement) for reverberation times between 0.4 and 1 s, but was reduced with a longer reverberation time of 2.5 s \citep{Zahorik2016}. Additionally, in order for the exposure to the room acoustics to be effective, listeners had to listen for a minimum of 850 ms \citep{Brandewie2013}. Listeners were insensitive to the exact location within the room, which strongly affects the spectral weighting of the response---a sort of a  ``room constancy'' effect (\citealp{Brandewie2018}; cf. \citealp{Weisser2004}). Unlike \citet{Watkins2005a}, binaural listening was usually found to improve performance \citep{Brandewie2010}, although the test design in this case relied on contralateral presentation of speech and noise, which makes it difficult to generalize \citep{Zahorik2019}. In comparative behavioral tests of gerbils, which employed startle reflex responses as proxy for the threshold of sinusoidal amplitude modulation (AM) of broadband noise, reverberation compensation was not established \citep{Lingner2013}. This is maybe due to the gerbils' smaller head size and lack of corresponding binaural cues. It was suggested that gerbils may have never evolved to deal with reverberation in the first place, which makes the direct comparison to humans less telling. 

Several physiological studies in animals have also demonstrated reverberation compensation in how AM signals are coded at the level of the inferior colliculus (IC), yet no compensation in the ventral cochlear nucleus (VCN). Single IC unit recordings of unanesthetized rabbits found less synchronized responses to sinusoidal AM narrowband noise in reverberant conditions compared to anechoic responses \citep{Kuwada2012}. Synchrony, spike rate, and neural gain were about constant for source distances of up to 40 cm, and degraded at larger distances, even after controlling for level, in what appears to be the result of growing interaural decorrelation (i.e., spatial decoherence). The same neurons usually responded to azimuth as well. However, when examined in individual azimuths, higher synchrony and gain were recorded than would be expected from the loss of modulation depth, suggesting again a compensatory mechanism for reverberation \citep{Kuwada2014}. \citet{Slama2015} replicated and refined these results, by observing that the greatest reverberation compensation occurred for IC neurons that exhibited the most modulation depth compression, which itself may have had a cochlear origin. While envelope distortion, spectral coloration, interaural envelope disparities, and average interaural coherence could not explain the results, a subset of the neurons did respond to the interaural cues. These results are contrasted with recordings in the anesthetized guinea-pig VCN, where all unit types (primary, sustained/onset choppers) showed reduced temporal synchrony to pitch (speech intonation) in reverberant conditions \citep{Sayles2008, Sayles2015}. 

\subsection{Synthesis}
The main effect of reverberation is to decohere the signal, so only the direct sound (the early portion of the signal) may have its phase information relatively intact, to the extent that it can be detected coherently. The indirect, or reverberant, sound that arrives later is either partially coherent or incoherent, so its phase function cannot be recovered. For this portion of the signal, noncoherent detection or incoherent imaging are necessary, whereas a strictly coherent detection may  be unnecessarily noisy. The speech signal itself poses a similar challenge in detection, since it contains a complex mixture of coherent and incoherent sound elements, which renders the speech partially coherent. Any loss of modulation depth can be directly associated with the impulse function of the room, and hence with its reverberation time \citep{Schroeder1981}, and effectively, with the degree of decoherence. Therefore, from the system engineering point of view, it makes sense that the auditory system adjusts its detection according to the specific combination of signal and acoustics, so that its detected product at the output can be used to extract the desirable information from the signal. The two interpretations reviewed in the previous section---that the system adapts directly to reverberation, or to the modulation depth changes---both refer to the same physics, but attribute the effect to different levels of explanation. We argue instead that attributing the changes to the degree of coherence is the most parsimonious level of explanation. This argument is very similar to the one that was made in the context of auditory depth of field in \cref{NonSimultDepth}. Either way, all interpretations require some form of time-variant signal processing that is not generally considered in classical models of hearing.

It can be concluded that insofar as the reverberation compensation effect exists, it is relatively small, based on the tests that have been reviewed above. Can this effect be thought of as auditory accommodation? The answer is a cautious yes, as it matches well the main requirement set forth above, as well as other features of ocular accommodation. First, it matches the main prediction of auditory accommodation set at the beginning of this subsection---it reacts to the amount of reverberation in the signal, which is related to the degree of coherence (\cref{CoherenceReverb}). Second, it takes some time to be engaged---about three times longer than ocular accommodation---which suggests a dynamic mechanism as well. Third, its effect is stronger for faster speech stimuli, which suggests that higher modulation frequencies are implicated more strongly. This can be related to the cutoff of the TMTF---the main proxy variable that determines sharpness. Fourth, compensation may take place somewhere between the VCN and the IC (or in parallel in the dorsal cochlear nucleus, DCN), which matches with some areas that are hypothetically involved in accommodation (see \cref{HypoAcco}). 

All that said, if the reverberation compensation effect constitutes the hypothetical auditory accommodation, then it is relatively elusive and small, and may not be nearly as significant as in vision. As such, it may not make a very strong case for accommodation as a whole. However, other realistic signals but speech that are not purely coherent or incoherent have not been investigated frequently in the literature, and some real-world dynamic listening phenomena may be either unmapped, or misclassified as something else (e.g., a complex effect of attention or release from masking). Hypothetically, such effects may be strong enough to merit the evolutionary investment in their improvement. Therefore, the subsequent discussion does not necessarily apply to the reverberation compensation effect as the only possible candidate for accommodation effects, or even as the ``it'' effect, and may apply to other phenomena as well, which remain somewhat vague at present. Combining the insight obtained from the reverberation adaptation effect and the auditory depth of field, we shall be aware of the possibility that various masking release or enhancement effects can be reframed as effects of accommodation. As such, they could provide much more impressive advantage in detection than for reverberation compensation alone, which may be easier to justify in terms of the evolutionary investment that has led to its existence.

\section{The hypothetical accommodating component(s)}
\label{HypoAcco}
The case for having an accommodating auditory imaging system will be laid out for all of the components that have 
been touched upon in this work, which are not necessarily independent. The most immediately attractive known system to explore is the olivocochlear efferent reflex, which shares many of the superficial properties of ocular accommodation. It will be analyzed in the context of the hypothetical time lens and the phase-locked loop that we associated with the organ of Corti. Although the analysis is speculative at this stage, it is valuable beyond the present analysis in further consolidating our understanding of the different system parts, how they interact, and why they matter. As such, this discussion will inspire, at least to some degree, the next chapter about hearing impairments. We shall primarily focus on upstream auditory areas that receive the stimulus before auditory retina, which is assumed to be the IC, and seem to be capable of adaptation. 

Before delving into the individual component accommodation hypotheses, it is going to be instructive to put forth a strong overarching hypothesis of what the accommodation system does: \textbf{Auditory accommodation calibrates the signal detection to produce an image that is optimally sharp and as noise-free as possible. As most natural signals are partially coherent, the system is poised to produce a certain mix between noncoherent and coherent detection. The relative weighting of coherent and noncoherent products may be accommodated by manipulation of phase-locking, noise (decoherence), dispersion, and gain at different points in the auditory pathways. The system determines its optimal degree of coherence from previous stimuli and updates its operating points continuously, also using input from attention.}

In interpreting results from literature, several key points should be made. First, the auditory system may not be geared to work with completely coherent or completely incoherent stimuli, so the influence of both strategies is always present to some extent in normal hearing. Separating the contributions of the two may not be trivial, since they often produce very similar images. Second, much of the data in literature exclusively utilizes broadband (incoherent) and tonal (coherent) stimuli. A few studies also utilize narrowband (partially coherent), including speech, or other informative stimuli\footnote{Based on the analysis in \cref{ExCohere}, we consider speech to generally be partially coherent, but also highly nonstationary. This means that different regions in time and frequency may be instantaneously incoherent or coherent, but they do not extend for very long.}. Therefore, some relevant evidence should be interpreted with care. Third, a related concern is that these non-informative stimuli are given meaning that is imposed by the experimenter, but may not correspond to what the system considers meaningful. For example, in masking experiments, it is common to treat the masker as noise and the probe as signal or target that the listener should detect. However, it has to be considered that the masker is an auditory object in its own right and may be taken to be more interesting for the system to detect. Strictly speaking, both types of stimuli contain almost no information, but might nevertheless resemble signals that have some significance to the listener in their everyday environment. 

As was noted in the introduction, we do not deal directly with gain accommodation, although inasmuch as it exists, it should work in concert with all other accommodating components. Cochlear gain accommodation was implied recently by \citet{Carney2018}, who argued that efferent control of the hair cells may optimize the effective level they operate in, so that envelope level variations across channels may be adequately coded in the auditory nerve. This level is thought to match typical speech levels of 55--65 dB SPL, which has to be coded despite limited dynamic range in several elements of the system. This should impact the ``fluctuation profiles'' in the IC---not unlike our polychromatic image---elements of which can then receive attentional focus. 

The section is organized according to the presumed order of signal processing of the auditory stimulus that is relevant to dispersion and coherence. As such, we shall attempt any direct analysis of the role of the middle ear reflexes. 

\subsection{Cochlear group-delay dispersion}
The cochlear group-delay dispersion (Figure \ref{Reu}) is just large enough to dominate over small variations in normal atmospheric conditions of up to about 1 km (Figure \ref{atmo}) and fluctuations in the group-delay dispersion of the ear canal (including its multiple sign changes, Figure \ref{earcanalGDphase}). If it were any smaller, it could have resulted in unstable imaging, especially in variable environmental conditions, in all but a small range of the audio spectrum. On the other hand, cochlear group-delay dispersion that is larger than estimated can only have a relatively small effect on the total defocus, unless it is combined with additional neural group-delay dispersive changes. Either way, a large part of the cochlear group-delay dispersion may be a direct result of the spectral analysis property of the cochlea, which produces dispersion by having the auditory channels distributed along its length. 

Therefore, hypothetical accommodation of the cochlear dispersion entails a rapid change to the global group-velocity dispersion in the cochlea, or a change in its spatial phase dependence---effectively a cochlear map change. The former may be accomplished by actively changing the velocity of the traveling wave, the viscosity of the cochlear perilymph and endolymph, or maybe the velocity of sound in the perilymph or endolymph \citep{Donaldson1996}. However, no such dynamic change processes are documented in literature at present, to the best knowledge of the author. Additionally, it is not at all clear that the necessary anatomical mechanisms to efficiently realize such changes are feasible. 

In conclusion, cochlear group-delay dispersion is an unlikely parameter for accommodation. A long-term (non-accommodating) change in cochlear mapping will be discussed in \cref{CochlearDispChanges} in the context of hearing impairments.

\subsection{Time-lens curvature and the phase-locked loop}
\label{MOCR}
It has been commonly theorized that the primary role of the outer hair cells (OHCs) is to provide nonlinear amplification (see \cref{ElementsHearing} and \cref{TheInnerEar}), but this function may not be universal among different animal clades with similar organs \citep{Peng2011}. For example, it was recently found that short hair cells\footnote{There are two types of hair cells in the avian cochlea---\term{tall hair cells} (THCs) that are innervated with afferents and hence are functionally homologous to the mammalian inner hair cells, and \term{short hair cells} (SHCs) that have motile properties in common with OHCs.} in chickens do not produce any measurable amplification or tuning on the traveling wave \citep{Xia2016}. The temporal imaging theory hypothesized two new functions of the organ of Corti and OHCs---phase-locking through the phase locked loop (PLL) and phase modulation through a time lens. At present, the two new roles are speculative and it is not clear whether they interact with amplification, with each other, or with other functions associated with the OHCs.

\subsubsection{The medial olivocochlear reflex}
Attractive for hypothesizing any kind of OHC-related accommodation, there is a neural reflex mechanism in place to facilitate it---the medial olivocochlear (MOC) reflex, or MOCR. The MOC system shares some notable similarities with ocular accommodation. First, both systems are primarily cholinergic, where only in the eye it is part of the parasympathetic nervous system, while in the auditory system there is limited parasympathetic innervation \citep{Linker2018}. Both are bilateral reflexes, whose most dominant effect can be seen within hundreds of milliseconds (\citealp{Backus2006, Charman2008}; but see \citealp{Salloom2023} for more recent evidence of shorter time constants in the MOCR). Both auditory and ocular reflexes may also be mediated by higher-level processing such as attention to particular targets. The MOC system is also found in a hierarchically similar place to ocular accommodation in the eye, since it is attached to the lens in both systems. However, in vision, the lens is activated from a midbrain nucleus whereas it is activated from the brainstem in hearing. Incidentally, it has been shown that visual working memory tasks modulate the MOC activity as well \citep{Marcenaro2021,Vicencio2021}.

Notable differences between the reflexes are that normal ocular accommodation is almost completely bilaterally symmetrical, whereas the MOC can have a marked asymmetry, which depends on the degree of symmetry of the stimulus that activates the monoaural ipsi- or contralateral reflex, or its binaural version \citep{Guinan2018}. Additionally, the auditory MOCR has two or three associated time constants with its operation, where the hundreds of milliseconds one would be considered medium \citep{Backus2006}. Finally, ocular accommodation is tightly coupled to vergence and pupil constriction, which have no clear parallels in hearing, although the independent middle-ear reflex can be activated in similar conditions to the MOCR in some listeners \citep{Mertes2020}. 

The role of the MOC system is not well understood. The different functions that have been attributed to it are controversial, especially since many aspects of normal hearing appear to be possible with severed MOC efferents \citep{Scharf1997}. The MOC efferents inhibit the OHC somatic electromotility and, therefore, generally reduce and linearize the amplification in the cochlea, which may entail dynamic range optimization in some conditions \citep{Kunzel2017,Guinan2018,Lopez2018,Jennings2021}. In humans, its effect is most prominent between 500 and 2000 Hz and medium sound pressure levels. Results from studies on humans tend to be inconsistent between methods and suffer from high noise where otoacoustic emission (OAE) techniques are employed as a proxy to its activity \citep{Guinan2018,Jennings2021}. This sets several functions that are hypothesized for the reflex on a shaky ground, such as improvement of speech-in-noise perception, localization, or release from masking effects that improve tone thresholds in low-level noise \citep{Lopez2018}. The interaction between the MOCR and masking has been particularly thoroughly investigated, although a conclusive understanding of this function still cannot be produced---especially when different experimental methods are contrasted \citep{Jennings2021}.

Aside from the broad range of results in literature, it is challenging to interpret the MOC data because any interpretation depends on secondary models that are themselves not necessarily free of controversy (e.g., various OAE measures, assumptions about the involved time constants of MOCR activation). Another serious difficulty in assigning a higher-level role for the reflex within the system as a whole is that the stimuli that are most frequently used in these tests do not carry any information. This category includes narrowband and broadband noise that elicits the reflex, as well as distortion products from pure tones that are used as signals. Thus, if the MOCR system has evolved to realize a certain process that is commonly encountered in naturally occurring circumstances, then pure tone(s), broadband noise, and dichotic stimuli are likely to be very poor representations of such circumstances.

\subsubsection{Time-lens curvature}
The auditory time lens directly affects the amount of defocus in the system, since it can counteract the chirping effect of the cochlear and neural dispersion (Eq. \ref{temporal_imaging_condition}). We obtained a broad range of curvatures, which nevertheless excludes the curvature range that is necessary to achieve a sharp focus. We have mostly relied on the large-curvature time lens estimates that produced the smallest defocus from the obtained range, but found relatively little impact on curvature variations within the range in all the phenomena we examined, except for the stretched octave effect (\cref{TransChromAb}). However, the small-curvature estimates we obtained based on two studies suggested a near zero effect of the time lens. Additionally, one of the two studies used to obtain the large-curvature estimates was directly based on change in excitation of the MOCR in the gerbil, which seemed to have a dramatic effect on curvature \citep{Guinan2008}. Therefore, there seems to be evidence for variable time lens curvature. 

We proposed that acoustic phase modulation can be theoretically achieved through modulation of the stiffness of the medium (\cref{StiffnessPM}), although we did not directly rely on this principle in estimating the time-lens curvature. Stiffness is also theoretically related to amplification by the OHCs. In vitro, the magnitude of the OHC stiffness decreases as a reaction to acetylcholine, which would imply facilitation of gain, rather than gain inhibition that the MOC is known to produce \citep{Dallos1997}. However, as Dallos et al. noted, in-vivo effects are generally more complex, which may account for these results that appear to contradict amplification. \citet{Cooper2003} suggested that the stiffness changes are slow (10--100 s) relative to the rapid changes associated with amplificative negative damping. As for phase modulation, any change in stiffness mediated by the efferents---even a small one---is expected to affect the time-lens curvature, although the slow changes are nothing like ocular accommodation that has a dynamic effect in vision over shorter time scales. 

It was mentioned in \cref{PhaseModEvidence} that when the MOC efferent to the OHC is not stimulated, the traveling wave of the basilar membrane (BM) exhibits a slow phase modulation over the first few cycles of a click response \citep{Guinan2008}. The effect could be switched off or diminished by stimulating the efferents. It suggests that the OHCs automatically apply phase modulation as part of their nonlinear response, as was hypothesized in \cref{StiffnessPM}.


If the curvature estimates in Figure \ref{Focaltimefromphase} are correct, then switching off the phase modulation can substantially decrease the time-lens term in the imaging equation (\ref{temporal_imaging_condition}), which is expected to bring it much closer to sharp focus, depending on the baseline phase modulation in the system (see \cref{ImagingDefocusFound}). Eliminating or reducing the defocus by changing the lens curvature entails a smaller separation between the coherent and incoherent imaging responses. In terms of the modulation transfer function (MTF), the theoretical incoherent focused MTF has a broader bandwidth than the coherent MTF. In a focused system, sources of different degrees of coherence are no longer distinguishable through their modulation bandwidths. The advantage in assuming such processing for sound may be twofold. First, if the source of interest tends to be incoherent, then the object contour can be better defined by letting the high-frequency modulation content go through, which may improve the demodulation (but it is subject to constraints of the sampling rate). Second, this may be useful if the system does not attempt to emphasize any coherent sounds over incoherent one, but rather make them sound qualitatively similar. Another way of putting it is that decrease of focus increases the auditory depth of field, which may eliminate perceptual cues that can be used to differentiate between objects of different coherence types (remember that, somewhat unintuitively, a shallow depth of field provides an effective way to distinguish between objects; \cref{NonSimultDepth}). Depending on the acoustic conditions, these functions may prove more or less useful in realistic listening situations.

\subsubsection{The phase-locked loop}
\paragraph{Motivation}
Similar considerations can be applied to the auditory PLL function as were applied in the time-lens analysis, except that its accommodation is considerably easier to justify. Accommodating the PLL can hypothetically enhance or degrade the phase locking performance that is achieved mechanically and is transformed to neural synchronization after transduction. Evidence of a relation between the MOC and phase locking is rather limited, but nevertheless consistent with the PLL theory of operation.

First, let us consider what might be achieved by accommodating the PLL. The main parameter that applies to all PLL orders is the loop gain, which determines its hold-in bandwidth (\cref{LinearizedPLL}). It is the product of the different gains in the loop---the phase detector, the low-pass filter, and the oscillator. In the auditory PLL, we specifically designated the somatic motility with the role of supplying additional power to the loop. The loop gain affects how efficiently the PLL locks onto a signal in noise, how much noise is rejected in the process, how quickly it takes to acquire the lock, how stable the PLL is, how broad the (pseudo) narrowband filtering appears, and how well the lock can be maintained with random instantaneous frequency modulation. 

The motivation of modifying the PLL loop gain should depend on the kind of input signal, its signal-to-noise ratio, and the ideal strategy that the system can use to demodulate it. If the signal is coherent so it has a nonrandom phase function, then a coherent detection strategy that incorporates the PLL may be warranted and the loop gain may have to be set accordingly. But if the signal is incoherent, then phase locking to it, if at all possible, may provide little advantage. Worse, coherent detection of incoherent signals may require more computational resources and may result in excessive phase noise at the output. In this case, noncoherent detection may be advantageous and the PLL may be either bypassed or its contribution to the received signal reduced. 

It should be noted that in higher-order PLLs, additional parameters may be tunable, which determine the filter properties. However, given how little is known at present about this system, we will not speculate about other parameters beside the loop gain.

\paragraph{Empirical evidence}
The effect of the MOCR on phase locking on the level of the auditory nerve was physiologically estimated only a handful of times, using direct stimulation of efferent nerves that innervate the OHCs. In the cat, when the MOC was directly stimulated, the saturation point of the synchronization index to tones near and below the characteristic frequency was found to increase in level by 2--14 dB for on-frequencies \citep[Figure 4]{Gifford1983}, and by 0--16 dB for off-frequencies \citep[Figure 5]{Stankovic2000}. For the on-frequency tones, levels as high as 40 dB SPL achieve maximum synchronization, but these levels increase for off-frequencies, as they do not excite the fiber as much as on-frequencies. Using the same stimuli, the saturation levels were generally higher if measured using the spiking rates rather than synchronization. These increases were correlated with lowering of the operation point of the spiking rate in the auditory nerve and shift to considerably higher levels, which suggest a reduction of amplification and loss of compression (or linearization) and a release from neural adaptation. The only other relevant measurement found, somewhat trivially, that a contralateral stimulation by the same tone as the ipsilateral ear does not affect synchronization \citep{WarrenIII1989II}. As with the vast majority of auditory phase locking measurements, none of the above presented the time course of the lock acquisition or the tracking capabilities to nonstationary signals, which would have been essential to evaluate some of the PLL most important features (see \cref{Corollaries}). 

A different effect of efferent stimulation on the OHC was found in in-vitro samples of the bullfrog's sacculus\footnote{The sacculus is part of the vestibular system that is found in all vertebrates. It contains a sensory epithelium with hair cells and supporting cells, as well as afferent and efferent innervation. The hair cells are sensitive to low-frequency vibrations and sound and are similar to those found in the auditory system \citep{Fritzsch2013}.}, where the hair bundle spontaneous frequency (in the range of 10--80 Hz) changed from its baseline, and its phase locking to external tones significantly deteriorated \citep{Lin2020,Bozovic2021}. According to our PLL model, the hair bundle phase locking is the main precursor for neural phase locking. Thus, if the results from the frog's hair bundle translate to the mammalian OHCs, we would expect to see a drop in phase locking in the auditory nerve. The data from \citet{Lin2020}, however, cannot be straightforwardly compared to the mammalian data, so generalizing these results for acoustic stimuli at different levels and higher frequencies in mammalian OHCs requires more research. 

According to the PLL theory, the phase detector of the PLL is represented by the quadratic $f_2-f_1$ distortion product that is emitted by the OHCs (\cref{PLLcomponents}). Therefore, the $f_2-f_1$ level may be an indication of the phase detector sensitivity $K_m$ (\cref{LinearizedPLL}) and its general function. The effect of contralateral MOCR is known to either suppress or enhance the quadratic product, depending on the stimulus properties, but it hardly affects the more dominant cubic distortion product $2f_2-f_1$ \citep{Brown1988,Kirk1993, Althen2012}. Phase changes to $f_2-f_1$ are also observed during suppression \citep{Wittekindt2009}. The effects are usually dependent on the frequencies of the primary tones and the contralateral stimulus. Large ($\pm11$ dB) and spectrally non-specific effects of both suppression and enhancement of the DPOAE products can be triggered on a cortical level, which ultimately controls to the MOC through the corticofugal efferent network \citep{Jager2016}. 

A possibly PLL-related effect that appears to be triggered by the MOC efferents is a slight broadening of the auditory filters. It was indirectly found in humans that by stimulating the contralateral ear with broadband noise, which caused the ipsilateral ear's bandwidth to decrease by a small amount, as the measured delay of evoked otoacoustic emissions (OAE) during reflex activation decreased by 5\% at 500--2000 Hz, compared to baseline \citep{Francis2010}. Filter broadening in human was also demonstrated by measuring the filter shape using the notched-noise masking thresholds for tones, when the contralateral ear was stimulated with pink noise or narrowband noise (centered on the same frequency of 1000 or 2000 Hz; \citealp{Wicher2014}). It was found that the tip of the filter was the same as the quiet condition, but in the pink noise conditions the filter broadened by about 17\% at the 2000 Hz and by a smaller amount at 1000 Hz, which was statistically insignificant. The results were cross-validated by testing the distortion-product otoacoustic emission (DPOAE) that changed by up to 2 dB only in the pink noise condition. These trends generally disclose small effects that are not always consistent with similar studies that employed somewhat different methods \citep{Wicher2014}. 

Higher-level measures in humans usually attempted to estimate whether the MOCR has any effect on speech intelligibility. A recent study found that in a lexical task\footnote{The lexical task paradigm entails the identification of whether a spoken token is a word or a non-word.} the MOCR was strongly activated during a vocoded-speech presentation, much more than during speech-in-speech-shaped noise and speech-in-babble noise conditions, while during natural speech in quiet it was only moderately activated \citep{Hernandez2021}. The MOCR activity was recorded by monitoring the contralateral-ear click OAEs. Its effect was neurally modeled using the original speech tokens, which indicated that the received envelope function in the vocoded speech is closer to the natural speech than during the other conditions. 

The significance of the MOCR for speech intelligibility in difficult listening conditions is roughly in agreement between two studies. The first one found that recognition of monosyllabic words in noise with unilateral vestibular neurotomized patients with sectioned efferents had better performance (up to 20\%) in their healthy ear with intact reflex than in the operated ear, when presented with broadband noise to their the contralateral ear \citep{Giraud1997}. Similar results were reported by \citet[Figure 7]{Zeng2000}, although they were confounded by the hearing loss of the de-efferenated subjects. The speech reception threshold difference was 1--8 dB worse in the operated ears with large individual differences among four subjects. 

\paragraph{Synthesis}
There are at least two standard ways to synthesize the above findings and the role of the MOCR. One standard interpretation is that the auditory system manages the dynamic range and achieves release from neural adaptation that can happen after prolonged exposure to noise, which can lead to loss of fidelity \citep[e.g.,][]{ClarkBrown2012,Kunzel2017}. This model class does not explain why this reflex should depend on the contralateral stimulus, or why the uncompressed signal and noise should be perceived any more clearly than the compressed signal in noise, especially given that the system reacts to broadband noise that does not necessarily affect the signal passband. However, especially when combined with the function of the middle ear reflex, the model does point to that the auditory system may strive to process signals at a medium level, perhaps to maintain a convenient operating point \citep[e.g.,][]{Carney2018}. An alternative explanation is that the auditory system improves inputs at positive signal-to-noise ratio (SNR) by linearizing the input-output characteristics through gain reduction, which comes at the expense of negative SNR inputs---as may be gathered from masking experiments that are interpreted using the power-spectrum model \citep{Jennings2021}. Once again, the design logic from the system point of view is somewhat unclear, since the low SNR situations may become hopelessly inaudible, and the role of contralateral stimulus in activating this feature is still not well motivated.

An alternative, or perhaps a parallel, explanation to the MOCR advantage is that the auditory system sets its PLL (loop) gain according to the degree of coherence of the (preceding) stimuli. As phase locking may be less effective with the MOC inhibiting the OHCs ability to synchronize, it can imply that only the highest-level portions of the signals will be phase-locked, whereas the lowest portion will not, and appear more blurry. This coincides with the modeling in \citet{Hernandez2021}, which found that envelope fidelity---information that can be extracted without phase locking using noncoherent detection---is enhanced when the MOCR is engaged. It also coincides with the dual-processing model by \citet{Shamma2013}, who proposed somewhat different signal processing paths for the envelope detection and temporal fine structure (TFS) detection, which we have associated with noncoherent and coherent detections. In their TFS detection, the signal is saturated, so that no envelope cues remain in the spiking pattern---only temporal cues. Thus, the effect we saw that the MOCR stimulation caused in the saturation point of synchronization and spiking rate \citep{Stankovic2000} may reflect exactly that---enhancement of the dynamic range and reduction of synchronization improves the noncoherent detection at the expense of coherent detection. 

As speech signals are only partially coherent, either type of detection (coherent or noncoherent) can be used to detect it, as has been shown in the auditory literature in reference to envelope and TFS processing of speech (\citealp{Lorenzi2006,Paliwal2011}; see \cref{EnvelopeTFS}). Given that the quadratic DPOAE amplitude can be either enhanced or suppressed with the MOCR, such a control system may have some flexibility in setting the proportion of noncoherent to coherent detection, especially at medium levels. The large dynamic range of the DPOAE effects observed through corticofugal activation suggests that attentional mechanisms can indirectly control the detection strategy as well. 

The PLL explanation can imply that the contralateral ear reflex has something to do with the signal interaural correlation, or rather, with an internal estimate of the spatial coherence in the system (\cref{InterauralCoherence}). Broadband noise that is completely incoherent is known to trigger the contralateral reflex, whereas partially coherent (narrowband) noise does so partially, and coherent signals (tones) often do not (but see \citealp{Althen2012}). The same goes also for the ipsilateral signals, as they are usually tones that are incoherent, partially coherent, or coherent with the contralateral stimulus, respectively\footnote{See also \citet{Lilaonitkul2009} for a systematic comparison between ipsi-, contra-, and bilateral elicitor bandwidth effects.}. When an uncorrelated stimulus is presented to the contralateral ear, the system may register it automatically as an important signal, rather than ``noise'' per se. Then, enhancing the most general signal processing---noncoherent detection---may constitute a more robust heuristics to deal with such stimuli. 

All in all, given the scant evidence for phase-locking effects in the MOC system, as well as the unavailable parameters of the hypothetical auditory PLL, the accommodation of the loop gain in the PLL is speculative, at present. Nevertheless, there appears to be merit in such a process and there are several strong findings that support such an effect in humans and animals. What remains relatively unclear at this point is whether the low-level shift that was observed in the synchronization index in the cat is relevant to humans and would it have any measurable effect in realistic listening conditions. Similarly, the bullfrog hair bundle synchronization data relevance remain to be seen in mammals. 

\subsubsection{PLL and time-lens accommodation}
Perhaps unsurprisingly, the hypothetical accommodations of the time-lens curvature and the PLL that were discussed above can serve the same purpose in sound processing. The time lens accommodation was tied to an apparent reduction in defocus, as the incoherent response gets closer to that of the coherent one and the depth of field increases. In such a design, it makes perfect sense that no phase locking should take place, as incoherent sounds may be detected noncoherently, only using their envelope. The reduction in the loop gain of the PLL also biases the system for more noncoherent detection than coherent detection. Putting the two together, it seems that the MOCR is designed to reduce the level of coherent detection and increase the noncoherent envelope detection. This explanation is supported by the modeling and data from \citet{Hernandez2021}. However, it may not match some of the hypotheses about the MOCR function that are found in the literature---none of which is in consensus at present \citep{Lopez2018}. Data regarding unmasking of speech in noise has been particularly inconsistent between studies and methods \citep{Smith2021}. Given that speech is partially coherent and may be almost as equally well-recognized with coherent as with noncoherent detection, the comment made by \citet{Lauer2021} about the MOC function may be apt: ``\textit{In some cases, these effects may not be apparent because compensatory or redundant processes are likely in play.}''

\subsection{The temporal aperture duration and shape}
The aperture time was expressed earlier (Eq. \ref{t0xindp}) as a function of all three dispersive elements (cochlear, time lens, neural), because of physical and mathematical constraints. However, the aperture may be set independently of these constraints, if it turns out to depend on a purely neural mechanism. A simple neural correlate, for example, is the action potential in the auditory nerve, which factors into wave I of the auditory brainstem response. The typical width of the largest peak of the action potential is just under 0.5 ms long \citep[e.g.,][]{Yoshie1968,Picton1974}, which is approximately the same as in the high-frequency channels computed of 4--8 kHz (table \ref{t0opts}). This correspondence fails at low frequencies, though, where the temporal aperture may not be purely neural and the aperture stop may depend on the cochlear channel (\cref{LowFreqCorr}). 

No less important than the duration of the aperture is its corresponding shape, as is expressed by the pupil function---effectively, a temporal window. For practical reasons, it was approximated here as a Gaussian function, which has a single parameter---width. However, although it was found to be a very faithful representation of the physiologically measured Chinchilla's pupil function, the measurement showed an asymmetrical fat tail on one side that deviates from the Gaussian (\cref{ChinAperture}). We know from modulation transfer function theory in optics that the pupil function essentially defines the image properties---its contrast, sharpness, magnification, and aberrations. Therefore, an ability to accommodate the pupil shape should directly affect the image quality in some situations. Wave I itself is known to vary in shape between people, but its neural source may be mixed with contributions from nearly overlapping sources (e.g., the hair cell membrane potential) \citep{Kamerer2020}. The feasibility and plausibility of biophysically achieving such an accommodation are unknown. 

The advantage of being able to vary the aperture duration is significant, as it affects the modulation transfer function characteristics, in concert with the neural group dispersion. Accommodation of the aperture time alone is probably analogous to the pupil function in vision, rather than to lens accommodation. Inasmuch as we may be looking for an analog to the role of the pupil in vision that limits the amount of energy on the retina, then the middle-ear and MOC reflexes may be more suitable candidates. But if the pupil is required not only in restricting the amount of energy that is transduced, then direct adjustment of the auditory aperture may have some merit. This can be achieved, for example, through activation of the lateral olivocochlear (LOC) efferents that synapse to the auditory nerve dendrites to the inner hair cells. As the LOC efferents are unmyelinated, such an accommodation may be too slow to be useful in dynamic situations, though.

All in all, this potential candidate element for accommodation seems rather unlikely, at least as a parameter independent from the other dispersion parameters, or as something that can change quickly enough to dynamically track the signal. 

\subsection{Neural group-delay dispersion}
\label{NeuralAccommodation}
Neural group-delay dispersion characterizes a piecewise conduit of information transmission before forming an image on the auditory retina---probably the IC. While this part of the system is hardwired, there is ample evidence that the brainstem has a range of dynamic capabilities that enable selective and rapid adaptation to different signals and conditions. A detailed review and analysis of the processes that mediate neural plasticity is out of the scope of this work (but see for example, \citealp{Tzounopoulos, Irvine2018}). Instead, three very general properties of the auditory brain are mentioned below, which may function as short-term mechanisms that can constitute the plausible auditory neural machinery for group dispersion accommodation. 

The first property is wide innervation of the descending auditory network in the brainstem, which allows for top-down control by higher auditory centers, through the formation of feedback loops. Thus, there are widespread descending projections from the IC to the cochlear nucleus (CN) and superior olivary complex (SOC), and from the thalamus to the IC, CN and SOC, among many others (\citealp{Schofield2010}; see Figure \ref{BrainstemFig} for notable efferent projections). Some of the connections to the CN project directly from the acousticomotor area of the external nucleus of the IC (ICX), which is itself connected to the superior colliculus (SC), where it is coordinated with visual and tactile inputs \citep{Huffman1990}. The SC itself may be indirectly controlling ocular accommodation as well, via the Edinger-Westphal nucleus \citep{May2016}. The exact function of these networks is not well understood, although the range of possible functions is constrained by the type of connections, e.g., excitatory or inhibitory \citep{Milinkeviciute}. Recent electrophysiological measurements demonstrated how the addition of context can quickly modulate the response at the subcortical auditory nuclei level. This was cleverly shown by having naive participants listen to stimuli made of three sine waves before and after revealing to them that these stimuli represent, in fact, sparse speech \citep{Remez1981,Cheng2021}. Hearing the stimuli as recognizable speech rather than tonal noise had a quick and dramatic effect on the frequency following response (FFR) amplitude, which led the authors to conclude that the subcortical facilitation of auditory processing could only be mediated by the descending efferent network from the cortex. 

The second property of the auditory pathways is that neurons tend to have multiple receptor types, which allow for different neuromodulators that innervate the system to fine-tune the involved auditory functions and their associated signal processing in complex and nontrivial ways \citep{SchofieldOliver2018}. The plasticity that arises as a result can take place at different time scales and may broadly reflect behavioral arousal, environmental stress, attentional inputs (e.g., regarding salience), past experience, and even social situations. 

The third general property is a short-term synaptic plasticity that has been identified throughout the auditory pathways, which affects the amplitude, timing, and rate of neural discharges \citep{Friauf2015}. Synaptic plasticity can be observed in amplitude changes of postsynaptic responses to presynaptic activation and hence, it modulates the synaptic transmission efficacy (i.e., signal processing speed and capacity).  

As the dispersive properties of the auditory brainstem ultimately reflect its information transfer dynamics, it is reasonable to expect that the plasticity that is offered by processes as mentioned above can vary the neural dispersion in different pathways. When dispersion is changed in a frequency-dependent manner, the group-delay dispersion changes as well. Depending on the magnitude of this change, it has the potential to strongly affect temporal imaging. The effect of accommodating the neural dispersion can be substantial in setting the cutoff frequency of the low-pass response of the TMTF, which may be significant for tasks that require high-frequency modulation content. At this point, we are unable to tell how plausible it is that the group-delay dispersion is modulated at all, and if it is, then what its temporal dynamics is like and whether it can serve as auditory accommodation.

\subsection{Filter bandwidth}
\label{Fbandwidth}
The advantage of having variable-bandwidth filters is a degree of control of the received coherence in every channel: narrow channels produce more coherent outputs than wide channels (\cref{CoherentFiltering}). Therefore, the sound of an incoherent object will appear more coherent in a narrower channel, which entails less blur, at the expense of spectral information loss from the full broadband sound, which should be processed temporally in a wide channel. This can be thought of in the extreme case of a high-Q resonant filter that oscillates with random noise as input---the filtered oscillation is quasi-tonal and partially-coherent, as it maintains the instantaneous phase of the broadband input, which varies slowly around the center frequency. Bandwidth accommodation is also likely to affect how polychromatic images are fused across channels. Also, in situations where off-frequency masking is dominant, the channel bandwidth may be able to decrease or increase the degree of masking---effectively, to adjust the contrast in the complex polychromatic images.  

Evidence that the MOCR system affects the auditory filter bandwidth was brought up in \cref{MOCR} in the context of PLL loop gain accommodation. Another general mechanism that potentially affects the bandwidth involves the corticocortical and corticofugal descending auditory pathways, which have the capability to selectively modulate spectral, temporal, amplitudinal, and spatial responses of neurons, as was measured primarily in bats and mice \citep{Suga2020}. One of the common findings is of filter sharpening following electric stimulation, fear response, or a repeating tonal stimulation---mainly in the primary auditory cortex and the medial geniculate body \citep{Suga2008, Suga2020}, and in the IC \citep[e.g.,][]{Yan2005}. It was observed following the retuning of off-frequency neurons and the reacquisition of the best frequency that was most relevant to the task at hand. There is not much direct evidence for such bandwidth shifts more upstream, but excitatory descending projections found in the DCN of the mouse may suggest that similar changes in bandwidth may take place there \citep[e.g.,][]{Milinkeviciute}. Similarly, bandwidth tuning effects were found in chickens in response to inhibitory GABAergic inputs to the nucleus magnocellularis from the superior olivary nucleus (SON)---the avian nuclei that are analogous to the CN and superior olivary complex in mammals, respectively \citep{Fukui}. While the function of the SON is mainly associated with localization, other roles of this or other control pathways may induce similar tuning effects for processing other attributes. 

Another mechanism to take into consideration is lateral inhibition, which sharpens the frequency response by suppressing off-frequency components in adjacent filters. It was originally found in insect vision \citep{Hartline1956}, and has been documented throughout the auditory pathways \citep{Nomoto1964, Sachs1968}, but is probably better thought of as a universal sensory mechanism \citep{Bekesy1967}. In the auditory nerve, it likely reflects the nonlinearity of the cochlea \citep{Ruggero1992Popper}. In the CN it may either preserve the cochlear response or sharpen it further \citep{Rhode1986ventral, Rhode1994Encoding, Caspary1994,Kopp2002}. This general mechanism is invoked to explain improved hearing in noise, as it is used to increase spectral contrasts \citep{Kluender2003}. However, the real-time effects and dynamic properties of lateral inhibition may be difficult to extrapolate from these low-level examples.  

There are two caveats to bandwidth accommodation. First, it is not clear that the auditory filter bandwidth as is encompassed by steady-state auditory filter models, has a significant influence on temporal information processing that is relevant in imaging (see \cref{AudDefocus}). The different neural mechanisms mentioned are bound to produce instantaneous bandwidths that are context- and situation-dependent. Either way, be it the auditory filter, or the neural sampler, or any other temporal constriction---the narrowest one in the processing chain functions as the aperture stop, which then has the limiting effect on the temporal modulation transfer function, which may depend on the filter bandwidth only at low frequencies \cref{TempAperture}. Second, even if broader channels can let in faster modulation frequencies, they still have to be sampled appropriately in order to make use of the extra temporal information that they can hold. This is discussed below in \cref{SamplingAccommodation}. 

In summary, there are several general physiological mechanisms in place for sharpening or broadening the effective bandwidth of auditory channels. In the present state of knowledge, none of them stands out as a distinct mechanism that resembles accommodation, but this possibility will have to be considered in the future. Bandwidth manipulation midway in the auditory brainstem may be better thought of as analogous to spatial filtering in optics, which works to process the modulation band rather than the carrier band information. Therefore, more complex passband morphologies that are occasionally observed in the auditory brain may have to be reinterpreted accordingly. In contrast, vision performs spatial filtering only with the pupil, which is roughly a circular aperture that may have equivalent one-dimensional analogs in the auditory brainstem in the form of bandpass filters. 

\subsection{Sampling rate}
\label{SamplingAccommodation}
It was argued in \cref{TemporalSampling} that spikes in the auditory nerve correspond to discrete samples, and that each one may embody a narrowband image in its own right (\cref{NaturalSamp}). If this argument is accepted, then the entire ascending auditory pathway and the stochastic nature of neural firing are expected to encompass several points of imperfect resampling between synapses in the brainstem. Furthermore, it can contribute to the effective downsampling that is observable in the IC and further downstream---the gradual decrease of maximum spiking rates. It means that the highest modulation-frequency information, which is captured by the fastest sampling sequences, is lost in transmission before the IC, unless it is recoded (or is used, if it reaches its destination) earlier. However, the fact that the spiking is nonuniformly distributed suggests that an instantaneous cutoff frequency is a more correct way to approach sampling, instead of an exact Nyquist rate as in uniform sampling (\cref{nonuniformunder}). It brings about the possibility of generating more spikes in order to better sample the incoming stimulus, and thereby provide less opportunities for information loss in the resampling and downsampling process, on the ascending pathways to the IC. 

There are several findings that support the idea that the spiking rate can be dynamically set by the system. Neural adaptation is one obvious example where the spike rate is known to be variable, as it decays after the signal onset \citep{Kiang1965}. As was mentioned in \cref{MOCR}, the MOCR is known to reduce the operation point of the OHC process, which causes a corresponding drop in the saturation rate of the auditory nerve (also referred to as adaptation), which is interpreted as dynamic range management of the system \citep{Lopez2018,Guinan2018}. However, it also entails that higher-level signals can be coded with the same spiking rate as lower-level ones, which is equivalent to a reduction of sampling rate. This change will have no discernible effect on tones and does not appear to not affect their loudness \citep{Morand2002}, but more dynamic signals can be instantaneously undersampled as a result of MOC inhibition. 

In more dynamic signals such as speech, the onsets may not be as well defined as with synthetic stimuli, but changes that can appear as onsets are important cues in phonetic segmentation in general \citep{DelgutteHardcastle1999}. For example, changes between formants are marked with frequency glides that excite successive fibers in the auditory nerve. Each fiber reacts with an unadapted onset response at a higher spiking rate, enabling a more precise representation of the signal, which cannot be attributed only to intensity differences (that are more readily associated with rate changes) \citep{Delgutte1984IV}. At the level of the CN, onset responses characterize most of the cells and fewer cells exhibit sustained responses. Using data from behavioral tests with speech-like stimuli, it was suggested that such adaptation effects enhance contrasts in complex sounds and may have a role in segmenting coarticulated speech \citep{Kluender2003}. This is supported by results from the aliasing detection study presented in \cref{Aliasing} (Experiments 4 and 5), in which adaptation was suggested to cause an almost one order of magnitude increase, on average, in threshold of temporal discrimination between successive pulses. 

At higher levels of processing, sampling may be less directly relevant than in the auditory nerve and brainstem, as information downstream is coded more efficiently and sparsely, and at lower rates. Nevertheless, cortical firing rates have been shown to be modulated by focused attention, in hearing \citep{Miller1972}, vision \citep[e.g.,][]{Moran1985,Luck1997, Spitzer}, and somatosensory processing \citep{Hsiao1993}. 

In summary, there is no strong evidence that the sampling rate can be modulated en masse in the brainstem, although it is locally quite dynamic and may be affected by several subprocesses in different contexts. An open question is whether the increase in cortical firing rate ``trickles down'' to the brainstem, and if so, whether it requires focused or selective attention or whether other mechanisms can cause it. At the present level of knowledge, it does not appear as a very likely standalone mechanism to embody auditory accommodation.  

\subsection{Synchronization accommodation}
\label{NeuJitter}
\subsubsection{Phase-lock coupling strength and noise}
Synchronization accommodation would hypothetically involve controlling the precision of phase locking---the auditory response to the fine-structure of the stimulus. There are two general mechanisms that can control or modulate phase locking. They are the modification of the degree of coupling between the local oscillator and the signal path and the variation of the amount of sampling noise. The first process was considered in \cref{MOCR}, where it was argued that the MOCR affects the level of phase locking that is provided by the OHCs. Activating the MOCR decreases the level of phase locking in the auditory nerve, which is equivalent to decoupling the PLL from the main signal path. We do not know if an opposite effect is possible with the MOCR (i.e., enhancing the coupling strength between the external signal and the local oscillator). Unless there are additional neural PLLs downstream (\cref{NPLLs}), then decoupling/coupling accommodation may limited to the cochlea. 

\subsubsection{Jitter}
The second process that may accommodate synchronization is based on addition or removal of noise, which modulates the degree of coherence of the transduced signal. One way to add noise is to relax the sampling precision, as the spikes are synchronized to both carrier and envelope frequencies with finite precision. The small instantaneous deviations between the spikes and the incoming wave affect the level of transmitted coherence of the input, as they correspond to proportional phase error---phase noise or \term{jitter} in the sampling. If the precision is perfect (zero jitter), then within- and across-channel cochlear-level coherence is conserved in the transduction process. Otherwise, it is eroded. While zero jitter is physically impossible, being able to control the finite level of jitter would theoretically enable the modulation of the degree of coherence of incoming signals and, which selectively subjects them to the defocus and to decoherence. So, the more incoherent signals are, the more defocus applies to them and the less effective coherent detection gets in demodulating them. Upstream, jitter may apply only to carrier phase locking and have little to no impact on slower timing patterns that belong to the envelope and to its processing. Thus, if the temporal precision of the envelope is also conserved downstream, then it may be used in the formation of auditory streams that can be attended to at a higher processing level \citep{Singer1999,Niebur2002,Elhilali2009,Shamma2011}.

Jitter exists during the transduction at the ribbon synapse between the IHC and the auditory nerve \citep{Rutherford2021}. This synapse reacts relatively slowly to high-frequency inputs, where jitter can limit the coding fidelity. It is considered a property of fiber type, i.e. low-, medium-, or high-spontaneous rate fiber. Whether this property can be accommodated to---say, using the slow LOC efferents---is unknown at present.

In the neural domain, there are several discontinuities between the auditory nerve and the CN in terms of synchronization, which varies between cell types and between the DCN and the VCN. The DCN appears to have poor phase locking capability, yet a spectrally fine-tuned response. In comparison, the VCN provides an excellent phase locking response, which is primarily projected to the SOC (associated mainly with localization processing) \citep{Rhode1986ventral}, but not exclusively, as it also projects ipsilaterally via the trapezoid body and the ventral nucleus of the lateral lemniscus (VNLL) to the IC \citep{Oertel2002}. Particularly in the anteroventral cochlear nucleus (AVCN), phase locking is thought to be the result of multiple inputs that feed into ``high-sync'' bushy cells, which yield an improvement over the initial synchronization observed in the auditory nerve \citep{Joris1994}. The bushy cells are exceptionally well-designed for precise temporal coding on all cell levels (synapse, membrane, action potential, high frequency operation, quick repolarization, sustain operation; \citealp{Kuenzel2019}). Furthermore, these are multipolar cells with multiple inputs that have extensive neuromodulatory capabilities that can likely fine-tune these well-calibrated phase-locking features (see \cref{NeuralAccommodation}). While the exact function of such a neuromodulation may not be well-understood at present, it is possible that it has a role in accommodating the degree of synchronization in some or all of the subnuclei of the CN. The overall design here appears to be promoting decoherence in the DCN, which may be suitable for noncoherent detection, and coherence in the VCN, which may be more suitable for coherent detection. Their combined product at the IC is partially coherent to a degree that is determined by their individual contributions. 

The analogous operation of jittering in continuous media is diffusion. Therefore, in optics and acoustics, it is sometimes achieved with diffusers. 

\subsubsection{Dither}
The alternative to jitter is to add noise directly to the signal, which is sometimes referred to as dither. It has been discussed in the context of mechanical motion of the hair bundles, as it was found in hair bundles of the frog's sacculus in vitro that a small amount of noise that can be generated from Brownian motion, for example, can actually improve the SNR by up to 3 dB at low-SNR conditions, owing to a phenomenon called \term{stochastic resonance} \citep{Jaramillo1998,Indresano2003,Benzi1981}. However, while it can be assumed with some confidence that these findings apply to audio frequencies in the cochlea as well, the effect was demonstrated using white noise and pure tones, so it may be difficult to generalize. Also, we do not know whether the source of noise can be controlled internally within the cochlea, or if it strictly depends on external supply. The effect appears to contribute to coherent detection, but it may be inverted to do the opposite at different input SNRs. For example, since phase locking is known to diminish at high frequencies even though the OHC architecture seems to be the same along the entire spectrum, the possibility that the OHCs are designed to add dither at high frequencies through random movements---rather than phase lock---may be worth exploring. 

Dither may be more commonly explored within the neural domain, where the amount of spontaneous spiking directly corresponds to noise. Spontaneous spikes were shown to have a desirable dithering effect on envelope detection in auditory nerve axons \citep{Yamada1999}. Spontaneous activity in the central auditory system was shown to depend on the stimulus and may or may not relate to intended manipulation of the noise floor. For example, spontaneous activity in the DCN, VCN, and IC widely differs between different cell types and some are inhibited with the increase of contralateral input level or after onset \citep[e.g.,][]{Syka1984,Rhode1986dorsal,Rhode1986ventral,Joris1994}.  Promisingly, early deep-neural-network simulations of the simplified DCN processing suggest that the addition of Gaussian noise at that stage can improve speech recognition accuracy after cochlear hearing loss \citep{Schilling2022}. Nonetheless, we do not know at present whether any of those internal noise floor variations can be modulated or have a desirable effect on detection---whether through decoherence or other signal processing tricks.

\subsubsection{Relation to tinnitus?}
Another perspective on synchronization may be gathered from recent findings and models of \textbf{tinnitus}---the perception of phantom sounds that do not correspond to external acoustic sources. In tinnitus, cortical synchrony is a robust correlate to the perceived phantom sound \citep{Eggermont1984,Eggermont2012a}. An accommodation-relevant hypothesis is that (some forms of) tinnitus is an extreme manifestation of a naturally occurring process in the auditory system, whose normal function is to accommodate to the target stimulus. When such a system receives a complex signal that is composed of a mix of partially coherent stimuli, accommodation may be geared to selectively enhance synchronization in order to cohere and sharpen certain bands, or to decohere and to blur others. When a real stimulus is being processed, the synchronization accommodation is subtle and automatic (unconscious). In contrast, when nothing but noise is being detected (i.e., only weak cochlear activity, or spontaneous activity in the auditory nerve), then the system coheres to random patterns. If in addition there is sufficient central gain that is applied to the noisy channel, it may be perceived as audible noise with pitch strength that is inversely proportional to the bandwidth involved (i.e., narrow bandwidth---high pitch strength). In the normal system, this adaptation would seldom ``glitch'', but when it does it may be experienced as a spontaneous \textbf{transient tinnitus} \citep{Flottorp1953}---an obscurity even within the tinnitus literature that is excluded from most surveys and is sometimes attributed to spurious activity of the OHCs \citep[e.g.,][pp. 3 and 15]{Eggermont2012a}. In hearing impaired listeners with mild tinnitus, an advantage is occasionally reported in speech-in-noise measures, compared to listeners with hearing loss without tinnitus \citep{Husain2023}. If accommodative synchronization indeed exists, then the synchronization process must include a control signal that modulates the relevant parameters of accommodation, as controlled by attention, for example. This hypothesis combines at least two of the prominent theories of tinnitus---elevated synchronization and central gain (see \citealp{Henry2014Tin} for a concise review)---but has a higher explanatory power, because it is less arbitrary as far as the complete system function is considered. 

For this dual-hypothesis---that tinnitus is an abnormal form of synchronization accommodation---to be correct, two conditions have to be met. First, the hypothetical synchronization accommodation has to be applied before the imaging stage at the IC, i.e., while in the brainstem, or even in the cochlea. A related clue may be found in one prominent tinnitus theory, which traces its generation, but not its maintenance, to the DCN \citep[e.g.,][]{Henton2021}. Fusiform cells in the DCN of guinea pigs were found to be hyperactive (with elevated spontaneous firing rate) with increased spontaneous synchrony both in noise- and in drug-induced tinnitus \citep{Martel2019}. However, this kind of plasticity is very slow in comparison with any useful reaction time for accommodation. Hence, the second condition is that the ability to induce a change in the amount of jitter or dither has to take place over short time durations. Tinnitus research has looked into long-term plasticity measured over days or longer, so there is no direct indication for this, to the best knowledge of the author. However, tinnitus studies generally exclude the normal, non-pathological, operation of the DCN, which may include some of the neuromodulatory mechanisms (of the kind mentioned in \cref{NeuralAccommodation} and \cref{Fbandwidth}), which may be able to achieve synchronization accommodation over shorter time constants \citep{Kuenzel2019}.

It should be noted that it may be impossible to treat synchronization completely independently from sampling rate. In the VCN, it has been emphasized that the relatively low firing rates in comparison with the auditory nerve may pay off as a processing strategy that achieves an increased temporal coding precision (\citealp{Keine2017,Dehmel2010,Kuenzel2011}; for similar findings in trapezoid body, see \citealp{Wei2017}). The same cannot be said about the auditory nerve, where synchronization and spiking rate seem to be largely independent factors \citep{Johnson1980}, 

Another clue that tinnitus may be related to an early-stage accommodation of synchronization is from recent auditory brainstem response (ABR) findings in tinnitus patients and normal-hearing controls. Synchronization can only be considered for signals that are coherent or partially coherent, but not for incoherent signals. Therefore, we would hypothesize to observe differential processing of different types of signals in the brainstem, depending on their degree of coherence. It was found by \citet{Tan2023} that in tinnitus patients the interpeak interval (latency) between waves I and V were shorter only with chirp stimuli (coherent), but not with (incoherent) clicks. The reduced interpeak intervals with the chirps at 45 dB (normal hearing level) were also correlated with lower speech intelligibility scores in those patients at 85 dB SPL. \citet{Tan2023} proposed that these results might be related to cochlear synaptopathy, but there was limited supportive evidence for this in the subject group of the study. 

All considered, accommodation through synchronization may be an attractive feature, but does not have strong evidence to support it at present. The exact synchronization processes in normal hearing---especially fast ones that can respond dynamically (at the scale of hundreds of milliseconds)---are completely uncharted. Additionally, it is unlikely that such accommodation would work completely independently from the accommodation of the firing rate, gain, and channel bandwidth. 

A lengthier discussion about tinnitus and accommodation will be deferred to the next chapter (\cref{TinnitusAccom}), along with some interesting analogies of related disorders, which will make the case for synchronization accommodation more compelling.

\subsection{Coherent and incoherent stream mixing}
\label{StreamMixing}
The imaging theory in this work has aggregated the neural pathways in the brainstem to a single parameter---the neural group-delay dispersion. In reality, each pathway may have somewhat different dispersive characteristics, as well as different internal noise characteristics, which are suitable for various kinds of processing. As an example that was mentioned earlier, depending on the particular cell type, in two out of the main three divisions of the CN, the DCN tends to have better spectral and worse temporal resolution, whereas the VCN has it the other way round \citep{Rhode1986ventral,Rhode1986dorsal,Joris1998}. These nuclei project both ipsilaterally and contralaterally to the IC. This means that the IC receives multiple versions of the same stimulus (e.g., \citealp[p. 264]{Ehret1997} and \citealp[p. 26]{Malmierca})---each one is potentially characterized by different defocusing and a different degree of coherence. Noise that is applied in either one of the pathways would reduce the degree of coherence selectively only to that processing path. Hypothetically, the IC can selectively weight the contributions of both inputs in order to optimize the received partial coherence, and therefore bring objects in and out of focus in the final mix that is then propagated to the auditory cortex. 

A close variation of this idea is that the two main pathways of the DCN and VCN specialize in noncoherent and coherent detection (\cref{CommunicationTheory}, \cref{DetectionSchemes}), respectively. The output of these two detectors may be optimal for some signals, but not for others. Weighting their contributions to the complete image may endow the system with a broad range of hearing strategies that can be optimal for arbitrary inputs. A similar design to this was hypothesized to exist in the avian auditory system \citep{Sullivan1984,Warchol1990} (see \cref{CentralNeuroanatomy}). It was also explored in some depth in \cref{DetectionSchemes} in relation to the PLL and with some evidence reviewed to support it---mainly from frequency-following response (FFR) studies in humans.

Compared to a few of the hypothetical accommodation mechanisms discussed above, these two hypotheses are attractive, as they seem relatively straightforward to implement. This is the case, especially given that there are already parallel signal processing pathways in the brainstem that converge in the IC with unclear role division. Nevertheless, the specialization of the VCN and DCN (or their avian analogs) serve only as circumstantial evidence, and at present there is no concrete proof that either variation of the mixing hypothesis holds. Furthermore, considering the cellular and functional diversity within the VCN and DCN, this role division may be too crude a classification. However, given the known capabilities of the auditory system to produce responses based either on coherent or on noncoherent detection, we speculate that some version of the second variation can describe the main split in the mammalian brainstem, which will be used as a hypothetical building blocks in the complete auditory model that is proposed in \cref{CompleteModel}. Even so, the hypothetical mixing feature may not count as accommodation proper, if only because its assumed function does not resemble ocular accommodation or any of the other two ocular reflexes. Instead, the mixing of the two pathways may be a much more fundamental feature of the auditory system writ large. In this respect, the precise role of the third branch of the cochlear nucleus (the posteroventral cochlear nucleus, PVCN), which is not always easy to discern in the human CN, remains to be defined more closely.

\section{What informs auditory accommodation?}
Regardless of the specific accommodated variable, the hypothetical accommodation system has to infer from the signal itself how to accommodate. Of the different accommodation mechanisms considered, the various functions of the MOCR are probably the only ones that have clear triggering. Both ipsi- and contralateral reflexes appear to depend most strongly on the bandwidth of the elicitor signal, and on its absolute level \citep{Lilaonitkul2009}. As the different studies tend to use broadband or narrowband white or pink noise to trigger the reflex, one may wonder whether periodic broadband, or other types of signals may elicit a similar efferent response. Triggering by the contralateral ear may indicate that the system strives to have some symmetry in processing between the two ears, perhaps to maintain some continuity of localization processing. Alternatively, a common mechanism of ipsilateral coherence detection between bands, or contralateral spatial coherence (interaural cross-correlation) has to be considered too, which can potentially inform the system about the optimal strategy to detect the image of interest for the listener. Such mechanisms may be realized using broadband coincidence detectors, which are thought to exist in the PVCN \citep{Oertel2000,Lu2018} that, in turn, projects to the SOC, where the MOC bundle starts. A problem with this hypothesis is that it appears that the MOC input comes from chopper units that provide sustained output \citep{Brown2003}, whereas the broadband coincidence detection properties are notable mainly in the octopus cells and are less quick and precise in the chopper cells \citep{Lu2018}. Either way, less speculative explanations can be established only after the accommodation function is elucidated.

\section{Listening effort and accommodation}
The term \term{listening effort} was introduced by \citet{Downs} in order to operationalize the added cognitive difficulty that aided hearing-impaired listeners experience, which may not be captured by their speech intelligibility scores alone. Listening effort has been recently defined in a consensus paper as ``\textit{the deliberate allocation of mental resources to overcome obstacles in goal pursuit when carrying out a} [listening] \textit{task}'' \citep{Pichora}. The usefulness of having a concept of effort lies in the ability of listeners to relate to it \citep[e.g.,][]{Luts2010}. Also, it differentiates certain tasks and listener groups in a way that may be more sensitive than other measures such as speech intelligibility, especially in ceiling-performance conditions. However, listening effort requires a model to base objective measures on and is difficult to pin down precisely, because it is correlated and potentially confounded with attention, fatigue, motivation, and other high-level cognitive variables, which are themselves not necessarily clearly defined. Listening fatigue, for example, is thought to combine emotional, cognitive, and peripheral components \citep{Hornsby2016}. Or, in some models, the effects of cochlear hearing loss are thought to be compensated by cognitive and executive functions \citep{Peelle}. These complex relations within the concept of listening effort make it a particularly different to measure and pin down reliably \citep{Shields2023}.

While the analogous \term{visual effort} is not well-defined either, it is less abstract to speak of visual strain or fatigue, or accommodation and vergence efforts, which directly relate to ocular muscle actions, at least in the peripheral processing stage \citep[e.g.,][]{Toates}. For example, accommodation effort was found in situations where the visual system attempts to maintain a sharp focus despite low-light conditions. In extreme low light, it retains a fixed focus and exhibits decreased resolution (\term{night myopia}; \citealp[pp. 1.34--1.35]{Johnson1976,CharmanBass3}). Vergence effort has been specifically shown to be associated with visual fatigue, or eye strain, for people spending much time in front of a computer screen \citep{Tyrrell1990}. Although accommodation and vergence efforts are not the cause of the so-called \term{computer vision syndrome} (characterized by headaches, eye strain, image blur, neck pain, and more), they are typical in lengthy viewing of visually demanding targets, such as computer or smartphone screens \citep{Rosenfield}. Accommodation effort is not completely automatic since the ``effort-to-see'' can be mediated by attention \citep{Francis2003}. Accommodation may be voluntarily and effortfully controlled in some conditions, which in turn affects vergence too, as part of the accommodation reflex (see \cref{TheHumanEye}) \citep{McLin}. Another compensatory voluntary action is eyelid squinting, which sharpens vision by reducing the aperture and field stops that leads to mitigation of refractive errors, but which may cause eye strain as well \citep{Sheedy2003}. This voluntary action circumvents the inability to control the involuntary pupil constriction that normally determines the aperture stop of the eye.

The anatomical section of the eye that is behind the lens is strictly optical and unmistakably peripheral. In the auditory system, past the auditory nerve, processing is done in the central nervous system and may therefore be considered more firmly integrated with cognitive functions than the analogous parts of the eye. Most of the eye muscles have no analogs in (human) hearing, so an interoceptive monitoring of auditory accommodation---through sensation of strain, fatigue, and pain---is unavailable to auditory circuits. Physical fatigue of the eye may also tap into muscle fatigue more precisely, which is defined in some contexts as ``\textit{an exercise-induced reduction in the ability of muscle to produce force or power whether or not the task can be sustained}'', while noting that the reduction is task dependent \citep{Enoka2008}. 

Inasmuch as forming a sharp image of attended objects is a goal of both the visual and the auditory systems, having analogous low-level accommodation processes that maintain sharpness may help with the disentanglement of low- and high-level components of listening effort and demystify some of its inner workings. Therefore, it is proposed here that listening effort is, in fact, \term{auditory accommodation effort}---it measures the activity of the process that has to be dynamically performed by the brainstem (perhaps along with other areas and with the mediation of attention, or even fluctuations in the metabolic demands within the organ of Corti) to maintain an optimal image quality, using the various mechanisms considered above.  

\section{Discussion}
Although the existence and function of auditory accommodation remains hypothetical at this stage, this is nevertheless the first systematic exploration of the possibility of accommodation in hearing, to the best knowledge of the author. While the number of speculations in this chapter may appear prohibitive, it is maybe comforting to know that much uncertainty had characterized the understanding of ocular accommodation as well. For a long time after it was first discovered by Christoph Scheiner in 1619, it had been debated which one of several possible anatomical mechanisms is at the root of accommodation \citep{Charman2008}. Only in 1801 did Thomas Young prove that it is caused by changes in the focus of the crystalline lens. Given that a relatively larger portion of the auditory system is neural than the in visual system, hearing does not lend itself as conveniently to the analysis that vision received using optical principles alone. At the same time, the involved circuits in the brainstem are relatively compact, so substituting some of their actions with more intuitive closed-form operations may be possible, despite their complexity. 

Of all the mechanisms that were reviewed, the MOCR stands out as the one that can be most readily analogized to accommodation in vision in terms of anatomy. Its existence itself is unmistakeable and its prevalence in different forms in all vertebrates is indicative that its function has to be biologically useful, perhaps more than has been gathered from results of various behavioral studies to date. Its main function, however, remains  opaque according to the current literature, and we hypothesized that its role is to control the phase locking precision, which can preferentially enhance noncoherent (intensity envelope-based) detection at the expense of coherent detection. This function seems to be tied with the accommodation of the time-lens curvature, which controls the amount of defocus that determines the separation between coherent and incoherent parts of the stimulus. Consequently, such a change may impact the depth of field too, where it is dictated by the different degrees of coherence of the various elements in the acoustic scene.

The psychoacoustical and physiological evidence presented in \cref{PsychAcco} for what appears to be adaptation to reverberant fields does not clearly coincide with the effects that characterize the MOCR. One difference is their different reaction times, which is about 100 ms for the MOCR and almost 1 s for the reverberation adaption. Therefore, the latter may be a result of one of the other mechanisms that were discussed, or a combination thereof, which is placed somewhere in the brainstem. This can relate either to neural group-delay dispersion, synchronization, or coherent/incoherent mixing accommodation.

The accommodation analysis attempted as much as possible to tease apart the various mechanisms as parameters that can be manipulated independently of one another. However, in all likelihood, as was occasionally implied, sampling rate (rate coding), neural synchrony, neural group dispersion, and filter bandwidth are all tied together somehow. This was seen in the MOCR that affected bandwidth, sampling rate, phase locking, phase curvature, degree of coherence, and gain in different amounts. Or, it can be inferred from the fact that the increase in the VCN synchronization rate seems to come at the expense of high firing rates. If a combination of such mechanisms turns out to be mostly working in tandem, then the analogy of auditory accommodation may be expanded to include an entire ``auditory accommodation reflex'' as in the near triad of vision.

By eliminating and refining some of the hypothetical accommodation mechanisms, more clarity may be obtained that can eventually reduce the overall apparent complexity of the auditory system, rather than exacerbate it. This may be obtained by creating more rigorous connections to higher-level phenomena such as listening effort and tinnitus, as well as by unifying parallel concepts from vision.

\chapter{Dispersive and synchronization hearing impairments}
\label{impairments}
\section{Introduction}
Hearing and vision are the two foremost senses in humans that gather information about their environment. Impairment in either sense can lead to a substantial loss in quality of living. In comparison with prescription glasses, contact lenses, and corrective surgeries that passively correct the most prevalent refractive errors type of eye disorders, sensorineural hearing impairments are not as amenable to treatment using today's technology that is centered on restoring audibility using active circuitry. However, the comparison between the most prevalent visual and auditory disorders may be fundamentally misleading, because the visual ones are refractive where visibility is not a problem (except for cataract), whereas the auditory ones are often dominated by audibility issues. But the present work begs the question whether dispersive and synchronization disorders of hearing exist, in a manner that degrades the auditory image and its perception, similarly to refractive problems in vision. In other words, it is argued that there can be conditions under which listeners experience abnormal loss of sharpness (or blur) of auditory images. 

The comparison between auditory and visual impairments may also appear somewhat naive and even unfair, because of the heightened intricacy of structure, functional complexity, and enclosed anatomy of the hearing organ compared to the eye. The acoustic signal reaching the outer ear is transduced multiple times before it is neurally coded, whereas the optical signal is carried electromagnetically throughout the periphery and its transduction is confined to the photoreceptors and a subset of nonimaging photosensitive retinal ganglion cells \citep{BurnsChalupa2003, Hoang2010}. Furthermore, as this work has attempted to show, the analogous image processing that takes place within the visual periphery is realized centrally within the auditory system. Other technical factors challenge hearing theory in unique ways compared to vision: a much larger total relative bandwidth, overlapping modulation and carrier spectra, strong nonlinearity, folded dimensionality, and highly concealed and encoded image. These factors undoubtedly hamper the attempts to fully account for some of the more complex hearing impairments to a degree that enables engineering a fully satisfactory treatment for them.

This chapter aims to explore what can go wrong in auditory temporal imaging and whether it can be a root cause for any hearing disorders. There are three approaches that such an analysis can follow. A bottom-up approach may be pursued to systematically examine every component in the imaging system, test its likelihood to break down, and analyze the effects it may have on the image. In general, this should include drift in the main parameter values of the system (e.g., cochlear dispersion, time-lens curvature), excessive aberrations, and impairments in the accommodative function. A second top-down approach may be pursued to survey known hearing impairments and find out whether they can be reinterpreted using the new theory. In the third approach, known impairments in vision can be analogized to hearing in search for known disorders. Ideally, these approaches should converge and help demystify some of the most persistent and lesser-understood aspects of certain hearing disorders. The exploration space is therefore very large, so only select cases will be examined, where the degree of speculation is also deemed relatively modest. In that, an ulterior intention of this chapter is to open the door to other interpretations of hearing disorders in light of an imaging theory---most likely by incorporating new data and methods. 

The analyses presented in the chapter are organized according to the main dispersive system components, where available evidence for changes in each parameter is considered---often with respect to specific disorders. Therefore, several conditions, such as age-related hearing loss, are featured in several sections throughout the chapter. We systematically explore possible effects and evidence to changes in to the cochlear and neural group-delay dispersion, the time-lens curvature, problems in phase locking, enhanced aberrations, and disorders of accommodation. The result is not a complete map of possible dispersive disorders, but it is deemed sufficient to encourage further explorations that will hopefully be less speculative in nature. 

\section{The missing dimensions of hearing impairments}
While there is no specific test at present that assesses auditory sharpness, it is clear that information about it cannot be possibly obtained in audibility tests such as the pure-tone audiogram. Indeed, the limitation of the audiogram as a diagnostic tool for sensorineural losses has been occasionally noted, as it does not necessarily correlate with suprathreshold listening performance where audibility is not an issue \citep[e.g.,][]{Plomp1978,Papakonstantinou,Parker2020}. Several extra factors have been proposed that may constitute the missing dimensions that describe the space of hearing impairments more completely---often involving the outer-hair cell (OHC) function and the auditory nerve \citep[e.g.,][]{Parker2020}. Most recently, the degeneration of the ribbon synapse of the auditory nerve, or \term{cochlear synaptopathy}, has been implicated with a host of ``\term{hidden hearing loss}'' problems \citep{Schaette2011}. Definitively demonstrating it in humans and rigorously associating it with hearing impairments have been elusive \citep{Bramhall2019,Ripley2022}, but at least on one occasion it was suggested that a loss of spectrotemporal details in hidden hearing loss may be analogous to visual blur of object edge details \citep{Shinn2017Kraus}. Another prominent line or research attempting to identify the missing dimensions in hearing loss has focused on the association between degraded sensitivity either to temporal-fine structure (TFS) or to envelope cues. Often these studies also estimate the broadening of the auditory filters, or the distortion product otoacoustic emissions (DPOAE) to monitor the function of the OHCs. Either one of these factors could be considered an additional dimension to hearing loss beyond audibility. 

As has been contended repeatedly in this work, in order to be able to make sense of envelope and TFS cues, it is essential to consider the degree of coherence of the incoming signal (and masker), as well as its downstream incarnations. Perfectly coherent and incoherent signals may be relatively easy to analyze, but in reality, sounds (like speech) are partially coherent, which requires a more involved analysis. This is not only because partial coherence is more complex mathematically, but also because the degree of coherence is often unknown and fluctuates in time, which has to be considered in light of multiple time constants throughout the ascending auditory system. This means that the available part of the signal that contains usable TFS information---that may be extracted using coherent detection---fluctuates both extrinsically and intrinsically to the auditory processing. Furthermore, once the signal is transduced, the auditory system is most certainly dynamically engaged in optimizing and accommodating to the specific signals (at least during selective attention), using a complex network of feedback processes. The very property of phase locking was speculated to be one such process, which we hypothesized arises in the auditory phase-locked loop (PLL) of the OHCs and is propagated downstream from there. The normal operation of the PLL itself may come across as a narrowband filter, so phase-locking and frequency selectivity can appear to correlate using certain measurement methods. Finally, the usefulness of phase locking itself depends on its retention downstream, so that different sites of lesion can degrade its quality between the OHCs and the auditory cortex, and then contribute to a measurable loss of coherence. 

Keeping in mind these complex interrelationships between the different components of the auditory system, the potential role of dispersive losses as a possible cause for hearing impairment will be explored below. Also, phase locking as a function that depends on the health of the OHCs---and hence on the auditory PLL---will also be discussed. 

\section{Evidence for dispersive shifts in hearing impairments}
This section systematically reviews the different elements of the auditory imaging system using the bottom-up approach spelled out in the beginning of the chapter. It highlights data from literature that may point to significant drifts in the values of the dispersive variables in hearing-impaired listeners. The methods considered are closely related to those that were originally used to estimate the normal-hearing dispersive values of the system (\cref{paramestimate}).

\subsection{Changes in the total group-velocity dispersion}
\label{TotalImpairedChange}
Temporal auditory imaging includes three gross dispersive elements: the cochlear group-delay dispersion $u$, the time-lens curvature of the outer hair cells (OHCs) $s$, and the neural group-delay dispersion $v$ from the inner hair cells (IHCs) through to the auditory nerve and all the way to the inferior colliculus (IC). The cascade of the three is observable as a characteristic chirping effect, which constitutes a defocus that differentially affects coherent and incoherent signals. Earlier, psychoacoustic results of the phase curvature measurements by \citet{OxenhamDau} were used to test the temporal imaging theory and derive the temporal aperture of the system (\cref{CurvatureModeling}). Additional studies are reviewed below that indicate whether there are significant differences between normal-hearing and hearing-impaired subjects with regards to the total curvature. 

\subsubsection{Schroeder phase curvature drift}
Only few psychoacoustic studies used the Schroeder phase method to specifically test hearing-impaired subjects \citep{Summers1998,Summers2000,Oxenham2004}. In general, there are differences between sensorineural hearing-impaired subjects and normal-hearing controls mainly in the level of masking achieved and, occasionally, in the curvature as well. \citet{Oxenham2004} provided detailed curvature data at two frequency bands with the same stimuli as in \citet{OxenhamDau}---a 250 Hz tone masked with Schroeder complex with $f_0=12.5$ Hz and a 1000 Hz tone masked with $f_0=100$ Hz (as well as $f_0=12.5$ Hz, which was not used in \cref{CurvatureModeling}). The two frequencies that were measured produced mixed results, though \citep[Figure 1]{Oxenham2004}. While all 12 hearing-impaired subjects showed a significantly reduced masking effect, the observed curvature at 250 Hz shifted from $C=0$ (see Eq. \ref{schphase}) only in two subjects (HI10 to $C=0.125$ and HI12 to $C=0.5$), who also had the largest audibility loss at that frequency (55 dB HL). At 1000 Hz ($f_0=100$ Hz), the three normal-hearing subjects had curvatures of $0.5 \leq C \leq 1$. At the same frequency, the hearing-impaired curvature shifted to $C \approx 0$ in two cases (HI2 to $C=0.125$ and HI4 to $C=0.25$; both with no exceptional pure-tone audiograms), and was about the same as for the normal hearing in two other cases (HI10 and HI6), but it was unmeasurable in the other eight cases with no clear peaks. As \citet{Oxenham2004} noted, a key assumption in this test method is that the curvature is locally constant (i.e., within the paratonal approximation). However, curvature constancy may not be true in the case of hearing loss. Also, a complex frequency-dependent change to the impaired cochlear phase response may have caused some shifts. The authors tried to correlate the findings with filter broadening and found a significant but weak effect, which can be taken as indicative of a loss of OHC function that causes a loss of compression. Similar results were obtained at 1, 2, and 4 kHz of by \citet{Summers1998}, but with curvature values limited to $C = 0,\pm 1$.

\subsubsection{Derived-band auditory brainstem response changes}
An alternative method to evaluate the possible effect of hearing impairment on the total group-delay dispersion relies on the derived-band evoked auditory brainstem response (ABR) \citep{Don1978}. This method employs unfiltered broadband clicks that are partially masked by high-pass filtered pink noise. By successively subtracting the adjacent-band ABR responses, it is possible to recover octave-wide narrowband ABR responses. The narrowband responses are primarily attributed to changes in the cochlear function, but include all dispersive contributions downstream \citep{Don1998}. Changes to the total group-delay dispersion may be appreciated from changes in the frequency-dependent slope of the ABR latencies. \citet{Strelcyk2009} compared the derived-band ABR in five frequency bands of five normal-hearing and 12 hearing-impaired subjects with mild-to-moderate losses. Wave-V latency curves were obtained for all subjects, but had larger variability in the hearing-impaired data. Using psychoacoustic frequency selectivity measures at 2 kHz, the authors concluded that the difference in latencies between the two groups is most likely due to auditory filter broadening, and not due to synaptic or central differences. This is in accordance with linear filter theory, which predicts slower filter response (larger group delay) the narrower the bandwidth gets. The results agree with previous ABR data where filter bandwidth was not measured directly \citep{Don1998}.

As in other studies (\cref{NeuralDispEst}), the latency data in \citet{Strelcyk2009} were fitted according to a power law of a similar form to Eq. \ref{eqNeely}
\begin{equation}
	\tau(f,i) = a + bc^{0.93-i}f^{-d}
	\label{DerABRPower}
\end{equation}
where $b$ is a subject specific parameter, $a$ is a constant delay, $i$ is the click intensity normalized to 100 dB, and the frequency $f$ is in kHz. The equation did not have a significantly different frequency-dependent slope $d$ for the two subject groups, but had a significantly steeper intensity-dependent slope $c$ for the hearing-impaired group. 

\begin{figure} 
		\centering
		\includegraphics[width=1\linewidth]{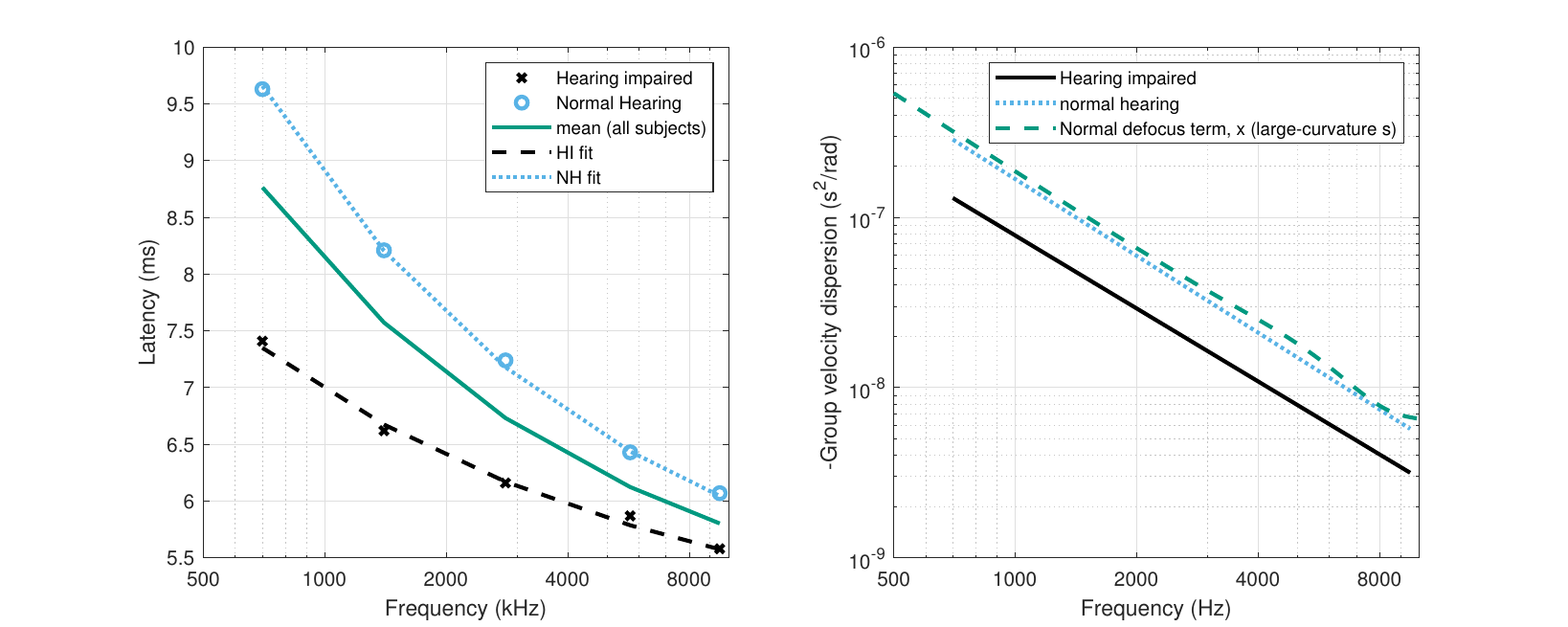}	
		\caption{Comparison of the derived-ABR latency and group-velocity dispersion of a normal-hearing and a hearing-impaired subject from \citet{Strelcyk2009}, Figure 3. The original normal-hearing data is plotted in black asterisks and the hearing-impaired subject in blue circles. \textbf{Left:} The latency measurements and individual fits to Eq. \ref{DerABRPower}, as well as the average across all subjects (hearing impaired and normal hearing). \textbf{Right:} The corresponding group-velocity dispersion, compared to the defocus term $x$ obtained earlier in \cref{CurvatureModeling}, obtained for the large-curvature time lens (both broad- and narrow-filter estimates) that we have used in most chapters (\cref{TimeLensExtrapolation}), which produces a close prediction to the normal ABR.}
		\label{DerABRcomp}
\end{figure}

In order to estimate the range of variation between impaired and normal hearing, two corner ABR responses (at 93 dB SPL) from \citet[Figure 3]{Strelcyk2009} are replotted in Figure \ref{DerABRcomp}, left. The data points were iteratively fitted to Eq. \ref{DerABRPower} from the initial mean values provided in \citet{Strelcyk2009}, where $b$ was matched individually and $d$ only for the hearing-impaired subject. The group-delay dispersion can be readily calculated from the power law, by differentiating $\tau$ with respect to $\omega$ and dividing by 2. The fits were then compared to the defocus parameter, as was obtained earlier from Eq. \ref{xdefocus}, where it was defined as $x = u + sv/(s+v)$. The normal-hearing subject response from \citet{Strelcyk2009} is relatively close to the $x$  value we obtained earlier (smaller by a factor of 0.8, approximately), whereas the hearing-impaired response is half as large, on average (Figure \ref{DerABRcomp}, right). 

The decrease in latency and, consequently, in total group-delay dispersion data analyzed here suggests that the defocus $x$ must decrease accordingly, through an effect on $u$, $v$, or $s$. As the magnification is close to unity, $M \approx 1$, then $x$ can be rewritten as $x = u + v/M \approx u + v$. Therefore, the causes for the measurable change may stem from a change in $u$ or in $v$. However, the two parameters are close in magnitude (Figure \ref{imagingcond}, right), so a decrease of $x$ by half, as was found above, cannot be achieved by changing a single parameter. Nonetheless, a filter broadening explanation may be stretched to account also for the interdependence of $u$ and $v$ that leads to covariation when the filter is broadened. Hence, on the cochlear side, the mechanical activity of the OHCs extends to the basilar membrane region also basal to the characteristic frequency (CF), which impacts the cochlear group delay that belongs to $u$. On the neural side, the group delay of the auditory filter and its associated IHC belong to $v$ and depend on its broadened bandwidth (Figure \ref{DispStages}). This functional division makes the differentiation between $u$ and $v$ challenging (see also \citealp{Eggermont1979}). We will look into contributions of the specific components below (\cref{CochlearDispChanges} and \cref{GroupDelayDisp}), using data that pertain to where differentiating between them is more clear-cut.

\subsubsection{Filter broadening and group-delay dispersion changes}
Regardless of the exact cause of the differences between the two listener groups (filter broadening or other), the implications of the drop in the total group-delay dispersion entail a decrease in the maximum amount of defocusing that the system can provide (not taking into account accommodative effects on the time lens that may dynamically vary the defocus). It means that the ability to segregate signals based on their degree of coherence---achieving optimal sharpness and depth of field---may be impaired. This may translate to difficulties in hearing speech in noise, which are common in hearing-impaired populations. However, a direct connection between loss of defocus and speech intelligibility cannot be established at present. Still, it is instructive to briefly review some of the effects associated with filter broadening on hearing. 

In literature, the effect of increased bandwidth on hearing is considered a function of degraded frequency selectivity, which diminishes the spectral contrasts of speech in noise \citep[e.g,][pp. 214--219]{Moore2007}. In cochlear hearing-impaired listeners, the role of frequency selectivity may be difficult to test independently of other factors, and the results have been mixed. For example, a principal component analysis was used to account for the performance of hearing-impaired subjects in an extensive test battery \citep{Festen1983}. Two components accounted for 65\% of the variance, and frequency selectivity was clustered along with speech-in-noise performance, whereas hearing threshold was clustered with speech in quiet. Similar test battery correlations were found in \citet{Horst1987} and \citet{DreschlerII1985}, where temporal as well as spectral resolution was found to be strongly correlated with performance. In contrast, smearing the spectral envelope of speech for both normal- and impaired-hearing listeners was only weakly correlated with speech reception thresholds, indicating that spectral resolution per se may not be critically important for speech recognition \citep{terKeurs1993b}. Similarly, in another test battery reported by \citet{StrelcykDau}, no significant correlations were found between frequency selectivity and speech reception, while temporal measures proved to be more strongly correlated. Interestingly, \citet{StrelcykDau} included two subjects with normal audiograms, who had poor speech-in-noise performance (``obscure dysfunction''), whose frequency selectivity at 750 Hz was comparable to that of hearing-impaired subjects. The two subjects' binaural masked detection performance was also poorer than normal-hearing subjects. 

One method to tease apart the effect of frequency selectivity from elevated threshold has been through simulation. In normal hearing, lower speech intelligibility was found in noise, but not in quiet, in simulated filter broadening experiments \citep{terKeurs1992,Baer1993}. The broadening or spectral smearing was achieved by convolving a Gaussian filter of the desired bandwidth with the envelope of the short-time Fourier transform of the signal, and then resynthesizing it \citep{terKeurs1992}. The performance worsened with poorer signal-to-noise ratio (SNR) and broader filters. However, with competing speech as masker a poorer SNR was required to degrade the intelligibility to that of broadband noise with better SNR (\citealp{terKeurs1993a, Baer1994}; see also, \citealp{Duquesnoy1983}). This was the case for normal bandwidths and was further exacerbated with broadened filters. The difference was attributed to the increased capability to listen in the dips at equal SNRs with speech masker that is naturally amplitude modulated. Higher-level stream segregation may have a role in this difference as well. 

Any filter broadening effects that are captured in these findings appear to be too general to be able to tie directly to dispersion. However, a more mundane explanation for the deterioration in performance has to do with the shift in the dual spectrum of speech (\cref{TwoSpectra}), as components that are normally resolved in adjacent filters when they are narrow have to be processed temporally within the broadened channels. However, the temporal modulation transfer function (TMTF) cutoff frequency is constrained by the sampling rate of the system, which might not keep up with the increased modulation frequencies in the shifted dual-spectrum. Effects of changes to the TMTF are discussed in \cref{GroupDelayDisp} and \cref{LossCoherenceAmp}. 

\subsection{Changes in cochlear dispersion as a result of impairment}
\label{CochlearDispChanges}
Cochlear insults that impair the OHC mechanics lead to changes in the response of the basilar membrane (BM), primarily around the CF. By selectively disrupting the OHC motility using furosemide in the chinchilla's cochlear base around CF$=9$ kHz, \citet{Ruggero1991} found that the phase response is both frequency- and intensity-dependent in an area confined to the CF range, but is unaffected at lower frequencies than the CF (1 and 5 kHz). In another physiological study of noise-induced hearing-impaired chinchillas, the relative delay between channel CFs separated by up to half an octave was found to change relatively to normal hearing animals, in a frequency-dependent manner \citep{HeinzLopez2010}. These and similar findings suggest that it may be impossible to completely disentangle changes in the OHC curvature (associated mainly with $s$) from the cochlear dispersion $u$. 

One global cochlear change that relates more specifically to the BM mechanics is an abnormal rise in fluid pressure in the scala media. This condition is referred to as \term{endolymphatic hydrops} and is usually associated with Meni\`ere's disease. Endolymphatic hydrops cause the BM stiffness to increase and, consequently, for the cochlear traveling wave velocity to increase as well \citep{Donaldson1996}. This velocity does not change with cochlear hearing loss without hydrops (but see \citealp{Eggermont1979}). Derived-band ABR may be used to compare the traveling wave velocity, using the inverse of the latency function \citep{Donaldson1993}. This procedure assumes a negligible contribution of the OHCs \citep{Don1998}, which may not be entirely justifiable given that a subset of Meni\`ere's disease patients exhibit OHC pathology as well \citep{Cianfrone}. Also, OHC motility depends on cochlear fluid changes that tend to cause hearing fluctuation---a characteristic of the disease \citep{Dulon1987}. However, the presence of endolymphatic hydrops has been attributed to changes in the traveling-wave velocity, based on derived-band ABR measurements \citep[e.g.,][]{Kim1994,Donaldson1996,Claes}. Abnormal data from  of Meni\`ere's patients suggest that even when the ABR latency increases, its frequency dependence remains close to normal, as the traveling wave velocity slope is unchanged on a log scale \citep[subjects GM, VF, and TH]{Donaldson1996}. All other seven patients in that study had normal latency curves, as did the noise-induced hearing-impaired subjects in the study. This means that the cochlear group-delay dispersion $u$ does not always change with the velocity in Meni\`ere's patients, but if it does, then it increases by a constant factor. As a consequence of Meni\`ere's disease, patients normally have poorer speech intelligibility in quiet than normal-hearing listeners \citep{Bosmana,Tonndorf1976}. In contrast, they exhibit no unusual sensitivity to noise compared to normal-hearing subjects---contrary to listeners with noise-induced and acquired impairments \citep{Bosmana}. Once again, it is not clear that the dispersive shift per se should have a significant effect that can lead to this peculiar pattern. In principle, it may be that speech in quiet relies more on coherent imaging, where dispersion can be more dominant to the different coherent and incoherent components of speech, so it might be impacted. In contrast, speech in noise relies more heavily on incoherent imaging, where phase locking makes relatively little difference, so that changes to $u$ also do not matter much. 

Changes in the apparent cochlear dispersion can be theoretically caused by conductive losses as well. However, these losses tend to be flat and using broadband evoked ABR measurements reveals only small mean differences in latency, with high variability, between impaired and normal ears \citep{Mackersie1994,Ferguson1998}. These findings are not suggestive of great dispersive changes to the cochlear group-delay dispersion $u$, but derived-band measures will be needed to obtain more refined association.

\subsection{Static changes in the OHC (time lens) curvature}
\label{OHCimpair}
We hypothesized two new roles for the OHCs in this work: a phase locked loop (PLL) and a time lens. The speculative nature of these ideas notwithstanding, disentangling possible impairments in them may not always be possible, especially considering the traditional functions of the OHCs---to provide amplification and dynamic range compression, sharpening of the filters, generation of distortion products, and therefore the production of otoacoustic emissions (OAEs). Below, we focus on the time lensing feature of the OHC and return to the PLL in \ref{NeuSyncPL}.

We have argued that the time lens with its estimated curvature has a role in setting the degree of defocus in the image and its accompanying depth of field. Having a time lens also enables the temporal image to retain its orientation and not become time-reversed as would be the case with two negative group-delay dispersion segments--- more in line with current knowledge of hearing operation. We look at two of its functions that were explored earlier. The olivocochlear reflex effects that may be related to the time-lens impairment are examined in \cref{AccommodationImpairment}.

\subsubsection{Forward masking and depth of field}
It was argued in \cref{NonSimultDepth} that the pattern of forward-masking variation for different degrees of coherence between temporally adjacent stimuli can serve as the auditory analog for the spatial depth of field. As nonsimultaneous masking interacts with cochlear hearing impairment, it is interesting to find out if any effects may be attributed directly to the focal time of the lens, which by way of analogy is thought to control the depth of field, at least in part. Another significant part of the depth of field is most likely controlled centrally, so the exact cause may be difficult to pin down in most of the observations below.  

In general, the release from forward masking changes with the hearing impairment, due to the elevated absolute threshold. When the masking thresholds are compared at equal sound pressure level, then the baseline simultaneous level is the same in both groups, but the forward masking decays much more slowly for the hearing-impaired than the normal-hearing listeners \citep{Glasberg1987}. This effect can be accounted for using the loss of compression in the impaired cochlea \citep{Oxenham1997}. If instead the comparison is done at equal sensation level, so that the relative levels above the normal and elevated absolute thresholds are compared, then the masking decay is better matched between the two groups, although still somewhat slower for the hearing-impaired listeners. These results hold for both on-frequency and off-frequency masking \citep{Nelson1989}. The ability to benefit from release from forward masking may not be always equal between the groups, especially at high frequencies, where hearing-impaired listeners may benefit less than normal-hearing listeners from various manipulations in the masker \citep{Svec2016}. As a consequence, old normal-hearing and hearing-impaired listeners tend to score more poorly than young normal-hearing controls in vowel-consonant or consonant-vowel pair identification, as a function of noise and the delay between the phonemes, which may potentially translate to poorer speech intelligibility in complex conditions \citep{Fogerty2017}.

It is difficult to draw direct conclusions from these data regarding the time lens. It is not impossible that the nonlinearity of the cochlea and loss of compression also entails time-lens curvature reduction, as they are all thought to depend on the healthy operation of the OHCs. However, at present we have no direct information to support this, based on forward masking patterns alone. Given that the forward masking decay is slow and long, elongation of its decay pattern may be a result of central changes.

Regardless of the specific cause for it, there appears to be substantial changes to the extent of depth of field as experienced by hearing impaired listeners, primarily due to the change in the decay function cause by masking.

\subsubsection{Binaural diplacusis}
As the lens curvature dominates the expression of magnification, it enabled a rough prediction of the stretched octave effect, which is qualitatively similar to published data. If this effect is indeed related to the time lens, then it implies that abnormal pitch height perception may correspond to static shifts in the time-lens curvature.

\term{Binaural diplacusis} is a hearing disorder in which the same stimulus is perceived as having a different pitch in the two ears \citep{Shambaugh}. It may also be accompanied by a loss of tone quality and increase of distortion \citep{Ward1955}. Slight diplacusis ($< 1.5/16$ octave difference, which is 112 cents---a slightly sharpened semitone) is not uncommon in normal hearing people, but it is quite common with unilateral hearing-impaired loss, where differences may be up to a few tones in extreme cases \citep{Colin2016}. Because of its small effect, it is usually not a dominant source of dissatisfaction even among professional musicians \citep{Jansen2009}. In sloping high-frequency hearing loss, the pitch difference tends to be larger in the impaired range compared to the unimpaired range, and the pitch is usually perceived as higher than the reference pitch of the better ear. 

Binaural diplacusis is particularly challenging for pitch theory, because the direction of pitch change is inconsistent among studies and listeners, even when they have similar losses (i.e., pure-tone audiograms) f. Still, the most common explanation so far has been due to place theory, as impaired cochlea results in a shift in the frequency-position mapping of the BM, so that places that are normally maximally excited by a frequency $f$ move apically toward less impaired areas, which are normally excited by a lower frequency $f' < f$ \citep{Muller2010,Colin2016}. However, this explanation is at odds with the few cases that show pitch shifts in the other direction, sometimes combined with intensity-dependent effects that change the shift direction too \citep{Burns1986}. Standard temporal theory of pitch predicts correspondence to the stimulus period independently of place, but it also does not offer an obvious explanation for the inconsistencies. These considerations suggest that a combined place-time pitch perception may be the most adequate to account for the various observations \citep{Colin2016,Moore2007}.

The temporal imaging theory may be able to resolve the inconsistency of pitch perception in binaural diplacusis since it scales the frequency by the magnification factor. The scaling operation is both place- and time-dependent and depends almost exclusively on the OHC curvature $s$ and on the neural dispersion $v$. Any shift in cochlear mapping or impairment to the degree of phase modulation of the healthy OHCs are therefore expected to affect the perceived pitch. When the time-lens curvature is unimpaired and the frequency-position remapping is the only physiological change, the frequency scaling will change along the magnification curve of Figure \ref{FigStretched} (right). If the pitch that is normally associated with frequency is scaled, or stretched according to $f \rightarrow M_f f $, then the impaired pitch, under cochlear remapping, would be associated with the place and magnification of $ f' < f $, so $f \rightarrow M_{f'} f$. However, as was concluded in \cref{TotalImpairedChange}, the cochlear impairment also causes a decrease in $v$, which leads to an increase in magnification $M^*>M$. The compounded effect of remapping and magnification shift may cancel out, as $f \rightarrow M^*_{f'} f$, where $M^*_{f'}$ may be either larger or smaller than $M_f$. As magnification is a function of frequency, the effect will be noticeable mostly where the magnification function bends, above 1 kHz, approximately (Figure \ref{FigStretched}). In case that the OHC dysfunction changes the value of $s$, deleterious changes in pitch may take place as well.

These hypothetical explanations of pitch changes also have to take into account the existence of dead regions in the cochlea, where severe diplacusis has been occasionally reported \citep{Huss2005a}. Such cases may be a result of severe aberration that distorts the signal that is coded off-frequency in active channels at the edge of the dead region (see \cref{HigherAbbImpair}).



\subsection{Impairments in neural group-delay dispersion}
\label{GroupDelayDisp}
The neural group-delay dispersion includes all auditory pathways past the OHCs: the peripheral IHCs and the auditory nerve, as well as the different subcortical circuits, at least down to the IC. As such, there are several sites that can be impaired \citep{Rance2015} and may cause the signal to be delayed in a frequency-dependent manner. Accordingly, multiple disorders have been identified, hypothesized, and associated with the peripheral and central parts of these pathways---most of which are not well understood at present. The dispersive point-of-view developed here assumes only a single time lens per auditory channel, so all pathways downstream from the auditory nerve are readily combined into a single frequency-dependent variable $v$, although nonuniform resampling artifacts and different noise regiments may apply. Changed parameters may include the neural group-delay dispersion (e.g., frequency-dependent latency), changes in the internal coherence (phase locking) properties, and changes in sampling rate. In some situations, the latter two may be interpreted as the addition of noise to the signal (see \cref{NeuJitter}). Additionally, changes in the pupil function, or the role of aperture stop, may be considered as well, although it is likely dependent on the other dispersive changes. Evidence for some of these changes is presented below.

\subsubsection{Direct evidence for neural group dispersive changes}
Based on data by \citet{Strelcyk2009}, it was concluded in \cref{TotalImpairedChange} that changes in neural group-delay dispersion may be inevitable in listeners whose ABR latency dropped to about half its normal value. Otherwise, it was impossible to arrive at the dispersed ABR figures of some hearing-impaired individuals, given the baseline values used throughout this work. \citet{Strelcyk2009} concluded that the drop in ABR latency is very likely a direct result of auditory filter broadening, which causes a reduction in the filter group delay. However, it is not difficult to come up with other mechanisms in the periphery or in the brainstem that may cause frequency-dependent delay. When such hypothetical changes cannot be represented as a linear function of frequency, then the changes to the group delay are also frequency-dependent, which entails a change in group-delay dispersion $v$. 

Theoretically, channel-specific damage to any of its neural dispersive elements can be either a result of within-channel changes to the filter group delay, or of dominant off-frequency response in place of damaged channels. The latter case is a form of aberration and was discussed in \cref{HigherOrderAb} in the context of dead regions and is further analyzed in \cref{HigherAbbImpair}. The former case is more difficult to establish directly based on available data. For example, the IHC synaptic delay is considered to be independent of the CF \citep[e.g.,][]{Palmer1986}. In the elusive condition of cochlear synaptopathy, the loss of IHC synapses is thought to impair the auditory function, but the minimum amount of loss necessary to cause perceptible impairment in humans is unknown at present and appears inconsistent between studies \citep{Bramhall2019}. Additionally, it has not been reported in this context, to the best knowledge of the author, whether synaptic impairment that does not completely incapacitate the synaptic activity can result in a shift to normal synaptic delay values. At the same time, in-vitro experiments indicate that the IHC synaptic delay is plastic under some conditions \citep{Goutman2011,Cho2014}---a fact whose impact on hearing is also unknown at present. 

Neural conductive abnormalities also characterize acoustic tumors (\term{acoustic neuroma}), which can be indirectly detected using derived-band methods \citep{Eggermont1986,Don1997,Philibert}, but may be also gathered from an increase in ABR latency and decrease in amplitude (rather than complete absence of ABR; \citealp{Telian}). Regardless of the specific causes, if any such changes to neural conduction are selectively applied in some channels but not in others, then a respective change in neural group-delay dispersion will occur as well. Depending on the magnitude of the change, it may not be observable in broadband ABR measurements. 

\subsubsection{Effects on the temporal modulation transfer function (TMTF)}
Effects of a degraded contrast sensitivity function (CSF) (either in a limited band or broadband range of the spatial frequency spectrum) in vision are well-known to reflect a blurred image perception \citep[e.g.][]{Bour1996,Woods2000}. Thus, continuing the space-time analogy between vision and hearing, it is naturally to seek for various correlates between degraded TMTFs and various indicators of hearing loss. 

A significant change in neural group-delay dispersion $v$ can have different effects on the perceived image quality. If the system were completely continuous, then a smaller $v$ should lead to a larger bandwidth of both the coherent and the incoherent TMTFs. This is also obtained by considering an increased bandwidth from the broadened filters, if they function as the aperture stop (something we suspect normally occurs only at low frequencies, \cref{LowFreqCorr}). In this case, the effect of the modulation bandwidth increase will be compounded by the elongation of the temporal aperture as well. In reality, it has been repeatedly found that the TMTF bandwidth under cochlear hearing loss, once controlled for audibility, is about the same as in normal-hearing subjects, with rare cases that exhibit broader than normal TMTFs (\citealp{Bacon1992,Moore1992,Moore2001,Fullgrabe2003}, but see \citealp{Grant1998}).

Physiological data from the auditory nerve of chinchillas with noise-induced loss confirmed the psychoacoustical data, as relatively few fibers had an extended TMTF bandwidth, despite the broadened tuning of the carrier frequency \citep{Kale2012}. This was explained using related data that found no increase in amplitude modulation (AM) synchronization strength as a result of noise-induced loss in chinchillas \citep{Kale2010}, so the availability of extra high-frequency modulation information could not be coded by the auditory nerve \citep{Kale2012}. A followup study that employed nonlinear analysis (Wiener kernel) of firing patterns of the chinchilla auditory nerve fibers with noise-induced cochlear impairment revealed changes in the TMTF shape and a 20--30\% increase in frequency cutoff \citep{HenryKale2014}. At the same time, the auditory tuning curves broadened by 100--200\%. Therefore, in line with previous conclusions \citep{Kale2010,Kale2012}, the increased modulation bandwidth was not fully matched by a corresponding firing rate, perhaps due to temporal limitations stemming from adaptation or refratoriness in the auditory nerve \citep{HenryKale2014}.

It is possible to reframe the above findings in sampling terminology: high modulation frequency from the broadened cochlear filters may be neurally undersampled. This could have resulted in what appears like occasional aliasing (\citealp[Figures 2D, 5D and 5F]{Kale2012}; see \cref{nonuniformunder} for a further discussion about aliasing with additional examples). However, as neural sampling is nonuniform, aliasing might be only a spurious concern, as the extra modulation information will be converted to noise, unless it is removed by an anti-aliasing filter (\cref{nonuniformunder}). Whether undersampling is perceived as noise or aliasing, its effect may be detrimental to hearing.

If certain species or situations exist in which undersampling due to broadened filters is not a concern, increased TMTF bandwidth still entails risk of losing defocus---as the contrast between coherent and incoherent stimuli erodes when it can only differentiate very high---possibly irrelevant---modulation frequencies\footnote{For example, contrast the amplitude-modulation spectra surveyed in \citealp{Singh2003} and the spectrum necessary for full speech intelligibility in \citet{Drullman1994a} and \citet{Drullman1994b}.}. A somewhat similar mechanism has been hypothesized without reference to imaging, based on extra gain that was seen to be internally applied to the TMTF of unilaterally hearing-impaired human listeners \citep{Moore1996}\footnote{Anecdotally, \citet{Moore1996} referred to the enhanced modulation depth as ``magnification'', but with no relation to imaging and no scaling implied for the imaged modulation frequencies. Modulation depth is exactly analogous to contrast, though, which is only indirectly dependent on magnification in standard imaging.} and in chinchillas \citep{Kale2010}. It was suggested by \citet{Kale2010} that this gain may be excessive to the extent that hearing speech in fluctuating masker (e.g., competing speech) gets more difficult with cochlear hearing loss, because the received modulation depth of both target and masker is enhanced. 

Theoretically, as a limiting case, a broadened filter causes the temporal aperture to significantly shorten and approach the dispersion limit, not only at low frequencies (below 660 Hz), where the filters are thought to function as the aperture stop (\cref{LowFreqCorr}). It was seen in \cref{PinholeArch} that the system is biased toward geometrical blur, rather than being dispersion-limited. However, with much shorter aperture stop this bias can change and can lead to dispersive blur that might affect coherent objects too. This possibility is currently uncharted. 

While the loss of defocus does not imply enhanced gain, it entails increase in contrast and loss of blur for high-frequency modulations, as fluctuating noise and speech, for example, may be perceived as equally sharp with respect to their modulation spectrum. This may be a component in the loss of speech intelligibility with hearing impairments, as it can impact the depth of field. However, more precise estimation of the significance of this effect alone is lacking.

\subsubsection{Auditory neuropathy}
Other hearing impairments may result in an increase in $v$ due to abnormal latency patterns, which are not directly related to the cochlear filter bandwidth. In such cases, $v$ may change independently of the temporal aperture and cochlear dispersion. \term{Auditory neuropathy} can be diagnosed when listeners have normal OAEs (undamaged OHCs) and cochlear microphonic response, but missing or abnormal ABR \citep{Berlin2010}. Documented neuropathy cases can have almost any possible audiogram from normal to profound, and the effect on their hearing also varies considerably between cases. Auditory neuropathy can have different etiologies that manifest in loss of IHCs, in demyelination or deafferenation of the auditory nerve, and/or the dysfunction of the synapse between the IHC and the auditory nerve (synaptopathy). Central lesions (e.g., acoustic neuroma) are excluded from this family of disorders. Auditory neuropathy can either cause reduced ability to phase lock and/or lead to poorer conductivity of information because of deafferenation \citep{Rance2015}. Common temporal symptoms of auditory neuropathy are a decrease in TMTF cutoff frequency and an increase in gap detection \citep{Zeng1999,Zeng2005}. Additionally, if not altogether absent, there is an increase in click-evoked ABR wave-I to wave-V latency of a couple of milliseconds \citep{Starr2001,McMahon2008}. These features are fundamentally different from temporal processing in cochlear hearing loss, where adequate audibility can largely bring the performance to near-normal levels and the ABR latency decreases. 

There is no published derived-band or evoked-burst-tone ABR that can clarify which channels dominate the latency increase in click-evoked ABR of auditory neuropathy, which can be indicative of the direction of change of $v$.  Indirectly, it may be gathered from some frequency-dependent measures (e.g., frequency discrimination and beat detection) that the largest differences between auditory-neuropathy and normal-hearing patients manifest at low frequencies, whereas performance is near normal at high frequencies \citep{Zeng2005}. However, there are documented cases of auditory neuropathy with high-frequency loss and near normal speech audiometry, but no brainstem response beyond the cochlear microphonics \citep[e.g.,][Patient C]{Berlin2010}. An excessive rise in $v$ should lead to blur and loss of intelligibility in quiet and in noise for signals with significant high-modulation content, because of poor modulation response. This may have been the case for the few auditory neuropathy patients that were able to complete a speech intelligibility test (25 out of 95 tested; \citealp[Table 5]{Berlin2010}). A more general option is that both coherent and noncoherent detection circuits are impaired (e.g., both the DCN and the VCN) in these cases, and create distorted or blurry images that do not converge properly at the level of the auditory retina (the IC). 

\section{Phase locking and coherent detection impairments}
\label{NeuSyncPL}
Throughout this work, we have emphasized the dichotomy between coherent and incoherent imaging and, in parallel, the one between coherent and noncoherent detection. Coherent detection (and imaging) ultimately depends on the auditory system being able to track the phase of the incoming signal, which is achieved through synchronization that is observable as phase locking. Therefore, a system with impaired phase locking can be expected to have impaired coherent detection and possibly become disproportionately dependent on noncoherent detection---on the slowly-varying temporal envelope. The effect of this kind of degradation may become noticeable in signals that are naturally coherent, such as sustained periodic sounds as are found in music. It may become also critical in localization based on interaural time difference. 

\subsection{Loss of balance between coherent and noncoherent detection}
\label{LossofCohNonCoh}
The auditory defocus was introduced using the imaging analysis, but it is useful also in exaggerating the differences between coherent and noncoherent detection. Given a signal that is not completely incoherent, strong phase locking between two system elements suggests that coherence is conserved and sharp imaging is possible, whereas poor phase locking entails that some coherent stimuli may become decohered and effectively blurred. Where the normal-heard image comprises a certain balance of coherent and incoherent contributions, this balance may shift when the coherent contribution is impaired. Therefore, if impaired conditions can be identified where phase locking is degraded, they may correspond to increased blur of coherent or partially coherent signals, which can make them difficult to tell apart from incoherent maskers. For example, this can manifest as an increase of signal-in-noise threshold. Opposite conditions, where phase locking is excessive, may generate an auditory perception of pseudo-coherent images originating from incoherent stimuli, noise, or partially coherent narrowband sounds. This can give rise to coherent noise in some conditions rather than enhance the image sharpness. Overall, it may also entail a loss of the defocus effectiveness, as it is no longer possible to harness the signal coherence as a cue to differentially process signals according to their phase function. 

Degraded TFS sensitivity---a proxy for phase locking---has been correlated with poor speech reception in modulated and unmodulated noise in sensorineural hearing-impaired listeners (\citealp[e.g.,][]{Buss2004,Lorenzi2006,StrelcykDau,Hopkins2011}; see \citealp{Moore2014,Moore2019} for recent reviews). However, the degradation in TFS processing is correlated with old age as well, independently of the key factors of filter broadening and audibility (\citealp{Lorenzi2006,Hopkins2011,Fullgrabe2015}; see also \cref{LossCoherenceAmp} for a brief review of presbycusis). Using a nonlinear Wiener-kernel technique\footnote{See reservations about the Wiener-kernel method in \cref{WienerKernel}.} to analyze single auditory-nerve units, \citet{Henry2019} obtained the phase locking and envelope coding amplitudes in normal-hearing and noise-induced hearing-impaired chinchillas. Additionally, the normal-hearing chinchillas were treated with furosemide, which induces a temporary and reversible metabolic hearing loss that serves as a model for the endocochlear potential reduction effect that is characteristic in aging. Both impaired groups had a downward-shifted tuning of the TFS, which peaked at lower than normal CFs, in a manner that corresponded to a broadening of the tuning curves in the respective channels. The noise-induced animal group had the mistuning appearing at lower levels than the metabolic-loss group, and it also exhibited a mistuned envelope response, which was normal in the metabolic impairment group. These single-unit results also carried over to frequency-following response (FFR) measurements (see \cref{DetectionSchemes}), where the balance between envelope and TFS power was distorted in noise-induced hearing-impaired relatively to normal-hearing chinchillas \citep{Parida2021}. The TFS power was higher in hearing-impaired animals as a result of the low-frequency energy that entered through the mistuned channels. It was suggested that such a response pattern in humans may cause an upward spread of masking that may make speech intelligibility worse. In general, however, it has been difficult to establish a consistent trend of the effect of hearing impairment on the envelope and TFS FFR \citep{Jacxsens2024}.

Listening in reverberation is a particularly interesting situation for the dual detection (coherent/noncoherent) strategy, since, realistically, only direct sounds can be coherently detected using TFS (\cref{DetectionSchemes}). In the limit of a diffuse sound field (\cref{CoherenceReverb}), the reverberant sound is incoherent and has to be detected noncoherently using envelope cues. In realistic reverberant fields, there may be an advantage in partially-coherent detection, which can target the direct signal and blur the indirect field. In another FFR study, listeners had to identify reverberant spoken digits from the front that were being masked by similarly spoken digits arriving from the sides, which forced the listeners to use interaural time difference (ITD) cues \citep{Ruggles2012}. Envelope cues, as were quantified by $\FFR_{ENV}$, were more robust to reverberation than TFS cues, but they were more effectively used by younger listeners. Older listeners tended to rely more on TFS cues that were degraded, which lowered their relative digit identification performance. In a different study, reverberant speech intelligibility performance of older listeners correlated only with $\FFR_{TFS}$ of several harmonics around the first formant, whereas envelope cues were not significantly correlated with performance \citep{Fujihira2015}. 

These two studies underline the importance of TFS cues that are implicated by their erosion with age and, predictably, reverberation. However, they are curious in that envelope detection---the most immediate and less costly strategy for detection in reverberant fields---is weaker in older listeners. This may entail a decrement in the DCN, or other subcortical pathways that are normally responsible for noncoherent detection. This indeed has been found in aged rats that had lower synchronization to sinusoidal AM stimuli and loss of tuning to modulation frequencies (in units with bandpass response) than young rats \citep{Schatteman2008}. Comparable findings about the VCN are not as suggestive, though---they show some age-related decrement, whose consequences are not necessarily detrimental---at least not at low CFs \citep[e.g.,][]{Wang2006,Wang2019}. These findings may provide some credence to the hypotheses that the DCN is responsible for noncoherent detection and that it may become impaired with age, whereas the VCN that is specialized in coherent detection is more resistant to aging-related deficits. 

\subsection{OHC impairment}
\label{OHCimpair}
Earlier in this work, we tied the phase locking that is measurable on the auditory nerve and higher levels to a hypothetical auditory PLL. It was argued that different parts of the OHC comprise the necessary elements for a generic PLL: a phase detector that is responsible for the $f_2-f_1$ distortion product generation (the mechanoelectric transduction channels), a (low-pass) loop filter (the OHC cell membrane), and an oscillator (the hair bundle) (\cref{CortiPLL}). The somatic electromotility then provides the loop gain for the closed feedback loop of the PLL. Therefore, it is interesting to find out whether an impairment in the OHCs may lead to degraded phase locking. However, establishing a direct link between normal operation of phase locking and the OHCs has not been done at a level that provides an unequivocal support for the auditory PLL hypothesis, but several interesting findings are indicative that such a connection does exist. 

Physiological evidence of degradation in phase locking with hearing loss has been mixed and usually not directly associated with the OHCs, sometimes due to questionable experimental methods or control. \citet{Harrison1979} were the first ones to test this connection using pure tones as stimuli. They found no impact of selective kanamycin-induced OHC degeneration in guinea pigs on the synchronization as measured in the auditory nerve. These results were contested by \citet{Woolf1981}, who pointed at the unprecedented high synchronization index ($> 0.45$) obtained in the tested animals at frequencies of 5--7 kHz \citep[Figure 2]{Harrison1979}. In findings from similarly treated chinchillas, the phase locking in channels with damaged OHCs showed a decreased frequency range in which phase locking could be observed in measurements of ventral cochlear nucleus (VCN) neurons \citep{Woolf1981}. In that study, though, it is not clear how well the presentation levels in the impaired animals were matched to the normal-hearing controls \citep[p. 20]{Moore2014}. Another study on noise-induced hearing-impaired cats tested the phase locking to tones and to the vowel /$\epsilon$/ embedded in a synthesized syllable \citep{Miller1997}. Using a somewhat nonstandard ``synchronization coefficient'' measure, the phase locking to tones had about the same average in the impaired animals, but with a larger corresponding variance, in comparison to normal-hearing controls. In contrast, there was a substantial degradation in the coding of the F2 (1700 Hz) and F3 (2500 Hz) formants of the vowel in the impaired animals. This was the case as the channels became broader and let through a larger number of spectral components that the auditory nerve had to synchronize to, but at a lower power. However, as the tuning curves and frequency selectivity data revealed, the hearing loss had affected both the IHCs and the OHCs, which made it impossible to implicate the drop in phase locking with one hair cell type or the other. 

Better methods are found in more recent studies that indirectly tested the relation between phase locking and OHC function. In noise-induced hearing-impaired chinchillas, the maximum phase locking frequency dropped by about 500 Hz, so the envelope became more dominant at lower frequencies \citep{Kale2010}. The authors observed that the phase locking capability was independent of the auditory fibers themselves, but since no decrease was measured in frequency selectivity, it was suggested that the shift in the phase locking range was caused by damage to the IHCs and not to the OHCs. In another noise-induced hearing-impaired chinchilla study, phase locking dropped only for tones in noise but not in quiet \citep{Henry2012}. The amount of decrease depended on the SNR and in some cases brought the impaired synchronization to negligible levels\footnote{The drop in synchronization strength is less than 0.1 at -20 dB SNR, according to Figure 2 in \citet{Henry2012}. However, it is displayed as a constant average drop, which in relative terms can be anything from 10\% to 100\% of the baseline synchronization.}. Auditory filter broadening was shown to negatively correlate with the degree of synchronization strength, which indicates that phase locking is at least partially determined by the periphery, and possibly by the OHCs rather than to the IHCs. 

The interpretation of the results that point to a peripheral source of phase locking is made complicated because of the tonotopic mapping distortion and the change in envelope to TFS power balance that has been observed in animals with hearing loss \citep{Henry2016,Henry2019,Parida2021}. Nevertheless, the shifts in phase locking were thought to be related to loss of OHC gain in one study \citep{Henry2019} and in another study to OHC function---as can be seen by a weak DPOAE response \citep{Parida2021}. In human subjects too, the TFS FFR was weakly correlated with DPOAEs in young and old normal-hearing listeners \citep{Anderson2021}. In contrast, auditory nerve afferent function was quantified using the ABR wave-I amplitude, but it was not significantly correlated with phase locking and did not interact with the DPOAE (\citealp{Anderson2021}; in line with \citealp{Parker2020}). Contrary to these findings, neither phase-locking nor envelope synchronization degradation was found in aged gerbils, but rather an increase in spontaneous rates, possibly due to auditory nerve fiber or synapse loss, independent of filter broadening \citep{Heeringa2020}. 

Integrating these results is somewhat puzzling, because of their inconsistency. According to the PLL model we indeed expect to see a drop in phase locking with a drop in loop gain that is taken to be caused by OHC function decrement. This may appear as loss of cochlear gain, irrespective of phase locking, though. Additionally, loss of loop gain in the PLL should cause a decrease in its pull-in range, so that off-frequencies relative to the CF may not lock in as effectively and they can lose some of their ability to reject noise. These predictions are roughly in accord with a subset of the studies \citep{Woolf1981,Miller1997, Henry2012}. Loss of phase-detection gain is not necessarily dependent on the DPOAE, which is quantified using the cubic distortion product $2f_2-f_1$, rather than the quadratic $f_2-f_1$---the critical distortion product for the PLL. However, if the two product types co-vary, then the weakening of the phase detection product can be deduced from both \citep{Parida2021,Anderson2021}. Qualitatively, this appears to be the case in normal-hearing listeners from data in \citet{Baiduc2018}, but no corresponding hearing-impaired data are available to corroborate this reasoning. The results in the gerbil from \citet{Heeringa2020} could not be explained using the auditory PLL framework, unless the aged animals had intact OHC gain that did not disrupt the PLL function, unlike the conclusion that we drew from human data by \citet{Anderson2021}. Finally, we generally assume that the type of animal tested is a relatively insignificant factor, although we have to allow for this assertion to be wrong if future data will say otherwise. 

\subsection{Neural impairment}
In addition to the cochlear PLL, we hypothesized that neural PLLs might exist more centrally than the cochlea. While it is in line with a few modality-general neural models, little evidence could be produced to illustrate it in hearing (\cref{NPLLs}). 

When looking at hearing impairments, the lack of substantiating data is even more stark, and only one study was found that specifically tested synchronization impairment in a brainstem tissue as a function of impairment. The robustness of phase locking between the auditory nerve and the bushy cells of the anteroventral cochlear nucleus (AVCN) was studied in brain slice preparations of two strains of young and aged mice with and without age-related hearing impairment \citep{Wang2006}. It was found that while the auditory nerve excitatory postsynaptic potential jitter pattern was unaffected by aging, the spike threshold in the endbuld of Held of the bushy cells was elevated at high CFs in the (aged) hearing-impaired mice. While this finding might have implications primarily for temporal-based localization tasks, it might also affect other coherent detection processing tasks, which rely on the excellent temporal precision of the AVCN for processing. This can theoretically point to loss of phase locking of central origin. All considered, though, the VCN does not appear to be strongly involved in hearing impairment effects, as was also suggested in \cref{LossofCohNonCoh}. 

\subsection{Excessive phase locking}
In principle, phase locking precision may also be higher than is required for hearing a particular stimulus. As phase locking (or coherent detection more generally) employs a local oscillator, this may be the case if the output of the oscillator is too dominant in the signal chain due to excessive coupling, if it is locked in a self-oscillating mode, or if it is applied to incoherent inputs that do not benefit from coherent detection. These hypothetical dysfunctions are most readily associated with tinnitus. 

Heightened phase locking that is associated with tinnitus has been studied primarily in cortical areas \citep[e.g.,][]{Norena2003,Weisz2007}, whereas subcortical theories of tinnitus relate its existence to hyperactivity---possibly combined with central gain---that usually corresponds to the edge frequencies of cochlear impairment \citep[e.g.,][]{Henry2014Tin}. In and of themselves, these cortical and subcortical mechanisms are not contradictory, but they seem to constitute a systemic abnormal behavior, rather than a specific impairment of the coherent processing of the system. 

An alternative point of view using auditory accommodation is presented in \cref{TinnitusAccom}. 

\section{Sampling rate and clocking impairments}
The maximum sampling rate of the system---auditory nerve spiking that is initially encoded in the cochlear nucleus (CN)---has to correspond to the highest modulation frequency in the stimulus, which is itself determined by the smallest aperture in the system (\cref{TemporalSampling}). Too low a sampling rate may cause high modulation components to be undersampled and potentially aliased, leading to blur. Too high a sampling rate can lead to over-detection of high-modulation frequencies that cannot be further recoded and processed by the system, and may be perceived as noise. Alternatively, if a high sampling rate can be detected, it may lead to sensory information overload, which might lead to non-auditory-specific adverse psychological and cognitive effects \citep[p. 180]{WeisserPhD}. The overall sampling rate may indirectly decrease as a result of cochlear impairment, synaptic changes, or deafferenation, as they may all entail a lower-energy transduced product, and therefore a reduced spiking probability either peripherally or centrally. The author is not aware of any documented evidence to a global change in sampling rates in the system---perhaps in accord with the fact that the auditory system is not centrally clocked---as most changes appear to be instantaneous and stimulus driven. Nevertheless, this hypothetical impairment is presented here because long-term global changes to in the spiking rate most are most certainly expected to have perceptible impact on the quality of auditory imaging and, perhaps, on other cognitive variables.





\section{Excessive aberrations}
\label{ImpairAberrations}
Optical aberrations are the primary cause for some of the most prevalent eye disorders, which in this context are referred to as refractive errors. Notably, myopia (short-sightedness) and hyperopia (far-sightedness) are caused by defocus and can be treated by wearing eyeglasses, contact lenses, or sometimes by going through a corrective surgery. In contrast, hearing disorders are primarily driven by loss of audibility, whose closest, most prevalent, analogous visual impairment may be \term{cataract}---a leading cause of blindness worldwide---the opacification (clouding; attenuation of light) of the crystalline lens that results in loss of image quality \citep{ChouBass3}. Several types of aberrations were discussed in \cref{HigherOrderAb} and \cref{ChromaticA} as part of the normal operation of the ear, which may have relevance to impaired operation as well, although probably not as critical as aberrations in vision. 

\subsection{Excessive transverse chromatic aberration}
Transverse chromatic aberration was indirectly discussed in the context of binaural diplacusis and abnormal curvature of the OHCs (\cref{OHCimpair}). Different magnification of the two ears combined with a remapped cochlea due to impairment can lead to a different pitch perception of the same stimulus frequency. This explanation may be also applied to a single ear, where pitch is stretched in normal-hearing people (\cref{TransChromAb}). Thus, an abnormal magnification may skew the pitch stretching curve. If the loss is symmetric, then the effect will be bilateral and will not give rise to an obvious diplacusis. While the stretching effect is generally much more pronounced for pure tones than for complex tones \citep{Jaatinen}, such an aberration may potentially impair the perception of musical intervals, melody, and harmony, as well as music enjoyment as a whole. 

There is only limited evidence that can attest to an abnormal transverse chromatic aberration in hearing impaired listeners. Two trained musicians with high-frequency moderate-severe SNHL performed a monotic (same-ear) octave matching test for the same ear with pure tones of a fundamental frequency of 125--2000 Hz \citep[Table 1]{Arehart1999}. The matching became more off-tune closer to the spectral region of loss and at 2 kHz it was about 1.5 semitones up for one subject and 1.5 semitones down for the other. In another study, \citet{Huss2005b} tested monotic octave matching in listeners with dead regions. The performance was mixed and often erratic, but in general, tones whose frequencies fell within the dead region range were perceived to be of higher pitch than normal. This can be translated to an enhanced stretching effect, and may suggest lower magnification, as may be implied by filter broadening (that leads to a drop in neural group-delay dispersion $v$ and, consequently, in the magnification $M$). However, in another octave-matching study where the psychophysical tuning curves were evaluated (dead regions not reported), it was noted that there was no clear correspondence between changes in the location of the channel tuning curve peaks and the pitch shifts \citep{Turner1983}. 

Overall, the limited available data suggest that pitch perception may be impaired with cochlear hearing loss, in a way that can well correspond to the magnification model, perhaps following from a curvature change of the OHCs, or a change in the filter bandwidth. However the involved changes may not be particularly debilitating. 

\subsection{Neural syncrhony: Excessive temporal chromatic aberration}
\label{ChromImpair}
If conditions exist in which the neural group-delay dispersion is excessively long, it may result in an audible temporal chromatic aberration, beyond the natural depth of field that was assumed tolerable earlier (\cref{AudDepth}). Such may be the case when the evoked-ABR wave V appears completely absent. In theory, if the neural signal is very dispersed, then the different channels cannot synchronize to produce what appears as a single peak\footnote{Channel synchrony---a term that is frequency used in this context---should not be confused with a lack of synchronized phase locking within channels, which has a very different effect (see \cref{NeuSyncPL}).}. Instead, the energy from the different channels will be smeared in time. For example, this appears to be the case in the ``poor morphology'' patients in \citet{McMahon2008}, whose wave-V responses were severely smeared. This effect is the neural dyssynchrony component of the auditory neuropathy/dyssynchrony disorder \citep{McMahon2008,Berlin2010}. However, the effects of neural dyssynchrony and poor conductance are often conjoined, so there are no available accounts of the perceptual effects of the dyssynchrony component alone in the experience of listeners who suffer from it.

Other auditory disorders may be less sensitive to neural dyssynchrony. On one extreme of the latency changes, the evoked-ABR of Meni\`ere's disease patients exhibits significantly longer latency than normal \citep{Donaldson1996}, which may also involve dispersive changes (\cref{CochlearDispChanges}). On the other extreme, the effect may be much smaller in moderate cochlear hearing loss, where the ABR tends to have shorter latency than normal, judging from the derived-band ABR in \citet{Strelcyk2009} that was reviewed in \cref{TotalImpairedChange}. If across-channel wave-V latency is on the order of 1--2 ms, then in normal-hearing people it is a difference that may be just noticeable (\cref{AudDepth}), but probably of little consequence \citep{Moller2007}. \citet{StoneMoore2003} tested the effects of frequency-dependent delays, which were applied uniformly to the low-mid frequency bands, while frequencies above 2 kHz were not delayed. Subjects with flat or sloping moderate sensorineural hearing loss were tested aided (with amplification) with closed-fitting and a linear amplification program. When a delay of 4 ms was applied, subjects were only slightly more annoyed with their own-voice sound, and experienced negligible consonant/vowel recognition drop. The effects of delay became more adverse the longer it was, but overall were never dramatically bad.

\subsection{Higher-order aberrations and adjacent-filter phase mismatch}
\label{HigherAbbImpair}
Spherical aberration and coma were the two major higher-order aberrations that were hypothesized to be relevant in hearing. They both entail changes in the phase function away from the center frequency of the channel. Dead regions that characterize profound cochlear loss have already been discussed in \cref{HigherOrderAb}, where their effects on pitch perception were used as possible evidence for the existence of these aberrations. However, as far as the impairment itself is concerned, the reverse causation is probably more useful: severe aberrations at the flanks of the auditory filter cause some of the abnormal sensations to pure tones and other stimuli that were reported by listeners with dead regions \citep{Huss2005a, Huss2005b}.  

A corollary of off-frequency aberrations may have general relevance to hearing-impaired listeners with broadened filters. In many of the analyses above and in all accounts of cochlear hearing impairments, the effect of filter broadening due to OHC dysfunction is pervasive. However, having narrow auditory filters in the normal hearing system can be justified from the system design perspective to minimize higher-order aberrations, and thereby to retain operation within the paratonal approximation. In general, the broader the channel is (relative to its center frequency), the more dominant its aberrating flanks will be with respect to arbitrary stimuli. This should have two main effects on imaging. First, signals that are normally resolved are processed in the same filter, creating modulated patterns that are perceptually unlike the unresolved image. For example, two previously resolved tones may beat together in a broadened channel. In this case, though, if the modulation frequency is not prohibitively high, then the information about the existence of the two tones is conserved, but transformed to a qualitatively different percept (\cref{TwoSpectra}). Therefore, the filter broadening effect on envelope cues (amplitude effects only) is not necessarily detrimental to hearing, if the listener can adapt to the transformed images. This effect is not related directly to off-frequency aberrations. 

The second effect on imaging is a direct result of the aberration, once we consider complex envelope cues that include temporal fine-structure. Here, signals whose instantaneous frequency is well within the broadened filter bandwidth may undergo severer aberration than is encountered in the normal cochlea. As was argued in \cref{HigherOrderAb}, the effect of aberration is to distort the phase function of the image and therefore produce blur. Furthermore, if the signal is spectrally positioned so it is analyzed by two filters simultaneously (see the audio examples given in \cref{HigherOrderAb}), then there is high likelihood that the two filters do not have matching phase response that extends to the broadened flanks, and further degradation of the phase information will occur upon combining the images between channels (e.g., during a sweep). This means that with broadened auditory filters, the original information contained in the instantaneous phase spectrum may no longer be available to the listener due to the effect of higher-order aberrations. 

Although it is mathematically impossible to truly isolate the effects of envelope and temporal fine structure in arbitrary signals (\cref{AudChallenges}), there is some data that can support the two aberrating effects above. \citet{Lorenzi2006} showed that young and elderly listeners with moderate flat cochlear hearing loss performed poorly ($<20\%$) in a vocoded consonant identification task when they had to rely on TFS cues only, unlike normal hearing subjects ($>90\%$). In contrast, all subjects had no difficulty to use envelope cues only ($>90\%$). All stimuli were processed in a similar manner, by filtering the speech signals into broad filters, and using the Hilbert transform to extract the envelope magnitude and phase. Using the TFS cues without the disambiguating envelope cues may be severely affected by higher-order aberrations. If the hearing-impaired listeners have broader filters, then their received TFS image may be distorted across channels. Note that these patterns appear to be opposite to findings that older listeners tend to rely on TFS cues more than envelope cues in reverberation, where envelope cues should be more accessible (\citealp{Fujihira2015} and \cref{NeuSyncPL}).

The relation between TFS sensitivity and frequency selectivity appears to be more complex. For example, young and old normal-hearing and hearing-impaired subjects were tested for their frequency selectivity, TFS, and speech processing \citep{Hopkins2011}. When the effect of elevated threshold was removed, there was no significant correlation between the TFS performance and frequency selectivity, as determined by the filters bandwidth. This led the authors to the conclusion that the impaired TFS response is driven by other factors such as loss of fibers, or weakening of phase locking capability. In the same test, monaural TFS scores were correlated with the speech scores in modulated noise masker, while frequency selectivity was correlated with speech scores in unmodulated noise. These results are similar to those found in \citet{StrelcykDau}, who found correlation between TFS measures and speech-on-speech performance, but not in other masker conditions, including modulated noise. Additionally, no correlation was found between TFS and frequency selectivity. These results may be explained by better ability of subjects to listen in the dips when the masker is modulated, which requires using temporal cues. If aberration is key in any of these tests, then there may be large individual variability between subjects, which may be independent of hearing status, but could be exacerbated by age. In analogy to vision, this would be entirely normal, as higher-order ocular aberrations show large individual variations that hardly affect the image quality when the pupil is relatively small, but have a much larger effect when the pupil is open and lets in higher modulation frequencies \citep{Liang} 

Physiological data using Wiener-kernel methods on chinchillas with noise-induced loss reveal that TFS cochlear mapping becomes dissociated from the normal envelope mapping at high frequencies, so their tuning does not match \citep{Henry2016, Henry2019, Parida2021}. It suggests that basal place coding becomes unreliable. However, at low frequencies, the tonotopic mapping remains normal, despite broadened filters. More recent data associated this tonotopic distortion with more energy being admitted to the channels off-frequency in animals with noise-induced loss, compared to normal-hearing ones---an effect which was not captured by the traditional filter broadening measure ($Q_{10}$) \citep{Parida2022}. While these data are not presented as phase response, it is possible that they can be reinterpreted as an asymmetrical coma-like aberration in basal sites. 

This small sample of studies reveals a complex interrelationship between the basic elements of hearing that are not well understood today. While higher-order aberrations may seem as a far-fetched hypothesis at the moment, it is entirely consistent even with standard filter theory, if phase distortion is considered with respect to TFS. Still, more work will be needed to elucidate these issues and, hopefully, to come up with more succinct explanations of these hearing impairments and their effects. 

\section{Impairments of auditory accommodation}
\label{AccommodationImpairment}
The final class of impairments that is considered here is based on the existence of auditory accommodation and its dysfunction. The role of accommodation and possible mechanisms that can facilitate it were discussed in \cref{accommodation}. One effect that was hypothesized in \cref{PsychAcco} as a possible behavioral correlate of accommodation is compensation for the level of reverberation when the acoustical context of the environment is given before the target, which was measured in speech intelligibility tasks. When this experiment was repeated with a mixed-age group of listeners with mild--moderate sensorineural hearing loss, the compensatory effect seemed to be intact, despite poorer speech intelligibility performance overall \citep{Zahorik2011}. However, in general, old hearing-impaired listeners are less sensitive to different reverberation durations \citep{Reinhart2018}, and even old normal-hearing listeners show degraded listening in reverberation---even when audibility is controlled for \citep{Nabelek1982,Helfer1990,Harris1985,Gordon1993,Gordon1995, Ruggles2012,Fujihira2015}. We reviewed earlier the possibility that these results have a peripheral cause in the OHCs, but a central cause may be possible too, in line with animal studies that found subcortical correlates in reverberation encoding in the IC (\cref{PsychAcco} and \citealp{Heeringa2020}).  

Two different varieties of accommodation impairments are explored below. The discussion is more inspired by analogy to vision than in previous sections, although some evidence from hearing research is presented as well. The motivation for this is to underscore that ocular accommodation, while a relatively obvious feature of the eye in contemporary perspective, is not at all a trivial process given that it involves optical imaging considerations, a biomechanical variable lens system, a neural control and feedback system, constraints to the pupil and binocular vergence (accommodation) reflex, as well as non-visual factors such as attention. The starting point for understanding accommodation, though, must be image sharpness, so it is difficult to imagine making sense of such a system in the absence of the very notion of the imaging function of the eye. When considering accommodative impairments, this logical chain of processes can only become more opaque if it has serious holes in its understanding. Therefore, the main purpose of this section is to use the concept of auditory imaging developed in this work and try to unravel potential elements in the rest of the accommodative system that could lead to impairments. Despite the speculative nature of this analysis, it may be more valuable to pursue it than to overlook it completely, in hope that it can advance the understanding of these impairments in the future.  

It is noted again that gain accommodation has been largely neglected in this work on purpose, although its hypothetical existence may be well-incorporated into the auditory accommodation framework. Compression is known to be reduced with hearing loss and it is tightly associated with the OHCs, which we used for the time lens and PLL models that are not necessarily independent of compression. Thus, dynamic gain manipulation in the central auditory system, or as is mediated peripherally by the OHCs, will eventually have to be analyzed as a possible mechanism whose range decreases in presbycusis. 

\subsection{Neural synchrony: Loss of defocus amplitude}
\label{LossCoherenceAmp}

\subsubsection{Presbyopia and prebycusis}
The general findings about age-related effects in reverberation and noise present a notional analogy between presbycusis and presbyopia, which goes beyond the inevitable age-related decline in both auditory and visual acuities.  

\term{Presbyopia} is the most common disorder related to accommodation and is defined as age-dependent subjective loss of the accommodation amplitude from a maximum of about 14 diopters at young age to below three diopters \citep{Charman2008}. The perceived subjective loss is a combination of objective loss related to the biomechanical changes in accommodation, as well as depth-of-field effects, which can mitigate the associated blur (e.g., by constricting the pupil). The typical onset of subjective presbyopia occurs around the age of 40, but the amplitude (i.e., the dioptric power range of accommodation) is continuously lost with age even when it is not subjectively noticed. These changes are correlated with other age-related changes to the lens, ciliary muscles, zonules, and other parts of the eye. However, there is no consensus at present as for which combination of degradations is the root cause of presbyopia. This impairment often manifests as hyperopia (far-sightedness) that can be corrected for with glasses, at least up to a certain degree of satisfaction.

\term{Prebycusis} is broadly defined as age-related hearing impairment that can emerge from peripheral, central, and/or cognitive causes \citep{Gates2005,Humes2012}. Co-occurrence of various symptoms of decline in all three types of causes is very common in old age, so several causes may coalesce in actual cases of prebycusis. The peripheral component famously manifests as a sloping high-frequency cochlear loss, which can be measured even in animals that are reared in quiet. The main peripheral effect (called \term{strial presbycusis}) is caused by age-related metabolic degeneration of the stria vascularis in the cochlea and, consequently, a drop in the endolymphatic voltage that is required for normal OHC amplification. However, the complete impact of presbycusis may be a cumulative effect of noise trauma, ototoxicity, genetic susceptibility, and other factors, so the effect of aging alone cannot always be easily isolated. Presbycusis leads to decline in speech intelligibility in noise and in quiet. Additional temporal processing deficits (e.g., poorer gap detection, time-compressed speech intelligibility, localization) are characteristic in presbycusis as well.
 
The main question we would like to answer is whether there is an analog in presbycusis to the drop in accommodation amplitude of presbyopia. Whereas ocular accommodation reacts to distance---the main parameter that determines the focal length of the lens---hearing is sensitive to coherence, so we can expect that auditory accommodation is designed to react to it as well. Two mechanisms are particularly pertinent for coherence processing: defocus and phase locking. Defocus is the main imaging property that has been discussed that differentiates acoustic objects according to their relative degree of coherence. Phase locking to the carrier is required for coherent imaging and detection, which can lead to eventual demodulation of the complex envelope. If either the defocus or the ability to phase lock deteriorates as a result of aging, then the analogy to a drop in the accommodation amplitude may be valid. While some inherent decoherence (noise) in the system was hypothesized as a plausible mechanism of accommodation (\cref{NeuJitter}), uncontrolled application of internal decoherence can have effects that interact with both defocus and phase locking and as such may adversely impact hearing. 

\subsubsection{Hypothetical effects of medial olivocochlear reflex impairment}
We speculated in \cref{MOCR} that the medial olivocochlear (MOC) reflex (MOCR) has two complementary functions that work to weaken the coherent detection. One is a decrement in phase-locking strength, which entails a weakening of downstream TFS representation. It does not necessarily mean that the envelope representation is enhanced, though, although it seems like a reasonable processing strategy. The second MOCR effect is a reduction (or change) of the time-lens curvature that can decrease the defocus of the imaging system that differentiates incoherent and coherent objects. If the curvature $s$ is independent of other dispersive shifts (i.e., of $u$ and $v$), then it is expected to improve the incoherent imaging without affecting the coherent imaging. This is because with a smaller defocus, the absolute value of the defocus term $W_d = \frac{1}{u} + \frac{1}{v} + \frac{1}{s}$ in Eq. \ref{winc} becomes smaller, which makes the theoretical incoherent MTF closer to the coherent one, in terms of its cutoff frequency (see Figure \ref{OTAcutoff}). Aging-related accommodation impairment, therefore, can potentially degrade the effectiveness of the time-lens curvature and phase-locking MOCR inhibitory role, which may entail poor adaptation to different degrees of coherence. Such an impairment may be accompanied by a static change to the time lens or the other dispersive elements, which leads to a change in the operating point of the dynamic system.

We therefore predict that auditory accommodation should be associated with loss of defocus amplitude, which can manifest also in strengthening of envelope processing. Furthermore, loss of phase locking may reduce the defocus amplitude from the coherent side. These changes may be driven by decline in the MOC function, although other processes may contribute to it. 

\subsubsection{Medial olivocochlear reflex and aging}
Even though its exact role is disputed (see \cref{MOCR}), the interaction of the MOC system with the aged hearing function is robust, which constitutes the most superficial similarity required between presbycusis and presbyopia. For example, changes in the endolymphatic electric potential have a global effect following excitation of the crossed olivocochlear bundle reflex, which directly implicates the OHCs and the amount of intermodulation distortion in the cochlea (\cref{MOCR}), at least in the vicinity of the impacted channels \citep{Mountain1980}. More recently, it has been shown that in aged mice with enhanced cholinergic receptors (of the type that exists in the efferent synapses with the OHCs) the auditory system hardly shows any of the normal age-related deterioration that is found in wild type mice \citep{Boero2020}. This was shown using normal ABR measures and OHC operation (DPOAE amplitudes), as well as only a minimal loss of afferent synaptopathy---all of which were comparable to young mouse controls. All mouse groups were raised in soundproof conditions, so noise-induced effects were ruled out. These results reinforce earlier findings that demonstrated that age-induced decline occurs in parallel in the DPOAE and the ABR of mice and humans \citep{Kim2002,Jacobson2003}. Also, it was found that a loss of MOC function (suppression of DPOAEs in middle-age groups) preceded the deterioration of the OHCs themselves (loss of DPOAE amplitude). This early MOC loss was more prominent at high frequencies ($>4$ kHz) in both normal-hearing humans and mice. In humans, the decline of the contralateral MOCR was evaluated from the weaker DPOAE and click-evoked OAE responses in aged listeners (41--60 years old) \citep{Lisowska2014}. While the MOCR effect is small (typically 0.5--2 dB), the DPOAE baseline was lower on average, suppression effects were smaller, and also the proportion of DPOAE enhancement events due to contralateral broadband noise increased in the aged group\footnote{Enhancement rather than inhibition of the DPOAE is sometimes observed as a result of the MOC activation, although at a lower frequency than inhibition. It has been interpreted as a consequence of constructive interference of the DPOAE and reflection products in the ear \citep{Abdala2009}.}. All effects were stronger at high frequencies (2--3 kHz). Nevertheless, inasmuch as any of those effects translates to speech reception, there seems to be only a negligible MOCR involvement in the improvement of speech intelligibility measures in noise, at least in studies that employed OAEs as a proxy for the detection of MOCR inhibition \citep{Gafoor2023a,Gafoor2023b}.

\subsubsection{Aging-related hearing and accommodation impairments}
Aging effects are best observed in studies that separately controlled for both hearing status and aging. Hearing status and age interact in conventional hearing performance measures---speech intelligibility in noise and reverberation, and interrupted speech intelligibility \citep{Gordon1993,Gordon1995}, but not in temporal tasks, such as duration discrimination and gap detection where age alone is the dominant factor \citep{Fitzgibbons1994,Fitzgibbons1996,Fitzgibbons2001,Snell1997,Snell2000}. 

In a different series of studies, the midbrain (FFR) and cortical (magnetoencephalography, MEG) responses to speech in quiet and speech in noise of young normal-hearing, older normal-hearing, and older hearing-impaired listeners were compared \citep{Presacco2016,Presacco2019}. As in the psychoacoustic studies, speech intelligibility scores were correlated with the hearing status of all groups. It was found that the FFR strength for the /da/ syllable---both its transient and its steady-state portions---was higher in young listeners than in old listeners, independently of their hearing status and whether it was measured in quiet or in noise. Interestingly, the low-frequency envelope (1--8 Hz bandwidth) reconstruction fidelity based on the MEG data was higher for the older listener groups. This effect may be caused by an enhanced cortical response that is observed in old age listeners \citep[e.g., an increase in the cortical N1 amplitude, ][]{Tremblay2003}. Although we expected an improvement in incoherent processing at a lower subcortical processing level, these cortical changes may be an indication of a downstream effect of such a change that begins with adaptation in upstream auditory processing. In this overarching interpretation, these results could be in accord with accommodative MOCR impairment prediction, which has not been measured in the cited literature\footnote{A recent analysis of old normal-hearing and hearing-impaired listener data suggested that ``\textit{patients with impaired audibility cannot adjust the corresponding proportion of the envelope and TFS under noise conditions the way that individuals with NH} [normal hearing] \textit{are able to}'' \citep{Hao2018}. While our present analysis is mainly concerned with aging effects and not with cochlear hearing loss, these ideas are close to ours.}.

\begin{figure} 
		\centering
		\includegraphics[width=0.5\linewidth]{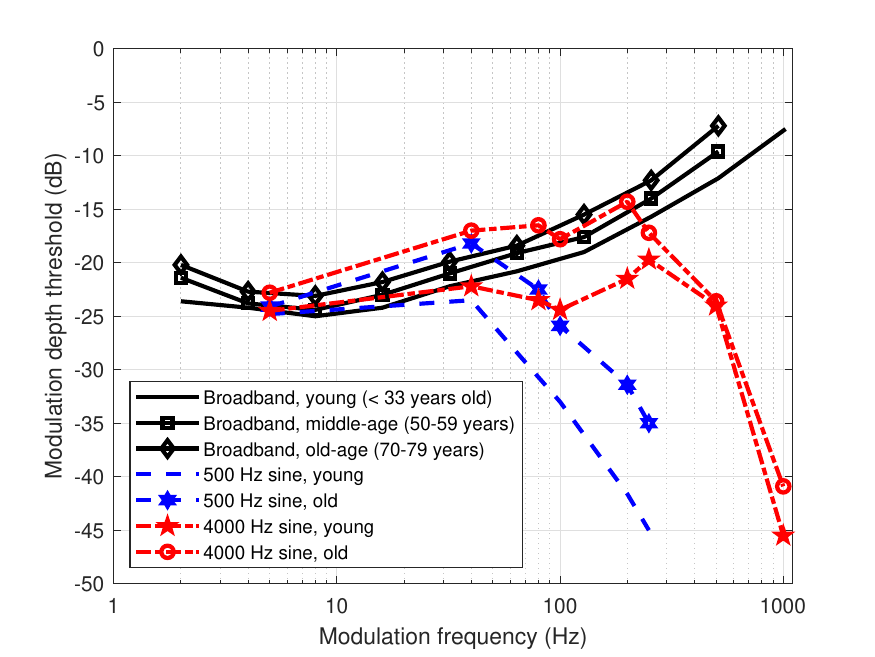}	
		\caption{Comparison of young and old normal-hearing TMTFs based on broadband carrier data by \citet{Takahashi} and sinusoidal carrier data by \citep{He2008}.}.
		\label{oldyoung}
\end{figure}

The loss in defocus amplitude that leads to a reduction in phase-locking fidelity and coherent detection, as well as to changes in the incoherent imaging fidelity, can be thought of as resulting from extension of TMTF bandwidth. Whether this extension has any effect on the image sharpness depends in part on the modulation spectral content of the object and on its degree of coherence. Reduction in coherent detection contrast might be gathered from available studies that measured the differences between young and old normal hearing listeners in broadband and tonal TMTFs, summarized below and in Figure \ref{oldyoung}.

Tonal TMTFs were measured by \citet{He2008} in young and old normal-hearing groups. The TMFTs were found to be similar in shape in the two groups, but less sensitive overall in the older listeners. The two subject groups had normal audiograms below 4 kHz, but the older group (60--80 years old) had mild losses at 6--8 kHz. Carriers tested were 500 and 4000 Hz---two spectral regions that differently probe the temporal sensitivity of the listeners. While both thresholds were elevated by 5--10 dB in the older group throughout the modulation frequency range, the cutoff frequencies, where the modulation are spectrally resolved in adjacent auditory filters, were the same in both groups. However, the threshold at 500 Hz was elevated also in the resolved region (blue curves in Figure \ref{oldyoung}), while at 4000 Hz the thresholds of both groups were the same once the signal was resolved (red curves in Figure \ref{oldyoung}). Therefore, coherent stimuli suffered a loss of contrast in older listeners, which was not necessarily associated with a loss of sharpness. These findings indicate a deterioration in the ability of old-age listeners to use temporal cues. Age-related decline in phase locking that leads to degradation of coherent detection might have impacted the performance, since the modulated stimulus was coherent. Similar studies did not always find consistent results. For example and additional references see \citet{Carcagno2021}, who described a small age effect on loss of contrast in coherent AM detection (with masking noise), as well as less conclusive overall results on the impact of age-related temporal processing degradation.

\citet{Takahashi} compared normal-hearing subjects of four age groups: young ($\le33$ years), 50--59, 60--69, and 70--79 years old. The older groups had typical presbycusis sloping audiograms with mild losses at 6 and 8 kHz. The incoherent broadband TMTFs of the older subject groups were slightly less sensitive than the young group's TMTF, but the difference was insignificant and may have resulted from higher threshold due to their mild hearing losses (black curves in Figure \ref{oldyoung}). Similar conclusions were obtained in two other experiments carried out in this study of modulation masking and speech intelligibility in modulated and unmodulated noise: while the results were generally poorer for the older groups, an age effect per se could not be statistically established. Either way, unlike what can be predicted by a loss of defocus---not only was an improvement in the incoherent MTF bandwidth not observed in this measurement, but also the sensitivity itself decreased slightly. Nonetheless, as the TMTF broadband depends on across-channel integration, it may not be the right way to tap into channel specific processes, which may require narrowband stimuli that are partially coherent. Still, both aged-listener TMTF studies fit the hypothesis of loss of phase locking precision as a result of aging. This conclusions may be bolstered by a study by \citet{Mamo2016}. An isolated effect of degraded phase locking that was simulated by artificially adding jitter to a /da/ syllable and comparing its evoked-speech ABR in young and old normal-hearing listeners (normal pure-tone audiograms below 4 kHz). Adding the jitter effectively equalized the ABR amplitudes (envelope FFR) of the young group to the old group, which was not affected by the additional jitter.  

In order for the presbycusis-presbyopia analogy to work, the auditory accommodation hypothesis---that accommodation exists and it relies on adaptation of phase locking (\cref{NeuJitter}) and maybe on adaptive mixing of coherent and incoherent streams (\cref{StreamMixing})---will have to be substantiated with more evidence. Especially, evidence will have to be establisehd for the causal effects of MOC degradation to phase locking and loss of defocus amplitude. 

\subsection{Loss of depth-of-field control}
In vision, accommodation impacts the depth of field of the image by changing the focal length. While pupil constriction would be expected to impact it too, aberrations and directional effects that are caused by the pupil edge tend to cancel out most of the depth-of-field changes, at least at large pupil diameters \citep[e.g.,][]{Marcos1999}. As the visual system reacts to the effective image sharpness, accommodation can take advantage of the depth of field and get away with perfect focusing, as long as sufficient sharpness is obtained \citep{Bernal2014}. 

Since the depth of field determines the extent of the defocus effect, we considered the possibility that the auditory accommodation function is to directly manipulate the depth of field, by controlling the focal time of the lens. This form of release from masking (or \term{antimasking}, as it is normally referred to in this context) should be then measurable for a combination of probe and masker with or without their MOCR activated. However, results to this effect have been inconsistent in both normal-hearing and hearing-impaired listeners and cannot be accounted for using cochlear gain changes alone \citep{Jennings2021}. In animal studies, the antimasking effect is to reduce the firing rate of a tone in noise, which provides a release from adaptation, but not to the extent that the MOCR masking threshold recovers the tone threshold in quiet. A major difficulty in identifying antimasking conclusively, though, is that physiological effects of reduction in suprathreshold masking in animals do not necessarily map to psychoacoustical effects in humans, where the MOCR stimulation is inferred indirectly using OAEs. Another difficulty is that listeners whose MOC was sectioned do not necessarily have hearing performance that clearly shows degraded forward masking performance. For example, one subject with presumed severed efferents had relatively normal tone-on-tone masking recovery \citep{Zeng2000}. Four subjects in the same study had reduced simultaneous masking overshoot effect in their operated ears, in which the broadband masking threshold of a probe that appears immediately after onset tends to be higher \citep{Zwicker1965}. In contrast, cochlear implant users, whose OHCs and hence MOCR are bypassed by directly stimulating the auditory nerve, exhibited normal overshoot effect of sinusoidal AM tones in noise, which suggests that the masking effect is caused by a central mechanism \citep{Marrufo2019}.

Thus, unfortunately, at present we cannot rely on any specific data that may confirm or disconfirm our speculation that the auditory depth of field may be accommodated to by the MOCR. Like several other hypotheses brought up here, it is in itself linked to several other strong hypotheses, which will have to be confirmed or rejected first, regarding the time lensing in the auditory system and the validity of the MOCR in accommodation. 

\subsection{Tinnitus as an accommodative impairment}
\label{TinnitusAccom}
This section is an extension of the hypothesis put forth in \cref{NeuJitter}---that fast and selective adaptation of the degree of synchronization to different elements in the stimulus is a plausible mechanism to achieve accommodation. It was further suggested that some forms of tinnitus (mainly transient) may be an excessive manifestation of this kind of accommodation. 

Tinnitus research is too vast to allow for all but a brief overview in the present work, especially given the uncertainties involved and the number of proposed theories, but see \citet{Moller2011}, \citet{Eggermont2012a}, \citet{Eggermont2012b},\citet{Baguley2013}, \citet{Henry2014Tin}, \citet{Roberts2019}, and \citet{Henton2021} for recent reviews. After a very brief introduction to tinnitus, the hypothetical connection with accommodation will be explored to include additional processes, by way of analogy to vision.  

Subjective tinnitus is characterized by the perception of phantom sounds, which are the result of abnormal neural activity that has no acoustic correlates inside or outside the body \citep{Eggermont2012a}. Subjects with tinnitus have enhanced synchronized activity between various circuits in the thalamus and cortex, in response to an increase in spontaneous firing rate that is caused by a spontaneous activity along the auditory pathways \citep[Chapter 9]{Eggermont2012a}. There are many possible causes for tinnitus (e.g., noise- or music-induced hearing loss, cochlear insults, ototoxicity), but none that is either necessary or sufficient for it to develop. Therefore, many tinnitus cases have an unknown etiology \citep{KleinjungMoller2011,Eggermont2012a}. Nevertheless, it is commonly thought that an initial degree of cochlear or neural impairment is the underlying cause, even if it goes undetected with normal clinical techniques \citep{Norena2011}. Although ``ringing'' is the most common descriptor of the perceived tinnitus, the full range of sounds that can be perceived by tinnitus patients is very broad and no universal perception exists \citep{Stouffer}. Accordingly, the correspondence between the nature of the phantom sounds and the etiology is not universal either, but for hearing-impaired listeners the tinnitus pitch spectrally corresponds to the impaired range \citep{Norena2002}.

In general, tinnitus may have either a peripheral or a central origin, but it appears that it is essentially a central response to an impairment in the periphery, in the inner ear (some injury to the OHCs, IHCs, and/or the auditory nerve) \citep{Jastreboff1990,Schaette2006,Norena2011}. According to this model, several central gain mechanisms that exist along the auditory system can compensate for a lack of activity in the impaired channels by a gain increase---itself a result of homeostatic plasticity \citep[e.g.,][]{Hutchison2023} that maintains a stable level of neural activity and efficient coding through various neuroplastic mechanisms. Tinnitus is then a side-effect of spontaneous activity amplification by the hyperactive system, which is perceived as noise. However, the gain hypothesis has been recently challenged as it was observed that gain is correlated with age and not with tinnitus. so that not all people with high gain experience tinnitus \citep{Johannesen2021}. In another line of models, the dorsal cochlear nucleus (DCN) has been repeatedly implicated as the specific source of tinnitus, although it most likely does not sustain it \citep{Henton2021}. Tinnitus-related hyperactivity has been observed in the VCN as well \citep{Norena2011, Henry2014Tin}. The IC, medial geniculate body (MGB), and primary auditory cortex (A1) may all be implicated in a cascade or feedback process that can lead to tinnitus, which appears to emerge at lower processing levels and involve some degree of neural hyperactivity \citep{Henton2021}.

Just as accommodation normally functions with a near-triad that (nearly) reflexively combines vergence and pupil constriction (see \cref{OcularAcco}), so does synchronization accommodation may work in tandem with other mechanisms. Because these processes are interconnected, the failure of one can lead to the failure of the others in unobvious ways. This can be illustrated with an analogy to two ocular accommodation disorders that are caused by incorrect central control of the accommodative reflex muscles. The first disorder is \term{refractive accommodative esotropia}---a form of \term{strabismus} (crossed eye), which is itself the result of an uncorrected hyperopia (far-sightedness) \citep{Lembo2019}. This condition occurs when the visual system tries to compensate for the defocus by using focal accommodation, but then also involuntarily converges because of the accommodation reflex. This problem is prevalent in infants and can usually be corrected with glasses for hyperopia or, if it fails, with a surgery. The second disorder is the \term{spasm of accommodation}---the permanent contraction of the ciliary muscles, which blocks any possibility for accommodation and thereby leaves the patient effectively myopic or hyperopic \citep{Goldstein1996}. Once again, reflexive side-effects exist in the form of pupil constriction and/or vergence, all are found in different and inconsistent amounts among patients. Both disorders reveal nontrivially coupled symptoms, which require the joint understanding of the refractive problem and the accommodation system in order to be diagnosed and treated successfully. 

The analogy between these two ocular disorders to tinnitus is not an obvious one, because the accommodation principles of the temporal and spatial imaging systems are fundamentally different. However, to the extent that tinnitus points to an impaired signal processing in the auditory system that is also part of a reflexive system, then the accommodative impairments of the eye may provide some inspiration for understanding the causal chain of auditory impairments. For example, tinnitus is often comorbid with cochlear impairments, and central gain was proposed as a compensatory mechanism for the loss of signal strength. But this may also lead to \term{hyperacusis}---the abnormal sensitivity to loudness in otherwise normal hearing---whereas tinnitus itself also involves excessive synchronization. Thus, synchronization and gain may be two processes that are tied by a reflex, which may also include any of the other mechanisms discussed in \cref{HypoAcco}. The two may or may not be tied also to the MOCR. 

In \cref{LossCoherenceAmp}, the effect of presbycusis on accommodation was hypothesized to be the loss of defocus amplitude, which is caused by a reduction in phase locking precision. In this subsection the possibility that tinnitus is uncontrolled phase-locking synchronization has been discussed. The two impairments need not be mutually exclusive. If the defocus amplitude is decreased, tinnitus may be a compensatory makeup shift in the available defocus range, which recalibrates the operating point from which synchronization can be accommodated (see Figure \ref{operatingpoint} for a cartoon illustration of this operating point). Indeed, tinnitus and presbycusis typically occur together. Histopathological analysis of the temporal bones of two groups of presbycusis patients with and without tinnitus revealed that the tinnitus group had a greater loss of OHC in the basal and middle turns of the cochlea and more atrophied stria vascularis in the basal turn \citep{Terao2011}. Thus, it may support the hypothesis that tinnitus may be caused by a central mechanism in order to compensate for a deficiency in the peripheral signal. If the OHC function is to produce gain, then central gain compensation may be indeed likely. Our interpretation of the OHC as the starting point of phase locking would imply that phase-locking compensation may be produced centrally, which can then appear as self-oscillations that produce tinnitus. 

Finally, it is curious to note that several obscure disorders may point at systemic connections between the assumed auditory and the known visual accommodative reflexes. For example, tinnitus can interact with vergence \citep{Yang2010} and saccadic movements \citep{Lang2013}, while some hyperacusis cases are coupled with hypersensitivity to light \citep{Philips1982,Woodhouse1993}. Additionally, there are documented cases of tonal tinnitus that are induced by gaze that is shifted away from the center, which occur following surgical deafferenation of the auditory periphery after the excision of lesions from the cerebellopontine angle (the area between the pons and the cerebellum; \citealp{Cacace1994,Mennink2020}). These obscure impairments may point to deeper, albeit weaker, connections between the various elements in the auditory and visual systems.

\section{Conclusion}
In this chapter we explored the hypothetical connections between known hearing impairments and several dispersive, phase-locking, and accommodative dysfunctions. Given the amount of unknowns about temporal imaging and the relatively limited empirical data that were found to test against the predictions, some of the hypothesized disorders are highly speculative and the overall picture is somewhat patchy. Additionally, the relevant analyses have not been consistent throughout and it appears that some disorders may be more plausible than others. For example, a certain form of age-related accommodation impairment seems to be relatively compelling, given the known age effects on the MOCR and on the envelope and TFR processing balance changes. Excessive higher-order aberrations for broad auditory channels seem compelling as well, as they can cause various undesirable distortions, but at this point they are much more difficult to establish. Other dispersive changes in the cochlear and neural parts of the auditory system are not clearly implicated in great changes to performance, so they may be difficult to establish as well. Finally, dysfunction of the OHCs has already been implicated in a large range of hearing impairments. On top of these we stacked changes to phase locking and time lensing, which can have complex effects that likely interact with the global processing strategy of the system and are not straightforward to probe. 

Even though the above analysis may appear premature given the current state of knowledge, it also seems worthwhile. The intention behind this speculative chapter has been to apply the ideas that were developed in the preceding chapters in hope to shed some light on hearing impairments that have resisted a compelling theoretical account, and hence---a satisfactory treatment. Ideally, this discussion can motivate further explorations of these topics that can contribute to the eventual resolution of at least some of the most persistent hearing disorders.

\chapter{General model and discussion}
\label{GeneralDisc}
\section{Introduction}
Several ideas that are new to hearing science have been introduced in this work. While a suitable imaging theory for the auditory system has been the centerpiece of this work, a prominent role has been given to both communication and coherence theories, which have been considered to be indispensable stepping stones for the development of a comprehensive auditory theory. The inclusion of these perspectives within the acoustic signal path that is received by the auditory system had to be motivated by a de-idealized presentation of real-world acoustics compared to what has been traditionally considered in hearing science texts. However, this realistic and more complex account is rewarded by being able to readily produce hypotheses for explanations of a wide range of auditory effects that have so far been largely left out of standard theory. This repackaging of the auditory narrative also produced promising links to both vision and neuroscience---either in way of analogy (to the eye) or through the integration of analog and neural information processing paths (the auditory brainstem and inferior colliculus). Several speculative elements and notable uncertainties have been foundational for the development of this work, but they will eventually have to be rigorously tested, in order to obtain confidence in the more advanced aspects of the new theory. 

In this concluding chapter, the full functional model of the auditory system is considered, as aggregated from the previous chapters. It will facilitate the discussion of some of the strengths and weaknesses of this work, which will lead us to make a few suggestions for improvements and further investigations. The chapter concludes with a short discussion of some of the overarching themes that go beyond hearing alone, but can be impacted by the theory. 

\section{A provisional functional model of the mammalian\\auditory system}
\label{CompleteModel}
\subsection{Introduction}
Various functions of the auditory system, which have not been considered previously in standard theory, have been presented in this work. Given the high number of known and hypothesized functions of the auditory system, it will be useful to try and put together the newfound elements along with the familiar ones in a single model. This should help us synthesize a more cogent understanding of what the auditory system actually does and identify the strengths and weaknesses of the proposed theory. 

There are two major difficulties in coming up with a model that encompasses the entire auditory system, at least up to the midbrain. First, the operational details of the cochlea are not uncontroversial. Especially, the organ of Corti and the outer hair cells (OHCs) are implicated in several functions---amplification is chief among them---whose functional inner workings are not entirely proven empirically and whose intricate mechanics is still being studied. It affects other important hearing characteristics, such as the auditory filter sharpness and distortion product generation. Second, the exact roles of the auditory brainstem remain fuzzy. The various nuclei are correlated with the detection of multiple acoustic features and specialized responses, which do not necessarily add up to clear, reducible, or nameable functions. Therefore, discussions about this critical part of the system tend to be somewhat vague so much so that entire nuclei and pathways have little more than conjectured roles associated with them (see \cref{CentralNeuroanatomy}). 

There are three dominant classes of auditory models in literature. The first class treats the acoustic input as a time signal and uses standard signal-processing techniques---both linear and nonlinear---to account for the various transformations that the signal undergoes. Typically, the processing gives rise to some desirable auditory feature extraction or a specific response, which can also be cross-validated psychoacoustically or electrophysiologically. This model class does not always adhere to known physiological correlates, which should carry out the signal processing operations in reality. Nevertheless, it is useful in simulating and predicting different experimental responses and can be highly insightful in forming hypotheses about the physiological roles of the different auditory organs. Occasionally, the simplest models in this class have no particular processing that can be attributed to the brainstem (e.g., a ``decision making device'' that immediately follows the rectification of the transduced signal), or to any other auditory nuclei for that matter. In this and similar cases, these nuclei are occasionally (and implicitly) explained away as ``relay neurons''. 

The second class of models usually traces the various neural pathways and their connections and tries to infer what their function is. These models emphasize the excitatory or inhibitory, and ascending or descending nature of of the projections. Occasionally, there would be additional emphases on the types of cells and neurotransmitters involved, and on their relative topological positions. In their most basic form, these models are nothing more than a connection diagram, which tends to be both vague and complex, due to multiple pathways and cells with no clear role in the emergent system function. A subset of these models hypothesize particular circuits that assume a desirable operation (e.g., acoustic feature extraction), which comprises specific neuron types with known responses to particular stimuli. In these models, the circuit interconnections and complete operations can be hypothetical, or based on partially known correlates that are available from physiological data. Yet more advanced physiological models relate directly to the biophysical properties of the cells, such as the different time constants of presynaptic and postsynaptic potentials, depolarization thresholds, and complex timing delays. This approach enables precise simulation of the neural signal between selected parts of the system, which can mimic the response of known acoustic signal types. 

The third class of models combines elements from both signal processing and physiology and attempts to provide a phenomenological account of hearing. Such models can be highly effective, although they do not necessarily bring out any intuition as for why the system architecture and signal processing is the way it is. 

\subsection{The model}
The explanation level that has been sought after in the present work lies somewhere in between these three model classes. Its focus is on the purpose of the auditory system as a whole and the functions that are required to realize it. The functional realization, though, is not necessarily unique, and more than one signal processing or physiological configuration can be conceived that can plausibly perform it. Therefore, the model grossly localizes some functions in the system physiology, but does not commit to the details of operation. Furthermore, because of the relative poverty of parameters that were computed here with certainty, this kind of system description does not yet offer a clear path to simulation and remains high-level. 

\begin{figure}
		\centering
		\includegraphics[width=1\linewidth]{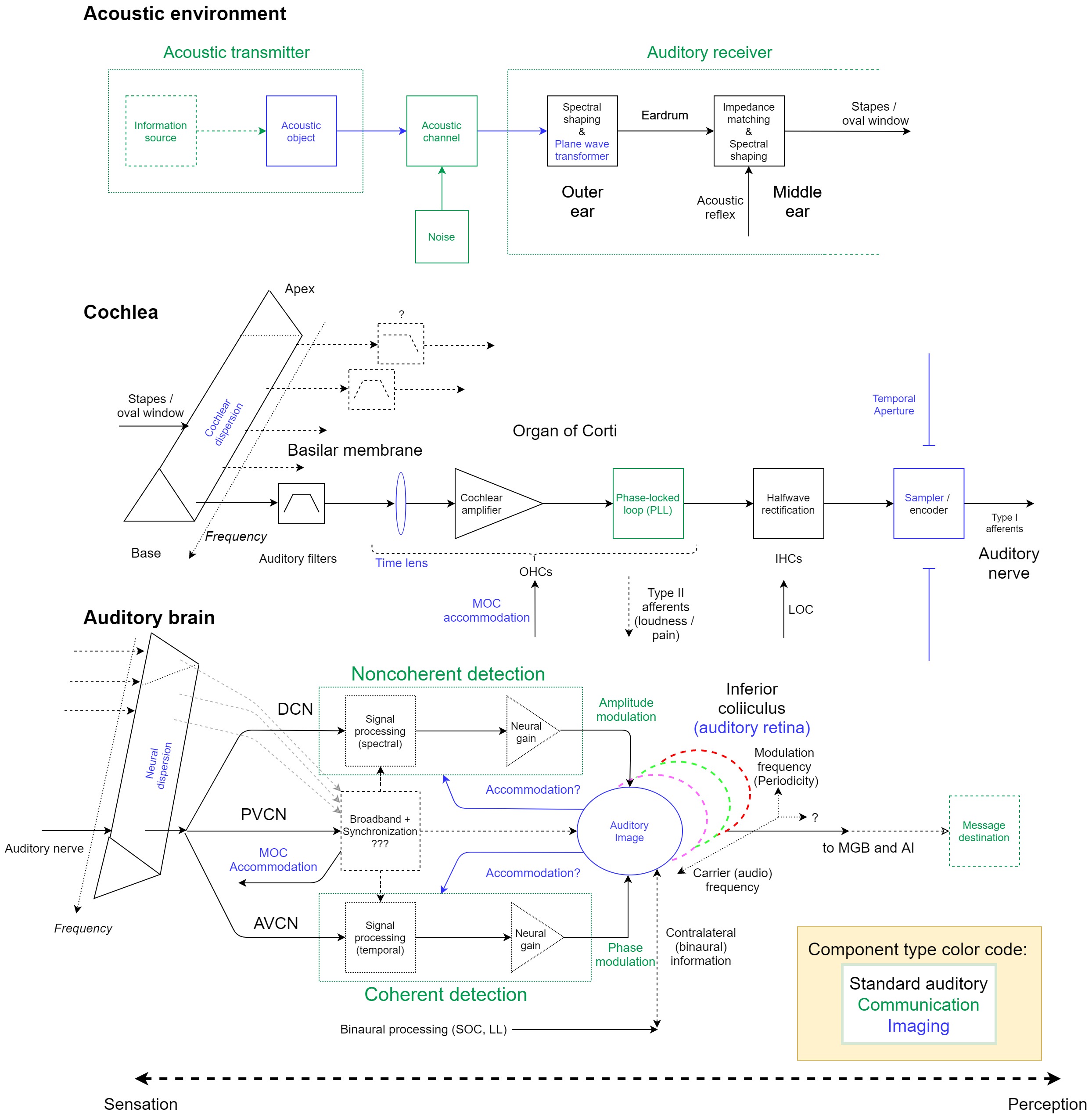}	
		\caption{A functional diagram of the provisional monaural auditory system, which contains standard elements (in black), as well as communication elements (in green) and imaging elements (in blue). The dotted lines around the message source and destination indicate that they are optional. See text for details.}
		\label{AuditoryCoherence}
\end{figure}

This section briefly presents how the various old and new components form a single auditory system. The provisional model is illustrated in Figure \ref{AuditoryCoherence} and includes both well-known components (in black), as well as new ones that are hypothesized in this work (communication in green and imaging in blue). There are still some uncertainties, which are most prominent in the brainstem, where the exact function is less obvious. Below is a general description of the model.

The top section of Figure \ref{AuditoryCoherence} includes the standard communication and information system building blocks (see Figure \ref{ThreeParadigms}). The ``acoustic transmitter'' is made from an optional information source that modulates the acoustic object. The object radiates energy into a noisy acoustic channel and is then received by the ear---the ``auditory receiver''. The channel additionally introduces various distortions to the signal, as was described in \cref{acoustenv}. The auditory receiver input port is the outer ear, which is responsible for spectral shaping of the acoustic wave caused by the scattering from the pinna and the concha, and for emphasis of canal resonance frequencies. Additionally, the ear canal allows only plane-wave propagation (below 4 kHz, in humans), which is a key requirement for imaging. The middle ear receives the vibrations from the eardrum and acts as an acoustic impedance transformer, yet unlike standard transformers, it is frequency dependent and produces a reactive component in its output. This linear mechanical structure injects significant power to the signal. The middle ear also receives control input from the acoustic reflex efferents, whose circuit is omitted from the diagram.

The middle section of Figure \ref{AuditoryCoherence} continues from the stapes / oval window, where it is coupled to the cochlear perilymph. The vibrations are transformed to a traveling wave in the basilar membrane (BM) from the base to the apex. A traveling wave excites the tectorial membrane as well, which is also omitted from the figure. Cochlear group-delay dispersion is indicated by the uncoiled BM that is shaped as a prism (in the diagram), which also exhibits the basic (broad) frequency selectivity---indicated with a bandpass filter bank. Only the most basal channel is displayed all the way through, to avoid clutter in the diagram. In the apex, there is some uncertainty about the bandpass nature of the filters, which is illustrated with a low-pass filter. 

The hypothetical time lens is due to the phase modulation inside the organ of Corti. It is followed by the standard but not yet empirically settled cochlear amplifier, which is a function of the OHCs. The novel phase-locked loop (PLL) function is also associated with the OHCs and drives the inner hair cells (IHCs), which transduce and rectify the stimulus (for details about the hypothetical PLL circuit itself, see \cref{CortiPLL}). If these three functions (time lens, amplifier, PLL) are indeed realized by the organ of Corti, then it is possible that they are either nested or occur in parallel, and not in series, as is displayed in the figure. So, a gain stage was assumed to be part of the auditory PLL, but it may also operate standalone.

Different efferent inputs to the OHCs are indicated on the diagram with no elaboration (medial olivocochlear, MOC) and IHCs (lateral olivocochlear, LOC). Slow Type-II afferents from the OHCs are indicated as well. The main output from the cochlea to the auditory nerve is through fast Type-I afferents that receive the rectified input from the IHCs. Each neural firing represents a sample of the continuous stimulus (or the deflection of the IHCs) and its temporal window constitutes the temporal aperture stop of the system, which is drawn as a vertical opening. Implicit in the figure is the high number of auditory nerve fibers for every hair cell (\cref{ComparativeCochlea}). In the brain, (nonuniform) sampling is always accompanied with encoding (see \cref{TemporalSampling}), which appears in the same block. 

The bottom section of Figure \ref{AuditoryCoherence} roughly illustrates the auditory brain function. First, neural group-delay dispersion is placed at the input, but should be distributed in the entire brainstem, in a manner that is unknown at present (with the exception of data by \citealp{Morimoto2019} of differences between waves I and V, see \cref{NeuDispExist}). From the auditory nerve, the neural pathway splits to three branches (acoustic stria): the dorsal cochlear nucleus (DCN), the posteroventral cochlear nucleus (PVCN), and the anteroventral cochlear nucleus (AVCN). 

This work adopted the communication engineering differentiation between coherent and noncoherent detection. Combining it with various findings about the cochlear nucleus in mammals and their avian analogs, the following rough functions are hypothesized for the two main branches of the DCN and AVCN. The DCN is hypothesized to primarily realize noncoherent detection functions, which mainly imply spectral and (real) envelope processing. The AVCN is hypothesized to realize the coherent detection functions, which usually imply temporal processing and also include the binaural processing of the system (not shown in the figure). Hypothetical neural gain function is indicated in both branches, which may be varied using accommodation efferent input from the inferior colliculus. Neural gain---or really, attenuation---may apply the relative weighting of the coherent and the noncoherent detected products that arrive to the midbrain to form a partially coherent image. Auditory accommodation may modulate other circuits beside gain as well, so it is displayed with no clear target. The function of the PVCN is not entirely clear, although it is known to involve broadband processing and provide synchronization to periodic stimuli. This is realized using coincidence detectors with variable delays from different parts of the cochlea---effectively canceling out some of the dispersion from earlier processing. The PVCN is also the source of the MOC reflex, which we considered to be part of auditory accommodation as well. While not shown in the diagram, the neural system continues to nonuniformly sample the stimulus beyond the auditory nerve. 

We have referred to the inferior colliculus (IC) as the auditory retina in this work, since it is where the image appears and the different processes converge, in analogy to the retina that contains extensive convergent neural circuitry within vision (\cref{AnaPhysioComp}). Its structure is laminar and each lamina processes a narrow band of carrier frequencies. In addition to tonotopy, periodicity---or rather modulation frequency that is more applicable universally---seems to be at least one additional orthogonal dimension of the image. In this model, only the coherent, noncoherent, binaural, and broadband inputs are shown, although almost all lower auditory nuclei project to the IC (Figure \ref{BrainstemFig}). As a communication target, the image is also where demodulation takes place (as detection). Like the retina, there is considerable signal processing and information compression that may be taking place in the IC. Given that in some mammals it was found that specific IC cells are sensitive to primitive (tuned) stimuli types, it suggests that the analogy between the IC and the retina must not be oversimplified to cast the IC as a simple screen and a detector. 

The main output from the IC is to the medial geniculate body (MGB) and it continues from there to the primary auditory cortex (A1) and the rest of the cortex, where perception is thought to emerge. Potentially, it is the destination of where a message that was acoustically sent is being received, which complements the optional information transmission on the other side of the processing chain. As was discussed in \cref{AcouAudComm}, the communication system can work more or less independently of the potential intent to use it to send messages. This destination resides around the same areas where perception may be taking place. In contrast, auditory sensation may encroach into the auditory brainstem. Therefore, the borderline between the two is not well defined and we do not imply that perception can be truly localized to one area of the brain.

The auditory model in Figure \ref{AuditoryCoherence} represents the main general functional blocks in the system, but remains relatively vague or agnostic regarding several key areas---mainly of the auditory brainstem. Some of the gaps in the model highlight the gaps in the present theory on the whole, whereas others reflect more general gaps in knowledge that are inherited from hearing and brain research, as well as the coarse grained perspective of this model. Either way, it contains certain speculations that will have to be tested in experiment. These gaps and weaknesses are reviewed in the next section.

\section{Known gaps and weaknesses in the proposed theory}
There are several aspects of the theory that have been left out of this work, either by design, or because of lack of sufficient knowledge and available methods at present. Below are several known gaps and weaknesses in the theory, as it stands at the time of writing. Some of these gaps can be appreciated directly from the model in Figure \ref{AuditoryCoherence}.

\subsection{Analytical approximations of the dispersive paratonal equation}
The dispersive paratonal equation that was originally derived by \citet{Akhmanov1968,Akhmanov1969} has been adopted from optics without changes (except for renaming it). While both acoustical and electromagnetic fields are assumed to be scalar, the acoustical problem deals with the audio frequency range that is many orders of magnitude lower than light. This is correct as long as the slowly-varying envelope condition is satisfied, which ensures that the group-velocity dispersion can be expanded around the carrier frequency and is well-approximated up to second-order. These conditions are synonymous with how we defined the paratonal approximation.

Two additional strong assumptions were made to be able to work with a tractable equation that may have inadvertently skewed some of the predictions. First, absorption curvature (or gain dispersion) was neglected, as is customary in optical imaging. The somewhat less common linear absorption (\cref{airtravel}) was also not considered. Linear absorption is known to have phase effects that are similar to dispersion \citep[pp. 358--359]{Siegman} and can be similarly incorporated into a complex group velocity (Eq. \ref{eq:vg}), which is a physically unintuitive concept and was not pursued here. However, even though the data about absorption are unavailable, we cannot rule out that the absorption curvature is of the same order of magnitude as the dispersion. This uncertainty is exacerbated by the abnormal low-frequency results we obtained in the cochlear curvature modeling (\cref{CurvatureModeling}), where the channels may no longer be narrowband. Also, the strict psychoacoustical modeling (\cref{PsychoEstimation}) has turned out complex parameter predictions that are only partially consistent with the mixed physiological modeling of \cref{paramestimate}. Consistency between the data sets may have to eventually incorporate absorption curvature. The uncertainty in the parameters is also compounded by the lack of distributions and confidence intervals for the obtained predictions (both physiological and psychoacoustical). 

The second strong assumption we made was of medium linearity. A level-dependent group-delay dispersion would require a nonlinear paratonal equation that takes the form of the nonlinear Schr\"odinger equation. It is sometimes applied to model the amplifying medium that is placed inside the resonator of a laser device, and it is arguable, therefore, that the active parts of the cochlea should be similarly modeled. While this equation has a closed-form solution \citep{Zakharov1972}, applying it here would have greatly complicated the analysis at this initial exploratory stage and would rest on shaky grounds with respect to the modeled cochlear medium. Most of the dispersive physiological data were obtained from low-level measurements at 40 dB SPL, which could warrant a linear treatment, but these results were then mixed with others that were obtained with observations done at higher levels. It should be noted that the level dependence of the phase may be negligible around the characteristic frequency (CF) in the cochlea and the auditory nerve \citep{Geisler1982}. The extent of the nonlinear effect in more central auditory areas is unknown. 

Adding level effects to the simple imaging equations should make it easier to incorporate compression into the theory, which has been purposefully avoided in order to simplify the analysis. The exclusion of level effects also entails that hearing threshold effects are not considered. As the spike rate in the auditory nerve is lower with low-level inputs, the sampling operation will be affected (degraded) by level as well, as reconstruction (should it exist) depends on less spikes---fewer samples. 

\subsection{Multiple roles assigned to the organ of Corti}
Two new roles were hypothesized for the organ of Corti and the OHCs: PLL and time lens. That these two functions are not in contradiction to one another, or to the amplification function that is often associated with the OHCs, is uncertain. 

The PLL model is framed in general terms and has a relatively strong empirical basis to it---both in supportive physiological and psychoacoustical evidence, and in the known properties of the PLL as a general model for a nonlinear oscillator (\cref{PLLChapter}). However, the associations and details of the different components that are hypothesized in the complete loop (i.e., the phase detector, filter, oscillator, and feedback coupling) may require a future update when new observations are made available. Perhaps the most daring speculation in the auditory PLL model is assigning a central role to the quadratic intermodulation distortion product, and assuming a role for internal infrasound information, which is currently not backed by direct evidence (\cref{LocalOsc}). It also assumes a critical role for the self-oscillations of the OHC hair bundles that can synchronize to external coherent stimuli. Despite the relative confidence in the PLL model, none of its parameters were estimated in the present work and none are available in literature about phase locking. Elementary specifications such as the order of the PLL, pull-in range, lock-in time are all missing. These parameters affect, for example, the PLL ability to retain lock of a linear chirp. 

The time lens function has been derived on a more phenomenological basis (\cref{lenscurve}). While the physical mechanism that was presented for acoustic phase modulation should not be particularly controversial or complex, the existence and derivation of its magnitude has been based on four studies that yielded two clustered sets of curvature values. All four required us to invoke the cochlear frequency scaling assumption to convert the data to humans, on top of the uncertain time lensing mechanism, and a very large gap between the estimates from the two animals. The analysis also intersected with the controversy regarding the true bandwidth of the auditory filter in humans, which has not been settled yet. This led us to repeatedly consult a broad range of curvatures throughout this work. Most effects were rather insensitive to curvature changes, but most appeared more adequate with the large curvature estimate---either based on broad-broad or narrow-band auditory filters. The uncertainty is compounded by the high likelihood that the time-lens curvature is adaptive, by virtue of the auditory accommodation---the MOC reflex (MOCR). Hypothetically, this mechanism could account for the existence of the small-curvature estimates of the time lens. The connection between the time lensing and the MOCR was deemed attractive in that it may be able to accommodate the auditory depth of field, although no direct proof was available to corroborate this idea.

It should be remembered that temporal imaging can be achieved without a lens, using a pinhole camera design \citep{Kolner1997}. The very short auditory pupil function may be analogized to a pinhole, so such a design is not completely far-fetched. However, a stronger constraint that motivated the acceptance of the time lens existence is the common-sense requirement that the magnification of the auditory system should be positive and close to unity. A negative magnification would entail a time-reversal of each sample of the envelope. Unless the cochlear and neural dispersions were misestimated in both magnitude and sign, unity positive magnitude cannot be achieved without a lens. The only other test we identified for directly assessing the time lens has been to model the psychoacoustical octave stretching effect using the image magnification, which is dominated by the time lens magnitude (\cref{TransChromAb}). This approach seemed promising but highly imprecise using the available data. All in all, the status of the time lens remains less than conclusive. 

\subsection{Neural group-delay dispersion}
The neural group-delay dispersion was computed from the difference between simultaneous auditory brainstem responses and evoked otoacoustic responses of the same subjects (\cref{NeuralDisp}). This calculation cannot differentiate the different neural paths that exist between the auditory nerve and the IC and therefore it imposes a blanket dispersion to all the different parts of the auditory brainstem. Although it is not impossible that the different pathways have about similar dispersion dependence, they are unlikely to all be exactly the same. Initial data on the chirp slope that is required to maximize either wave I or wave V of the auditory brainstem responses, suggested that the auditory nerve and brainstem have different dispersions \citep{Morimoto2019}. In the general hypothetical model of the complete auditory brainstem of Figure \ref{AuditoryCoherence}, the different pathways are specifically assumed to have different functions---coherent and noncoherent detection. In the imaging theory, each one corresponds to another extreme solution using the same neural dispersion. However, in reality, the parametrization of the different path dispersions may produce different effects. Supporting evidence to this effect may be found in frequency-following responses (FFR) of amplitude modulated tones in humans, which were associated with different group delays and brain areas that generated the envelope synchronization and phase locking to temporal fine structure \citep{King2016}.

\subsection{Modulation filtering}
Throughout the auditory brain, there are units that are tuned either to carrier or to modulation frequencies, or to both. The existence of bandpass modulation-frequency filters is not accounted for by the present theory. Such filtering can be analyzed using the various modulation domain transfer functions. The analog in optics would be of spatial filtering, which is achieved in the inverse domain and affects the modulation band---the spatial frequencies. This optical technique provides relatively simple image processing capabilities, which can be implemented also in analog means (e.g., by constructing a special pupil function). The exact roles of these filters in the auditory brain may be specific to the species and its acoustic ecology. The degree to which these filters are eventually perceived may vary, especially with consideration to integration across channels.

\subsection{The role of the PVCN}
A particular challenge in the present model is the third pathway from the auditory nerve---the PVCN---whose function is not very clear. Neurophysiological research has mainly dealt with the DCN and AVCN and in the past has not always distinguished between the AVCN and the PVCN. Interestingly, the avian auditory system has analogs to all the main auditory organs in the brain, but has only two pathways split from the auditory nerve, which correspond to the AVCN and to a combination of the DCN and PVCN (see \cref{CentralNeuroanatomy}). The roles of the two pathways can fall in the purview of the assumed coherent and noncoherent blocks from the communication model. However, despite several anatomical and morphological differences between humans and other mammals, there is little difference in their hearing in terms of general mechanisms and responses. 

The literature suggests three main differences between avian (mainly songbirds) and human hearing perception \citep{Dooling2000Dooling}: 1. Lack of avian high-frequency hearing, with a bandwidth that is usually cut off below 10 kHz. 2. More acute temporal processing (shorter time constants). 3. Higher sensitivity to small and fast details within sound sequences, which go unnoticed by human listeners. Although these differences (especially 1) may be attributed to anatomical differences in the cochlear and hair cell structure, they are most likely related to higher-level mechanisms. What is particularly baffling about 2 and 3 is that they exhibit superior performance to humans\footnote{Superior performance here means that less information is lost. Alternatively, if information loss or integration of information over long time frames is key \citep{Weisser2019}, then the avian hearing superiority may be questioned.}, despite an absence of an entire auditory pathway. Therefore, a more refined model may be helped by comparative research between mammals and birds.

\subsection{Accommodation}
Accommodation was introduced in \cref{accommodation} as a plausible feature that the auditory system likely has, in analogy to vision, except that it is more deeply embedded in neural mechanisms. Several possible mechanisms were considered, but they were all speculative to different degrees. Given the amount of efferent connections in the system, it is not hard to imagine that some particular circuits can be recast as accommodation. However, the present theory remains uncertain as for which one it is. Perhaps the most likely hypothesis is related to the degree of coherence of the image, which is a function of the source coherence, and the products of the coherent and noncoherent detection pathways. This can then corroborate an interpretation of the MOCR system as one that adapts the time-lens curvature and the PLL gain to dynamically skew the processing to be more coherent or noncoherent, depending on the stimulus and situation, and thereby also affect the auditory depth of field. In addition, this dual processing logic may entail modulation of the relative neural gain of the pathways that are responsible for each kind of processing. Other mechanisms may exist that accommodate almost every other parameter in the system, either as part of an accommodation reflex, or independently. In all cases, the number of unknowns stands in the way of a comprehensive understanding of the system operation. 

\subsection{Coherence model}
Throughout the analysis, we have appealed to coherence considerations to motivate a large number of effects, but it was done without specifying a particular model to compute auditory coherence. Although it should be possible to use the various formulas for continuous coherence in the traveling-wave domain, they may lose validity in central processing. Also, the nonstationarity that characterizes typical acoustic signals has to be matched with proper time constants that may be longer than the aperture time---perhaps depending on the processing stage. Furthermore, there are clearly two dimensions of coherence---both within channel that is applicable to classical imaging and communication, and across channel where object coherence (in its psychoacoustic sense) comes into play. 

Another important aspect of coherence that we relied on is that the weighted sum of the coherent and incoherent images gives rise to the partially coherent image that ``appears'' on the IC. The details of the specific images, their weightings, and how they are accommodated are missing too, although the logic of this operation is fairly straightforward. However, the combination of these two images has been a central point of this theory and will have to be better formalized. 

\subsection{No simulation}
No simulation of the complete auditory imaging system has been provided in this work. Such a simulation will have to be attempted when the dispersion parameters and pupil function of the system are known with more certainty. This simulation requires numerical computation that has several hurdles to overcome, depending on the method used \citep[pp. 121--153]{Goodman}. First, discretized quadratic transformations (as defocusing implies) require computational oversampling to avoid aliasing of the output. Second, nonuniform sampling must be implemented as well, to adequately represent the neural transduction and spiking, yet this is not a standard element of the optical methods. The auralization of such transformations may have poor sound quality due to the apparent periodicity pitch of the low-frequency sampled pulses, or noisiness, depending on the sampling regularity and the type of detection assumed. This means that an anti-aliasing filter has to be implemented, unless aliasing can be converted to noise through nonuniform sampling, as theory predicts. A third issue has to do with the broadband configuration of the entire set of auditory filters, which have to be combined in a way that is able to preserve features of both amplitude and intensity imaging.  

A related problem in simulation can take place in the hypothetical reconstruction stage of the sampled signals, which requires using unmodulated carriers that correspond to the CF of the auditory filter, especially in coherent detection and imaging. It is not unlikely that the combination of the inputs from multiple fibers and channels is essential for the production of a perceptually seamless and continuous sound. It should be remembered, though, that the auralized response may not sound good without appropriate decoding that is not entirely known. Perception deals with an auditory code that is recoded several times before the smooth tonal experience is perceived, downstream. The pitch percept, for example, may be likened to color, in the sense that it is not directly caused by the carrier frequency, but is partly determined by place-coding that is transformed to a higher-level representation in the cortex, which corresponds to pitch. 

\subsection{Across-channel response}
Even though the significance of the polychromatic image has been emphasized, the present theory is based on single-channel analysis and it remains uncertain about how across-channel integration and broadband response is achieved in reality. Physiologically, it can be achieved in a few ways. Broadband units (e.g., the octopus cells) are the simplest solution that has been proposed so far, but there are few observations that demonstrate innervation by multiple channels, which can facilitate the cross-channel integration (for example, \citealp{McGinley2012}). Another option is that the channels are interconnected, say, by the interneurons in the brainstem, and have the ability to mode-lock, which gives rise to enhanced harmonicity sensitivity (\cref{Phaselocking}). This may work in tandem with the helical geometrical structure of the dorsal lateral lemniscus, as was hypothesized by \citet{Langner2015}. In the same work, \citet{Langner2015} argued for a powerful neural model that combines units from the DCN, PVCN, and VCN, which together detect the broadband signal periodicity by synchronizing to its envelope phase. The resultant periodotopy on the IC laminae is orthogonal to the carrier frequency tonotopy. In principle, Langner's model may  apply to aperiodic or quasiperiodic modulation and also to unresolved periodicity, which is sometimes referred to as residue pitch. However compelling this model may be, it is unknown at present whether the exact circuitry that it hypothesizes exists in the brainstem. 

\subsection{Lateral inhibition}
Lateral inhibition is crucial in both vision and hearing, but has not been treated in this work and, rather, it has been implied throughout, as is sometimes done in auditory models. It should be distinguished from lateral suppression in the cochlea that is more automatic and perhaps less goal-oriented. Because of the substantial overlap between the auditory filters, it is important to consider the effect of lateral inhibition that defines stable channels and affects the integration of information between channels. Linearly frequency-modulated (FM) stimuli, which have been repeatedly considered as the staple signals in temporal imaging, are particularly prone to excite multiple auditory channels over a short time span (depending on their slope). Thus, these signals inherently challenge the simplistic single-channel models that have been presented in this work. On the one hand, we know that the auditory system ``takes care'' of the integration, so that these stimuli always sound seamless, even as they cross several channels. On the other hand, it requires an active weighting process across channels, which has to include the signal phase as well. In any case, inhibition should play a role in modeling these processes. 

\subsection{Binaural hearing}
A major topic that has been deliberately left out of the analysis is binaural integration of monaural images. The binaural system in hearing has often been studied as an independent subfield, partly due to what appears to be dedicated pathways from the superior olivary complex to the auditory cortex and beyond. The fact that almost all auditory pathways converge in the IC should provide some clues for an integrated monaural-binaural model. 

All that said, binaural perception is readily included within a general coherence theory of hearing, since coherence has been long used in binaural research. According to this reformulation, interaural cross-correlation is really a spatial coherence measurement (\cref{InterauralCoherence}). Additionally, the familiar interaural time and level differences are also readily seen as special cases of the complex degree of coherence measurement that are sensitive to different cues of partial coherence. The entire structure of the binaural system is reminiscent of an interferometer, as was first noted by \citet{Cohen1988}, which is the kind of measurement device that produces the most precise data in physics. Therefore, interferometry science may have something to say about the ears as well, although this direction has not been consulted until very recently \citep{Dietz2021Litovsky}. 

\subsection{Neuromodulation}
The topic of neuromodulation was briefly invoked in \cref{accommodation} as a possible general mechanism to realize auditory accommodation. Mounting evidence suggests that various neuromodulatory pathways in the auditory system, including the brainstem and midbrain, can have significant effects on sound processing that take place over different time scales. At present, these effects are excluded from all standard auditory models, which may create the false impression that the signal processing of the auditory system is static, or that plasticity occurs only on a high and slow processing level. Undoubtedly, inclusion of these effects greatly complicates the understanding of this system, but it also has the potential to bring its modeling much closer to reality, by introducing various points of calibration and plasticity that can be individually matched. 

\subsection{Multiple time lenses and PLLs}
We modeled the auditory channel as though it contains a dedicated time lens and PLL, which were associated with the apparent uniformity of the hair cells and supporting cell tissues in the organ of Corti. This implicit assumption is unsubstantiated at present. The PLL, in particular, depends on a local oscillator for its function, which we took as the hair bundle self-oscillations, knowing that in vitro, hair bundles of different hair cell types tend to oscillate \citep{Martin1999}. However, inasmuch as the spontaneous otoacoustic emissions (SOAEs) represent existing oscillators in the cochlea, they portray a rather limited picture---only a subset of subjects have measurable SOAEs and their retuning is not as flexible as we would like to see in a general-purpose PLL. A charitable interpretation would then be that the oscillations are usually too faint to be measurable, that they are suppressed by their neighboring oscillators, or that they remain active only during supra-threshold stimulation that does not allow for external measurement in a favorable signal-to-noise ratio. Furthermore, we do not know what the effect is of chaining parallel PLLs that may all couple and synchronize to one another. Can they be isolated? How do they suppress one another? For a treatment of a related problem of coupled nonlinear oscillators in the cochlea, see \citet{Wit2017}.

Another difficulty in the theory is that mechanical phase locking does not work much beyond 4 kHz in mammals, whereas the OHCs extend to much higher frequencies. There, the PLL function cannot work, unless it is somehow able to synchronize directly to the envelope. Another more remote option is that the OHCs are desynchronized at high frequencies to the extent that they decohere coherent signals and gradually facilitate their power amplification and noncoherent detection.

To the extent that they are independent of PLLs, multiple parallel time lenses may also give rise to complex nonlinear effects, although less consequential, perhaps. The layout of small and channel-specific time lenses might be compared to compound eyes of insects and crustaceans, which do not have a single crystalline lens, but rather numerous small optical units (\term{ommatidia})---complete with their own lenses that are directly connected to a few photoreceptor cells \citep[e.g.,][]{Nilsson1989Stavenga}. Unlike the ear, the ommatidia are not coupled, or at least not in such a complex way. But maybe somewhat like the many channels of the ear, the images from every ommatidium have to be integrated into a whole in the animal's brain. This interesting association has not been explored any further.

\section{Novel contributions to auditory theory}
Despite its shortcomings, it is important to not lose sight of the insights that the temporal imaging theory has to offer to the understanding of auditory theory more generally. The primary ones are briefly discussed below.

\subsection{Temporal auditory imaging}
The temporal imaging theory, as was introduced into optics by \citet{KolnerNazarathy} and \citet{Kolner}, has been applied to the mammalian auditory system (\crefrange{temporaltheory}{ImpFun}). The theory treats complex-envelope pulses as the objects to be imaged and operates in the realm of dispersive media. Such a foundation makes little sense in an acoustic world that comprises mostly pure and complex tones and where AM-FM modulation is a rarity. However, once we let realistic curvature-ridden signals and environments replace the idealized stimuli---sounds that had been inherited from an idealized musical world and may have occupied the collective auditory scientific spotlight as a consequence---a dispersive point of view becomes much less puzzling. This is the domain where paratonal sound processing resides. 

The introduction of a rigorous imaging framework to hearing enables us to make intuitive predictions that are based on analogies from optics and vision, as long as two substitutions are made: the spatial envelope (the monochromatic object) should be replaced with the temporal envelope, and, in some cases, the distance from focus of the optical object should be substituted with the degree of coherence of the acoustic signal. These substitutions open the door for auditory focus, sharpness, blur, depth of field, aberrations, and accommodation as valid and useful concepts in hearing. A third substitution between color and pitch is more complex, though, because the relative bandwidth of hearing is considerably larger than vision that enables harmonic perception, and auditory phase perception allows for coherent (interference) phenomena that do not exist in vision that is strictly incoherent. 

Temporal imaging requires several second-order phenomena in the auditory system to assume relatively prominent roles in the signal processing chain. Predominantly, these phenomena include the group-delay dispersion of the cochlea and the auditory brainstem pathways, the exact temporal window shape that is achieved in transduction and works as a temporal aperture, and more speculatively, the phase modulation in the organ of Corti that works as a time lens. In current literature, these parameters have been either neglected or presented as physiological happenstance rather than essential to the signal processing of the ear. This neglect is not coincidental and is unsurprising in the case of unmodulated tones or signals with low-frequency modulation. Such stimuli are usually unaffected by the typical low-pass character of the modulation response and they do not disperse easily. This is also because the associated curvatures have very small magnitude, which has little influence on sustained narrowband signals. Additionally, the auditory filters work well within the narrowband approximation in normal-hearing conditions, so low-frequency modulation response generally dominates unresolved auditory stimuli. In contrast, high-frequency modulation that is more prone to dispersion gets resolved in adjacent filters and may become noticeable only in broadened filters and through a hearing impairment. Therefore, relatively imaginative experiments may be needed to uncover exceptional cases in this system. 

Another interesting point, which has not been dwelt on in too much depth in the text is the significance of the pupil function in imaging. In optics, the imaging system is completely described by the generalized pupil function, which contains information about resolution, blur, and aberrations. In case we are interested in the demodulated product of the hearing system, then the auditory pupil function should be equally important. In the text we have used a Gaussian pupil that is analytically tractable and seemed to have yielded several useful results and insights, despite its oversimplified nature. 

\subsection{Sampling}
An additional layer of theory that is required for the temporal imaging to work is the treatment of neural transduction as a sampling process that may or may not be amalgamated with the neural encoding operation. This is not a new idea in hearing theory, as various psychoacoustic and physiological models assumed the discretization of the stimulus (see \cref{SamplingInTheory} and \cref{AliasIntro}). However, sampling has not been embraced as a standard part of the auditory signal processing chain and is rarely considered a mandatory function of the auditory neural code. Whenever a discrete representation of sound is employed, sampling is usually taken for granted and sampling theory is not consulted. This means that several interesting sampling-related effects have not been traditionally considered in hearing such as aliasing, the instantaneous sampling rate effect on temporal resolution, or the tradeoff between aliasing distortion and noise that can be achieved with nonuniform sampling (as was demonstrated spatially in the retina; see \cref{nonuniformunder}). 

A related effect that is under-explored in hearing is the possible interaction, or even downright equivalence, of neural adaptation and the notion of having a variable sampling rate (\cref{Aliasing}). Nevertheless, it should be plausible and even intuitive that the auditory system densely samples the signal around the onset---the temporal analog of spatial edges---to provide higher-resolution information about the points where the signal exhibits the most variation due to transient effects.

\subsection{Modulation domain}
A central point that this work emphasized is the conditional independence of the carrier and the modulation domains---independence that applies only in strict narrowband and stationary conditions that are not particularly realistic. The idea of a separate modulation spectrum and periodicity maps is not new and can be traced back to \citet{Licklider1951}. But even when it is discussed at length, the subtext in hearing theory has been that the significance of the modulation domain is secondary to the carrier domain. This is mathematically, physically, and perceptually misleading and yet it is not a far-fetched impression that one may get by exclusively considering unmodulated (stationary) stimuli. The present theory attempts to dispel some of the confusion regarding modulation in hearing by maintaining consistent representation of the acoustic wave (the object), and the communicated signal, which is eventually perceived as an image. For the price of restricting the usage of Fourier frequencies that are unvarying and extend to infinity mainly to carrier frequencies, we obtain a unifying mathematical framework that encompasses several theories that are relevant to hearing. A corollary of this move is the recognition that the envelope and temporal fine structure are interdependent, so that AM-FM signals are inevitable, which reflects a signal representation that has a carrier and a complex envelope, rather than a frequency-modulated carrier and an amplitude-modulated envelope, separately. This representation also facilitates synthesis with coherence and PLL theories. 

This point of view also highlights the informational content of the sound source object. In hearing, the acoustic signal contains useful information both in the carrier and in the modulation domains. Due to the finite bandwidth of the auditory filters, there is some fluidity between the carrier and modulation domains for signals with high modulation rates that are on the limit of being resolved by adjacent filters. This means that temporal scaling of the signal, or changes in the filter bandwidth can result in information moving from one domain to the other. Unresolved high-frequency modulation is theoretically more prone to phase distortion (over-modulation)---that is, to image aberration. Unresolved high-frequency modulation is also constrained by undersampling errors, which can potentially make the quality of received information lower than the resolved version of the same signal. Aberration, over-modulation, undersampling, and movement between carrier and modulation domains have rarely if ever been treated in hearing theory, so it is hoped that they may inspire more holistic ways of thinking about normal and impaired auditory processing. 

\subsection{Coherence and defocus}
\label{CoherenceAndDefous}
The key parameter that repeated in the context all all theories that were synthesized here is the degree of coherence of the signal. Even though coherence has been used occasionally in hearing theory, its application has been inconsistent and has lacked rigorous ties to scalar wave coherence theory. Along with the static frequency interpretation of acoustic time signals, this may have underpinned an idealized point of view in which coherence is an auxiliary signal processing concept, rather than a dynamic property of the acoustic source and transmitted signal. In reality, signals tend to be neither coherent nor incoherent, but rather, partially coherent.

The differential processing of incoherent and coherent signals under defocus can readily explain the defocus unmistakable existence within the auditory system. With a degree of coherence that can vary over the entire range between completely incoherent to completely coherent, it is reasonable that the auditory system takes advantage of its defocusing property to further differentiate these two signal types, as a key to segregate different sources. As signal informativeness is often implied by its coherence, an ability to manipulate signals accordingly may provide a considerable advantage for hearing in noisy, reverberant, and other distorting conditions. Moreover, coherence plays a dominant role in imaging theory and readily distinguishes between coherent and incoherent modulation transfer functions---something that has not been previously accounted for in auditory theory (re broadband and tonal temporal modulation transfer functions, TMTFs). 

\subsection{Coherent and noncoherent detection}
\label{CoherentNonCoherent}
Once we accept that information is carried by signals that are not mathematically idealized and that coherence is a field property that propagates in space, hearing becomes amenable to be treated as a communication detection problem. Known signal detection techniques are distinguished as either coherent or noncoherent, which intuitively refers to the similarly worded imaging optics distinction, but is not exactly the same. Coherent detection preserves the carrier phase and generally relies on a local oscillator in the receiver, whereas noncoherent detection does not. Both detection types are advantageous in different conditions. Coherent detection is ideal for signals in noise and for carrier synchronization applications, whereas noncoherent detection is much easier to implement and less prone to failure as a result. Each method is capable to suppress noise from the other kind: noncoherent detection can provide immunity against coherent noise, and vice versa. Noncoherent detection strongly resembles the traditional power spectrum (and phase deaf) hearing model that has been dominant and effective in many applications, despite accumulating cracks that have been revealed in this approach (\cref{AuditoryPhaseReview}). Many of the countervailing observations against the power spectrum model have to do with various phase (temporal fine structure) effects that are thought to be related to the phase locking property of the auditory system. 

The unifying perspective proposed here is that the auditory system has a dual detection capability, which can universally receive arbitrary signals. Such a system should dynamically optimize its detection according to the coherence of the signal, its environment, and various internal factors of the listener. On a high level, it functionally calls for auditory accommodation system that is somewhat analogous to vision. We were able to identify a range of processes that can fall into the purview of accommodation, but much still remains to be revealed. In particular, the idea that the organ of Corti functions as a PLL that can facilitate coherent detection and perhaps respond to accommodation---probably in unison with its effect on the time lens---has the potential to clarify several murky points in the present understanding of the system. These ideas may be a step toward understanding the interrelationship between phase locking, envelope synchronization, temporal fine structure, and the elusive duality in processing (i.e., spectral and temporal) that the system seems to manifest in almost any task. 

Nonetheless, contrary to the traditional distinction between coherent and noncoherent detections, it appears that the system eventually combines the two detected products and produces a partially coherent output. This strategy is accounted for much more readily with principles from imaging optics, rather than communication engineering alone. 

\section{Missing experimental data}
Part of what makes the theoretical claims in this work speculative is the relative poverty of direct data that could be harvested from literature to substantiate them. This is understandable given the small magnitude involved in the associated quantities, their obscure function, and a general attendance to other auditory properties that have been better motivated. In this section, we highlight some types of data that are most sorely missing for this theory. 

\subsection{Human dispersion parameters}
At this stage, the auditory temporal imaging theory rests on insufficient objective/physiological data that were assorted from several sources---usually in different measurement conditions (i.e., species, stimulus level). We also generalized from data measured on very small populations, although we do not know what the spread of the dispersion parameters is in the normal population. Therefore, better estimates of the frequency-dependent cochlear and neural group-delay dispersions are needed---no matter how small in magnitude they appear to be. 

Possibly, alternative methods should be conceived to obtain these estimates of the dispersion parameters that require less transformations or corrections (e.g., from animal models to humans, from cadavers to living, or from high to low frequencies). In the present estimates, there was no way to altogether avoid employing some methods that have not been mired in some controversy. Therefore, some dataset choices (e.g., picking one dataset out of the four derived neural group-delay dispersion calculated that are physically plausible, \cref{NeuralDispEst}) were preferred because they provided a somewhat superior prediction to independently measured data (e.g., the psychoacoustic curvature measurements \cref{CurvatureModeling}). While the differences have usually been small and the results rather stable across several datasets, more certain data are clearly needed to establish sufficient confidence in the theory and predictions. 

Perhaps the most uncertain parameter is the time-lens curvature when the lens is ``relaxed''. Then, if it indeed turns out that the curvature can be accommodated, then its bounds in humans should be clearly established, as well as the conditions that lead the system to actively accommodate to. 

Another parameter that has been largely missing is the pupil function---its shape and effective temporal aperture and their frequency dependence. Importantly, any noninvasive methods that can be developed to directly measure this function in humans can go a long way to simplify the derivation of individualized transfer functions. Should dispersive pathologies exist in the listener's hearing, they should appear in the generalized pupil function. 

Both auditory brainstem response (ABR) and otoacoustic emissions (OAE) are particularly attractive candidate methods for noninvasively determining the various parameters in humans. However, the underlying theory---especially of OAE---must be settled before they can be harnessed for these tasks. Furthermore, both ABR and OAE may require some corrections to precisely segment the desirable dispersive element without contributions from the adjacent dispersive segments. 

Ultimately, superior estimates of the various dispersive parameters should decisively determine what the correct configuration of the auditory system is and whether it matches the one that has been championed in this work. This means that if either the cochlear or neural dispersive path curvature is measurably zero, then the possible image is going to be different and may entail that some of our conclusions are erroneous. If an active phase modulation capability producing the time lens effect is not found, then we would be dealing with an auditory pinhole camera design, which is similar but not identical to our model imaging system and could offer less flexibility in hearing. The combination of these three components determine the focusing capability of the system, which, for all we know, may be configured completely differently between different animal species. Therefore, generalizations between species must be done with care.  

\subsection{Behavioral data}
A critical section in this work was based on high-quality cochlear phase curvature data by \citet{OxenhamDau}, which allowed us to validate the physiological predictions and derive estimates for the aperture time. However, these results are based on relatively few subjects ($N=4$) and are extremely demanding for listeners. A modified method that is based on B\'ek\'esy's tracking technique was demonstrated, which could reduce the measurement time per frequency from 45 minutes to 8 minutes only (\citealp{Rahmat2015}; see also \citealp{Klyn2015}). While an even shorter measurement duration will be preferable, this is undoubtedly a promising direction. 

Ideally, the individual time-lens curvature can be measured in a more robust way than by using the stretched octave effect, which is also a tedious measurement. More importantly, if time-lens accommodation is confirmed, then the adequate ``relaxed'' curvature should be established (analogous to the emmetropic state of the eye). Then, the limits of the smallest and the largest curvatures should be established across the population along with the required conditions to elicit them.

Another perspective that can be rather readily tested is how normal-hearing and hearing-impaired listeners rate or respond to stimuli that can be classified as sharp or blurry. A continuum can be generated between the two extremes, based on different principles of (de)coherence, which listeners can judge. Listeners' willingness to apply this measure and their consistency may be revealing. Ideally, such results can be correlated to the acoustical features of the signals. This may provide additional data about possible dispersive impairments that some subjects may experience.

Better controlled data of various psychoacoustic effects may be useful to establish whether a completely behavioral test battery can be used to fully characterize the dispersion parameters of a listener. This was attempted in \cref{PsychoEstimation} with mixed results that were not straightforward to interpret, but which could have benefited from better controlled data. This refers to octave stretching, frequency-dependent temporal acuity, and beating data, in addition to the phase curvature using Schroeder phase complex.

It will also be important to find out what the individual spread is in the normal hearing population. We relied mostly on (small) population averages, but it is unknown how relevant these values are for individual listeners.

\subsection{Acoustics}
On the physical acoustics front, there is a total lack of literature on the group-delay dispersion in various media, as well as on methods of estimating it. The straightforward way is to directly obtain it from the second-derivative of the frequency-dependent phase function. However, this quantity may be generally very small, so measurements should be done carefully due to possible sensitivity to small changes in position, boundary conditions, equipment, etc. These data should be complemented with the group-delay absorption to find out how it compares to dispersion. Additionally, it can be useful to confirm that the Kramers-Kronig relations are indeed applicable in typical media using bandlimited measurements. 

Group-delay dispersion measurements should include the phase response of the outer ear canal, which were estimated to be negligible, yet tend to erratically fluctuate between positive and negative values (\cref{outerear}). Most curious may be testing for the possibility of dispersion distortion, which is caused when the same modulation information is carried by different modes with different group velocities and spatial distributions. This is expected to take place in the vicinity of the eardrum at frequencies above 4 kHz, but the size of this effect is unknown---especially given the small acoustical path involved relative to the wavelength. This is a new distortion to acoustics, which can have important implications on information transfer fidelity. 

Acoustic phase modulation constitutes the time-lens operation (in the quadratic approximation), but has not been investigated in acoustic research. However, there should be multiple ways to achieve it, depending on which medium parameters are being modulated (e.g., density, compressibility) and over what spatial and temporal extent. This is an uncharted territory that may produce interesting theoretical and technological insights in acoustics. 

Finally, detailed coherence data that describe the range of natural and synthetic stimuli are missing. The data provided in \cref{ExCohere} are limited in scope and were mostly presented to get an idea of the range of values that are involved and the effects of room acoustics, nonstationarity, and analysis filter selection, among others. Such data would be required in order to establish a more rigorous coherence theory for the auditory system itself. 

\subsection{Cross-disciplinary digging}
Communication theory and imaging Fourier optics have been used as the central sources of inspiration and analogy for the present theory. However, the analogy breaks down due to the idiosyncrasies of the auditory system, such as the large number of overlapping channels, the dual nature of the system that leads to partially coherent images, the mixed acoustic-neural domains, the multi-purposedness of hearing, and the mechanical complexity of the cochlea. Still, both communication and optics have a wealth of theory, methods, and designs, along with extensive empirical records, which undoubtedly contain additional clues that are relevant for the hearing system. For example, the topic of ultrawide-bandwidth transmission and reception is relatively new in communication engineering and it matches the auditory system bandwidth, by definition. This topic has not been consulted in this work more than at the most superficial level, and is expected to harbor ideas that are relevant to hearing.

\section{Overarching themes}
Given the interdisciplinary nature of this work, some of its assertions may impact closely related disciplines outside of hearing. This chapter concludes with a short discussion of a few of them. 

\subsection{Imaging as a unifying sensory principle}
This work has demonstrated how mammalian hearing can be reframed as a temporal imaging system, which can be formulated with the analytical tools developed for light and spatial imaging optics. While hearing and vision have been repeatedly juxtaposed and contrasted throughout history (\cref{HearingTheoryVision}), the present theory provides the most comprehensive account to date for the extent of this analogy. Specifically, an auditory image has been rigorously shown that shares the basic conceptual properties of the visual image such as focus and blur. However, the dominant dimensionality of the auditory image is ``antonymous'' to that of vision---temporal instead of spatial---and the number of carrier channels involved is orders of magnitude larger than in human vision. The physical signal that is applicable to hearing is also associated with completely different frequency and wavelength ranges, which results in a different operating principle than vision---coherence-dependent focus, instead of distance dependence. Therefore, the resemblance and analogy between the two image types is nontrivial. However, there are enough parallels between the two, which can lead to some form of theoretical union. It implies that some general theoretical principles can be studied only once and be applied to both vision and hearing as special cases.

This begs the question---how universal is the imaging operation? Hearing requires complex energy transformations, but seems to have evolved a similar mathematical solution as vision, at least in humans and many mammals. It is also known that in some species of lizards and tautara (see Figure \ref{EarEvo}), the pineal complex has an eye-like organ---the parietal eye that has a role in thermoregulation---complete with a lens, cornea, retina, and photoreceptors with sensitivity overlapping the visual portion of the electromagnetic spectrum \citep{Tosini1997}. Does that imply that other sensory channels may operate using similar principles? On the face of it, it seems highly unlikely, as the vast majority of senses are not based on radiated waves, as hearing and vision are, where the dimensional transformation is relatively neat. It may be that other sensory modalities require transformations that are not at all trivial and are intuitively challenging, even though they become comparable on the level of the cortex and assume similar mathematics to imaging along their sensory pathways. 

\subsection{Perception and sensation}
Perception and sensation are usually treated together in psychology and philosophy texts, although the emphasis on perception is significantly greater. Visual perception is typically taken as the gold standard for perception in general (\cref{AuditoryImageDiscusion}). This approach mirrored in the present work for practical reasons, as the auditory system has turned out to resemble vision in some unexpected ways. But we have argued throughout that the auditory periphery encroaches well into the central nervous system, much like the eye, since the optic nerve is part of the central nervous system. It means that the borderline between auditory sensation and perception is fuzzier than in the visual system. It is generally understood that sensation operates on a low level of processing of the peripheral nervous system, which implies that it is automatic and unconscious. The sensory input is therefore conveniently fed into perception, which occurs somewhere in the brain and is typically associated with the cortex. 

However, in the present theory we have a system for which this usual role division fails, inasmuch as the auditory physiology reveals either the sensory or the perceptual roles. The evidence for this is that the same operation that the visual system performs in optical periphery (diffraction), hearing performs in the brainstem (dispersion). It does not imply that there is consciousness that is associated with the high-level sensory perception. But it does raise the question of whether perception is produced incrementally and only manifests fully in the cortex. Our theory is therefore reminiscent of the ancient Greek philosophers' view, who did not have separate notions for sensation and perception \citep{Hamlyn1961}. 

\subsection{Analog and digital computation in service of sensation}
Another facet of sensation is related to biological computation. We obtained a mathematical analogous function in the auditory neural domain and in the visual peripheral (ocular) domain. Traditional views \citep[e.g.,][]{Marr2010} take for granted the analog information processing of the eye, prior to the retina, which is responsible for the sharp visual image. In contrast, the analogous auditory brainstem function is part of the signal processing of the system. We have a fully neural system that realizes a function (group-delay dispersion) that the eye does with the vitreous humor (the eyeball)---diffraction. In strict information processing terms, it means that the eye works as an analog computer, whereas the ear is neural (if we shy away from using the word ``digital'' in the context of the brain). 

The inclusion of analog computation in the biological toolkit is not a subtle narrative change, because it demonopolizes the neural system from being the only biological element that can process information. At least in hearing and vision, biology has been more accommodating in utilizing diverse physical information processing principles that were available through evolution to achieve the desirable computational goal. 

See \citet{Sarpeshkar1998} for further ideas about combining analog and digital information processing in the brain and also \citet{Maclennan2007} for a discussion about analog computation in biology.

\subsection{Fast and slow processing}
Our analysis has made explicit a dual processing strategy of the auditory system, which has been in plain sight for a long time, although it has not been called by name (\cref{DetectionSchemes}). In the simplest terms, the ear combines coherently and incoherently detected images into a partially coherent image. One of the main differences between the two are the time scales. Coherent processing requires phase locking to carrier frequencies that may have to be matched with high sampling rates, whereas incoherent processing can track the slow envelope, thereby discarding much of the phase information with no apparent loss of auditory performance (i.e., speech intelligibility). This kind of processing may be simpler to accomplish, as it aggregates signals over several channels and is not particularly ``fussy'' about instantaneous changes of the signal. It likely requires slower sampling rates and can be performed with much more relaxed tolerance, which may free up brain resources to other tasks, if necessary. 

This fast and slow processing is remarkable in that it emerges at such a low level in auditory processing (in the cochlear nuclei). Such ideas have been popularized in various guises of brain processing, but never at such a basic processing level. For example, according to \citet{Stanovich2000}, reasoning may be attributed either to ``\term{System 1}'', which is inherently fast and requires heuristics that can lead to wrong conclusions, or to ``\term{System 2}'', which is slow and effortful, but is more logical and potentially precise. In another influential brain theory, \citet{McGilchrist2009} associated processing of details with the left hemisphere and processing of the whole with the right hemisphere, so the two operate on two wildly different time scales.

If the dual auditory processing hypothesis will be confirmed, then it may have the potential to show that the distinction between fine-grained and coarse-grained processing is a fundamental processing feature of our brain. It may even work as a recursive principle that can be repeatedly applied at different levels of processing.

\subsection{The auditory literature}
A large portion of this work involved the harvest of existing auditory research literature for available data. Having reviewed what feels like innumerable papers for the sake of this research, I have no doubt that there is much relevant data that wait to be utilized, which can potentially save resources for truly novel research. Many of these publications have shown an immense level of accomplishment and creativity, which I can only be grateful for that I could access. By embracing more literature from the past, I believe that it should be possible to unify additional concepts that appear disparate today. This process may also be used to rule out long-standing hypotheses that no longer have merit. The alternative may entail endless divergence with growing challenge to close off loose ends and settle open questions that may otherwise be left unanswered. 

%

\begin{appendices}
\include{coherence_app}
\include{Waves}
\include{LCT_appendix}
\include{rectimpulse}
\include{Aliasing}
\include{psychoacoustic_params}

\end{appendices}
\clearpage
\cleardoublepage
\phantomsection
\addcontentsline{toc}{chapter}{Bibliography}
\bibliography{Hearing_references}
\end{document}

%% file: lists.tex
\chapter*{List of acronyms}
\vspace{-1em} 
\begin{longtable}{ll}
\textbf{A1} & Primary auditory cortex\\
\textbf{A2} & Secondary auditory cortex\\
\textbf{A3} & Tertiary auditory cortex\\
\textbf{ABR} & Auditory brainstem response\\
\textbf{AC} & Alternating current\\
\textbf{AM} & Amplitude modulation\\
\textbf{ATF} & Amplitude transfer function\\
\textbf{AVCN} & Anteroventral cochlear nucleus\\
\textbf{BM} & Basilar membrane\\
\textbf{CCD} & Charge-coupled device\\
\textbf{CF} & Characteristic frequency\\
\textbf{CN} & Cochlear nucleus\\
\textbf{CMR} & Comodulation masking release\\
\textbf{cPSF} & coherent Point spread function\\
\textbf{CSF} & Contrast sensitivity function\\
\textbf{DC} & Direct current\\
\textbf{DCN} & Dorsal cochlear nucleus\\
\textbf{DNLL} & Dorsal nucleus of the lateral lemniscus\\
\textbf{DPOAE} & Distortion-product otoacoustic emission\\
\textbf{EEOAE} & Electrically evoked otoacoustic emission\\
\textbf{EOAE} & Evoked otoacoustic emission\\
\textbf{ERB} & Equivalent rectangular bandwidth\\
\textbf{FEM} & Finite-element method\\
\textbf{FFR} & Frequency-following response\\
\textbf{FM} & Frequency modulation\\
\textbf{fMRI} & functional Magnetic resonance imaging\\
\textbf{FWHM} & Full-width half maximum\\
\textbf{GDD} & Group-delay dispersion\\
\textbf{GVD} & Group-velocity dispersion\\
\textbf{IACC} & Interaural acoustic cross-correlation\\
\textbf{IC} & Inferior colliculus\\
\textbf{ICC} & Central nucleus of the inferior colliculus\\
\textbf{ICD} & Dorsal cortex of the inferior colliculus\\
\textbf{ICX} & External nucleus of the inferior colliculus\\
\textbf{IHC} & Inner hair cell\\
\textbf{INLL} & Intermediate nucleus of the lateral lemniscus\\
\textbf{iPSF} & incoherent Point spread function\\
\textbf{LCT} & Linear canonical transform\\
\textbf{LED} & Light-emitting diode\\
\textbf{LGB} & Lateral geniculate body\\
\textbf{LL} & Lateral lemniscus\\
\textbf{LOC} & Lateral olivocochlear\\
\textbf{LSO} & Lateral superior olive\\
\textbf{MDI} & Modulation discrimination interference\\
\textbf{MEG} & Magnetoencephalography\\
\textbf{MET} & Mechanoelectrical transduction\\
\textbf{MGB} & Medial geniculate body\\
\textbf{MLR} & Middle latency response\\
\textbf{MNTB} & Medial nucleus of the trapezoid body\\
\textbf{MOC} & Medial olivocochlear\\
\textbf{MOCR} & Medial olivocochlear reflex\\
\textbf{MSO} & Medial superior olive\\
\textbf{MTF} & Modulation transfer function\\
\textbf{NLL} & Nuclei of the lateral lemniscus\\
\textbf{NPLL} & Neural phase-locked loop\\
\textbf{OAE} & Otoacoustic emission\\
\textbf{OHC} & Outer hair cell\\
\textbf{OTF} & Optical transfer function\\
\textbf{PLL} & Phase-locked loop\\
\textbf{PM-HLL} & Period-modulated harmonic locked loop \\
\textbf{PSF} & Point spread function\\
\textbf{PTF} & Phase transfer function\\
\textbf{PVCN} & Posteroventral cochlear nucleus\\
\textbf{QM} & Quadrature modulation\\
\textbf{RC} & Resistance capacitance\\
\textbf{RMS} & Root mean square\\
\textbf{SC} & Superior colliculus\\
\textbf{SFM} & Sinusoidal frequency modulation\\
\textbf{SFOAE} & Stimulus-frequency otoacoustic emissions\\
\textbf{SHC} & Short hair cell\\
\textbf{SNR} & Signal-to-noise ratio\\
\textbf{SOAE} & Spontaneous otoacoustic emission\\
\textbf{SOC} & Superior olivary complex\\
\textbf{SON} & Superior olivary nucleus\\
\textbf{SPL} & Sound pressure level\\
\textbf{SPN} & Superior paraolivary nucleus \\
\textbf{TOAE} & Transient otoacoustic emission\\
\textbf{TFS} & Temporal fine structure\\
\textbf{THC} & Tall hair cell\\
\textbf{TM} & Tectorial membrane\\
\textbf{TMTF} & Temporal modulation transfer function\\
\textbf{UWB} & Ultra-wideband\\
\textbf{V1} & Primary visual cortex\\
\textbf{VCN} & Ventral cochlear nucleus\\
\textbf{VCO} & Voltage controlled oscillator\\
\textbf{VNLL} & Ventral nucleus of the lateral lemniscus\\
\textbf{VNTB} & Ventral nucleus of the trapezoid body \\
\end{longtable}

\chapter*{List of symbols}
\begin{longtable}{ll}
$a,b,c$ & Constants in polynomial (\cref{PsychoEstimation}) \\
$a,b,c,d,i$ & Coefficients of power law of ABR or OAE (\cref{NeuralDispEst}, \cref{TotalImpairedChange}) \\
$a,b,d$ & Linear canonical transform (LCT) coefficients (\cref{AppLCT}) \\
$a$ & Complex temporal envelope \\
$a$ & Arbitrary signal (\cref{TemporalSampling}) \\
$A$ & Complex envelope spectrum \\
$a_g $ & Ideal geometrical image of envelope (\cref{ImpulseResponseDer}) \\
$a_n$ & Complex temporal envelope in neural domain (\cref{ImagingDerivation}) \\
$a_s$ & Sampled signal (\cref{TemporalSampling}) \\
$A_s$ & Sampled spectrum (\cref{TemporalSampling}) \\
$b_n$ & Random eigenfunction coefficient (\cref{SpectralCoherence}) \\
$B$ & Bandwidth \\
$B_n$ & Noise-equivalent bandwidth (\cref{LinearizedPLL}) \\
$c$ & Wave speed; Speed of light; Speed of sound \\
$C$ & Circle of confusion (\cref{ChapterImaging}) \\
$C$ & Channel capacity (\cref{InfoNutshell}) \\
$C$ & Schroeder phase curvature (\cref{ModelSchr}, \cref{TotalImpairedChange}) \\
$C$ & Real part of Fresnel integral (\cref{RectImp}) \\
$C_T$ & Linear canonical transform (LCT) kernel for operator $T$ (\cref{AppLCT}) \\
$d$ & Temporal acuity; Detectable gap (\cref{EmpiricalBB}, \cref{GapDetect}, \cref{PsychoEstimation}) \\
$d$ & Group-delay dispersion impulse response (\cref{ImagingEqs}, \cref{AudImpulseRes}) \\
$D$ & Aperture size (\cref{ChapterImaging}) \\
$D$ & Group-delay dispersion transfer function (\cref{ImagingEqs}, \cref{AudImpulseRes}) \\
$D$ & Pulse sequence duration (\cref{Aliasing}) \\
$D'$ & Pulse sequence duration minus the last pulse duration (\cref{Aliasing}) \\
$D_a$ & Acoustic dipole strength (Table \ref{soundvslight}) \\
$D_e$ & Electromagnetic dipole strength (Table \ref{soundvslight}) \\
$\bm{E}$ & Electric field (vector) (Table \ref{soundvslight}) \\
$E$ & Electric field (scalar) (\cref{ChapterImaging}) \\
$ERB$ & Equivalent rectangular bandwidth of auditory filter \\
$f$ & Frequency \\
$f$ & Focal length (\cref{ChapterImaging})\\
$f$ & Arbitrary function (\cref{LinAnaDisp})\\
$F$ & Fourier transform of arbitrary function $f$ (\cref{LinAnaDisp})\\
$\bm{F}$ & Arbitrary wave field (Table \ref{soundvslight})\\
$f'$ & Doppler-shifted frequency (\cref{ComplexSourceMod}) \\
$f'$ & Remapped imaged frequency under binaural diplacusis (\cref{OHCimpair}) \\
$f^\#$ & F-number (\cref{GeometricalOptics}) \\
$f_0$ & Fundamental frequency \\
$f_c$ & Carrier frequency \\
$f_m$ & Modulation frequency \\
$f_n$ & Cutoff frequency (Figure \ref{PLLdemos}) \\
$f_T$ & Focal time \\
$f_{T,gg}$ & Focal time estimate of the gerbil and guinea pig time lens\\
$f_{T,h}$ & Focal time estimate of the human time lens\\
$f_s$ & Sampling rate \\
$f_T^\#$ & Temporal imaging f-number (\cref{AudFNum}) \\
${\cal F}$ & Fourier transform \\
${\cal F}^{-1}$ & Inverse Fourier transform \\
$g$ & Arbitrary function (\cref{LinAnaDisp}) \\
$g$ & Complex envelope / Mapping function (\cref{CommunicationTheory}) \\
$g_1,g_2$ & Fresnel integral variables (\cref{RectImp}) \\
$g_T$ & Linear canonical transformed (LCT) function (\cref{AppLCT}) \\
$h$ & Impulse response function \\
$\tilde h $ & Impulse response function with reduced coordinate; Point spread function (\cref{ImpulseResponseDer}) \\
$\tilde h_d$ & Defocused impulse response function for Gaussian pupil (\cref{ImagingNotSatisfied}) \\
$h_{dr}$ & Defocused impulse response function for rectangular pupil (\cref{RectImp}) \\
$h_L$ & Time-lens impulse response function (\cref{TheTimeLens}, \cref{AudImpulseRes}) \\
$\bm{H}$ & Magnetic field (vector) (Table \ref{soundvslight}) \\
$H$ & (Shannon's) entropy (\cref{InfoNutshell}) \\
$H$ & Transfer function \\
$H_d$ & Amplitude transfer function (ATF) of generalized Gaussian pupil (\cref{ImpFun}) \\
$H_{dr}$ & Amplitude transfer function (ATF) of generalized rectangular pupil (\cref{ImpFun}) \\
$H_L$ & Time lens transfer function (\cref{TheTimeLens}) \\
${\cal H}$ & Hilbert transform (\cref{PhysicalSignals}) \\
${\cal H}^{-1}$ & Inverse Hilbert transform (\cref{PhysicalSignals}) \\
${\cal H}$ & Optical transfer function (\cref{ImpFun}) \\
${\cal H}_d$ & Defocused optical transfer function with Gaussian pupil (\cref{ImpFun}) \\
${\cal H}_{dr}$ & Defocused optical transfer function with rectangular pupil (\cref{ImpFun}) \\
$i$ & $\sqrt{-1}$ \\ 
$I$ & Intensity \\ 
$I$ & Frequency interval (\cref{TransChromAb}) \\ 
$I_{coh}$ & Coherent part of intensity (\cref{BasicCoherenceDeriv}) \\ 
$I_{inc}$ & Incoherent part of intensity (\cref{BasicCoherenceDeriv}) \\ 
$I_{max}$ & Maximum intensity (interference) \\ 
$I_{min}$ & Maximum intensity (interference) \\ 
$I_{tot}$ & Total intensity (\cref{BasicCoherenceDeriv}) \\ 
$I_0$ & Object intensity (\cref{ChapterImaging})\\ 
$I_1$ & Image intensity (\cref{ChapterImaging})\\ 
$I_r$ & Intensity (radial) (Table \ref{soundvslight}) \\ 
$I_\theta$ & Intensity (azimuth) (Table \ref{soundvslight}) \\ 
$I_\phi$ & Intensity (elevation) (Table \ref{soundvslight}) \\ 
$J_n$ & Bessel function of the first kind of order $n$\\
$k$ & Wavenumber; Spatial frequency \\
$k_x$ & Horizontal component of spatial frequency (Table \ref{tab:acousticvision}) \\
$k_y$ & Vertical component of spatial frequency (Table \ref{tab:acousticvision}) \\
$\bm{k}$ & Wavenumber vector\\
$k_c$ & Wavenumber at carrier frequency\\
$k_i$ & Imaginary-part of wavenumber function (absorption) (\cref{airtravel}) \\
$k_r$ & Real-part of wavenumber function (dispersion) (\cref{airtravel}) \\
$K$ & Gain (\cref{CommunicationTheory}) \\
$K$ & Stiffness (\cref{StiffnessPM}) \\
$K$ & Loop gain (\cref{PLLs}) \\
$K_{BM}$ & Passive basilar membrane stiffness (\cref{StiffnessPM})\\
$K_f$ & Filter DC gain (\cref{PLLs}) \\
$K_m$ & Phase detector sensitivity (\cref{PLLs}) \\
$K_v$ & Voltage controller oscillator gain (\cref{PLLs}) \\
$l$ & Integer \\
${\mathscr L}(T) $ & Linear canonical transform (LCT) for kernel operator $T$ (\cref{AppLCT}) \\
$m$ & Integer \\
$m$ & Message (baseband) function (\cref{CommunicationTheory}) \\
$m$ & Modulation depth (audibility, contrast) \\
$m$ & Frequency slope (chirpiness) (\cref{PulseCalc}) \\
$M$ & Magnification \\
$M'$ & Complex magnification (defocused system) (\cref{GaussImaging}) \\
$M^*, M_f, M_{f'}, M_{f'}^*$ & Distorted magnification under binaural diplacusis (\cref{OHCimpair})\\
$m_0$ & Chirp slope of object (\cref{ModelSchr})  \\
$M_0$ & Magnification $-v/u$ \\
$m_1$ & Chirp slope of image (\cref{ModelSchr}) \\
$m_r$ & Rectangular pulse chirp slope (\cref{PulseCalc}) \\
$n$ & Index of refraction \\
$n$ & Noise time-signal (\cref{CommunicationTheory}) \\
$n$ & Harmonic number \\
$n$ & Integer \\
$n_k$ & Neural group velocity dispersion coefficient (\cref{NeuralDispEst}) \\
$N$ & Noise power (\cref{InfoNutshell}) \\
$N$ & Number of samples \\
$N$ & Schroeder phase number of harmonics (\cref{ModelSchr}) \\
$N$ & Number of pulses in sequence (\cref{Aliasing}) \\
$NS$ & Nonstationarity index (\cref{NonstationarityMeasurement}) \\
$p$ & Pressure \\
$p$ & Probability (\cref{InfoNutshell}) \\
$P$ & Pressure (frequency domain) \\
$P$ & Pupil function \\
${\cal P}$ & Cauchy principal value (\cref{PhysicalSignals}) \\
${\cal P}$ & Generalized pupil function (\cref{ImagingNotSatisfied}) \\
$P_e$ & Entrance pupil (\cref{NaturalSamp}) \\
$P_g$ & Gaussian pupil function (\cref{ImagingNotSatisfied}) \\
$P_r$ & Rectangular pupil function (\cref{RectPupil}, \cref{RectImp}) \\
$q,q'$ & Linear canonical transform (LCT) variables (\cref{AppLCT}) \\
$Q$ & Q-factor of bandpass filter \\
$Q_{10}$ & Q-factor of bandpass filter at -10 dB points  (\cref{TimeLensExtrapolation}) \\
$Q_{ERB}$ & Q-factor of bandpass filter at the ERB bandwidth (\cref{TimeLensExtrapolation}) \\
$r$ & Displacement \\
$r$ & Received signal time-signal (\cref{CommunicationTheory}) \\
$r$ & Correlation coefficient (Pearson's r) \\
$\bm{r}$ & Position vector \\
$r_c$ & Critical distance (\cref{CoherenceReverb}) \\
$R$ & Fixed distance (\cref{InterauralCoherence}) \\
$R$ & Synchronization strength/index (\cref{SyncIndex}) \\ 
$R_{pp}$ & Autocorrelation function of $p$ (\cref{TempSpatCoh}) \\
$s$ & Time lens curvature \\
$s$ & Time signal \\
$s$ & Laplace-transform complex frequency (\cref{LinearizedPLL}) \\
$s$ & Sampler delta function array (comb function) (\cref{TemporalSampling}) \\
$\hat{s}$ & Estimate of time signal (\cref{ComplexSourceMod}) \\
$s_{T,gg}$ & Curvature estimate of the gerbil and guinea pig time lens\\
$\bm{S}$ & Poynting vector  (Table \ref{soundvslight}) \\
$S, S_I$ & Intensity/power spectrum \\
$S$ & Signal power (\cref{InfoNutshell}) \\
$S$ & Power spectral density / Spectrum (\cref{CoherenceTheory}) \\
$\mathit{S}$ & Surface area of room (\cref{CoherenceReverb}) \\
$S$ & Imaginary part of Fresnel integral (\cref{RectImp}) \\
$S_a$ & Amplitude spectrum (\cref{PowerMods}) \\
$s_h$ & Time-lens curvature in humans (\cref{TimeLensExtrapolation}) \\
$S_\omega$ & Acoustic point source strength (Table \ref{soundvslight}) \\
$t$ & Time \\
$t_0$ & Gaussian pulse width parameter of the object; Entrance pupil \\
$t_1$ & Gaussian pulse width parameter of the image; Exit pupil \\
$t'$ & Complex Gaussian pulse width parameter \\
$T$ & Period \\
$T$ & Integration time constant \\
$T$ & Complex tone standard (\cref{AudFNum}) \\
$T$ & Linear canonical transform (LCT) matrix operator (\cref{AppLCT}) \\
$\tilde T$ & Integration variable (\cref{ImpulseResponseDer}) \\
$T_{60}$ & Reverberation time \\
$T_a$ & Aperture time \\
$T_c$ & Carrier period \\
$T_s$ & Sampling rate period \\
$v$ & Velocity (\cref{ComplexSourceMod}) \\
$v$ & Distance between lens and screen (\cref{ChapterImaging})\\
$v$ & Neural group-delay dispersion \\
$V$ & Visibility (\cref{OpticalCoh}, \cref{CoherenceTheory}) \\
$V$ & Volume (\cref{CoherenceReverb}) \\
$V$ & Voltage (\cref{StiffnessPM}) \\
$v_g$ & Group velocity \\
$V_c$ & Output voltage from loop filter (\cref{LinearizedPLL}) \\
$V_d$ & Output voltage from phase detector (\cref{LinearizedPLL}) \\
$V_i$ & Input voltage to phase detector (\cref{LinearizedPLL}) \\
$V_o$ & Output voltage from voltage controller oscillator (\cref{LinearizedPLL}) \\
$v_p$ & Phase velocity \\
$v_{V-I}$ & Neural group-delay dispersion for the segment between wave I and wave V (\cref{NeuralDispEst}) \\
$u$ & Acoustic velocity (Table \ref{soundvslight})\\
$u$ & Distance between object and lens (\cref{ChapterImaging})\\
$u$ & Auditory input (cochlear) group-delay dispersion \\
$\bm{u}$ & Sound velocity (Table \ref{soundvslight}) \\
$u_e$ & External environment group-delay dispersion (\cref{paramestimate}) \\
$u_o$ & Outer-ear group-delay dispersion (\cref{paramestimate}) \\
$u_m$ & Middle-ear group-delay dispersion (\cref{paramestimate}) \\
$u_c$ & Cochlear group-delay dispersion (\cref{paramestimate}) \\
$w$ & Wave energy density (Table \ref{soundvslight}) \\
$W$ & Cross-spectral density (\cref{CoherenceTheory}) \\
$W_d$ & Reciprocal defocus parameter $=1/u + 1/v + 1/s$ (\cref{ImpFun}, \cref{PsychoEstimation}) \\
$W_p$ & Single pulse duration (\cref{Aliasing}) \\
$x$ & x-axis coordinate \\
$x$ & In-phase modulation (\cref{CommunicationTheory}) \\
$x$ & Real part of analytic signal; arbitrary time signal (\cref{PhysicalSignals}) \\
$x$ & Defocus parameter $=u+sv/(s+v)$ (\cref{ModelSchr}, \cref{TotalImpairedChange}) \\
$X$ & Fourier transform of real part of analytic signal $X$ (\cref{PhysicalSignals}) \\
$x_0$ & Arbitrary spatial position (\cref{DispersiveAn}) \\
$y$ & y-axis coordinate \\
$y$ & Quadrature modulation (\cref{CommunicationTheory}) \\
$x$ & Imaginary part of analytic signal (\cref{PhysicalSignals}) \\
$z$ & Distance along the z-axis (the optical axis) \\
$z$ & Analytic signal (\cref{PhysicalSignals}) \\
$X$ & Fourier transform of analytic signal $Z$ (\cref{PhysicalSignals}) \\
$Z_0$ & Characteristic impedance (Table \ref{soundvslight}) \\
&\\
$\alpha,\beta,\gamma$ & Direction cosines of wavenumber vector (\cref{DiffractionFourier}) \\
$\alpha$ & Instantaneous phase of complex degree of coherence $\gamma$ (\cref{CoherenceTheory}) \\
$\alpha$ & Total absorption ($=k_i$) (\cref{temporaltheory}, \cref{AppWaves}) \\
$\alpha$ & Gaussian shape factor (\cref{AperturePsych}) \\
$\alpha'$ & Phase-velocity absorption (\cref{AppWaves}) \\
$\alpha''$ & Group-velocity absorption (\cref{temporaltheory}, \cref{AppWaves}) \\
$\mathit{\alpha}$ & Absorption coefficient (\cref{CoherenceReverb}) \\
$\beta$ & Total dispersion ($=k_r$) (\cref{temporaltheory}, \cref{AppWaves}) \\
$\beta'$ & Phase-velocity absorption  (\cref{temporaltheory}, \cref{AppWaves}) \\
$\beta''$ & Group-velocity dispersion (\cref{temporaltheory}, \cref{paramestimate}, \cref{ImagingEqs}, \cref{AppWaves})\\
$\gamma$ & Frequency deviation (\cref{FreqInstFreq}) \\
$\gamma$ & Mutual coherence function (\cref{CoherenceTheory}) \\
$\gamma_{ii} $ & Self-coherence function of input signal (\cref{PLLCoherence}) \\
$\gamma_{oo} $ & Self-coherence function of output signal (\cref{PLLCoherence}) \\
$\gamma_{io} $ & Coherence function between input and output signals (\cref{PLLCoherence}) \\
$\Gamma$ & Complex degree of coherence (\cref{CoherenceTheory}, \cref{PLLCoherence}) \\
$\delta$ & Dirac delta function \\
$\delta$ & Linear phase of coherence function (\cref{CoherenceTheory}) \\
$\delta_{nm}$ & Kronecker delta \\
$\Delta f$ & Spectral bandwidth of narrowband signal \\
$\Delta f, \Delta f_{beat}$ & Frequency spacing between beating tones (\cref{PsychoEstimation}) \\
$\Delta f, \Delta f_{oct}$ & Octave stretch in cents (\cref{TransChromAb}, \cref{PsychoEstimation}) \\ 
$\Delta f'$ & Iterated octave stretch in cents (\cref{TransChromAb}) \\ 
$\Delta f_h$ & Bandwidth of the time-lens phase modulation in humans  (\cref{TimeLensExtrapolation}) \\
$\Delta k$ & Wavenumber difference between beating components (\cref{LinAnaDisp})\\
$\Delta l$ & Coherence length (\cref{CoherenceTimeLength}) \\
$\Delta n$ & Path change in index of refraction  (\cref{StiffnessPM}) \\
$\Delta t$ & Deviation from complex tone standard (\cref{AudFNum}) \\
$\Delta t_{opt}$ & Optimal ratio between geometrical and dispersive blurs (Table \ref{t0opts}, \cref{ImagingNotSatisfied}) \\
$\Delta \tau$ & Coherence time \\
$\Delta \omega$ & Frequency difference between beating components (\cref{LinAnaDisp})\\
$\Delta \omega$ & Bandwidth of narrowband signal \\
$\Delta \omega$ & Phase ramp (\cref{NarrowbandApp}) \\
$\Delta \omega$ & Phase ramp (\cref{NarrowbandApp}) \\
$\dot{\omega}$ & Frequency velocity (\cref{NarrowbandApp}) \\
$\ddot{\omega}$ & Frequency acceleration (\cref{NarrowbandApp}) \\
$\epsilon$ & Dielectric constant (permittivity) (Table \ref{soundvslight}) \\
$\epsilon_0$ & Dielectric constant in vacuum (permittivity) (Table \ref{soundvslight}) \\
$\zeta$ & z-axis coordinate in the traveling pulse coordinate system (\cref{temporaltheory},\cref{paramestimate}, \cref{ImagingDerivation}) \\
$\zeta$ & Damping factor (Figure \ref{PLLdemos}) \\
$\eta$ & Displacement function (\cref{PrimitiveSources}) \\
$\eta$ & Mask coordinate (\cref{DiffractionFourier}) \\
$\theta$ & Azimuth (Table \ref{soundvslight}) \\
$\theta$ & Angle from optical axis (\cref{ChapterImaging}) \\
$\theta_{min}$ & Angular resolution (\cref{DiffractionFourier}) \\
$\kappa$ & Adiabatic compressibility (Table \ref{soundvslight}) \\
$\lambda$ & Wavelength \\
$\Lambda$ & Triangle function \\
$\lambda_n$ & Eigenvalue (\cref{SpectralCoherence}) \\
$\mu$ & Magnetic permeability (Table \ref{soundvslight}) \\
$\mu$ & Spectral degree of coherence (\cref{CoherenceTheory}) \\
$\mu_0$ & Magnetic permeability in vacuum (Table \ref{soundvslight}) \\
$\xi$ & Mask coordinate (\cref{DiffractionFourier}) \\
$\rho$ & Fluid density (Table \ref{soundvslight}) \\
$\rho_{BM}$ & Basilar membrane density per unit length (\cref{StiffnessPM}) \\
$\sigma$ & Multiplicative factor (\cref{InitialCurveEstimates}, \cref{PsychoEstimation}) \\
$\tau$ & Time lag \\
$\tau$ & Time coordinate in the traveling pulse coordinate system, group delay \\
$\tilde\tau_0 $ & Reduced coordinate of the object time (\cref{ImpulseResponseDer}) \\
$\tau_0$ & Initial time of pulse in the traveling pulse coordinate system \\
$\tau_1,\tau_2$ & Time constants (Figure \ref{PLLdemos}) \\
$\tau_e$ & Effective duration (\cref{CoherentSources}, \cref{ExCohere}) \\
$\tau_g$ & Group delay \\
$\tau_p$ & Phase delay \\
$\varphi$ & Phase function (time domain) \\
$\phi$ & Phase function (frequency domain) \\
$\phi$ & Elevation (Table \ref{soundvslight}) \\
$\phi_e$ & Phase error (\cref{PLLs}) \\
$\phi_n$ & Schroeder phase for component $n$ (\cref{ModelSchr}) \\
$\phi_{\Delta f/2}$ & Time-lens phase at the cutoff frequencies of $Q_{10}$ (\cref{TimeLensExtrapolation}) \\
$\psi$ & Field function (\cref{LinAnaDisp})\\
$\Psi$ & Eigenfunction  (\cref{PrimitiveSources}, \cref{SpectralCoherence}) \\
$\omega$ & Angular frequency; Modulation frequency \\
$\bar{\omega}$ & First moment of frequency (\cref{CoherenceTheory}) \\
$\Omega$ & Integration domain including radiating source and receivers  (\cref{SpectralCoherence}) \\
$\omega_c$ & Carrier angular frequency \\
$\omega_{coh}$ & Coherent amplitude transfer function cutoff frequency (Gaussian pupil) (\cref{ImpFun}) \\
$\omega_{coh,r}$ & Coherent amplitude transfer function cutoff frequency (rectangular pupil) (\cref{ImpFun}) \\
$\omega_H$ & Hold-in range (\cref{PLLs}) \\
$\omega_{inc}$ & Incoherent modulation transfer function cutoff frequency (Gaussian pupil) (\cref{ImpFun}) \\
$\omega_L$ & Lock range (\cref{PLLs}) \\
$\omega_m$ & Modulation angular frequency \\
$\omega_O$ & Pull-out range (\cref{PLLs}) \\
$\omega_P$ & Pull-in range (\cref{PLLs}) \\

\end{longtable}

\chapter*{Mathematical conventions}
\label{MathConventions}
In this work, a forward-propagating wave is implicitly taken to be the real part of a signal of the form:
\begin{equation}
	p(x,t) = e^{i(\omega t - kx)}
\end{equation}
This matches the convention that is typically used in signal processing, where the argument that contains the angular frequency $\omega$ is positive. It is also the convention in most of the optics and communication texts cited here, as well as several others that include, notably, \citet{Haus,Siegman,Kolner, Cohen1995, FletcherRossing, Kinsler, Zverev, New2011, Couch} and \citet{Kuttruff}. Therefore, formulas that were taken from \citet{Morse,Jackson,Whitham,Born} and \citet{Goodman} sometimes had to be adopted by changing the sign as they originally appeared in the form $p(x,t) = e^{i(kx- \omega t)}$. 

~\\

The Fourier transform that is used in the text conforms to
\begin{equation}
	F(\omega) = \int_{-\infty}^\infty f(t) e^{-i\omega t} dt
\end{equation}
whereas the inverse Fourier transform is
\begin{equation}
	f(t) = \frac{1}{2\pi}\int_{-\infty}^\infty F(\omega) e^{i\omega t} d\omega
\end{equation}
Thus, throughout this work, lowercase signal functions are time domain and their spectral domain counterparts are capitalized.

%% file: coherence_app.tex
\chapter{Examples of realistic coherence functions of acoustic sources}
\label{ExCohere}

\section{Introduction}
This appendix supplements some of the statements made throughout the review of coherence theory that are relevant for acoustics and hearing (\cref{CoherenceTheory})---primarily in the sections about realistic acoustic sources (\cref{CoherentSources}) and room acoustics (\cref{CoherenceReverb}). The overarching goal of the examples in this appendix is to demonstrate that partial coherence is prevalent in relatively ordinary conditions---a claim that is repeatedly made throughout this work. As information supporting this claim could not be conveniently gathered from available literature, it was deemed instructive to generate the required data using available recordings. The interpretation of some of the effects is not always straightforward, especially since several critical factors go into every analysis---frequency-dependent coherence, nonstationarity of the sources, reverberation time, integration time of the coherence function, the analysis filter, and more. The challenge in interpretation is exacerbated because the relative weighting and exact interplay between these factors are unknown with respect to hearing. Nevertheless, it is the hope here that the figures below will serve the immediate purpose of showing how real-world acoustic fields are not black or white---they are neither coherent nor incoherent, but are generally a mixture of both: they are partially coherent.

The basic relation used throughout this work to obtain the coherence function is the correlation function:
\begin{equation}
	\Gamma(\bm{r_1},\bm{r_2},\tau) = \frac{1}{2T} \int_{-T}^T p^*(\bm{r_1},t)p(\bm{r_2},t+\tau)dt
	\label{AutoCorrRep2}
\end{equation}
which is integrated over a finite duration $2T$ that in the limit of  $T \rightarrow \infty$ converges to the ensemble average value, if stationarity is satisfied. When $\bm{r_1}=\bm{r_2}=\bm{r}$, $\Gamma(\bm{r,r},\tau)$ is referred to as the self-coherence function, and then Eq. \ref{AutoCorrRep2} assumes the form of a standard autocorrelation function. Eq. \ref{AutoCorrRep2} is normalized by the root mean square of both signals according to:
\begin{equation}
	\gamma(\bm{r_1},\bm{r_2},\tau) = \frac{\Gamma(\bm{r_1},\bm{r_2},\tau)}{\sqrt{\Gamma(\bm{r_1},\bm{r_1},0)}\sqrt{\Gamma(\bm{r_2},\bm{r_2},0)}}
	\label{Degreeofcoherence2}
\end{equation}
to obtain the complex degree of coherence, $\gamma(\bm{r_1},\bm{r_2},\tau)$.

In most of the following analyses, we are interested in the transition between coherence ($\gamma = 1$) and incoherence ($\gamma = 0$). In the room acoustic and psychoacoustic literature, it was indirectly assessed using the effective duration, which is defined as the duration for a 10 dB drop from the peak of the autocorrelation function, which is always coherent at $\tau = 0$ and drops for larger lags. This measure has been shown to correlate with various subjective measures, but the choice of 10 dB does not seem to have been justified in literature and appears to be somewhat arbitrary. In contrast, the coherence time as is defined in coherence theory (\cref{CoherenceTimeLength}) is computed at the half power point (-3 dB). This duration also corresponds to the point at which the coherent and incoherent components in the total (partial) coherence function are exactly equal, according to Eq. \ref{cohtoincohratio}: $I_{coh/}I_{incoh} = |\gamma(\bm{r_1},\bm{r_1},\tau)/|(1-|\gamma(\bm{r_1},\bm{r_1},\tau)|)$. Therefore, we shall make a distinction between the coherence time (-3 dB) and the effective duration (-10 dB). As a working hypothesis, we will assume that complete (subjective) coherence is obtained for delays shorter than the coherence time $\tau<\Delta \tau$ and subjective complete incoherence is obtained for delays longer than the effective duration $\tau > \tau_e$. The broad margin in between corresponds to subjective partially coherent sound. The exact cutoff in auditory relevant measures is, of course, unknown at present. 

A notable and known pattern that repeats in the following measurements is that the coherence time decreases with frequency. This is a corollary of the 1/f-like distribution of the power spectrum of different sources of physical noise, which also characterizes speech and music loudness and pitch fluctuations \citep{VanDerZiel1950,Voss1975}.

Several sound examples from the recordings that are analyzed throughout are found in the audio demo directory \textsc{/Appendix A - Coherence/}.

\section{White noise and the cohering effect of two filter types}
\label{WhiteCoherence}
Although the autocorrelation function is routinely applied in acoustics to broadband sources, the physical coherence function that carries information about interference between sources is well defined only for narrowband signals. In order to obtain a narrowband approximation that is required to get meaningful results out of the temporal-spatial coherence functions, it is necessary to bandpass-filter the various broadband acoustic signals. However, the choice of filter can significantly influence the output coherence function and is therefore not completely arbitrary (\cref{CoherentFiltering}). To illustrate this effect, we shall compare the white-noise coherence function for different Butterworth and gammachirp filters. The Butterworth filters are designed to maximize the flatness of the passband magnitude, but are characterized by relatively slow rise time and frequency-dependent group delay in the flanks \citep[109--117]{Zverev}. The gammachirp filters are the standard model for the auditory filters \citep{Patterson1987,Irino1997}. They are faster than the Butterworth filters and account for the asymmetry of the passband. However, they are modeled using stationary signals (pure tone and broadband maskers) and probably have an incorrect phase response \citep{Lentz2001}  that may not square well with some transient nonlinear characteristics. The passband in both cases is set according to the equivalent rectangular bandwidth of the auditory filters \citep{Glasberg1990}:
\begin{equation}
	ERB = 0.108f + 24.7 \,\,\, \Hz
	\label{ERBrep}
\end{equation}
for frequency $f$ in Hz. 

\begin{figure} 
		\centering
		\includegraphics[width=1\linewidth]{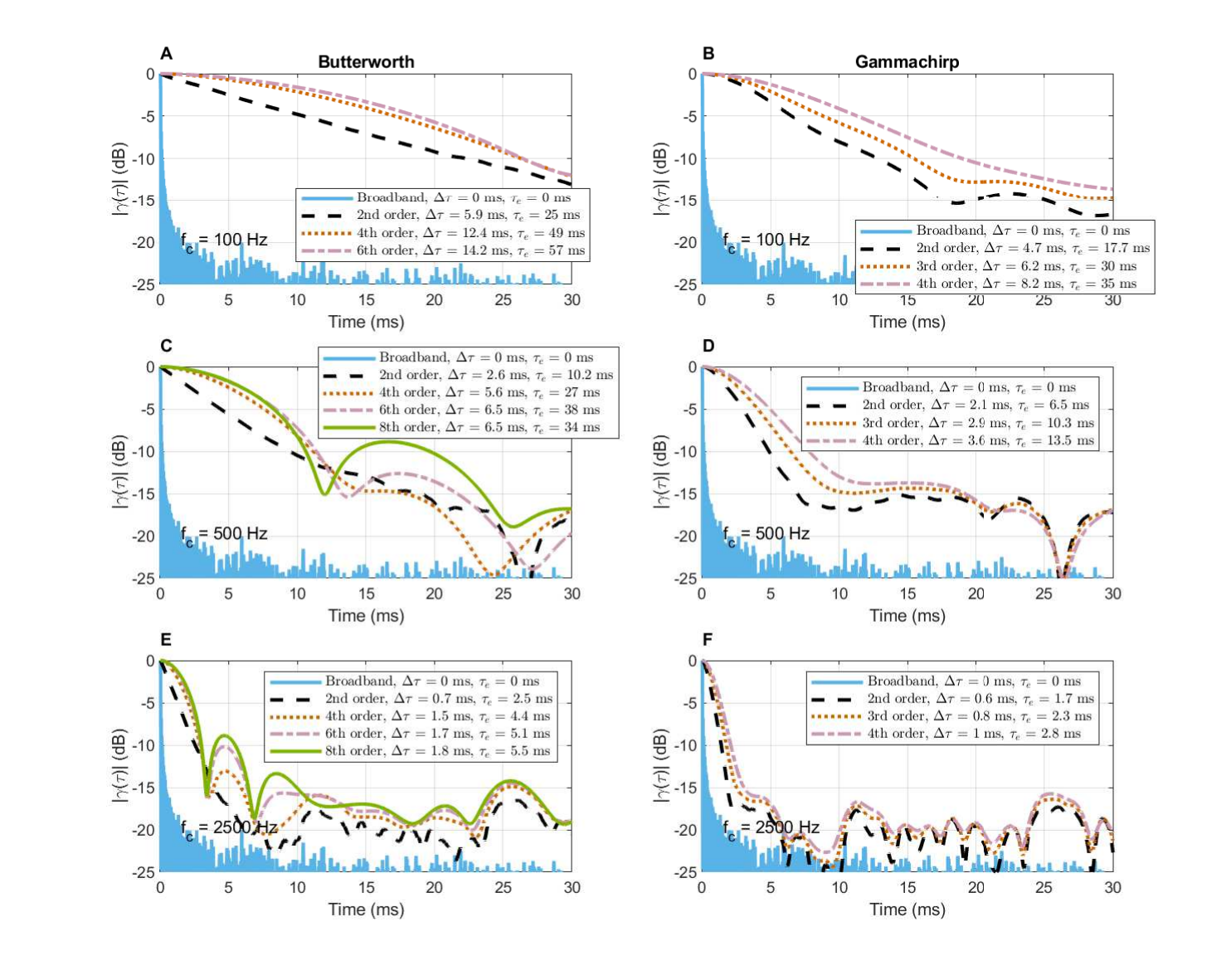}	
		\caption{Demonstration of the cohering effect of two different types of filters with bandwidths set according to the ERB in Eq.\ref{ERBrep}, in four orders each. On the right are band-pass Butterworth filters of 2nd, 4th, 6th, and 8th order (standard bandpass orders are always even). On the left are Gammachirp filters of 2nd, 3rd, and 4th orders. White Gaussian noise of 5 s was used as signal and the integration time was 100 ms, in all cases.} 
		\label{filtertype}
\end{figure}

White noise, in theory, has a zero coherence time in the limit of infinite bandwidth of the noise. However, as was proven in \cref{CoherentFiltering} and was originally shown in \citet{Jacobsen1987}, the choice of filter bandwidth affects the measurement, and hence, the apparent coherence of the source. This is illustrated in Figure \ref{filtertype}. The pre-filtered autocorrelation of a 5 s pseudo-random white noise is shown in all plots as the blue trace. The negative half plane of the autocorrelation is omitted due to its symmetry. The autocorrelation of Eq. \ref{AutoCorrRep} was calculated with $T=0.1$ s and 50\% overlap between analysis frames and then averaged over all frames. The envelope of the obtained autocorrelation that appears in each plot was extracted using the magnitude of the Hilbert transform of the average autocorrelation function on all frames \citep{Ando1989,Dorazio2011}. The coherence time $\Delta \tau$ (seen on the plot top right corners) was calculated by finding the duration it took the envelope to drop by 3 dB from its peak value. The effective duration was found by fitting a linear curve to the logarithmic envelope of the coherence function between 0 and -5 dB and extrapolating it to -10 dB, in accord with the standard definition of effective duration \citep{Ando1985}. 

In all plots, the coherence time $\Delta \tau$ and effective duration $\tau_e$ are smaller than 0.15 ms\footnote{The coherence time is defined as the measured at the half-maximum point of the main lobe, whereas the effective duration is measured when the function drops to -10 dB. However, the latter is evaluated through extrapolation from -5 dB. See \cref{FourSources} and \cref{TypicalSourceCoh} for further details.}. Butterworth filtering  (2nd-, 4th-, 6th-, and 8th-order) is displayed in plots A, C, and E for center frequencies of 100, 500, and 2500 Hz, respectively. In all cases, the coherence time (and effective duration) increases as the center frequency decreases. Both $\Delta \tau$ and $\tau_e$ are approximately doubled when the filter order is increased from 2nd- to 6th-order. The effect of an 8th-order filter appears more unpredictable, though. The gammachirp filtering (plots B, D, F) is more predictable and has much shorter times associated with it. However, the filter orders are not directly equivalent and (the standard) 4th-order gammachirp results in approximately the same coherence time and effective duration values as the 2nd-order Butterworth, in all cases. 

Regardless of the specific type and order of filter selected, the effect on coherence is unmistakable, as even white noise that enters as a completely incoherent signal, leaves the filter as partially coherent. This is an important result that will be used at several points in the main text.

The 6th-order Butterworth filter will be used throughout this appendix, as it has a known phase response and has sufficient frequency resolution to reflect features visible in the broadband spectrum. Its downside appears to be that at short coherence times and frequencies, the filter dominates the coherence function response and makes it excessively long---it likely exaggerates the low-frequency signal coherence time. Therefore, the values below should be interpreted in a comparative and relative sense rather in the strict absolute manner.

\section{Effect of integration time}
\label{IntegrationTimeConstant}
\begin{figure} 
		\centering
		\includegraphics[width=1\linewidth]{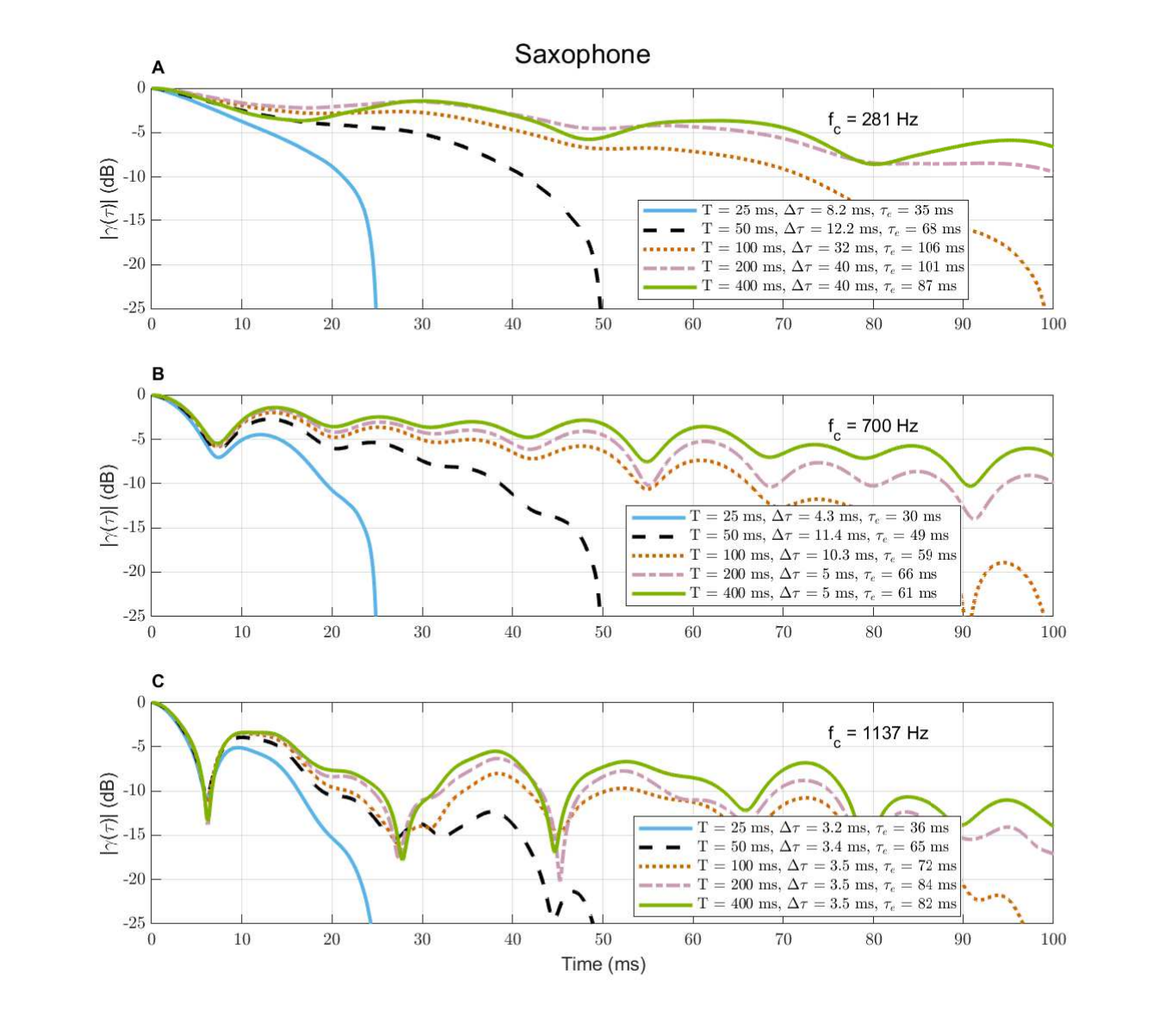}	
		\caption{The effect of the integration time $T$ on the same autocorrelation function as is in plots A, B, and C, for a 6th-order Butterworth filter, which has been used as a standard in this appendix. The coherence time $\Delta\tau$ and effective duration $\tau_e$ are displayed for all conditions.} 
		\label{SaxT}
\end{figure}

Examples of the effects of the integration time of the correlation integral (Eq. \ref{AutoCorrRep2}) are shown in Figure \ref{SaxT} for an acoustic source that is presented in \cref{PositionFrequency}---a saxophone that tends to produce long notes of high coherence. The effect of varying the integration constant duration $T$ between 25 and 400 ms is frequency dependent in the saxophone example, since longer $T$ results in longer coherence time at low frequencies (plot A), while it is maximized with $T=100$ ms at the midrange frequency (plot B), and is unaffected at the highest frequencies (plot C). The effective duration is somewhat different, as it is more sensitive to fluctuations. The shortest integration times (25 and 50 ms) are clearly too short to capture significant portions of the signal that can reveal relevant coherent patterns. The longest integration times (200 and 400 ms) do not necessarily add critical information to the analysis that is maybe more suitable for sustained and stationary sounds. 

\section{Stationarity and nonstationarity}
\label{NonstationarityMeasurement}
\begin{figure} 
		\centering
		\includegraphics[width=1\linewidth]{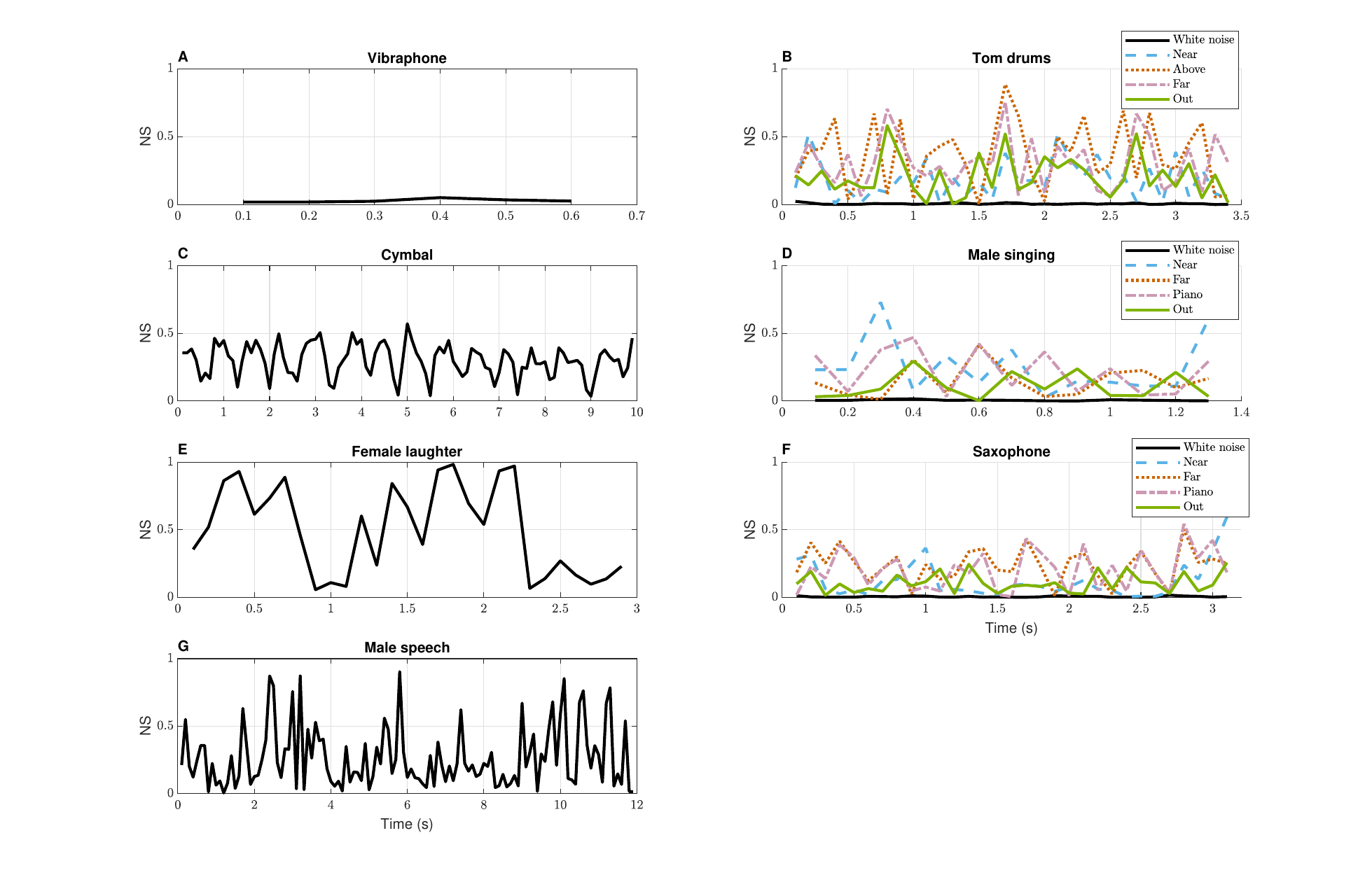}	
		\caption{The nonstationarity index of the seven sources used throughout the appendix, according to Eq. \ref{NonstationIndex}. The sources were analyzed in 100 ms consecutive frames. On the right hand side, each source is analyzed in all four microphone positions available, which generally showed similar patterns. For comparison, the black curves on the right-hand figures (B, D, and F) shows the nonstationarity index of pseudo-random white noise of identical duration, which approaches zero with $NS \approx 0.005$.} 
		\label{nonst}
\end{figure}

It was mentioned several times in the text that the assumption of stationarity does not hold for typical acoustic signals that function as auditory stimuli. Nonstationarity complicates the analysis and sets many of the results obtained using the stationarity assumption as limiting cases. In the figure below (\ref{nonst}), the nonstationarity of the seven sound samples that have been used in this appendix and are presented in the subsequent sections, is analyzed using a simple measure. \citet{Kapilow1999} proposed three stationarity indices to facilitate various algorithms of speech processing, which can produce severe artifacts if used to process sound segments that are nonstationary (mainly phonemic boundaries). The simplest index proposed in the paper ($C_n^1$ in \citealp{Kapilow1999}) is based on the root mean square (RMS) level difference between consecutive analysis windows of the sound:
\begin{equation}
	NS = \frac{|p_n - p_{n-1}|}{p_n + p_{n-1}}
	\label{NonstationIndex}
\end{equation}
where $p_n$ refers to the RMS level of the pressure in time frame $n$, which was set here to be 100 ms long. When the difference term in the numerator is 0, the signal does not vary in level, which entails stationarity. If it maximally varies, $NS$ is 1. Therefore, it seems appropriate to think of this measure as a nonstationarity index rather than a stationarity index. 

In Figure \ref{nonst}, the plots on the left refer to the four sources used in \cref{FourSources} and the plots on the right to the other three sources that are used in \cref{PositionFrequency}. The vibraphone note (plot A)---being tonal and sustained---appears stationary as $NS \approx 0$. In plots B, D, and F, the nonstationarity of pseudorandom white noise is displayed in black and shows zero nonstationarity, as expected. All other examples are nonstationary to different degrees, but musical sounds (vibraphone, saxophone, singing) appear to be more stationary than others. Speech and laughter have particularly erratic nonstationarity patterns. In the case of the sources on the right of Figure \ref{nonst}, alternative microphone positions were all plotted for comparison and exhibit reasonably close figures, which implies that moderate reverberation does not have a strong effect on the inherent stationarity of the signal. In fact, in some cases the remote positions appear more nonstationary on average than the near position (tom drum), whereas it is the opposite in other cases (singing).

The nonstationarity estimated in the various cases suggests that using stationary coherence function of the form of Eq. \ref{AutoCorrRep2} is not strictly correct. This equation assumes that the correlation operation is time invariant, so that the signals can be compared (correlated) at any two points $t_1$ and $t_2$ and produce the same results. This gives rise to the time-invariant time delay variable $\tau$. However, the nonstationarity entails time dependence, so relevant signals have to be evaluated with the two time coordinates instead. For simplicity, we do not pursue this approach here, but show only some examples of the time-dependent effective duration in the next section. 

\section{Four narrowband coherence-time examples}
\label{FourSources}
From  \cref{CoherentSources}, it appears that a relatively fragmentary sample of effective durations of various acoustic sources is provided in literature, and almost none of acoustic coherence times. Critically, the available reports have exclusively focused on the autocorrelation of the full (broadband) spectrum, which is useful in the discussion of pitch and broadband periodicity, but produces difficulties in interpreting nonstationary broadband sounds. Broadband autocorrelation peaks may correspond to different sounds associated with specific sources (e.g., certain musical instruments in an ensemble), or within sources when their spectra are complex and comprise multiple modes. As the auditory system decomposes the broadband signal to narrowband channels, there may be some information that cannot be garnered from such broadband analyses. Therefore, in order to anchor the subsequent discussions with relevant data, narrowband autocorrelation analysis is provided for four acoustic sources.

\begin{figure} 
		\centering
		\includegraphics[width=1\linewidth]{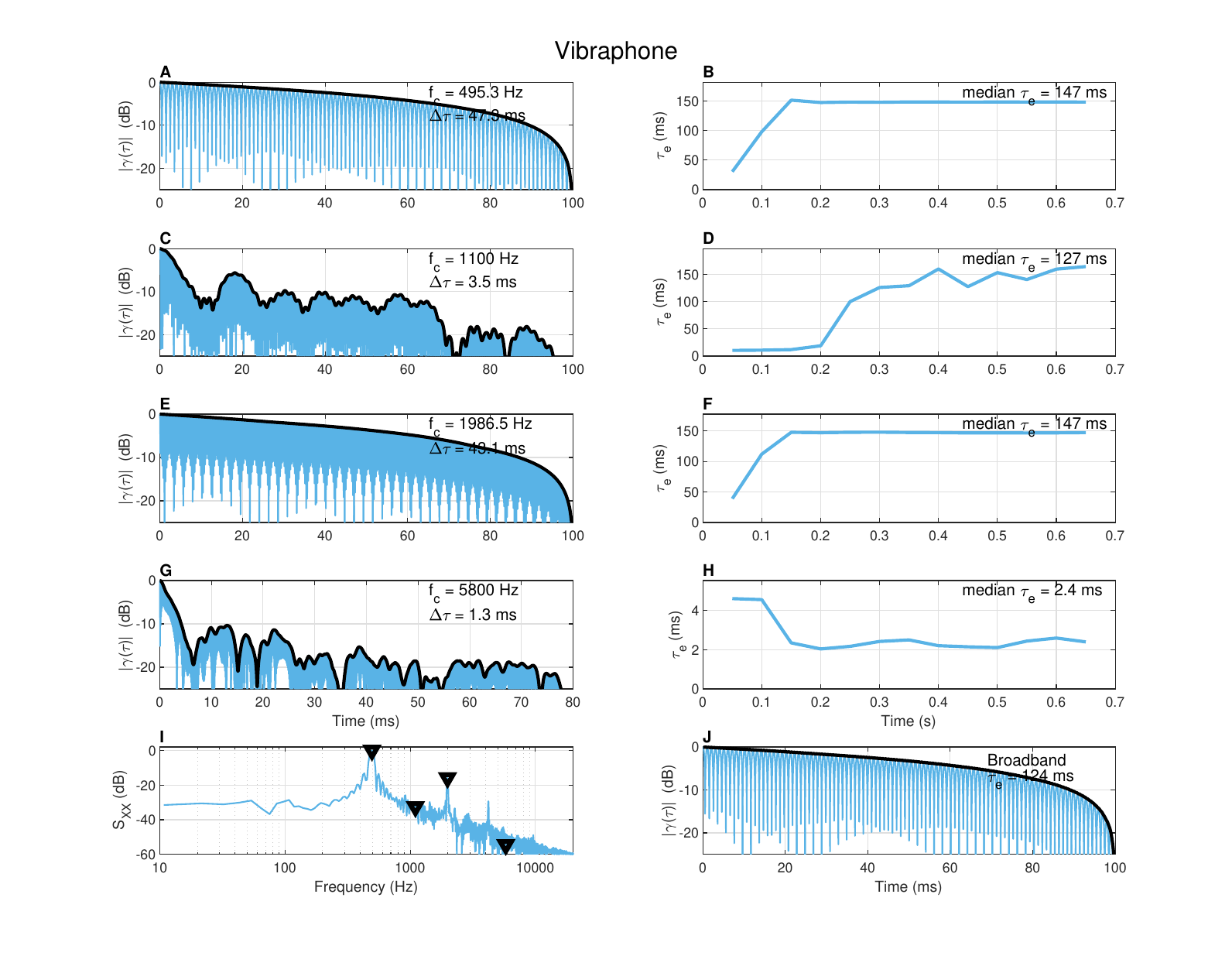}	
		\caption{Autocorrelation of a vibraphone note (B4 = 494 Hz) of duration 0.7 s, recorded in near-field in a dry room. The integration time is $T = 0.1$ s. The average autocorrelation curves are plotted on a dB scale. The coherence time ($\Delta \tau$) is the -3 dB point on the Hilbert envelope of the coherence function. The effective duration ($\tau_e$) is extrapolated to the -10 dB on the envelope (see text). Bandpass filtering was used to obtain narrowband autocorrelation curves with center frequency as marked in plots A, C, E, and G and bandwidth equal to the equivalent rectangular bandwidth of a corresponding auditory filter. The corresponding plots B, D, F, and H give the running effective duration and its median value for that frequency band. The frequencies were selected from the spectrum (plot I), where they are marked with triangles. The broadband autocorrelation function appears in plot J.}
		\label{Vibes}
\end{figure}

\begin{figure} 
		\centering
		\includegraphics[width=1\linewidth]{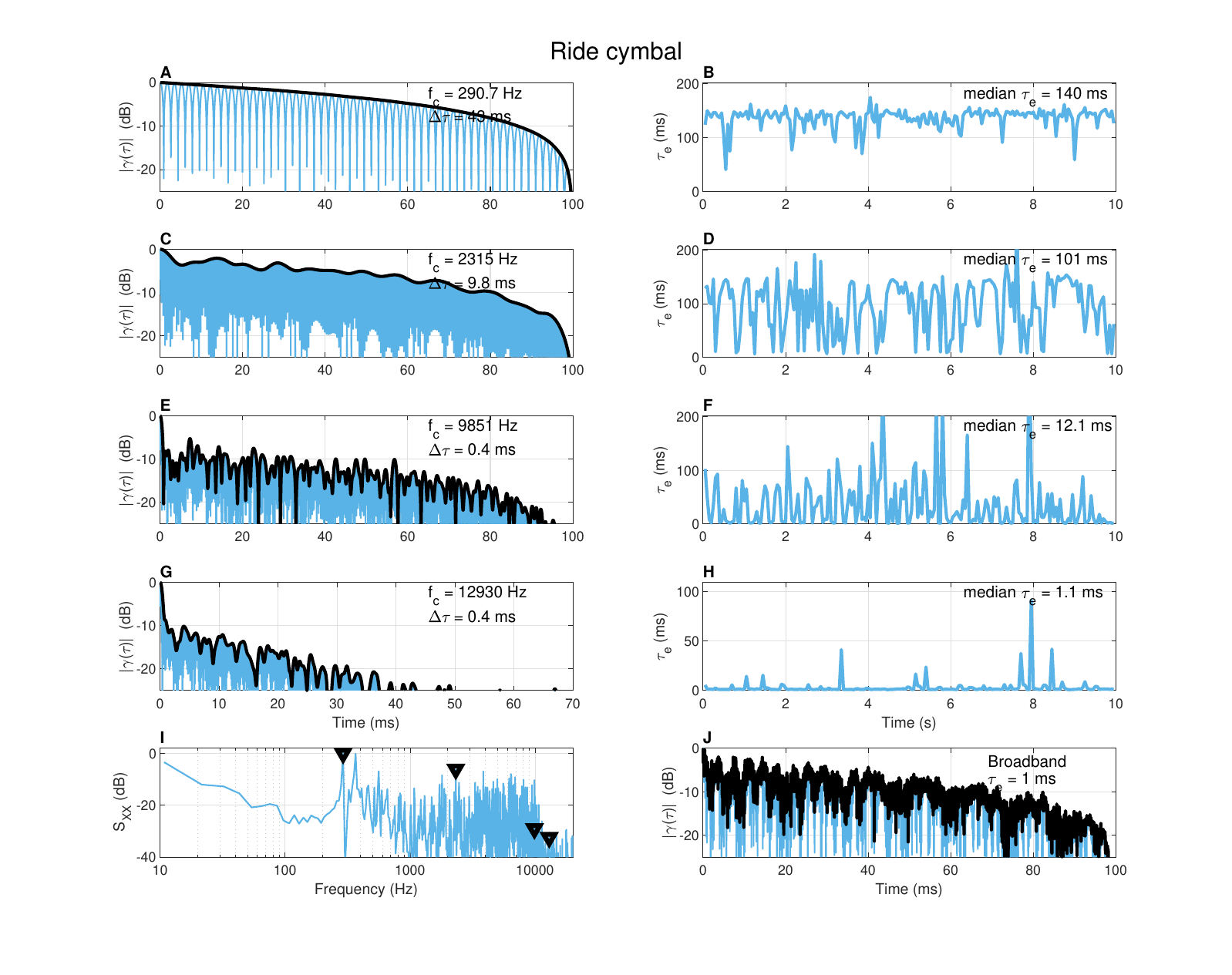}	
		\caption{Autocorrelation of a ride cymbal being played with a drum stick for 10 s, recorded with a close microphone in a dry studio. The measurement details are given in Figure \ref{Vibes} and in the text.}
		\label{Ride}
\end{figure}

\begin{figure} 
		\centering
		\includegraphics[width=1\linewidth]{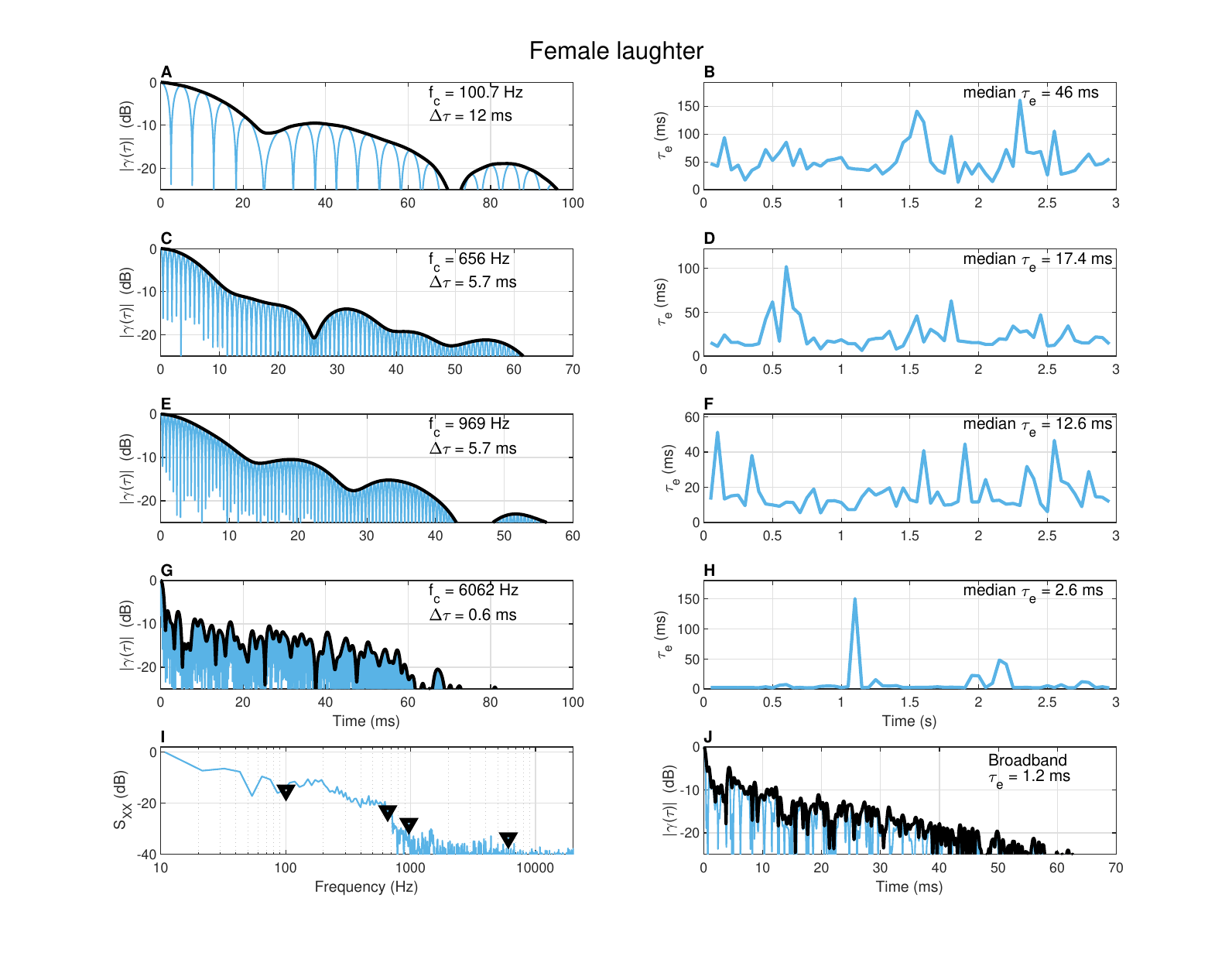}	
		\caption{Autocorrelation of female laughter of 3 s duration, recorded in a standardized audiometric booth. See measurement details in Figure \ref{Vibes} and in the text.}
		\label{Laughter}
\end{figure}

\begin{figure} 
		\centering
		\includegraphics[width=1\linewidth]{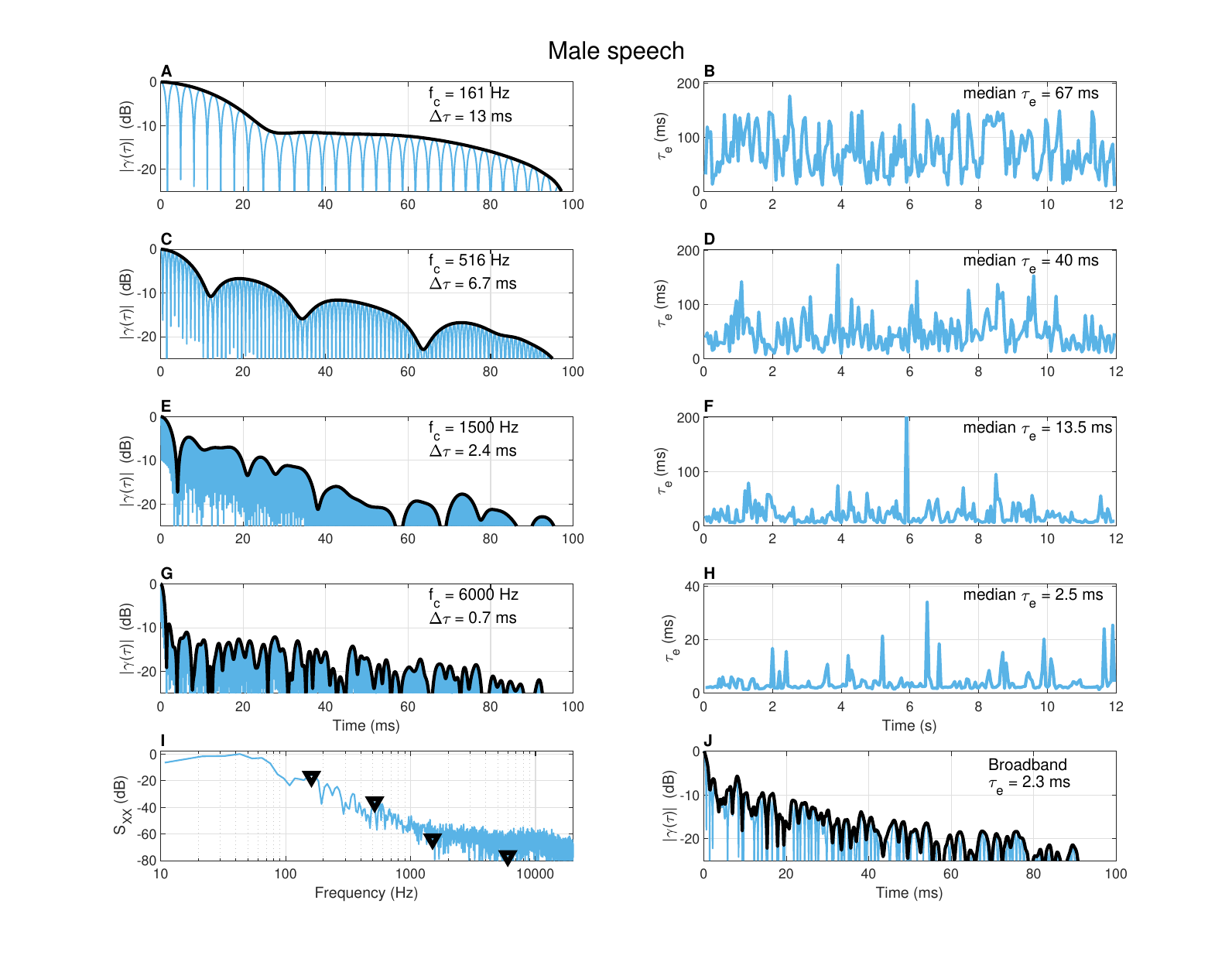}	
		\caption{Autocorrelation of male speech of 12 s duration, recorded in a standardized audiometric booth. See measurement details in Figure \ref{Vibes} and in the text.} 
		\label{MaleSpeech}
\end{figure}

 
The narrowband autocorrelation curves in the next four Figures (\ref{Vibes}, \ref{Ride}, \ref{Laughter}, and \ref{MaleSpeech}) were all computed using identical methods as in \cref{WhiteCoherence}, and are displayed in a uniform format. The center frequencies were selected by analyzing the spectrum (plots I). This was done in order to have examples that contrast different coherence time characteristics in different bands. The median value of the effective duration $\tau_e$ values are printed in plots B, D, F, where changes are shown along the duration of the recording. The corresponding narrowband filtering was achieved with 6th-order bandpass Butterworth filters, set with the equivalent rectangular bandwidth of auditory filters with the center frequency (Eq. \ref{ERBrep}).

Both $\Delta \tau$ and $\tau_e$ are dependent on the choice of the integration time $T$. The particular choice of $T=0.1$ s produced values that subjectively reflected the author's impression of coherence from listening to the sounds. However, different choices of $T$ would retain the relative differences between the different coherence functions, as was seen in \cref{IntegrationTimeConstant}. The spectrum of the sound was obtained using the same time frames, using fast-Fourier transform (FFT) with 4096 points (plots I). Finally, the broadband autocorrelation is displayed in plots J along with its associated effective duration. Corresponding audio demos to the samples that are presented in this analysis are found in the demo directory \textsc{/Appendix A - Coherence/Narrowband coherence/}.

Figure \ref{Vibes} shows the analysis of a sustained complex tone (B4 = 494 Hz, 0.7 s) of a vibraphone\footnote{A xylophone-like instrument with tuned metallic bars that produce tones of sharp attack and long sustain. The sound is produced with soft mallets.} in four frequency bands. The fundamental and the second octave overtone (1986 Hz) were selected as the most coherent modes (plots A--B and E--F, according to the spectrum in plot J), whereas two other frequencies (1100 and 5800 Hz) that are not associated with apparent overtones are plotted in C--D and G--H. The coherence times are an order of magnitude longer in overtones ($\sim$45 ms) than in non-overtone frequencies ($\sim$1--3 ms). The audio version of the fundamental sounds almost indistinguishable from a pure tone, whereas the overtone has a subtle timbral variation in its attack, despite a very small (4 ms) decrease in mean $\Delta \tau$ (but not in the running $\tau_e$) compared to the fundamental. For the 1100 Hz (plots C and D), the mean $\Delta \tau$ is much shorter (3.5 ms) despite a long $\tau_e$ (147 ms), which reveals a step in the corresponding effective duration curve (plot B). It is also audible in the faint recording, which starts with a clunky noise-like attack sound before it becomes more tonal. The last frequency (plots G and H) has much shorter $\Delta\tau$ and $\tau_e$ ($<2.5$ ms), which indeed sounds like noisy-clicky sound with no discernible pitch.

The second sound analyzed in Figure \ref{Ride} is musical but not tonal---a large (22'') ride cymbal mounted on a stand that was struck with a drumstick in an unsteady rhythm. It was recorded in near-field and dry studio conditions. Cymbals are characterized by high modal density and distinct high-frequency attack sound that are excited by the drumstick. The lowest eigenfrequency (290 Hz, plots A and B) gave coherence time that is comparable to the vibraphone modes (43 ms), which nevertheless sounds like a hollow tone---somewhat less clear and steady than a pure tone---which never decays significantly between hits. The next frequency band (2315 Hz, plots C and D) sounds dirtier and retains an audible indication for the drumstick hits with variable impact levels, although it is still largely sustained, with shorter coherence time (9.8 ms). The last two frequencies are of very short coherence time (0.4 ms) and are neither tonal nor sustained, as the impulsive hits are distinctly heard. The highest frequency (13 kHz, plots G and H) sounds like faint dirty clicks. In this case, the decrease in coherence time with frequency is also reflective of the decrease in the decay time of high frequency modes in cymbals \citep[652--655]{FletcherRossing}.

The third sound (Figure \ref{Laughter}) is a 3 s female-laughter recording done in an audiometric booth in near-field conditions. Its spectrum (plot I) does not reveal clear peaks and indeed none of the narrowband recordings resembles laughter, suggesting that laughter may be described more faithfully as a modulated broadband sound than using specific normal modes, or that the modal excitation is too nonstationary to be modeled in the way presented here (cf. \citealp{Bachorowski2001}). Relative agreement in the decrease rate of both coherence time and effective duration was obtained in all frequencies, where the coherence time dropped from about 12 ms at 100 Hz to 0.6 ms at 6 kHz. In general, it is evident that low frequencies have longer coherence times than high frequencies, although it depends also on the particular signal contents, which seem to be highly variable, judging from the running effective duration plots. 

The fourth and last sound (Figure \ref{MaleSpeech}) is a 12 s male-speech recording done in the same conditions as the female laughter from Figure \ref{Laughter}. The spectrum (plot I) is rich with resonances, but only few that are prominent. Therefore, an attempt was made to pick frequencies that display the largest variance in coherence time. At 161 Hz (plots A and B), the coherence time is 13 ms and, as can be heard in the recording, it is clearly amplitude- and frequency-modulated---the latter as a result of prosodic changes of the fundamental frequency. At 516 Hz (plots C and D), the filtered recording sounds more speech-like with amplitude-modulated noisy carrier that is hardly tonal. Here the mean coherence time (6.7 ms) is much shorter than the effective duration median (40 ms). Shorter coherence time may correspond to the atonal timbre of this sound. This impression is stronger at 1500 Hz (plots E and F), where no clear peaks are observed in the spectrum, the mean coherence time is very low (2.4 ms), and the recording sounds like a deeply amplitude-modulated narrowband buzz. This is exacerbated in the last band (6000 Hz, plots G and H), where the modulated buzz is less deep and more sustained, but sounds more like a guiro than speech. 

In all cases, the broadband autocorrelation (plots J in the figures) gives only partial information about the source, which masks any distinction between coherent and incoherent parts it may have. This is perhaps not surprising, because the autocorrelation in coherence theory is strictly a narrowband measure. So, in both speech recordings, the broadband coherence time estimate is conservative, in that it estimates the speech to be less coherent than it may actually be. This can be true especially for voiced phones. In contrast, the musical vibraphone note is estimated to be more coherent than not---effectively neglecting any noise-like incoherent components of the timbre.

Another recurrent observation about the data is that the running and median autocorrelation functions occasionally yield different coherence time estimates, which should be  $\Delta \tau \approx 3\tau_e/10$ for a `well-behaved' exponentially decaying self-coherence curve. While the nonstationary coherence function fine structure is captured by the running autocorrelation, it is often too erratic to be truly useful, whereas the mean and median values appear to correspond to higher-level perception of the overall sound. Either way, these differences underline the nonstationarity of even the simplest acoustic narrowband sources. 

\section{The effects of room acoustics on coherence}
It is instructive to examine a few concrete examples of sounds whose coherence properties are affected by reverberation. As was stressed throughout the main text, the degree of coherence of the acoustic source can be affected by a number of factors. The primary concern in the examples below is to explore how the relative degree of coherence varies between different sources, frequency bands, and recording positions. Three sets of recordings are tested throughout: a floor tom drum\footnote{The largest drum of the drum set, except for the bass drum.}, a tenor saxophone, and a male voice singing a three-word melody. Each set was recorded in four positions within the same large studio, with different high-quality microphones. The computational methods of the various functions are identical to those presented in \cref{FourSources}, but this time we present the mean coherence time and effective duration as a function of frequency. 

Unlike the samples in the previous section, the signal-to-noise ratio of the following samples was not always high. Thus, apparent incoherence may be the result of sounds that are truly unrelated to the source---either noise or altogether different sources. 

\subsection{Effect of position and frequency}
\label{PositionFrequency}
All the sound samples used below were recorded in a large studio ($V\approx 800 \,\, \m^3$) with medium reverberation time $T_{60} \approx 1$ s. The walls and ceiling were treated with heavy absorption and each instrument was recorded using various microphone positions. One was always close (the ``near'' position), one was positioned somewhere in the room to capture more of the reverberant field (``far''), and another was positioned in a much smaller room that was connected through an open doorway, but far away from the source (``out''). For the tom drum, there was another position that was not as close as the ``near'' position and picked up the radiation from above the drum (the '`above'' position). For the singing and saxophone, another microphone was placed right over an open grand-piano soundboard, close to the strings, which may have resonated with the ambient radiated sound. 

Audio demos that correspond to the samples in the analysis are found in ``\textbf{/Appendix A - Coherence/Room acoustics/}.''

\begin{figure}
		\centering
		\includegraphics[width=1\linewidth]{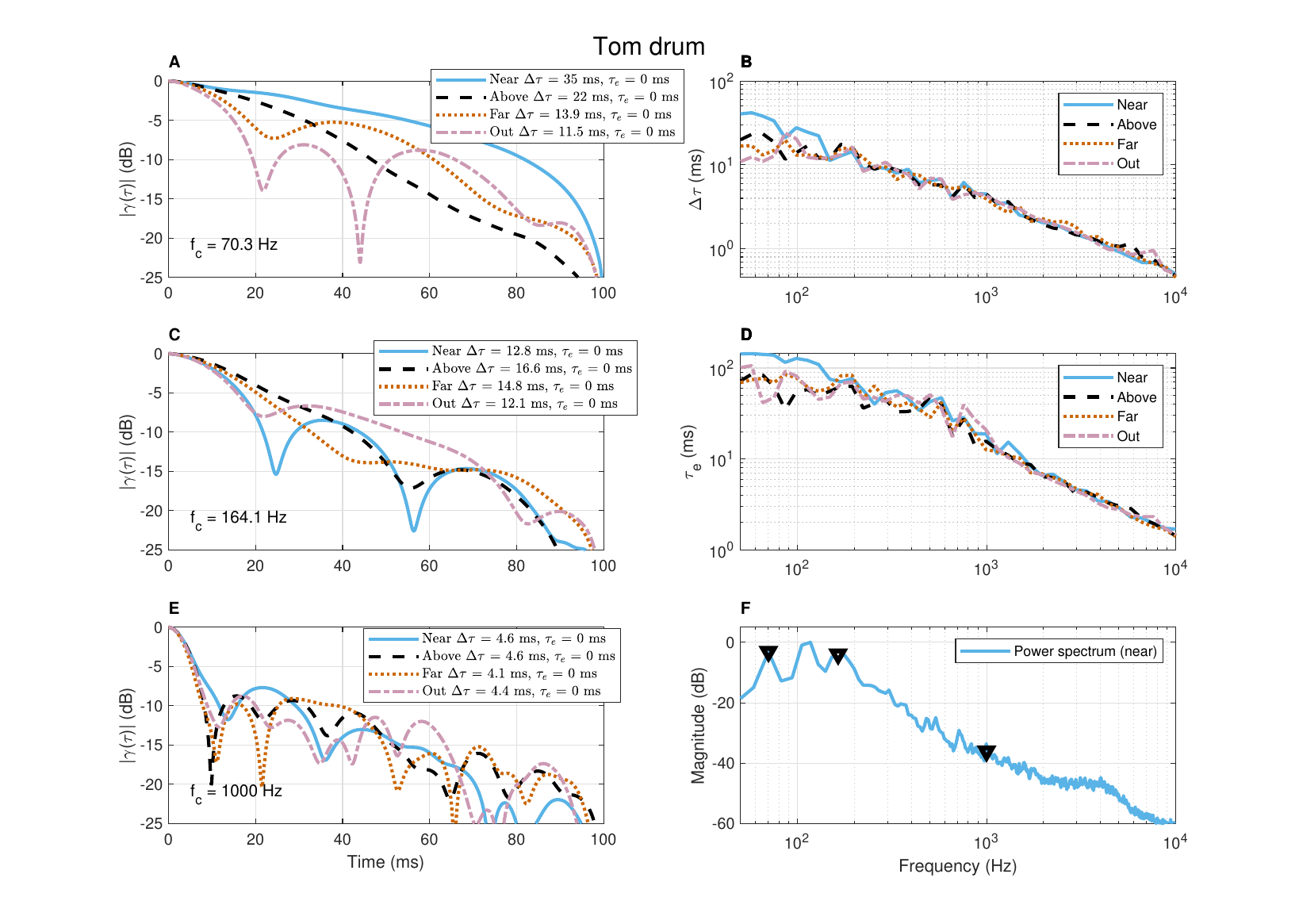}	
		\caption{The autocorrelation of a floor tom drum of a drum set being played for 3 s. The autocorrelation function envelopes of the same material recorded in four different positions in the room are compared at three different frequencies. The first at 70 Hz (plot A) and second at 164 Hz (C) are resonances of the drum, as is marked in the spectrum of the near position (plot F). At 1000 Hz (plot E) and in general above 200 Hz, there are no special spectral features and the sound may be largely incoherent at the source. Plots B and D are the frequency-dependent mean coherence time (at -10 dB) and median running coherence time, or effective duration, respectively.} 
		\label{Toms}
\end{figure}

Figure \ref{Toms} shows the autocorrelation of the floor toms that were played for 3.5 s. The hi-hat cymbal was being played throughout as well, as can be heard in the above position. However, it was less dominant in all other recordings and nearly inaudible in the near position. For clarity, only the autocorrelation envelope is shown in plots A, C, and E. As can be seen in the spectrum (plot F) of the near position, it has strong normal modes below 170 Hz, but not much energy and no special features are recognized above that. The coherence time of the near recording is distinctly longer below 170 Hz, but at higher frequencies all positions seem to approximately converge to similar coherence time. This can also be seen in the frequency-dependent coherence time and effective duration plots (B and D), which appear to be nearly merged above 1000 Hz. This suggests that the degree of coherence is almost the same at these frequencies in all room positions, which can be a result of the spectral content of the source combined with its incoherent nature.

\begin{figure} 
		\centering
		\includegraphics[width=1\linewidth]{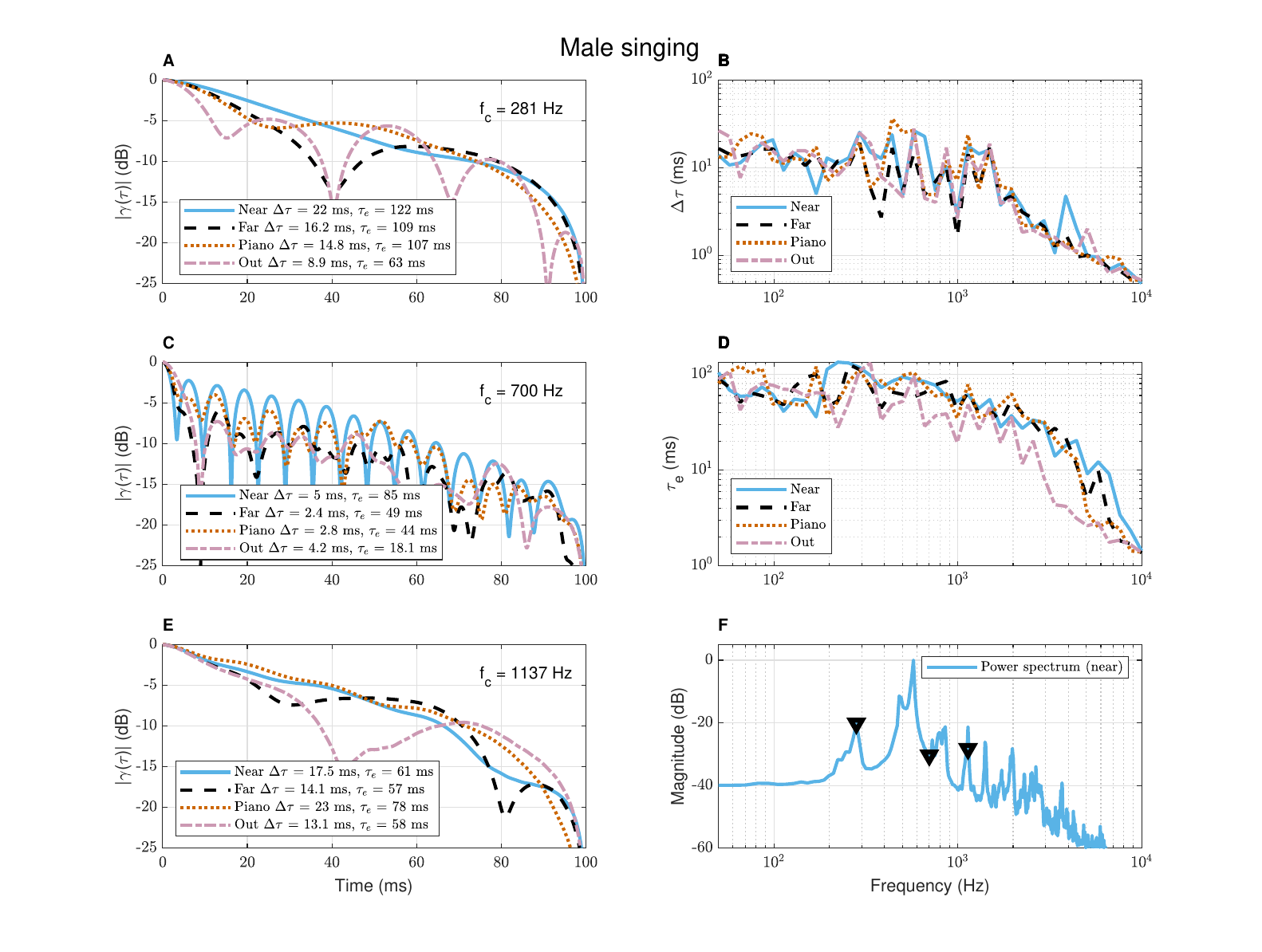}	
		\caption{Autocorrelation of male singing for 1.4 s. The autocorrelation function envelopes of the same material recorded simultaneously in four different positions in the room is compared at three different frequencies. The first at 281 Hz (plot A) is $f_0$ and the third at 1137 Hz (plot E) is a distinct resonance. The 700 Hz frequency (plot C) was chosen to illustrate the effect of no discernible resonance and low energy on the autocorrelation. The frequencies are marked in the spectrum (plot F). Plots B and D are the frequency-dependent mean coherence time (at -10 dB) and median running coherence time, or effective duration, respectively.} 
		\label{Head}
\end{figure}

\begin{figure} 
		\centering
		\includegraphics[width=1\linewidth]{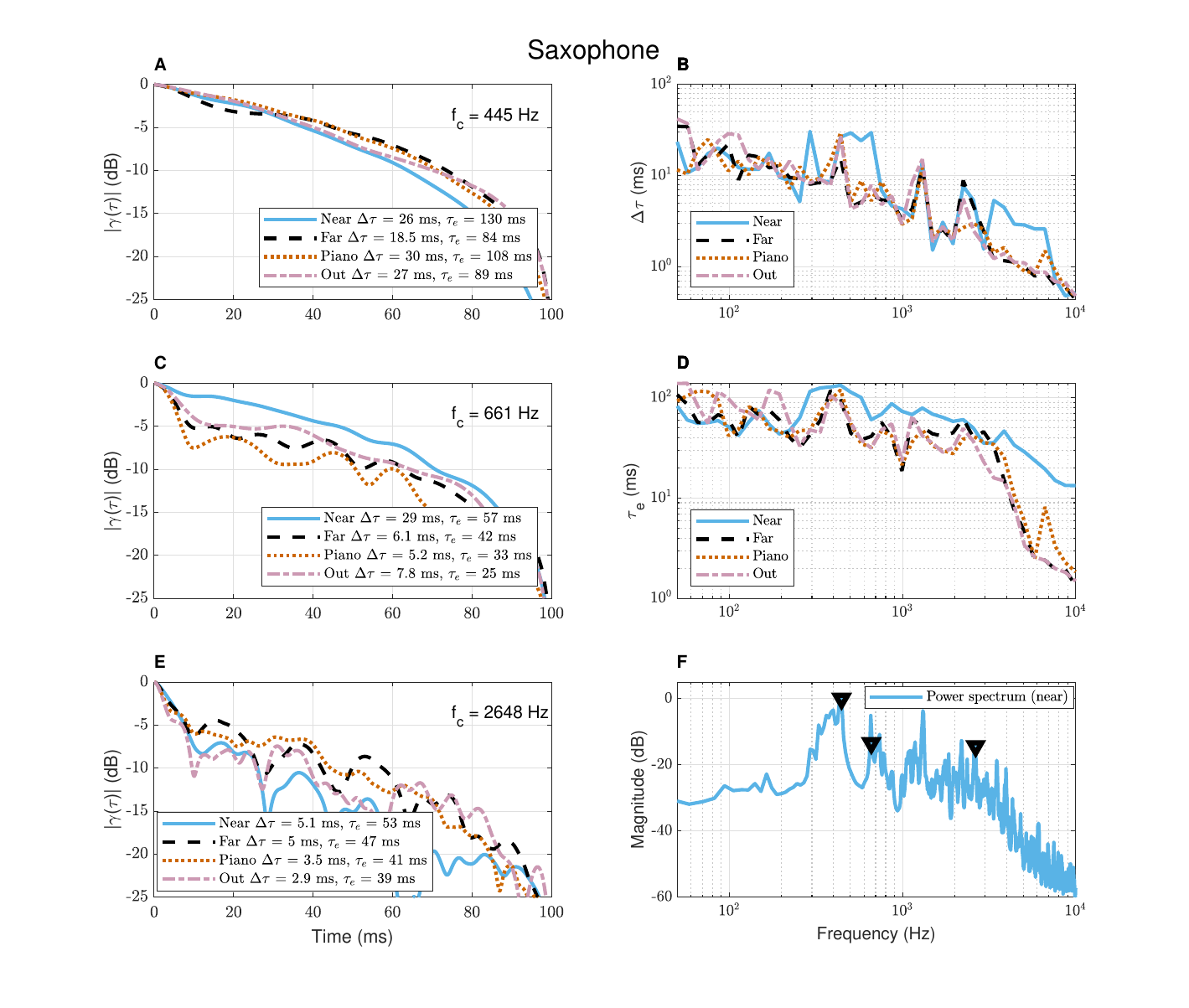}	
		\caption{Autocorrelation of tenor saxophone playing fast notes for 3 s, in the same conditions as the male singing example from Figure \ref{Head}. The frequencies in the analysis (plots A, C, E) were selected from peaks in the spectrum (plot F), but the near position tended to exhibit longer coherence time (plot B) and effective duration (D) than the other microphone positions.} 
		\label{Sax}
\end{figure}

In the second example (Figure \ref{Head}), male singing for 1.4 s is analyzed. Once again, the coherence time estimates were generally longer for the near position both at low frequencies below 150 Hz, and around a couple of distinct resonances (plot B). While the tendency for the coherence time to get shorter with higher frequencies is apparent here too, the additional energy in other spectral components makes the frequency dependence more erratic. This may be gathered by inspecting the second frequency (700 Hz), which was analyzed because it does not correspond to any clear peak. The coherence time at 700 Hz drops significantly to around 5 ms in the near position, or shorter in other positions (plot C). It is lower than the next resonance at 1137 Hz, which has 13--23 ms coherence time, depending on the positions (plot E). The different nature of this source coherence is evident in the variation of the coherence time and effective duration in frequency (plots B and D), which are not nearly as monotonically linear as that of the tom drum. 

In the third example (Figure \ref{Sax}), a short passage (3 s) of tenor saxophone fast notes was recorded in an identical setup to the singing. While the near-field recording has the longest coherence time, the rank order of the other recordings from the room is not always predictable. Especially in the connected room and the piano, there may have been specific room modes or soundboard resonances that enhanced some of the frequencies. In general, even though the notes that were played are short, they still consist of relatively long periodic portions that have high coherence at low frequencies. However, the relatively high coherence manifests primarily in near-field, as can be seen in the coherence time and effective duration curves (plots B and D). 

\subsection{Decoherence as a function of position and frequency}
\begin{figure} 
		\centering
		\includegraphics[width=1\linewidth]{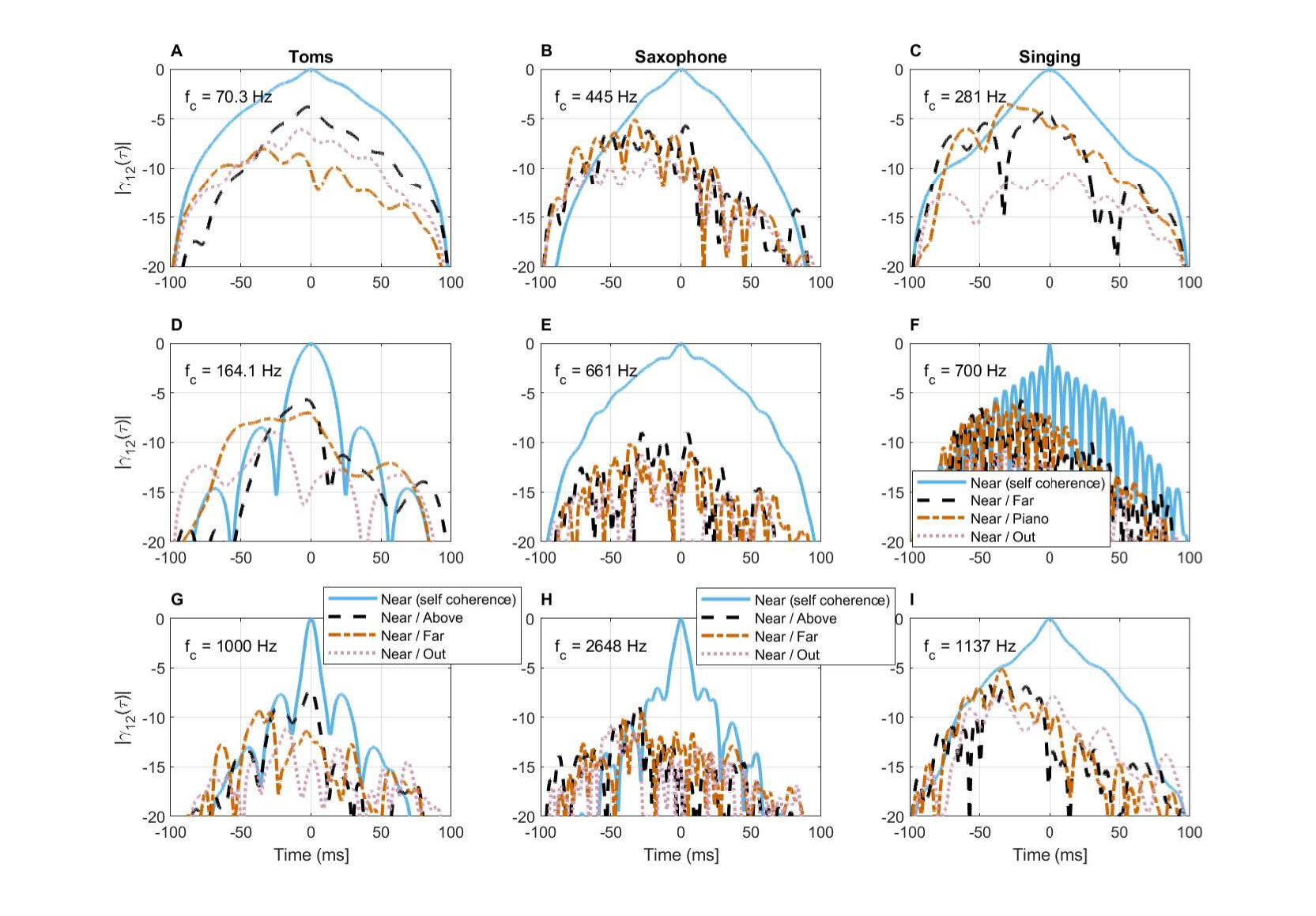}	
		\caption{Cross-correlation of pressure fields in four different positions in the same space for three different sound sources. Each sound source was analyzed around three frequencies, according to Figures \ref{Toms}, \ref{Head}, and \ref{Sax}.} 
		\label{crossrev}
\end{figure}

It has been shown that in reverberation the pressure field is decohered between two points with increasing distance and frequency, given that the sound is stationary and represents a perfectly diffuse field \citep{Cook1955}. How relevant is this effect to more realistic sounds that are nonstationary and are not placed within a perfectly diffuse field? Figure \ref{crossrev} shows, once again, the tom drum, saxophone, and male singing recordings---each analyzed at the same three frequencies as in \cref{PositionFrequency}. The coherence functions of pairs of recordings are displayed in comparison with the self-coherence (autocorrelation) function in the near position. It can be seen that in none of the cases is there complete (spatial) coherence between the close and any remote microphone, but rather moderate partial coherence (between -10 and -3 dB) over a narrow time interval, which includes the acoustic path delay. According to Eqs. \ref{roomdeocherence} and \ref{mugamma0}, at a distance of 0.8 m (the approximate distance between the ``near'' and ``above'' microphones according to the their broadband cross-correlation), the coherence of the tom drum at 70 Hz should have been -16 dB (close to the first zero of the sinc function), whereas we obtained -4 dB. At 164 Hz, for that same distance, the sinc function is past its first zero and the coherence is expected to be -9 dB, whereas the obtained value is -5.7 dB. In the case of the saxophone and singing, for a microphone distance of 1--1.5 m, the coherence of the lowest frequencies should have been around -15 dB in both cases, whereas we obtained -5 and -6 dB, respectively. Similar values are obtained from the coherence functions between the far-field microphones. 

Another interesting feature is that it is unpredictable which of the three remote positions is the most decohered one with respect to the near position. In one case it is the ``far'' position (70 Hz, tom), in some cases it is the ``out'' position of the small room (tom---164 and 1000 Hz; singing---281 and 700 Hz), and in yet other cases the three positions are about equal (singing---1137 Hz; all saxophone frequencies). Such an erratic behavior can be the result of interaction between the coherence function of the source with local modes and reflections, in addition to the different coherence at the source of the three instruments. 

All in all, from the small sample of recordings that was tested, the room acoustics in question cannot be considered a true diffuse field, because it indicates that there are extensive frequency ranges of relatively high degree of coherence. Therefore, the pressure fields in these positions may be better described as partially coherent, rather than completely incoherent. 

\subsection{Cross-spectral coherence}
\label{Crossspectral}
\begin{figure} 
		\centering
		\includegraphics[width=1\linewidth]{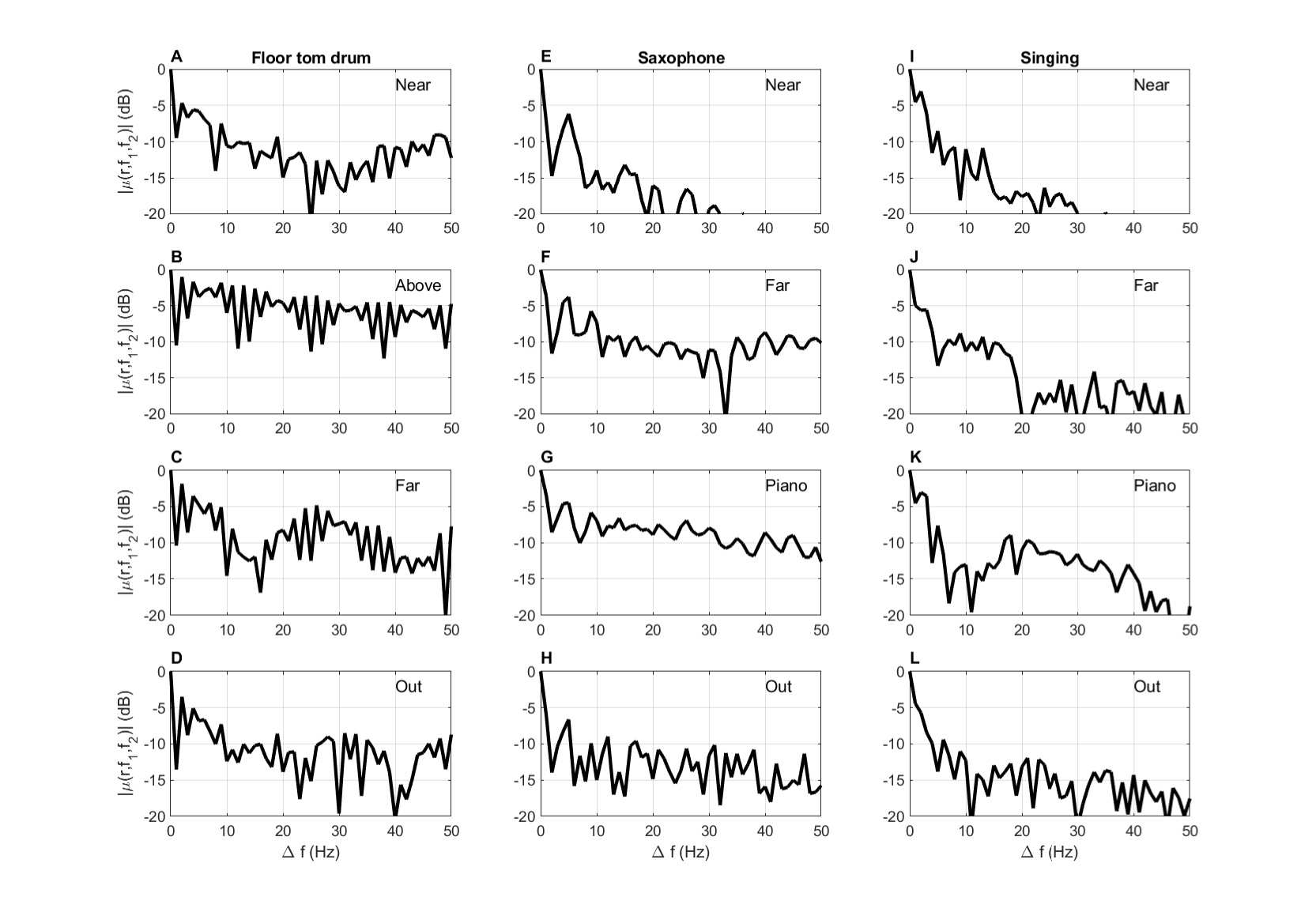}	
		\caption{The autocorrelation of the pressure frequency response of three sources in four different positions in space. See \cref{PositionFrequency} for further details about the sources and recording positions.} 
		\label{acfft}
\end{figure}

Correlation between frequency components of the pressure field is another effect that was theoretically predicted for diffuse fields \citep{Schroeder1962}, but has not been demonstrated empirically, to the best knowledge of the author. We demonstrate the effect for the same sound recordings, which we already know did not take place in a perfectly diffuse field. The autocorrelation functions of the (complex) fast Fourier transforms (non-negative frequencies only, 1 Hz resolution) of the broadband pressure fields are plotted in Figure \ref{acfft} for a range of 50 Hz. The data vary in how quickly neighboring frequency components become effectively decorrelated. Borrowing from the previous time-domain analysis, the drop from complete coherence to -3 dB is instantaneous---within $\Delta f = 1$ Hz. if we treat the -10 dB mark as a rough indicator for incoherence, then a broad range of responses is evident from the plots. In two cases---the ``above'' tom and `'piano'' saxophone positions---the sound barely decoheres over 50 Hz. Other cases can be anything from $\Delta f$ of 5 to 20 Hz, depending on how the different fluctuations are interpreted. In Schroeder's original paper, $\Delta f= 3$ Hz for $T_{60} = 1$ s led to a drop in coherence below -10 dB \citep[Figure 3]{Schroeder1962}. 

As larger frequency deviations than predicted by stationary coherence theory were obtained for the present examples, we can conclude that adjacent components are partially coherent, albeit weakly. This supports the general theory of nonstationary coherence, as was briefly reviewed in \cref{Nonstationarytheory}, which explicitly discusses coherence between different frequency components---something that violates the stationarity assumption in standard coherence theory (Eq. \ref{SpectralCorrelation1}).

\section{Conclusion}
This appendix presented the coherence functions of a small sample of real acoustic sources that were recorded indoors in different room acoustical conditions. The range of responses is wide and so are the values that are obtained from them. The correspondence between the theoretical results, which are based on the approximate diffuse room acoustics, to the present ones is not always predictable and is made more difficult because of nonstationarity. Still, these results demonstrate the applicability of the concept of partial coherence, either as a descriptor of the narrowband self-coherence of a sound, or of the spatial coherence between difference receiver positions. 

The collection of coherence functions above is by no means a representative sample of acoustic sources in general. However, given the dearth of published data about it in the literature, it provides a starting point for more comprehensive attempts to chart this large territory in the future.

%% file: Waves.tex
\chapter{Waves}
\label{AppWaves}
The main aim of this appendix is to expand the mathematical toolkit that can be used along with the temporal imaging solutions we obtained in \crefrange{temporaltheory}{ImpFun}. The main emphasis is on the properties of the group-velocity dispersion, which we have normally referred to as group-delay dispersion. This quantity has not made it to acoustics until now and it may require some effort to develop the appropriate intuition for it, insofar as it  can be applied for audio frequencies. Therefore, the derivations below are fairly constrained in their scope. It is suggested that the reader become familiarized with the section about dispersion (\cref{PhysicalWaves}) before reading this appendix.

\section{Group-velocity dispersion}
\begin{figure} 
		\centering
		\includegraphics[width=.85\linewidth]{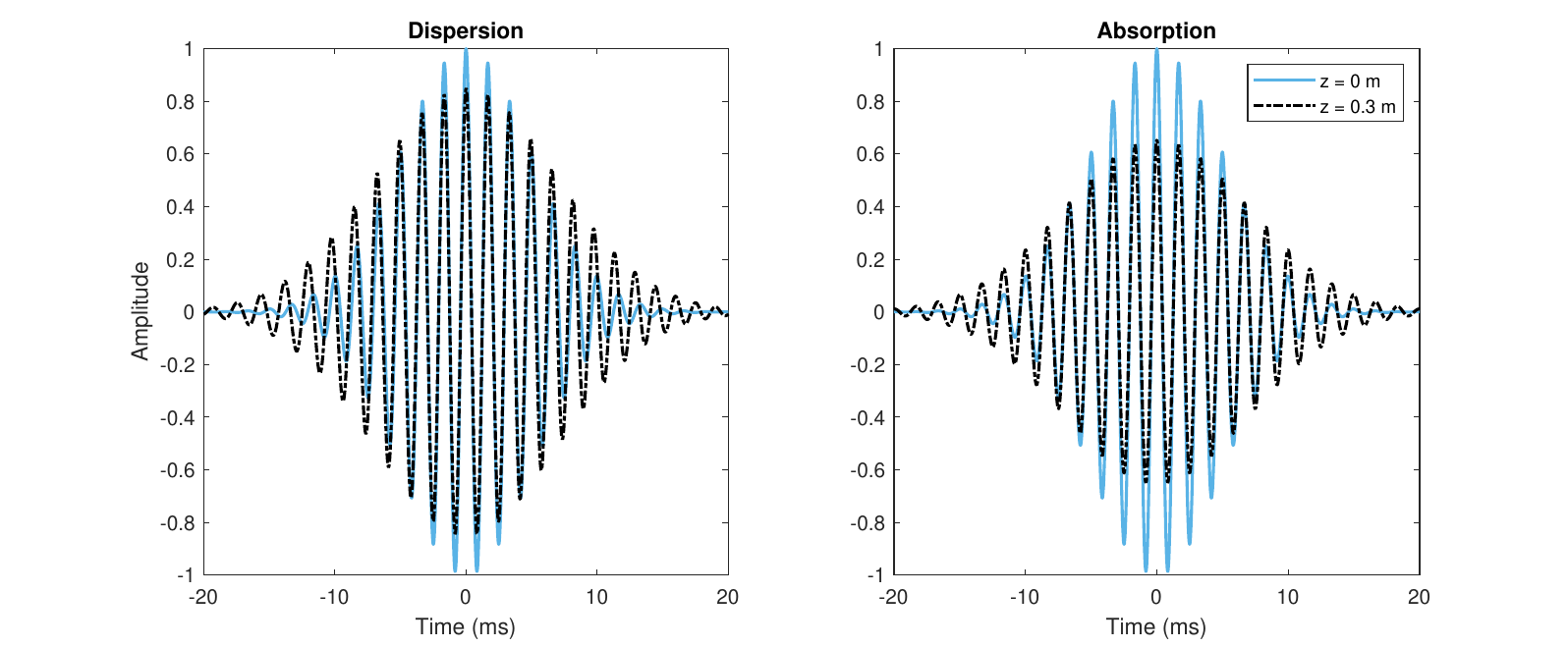}	
		\caption{An example of a Gaussian pulse at $z=0$ (solid blue) following dispersion (left) and absorption (right) after moving 30 cm in space (dashed black), according to Eqs. (\ref{pressureclear}) and (\ref{saturated}) (real part). The figures show all features of group dispersion discussed in the text: reduced amplitude, broader pulse, and chirped phase. The chirp can be clearly seen by comparing the (phase) peak locations before and after the pulse (envelope) peak. While the phase appears in sync at $t=0$, the periods are shorter at $t<0$ and longer at $t>0$. The parameters used for dispersion simulation are $f_c = 600$ Hz, $v_p = c = 343$ m/s, $t_0=5$ ms, $v_g=0.8c=274$ m/s, $\beta''=0.00008$ s$^2$/m rad. For the absorption simulation the same pulse was used with $\alpha_0 = -0.3i$ 1/m, and $\alpha'' = -0.00008i$ s$^2$/m rad.}
		\label{fig:dispersion}
\end{figure}

In the physical world, pure tones do not exist. Additionally, real physical media always exhibit some dispersion \citep[p. 3]{Brillouin1960}, which means that the modulations that go through them deform with time and will eventually vanish, as every frequency component that contributes to the envelope travels at a different phase velocity. Dispersion is also accompanied by absorption in any causal, physical medium (see \ref{airtravel}). In narrow frequency bands, where the group velocity is about constant and the medium is not highly absorptive, it is also equal to the \term{signal velocity} \citep[pp. 9--10]{Brillouin1960}---the velocity at which most of the energy of the signal travels, which cannot be higher than $c$ itself \citep[pp. 74--79]{Brillouin1960}\footnote{Note that the term signal velocity is used in the context of waveguide analysis in \citet[p. 479]{Morse}, but with a non-standard interpretation. For Morse \& Ingard the signal velocity is the velocity of the front of the wave, which has to be $c$. However, these are two separate quantities in \citet{Brillouin1960} and elsewhere, where the signal velocity is \textbf{not equal} to the \term{front velocity}, but to the group velocity, unless the medium is dispersionless.}. But, in general, the group velocity is frequency dependent as well. An approximation for the group velocity can be found by expanding the wavenumber $k$ around a narrow band, centered around center frequency $\omega_c$, using Taylor series\footnote{Note that sometimes the inverse is done, by expanding $\omega(k)$ around $k_c$ (Eq. \ref{FourSol} and \citealp[pp. 431--435]{Elmore}). We shall stick with the formalism found in optics, as is also advocated in \citet{Lighthill}.}:
\begin{equation}
k (\omega) = k_c + \frac{{dk }}{{d\omega }}(\omega- \omega_c)  + \frac{1}{2}\frac{{d^2 k }}{{d\omega ^2 }}(\omega- \omega_c)  ^2  + ... 
\label{beta_taylor1}
\end{equation}
where $k(\omega)$ is centered around the wavenumber of the center frequency, $k_c=\omega_c/c$. In general, $k$ is complex, so its real part represents the medium dispersion and its imaginary part the medium absorption:
\begin{equation}
k(\omega) = \beta(\omega) + i\alpha(\omega)
\label{complexk}
\end{equation}
and both parts form a Hilbert-transform pair---they are related through the Kramers-Kronig relations as long as the propagation medium is linear, time-invariant, and causal \citep{Toll}. The coefficient of the linear term is the inverse of the group velocity, $\Re\left(\frac{{dk}}{{d\omega }}\right)=\beta'= 1/v_g$. The real part of the coefficient of the quadratic term stands for the \term{group-velocity dispersion} (GVD)\footnote{This term has not been imported to acoustics, to the best knowledge of the author, except for acoustic measurements of tubes in \citet{Latif2000}. Another near-mention may have been in the context of acoustic dispersion in waveguides. In their analysis of the phase and group velocities of acoustic waveguides, \citet[pp. 477--478]{Morse} expounded on a similar problem as presented here of a Gaussian pulse propagating in one-dimension, over the fundamental (plane wave) mode of the tube. However, even though an absorptive ``frequency spread'' term appeared in their Taylor expansion of $k$ (as in Eq. \ref{beta_taylor1}), they unfortunately neglected it in their subsequent analysis and discussion, so GVD never came about. GVD is commonly used in optics and fiber optics \citep[e.g.,][]{Agrawal,New2011}.}. Taking a similar approach as \citet[pp. 120--121]{New2011} and \citet[pp. 335--338]{Siegman}, let us examine the effect of the GVD on an arbitrary narrowband signal, centered around $\omega_c$, while neglecting the effect of absorption, for the moment,
\begin{equation}
p(z,t) = a(z,t) e^{ i\varphi(z,t)} = a(z,t) e^{i\left[\omega t-k(\omega)z\right]}
\label{narrowband}
\end{equation}
We take a Gaussian pulse as a particular case, whose width is set by $t_0$ and is centered at $z=0$. Its envelope is
\begin{equation}
a(0,t) = ae^{-t^2/2t_0^2}
\label{envelopeT}
\end{equation}
From Eq. \ref{temptospectenv}, we can obtain the envelope spectrum using Fourier transform centered at $\omega_c$
\begin{equation}
A(0,\omega-\omega_c) = \int_{ - \infty }^\infty ae^{-t^2/2t_0^2}e^{-i(\omega-\omega_c)t}dt = \sqrt{2\pi} at_0\exp\left[-\frac{1}{2}(\omega-\omega_c)^2 t_0 ^2 \right]
\label{envelopeF}
\end{equation}
We can apply the dispersive propagation factor $k(\omega)$ on the initial spectrum---also a Gaussian---now propagated to $z$
\begin{equation}
A(z,\omega-\omega_c) = \sqrt{2\pi} at_0\exp\left[-\frac{1}{2}(\omega-\omega_c)^2 t_0 ^2 \right] \exp \left\{-iz\left[k_c + \frac{1}{v_g}(\omega- \omega_c)  + \frac{1}{2}\beta''(\omega- \omega_c)  ^2\right]\right\}
\label{envelopeFZ}
\end{equation}
where we set the GVD parameter to be $\beta'' = \frac{{d^2 k }}{{d\omega ^2 }}$. Finally, we apply the inverse Fourier transform to get back the time-signal envelope
\begin{equation}
a(z,t) = {\cal F}^{ - 1} \left[A(z,\omega-\omega_c) \right] = \frac{1}{2 \pi} \int_{ -\infty }^\infty A(z,\omega-\omega_c)e^{i(\omega - \omega_c) t} d(\omega -\omega_c)
\label{envelopetZ}
\end{equation}
The explicit solution can be computed using Siegman's lemma (Eq. \ref{SiegmansLemma})
\begin{equation}
	p(z,t) = a(z,t) e^{i\omega_c t} = \frac{at_0}{\sqrt{t_0^2+i\beta'' z}}\exp\left[i(\omega_c t-k_c z)\right] \exp\left[-\frac{1}{2} \left( t - \frac{z}{v_g}  \right)^2 \frac{1}{t_0^2+i\beta'' z} \right]
\label{pressurealmost}
\end{equation}
We introduce a traveling wave time coordinate that moves at group velocity, just like in Eq. \ref{traveltau}
\begin{equation}
	\tau = t - \frac{z}{v_g} 
\end{equation}
which is equivalent to the group delay of the envelope. Setting $u = \beta''z/2$, we can tidy up Eq. \ref{pressurealmost}
\begin{equation}
	p(z,t) = at_0\sqrt{\frac{t_0^2-2iu}{t_0^4+4u^2}}\exp\left[i(\omega_c t-k_c z)\right] \exp\left(
-\frac{t_0^2}{t_0^4+4u^2}\frac{\tau^2 }{2}\right) \exp \left( \frac{2iu}{t_0^4+4u^2}\frac{\tau^2 }{2} \right)
\label{pressureclear}
\end{equation}
Following \citet{New2011}, the quadratic term was rearranged so to bring it to the explicit form of a complex Gaussian, as will be standardized in \cref{PulseCalc}, which facilitates the discussion about a pulse defined by a real envelope and a linear chirp. The complete pressure expression of Eq. (\ref{pressureclear}) has four terms, all of which provide valuable insight about the process of pulse dispersion. First, the narrowband carrier $\omega_c$ is intact and moves at phase velocity $v_p = \omega_c/k_c$. All other terms are impacted by the GVD through the propagation variable $u$, which is proportional to the distance from $z=0$. The Gaussian pulse itself moves at group velocity and becomes broader with distance, because $t_0^{'2}=(t_0^4 + 4u^2)/t_0^2 > t_0^2$. The amplitude of the pulse decreases with distance approximately as $t_0/t_0'$, but also becomes phase shifted given the imaginary amplitude that depends on $u$. The last phase term is a linear chirp that is superimposed on the carrier---a quadratic phase term that quickly varies with $\tau$. The last term therefore causes the instantaneous frequency around $\omega_c$ to be time and space dependent, with a frequency slope of $m'= 2u/(t_0^4+4u^2)$ (see Eq. \ref{GeneralizedPhase}). All in all, the effect of group dispersion is to introduce both amplitude and frequency modulations, which generally deform and smear the original modulations. An instructive example of a reference pulse at $z=0$ compared with a dispersed version of itself a little later is drawn in the left plot of Figure \ref{fig:dispersion}, where all the effects mentioned above are visible. 

It is instructive to look at three limiting cases. First, when there is no group-velocity dispersion, so $\beta''=0$ and  $u=0$, the pulse shape and energy do not change or acquire a chirp. Second, critically, when we have a pure tone, then the time signal has an infinite width and energy. We can test the effect by setting $t_0 \rightarrow \infty$. Obviously, in this case there is no longer a pulse, but also no chirping takes place. Therefore, pure tones do not chirp. The third limiting case occurs when the pulse is instantaneous---a delta function---so $t_0 \rightarrow 0$ and the chirping effect happens instantaneously as a transient that depends only on $u$. This behavior resembles the effect of a single reflection, which is not dispersive by strict definition, but can cause pulse broadening and phase shifts much like propagation in a dispersive medium (\cref{Reflections}). 

It is also important to note that not only are pure tones unsuitable for demonstrating group-velocity dispersion, but also the entire concept of a single frequency is undetermined when dealing with real modulations. This is the reason why we preferred to refer to Akhmarov's original ``paraxial'' dispersion equation (Eq. \ref{eq:dispersionabs}) as \term{paratonal} instead. The prefix ``para'' is used in the sense of ``\textit{closely resembling: almost}\footnote{From Merriam-Webster's dictionary: \url{https://www.merriam-webster.com/dictionary/para}.}'', so paratonal preserves the identity that is associated with the channel---its carrier and its corresponding pure tone, and the relative narrow bandwidth that is required.

\section{Group-velocity absorption}
\label{GroupABS}
Similar effects to dispersion are observed by an analogous manipulation of the absorption \citep[p. 335]{Siegman}, which can be explored by examining a strictly imaginary dispersion relation $k(\omega) = i\left[\alpha_0 + \alpha'(\omega- \omega_c) + \frac{\alpha''}{2}(\omega- \omega_c)^2 \right]$. Due to the similarity, absorption is sometimes referred to as gain dispersion, but with opposite signs (\citealp[356-360]{Siegman}; \citealp[283-287]{Haus}), so when any of the $\alpha$'s are positive, there is gain. In most cases (e.g., air, saturated media in lasers) $\alpha'=0$ and only the quadratic effect is of interest. The solution to the paratonal equation \ref{eq:dispersionabs} with Eq. \ref{envelopeT} as input envelope is then a variation on Eq. \ref{pressurealmost}
\begin{equation}
p(z,t) = \frac{at_0}{\sqrt{t_0^2-\alpha'' z}}\exp\left[i(\omega_c t-k_c z)+\alpha_0 z\right] \exp\left(
-\frac{1}{2} \frac{t ^2 }{t_0^2-\alpha'' z} \right) \,\,\,\,\,\,\, \alpha'' < 0
\label{saturated}
\end{equation}
where the condition for negative $\alpha''$ is necessary to make the Fourier transform solution method of \ref{envelopetZ} tractable. The effect on the amplitude of the pulse and the width are apparent directly from the equation without any additional transformation. A uniform attenuation is applied to the entire pulse, which is proportional to the distance, when $\alpha_0<0$. A time/frequency-dependent loss broadens the pulse when $\alpha''<0$, which also scales the amplitude of the pulse accordingly. An example of absorption is given on the right of Figure \ref{fig:dispersion} for similar parameters to the above example. If $\alpha' \neq 0$ then the phase would be affected as well, and manifest as a subtler chirp than the linear chirp caused by the quadratic dispersive terms. However, this can be compounded into a complex group velocity instead, so its effect is less critical. Unlike dispersion, absorption is accompanied by energy loss. 

\section{Complex pulse calculus}
\label{PulseCalc}
This section presents several useful formulas for working with complex Gaussian functions, which are used throughout the text. Additionally, it aims to provide some intuition for these functions---whether they are used as signal envelopes or as filters.

The fundamental sound pulse used in this work---the sound atom, or the logon---is the complex Gaussian function that modulates an arbitrary carrier. Its real part in the exponential represents the energy and width or the pulse, whereas the imaginary part imparts the signal with quadratic phase\footnote{Note that unlike Gabor's logons \citep{Gabor}, the envelope used here is complex rather than real.}. Strictly speaking, the Gaussian makes for non-analytic signals, because it has a non-zero negative spectrum. Exactly the same functions are also used as filters in the imaging transforms that are employed throughout the text, which are similarly non-causal because they do not vanish at $t<0$. Nevertheless, the complex Gaussian function enables closed-form solutions for all the relevant equations and is characterized by the same defining features as other relevant signals in analysis \citep{Siegman,Zverev}. Critically, it is the simplest second-order curved signal/filter that has a characteristic duration and chirp (curvature), which enables instantaneous amplitude and frequency modulations, respectively. 

The general class of pulses we consider contains a chirp at the source, which is the simplest type of frequency modulation (FM). It is easy to see how it directly interacts with the group-velocity dispersion of the medium. Consider a signal with the complex envelope
\begin{equation}
	a(0,t) = a\exp \left( -\frac{t^2}{2t_0^2} + \frac{im_0t^2}{2}  \right)
	\label{BasicAMFM}
\end{equation}
where $m_0$ is the frequency velocity---the slope of the instantaneous frequency (Eq. \ref{GeneralizedPhase}) of the signal with envelope $a(0,t)$. As before, we can obtain a full solution to the dispersion problem. For this purpose, it is convenient to make the following substitution\footnote{A similar approach was developed by \citet[Chapter 9]{Siegman}, where a complex Gaussian parameter was defined as $\Gamma \equiv a-ib$, where $a=1/2t_0^2$ denotes the width and $b$ denotes frequency slope. A geometrical representation of $\Gamma$ is then investigated in the spectral domain, as a function of dispersion and the resultant pulse compression.} of a complex Gaussian width $t'$,
\begin{equation}
	\frac{1}{t^{'2}} = \frac{1}{t_0^2} - im_0 = \frac{1-im_0 t_0^2}{t_0^2} \,\,\,\, \Rightarrow \,\,\,\, t^{'2} = \frac{t_0^2}{1-im_0 t_0^2} = t_0^2\frac{1+im_0 t_0^2}{1+m_0^2 t_0^4}
\label{complext}
\end{equation}
By using $t'$ directly in Eq. \ref{pressurealmost}, we can obtain the full signal at $z$ after propagating through the dispersive medium
\begin{multline}
p(z,t) = \frac{t_0^2(1 + im_0t_0^2)}{1+m^2t_0^4} \sqrt{\frac{t_0^2(1 + um_0) - i(2um_0^2t_0^4+m_0 t_0^4+2u)}{(1+2m_0u)^2t_0^4+4u^2}}\exp\left[i(\omega_c t-k_c z)\right] \\
\cdot \exp\left[-\frac{t_0^2(1 + um_0)}{(1+2m_0u)^2t_0^4+4u^2}\frac{\tau^2 }{2}\right] \exp\left[\frac{i(2um_0^2t_0^4+m_0 t_0^4+2u)}{(1+2m_0u)^2t_0^4+4u^2}\frac{\tau^2 }{2}\right] 
\label{pressureclearchirp}
\end{multline}
which indicates that the dispersed chirp slope depends on $m_0$, $u$ and on $t_0$ itself.

The eventual filtering as is given in Eqs. \ref{pressureclearchirp} and \ref{saturated} is most readily treated as a multiplication of several Gaussian elements, which are then governed by relatively simple operations of scaling. This operation is also common in imperfect imaging, however, which is implicated by chirping as a result of defocus. Thus, we would like to find out how two arbitrary complex Gaussian functions of the form of Eq. \ref{BasicAMFM} interact under multiplication. This is encountered in the (Fourier-transformed) convolution of the input envelope spectrum with the imaging system transfer function, or equivalently, when a temporal envelope is processed by a time-domain phase modulator (i.e., a time lens). Let us look at two complex Gaussians with constant amplitudes, $a_1$ and $a_2$, characterized by parameters $t_1$ and $m_1$, and $t_2$ and $m_2$, respectively, or the complex widths $t'_1$ and $t'_2$. The product of the two complex Gaussians is
\begin{multline}
a_1(\tau)a_2(\tau) = a_1a_2\exp\left( -\frac{\tau^2}{2t_1^2} + \frac{im_1\tau^2}{2} \right)\exp\left( -\frac{\tau^2}{2t_2^2} + \frac{im_2\tau^2}{2} \right)\\
= a_1a_2\exp\left[ -\frac{(t_1^2 + t_2^2)}{2t_1^2 t_2^2}\tau^2 + \frac{i(m_1+m_2)}{2}\tau^2 \right] = a_1a_2\exp\left[ -\frac{(t_1^{'2} + t_2^{'2})}{2t_1^{'2} t_2^{'2}}\tau^2 \right]
\label{GaussProduct}
\end{multline}
Importantly, the third equality shows that the slopes of the linear chirps are additive under modulation, whereas for the pulse width, it is the reciprocal of the width squared that is additive. 

Another recurrent operation on the envelope is scaling, where the pulse is being magnified by a factor $M$ as a result of imaging. In this case, the envelope time variable is substituted by $\tau/M$ (see Eq. \ref{eq:image})
\begin{equation}
a\left(\frac{\tau}{M}\right) = \frac{a_0}{\sqrt{M}}\exp \left(-\frac{\tau^2}{2M^2t_0^{'2}} \right) = \frac{a_0}{\sqrt{M}}\exp\left( -\frac{\tau^2}{2M^2t_0^2} + \frac{im_0\tau^2}{2M^2} \right)
\end{equation}
which means that the pulse is magnified, while its chirp rate is demagnified (for $M>1$)
\begin{equation}
t_0 \xrightarrow{\text{M}} Mt_0
\end{equation}
\begin{equation}
m_0 \xrightarrow{\text{M}} \frac{m_0}{M^2}
\end{equation}
\begin{equation}
a_0 \xrightarrow{\text{M}} \frac{a_0}{\sqrt M}
\end{equation}
where the last relation follows from the first one. 

In some cases, the complex width $t_1'$ of a particular transform is available numerically, but it is necessary to obtain the underlying width and chirp factors. Bringing it to the form of \ref{complext}, this can be done using these expressions
\begin{equation}
t_1 = \left[\Re\left(\frac{1}{t^{'2}_1}\right)\right]^{-\frac{1}{2}} 
\end{equation}
\begin{equation}
m_1 = -\Im\left({\frac{1}{t_1^{'2}}}\right) 
\label{unknownpulseparams}
\end{equation}
A common and somewhat more realistic linear FM pulse that is often employed \citep[p. 57]{Levanon} has a rectangular envelope of width $T$
\begin{equation}
a(\tau) = \frac{a_0}{\sqrt{T}}\rect \left( \frac{t}{T} \right) \exp\left( i\pi m_r\tau^2 \right) \,\,\,\,\,\,\,\,\, m_r = \pm\frac{B}{T}
\label{rectchirp}
\end{equation}
where the frequency slope $m_r$ is by definition the quotient of the bandwidth $B$ and pulse width $T$. By comparing the coefficients of our complex pulse from Eq. \ref{BasicAMFM}, a simple transformation between the slopes can be obtained
\begin{equation}
m_0 = 2\pi m_r = 2\pi\frac{B}{T}
\label{slopeconv}
\end{equation}
This relation can provide some insight about the magnitude of $m_0$ of the Gaussian pulse. When the pulse is made narrow, $T \rightarrow 0$, its associated frequency velocity can be very large, depending on the bandwidth $B$. If in addition the spectral bandwidth is made very large, $B \rightarrow \infty$, the pulse functionally approximates a delta function in the temporal domain. Therefore, the closer the pulse is to a perfect impulse, the larger is its frequency velocity. If the impulse does not disperse much in propagation, then its impulse shape is retained, which means that it is experienced almost instantaneously across the spectrum. 

\begin{figure} 
		\centering
		\includegraphics[width=0.5\linewidth]{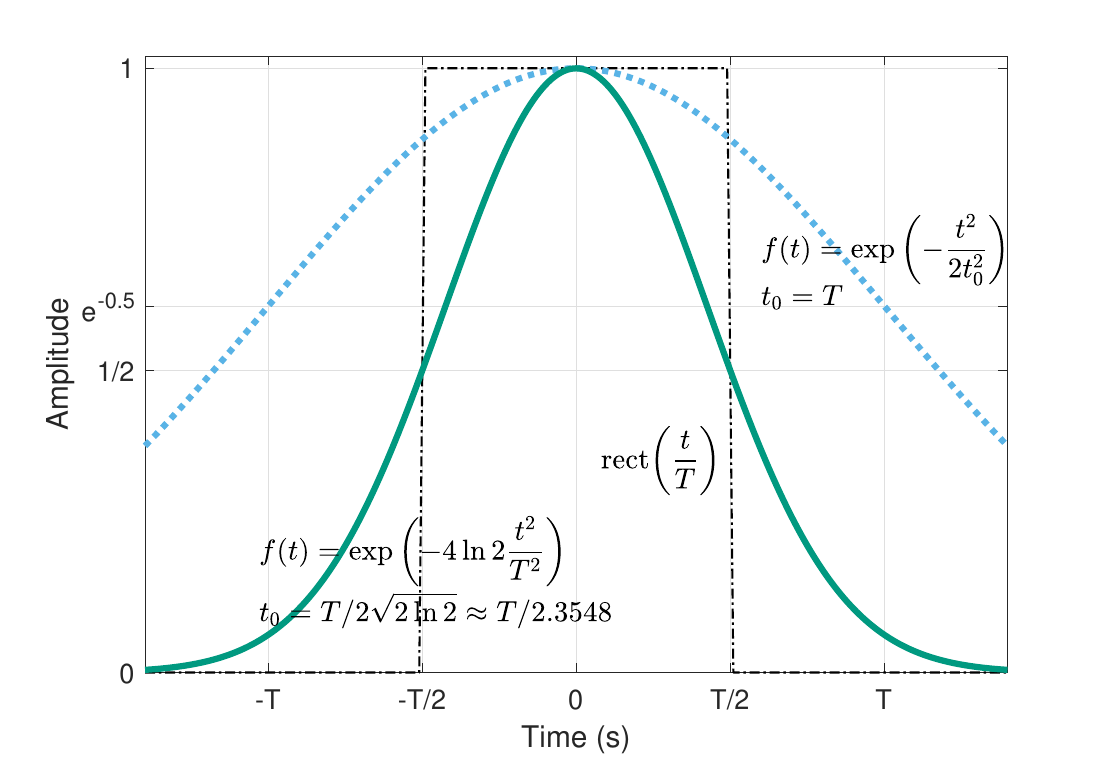}	
		\caption{A rectangular pulse (dash-dot black) of duration T and two Gaussian pulses of different widths $t_0$ relative to T. The amplitude and time are given in arbitrary units, so that the rectangular pulse has an area of T, whereas the Gaussian pulses are not normalized. The broad Gaussian in dotted blue has the standard width of $2t_0^2$ in the denominator of the exponent, which indicates that when $t=T$, the amplitude drops to $1/\sqrt{e}$ and the intensity to $1/e$. Typically, this is not useful when working with $t_0$, so it is preferable to convert it to equivalent rectangular duration, which is measured across both the negative and positive support of the pulse. The standard rectangular pulse and the Gaussian that intersects with it at $T/2$ have the same width. Then, the conversion between the two is given by the FWHM.}
		\label{fwhm} 
\end{figure}

Finally, depending on the application, the effective Gaussian pulse width has to be scaled to constrain its infinite support, but still preserve some of its characteristics. In photonics, pulses are customarily quantified using the full-width half maximum (FWHM) of the particular pulse function. It is defined with respect to the pulse power, using $t_0$ as a parameter, so at half the peak amplitude (quarter the peak power) $t = \FWHM \cdot t_{0}=2\sqrt{2\ln2}t_0$ \citep[e.g.,][p. 120]{New2011}. See Figure \ref{fwhm} for illustration.

\begin{figure} 
		\centering
		\includegraphics[width=0.5\linewidth]{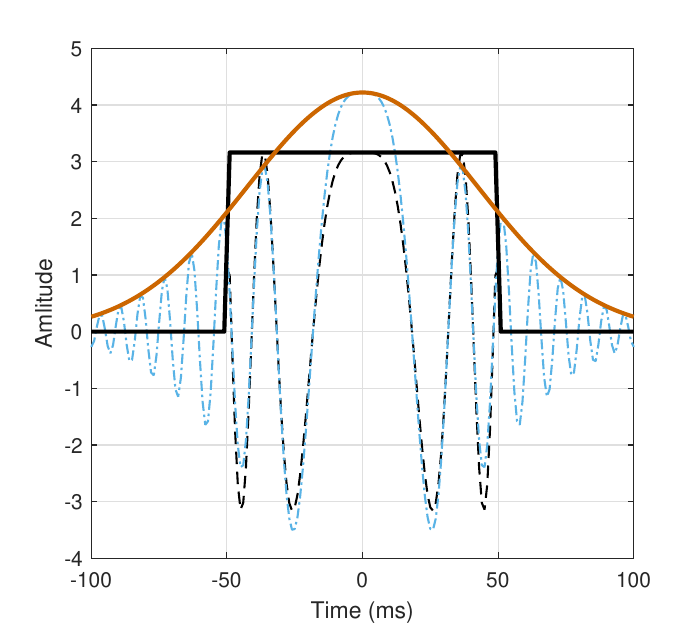}	
		\caption{Rectangular and Gaussian pulses that have the same instantaneous frequency in the overlapping duration, according to the substitutions in Eqs. \ref{rectchirp} and \ref{slopeconv}, where $B = 150$ Hz and $T = 0.1$ s (of the full rectangular width). The Gaussian has a $t_0 = T/(2 \sqrt{2\ln 2})$. In this example, the pulses are normalized to have equal total equal power.}
		\label{fig:rectgauss}
\end{figure}

Another example is plotted in Figure \ref{fig:rectgauss} of complex Gaussian and rectangular pulses that have the same instantaneous frequency in the overlapping duration of the rectangular pulse. The Gaussian width is set to intersect the rect function so that the two pulses have equal power. Note that it is impossible to have the total pulse power, the peak amplitude, and the half power simultaneously equalized. 


%% file: LCT_appendix.tex
\chapter{Linear canonical transform approach to the dispersion integral}
\label{AppLCT}

A general approach to a solution of the imaging transform in Eq. \ref{totalpath} is to reformulate the dispersion integral of the envelope as a \term{linear canonical transform} (LCT), similar to the Frensel-Kirchhoff diffraction integral in optics. In one-dimensional imaging over the time-frequency plane, this transform has the general kernel \citep[e.g., ][Chapter 9]{Wolf}
\begin{equation}
C_T(q,q') = \frac{1}{\sqrt{2\pi b}}e^{{-i\pi} \over {4}} \exp\left[ {{i} \over {2b}} (aq^2-2q'q+dq'^2)\right] \,\,\,\,\,\,\,\,\,\, b \neq 0
\label{LCT}
\end{equation}
From this, the LCT of function $g(q)$ is
\begin{equation}
g_T(q') = {\mathscr L}(T) g(q) \equiv \int_{-\infty}^\infty g(q)C_T(q,q')dq
\label{LCT2}
\end{equation}
where the generic variables $q$ and $q'$ can be assigned the roles of either $f$ or $t$. Depending on the coefficients $a$, $b$, and $d$, the transform rotates the function $g(q)$ in the time-frequency phase plane (as can be obtained by the Wigner-Ville distribution, for example) at an arbitrary angle. Many of the LCT properties are identical to the Fourier transform, which is a special case when the phase rotation angle is $\pi/2$, or $-\pi/2$ for the inverse Fourier transform. The advantage of conforming to the LCT formulation is that it offers a sophisticated toolkit based on operator and matrix algebra, in which a cascade of operations on the input (transforms) can be realized and interpreted without having to solve the integrals explicitly (see for example, \citealp[Chapters 4, 12, and Appendix A]{Shamir}). By comparing the coefficients of Eqs. (\ref{LCT}) and (\ref{totalpath}), the parameters $a$, $b$, and $d$ can be found
\begin{equation}
a = 2\left(\frac{uv}{s}+v+u \right)  \,\,\,\,\,\,\,\,\,\,\,
b = -\frac{v+s}{s} \,\,\,\,\,\,\,\,\,\,\,
d = -\frac{1}{2s}
\label{abd}
\end{equation}
assigning $q=\omega$ and $q'=\tau$. The transformation $T$ shall be defined as
\begin{equation}
T = \left[ \begin{array}{l}
a\,\,\,\,\,b\\
0\,\,\,\,\,d
\end{array} \right]
\label{Transformation}
\end{equation}
And Eq. (\ref{totalpath}) can be rewritten as an LCT, only correcting for the amplitude by $e^{{i\pi} \over {4}} /\sqrt{2\pi}$
\begin{equation}
a_n(\zeta _2 ,\tau) = \frac{e^{{i\pi} \over {4}}}{\sqrt{2\pi}}{\mathscr L}(T)\left[A(0,\omega ') \right]
\label{totalpathLCT}
\end{equation}
Such a general transformation can always be decomposed into three cascaded operations (but other decompositions exist): a fractional Fourier transform that rotates the phase space, a magnifying operation, and a quadratic-phase modulation \citep{Bastiaans}. 

%% file: rectimpulse.tex
\chapter{Impulse response of rectangular pupil with positive defocus}
\label{RectImp}
The rectangular pupil can be informative as a limiting case, where extreme time-limitation of the object support is required. It can be solved in a closed form following the formulas provided in \citet{Klauder1960}, but only for positive values of the defocus $W_d$. Let us define a rectangular pupil $P_r$ of width $T_a$, 
\begin{equation}
P_r(\tau ) = \rect\left(\frac{\tau}{T_a}\right)  = \left\{ \begin{array}{l}
1\,\,\,\,\,\,\,\,|\tau | < T_a/2\\
\frac{1}{2}\,\,\,\,\,\,\,|\tau | = T_a/2\\
0\,\,\,\,\,\,\,\,|\tau | > T_a/2
\end{array} \right.\
\end{equation}
where $T_a$ is equal here and in the Gaussian pupil, because of how we defined it in Eq. \ref{GaussPupil}. Using this pupil in Eq. \ref{impresponse4}, we obtain the following integral
\begin{equation}
\tilde h_{dr}(\tau-\tilde\tau_0) = \frac{1}{2\pi} \exp\left( \frac{i\omega_c\tau^2}{2Mf_T} \right) \int_{-T_a/2}^{T_a/2} \exp ( iv^2 W_d \tilde T^2) \exp \left[ { - i\tilde T (\tau  - \tilde\tau_0 )} \right] d\tilde T
\label{impresponseRect1}
\end{equation}
Then, the response would be given according to
\begin{equation}
\tilde h_{dr}(\tau-\tilde\tau_0) = \frac{1}{2\pi} \exp\left( \frac{i\omega_c\tau^2}{2Mf_T} \right) \sqrt{\frac{\pi}{2v^2 W_d}}\exp \left[-\frac{i(\tau-\tau_0)^2}{4\pi v^2 W_d} \right]\left[C(g_2) + iS(g_2) - C(g_1)-iS(g_1) \right]
\label{impresponseRect2}
\end{equation}
Where $C$ and $S$ are the real and imaginary parts of the complex Fresnel integrals, respectively, which are defined as
\begin{equation}
S(g) =  \int_0 ^g \sin (\psi^2) d\psi
\label{FresnelS}
\end{equation}
\begin{equation}
C(g) =  \int_0 ^g \cos (\psi^2) d\psi
\label{FresnelC}
\end{equation}
The variables $g_1$ and $g_2$ are defined as
\begin{equation}
g_1(\tau) = \frac{\tau-\tilde\tau_0}{\sqrt{2\pi v^2W_d}} + \sqrt{\frac{v^2W_dT_a^2}{2\pi}}
\label{FresnelG1}
\end{equation}
\begin{equation}
g_2(\tau) = \frac{\tau-\tilde\tau_0}{\sqrt{2\pi v^2W_d}} - \sqrt{\frac{v^2W_dT_a^2}{2\pi}}
\label{FresnelG1}
\end{equation}
A numerical solution exists only when $g$ is positive, which is unfortunately not the case in the auditory system using the values of $v$, $T_a$, and $W_d$ obtained in this work. 

%% file: Aliasing.tex
\chapter{Evidence of discrete sampling in hearing through aliasing of double- and triple-pulse sequences}
\label{Aliasing}

\textbf{Abstract}
\\
Three experiments are described in which listeners had to count the number of events in sequences of Gabor pulses at carrier frequencies of 6 and 8 kHz. The results were interpreted using a sampler model that allows for aliasing to take place. The model entails that the pulses are effectively sampled at an instantaneous sampling rate, which determines the maximum pulse rate that can be discriminated without ambiguity. Therefore, it provides a basis for perceived confusion between stimuli containing brief sequences of either two or three pulses, which is not readily explained using standard temporal integration models. The calculated instantaneous sampling rates are compared to known physiological spiking rates in the auditory nerve, which reveals an onset effect and temporal acuity adaptation. The addition of off-frequency notched broadband noise is shown to affect only a subset of the listeners.

\section{Introduction}
\label{AliasIntro}
The experience of sound perception is seamless. Tones, in particular, sound smooth with no breaks or gaps, which is in accord with the classical physical acoustical wave description of sound sources. However, beyond the cochlea, sound is transduced to neural spikes, which on their face appear discrete. Nevertheless, manys auditory models have treated sound as continuous also beyond the auditory nerve. Given the number of simultaneously active auditory nerve fibers in every auditory channel, an effectively continuous sound sensation may arguably have a physiological basis. At the same time, several temporal models suggested that a discrete description is more correct, which can also provide an intuitive explanation for apparent discontinuities in sounds, as are evident from gap-detection experiments. The basic question remains, though: is hearing continuous or discrete? If the auditory system is truly discrete, then certain sampling-theoretical constraints should apply, which have not been rigorously considered in the context of hearing, although they may have perceptual effects. 

\subsection{Continuous and discrete auditory temporal models}
\label{ModelsIntro}
A large class of continuous temporal auditory models is originally due to \citet{Munson1947} and \citet{Zwislocki1960,Zwislocki1969} and are variably referred to as sliding temporal window \citep{Penner1975} or leaky integrator models \citep{Viemeister1979}. While these models typically acknowledge that they simplify temporal effects that have neural origin, they do not point to a specific location where these effects take place within the central auditory system. These models can account well for some of the results in  temporal acuity experiments, which include temporal integration (forward masking threshold decay), brief increments or decrements in tone intensity, and (not so well) broadband temporal modulation transfer function cutoff frequencies \citep{Moore1988,Oxenham1994}. These continuous models generally include variations on four basic components: cochlear bandpass filtering, a nonlinearity (attributed to both dynamic range compression and neural transduction), low-pass filtering, and a decision device \citep[pp. 183--189]{Moore2013}. Another type of continuous-processing temporal model hypothesizes a central modulation filter bank that processes the auditory signals and can isolate specific temporal patterns within the individual filter bandwidths \citep{Dau1997a,Dau1997b}.

A second class of models hypothesizes a discrete sampling window rather than a continuous sliding window. \citet{Viemeister1991} proposed the multiple-look model that accounts for short pulses that are separated by more than 5 ms, which do not show apparent power integration between one another. According to this model, the auditory system acquires the samples and stores them in short-term memory, where they can be integrated using a longer time constant that is associated with the memory itself. The multiple-look model can account for some temporal integration effects, in addition to the gap detection experiments that are readily understood with a discrete framework \citep[e.g.,][]{Hofman1998,Hoglund2009}. However, the multiple-look model prescribes equal weighting for the looks and no differentiation of masker regularity and therefore was unsuccessful in predicting the effects of comodulation masking release \citep{Buus1999}, informational masking release \citep{Kidd2003}, and continuous or pulsed tonal stimuli masked by noise \citep{Wright2021}. Nevertheless, more general, successful auditory models include sampling in a way that does not claim to adhere to the particular multiple-look model \citep{Patterson1992,Lyon2018}.

Both continuous and discrete models are not universally successful in their original form partly because of the difficulty to know how to account for more complex stimuli. The discrete model has been especially problematic. For example, \citet{Buus1999} could not account for coherent comodulation masking release effects using the multiple-look model, which hypothesizes that the looks are incoherently summed as each sample is represented by its intensity only. In another case, the release from informational masking improved with the number of signal and masker bursts in the sequence, but deteriorated when the inter-burst intervals were increased \citep{Kidd2003}. In yet another experiment, unpredictability of the temporal structure of continuous or pulsed tonal stimuli masked by noise was poorly accounted for by the multiple-look model that should have been sensitive to the signal duration, whereas a continuous temporal integration yielded a much better prediction \citep{Wright2021}. These results could not be accounted for by the multiple-look model, which prescribes equal weighting for the looks and no differentiation of masker regularity. However, these failures of the model implementation for experiments that explored effects in the tens or hundreds of milliseconds ranges do not discredit the hypothesis that a discretized representation exists on the millisecond range. All these studies (including the original one by \citealp{Viemeister1991}) applied an incoherent and energetic summation of the looks that discards all phase information. Furthermore, complex processing and information management within and across channels for stimuli as complex as were presented in the informational masking test preclude the application of static temporal processing models, whether they are discrete or continuous, so the continuous temporal integration models are not expected to perform much better\footnote{Note that this conclusion is generalized here to the discrete model by \citet{HeilMatysiak2017}, as it shares a similar logic to the multiple-look model, although the examples below have not been tested against this more recent model.}.

Relatively few physiological models explicitly embraced the idea of discrete processing in the auditory brain. In fact, in the very first temporal processing model by \citet{Munson1947}, he explicitly modeled the loudness response as an integrated measure of auditory nerve spikes---each of which represents an ``elemental quantum'' of loudness: ``\textit{each pulse of the action potential... mediates a small elemental contribution to the magnitude of the sensation experienced, and that as time elapses after its advent, the effectiveness of the element diminishes.}'' More recently, \citet{HeilMatysiak2017} proposed a probabilistic model with some parallels to the multiple-look model, but also with more constraints. These include modeling the ``sensory event'' (spiking) detection of the signal envelope as a Poisson point process and considering the spontaneous activity in the auditory nerve with no acoustic input. This model can produce the same results as the classical temporal integration models, but also accounts for more complex threshold effects of several masking experiments in humans and animals.

Other approaches to signal processing in hearing applied the concept of sampling more centrally to modeling, usually by assuming sampling at the level of the auditory nerve. \citet{Lewis1995} and \citet{Yamada1999} referred to the noise from the high spontaneous rate auditory nerve fibers as performing dithering\footnote{Dithering is smoothing of sampling fluctuations, which are caused by the minimum quantization level (its finite resolution), through the addition of random low-level noise.}---a term that is normally used only in the context of sampling and conversion between digital and analog signal representations. A more specific mechanism of sampling was considered by \citet{HeilIrvine1997} and \citet{Heil2003}, where the auditory nerve coding of the onset of temporal envelopes was modeled as equivalent to point-by-point sampling of the envelope function, which tracks it at high resolution, limited by the spike/sampling rate. Another neural processing model makes use of the concept of stochastic undersampling to show how deafferentation of the auditory nerve is analogous to noise \citep{Poveda2013, Poveda2014}. This model has some parallels to the classical volley principle, whereby the acoustic input is adequately sampled (or even oversampled) by a population of neural fibers, each of which by itself undersamples the signal \citep{Wever1930Present}. 

Similar ideas were sometimes attributed to higher-level nuclei such as the brainstem. \citet{Warchol1990} suggested that high spontaneous rates in the avian auditory cochlear nucleus enable better sampling of the stimulus. In another signal processing auditory model, \citet{Yang1992} noted that the anteroventral cochlear nucleus (AVCN) receives inputs from the auditory nerve, which could be instantaneously mismatched and then lead to effective lateral inhibition. This perspective may be interpreted as another form of nonuniformity in the sampling that exists beyond the stochastic auditory nerve spiking pattern. Further downstream, \citet{Poeppel2003} suggested that the two auditory cortices work by asymmetrically sampling the incoming sound---the left hemisphere samples the auditory cortex at around 40 Hz, and the right hemisphere at 4--10 Hz. 
Additional auditory signal processing models exist that were inspired by nonuniform or irregular sampling of wavelet frames, but whose exact physiological correlate was not made explicit \citep{Yang1992,Benedetto1993}. Independently of the various auditory models, a recent paper attempted to find evidence for discrete auditory representation of sound in the brain \citep{VanRullen2014}. It concluded that hearing, unlike vision (see below), is not discrete on the subcortical levels, although it might be discrete on a cortical, or specifically attentional, level. However, the methods that were used to reach these conclusions were somewhat arbitrary and tended to conflate the carrier and modulation domains of broadband stimuli (including in the comparison between hearing and vision). These make the conclusion of excluding discrete subcortical mechanisms somewhat tenuous. 

Sampling in the spectral domain of the spectral envelope was also considered in the context of a model for vowel identification, which can be degraded when the harmonic content is rich and the fundamental frequency is high, because of spectral undersampling and resultant aliasing distortion \citep{deCheveigne1999}. The model was also formulated in the temporal domain using autocorrelation, which may have a physiological correlate. More generally, the model was applied for pitch perception as well \citep{Cheveigne}.

\begin{figure} 
		\centering
		\includegraphics[width=1\linewidth]{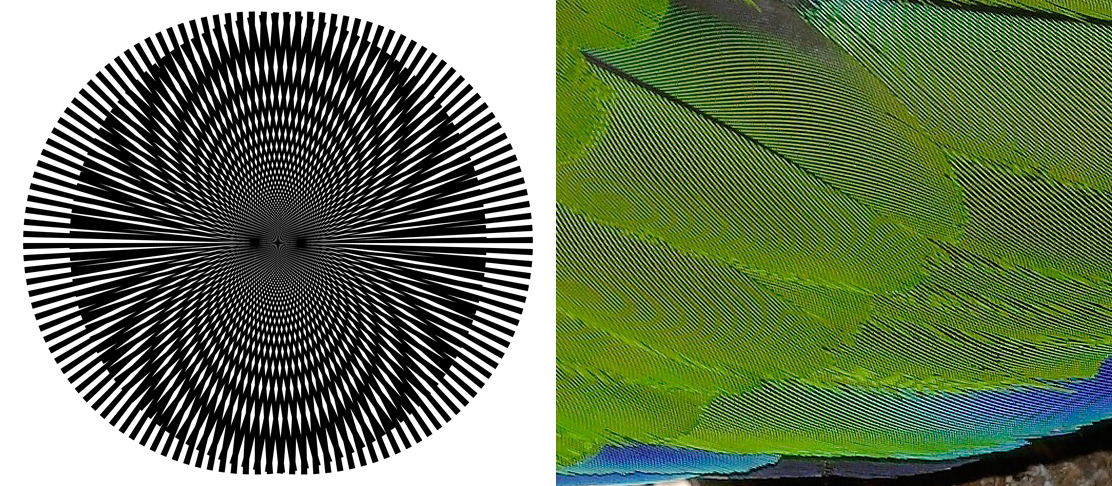}	
		\caption{Examples of Moir\'e patterns. \textbf{Left}: Curved Moir\'e pattern formed by two spoked wheels with different angular frequencies (image by SharkD, \url{https://en.wikipedia.org/wiki/Moir\%C3\%A9_pattern\#/media/File:Moire_Lines.svg}). \textbf{Right}: Closeup image of parrot feathers (image by Fir0002/Flagstaffotos, \url{https://en.wikipedia.org/wiki/Moir\%C3\%A9_pattern\#/media/File:Moire_on_parrot_feathers.jpg}).}
		\label{Moire1}
\end{figure}

The fundamental question about the sampling nature of sensation has received considerably more attention in vision, where sampling effects are modeled both in the spatial and in the temporal dimensions. In the spatial domain of vision, aliasing can be caused when the object contains high frequencies that are imaged by the photoreceptors of the retinal cone mosaic, which are separated by finite distances \citep[pp. 71--85]{Williams2003Chalupa,PackerShevell}. The maximum spatial frequencies that can be imaged may be calculated from the two-dimensional Nyquist rate of the mosaic. Additionally, neural mechanisms in the retina may not be capable of coding higher frequencies. Therefore, high spatial frequencies may be perceived in reality, but they cannot be resolved unambiguously. Aliasing can sometimes appear as a Moir\'e pattern, which can be understood as the reproduction of one grating (evenly spaced lines) by a different grating of a similar period that gives rise to a third pattern (\citealp{Rayleigh1874}; see also \citealp[pp. 48--50]{Amidror}). See Figure \ref{Moire1} for examples. 
In the temporal visual domain, an illusion of a continuously moving image can be generated by projecting sequences of still images at low frame rates. It suggests that the continuous perception in vision may be the result of processing of series of discrete snapshots. This idea is not universally accepted in vision \citep[e.g.,][]{Kline2008}, but has been repeatedly considered \citep[e.g.,][]{Andrews2005, Simpson2005, VanRullen2014}. When the frame rate of the moving image is slower than, or approximately equal to, the ``refresh rate'' of the visual system, a perceptual flicker occurs, which is a temporal modulation pattern superimposed on the image \citep{Kelly1972Jameson}. Some flicker types can be interpreted as temporal aliasing, in which the sampling generated by the visual system does not overlap the discontinuous objects presented to the eyes. The mismatching rates and the lack of anti-aliasing filtering\footnote{In system engineering, a well-designed analog-to-digital converter has to include an anti-aliasing (low-pass) filter, whose purpose is to remove high frequencies from the broadband input that are above half the sampling rate, which would otherwise be shifted downwards to frequencies within the passband, creating an aliased (and hence distorted) output (e.g., \citealp[pp. 389--391]{Proakis}).} or long-term image reconstruction mechanism in the visual system may cause noticeable gaps in the perceived images. An early discussion of the analogous idea of auditory flicker caused by tones that are amplitude-modulated at sufficiently high rates was given by \citet[pp. 408--416]{Wever1949}.


\subsection{The present study}
In the present work, we attempt to reexamine the nature of the auditory system---whether it is continuous or discrete---at a fine-grained level of the sampling mechanism, should it exist. First, we hypothesize that if the auditory signal is discrete, then under some conditions it may be possible to evoke sensory aliasing. We assume that there is no such anti-aliasing filter in the auditory system. Using a psychoacoustic counting task, it is possible to elicit an audible confusion in the number of short events (sequences of two and three pulses), which suggests that aliasing may be at play. We infer from the results the bounds of the effective sampling rates of the system, using the Shannon-Nyquist limit. We use these results to estimate what sampling rates are possible under different conditions and find relatively high rates at onsets, which significantly drop after a few milliseconds. These patterns are in agreement with known neural adaptation patterns in the auditory nerve. 

\section{Experiments}
A battery of several mini-experiments was administered in a single session that lasted about 45 minutes per subject. The results are reported as separate conditions of Experiments 1, 2 and 3, for clarity of presentation. The testing began with two training rounds that had the same structure as Experiments 1 and 3 (one round each, see below), but with correct / incorrect feedback for the subject. The testing order was set to Experiments 2, 1, 3, ending on the loud conditions of Experiments 3, 2, and 1 for half the subjects. The other half were tested on the loud conditions of Experiments 1, 2, and 3 first, and then on the normal-level conditions of Experiments 2, 1, and 3. 

\subsection{Experiment 1: Confusion between one, two, and three pulses}
\subsubsection{Introduction}
\begin{figure} 
		\centering
		\includegraphics[width=1\linewidth]{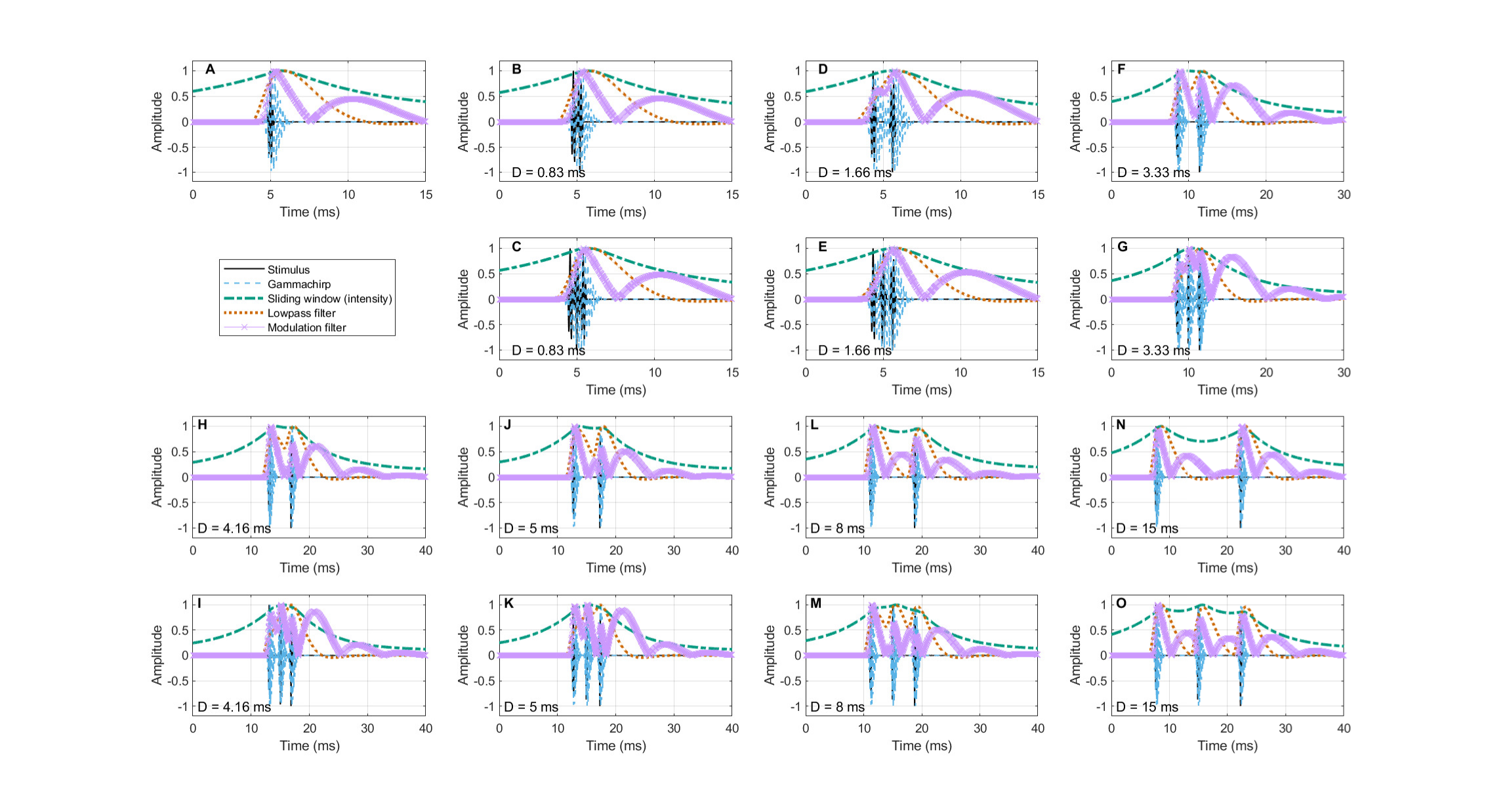}	
		\caption{The output of simple continuous temporal integration models of single, double, and triple Gabor pulse stimuli with 6000 Hz carrier and duration of $W_p = 0.45$ ms per pulse, with seven total durations (0.83, 1.66, 3.33, 4.16, 5, 8, and 15 ms) of both double- and triple-pulse sequences are displayed in plots B--O. The original stimuli are plotted in black solid curves. The blue dashed curves show the output from a fourth-order level-independent gammachirp filter that models the band-pass filtering in the cochlea \citep{Irino1997} that is typically used as the first stage of temporal models. These filters can easily track the pulse envelope for the durations and carrier tested. The green dash-dot curves show the output of the sliding temporal window following an additional nonlinear stage, based on an asymmetrical round-exponential window parameters reported in \citet{Oxenham1994}. They show a smeared response that appears as a single broad pulse in all durations, perhaps with the exception of the longest pulse in plot N that can be interpreted as a double-pulse. The dotted red curves are the output following half-wave rectification, squaring, and low-pass filtering (fourth-order Butterworth with 100 Hz cutoff). The choice of low-pass cutoff frequency determines the degree of smearing in the output of the triple-pulse, which appears as a single-pulse in the shortest durations (plot B--E \& G), ambiguous for some of the double-pulses (plots F, H--K \& M), while it retains the pulse sequence shapes in the longer / slower sequences (plots L, N \& O). The output of the half-wave rectified signal was also modulation-filtered (second-order Butterworth) and is plotted in purple crosses. These modulation-filtering parameters are based on \citet{Moore2009}, using the minimum estimated center frequency (74 Hz, centered logarithmically) and a narrow filter (Q = 1.23), which produces noticeable ringing and gives rise to ambiguous patterns in several cases, including for the single pulse stimulus (plot A).}
		\label{NoAliasing}
\end{figure}

While several studies have looked into the ability of listeners to differentiate between one and two pulses or clicks \citep{Exner1875,Gescheider1966,Williams1972}, none to date specifically examined differentiation between two and three pulses. The distinction between the two sequence types is important, because both continuous (low-pass filtered due to a sliding temporal window) and discrete processing (aliasing) would give rise to confusion between two pulses and a single pulse. However, adding one more pulse to the stimulus affords a more critical benchmark to the discrete processing, aliasing hypothesis. With aliasing, three pulses can be confused with two pulses (three-to-two confusion), whereas the effect of a sliding window (continuous processing) is to smear three pulses into a single broad one, causing a three-to-one confusion. This is illustrated for some of the stimuli used in Experiments 1 and 2 in Figure \ref{NoAliasing}, using approximate continuous temporal models and setting them to produce the most ambiguous output of a triple-pulse---one that might be confused with a double-pulse. In general, the output is either a smeared replica of the input, or a combined single large pulse, when the low-pass frequency cutoff is set sufficiently low and the sequence duration is short (See also \citealp[p. 186, Figure 5.11]{Moore2013}). 

An alternative to the simple sliding window models is to use a modulation filtering temporal model that is low frequency and relatively narrowband \citep{Moore2009}. This model can result in different pulse morphology due to filter ringing, which may be perceived as additional pulses in succession to the input pulses. With the correct timing of the ringing aligned with the periodicity of the pulse sequence, it may be interpreted as a double-pulse when the input is a triple-pulse (Figure \ref{NoAliasing}, G) and as an irregular triple-pulse when the input is a double-pulse (Figure \ref{NoAliasing}, F, H, and J). For these parameters, the single-, double-, and triple-pulses up to duration of 1.66 ms have almost identical morphology, of a single pulse followed by an additional low-energy pulse due to ringing. Pulse sequences of longer durations may sound ambiguous in terms of their numerosity, given their ambiguous morphology. For example, the 8 ms sequences (Figure \ref{NoAliasing}, L and M) might appear as a quadruple-pulse for both double- and triple-pulse inputs, depending on how the extra ringing pulses are perceived / counted. 

\begin{figure} 
		\centering
		\includegraphics[width=1\linewidth]{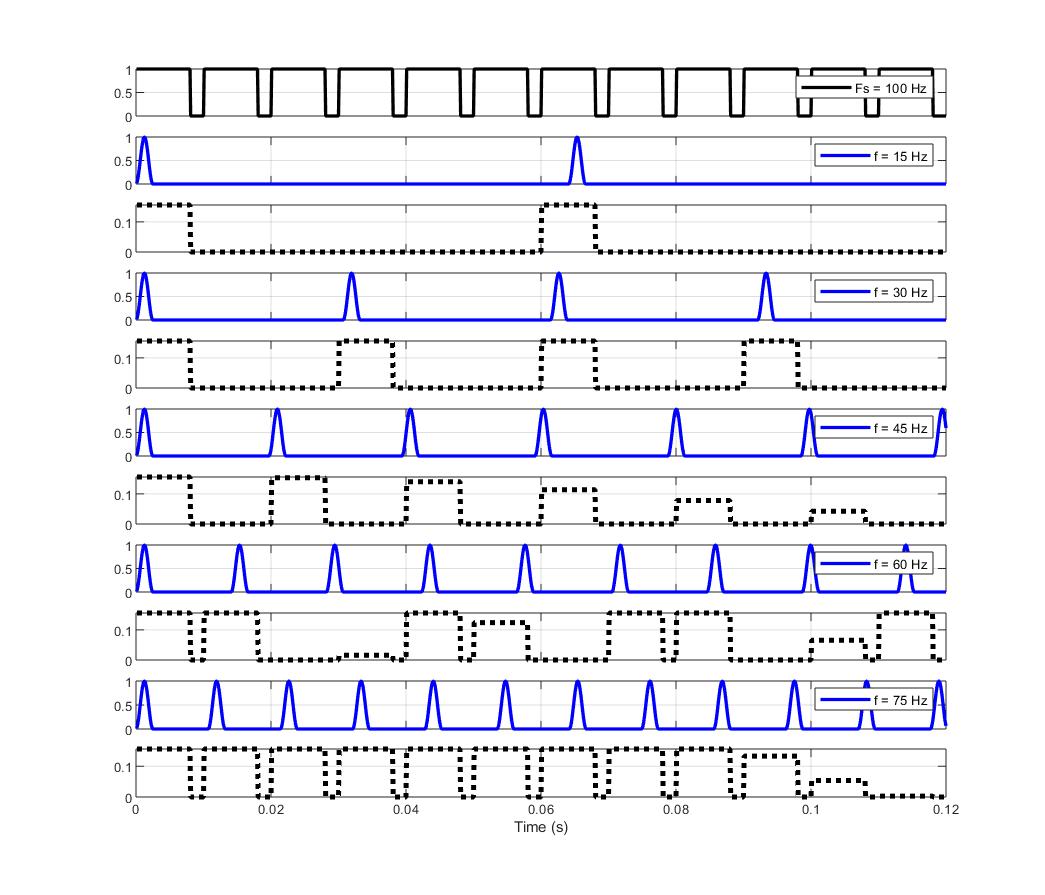}	
		\caption{A cartoon illustration of flat-top sampling (\cref{FlatTop}) using a 100 Hz sampling rate, and rectangular samples with 80\% duty cycle (top curve in solid black). The input are pulse trains at rates of 15, 30, 45, 60, 75 Hz (in solid blue) and their sampled response are in dot black, which illustrates different degrees of aliasing. Frequencies below half the sampling rate show no aliasing, whereas the two highest frequencies exhibit some aliasing, as the signals are undersampled and folded downwards, giving ambiguous (and on average lower) frequencies than the input.}
		\label{AliasDemo}
\end{figure}

Assuming that a sampling mechanism is responsible for capturing all incoming sound, let the instantaneous sampling rate be $f_s$. Using pulse trains as the simplest auditory multi-event available, let the pulse sequence periodicity be $f_p$. The pulse sequence periodicity can be accurately sampled as long as $f_p \leq f_s/2$, according to the sampling theorem \citep{Shannon1948}. If $f_p > f_s/2$ then aliasing will occur, as higher frequencies will appear at lower sampled frequency than their continuous version (after reconstruction). This is illustrated in a cartoon example in Figure \ref{AliasDemo}. 

In the experiments below, the pulse sequence duration $D$ is varied throughout the tests. In each sequence of total duration $D$, either $N=2$ or $3$ pulses can be fitted, which determines the periodicity of the pulse trains. The threshold of aliasing can be estimated by the inequality
\begin{equation}
		\frac{N-2}{D'} \leq \frac{f_s}{2} \leq \frac{N-1}{D'} \,\,\,\,\,\,\,\,\,\,\,\,\, N \geq 1 
		\label{aliasingeq}
\end{equation}
where the duration $D'$ is the duration $D$ corrected for the width of a single pulse $W_p$ that is positioned at the end of the sequence, $D' = D - W_p$. The larger the number of pulses per sequence $N$ is, the more precise are the bounds that contain $f_s$. However, it was determined in pilot testing that counting more than three pulses may be prohibitively difficult for untrained listeners, so in all the following experiments $N \leq 3$. 

In order to test the existence of aliasing, several psychoacoustic experiments were devised. The first experiment tested whether pulses with one, two, and three pulses (referred to throughout the text as single-, double-, and triple-pulses) are confused by listeners. The stimuli were pulses separated by silent gaps of different durations. Gabor pulses of constant carrier and Gaussian envelope were generated and employed to minimize the spectral and temporal smearing, by minimizing the uncertainty product of the very short signals \citep{Gabor}. Another condition was added with a higher carrier frequency, in order test whether the auditory channel center frequency has an effect on the observed sampling rate. The absolute level of the stimulus was modified in yet another condition. The motivation there was that the bandwidth of the auditory filters is known to increase as a function of stimulus intensity \citep{Glasberg2000}. This broadening should have a reciprocal effect in the temporal domain, making the sampling window narrower. The observed effect may be a sharper image of the pulses, where the gaps between them sound more distinct in case of near-aliasing at lower levels. Furthermore, it is known that the auditory nerve fires at a higher rate for inputs of higher intensity, at least when it is below its saturation level \citep{Kiang1965,Liberman1978}. 

\subsubsection{Methods}

\paragraph{Subjects}
Ten subjects with normal pure-tone audiograms ($<20$ dBHL up to 8 kHz, recently assessed) participated in the study---3 female and 7 male, of 23--46 years old. All subjects participated voluntarily after the procedure was explained to them.

\paragraph{Setup}
The experiment took place inside an anechoic chamber. 
A UFX RME sound card (RME Audio AG, Haimhausen, Germany) was used at a sampling rate 48 kHz. Stimuli were generated on MATLAB (The Mathworks, Inc., Natick, MA) and played diotically through Sennheiser HD-25 headphones (Sennheiser Electronic GmbH \& Co. KG, Wedemark, Germany), which were calibrated using a G.R.A.S. Ear Simulator RA0045 (G.R.A.S. Sound \& Vibration A/S, Holte, Denmark), connected to G.R.A.S. microphone preamplifier 26 AC, and Br\"uel \& Kj\ae r amplifier Type 2636 (Br\"uel \& Kj\ae r Sound \& Vibration Measurement A/S, N\ae rum, Denmark). Calibration gains were found for tones of 6 and 8 kHz at 60 dB SPL (root mean square, RMS, levels), but were then multiplied by $\sqrt{2}$ to determine a set value for the pulse amplitudes used in the experiment. The pulses of 6 and 8 kHz were level-equalized using these calibration values.

\paragraph{Stimuli}
Double- or triple-pulse sequences were contained in an initial stimulus length of $D=350$ ms, which could fit either two or three pulses. The pulses were 0.45 ms long (full-width half maximum), so they contained about three carrier periods of 6 or 8 kHz, regardless of the total stimulus duration (see examples in Figure \ref{NoAliasing}). The stimulus level was 60 dB SPL and the 6 kHz stimuli were also presented at 80 dB SPL. As Gaussian pulses minimize the uncertainty relations of $4\pi\Delta f \Delta t \geq 1$ \citep{Gabor}, the bandwidth of the pulse is about 338 Hz for both carriers. 
The carrier frequency was first generated for the entire stimulus duration before it was multiplied by the pulse envelopes (including the gaps), so to keep them in undisrupted phase at all onsets. This was found to yield continuous gap detection psychometric functions, as opposed to independent phase relations between pulses that made the psychometric function discontinuous \citep{Shailer1987}. In each test, single-, double-, or triple-pulse trains were presented in 13 fix total durations $D$ that varied between 0.8--200 ms. The shortest nominal duration entailed that the pulses had no gaps between them, so that stimuli are approximately 0.4, 0.8 and 1.2 ms long, for one, two and three pulse-trains respectively. Obviously, the single-pulse stimuli were identical across the test, regardless of their nominal durations. 

\paragraph{Procedure}
Single-, double-, and triple-pulses were presented to subjects that had to determine how many pulses they heard by pressing the respective digit (1, 2, or 3) on the computer keyboard. Prior to the measurement, there was a training round with correct / incorrect feedback, which was eliminated in the actual test. The presentation order was randomized with respect to the gap and number of pulses, so that each subject was tested once on every number of pulses and duration (total of 39 stimuli per carrier and level condition per subject). Note that the subjects were tested multiple times on the same single-pulse, as it was identical across durations.


\subsubsection{Results}
\label{Sec:Exp2Res}
The confusion matrix in Table \ref{tab:Confusions} summarizes the results of Experiment 1. Four distinct patterns are apparent in the responses, which depend primarily on the duration of the sequences (marked with different shades of gray in Table \ref{tab:Confusions}). When the inter-pulse gaps (the quiet parts of the stimulus between the Gabor pulses) are long, perfect counting is possible with both 6 and 8 kHz carriers. This is the case for pulses that are separated by gaps that are 200 ms at 6 kHz, or longer than 75 ms at 8 kHz. Another response pattern occurs when double- and triple-pulses are confused more or less equally, but are not mistaken for a single pulse. For 6 kHz it happens most clearly between 15 and 3.33 ms, whereas for 8 kHz between 50 and 8 ms. However, hearing a double-pulse instead of triple-pulse becomes more common the shorter the stimuli are. For stimulus durations of 1.66 ms, listeners no longer perceived three pulses, and heard mostly a single- instead of a triple-pulse, although they did sometimes get the double-pulses right. Finally, at the shortest stimulus durations, 0.83 ms, all pulses tend to fuse into one (and indeed there are no gaps in the signal between the Gaussian pulses used here---0.83 ms was the duration of the double-pulse, whereas the triple-pulse was 1.2 ms long), so almost all responses were of a single-pulse. The 80 dB SPL condition data are very similar to the 60 dB SPL data, with a slight tendency for less double- / single- and triple- / single-pulse confusions, which is observed mainly in the short sequence durations of 1.66 ms. 

\begin{table} 
\footnotesize\sf
\centering
\begin{tabular}{|c|c|ccc|ccc|ccc|}
\hline
&&\multicolumn{9}{c|}{\textbf{Pulses heard}}\\\hline
&&\multicolumn{6}{c|}{\textit{Experiment 1}}&\multicolumn{3}{c|}{\textit{Level Condition}}\\\hline
&&\multicolumn{3}{c|}{6 kHz, 60 dB}&\multicolumn{3}{c|}{8 kHz, 60 dB}&\multicolumn{3}{c|}{6 kHz, 80 dB}\\
$D$ (ms)&Stimulus&\textbf{1}&\textbf{2}&\textbf{3}&\textbf{1}&\textbf{2}&\textbf{3}&\textbf{1}&\textbf{2}&\textbf{3}\\\hline
\hline
\multirow{3}{*}{200}&\textbf{1}&\cellcolor{lightgray!90}10&\cellcolor{lightgray!90}0&\cellcolor{lightgray!90}0&\cellcolor{lightgray!90}10&\cellcolor{lightgray!90}0&\cellcolor{lightgray!90}0&\cellcolor{lightgray!90}10&\cellcolor{lightgray!90}0&\cellcolor{lightgray!90}0\\
&\textbf{2}&\cellcolor{lightgray!90}0&\cellcolor{lightgray!90}10&\cellcolor{lightgray!90}0&\cellcolor{lightgray!90}0&\cellcolor{lightgray!90}10&\cellcolor{lightgray!90}0&\cellcolor{lightgray!90}0&\cellcolor{lightgray!90}10&\cellcolor{lightgray!90}0\\
&\textbf{3}&\cellcolor{lightgray!90}0&\cellcolor{lightgray!90}0&\cellcolor{lightgray!90}10&\cellcolor{lightgray!90}0&\cellcolor{lightgray!90}0&\cellcolor{lightgray!90}10&\cellcolor{lightgray!90}0&\cellcolor{lightgray!90}0&\cellcolor{lightgray!90}10\\\hline
\multirow{3}{*}{100}&\textbf{1}&10&0&0&\cellcolor{lightgray!90}10&\cellcolor{lightgray!90}0&\cellcolor{lightgray!90}0&10&0&0\\
&\textbf{2}&2&8&0&\cellcolor{lightgray!90}0&\cellcolor{lightgray!90}10&\cellcolor{lightgray!90}0&0&10&0\\
&\textbf{3}&0&1&9&\cellcolor{lightgray!90}0&\cellcolor{lightgray!90}0&\cellcolor{lightgray!90}10&0&2&8\\\hline
\multirow{3}{*}{75}&\textbf{1}&10&0&0&\cellcolor{lightgray!90}10&\cellcolor{lightgray!90}0&\cellcolor{lightgray!90}0&10&0&0\\
&\textbf{2}&0&10&0&\cellcolor{lightgray!90}0&\cellcolor{lightgray!90}10&\cellcolor{lightgray!90}0&1&9&0\\
&\textbf{3}&0&3&7&\cellcolor{lightgray!90}0&\cellcolor{lightgray!90}0&\cellcolor{lightgray!90}10&0&2&8\\\hline
\multirow{3}{*}{50}&\textbf{1}&10&0&0&\cellcolor{lightgray!50}10&\cellcolor{lightgray!50}0&\cellcolor{lightgray!50}0&\cellcolor{lightgray!50}10&\cellcolor{lightgray!50}0&\cellcolor{lightgray!50}0\\
&\textbf{2}&0&9&1&\cellcolor{lightgray!50}0&\cellcolor{lightgray!50}6&\cellcolor{lightgray!50}4&\cellcolor{lightgray!50}0&\cellcolor{lightgray!50}9&\cellcolor{lightgray!50}1\\
&\textbf{3}&0&2&8&\cellcolor{lightgray!50}0&\cellcolor{lightgray!50}1&\cellcolor{lightgray!50}9&\cellcolor{lightgray!50}0&\cellcolor{lightgray!50}4&\cellcolor{lightgray!50}6\\\hline
\multirow{3}{*}{20}&\textbf{1}&10&0&0&\cellcolor{lightgray!50}10&\cellcolor{lightgray!50}0&\cellcolor{lightgray!50}0&\cellcolor{lightgray!50}10&\cellcolor{lightgray!50}0&\cellcolor{lightgray!50}0\\
&\textbf{2}&0&9&1&\cellcolor{lightgray!50}0&\cellcolor{lightgray!50}4&\cellcolor{lightgray!50}6&\cellcolor{lightgray!50}0&\cellcolor{lightgray!50}5&\cellcolor{lightgray!50}5\\
&\textbf{3}&1&1&8&\cellcolor{lightgray!50}0&\cellcolor{lightgray!50}2&\cellcolor{lightgray!50}8&\cellcolor{lightgray!50}0&\cellcolor{lightgray!50}3&\cellcolor{lightgray!50}7\\\hline
\multirow{3}{*}{15}&\textbf{1}&\cellcolor{lightgray!50}10&\cellcolor{lightgray!50}0&\cellcolor{lightgray!50}0&\cellcolor{lightgray!50}10&\cellcolor{lightgray!50}0&\cellcolor{lightgray!50}0&\cellcolor{lightgray!50}10&\cellcolor{lightgray!50}0&\cellcolor{lightgray!50}0\\
&\textbf{2}&\cellcolor{lightgray!50}0&\cellcolor{lightgray!50}6&\cellcolor{lightgray!50}4&\cellcolor{lightgray!50}1&\cellcolor{lightgray!50}7&\cellcolor{lightgray!50}2&\cellcolor{lightgray!50}0&\cellcolor{lightgray!50}4&\cellcolor{lightgray!50}6\\
&\textbf{3}&\cellcolor{lightgray!50}1&\cellcolor{lightgray!50}2&\cellcolor{lightgray!50}7&\cellcolor{lightgray!50}0&\cellcolor{lightgray!50}3&\cellcolor{lightgray!50}7&\cellcolor{lightgray!50}0&\cellcolor{lightgray!50}3&\cellcolor{lightgray!50}7\\\hline
\multirow{3}{*}{10}&\textbf{1}&\cellcolor{lightgray!50}10&\cellcolor{lightgray!50}0&\cellcolor{lightgray!50}0&\cellcolor{lightgray!50}10&\cellcolor{lightgray!50}0&\cellcolor{lightgray!50}0&\cellcolor{lightgray!50}10&\cellcolor{lightgray!50}0&\cellcolor{lightgray!50}0\\
&\textbf{2}&\cellcolor{lightgray!50}0&\cellcolor{lightgray!50}7&\cellcolor{lightgray!50}3&\cellcolor{lightgray!50}0&\cellcolor{lightgray!50}8&\cellcolor{lightgray!50}2&\cellcolor{lightgray!50}0&\cellcolor{lightgray!50}7&\cellcolor{lightgray!50}3\\
&\textbf{3}&\cellcolor{lightgray!50}0&\cellcolor{lightgray!50}3&\cellcolor{lightgray!50}7&\cellcolor{lightgray!50}0&\cellcolor{lightgray!50}3&\cellcolor{lightgray!50}7&\cellcolor{lightgray!50}0&\cellcolor{lightgray!50}2&\cellcolor{lightgray!50}8\\\hline
\multirow{3}{*}{8}&\textbf{1}&\cellcolor{lightgray!50}10&\cellcolor{lightgray!50}0&\cellcolor{lightgray!50}0&\cellcolor{lightgray!50}9&\cellcolor{lightgray!50}1&\cellcolor{lightgray!50}0&\cellcolor{lightgray!50}9&\cellcolor{lightgray!50}1&\cellcolor{lightgray!50}0\\
&\textbf{2}&\cellcolor{lightgray!50}0&\cellcolor{lightgray!50}6&\cellcolor{lightgray!50}4&\cellcolor{lightgray!50}1&\cellcolor{lightgray!50}2&\cellcolor{lightgray!50}7&\cellcolor{lightgray!50}0&\cellcolor{lightgray!50}8&\cellcolor{lightgray!50}2\\
&\textbf{3}&\cellcolor{lightgray!50}0&\cellcolor{lightgray!50}4&\cellcolor{lightgray!50}6&\cellcolor{lightgray!50}1&\cellcolor{lightgray!50}4&\cellcolor{lightgray!50}5&\cellcolor{lightgray!50}0&\cellcolor{lightgray!50}1&\cellcolor{lightgray!50}9\\\hline
\multirow{3}{*}{5}&\textbf{1}&\cellcolor{lightgray!50}10&\cellcolor{lightgray!50}0&\cellcolor{lightgray!50}0&10&0&0&\cellcolor{lightgray!50}10&\cellcolor{lightgray!50}0&\cellcolor{lightgray!50}0\\
&\textbf{2}&\cellcolor{lightgray!50}0&\cellcolor{lightgray!50}6&\cellcolor{lightgray!50}4&4&2&4&\cellcolor{lightgray!50}0&\cellcolor{lightgray!50}4&\cellcolor{lightgray!50}6\\
&\textbf{3}&\cellcolor{lightgray!50}0&\cellcolor{lightgray!50}3&\cellcolor{lightgray!50}7&1&6&3&\cellcolor{lightgray!50}0&\cellcolor{lightgray!50}3&\cellcolor{lightgray!50}7\\\hline
\multirow{3}{*}{4.16}&\textbf{1}&\cellcolor{lightgray!50}10&\cellcolor{lightgray!50}0&\cellcolor{lightgray!50}0&10&0&0&9&1&0\\
&\textbf{2}&\cellcolor{lightgray!50}0&\cellcolor{lightgray!50}7&\cellcolor{lightgray!50}3&3&5&2&0&8&2\\
&\textbf{3}&\cellcolor{lightgray!50}0&\cellcolor{lightgray!50}4&\cellcolor{lightgray!50}6&3&2&5&1&4&5\\\hline
\multirow{3}{*}{3.33}&\textbf{1}&\cellcolor{lightgray!50}9&\cellcolor{lightgray!50}1&\cellcolor{lightgray!50}0&9&0&1&10&0&0\\
&\textbf{2}&\cellcolor{lightgray!50}0&\cellcolor{lightgray!50}5&\cellcolor{lightgray!50}5&3&5&2&1&6&3\\
&\textbf{3}&\cellcolor{lightgray!50}1&\cellcolor{lightgray!50}7&\cellcolor{lightgray!50}2&2&7&1&1&7&2\\\hline
\multirow{3}{*}{1.66}&\textbf{1}&\cellcolor{lightgray!25}9&\cellcolor{lightgray!25}1&\cellcolor{lightgray!25}0&\cellcolor{lightgray!25}10&\cellcolor{lightgray!25}0&\cellcolor{lightgray!25}0&\cellcolor{lightgray!25}10&\cellcolor{lightgray!25}0&\cellcolor{lightgray!25}0\\
&\textbf{2}&\cellcolor{lightgray!25}7&\cellcolor{lightgray!25}3&\cellcolor{lightgray!25}0&\cellcolor{lightgray!25}9&\cellcolor{lightgray!25}1&\cellcolor{lightgray!25}0&\cellcolor{lightgray!25}4&\cellcolor{lightgray!25}6&\cellcolor{lightgray!25}0\\
&\textbf{3}&\cellcolor{lightgray!25}9&\cellcolor{lightgray!25}1&\cellcolor{lightgray!25}0&\cellcolor{lightgray!25}9&\cellcolor{lightgray!25}1&\cellcolor{lightgray!25}0&\cellcolor{lightgray!25}4&\cellcolor{lightgray!25}5&\cellcolor{lightgray!25}1\\\hline
\multirow{3}{*}{0.83*}&\textbf{1}&\cellcolor{lightgray!25}10&\cellcolor{lightgray!25}0&\cellcolor{lightgray!25}0&\cellcolor{lightgray!25}10&\cellcolor{lightgray!25}0&\cellcolor{lightgray!25}0&\cellcolor{lightgray!25}10&\cellcolor{lightgray!25}0&\cellcolor{lightgray!25}0\\
&\textbf{2}&\cellcolor{lightgray!25}10&\cellcolor{lightgray!25}0&\cellcolor{lightgray!25}0&\cellcolor{lightgray!25}10&\cellcolor{lightgray!25}0&\cellcolor{lightgray!25}0&\cellcolor{lightgray!25}10&\cellcolor{lightgray!25}0&\cellcolor{lightgray!25}0\\
&\textbf{3}&\cellcolor{lightgray!25}7&\cellcolor{lightgray!25}2&\cellcolor{lightgray!25}1&\cellcolor{lightgray!25}9&\cellcolor{lightgray!25}1&\cellcolor{lightgray!25}0&\cellcolor{lightgray!25}7&\cellcolor{lightgray!25}3&\cellcolor{lightgray!25}0\\\hline\hline
\multirow{3}{*}{Total}&\textbf{1}&128&2&0&128&1&1&128&2&0\\
&\textbf{2}&29&76&25&41&60&29&25&77&28\\
&\textbf{3}&27&35&68&35&30&65&22&40&68\\\hline

\end{tabular}
\caption{Confusion matrix of pulse sequences of variable durations ($D$) at 6 and 8 kHz and 60 dB SPL and 6 kHz at 80 dB SPL in Experiment 1. The number in each cell refers to the number of correct responses pooled over the 10 subjects (one condition per subject). *Note that the 0.83 ms stimulus contained no gaps between the pulses and was in fact 1.2 ms long in the triple pulse case.}
\label{tab:Confusions}
\end{table}

It is possible to crudely estimate the duration thresholds between the double- and triple-pulses and between single- and double-pulses, at least in the duration ranges where only two responses were confused and were not contaminated by a third one. Using Eq. \ref{aliasingeq}, it can be done by calculating the average between the two bounds in each duration point in the table that gives the individual sampling frequency. At 6 kHz, double/triple confusions begin to be common for durations between 5 and 8 ms, which gives effective sampling rates of 660 and 426 Hz, respectively. The confusions are virtually gone at 1.66 ms, which would have produced rates that are smaller than 2500 Hz. Similarly, the single/double confusion typically happens at around 1.66 ms, which corresponds to a 1250 Hz. These values are less distinct and longer in the 8 kHz stimuli, as they occur at 8--10 ms, corresponding to a sampling rate of 313--397 Hz. 

\subsubsection{Discussion}
Experiment 1 revealed a confusion pattern that may be in line with a discrete processing, but can be contrasted with predictions from continuous temporal models. First, the results can be compared with the sliding window and low-pass filtering in Figure \ref{NoAliasing}. The sliding window predicts that most stimuli of duration 15 ms or less would be perceived as a single-pulse or a double-pulse for the longest of these durations. This is clearly not the case, according to Table \ref{tab:Confusions}, which shows three-to-two confusions and not three-to-one or two-to-one confusions. Similarly, the low-pass filtering model predicts a single-pulse perception for stimuli of 5 ms or less, which is also not the case, as triple-pulses tended to be confused with double-pulses or identified correctly down to 3.33 ms. These models predict that the double-pulses would be correctly identified down to 3.33 ms and 4.16 ms. However, the observed identification rate of the double-pulses are not much better than the triple-pulses. 

The modulation-band filtering model produces more ambiguous results that may coincide with the observed patterns. First, it predicts that stimuli of durations 1.66 ms or less would be perceived identically to a single-pulse, which appears as a double-pulse due to ringing (Figure \ref{NoAliasing}). This is largely in accord with the results at the 60 dB SPL condition, whereas about half of the double-pulses were identified correctly at 1.66 ms at the 80 dB SPL condition. The pulse morphology predicted by the model at durations of 3.33 ms, 4.16 ms and 5 ms suggests that confusion between double- and triple-pulses may be possible, as was indeed the case in the results. With longer stimuli, the ringing appears more discernible, so that sequences may appear to consist of four or six pulses---answers that were not available as options in the alternative-forced choice test, even if listeners could hear and count them. However, if the ringing of the double-pulses at 3.33 ms and 4.16 ms could be indeed perceived as energetic enough to elicit confusions with triple-pulses, then one would expect such confusions to occur also between a single-pulse and a double-pulse, which was the case only in 1.5\% of the responses, whereas the vast majority of the single-pulse responses were never confused. Therefore, although the modulation filtering model does give rise to a certain ambiguity that can account for several pulse confusion patterns, it is not internally consistent with the entirety of the patterns observed. 

We note that the higher-frequency carrier measured (8 kHz) exhibited a lower and not as well-differentiated aliasing range as in comparison with the 6 kHz channel. 

It is impossible to know what cues made subjects discriminate the pulse sequences, especially in the limit of a single fused event. Additionally, the precision of this test is low as far as the sampling rate estimation goes, given the non-adaptive method of measurement. If aliasing indeed exists, the individual threshold of sampling rate may be estimated more precisely using an adaptive method. 

\subsection{Experiment 2: Within-channel adaptive two-three numerosity\\threshold}
\label{Experiment2}
\subsubsection{Introduction}
The confusion matrix from Experiment 1 revealed a broad range of effective sampling rates that may fit the aliasing working hypothesis. To determine the hypothetical sampling rate more accurately, an adaptive task was devised using the same stimuli for the double- and triple-pulses, but eliminated the option of a single-pulse. Hence, only one threshold at a time was explicitly tested here, instead of two. 

The threshold was tested in four conditions. The first condition utilizes the same double- and triple-pulse stimuli as in Experiment 1, only with total stimulus duration that is set adaptively. It was implicitly assumed in Experiment 1 that the stimuli are narrowband, as the spectral splatter using the Gabor pulses is thought to be minimal because of their relatively small bandwidth and high carrier frequencies, which should confine the performance more narrowly to a single channel. However, in gap detection experiments, when continuous pure tones stop and restart abruptly, it creates a spectral splatter that has been sometimes masked with a notched broadband noise that occupies the adjacent channels, which ensures that the adjacent auditory bands to the carrier will have too low a signal-to-noise ratio to be capable of detecting the pulses in the flanks of their passband \citep[e.g.,][]{Shailer1987}. A similar argument for using such noise is that it reduces the availability of off-frequency listening \citep{Patterson1980}. Therefore, the second condition was set find out whether the high sampling rate estimates could be a result of integration of information over more than a single auditory channel. If this is the case, then a lower effective sampling rate may be measurable by masking the response of off-frequency channels using notched broadband noise. In the third condition, the carrier frequency of the pulses was roved within intervals. This was done in attempt to reduce some of the predictability that may characterize the stimuli, which could make the test easier by cuing the listener to a specific carrier frequency throughout the tested condition. It was hypothesized that a less predictable carrier may require extra vigilance from the system in order to sample the stimulus (possibly mediated by a top-down mechanism), which may putatively bring it closer to its maximum sampling rate capability. In the fourth condition, the stimulus was also presented at a level of 80 dB SPL to find out if there are any intensity effects that are more significant in an adaptive task, compared to those seen in Experiment 1. 

\subsubsection{Methods}
\label{Exp2Methods}
\paragraph{Stimuli}
The stimuli in the first condition were identical to those from Experiment 1. The complete pulse sequence was contained in an initial stimulus length of 350 ms, which accommodated either the double- or triple-pulse. In the second condition, the pulses were presented with notched noise that masks off-frequency channels. The noise was filtered with a third-order Butterworth band-stop filter that removed the energy at the same equivalent rectangular bandwidth (ERB) as the carriers (\citealp{Glasberg1990}). The noise RMS level was set at -40 dB from the pulse peak amplitude and its duration was either 50 ms, or 1.5 times the total stimulus duration---the longest of the two---and the stimulus middle was aligned with the middle of the masker. The relative level of the noise was determined in pilot testing so that discrimination was made more difficult, but the target still sounded clear. In the third condition, the carrier frequency was randomized around the center frequency of 6 kHz. Within each pulse sequence, the carrier was determined using a uniform distribution between 5495 and 6504 Hz, which corresponds to 1.5 times the ERB of the 6 kHz channel, according to \citet{Glasberg1990}. Therefore, no two consecutive pulses had the same exact frequency. 

\paragraph{Procedure}
Either a double- or a triple-pulse sequence was played at random and subjects had to determine whether they heard two or three pulses by pressing the respective digit key (2 or 3) in a single-interval two-alternative forced-choice (2AFC) task. An adaptive three-down one-up procedure was used, so that the stimulus duration was halved if the number of pulses were identified correctly three times in a row, but increased by half its duration for each incorrect response. The 79.4\% threshold was determined after 14 reversals, by calculating the mean value of the last ten reversals \citep{Levitt1971,Schlauch1990}. 

\subsubsection{Results}
All log-transformed data sets were successfully tested for normality using the Jarque-Bera test at the $p>0.05$ level \citep{Jarque1987}. 

\begin{figure} 
		\centering
		\includegraphics[width=1\linewidth]{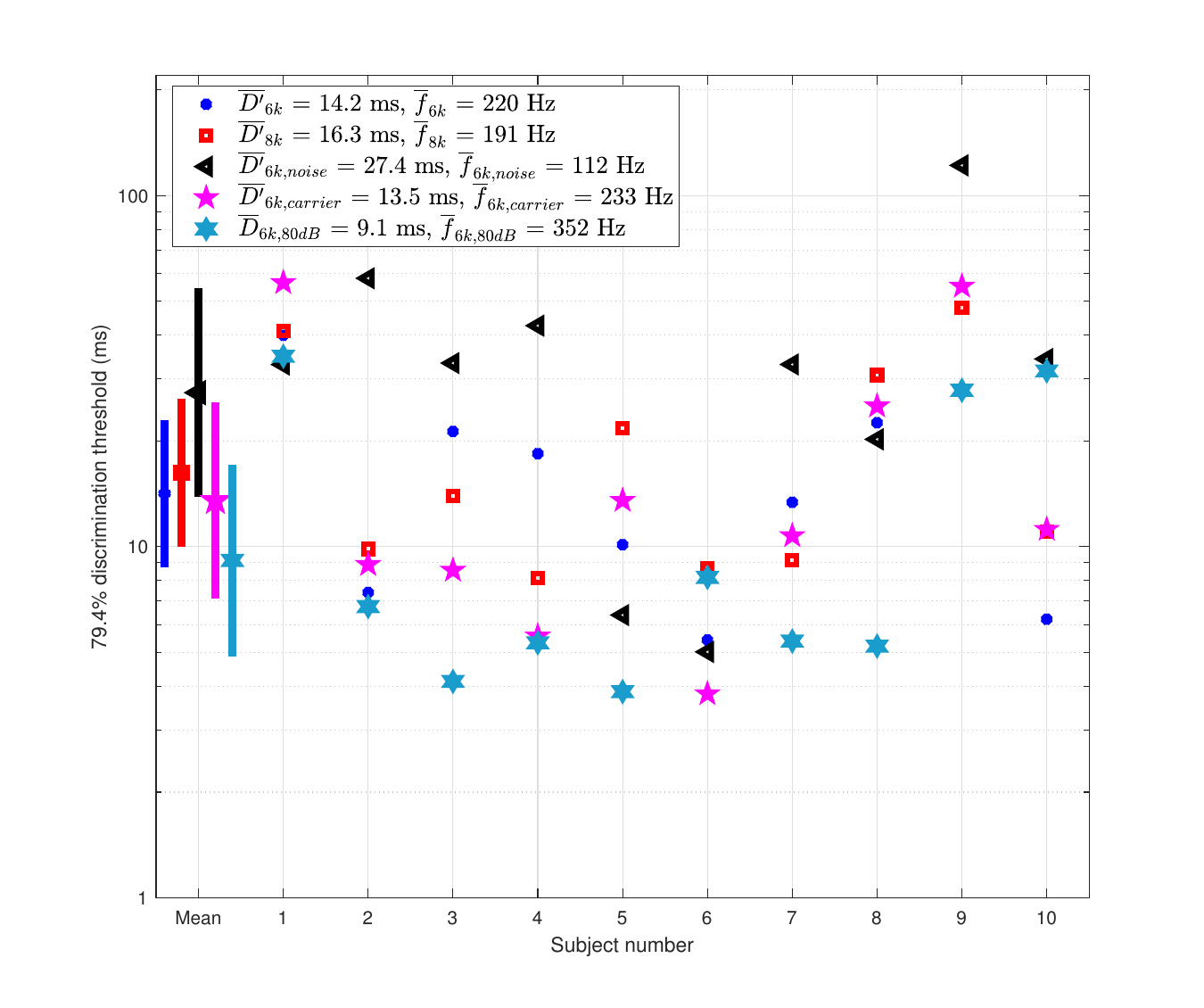}	
		\caption{Mean discrimination thresholds of double- and triple-Gabor pulse sequences for ten subjects (Experiment 2). The mean threshold values $D'$ (the duration of the total stimulus minus the duration of one pulse) for the group and their 95\% confidence intervals are displayed, as well as and the individual subject data. The 6 kHz thresholds are marked with blue asterisks and the 8 kHz with red squares. The broadband masking noise condition for the adjacent channels was added is indicated by black triangles. The roving carrier condition is plotted with magenta stars. The corresponding mean sampling rate estimates according to the bounds in Eq. \ref{aliasingeq} are given in the legend.}
		\label{withinchannel}
\end{figure}

The mean as well as the individual discrimination thresholds between double- and triple-pulse sequences are displayed in Figure \ref{withinchannel} for the 6 and 8 kHz pulse sequences, along with their 95\% confidence intervals. Sequences of mean durations of 14.2 ms at 6 kHz (95\% CI [8.7, 23.0] ms) and 16.3 ms at 8 kHz (95\% CI [10.0, 26.4] ms) were perceived correctly as containing either two or three pulses. On average, this is one event per 4.6--6.9 ms at 6 kHz, and 5.3--7.9 ms at 8 kHz. Note that the duration refers to $D$ from  Eq. \ref{aliasingeq}, but the event rates were computed using $D' = D - W_p$, to correct for the duration of the closing pulse in the sequence. 

The sampling rate for each subject was calculated by plugging their $D'$ threshold in the two bounds of Eq. \ref{aliasingeq}, and the arithmetic average of the two was computed individually. Then, the arithmetic average of the individual rates was taken as the group sampling rate estimate. The 6 kHz pulses appear to have been sampled at a higher rate than the 8 kHz pulses---220 Hz (95\% CI [133, 362] Hz) at 6 kHz and 191 Hz (95\% CI [115, 314] Hz) at 8 kHz. The two rates are not significantly different, though, according to a paired t-test of the individual sampling rates ($p=0.41$). In the condition where off-frequency masking noise was added, there was a dramatic effect on the performance of six out of the ten subjects. The respective t-test comparison between the noise and no noise (randomized carriers) was significant ($p=0.043$). Additionally, the mean sampling rate in the noise condition dropped by almost 50\% from the quiet condition to 112 Hz (95\% CI [56, 224] Hz). In contrast, the roving carrier condition did not produce performance that was significantly different from the fixed carrier condition (paired t-test between sampling frequencies of fixed and randomized carriers gave $p = 0.8$).

Interestingly, the responses in the noise condition reveal two clear clusters: subjects whose temporal performance was unaffected by the noise (subjects \#1, \#5, \#6 and \#8) and all the rest, whose performance significantly deteriorated with the noise. The split in the performance was independent of the individual sampling rate baseline values in the quiet condition. 

The high-intensity condition produced a sampling rate that was 132 Hz higher (352 Hz, 95\% CI [180, 679] Hz) at 80 dB SPL than at 60 dB SPL, yet this difference was found insignificant according to a paired t-test ($p = 0.19$).

\subsubsection{Discussion}
The rates obtained in this adaptive test were lower (slower) than those estimated using the confusion matrix data of Experiment 1. However, the adaptive task produces the 79.4\% threshold data, whereas Experiment 1 data that showed higher rates refer to the 50\% threshold, so the thresholds in Experiment 2 are unsurprisingly higher. 

We obtained a minimum temporal resolution threshold of 4.6--7.9 ms per event, which is higher than known broadband temporal acuity performance in humans, but is consistent with the sinusoidal gap detection 75\% thresholds of about 5 ms \citep[pp. 200--201]{Shailer1987,Moore2013}. The best performing subject's temporal threshold resolution of 1.6 ms was on par with the monoaural (broadband) resolution of 1.6 ms between one and two square 1-ms clicks reported in \citet{Gescheider1966} and 1.66 ms gap detection between to 3-ms tones \citep{Williams1972}. 


The addition of noise had a much stronger effect on a subset of six listeners, but not on the other four. This clustering does not appear to be related to the baseline sampling rates without noise. However, it suggests that off-frequency listening may not necessarily be the cause for the apparent higher sampling rates obtained in the quiet condition, since four of the subjects had stable performances despite the noise. Alternatively, it can indicate that the six affected subjects had significantly broadened auditory filters at 6 kHz. While this seems improbable, it was not tested and cannot be ruled out.

Roving the sequence carrier did not cause a significant change in performance, which suggests that spectral predictability between intervals may not contribute much to the sampling rate values that were observed. Similarly, no level effects were found in this task, or any level effects were too small to produce significant results given the level difference (20 dB) and relatively low power of this experiment. 


\subsection{Experiment 3: The possibility of nonuniform sampling}
\label{AliasingExp3}
\subsubsection{Introduction}
If the observations from the first two experiments reflect a physiological process in the auditory nerve or other auditory nuclei downstream, then we may expect to see adaptation effects as a result. Thus, we would like to find out if the results obtained in Experiments 1 and 2 can be generalized to longer sequences, or whether they represent rates that are too high to be sustained in more central nuclei that exhibit slower spiking rates \citep{Joris2004}. In other words, the observed responses may be relevant for the onset of the signals, but the corresponding rates may change later due to a nonuniform sampling strategy within the auditory system. The stimuli used in the first two experiments may have been short enough to always be processed as onsets by the auditory system. This may trigger adaptation effects in the auditory nerve, for example. On a higher level of processing, the listener knows that their attention is required for a brief moment at the onset, which may exert their finest detection capability only momentarily. Theoretically, an adaptive system may be geared to be frugal in its information processing resources \citep[pp. 143--162]{WeisserPhD}, so it may be designed to use a higher (or its highest) sampling rate just to monitor the stimulus onset. This can enable performance optimization on short temporal scales, similarly to improved spatial sampling in the central retinal areas in vision \citep{Yellott1983}. In contrast, if the system detects a stimulus that seems predictable beyond its onset, then a correspondingly unvarying (and maybe low) sampling rate can be generated that is sufficient to yield low resolution sampling of the input. 

A possible issue with the regularly spaced pulse trains may be an exaggerated predictability between trials, since the context pulses are evenly spaced in a way that could be learned, possibly through a top-down effect. Thus, predictable stimulus rates may cause the system to produce sampling rates that are just fast enough for adequately sampling them, but not to spend more resources on higher rates. Jittering the pulse timing may further decrease their predictability. If the sampling rate set by the system is a result of predictability and not only of adaptation immediately after the onset, then reducing the stimulus predictability may coax the system to maintain a higher instantaneous sampling rate.

\subsubsection{Methods}
In order to test the hypothesis that the sampling rate can vary adaptively, the double- / triple-pulse confusion paradigm was repeated in a context of longer pulse trains. The stimuli were either evenly-spaced pulse trains, or they contained a triple-pulse sequence within the duration of a double-pulse (Figure \ref{ModelComp2}, right). In other words, the pulse-sequence rhythm is momentarily disrupted, as a single pulse (the middle one of the triple-pulse) is presented at double the rate of the pulse train. The assumption here is that an evenly spaced pulse train can facilitate a timing prediction of when the next pulse should arrive after the onset. Thus, when the pulse train rate is half the instantaneous sampling rate of the system, a faster inter-stimulus pulse---hidden in between the regularly spaced pulses---may go undetected. Therefore, if the high temporal acuity performance of Experiments 1 and 2 were an onset-driven brief increase in the sampling rate, then the expected rate during a fast triple-pulse embedded in a slow pulse train context may be lower than when tested standalone, as in Experiment 2.

\paragraph{Procedure}
Subjects were presented with two consecutive pulse trains and were asked to determine whether they are the same or different, by pressing a corresponding key. The two pulse trains were separated by the larger of a 450 ms gap or 1.5 times the duration of the pulse train. As in Experiment 2, a three-down one-up adaptive procedure was used with 14 reversals, from which only the latter ten were used for calculating the mean. The pulse train presentation was randomized with respect to whether they are two identical evenly spaced sequences that effectively contains a double-pulse (Figure \ref{ModelComp2}), or one of the two contains a triple-pulse (Figure \ref{ModelComp2}). Two conditions were administered with the different carrier frequencies of 6 and 8 kHz, each resulting in its own threshold. Two further conditions were administered where the pulse train was made more irregular both temporally and spectrally. This was done using randomized inter-pulse gaps (jittering), with and without roving carriers, as was tested in Experiment 2 for the short stimuli (\ref{Exp2Methods}). A final condition was repeated for the pulse train set at a higher level (80 dB SPL) than the standard presentation.

\paragraph{Stimuli}
The pulses that were used to construct the pulse trains were identical to the ones of the previous experiments. The evenly spaced sequences contained eight pulses at 6 or 8 kHz (Figure \ref{ModelComp2}, left). The odd sequence contains nine pulses: eight are identical to the evenly-spaced sequences, but another one appears exactly halfway between two pulses somewhere within the six middle pulses (Figure \ref{ModelComp2}, right). Therefore, the double- or triple-pulse trains are inserted inside longer sequences of the same period as the double-pulse train. The position of the triple-pulse within the entire pulse train was randomized and could start anywhere between the second and the seventh pulse. To obtain jittering in the pulse train, the spacing between the pulses was randomized to be within $\pm10\%$ of its mean period value, following a uniform distribution. The roving frequency condition was generated using the randomized carrier frequencies as in Experiment 2, so that every pulse within a sequence had a somewhat different carrier frequency. 

\subsubsection{Results}
\begin{figure}  
		\centering
		\includegraphics[width=1\linewidth]{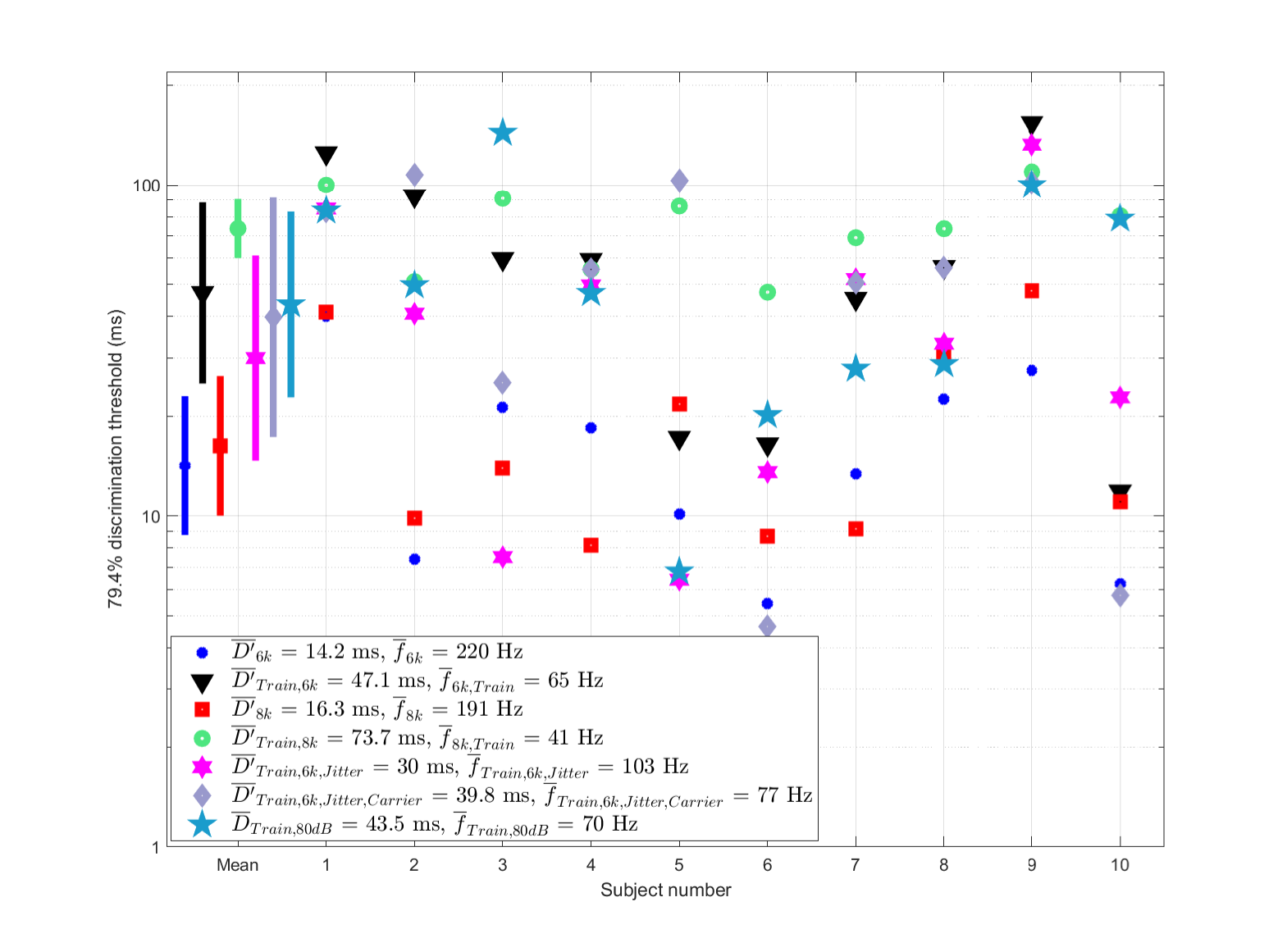}	
		\caption{Mean discrimination thresholds of evenly spaced pulse trains with a hidden pulse at 6 kHz (black triangles) and 8 kHz (green circles) carriers. The two conditions of jittered periodicity with and without roving carrier are shown in magenta stars and gray diamonds, respectively. For comparison, the thresholds of the standard double- and triple-pulse sequences from the previous experiment  are shown in blue asterisks (6 kHz) and red squares (8 kHz). The group mean values include the 95\% confidence intervals.}
		\label{pulsetrain1}
\end{figure}

The individual performances in all conditions are displayed in Figure \ref{pulsetrain1}, along with the main results from Experiment 2. The comparison between the pulse train and short stimuli reveals a noticeable drop in temporal acuity, confirmed by paired t-tests between the standalone sequences and those in the context of longer pulse trains are significant at the $p<0.05$ level in both conditions ($p=8\cdot10^{-5}$ for 6 kHz and $p=5\cdot10^{-6}$ for 8 kHz). The mean sampling rates dropped by 3.4 times for the 6 kHz pulses to 65 Hz (95\% CI [34, 122] Hz) and by 4.6 times on average for the 8 kHz pulses to 41 Hz (95\% CI [33, 50] Hz) compared to the rates measured in Experiment 2. Despite some lowering of the thresholds of the pulse trains with jitter with and without frequency roving, there was no significant improvement in the sampling rate of either condition, as paired t-tests of data at the 0.05 level revealed ($p=0.08$ for the jittered condition and $p=0.52$ for the jittered and roving condition). The high-intensity condition produced a rate that was only 5 Hz higher for the pulse train sequence at 80 dB SPL than the 60 dB SPL condition (70 Hz (95\% CI [36, 134] Hz)). This difference is insignificant using a paired t-test ($p=0.77$). 

\subsubsection{Discussion}
\begin{figure}  
		\centering
		\includegraphics[width=1\linewidth]{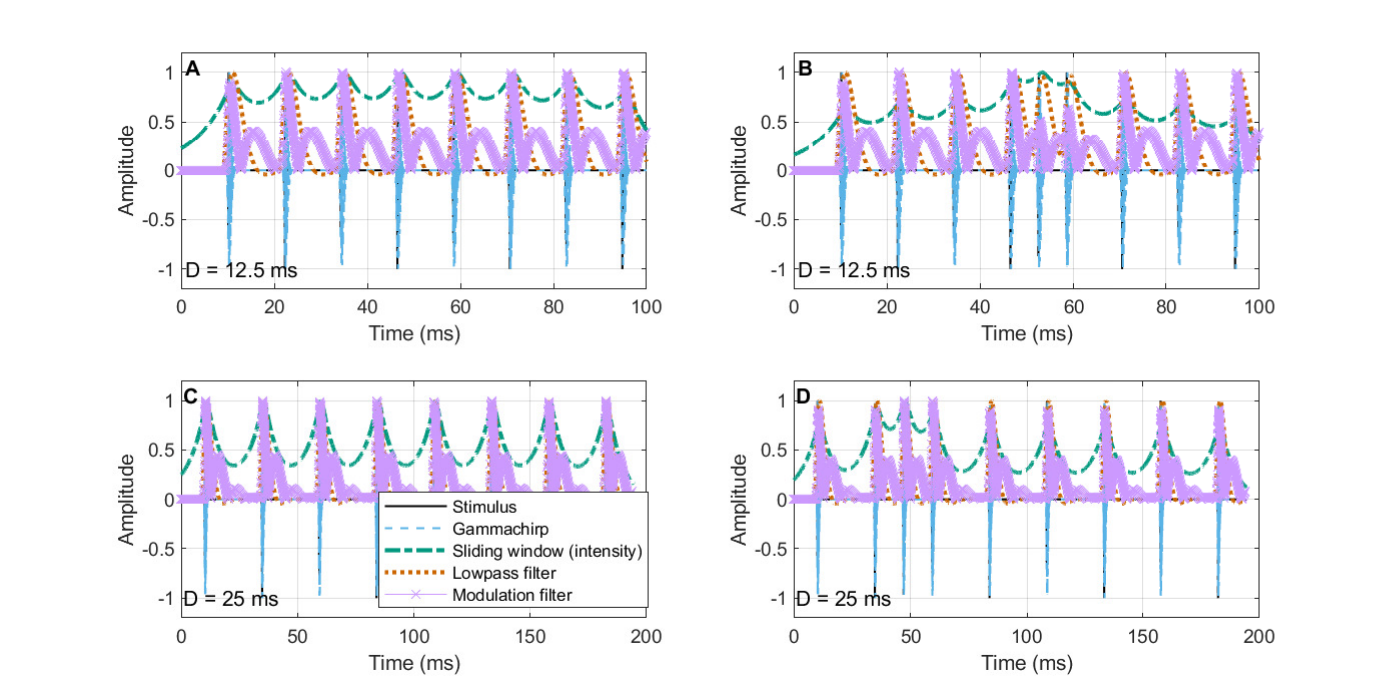}	
		\caption{Comparison of pulse train stimuli with hidden pulse of two durations of the pulse period: 12.5 and 25 ms. See Figure \ref{NoAliasing} for details about the models.}
		\label{ModelComp2}
\end{figure}

When the (hidden) triple-pulse is inserted in a context of a periodic pulse train, no subject could discriminate the middle pulse out of the three as well as they did in the standalone condition of the same triple-pulse without the pulse-train context (Experiment 2). It suggests that the temporal resolution at the onset may be indeed higher than during the sequence. As before, we can compare the prediction of simple continuous models for the pulse trains at the thresholds achieved in the experiment or even lower. In Experiment 1, it was found that modulation band filtering can sometimes give rise to ambiguous outputs in terms of numerosity, albeit in an manner that is inconsistent with all the results. Since the thresholds in the present test are at least twice as high, there is no ambiguity that is implied by the modulation filtering model for a duration shorter (25 ms) than the threshold achieved in the best condition (30 ms), as can be seen in Figure \ref{ModelComp2} (bottom). A shorter duration stimulus of 12.5 ms yields something that may be considered morphologically ambiguous (Figure \ref{ModelComp2}, top), which may have given rise to a ``different'' response in the test. However, the ambiguity is questionable in terms of its associated pulse numerosity. 

If the results are instead interpreted as the output of a sampler with an adaptive sampling rate, then the triple-pulse can be thought to ``fool'' the sampler into missing a fast hidden event. An adaptation-perspective interpretation suggests that a prediction mechanism may be employed to detect the periodicity of the stimulus, which generates samples around the expected events. Alternatively, it could have merely been the effect of being far enough from the stimulus onset that led to lower-resolution sampling. A continuous model such as the one that was tested against above may have to include an adaptive stage with variable filter parameters in order to capture these results. A more surgical test that targets adaptation effects may require separating the placement randomization of the extra pulse to conditions of early and late placements, which may enable the analysis of the detection sensitivity decrease after onset. 

Another alternative explanation of the data may involve forward masking by the initial pulses that has a cumulative effect by the time the hidden pulse appears. Such an effect is expected to be negligible because of the very short pulse duration, the fact that the pulses were presented at equal levels, and the fact that the thresholds of a couple of subjects changed relatively little as a result of the lengthening of the sequences. 

Adding a subtle jitter to the pulse trains had an insignificant effect on the group level. It is possible that the jitter amplitude of 10\% of the period was not large enough to elicit a substantial shift in performance. The addition of carrier roving did little to improve the performance and may have even canceled out the effect of jitter, just as Experiment 2 showed for the shorter stimuli. Unfortunately, the carrier randomization effect was not tested in a separate condition with the long pulse trains. Thus, the onset effect hypothesized may have been a more dominant factor than does a relatively small amount of jitter on listeners' temporal acuity.

It should be noted that the two tasks compared---one-interval forced choice in Experiment 2 and same-different in Experiment 3---are not identical and they may not be reflect the exact same psychometric function. The same-different task of Experiment 3 may have been more prone to response bias of the ``same'' type, which could have resulted in a worse sensitivity and therefore higher threshold---lower sampling rate \citep[pp. 217--218]{Macmillan2005}. Even then, however, the highly significant difference between the results of the two experiments seems to point at a true underlying difference. 

\section{General discussion}
\label{GenDisAlias}
In a series of experiments, the hypothesis of a discrete auditory sampling system was explored using adaptive and non-adaptive psychoacoustic methods. The resultant patterns are generally in agreement with theoretical predictions from temporally discrete auditory processing that includes adaptation effects. 


\subsection{Relationship to continuous temporal processing models}
While we did not test against the continuous temporal auditory models specifically, we used them as conceptual benchmarks. As was illustrated in Figures \ref{NoAliasing} and \ref{ModelComp2}, none of the continuous models can consistently describe our data. Application of modulation filtering was the only model that could sometimes give rise to ambiguity between double- and triple-pulses, but in a way which was inconsistent with the results of similar pulse morphologies, as well as with adaptive effects that can only be accounted for by dynamically changing the modulation filter parameters---something that was not attempted here.

We did not test the results against discrete auditory models, which were not formulated on sampling theoretic basis (see \ref{ModelsIntro}). Furthermore, the cited auditory models (both discrete and auditory) specifically tackled audibility thresholds, whereas the present study explored the perception of numerosity, which may or may not logically precede processing of long-term integration and loudness perception. Nevertheless, the results point to the validity of the general hypothesis of discrete auditory sensation, which has to give rise to sampling-theoretical artifacts like aliasing, in line with similar finding from vision (see \ref{ModelsIntro}).

\subsection{Physiological correlates of the results}
The physiological correlate of the auditory sampler has not been discussed in the psychoacoustic literature that hypothesized its existence \citep{Viemeister1991,Patterson1992,Lyon2018}, with the exception of \citet{HeilMatysiak2017} that attributed it to general neural spiking distributions in the auditory system. The neural spike production in the auditory nerve is probably the most immediate candidate mechanism, since it is the first point of a discrete mechanism past the continuous cochlear mechanics.

The auditory nerve exhibits neural (spiking rate) adaptation, which may explain the difference between the responses to short pulse sequences and pulse trains. In general, onset response in the auditory nerve is characterized by a faster spiking rate \citep{Galambos1943} and was recognized in early temporal models to be an important factor in modeling the temporal response of the system \citep[e.g.,][]{Zwislocki1960}. Rates that were electrophysiologically recorded in the auditory nerve of the gerbil are larger than those obtained here \citep{Westerman1984}. In the gerbil, a fast instantaneous rate (``the rapid component'' that includes the initial 1 ms of the onset) was measured until about 10 ms from the onset and had an average rate of 642 spikes/s \citep{Westerman1984}. These values are almost three times faster than the 79.4\% threshold of Experiment 2, and less than double those measured with 80 dB SPL pulses. They are within the range of the coarse double-triple threshold from Experiment 1. The next, ``short-term'', component in the gerbil's response had a lower rate of 35-261 spikes/s---a broad range that contains the rates observed in the majority of the conditions in Experiments 2 and 3, with the exception of the 80 dB SPL short-pulse sequence measurement of Experiment 2 (Figure \ref{withinchannel}). 

An alternative physiological correlate to the sampling operation is the T-Stellate type of neurons (choppers) that are mostly present in the cochlear nucleus (CN) \citep{Oertel}. The average spiking rates vary significantly between different cell types and as a function of level and frequency and can be several hundreds of spikes per second \citep{Rhode1994Temporal}, which may correspond to the observed rates here. 

Both hypothetical correlates---the auditory nerve and the chopper cells---appear rather similar and it is possible that the chopper cells achieve downsampling, which means that the eventual perceived effect is a combination of both, and possibly further sampling units downstream. 

One difficulty about using the neural correlates along with sampling theory is that spikes, in general, are not generated in the zeros of the incoming waves. However, sampling requires the generation of information also about low-level inputs, in order to be able to correctly reconstruct the arbitrary input. The sampling rate estimates were based on uniform sampling that includes the envelope zeros, contrary to normal auditory coding observations---it can be seen that any peristimulus spiking histogram traces the envelopes of periodic signals, but the spiking is distributed around the peaks \citep[e.g.,][]{Heil1997II}. Mathematically, of course, it is necessary to distinguish between an interpolated (reconstructed) signal that is constant and one that is periodic, which can only be done by sampling at double the stimulus rate. It should be noted, however, that we used the sampling theorem assuming uniform sampling rate. Though, uniform sampling appears to be false in the auditory system for longer stimuli, so nonuniform sampling may have to be considered instead. Therefore, the computed sampling rates can only refer to instantaneous sampling rates, which could lead to instantaneous aliasing in the system. 

In the motivation for the longer the pulse train stimuli in Experiment 3, we repeatedly invoked the design goal of making the stimulus less predictable, where the random appearance of the hidden pulse in the pulse train may have contributed to the lower observed threshold. However, the association of predictability with a low-level physiological correlate such as auditory-nerve or brainstem spiking rate adaptation may be controversial. Predictability in auditory processing---or rather, predictive coding that is geared to minimize uncertainty in future stimuli \citep[e.g.][]{Clark2013}---has been explored mostly in cortical processing and, to a more limited degree, in the midbrain and thalamus \citep{Heilbron2018,Carbajal2018}. While predictive coding is thought to be recurring hierarchically in the brain, a more localized relation between adaptation and predictive coding on the subcortical level (midbrain and thalamus) has only recently been proposed, but it was tested with stimuli that are up to two orders of magnitude longer than the stimuli we employed in the present study \citep{Tabas2020,Tabas2021}. Therefore, the interpretation we have provided here that ties together adaption and prediction may be considered speculative at present and has to be further investigated to find out whether it is warranted.

\subsection{Individual variation and resemblance to informational masking}
The large individual variability observed in all measurements suggests that the temporal properties of the auditory system may be age-dependent, learned, or dependent on other hidden parameters. The broad range of individual performances indicates that the auditory mechanics may be capable of transmitting the full information to the auditory nerve, but it may not always be processed with corresponding resolution in the central auditory circuitry. This may be a function of local neural circuits, of their instantaneous synchronization, or even of attentional resources that mediate early processing stages in the system. 

The clustering into two subject groups according to their responses to the noise masker in Experiment 2 is reminiscent of the clustering observed in informational masking studies between ``high-threshold'' and ``low-threshold'' listeners \citep{Neff1993}. Approximately, only half of every random sample of test subjects (the high-threshold group) appears to be sensitive to informational masking effects, when the target is a tone that is masked by other tones that are well-separated in frequency. The masking remains high even when the separation is increased and is uncorrelated with the bandwidth of the auditory filter that corresponds to the target, when measured using a notched broadband noise task. \citet{Neff1993} suggested that the difference may be a result of different attentional filtering and is anyway not peripheral in origin. The main difference between this paradigm and the present study (Experiment 2) is that the masking noise was broadband and not tonal. Also, the task was temporal and not spectral. However, the explanation in which attention is divided in the subjects with high threshold in noise may work here as well, as it may seem improbable that the six other subjects all have broadened auditory filters. 

An alternative explanation for this split in informational masking listener sensitivity was proposed in \cref{SupraMasking} based on coherent imaging theory. It was speculated there that the sensitivity to the kind of events that are normally presented as stimuli in informational masking tests (i.e., short tone bursts) may be interpreted by the auditory system as coherent noise, in analogy to speckle noise in optics (e.g., noise that appear like dust or small imperfections in the illuminated object). Then, the sensitivity may depend on the internal weighting of the auditory system to its coherent and incoherent images (roughly corresponding to the temporal fine structure and envelope processing, respectively) that makes the eventual perceived partially-coherent image. As the stimuli applied in the present experiment are of a similar nature to those found in informational masking studies, this explanation may apply here as well. Namely, listeners who tend to give higher weight to incoherent imaging are more affected by the masking noise that triggers a smoother listening, whereas listeners who give higher weight to coherent imaging do not smooth out the short speckle-like pulses. 

\subsection{No intensity effects}
If there is a level effect at play on the temporal acuity of short pulse trains, then it may have to be revealed with more exhaustive testing, or at lower absolute intensity than was applied here. The auditory nerve spiking rate is also dependent on level, although it saturates at high levels \citep{Kiang1965,Liberman1978}. However, the higher rates measured at 80 dB SPL compared to 60 dB SPL (Figure \ref{withinchannel}) were insignificant in our study, which makes the direct link to the auditory nerve somewhat less certain. These presentation levels were chosen for comfort and audibility, before adaptation effects in the auditory nerve were considered from the results. Also, inasmuch as the change in the auditory filter bandwidth should have an effect on temporal acuity, the 20 dB input level difference may not be large enough to elicit it, and a larger difference in stimulus intensity may be required \citep{Glasberg2000}. Given the trend of the insignificant results from Experiment 1, it is possible that a test with higher statistical power could reveal an intensity effect.

\subsection{Discrete or continuous sampling in the auditory system}
Even after characterizing the auditory system input as sampled by an hypothetical adaptive detector, there would still remain many questions unanswered, should the system be understood more fully as a sampler. Some of these questions may be understood if the auditory nerve spiking is the responsible mechanism for sampling. For example, if hearing is discrete, then how are the samples triggered? Is there a regular clock, or an ad-hoc oscillator that triggers the sampling process? What is the sampling rate? Is it uniform or nonuniform? Other questions remain much more opaque: What is the shape of the sampling window? Is there a duty cycle? Are these parameters fixed or variable (i.e., adaptive)? Yet other questions require a more elaborate integration of the auditory system function: At what level(s) of the auditory system does sampling take place? Is reconstruction of the continuous signal from discrete samples a relevant concept physiologically and perceptually? If so, how and where are samples reconstructed to give the listener the experience of continuous perception? Are there any artifacts as a result of discretization and perceptual reconstruction (e.g., aliasing or spectral effects of windowing)? Can the processing be considered truly digital, once it is in discrete form?

\section{Conclusion}
The results from the above psychoacoustic experiments appear to reflect more closely well-studied features of the neural response to sound, which have not been directly integrated with behavioral data related to temporal acuity. Despite the wealth of knowledge about the nature of the auditory nerve and its coding, its sampling-theoretic considerations have been largely neglected, which means that any insights that may be garnered from sampling theory have been largely left uncharted to date. The above experiments were readily interpreted using a discrete sampling theoretic model that uses the notion of aliasing and non-uniform adaptive sampling rate, where the samples may have been triggered by the auditory nerve spiking. 

%% file: psychoacoustic_params.tex
\chapter{Dispersion parameter estimation from psychoacoustic data}
\label{PsychoEstimation}
The main aim of this appendix is to find out whether it is possible to re-derive the dispersion parameters of the auditory system using non-invasive psychoacoustical methods. The answer is a hesitant ``yes'', as the computed parameters are relatively consistent with the physiological estimates, although they are generally complex valued, which is not physically impossible. They also shed some more light on the anomalous low-frequency behavior that was noted in the text and in several places in literature. However, as the data used for this appendix were not controlled for level and individual listeners, they add significant uncertainty to the estimates. 

\section{Introduction}
The quantitative results in this work are based on the frequency-dependent values of the group-delay dispersion of the cochlea ($u$), the curvature of the time lens ($s$), the neural group-delay dispersion ($v$), and the temporal aperture ($T_a$). Obtaining estimates for their values earlier in \cref{paramestimate} involved a combination of animal and human physiological measurements, whose validity could not always be ascertained, due to the paucity of relevant data and methods, as well as uncertainty in interpreting some of the data from literature. This was particularly problematic in the estimation of the time lens curvature, which was based on data from a number of animal studies that yielded two value clusters (small- and large-curvature estimates) that were not trivially transformed to human values. However, throughout this work, several psychoacoustic effects were modeled reasonably well using these parameters, despite the uncertainty that was mainly evident at low frequencies. This raises the possibility that the modeling may be inverted, so that the entire dispersion parameter set could be obtained using these known psychoacoustic effects, without resorting to animal and physiological data. This appendix therefore attempts to obtain $u$, $s$, $v$, and $T_a$ using published data from tone beating, phase curvature, and stretched octave testing. Gap detection threshold data are required as well and had to be obtained through a self-administered test of the author. Together they constitute a system of four equations with four unknowns. 

For simplicity, we will stick to the seven octave frequencies of the curvature data from \citet{OxenhamDau} of 125--8000 Hz.  

In the following, ``psychoacoustic'' data or estimates refer to the present modeling, whereas the ``physiological'' estimates refer to the previous methods that were discussed in \cref{paramestimate}.

\subsection{Beating}
From the tone beating analysis presented in \cref{CohereBeating} and \cref{BeatingCurves}, we can obtain the temporal aperture, using the mean frequency difference $\Delta f_{beat}$ at which tones are spectrally resolved in two separate filters:
\begin{equation}
	T_a = \frac{2\ln2 \,\,}{\pi \Delta f_{beat}} \approx \frac{0.441}{\Delta f_{beat}}
	\label{beatingrelations}
\end{equation}
The most systematic data of beating perception were replotted in Figure \ref{bearingfig}, based on published results. Each dataset leads to slightly different dispersion predictions. Here, we used the data from \citet{PlompSteen1968}, who explicitly asked subjects to rate ``\textit{just the absence of interference}'' between two pure tones at 60 phons. The data is very similar to psychoacoustic data using other methods, except for frequencies above 4000 Hz, where the slope changed. The predictions that are obtained below 4000 Hz using this dataset are almost identical to those obtained had the data from \citet{Plomp1964} been used instead. 

\subsection{Phase curvature}
The next effect relates to the internal phase curvature that the auditory system exhibits, which in the text was associated with a permanent defocus. A closed-form solution was obtained by assuming that the information of the continuous linear modulation is contained in a single Gaussian pulse that has the duration of the temporal aperture. The explicit solution to the curvature equation (\ref{quad1}) is 
\begin{equation}
m_0^2 + \frac{1}{2x}m_0 + \frac{1}{t_0^4}=0 \,\,\,\,\,\,\,\, x,t_0 \ne 0
\label{quad2}
\end{equation}
with $m_0$ being the slope of the instantaneous frequency of the stimulus that is needed to obtain minimal masking thresholds, and $t_0$ is the Gaussian width that is determined by the temporal aperture $T_a$ through the full-width half maximum (FWHM), so that $t_0 \approx T_a/2.355$. The curvature of $x$ is defined as
\begin{equation}
x = u + \frac{vs }{v + s}  = u + \frac{v}{M}
\label{xdefocusrep}
\end{equation}
with $M$ being the magnification of the system. If we have the $t_0$ values from the beating data and the $m_0$ from the psychoacoustic curvature data, then $x$ can be isolated from Eq. \ref{quad2}
\begin{equation}
x = -\frac{m_0 t_0^4}{2(m_0^2 t_0^4 + 1)}
\label{xalt}
\end{equation}
An estimate for $x$ can be therefore obtained independently of $v$, $u$, and $s$, using $T_a$ from the beating data. In their comprehensive study, \citet{OxenhamDau} produced two very similar datasets of psychoacoustic curvature data, which varied mainly at 125 and 250 Hz. The authors considered the data from their Method 2 to be more accurate. Therefore, we used their Method 2 data, which were measured for a Schroeder phase masker level of 75 dB SPL and individual components at 64 dB SPL (see also \cref{ModelSchr}). 

Additionally, the following constraint was obtained from solving for $t_0$ in the text, which is required in order to cancel out the input chirp
\begin{equation}
	t_0 \geq 2\sqrt{|x|}
\label{msolutionINEQrep}
\end{equation}
This was satisfied in \cref{InitialCurveEstimates} after optimization by turning the inequality to equality with the factor 2.116 instead of 2, which indicated that the solution may not be optimal (as it indicates two solutions to \ref{quad1} instead of one). Using the beating and Method 2 data completely determines the multiplicative constant in the inequality of Eq. \ref{msolutionINEQrep} (if different from 2), which ensures that the chirping is completely dispersive (real). While the inequality is maintained automatically using the $t_0$ and $m_0$ values, it is frequency dependent at low frequencies, unlike the analysis in \cref{InitialCurveEstimates}. Using both the beating and curvature data, the value of $\sigma = t_0 / (2\sqrt{|x|})$ is almost at its physical limit of 1 at 8000 Hz, drops slightly down to 1000 Hz and then jumps to 1.1172 (500 Hz) and up to much larger values of 1.6264 (250 Hz) and 6.8482 (125 Hz). This is in accord with the anomalous behavior originally noticed below 1000 Hz in the original paper and addressed earlier in \cref{TempAperture}, and will be seen also below. The values of the multiplicative factor that represent the difference from equality in \ref{msolutionINEQrep} to inequality are given in Table \ref{MultFactor}.

\begin{table} 
\centering
\scriptsize
\begin{tabular}{|c|c|c|c|c|c|c|c|}
\hline
\textbf{Multiplicative factor}&\textbf{125 Hz}&\textbf{250 Hz}&\textbf{500 Hz}&\textbf{1000 Hz}&\textbf{2000 Hz}&\textbf{4000 Hz}&\textbf{8000 Hz}\\\hline
$\sigma$ & 6.8482   & 1.6264 & 1.1172 & 1.0019 & 1.0028 & 1.0003 & 1  \\
\hline\end{tabular}
  \caption{The multiplicative factor that describes the inequality \ref{msolutionINEQrep}, so that $t_0 = 2\sigma\sqrt{|x|}$}
	\label{MultFactor}
\end{table}

\subsection{Stretched octave}
The next expression comes from data of stretched octave perception in pure tone intervals---the amount of frequency detuning that is necessary to perceive two sequential sounds as though they are exactly one octave apart. The following relation between two frequencies was discussed in \cref{TransChromAb}:
\begin{equation}
	\frac{f_2}{2f_1} = \frac{M(f_1)}{M(f_2)} 
	\label{MagStretch9}
\end{equation}
where $M(f_1)$ and $M(f_2)$ are the auditory magnifications that are associated with the interval frequencies $f_1$ and $f_2$, respectively. We define $\Delta f_{oct}$ as the perceived frequency deviation from a pure octave, so that $\Delta f_{oct} = f_2 - 2f_1$, so that $\Delta f_{oct}=0$ only when $f_2=2f_1$. Using Eq. \ref{MagnificationDef}, this expression can be rewritten as:
\begin{equation}
	\frac{M(f_1)}{M(f_2)} = \frac{s_2(v_1 + s_1)}{s_1(v_2 + s_2)} = 1 + \frac{\Delta f_{oct}}{2 f_1}
	\label{MagStretch10}
\end{equation}
We shall use this expression in both forms---for $M$ as a parameter, and once $v$ is known, explicitly to solve for $s$. A necessary approximation that we have to do in order to make the equations tractable is that $M(2f_1) \approx M(f_2)$. Furhtermore, in order to solve for the magnification $M = (s+v)/s$, it is necessary to have an initial condition---$M$ at one frequency. Based on the pure tone data from literature compiled in \citet{Jaatinen} (and replotted in Figure \ref{FigStretched}), we only have a single pure-tone datum (without confidence intervals) at 100 Hz and no other low-frequency measurements. In contrast, the complex tone (of real instruments) stretched octave data from \citet{Jaatinen} suggest that there is a large spread of values of positive and negative detuning at low frequencies, which averages to a negative trend. Hence, above 200 Hz, the pure tone curve was used as a reference, whereas the complex tone curve was used as a reference below 200 Hz. It was also used for the magnification estimate at 62.5 Hz, which provides the initial point necessary to obtain the magnifications from 125 Hz and above. Note that the sound pressure levels of the measured stimuli were not controlled across the original experiments that were used to collect the stretched octave data in \citet{Jaatinen}. 

\subsection{Double-pulse gap detection}
\label{DoublePulseGap}
The last measurement relates to the temporal resolution of the system, which was modeled using the temporal-imaging amplitude transfer function (ATF; a synonym for impulse response function) of a single channel (Eq. \ref{GenGaussPupil}). The detection of a gap between two clicks was introduced in \cref{GapDetect} (Eq. \ref{TemporalRes}), which was based on ideally-tuned impulses that, strictly speaking, violates the uncertainty principle and is physically impossible. Therefore, we would like to replace the impulses with Gabor pulses---short tone bursts with a Gaussian envelope that minimizes the uncertainty relations product \citep{Gabor}. The response to these pulses can be directly computed and is dependent on all four unknown parameters. The envelope of the pulse is
\begin{equation}
	a_i(t) = \exp\left( -4\ln2 \,\, \frac{t^2}{T_1^2}\right)
\end{equation}
where $T_1$ is the FWHM of the pulse. Omitting all multiplicative constants, the Fourier transform of the pulse envelope is
\begin{equation}
	A_i(\omega) = \exp\left( -\frac{T_1^2 \omega^2}{16\ln2 \,\,}\right)
\end{equation}
The dispersive response to this pulse can be obtained by multiplying this expression with the defocused ATF of the system (Eq. \ref{GenGaussPupil}) and taking its inverse Fourier transform. Again, omitting the multiplicative constant, this gives
\begin{equation}
	a_o(t) = \exp\left[ -\frac{t^2}{4}\frac{\frac{T_1^2}{16\ln2 \,\,} + \frac{16\ln2 \,\, v^2}{T_a^2} + iW_d v^2 }{\left(\frac{T_1^2}{16\ln2 \,\,} + \frac{16\ln2 \,\, v^2}{T_a^2}\right)^2 +v^4 W_d^2}    \right]
\end{equation}
where $W_d = 1/u + 1/v + 1/s$. We are looking at time $-d/2$ in which the intensity of the pulse is 1/4 of the peak, so when it is incoherently added to another pulse at $d/2$, the total level is half of the peak. The phase cancels out in the squaring. From this we can obtain the detectable gap $d$
\begin{equation}
	d = \sqrt{  T_1^2 + \left(\frac{16\ln2 \,\, v}{T_a}\right)^2  + \frac{16\ln2 \,\, W_d^2 v^4}{\frac{16\ln2 \,\, v^2}{T_a^2} + \frac{T_1^2}{16 \ln2 \,\,} }  }
	\label{GapGaussThers}
\end{equation}
When $T_1 = 0$, this expression reduces to Eq. \ref{TemporalRes}. Note that when $T_1$ is very short, the gap obtained using the Gabor pulses may be smaller than that obtained with the delta function.

The main challenge in applying this expression in practice is that the uncertainty relations of the Fourier transform simultaneously constrain the temporal and spectral resolution of the stimulus. In order to best observe the dispersive effects, the Gaussian must be as short as possible, whereas in order to spectrally localize the pulse on the cochlea, it must be made as long as possible. Additionally, neural sampling limitations may be constraining the achievable resolution that can be achieved for very short gaps. At low frequencies this is particularly difficult. Another practical challenge is to psychoacoustically capture the correct point of where the two pulses are just noticeable and align it with the computed half the maximum level. Different gradations of ``two-pulse-ness'' are audible when listening to such stimuli, which may map to a continuous psychometric curve. 

As the two-pulse ``gap detection'' data were not available from literature, the data used are the author's own self-administered measurements. Different Gabor pulse widths produce somewhat different results, which may all be useful at different frequencies, yet are not completely consistent. The shortest pulses tested had a width of 0.3 of the carrier period. This gave a faint but unmistakable sense of pitch, which is thought to roughly localize to a cochlear place, unlike an infinitesimally short impulse. The shortness of the pulse allows for some differentiation of the dispersion effect at very low frequencies. 

The test method employed here was chosen for its simplicity for the author's self testing, but is not rigorous. The presentation level was comfortable, yet uncontrolled. All stimuli were produced using MATLAB (The Mathworks Inc., Natick, MA) at 44.1 kHz and 24 bits and delivered to Sennheiser HD-25 closed headphones (Sennheiser Electronic GmbH \& Co. KG, Wedemark, Germany), using a Focusrite Scarlett 18i20 sound card (Focusrite Plc, High Wycombe, UK). The threshold was found manually using the adjustment method. The gap frequencies were presented at a random order and two sets were measured for two pulse widths (FWHM)---one of a single carrier period and one that was 0.3 of the period. The thresholds obtained for the single period pulses are given Table \ref{GaborPair}. 

\begin{table} 
\centering
\scriptsize
\begin{tabular}{|c|c|c|c|c|c|c|c|}
\hline
\textbf{Pulse width (periods)}&\textbf{125 Hz}&\textbf{250 Hz}&\textbf{500 Hz}&\textbf{1000 Hz}&\textbf{2000 Hz}&\textbf{4000 Hz}&\textbf{8000 Hz}\\\hline
1   & 4.8 & 4.1 & 3.1 & 2.7 & 1.4 & 1.5 & 1.6 \\
0.3 & 4.5 & 3.5 & 3.2 & 2.4 & 1.2 & 1.5 & 1.5 \\
\hline\end{tabular}
  \caption{Gap detection thresholds $d$ with Gabor pulses for a single subject. All threshold durations are in milliseconds and represent the average of two repetitions for each pulse duration set.}
	\label{GaborPair}
\end{table}

\subsection{Solving for $v$}
Once we have $T_a$ (or rather $t_0 = T_a/\FWHM$), $x$, and $M$, we can solve for $v$ in Eq. \ref{GapGaussThers}, by noting that 
\begin{equation}
	W_d = \frac{1}{u} + \frac{1}{v} + \frac{1}{s} = \frac{1}{u} + \frac{M}{v}
\end{equation}
Isolating $u$ from Eq. \ref{xdefocusrep} and using in $W_d$ we obtain, after some manipulation
\begin{equation}
	W_d = \frac{M^2x}{v(Mx - v)}
\end{equation}
Using this expression in Eq. \ref{GapGaussThers} turns it to an equation with one unknown. Unfortunately, it is a sixth-order polynomial equation that cannot be simplified. After some work, this unseemly expression is obtained for $v$
\begin{multline}
	-\frac{a^3}{T_a^2}v^6 + \frac{2a^3Mx}{T_a^2}v^5 + \left( ab - \frac{a^3M^2x^2}{T_a^2} - a^2c\right)v^4  + (2a^2cMx - 2abMx )v^3 \\
	 + ( abM^2x^2 - a^2cM^2x^2 - aM^4T_a^2x^2 + bcT_a^2 )v^2 - (2bcMxT_a^2) v + bcM^2T_a^2x^2 = 0
	\label{sixrootv}
\end{multline}
with the following auxiliary constants defined
\begin{equation}
	a = 16\ln 2 \,\,\,\,\,\,\,\,\,\, b = d^2 - T_1^2 \,\,\,\,\,\,\,\,\,\,\, c = \frac{T_1^2}{a}
\end{equation}
This equation can be solved numerically.

\section{Results and discussion}
\subsection{Root selection}
There are several degrees of freedom in running the above model and solving for the six roots of $v$. The solutions change with slightly different choices of psychoacoustic data, but often not significantly, for a given root of Eq. \ref{sixrootv}. Initially, it was necessary to eliminate as many unphysical and improbable solutions as possible.

Elimination of irrelevant solutions of $v$ turned out to be a nontrivial task with potentially far-reaching implications on the temporal imaging theory, as presented throughout this work. Ideally, we would have liked choose between real roots and discard any complex ones. Depending on the exact temporal aperture estimates to some extent, there was at most one real solution at all frequencies, which has the wrong sign: the cochlear group-delay dispersion must be negative to produce a group delay that decreases with frequency. According to this solution, the neural group-delay dispersion has a matching negative sign to the physiological estimates, but is two orders of magnitude higher than the physiological estimates. The same thing goes for the time lens curvature, although we are anyway less certain of its physiologically estimated values. This solution also produces a modulation transfer function that does not correspond to known auditory responses as it has lower coherent than incoherent cutoff frequency across the spectrum. Therefore, this solution is ruled out because of poor match to the empirical findings. 

\begin{figure} 
		\centering
		\includegraphics[width=1\linewidth]{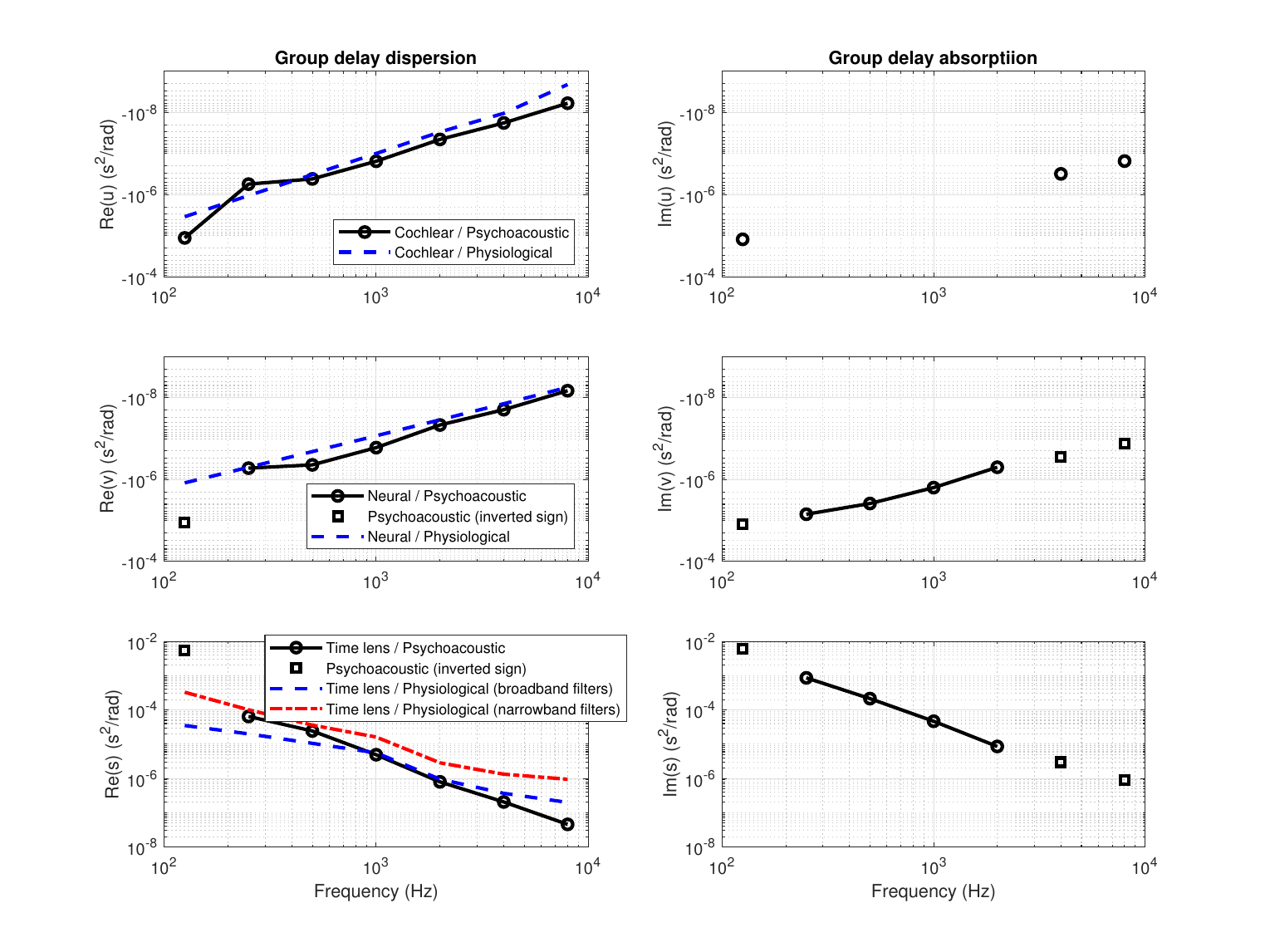}	
		\caption{The cochlear ($u$) and neural ($v$) group delay-dispersions and the time-lens curvature ($s$) based on the psychoacoustic model of beating threshold (Eq. \ref{beatingrelations}), phase curvature data (Eq. \ref{quad2}), stretched octave (Eq. \ref{MagStretch10}), and double Gabor pulse gap detection (Eq. \ref{GapGaussThers}), plotted in black circles / solid lines. The real parts are displayed on the left. At the lowest frequency (125 Hz), the values of $v$ and $s$ have inverted sign and are presented as black squares. For comparison, the estimates from the main text that are based on electrophysiological human and physiological animal data (\cref{paramestimate}) are displayed in blue dashed lines on the left. In addition, the two most used physiological estimates of the time lens curvature---the broad-filter (blue dash) and narrow-filter (red dash-dot) large-curvatures are displayed on the bottom left. The imaginary parts that correspond to group-delay absorption are displayed on the right (black circles / solid lines) without a physiological counterpart. Between 250 and 2000 Hz, $u$ is real. Inverted-sign values in both $v$ and $s$ are marked with the disconnected black squares.}
		\label{psychmodel1}
\end{figure}

\begin{figure} 
		\centering
		\includegraphics[width=0.5\linewidth]{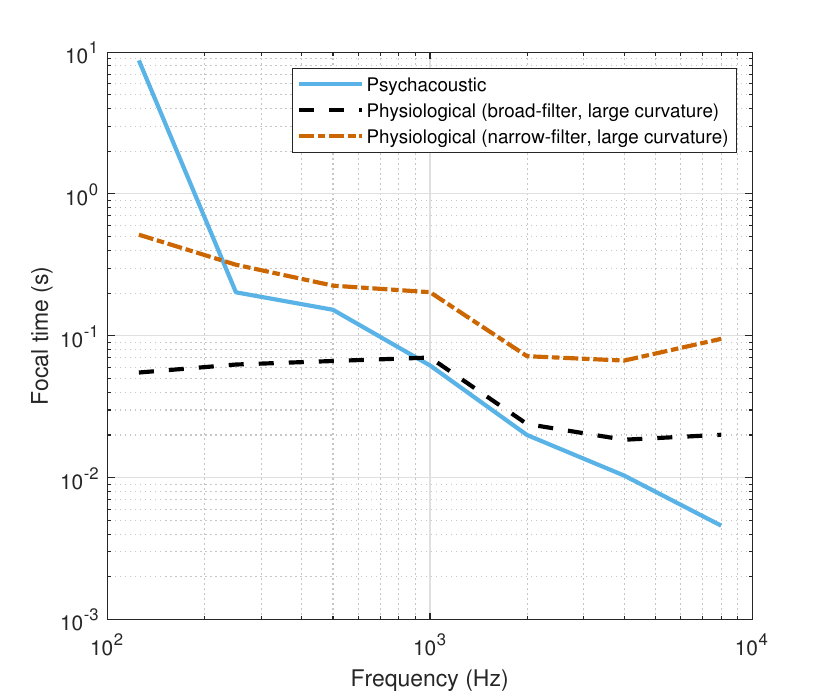}	
		\caption{The focal time $f_T = 2\omega_c s$ corresponding to the (real part of the) time-lens curvature $s$. The present psychoacoustic estimate is compared with the two limiting physiological estimates of the broad-filter curvature and the narrow-filter large-curvature time lenses.}
		\label{psychmodel}
\end{figure}

This leaves us with complex roots (or mixed real and complex) to choose from. One of them is more attractive than the rest, because its real part is relatively close to the physiological estimates. Critically, this solution entails that the lowest frequency (125 Hz) has an opposite sign in the neural group-delay dispersion $v$ and the time-lens curvature $s$. The solutions for the three dispersion parameters are displayed on the left-hand side of Figure \ref{psychmodel1}, where they are compared to the physiological estimates from the text (\cref{paramestimate}). The focal time of the time lens is displayed in Figure \ref{psychmodel}. 

\subsection{Complex solutions?}
The biggest challenge in the proposed solution is that it produces complex values in all parameters and frequencies, except for three frequencies of $u$. The imaginary absorptive parts of the parameters are of about one order of magnitude larger than the respective real dispersive parts of all three parameters (Figure \ref{psychmodel1}, right-hand plots). Their existence, even if troubling at first sight, is physically appropriate, given the causality/dispersion relations (Kramers-Kronig relations), which tie together the real and imaginary parts of any causal, linear, time-invariant medium (\cref{airtravel}). While absorption was mentioned early in the original derivation in the text (\cref{temporaltheory}), it was conveniently neglected (as is the convention in imaging optics) and did not seem to be necessary to obtain good fit of the parameters. Additionally, positive absorption terms of the form $\exp(\alpha''\zeta \omega^2/2)$, where $\alpha''\zeta > 0$ (unlike the imaginary values in Figure \ref{psychmodel1}, including the time-lens curvature), would make the expressions intractable in closed-form if they are included in integrals (\cref{GroupABS}). Therefore, with the present knowledge of the system and theory, it is uncertain whether the various expressions derived for the psychoacoustic effects (mainly the phase curvature and gap detection) should hold in general for complex parameters. In principle, large group-delay absorption may further rotate the signal in the phase space, when it is combined with dispersion\footnote{Note that first-order absorption coefficient that is linearly dependent on frequency has an effect on phase similar to dispersion, but is not considered in this work.}. 

\begin{figure} 
		\centering
		\includegraphics[width=1\linewidth]{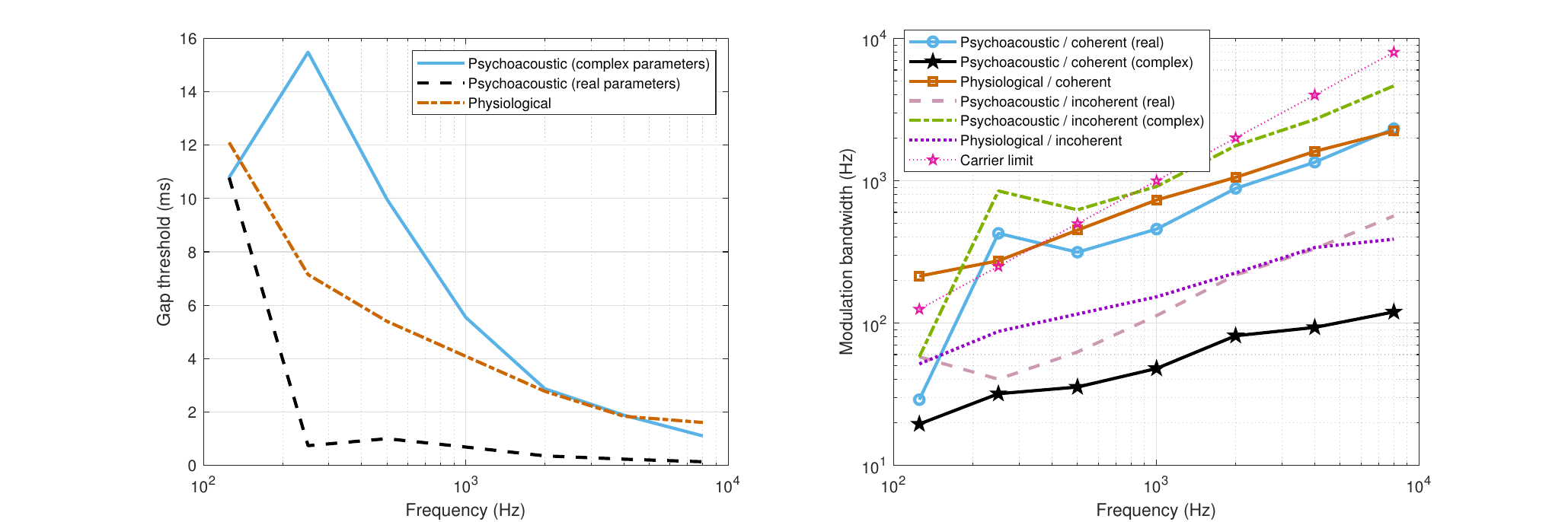}	
		\caption{Comparison of dispersive computations using real and complex parameters. \textbf{Left}: Double-click audibility threshold according to three estimates, using Eq. \ref{TemporalRes}. The solid blue curve was computed using the complex parameter values shown in Figure \ref{psychmodel1}, whereas the black dashed curve used their real parts only. The physiological model lies in the middle in dash-dot red. \textbf{Right}: Predicted cutoff frequencies of coherent and incoherent temporal modulation transfer functions (TMTFs). The physical bound of over-modulation on the carrier is marked with dotted purple stars. }
		\label{doublclickthres}
\end{figure}

One way to test the significance of the imaginary part on the system parameters is by comparing real-numbered predictions done with the complex valued solutions and those done with their real parts only. Two examples that were used in the text are repeated here and are displayed in Figure \ref{doublclickthres}. On the left, the double-click gap detection predictions of Eq. \ref{TemporalRes} (same as Eq. \ref{GapGaussThers} with $T_1=0$), led earlier to plausible temporal thresholds, at least between 1000 and 8000 Hz, where the predicted range was approximately as suggested by the rule of thumb of 2--3 ms (\cref{GapDetect}). Using only the real part of the new estimates of the dispersion parameters, the gap detection thresholds at all frequencies turn out shorter than with the physiological estimates, except for at 125 Hz. In contrast, using the full complex values leads to an increase and overestimation of all thresholds except for the 125 Hz value. 

The temporal modulation transfer function (TMTF) cutoff frequencies are compared on the right of Figure \ref{doublclickthres}, similarly to Figure \ref{OTAcutoff}. Both coherent and incoherent real-parameter psychoacoustic estimates are of lower frequencies than the physiological estimate. The complex-parameter incoherent estimate is implausible, since it predicts cutoff frequencies that are much higher than the coherent limits. Interestingly, the real coherent estimate at two out of three low carrier frequencies dodges the over-modulation problem that was highlighted in \cref{LowFreqCorr}, since its predicted bandwidth is much lower than the carrier frequency by a factor of five. 

\subsection{Low-frequency inversion}
The group-delay dispersion sign changes in $v$ and $s$ account for the physiological curvature anomalous behavior in the \citet{OxenhamDau} lowest frequency measurement. Moreover, it may constitute a parallel to the cat's auditory nerve results in \citet{Carney1999}, who found that at low frequencies the impulse response glide slope changed sign from rising at high frequencies ($>1500$ Hz) to approximately flat between 750 and 1500 Hz, and to falling at low frequencies ($<750$ Hz). These frequency ranges of the cat can be scaled to human above 530 Hz (high), 260-530 Hz (flat), and below 260 Hz (low) \citep{Greenwood1990}. However, no sign inversion was identified in humans in the psychoacoustic curvature experiments by \citet{OxenhamDau}. The difference might lie in the extra neural dispersion that was excluded by \citet{Carney1999}, as they tapped the signal in the auditory nerve and not in the inferior colliculus. The sign inversion may mean that at some point in the apical region the signal is in sharp focus.

Alternatively, the sign inversion in $v$ and $s$ may be erroneous. The inversion may be a result of imprecise low-frequency data of one of the parameters used in the test, or caused by another inconsistency. Such numerical instability may be the case as with some choices of psychoacoustic data (e.g., different beating thresholds), if it is $u$ that is inverted, rather than $v$ and $s$. However, it is unlikely that $u$ changes sign, because this would imply that the group delay of the cochlea decreases at very low frequencies, which seems unreasonable, if energy to the inner hair cells is transmitted sequentially through the traveling wave. Sign changes in neural dispersion also seem somewhat less easy to justify. In contrast, it is perhaps possible that the time lens loses its phase-modulation capability close to the helicotrema, where the wide basilar membrane may become less compliant and is subjected to different boundary conditions than in the basal turns (note that a large value of $s$ entails a \textbf{small} effect of the lens). 

\subsection{Magnification}
The magnification has two alternative expressions that are exactly equal only when the system is in sharp focus, $M = \frac{s+v}{s}$ (the actual magnification factor) and $M_0 = -\frac{v}{u}$ (real parts assumed everywhere). Given the phase curvature as was measured by \citet{OxenhamDau} and others, we know that the system has to be permanently defocused, which means that $M \neq M_0$ (\cref{ImagingDefocusFound}). This is the case in both the psychoacoustic and the physiological estimates. There are three main differences between the two models. First, $M$ is much closer to unity in the new estimate, similar to the two large-curvature time lens physiological estimates, which perfectly explains (by design) the stretched octave data and the musical range. Second, as a result, $M_0$ changes sign around 160--200 Hz---with an unknown effect. It should be noted that two complex roots of Eq. \ref{sixrootv} were ruled out partly because of their estimated $M_0 \approx 0$, which could not be justified. Finally, the new $M_0$ is much closer to -1 and is fairly constant, given that $u$ and $v$ are of very similar magnitudes. $M_0$ was used in \cref{ImpulseResponseDer} to approximate time-invariant impulse response. 

\begin{figure} 
		\centering
		\includegraphics[width=0.65\linewidth]{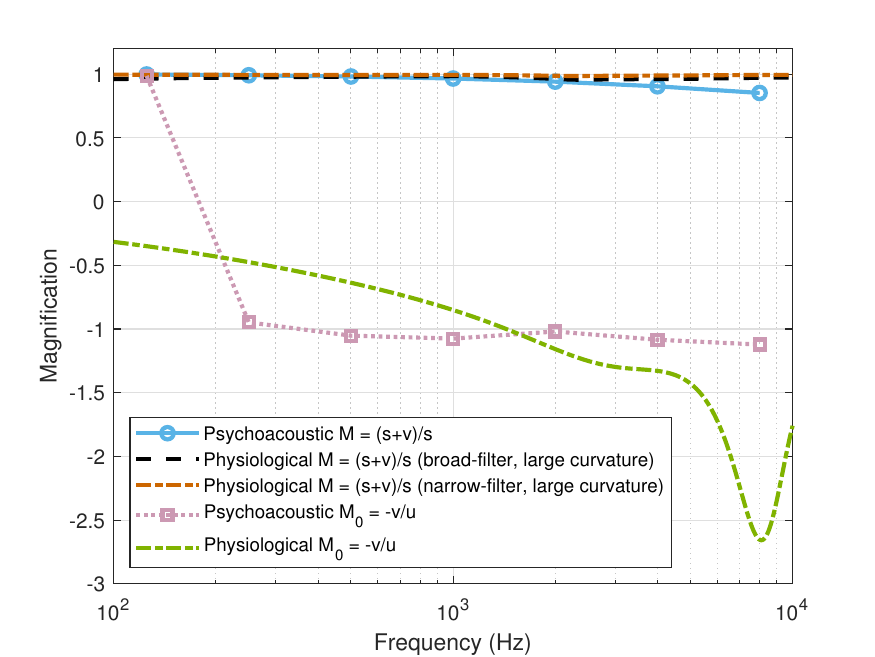}	
		\caption{Different estimates for the two magnification definitions of the imaging system, $M = \frac{s+v}{s}$ and $M_0 = -\frac{v}{u}$. When the system is in sharp focus $M=M_0$. In both psychoacoustic and physiological estimates $M \approx 1 $, especially in the large-curvature estimate. According to the psychoacoustic estimate $M_0 \approx -1$, but since $v$ changes sign at low frequencies, then $M_0$ changes sign at around 160--200 Hz as well.}
		\label{psychmodelmag}
\end{figure}

\section{General discussion}
The psychoacoustic solution proposed here is almost completely independent of the previous solution that was based on physiological data, except for the phase curvature measurements that were used in both methods. At the level of rigor obtained here, it provides only a partial validation to the temporal imaging theory in its current formulation. First, the only plausible solutions are complex, which may or may not relate to physical absorptive attributes of the auditory medium. At this stage, there is no method to map it to any known measurements. Then, using the real part only is not always justified, as some derivative figures (e.g., double-click detection thresholds) are highly sensitive to the imaginary part as well---probably because of its significant order of magnitude relative to dispersion. That said, most real-valued results are relatively consistent with the physiological data. The one exception is at the lowest frequency band, which shows a sign change in one or more parameters. This might be consistent with the anomalous response observed in the \citet{OxenhamDau} measurements and in cat data from \citet{Carney1999}, although we would expect to see sign inversion taking place at a higher frequency---something which was not seen in the data. It suggests that the low-frequency range may be governed by somewhat different equations than were used in this work. 

Another interesting result from this work relates specifically to the time-lens curvature. Its physiological estimation process (\cref{lenscurve}) required some speculative assumptions that nevertheless relied on five studies with what appeared to be unmistakable phase modulation in the organ of Corti. Assuming scaling between animal data and humans, we obtained wide bounds of the curvature, but picked mainly those that correspond to either broad or narrow auditory filter response in humans. Predictions that were based on these values were sometimes closer using one curvature (small or large) than with the other, but were generally similar. The psychoacoustic curvature estimate is somewhere in between the two, as it is closer to the narrow-filter physiological estimate at low frequencies and to the broad-filter curvature above 1000 Hz. Incidentally, the psychoacoustic estimate is far from the small-curvature estimate that was ruled out in much of the analyses throughout this work, for producing all sorts of outliers and unlikely sign inversions. However, as noted earlier, it is more likely than not that the time-lens curvature is adaptive by virtue of the medial olivocochlear reflex accommodation. In this case, small-curvature values may better fit conditions that have not been explored here. 

Should the above model and equations prove correct over time, they suggest an indirect way to estimate the dispersion parameters noninvasively. The estimates may even reveal individual differences, analogous to refractive errors in vision. Beating, Gabor pulse detection, and stretched octave tests are relatively easy to administer, even if not particularly easy to obtain stable results from. But the phase curvature testing procedure, as administered by \citet{OxenhamDau}, is very tedious and will have to be made shorter and simpler in order to be suitable for human data collection on a larger test sample. An alternative technique  has been recently proposed by \citet{Rahmat2015} and cuts the average test time by a factor of 5.5. Importantly, administering such a comprehensive test battery should be done at a uniform sound pressure level---a control that was not available to us here and has undoubtedly detracted from the overall accuracy of the results.